\documentclass[twocolumn,aps,showpacs]{revtex4}

\begin{document}


\title{Bibliographic guide to the foundations of quantum
mechanics and \\
quantum information}
\author{Ad\'{a}n Cabello}
\email{adan@us.es}
\affiliation{Departamento de F\'{\i}sica
Aplicada II, Universidad de Sevilla, 41012 Sevilla, Spain}
\date{\today}
\maketitle
\begin{quotation}
``[T]here's much more difference (\ldots)
between a human being who knows quantum mechanics and one that
doesn't than between one that doesn't and the other great apes.''
\end{quotation}
\begin{flushright}
M. Gell-Mann \\
at the annual meeting of the American Association for\\
the Advancement of Science, Chicago 11 Feb. 1992.\\
Reported in {\bf [Siegfried 00]}, pp.~177-178.
\end{flushright}

\begin{quotation}
``The Copenhagen interpretation {\em is} quantum mechanics.''
\end{quotation}
\begin{flushright}
R. Peierls. \\
Reported in {\bf [Khalfin 90]}, p.~477.
\end{flushright}

\begin{quotation}
``Quantum theory needs no `interpretation'.''
\end{quotation}
\begin{flushright}
C. A. Fuchs and A. Peres. \\
Title of {\bf [Fuchs-Peres 00 a]}.
\end{flushright}

\begin{quotation}
``Unperformed experiments have no results.''
\end{quotation}
\begin{flushright}
A. Peres. \\
Title of {\bf [Peres 78 a]}.
\end{flushright}


\section*{Introduction}
This is a collection of references (papers, books, preprints, book
reviews, Ph.\ D. thesis, patents, web sites, etc.), sorted
alphabetically and (some of them) classified by subject, on
foundations of quantum mechanics and quantum information.
Specifically, it covers hidden variables (``no-go'' theorems,
experiments), ``interpretations'' of quantum mechanics,
entanglement, quantum effects (quantum Zeno effect, quantum
erasure, ``interaction-free'' measurements, quantum
``non-demolition'' measurements), quantum information
(cryptography, cloning, dense coding, teleportation), and quantum
computation. For a more detailed account of the subjects covered,
please see the table of contents in the next pages.

Most of this work was developed for personal use, and is therefore
biased towards my own preferences, tastes and phobias. This means
that the selection is incomplete, although some effort has been
made to cover some gaps. Some closely related subjects such as
quantum chaos, quantum structures, geometrical phases,
relativistic quantum mechanics, or Bose-Einstein condensates have
been deliberately excluded.

Please note that this guide has been directly written in LaTeX
(REVTeX4) and therefore a corresponding BibTeX file does not
exist, so do not ask for it.

Please e-mail corrections to adan@us.es (under subject: Error).
Indicate the references as, for instance, {\bf [von Neumann 31]},
not by its number (since this number may have been changed in a
later version). Suggestions for additional (essential) references
which ought to be included are welcome (please e-mail to
adan@us.es under subject: Suggestion).

\section*{Acknowledgments}
The author thanks those who have pointed out errors, made
suggestions, and sent copies of papers, lists of personal
publications, and lists of references on specific subjects.
Special thanks are given to J.~L.~Cereceda, R.~Onofrio, A.~Peres,
E.~Santos, C.~Serra, M.~Simonius, R.~G.~Stomphorst, and
A.~Y.~Vlasov for their help on the improvement of this guide. This
work was partially supported by the Universidad de Sevilla grant
OGICYT-191-97, the Junta de Andaluc\'{\i}a grants FQM-239 (1998,
2000, 2002), and the Spanish Ministerio de Ciencia y
Tecnolog\'{\i}a grants BFM2000-0529, BFM2001-3943, and
BFM2002-02815.

\newpage


\tableofcontents


\newpage





\section{Hidden variables}


\subsection{Von Neumann's impossibility proof}


{\bf [von Neumann 31]},
{\bf [von Neumann 32]} (Sec.~IV.~2),
{\bf [Hermann 35]},
{\bf [Albertson 61]},
{\bf [Komar 62]},
{\bf [Bell 66, 71]},
{\bf [Capasso-Fortunato-Selleri 70]},
{\bf [Wigner 70, 71]},
{\bf [Clauser 71 a, b]},
{\bf [Gudder 80]} (includes an example in two dimensions showing
that the expected value cannot be additive),
{\bf [Selleri 90]} (Chap.~2),
{\bf [Peres 90 a]} (includes an example in two dimensions showing
that the expected value cannot be additive),
{\bf [Ballentine 90 a]} (in pp.~130-131 includes an example in four
dimensions showing that the expected value cannot be additive),
{\bf [Zimba-Clifton 98]},
{\bf [Busch 99 b]} (resurrection of the theorem),
{\bf [Giuntini-Laudisa 01]}.


\subsection{Einstein-Podolsky-Rosen's argument of incompleteness of QM}


\subsubsection{General}


{\bf [Anonymous 35]},
{\bf [Einstein-Podolsky-Rosen 35]},
{\bf [Bohr 35 a, b]} (see \ref{subsubBohrsreply}),
{\bf [Schr\"{o}dinger 35 a, b, 36]},
{\bf [Furry 36 a, b]},
{\bf [Einstein 36, 45]}
(later Einstein's arguments of incompleteness of QM),
{\bf [Epstein 45]},
{\bf [Bohm 51]} (Secs.~22. 16-19.
Reprinted in {\bf [Wheeler-Zurek 83]}, pp.~356-368;
simplified version of the EPR's example with two spin-$\frac{1}{2}$
atoms in the singlet state),
{\bf [Bohm-Aharonov 57]} (proposal of an experimental test with photons
correlated in polarization. Comments:),
{\bf [Peres-Singer 60]},
{\bf [Bohm-Aharonov 60]};
{\bf [Sharp 61]},
{\bf [Putnam 61]},
{\bf [Breitenberger 65]},
{\bf [Jammer 66]} (Appendix B; source of additional bibliography),
{\bf [Hooker 70]} (the quantum approach does not ``solve'' the paradox),
{\bf [Hooker 71]},
{\bf [Hooker 72 b]} (Einstein vs. Bohr),
{\bf [Krips 71]},
{\bf [Ballentine 72]} (on Einstein's position toward QM),
{\bf [Moldauer 74]},
{\bf [Zweifel 74]} (Wigner's theory of measurement solves the paradox),
{\bf [Jammer 74]} (Chap.~6, complete account of the historical development),
{\bf [McGrath 78]} (a logic formulation),
{\bf [Cantrell-Scully 78]} (EPR according QM),
{\bf [Pais 79]} (Einstein and QM),
{\bf [Jammer 80]} (includes photographs of Einstein, Podolsky, and
Rosen from 1935, and the New York Times article on EPR, {\bf [Anonymous 35]}),
{\bf [Ko\c{c} 80, 82]},
{\bf [Caser 80]},
{\bf [M\"{u}ckenheim 82]},
{\bf [Costa de Beauregard 83]},
{\bf [Mittelstaedt-Stachow 83]} (a logical and relativistic formulation),
{\bf [Vujicic-Herbut 84]},
{\bf [Howard 85]} (Einstein on EPR and other later arguments),
{\bf [Fine 86]} (Einstein and realism),
{\bf [Griffiths 87]} (EPR experiment in the consistent histories interpretation),
{\bf [Fine 89]} (Sec.~1, some historical remarks),
{\bf [Pykacz-Santos 90]} (a logical formulation with axioms derived from experiments),
{\bf [Deltete-Guy 90]} (Einstein and QM),
(Einstein and the statistical interpretation of QM:)
{\bf [Guy-Deltete 90]},
{\bf [Stapp 91]},
{\bf [Fine 91]};
{\bf [Deltete-Guy 91]} (Einstein on EPR),
{\bf [H\'{a}jek-Bub 92]} (EPR's argument is ``better'' than later arguments
by Einstein, contrary to Fine's opinion),
{\bf [Combourieu 92]} (Popper on EPR, including a letter by Einstein from 1935
with containing a brief presentation of EPR's argument),
{\bf [Bohm-Hiley 93]} (Sec.~7.~7, analysis of the EPR experiment according
to the ``causal'' interpretation),
{\bf [Schatten 93]} (hidden-variable model for the EPR experiment),
{\bf [Hong-yi-Klauder 94]} (common eigenvectors of relative position and
total momentum of a two-particle system, see also
{\bf [Hong-yi-Xiong 95]}),
{\bf [De la Torre 94 a]} (EPR-like argument with two components of position
and momentum of a single particle),
{\bf [Dieks 94]} (Sec.~VII, analysis of the EPR experiment according
to the ``modal'' interpretation),
{\bf [Eberhard-Rosselet 95]} (Bell's theorem based on a generalization of EPR
criterion for elements of reality which includes values predicted with almost certainty),
{\bf [Paty 95]} (on Einstein's objections to QM),
{\bf [Jack 95]} (easy-reading introduction to the EPR and Bell arguments,
with Sherlock Holmes).


\subsubsection{Bohr's reply to EPR}
\label{subsubBohrsreply}


{\bf [Bohr 35 a, b]},
{\bf [Hooker 72 b]} (Einstein vs. Bohr),
{\bf [Ko\c{c} 81]} (critical analysis of Bohr's reply to EPR),
{\bf [Beller-Fine 94]} (Bohr's reply to EPR),
{\bf [Ben Menahem 97]} (EPR as a debate between two possible interpretations of
the uncertainty principle: The {\em weak} one---it is not possible to measure or
prepare states with well defined values of conjugate observables---, and the {\em strong}
one ---such states do not even exist---. In my opinion, this paper is extremely useful
to fully understand Bohr's reply to EPR),
{\bf [Dickson 01]}
(Bohr's thought experiment is a reasonable realization of EPR's argument),
{\bf [Halvorson-Clifton 01]} (the claims that the point in Bohr's reply is a
radical positivist are unfounded).


\subsection{Gleason theorem}


{\bf [Gleason 57]}, {\bf [Piron 72]}, simplified unpublished proof
by Gudder mentioned in {\bf [Jammer 74]} (p.~297), {\bf [Krips 74,
77]}, {\bf [Eilers-Horst 75]} (for non-separable Hilbert spaces),
{\bf [Piron 76]} (Sec.~4.~2), {\bf [Drisch 79]} (for non-separable
Hilbert spaces and without the condition of positivity), {\bf
[Cooke-Keane-Moran 84, 85]}, {\bf [Redhead 87]} (Sec.~1.~5), {\bf
[Maeda 89]}, {\bf [van Fraassen 91 a]} (Sec.~6.~5), {\bf [Hellman
93]}, {\bf [Peres 93 a]} (Sec.~7.~2), {\bf [Pitowsky 98 a]}, {\bf
[Busch 99 b]}, {\bf [Wallach 02]} (an ``unentangled'' Gleason's
theorem), {\bf
[Hrushovski-Pitowsky 03]} (constructive proof of Gleason's
theorem, based on a generic, finite, effectively generated set of
rays, on which every quantum state can be approximated), {\bf
[Busch 03 a]} (the idea of a state as an expectation value
assignment is extended to that of a generalized probability
measure on the set of all elements of a POVM. All such generalized
probability measures are found to be determined by a density
operator. Therefore, this result is a simplified proof and, at the
same time, a more comprehensive variant of Gleason's theorem),
{\bf [Caves-Fuchs-Manne-Renes 04]} (Gleason-type
derivations of the quantum probability rule for POVMs).


\subsection{Other proofs of impossibility of hidden variables}


{\bf [Jauch-Piron 63]},
{\bf [Misra 67]},
{\bf [Gudder 68]}.


\subsection{Bell-Kochen-Specker theorem}
\subsubsection{The BKS theorem}


{\bf [Specker 60]},
{\bf [Kochen-Specker 65 a, 65 b, 67]},
{\bf [Kamber 65]},
{\bf [Zierler-Schlessinger 65]},
{\bf [Bell 66]},
{\bf [Belinfante 73]} (Part~I, Chap.~3),
{\bf [Jammer 74]} (pp.~322-329),
{\bf [Lenard 74]},
{\bf [Jost 76]} (with 109 rays),
{\bf [Galindo 76]},
{\bf [Hultgren-Shimony 77]} (Sec.~VII),
{\bf [Hockney 78]} (BKS and the ``logic'' interpretation of QM proposed by Bub; see
{\bf [Bub 73 a, b, 74]}),
{\bf [Alda 80]} (with 90 rays),
{\bf [Nelson 85]} (pp.~115-117),
{\bf [de Obaldia-Shimony-Wittel 88]} (Belinfante's proof requires 138 rays),
{\bf [Peres-Ron 88]} (with 109 rays),
unpublished proof using 31 rays by Conway and Kochen (see
{\bf [Peres 93 a]}, p.~114, and
{\bf [Cabello 96]} Sec.~2.~4.~d.),
{\bf [Peres 91 a]} (proofs with 33 rays in dimension 3 and 24 rays in dimension 4),
{\bf [Peres 92 c, 93 b, 96 b]},
{\bf [Chang-Pal 92]},
{\bf [Mermin 93 a, b]},
{\bf [Peres 93 a]} (Sec.~7.~3),
{\bf [Cabello 94, 96, 97 b]},
{\bf [Kernaghan 94]} (proof with 20 rays in dimension 4),
{\bf [Kernaghan-Peres 95]} (proof with 36 rays in dimension 8),
{\bf [Pagonis-Clifton 95]} [why Bohm's theory eludes BKS theorem; see also
{\bf [Dewdney 92, 93]}, and
{\bf [Hardy 96]} (the result of a measurement in Bohmian mechanics depends not
only on the context of other simultaneous measurements but also on how
the measurement is performed)],
{\bf [Bacciagaluppi 95]} (BKS theorem in the modal interpretation),
{\bf [Bell 96]},
{\bf [Cabello-Garc\'{\i}a Alcaine 96 a]} (BKS proofs in dimension $n \ge 3$),
{\bf [Cabello-Estebaranz-Garc\'{\i}a Alcaine 96 a]} (proof with 18 rays in dimension 4),
{\bf [Cabello-Estebaranz-Garc\'{\i}a Alcaine 96 b]},
{\bf [Gill-Keane 96]},
{\bf [Svozil-Tkadlec 96]},
{\bf [DiVincenzo-Peres 96]},
{\bf [Garc\'{\i}a Alcaine 97]},
{\bf [Calude-Hertling-Svozil 97]} (two geometric proofs),
{\bf [Cabello-Garc\'{\i}a Alcaine 98]}
(proposed {\em gedanken} experimental test on the existence
of non-contextual hidden variables),
{\bf [Isham-Butterfield 98, 99]},
{\bf [Hamilton-Isham-Butterfield 99]},
{\bf [Butterfield-Isham 01]}
(an attempt to construct a realistic contextual interpretation of QM),
{\bf [Svozil 98 b]} (book),
{\bf [Massad 98]} (the Penrose dodecahedron),
{\bf [Aravind-Lee Elkin 98]} (the 60 and 300 rays corresponding respectively to
antipodal pairs of vertices of the 600-cell 120-cell ---the two most complex of the
four-dimensional regular polytopes--- can both be used to prove BKS theorem in
four dimensions. These sets have critical non-colourable subsets with 44 and 89 rays),
{\bf [Clifton 99, 00 a]}
(KS arguments for position and momentum components),
{\bf [Bassi-Ghirardi 99 a, 00 a, b]}
(decoherent histories description of reality cannot be considered satisfactory),
{\bf [Griffiths 00 a, b]}
(there is no conflict between consistent histories and Bell and KS theorems),
{\bf [Michler-Weinfurter-\.{Z}ukowski 00]}
(experiments),
{\bf [Simon-\.{Z}ukowski-Weinfurter-Zeilinger 00]}
(proposal for a {\em gedanken} KS experiment),
{\bf [Aravind 00]} (Reye's configuration and the KS theorem),
{\bf [Aravind 01 a]} (the magic tesseracts and Bell's theorem),
{\bf [Conway-Kochen 02]},
{\bf [Myrvold 02 a]} (proof for position and momentum),
{\bf [Cabello 02 k]} (KS theorem for a single qubit),
{\bf [Pavi\v{c}i\'{c}-Merlet-McKay-Megill 04]} (exhaustive construction of all proofs of the KS theorem;
the one in {\bf [Cabello-Estebaranz-Garc\'{\i}a Alcaine 96 a]} is the smallest).


\subsubsection{From the BKS theorem to the BKS with locality theorem}


{\bf [Gudder 68]},
{\bf [Maczy\'{n}ski 71 a, b]},
{\bf [van Fraassen 73, 79]},
{\bf [Fine 74]},
{\bf [Bub 76]},
{\bf [Demopoulos 80]},
{\bf [Bub 79]},
{\bf [Humphreys 80]},
{\bf [van Fraassen 91 a]} (pp.~361-362).


\subsubsection{The BKS with locality theorem}


Unpublished work by Kochen from the early 70's,
{\bf [Heywood-Redhead 83]},
{\bf [Stairs 83 b]},
{\bf [Krips 87]} (Chap.~9),
{\bf [Redhead 87]} (Chap.~6),
{\bf [Brown-Svetlichny 90]},
{\bf [Elby 90 b, 93 b]},
{\bf [Elby-Jones 92]},
{\bf [Clifton 93]},
(the Penrose dodecahedron and its sons:),
{\bf [Penrose 93, 94 a, b, 00]},
{\bf [Zimba-Penrose 93]},
{\bf [Penrose 94 c]} (Chap.~5),
{\bf [Massad 98]},
{\bf [Massad-Aravind 99]};
{\bf [Aravind 99]} (any proof of the BKS can be converted into
a proof of the BKS with locality theorem).


\subsubsection{Probabilistic versions of the BKS theorem}


{\bf [Stairs 83 b]} (pp.~588-589),
{\bf [Home-Sengupta 84]} (statistical inequalities),
{\bf [Clifton 94]} (see also the comments),
{\bf [Cabello-Garc\'{\i}a Alcaine 95 b]}
(probabilistic versions of the BKS theorem and proposed experiments).


\subsubsection{The BKS theorem and the existence of dense
``KS-colourable'' subsets of projectors}


{\bf [Godsil-Zaks 88]} (rational unit vectors in $d=3$ do not admit a ``regular colouring''),
{\bf [Meyer 99 b]}
(rational unit vectors are a dense KS-colourable subset in dimension $3$),
{\bf [Kent 99 b]} (dense colourable subsets of projectors
exist in any arbitrary finite dimensional real or complex Hilbert space),
{\bf [Clifton-Kent 00]}
(dense colourable subsets of projectors exist with
the remarkable property that
every projector belongs to only one resolution of the identity),
{\bf [Cabello 99 d]},
{\bf [Havlicek-Krenn-Summhammer-Svozil 01]}, {\bf [Mermin 99 b]},
{\bf [Appleby 00, 01, 02, 03 b]},
{\bf [Mushtari 01]} (rational unit vectors do not admit a ``regular colouring'' in $d=3$ and $d \ge 6$, but do admit
a ``regular colouring'' in
$d=4$ ---an explicit example is presented--- and $d=5$ ---result announced by P. Ovchinnikov---),
{\bf [Boyle-Schafir 01 a]},
{\bf [Cabello 02 c]} (dense colourable subsets cannot simulate QM
because most of the many possible colourings of these sets must be statistically
irrelevant in order to reproduce some of the statistical predictions of QM,
and then, the remaining statistically relevant colourings cannot reproduce
some different predictions of QM),
{\bf [Breuer 02 a, b]} (KS theorem for unsharp spin-one observables),
{\bf [Peres 03 d]},
{\bf [Barrett-Kent 04]}.


\subsubsection{The BKS theorem in real experiments}


{\bf [Simon-\.{Z}ukowski-Weinfurter-Zeilinger 00]} (proposal),
{\bf [Simon-Brukner-Zeilinger 01]},
{\bf [Larsson 02 a]} (a KS inequality),
{\bf [Huang-Li-Zhang-(+2) 03]}
(realization of all-or-nothing-type KS experiment with single photons).


\subsection{Bell's inequalities}


\subsubsection{First works}


{\bf [Bell 64, 71]},
{\bf [Clauser-Horne-Shimony-Holt 69]},
{\bf [Clauser-Horne 74]},
{\bf [Bell 87 b]} (Chaps.~7, 10, 13, 16),
{\bf [d'Espagnat 93]} (comparison between the assumptions in
{\bf [Bell 64]} and in
{\bf [Clauser-Horne-Shimony-Holt 69]}).


\subsubsection{Bell's inequalities for two spin-$s$ particles}


{\bf [Mermin 80]} (the singlet state of two spin-$s$ particles violates a particular Bell's inequality
for a range of settings that vanishes as ${1 \over s}$ when $s \rightarrow \infty$)
{\bf [Mermin-Schwarz 82]} (the
${1 \over s}$ vanishing might be peculiar to the particular inequality used in {\bf [Mermin 80]}),
{\bf [Garg-Mermin 82, 83, 84]} (for some Bell's inequalities the range of settings does not diminish
as $s$ becomes arbitrarily large),
{\bf [\"{O}gren 83]} (the range of settings for which quantum mechanics violates
the original Bell's inequality is the same magnitude, at least for small $s$),
{\bf [Mermin 86 a]},
{\bf [Braunstein-Caves 88]},
{\bf [Sanz-S\'{a}nchez G\'{o}mez 90]},
{\bf [Sanz 90]} (Chap.~4),
{\bf [Ardehali 91]} (the range of settings vanishes as ${1 \over s^2}$),
{\bf [Gisin 91 a]} (Bell's inequality holds for all non-product states),
{\bf [Peres 92 d]},
{\bf [Gisin-Peres 92]} (for two spin-$s$ particles in the singlet state
the violation of the CHSH inequality is {\em constant} for any $s$;
large $s$ is no guarantee of classical behavior)
{\bf [Geng 92]} (for two different spins),
{\bf [W\'{o}dkiewicz 92]},
{\bf [Peres 93 a]} (Sec.~6.~6),
{\bf [Wu-Zong-Pang-Wang 01 a]} (two spin-$1$ particles),
{\bf [Kaszlikowski-Gnaci\'{n}ski-\.{Z}ukowski-(+2) 00]}
(violations of local realism by two entangled $N$-dimensional systems
are stronger than for two qubits),
{\bf [Chen-Kaszlikowski-Kwek-(+2) 01]}
(entangled three-state systems violate local realism
more strongly than qubits: An analytical proof),
{\bf [Collins-Gisin-Linden-(+2) 01]} (for arbitrarily high
dimensional systems),
{\bf [Collins-Popescu 01]}
(violations of local realism by two entangled quNits),
{\bf [Kaszlikowski-Kwek-Chen-(+2) 02]}
(Clauser-Horne inequality for three-level systems),
{\bf [Ac\'{\i}n-Durt-Gisin-Latorre 02]}
(the state ${1 \over \sqrt{2+\gamma^2}}(|00\rangle+\gamma|11\rangle+|22\rangle)$,
with $\gamma =(\sqrt{11}-\sqrt{3})/2 \approx 0.7923$, can violate the
Bell inequality in {\bf [Collins-Gisin-Linden-(+2) 01]}
more than the state with $\gamma =1$),
{\bf [Thew-Ac\'{\i}n-Zbinden-Gisin 04]} (Bell-type test of energy-time entangled qutrits).


\subsubsection{Bell's inequalities for two particles and more
than two observables per particle}


{\bf [Braunstein-Caves 88, 89, 90]} (chained Bell's inequalities, with more than two
alternative observables on each particle),
{\bf [Gisin 99]},
{\bf [Collins-Gisin 03]} (for three possible two-outcome measurements per qubit, there
is only one inequality which is inequivalent to the CHSH inequality; there are states which
violate it but do not violate the CHSH inequality).


\subsubsection{Bell's inequalities for $n$ particles}


{\bf [Greenberger-Horne-Shimony-Zeilinger 90]} (Sec.~V),
{\bf [Mermin 90 c]},
{\bf [Roy-Singh 91]},
{\bf [Clifton-Redhead-Butterfield 91 a]} (p.~175),
{\bf [Hardy 91 a]} (Secs. 2 and 3),
{\bf [Braunstein-Mann-Revzen 92]},
{\bf [Ardehali 92]},
{\bf [Klyshko 93]},
{\bf [Belinsky-Klyshko 93 a, b]},
{\bf [Braunstein-Mann 93]},
{\bf [Hnilo 93, 94]},
{\bf [Belinsky 94 a]},
{\bf [Greenberger 95]},
{\bf [\.{Z}ukowski-Kaszlikowski 97]}
(critical visibility for $n$-particle GHZ
correlations to violate local realism),
{\bf [Pitowsky-Svozil 00]} (Bell's inequalities for the
GHZ case with two and three local observables),
{\bf [Werner-Wolf 01 b]},
{\bf [\.{Z}ukowski-Brukner 01]},
{\bf [Scarani-Gisin 01 b]} (pure entangled states may exist
which do not violate Mermin-Klyshko inequality),
{\bf [Chen-Kaszlikowski-Kwek-Oh 02]}
(Clauser-Horne-Bell inequality for three three-dimensional systems),
{\bf [Brukner-Laskowski-\.{Z}ukowski 03]}
(multiparticle Bell's inequalities involving many measurement settings:
the inequalities
reveal violations of local realism for some states for which the two
settings-per-local-observer inequalities fail in this task),
{\bf [Laskowski-Paterek-\.{Z}ukowski-Brukner 04]}.


\subsubsection{Which states violate Bell's inequalities?}


(Any pure entangled state does violate Bell-CHSH inequalities:)
{\bf [Capasso-Fortunato-Selleri 73]},
{\bf [Gisin 91 a]} (some corrections in
{\bf [Barnett-Phoenix 92]}),
{\bf [Werner 89]} (one might naively think that as in the case
of pure states, the only mixed states which do not violate Bell's inequalities
are the mixtures of product states, i.e. separable states.
Werner shows that this conjecture is false),
(maximum violations for pure states:)
{\bf [Popescu-Rohrlich 92]},
(maximally entangled states violate maximally Bell's inequalities:)
{\bf [Kar 95]},
{\bf [Cereceda 96 b]}.
For mixed states:
{\bf [Braunstein-Mann-Revzen 92]} (maximum violation for mixed states),
{\bf [Mann-Nakamura-Revzen 92]},
{\bf [Beltrametti-Maczy\'{n}ski 93]},
{\bf [Horodecki-Horodecki-Horodecki 95]} (necessary and sufficient condition
for a mixed state to violate the CHSH inequalities),
{\bf [Aravind 95]}.


\subsubsection{Other inequalities}


{\bf [Baracca-Bergia-Livi-Restignoli 76]} (for non-dichotomic observables),
{\bf [Cirel'son 80]} (while Bell's inequalities give limits for the correlations
in local hidden variables theories,
Cirel'son inequality gives the upper limit for {\em quantum} correlations and,
therefore, the highest possible violation of Bell's inequalities according to QM;
see also
{\bf [Chefles-Barnett 96]}),
{\bf [Hardy 92 d]},
{\bf [Eberhard 93]},
{\bf [Peres 98 d]} (comparing the strengths of various Bell's inequalities)
{\bf [Peres 98 f]} (Bell's inequalities for any number of observers, alternative
setups and outcomes).


\subsubsection{Inequalities to detect genuine $n$-particle nonseparability}


{\bf [Svetlichny 87]},
{\bf [Gisin-Bechmann Pasquinucci 98]},
{\bf [Collins-Gisin-Popescu-(+2) 02]},
{\bf [Seevinck-Svetlichny 02]},
{\bf [Mitchell-Popescu-Roberts 02]},
{\bf [Seevinck-Uffink 02]} (sufficient conditions for three-particle
entanglement and their tests in recent experiments),
{\bf [Cereceda 02 b]},
{\bf [Uffink 02]} (quadratic Bell inequalities which
distinguish, for systems of $n>2$ qubits, between fully entangled states
and states in which at most $n-1$ particles are entangled).


\subsubsection{Herbert's proof of Bell's theorem}


{\bf [Herbert 75]},
{\bf [Stapp 85 a]},
{\bf [Mermin 89 a]},
{\bf [Penrose 89]} (pp.~573-574 in the Spanish version),
{\bf [Ballentine 90 a]} (p.~440).


\subsubsection{Mermin's statistical proof of Bell's theorem}


{\bf [Mermin 81 a, b]},
{\bf [Kunstatter-Trainor 84]} (in the context of the statistical interpretation of QM),
{\bf [Mermin 85]} (see also the comments ---seven---),
{\bf [Penrose 89]} (pp.~358-360 in the Spanish version),
{\bf [Vogt 89]},
{\bf [Mermin 90 e]} (Chaps.~10-12),
{\bf [Allen 92]},
{\bf [Townsend 92]} (Chap.~5, p.~136),
{\bf [Yurke-Stoler 92 b]} (experimental proposal with two independent sources of particles),
{\bf [Marmet 93]}.


\subsection{Bell's theorem without inequalities}


\subsubsection{Greenberger-Horne-Zeilinger's proof}


{\bf [Greenberger-Horne-Zeilinger 89, 90]},
{\bf [Mermin 90 a, b, d, 93 a, b]},
{\bf [Greenberger-Horne-Shimony-Zeilinger 90]},
{\bf [Clifton-Redhead-Butterfield 91 a, b]},
{\bf [Pagonis-Redhead-Clifton 91]} (with $n$ particles),
{\bf [Clifton-Pagonis-Pitowsky 92]},
{\bf [Stapp 93 a]},
{\bf [Cereceda 95]} (with $n$ particles),
{\bf [Pagonis-Redhead-La Rivi\`{e}re 96]},
{\bf [Belnap-Szab\'{o} 96]},
{\bf [Bernstein 99]} (simple version of the GHZ argument),
{\bf [Vaidman 99 b]} (variations on the GHZ proof),
{\bf [Cabello 01 a]} (with $n$ spin-$s$ particles),
{\bf [Massar-Pironio 01]} (GHZ for position and momentum),
{\bf [Chen-Zhang 01]} (GHZ for continuous variables),
{\bf [Khrennikov 01 a]},
{\bf [Kaszlikowski-\.{Z}ukowski 01]} (GHZ for $N$ qu$N$its),
{\bf [Greenberger 02]} (the history of the GHZ paper),
{\bf [Cerf-Massar-Pironio 02]} (GHZ for many qudits).


\subsubsection{Peres' proof of impossibility of recursive elements of reality}


{\bf [Peres 90 b, 92 a]},
{\bf [Mermin 90 d, 93 a, b]},
{\bf [Nogueira-dos Aidos-Caldeira-Domingos 92]},
(why Bohm's theory eludes Peres's and Mermin's proofs:)
{\bf [Dewdney 92]},
{\bf [Dewdney 92]} (see also
{\bf [Pagonis-Clifton 95]}),
{\bf [Peres 93 a]} (Sec.~7.~3),
{\bf [Cabello 95]},
{\bf [De Baere 96 a]} (how to avoid the proof).


\subsubsection{Hardy's proof}


{\bf [Hardy 92 a, 93]},
{\bf [Clifton-Niemann 92]} (Hardy's argument with two spin-$s$ particles),
{\bf [Pagonis-Clifton 92]} (Hardy's argument with $n$ spin-$\frac{1}{2}$ particles),
{\bf [Hardy-Squires 92]},
{\bf [Stapp 92]} (Sec.~VII),
{\bf [Vaidman 93]},
{\bf [Goldstein 94 a]},
{\bf [Mermin 94 a, c, 95 a]},
{\bf [Jordan 94 a, b]}, (nonlocality of a single photon:)
{\bf [Hardy 94, 95 a, 97]};
{\bf [Cohen-Hiley 95 a, 96]},
{\bf [Garuccio 95 b]},
{\bf [Wu-Xie 96]} (Hardy's argument for three spin-$\frac{1}{2}$ particles),
{\bf [Pagonis-Redhead-La Rivi\`{e}re 96]},
{\bf [Kar 96]},
{\bf [Kar 97 a, c]} (mixed states of three or more spin-$\frac{1}{2}$ particles
allow a Hardy argument),
{\bf [Kar 97 b]} (uniqueness of the Hardy state for a fixed choice of observables),
{\bf [Stapp 97]}, {\bf [Unruh 97]},
{\bf [Boschi-Branca-De Martini-Hardy 97]} (ladder argument),
{\bf [Schafir 98]} (Hardy's argument in the many-worlds and consistent
histories interpretations),
{\bf [Ghosh-Kar 98]} (Hardy's argument for two spin $s$ particles),
{\bf [Ghosh-Kar-Sarkar 98]} (Hardy's argument for three spin-$\frac{1}{2}$ particles),
{\bf [Cabello 98 a]} (ladder proof without probabilities for two spin $s \geq 1$ particles),
{\bf [Barnett-Chefles 98]} (nonlocality without inequalities for all pure entangled states
using generalized measurements which perform unambiguous state discrimination between
non-orthogonal states),
{\bf [Cereceda 98, 99 b]} (generalized probability for Hardy's nonlocality contradiction),
{\bf [Cereceda 99 a]} (the converse of Hardy's theorem),
{\bf [Cereceda 99 c]} (Hardy-type experiment for maximally entangled states
and the problem of subensemble postselection),
{\bf [Cabello 00 b]} (nonlocality without inequalities has not
been proved for maximally entangled states),
{\bf [Yurke-Hillery-Stoler 99]} (position-momentum Hardy-type proof),
{\bf [Wu-Zong-Pang 00]} (Hardy's proof for GHZ states),
{\bf [Hillery-Yurke 01]} (upper and lower bounds on maximal violation of local realism
in a Hardy-type test using continuous variables),
{\bf [Irvine-Hodelin-Simon-Bouwmeester 04]} (realisation of {\bf [Hardy 92 a]}).


\subsubsection{Bell's theorem without inequalities for EPR-Bohm-Bell states}


{\bf [Cabello 01 c, d]},
{\bf [Nistic\`{o} 01]} (GHZ-like proofs
are impossible for pairs of qubits),
{\bf [Aravind 02, 04]},
{\bf [Chen-Pan-Zhang-(+2) 03]} (experimental implementation).


\subsubsection{Other algebraic proofs of no-local hidden variables}


{\bf [Pitowsky 91 b, 92]},
{\bf [Herbut 92]},
{\bf [Clifton-Pagonis-Pitowsky 92]},
{\bf [Cabello 02 a]}.


\subsubsection{Classical limits of no-local hidden variables proofs}


{\bf [Sanz 90]} (Chap.~4),
{\bf [Pagonis-Redhead-Clifton 91]} (GHZ with $n$ spin-$\frac{1}{2}$ particles),
{\bf [Peres 92 b]},
{\bf [Clifton-Niemann 92]} (Hardy with two spin-$s$ particles),
{\bf [Pagonis-Clifton 92]} (Hardy with $n$ spin-$\frac{1}{2}$ particles).


\subsection{Other ``nonlocalities''}


\subsubsection{``Nonlocality'' of a single particle}


{\bf [Grangier-Roger-Aspect 86]},
{\bf [Grangier-Potasek-Yurke 88]},
{\bf [Tan-Walls-Collett 91]},
{\bf [Hardy 91 a, 94, 95 a]},
{\bf [Santos 92 a]},
{\bf [Czachor 94]},
{\bf [Peres 95 b]},
{\bf [Home-Agarwal 95]},
{\bf [Gerry 96 c]},
{\bf [Steinberg 98]} (single-particle nonlocality and conditional measurements),
{\bf [Resch-Lundeen-Steinberg 01]}
(experimental observation of nonclassical effects on
single-photon detection rates),
{\bf [Bj{\o}rk-Jonsson-S\'{a}nchez Soto 01]}
(single-particle nonlocality and entanglement with the vacuum),
{\bf [Srikanth 01 e]},
{\bf [Hessmo-Usachev-Heydari-Bj\"{o}rk 03]}
(experimental demonstration of single photon ``nonlocality'').


\subsubsection{Violations of local realism exhibited in sequences of measurements
(``hidden nonlocality'')}


{\bf [Popescu 94, 95 b]} (Popescu notices that the LHV model proposed in
{\bf [Werner 89]} does not work for sequences of measurements),
{\bf [Gisin 96 a, 97]} (for two-level systems nonlocality can be revealed
using filters),
{\bf [Peres 96 e]} (Peres considers collective tests on Werner states and uses
consecutive measurements to show the impossibility of constructing LHV models
for some processes of this kind),
{\bf [Berndl-Teufel 97]},
{\bf [Cohen 98 b]} (unlocking hidden entanglement with classical information),
{\bf [\.{Z}ukowski-Horodecki-Horodecki-Horodecki 98]},
{\bf [Hiroshima-Ishizaka 00]}
(local and nonlocal properties of Werner states),
{\bf [Kwiat-Barraza L\'{o}pez-Stefanov-Gisin 01]}
(experimental entanglement distillation and `hidden'
non-locality), {\bf [Wu-Zong-Pang-Wang 01 b]}
(Bell's inequality for Werner states).


\subsubsection{Local immeasurability or indistinguishability (``nonlocality without entanglement'')}


{\bf [Bennett-DiVincenzo-Fuchs-(+5) 99]}
(an unknown member of a product basis cannot be reliably
distinguished from the others by local measurements and classical
communication),
{\bf [Bennett-DiVincenzo-Mor-(+3) 99]},
{\bf [Horodecki-Horodecki-Horodecki 99 d]}
(``nonlocality without entanglement'' is an EPR-like incompleteness
argument rather than a Bell-like proof),
{\bf [Groisman-Vaidman 01]}
(nonlocal variables with product states eigenstates),
{\bf [Walgate-Hardy 02]},
{\bf [Horodecki-Sen De-Sen-Horodecki 03]}
(first operational method for checking indistinguishability of
orthogonal states by LOCC;
any full basis of an arbitrary number of systems is
not distinguishable, if at least one of the vectors is entangled),
{\bf [De Rinaldis 03]}
(method to check the LOCC
distinguishability of a complete product bases).


\subsection{Experiments on Bell's theorem}


\subsubsection{Real experiments}


{\bf [Kocher-Commins 67]},
{\bf [Papaliolios 67]},
{\bf [Freedman-Clauser 72]} (with photons correlated in polarizations after the decay
$J=0 \rightarrow 1 \rightarrow 0$ of Ca atoms; see also
{\bf [Freedman 72]},
{\bf [Clauser 92]}),
{\bf [Holt-Pipkin 74]} (id. with Hg atoms; the results of this experiment agree
with Bell's inequalities),
{\bf [Clauser 76 a]},
{\bf [Clauser 76 b]} (Hg),
{\bf [Fry-Thompson 76]} (Hg),
{\bf [Lamehi Rachti-Mittig 76]} (low energy proton-proton scattering),
{\bf [Aspect-Grangier-Roger 81]} (with Ca photons and one-channel polarizers; see also
{\bf [Aspect 76]}),
{\bf [Aspect-Grangier-Roger 82]} (Ca and two-channel polarizers),
{\bf [Aspect-Dalibard-Roger 82]} (with optical devices that change the
orientation of the polarizers during the photon's flight; see also
{\bf [Aspect 83]}),
{\bf [Perrie-Duncan-Beyer-Kleinpoppen 85]} (with correlated photons simultaneously
emitted by metastable deuterium),
{\bf [Shih-Alley 88]} (with a parametic-down converter),
{\bf [Rarity-Tapster 90 a]} (with momentum and phase),
{\bf [Kwiat-Vareka-Hong-(+2) 90]} (with photons emitted by a non-linear
crystal and correlated in a double interferometer; following Franson's proposal
{\bf [Franson 89]}),
{\bf [Ou-Zou-Wang-Mandel 90]} (id.),
{\bf [Ou-Pereira-Kimble-Peng 92]} (with photons correlated in amplitude),
{\bf [Tapster-Rarity-Owens 94]} (with photons in optical fibre),
{\bf [Kwiat-Mattle-Weinfurter-(+3) 95]}
(with a type-II parametric-down converter),
{\bf [Strekalov-Pittman-Sergienko-(+2) 96]},
{\bf [Tittel-Brendel-Gisin-(+3) 97, 98]}
(testing quantum correlations with photons 10 km apart in optical fibre),
{\bf [Tittel-Brendel-Zbinden-Gisin 98]} (a Franson-type test of Bell's
inequalities by photons 10,9 km apart),
{\bf [Weihs-Jennewein-Simon-(+2) 98]}
(experiment with strict Einstein locality conditions, see also {\bf [Aspect 99]}),
{\bf [Kuzmich-Walmsley-Mandel 00]},
{\bf [Rowe-Kielpinski-Meyer-(+4) 01]} (experimental violation of a Bell's
inequality for two beryllium ions with nearly perfect detection efficiency),
{\bf [Howell-Lamas Linares-Bouwmeester 02]}
(experimental violation of a spin-$1$ Bell's
inequality using maximally-entangled four-photon states),
{\bf [Moehring-Madsen-Blinov-Monroe 04]}
(experimental Bell inequality violation with an atom and a photon; see also
{\bf [Blinov-Moehring-Duan-Monroe 04]}).


\subsubsection{Proposed {\em gedanken} experiments}


{\bf [Lo-Shimony 81]} (disotiation of a metastable molecule),
{\bf [Horne-Zeilinger 85, 86, 88]} (particle interferometers),
{\bf [Horne-Shimony-Zeilinger 89, 90 a, b]} (id.) (see also
{\bf [Greenberger-Horne-Zeilinger 93]},
{\bf [Wu-Xie-Huang-Hsia 96]}),
{\bf [Franson 89]} (with position and time), with observables with a discrete
spectrum and ---simultaneously--- observables with a continuous spectrum
{\bf [\.{Z}ukowski-Zeilinger 91]} (polarizations and momentums),
(experimental proposals on Bell's inequalities without additional assumptions:)
{\bf [Fry-Li 92]},
{\bf [Fry 93, 94]},
{\bf [Fry-Walther-Li 95]},
{\bf [Kwiat-Eberhard-Steinberg-Chiao 94]},
{\bf [Pittman-Shih-Sergienko-Rubin 95]},
{\bf [Fern\'{a}ndez Huelga-Ferrero-Santos 94, 95]}
(proposal of an experiment with photon pairs and detection of the recoiled atom),
{\bf [Freyberger-Aravind-Horne-Shimony 96]}.


\subsubsection{EPR with neutral kaons}


{\bf [Lipkin 68]},
{\bf [Six 77]},
{\bf [Selleri 97]},
{\bf [Bramon-Nowakowski 99]},
{\bf [Ancochea-Bramon-Nowakowski 99]}
(Bell-inequalities for $K^0 \bar{K^0}$ pairs from $\Phi$-resonance
decays),
{\bf [Dalitz-Garbarino 00]}
(local realistic theories for the two-neutral-kaon system),
{\bf [Gisin-Go 01]} (EPR with photons and kaons: Analogies),
{\bf [Hiesmayr 01]} (a generalized Bell's inequality for
the $K^0 \bar{K^0}$ system),
{\bf [Bertlmann-Hiesmayr 01]} (Bell's inequalities for entangled kaons
and their unitary time evolution),
{\bf [Garbarino 01]},
{\bf [Bramon-Garbarino 02 a, b]}.


\subsubsection{Reviews}


{\bf [Clauser-Shimony 78]},
{\bf [Pipkin 78]},
{\bf [Duncan-Kleinpoppen 88]},
{\bf [Chiao-Kwiat-Steinberg 95]} (review of the experiments proposed by these authors
with photons emitted by a non-linear crystal after a parametric down conversion).


\subsubsection{Experimental proposals on GHZ proof, preparation of GHZ states}


{\bf [\.{Z}ukowski 91 a, b]},
{\bf [Yurke-Stoler 92 a]} (three-photon GHZ states can be
obtained from three spatially separated sources of one photon),
{\bf [Reid-Munro 92]},
{\bf [W\'{o}dkiewicz-Wang-Eberly 93]} (preparation of a GHZ state with a four-mode
cavity and a two-level atom),
{\bf [Klyshko 93]},
{\bf [Shih-Rubin 93]},
{\bf [W\'{o}dkiewicz-Wang-Eberly 93 a, b]},
{\bf [Hnilo 93, 94]},
{\bf [Cirac-Zoller 94]} (preparation of singlets and GHZ states with two-level
atoms and a cavity),
{\bf [Fleming 95]} (with only one particle),
{\bf [Pittman 95]} (preparation of a GHZ state with four
photons from two sources of pairs),
{\bf [Haroche 95]},
{\bf [Lalo\"{e} 95]},
{\bf [Gerry 96 b, d, e]} (preparations of a GHZ state using cavities),
{\bf [Pfau-Kurtsiefer-Mlynek 96]},
{\bf [Zeilinger-Horne-Weinfurter-\.{Z}ukowski 97]} (three-particle GHZ
states prepared from two entangled pairs),
{\bf [Lloyd 97 b]} (a GHZ experiment with mixed states),
{\bf [Keller-Rubin-Shih-Wu 98]},
{\bf [Keller-Rubin-Shih 98 b]},
{\bf [Laflamme-Knill-Zurek-(+2) 98]} (real experiment to
produce three-particle GHZ states using nuclear magnetic resonance),
{\bf [Lloyd 98 a]}
(microscopic analogs of the GHZ experiment),
{\bf [Pan-Zeilinger 98]} (GHZ states analyzer),
{\bf [Larsson 98 a]} (necessary and sufficient conditions on detector
efficiencies in a GHZ experiment),
{\bf [Munro-Milburn 98]} (GHZ in nondegenerate parametric oscillation via
phase measurements),
{\bf [Rarity-Tapster 99]} (three-particle entanglement obtained
from entangled photon pairs and a weak coherent state),
{\bf [Bouwmeester-Pan-Daniell-(+2) 99]} (experimental observation
of polarization entanglement for three spatially separated
photons, based on the idea of
{\bf [Zeilinger-Horne-Weinfurter-\.{Z}ukowski 97]}),
{\bf [Watson 99 a]},
{\bf [Larsson 99 b]} (detector efficiency in the GHZ experiment),
{\bf [Sakaguchi-Ozawa-Amano-Fukumi 99]} (microscopic analogs of the GHZ
experiment on an NMR quantum computer),
{\bf [Guerra-Retamal 99]}
(proposal for atomic GHZ states
via cavity quantum electrodynamics),
{\bf [Pan-Bouwmeester-Daniell-(+2) 00]} (experimental test),
{\bf [Nelson-Cory-Lloyd 00]} (experimental GHZ correlations using NMR),
{\bf [de Barros-Suppes 00 b]}
(inequalities for dealing with detector inefficiencies in GHZ experiments),
{\bf [Cohen-Brun 00]}
(distillation of GHZ states by selective information manipulation),
{\bf [\.{Z}ukowski 00]} (an analysis of the ``wrong'' events in the Innsbruck
experiment shows that they cannot be described using a local realistic model),
{\bf [Sackett-Kielpinski-King-(+8) 00]}
(experimental entanglement of four ions: Coupling between the ions
is provided through their collective motional degrees of freedom),
{\bf [Zeng-Kuang 00 a]} (preparation of GHZ states via Grover's algorithm),
{\bf [Ac\'{\i}n-Jan\'{e}-D\"{u}r-Vidal 00]}
(optimal distillation of a GHZ state),
{\bf [Cen-Wang 00]} (distilling a GHZ state from
an arbitrary pure state of three qubits),
{\bf [Zhao-Yang-Chen-(+2) 03 b]} (nonlocality with a polarization-entangled four-photon GHZ state).


\subsubsection{Experimental proposals on Hardy's proof}


{\bf [Hardy 92 d]} (with two photons in overlapping optical interferometers),
{\bf [Yurke-Stoler 93]} (with two identical fermions in overlapping interferometers and
using Pauli's exclusion principle),
{\bf [Hardy 94]} (with a source of just one photon),
{\bf [Freyberger 95]} (two atoms passing through two cavities),
{\bf [Torgerson-Branning-Mandel 95]},
{\bf [Torgerson-Branning-Monken-Mandel 95]} (first real experiment,
measuring two-photon coincidence),
{\bf [Garuccio 95 b]} (to extract conclusions
from experiments like the one by Torgerson et al. some inequalities must be derived),
{\bf [Cabello-Santos 96]} (criticism of the conclusions
of the experiment by Torgerson et al.),
{\bf [Torgerson-Branning-Monken-Mandel 96]} (reply),
{\bf [Mandel 97]} (experiment),
{\bf [Boschi-De Martini-Di Giuseppe 97]},
{\bf [Di Giuseppe-De Martini-Boschi 97]} (second real experiment),
{\bf [Boschi-Branca-De Martini-Hardy 97]}
(real experiment based on the ladder version of Hardy's argument),
{\bf [Kwiat 97 a, b]},
{\bf [White-James-Eberhard-Kwiat 99]} (nonmaximally entangled states:
Production, characterization, and utilization),
{\bf [Franke-Huget-Barnett 00]}
(Hardy state correlations for two trapped ions),
{\bf [Barbieri-De Martini-Di Nepi-Mataloni 04]}
(experiment of Hardy's ``ladder theorem''
without ``supplementary assumptions''),
{\bf [Irvine-Hodelin-Simon-Bouwmeester 04]} (realisation of {\bf [Hardy 92 a]}).


\subsubsection{Some criticisms of the experiments on Bell's inequalities. Loopholes}


{\bf [Marshall-Santos-Selleri 83]} (``local realism has not been refuted by atomic
cascade experiments''),
{\bf [Marshall-Santos 89]},
{\bf [Santos 91, 96]},
{\bf [Santos 92 c]} (local hidden variable model which agree with the predictions
of QM for the experiments based on photons emitted by atomic cascade,
like those of Aspect's group),
{\bf [Garuccio 95 a]} (criticism for the experiments with photons emitted
by parametric down conversion),
{\bf [Basoalto-Percival 01]}
(a computer program for the Bell detection loophole).


\section{``Interpretations''}


\subsection{Copenhagen interpretation}


{\bf [Bohr 28, 34, 35 a, b, 39, 48, 49, 58 a, b, 63, 86, 96, 98]}
({\bf [Bohr 58 b]} was regarded by Bohr as his clearest presentation of the
observational situation in QM. In it he asserts that QM cannot exist without
classical mechanics: The classical realm is an essential
part of any proper measurement, that is, a measurement whose results can be
communicated in plain language. The wave function represents, in Bohr's words, ``a
purely symbolic procedure, the unambiguous physical interpretation of which in the last
resort requires a reference to a complete experimental arrangement''),
{\bf [Heisenberg 27, 30, 55 a, b, 58, 95]}
({\bf [Heisenberg 55 a]} is perhaps Heisenberg's most
important and complete statement of his views: The wave function is ``objective'' but it
is not ``real'', the cut between quantum and classical realms cannot be pushed so far that
the entire compound system, including the observing apparatus, is cut off from the rest of
the universe. A connection with the external world is essential. Stapp points out in
{\bf [Stapp 72]} that ``Heisenberg's writings are more direct [than Bohr's].
But his way of speaking suggests a subjective interpretation that appears
quite contrary to the apparent intention of Bohr''.
See also more precise differences between Bohr and Heisenberg's
writings pointed out in
{\bf [DeWitt-Graham 71]}),
{\bf [Fock 31]} (textbook),
{\bf [Landau-Lifshitz 48]} (textbook),
{\bf [Bohm 51]} (textbook),
{\bf [Hanson 59]},
{\bf [Stapp 72]} (this reference is described in
{\bf [Ballentine 87 a]}, p.~788 as follows: `In attempting to save ``the Copenhagen
interpretation'' the author radically revises what is often, rightly or wrongly,
understood by that term. That interpretation in which Von Neumann's ``reduction''
of the state vector in measurement forms the core is rejected, as
are Heisenberg's subjectivistic statements. The very ``pragmatic'' (one could also say
``instrumentalist'') aspect of the interpretation is emphasized.'),
{\bf [Faye 91]} (on Bohr's interpretation of QM),
{\bf [Zeilinger 96 b]} (``It is suggested that the objective
randomness of the individual quantum event is a necessity of a description of the world
(\ldots). It is also suggested that the austerity of the Copenhagen interpretation should serve
as a guiding principle in a search for deeper understanding.''),
{\bf [Zeilinger 99 a]} (the quotations are not in their original order, and
some italics are mine:
``We have knowledge, i.e., information, of an
object only through observation (\ldots).
Any physical object can be described by a set of true propositions (\ldots).
[B]y proposition we mean something which can be verified directly by experiment (\ldots).
In order to analyze the information content of elementary systems,
we (\ldots) decompose a system (\ldots) into constituent systems (\ldots). [E]ach
such constituent systems will be represented by fewer propositions. How far, then,
can this process of subdividing a system go? (\ldots). [T]he limit is reached when an
individual system finally represents the truth value to one single proposition {\em only}.
Such a system we call an elementary system. We thus suggest a principle of quantization
of information as follows: {\em An elementary system represents the truth value of one
proposition}. [This is what Zeilinger proposes as the foundational principle
for quantum mechanics. He says that he
personally prefers the Copenhagen interpretation because of its extreme
austerity and clarity. However, the purpose of this paper is to attempt to go
significantly beyond previous interpretations] (\ldots).
The spin of [a spin-1/2] (\ldots) particle carries the answer to
one question {\em only}, namely, the question What is its spin along the $z$-axis? (\ldots). Since
this is the only information the spin carries, measurement along any other
direction must necessarily contain an element of randomness (\ldots). We have thus
found a reason for the irreducible randomness in quantum measurement. It is the simple
fact that an elementary system cannot carry enough information to provide
definite answers to all questions that could be asked experimentally (\ldots). [After
the measurement, t]he new information the system now represents
has been spontaneously {\em created}
in the measurement itself (\ldots). [The information carried by composite systems
can be distributed in different ways: E]ntanglement results if all possible
information is exhausted in specifying {\em joint} (\ldots)
[true propositions] of the constituents''. See \ref{itfrombit}),
{\bf [Fuchs-Peres 00 a, b]} (quantum theory needs no ``interpretation'').


\subsection{De Broglie's ``pilot wave'' and Bohm's ``causal'' interpretations}


\subsubsection{General}


{\bf [Bohm 52]},
{\bf [de Broglie 60]},
{\bf [Goldberg-Schey-Schwartz 67]}
(computer-generated motion pictures of one-dimensional
quantum-mechanical transmission and reflection phenomena),
{\bf [Philippidis-Dewdney-Hiley 79]} (the
quantum potential and the ensemble of particle trajectories are computed and illustrated
for the two-slit interference pattern),
{\bf [Bell 82]},
{\bf [Bohm-Hiley 82, 89]},
{\bf [Dewdney-Hiley 82]},
{\bf [Dewdney-Holland-Kyprianidis 86, 87]},
{\bf [Bohm-Hiley 85]},
{\bf [Bohm-Hiley-Kaloyerou 87]},
{\bf [Dewdney 87, 92, 93]},
{\bf [Dewdney-Holland-Kyprianidis-Vigier 88]},
{\bf [Holland 88, 92]},
{\bf [Englert-Scully-S\"{u}ssmann-Walther 93 a, b]}
({\bf [D\"{u}rr-Fusseder-Goldstein-Zangh\`{\i} 93]})
{\bf [Albert 92]} (Chap.~7),
{\bf [Dewdney-Malik 93]},
{\bf [Bohm-Hiley 93]} (book),
{\bf [Holland 93]} (book),
{\bf [Albert 94]},
{\bf [Pagonis-Clifton 95]},
{\bf [Cohen-Hiley 95 b]} (comparison between Bohmian mechanics, standard QM and
consistent histories interpretation),
{\bf [Mackman-Squires 95]} (retarded Bohm model),
{\bf [Berndl-D\"{u}rr-Goldstein-Zangh\`{\i} 96]},
{\bf [Goldstein 96, 99]},
{\bf [Cushing-Fine-Goldstein 96]} (collective book),
{\bf [Garc\'{\i}a de Polavieja 96 a, b, 97 a, b]} (causal interpretation in phase
space derived from the coherent space representation of the Schr\"{o}dinger equation),
{\bf [Kent 96 b]} (consistent histories and Bohmian mechanics),
{\bf [Rice 97 a]},
{\bf [Hiley 97]},
{\bf [Deotto-Ghirardi 98]} (there are infinite theories similar to Bohm's ---with
trajectories--- which reproduce the predictions of QM),
{\bf [Dickson 98]},
{\bf [Terra Cunha 98]},
{\bf [Wiseman 98 a]}
(Bohmian analysis of momentum transfer in welcher Weg measurements),
{\bf [Blaut-Kowalski Glikman 98]},
{\bf [Brown-Sj\"{o}qvist-Bacciagaluppi 99]}
(on identical particles in de Broglie-Bohm's theory),
{\bf [Leavens-Sala Mayato 99]},
{\bf [Griffiths 99 b]} (Bohmian mechanics and consistent histories),
{\bf [Maroney-Hiley 99]} (teleportation understood through the Bohm
interpretation),
{\bf [Belousek 00 b]},
{\bf [Neumaier 00]} (Bohmian mechanics contradict quantum mechanics),
{\bf [Ghose 00 a, c, d, 01 b]}
(incompatibility of the de Broglie-Bohm theory with quantum mechanics),
{\bf [Marchildon 00]} (no contradictions between Bohmian and quantum mechanics),
{\bf [Barrett 00]} (surreal trajectories),
{\bf [Nogami-Toyama-Dijk 00]},
{\bf [Shifren-Akis-Ferry 00]},
{\bf [Ghose 00 c]} (experiment to distinguish between de
Broglie-Bohm and standard quantum mechanics),
{\bf [Golshani-Akhavan 00, 01 a, b, c]}
(experiment which distinguishes between the
standard and Bohmian quantum mechanics),
{\bf [Hiley-Maroney 00]}
(consistent histories and the Bohm approach),
{\bf [Hiley-Callaghan-Maroney 00]},
{\bf [Gr"{o}ssing 00]} (book; extension of the de
Broglie-Bohm interpretation into the relativistic
regime for the Klein-Gordon case),
{\bf [D\"{u}rr 01]} (book),
{\bf [Marchildon 01]} (on Bohmian trajectories
in two-particle interference devices),
{\bf [John 01 a, b]} (modified de Broglie-Bohm theory closer to
classical Hamilton-Jacobi theory),
{\bf [Bandyopadhyay-Majumdar-Home 01]},
{\bf [Struyve-De Baere 01]},
{\bf [Ghose-Majumdar-Guha-Sau 01]} (Bohmian trajectories for photons),
{\bf [Shojai-Shojai 01]} (problems raised by the relativistic form of
de Broglie-Bohm theory),
{\bf [Allori-Zangh\`{\i} 01 a]},
(de Broglie's pilot wave theory for the Klein-Gordon equation:)
{\bf [Horton-Dewdney 01 b]}, {\bf [Horton-Dewdney-Ne'eman 02]};
{\bf [Ghose-Samal-Datta 02]}
(Bohmian picture of Rydberg atoms),
{\bf [Feligioni-Panella-Srivastava-Widom 02]},
{\bf [Gr\"{u}bl-Rheinberger 02]},
{\bf [Dewdney-Horton 02]} (relativistically invariant extension),
{\bf [Allori-D\"{u}rr-Goldstein-Zangh\`{\i} 02]},
{\bf [Bacciagaluppi 03]}
(derivation of the symmetry postulates for identical particles from
pilot-wave theories),
{\bf [Tumulka 04 a]}.


\subsubsection{Tunneling times in Bohmian mechanics}


{\bf [Hauge-Stovneng 89]} (TT: A critical review),
{\bf [Spiller-Clarck-Prance-Prance 90]},
{\bf [Olkhovsky-Recami 92]} (recent developments in TT),
{\bf [Leavens 93, 95, 96, 98]},
{\bf [Leavens-Aers 93]},
{\bf [Landauer-Martin 94]} (review on TT),
{\bf [Leavens-Iannaccone-McKinnon 95]},
{\bf [McKinnon-Leavens 95]},
{\bf [Cushing 95 a]}
(are quantum TT a crucial
test for the causal program?; reply: {\bf [Bedard 97]}),
{\bf [Oriols-Mart\'{\i}n-Su\~{n}e 96]}
(implications of the noncrossing property of Bohm
trajectories in one-dimensional tunneling configurations),
{\bf [Abolhasani-Golshani 00]}
(TT in the Copenhagen interpretation; due to experimental
limitations, Bohmian mechanics leads to same TT),
{\bf [Majumdar-Home 00]}
(the time of decay measurement in the Bohm model),
{\bf [Ruseckas 01]} (tunneling time determination in standard QM),
{\bf [Stomphorst 01, 02]},
{\bf [Chuprikov 01]}.


\subsection{``Relative state'', ``many worlds'', and ``many minds'' interpretations}


{\bf [Everett 57 a, b, 63]},
{\bf [Wheeler 57]},
{\bf [DeWitt 68, 70, 71 b]},
{\bf[Cooper-Van Vechten 69]} (proof of the unobservability of the splits),
{\bf [DeWitt-Graham 73]},
{\bf [Graham 71]},
{\bf [Ballentine 73]} (the definition of the ``branches'' is dependent
upon the choice of representation; the assumptions of the many-worlds interpretation are
neither necessary nor sufficient to derive the Born statistical formula),
{\bf [Clarke 74]}
(some additional structures must be added in order to determine which states will
determine the ``branching''),
{\bf [Healey 84]} (critical discussion),
{\bf [Geroch 84]},
{\bf [Whitaker 85]},
{\bf [Deutsch 85 a, 86]} (testable split observer experiment),
{\bf [Home-Whitaker 87]} (quantum Zeno effect in the many-worlds interpretation),
{\bf [Tipler 86]},
{\bf [Squires 87 a, b]} (the ``many-views'' interpretation),
{\bf [Whitaker 89]} (on Squires' many-views interpretation),
{\bf [Albert-Loewer 88]},
{\bf [Ben Dov 90 b]},
{\bf [Kent 90]},
{\bf [Albert-Loewer 91 b]} (many minds interpretation),
{\bf [Vaidman 96 c, 01 d]},
{\bf [Lockwood 96]} (many minds),
{\bf [Cassinello-S\'{a}nchez G\'{o}mez 96]} (and
{\bf [Cassinello 96]}, impossibility of deriving the probabilistic postulate using a
frequency analysis of infinite copies of an individual system),
{\bf [Deutsch 97]} (popular review),
{\bf [Schafir 98]} (Hardy's argument in the many-worlds and in the consistent
histories interpretations),
{\bf [Dickson 98]},
{\bf [Tegmark 98]} (many worlds or many words?),
{\bf [Barrett 99 a]},
{\bf [Wallace 01 b]},
{\bf [Deutsch 01]} (structure of the multiverse),
{\bf [Butterfield 01]},
{\bf [Bacciagaluppi 01 b]},
{\bf [Hewitt-Horsman 03]} (status of the uncertainty relations in the many worlds interpretation).


\subsection{Interpretations with explicit collapse or dynamical reduction theories
(spontaneous localization, nonlinear
terms in Schr\"{o}dinger equation, stochastic theories)}


{\bf [de Broglie 56]},
{\bf [Bohm-Bub 66 a]},
{\bf [Nelson 66, 67, 85]},
{\bf [Pearle 76, 79, 82, 85, 86 a, b, c, 89, 90, 91, 92, 93, 99 b, 00]},
{\bf [Bialynicki Birula-Mycielski 76]} (add a nonlinear term to the
Schr\"{o}dinger equation
in order to keep wave packets from spreading beyond any limit.
Experiments with neutrons,
{\bf [Shull-Atwood-Arthur-Horne 80]} and
{\bf [G\"{a}hler-Klein-Zeilinger 81]}, have resulted in such small upper
limits for a possible nonlinear term of a kind that some quantum features would
survive in a macroscopic world),
{\bf [Dohrn-Guerra 78]},
{\bf [Dohrn-Guerra-Ruggiero 79]}
(relativistic Nelson stochastic model),
{\bf [Davidson 79]} (a generalization of the Fenyes-Nelson stochastic
model),
{\bf [Shimony 79]} (proposed neutron interferometer test of some
nonlinear variants),
{\bf [Bell 84]},
{\bf [Gisin 84 a, b, 89]},
{\bf [Ghirardi-Rimini-Weber 86, 87, 88]},
{\bf [Werner 86]},
{\bf [Primas 90 b]},
{\bf [Ghirardi-Pearle-Rimini 90]},
{\bf [Ghirardi-Grassi-Pearle 90 a, b]},
{\bf [Weinberg 89 a, b, c, d]} (nonlinear variant),
{\bf [Peres 89 d]} (nonlinear variants violate the second law
of thermodynamics),
(in Weinberg's attempt faster than light communication is possible:)
{\bf [Gisin 90]},
{\bf [Polchinski 91]}, {\bf [Mielnik 00]};
{\bf [Bollinger-Heinzen-Itano-(+2) 89]}
(tests Weinberg's variant),
{\bf [W\'{o}dkiewicz-Scully 90]}),
{\bf [Ghirardi 91, 95, 96]},
{\bf [Jordan 93 b]} (fixes the Weinberg variant),
{\bf [Ghirardi-Weber 97]},
{\bf [Squires 92 b]} (if the collapse is a physical phenomenon it would be possible to
measure its velocity),
{\bf [Gisin-Percival 92, 93 a, b, c]},
{\bf [Pearle-Squires 94]} (nucleon decay experimental results could be considered
to rule out the collapse models, and support a version in which the rate of collapse
is proportional to the mass),
{\bf [Pearle 97 a]} {explicit model of collapse, ``true collapse'', versus interpretations with
decoherence, ``false collapse''),
{\bf [Pearle 97 b]} (review of Pearle's own contributions),
{\bf [Bacciagaluppi 98 b]} (Nelsonian mechanics),
{\bf [Santos-Escobar 98]},
{\bf [Ghirardi-Bassi 99]},
{\bf [Pearle-Ring-Collar-Avignone 99]},
{\bf [Pavon 99]} (derivation of the wave function collapse in the context of
Nelson's stochastic mechanics),
{\bf [Adler-Brun 01]}
(generalized stochastic Schr\"{o}dinger equations for state vector collapse),
{\bf [Brody-Hughston 01]}
(experimental tests for stochastic reduction models).


\subsection{Statistical (or ensemble) interpretation}


{\bf [Ballentine 70, 72, 86, 88 a, 90 a, b, 95 a, 96, 98]},
{\bf [Peres 84 a, 93]},
{\bf [Pavi\v{c}i\'{c} 90 d]} (formal difference between the Copenhagen and the statistical interpretation),
{\bf [Home-Whitaker 92]}.


\subsection{``Modal'' interpretations}


{\bf [van Fraassen 72, 79, 81, 91 a, b]},
{\bf [Cartwright 74]},
{\bf [Kochen 85]},
{\bf [Healey 89, 93, 98 a]},
{\bf [Dieks 89, 94, 95]},
{\bf [Lahti 90]} (polar decomposition and measurement),
{\bf [Albert-Loewer 91 a]} (the Kochen-Healey-Dieks interpretations do not
solve the measurement problem),
{\bf [Arntzenius 90]},
{\bf [Albert 92]} (appendix),
{\bf [Elby 93 a]},
{\bf [Bub 93]},
{\bf [Albert-Loewer 93]},
{\bf [Elby-Bub 94]},
{\bf [Dickson 94 a, 95 a, 96 b, 98]},
{\bf [Vermaas-Dieks 95]} (generalization of
the MI to arbitrary density operators),
{\bf [Bub 95]},
{\bf [Cassinelli-Lahti 95]},
{\bf [Clifton 95 b, c, d, e, 96, 00 b]},
{\bf [Bacciagaluppi 95, 96, 98 a, 00]},
{\bf [Bacciagaluppi-Hemmo 96, 98 a, 98 b]},
{\bf [Vermaas 96]},
{\bf [Vermaas 97, 99 a]} (no-go theorems for MI),
{\bf [Zimba-Clifton 98]},
{\bf [Busch 98 a]},
{\bf [Dieks-Vermaas 98]},
{\bf [Dickson-Clifton 98]} (collective book),
{\bf [Bacciagaluppi-Dickson 99]} (dynamics for MI),
{\bf [Dieks 00]} (consistent histories and relativistic invariance
in the MI),
{\bf [Spekkens-Sipe 01 a, b]},
{\bf [Bacciagaluppi 01 a]} (book),
{\bf [Gambetta-Wiseman 04]} (modal dynamics extended to include POVMs).


\subsection{``It from bit''}
\label{itfrombit}


{\bf [Wheeler 78, 81, 95]} (the measuring process creates a ``reality''
that did not exist objectively before the intervention),
{\bf [Davies-Brown 86]} (``the game of the 20 questions'', pp.~23-24
[pp.~38-39 in the Spanish version], Chap.~4),
{\bf [Wheeler-Ford 98]} ([p.~338:] ``A measurement, in this context,
is an irreversible act in which uncertainty collapses to
certainty. It is the link between the quantum and the classical
worlds, the point where what {\em might} happen (\dots) is replaced by
what {\em does} happen (\dots)''. [p.~338:] ``No elementary
phenomenon, he [Bohr] said, is a phenomenon until it is a
registered phenomenon''. [pp.~339-340:] ``Measurement, the act of
turning potentiality into actuality, is an act of choice, choice
among possible outcomes''. [pp.~340-341:] ``Trying to wrap my
brain around this idea of information theory as the basis of
existence, I came up with the phrase ``it from bit.'' The universe
an all that it contains (``it'') may arise from the myriad yes-no
choices of measurement (the ``bits''). Niels Bohr wrestled for
most of his life with the question of how acts of measurement (or
``registration'') may affect reality. It is registration (\dots)
that changes potentiality into actuality. I build only a little on
the structure of Bohr's thinking when I suggest that we may never
understand this strange thing, the quantum, until we understand
how information may underlie reality. Information may not be just
what we {\em learn} about the world. It may be what {\em makes}
the world.

An example of the idea of it from bit: When a photon is absorbed,
and thereby ``measured''---until its absortion, it had no true
reality---an unsplittable bit of information is added to what we
know about the word, {\em and}, at the same time that bit of
information determines the structure of one small part of the
world. It {\em creates} the reality of the time and place of that
photon's interaction'').


\subsection{``Consistent histories'' (or ``decoherent histories'')}


{\bf [Griffiths 84, 86 a, b, c, 87, 93 a, b, 95, 96, 97, 98 a, b, c, 99, 01]},
{\bf [Omn\`{e}s 88 a, 88 b, 88 c, 89, 90 a, b, 91, 92, 94 a, b, 95, 97, 99 a, b, 01, 02]},
{\bf [Gell-Mann-Hartle 90 a, 90 b, 91, 93, 94]},
{\bf [Gell-Mann 94]} (Chap.~11),
{\bf [Halliwell 95]} (review),
{\bf [Di\'{o}si-Gisin-Halliwell-Percival 95]},
{\bf [Goldstein-Page 95]},
{\bf [Cohen-Hiley 95 b]} (in comparation
with standard QM and causal de Broglie-Bohm's interpretation),
{\bf [Cohen 95]} (CH in pre- and post-selected systems),
{\bf [Dowker-Kent 95, 96]},
{\bf [Rudolph 96]} (source of critical references),
{\bf [Kent 96 a, b, 97 a, 98 b, c, 00 b]} (CH approach allows
contrary inferences to be made from the same data),
{\bf [Isham-Linden-Savvidou-Schreckenberg 97]},
{\bf [Griffiths-Hartle 98]},
{\bf [Brun 98]},
{\bf [Schafir 98 a]} (Hardy's argument in the many-world and CH interpretations),
{\bf [Schafir 98 b]},
{\bf [Halliwell 98, 99 a, b, 00, 01, 03 a, b]},
{\bf [Dass-Joglekar 98]},
{\bf [Peruzzi-Rimini 98]}
(incompatible and contradictory retrodictions in the CH approach),
{\bf [Nistic\`{o} 99]}
(consistency conditions for probabilities of quantum histories),
{\bf [Rudolph 99]} (CH and POV measurements),
{\bf [Stapp 99 c]} (nonlocality, counterfactuals, and CH),
{\bf [Bassi-Ghirardi 99 a, 00 a, b]}
(decoherent histories description of reality cannot be considered satisfactory),
{\bf [Griffiths-Omn\`{e}s 99]},
{\bf [Griffiths 00 a, b]} (there is no conflict between CH and Bell, and
Kochen-Specker theorems),
{\bf [Dieks 00]} (CH and relativistic invariance
in the modal interpretation),
{\bf [Egusquiza-Muga 00]} (CH and quantum Zeno effect),
{\bf [Clarke 01 a, b]},
{\bf [Hiley-Maroney 00]} (CH and the Bohm approach),
{\bf [Sokolovski-Liu 01]},
{\bf [Raptis 01]},
{\bf [Nistic\`{o}-Beneduci 02]},
{\bf [Bar-Horwitz 02]},
{\bf [Brun 03]},
{\bf [Nistic\`{o} 03]}.


\subsection{Decoherence and environment induced superselection}


{\bf [Simonius 78]} (first explicit treatment of decoherence due to the
environment and the ensuing symmetry breaking and
``blocking'' of otherwise not stable states),
{\bf [Zurek 81 a, 82, 91 c, 93, 97, 98 a, 00 b, 01, 02, 03 b, c]},
{\bf [Joos-Zeh 85]},
{\bf [Zurek-Paz 93 a, b, c]},
{\bf [Wightman 95]} (superselection rules),
{\bf [Elby 94 a, b]},
{\bf [Giulini-Kiefer-Zeh 95]} (symmetries, superselection rules, and decoherence),
{\bf [Giulini-Joos-Kiefer-(+3) 96]} (review, almost
exhaustive source of references,
{\bf [Davidovich-Brune-Raimond-Haroche 96]},
{\bf [Brune-Hagley-Dreyer-(+5) 96]}
(experiment, see also
{\bf [Haroche-Raimond-Brune 97]}),
{\bf [Zeh 97, 98, 99]},
{\bf [Yam 97]} (non-technical review),
{\bf [Dugi\'c 98]} (necessary conditions for the occurrence of the
``environment-induced'' superselection rules),
{\bf [Habib-Shizume-Zurek 98]} (decoherence, chaos and the
correspondence principle),
{\bf [Kiefer-Joos 98]} (decoherence: Concepts and examples),
{\bf [Paz-Zurek 99]} (environment induced superselection of energy eigenstates),
{\bf [Giulini 99, 00]},
{\bf [Joos 99]},
{\bf [Bene-Borsanyi 00]} (decoherence within a single atom),
{\bf [Paz-Zurek 00]},
{\bf [Anastopoulos 00]} (frequently asked questions about decoherence),
{\bf [Kleckner-Ron 01]},
{\bf [Braun-Haake-Strunz 01]},
{\bf [Eisert-Plenio 02 b]} (quantum Brownian motion does not necessarily create entanglement
between the system and its environment;
the joint state of the system and its environment
may be separable at all times).


\subsection{Time symetric formalism, pre- and post-selected systems, ``weak''
measurements}


{\bf [Aharonov-Bergman-Lebowitz 64]},
{\bf [Albert-Aharonov-D'Amato 85]},
{\bf [Bub-Brown 86]} (comment:
{\bf [Albert-Aharonov-D'Amato 86]}),
{\bf [Vaidman 87, 96 d, 98 a, b, e, 99 a, c, d, 03 b]},
{\bf [Vaidman-Aharonov-Albert 87]},
{\bf [Aharonov-Albert-Casher-Vaidman 87]},
{\bf [Busch 88]},
{\bf [Aharonov-Albert-Vaidman 88]} (comments:
{\bf [Leggett 89]},
{\bf [Peres 89 a]}; reply:
{\bf [Aharonov-Vaidman 89]}),
{\bf [Golub-G\"{a}hler 89]},
{\bf [Ben Menahem 89]},
{\bf [Duck-Stevenson-Sudarshan 89]},
{\bf [Sharp-Shanks 89]},
{\bf [Aharonov-Vaidman 90, 91]},
{\bf [Knight-Vaidman 90]},
{\bf [Hu 90]},
{\bf [Zachar-Alter 91]},
{\bf [Sharp-Shanks 93]} (the rise and fall of time-symmetrized quantum
mechanics; counterfactual interpretation of the ABL rule leads to results that
disagree with standard QM; see also
{\bf [Cohen 95]}),
{\bf [Peres 94 a, 95 d]} (comment:
{\bf [Aharonov-Vaidman 95]}),
{\bf [Mermin 95 b]} (BKS theorem puts limits to the ``magic'' of retrodiction),
{\bf [Cohen 95]} (counterfactual use of the ABL rule),
{\bf [Cohen 98 a]},
{\bf [Reznik-Aharonov 95]},
{\bf [Herbut 96]},
{\bf [Miller 96]},
{\bf [Kastner 98 a, b, 99 a, b, c, 02, 03]},
{\bf [Lloyd-Slotine 99]},
{\bf [Metzger 00]},
{\bf [Mohrhoff 00 d]},
{\bf [Aharonov-Englert 01]},
{\bf [Englert-Aharonov 01]},
{\bf [Aharonov-Botero-Popescu-(+2) 01]} (Hardy's paradox and weak values),
{\bf [Atmanspacher-R\"{o}mer-Walach 02]}.


\subsection{The transactional interpretation}


{\bf [Cramer 80, 86, 88]}, {\bf [Kastner 04]}.


\subsection{The Ithaca interpretation: Correlations without correlata}


{\bf [Mermin 98 a, b, 99 a]},
{\bf [Cabello 99 a, c]}, {\bf [Jordan 99]},
{\bf [McCall 01]},
{\bf [Fuchs 03 a]} (Chaps.~18, 33),
{\bf [Plotnitsky 03]}.


\section{Composite systems, preparations, and measurements}


\subsection{States of composite systems}


\subsubsection{Schmidt decomposition}


{\bf [Schmidt 07 a, b]},
{\bf [von Neumann 32]} (Sec.~VI.~2),
{\bf [Furry 36 a, b]},
{\bf [Jauch 68]} (Sec. 11.\ 8),
{\bf [Ballentine 90 a]} (Sec.~8.~3),
{\bf [Albrecht 92]} (Secs.~II, III and Appendix),
{\bf [Barnett-Phoenix 92]},
{\bf [Albrecht 93]} (Sec.~II and Appendix),
{\bf [Peres 93 a]} (Chap.~5),
{\bf [Elby-Bub 94]} (uniqueness of triorthogonal decomposition of pure states),
{\bf [Albrecht 94]} (Appendix),
{\bf [Mann-Sanders-Munro 95]},
{\bf [Ekert-Knight 95]},
{\bf [Peres 95 c]} (Schmidt decomposition of higher order),
{\bf [Aravind 96]},
{\bf [Linden-Popescu 97]}
(invariances in Schmidt decomposition under local transformations),
{\bf [Ac\'{\i}n-Andrianov-Costa-(+3) 00]}
(Schmidt decomposition and classification of three-quantum-bit pure states),
{\bf [Terhal-Horodecki 00]} (Schmidt number for density matrices),
{\bf [Higuchi-Sudbery 00]},
{\bf [Carteret-Higuchi-Sudbery 00]}
(multipartite generalisation of the Schmidt decomposition),
{\bf [Pati 00 c]} (existence of the Schmidt decomposition for tripartite
system under certain condition).


\subsubsection{Entanglement measures}


{\bf [Barnett-Phoenix 91]} (``index of correlation''),
{\bf [Shimony 95]},
{\bf [Bennett-DiVincenzo-Smolin-Wootters 96]} (for a mixed state),
{\bf [Popescu-Rohrlich 97 a]},
{\bf [Schulman-Mozyrsky 97]},
{\bf [Vedral-Plenio-Rippin-Knight 97]},
{\bf [Vedral-Plenio-Jacobs-Knight 97]},
{\bf [Vedral-Plenio 98 a]},
{\bf [DiVincenzo-Fuchs-Mabuchi-(+3) 98]},
{\bf [Belavkin-Ohya 98]},
{\bf [Eisert-Plenio 99]} (a comparison of entanglement measures),
{\bf [Vidal 99 a]} (a measure of entanglement is defended which quantifies the
probability of success in an optimal local conversion from a single copy of a pure
state into another pure state),
{\bf [Parker-Bose-Plenio 00]} (entanglement quantification and purification
in continuous-variable systems),
{\bf [Virmani-Plenio 00]} (various entanglement measures do not give
the same ordering for all quantum states),
{\bf [Horodecki-Horodecki-Horodecki 00 a]}
(limits for entanglement measures),
{\bf [Henderson-Vedral 00]} (relative entropy of entanglement and irreversibility),
{\bf [Benatti-Narnhofer 00]} (on the additivity of entanglement formation),
{\bf [Rudolph 00 b]},
{\bf [Nielsen 00 c]} (one widely used method for defining measures of entanglement violates
that dimensionless quantities do not depend on the system of units being used),
{\bf [Brylinski 00]} (algebraic measures of entanglement),
{\bf [Wong-Christensen 00]},
{\bf [Vollbrecht-Werner 00]}
(entanglement measures under symmetry),
{\bf [Hwang-Ahn-Hwang-Lee 00]} (two mixed states such that their ordering
depends on the choice of entanglement measure
cannot be transformed, with unit efficiency,
to each other by any local operations),
{\bf [Audenaert-Verstraete-De Bie-De Moor 00]},
{\bf [Bennett-Popescu-Rohrlich-(+2) 01]}
(exact and asymptotic measures of multipartite pure state
entanglement),
{\bf [Majewski 01]},
{\bf [\.{Z}yczkowski-Bengtsson 01]}
(relativity of pure states entanglement),
{\bf [Abouraddy-Saleh-Sergienko-Teich 01]}
(any pure state of two qubits may be decomposed
into a superposition of a maximally entangled state and an orthogonal
factorizable one. Although there are many such decompositions, the weights of
the two superposed states are unique),
{\bf [Vedral-Kashefi 01]} (uniqueness of entanglement measure and thermodynamics),
{\bf [Vidal-Werner 02]} (a computable measure of entanglement),
{\bf [Eisert-Audenaert-Plenio 02]},
{\bf [Heydari-Bj\"{o}rk-S\'{a}nchez Soto 03]} (for two qubits),
{\bf [Heydari-Bj\"{o}rk 04 a, b]} (for two and $n$ qudits of different dimensions).


\subsubsection{Separability criteria}


{\bf [Peres 96 d, 97 a, 98 a]} (positive partial transposition (PPT) criterion),
{\bf [Horodecki-Horodecki-Horodecki 96 c]},
{\bf [Horodecki 97]},
{\bf [Busch-Lahti 97]},
{\bf [Sanpera-Tarrach-Vidal 97, 98]},
{\bf [Lewenstein-Sanpera 98]} (algorithm to obtain the best separable
approximation to the density matrix of a composite system.
This method gives rise to a condition of separability and to a measure
of entanglement),
{\bf [Cerf-Adami-Gingrich 97]},
{\bf [Aravind 97]},
{\bf [Majewski 97]},
{\bf [D\"{u}r-Cirac-Tarrach 99]}
(separability and distillability of multiparticle systems),
{\bf [Caves-Milburn 99]} (separability of various states for $N$ qutrits),
{\bf [Duan-Giedke-Cirac-Zoller 00 a]}
(inseparability criterion for continuous variable systems),
{\bf [Simon 00 b]} (Peres-Horodecki separability criterion for
continuous variable systems),
{\bf [D\"{u}r-Cirac 00 a]} (classification of multiqubit mixed states:
Separability and distillability properties),
{\bf [Wu-Chen-Zhang 00]} (a necessary and sufficient criterion for
multipartite separable states),
{\bf [Wang 00 b]},
{\bf [Karnas-Lewenstein 00]}
(optimal separable approximations),
{\bf [Terhal 01]} (review of the criteria for separability),
{\bf [Chen-Liang-Li-Huang 01 a]}
(necessary and sufficient condition of separability of any system),
{\bf [Eggeling-Vollbrecht-Wolf 01]}
({\bf [Chen-Liang-Li-Huang 01 a]} is a reformulation of the
problem rather than a practical criterion;
reply: {\bf [Chen-Liang-Li-Huang 01 b]}),
{\bf [Pittenger-Rubin 01]},
{\bf [Horodecki-Horodecki-Horodecki 01 b]}
(separability of $n$-particle mixed states),
{\bf [Giedke-Kraus-Lewenstein-Cirac 01]}
(separability criterion for all bipartite Gaussian states),
{\bf [Kummer 01]} (separability for two qubits),
{\bf [Albeverio-Fei-Goswami 01]}
(separability of rank two quantum states),
{\bf [Wu-Anandan 01]} (three necessary separability criteria for
bipartite mixed states),
{\bf [Rudolph 02]},
{\bf [Doherty-Parrilo-Spedalieri 02, 04]},
{\bf [Fei-Gao-Wang-(+2) 02]},
{\bf [Chen-Wu 02]} (generalized partial transposition
criterion for separability of multipartite quantum states).


\subsubsection{Multiparticle entanglement}


{\bf [Elby-Bub 94]} (uniqueness of triorthogonal decomposition of pure states),
{\bf [Linden-Popescu 97]},
{\bf [Clifton-Feldman-Redhead-Wilce 97]},
{\bf [Linden-Popescu 98 a]},
{\bf [Thapliyal 99]} (tripartite pure-state entanglement),
{\bf [Carteret-Linden-Popescu-Sudbery 99]},
{\bf [Fivel 99]},
{\bf [Sackett-Kielpinski-King-(+8) 00]}
(experimental four-particle entanglement),
{\bf [Carteret-Sudbery 00]} (three-qubit pure states are classified by
means of their stabilizers in the group of local unitary transformations),
{\bf [Ac\'{\i}n-Andrianov-Costa-(+3) 00]}
(Schmidt decomposition and classification of three-qubit pure states),
{\bf [Ac\'{\i}n-Andrianov-Jan\'{e}-Tarrach 00]}
(three-qubit pure-state canonical forms),
{\bf [van Loock-Braunstein 00 b]}
(multipartite entanglement for continuous variables),
{\bf [Wu-Zhang 01]} (multipartite pure-state entanglement and the
generalized GHZ states),
{\bf [Brun-Cohen 01]}
(parametrization and distillability of three-qubit entanglement).


\subsubsection{Entanglement swapping}


{\bf [Yurke-Stoler 92 a]} (entanglement from independent particle sources),
{\bf [Bennett-Brassard-Cr\'{e}peau-(+3) 93]} (teleportation),
{\bf [\.{Z}ukowski-Zeilinger-Horne-Ekert 93]} (event-ready-detectors),
{\bf [Bose-Vedral-Knight 98]} (multiparticle generalization of ES),
{\bf [Pan-Bouwmeester-Weinfurter-Zeilinger 98]} (experimental ES:
Entangling photons that have never interacted),
{\bf [Bose-Vedral-Knight 99]} (purification via ES),
{\bf [Peres 99 b]} (delayed choice for ES),
{\bf [Kok-Braunstein 99]} (with the current state of technology,
event-ready detections cannot be performed with the experiment of
{\bf [Pan-Bouwmeester-Weinfurter-Zeilinger 98]}),
{\bf [Polkinghorne-Ralph 99]} (continuous variable ES),
{\bf [\.{Z}ukowski-Kaszlikowski 00 a]} (ES with parametric down
conversion sources),
{\bf [Hardy-Song 00]} (ES chains for general pure states),
{\bf [Shi-Jiang-Guo 00 c]}
(optimal entanglement purification via ES),
{\bf [Bouda-Bu\v{z}zek 01]} (ES between multi-qudit systems),
{\bf [Fan 01 a, b]},
{\bf [Son-Kim-Lee-Ahn 01]}
(entanglement transfer from continuous variables to qubits),
{\bf [Karimipour-Bagherinezhad-Bahraminasab 02 a]}
(ES of generalized cat states),
{\bf [de Riedmatten-Marcikic-van Houwelingen-(+3) 04]}
(long distance ES with photons from separated sources).


\subsubsection{Entanglement distillation (concentration and purification)}


(Entanglement concentration: How to create, using only LOCC, maximally entangled
pure states from not maximally entangled ones.
Entanglement purification: How to distill pure maximally entangled states out of mixed entangled states.
Entanglement distillation means both concentration or purification)
{\bf [Bennett-Bernstein-Popescu-Schumacher 95]}
(concentrating partial entanglement by local operations),
{\bf [Bennett 95 b]},
{\bf [Bennett-Brassard-Popescu-(+3) 96]},
{\bf [Deutsch-Ekert-Jozsa-(+3) 96]},
{\bf [Murao-Plenio-Popescu-(+2) 98]}
(multiparticle EP protocols),
{\bf [Rains 97, 98 a, b]},
{\bf [Horodecki-Horodecki 97]} (positive maps
and limits for a class of protocols of entanglement distillation),
{\bf [Kent 98 a]} (entangled mixed states and local purification),
{\bf [Horodecki-Horodecki-Horodecki 98 b, c, 99 a]},
{\bf [Vedral-Plenio 98 a]} (entanglement measures and EP procedures),
{\bf [Cirac-Ekert-Macchiavello 99]} (optimal purification of single qubits),
{\bf [D\"{u}r-Briegel-Cirac-Zoller 99]} (quantum repeaters based on EP),
{\bf [Giedke-Briegel-Cirac-Zoller 99]} (lower bounds for attainable fidelity in EP),
{\bf [Opatrn\'{y}-Kurizki 99]} (optimization approach to entanglement distillation),
{\bf [Bose-Vedral-Knight 99]} (purification via entanglement swapping),
{\bf [D\"{u}r-Cirac-Tarrach 99]}
(separability and distillability of multiparticle systems),
{\bf [Parker-Bose-Plenio 00]} (entanglement quantification and EP
in continuous-variable systems),
{\bf [D\"{u}r-Cirac 00 a]} (classification of multiqubit mixed states:
Separability and distillability properties),
{\bf [Brun-Caves-Schack 00]} (EP of unknown quantum states),
{\bf [Ac\'{\i}n-Jan\'{e}-D\"{u}r-Vidal 00]}
(optimal distillation of a GHZ state),
{\bf [Cen-Wang 00]} (distilling a GHZ state from
an arbitrary pure state of three qubits),
{\bf [Lo-Popescu 01]}
(concentrating entanglement by local actions--beyond mean values),
{\bf [Kwiat-Barraza L\'{o}pez-Stefanov-Gisin 01]}
(experimental entanglement distillation),
{\bf [Shor-Smolin-Terhal 01]}
(evidence for nonadditivity of bipartite distillable entanglement),
{\bf [Pan-Gasparoni-Ursin-(+2) 03]}
(experimental entanglement purification of arbitrary unknown states, {\em Nature}).


\subsubsection{Disentanglement}


{\bf [Ghirardi-Rimini-Weber 87]} (D of wave functions),
{\bf [Chu 98]} (is it possible to disentangle an entangled
state?),
{\bf [Peres 98 b]} (D and computation),
{\bf [Mor 99]} (D while preserving all local properties),
{\bf [Bandyopadhyay-Kar-Roy 99]}
(D of pure bipartite quantum states by local cloning),
{\bf [Mor-Terno 99]} (sufficient conditions for a D),
{\bf [Hardy 99 b]} (D and teleportation),
{\bf [Ghosh-Bandyopadhyay-Roy-(+2) 00]}
(optimal universal D for two-qubit states),
{\bf [Bu\v{z}ek-Hillery 00]} (disentanglers),
{\bf [Zhou-Guo 00 a]} (D and inseparability correlation
in a two-qubit system).


\subsubsection{Bound entanglement}


{\bf [Horodecki 97]},
{\bf [Horodecki-Horodecki-Horodecki 98 b, 99 a]} (a BE state is an entangled
mixed state from which no pure entanglement can be distilled),
{\bf [Bennett-DiVincenzo-Mor-(+3) 99]}
(unextendible incomplete product bases provide a systematic way of
constructing BE states),
{\bf [Linden-Popescu 99]} (BE and teleportation),
{\bf [Bru\ss-Peres 00]} (construction of quantum states with BE),
{\bf [Shor-Smolin-Thapliyal 00]},
{\bf [Horodecki-Lewenstein 00]} (is BE for continuous variables a rare
phenomenon?),
{\bf [Smolin 01]} (four-party unlockable BE state,
$\rho_S={1 \over 4} \sum_{i=1}^4 |\phi_i\rangle\langle\phi_i| \otimes |\phi_i\rangle\langle\phi_i|$,
where $\phi_i$ are the Bell states),
{\bf [Murao-Vedral 01]}
(remote information concentration ---the reverse process to quantum telecloning---
using Smolin's BE state),
{\bf [Gruska-Imai 01]} (survey, p.~57),
{\bf [Werner-Wolf 01 a]} (BE Gaussian states),
{\bf [Sanpera-Bru\ss-Lewenstein 01]} (Schmidt number witnesses
and BE),
{\bf [Kaszlikowski-\.{Z}ukowski-Gnaci\'{n}ski 02]} (BE admits a local realistic description),
{\bf [Augusiak-Horodecki 04]} (some four-qubit bound entangled states can maximally violate two-setting Bell inequality;
this entanglement does not allow for secure key distillation, so neither entanglement
nor violation of Bell inequalities implies quantum security; it is also pointed out how that kind of bound entanglement
can be useful in reducing communication complexity),
{\bf [Bandyopadhyay-Ghosh-Roychowdhury 04]} (systematic method for generating bound entangled states in any bipartite system),
{\bf [Zhong 04]}.


\subsubsection{Entanglement as a catalyst}


{\bf [Jonathan-Plenio 99 b]}
(using only LOCC one cannot transform $|\phi_1\rangle$ into $|\phi_2\rangle$, but with the assistance of an
appropriate entangled state $|\psi \rangle$ one can transform $|\phi_1\rangle$ into $|\phi_2\rangle$
using LOCC in such a way that the state $|\psi \rangle$ can be returned back after the process:
$|\psi \rangle$ serves as a catalyst for otherwise impossible transformation),
{\bf [Barnum 99]} (quantum secure identification using entanglement and
catalysis),
{\bf [Jensen-Schack 00]}
(quantum authentication and key distribution using catalysis),
{\bf [Zhou-Guo 00 c]} (basic limitations for entanglement catalysis),
{\bf [Daftuar-Klimesh 01 a]} (mathematical structure of entanglement catalysis),
{\bf [Anspach 01]} (two-qubit catalysis in a four-state pure bipartite
system).


\subsection{State determination, state discrimination,
and measurement of arbitrary observables}


\subsubsection{State determination, quantum tomography}


{\bf [von Neumann 31]},
{\bf [Gale-Guth-Trammell 68]} (determination of the quantum state),
{\bf [Park-Margenau 68]},
{\bf [Band-Park 70, 71, 79]},
{\bf [Park-Band 71, 80, 92]},
{\bf [Brody-Meister 96]} (strategies for measuring identically prepared particles),
{\bf [Hradil 97]} (quantum state estimation),
{\bf [Raymer 97]} (quantum tomography, review),
{\bf [Freyberger-Bardroff-Leichtle-(+2) 97]} (quantum tomography, review),
{\bf [Chefles-Barnett 97 c]}
(entanglement and unambiguous discrimination between non-orthogonal states),
{\bf [Hradil-Summhammer-Rauch 98]} (quantum tomography as normalization of
incompatible observations).


\subsubsection{Generalized measurements, positive operator-valued measurements (POVMs),
discrimination between non-orthogonal states}


{\bf [Neumark 43, 54]} (representation of a POVM by a
projection-valued measure ---a von Neumman measure---
in an extended higher dimensional Hilbert space; see also {\bf [Nagy 90]}),
{\bf [Berberian 66]} (mathematical theory of POVMs),
{\bf [Jauch-Piron 67]} (POVMs are used in a generalized analysis of the localizability of quantum systems),
{\bf [Holevo 72, 73 c, 82]},
{\bf [Benioff 72 a, b, c]},
{\bf [Ludwig 76]} (POVMs),
{\bf [Davies-Lewis 70]} (analysis of quantum observables in terms of POVMs),
{\bf [Davies 76, 78]},
{\bf [Helstrom 76]},
{\bf [Ivanovic 81, 83, 93]},
{\bf [Ivanovic 87]} (how discriminate {\em unambiguously}
between a pair of non-orthogonal pure states ---the procedure has less than unit
probability of giving an answer at all---),
{\bf [Dieks 88]},
{\bf [Peres 88 b]} (IDP: Ivanovic-Dieks-Peres measurements),
{\bf [Peres 90 a]} (Neumark's theorem),
{\bf [Peres-Wootters 91]} (optimal detection of quantum information),
{\bf [Busch-Lahti-Mittelstaedt 91]},
{\bf [Bennett 92 a]} (B92 quantum key distribution scheme: Using two nonorthogonal states),
{\bf [Peres 93 a]} (Secs.~9.~5 and~9.~6),
{\bf [Busch-Grabowski-Lahti 95]},
{\bf [Ekert-Huttner-Palma-Peres 94]} (application of IDP to eavesdropping),
{\bf [Massar-Popescu 95]} (optimal measurement procedure for an {\em infinite}
number of identically prepared two-level systems: Construction of an infinite POVM),
{\bf [Jaeger-Shimony 95]} (extension of the IDP analysis to two states
with {\em a priori} unequal probabilities),
{\bf [Huttner-Muller-Gautier-(+2) 96]} (experimental unambiguous
discrimination of nonorthogonal states),
{\bf [Fuchs-Peres 96]},
{\bf [L\"{u}tkenhaus 96]} (POVMs and eavesdropping),
{\bf [Brandt-Myers 96, 99]} (optical POVM receiver for quantum cryptography),
{\bf [Grossman 96]} (optical POVM; see appendix A of {\bf [Brandt 99 b]}),
{\bf [Myers-Brandt 97]} (optical implementations of POVMs),
{\bf [Brandt-Myers-Lomonaco 97]} (POVMs and eavesdropping),
{\bf [Fuchs 97]} (nonorthogonal quantum states maximize classical information capacity),
{\bf [Biham-Boyer-Brassard-(+2) 98]} (POVMs and eavesdropping),
{\bf [Derka-Bu\v{z}ek-Ekert 98]} (explicit construction of
an optimal {\em finite} POVM for two-level systems),
{\bf [Latorre-Pascual-Tarrach 98]} (optimal, finite, {\em minimal} POVMs for the
cases of two to seven copies of a two-level system),
{\bf [Barnett-Chefles 98]} (application of the IDP to construct a
Hardy type argument for maximally entangled states),
{\bf [Chefles 98]} (unambiguous discrimination between multiple quantum states),
{\bf [Brandt 99 b]} (review),
{\bf [Nielsen-Chuang 00]},
{\bf [Chefles 00 b]}
(overview of the main approaches to quantum state discrimination),
{\bf [Sun-Hillery-Bergou 01]}
(optimum unambiguous discrimination between linearly independent
nonorthogonal quantum states),
{\bf [Sun-Bergou-Hillery 01]}
(optimum unambiguous discrimination
between subsets of non-orthogonal states),
{\bf [Peres-Terno 02]}.


\subsubsection{State preparation and measurement of arbitrary observables}


{\bf [Fano 57]},
{\bf [Fano-Racah 59]},
{\bf [Wichmann 63]} (density matrices arising from incomplete measurements),
{\bf [Newton-Young 68]} (measurability of the spin density matrix),
{\bf [Swift-Wright 80]} (generalized Stern-Gerlach experiments for the
measurement of arbitrary spin operators),
{\bf [Vaidman 88]} (measurability of nonlocal states),
{\bf [Ballentine 90 a]} (Secs. 8. 1-2, state preparation and determination),
{\bf [Phoenix-Barnett 93]},
{\bf [Popescu-Vaidman 94]} (causality constraints on nonlocal measurements),
{\bf [Reck-Zeilinger-Bernstein-Bertani 94 a, b]} (optical realization of
any discrete unitary operator),
{\bf [Cirac-Zoller 94]} (theoretical preparation of two
particle maximally entangled states and GHZ states with atoms),
{\bf [\.{Z}ukowski-Zeilinger-Horne 97]} (realization of any photon observable,
also for composite systems),
{\bf [Weinacht-Ahn-Bucksbaum 99]} (real experiment to control
the shape of an atomic electron's wavefunction),
{\bf [Hladk\'{y}-Drobn\'{y}-Bu\v{z}ek 00]} (synthesis of arbitrary unitary operators),
{\bf [Klose-Smith-Jessen 01]} (measuring the state of a large angular momentum).


\subsubsection{Stern-Gerlach experiment and its successors}


{\bf [Gerlach-Stern 21, 22 a, b]}, (SGI: Stern-Gerlach interferometer; a SG followed
by an inverted SG:)
{\bf [Bohm 51]} (Sec.~22.~11),
{\bf [Wigner 63]} (p.~10),
{\bf [Feynman-Leighton-Sands 65]} (Chap.~5);
{\bf [Swift-Wright 80]} (generalized SG experiments
for the measurement of arbitrary spin operators),
(coherence loss in a SGI:)
{\bf [Englert-Schwinger-Scully 88]},
{\bf [Schwinger-Scully-Englert 88]},
{\bf [Scully-Englert-Schwinger 89]};
{\bf [Summhammer-Badurek-Rauch-Kischko 82]}
(experimental ``SGI'' with polarized neutrons),
{\bf [Townsend 92]} (SG, Chap.~1, SGI, Chap.~2),
{\bf [Platt 92]} (modern analysis of a SG),
{\bf [Martens-de Muynck 93, 94]} (how to measure the spin of the electron),
{\bf [Batelaan-Gay-Schwendiman 97]} (SG for electrons),
{\bf [Venugopalan 97]}
(decoherence and Schr\"{o}dinger's-cat states in a SG experiment),
{\bf [Patil 98]} (SG according to QM),
{\bf [Hannout-Hoyt-Kryowonos-Widom 98]} (SG and quantum measurement theory),
{\bf [Shirokov 98]} (spin state determination using a SG),
{\bf [Garraway-Stenholm 99]} (observing the spin of a free electron),
{\bf [Amiet-Weigert 99 a, b]}
(reconstructing the density matrix of a spin $s$ through SG measurements),
{\bf [Reinisch 99]}
(the two output beams of a SG for spin 1/2 particles should not show interference
when appropriately superposed because an entanglement between
energy level and path selection occurs),
{\bf [Schonhammer 00]} (SG measurements with arbitrary spin),
{\bf [Gallup-Batelaan-Gay 01]} (analysis of the propagation of electrons through an
inhomogeneous magnetic field with axial symmetry:
A complete spin polarization of the beam is demonstrated,
in contrast with the semiclassical situation, where the spin splitting is blurred),
{\bf [Berman-Doolen-Hammel-Tsifrinovich 02]}
(static SG effect in magnetic force microscopy),
{\bf [Batelaan 02]}.


\subsubsection{Bell operator measurements}


{\bf [Michler-Mattle-Weinfurter-Zeilinger 96]}
(different interference effects produce three different results,
identifying two out of the four Bell states with the other two states giving the same
third measurement signal),
{\bf [L\"{u}tkenhaus-Calsamiglia-Suominen 99]} (a never-failing measurement of the
Bell operator of a two two-level bosonic system is impossible with beam splitters, phase
shifters, delay lines, electronically switched linear elements, photo-detectors, and
auxiliary bosons),
{\bf [Vaidman-Yoran 99]},
{\bf [Kwiat-Weinfurter 98]}
(``embedded'' Bell state analysis:
The four polarization-entangled Bell states can be discriminated if, simultaneously, there
is an additional entanglement in another degree of freedom ---time-energy or momentum---),
{\bf [Scully-Englert-Bednar 99]}
(two-photon scheme for detecting the four polarization-entangled Bell states
using atomic coherence),
{\bf [Paris-Plenio-Bose-(+2) 00]} (nonlinear interferometric setup
to unambiguously
discriminate the four polarization-entangled EPR-Bell photon pairs),
{\bf [DelRe-Crosignani-Di Porto 00]},
{\bf [Vitali-Fortunato-Tombesi 00]} (with a Kerr nonlinearity),
{\bf [Andersson-Barnett 00]}
(Bell-state analyzer with channeled atomic particles),
{\bf [Tomita 00, 01]} (solid state proposal),
{\bf [Calsamiglia-L\"{u}tkenhaus 01]}
(maximum efficiency of a linear-optical Bell-state analyzer),
{\bf [Kim-Kulik-Shih 01 a]}
(teleportation experiment of an unknown arbitrary polarization state
in which nonlinear interactions are used for the Bell state measurements
and in which all four Bell states can be
distinguished),
{\bf [Kim-Kulik-Shih 01 b]}
(teleportation experiment with a complete Bell state measurement
using nonlinear interactions),
{\bf [O'Brien-Pryde-White-(+2) 03]} (experimental all-optical quantum CNOT gate),
{\bf [Gasparoni-Pan-Walther-(+2) 04]}
(quantum CNOT with linear optics and previous entanglement),
{\bf [Zhao-Zhang-Chen-(+4) 04]}
(experimental demonstration of a non-destructive quantum CNOT for two independent photon-qubits).


\section{Quantum effects}


\subsubsection{Quantum Zeno and anti-Zeno effects}


{\bf [Misra-Sudarshan 77]},
{\bf [Chiu-Sudarshan-Misra 77]},
{\bf [Peres 80 a, b]},
{\bf [Joos 84]},
{\bf [Home-Whitaker 86, 92 b, 93]},
{\bf [Home-Whitaker 87]} (QZE
in the many-worlds interpretation),
{\bf [Bollinger-Itano-Heinzen-Wineland 89]},
{\bf [Itano-Heinzen-Bollinger-Wineland 90]},
{\bf [Peres-Ron 90]} (incomplete collapse and
partial QZE),
{\bf [Petrosky-Tasaki-Prigogine 90]},
{\bf [Inagaki-Namiki-Tajiri 92]} (possible observation of the
QZE by means of neutron spin-flipping),
{\bf [Whitaker 93]},
{\bf [Pascazio-Namiki-Badurek-Rauch 93]} (QZE with
neutron spin),
{\bf [Agarwal-Tewori 94]} (an optical realization),
{\bf [Fearn-Lamb 95]},
{\bf [Presilla-Onofrio-Tambini 96]},
{\bf [Kaulakys-Gontis 97]} (quantum anti-Zeno effect),
{\bf [Beige-Hegerfeldt 96, 97]},
{\bf [Beige-Hegerfeldt-Sondermann 97]},
{\bf [Alter-Yamamoto 97]} (QZE and the impossibility of
determining the quantum state of a single system),
{\bf [Kitano 97]}, {\bf [Schulman 98 b]},
{\bf [Home-Whitaker 98]},
{\bf [Whitaker 98 b]} (interaction-free measurement and the QZE),
{\bf [Gontis-Kaulakys 98]},
{\bf [Pati-Lawande 98]},
{\bf [\'{A}lvarez Estrada-S\'{a}nchez G\'{o}mez 98]} (QZE in relativistic quantum
field theory),
{\bf [Facchi-Pascazio 98]}
(quantum Zeno time of an excited state of the hydrogen atom),
{\bf [Wawer-Keller-Liebman-Mahler 98]} (QZE in composite systems),
{\bf [Mensky 99]},
{\bf [Lewenstein-Rzazewski 99]} (quantum anti-Zeno effect),
{\bf [Balachandran-Roy 00, 01]} (quantum anti-Zeno paradox),
{\bf [Egusquiza-Muga 00]} (consistent histories and QZE),
{\bf [Facchi-Gorini-Marmo-(+2) 00]},
{\bf [Kofman-Kurizki-Opatrn\'{y} 00]} (QZE and anti-Zeno effects
for photon polarization dephasing),
{\bf [Horodecki 01 a]},
{\bf [Wallace 01 a]} (computer model for the QZE),
{\bf [Kofman-Kurizki 01]},
{\bf [Militello-Messina-Napoli 01]} (QZE in trapped ions),
{\bf [Facchi-Nakazato-Pascazio 01]},
{\bf [Facchi-Pascazio 01]} (QZE: Pulsed versus
continuous measurement),
{\bf [Fischer-Guti\'{e}rrez Medina-Raizen 01]},
{\bf [Wunderlich-Balzer-Toschek 01]},
{\bf [Facchi 02]}.


\subsubsection{Reversible measurements, delayed choice and quantum erasure}


{\bf [Jaynes 80]},
{\bf [Wickes-Alley-Jakubowicz 81]} (DC experiment),
{\bf [Scully-Dr\"{u}hl 82]},
{\bf [Hillery-Scully 83]},
{\bf [Miller-Wheeler 84]} (DC),
{\bf [Scully-Englert-Schwinger 89]},
{\bf [Ou-Wang-Zou-Mandel 90]},
{\bf [Scully-Englert-Walther 91]} (QE, see also
{\bf [Scully-Zubairy 97]}, Chap.~20),
{\bf [Zou-Wang-Mandel 91]},
{\bf [Zajonc-Wang-Zou-Mandel 91]} (QE),
{\bf [Kwiat-Steinberg-Chiao 92]} (observation of QE),
{\bf [Ueda-Kitagawa 92]} (example of a ``logically reversible'' measurement),
{\bf [Royer 94]} (reversible measurement on a spin-$\frac{1}{2}$ particle),
{\bf [Englert-Scully-Walther 94]} (QE, review),
{\bf [Kwiat-Steinberg-Chiao 94]} (three QEs),
{\bf [Ingraham 94]} (criticism in
{\bf [Aharonov-Popescu-Vaidman 95]}),
{\bf [Herzog-Kwiat-Weinfurter-Zeilinger 95]} (complementarity and QE),
{\bf [Watson 95]},
{\bf [Cereceda 96 a]} (QE, review),
{\bf [Gerry 96 a]},
{\bf [Mohrhoff 96]} (the Englert-Scully-Walther's experiment is a
`DC' experiment only in a semantic sense),
{\bf [Griffiths 98 b]} (DC experiments in the consistent
histories interpretation),
{\bf [Scully-Walther 98]} (an operational analysis of QE and DC),
{\bf [D\"{u}rr-Nonn-Rempe 98 a, b]} (origin of quantum-mechanical
complementarity probed by a ``which way'' experiment in an atom interferometer,
see also {\bf [Knight 98]}, {\bf [Paul 98]}),
{\bf [Bj{\o}rk-Karlsson 98]} (complementarity and QE in welcher Weg experiments),
{\bf [Hackenbroich-Rosenow-Weidenm\"{u}ller 98]} (a mesoscopic QE),
{\bf [Mohan-Luo-Kr\"{o}ll-Mair 98]} (delayed single-photon self-interference),
{\bf [Luis-S\'{a}nchez Soto 98 b]} (quantum phase difference is used to analyze
which-path detectors in which the loss of interference predicted by
complementarity cannot be attributed to a momentum transfer),
{\bf [Kwiat-Schwindt-Englert 99]} (what does a quantum eraser really erase?),
{\bf [Englert-Scully-Walther 99]} (QE in double-slit interferometers
with which-way detectors, see {\bf [Mohrhoff 99]}),
{\bf [Garisto-Hardy 99]} (entanglement of projection and a new class of QE),
{\bf [Abranyos-Jakob-Bergou 99]}
(QE and the decoherence time of a measurement process),
{\bf [Schwindt-Kwiat-Englert 99]} (nonerasing QE),
{\bf [Kim-Yu-Kulik-(+2) 00]} (a DC QE),
{\bf [Tsegaye-Bj\"{o}rk-Atat\"{u}re-(+3) 00]}
(complementarity and QE with entangled-photon states),
{\bf [Souto Ribeiro-P\'{a}dua-Monken 00]} (QE by transverse indistinguishability),
{\bf [Elitzur-Dolev 01]} (nonlocal effects of partial measurements and QE),
{\bf [Walborn-Terra Cunha-P\'{a}dua-Monken 02]} (a double-slit QE),
{\bf [Kim-Ko-Kim 03 b]} (QE experiment with frequency-entangled photon pairs).


\subsubsection{Quantum nondemolition measurements}


{\bf [Braginsky-Vorontsov 74]},
{\bf [Braginsky-Vorontsov-Khalili 77]},
{\bf [Thorne-Drever-Caves-(+2) 78]},
{\bf [Unruh 78, 79]},
{\bf [Caves-Thorne-Drever-(+2) 80]},
{\bf [Braginsky-Vorontsov-Thorne 80]},
{\bf [Sanders-Milburn 89]} (complementarity
in a NDM),
{\bf [Holland-Walls-Zller 91]} (NDM of photon number by atomic-beam deflection),
{\bf [Braginsky-Khalili 92]} (book),
{\bf [Werner-Milburn 93]} (eavesdropping using NDM),
{\bf [Braginsky-Khalili 96]} ({\em Rev. Mod. Phys.}),
{\bf [Friberg 97]} ({\em Science}),
{\bf [Ozawa 98 a]} (nondemolition monitoring of universal quantum computers),
{\bf [Karlsson-Bj{\o}rk-Fosberg 98]}
(interaction-free and NDM),
{\bf [Fortunato-Tombesi-Schleich 98]} (non-demolition endoscopic tomography),
{\bf [Grangier-Levenson-Poizat 98]}
(quantum NDM in optics, review article in {\em Nature}),
{\bf [Ban 98]}
(information-theoretical properties of a sequence of NDM),
{\bf [Buchler-Lam-Ralph 99]} (NDM with
an electro-optic feed-forward amplifier),
{\bf [Watson 99 b]}.


\subsubsection{``Interaction-free'' measurements}


{\bf [Reninger 60]} (is the first one to speak of ``negative result measurements'')
{\bf [Dicke 81, 86]} (investigates the change in the wave function of an atom due
to the non-scattering of a photon),
{\bf [Hardy 92 c]} (comments:
{\bf [Pagonis 92]},
{\bf [Hardy 92 e]}),
{\bf [Elitzur-Vaidman 93 a, b]},
{\bf [Vaidman 94 b, c, 96 e, 00 b, 01 a, c]},
{\bf [Bennett 94]},
{\bf [Kwiat-Weinfurter-Herzog-(+2) 95 a, b]},
{\bf [Penrose 95]} (Secs. 5.\ 2, 5.\ 9),
{\bf [Krenn-Summhammer-Svozil 96]},
{\bf [Kwiat-Weinfurter-Zeilinger 96 a]} (review),
{\bf [Kwiat-Weinfurter-Zeilinger 96 b]},
{\bf [Paul-Pavi\v{c}i\'{c} 96, 97, 98]},
{\bf [Pavi\v{c}i\'{c} 96 a]},
{\bf [du Marchie van Voorthuysen 96]},
{\bf [Karlsson-Bj{\o}rk-Fosberg 97, 98]} (investigates the transition from IFM
of classical objects like bombs to IFM of quantum objects; in that case they are
called ``non-demolition measurements''),
{\bf [Hafner-Summhammer 97]} (experiment with neutron interferometry),
{\bf [Luis-S\'{a}nchez Soto 98 b, 99]},
{\bf [Kwiat 98]},
{\bf [White-Mitchell-Nairz-Kwiat 98]} (systems that allow us to obtain
images from photosensible objects, obtained by absorbing or scattering
fewer photons than were classically expected),
{\bf [Geszti 98]},
{\bf [Noh-Hong 98]},
{\bf [Whitaker 98 b]} (IFM and the quantum Zeno effect),
{\bf [White-Kwiat-James 99]},
{\bf [Mirell-Mirell 99]} (IFM
from continuous wave multi-beam interference),
{\bf [Krenn-Summhammer-Svozil 00]}
(interferometric information gain versus IFM),
{\bf [Simon-Platzman 00]} (fundamental limit on IFM),
{\bf [Potting-Lee-Schmitt-(+3) 00]}
(coherence and IFM),
{\bf [Mitchison-Jozsa 01]} (IFM can be regarded as counterfactual
computations),
{\bf [Horodecki 01 a]} (interaction-free interaction),
{\bf [Mitchison-Massar 01]}
(IF discrimination between semi-transparent
objects),
{\bf [S\'{a}nchez Soto 00]} (IFM and the quantum Zeno effect, review),
{\bf [Kent-Wallace 01]} (quantum interrogation and the safer X-ray),
{\bf [Zhou-Zhou-Feldman-Guo 01 a, b]} (``nondistortion quantum
interrogation''),
{\bf [Zhou-Zhou-Guo-Feldman 01]}
(high efficiency nondistortion quantum interrogation
of atoms in quantum superpositions),
{\bf [Methot-Wicker 01]} (IFM applied to quantum computation:
A new CNOT gate),
{\bf [DeWeerd 02]}.


\subsubsection{Other applications of entanglement}


{\bf [Wineland-Bollinger-Itano-(+2) 92]}
(reducing quantum noise in spectroscopy using correlated ions),
{\bf [Boto-Kok-Abrams-(+3) 00]}
(quantum interferometric optical lithography: Exploiting entanglement
to beat the diffraction limit),
{\bf [Kok-Boto-Abrams-(+3) 01]} (quantum lithography:
Using entanglement to beat the diffraction limit),
{\bf [Bj{\o}rk-S\'{a}nchez Soto-S{\o}derholm 01]}
(entangled-state lithography: Tailoring any pattern with a single state),
{\bf [D'Ariano-Lo Presti-Paris 01]}
(using entanglement improves the precision of quantum measurements).


\section{Quantum information}


\subsection{Quantum cryptography}


\subsubsection{General}


{\bf [Wiesner 83]} (first description of quantum coding, along
with two applications: making money that is in principle
impossible to counterfeit, and multiplexing two or three messages
in such a way that reading one destroys the others),
{\bf [Bennett 84]},
{\bf [Bennett-Brassard 84]} (BB84 scheme for quantum key distribution (QKD)),
{\bf [Deutsch 85 b, 89 b]},
{\bf [Ekert 91 a, b, 92]} (E91 scheme: QKD using EPR pairs),
{\bf [Bennett-Brassard-Mermin 92]} (E91 is in practice equivalent to BB84:
Entanglement is not essential for QKD, and Bell's inequality is
not essential for the detection of eavesdropping),
{\bf [Bennett-Brassard-Ekert 92]},
{\bf [Bennett 92 a]} (B92 scheme: Using two nonorthogonal states),
{\bf [Ekert-Rarity-Tapster-Palma 92]},
{\bf [Bennett-Wiesner 92]},
{\bf [Phoenix 93]},
{\bf [Muller-Breguet-Gisin 93]},
{\bf [Franson 93]},
(one-to-any QKD:) {\bf [Townsend-Smith 93]}, {\bf [Townsend-Blow 93]},
{\bf [Townsend-Phoenix-Blow-Barnett 94]};
(any-to-any QKD:)
{\bf [Barnett-Phoenix 94]},
{\bf [Phoenix-Barnett-Townsend-Blow 95]};
{\bf [Barnett-Loudon-Pegg-Phoenix 94]},
{\bf [Franson-Ilves 94 a]},
{\bf [Huttner-Peres 94]},
{\bf [Breguet-Muller-Gisin 94]},
{\bf [Ekert-Palma 94]},
{\bf [Townsend-Thompson 94]},
{\bf [Rarity-Owens-Tapster 94]},
{\bf [Huttner-Ekert 94]},
{\bf [Huttner-Imoto-Gisin-Mor 95]},
{\bf [Hughes-Alde-Dyer-(+3) 95]} (excellent review),
{\bf [Phoenix-Townsend 95]},
{\bf [Ardehali 96]} (QKD based on delayed choice),
{\bf [Koashi-Imoto 96]} (using two mixed states),
{\bf [Hughes 97]},
{\bf [Townsend 97 a, 99]}
(scheme for QKD for several users by means of an optical fibre network),
{\bf [Biham-Mor 97]} (security of QC against collective attacks),
{\bf [Klyshko 97]},
{\bf [Fuchs-Gisin-Griffiths-(+2) 97]},
{\bf [Brandt-Myers-Lomonaco 97]},
{\bf [Hughes 97 b]} (relevance of quantum computation for crytography),
{\bf [L\"{u}tkenhaus-Barnett 97]},
{\bf [Tittel-Ribordy-Gisin 98]} (review),
{\bf [Williams-Clearwater 98]} (book with a chapter on QC),
{\bf [Mayers-Yao 98]},
{\bf [Slutsky-Rao-Sun-Fainman 98]} (security against individual attacks),
{\bf [Lo-Chau 98 b, c, 99]},
{\bf [Ardehali-Chau-Lo 98]} (see also {\bf [Lo-Chau-Ardehale 00]}),
{\bf [Zeng 98 a]},
{\bf [Molotkov 98 c]} (QC based on photon ``frequency'' states),
{\bf [Lomonaco 98]} (review),
{\bf [Lo 98]} (excellent review on quantum {\em cryptology} ---the art of secure
communications using quantum means---, both from the perspective of
quantum {\em cryptography} ---the art of quantum code-making--- and
quantum {\em cryptoanalysis} ---the art of quantum code-breaking---),
{\bf [Ribordy-Gautier-Gisin-(+2) 98]}
(automated `plug \& play' QKD),
{\bf [Mitra 98]}, (free-space practical QC:)
{\bf [Hughes-Nordholt 99]},
{\bf [Hughes-Buttler-Kwiat-(+4) 99]},
{\bf [Hughes-Buttler-Kwiat-(+5) 99]};
{\bf [L\"{u}tkenhaus 99]} (estimates for practical QC),
{\bf [Guo-Shi 99]} (QC based on interaction-free measurements),
{\bf [Czachor 99]} (QC with polarizing interferometers),
{\bf [Kempe 99]} (multiparticle entanglement and its applications to QC),
{\bf [Sergienko-Atat\"{u}re-Walton(+3) 99]}
(QC using parametric down-conversion),
{\bf [Gisin-Wolf 99]}
(quantum versus classical key-agreement protocols),
{\bf [Zeng 00]} (QKD based on GHZ state),
{\bf [Zeng-Wang-Wang 00]}
(QKD relied on trusted information center),
{\bf [Zeng-Guo 00]} (authentication protocol),
{\bf [Ralph 00 a]} (continuous variable QC),
{\bf [Hillery 00]} (QC with squeezed states),
{\bf [Zeng-Zhang 00]} (identity verification in QKD),
{\bf [Bechmann Pasquinucci-Peres 00]} (QC with 3-state systems),
{\bf [Cabello 00 c]} (QKD without alternative measurements using
entanglement swapping, see also {\bf [Zhang-Li-Guo 01 a]},
{\bf [Cabello 01 b, e]}),
{\bf [Bouwmeester-Ekert-Zeilinger 00]} (book on quantum information),
{\bf [Brassard-L\"{u}tkenhaus-Mor-Sanders 00]}
(limitations on practical QC),
{\bf [Phoenix-Barnett-Chefles 00]} (three-state QC),
{\bf [Nambu-Tomita-Chiba Kohno-Nakamura 00]}
(QKD using two coherent states
of light and their superposition),
{\bf [Cabello 00 f]} (classical capacity of a quantum channel
can be saturated with secret information),
{\bf [Bub 01 a]} (QKD using a pre- and postselected states).
{\bf [Xue-Li-Guo 01, 02]}
(efficient QKD with nonmaximally entangled states),
{\bf [Guo-Li-Shi-(+2) 01]} (QKD
with orthogonal product states),
{\bf [Beige-Englert-Kurtsiefer-Weinfurter 01 a, b]},
{\bf [Gisin-Ribordy-Tittel-Zbinden 02]} (review),
{\bf [Long-Liu 02]}
(QKD in which each EPR pair carries 2 bits),
{\bf [Klarreich 02]} (commercial QKD: {\em ID Quantique},
{\em MagiQ Technologies}, {\em BBN Technologies}),
{\bf [Buttler-Torgerson-Lamoreaux 02]}
(new fiber-based quantum key distribution schemes).


\subsubsection{Proofs of security}


{\bf [Lo-Chau 99]},
{\bf [Mayers 96 b, 01, 02 a]},
{\bf [Biham-Boyer-Boykin-(+2) 00]},
{\bf [Shor-Preskill 00]} (simple proof of security of the BB84),
{\bf [Tamaki-Koashi-Imoto 03 a, b]} (B92),
{\bf [Hwang-Wang-Matsumoto-(+2) 03 a]} (Shor-Preskill type security-proof
without public announcement of bases),
{\bf [Tamaki-L\"{u}tkenhaus 04]} (B92
over a lossy and noisy channel),
{\bf [Christandl-Renner-Ekert 04]}
(A generic security proof for QKD which can be
applied to a number of different protocols. It relies on the fact that privacy amplification is equally secure when an
adversary's memory for data storage is quantum rather than classical),
{\bf [Hupkes 04]} (extension of the first proof for the unconditional security of
the BB84 by Mayers, without
the constraint that a perfect source is required).


\subsubsection{Quantum eavesdropping}


{\bf [Werner-Milburn 93]},
{\bf [Barnett-Huttner-Phoenix 93]} (eavesdropping strategies),
{\bf [Ekert-Huttner-Palma-Peres 94]},
{\bf [Huttner-Ekert 94]},
{\bf [Fuchs-Gisin-Griffiths-Niu-Peres 97]},
{\bf [Brandt-Myers-Lomonaco 97]},
{\bf [Gisin-Huttner 97]},
{\bf [Griffiths-Niu 97]},
{\bf [Cirac-Gisin 97]},
{\bf [L\"{u}tkenhaus-Barnett 97]},
{\bf [Bru\ss\, 98]},
{\bf [Niu-Griffiths 98 a]} (optimal copying of one qubit),
{\bf [Zeng-Wang 98]} (attacks on BB84 protocol),
{\bf [Zeng 98 b]} (id.),
{\bf [Bechmann Pasquinucci-Gisin 99]},
{\bf [Niu-Griffiths 99]}
(two qubit copying machine for economical quantum eavesdropping),
{\bf [Brandt 99 a]} (eavesdropping optimization
using a positive operator-valued measure),
{\bf [L\"{u}tkenhaus 00]}
(security against individual attacks for realistic QKD),
{\bf [Hwang-Ahn-Hwang 01 b]} (eavesdropper's optimal information
in variations of the BB84 in the coherent attacks).


\subsubsection{Quantum key distribution with orthogonal states}


{\bf [Goldenberg-Vaidman 95 a]} (QC with orthogonal states)
({\bf [Peres 96 f]},
{\bf [Goldenberg-Vaidman 96]}),
{\bf [Koashi-Imoto 97, 98 a]},
{\bf [Mor 98 a]} (if the individual systems go one after another, there
are cases in which even orthogonal states cannot be cloned),
{\bf [Cabello 00 f]} (QKD in the Holevo limit).


\subsubsection{Experiments}


{\bf [Bennett-Bessette-Brassard-(+2) 92]} (BB84 over $32$ cm through air),
{\bf [Townsend-Rarity-Tapster 93 a, b]},
{\bf [Muller-Breguet-Gisin 93]}
(B92 through more than 1 km of optical fibre),
{\bf [Townsend 94]},
{\bf [Muller-Zbinden-Gisin 95]} (B92 through $22.8$ km of optical fibre),
{\bf [Marand-Townsend 95]} (with phase-encoded photons
over $30$ km),
{\bf [Franson-Jacobs 95]},
{\bf [Hughes-Luther-Morgan-(+2) 96]} (with phase-encoded photons),
{\bf [Muller-Zbinden-Gisin 96]} (real experiment through $26$ km of optical fibre),
{\bf [Zbinden 98]} (review of different experimental setups based on optical fibres),
(`plug and play' QKD:) {\bf [Muller-Herzog-Huttner-(+3) 97]},
{\bf [Ribordy-Gautier-Gisin-(+2) 98]};
(quantum key transmision through 1 km of atmosphere:)
{\bf [Buttler-Hughes-Kwiat-(+6) 98]},
{\bf [Buttler-Hughes-Kwiat-(+5) 98]},
{\bf [Hughes-Buttler-Kwiat-(+4) 99]},
{\bf [Hughes-Nordholt 99]} (B92 at a rate of $5$ kHz and over $0.5$ km in broad
daylight and free space, with polarized photons),
{\bf [Gisin-Brendel-Gautier-(+5) 99]},
{\bf [M\'{e}rolla-Mazurenko-Goedgebuer-(+3) 99]}
(quantum cryptographic device using single-photon phase modulation),
{\bf [Hughes-Morgan-Peterson 00]} ($48$ km),
{\bf [Buttler-Hughes-Lamoreaux-(+3) 00]}
(daylight quantum key distribution over $1.6$ km),
{\bf [Jennewein-Simon-Weihs-(+2) 00]}
(E91 with individual photons entangled in polarization),
{\bf [Naik-Peterson-White-(+2) 00]}
(E91 with individual photons entangled in polarization from parametric down-conversion),
{\bf [Tittel-Brendel-Zbinden-Gisin 00]}
(with individual photons in energy-time Bell states),
{\bf [Ribordy-Brendel-Gautier-(+2) 01]}
(long-distance entanglement-based QKD),
{\bf [Stucki-Gisin-Guinnard-(+2) 02]}
(over $67$ km with a plug \& play system),
{\bf [Hughes-Nordholt-Derkacs-Peterson 02]}
(over $10$ km in daylight and at night),
{\bf [Kurtsiefer-Zarda-Halder-(+4) 02]}
(over a free-space path of $23.4$ km between the summit of
Zugspitze and Karwendelspitze, {\em Nature}),
{\bf [Waks-Inoue-Santori-(+4) 02]} (quantum cryptography with a photon turnstile,
{\em Nature}).


\subsubsection{Commercial quantum cryptography}


{\bf [ID Quantique 01]},
{\bf [MagiQ Technologies 02]},
{\bf [QinetiQ 02]},
{\bf [Telcordia Technologies 02]},
{\bf [BBN Technologies 02]}.


\subsection{Cloning and deleting quantum states}


{\bf [Wootters-Zurek 82]} (due to the linearity of QM, there is no
{\em universal quantum cloner} ---a device for producing two copies from an
arbitrary initial state--- with fidelity 1),
{\bf [Dieks 82]},
{\bf [Herbert 82]} (superluminal communication would be possible with a
perfect quantum cloner),
{\bf [Barnum-Caves-Fuchs-(+2) 96]}
(noncomuting mixed states cannot be broadcast),
{\bf [Bu\v{z}ek-Hillery 96]} (it is possible to build a cloner which
produces two {\em approximate} copies of an arbitrary
initial state, the maximum fidelity for that process is $\frac{5}{6}$),
{\bf [Hillery-Bu\v{z}ek 97]} (fundamental inequalities in quantum copying),
{\bf [Gisin-Massar 97]}
(optimal cloner which makes $m$ copies from $n$ copies of the original state),
{\bf [Bru\ss-DiVincenzo-Ekert-(+2) 98]}
(the maximum fidelity of a universal quantum cloner is $\frac{5}{6}$),
{\bf [Moussa 97 b]} (proposal for a cloner based on QED),
{\bf [Bru\ss-Ekert-Macchiavello 98]},
{\bf [Gisin 98]} ($\frac{5}{6}$ is the maximum fidelity of a
universal quantum cloner, supposing that it cannot
serve for superluminial transmission of information),
{\bf [Mor 98 a]} (if the individual systems go one after another, there
are cases in which even orthogonal states cannot be cloned),
{\bf [Koashi-Imoto 98 a]} (necessary and sufficient condition for two pure
entangled states to be clonable by sequential access to both systems),
{\bf [Westmoreland-Schumacher 98]},
{\bf [Mashkevich 98 b, d]},
{\bf [van Enk 98]} (no-cloning and superluminal signaling),
{\bf [Cerf 98 b]} (generalization of the cloner proposed by Hillery and
Bu\v{z}ek in case that the two copies are not identical; the inequalities
that govern the fidelity of this process),
{\bf [Werner 98]} (optimal cloning of pure states),
{\bf [Zanardi 98 b]} (cloning in $d$ dimensions),
{\bf [Cerf 98 c]} (asymmetric cloning),
{\bf [Duan-Guo 98 c, f]} (probabilistic cloning),
{\bf [Keyl-Werner 98]} (judging single clones),
{\bf [Bu\v{z}ek-Hillery 98 a, b]}
(universal optimal cloning of qubits and quantum registers),
{\bf [Bu\v{z}ek-Hillery-Bednik 98]},
{\bf [Bu\v{z}ek-Hillery-Knight 98]},
{\bf [Chefles-Barnett 98 a, b]},
{\bf [Masiak-Knight 98]}
(copying of entangled states and the degradation of correlations),
{\bf [Niu-Griffiths 98]}
(two qubit copying machine for economical quantum eavesdropping),
{\bf [Bandyopadhyay-Kar 99]},
{\bf [Ghosh-Kar-Roy 99]} (optimal cloning),
{\bf [Hardy-Song 99]} (no signalling and probabilistic quantum
cloning),
{\bf [Murao-Jonathan-Plenio-Vedral 99]}
(quantum telecloning: a process combining quantum teleportation and
optimal quantum cloning from one input to $M$ outputs),
{\bf [D\"{u}r-Cirac 00 b]} (telecloning from $N$ inputs to $M$ outputs),
{\bf [Albeverio-Fei 00 a]} (on the optimal cloning of an $N$-level quantum
system),
{\bf [Macchiavello 00 b]}
(bounds on the efficiency of cloning for two-state quantum systems),
{\bf [Zhang-Li-Wang-Guo 00]} (probabilistic quantum cloning via GHZ states),
{\bf [Pati 00 a]}
(assisted cloning and orthogonal complementing of an unknown state),
{\bf [Pati-Braunstein 00 a]} (impossibility of deleting an unknown quantum
state: If two photons are in the same initial polarization state, there is
no mechanism that produces one photon in the same initial state and another
in some standard polarization state),
{\bf [Simon-Weihs-Zeilinger 00 a, b]}
(optimal quantum cloning via stimulated emission),
{\bf [Cerf 00 a]} (Pauli cloning),
{\bf [Pati 00 b]},
{\bf [Zhang-Li-Guo 00 b]} (cloning for $n$-state system),
{\bf [Cerf-Ipe-Rottenberg 00]}
(cloning of continuous variables),
{\bf [Cerf 00 b]} (asymmetric quantum cloning in any dimension),
{\bf [Kwek-Oh-Wang-Yeo 00]} (Bu\v{z}ek-Hillery cloning revisited
using the bures metric and trace norm),
{\bf [Galv\~{a}o-Hardy 00 b]}
(cloning and quantum computation),
{\bf [Kempe-Simon-Weihs 00]} (optimal photon cloning),
{\bf [Cerf-Iblisdir 00]} (optimal $N$-to-$M$ cloning of
conjugate quantum variables),
{\bf [Fan-Matsumoto-Wadati 01 b]}
(cloning of $d$-level systems),
{\bf [Roy-Sen-Sen 01]}
(is it possible to clone using an arbitrary blank state?),
{\bf [Bru\ss-Macchiavello 01 a]}
(optimal cloning for two pairs of orthogonal states),
{\bf [Fan-Matsumoto-Wang-(+2) 01]}
(a universal cloner allowing the input to be arbitrary states in symmetric subspace),
{\bf [Fan-Wang-Matsumoto 02]}
(a quantum-copying machine for equatorial qubits),
{\bf [Rastegin 01 a, b, 03 a]} (some bounds for quantum copying),
{\bf [Cerf-Durt-Gisin 02]} (cloning a qutrit),
{\bf [Segre 02]} (no cloning theorem versus the second law of thermodynamics),
{\bf [Feng-Zhang-Sun-Ying 02]}
(universal and original-preserving quantum copying is impossible),
{\bf [Qiu 02 c]} (non-optimal universal quantum deleting machine),
{\bf [Ying 02 a, b]},
{\bf [Han-Zhang-Guo 02 b]} (bounds for state-dependent quantum cloning),
{\bf [Rastegin 03 b]} (limits of state-dependent cloning of mixed states),
{\bf [Pati-Braunstein 03 b]} (deletion of unknown quantum state against a copy
can lead to superluminal signalling, but erasure of unknown quantum state does
not imply faster than light signalling),
{\bf [Horodecki-Horodecki-Sen De-Sen 03]}
(no-deleting and no-cloning principles as consequences of conservation of
quantum information),
{\bf [Horodecki-Sen De-Sen 03 b]} (orthogonal pure states can be cloned and deleted.
However, for orthogonal mixed states deletion is forbidden and cloning necessarily produces an
irreversibility, in the form of leakage of information into the environment),
{\bf [Peres 02]} (why wasn't the no-cloning theorem discovered fifty years earlier?).


\subsection{Quantum bit commitment}


{\bf [Brassard-Cr\'{e}peau-Jozsa-Langlois 93]},
{\bf [Mayers 97]}
(unconditionally secure QBC is impossible),
{\bf [Brassard-Cr\'{e}peau-Mayers-Salvail 97]}
(review on the impossibility of QBC),
{\bf [Kent 97 b, 99 a, c, d, 00 a, 01 a, b]},
{\bf [Lo-Chau 96, 97, 98 a, d]},
{\bf [Brassard-Cr\'{e}peau-Mayers-Salvail 98]}
(defeating classical bit commitments with a quantum computer),
{\bf [Hardy-Kent 99]} (cheat sensitive QBC),
{\bf [Molotkov-Nazin 99 c]}
(unconditionally secure relativistic QBC),
{\bf [Bub 00 b]},
{\bf [Yuen 00 b, c, 01 a, c]} (unconditionally secure QBC is possible),
{\bf [Nambu-Chiba Kohno 00]}
(information-theoretic description of no-go theorem of a QBC),
{\bf [Molotkov-Nazin 01 b]} (relativistic QBC)
{\bf [Molotkov-Nazin 01 c]} (QBC in a noisy channel),
{\bf [Li-Guo 01]},
{\bf [Spekkens-Rudolph 01 a]} (degrees of concealment and bindingness in
QBC protocols),
{\bf [Spekkens-Rudolph 01 b]} (optimization of coherent attacks in
generalizations of the BB84 QBC protocol),
{\bf [Cheung 01]} (QBC can be unconditionally secure),
{\bf [Srikanth 01 f]},
{\bf [Bub 01 b]} (review),
{\bf [Shimizu-Imoto 02 a]}
(fault-tolerant simple QBC unbreakable by individual attacks),
{\bf [Nayak-Shor 03]} (bit-commitment-based quantum coin flipping),
{\bf [Srikanth 03]}.


\subsection{Secret sharing and quantum secret sharing}


{\bf [\.{Z}ukowski-Zeilinger-Horne-Weinfurter 98]},
{\bf [Hillery-Bu\v{z}ek-Berthiaume 99]}
(one- to two-party SS and QSS using three-particle
entanglement, and one- to three-party SS using four-particle entanglement),
{\bf [Karlsson-Koashi-Imoto 99]} (one- to two-party SS using two-particle entanglement,
and QSS using three-particle entanglement),
{\bf [Cleve-Gottesman-Lo 99]} (in a ($k$, $n$) threshold scheme, a
secret quantum state is divided into $n$ shares such that any $k$
shares can be used to reconstruct the secret, but any set of $k-1$
shares contains no information about the secret. The ``no-cloning
theorem'' requires that $n<2k$),
{\bf [Tittel-Zbinden-Gisin 99]} (QSS using pseudo-GHZ states),
{\bf [Smith 00]} (QSS for general access structures),
{\bf [Bandyopadhyay 00 b]},
{\bf [Gottesman 00 a]} (theory of QSS),
{\bf [Karimipour-Bagherinezhad-Bahraminasab 02 b]} (SS).


\subsection{Quantum authentication}


{\bf [Ljunggren-Bourennane-Karlsson 00]} (authority-based user
authentication in QKD), {\bf [Zeng-Guo 00]} (QA protocol), {\bf
[Zhang-Li-Guo 00 c]} (QA using entangled state), {\bf
[Jensen-Schack 00]} (QA and QKD using catalysis), {\bf
[Shi-Li-Liu-(+2) 01]} (QKD and QA based on entangled state), {\bf
[Guo-Li-Guo 01]} (non-demolition measurement of nonlocal variables
and its application in QA), {\bf [Curty-Santos 01 a, c]}, {\bf
[Barnum 01]} (authentication codes), {\bf
[Curty-Santos-P\'{e}rez-Garc\'{\i}a Fern\'{a}ndez 02]},
{\bf [Kuhn 03]} (QA using entanglement and
symmetric cryptography),
{\bf [Curty 04]}.


\subsection{Teleportation of quantum states}


\subsubsection{General}


{\bf [Bennett-Brassard-Cr\'{e}peau-(+3) 93]},
{\bf [Sudbery 93]} (News and views, {\em Nature}),
{\bf [Deutsch-Ekert 93]},
{\bf [Popescu 94]},
{\bf [Vaidman 94 a]},
{\bf [Davidovich-Zagury-Brune-(+2) 94]},
{\bf [Cirac-Parkins 94]},
{\bf [Braunstein-Mann 95]},
{\bf [Vaidman 95 c]},
{\bf [Popescu 95]},
{\bf [Gisin 96 b]},
{\bf [Bennett-Brassard-Popescu-(+3) 96]},
{\bf [Horodecki-Horodecki-Horodecki 96 b]},
{\bf [Horodecki-Horodecki 96 b]},
{\bf [Taubes 96]},
{\bf [Braunstein 96 a]},
{\bf [Home 97]} (Sec.~4.~4),
{\bf [Moussa 97 a]},
{\bf [Nielsen-Caves 97]}
(reversible quantum operations and their application to T),
{\bf [Zheng-Guo 97 a, b]},
{\bf [Watson 97 b]},
{\bf [Anonymous 97]},
{\bf [Williams-Clearwater 98]} (book with a chapter on T),
{\bf [Brassard-Braunstein-Cleve 98]} (T as a quantum computation),
{\bf [Braunstein-Kimble 98 a]} (T of continuous quantum variables),
{\bf [Collins 98]} ({\em Phys. Today}),
{\bf [Pan-Bouwmeester-Weinfurter-Zeilinger 98]},
{\bf [Garc\'{\i}a Alcaine 98 a]} (review),
{\bf [Klyshko 98 c]} (on the realization and meaning of T),
{\bf [Molotkov 98 a]} (T of a single-photon wave packet),
{\bf [de Almeida-Maia-Villas B\^{o}as-Moussa 98]}
(T of atomic states with cavities),
{\bf [Ralph-Lam 98]} (T with bright squeezed light),
{\bf [Horodecki-Horodecki-Horodecki 99 c]}
(general T channel, singlet fraction and quasi-distillation),
{\bf [Vaidman 98 c]} (review of all proposals and experiments, and
T in the many-worlds interpretation),
{\bf [Zubairy 98]} (T of a field state),
{\bf [Nielsen-Knill-Laflamme 98]}
(complete quantum T using nuclear magnetic resonance),
{\bf [Stenholm-Bardroff 98]} (T of $N$-dimensional states),
{\bf [Karlsson-Bourennane 98]} (T using three-particle entanglement),
{\bf [Plenio-Vedral 98]} (T, entanglement and thermodynamics),
{\bf [Ralph 98]} (all optical quantum T),
{\bf [Maierle-Lidar-Harris 98]} (T of superpositions of chirial amplitudes),
{\bf [Vaidman-Yoran 99]} (methods for reliable T),
{\bf [L\"{u}tkenhaus-Calsamiglia-Suominen 99]} (a never-failing measurement of the
Bell operator in a two two-level bosonic system is impossible with beam splitters, phase
shifters, delay lines, electronically switched linear elements, photo-detectors, and
auxiliary bosons),
{\bf [Linden-Popescu 99]} (bound entanglement and T),
{\bf [Molotkov-Nazin 99 b]} (on T of continuous variables),
{\bf [Tan 99]}
(confirming entanglement in continuous variable quantum T),
{\bf [Villas B\^{o}as-de Almeida-Moussa 99]} (T of a zero- and one-photon
running-wave state by projection synthesis),
{\bf [van Enk 99]} (discrete formulation of T of continuous variables),
{\bf [Milburn-Braunstein 99]} (T with squeezed vacuum states),
{\bf [Ryff 99]},
{\bf [Koniorczyk-Janszky-Kis 99]} (photon number T),
{\bf [Bose-Knight-Plenio-Vedral 99]}
(proposal for T of an atomic state via cavity decay),
{\bf [Ralph-Lam-Polkinghorne 99]} (characterizing T in optics),
{\bf [Maroney-Hiley 99]} (T understood through the Bohm
interpretation), {\bf [Hardy 99 b]} (a toy local theory in which
cloning is not possible but T is),
{\bf [Parkins-Kimble 99]} (T of the wave function of a massive
particle),
{\bf [Marinatto-Weber 00 b]} (which kind of two-particle states can be teleported
through a three-particle quantum channel?),
{\bf [Bouwmeester-Pan-Weinfurter-Zeilinger 00]}
(high-fidelity T of independent qubits),
{\bf [Zeilinger 00 c]},
{\bf [van Loock-Braunstein 00 a]} (T of continuous-variable entanglement),
{\bf [Banaszek 00]} (optimal T with an arbitrary pure state),
{\bf [Opatrn\'{y}-Kurizki-Welsch 00]}
(improvement on T of continuous variables
by photon subtraction via conditional measurement),
{\bf [Horoshko-Kilin 00]} (T using quantum nondemolition technique),
{\bf [Murao-Plenio-Vedral 00]} (T of quantum information to $N$ particles),
{\bf [Li-Li-Guo 00]} (probabilistic T and entanglement matching),
{\bf [Cerf-Gisin-Massar 00]} (classical T of a qubit),
{\bf [DelRe-Crosignani-Di Porto 00]} (scheme for total T),
{\bf [Kok-Braunstein 00 a]} (postselected versus nonpostselected T
using parametric down-conversion),
{\bf [Bose-Vedral 00]} (mixedness and T),
{\bf [van Loock-Braunstein 00 b]}
(multipartite entanglement for continuous variables: A quantum T network),
{\bf [Braunstein-D'Ariano-Milburn-Sacchi 00]}
(universal T with a twist),
{\bf [Bouwmeester-Ekert-Zeilinger 00]} (book on quantum information),
{\bf [D\"{u}r-Cirac 00 b]} (multiparty T),
{\bf [Henderson-Hardy-Vedral 00]} (two-state T),
{\bf [Motoyoshi 00]} (T without Bell measurements),
{\bf [Vitali-Fortunato-Tombesi 00]}
(complete T with a Kerr nonlinearity),
{\bf [Galv\~{a}o-Hardy 00 a]}
(building multiparticle states with T),
{\bf [Banaszek 00 a]} (optimal T with an arbitrary pure state),
{\bf [Lee-Kim 00]} (entanglement T via Werner states),
{\bf [Lee-Kim-Jeong 00]} (transfer of nonclassical
features in T via a mixed quantum channel),
{\bf [\.{Z}ukowski 00 b]} (Bell's theorem for the nonclassical part of
the T process),
{\bf [Clausen-Opatrn\'{y}-Welsch 00]}
(conditional T using optical squeezers),
{\bf [Grangier-Grosshans 00 a]} (T criteria for continuous variables),
{\bf [Koniorczyk-Kis-Janszky 00]},
{\bf [Gorbachev-Zhiliba-Trubilko-Yakovleva 00]}
(T of entangled states and dense coding
using a multiparticle quantum channel),
{\bf [van Loock-Braunstein 00 d]}
(telecloning and multiuser quantum channels
for continuous variables),
{\bf [Hao-Li-Guo 00]} (probabilistic dense coding and T),
{\bf [Zhou-Hou-Zhang 01]}
(T of $S$-level pure states by two-level EPR states),
{\bf [Trump-Bru\ss-Lewenstein 01]}
(realistic T with linear optical elements),
{\bf [Werner 01 a]} (T and dense coding schemes),
{\bf [Ide-Hofmann-Kobayashi-Furusawa 01]}
(continuous variable T of single photon states),
{\bf [Wang-Feng-Gong-Xu 01]} (atomic-state T by using a quantum switch),
{\bf [Braunstein-Fuchs-Kimble-van Loock 01]}
(quantum versus classical domains for T with continuous variables),
{\bf [Bowen-Bose 01]} (T as a depolarizing quantum channel),
{\bf [Shi-Tomita 02]} (T using a W state),
{\bf [Agrawal-Pati 02]} (probabilistic T),
{\bf [Yeo 03 a]} (T using a three-qubit W state),
{\bf [Peres 03 b]} (it includes a narrative of how Peres remembers that T was conceived).


\subsubsection{Experiments}


{\bf [Boschi-Branca-De Martini-(+2) 98]} (first experiment),
{\bf [Bouwmeester-Pan-Mattle-(+3) 97]} (first published experiment),
{\bf [Furusawa-S{\o}rensen-Braunstein-(+3) 98]},
(first T of a state that describes a light field, see also
{\bf [Caves 98 a]}),
{\bf [Sudbery 97]} (News and views, {\em Nature}), (Comment:
{\bf [Braunstein-Kimble 98 b]}, Reply:
{\bf [Bouwmeester-Pan-Daniell-(+3) 98]}), (discussion on which
group did the first experiment:)
{\bf [De Martini 98 a]},
{\bf [Zeilinger 98 a]};
{\bf [Koenig 00]} (on Vienna group's experiments on T),
{\bf [Kim-Kulik-Shih 01 a]}
(T experiment of an unknown arbitrary polarization state
in which nonlinear interactions are used for the Bell state measurements
and in which all four Bell states can be
distinguished),
{\bf [Pan-Daniell-Gasparoni-(+2) 01]} (four-photon entanglement
and high-fidelity T),
{\bf [Lombardi-Sciarrino-Popescu-De Martini 02]}
(T of a vacuum--one-photon qubit),
{\bf [Kim-Kulik-Shih 02]} (proposal for an
experiment for T with a complete Bell state measurements using nonlinear interactions),
{\bf [Marcikic-de Riedmatten-Tittel-(+2) 03]}
(experimental
probabilistic quantum teleportation: Qubits
carried by photons of 1.3 mm wavelength are teleported onto
photons of 1.55 mm wavelength from one laboratory to another,
separated by 55 m but connected by 2 km of standard telecommunications
fibre, {\em Nature}),
{\bf [Pan-Gasparoni-Aspelmeyer-(+2) 03]} ({\em Nature}).


\subsection{Telecloning}


{\bf [Murao-Jonathan-Plenio-Vedral 99]}
(quantum telecloning: a process combining quantum teleportation and
optimal quantum cloning from one input to $M$ outputs),
{\bf [D\"{u}r-Cirac 00 b]} (telecloning from $N$ inputs to $M$ outputs),
{\bf [van Loock-Braunstein 00 d]}
(telecloning and multiuser quantum channels
for continuous variables),
{\bf [van Loock-Braunstein 01]}
(telecloning of continuous quantum variables),
{\bf [Ghiu 03]} (asymmetric quantum telecloning of $d$-level systems),
{\bf [Ricci-Sciarrino-Sias-De Martini 03 a, b]} (experimental results),
{\bf [Zhao-Chen-Zhang-(+3) 04]} (experimental demonstration of five-photon entanglement and
open-destination teleportation),
{\bf [Pirandola 04]}
(the standard, non cooperative, telecloning protocol can be outperformed by a cooperative one).


\subsection{Dense coding}


{\bf [Bennett-Wiesner 92]} (encoding $n^2$ values in a $n$-level system),
{\bf [Deutsch-Ekert 93]} (popular review),
{\bf [Barnett-London-Pegg-Phoenix 94]} (communication using quantum states),
{\bf [Barenco-Ekert 95]} (the Bennett-Wiesner scheme for DC based
on the discrimination of the four Bell states is the optimal one,
i.e. it maximizes the mutual information, even if the initial
state is not a Bell state but a non-maximally entangled state),
{\bf [Mattle-Weinfurter-Kwiat-Zeilinger 96]}
(experimeltal transmission of a ``trit'' using a two-level quantum system,
with photons entangled in polarization),
{\bf [Huttner 96]} (popular review of the MWKZ experiment),
{\bf [Cerf-Adami 96]}
(interpretation of the DC in terms of negative information),
{\bf [Bose-Vedral-Knight 99]}
(Sec.~V.~B, generalization with several particles and several transmitters),
{\bf [Bose-Plenio-Vedral 98]} (with mixed states),
{\bf [Shimizu-Imoto-Mukai 99]} (DC in photonic quantum communication
with enhanced information capacity),
{\bf [Ban 99 c]} (DC via two-mode squeezed-vacuum state),
{\bf [Bose-Plenio-Vedral 00]} (mixed state DC and its
relation to entanglement measures),
{\bf [Fang-Zhu-Feng-Mao-Du 00]} (experimental implementation of DC using
nuclear magnetic resonance),
{\bf [Braunstein-Kimble 00]} (DC for continuous variables),
{\bf [Ban 00 b, c]} (DC in a noisy quantum channel),
{\bf [Gorbachev-Zhiliba-Trubilko-Yakovleva 00]}
(teleportation of entangled states and DC
using a multiparticle quantum channel),
{\bf [Hao-Li-Guo 00]}
(probabilistic DC and teleportation),
{\bf [Werner 01 a]} (teleportation and DC schemes),
{\bf [Hiroshima 01]} (optimal DC with mixed state entanglement),
{\bf [Bowen 01 a]} (classical capacity of DC),
{\bf [Hao-Li-Guo 01]} (DC using GHZ),
{\bf [Cereceda 01 b]} (DC using three qubits),
{\bf [Bowen 01 b]},
{\bf [Li-Pan-Jing-(+3) 01]} (DC exploiting bright EPR beam),
{\bf [Liu-Long-Tong-Li 02]} (DC between multi-parties),
{\bf [Grudka-W\'{o}jcik 02 a]} (symmetric DC between multiparties),
{\bf [Lee-Ahn-Hwang 02]},
{\bf [Ralph-Huntington 02]} (unconditional continuous-variable DC),
{\bf [Mizuno-Wakui-Furusawa-Sasaki 04]}
(experimental demonstration of DC using entanglement of
a two-mode squeezed vacuum state),
{\bf [Schaetz-Barrett-Leibfried-(+6) 04]} (experimental DC with atomic qubits).


\subsection{Remote state preparation and measurement}


(In remote state preparation Alice knows the state which is to be remotely
prepared in Bob's site without sending him the qubit or the complete classical
description of it. Using one bit and one ebit Alice
can remotely prepare a qubit (from an special ensemble) of her choice at Bob's
site. In remote state measurement Alice asks Bob to simulate any single
particle measurement statistics on an arbitrary qubit}
{\bf [Bennett-DiVincenzo-Smolin-(+2) 01]},
{\bf [Pati 01 c, 02]},
{\bf [Srikanth 01 c]},
{\bf [Zeng-Zhang 02]},
{\bf [Berry-Sanders 03 a]} (optimal RSP),
{\bf [Agrawal-Parashar-Pati 03]} (RSP for multiparties),
{\bf [Bennett-Hayden-Leung-(+2) 02]} (general
method of remote state preparation for arbitrary states of many qubits, at a
cost of 1 bit of classical communication and 1 bit of entanglement per qubit
sent),
{\bf [Shi-Tomita 02 c]} (RSP of an entangled state),
{\bf [Abeyesinghe-Hayden 03]} (generalized RSP),
{\bf [Ye-Zhang-Guo 04]},
{\bf [Berry 04]} (resources required for exact RSP).


\subsection{Classical information capacity of quantum channels}


(A quantum channel is defined by the action of
sending one of $n$ possible messages, with different {\em a priori}
probabilities, to a receiver in the form of one of $n$ distinct
density operators. The receiver can perform any generalized
measurement in an attempt to discern which message was sent.)
{\bf [Gordon 64]},
{\bf [Levitin 69, 87, 93]},
{\bf [Holevo 73 a, b, 79, 97 a, b, 98 a, b, c]},
{\bf [Yuen-Ozawa 93]},
{\bf [Hall-O'Rourke 93]},
{\bf [Jozsa-Robb-Wootters 94]} (lower bound for accessible information),
{\bf [Fuchs-Caves 94]} (simplification of the Holevo upper bound of the
maximum information extractable in a quantum channel, and upper and lower bounds for
binary channels),
{\bf [Hausladen-Schumacher-Westmoreland-Wootters 95]},
{\bf [Hausladen-Jozsa-Schumacher-(+2) 96]},
{\bf [Schumacher-Westmoreland-Wootters 96]}
(limitation on the amount of accessible information in a quantum channel),
{\bf [Schumacher-Westmoreland 97]}.


\subsection{Quantum coding, quantum data compression}


{\bf [Schumacher 95]} (optimal compression of quantum
information carried by ensembles of pure states),
{\bf [Lo 95]} (quantum coding theorem for mixed states),
{\bf [Horodecki 98]} (limits for compression of quantum
information carried by ensembles of mixed states),
{\bf [Horodecki-Horodecki-Horodecki 98 a]}
(optimal compression of quantum information for one-qubit source
at incomplete data),
{\bf [Barnum-Smolin-Terhal 97, 98]},
{\bf [Jozsa-Horodecki-Horodecki-Horodecki 98]}
(universal quantum information compression),
{\bf [Horodecki 00]}
(toward optimal compression for mixed signal states),
{\bf [Barnum 00]}.


\subsection{Reducing the communication complexity with quantum entanglement}


{\bf [Yao 79]},
{\bf [Cleve-Buhrman 97]} (substituting quantum entanglement for communication),
{\bf [Cleve-Tapp 97]},
{\bf [Grover 97 a]},
{\bf [Buhrman-Cleve-van Dam 97]} (two-party communication complexity problem:
Alice receives a string $x=(x_0, x_1)$ and Bob a string $y=(y_0, y_1)$.
Each of the strings is a combination of two bit values: $x_0,y_0 \in \{0,1\}$
and $x_1, y_1 \in \{-1,1\}$. Their common goal is to compute the
function $f (x,y)= x_1 y_1 (-1)^{x_0 y_0}$,
with as high a probability as possible, while exchanging
altogether only 2 bits of information.
This can be done with a probability of success
of $0.85$ if the two parties share two qubits in a
maximally entangled state, whereas with shared random
variables but without entanglement, this probability cannot exceed $0.75$.
Therefore, in a classical protocol 3 bits of information
are necessary to compute $f$ with a probability of at least
$0.85$, whereas with the use of entanglement 2 bits of
information are sufficient to compute $f$ with the same
probability),
{\bf [Buhrman-van Dam-H\o{}yer-Tapp 99]} (reducing the communication complexity
in the ``guess my number'' game using a GHZ state, see
also {\bf [Steane-van Dam 00]} and {\bf [Gruska-Imai 01]} (p.~28)),
{\bf [Raz 99]} (exponential separation of quantum and classical communication complexity),
{\bf [Galv\~{a}o 00]} (experimental requirements for quantum communication complexity protocols),
{\bf [Lo 00 a]} (classical-communication cost in distributed quantum-information
processing: A generalization of quantum-communication complexity),
{\bf [Klauck 00 b, 01 a]},
{\bf [Brassard 01]} (survey),
{\bf [H\o{}yer-de Wolf 01]} (improved quantum communication complexity bounds for disjointness and equality),
{\bf [Xue-Li-Zhang-Guo 01]}
(three-party quantum communication complexity via entangled tripartite pure states),
{\bf [Xue-Huang-Zhang-(+2) 01]}
(reducing the communication complexity with quantum entanglement),
{\bf [Brukner-\.{Z}ukowski-Zeilinger 02]}
(quantum communication complexity protocol with two entangled qutrits),
{\bf [Galv\~{a}o 02]} (feasible quantum communication complexity protocol),
{\bf [Massar 02]} (closing the detection loophole and communication complexity),
{\bf [Brukner-\.{Z}ukowski-Pan-Zeilinger 04]}
(violation of Bell's inequality: Criterion for quantum communication complexity advantage).


\subsection{Quantum games and quantum strategies}


{\bf [Meyer 99 a]}
(comment: {\bf [van Enk 00]};
reply: {\bf [Meyer 00 a]}),
{\bf [Eisert-Wilkens-Lewenstein 99]}
(comment: {\bf [Benjamin-Hayden 01 b]}),
{\bf [Marinatto-Weber 00 a]} (comment:
{\bf [Benjamin 00 c]}; reply:
{\bf [Marinatto-Weber 00 c]}),
{\bf [Eisert-Wilkens 00 b]},
{\bf [Li-Zhang-Huang-Guo 00]}
(quantum Monty Hall problem),
{\bf [Du-Xu-Li-(+2) 00]}
(Nash equilibrium in QG),
{\bf [Du-Li-Xu-(+3) 00]}
(multi-player and multi-choice QG),
{\bf [Du-Xu-Li-(+3) 00]}
(quantum strategy without entanglement),
{\bf [Wang-Kwek-Oh 00]}
(quantum roulette: An extended quantum strategy),
{\bf [Johnson 01]} (QG with a corrupted source),
{\bf [Benjamin-Hayden 01 a]},
{\bf [Du-Xu-Li-(+2) 01]}
(entanglement playing a dominating role in QG),
{\bf [Du-Li-Xu-(+3) 01 a]}
(quantum battle of the sexes),
{\bf [Kay-Johnson-Benjamin 01]}
(evolutionary QG),
{\bf [Parrondo 01]},
{\bf [Iqbal-Toor 01 a, b, c, 02 a, b, c, d, e]},
{\bf [Du-Li-Xu-(+4) 01]} (experimental realization of
QG on a quantum computer),
{\bf [Piotrowski-Sladkowski 01]} (bargaining QG),
{\bf [Nawaz-Toor 01 a]} (strategies in quantum Hawk-Dove game),
{\bf [Klarreich 01]} ({\em Nature}),
{\bf [Nawaz-Toor 01 b]}
(worst-case payoffs in quantum battle of sexes game),
{\bf [Du-Li-Xu-(+3) 01 b]},
{\bf [Flitney-Ng-Abbott 02]}
(quantum Parrondo's games),
{\bf [D'Ariano-Gill-Keyl-(+3) 02]}
(quantum Monty Hall problem),
{\bf [Chen-Kwek-Oh 02]}
(noisy QG),
{\bf [Flitney-Abbott 02]}
(quantum version of the Monty Hall problem),
{\bf [Han-Zhang-Guo 02 a]}
(GHZ and W states in quantum three-person prisoner's dilemma),
{\bf [Protopopescu-Barhen 02]}
(solving continuous global
optimization problems using quantum algorithms),
{\bf [van Enk-Pike 02]}
(classical rules in quantum games),
{\bf [Ma-Long-Deng-(+2) 02]}
(cooperative three- and four-player quantum games),
{\bf [Meyer 02]},
{\bf [Du-Li-Xu-(+3) 02]}
(entanglement enhanced multiplayer quantum games);
{\bf [Li-Du-Massar 02]}
(continuous-variable quantum games),
{\bf [Lee-Johnson 02 b]} (review),
{\bf [Guinea-Mart\'{\i}n Delgado 03]}
(quantum chinos game),
{\bf [Chen-Hogg-Beausoleil 03]} (quantum $n$-player public goods game),
{\bf [Du-Xu-Li-(+2) 02]}
(playing prisoner's dilemma with quantum rules),
{\bf [Gravier-Jorrand-Mhalla-Payan 03]},
{\bf [\"{O}zdemir-Shimamura-Morikoshi-Imoto 02]}
(samaritan's dilemma),
{\bf [Shimamura-\"{O}zdemir-Morikoshi-Imoto 03]},
{\bf [\"{O}zdemir-Shimamura-Imoto 04]},
{\bf [Kargin 04]} (coordination games).


\subsection{Quantum clock synchronization}


{\bf [Chuang 00]},
{\bf [Jozsa-Abrams-Dowling-Williams 00]},
{\bf [Burt-Ekstrom-Swanson 00]},
{\bf [Genovese-Novero 00 c]}
(QCS based on entangled photon pairs transmission),
{\bf [Shahriar 00]},
{\bf [Preskill 00 b]} (QCS and quantum error correction),
{\bf [Hwang-Ahn-Hwang-Han 00]}
(entangled quantum clocks for measuring proper-time difference),
{\bf [Giovannetti-Lloyd-Maccone 01 a, 02 a]},
{\bf [Harrelson-Kerenidis 01]},
{\bf [Giovannetti-Lloyd-Maccone-Wong 01]},
{\bf [Janzing-Beth 01 c]} (quasi-order of clocks and synchronism and
quantum bounds for copying timing information),
{\bf [Yurtsever-Dowling 02]},
{\bf [Giovannetti-Lloyd-Maccone-Wong 02]},
{\bf [Giovannetti-Lloyd-Maccone-Shahriar 02]}
(limits to QCS induced by completely dephasing communication channels),
{\bf [Kr\v{c}o-Paul 02]} (a multi-party protocol),
{\bf [Valencia-Scarcelli-Shih 04]}.


\section{Quantum computation}


\subsection{General}


{\bf [Benioff 80, 81, 82 a, b, c, 86, 95, 96, 97 a, b, 98 a, c, d]},
{\bf [Feynman 82]} (Feynman asked whether or not the behavior of every physical
system can be simulated by a computer, taking no more time than the physical
system itself takes to produce the observed behavior.
Feynman suggests that it may not be possible to
simulate a quantum system in real time by a classical computer whereas it may be
possible with a quantum computer.
So if Feynman's suggestion is correct it implies there are
tasks a QC can perform far more efficiently than a classical computer),
{\bf [Deutsch 85 b]} (quantum equivalent of a Turing machine),
{\bf [Feynman 85, 86]} (physical limitations of classical computers),
{\bf [Deutsch 89]} (QC networks),
{\bf [Deutsch 92]},
{\bf [Deutsch-Jozsa 92]},
{\bf [Bennett 93]},
{\bf [Brown 94]} (popular review)
{\bf [Sleator-Weinfurter 95]},
{\bf [Bennett 95 a]} (review, see for more references),
{\bf [Lloyd 93, 94 a, b, 95 a, b]},
{\bf [Shor 95]} (how to reduce decoherence in QC memory),
{\bf [Dove 95]},
{\bf [Pellizzari-Gardiner-Cirac-Zoller 95]} (how to reduce decoherence in
a QC based on cavities by continuous observation),
{\bf [Chuang-Yamamoto 95]} (a simple QC),
{\bf [Glanz 95 a]},
{\bf [Plenio-Vedral-Knight 96]} (review),
{\bf [Barenco 96]} (review),
{\bf [Barenco-Ekert-Macchiavello-Sampera 96]} (review),
{\bf [Haroche-Raimond 96]} (review),
{\bf [Deutsch 97]} (review),
{\bf [Myers 97]} (can a QC be fully quantum?),
{\bf [Grover 97 a]} (quantum telecomputation),
{\bf [Bennett-Bernstein-Brassard-Vazirani 97]} (strengths
and weaknesses of QC),
{\bf [Warren-Gershenfeld-Chuang 97]}
(the usefulness of NMR QC),
{\bf [Williams-Clearwater 98]} (book),
{\bf [Hughes 98]} (relevance of QC for cryptography),
{\bf [Preskill 98 a, b]} (pros and cons of QC),
{\bf [Lo-Spiller-Popescu 98]} (book),
{\bf [Berman-Doolen-Mainieri-Tsifrinovich 98]} (book),
{\bf [Gram\ss\, 98]} (book),
{\bf [Milburn 98]} (book),
{\bf [Steane 98 b]} (review),
{\bf [Farhi-Gutmann 98 a]} (analog analogue of a digital QC),
{\bf [Loss-DiVincenzo 98]},
{\bf [Schack 98]} (using a QC to
investigate quantum chaos),
{\bf [Vedral-Plenio 98 b]} (review),
{\bf [Buhrman-Cleve-Wigderson 98]}
(classical vs. quantum communication and QC),
{\bf [Ekert-Fern\'{a}ndez Huelga-Macchiavello-Cirac 98]}
(using entangled states to make
computations between distant nodes of a quantum network),
{\bf [Deutsch-Ekert 98]} (review),
{\bf [Scarani 98]} (review),
{\bf [Privman-Vagner-Kventsel 98]} (QC based
on a system with quantum Hall effect),
{\bf [Gershenfeld-Chuang 98]} (QC with molecules, review),
{\bf [DiVincenzo 98 a]},
{\bf [Kane 98]} (QC based on silicon and on RMN),
{\bf [Farhi-Gutmann 98 b]} (decision trees),
{\bf [Linden-Fremann 98 b]}
(Deutsch-Jozsa algorithm on a three-qubit NMR QC),
{\bf [Collins-Kim-Holton 98]}
(Deutsch-Jozsa algorithm as a test of QC),
{\bf [Terhal-Smolin 98]} (single quantum querying of a database),
{\bf [Rieffel-Polak 98]} (introduction for non-physicists),
{\bf [Zalka 98 d]} (an introduction to QC),
{\bf [Luo-Zeng 98]}
(NMR QC with a hyperpolarized nuclear spin bulk),
{\bf [Gruska 99]} (book),
{\bf [Braunstein-Caves-Jozsa-Linden-Popescu-Schac 99]}
(separability of very noisy mixed states and implications for NMR QC),
{\bf [Brun-Schac 99]},
{\bf [Braunstein 99]} (book),
{\bf [Brooks 99]} (book),
{\bf [Williams 99]} (book),
{\bf [DiVincenzo-Loss 99]},
{\bf [Sanders-Kim-Holton 99]},
{\bf [Gottesman-Chuang 99]} (QC
using teleportation and single-qubit operations),
{\bf [Preskill 99 d]} (Chap.~6),
{\bf [Macchiavello-Palma-Zeilinger 00]} (book of collected papers),
{\bf [Lloyd 00 a]} (quantum search without entanglement),
{\bf [Cirac-Zoller 00]} (scalable QC with ions in an array of
microtraps),
{\bf [Bouwmeester-Ekert-Zeilinger 00]} (book on quantum information),
{\bf [Bennett-DiVincenzo 00]} (review in {\em Nature} on
quantum information and QC),
{\bf [Nielsen-Chuang 00]} (book),
{\bf [Bacon-Kempe-Lidar-Whaley 00]}
(universal fault-tolerant QC
on decoherence-free subspaces),
{\bf [Beige-Braun-Tregenna-Knight 00]}
(QC using dissipation to remain
in a decoherence-free subspace),
{\bf [Osborne 00 d]},
{\bf [Georgeot-Shepelyansky 00]} (in the quantum chaos regime, an ideal
state quickly disappears, and exponentially many states become mixed;
below the quantum chaos border an ideal state can survive for long times,
and an be used for QC),
{\bf [Knill-Nielsen 00 a]}
(theory of QC),
{\bf [Ekert-Hayden-Inamori 00]} (basic concepts in QC),
{\bf [Ekert-Hayden-Inamori-Oi 01]} (what is QC),
{\bf [Knill-Laflamme-Milburn 01]}
(scheme for efficient QC with linear optics),
{\bf [Linden-Popescu 01]}
(entanglement is necessary for QC),
{\bf [Hardy-Steeb 01]} (book),
{\bf [Kitaev-Shen-Vyalyi 02]} (book),
{\bf [Lomonaco 02]} (book),
{\bf [Lomonaco-Brandt 02]} (book),
{\bf [Zalka 02]} (lectures on QC),
{\bf [Biham-Brassard-Kenigsberg-Mor 03]}
(the Deutsch-Jozsa problem and the Simon problem
can be solved using a separable state).


\subsection{Quantum algorithms}


\subsubsection{Deutsch-Jozsa's and Simon's}


{\bf [Deutsch 85 b]},
{\bf [Deutsch-Jozsa 92]},
{\bf [Simon 94, 97]},
{\bf [Cleve-Ekert-Macchiavello-Mosca 98]},
{\bf [Chi-Kim-Lee 00 a, 01]}
(initialization-free generalized DJ algorithm),
{\bf [Vala-Amitay-Zhan-(+2) 02]}
(experimental implementation of the DJ algorithm for
three-qubit functions using rovibrational molecular wave packets
representation),
{\bf [Gulde-Riebe-Lancaster-(+6) 03]} (implementation of the
DJ algorithm on an ion-trap quantum computer, {\em Nature}),
{\bf [Brazier-Plenio 03]}
(the DJ algorithm is surprisingly good as the problem becomes less
structured and is always better than the van Dam algorithm for low numbers of
queries),
{\bf [Ermakov-Fung 03]} (NMR implementation of the DJ algorithm
using different initial states),
{\bf [Bianucci-Muller-Shih-(+3) 04]}
(experimental realization of the one qubit DJ algorithm in a
quantum dot),
{\bf [Cereceda 04 c]}
(generalization of the DJ algorithm using two qudits).


\subsubsection{Factoring}


{\bf [Shor 94, 97]} (the number of steps any classical computer requires in order
to find the prime factors of an $l$-digit integer increases exponentially with $l$,
at least using algorithms known at present. Factoring large integers
is therefore conjectured to be intractable classically, an observation
underlying the security of widely used cryptographic codes.
Quantum computer, however, could factor
integers in only polynomial time, using Shor's quantum factoring algorithm),
{\bf [Ekert-Jozsa 96]} ({\em Rev. Mod. Phys.}),
{\bf [Plenio-Knight 96]}
(realistic lower bounds for the factorization time of large
numbers),
{\bf [Zalka 98 c]} (fast versions of Shor's factoring algorithm),
{\bf [Berman-Doolen-Tsifrinovich 00]}
(influence of superpositional wave function oscillations on Shor's
algorithm),
{\bf [Lomonaco 00 b]} (Shor's quantum factoring algorithm),
{\bf [McAnally 01]}
{\bf [Vandersypen-Steffen-Breyta-(+3) 01]}
(experimental realization of Shor's quantum factoring algorithm
using nuclear magnetic resonance, {\em Nature}),
{\bf [Lavor-Manssur-Portugal 03]} (review of Shor's factoring algorithm).


\subsubsection{Searching}


{\bf [Grover 96 b, 97 b, c, 98 a, b, c, d, 00 c, 02 b, c]} (a QA for a
quicker search of an item in a non-ordered $n$ items database: While a classical
algorithm requires $\frac{n}{2}$ steps to obtain a 50\% probability of success,
Grover's algorithm obtains 100\% success with $\frac{\pi \sqrt{n}}{4}$ steps),
{\bf [Brassard 97]} (on Grover's algorithm),
{\bf [Boyer-Brassard-H\o{}yer-Tapp 96, 98]} (optimal
number of iterations for the amplitude of the solution state in Grover's algorithm),
{\bf [Collins 97]} (on Grover's algorithm and other advances in quantum computation),
{\bf [Terhal-Smolin 97]} (searching algorithms),
{\bf [Biron-Biham-Biham-(+2) 98]} (generalized Grover's algorithm),
{\bf [Chuang-Gershenfeld-Kubinec 98]}
(experimental implementation of quantum fast search),
{\bf [Ross 98]}
(a modification of Grover's algorithm as a fast database search),
{\bf [Carlini-Hosoya 98]} (an alternative algorithm for database search),
{\bf [Buhrman-de Wolf 98]} (lower bounds for a quantum search),
{\bf [Roehrig 98]} (an upper bound for searching in an ordered list),
{\bf [Zalka 99 a]} (Grover's algorithm is optimal),
{\bf [Jozsa 99]} (searching in Grover's algorithm),
{\bf [Long 01]} (Grover algorithm with zero theoretical failure rate),
{\bf [Patel 01 a]},
{\bf [Li-Li 01]} (a general quantum search algorithm),
{\bf [Murphy 01]},
{\bf [Grover 01]} (pedagogical article describing the
invention of the quantum search algorithm),
{\bf [Bae-Kwon 01]},
{\bf [Miao 01 a]} (construction for the unsorted quantum search
algorithms),
{\bf [Collins 02]}.


\subsubsection{Simulating quantum systems}


{\bf [Feynman 82]} (Feynman asked whether or not the behavior of every physical
system can be simulated by a computer, taking no more time than the physical
system itself takes to produce the observed behavior.
Feynman suggests that it may not be possible to
simulate a quantum system in real time by a classical computer whereas it may be
possible with a quantum computer.
So if Feynman's suggestion is correct it implies there are
tasks a QC can perform far more efficiently than a classical computer),
{\bf [Feynman 86]},
{\bf [Lloyd 96]} (Feynman's 1982 conjecture, that quantum computers can be
programmed to simulate any local quantum system, is shown to be correct),
{\bf [Wiesner 96]} (simulations of many-body quantum systems),
{\bf [Meyer 96 a, 97]},
{\bf [Lidar-Biham 97]},
{\bf [Abrams-Lloyd 97]}
(simulation of many-body Fermi systems on a universal quantum computer),
{\bf [Zalka 98 a, b]},
{\bf [Boghosian-Taylor 98 a, b]},
{\bf [Schack 98]} (using a quantum computer to investigate quantum chaos),
{\bf [Somaroo-Tseng-Havel-(+2) 99]}
(quantum simulations on a quantum computer),
{\bf [Terhal-DiVincenzo 00]},
{\bf [Leung 01 a]},
{\bf [Ortiz-Gubernatis-Knill-Laflamme 02]} (simulating fermions on a quantum computer),
{\bf [Somma-Ortiz-Gubernatis-(+2) 02]},
{\bf [Berman-Ezhov-Kamenev-Yepez 02]},
{\bf [Chepelianskii-Shepelyansky 02 a, b]},
{\bf [Wocjan-Rotteler-Janzing-Beth 02]} (simulating Hamiltonians in quantum networks: Efficient schemes and
complexity bounds),
{\bf [Jan\'{e}-Vidal-D\"{u}r-(+2) 03]} (simulation of quantum dynamics with quantum optical systems),
{\bf [Kraus-Hammerer-Giedke-Cirac 03]} (Hamiltonian simulation in continuous-variable
systems).


\subsubsection{Quantum random walks}


{\bf [Aharonov-Davidovich-Zagury 93]},
{\bf [Meyer 96 a, b]},
{\bf [Nayak-Vishwanath 00]},
{\bf [Watrous 01]},
{\bf [Aharonov-Ambainis-Kempe-Vazirani 01]},
{\bf [Ambainis-Bach-Nayak-(2) 01]},
{\bf [Travaglione-Milburn 02 a]},
{\bf [Konno 02]} (QRW in one dimension),
{\bf [Kempe 03]} (an introductory overview),
{\bf [Brun-Carteret-Ambainis 03 a, b, c]},
{\bf [Grimmett-Janson-Scudo 03]} (weak limits for quantum random walks),
{\bf [Bracken-Ellinas-Tsohantjis 04]}.


\subsubsection{General and others}


{\bf [Durr-H\o{}yer 96]} (a QA for finding the minimum),
{\bf [Cockhott 97]} (databases),
{\bf [Ekert-Macchiavello 98]},
{\bf [Cleve-Ekert-Macchiavello-Mosca 98]},
{\bf [Hogg 98 a, b]},
{\bf [Hogg-Yanik 98]} (local searching methods),
{\bf [Ekert-Jozsa 98]},
{\bf [Pati 98 c]},
{\bf [Pittenger 99]} (book on QA),
{\bf [Abrams-Lloyd 99]} (algorithm for finding eigenvalues and eigenvectors),
{\bf [Ahuja-Kapoor 99]} (algorithm for finding the maximum),
{\bf [Watrous 00]} (QA for solvable groups),
{\bf [Vandersypen-Steffen-Breyta-(+3) 00]}
(experimental realization of an order-finding algorithm with
an NMR quantum computer),
{\bf [Ivanyos-Magniez-Santha 01]} (QA for some instances
of the non-Abelian hidden subgroup problem),
{\bf [Alber-Beth-Horodecki-(+6) 01]} (Chap.~4),
{\bf [Galindo-Mart\'{\i}n Delgado 02]} (review),
{\bf [Shor 02 b]} (introduction to QA),
{\bf [Klappenecker-R\"{o}tteler 03]}.


\subsection{Quantum logic gates}


{\bf [Deutsch 89 a]} (a set of gates is {\em universal} if any unitary
action can be decomposed into a product of successive actions of these gates
on different subsets of the input qubits; the {\em Deutsch gate} is a three-qubit
universal gate),
{\bf [Barenco 95]} (almost any two-qubit gate is universal),
{\bf [DiVincenzo 95 b]}
(two-qubit gates are universal for quantum computation; its classical analog is not true:
classical reversible two-bit gates are not universal),
{\bf [Barenco-Bennett-Cleve-(+6) 95]} (one-qubit gates plus the CNOT gate are enough
for quantum computation),
{\bf [Cirac-Zoller 95]} (proposal for a quantum computer with ions),
{\bf [Monroe-Meekhof-King-(+2) 95]} (ions in a radiofrecuency trap),
{\bf [Domokos-Raimond-Brune-Haroche 95]}
(they control atoms using photons trapped in superconductor cavities),
{\bf [Barenco-Deutsch-Ekert-Jozsa 95]} (quantum logic gates),
{\bf [Schwarzschild 96]} (experimental quantum logic gates),
{\bf [Cory-Fahmy-Havel 97]} (NMR),
{\bf [Gershenfeld-Chuang 97]} (NMR),
(Los Alamos experiment with trapped ions:)
{\bf [Hughes-James-G\'{o}mez-(+12) 98]},
{\bf [Wineland-Monroe-Itano-(+5) 98]},
{\bf [James-Gulley-Holzscheiter-(+10) 98]};
{\bf [Stevens-Brochard-Steane 98]}
(experimental methods for processors with trapped ions),
{\bf [Brennen-Caves-Jessen-Deutsch 98]} (optical),
{\bf [Wei-Xue-Morgera 98]},
{\bf [Linden-Barjat-Carbajo-Freeman 98]}
(pulse sequences for NMR quantum computers: How to
manipulate nuclear spins while freezing the motion of coupled neighbours),
{\bf [Fuji 01]},
{\bf [Schmidt Kaler-Haffner-Riebe-(+7) 03]} (experimental Cirac-Zoller
CNOT quantum gate, {\em Nature}),
{\bf [O'Brien-Pryde-White-(+2) 03]} (experimental all-optical quantum CNOT gate),
{\bf [Gasparoni-Pan-Walther-(+2) 04]}
(quantum CNOT with linear optics and previous entanglement),
{\bf [Zhao-Zhang-Chen-(+4) 04]}
(experimental demonstration of a non-destructive quantum CNOT for two independent photon-qubits).


\subsection{Schemes for reducing decoherence}


{\bf [Briegel-D\"{u}r-Cirac-Zoller 98]}
(quantum repeaters for communication),
{\bf [Duan-Guo 98 a, b, d, h]} (reducing decoherence),
{\bf [Viola-Lloyd 98]}(dynamical suppression of decoherence in two-state
quantum systems),
{\bf [DiVincenzo-Terhal 98]} (decoherence: The obstacle to
quantum computation, review).


\subsection{Quantum error correction}


{\bf [Shor 95, 96]} (9:1),
{\bf [Steane 96 a, b, c, 98 d, e]} (QEC codes) (7:1),
{\bf [Calderbank-Shor 96]} (QEC),
{\bf [Gottesman 96]},
{\bf [DiVincenzo-Shor 96]},
{\bf [Bennett-DiVincenzo-Smolin-Wootters 96]} (5:1),
{\bf [Laflamme-Miquel-Paz-Zurek 96]} (perfect QEC code),
{\bf [Ekert-Macchiavello 96]},
{\bf [Schumacher-Nielsen 96]},
{\bf [Calderbank-Rains-Shor-Sloane 96, 97]},
{\bf [Chau 97 a, b]},
{\bf [Cleve-Gottesman 97]},
{\bf [Cerf-Cleve 97]} (information-theoretic interpretation of QEC codes),
{\bf [Knill-Laflamme 97]} (QEC codes),
{\bf [Plenio-Vedral-Knight 97 a, b]}
(QEC in the presence of spontaneous emission),
{\bf [Vedral-Rippin-Plenio 97]},
{\bf [Chuang-Yamamoto 97]},
{\bf [Braunstein-Smolin 97]}
(perfect QEC coding in 24 laser pulses),
{\bf [Braunstein 98 a, b]},
{\bf [Knill-Laflamme-Zurek 98 a, b]}
(arbitrarly high efficiency QEC codes),
{\bf [Gottesman 98 a, b]} (fault tolerant quantum computation),
{\bf [Preskill 98 c]} (brief history of QEC codes),
{\bf [Preskill 98 d]} (fault tolerant quantum computation),
{\bf [Cory-Price-Mass-(+5) 98]} (experimental QEC),
{\bf Kak 98]},
{\bf [Steinbach-Twamley 98]} (motional QEC),
{\bf [Koashi-Ueda 99]}
(reversing measurement and probabilistic QEC),
{\bf [Chau 99]},
{\bf [Kanter-Saad 99]}
(error-correcting codes that nearly saturate Shannon's bound),
{\bf [Preskill 99 d]} (Chap.~7),
{\bf [Knill-Laflamme-Viola 00]}
(theory of QEC for general noise),
{\bf [Barnes-Warren 00]} (automatic QEC),
{\bf [Nielsen-Chuang 00]} (Chap.~10),
{\bf [Knill-Laflamme-Martinez-Negrevergne 01]}
(implementation of the five qubit error correction benchmark),
{\bf [Schumacher-Westmoreland 01 b]}
(approximate quantum error correction),
{\bf [Korepin-Terilla 02]},
{\bf [Yang-Chu-Han 02]},
{\bf [Gottesman 02]} (introduction to QEC),
{\bf [Ahn-Wiseman-Milburn 03]} (QEC
for continuously detected errors),
{\bf [Pollatsek-Ruskai 03]}
(permutationally invariant codes for quantum error correction).


\subsection{Decoherence-free subspaces and subsystems}


{\bf [Palma-Suominen-Ekert 96]},
{\bf [Duan-Guo 97 a, 98 a, e]},
{\bf [Zanardi-Rasetti 97 a, b]},
{\bf [Zanardi 97, 98, 99]},
{\bf [Lidar-Chuang-Whaley 98]} (DFS for quantum computation),
{\bf [Lidar-Bacon-Whaley 99]},
{\bf [Bacon-Kempe-Lidar-Whaley 00]}
(universal fault-tolerant quantum computation on DFS),
{\bf [Lidar-Bacon-Kempe-Whaley 00, 01 a, b]},
{\bf [Kempe-Bacon-Lidar-Whaley 00]},
{\bf [Beige-Braun-Tregenna-Knight 00]}
(quantum computation using dissipation to remain in a DFS),
{\bf [Kwiat-Berglund-Altepeter-White 00]}
(experimental preparation a two-photon polarization-entangled singlet state and
demonstration of its invariance under collective decoherence),
{\bf [Kielpinski-Meyer-Rowe-(+4) 01]}
(experimental demonstration of the protection of a qubit against collective dephasing by
encoding it in two trapped ions),
{\bf [Viola-Fortunato-Pravia-(+3) 01]}
(experimental demonstration of the protection of a qubit against collective decoherence by
encoding it in a DF subsystem of three NMR qubits),
{\bf [Fortunato-Viola-Hodges-(+2) 02]}
(experimental demonstration of the protection of a qubit against collective dephasing by
encoding it two NMR qubits),
{\bf [Foldi-Benedict-Czirjak 02]}
(preparation of DF, subradiant states in a cavity),
{\bf [Feng-Wang 02 a]}
(quantum computing with four-particle DF states in an ion trap),
{\bf [Wu-Lidar 02 b]} (creating DFS using strong and fast pulses),
{\bf [Cabello 02 m]} (four-qubit DFS),
{\bf [Satinover 02 a]} (DFS in supersymmetric oscillator networks),
{\bf [Satinover 02 b]},
{\bf [Lidar-Whaley 03]} (review),
{\bf [Brown-Vala-Whaley 03]} (scalable ion trap quantum computation in decoherence-free subspaces with
pairwise interactions only),
{\bf [Ollerenshaw-Lidar-Kay 03]} (Grover's search algorithm on a NMR computer in which two qubits are
protected from a special kind of errors by encoding them in four qubits),
{\bf [Fonseca Romero-Mokarzel-Terra Cunha-Nemes 03]},
{\bf [Walton-Abouraddy-Sergienko-(+2) 03 b]} (DFS in QKD),
{\bf [Boileau-Gottesman-Laflamme-(+2) 04]} (B92 with double singlets).


\subsection{Experiments and experimental proposals}


(Implementation of an algorithm for solving the two-bit Deutsch problem with NMR:) {\bf
[Chuang-Vandersypen-Zhou-(+2) 98]}, {\bf [Jones-Mosca 98]};
{\bf [Jones-Mosca-Hansen 98]}
(implementation of Grover's quantum search algorithm with NMR),
{\bf [Nakamura-Pashkin-Tsai 99]}
(coherent control of macroscopic quantum states in a single-Cooper-pair box),
{\bf [Fu-Luo-Xiao-Zeng 99]}
(experimental realization of a discrete Fourier transformation on an NMR QC),
{\bf [Kwiat-Mitchell-Schwindt-White 99]} (Grover's search algorithm: An optical approach),
{\bf [Marx-Fahmy-Myers-(+2) 99]}
(realization of a 5-bit NMR QC using a new molecular architecture),
{\bf [Yannoni-Sherwood-Vandersypen-(+3) 99]} (NMR using liquid crystal solvents),
{\bf [Vandersypen-Steffen-Sherwood-(+3) 00]}
(first implementation of a three qubit Grover's algorithm),
{\bf [Jones 00 a, b]} (NMR QC: A critical evaluation),
{\bf [Vrijen-Yablonovitch-Wang-(+5) 00]} (electron spin resonance
transistors for quantum computing in silicon-germanium heterostructures),
{\bf [Cory-Laflamme-Knill-(+13) 00]}
(NMR based quantum information processing: Achievements and
prospects),
{\bf [Deutsch-Brennen-Jessen 00]}
(QC with neutral atoms in an optical lattice),
{\bf [DiVincenzo 00]},
{\bf [Kane 00]} (silicon-based QC),
{\bf [Opatrn\'{y}-Kurizki 00]} (QC based on photon exchange
interactions),
{\bf [Kielpinski-Ben Kish-Britton-(+6) 01]} (trapped-ion QC),
{\bf [Vandersypen-Steffen-Breyta-(+3) 01]}
(experimental realization of Shor's quantum factoring algorithm
using nuclear magnetic resonance, {\em Nature}).
{\bf [Gulde-Riebe-Lancaster-(+6) 03]} (implementation of the
Deutsch-Jozsa algorithm on an ion-trap quantum computer, {\em Nature}),
{\bf [Steffen-van Dam-Hogg-(+2) 03]}
(implementation of an adiabatic quantum optimization
algorithm),
{\bf [Ermakov-Fung 03]} (NMR implementation of the DJ algorithm
using different initial states),
{\bf [Brainis-Lamoureux-Cerf-(+3) 03]}
(fiber-optics implementation of the DJ and Bernstein-Vazirani
quantum algorithms with three qubits).


\section{Miscellaneous}


\subsection{Textbooks}


{\bf [Dirac 30]},
{\bf [Fock 31]},
{\bf [von Neumann 32]},
{\bf [Born 33]},
{\bf [Landau-Lifshitz 48]},
{\bf [Schiff 49]},
{\bf [Bohm 51]},
{\bf [Messiah 58]},
{\bf [Merzbacher 61]},
{\bf [Feynman-Hibbs 65]},
{\bf [Feynman-Leighton-Sands 65]},
{\bf [Sakurai 67, 85]},
{\bf [Cohen Tannoudji-Diu-Lalo\"{e} 73]},
{\bf [Galindo-Pascual 78]},
{\bf [Bohm 79]},
{\bf [Bransden-Joachain 89]},
{\bf [Greiner 89]},
{\bf [Pauli-Achuthan-Venkatesan 90]},
{\bf [Ballentine 90 a, 98]},
{\bf [Peres 93 a]},
{\bf [Isham 95]},
{\bf [Hecht 00]},
{\bf [Schwinger 01]},
{\bf [Bes 04]}.


\subsection{History of quantum mechanics}


{\bf [Jammer 66]} (the conceptual development of QM until 1927),
{\bf [van der Waerden 67]} (17 papers translated to English from 1916 to 1926),
{\bf [Kuhn-Heilbron-Forman-Allen 67]} (sources for history of QM),
{\bf [Hermann 71]} (1899-1913),
{\bf [Kangro 72]} (original on QM papers translated to English),
{\bf [Jammer 74]} (the philosophy of QM),
{\bf [Holton 80]}
(133 informally collected ``classic'' papers in quantum physics),
{\bf [Mehra-Rechenberg 82 a-e, 87 a, b, 00 a, b]}
(historical development of QM, 1900-1941),
{\bf [Jammer 85]} (the EPR problem in its historical development),
{\bf [Howard 85]} (Einstein, locality and separability),
{\bf [Pais 86]} (history of nuclear physics, quantum field theories, and subatomic
particles, 1927-1983),
{\bf [Icaza 91]} (historical development, 1925-1927),
{\bf [Marage-Wallenborn 95]} (the Solvay conferences),
{\bf [S\'{a}nchez Ron 01]} (1860-1926),
{\bf [Friedrich-Herschbach 03]} (Stern and Gerlach).


\subsection{Biographs}


{\bf [Planck 48]} (autobiography),
{\bf [Gerlach 48]} (Planck),
{\bf [Born 75]} (autobiography),
{\bf [Heims 80]} (von Neumann),
{\bf [Pais 82]} (a scientific biography of Einstein),
{\bf [Heilbron 86]} (Planck),
{\bf [Moore 89]} (Schr\"{o}dinger),
{\bf [Jammer 88]} (paper on Bohm),
{\bf [Bernstein 89]} (interview with Bell),
{\bf [Jammer 90, 93]} (papers on Bell),
{\bf [Kragh 90]} (Dirac),
{\bf [Pais 91]} (Bohr),
{\bf [MacRae 91]} (von Neumann),
{\bf [Cassidy 92]} (Heisenberg),
{\bf [Pines 93]} (Bohm's obituary),
{\bf [Israel-Gasca 95]} (von Neumann),
{\bf [Peres 96 a, b]} (Nathan Rosen 1909-95),
{\bf [Bergmann-Merzbacher-Peres 96]}
(Obituary: Nathan Rosen),
{\bf [Israelit 96]} (Nathan Rosen: 1909-1995),
{\bf [Peat 97]} (Bohm),
{\bf [Laurikainen 97]} (essays on Pauli),
{\bf [Wheeler-Ford 98]} (Wheeler's autobiography),
{\bf [Goddard 98]} (Dirac),
{\bf [Whitaker 98 a]} (Bell),
{\bf [Mehra 99]} (Einstein),
{\bf [Pais 00]} (biographical portraits of Bohr, Born, Dirac, Einstein,
von Neumann, Pauli, Uhlenbeck, Wigner and others),
{\bf [Holton 00]} (Heisenberg and Einstein),
{\bf [Aspect 00]} (contains a photograph of J. S. Bell and A. Aspect about 1986 in Paris),
{\bf [Jackiw-Shimony 02]} (Bell),
{\bf [Bell 02]} (Bell's wife reminiscences),
{\bf [Whitaker 02]} (Bell in Belfast: Early years and education),
{\bf [d'Espagnat 02]} (Bell),
{\bf [Enz 02]} (Pauli),
{\bf [Schroer 03]} (Jordan),
{\bf [Fern\'{a}ndez Ra\~{n}ada 04]} (Heisenberg),
{\bf [Lahera 04]} (Bohr).


\subsection{Philosophy of the founding fathers}


{\bf [Petersen 63]} (Bohr's philosophy),
{\bf [Heelan 65, 75]} (Heisenberg's philosophy),
{\bf [Hall 65]} (philosophical basis of Bohr's interpretation of quantum
mechanics),
{\bf [Folse 85]} (Bohr's philosophy),
{\bf [Laurikainen 85, 88]} (Pauli's philosophy),
{\bf [Fine 86]} (Einstein and QM),
{\bf [Honner 87]} (Bohr's philosophy),
{\bf [Murdoch 87]} (Bohr's philosophy),
{\bf [Faye 91]} (on Bohr's interpretation of QM),
{\bf [Faye-Folse 94]} (Bohr and philosophy),
{\bf [Bohr 98]} (collected writings beyond physics:
attempts to prove that biology
cannot be reduced to physics,
essays on the influence on his work of philosopher Hoffding),
{\bf [Jammer 99]} (Einstein and religion).


\subsection{Quantum logic}


{\bf [Birkhoff-von Neumann 36]} (first QL),
{\bf [Reichenbach 44]} (first three-valued QL),
{\bf [Putnam 57]} (three-valued QL),
{\bf [Mackey 63]},
{\bf [Finkelstein 69, 72]},
{\bf [Putnam 69, 74, 81]},
{\bf [Piron 72, 76]},
{\bf [van Fraassen 73, 74 b]},
{\bf [Scheibe 73]},
{\bf [Jammer 74]} (Chap.~8, historical account),
{\bf [Hooker 75, 79]} (collections of original papers),
{\bf [Suppes 76]} (collective book),
{\bf [Friedman-Putnam 78]},
{\bf [Stairs 78, 82, 83 a, b]},
{\bf [Greechie 78]} (a nonstandard QL),
{\bf [Beltrametti-Cassinelli 79]} (collective book),
{\bf [Beltrametti-Cassinelli 81]} (book),
{\bf [Beltrametti-van Fraassen 81]},
{\bf [Hughes 81]} (paper in {\em Sci. Am.}),
{\bf [Holdsworth-Hooke 83]} (a critical survey of QL),
{\bf [Redhead 87]} (Chap.~7),
{\bf [Pitowsky 89 a]} (book),
{\bf [Hughes 89]} (Chap.~7),
{\bf [Pykacz-Santos 90, 91, 95]},
{\bf [Pavi\v{c}i\'{c} 92 b]} (bibliography on quantum logics and related structures),
{\bf [R\'{e}dei 98]} (book),
{\bf [Svozil 98 b]} (book),
{\bf [Pykacz 98]},
{\bf [Coecke-Moore-Wilce 00]},
{\bf [McKay-Megill-Pavi\v{c}i\'{c} 00]}
(algorithms for Greechie diagrams),
{\bf [Dalla Chiara-Giuntini 01]}.


\subsection{Superselection rules}


{\bf [Wick-Wightman-Wigner 52]},
{\bf [Galindo-Pascual 78]},
{\bf [Gilmore-Park 79 a, b]},
{\bf [Mirman 79]},
{\bf [Wan 80]} (superselection rules, quantum measurement and
Schr\"{o}dinger's cat),
{\bf [Zurek 82]},
{\bf [Hughes-van Fraassen 88]}
(can the measurement problem be solved by superselection rules?),
{\bf [Giulini-Kiefer-Zeh 95]},
{\bf [Wightman 95]},
{\bf [Dugi\'c 98]},
{\bf [Cisneros-Mart\'{\i}nez y Romero-N\'{u}\~{n}ez Y\'{e}pez-Salas Brito 98]},
{\bf [Giulini 99, 00]} (the distinction between `hard' ---i.e.,
those whose existence is demonstrated by means of symmetry principles---
and `soft' ---or `environment-induced'--- superselection rules
is not well founded),
{\bf [Mayers 02 b]} (a charge superselection rule implies no restriction
on the operations that can be executed on any individual qubit),
{\bf [Kitaev-Mayers-Preskill 03]} (superselection rules do not enhance
the information-theoretic security of quantum cryptographic protocols),
{\bf [Verstraete-Cirac 03 a]} (nonlocality in the presence of superselection rules and data hiding
protocols),
{\bf [Schuch-Verstraete-Cirac 04 a, b]} (entanglement in the presence of superselection rules),
{\bf [Wiseman-Vaccaro 03]} (entanglement of indistinguishable particles shared between two parties),
{\bf [Wiseman-Bartlett-Vaccaro 03]} (entanglement constrained by
generalized superselection rules).


\subsection{Relativity and the instantaneous change of the quantum
state by local interventions}


{\bf [Bloch 67]}, {\bf [Aharonov-Albert 80, 81, 84]}, {\bf
[Herbert 82]} (superluminal communication would be possible with a
perfect quantum cloner), {\bf [Pearle 86 a]} (stochastic dynamical
reduction theories and superluminal communication), {\bf [Squires
92 b]} (explicit collapse and superluminal signals), {\bf [Peres
95 a, 00 b]}, {\bf [Garuccio 96]}, {\bf [Svetlichny 98]} (quantum
formalism with state-collapse and superluminal communication),
{\bf [Aharonov-Reznik-Stern 98]} (quantum limitations on
superluminal propagation), {\bf [Mittelstaedt 98]} (can
EPR-correlations be used for the transmission of superluminal
signals?), {\bf [Westmoreland-Schumacher 98]} (entanglement and
the nonexistence of superluminal signals; comments: {\bf
[Mashkevich 98 b]}, {\bf [van Enk 98]}), {\bf [Shan 99]} (quantum
superluminal communication does not result in the causal loop),
{\bf [Aharonov-Vaidman 01]}, {\bf [Svozil 01]}, {\bf
[Zbinden-Brendel-Tittel-Gisin 01]} (experimental test of
relativistic quantum state collapse with moving reference frames),
{\bf [Buhrman-Massar 04]} (any correlations more ``non local''
than those achievable in an EPR-Bell type experiment necessarily
allow generation of entanglement; in {\bf
[Bennett-Harrow-Leung-Smolin 03]} it is shown that any unitary
that can generate entanglement necessarily also allows signaling).


\subsection{Quantum cosmology}


{\bf [Clarke 74]} (quantum theory and cosmology),
{\bf [Hartle-Hawking 83]} (the wave function of the universe),
{\bf [Tipler 86]} (the many-worlds interpretation of quantum
mechanics in quantum cosmology),
{\bf [Hawking 87]},
{\bf [S\'{a}nchez G\'{o}mez 96]},
{\bf [Percival 98 b]} (cosmic quantum measurement).


\newpage




\section{Bibliography}


\begin{enumerate}


\subsection{}


\item {\bf [Aaberge 94]}:
T. Aaberge,
``On the size of the state-space for systems of quantum particles with spin'',
{\em Helv. Phys. Acta} {\bf 67}, 2, 127-143 (1994).

\item {\bf [Aaronson-Gottesman 04]}:
S. Aaronson, \& D. Gottesman,
``Improved simulation of stabilizer circuits'',
{\em Phys. Rev. A};
quant-ph/0406196.

\item {\bf [Abdel Aty-Abdalla-Obada 01]}:
M. Abdel-Aty, M. S. Abdalla, \& A.-S. F. Obada,
``Quantum information and entropy squeezing of a
two-level atom with a non-linear medium'',
{\em J. Phys. A} {\bf 34}, 43, 9129-9142 (2001).
Erratum: {\em J. Phys. A} {\bf 34}, 47, 10333 (2001).

\item {\bf [Abe-Rajagopal 99]}:
S. Abe, \& A. K. Rajagopal,
``Quantum entanglement inferred by the principle of maximum nonadditive entropy'',
{\em Phys. Rev. A} {\bf 60}, 5, 3461-3466 (1999);
quant-ph/9904088.

\item {\bf [Abe-Rajagopal 00]}:
S. Abe, \& A. K. Rajagopal,
``Towards nonadditive quantum information theory'',
{\em Int. Workshop on Classical and Quantum Complexity and
Nonextensive Thermodynamics (Denton, Texas, 2000)};
quant-ph/0003145.

\item {\bf [Abe-Rajagopal 01]}:
S. Abe, \& A. K. Rajagopal,
``Nonadditive conditional entropy and its significance for local realism'',
{\em Physica A} {\bf 289}, 1-2, 157-164 (2001);
quant-ph/0001085.

\item {\bf [Abe 02]}:
S. Abe,
``Nonadditive information measure and quantum entanglement in a class of mixed
states of an $N^n$ system'',
{\em Phys. Rev. A} {\bf 65}, 5, 052323 (2002);
quant-ph/0104135.

\item {\bf [{\AA}berg 03 a]}:
J. {\AA}berg,
``Subspace preserving completely positive maps'',
quant-ph/0302180.

\item {\bf [{\AA}berg 03 b]}:
J. {\AA}berg,
``Subspace local quantum channels'',
quant-ph/0302181.

\item {\bf [{\AA}berg 03 c]}:
J. {\AA}berg,
``Gluing of completely positive maps'',
quant-ph/0302182.

\item {\bf [{\AA}berg 04]}:
J. {\AA}berg,
``Operations and single-particle interferometry'',
{\em Phys. Rev. A} {\bf 70}, 1, 012103 (2004).

\item {\bf [Abbott 03]}:
D. Abbott,
``Dreams versus reality: Plenary debate session on quantum computing'',
quant-ph/0310130.

\item {\bf [Abe 03]}:
S. Abe,
``Nonadditive generalization of the quantum Kullback-Leibler divergence for
measuring the degree of purification'',
{\em Phys. Rev. A} {\bf 68}, 3, 032302 (2003).

\item {\bf [Abeyesinghe-Hayden 03]}:
A. Abeyesinghe, \& P. Hayden,
``Generalized remote state preparation:
Trading cbits, qubits, and ebits in quantum communication'',
{\em Phys. Rev. A} {\bf 68}, 6, 062319 (2003);
quant-ph/0308143.

\item {\bf [Abeyesinghe-Hayden-Smith-Winter 04]}:
A. Abeyesinghe, P. Hayden, G. Smith, \& A. Winter,
``Optimal superdense coding of entangled states'',
quant-ph/0407061.

\item {\bf [Abdel Aty 03]}:
M. Abdel-Aty,
``An investigation of entanglement and quasiprobability distribution in a
generalized Jaynes-Cummings model'',
{\em J. Math. Phys.} {\bf 44}, ?, 1457-? (2003).

\item {\bf [Abolhasani-Golshani 00]}:
M. Abolhasani, \& M. Golshani,
``Tunneling times in the Copenhagen interpretation
of quantum mechanics'',
{\em Phys. Rev. A} {\bf 62}, 1, 012106 (2000);
quant-ph/9906047.

\item {\bf [Abouraddy-Nasr-Saleh-(+2) 01]}:
A. F. Abouraddy, M. B. Nasr, B. E. A. Saleh,
A. V. Sergienko, \& M. C. Teich,
``Demonstration of the complementarity of one-
and two-photon interference'',
{\em Phys. Rev. A} {\bf 63}, 6, 063803 (2001);
quant-ph/0112065.

\item {\bf [Abouraddy-Saleh-Sergienko-Teich 01 a]}:
A. F. Abouraddy, B. E. A. Saleh, A. V. Sergienko, \& M. C. Teich,
``Degree of entanglement for two qubits'',
{\em Phys. Rev. A} {\bf 64}, 5, 050101(R) (2001);
quant-ph/0109081.

\item {\bf [Abouraddy-Saleh-Sergienko-Teich 01 b]}:
A. F. Abouraddy, B. E. A. Saleh, A. V. Sergienko, \& M. C. Teich,
``Role of entanglement in two-photon imaging'',
{\em Phys. Rev. Lett.} {\bf 87}, 12, 123602 (2001);
quant-ph/0108124.

\item {\bf [Abouraddy-Sergienko-Saleh-Teich 01]}:
A. F. Abouraddy, A. V. Sergienko, B. E. A. Saleh, \& M. C. Teich,
``Quantum entanglement and the two-photon stokes parameters'',
{\em Opt. Comm.};
quant-ph/0110172.

\item {\bf [Abouraddy-Saleh-Sergienko-Teich 01]}:
A. F. Abouraddy, B. E. A. Saleh, A. V. Sergienko, \& M. C. Teich,
``Entangled-photon Fourier optics'',
{\em J. Opt. Soc. Am. B};
quant-ph/0111054.

\item {\bf [Abouraddy-Toussaint-Sergienko-(+2) 02]}:
A. F. Abouraddy, K. C. Toussaint, Jr., A. V. Sergienko,
B. E. A. Saleh, \& M. C. Teich,
``Entangled-photon ellipsometry'',
{\em J. Opt. Soc. Am. B};
quant-ph/0202088.

\item {\bf [Abouraddy-Stone-Sergienko-(+2) 03]}:
A. F. Abouraddy, P. R. Stone, A. V. Sergienko,
B. E. A. Saleh, \& M. C. Teich,
``Entangled-photon imaging of a pure phase object'',
quant-ph/0311147.

\item {\bf [Abrams-Lloyd 97]}:
D. S. Abrams, \& S. Lloyd,
``Simulation of many-body Fermi systems on a universal quantum computer'',
{\em Phys. Rev. Lett.} {\bf 79}, 13, 2586-2589 (1997);
quant-ph/9703054.
See {\bf [Wiesner 96]}.

\item {\bf [Abrams-Lloyd 98]}:
D. S. Abrams, \& S. Lloyd,
``Nonlinear quantum mechanics implies polynomial-time solution for
$NP$-complete and $\#P$ problems'',
{\em Phys. Rev. Lett.} {\bf 81}, 18, 3992-3995 (1998);
quant-ph/9801041.

\item {\bf [Abrams-Lloyd 99]}:
D. S. Abrams, \& S. Lloyd,
``Quantum algorithm providing exponential speed increase
for finding eigenvalues and eigenvectors'',
{\em Phys. Rev. Lett.} {\bf 83}, 24, 5162-5165 (1999);
quant-ph/9807070.

\item {\bf [Abrams-Williams 99]}:
D. S. Abrams, \& C. P. Williams,
``Fast quantum algorithms for numerical integrals and stochastic
processes'',
quant-ph/9908083.

\item {\bf [Abramsky-Coecke 03]}:
S. Abramsky, \& B. Coecke,
``Physical traces: Quantum vs. classical information processing'',
{\em Electronic Notes in Theor. Comp. Sci.} {\bf 69}, ?-? (2003);
cs.CG/0207057.

\item {\bf [Abranyos-Jakob-Bergou 99]}:
Y. Abranyos, M. Jakob, \& J. Bergou,
``Quantum eraser and the decoherence time of a measurement process'',
{\em Phys. Rev. A} {\bf 60}, 4, R2618-R2621 (1999).

\item {\bf [Accardi-Regoli 00 a]}:
L. Accardi, \& M. Regoli,
``Locality and Bell's inequality'',
quant-ph/0007005.

\item {\bf [Accardi-Regoli 00 b]}:
L. Accardi, \& M. Regoli,
``Non-locality and quantum theory: New experimental evidence'',
quant-ph/0007019.

\item {\bf [Accardi-Regoli 00]}:
L. Accardi, \& R. Sabbadini,
``A generalization of Grover's algorithm'',
quant-ph/0012143.

\item {\bf [Accardi-Regoli 01]}:
L. Accardi, \& M. Regoli,
``The EPR correlations and the chameleon effect'',
quant-ph/0110086.

\item {\bf [Accardi-Imafuku-Regoli 01]}:
L. Accardi, K. Imafuku, \& M. Regoli,
``On the physical meaning of the EPR--chameleon experiment'',
quant-ph/0112067.

\item {\bf [Achilles-Silberhorn-Sliwa-(+2) 03]}:
D. Achilles, C. Silberhorn, C. Sliwa,
K. Banaszek, \& I. A. Walmsley,
``Fiber-assisted detection with photon number resolution'',
quant-ph/0305191.

\item {\bf [Achilles-Silberhorn-Sliwa-(+6) 03]}:
D. Achilles, C. Silberhorn, C. Sliwa,
K. Banaszek, I. A. Walmsley, M. J. Fitch, B. C. Jacobs,
T. B. Pittman, \& J. D. Franson,
``Photon number resolving detection using time-multiplexing'',
quant-ph/0310183.

\item {\bf [Ac\'{\i}n-Latorre-Pascual 00]}:
A. Ac\'{\i}n, J. I. Latorre, \& P. Pascual,
``Optimal generalized quantum measurements
for arbitrary spin systems'',
{\em Phys. Rev. A} {\bf 61}, 2, 022113 (2000).

\item {\bf [Ac\'{\i}n-Andrianov-Costa-(+3) 00]}:
A. Ac\'{\i}n, A. Andrianov, L. Costa, E. Jan\'{e},
J. I. Latorre, \& R. Tarrach,
``Schmidt decomposition and classification of three-quantum-bit
states'',
{\em Phys. Rev. Lett.} {\bf 85}, 7, 1560-1563 (2000);
quant-ph/0003050.

\item {\bf [Ac\'{\i}n-Tarrach-Vidal 00]}:
A. Ac\'{\i}n, R. Tarrach, \& G. Vidal,
``Optimal estimation of two-qubit pure-state entanglement'',
{\em Phys. Rev. A} {\bf 61}, 6, 062307 (2000);
quant-ph/9911008.

\item {\bf [Ac\'{\i}n-Jan\'{e}-D\"{u}r-Vidal 00]}:
A. Ac\'{\i}n, E. Jan\'{e}, W. D\"{u}r, \& G. Vidal,
``Optimal distillation of a GHZ state'',
{\em Phys. Rev. Lett.} {\bf 85}, 22, 4811-4814 (2000);
quant-ph/0007042.

\item {\bf [Ac\'{\i}n-Latorre-Pascual 01]}:
A. Ac\'{\i}n, J. I. Latorre, \& P. Pascual,
``Three-party entanglement from positronium'',
{\em Phys. Rev. A} {\bf 63}, 4, 042107 (2001);
quant-ph/0007080.

\item {\bf [Ac\'{\i}n-Jan\'{e}-Vidal 01]}:
A. Ac\'{\i}n, E. Jan\'{e}, \& G. Vidal,
``Optimal estimation of quantum dynamics'',
{\em Phys. Rev. A} {\bf 64}, 5, 050302(R) (2001);
quant-ph/0012015.

\item {\bf [Ac\'{\i}n 01 a]}:
A. Ac\'{\i}n,
``Entrelazado cu\'{a}ntico y estimaci\'{o}n de estados
cu\'{a}nticos. Quantum entanglement and quantum state
estimation'',
Ph.\ D. thesis, Universitat de Barcelona, 2001.
See {\bf [Ac\'{\i}n-Andrianov-Costa-(+3) 00]},
{\bf [Ac\'{\i}n-Andrianov-Jan\'{e}-Tarrach 01]},
{\bf [Ac\'{\i}n-Latorre-Pascual 00, 01]},
{\bf [Ac\'{\i}n-Bru\ss-Lewenstein-Sanpera 01]},
{\bf [Ac\'{\i}n-Tarrach-Vidal 00]}.

\item {\bf [Ac\'{\i}n-Bru\ss-Lewenstein-Sanpera 01]}:
A. Ac\'{\i}n, D. Bru\ss, M. Lewenstein, \& A. Sanpera,
``Classification of mixed three-qubit states'',
{\em Phys. Rev. Lett.} {\bf 87}, 4, 040401 (2001);
quant-ph/0103025.

\item {\bf [Ac\'{\i}n-Andrianov-Jan\'{e}-Tarrach 01]}:
A. Ac\'{\i}n, A. Andrianov, E. Jan\'{e}, \& R. Tarrach,
``Three-qubit pure-state canonical forms'',
in S. Popescu, N. Linden, \& R. Jozsa (eds.),
{\em J. Phys. A} {\bf 34}, 35
(Special issue: Quantum information and computation), 6725-6239 (2001);
quant-ph/0009107.

\item {\bf [Ac\'{\i}n 01 b]}:
A. Ac\'{\i}n,
``Statistical distinguishability between unitary operations'',
{\em Phys. Rev. Lett.} {\bf 87}, 17, 177901 (2001);
quant-ph/0102064.

\item {\bf [Ac\'{\i}n-Scarani 01]}:
A. Ac\'{\i}n, \& V. Scarani,
``Violation of Bell's inequalities implies distillability for $N$ qubits'',
quant-ph/0112102.

\item {\bf [Ac\'{\i}n 02]}:
A. Ac\'{\i}n,
``Distillability, Bell inequalities, and multiparticle bound entanglement'',
{\em Phys. Rev. Lett.} {\bf 88}, 2, 027901 (2002);
quant-ph/0108029.

\item {\bf [Ac\'{\i}n-Durt-Gisin-Latorre 02]}:
A. Ac\'{\i}n, T. Durt, N. Gisin, \& J. I. Latorre,
``Quantum nonlocality in two three-level systems'',
{\em Phys. Rev. A} {\bf 65}, 5, 052325 (2002);
quant-ph/0111143.

\item {\bf [Ac\'{\i}n-Vidal-Cirac 02]}:
A. Ac\'{\i}n, G. Vidal, \& J. I. Cirac,
``On the structure of a reversible entanglement
generating set for three--partite states'',
quant-ph/0202056.

\item {\bf [Ac\'{\i}n-Scarani-Wolf 02]}:
A. Ac\'{\i}n, V. Scarani, \& M. M. Wolf,
``Bell's inequalities and distillability in $N$-quantum-bit systems'',
{\em Phys. Rev. A} {\bf 66}, 4, 042323 (2002);
quant-ph/0206084.

\item {\bf [Ac\'{\i}n-Scarani-Wolf 03]}:
A. Ac\'{\i}n, V. Scarani, \& M. M. Wolf,
``Violation of Bell's inequalities and distillability for $N$ qubits'',
{\em J. Phys. A} {\bf 36}, 2, L21-L29 (2003).

\item {\bf [Ac\'{\i}n-Gisin-Scarani 03]}:
A. Ac\'{\i}n, N. Gisin, \& V. Scarani,
``Security bounds in quantum cryptography using $d$-level systems'',
{\em Quant. Inf. Comp.} {\bf 3}, 6, 563-? (2003);
quant-ph/0303009.

\item {\bf [Ac\'{\i}n-Masanes-Gisin 03]}:
A. Ac\'{\i}n, L. Masanes, \& N. Gisin,
``Equivalence between two-qubit entanglement and secure key distribution'',
{\em Phys. Rev. Lett.} {\bf 91}, 16, 167901 (2003);
quant-ph/0303053.

\item {\bf [Ac\'{\i}n-Gisin-Scarani 04]}:
A. Ac\'{\i}n, N. Gisin, \& V. Scarani,
``Coherent-pulse implementations of quantum cryptography protocols resistant to photon-number-splitting attacks'',
{\em Phys. Rev. A} {\bf 69}, 1, 012309 (2004);
quant-ph/0302037.

\item {\bf [Ac\'{\i}n-Gisin 03]}:
A. Ac\'{\i}n, \& N. Gisin,
``Quantum correlations and secret bits'',
quant-ph/0310054.

\item {\bf [Ac\'{\i}n-Gisin-Masanes-Scarani 03]}:
A. Ac\'{\i}n, N. Gisin, L. Masanes, \& V. Scarani,
``Bell's inequalities detect efficient entanglement
{\em Proc.\ of EQIS'03 (Kyoto, 2003)};
quant-ph/0310166.

\item {\bf [Ac\'{\i}n-Cirac-Masanes 04]}:
A. Ac\'{\i}n, J. I. Cirac, \& L. Masanes,
``Multipartite bound information exists and can be activated'',
{\em Phys. Rev. Lett.} {\bf 92}, 10, 107903 (2004);
quant-ph/0311064.

\item {\bf [Ac\'{\i}n-Chen, N. Gisin-(+4) 04]}:
A. Ac\'{\i}n, J. L. Chen, N. Gisin,
D. Kaszlikowski, L. C. Kwek, C. H. Oh, \& M. \.{Z}ukowski,
``Coincidence Bell inequality for three three-dimensional systems'',
{\em Phys. Rev. Lett.} {\bf 92}, 25, 250404 (2004).

\item {\bf [Ac\'{\i}n-Bagan-Baig-(+2) 04]}:
A. Ac\'{\i}n, E. Bagan, M. Baig,
L. Masanes, \& R. Mu\~{n}oz-Tapia,
``Bayesian approach to multiple copy 2-state discrimination'',
quant-ph/0410097.

\item {\bf [Ac\'{\i}n-Bae-Bagan-(+3) 04]}:
A. Ac\'{\i}n, J. Bae, E. Bagan,
M. Baig, L. Masanes, \& R. Mu\~{n}oz-Tapia,
``Secrecy content of two-qubit states'',
quant-ph/0411092.

\item {\bf [Aczel 02]}:
A. D. Aczel,
{\em Entanglement: The greatest mystery in physics},
Four Walls Eight Windows, New York, 2002.
Spanish version:
{\em Entrelazamiento: El mayor misterio de la f\'{\i}sica},
Cr\'{\i}tica, Barcelona, 2004.
Reviews: {\bf [Vedral 02 b]}, {\bf [van Dam 03 a]}.

\item {\bf [Adami-Cerf 97]}:
C. Adami, \& N. J. Cerf,
``Von Neumann capacity of noisy quantum channels'',
{\em Phys. Rev. A} {\bf 56}, 5, 3470-3483 (1997).

\item {\bf [Adami-Cerf 98 a]}:
C. Adami, \& N. J. Cerf,
``What information theory can tell us about quantum reality'',
in C. P. Williams (ed.),
{\em 1st NASA Int.\ Conf.\ on Quantum Computing and Quantum Communications
(Palm Springs, California, 1998)},
{\em Lecture Notes in Computer Science} {\bf 1509},
Springer-Verlag, New York, 1999, pp.~?-?;
quant-ph/9806047.

\item {\bf [Adami-Cerf 98 b]}:
C. Adami, \& N. J. Cerf,
``Quantum computation with linear optics'',
in C. P. Williams (ed.),
{\em 1st NASA Int.\ Conf.\ on Quantum Computing and Quantum Communications
(Palm Springs, California, 1998)},
{\em Lecture Notes in Computer Science} {\bf 1509},
Springer-Verlag, New York, 1999, pp.~?-?;
quant-ph/9806048.

\item {\bf [Adami-Dowling 02]}:
C. Adami, \& J. P. Dowling,
``Quantum computation--The ultimate frontier'',
in P. Kervin, L. Bragg, \& S. Ryan (eds.),
{\em Proc.\ AMOS 2001 Technical Conf.\ (Wailea, Maui, Hawaii, 2002)};
quant-ph/0202039.

\item {\bf [Adami 04]}:
C. Adami,
``The physics of information'',
quant-ph/0405005.

\item {\bf [Adamyan-Kryuchkyan 04]}:
H. H. Adamyan, \& G. Y. Kryuchkyan,
``Continuous variable entanglement of phase locked light beams'',
{\em Phys. Rev. A} {\bf 69}, 5, 053814 (2004);
quant-ph/0406128.

\item {\bf [Adcock-Cleve 01]}:
M. Adcock, \& R. Cleve,
``A quantum Goldreich-Levin theorem with cryptographic applications'',
quant-ph/0108095.

\item {\bf [Adenier 00]}:
G. Adenier,
``Refutation of Bell's theorem'',
quant-ph/0006014.

\item {\bf [Adenier 01]}:
G. Adenier,
``Representation of joint measurement in quantum mechanics.
A refutation of quantum teleportation'',
quant-ph/0105031.

\item {\bf [Adenier-Khrennikov 03]}:
G. Adenier, \& A. Y. Khrennikov,
``Experimental scheme to test the fair sampling assumption in EPR-Bell experiments'',
quant-ph/0309010.

\item {\bf [Adesso-Illuminati-De Siena 03]}:
G. Adesso, F. Illuminati, \& S. De Siena,
``Characterizing entanglement with global and marginal entropic measures'',
{\em Phys. Rev. A} {\bf 68}, 6, 062318 (2003);
quant-ph/0307192.

\item {\bf [Adesso-Serafini-Illuminati 04]}:
G. Adesso, A. Serafini, \& F. Illuminati,
``Determination of continuous variable entanglement by purity measurements'',
{\em Phys. Rev. Lett.} {\bf 92}, 8, 087901 (2004);
quant-ph/0310150.

\item {\bf [Adesso-Serafini-Illuminati 04 a]}:
G. Adesso, A. Serafini, \& F. Illuminati,
``Extremal entanglement and mixedness in continuous variable systems'',
{\em Phys. Rev. A};
quant-ph/0402124.

\item {\bf [Adesso-Serafini-Illuminati 04 b]}:
G. Adesso, A. Serafini, \& F. Illuminati,
``Quantification and scaling of multipartite entanglement in continuous
variable systems'',
{\em Phys. Rev. Lett.};
quant-ph/0406053.

\item {\bf [Adesso-Illuminati 04]}:
G. Adesso, \& F. Illuminati,
``Multipartite entanglement and its polygamy in continuous variable
systems'',
quant-ph/0410050.

\item {\bf [Adler 00]}:
S. L. Adler,
``Probability in orthodox quantum mechanics: Probability as a
postulate versus probability as an emergent phenomenon'',
submitted to {\em Proc.\ of the Ischia Conf.\ on ``Chance in Physics:
Foundations and Perspectives''};
quant-ph/0004077.

\item {\bf [Adler-Brun 01]}:
S. L. Adler, \& Todd A. Brun,
``Generalized stochastic Schr\"{o}dinger equations for state vector collapse'',
{\em J. Phys. A} {\bf 34}, 23, 4797-4809 (2001);
quant-ph/0103037.

\item {\bf [Adler-Brody-Brun-Hughston 01]}:
S. L. Adler, D. C. Brody, T. A. Brun, \& L. P. Hughston,
``Martingale models for quantum state reduction'',
{\em J. Phys. A} {\bf 34}, 42, 8795-8820 (2001);
quant-ph/0107153.

\item {\bf [Adler 01 a]}:
S. L. Adler,
``Environmental influence on the measurement process
in stochastic reduction models'',
quant-ph/0109029.

\item {\bf [Adler 01 b]}:
S. L. Adler,
``Why decoherence has not solved the measurement problem:
A response to P. W. Anderson'',
quant-ph/0112095.

\item {\bf [Aerts 95]}:
D. Aerts,
``Quantum structures: An attempt to explain
the origin of their appearance in nature'',
{\em Int. J.Theor. Phys.} {\bf 34}, ?, 1165-? (1995);
quant-ph/0111071.

\item {\bf [Aerts-Coecke-D'Hooghe-Valckenborgh 97]}:
D. Aerts, B. Coecke, B. D'Hooghe, \& F. Valckenborgh,
``A mechanistic macroscopic physical entity
with a three-dimensional Hilbert space description'',
{\em Helv. Phys. Acta} {\bf 70}, ?, 793-? (1997);
quant-ph/0111074.

\item {\bf [Aerts 98]}:
D. Aerts,
``The hidden measurement formalism: What can be explained and where
quantum paradoxes remain'',
{\em Int. J. Theor. Phys.} {\bf 37}, 1, 291-304 (1998).

\item {\bf [Aerts-Colebunders-Van der Voorde-Van Steirteghem 99]}:
D. Aerts, E. Colebunders, A. Van der Voorde, \& B. Van Steirteghem,
``State property systems and closure spaces: A
study of categorical equivalence'',
{\em Int. J. Theor. Phys.} {\bf 39}, ?, 259-? (1999);
quant-ph/0105108.

\item {\bf [Aerts 99 a]}:
D. Aerts,
``Foundations of quantum physics: A general
realistic and operational approach'',
{\em Int. J. Theor. Phys.} {\bf 39}, ?, 289-? (1999);
quant-ph/0105109.

\item {\bf [Aerts 99 b]}:
D. Aerts,
``The description of joint quantum entities and
the formulation of a paradox'',
{\em Int. J. Theor. Phys.} {\bf 39}, ?, 483-? (1999);
quant-ph/0105106.

\item {\bf [Aerts-Van Steirteghem 99]}:
D. Aerts, \& B. Van Steirteghem,
``Quantum axiomatics and a theorem of M.P. Soler'',
{\em Int. J. Theor. Phys.} {\bf 39}, ?, 497-? (1999);
quant-ph/0105107.

\item {\bf [Aerts-Broekaert-Smets 99]}:
D. Aerts, J. Broekaert, \& S. Smets,
``A quantum structure description of the liar paradox'',
{\em Int. J. Theor. Phys.} {\bf 38}, ?, 3231-3239 (1999);
quant-ph/0106131.

\item {\bf [Aerts 99 c]}:
D. Aerts,
``The stuff the world is made of: Physics and reality'',
in D. Aerts, J. Broekaert, \& E. Mathijs (eds.),
{\em Einstein meets Magritte: An interdisciplinary reflection},
Kluwer Academic, Dordrecht, Holland, 1999;
quant-ph/0107044.

\item {\bf [Aerts-Aerts-Broekaert-Gabora 00]}:
D. Aerts, S. Aerts, J. Broekaert, \& L. Gabora,
``The violation of Bell inequalities in the macroworld'',
{\em Found. Phys.} {\bf 30}, 10, 1387-1414 (2000);
quant-ph/0007044.

\item {\bf [Aerts-D'Hondt-Gabora 00]}:
D. Aerts, E. D'Hondt, \& L. Gabora,
``Why the disjunction in quantum logic is not classical'',
{\em Found. Phys.} {\bf 30}, 10, 1473-1480 (2000);
quant-ph/0007041.

\item {\bf [Aerts 01]}:
D. Aerts,
``Quantum mechanics: Structures, axioms and paradoxes'',
in D. Aerts, \& J. Pykacz (eds.),
{\em Quantum structures and the nature of
reality: the indigo book of the Einstein meets Magritte series},
Kluwer Academic, Dordrecht, Holland, 2001;
quant-ph/0106132.

\item {\bf [Aerts-Deses 04]}:
D. Aerts, \& D. Deses,
``State property systems and closure spaces: Extracting the classical and
nonclassical parts'',
in D. Aerts, M. Czachor, \& T. Durt,
{\em Probing the structure of quantum
mechanics: Nonlinearity, nonlocality, probability and axiomatics},
World Scientific, Singapore, 2004;
quant-ph/0404070.

\item {\bf [Aerts-Deses-Van der Voorde 04]}:
D. Aerts, D. Deses, \& A. Van der Voorde,
``Classicality and connectedness for state property systems and closure
spaces'',
{\em Int. J. Theor. Phys.};
quant-ph/0404071.

\item {\bf [Aerts-Kwiat-Larsson-\.{Z}ukowski 99]}:
S. Aerts, P. G. Kwiat, J.-\AA. Larsson, \& M. \.{Z}ukowski,
``Two-photon Franson-type experiments and local realism'',
{\em Phys. Rev. Lett.} {\bf 83}, 15, 2872-2875 (1999);
quant-ph/9912064.
Comment: {\bf [Ryff 01]}.
Reply: {\bf [Aerts-Kwiat-Larsson-\.{Z}ukowski 01]}.

\item {\bf [Aerts-Kwiat-Larsson-\.{Z}ukowski 01]}:
S. Aerts, P. G. Kwiat, J.-\AA. Larsson, \& M. \.{Z}ukowski,
``Reply: Aerts {\em et al.}'',
{\em Phys. Rev. Lett.} {\bf 86}, 9, 1909 (2001).
Reply to {\bf [Ryff 01]}.
See {\bf [Aerts-Kwiat-Larsson-\.{Z}ukowski 99]}.

\item {\bf [Aerts 01]}:
D. Aerts,
``The hidden measurement formalism: what can be
explained and where quantum paradoxes remain'',
{\em Int. J. Theor. Phys.} {\bf 37}, ?, 291-? (2001);
quant-ph/0105126.

\item {\bf [Afriat-Selleri 99]}:
A. Afriat, \& F. Selleri,
{\em The Einstein, Podolsky, and Rosen paradox in atomic, nuclear, and particle physics},
Plenum Press, New York, 1999.

\item {\bf [Afriat 03 a]}:
A. Afriat,
``Bell's inequality with times rather than angles'',
in M. Ferrero (ed.),
{\em Proc. of Quantum Information: Conceptual Foundations,
Developments and Perspectives (Oviedo, Spain, 2002)},
{\em J. Mod. Opt.} {\bf 50}, 6-7, 1063-1069 (2003).

\item {\bf [Afriat 03 b]}:
A. Afriat,
``Altering the remote past'',
{\em Found. Phys. Lett.} {\bf 16}, 3, 293-301 (2003).

\item {\bf [Agaian-Klappenecker 02]}:
S. S. Agaian, \& A. Klappenecker,
``Quantum computing and a unified approach to fast unitary transforms'',
{\em Image Processing: Algorithms and Systems, Electronic Imaging 2002},
San Jose, SPIE, 2002.

\item {\bf [Agarwal-Home-Schleich 92]}:
G. S. Agarwal, D. Home, \& W. Schleich,
``EPR correlation---parellelism between the Wigner function and the local
hidden variable approaches'',
{\em Phys. Lett. A} {\bf 170}, 5, 359-362 (1992).

\item {\bf [Agarwal 93]}:
G. S. Agarwal,
``Perspective of Einstein-Podolsky-Rosen spin correlations in the
phase-space formulation for arbitrary values of the spin'',
{\em Phys. Rev. A} {\bf 47}, 6, 4608-4615 (1993).

\item {\bf [Agarwal-Tewori 94]}:
G. S. Agarwal, \& S. P. Tewori,
``An all-optical realization of the quantum Zeno effect'',
{\em Phys. Lett. A} {\bf 185}, 2, 139-142 (1994).

\item {\bf [Agarwal 98]}:
G. S. Agarwal,
``State reconstruction for a collection of two-level systems'',
{\em Phys. Rev. A} {\bf 57}, 1, 671-673 (1998).

\item {\bf [Agarwal-Scully-Walther 01]}:
G. S. Agarwal, M. O. Scully, \& H. Walther,
``Inhibition of decoherence due to decay in a continuum'',
quant-ph/0101014.

\item {\bf [Agarwal-Ou 01]}:
G. S. Agarwal, \& Z. Y. Ou,
``Obituary: Leonard Mandel (1927-2001)'',
{\em Nature} {\bf 410}, 6828, 538 (2001).

\item {\bf [Agarwal-Puri-Singh 01]}:
G. S. Agarwal, R. R. Puri, \& R. P. Singh,
`Comment on ``One-step synthesis of multiatom Greenberger-Horne-Zeilinger states''
[Shi-Biao Zheng, Phys. Rev. Lett. {\bf 87}, 230404 (2001)]',
quant-ph/0112163.
Comment on {\bf [Zheng 01]}.

\item {\bf [Agarwal 02]}:
G. S. Agarwal,
``Heisenberg's uncertainty relations and quantum optics'',
{\em 100 Years of Werner Heisenberg--Works and Impact (Bamberg, Germany, 2001)},
{\em Fortschr. Phys.};
quant-ph/0201098.

\item {\bf [Agarwal-von Zanthier-Skornia-Walther 02]}:
G. S. Agarwal, J. von Zanthier, C. Skornia, \& H. Walther,
``Intensity-intensity correlations as a probe of interferences under
conditions of noninterference in the intensity'',
{\em Phys. Rev. A} {\bf 65}, 5, 053826 (2002).

\item {\bf [Agarwal-Pathak-Scully 02]}:
G. S. Agarwal, P. K. Pathak, \& M. O. Scully,
``Single-atom and two-atom Ramsey interferometry with quantized fields'',
{\em Phys. Rev. A} {\bf 67}, 4, 043807 (2003).

\item {\bf [Agarwal-Ariunbold-Zanthier-Walther 04]}:
G. S. Agarwal, G. O. Ariunbold, J. V. Zanthier, \& H. Walther,
``Nonclassical imaging for a quantum search of ions in a linear trap'',
quant-ph/0401141.

\item {\bf [Agarwal-Lougovski-Walther 04]}:
G. S. Agarwal, P. Lougovski, \& H. Walther,
``Quantum holography and multiparticle entanglement using ground-state
coherences'',
quant-ph/0402116.

\item {\bf [Agrawal-Pati 02]}:
P. Agrawal, \& A. K. Pati,
``Probabilistic quantum teleportation'',
{\em Phys. Lett. A} {\bf 305}, 1-2, 12-17 (2002).

\item {\bf [Agrawal-Parashar-Pati 03]}:
P. Agrawal, P. Parashar, \& A. K. Pati,
``Exact remote state preparation for multiparties'',
quant-ph/0304006.

\item {\bf [Aharonov-Kitaev-Nisan 98]}:
D. Aharonov, A. Y. Kitaev, \& N. Nisan,
``Quantum circuits with mixed states'',
in {\em Proc.\ of the 30th Annual ACM Symp.\ on the Theory of
Computing (El Paso, Texas, 1998)}, ACM Press, New York, 1998,
pp.~20-30;
quant-ph/9806029.

\item {\bf [Aharonov-Ben Or 99]}:
D. Aharonov, \& M. Ben-Or,
``Fault-tolerant quantum computation with constant error rate'',
submitted to {\em SIAM J. Comput.};
quant-ph/9906129.

\item {\bf [Aharonov-Ta Shma-Vazirani-Yao 00]}:
D. Aharonov, A. Ta-Shma, U. Vazirani, \& A. Yao,
``Quantum bit escrow'',
in
{\em Proc.\ of the 32nd Annual ACM
Symp.\ on Theory of Computing (2000)};
quant-ph/0004017.

\item {\bf [Aharonov 00]}:
D. Aharonov,
``Quantum to classical phase transition in noisy quantum computers'',
{\em Phys. Rev. A} {\bf 62}, 6, 062311 (2000).

\item {\bf [Aharonov-Ambainis-Kempe-Vazirani 01]}:
D. Aharonov, A. Ambainis, J. Kempe, \& U. Vazirani,
``Quantum walks on graphs'',
in {\em Proc. 33th STOC}, ACM, New York, 2001, 50-59.

\item {\bf [Aharonov-Ta Shma 03]}:
D. Aharonov, \& A. Ta-Shma,
``Adiabatic quantum state generation and statistical zero knowledge'',
quant-ph/0301023.

\item {\bf [Aharonov 03]}:
D. Aharonov,
``A simple proof that Toffoli and Hadamard are quantum universal'',
quant-ph/0301040.

\item {\bf [Aharonov-van Dam-Kempe-(+3) 04]}:
D. Aharonov, W. van Dam, J. Kempe,
Z. Landau, S. Lloyd, \& O. Regev,
``Adiabatic quantum computation is equivalent to standard quantum
computation'',
quant-ph/0405098.

\item {\bf [Aharonov-Bergman-Lebowitz 64]}:
Y. Aharonov, P. G. Bergman, \& J. L. Lebowitz,
``Time symmetry in the quantum process of measurement'',
{\em Phys. Rev.} {\bf 134}, 6B, B1410-B1418 (1964).
Reprinted in {\bf [Wheeler-Zurek 83]}, pp.~680-686.

\item {\bf [Aharonov-Albert 80]}:
Y. Aharonov, \& D. Z. Albert,
``States and observables in relativistic quantum field theories'',
{\em Phys. Rev. D} {\bf 21}, 12, 3316-3324 (1980).

\item {\bf [Aharonov-Albert-Au 81]}:
Y. Aharonov, D. Z. Albert, \& C. K. Au,
``New interpretation of the scalar product in Hilbert space'',
{\em Phys. Rev. Lett.} {\bf 47}, 15, 1029-1031 (1981).
Erratum: {\em Phys. Rev. Lett.} {\bf 47}, 24, 1765 (1981).

\item {\bf [Aharonov-Albert 81]}:
Y. Aharonov, \& D. Z. Albert,
``Can we make sense of the measurement process in
relativistic quantum mechanics?'',
{\em Phys. Rev. D} {\bf 24}, 2, 359-370 (1981).

\item {\bf [Aharonov-Albert 84]}:
Y. Aharonov, \& D. Z. Albert,
``Is the usual notion of time evolution adequate
for quantum-mechanical systems? II. Relativistic considerations'',
{\em Phys. Rev. D} {\bf 29}, 2, 228-234 (1984).

\item {\bf [Aharonov-Albert-Vaidman 86]}:
Y. Aharonov, D. Z. Albert, \& L. Vaidman,
``Measurement process in relativistic quantum theory'',
{\em Phys. Rev. D} {\bf 34}, 6, 1805-1813 (1986).

\item {\bf [Aharonov-Albert 87]}:
Y. Aharonov, \& D. Z. Albert,
``The issue of retrodiction in Bohm's theory'',
in {\bf [Hiley-Peat 87]}, pp.~224-226.

\item {\bf [Aharonov-Albert-Casher-Vaidman 87]}:
Y. Aharonov, D. Z. Albert, A. Casher, \& L. Vaidman,
``Surprising quantum effects'',
{\em Phys. Lett. A} {\bf 124}, 4-5, 199-203 (1987).
See {\bf [Busch 88]}.

\item {\bf [Aharonov-Albert-Vaidman 88]}:
Y. Aharonov, D. Z. Albert, \& L. Vaidman,
``How the result of a measurement of a component of the spin of a
spin-$\frac{1}{2}$ particle can turn out to be 100'',
{\em Phys. Rev. Lett.} {\bf 60}, 14, 1351-1354 (1988).
Comments: {\bf [Leggett 89]}, {\bf [Peres 89 a]}.
Reply: {\bf [Aharonov-Vaidman 89]}.
See {\bf [Duck-Stevenson-Sudarshan 89]}.

\item {\bf [Aharonov-Vaidman 89]}:
Y. Aharonov, \& L. Vaidman,
``The reply to Leggett and to Peres'',
{\em Phys. Rev. Lett.} {\bf 62}, 19, 2327 (1989).
Reply to {\bf [Leggett 89]}, {\bf [Peres 89 a]}.

\item {\bf [Aharonov-Anandan-Popescu-Vaidman 90]}:
Y. Aharonov, J. Anandan, S. Popescu, \& L. Vaidman,
``Superpositions of time evolutions of a quantum system
and a quantum time-translation machine'',
{\em Phys. Rev. Lett.} {\bf 64}, 25, 2965-2968 (1990).

\item {\bf [Aharonov-Vaidman 90]}:
Y. Aharonov, \& L. Vaidman,
``Properties of a quantum system during the time
interval between two measurements'',
{\em Phys. Rev. A} {\bf 41}, 1, 11-20 (1990).

\item {\bf [Aharonov-Vaidman 91]}:
Y. Aharonov, \& L. Vaidman,
``Complete description of a quantum system at a given time'',
{\em J. Phys. A} {\bf 24}, 10, 2315-2328 (1991).
See {\bf [Kirkpatrick 03 a]}.

\item {\bf [Aharonov-Anandan-Vaidman 93]}:
Y. Aharonov, J. Anandan, \& L. Vaidman,
``Meaning of the wave function'',
{\em Phys. Rev. A} {\bf 47}, 6, 4616-4626 (1993).

\item {\bf [Aharonov-Davidovich-Zagury 93]}:
Y. Aharonov, L. Davidovich, \& N. Zagury,
``Quantum random walks'',
{\em Phys. Rev. A} {\bf 48}, 2, 1687–1690 (1993).

\item {\bf [Aharonov-Vaidman 94]}:
Y. Aharonov, \& L. Vaidman,
``Protective measurements of two-state vectors'',
hep-th/9411196.

\item {\bf [Aharonov 95]}:
Y. Aharonov,
``A new formulation of quantum mechanics'',
in J. S. Anandan, \& J. L. Safko (eds.),
{\em Quantum coherence and reality.
In celebration of the 60th birthday of Yakir Aharonov.
Int.\ Conf.\
on Fundamental Aspects of Quantum Theory (?, ?)},
World Scientific, Singapore, 1995, pp.~?-?.

\item {\bf [Aharonov-Vaidman 95]}:
Y. Aharonov, \& L. Vaidman,
``Comment on `Time asymmetry in quantum mechanics:
A retrodiction paradox'\,'',
{\em Phys. Lett. A} {\bf 203}, 2-3, 148-149 (1995);
quant-ph/9501002.
See {\bf [Peres 94 a, 95 d]}.

\item {\bf [Aharonov-Popescu-Vaidman 95]}:
Y. Aharonov, S. Popescun, \& L. Vaidman,
``Causality, memory erasing, and delayed-choice experiments'',
{\em Phys. Rev. A} {\bf 52}, 6, 4984-4985 (1995);
quant-ph/9501006.
Comment on {\bf [Ingraham 94]}.

\item {\bf [Aharonov-Vaidman 96]}:
Y. Aharonov, \& L. Vaidman,
``About position measurements which do not show
the Bohmian particle position'',
in {\bf [Cushing-Fine-Goldstein 96]};
quant-ph/9511005.

\item {\bf [Aharonov-Anandan-Vaidman 96]}:
Y. Aharonov, J. Anandan, \& L. Vaidman,
``The meaning of protective measurements'',
in A. Mann, \& M. Revzen (eds.),
{\em The dilemma of Einstein, Podolsky and Rosen -- 60 years
later. An international symposium in honour of Nathan Rosen
(Haifa, Israel, 1995)},
{\em Ann. Phys. Soc. Israel} {\bf 12}, 296-304 (1996).

\item {\bf [Aharonov-Oppenheim-Popescu-(+2) 98]}:
Y. Aharonov, J. Oppenheim, S. Popescu,
B. Reznik, \& W. G. Unruh,
``Measurement of time-of-arrival in quantum mechanics'',
{\em Phys. Rev. A} {\bf 57}, 6, 4130-4139 (1998);
quant-ph/9709031.

\item {\bf [Aharonov-Anandan 98]}:
Y. Aharonov, \& J. Anandan,
``Meaning of the density matrix'',
quant-ph/9803018.
See {\bf [d'Espagnat 98 b]}.

\item {\bf [Aharonov-Vaidman 97]}:
Y. Aharonov, \& L. Vaidman,
``Protective measurements of two-state vectors'',
in {\bf [Cohen-Horne-Stachel 97 b]};
quant-ph/9602009.

\item {\bf [Aharonov-Reznik-Stern 98]}:
Y. Aharonov, B. Reznik, \& A. Stern,
``Quantum limitations on superluminal propagation'',
{\em Phys. Rev. Lett.} {\bf 81}, 11, 2190-2193 (1998);
quant-ph/9804018.

\item {\bf [Aharonov-Englert-Scully 99]}:
Y. Aharonov, B.-G. Englert, \& M. O. Scully,
``Protective measurements and Bohm trajectories'',
{\em Phys. Lett. A} {\bf 263}, 3, 137-146 (1999).
Erratum: {\em Phys. Lett. A} {\bf 266}, 2-3, 216-217 (2000).

\item {\bf [Aharonov-Reznik 00]}:
Y. Aharonov, \& B. Reznik,
`\,``Weighing'' a closed system and the time-energy
uncertainty principle',
{\em Phys. Rev. Lett.} {\bf 84}, 7, 1368-1370 (2000);
quant-ph/9906030.

\item {\bf [Aharonov-Vaidman 00]}:
Y. Aharonov, \& L. Vaidman,
``Nonlocal aspects of a quantum wave'',
{\em Phys. Rev. A} {\bf 61}, 5, 052108 (2000);
quant-ph/9909072.

\item {\bf [Aharonov-Englert 01]}:
Y. Aharonov, \& B.-G. Englert,
``The mean king's problem: Spin 1'',
{\em Zeitschrift f\"{u}r Naturforschung};
quant-ph/0101065.
See {\bf [Englert-Aharonov 01]}.

\item {\bf [Aharonov-Vaidman 01]}:
Y. Aharonov, \& L. Vaidman,
``Sending signals to space-like separated regions'',
{\em Mysteries and Paradoxes in Quantum Mechanics (Gargnano,
Italy, 2000)},
{\em Zeitschrift f\"{u}r Naturforschung};
quant-ph/0102083.

\item {\bf [Aharonov-Vaidman 01]}:
Y. Aharonov, \& L. Vaidman,
``The two-state vector formalism of quantum mechanics'',
in J. G. Muga, R. Sala Mayato, \& I. L. Egusquiza (eds.),
{\em Time in quantum mechanics},
to appear;
quant-ph/0105101.

\item {\bf [Aharonov-Reznik 01]}:
Y. Aharonov, \& B. Reznik,
``How macroscopic properties dictate microscopic probabilities'',
quant-ph/0110093.

\item {\bf [Aharonov-Erez-Reznik 01]}:
Y. Aharonov, N. Erez, \& B. Reznik,
``Superoscillations and tunneling times'',
quant-ph/0110104.

\item {\bf [Aharonov-Botero-Popescu-(+2) 02]}:
Y. Aharonov, A. Botero, S. Popescu,
B. Reznik, \& J. Tollaksen,
``Revisiting Hardy's Paradox: Counterfactual statements,
real measurements, entanglement and weak values'',
{\em Phys. Lett. A} {\bf 301}, 3-4, 130-138 (2002);
quant-ph/0104062.

\item {\bf [Aharonov-Massar-Popescu 02]}:
Y. Aharonov, S. Massar, \& S. Popescu,
``Measuring energy, estimating Hamiltonians, and the time-energy uncertainty
relation'',
{\em Phys. Rev. A} {\bf 66}, 5, 052107 (2002).

\item {\bf [Aharonov-Erez-Reznik 03]}:
Y. Aharonov, N. Erez, \& B. Reznik,
``Superluminal tunnelling times as weak values'',
in M. Ferrero (ed.),
{\em Proc. of Quantum Information: Conceptual Foundations,
Developments and Perspectives (Oviedo, Spain, 2002)},
{\em J. Mod. Opt.} {\bf 50}, 6-7, 1139-1149 (2003).

\item {\bf [Aharonov-Popescu-Reznik-Stern 04]}:
Y. Aharonov, S. Popescu, B. Reznik, \& A. Stern,
``Classical analog to topological nonlocal quantum interference effects'',
{\em Phys. Rev. Lett.} {\bf 92}, 2, 020401 (2004);
quant-ph/0311155.

\item {\bf [Ahlswede-Loeber 99]}:
R. Ahlswede, \& P. Loeber,
``Quantum data processing'',
quant-ph/9907081.

\item {\bf [Ahlswede-Winter 00]}:
R. Ahlswede, \& A. Winter,
``Strong converse for identification via quantum channels'',
quant-ph/0012127.

\item {\bf [Ahn-Doherty-Landahl 02]}:
C. Ahn, A. C. Doherty, \& A. J. Landahl,
``Continuous quantum error correction via quantum feedback control'',
{\em Phys. Rev. A} {\bf 65}, 4, 042301 (2002);
quant-ph/0110111.

\item {\bf [Ahn-Wiseman-Milburn 03]}:
C. Ahn, H. M. Wiseman, \& G. J. Milburn,
``Quantum error correction for continuously detected errors'',
{\em Phys. Rev. A} {\bf 67}, 5, 052310 (2003);
quant-ph/0302006.
See {\bf [Ahn-Wiseman-Jacobs 04]}.

\item {\bf [Ahn-Wiseman-Jacobs 04]}:
C. Ahn, H. M. Wiseman, \& K. Jacobs,
``Quantum error correction for continuously detected errors with any
number of error channels per qubit'',
quant-ph/0402067.
See {\bf [Ahn-Wiseman-Milburn 03]}.

\item {\bf [Ahn-Doherty-Hayden-Winter 04]}:
C. Ahn, A. Doherty, P. Hayden, \& A. Winter,
``On the distributed compression of quantum information''
quant-ph/0403042.

\item {\bf [Ahn-Oh-Kimm-Hwang 00]}:
D. Ahn, J. H. Oh, K. Kimm, \& S. W. Hwang,
``Time-convolutionless reduced-density-operator theory of a noisy
quantum channel: Two-bit quantum gate for quantum-information
processing'',
{\em Phys. Rev. A} {\bf 61}, 5, 052310 (2000);
quant-ph/9907074.

\item {\bf [Ahn-Lee-Kim-Hwang 02]}:
D. Ahn, J. Lee, M. S. Kim, \& S. W. Hwang,
``Self-consistent non-Markovian theory of a quantum-state evolution for
quantum-information processing'',
{\em Phys. Rev. A} {\bf 66}, 1, 012302 (2002).

\item {\bf [Ahn-Lee-Moon-Hwang 03]}:
D. Ahn, H.-J. Lee, Y. H. Moon, \& S. W. Hwang,
``Relativistic entanglement and Bell's inequality'',
{\em Phys. Rev. A} {\bf 67}, 1, 012103 (2003).

\item {\bf [Ahn-Lee-Hwang 03]}:
D. Ahn, H.-J. Lee, \& S. W. Hwang,
``Lorentz-covariant reduced-density-operator theory for
relativistic-quantum-information processing'',
{\em Phys. Rev. A} {\bf 67}, 3, 032309 (2003).

\item {\bf [Ahn-Weinacht-Bucksbaum 00]}:
J. Ahn, T. C. Weinacht, \& P. H. Bucksbaum,
``Information storage and retrieval
through quantum phase'',
{\em Science} {\bf 287}, 5452, 463-465 (2000).
Comments: {\bf [Meyer 00 c]}, {\bf [Kwiat-Hughes 00]}.
Reply: {\bf [Bucksbaum-Ahn-Weinacht 00]}.
See {\bf [Knight 00]}.

\item {\bf [Ahn-Rangan-Hutchinson-Bucksbaum 02]}:
J. Ahn, C. Rangan, D. N. Hutchinson, \& P. H. Bucksbaum,
``Quantum-state information retrieval in a Rydberg-atom data register'',
{\em Phys. Rev. A} {\bf 66}, 2, 022312 (2002).

\item {\bf [Ahnert-Payne 04 a]}:
S. E. Ahnert, \& M. C. Payne,
``Nonorthogonal projective positive-operator-value measurement of photon
polarization states with unit probability of success'',
{\em Phys. Rev. A} {\bf 69}, 1, 012312 (2004).

\item {\bf [Ahnert-Payne 04 b]}:
S. E. Ahnert, \& M. C. Payne,
``Weak measurement of the arrival times of single photons and pairs of
entangled photons'',
{\em Phys. Rev. A} {\bf 69}, 4, 042103 (2004);
quant-ph/0405156.

\item {\bf [Ahnert-Payne 04 c]}:
S. E. Ahnert, \& M. C. Payne,
``Linear optics implementation of weak values in Hardy's paradox'',
{\em Phys. Rev. A};
quant-ph/0408153.

\item {\bf [Ahnert-Payne 04 d]}:
S. E. Ahnert, \& M. C. Payne,
``A general implementation of all possible positive operator value
measurements of single photon polarization states'',
quant-ph/0408011.

\item {\bf [Ahuja-Kapoor 99]}:
A. Ahuja, \& S. Kapoor,
``A quantum algorithm for finding the maximum'',
quant-ph/9911082.

\item {\bf [Akguc-Reichl-Shaji-Snyder 04]}:
G. B. Akguc, L. E. Reichl, A. Shaji, \& M. G. Snyder,
``Bell states in a resonant quantum waveguide network'',
{\em Phys. Rev. A} {\bf 69}, 4, 042303 (2004).

\item {\bf [Akhavan-Rezakhani 03]}:
O. Akhavan, \& A. T. Rezakhani,
`Comment II on ``Dense coding in entangled states''\,',
{\em Phys. Rev. A} {\bf 68}, 1, 016302 (2003).
quant-ph/0306148.
Comment on {\bf [Lee-Ahn-Hwang 02]}.

\item {\bf [Akhtarshenas-Jafarizadeh 03]}:
S. J. Akhtarshenas, \& M. A. Jafarizadeh,
``Separability criterion induced from cross norm is not equivalent to positive partial transpose'',
{\em J. Phys. A} {\bf 36}, 5, 1509-1513 (2003).

\item {\bf [Akhavan-Rezakhani-Golshani 03]}:
O. Akhavan, A. T. Rezakhani, \& M. Golshani,
``Quantum dense coding by spatial state entanglement'',
{\em Phys. Lett. A} {\bf 313}, 4, 261-266 (2003).

\item {\bf [Akis-Ferry 01]}:
R. Akis, \& D. K. Ferry,
``Quantum waveguide array generator for performing Fourier transforms:
Alternate route to quantum computing'',
{\em Appl. Phys. Lett.} {\bf 79}, 17, 2823-? (2001).
Comment: {\bf [Lidar 02]}.
Reply: {\bf [Akis-Ferry 02]}.

\item {\bf [Akis-Ferry 02]}:
R. Akis, \& D. K. Ferry,
`Response to ``Comment on `Quantum waveguide array generator for performing
Fourier transforms: Alternate route to quantum computing'\,'' [Appl. Phys.
Lett. {\bf 80}, 2419 (2002).]',
{\em Appl. Phys. Lett.} {\bf 80}, 13, 2420-? (2002).
Reply to {\bf [Lidar 02]}.
See {\bf [Akis-Ferry 01]}.

\item {\bf [Akulin-Gershkovich-Harel 01]}:
V. M. Akulin, V. Gershkovich, \& G. Harel,
``Nonholonomic quantum devices'',
{\em Phys. Rev. A} {\bf 64}, 1, 012308 (2001).

\item {\bf [Aicardi-Borsellino-Ghirardi-Grassi 91]}:
F. Aicardi, A. Borsellino, G.-C. Ghirardi, \& R. Grassi,
``Dynamical models for state-vector reduction: Do they
ensure that measurements have outcomes?'',
{\em Found. Phys. Lett. A} {\bf 4}, 1, 109-128 (1991).

\item {\bf [Aichele-Lvovsky-Schiller 02]}:
T. Aichele, A. I. Lvovsky, \& S. Schiller,
``Optical mode characterization of single photons prepared
by means of conditional measurements on a biphoton state'',
{\em Eur. Phys. J. D} {\bf 18}, 2 (Special issue:
{\em Quantum interference and cryptographic keys:
Novel physics and advancing technologies (QUICK) (Corsica, 2001)}, 237-245 (2002).

\item {\bf [Aiello-De Martini-Giangrasso-Mataloni 95]}:
A. Aiello, F. De Martini, M. Giangrasso, \& P. Mataloni,
``Transverse quantum correlations, stimulated emission and
Einstein causality in the active optical microcavity'',
{\em Quantum Semiclass. Opt.} {\bf 7}, 4, 677-691 (1995).

\item {\bf [Al Amri-Babiker 04]}:
M. Al-Amri, \& M. Babiker,
``Quantum shifts for qubits in heterostructures'',
{\em Phys. Rev. A} {\bf 69}, 6, 065801 (2004).

\item {\bf [Albanese-Christandl-Datta-Ekert 04]}:
C. Albanese, M. Christandl, N. Datta, \& A. K. Ekert,
``Mirror inversion of quantum states in linear registers'',
quant-ph/0405029.

\item {\bf [Alber 99]}:
G. Alber,
``Entanglement and the linearity of quantum mechanics'',
{\em Proc.\ of the 10th Int.\ Symp.\ on Theoretical
Electrical Engineering, ISTET 99, Magdeburg},
quant-ph/9907104.

\item {\bf [Alber-Delgado-Jex 00]}:
G. Alber, A. Delgado, \& I. Jex,
``Optimal universal two-particle entanglement processes
in arbitrary dimensional Hilbert spaces'',
quant-ph/0006040.

\item {\bf [Alber-Delgado-Gisin-Jex 00]}:
G. Alber, A. Delgado, N. Gisin, \& I. Jex,
``Generalized quantum XOR-gate for quantum teleportation
and state purification in arbitrary dimensional Hilbert spaces'',
quant-ph/0008022.

\item {\bf [Alber-Beth-Charnes-(+3) 01]}:
G. Alber, T. Beth, C. Charnes, A. Delgado,
M. Grassl, \& M. Mussinger,
``Stabilizing distinguishable qubits against spontaneous decay
by detected-jump correcting quantum codes'',
{\em Phys. Rev. Lett.} {\bf 86}, 19, 4402-4405 (2001).

\item {\bf [Alber-Delgado-Gisin-Jex 01]}:
G. Alber, A. Delgado, N. Gisin, \& I. Jex,
``Efficient bipartite quantum state purification in
arbitrary dimensional Hilbert spaces'',
{\em J. Phys. A} {\bf 34}, 42, 8821-8834 (2001);
quant-ph/0102035.

\item {\bf [Alber-Beth-Horodecki-(+6) 01]}:
G. Alber, T. Beth, M. Horodecki, P. Horodecki, R. Horodecki,
M. R\"{o}tteler, H. Weinfurter, R. F. Werner, \& A. Zeilinger,
{\em Quantum information. An introduction to basic theoretical
concepts and experiments},
Springer-Verlag, Berlin, 2001.
See {\bf [Horodecki-Horodecki-Horodecki 01 c]}.

\item {\bf [Alber-Beth-Charnes-(+3) 03]}:
G. Alber, T. Beth, C. Charnes,
A. Delgado, M. Grassl, \& M. Mussinger,
``Detected-jump-error-correcting quantum codes, quantum error designs, and
quantum computation'',
{\em Phys. Rev. A} {\bf 68}, 1, 012316 (2003).

\item {\bf [Albert-Aharonov-D'Amato 85]}:
D. Z. Albert, Y. Aharonov, \& S. D'Amato,
``Curious new statistical prediction of quantum mechanics'',
{\em Phys. Rev. Lett.} {\bf 54}, 1, 5-7 (1985).
See {\bf [Bub-Brown 86]}, {\bf [Albert-Aharonov-D'Amato 86]},
{\bf [Hu 90]}, {\bf [Zachar-Alter 91]}.

\item {\bf [Albert-Aharonov-D'Amato 86]}:
D. Z. Albert, Y. Aharonov, \& S. D'Amato,
``Comment on `Curious properties of quantum ensembles which have been
preselected and postselected'\,'',
{\em Phys. Rev. Lett.} {\bf 56}, 22, 2427 (1986).
See {\bf [Albert-Aharonov-D'Amato 85]}, {\bf [Bub-Brown 86]},
{\bf [Hu 90]}, {\bf [Zachar-Alter 91]}.

\item {\bf [Albert-Loewer 88]}:
D. Z. Albert, \& B. Loewer,
``Interpreting the many-worlds interpretation'',
{\em Synthese} {\bf 77}, ?, 195-213 (1988).

\item {\bf [Albert-Vaidman 89 a]}:
D. Z. Albert, \& L. Vaidman,
``On a proposed postulate of state-reduction'',
{\em Phys. Lett. A} {\bf 139}, 1-2, 1-4 (1989).

\item {\bf [Albert-Vaidman 89 b]}:
D. Z. Albert, \& L. Vaidman,
``On a theory of the collapse of the wave function'',
in M. Kafatos (ed.),
{\em Bell's theorem, quantum
theory, and conceptions of the universe.
Proc.\ of a workshop (George Mason University, 1988)},
Kluwer Academic, Dordrecht, Holland, 1989, pp.~7-16.

\item {\bf [Albert 90]}:
D. Z. Albert,
``On the collapse of the wave-function'',
in A. I. Miller (ed.),
{\em Sixty-two years of uncertainty:
Historical, philosophical and physical inquiries into the foundations of quantum
mechanics.
Proc.\ Int. School of History of Science (Erice, Italy, 1989)},
Plenum Press, New York, 1990, pp.~153-165.

\item {\bf [Albert-Loewer 91 a]}:
D. Z. Albert, \& B. Loewer,
``The measurement problem: Some `solutions'\,'',
{\em Synthese} {\bf 86}, ?, 87-98 (1991).

\item {\bf [Albert-Loewer 91 b]}:
D. Z. Albert, \& B. Loewer,
``Wanted dead or alive: Two attempts to solve Schr\"{o}dinger's paradox'',
in A. I. Fine, M. Forbes, \& L. Wessels (eds.),
{\em Proc.\ of the 1990 Biennial Meeting of the Philosophy of Science
Association}, East Lansing, Michigan, 1991, vol. 1, pp.~277-285.
See {\bf [Ruetsche 95]}.

\item {\bf [Albert 91]}:
D. Z. Albert,
``The quantum mechanics of self-measurement'',
in {\bf [Zurek 90]}, pp.~471-476.

\item {\bf [Albert 92]}:
D. Z. Albert,
{\em Quantum mechanics and experience},
Harvard University Press, Cambridge, Massachusetts, 1994.

\item {\bf [Albert-Loewer 93]}:
D. Z. Albert, \& B. Loewer,
``Non-ideal measurements'',
{\em Found. Phys. Lett.} {\bf 6}, 4, 297-305 (1993).
See {\bf [Healey 93]}.

\item {\bf [Albert 94]}:
D. Z. Albert,
``Bohm's alternative to quantum mechanics'',
{\em Sci. Am.} {\bf 270}, 5, 32-39 (1994).
Spanish version: ``Teor\'{\i}a alternativa
de Bohm a la mec\'{a}nica cu\'{a}ntica'',
{\em Investigaci\'{o}n y Ciencia} 214, 20-27 (1994).
Reprinted in {\bf [Cabello 97 c]}, pp.~57-64.

\item {\bf [Albertson 61]}:
J. Albertson,
``Von Neumann's hidden-parameter proof'',
{\em Am. J. Phys.} {\bf 29}, 8, 478-484 (1961).

\item {\bf [Albeverio-Fei 00 a]}:
S. Albeverio, \& S.-M. Fei,
``A remark on the optimal cloning of an $N$-level quantum system'',
{\em Eur. Phys. J. B} {\bf 14}, ?, 669-672 (2000);
quant-ph/9912038.

\item {\bf [Albeverio-Fei 00 b]}:
S. Albeverio, \& S.-M. Fei,
``Teleportation of general finite-dimensional quantum systems '',
{\em Phys. Lett. A} {\bf 276}, 1-4, 8-11 (2000);
quant-ph/0012035.

\item {\bf [Albeverio-Fei 01]}:
S. Albeverio, \& S.-M. Fei,
``A note on invariants and entanglements'',
{\em J. Opt. B: Quantum Semiclass. Opt.} {\bf 3}, 4, 223-227 (2001);
quant-ph/0109073.

\item {\bf [Albeverio-Fei-Goswami 01]}:
S. Albeverio, S.-M. Fei, \& D. Goswami,
``Separability of rank two quantum states'',
{\em Phys. Lett. A} {\bf 286}, 2-3, 91-96 (2001);
quant-ph/0109089.

\item {\bf [Albeverio-Fei-Yang 02]}:
S. Albeverio, S.-M. Fei, \& W.-L. Yang,
``Optimal teleportation based on Bell measurements'',
{\em Phys. Rev. A} {\bf 66}, 1, 012301 (2002).

\item {\bf [Albeverio-Fei-Parashar-Yang 03]}:
S. Albeverio, S.-M. Fei, P. Parashar, \& W.-L. Yang,
``Nonlocal properties and local invariants for bipartite systems'',
{\em Phys. Rev. A} {\bf 68}, 1, 010313 (2003);
quant-ph/0307164.

\item {\bf [Albeverio-Chen-Fei 03]}:
S. Albeverio, K. Chen, \& S. M. Fei,
``Generalized reduction criterion for separability of quantum states'',
{\em Phys. Rev. A} {\bf 68}, 6, 062313 (2003);
quant-ph/0312185.

\item {\bf [Albeverio-Fei-Yang 03]}:
S. Albeverio, S. M. Fei, \& W. L. Yang,
``Quantum teleportation: From pure to mixed states and standard to optimal'',
{\em Found. Prob. Phys.},
V\"{a}xj\"{o} University Press, V\"{a}xj\"{o}, Sweden, 2003, pp.~37-56;
quant-ph/0308009.

\item {\bf [Albrecht 92]}:
A. Albrecht,
``Investigating decoherence in a simple system'',
{\em Phys. Rev. D} {\bf 46}, 12, 5504-5520 (1992).

\item {\bf [Albrecht 93]}:
A. Albrecht,
`Following a ``collapsing'' wave function',
{\em Phys. Rev. D} {\bf 48}, 8, 3768-3778 (1993).

\item {\bf [Albrecht 94]}:
A. Albrecht,
``Some remarks on quantum coherence'',
in S. M. Barnett, A. K. Ekert, \& S. J. D. Phoenix (eds.),
{\em J. Mod. Opt.} {\bf 41}, 12 (Special issue: Quantum
communication), 2467-2482 (1994).

\item {\bf [Albrecht 01]}:
A. Albrecht,
``Quantum ripples in chaos'',
{\em Nature} {\bf 412}, 6848, 687-688 (2001).
See: {\bf [Zurek 01 a]}.

\item {\bf [Alcaraz-Tsallis 02]}:
F. C. Alcaraz, \& C. Tsallis,
``Frontier between separability and quantum entanglement in a many spin system'',
{\em Phys. Lett. A} {\bf 301}, 3-4, 105-111 (2002);
quant-ph/0110067.

\item {\bf [Alcaraz-Saguia-Sarandy 04]}:
F. C. Alcaraz, A. Saguia, \& M. S. Sarandy,
``Entanglement and quantum phases in the anisotropic ferromagnetic
Heisenberg chain in the presence of domain walls'',
quant-ph/0312039.

\item {\bf [Alda 80]}:
V. Alda,
``On 0-1 measure for projectors'',
{\em Aplikace Matematiky} {\bf 25}, ?, 373-374 (1980).
See {\bf [Alda 81]} (II).

\item {\bf [Alda 81]}:
V. Alda,
``On 0-1 measure for projectors. II'',
{\em Aplikace Matematiky} {\bf 26}, ?, 57-58 (1981).
See {\bf [Alda 80]} (I).

\item {\bf [Alessandrini-Mancini-Tombesi 04]}:
F. Alessandrini, S. Mancini, \& P. Tombesi,
``Validation of entanglement purification by continuous variable
polarization'',
{\em J. Opt. B: Quantum Semiclass. Opt.};
quant-ph/0401169.

\item {\bf [D'Alessandro 03]}:
D. D'Alessandro,
``On quantum state observability and measurement'',
{\em J. Phys. A} {\bf 36}, 37, 9721–9735 (2003).

\item {\bf [Alexanian-Bose 02]}:
M. Alexanian, \& S. K. Bose,
`\,``Macroscopic'' quantum superpositions: Atom-field entangled and steady
states by two-photon processes',
{\em Phys. Rev. A} {\bf 65}, 3, 033819 (2002).

\item {\bf [Alexanian 03]}:
M. Alexanian,
``Cavity coherent-state cloning via Raman scattering'',
{\em Phys. Rev. A} {\bf 67}, 3, 033809 (2003).

\item {\bf [Alfinito-Viglione-Vitiello 00]}
E. Alfinito, R. G. Viglione, \& G. Vitiello,
``The decoherence criterion'',
quant-ph/0007020.

\item {\bf [Ali Can-Klyachko-Shumovsky 02]}:
M. Ali Can, A. A. Klyachko, \& A. S. Shumovsky,
``Entanglement and the SU(2) phase states in atomic systems'',
{\em Phys. Rev. A} {\bf 66}, 2, 022111 (2002).

\item {\bf [Ali Khan-Howell 04]}:
I. Ali Khan, \& J. C. Howell,
``Hong-Ou-Mandel cloning: Quantum copying without an ancilla'',
{\em Phys. Rev. A} {\bf 70}, 1, 010303 (2004).

\item {\bf [Alibart-Tanzilli-Ostrowsky-Baldi 04]}:
O. Alibart, S. Tanzilli, D. B. Ostrowsky, \& P. Baldi,
``Guided wave technology for a telecom wavelength heralded single photon
source'',
quant-ph/0405075.

\item {\bf [Alicki 00 a]}:
R. Alicki,
``Can quantum computer perform better than classical?'',
quant-ph/0006018.

\item {\bf [Alicki 00 b]}:
R. Alicki,
``On non efficiency of quantum computer'',
quant-ph/0006080.

\item {\bf [Alicki 01]}:
R. Alicki,
``A search for a border between classical and quantum worlds'',
quant-ph/0105089.

\item {\bf [Alicki-Horodecki-Horodecki-Horodecki 02]}:
R. Alicki, M. Horodecki, P. Horodecki, \& R. Horodecki,
``Dynamical description of quantum computing: Generic nonlocality of quantum
noise'',
{\em Phys. Rev. A} {\bf 65}, 6, 062101 (2002);
quant-ph/0105115.

\item {\bf [Alicki 02]}:
R. Alicki,
``Information-theoretical meaning of quantum-dynamical entropy'',
{\em Phys. Rev. A} {\bf 66}, 5, 052302 (2002).

\item {\bf [Alicki-Horodecki-Horodecki-(+3) 04]}:
R. Alicki, M. Horodecki, P. Horodecki,
R. Horodecki, L. Jacak, \& P. Machnikowski,
``Optimal strategy for a single-qubit gate and the trade-off between opposite types of decoherence'',
{\em Phys. Rev. A} {\bf 70}, 1, 010501 (2004).

\item {\bf [Alicki-Horodecki-Horodecki-Horodecki 04]}:
R. Alicki, M. Horodecki, P. Horodecki, \& R. Horodecki,
``Thermodynamics of quantum informational systems -- Hamiltonian
description'',
quant-ph/0402012.

\item {\bf [Alicki-Fannes 04]}:
R. Alicki, \& M. Fannes,
``Note on multiple additivity of minimal entropy output of extreme
$SU(d)$-covariant channels'',
quant-ph/0407033.

\item {\bf [Aliferis-Leung 04]}:
P. Aliferis, \& D. W. Leung,
``Computation by measurements: A unifying picture'',
quant-ph/0404082.

\item {\bf [Allahverdyan-Saakian 98]}:
A. E. Allahverdyan, \& D. B. Saakian,
``About optimal measurements in quantum hypothesizes testing'',
quant-ph/9810057.

\item {\bf [Allahverdyan-Saakian 00]}:
A. E. Allahverdyan, \& D. B. Saakian,
``Broadcast of classical information through a quantum channel'',
{\em Europhys. Lett.} {\bf 50}, 6, 718-723 (2000).

\item {\bf [Alleaume-Treussart-Messin-(+6) 04]}:
R. Alleaume, F. Treussart, G. Messin,
Y. Dumeige, J.-F. Roch, A. Beveratos,
R. Brouri-Tualle, J.-P. Poizat, \& P. Grangier,
``Experimental open air quantum key distribution with a single photon
source'',
quant-ph/0402110.

\item {\bf [Allen 92]}:
A. D. Allen,
``Debunking the Mermin contraption'',
{\em Phys. Essays} {\bf 5}, 2, 178-179 (1992).
Comment on {\bf [Mermin 81 a, b, 85]}.

\item {\bf [Alley-Shih 86]}:
C. O. Alley, \& Y. H. Shih,
``A new type of EPR experiment'',
in ? (ed.),
{\em Proc.\ of the 2nd Int.\ Symp.\ on the
Foundations of Quantum Mechanics in the
Light of the New Technology (Tokyo, 1986)},
Physical Society of Japan, Tokyo, 1986, pp.~47-52.

\item {\bf [Allori-Zangh\`{\i} 01 a]}:
V. Allori, \& N. Zangh\`{\i},
{\em Biannual IQSA Meeting (Cesena, Italy, 2001)};
``What is Bohmian mechanics'',
quant-ph/0112008.

\item {\bf [Allori-Zangh\`{\i} 01 b]}:
V. Allori, \& N. Zangh\`{\i},
{\em Biannual IQSA Meeting (Cesena, Italy, 2001)};
``On the classical limit of quantum mechanics'',
quant-ph/0112009.

\item {\bf [Allori-D\"{u}rr-Goldstein-Zangh\`{\i} 02]}:
V. Allori, D. D\"{u}rr, S. Goldstein, \& N. Zangh\`{\i},
``Seven steps towards the classical world'',
in R. Bonifacio, \& D. Vitali (eds.),
{\em Mysteries, Puzzles and Paradoxes in Quantum Mechanics IV:
Quantum Interference Phenomena (Gargnano, Italy, 2001)},
{\em J. Opt. B: Quantum Semiclass. Opt.} {\bf 4}, 4, S273-S276 (2002).

\item {\bf [de Almeida-Maia-Villas B\^{o}as-Moussa 98]}:
N. G. de Almeida, L. P. Maia, C. J. Villas-B\^{o}as, \& M. H. Y. Moussa,
``One-cavity scheme for atomic-state teleportation through GHZ states'',
{\em Phys. Lett. A} {\bf 241}, 4-5, 213-217 (1998).

\item {\bf [de Almeida-Villas B\^{o}as-Solano-Moussa 98]}:
N. G. de Almeida, C. J. Villas-B\^{o}as, E. Solano, \& M. H. Y. Moussa,
``Reliable teleportation of ionic motional states through a mapping
process'',
quant-ph/9909036.

\item {\bf [de Almeida-Napolitano-Moussa 00]}:
N. G. de Almeida, R. Napolitano, \& M. H. Y. Moussa,
``Accuracy of a teleported cavity-field state'',
{\em Phys. Rev. A} {\bf 62}, 1, 010101(R) (2000).

\item {\bf [de Almeida-Serra-Villas Boas-Moussa 04]}:
N. G. de Almeida, R. M. Serra, C. J. Villas-Boas, \& M. H. Y. Moussa,
``Engineering squeezed states in high-Q cavities'',
{\em Phys. Rev. A} {\bf 69}, 3, 035802 (2004);
quant-ph/0308025.

\item {\bf [Alon-Moiseyev-Peres 95]}:
O. E. Alon, N. Moiseyev, \& A. Peres,
``Infinite matrices may violate the associative law'',
{\em J. Phys. A} {\bf 28}, 6, 1765-1770 (1995);
quant-ph/9412007.

\item {\bf [Alonso-Muga-Sala Mayato 01]}:
D. Alonso, J. G. Muga, \& R. Sala Mayato,
`Comment on ``Foundations of quantum mechanics:
Connection with stochastic processes''\,',
{\em Phys. Rev. A} {\bf 64}, 1, 016101 (2001);
quant-ph/0011038.
Comment on {\bf [Olavo 00]}.

\item {\bf [Alonso-De Vincenzo-Gonz\'{a}lez D\'{\i}az 01]}:
V. Alonso, S. De Vincenzo, \& L. Gonz\'{a}lez D\'{\i}az,
`Ehrenfest's theorem and Bohm's quantum potential
in a ``one-dimensional box''\,',
{\em Phys. Lett. A} {\bf 287}, 1-2, 23-30 (2001).

\item {\bf [Alsing-Milburn 02]}:
P. M. Alsing, \& G. J. Milburn,
``On entanglement and Lorentz transformations'',
{\em Quant. Inf. Proc.} {\bf 2}, 6, 487-512 (2002);
quant-ph/0203051.

\item {\bf [Alsing-Milburn 03]}:
P. M. Alsing, \& G. J. Milburn,
``Teleportation with a uniformly accelerated partner'',
{\em Phys. Rev. Lett.} {\bf 91}, 18, 180404 (2003);
quant-ph/0302179.
See {\bf [Alsing-McMahon-Milburn 03]}.

\item {\bf [Alsing-McMahon-Milburn 03]}:
P. M. Alsing, D. McMahon, \& G. J. Milburn,
``Teleportation in a non-inertial frame'',
quant-ph/0311096.
See {\bf [Alsing-Milburn 03]}.

\item {\bf [Altafini 02]}:
C. Altafini,
``On the generation of sequential unitary gates from
continuous time Schr\"{o}dinger equations driven by external fields'',
{\em Quant. Inf. Proc.} {\bf 1}, 3, 207-224 (2002);
quant-ph/0203005.

\item {\bf [Altafini 04]}:
C. Altafini,
``Tensor of coherences parametrization of multiqubit density operators for entanglement characterization'',
{\em Phys. Rev. A} {\bf 69}, 1, 012311 (2004).

\item {\bf [Altaisky 01]}:
M. V. Altaisky,
``Quantum neural network'',
quant-ph/0107012.

\item {\bf [Altepeter-Branning-Jeffrey-(+6) 03]}:
J. B. Altepeter, D. Branning, E. Jeffrey, T. C. Wei, P. G. Kwiat,
R. T. Thew, J. L. O'Brien, M. A. Nielsen, \& A. G. White,
``Ancilla-assisted quantum process tomography'',
{\em Phys. Rev. Lett.} {\bf 90}, 19, 193601 (2003);
quant-ph/0303038.

\item {\bf [Altepeter-Hadley-Wendelken-(+2) 03]}:
J. B. Altepeter, P. G. Hadley, S. M. Wendelken,
A. J. Berglund, \& P. G. Kwiat,
``Experimental investigation of a two-qubit decoherence-free subspace'',
{\em Phys. Rev. Lett.} {\bf 92}, 14, 147901 (2004).

\item {\bf [Alter-Yamamoto 97]}:
O. Alter, \& Y. Yamamoto,
``Quantum Zeno effect and the impossibility of determining
the quantum state of a single system'',
{\em Phys. Rev. A} {\bf 55}, 4, R2499-R2502 (1997).

\item {\bf [Alter-Yamamoto 98]}:
O. Alter, \& Y. Yamamoto,
``Impossibility of determining the unknown quantum wavefunction of a single system:
Quantum non-demolition measurements, measurements without entanglement and
adiabatic measurements'',
{\em Fortschr. Phys.} {\bf 46}, 6-8, 817-827 (1998).

\item {\bf [Alter-Yamamoto 01]}:
O. Alter, \& Y. Yamamoto,
{\em Quantum measurement of a single system},
John Wiley \& Sons, New York, 2001.
Review: {\bf [Ballentine 03]}.

\item {\bf [Altewischer-van Exter-Woerdman 02]}:
E. Altewischer, M. P. van Exter, \& J. P. Woerdman,
``Plasmon-assisted transmission of entangled photons'',
{\em Nature} {\bf 418}, 6896, 304-306 (2002);
quant-ph/0203057.

\item {\bf [Altman 03]}:
C. Altman,
``Quantum state engineering with the rf-SQUID: A brief introduction'',
quant-ph/0307101.

\item {\bf [Altschul-Altschul 01]}:
M. S. Altschul, \& B. Altschul,
``Spontaneous parametric down-conversion
experiment to measure both photon trajectories and double-slit
interference'',
quant-ph/0106113.

\item {\bf [Alvarenga-Batista-Fabris-Goncalves 02]}:
F. G. Alvarenga, A. B. Batista, J. C. Fabris, \& S. V. B. Goncalves,
``Anisotropic quantum cosmological models:
A discrepancy between many-worlds and dBB interpretations'',
gr-qc/0202009.

\item {\bf [\'{A}lvarez-Garc\'{\i}a Alcaine 87]}:
G. \'{A}lvarez, \& G. Garc\'{\i}a Alcaine,
``Some remarks on realism, contextuality and quantum mechanics'',
{\em Anales de F\'{\i}sica A} {\bf 83}, 3, 247-253 (1987).

\item {\bf [\'{A}lvarez Estrada-S\'{a}nchez G\'{o}mez 98]}:
R. F. \'{A}lvarez Estrada, \& J. L. S\'{a}nchez G\'{o}mez,
``Time evolution and Zeno effect in relativistic quantum field theory'',
quant-ph/9807084.

\item {\bf [Alves-Jaksch 04]}:
C. M. Alves, \& D. Jaksch,
``Multipartite entanglement detection in bosons'',
{\em Phys. Rev. Lett.} {\bf 93}, 11, 110501 (2004);
quant-ph/0409036.

\item {\bf [Ambainis-Schulman-Vazirani 00]}:
A. Ambainis, L. J. Schulman, \& U. Vazirani,
``Computing with highly mixed states'',
to be published in {\em Proc.\ STOC'00};
quant-ph/0003136.

\item {\bf [Ambainis-Kikusts-Valdats 00]}:
A. Ambainis, A. Kikusts, \& M. Valdats,
``On the class of languages recognizable by 1-way quantum finite
automata'',
quant-ph/0009004.
Extends {\bf [Valdats 00]}.

\item {\bf [Ambainis-Wolf 01]}:
A. Ambainis, \& R. de Wolf,
``Average-case quantum query complexity'',
in S. Popescu, N. Linden, \& R. Jozsa (eds.),
{\em J. Phys. A} {\bf 34}, 35
(Special issue: Quantum information and computation), 6741-6754 (2001).

\item {\bf [Ambainis-Bach-Nayak-(2) 01]}:
A. Ambainis, E. Bach, A. Nayak,
A. Vishwanath, \& J. Watrous,
``One-dimensional quantum walks'',
in {\em Proc. 33th STOC},
ACM, New York, 2001, 60–69.

\item {\bf [Ambainis-Yang 01]}:
A. Ambainis, \& K. Yang,
``Extracting quantum entanglement
(general entanglement purification protocols)'',
quant-ph/0110011.

\item {\bf [Ambainis-Schulman-Ta Shma-(+2) 03]}:
A. Ambainis, L. J. Schulman, A. Ta-Shma,
U. Vazirani, \& A. Wigderson,
``The quantum communication complexity of sampling'',
{\em Siam J. Comput.} {\bf 32}, 1570-? (2002).

\item {\bf [Ambainis 02 a]}:
A. Ambainis,
``A new protocol and lower bounds for quantum coin flipping'',
quant-ph/0204022.

\item {\bf [Ambainis 02 b]}:
A. Ambainis,
``Lower bound for a class of weak quantum coin flipping protocols'',
quant-ph/0204063.

\item {\bf [Ambainis-Buhrman-Dodis-Roehrig 03]}:
A. Ambainis, H. Buhrman, Y. Dodis, \& H. Roehrig,
``Multiparty quantum coin flipping'',
quant-ph/0304112.

\item {\bf [Ambainis-Gottesman 03]}:
A. Ambainis, \& D. Gottesman,
``The minimum distance problem for two-way entanglement purification'',
quant-ph/0310097.

\item {\bf [Ambainis-Kempe-Rivosh 04]}:
A. Ambainis, J. Kempe, \& A. Rivosh,
``Coins make quantum walks faster'',
quant-ph/0402107.

\item {\bf [Ambainis 04]}:
A. Ambainis,
``Polynomial degree vs. quantum query complexity'',
quant-ph/0305028.

\item {\bf [Amico-Osterloh-Plastina-(+2) 04]}:
L. Amico, A. Osterloh, F. Plastina, R. Fazio, \& G. M. Palma,
``Dynamics of entanglement in one-dimensional spin systems'',
{\em Phys. Rev. A} {\bf 69}, 2, 022304 (2004);
quant-ph/0307048.

\item {\bf [Amiet-Weigert 99 a]}:
J.-P. Amiet, \& S. Weigert,
``Reconstructing a pure state of a spin s through
three Stern-Gerlach measurements
{\em J. Phys. A} {\bf 32}, 15, 2777-2784 (1999);
quant-ph/9809018.
See {\bf [Amiet-Weigert 99 b]} (II).

\item {\bf [Amiet-Weigert 99 b]}:
J.-P. Amiet, \& S. Weigert,
``Reconstructing the density matrix of a spin s through
Stern-Gerlach measurements: II'',
{\em J. Phys. A} {\bf 32}, 25, L269-L274 (1999).
See {\bf [Amiet-Weigert 99 a]} (I).

\item {\bf [Amin-Smirnov-van den Brink 03]}:
M. H. S. Amin, A. Y. Smirnov, \& A. M. van den Brink,
``Josephson-phase qubit without tunneling'',
{\em Phys. Rev. B} {\bf 67}, 10, 100508 (2003).

\item {\bf [Amlani-Orlov-Toth-(+3) 99]}:
I. Amlani, A. O. Orlov, G. Toth,
G. H. Bernstein, C. S. Lent, \& G. L. Snider,
``Digital logic gate using quantum-dot cellular automata'',
{\em Science} {\bf 284}, 5412, 289-291 (1999).

\item {\bf [Amosov-Holevo 01]}:
G. G. Amosov, \& A. S. Holevo (Kholevo),
``On the multiplicativity conjecture for quantum channels'',
math-ph/0103015.

\item {\bf [An 03]}:
N. B. An,
``Teleportation of coherent-state superpositions within a network'',
{\em Phys. Rev. A} {\bf 68}, 2, 022321 (2003).

\item {\bf [An 04 a]}:
N. B. An,
``Optimal processing of quantum information via $W$-type entangled coherent states'',
{\em Phys. Rev. A} {\bf 69}, 2, 022315 (2004).

\item {\bf [An 04 b]}:
N. B. An,
``Quantum dialogue'',
{\em Phys. Lett. A} {\bf 328}, ?, 6-? (2004);
quant-ph/0406130.

\item {\bf [Anandan 99]}:
J. Anandan,
``The quantum measurement problem and the possible
role of the gravitational field'',
{\em Found. Phys.} {\bf 29}, 3, 333-348 (1999).

\item {\bf [Anandan-Aharonov 99]}:
J. Anandan, \& Y. Aharonov,
``Meaning of the density matrix'',
{\em Found. Phys. Lett.} {\bf 12}, 6, 571-578 (1999).

\item {\bf [Anandan-Stodolsky 00]}:
J. Anandan, \& L. Stodolsky,
``On the relation between quantum mechanical and classical parallel
transport'',
{\em Phys. Lett. A} {\bf 266}, 2-3, 95-97 (2000).

\item {\bf [Anandan 02]}:
J. Anandan,
``Causality, symmetries and quantum mechanics'',
{\em Found. Phys. Lett.} {\bf 15}, 6, 415-438 (2002).

\item {\bf [Anandan 03]}:
J. Anandan,
``Laws, symmetries, and reality'',
based on the invited talk given by the author at the {\em Wigner Centennial Conf.\ (Pecs, Hungary, 2002)};
quant-ph/0304109.

\item {\bf [Anastopoulos 00]}:
C. Anastopoulos,
``Frequently asked questions about decoherence'',
quant-ph/0011123.

\item {\bf [Anastopoulos 01]}:
C. Anastopoulos,
``Quantum theory without Hilbert spaces'',
{\em Found. Phys.} {\bf 31}, 11, 1545-1580 (2001).

\item {\bf [Ancochea-Bramon-Nowakowski 99]}:
B. Ancochea, A. Bramon, \& M. Nowakowski,
``Bell-inequalities for $K^0 \bar{K^0}$ pairs from $\Phi$-resonance
decays'',
{\em Phys. Rev. D} {\bf 60}, 9, 094008 (1999);
hep-ph/9811404.

\item {\bf [And\aa s 92]}:
H. E. And\aa s,
``Bell's inequalities for quantum mechanics'',
{\em Phys. Lett. A} {\bf 167}, 1, 6-10 (1992).

\item {\bf [Andersen-Buchhave 03]}:
U. L. Andersen, \& P. Buchhave,
``Squeezing and entanglement in doubly resonant, type II, second-harmonic
generation'',
{\em J. Opt. Soc. Am. B} {\bf 20}, 1947-? (2003).

\item {\bf [Andersen-Gl\"{o}ckl-Lorenz-(+2) 04]}:
U. L. Andersen, O. Gl\"{o}ckl, S. Lorenz,
G. Leuchs, \& R. Filip,
``Experimental demonstration of continuous variable quantum erasing'',
{\em Phys. Rev. Lett.};
quant-ph/0409005.

\item {\bf [Anderson 94]}:
P. W. Anderson,
``Shadows of doubt'',
{\em Nature} {\bf 372}, 6503, 288-289 (1994).
Review of {\bf [Penrose 94 b]}.

\item {\bf [Anderson 97]}:
P. W. Anderson,
``Mind over matter'',
{\em Nature} {\bf 386}, 6624, 456 (1997).
Review of {\bf [Penrose 97 a]}.

\item {\bf [Anderson-van Enk-Rudolph 03]}:
X. Anderson, S. J. van Enk, \& T. Rudolph,
``Quantum communication protocols using the vacuum'',
quant-ph/0302091.

\item {\bf [Andersson-Barnett 00]}:
E. Andersson, \& S. M. Barnett,
``Bell-state analyzer with channeled atomic particles'',
{\em Phys. Rev. A} {\bf 62}, 5, 052311 (2000).

\item {\bf [Andersson 01]}:
E. Andersson,
``Generalized measurements on atoms in microtraps'',
{\em Phys. Rev. A} {\bf 64}, 3, 032303 (2001).

\item {\bf [Andersson-Barnett-Gilson-Hunter 02]}:
E. Andersson, S. M. Barnett, C. R. Gilson, \& K. Hunter,
``Minimum-error discrimination between three mirror-symmetric states'',
{\em Phys. Rev. A} {\bf 65}, 5, 052308 (2002);
quant-ph/0201074.

\item {\bf [Andersson-Jex-Barnett 03]}:
E. Andersson, I. Jex, \& S. M. Barnett,
``Comparison of unitary transforms'',
{\em J. Phys. A} {\bf 36}, 9, 2325-2338 (2003).

\item {\bf [Andr\'{e}-Lukin 02 a]}:
A. Andr\'{e}, \& M. D. Lukin,
``Atom correlations and spin squeezing near the Heisenberg limit: Finite-size
effect and decoherence'',
{\em Phys. Rev. A} {\bf 65}, 5, 053819 (2002).

\item {\bf [Andr\'{e}-Lukin 02 b]}:
A. Andr\'{e}, \& M. D. Lukin,
``Manipulating light pulses via dynamically controlled photonic band gap'',
{\em Phys. Rev. Lett.} {\bf 89}, 14, 143602 (2002).

\item {\bf [Andre-S{\o}rensen-Lukin 04]}:
A. Andr\'{e}, A. S. S{\o}rensen, \& M. D. Lukin,
``Stability of atomic clocks based on entangled atoms'',
{\em Phys. Rev. Lett.} {\bf 92}, 23, 230801 (2004);
quant-ph/0401130.

\item {\bf [Andrecut-Ali 04]}:
M. Andrecut, \& M. K. Ali,
``Entanglement dynamics in quantum cellular automata'',
{\em Phys. Lett. A} {\bf 326}, 5-6, 328-332 (2004).

\item {\bf [Andreev-Man'ko 01]}:
V. A. Andreev, \& V.I. Man'ko,
``The classification of two-particle spin states and generalized Bell
inequalities'',
{\em Phys. Lett. A} {\bf 281}, 5-6, 278-288 (2001).

\item {\bf [Angelakis-Beige-Knight-(+2) 01]}:
D. G. Angelakis, A. Beige, P. L. Knight,
W. J. Munro, \& B. Tregenna,
``Verifying atom entanglement schemes by testing Bell's
inequality'',
{\em Mysteries and Paradoxes in Quantum Mechanics (Gargnano, Italy,
2000)},
{\em Zeitschrift f\"{u}r Naturforschung};
quant-ph/0102079.

\item {\bf [Angelakis-Knight 02]}:
D. G. Angelakis, \& P. L. Knight,
``Testing Bell inequalities in photonic crystals'',
{\em Eur. Phys. J. D} {\bf 18}, 2 (Special issue:
{\em Quantum interference and cryptographic keys:
Novel physics and advancing technologies (QUICK) (Corsica, 2001)}, 247-250 (2002).

\item {\bf [Angelakis-Santos-Yannopapas-Ekert 04]}:
D. G. Angelakis, M. F. Santos, V. Yannopapas, \& A. K. Ekert,
``Quantum computation in photonic crystals'',
quant-ph/0410189.

\item {\bf [Angelidis 83]}:
T. D. Angelidis,
``Bell's theorem: Does the Clauser-Horne inequality
hold for all local theories?'',
{\em Phys. Rev. Lett.} {\bf 51}, 20, 1819-1822 (1983).
Comments: {\bf [Garg-Leggett 84]}, {\bf [Barut-Meystre 84 b]},
{\bf [Horne-Shimony 84]}, {\bf [Cushing 85]}.

\item {\bf [Angelidis-Popper 85]}:
T. D. Angelidis, \& K. R. Popper,
``Towards a local explanatory theory of the
Einstein-Podolsky-Rosen-Bohm experiment'',
in P. J. Lahti, \& P. Mittelstaedt (eds.),
{\em Proc.\ Symp.\ on the Foundations of Modern
Physics: 50 Years of the Einstein-Podolsky-Rosen Experiment
(Joensuu, Finland, 1985)},
World Scientific, Singapore, 1985, pp.~37-50.

\item {\bf [D'Angelo-Kim-Kulik-Shih 04]}:
M. D'Angelo, Y.-H. Kim, S. P. Kulik, \& Y. Shih,
``Identifying entanglement using quantum ghost interference and imaging'',
{\em Phys. Rev. Lett.} {\bf 92}, 23, 233601 (2004).

\item {\bf [Anglin-Laflamme-Paz-Zurek 95]}:
J. R. Anglin, R. Laflamme, J. P. Paz, \& W. H. Zurek,
``Decoherence and recoherence in an analogue of the
black hole information paradox'',
{\em Phys. Rev. D} {\bf 52}, 4, 2221-2231 (1995).

\item {\bf [Anglin-Zurek 96]}:
J. R. Anglin, \& W. H. Zurek,
``Decoherence of quantum fields:
Pointer states and predictability'',
{\em Phys. Rev. D} {\bf 53}, 12, 7327-7335 (1996).

\item {\bf [Anglin-Paz-Zurek 97]}:
J. R. Anglin, J. P. Paz, \& W. H. Zurek,
``Deconstructing decoherence'',
{\em Phys. Rev. A} {\bf 55}, 6, 4041-4053 (1997).

\item {\bf [Anonymous 35]}:
Anonymous,
``Einstein attacks quantum theory'',
{\em The New York Times}, 4 May, 1935.
Reprinted in {\bf [Jammer 80]}, p.~514.

\item {\bf [Anonymous 78]}:
Anonymous,
``?'',
{\em Sci. Am.} {\bf 242}, 8 (?), ?-? (1978).
Spanish version:
``Ciencia y sociedad: Esse est percipi'',
{\em Investigaci\'{o}n y Ciencia} 26, 40-142 (1978).

\item {\bf [Anonymous 96]}:
Anonymous,
``The weirdest computer of all'',
{\em The Economist} {\bf 340}, 7985, 105-107 (28 Sept. 1996).

\item {\bf [Anonymous 97]}:
Anonymous,
``Don't look now, I'm teleporting'',
{\em The Economist} {\bf 345}, 8047, 90 (13 Dec. 1997).

\item {\bf [Ansari 97]}:
N. A. Ansari,
``Violation of Bell's inequality in a driven
three-level cascade atomic system'',
{\em Phys. Rev. A} {\bf 55}, 3, 1639-1646 (1997).

\item {\bf [Anselmi-Chefles-Plenio 04]}:
F. Anselmi, A. Chefles, \& M. B. Plenio,
``Local copying of orthogonal entangled quantum states'',
quant-ph/0407168.

\item {\bf [Anspach 01]}:
P. H. Anspach,
``Two-qubit catalysis in a four-state pure bipartite
system'',
quant-ph/0102067.

\item {\bf [Antoci-Mihich 00]}:
S. Antoci, \& L. Mihich,
``Violating' Clauser-Horne inequalities within classical
mechanics'',
{\em Nuovo Cimento B} {\bf 115}, ? 459-466 (2000);
quant-ph/0004037.

\item {\bf [Anwar-Blazina-Carteret-(+5) 04]}:
M. S. Anwar, D. Blazina, H. A. Carteret,
S. B. Duckett, T. K. Halstead, J. A. Jones,
C. M. Kozak, \& R. J. K. Taylor,
``Preparing high purity initial states for nuclear magnetic resonance quantum computing'',
{\em Phys. Rev. Lett.} {\bf 93}, 4, 040501 (2004);
quant-ph/0312014.

\item {\bf [Anwar-Blazina-Carteret-(+2) 04]}:
M. S. Anwar, D. Blazina, H. Carteret,
S. B. Duckett, \& J. A. Jones,
``Implementing Grover's quantum search on a para-hydrogen based pure state
NMR quantum computer'',
quant-ph/0407091.

\item {\bf [Aoki-Takei-Yonezawa-(+4) 03]}:
T. Aoki, N. Takei, H. Yonezawa,
K. Wakui, T. Hiraoka, A. Furusawa, \& P. van Loock,
``Experimental creation of a fully inseparable tripartite continuous-variable
state'',
{\em Phys. Rev. Lett.} {\bf 91}, 8, 080404 (2003);
quant-ph/0304053.

\item {\bf [Appleby 99 a]}:
D. M. Appleby,
``Generic Bohmian trajectories of an isolated particle'',
{\em Found. Phys.} {\bf 29}, 12, 1863-1884 (1999);
quant-ph/9905003.

\item {\bf [Appleby 99 b]}:
D. M. Appleby,
``Bohmian trajectories post-decoherence'',
{\em Found. Phys.} {\bf 29}, 12, 1885-11916 (1999);
quant-ph/9908029.

\item {\bf [Appleby 00]}:
D. M. Appleby,
``Contextuality of approximate measurements'',
quant-ph/0005010.
See {\bf [Meyer 99 b]}, {\bf [Kent 99 b]},
{\bf [Clifton-Kent 00]}, {\bf [Cabello 99 d, 02 c]},
{\bf [Havlicek-Krenn-Summhammer-Svozil 01]}, {\bf [Mermin 99 b]},
{\bf [Appleby 01, 02]}, {\bf [Boyle-Schafir 01 a]}.

\item {\bf [Appleby 01]}:
D. M. Appleby,
``Nullification of the nullification'',
quant-ph/0109034.
See {\bf [Meyer 99 b]}, {\bf [Kent 99 b]},
{\bf [Clifton-Kent 00]}, {\bf [Cabello 99 d, 02 c]},
{\bf [Havlicek-Krenn-Summhammer-Svozil 01]}, {\bf [Mermin 99 b]},
{\bf [Appleby 00, 02]}, {\bf [Boyle-Schafir 01 a]}.

\item {\bf [Appleby 02]}:
D. M. Appleby,
``Existential contextuality and the models of Meyer, Kent, and Clifton'',
{\em Phys. Rev. A} {\bf 65}, 2, 022105 (2002);
quant-ph/0005056.
See {\bf [Meyer 99 b]}, {\bf [Kent 99 b]},
{\bf [Clifton-Kent 00]}, {\bf [Cabello 99 d, 02 c]},
{\bf [Havlicek-Krenn-Summhammer-Svozil 01]}, {\bf [Mermin 99 b]},
{\bf [Appleby 00, 01]}, {\bf [Boyle-Schafir 01 a]}.

\item {\bf [Appleby 03 a]}:
D. M. Appleby,
``The Hess-Philipp model is non-local'',
{\em Int. J. Quant. Inf.} (2003);
quant-ph/0210145.
Comment on {\bf [Hess-Philipp 01 a, b, c, 02 a]}.
Reply: {\bf [Hess-Philipp 02 f]}.
See {\bf [Mermin 02 e]}, {\bf [Myrvold 02 b]}.

\item {\bf [Appleby 03 b]}:
D. M. Appleby,
``The Bell-Kochen-Specker theorem'',
quant-ph/0308114.

\item {\bf [Appleby 04 a]}:
D. M. Appleby,
``Facts, values and quanta'',
quant-ph/0402015.

\item {\bf [Appleby 04 b]}:
D. M. Appleby,
``Probabilities are single-case, or nothing'',
quant-ph/0408058.

\item {\bf [Araki-Yanase 60]}:
H. Araki, \& M. M. Yanase,
``Measurement of quantum mechanical operators'',
{\em Phys. Rev.} {\bf 120}, 2, 622-626 (1960).

\item {\bf [Aravind 95]}:
P. K. Aravind,
``To what extent do mixed states violate the Bell inequalities?'',
{\em Phys. Lett. A} {\bf 200}, 5, 345-349 (1995).
Erratum: {\em Phys. Lett. A} {\bf 204}, 5-6, 431 (1995).
See {\bf [Horodecki-Horodecki-Horodecki 95]}.

\item {\bf [Aravind 96]}:
P. K. Aravind,
``Geometry of the Schmidt decomposition and Hardy's theorem'',
{\em Am. J. Phys.} {\bf 64}, 9, 1143-1150 (1996).

\item {\bf [Aravind 97 a]}:
P. K. Aravind,
`The ``twirl'', stella octangula and mixed state entanglement',
{\em Phys. Lett. A} {\bf 233}, 1-2, 7-10 (1997).
See {\bf [Bennett-Brassard-Popescu-(+3) 96]}.

\item {\bf [Aravind 97 b]}:
P. K. Aravind,
``Borromean entanglement of the GHZ state'',
in {\bf [Cohen-Horne-Stachel 97 b]}.

\item {\bf [Aravind 98 a]}:
P. K. Aravind,
``How many bits are needed to transmit a qubit reliaby?'',
{\em Phys. Lett. A} {\bf 241}, 4-5, 207-212 (1998).

\item {\bf [Aravind 98 b]}:
P. K. Aravind,
``General radiation states and Bell's inequalities'',
quant-ph/9806085.

\item {\bf [Aravind-Lee Elkin 98]}:
P. K. Aravind, \& F. Lee-Elkin,
``Two non-colourable configurations in four dimensions
illustrating the Kochen-Specker theorem'',
{\em J. Phys. A} {\bf 31}, 49, 9829-9834 (1998).

\item {\bf [Aravind 99]}:
P. K. Aravind,
``Impossible colorings and Bell's theorem'',
{\em Phys. Lett. A} {\bf 262}, 4-5, 282-286 (1999).

\item {\bf [Aravind 00]}:
P. K. Aravind,
``How Reye's configuration helps in proving
the Bell-Kochen-Specker theorem'',
{\em Found. Phys. Lett.} {\bf 13}, 6, 499-519 (2000).

\item {\bf [Aravind 01 a]}:
P. K. Aravind,
``The magic tesseracts and Bell's theorem'',
{\em Am. J. Phys.} {\bf 69}, 3, 348-354 (2001).

\item {\bf [Aravind 01 b]}:
P. K. Aravind,
``Mermin's pentagram and Bell's theorem'',
quant-ph/0104138.

\item {\bf [Aravind 01 c]}:
P. K. Aravind,
``Optimal quantum measurements for spin-1 and spin-3/2
particles'',
{\em Quant. Inf. Comp.} {\bf 1}, 3, 52-61 (2001);
quant-ph/0105037.

\item {\bf [Aravind 02 a]}:
P. K. Aravind,
``Bell's theorem without inequalities and only two
distant observers'',
{\em Found. Phys. Lett.} {\bf 15}, 4, 397-405 (2002);
quant-ph/0104133.

\item {\bf [Aravind 02 b]}:
P. K. Aravind,
``Solution to the King's problem in prime power dimensions'',
quant-ph/0210007.

\item {\bf [Aravind 03 a]}:
P. K. Aravind,
``Generalized Kochen-Specker theorem'',
{\em Phys. Rev. A} {\bf 68}, 5, 052104 (2003);
quant-ph/0301074.
See {\bf [Cabello 03 c]}.

\item {\bf [Aravind 03 b]}:
P. K. Aravind,
``Best conventional solutions to the king's problem'',
{\em Zeitschrift f\"{u}r Naturforschung A} {\bf 58}, 682-690 (2003);
quant-ph/0306119.

\item {\bf [Aravind 04]}:
P. K. Aravind,
``Quantum mysteries revisited again'',
{\em Am. J. Phys.} {\bf 72}, 10, 1303-1307 (2004).

\item {\bf [Ardehali 91]}:
M. Ardehali,
``Hidden variables and quantum-mechanical
probabilities for generalized spin-$s$ systems'',
{\em Phys. Rev. D} {\bf 44}, 10, 3336-3341 (1991).

\item {\bf [Ardehali 92]}:
M. Ardehali,
``Bell inequalities with a magnitude of
violation that grows exponentially with the number of particles'',
{\em Phys. Rev. A} {\bf 46}, 9, 5375-5378 (1992).

\item {\bf [Ardehali 93 a]}:
M. Ardehali,
``Experimental implications of Bell's theorem'',
{\em Phys. Lett. A} {\bf 176}, 5, 285-291 (1993).

\item {\bf [Ardehali 93 b]}:
M. Ardehali,
``Experimental consequences of Einstein-Podolsky-Rosen-Bohm-Bell
gedanken experiment'',
{\em Phys. Rev. A} {\bf 47}, 3, 1633-1638 (1993).
Erratum: {\em Phys. Rev. A} {\bf 49}, 6, 5158 (1994).

\item {\bf [Ardehali 93 c]}:
M. Ardehali,
``Nonlocality in quantum mechanics and
atomic-cascade photons'',
{\em Phys. Lett. A} {\bf 181}, 3, 187-192 (1993).

\item {\bf [Ardehali 94]}:
M. Ardehali,
``Optical test of local hidden-variable with
two-channel polarizers'',
{\em Phys. Rev. A} {\bf 49}, 5, Part A, R3143-R3146 (1994).

\item {\bf [Ardehali 96]}:
M. Ardehali,
``Quantum key distribution based on delayed choice'';
{\em Phys. Lett. A} {\bf 217}, ?, 301-? (1996);
quant-ph/9508008.

\item {\bf [Ardehali 97]}:
M. Ardehali,
``A Bell inequality which can be used to test
locality more simply than Clauser-Horne-Shmony-Holt inequality'',
quant-ph/9709043.

\item {\bf [Ardehali 98 a]}:
M. Ardehali,
``Clauser-Horne-Shimony-Holt correlation
and Clauser-Horne correlation do not lead to the
largest violations of Bell's inequlity'',
{\em Phys. Rev. A} {\bf 57}, 1, 114-119 (1998).

\item {\bf [Ardehali-Chau-Lo 98]}:
M. Ardehali, H. F. Chau, \& H.-K. Lo,
``Efficient quantum key distribution'',
quant-ph/9803007.
See {\bf [Lo-Chau 98 c]}, {\bf [Lo-Chau-Ardehali 00]}.

\item {\bf [Ardehali 98 b]}:
M. Ardehali,
``A Bell inequality which has a larger signal to noise ratio
than Clauser-Horne-Shimony-Holt inequality and which can be used to
test locality more simply than Clauser-Horne inequality'',
quant-ph/9805034.

\item {\bf [Ardehali 98 c]}:
M. Ardehali,
``Maximal violation of Bell's inequality in the case of real experiments',
quant-ph/9810010.

\item {\bf [Ardehali 98 d]}:
M. Ardehali,
``Bell's theorem, quantum mechanical non-locality and atomic cascade photons'',
quant-ph/9810058.

\item {\bf [Ardehali 98 c]}:
M. Ardehali,
``Quantum oblivious transfer protocol based on entangled states'',
quant-ph/9806040.

\item {\bf [Arecchi-Montina 00]}:
F. T. Arecchi, \& A. Montina,
``Quantum interference between macroscopically distinct optical
states'',
{\em Fortschr. Phys.} {\bf 48}, 5-7, 423-429 (2000).

\item {\bf [Arens-Varadarajan 00]}:
R. Arens, \& V. S. Varadarajan,
``On the concept of EPR states and their structure'',
{\em J. Math. Phys.} {\bf 41}, 2, 638-651 (2000);
quant-ph/9910066.

\item {\bf [D'Ariano-Yuen 96]}:
G. M. D'Ariano, \& H. P. Yuen,
``Impossibility of measuring the wave function of a single quantum
system'',
{\em Phys. Rev. Lett.} {\bf 76}, 16, 2832-2835 (1996).

\item {\bf [D'Ariano-Paris 99]}:
G. M. D'Ariano, \& M. G. A. Paris,
``Adaptive quantum homodyne tomography'',
{\em Phys. Rev. A} {\bf 60}, 1, 518-528 (1999).

\item {\bf [D'Ariano-Kumar-Macchiavello-(+2) 99]}:
G. M. D'Ariano, P. Kumar, C. Macchiavello, L. Maccone, \& N. Sterpi,
``Test of the state reduction rule'',
{\em Phys. Rev. Lett.} {\bf 83}, 13, 2490-2493 (1999).

\item {\bf [D'Ariano-Maccone-Sacchi-Garuccio 99]}:
G. M. D'Ariano, L. Maccone, M. F. Sacchi, \& A. Garuccio,
``Tomographic test of Bell's inequality'',
{\em J. Opt. B: Quantum Semiclass. Opt.} {\bf 1}, 5, 576-579
(1999);
quant-ph/9907091.

\item {\bf [D'Ariano 00 a]}:
G. M. D'Ariano,
``Universal quantum estimation'',
{\em Phys. Lett. A} {\bf 268}, 3, 151-157 (2000).

\item {\bf [D'Ariano-Macchiavello-Maccone 00]}:
G. M. D'Ariano, C. Macchiavello, \& L. Maccone,
``Quantum computations with polarized photons'',
{\em Fortschr. Phys.} {\bf 48}, 5-7, 573-577 (2000).

\item {\bf [D'Ariano 00 b]}:
G. M. D'Ariano,
``Quantum tomography: General theory and new experiments'',
{\em Fortschr. Phys.} {\bf 48}, 5-7, 579-588 (2000).

\item {\bf [D'Ariano-Rubin-Shih 00]}:
G. M. D'Ariano, M. Rubin, M. F. Sacchi, \& Y. H. Shih,
``Quantum tomography of the GHZ state'',
{\em Fortschr. Phys.} {\bf 48}, 5-7, 599-603 (2000);
quant-ph/9906067.

\item {\bf [D'Ariano-Maccone-Paris-Sacchi 00]}:
G. M. D'Ariano, L. Maccone, M. G. A. Paris, \& M. F. Sacchi,
``State preparation by photon filtering'',
{\em Fortschr. Phys.} {\bf 48}, 5-7, 671-? (2000);
quant-ph/9906077.

\item {\bf [D'Ariano-Maccone-Paris 00]}:
G. M. D'Ariano, L. Maccone, \& M. G. A. Paris,
``Orthogonality relations in quantum tomography'',
{\em Phys. Lett. A} {\bf 276}, 1-4, 25-30 (2000);
quant-ph/0005111.

\item {\bf [D'Ariano-Presti-Sacchi 00]}:
G. M. D'Ariano, P. L. Presti, \& M. F. Sacchi,
``Bell measurements and observables'',
{\em Phys. Lett. A} {\bf 272}, 1-2, 32-38 (2000);
quant-ph/0005121.

\item {\bf [D'Ariano-Sacchi 00]}:
G. M. D'Ariano, \& M. F. Sacchi,
``Quantum cloning optimal for joint measurements'',
in O. Hirota, \& P. Tombesi (eds.),
{\em Quantum Communication, Measurement and Computing 3},
Kluwer Academic/Plenum Press, New York, 2000, pp.~?-?;
quant-ph/0009080.

\item {\bf [D'Ariano-Macchiavello-Sacchi 01]}:
G. M. D'Ariano, C. Macchiavello, \& M. F. Sacchi,
``Joint measurements via quantum cloning'',
{\em J. Opt. B: Quantum Semiclass. Opt.} {\bf 3}, 2, 44-50 (2001);
quant-ph/0007062.

\item {\bf [D'Ariano-Lo Presti 01 a]}:
G. M. D'Ariano, \& P. Lo Presti,
``Quantum tomography for measuring experimentally the matrix
elements of an arbitrary quantum operation'',
{\em Phys. Rev. Lett.} {\bf 86}, 19, 4195-4198 (2001);
quant-ph/0012071.

\item {\bf [D'Ariano-Maccone-Paris 01]}:
G. M. D'Ariano, L. Maccone, \& M. G. A. Paris,
``Quorum of observables for universal quantum estimation'',
{\em J. Phys. A} {\bf 34}, 1, 93-104 (2001);
quant-ph/0006006.

\item {\bf [D'Ariano-De Martini-Sacchi 01]}:
G. M. D'Ariano, F. De Martini, \& M. F. Sacchi,
``Continuous variable cloning via network of parametric gates'',
{\em Phys. Rev. Lett.} {\bf 86}, 5, 914-917 (2001);
quant-ph/0012025.

\item {\bf [D'Ariano-De Martini-Macchiavello 01]}:
G. M. D'Ariano, F. De Martini, \& C. Macchiavello,
``Quantum statistics of photon cloning machines'',
{\em Phys. Lett. A} {\bf 238}, 1-2, 15-19 (2001).

\item {\bf [D'Ariano-Lo Presti 01 b]}:
G. M. D'Ariano, \& P. Lo Presti,
``Optimal nonuniversally covariant cloning'',
{\em Phys. Rev. A} {\bf 64}, 4, 042308 (2001);
quant-ph/0101100.

\item {\bf [D'Ariano-Lo Presti-Paris 01]}:
G. M. D'Ariano, P. Lo Presti, \& M. G. A. Paris,
``Using entanglement improves the precision of quantum measurements'',
{\em Phys. Rev. Lett.} {\bf 87}, 27, 270404 (2001);
quant-ph/0109040.

\item {\bf [D'Ariano-Paris-Perinotti 01]}:
G. M. D'Ariano, M. G. A. Paris, \& P. Perinotti,
``Improving quantum interferometry by using entanglement
(to take a decision you'd better use entanglement)'',
quant-ph/0110105.

\item {\bf [D'Ariano-Lo Presti-Sacchi 02]}:
G. M. D'Ariano, P. Lo Presti, \& M. F. Sacchi,
``A quantum measurement of the spin direction'',
{\em Phys. Lett. A} {\bf 292}, 4-5, 233-237 (2002);
quant-ph/0110065.

\item {\bf [D'Ariano-Lo Presti-Paris 02]}:
G. M. D'Ariano, P. Lo Presti, \& M. G. A. Paris,
``Improved discrimination of unitary transformations by entangled probes'',
in R. Bonifacio, \& D. Vitali (eds.),
{\em Mysteries, Puzzles and Paradoxes in Quantum Mechanics IV:
Quantum Interference Phenomena (Gargnano, Italy, 2001)},
{\em J. Opt. B: Quantum Semiclass. Opt.} {\bf 4}, 4, S273-S276 (2002);
quant-ph/0204050.

\item {\bf [D'Ariano-Lo Presti-Perinotti 02]}:
G. M. D'Ariano, P. Lo Presti, \& P. Perinotti,
``To take a (binary) decision you'd better use entanglement'',
in R. Bonifacio, \& D. Vitali (eds.),
{\em Mysteries, Puzzles and Paradoxes in Quantum Mechanics IV:
Quantum Interference Phenomena (Gargnano, Italy, 2001)},
{\em J. Opt. B: Quantum Semiclass. Opt.} {\bf 4}, 4, S277-S280 (2002).

\item {\bf [D'Ariano 02]}:
G. M. D'Ariano,
``Universal quantum observables'',
{\em Phys. Lett. A} {\bf 300}, 1, 175-181 (2002).

\item {\bf [D'Ariano-Gill-Keyl-(+3) 02]}:
G. M. D'Ariano, R. D. Gill, M. Keyl,
B. Kuemmerer, H. Maassen, \& R. F. Werner,
``The quantum Monty Hall problem'',
quant-ph/0202120.

\item {\bf [D'Ariano-Lo Presti 03]}:
G. M. D'Ariano, \& P. Lo Presti,
``Classical and quantum noise in measurements and transformations'',
quant-ph/0301110.
Superseded by {\bf [D'Ariano-Maccone-Lo Presti 04]}.

\item {\bf [D'Ariano-Macchiavello 03]}:
G. M. D'Ariano, \& C. Macchiavello,
``Optimal phase-covariant cloning for qubits and qutrits'',
{\em Phys. Rev. A} {\bf 67}, 4, 042306 (2003);
quant-ph/0301175.

\item {\bf [D'Ariano-Paris-Sacchi 03]}:
G. M. D'Ariano, M. G. A. Paris, \& M. F. Sacchi,
``Quantum tomography'',
in {\em Advances in imaging and electron physics};
quant-ph/0302028.

\item {\bf [D'Ariano-Macchiavello-Paris 03]}:
G. M. D'Ariano, C. Macchiavello, \& M. G. A. Paris,
``Local observables for entanglement detection of depolarized states'',
{\em Phys. Rev. A} {\bf 67}, 4, 042310 (2003);
quant-ph/0211146.

\item {\bf [D'Ariano-Sacchi 03]}:
G. M. D'Ariano, \& M. F. Sacchi,
``Protocols for entanglement transformations of bipartite pure states'',
{\em Phys. Rev. A} {\bf 67}, 4, 042312 (2003);
quant-ph/0209166.

\item {\bf [D'Ariano-Perinotti-Sacchi 03 a]}:
G. M. D'Ariano, P. Perinotti, \& M. F. Sacchi,
``Optimization of quantum universal detectors'',
{\em Proc. 8th Int. Conf. on Squeezed States and Uncertainty Relations ICSSUR' 2003 (Puebla, Mexico, 2003)},
quant-ph/0309161.

\item {\bf [D'Ariano-Perinotti-Sacchi 03 b]}:
G. M. D'Ariano, P. Perinotti, \& M. F. Sacchi,
``Informationally complete measurements and groups representation'',
quant-ph/0310013.

\item {\bf [D'Ariano-Lo Presti 03]}:
G. M. D'Ariano, \& P. Lo Presti,
``Imprinting complete information about a quantum channel on its output state'',
{\em Phys. Rev. Lett.} {\bf 91}, 4, 047902 (2003).

\item {\bf [D'Ariano 04]}:
G. M. D'Ariano,
``Extremal covariant quantum operations and POVM's'',
{\em J. Math. Phys.};
quant-ph/0310024.

\item {\bf [D'Ariano-Perinotti 04]}:
G. M. D'Ariano, \& P. Perinotti,
``On the realization of Bell observables'',
{\em Phys. Lett. A} {\bf 329}, 3, 188-192 (2004);
quant-ph/0406077.

\item {\bf [D'Ariano-Lo Presti-Perinotti 04]}:
G. M. D'Ariano, P. Lo Presti, \& P. Perinotti,
``Classical randomness in quantum measurements'',
quant-ph/0408115.

\item {\bf [D'Ariano-Maccone-Lo Presti 04]}:
G. M. D'Ariano, L. Maccone, \& P. Lo Presti,
``Quantum calibration of measuring apparatuses'',
quant-ph/0408116.

\item {\bf [D'Ariano-Perinotti 04]}:
G. M. D'Ariano, \& P. Perinotti,
``Efficient universal programmable quantum measurements'',
quant-ph/0410169.

\item {\bf [Arias-Gheondea-Gudder 02]}:
A. Arias, A. Gheondea, \& S. P. Gudder,
``Fixed points of quantum operations'',
{\em J. Math. Phys.} {\bf 43}, ?, 5872-? (2002).

\item {\bf [Arndt-Nairz-Vos Andreae-(+3) 99]}:
M. Arndt, O. Nairz, J. Vos Andreae, C. Keller,
G. van der Zouw, \& A. Zeilinger,
``Wave-particle duality of $C_{60}$ molecules'',
{\em Nature} {\bf 401}, 6754, 680-682 (1999).
See {\bf [Rae 99]}.

\item {\bf [Arndt-Nairz-Zeilinger 02]}:
M. Arndt, O. Nairz, \& A. Zeilinger,
``Interferometry with macromolecules:
Quantum paradigms tested in the mesoscopic world'',
in {\bf [Bertlmann-Zeilinger 02]}, pp.~333-350.

\item {\bf [Arnesen-Bose-Vedral 01]}:
M. C. Arnesen, S. Bose, \& V. Vedral,
``Natural thermal and magnetic entanglement in the 1D Heisenberg model'',
{\em Phys. Rev. Lett.} {\bf 87}, 1, 017901 (2001).

\item {\bf [Arntzenius 91]}:
F. Arntzenius,
``Kochen's interpretation of quantum mechanics'',
in A. I. Fine, M. Forbes, \& L. Wessels (eds.),
{\em Proc.\ of the 1990
Biennial Meeting of the Philosophy of Science Association},
East Lansing, Michigan, 1991, vol.~2, pp.~241-249.

\item {\bf [Arrighi-Patricot 03 a]}:
P. Arrighi, \& C. Patricot,
``A note on the correspondence between qubit quantum operations and
special relativity'',
{\em J. Phys. A} {\bf 36}, 20, L287-L296 (2003);
quant-ph/0212135.

\item {\bf [Arrighi-Patricot 03 b]}:
P. Arrighi, \& C. Patricot,
``Conal representation of quantum states and non-trace-preserving quantum
operations'',
{\em Phys. Rev. A} {\bf 68}, 4, 042310 (2003).

\item {\bf [Arrighi 03]}:
P. Arrighi,
``Quantum computation explained to my mother'',
quant-ph/0305045.

\item {\bf [Arsenovi\'{c}-Bo\v{z}i\'{c}-Vuskovi\'{c} 02]}:
D. Arsenovi\'{c}, M. Bo\v{z}i\'{c}, \& L. Vuskovi\'{c},
``Is the classical law of the addition of probabilities violated in quantum interference?'',
in R. Bonifacio, \& D. Vitali (eds.),
{\em Mysteries, Puzzles and Paradoxes in Quantum Mechanics IV:
Quantum Interference Phenomena (Gargnano, Italy, 2001)},
{\em J. Opt. B: Quantum Semiclass. Opt.} {\bf 4}, 4, S358-S365 (2002).

\item {\bf [Artiles-Gill-Guta 04]}:
L. M. Artiles, R. D. Gill, \& M. I. Guta,
``An invitation to quantum tomography (II)'',
math.ST/0405595.
See {\bf [Gill-Guta 04]} (I).

\item {\bf [Arul 01]}:
A. J. Arul,
``Impossibility of comparing and sorting quantum states'',
quant-ph/0107085.

\item {\bf [Arvind-Mallesh-Mukunda 99]}:
Arvind, K. S. Mallesh, \& N. Makunda,
``A generalized Pancharatnam geometric phase formula for three-level quantum systems'',
{\em J. Phys. A} {\bf 30}, 7, 2417-2431 (1997).

\item {\bf [Arvind-Mukunda 99]}:
Arvind, \& N. Mukunda,
``Bell's inequalities, multiphoton states and phase space
distributions'',
{\em Phys. Lett. A} {\bf 259}, 6, 421-426 (1999);
quant-ph/9806049.

\item {\bf [Arvind-Dorai-Kumar 99]}:
Arvind, K. Dorai, \& A. Kumar,
``Quantum entanglement in the NMR implementation of
the Deutsch-Jozsa algorithm'',
quant-ph/9909067.

\item {\bf [Arvind-Mukunda 00]}:
Arvind, \& N. Mukunda,
``A two-qubit algorithm involving quantum entanglement'',
quant-ph/0006069.

\item {\bf [Arvind 01]}:
Arvind,
``Quantum entanglement and quantum computational algorithms'',
{\em Pranama J. Phys.} {\bf 56}, 2-3, 357-365 (2001);
quant-ph/0012116.

\item {\bf [Arvind-Collins 03]}:
Arvind, \& D. Collins,
``Scaling issues in ensemble implementations of the Deutsch-Jozsa
algorithm'',
{\em Phys. Rev. A} {\bf 68}, 5, 052301 (2003);
quant-ph/0307153.

\item {\bf [Aschauer-Briegel 00]}:
H. Aschauer, \& H.-J. Briegel,
``Secure quantum communication over arbitrary distances'',
quant-ph/0008051.

\item {\bf [Aschauer-Briegel 02 a]}:
H. Aschauer, \& H. J. Briegel,
``Private entanglement over arbitrary distances, even using noisy apparatus'',
{\em Phys. Rev. Lett.} {\bf 88}, 4, 047902 (2002).

\item {\bf [Aschauer-Briegel 02 b]}:
H. Aschauer, \& H. J. Briegel,
``Entanglement purification with noisy apparatus
can be used to factor out an eavesdropper'',
{\em Eur. Phys. J. D} {\bf 18}, 2 (Special issue:
{\em Quantum interference and cryptographic keys:
Novel physics and advancing technologies (QUICK) (Corsica, 2001)}, 171-177 (2002);
quant-ph/0108060.
See {\bf [Aschauer-Briegel 02 c]}.

\item {\bf [Aschauer-Briegel 02]}:
H. Aschauer, \& H. J. Briegel,
``Security proof of quantum cryptography based entirely on entanglement
purification'',
{\em Phys. Rev. A} {\bf 66}, 3, 032302 (2002);
quant-ph/0111066.
See {\bf [Aschauer-Briegel 02 b]}.

\item {\bf [Aschauer-Calsamiglia-Hein-Briegel 03]}:
H. Aschauer, J. Calsamiglia, M. Hein, \& H. J. Briegel,
``Local invariants for multi-partite entangled states, allowing for a
simple entanglement criterion'',
quant-ph/0306048.

\item {\bf [Aschauer-D\"{u}r-Briegel 04]}:
H. Aschauer, W. D\"{u}r, \& H.-J. Briegel,
``Multiparticle entanglement purification for two-colorable graph states'',
quant-ph/0405045.

\item {\bf [Ashikhmin-Barg-Knill-Litsyn 99 a]}:
A. Ashikhmin, A. Barg, E. Knill, \& S. Litsyn,
``Quantum error detection I: Statement of the problem'',
quant-ph/9906126.
See {\bf [Ashikhmin-Barg-Knill-Litsyn 99 b]} (II).

\item {\bf [Ashikhmin-Barg-Knill-Litsyn 99 b]}:
A. Ashikhmin, A. Barg, E. Knill, \& S. Litsyn,
``Quantum error detection II: Bounds'',
quant-ph/9906131.
See {\bf [Ashikhmin-Barg-Knill-Litsyn 99 a]} (I).

\item {\bf [Ashikhmin-Knill 00]}:
A. Ashikhmin, \& E. Knill,
``Nonbinary quantum stabilizer codes'',
quant-ph/0005008.

\item {\bf [Ashikhmin-Litsyn-Tsfasman 01]}:
A. Ashikhmin, S. Litsyn, \& M. A. Tsfasman,
``Asymptotically good quantum codes'',
{\em Phys. Rev. A} {\bf 63}, 3, 032311 (2001);
quant-ph/000606.

\item {\bf [Asoudeh-Karimipour-Memarzadeh-Rezakhani 03]}:
M. Asoudeh, V. Karimipour, L. Memarzadeh, \& A. T. Rezakhani,
``Symmetrization and entanglement of arbitrary states of qubits'',
{\em Phys. Lett. A} {\bf 313}, 5-6, 330-337 (2003).

\item {\bf [Aspect 76]}:
A. Aspect,
``Proposed experiment to test the nonseparability of quantum mechanics'',
{\em Phys. Rev. D} {\bf 14}, 8, 1944-1951 (1976).
Reprinted in {\bf [Wheeler-Zurek 83]}, pp.~435-442.

\item {\bf [Aspect-Imbert-Roger 80]}:
A. Aspect, C. Imbert, \& G. Roger,
``Absolute measurement of an atomic cascade rate using a two-photon coincidence
technique. Application to the (4p2 1S0)->(4s4p 1P1)->(4s2 1S0) cascade of calcium
exited by a two photon absorption'',
{\em Opt. Comm.} {\bf 34}, ?, 46-? (1980).

\item {\bf [Aspect-Grangier-Roger 81]}:
A. Aspect, P. Grangier, \& G. Roger,
``Experimental tests of realistic local theories via Bell's theorem'',
{\em Phys. Rev. Lett.} {\bf 47}, 7, 460-463 (1981).

\item {\bf [Aspect-Grangier-Roger 82]}:
A. Aspect, P. Grangier, \& G. Roger,
``Experimental realization of Einstein-Podolsky-Rosen-Bohm
{\em Gedankenexperiment:} A new violation of Bell's inequalities'',
{\em Phys. Rev. Lett.} {\bf 49}, 1, 91-94 (1982).

\item {\bf [Aspect-Dalibard-Roger 82]}:
A. Aspect, J. Dalibard, \& G. Roger,
``Experimental test of Bell's inequalities using time-varying analyzers'',
{\em Phys. Rev. Lett.} {\bf 49}, 25, 1804-1807 (1982).
Reprinted in {\bf [Stroke 95]}, pp.~1247-1250.

\item {\bf [Aspect 83]}:
A. Aspect,
``Trois tests exp\'{e}rimentaux
des in\'{e}galit\'{e}s de Bell par
mesure de correlation de polarisation de photons'',
th\`{e}se, Universit\'{e} Paris-Sud, Centre d'Orsay, 1983.

\item {\bf [Aspect 85]}:
A. Aspect, in ``Reality and the quantum theory'',
{\em Phys. Today} {\bf 38}, 11, 9 (1985).
Comment on {\bf [Mermin 85]}.

\item {\bf [Aspect-Grangier 85]}:
A. Aspect, \& P. Grangier,
``Tests of Bell's inequalities with pairs of low energy correlated photons:
An experimental realization of Einstein-Podolsky-Rosen-type correlations'',
in P. J. Lahti, \& P. Mittelstaedt (eds.),
{\em Proc.\ Symp.\ on the Foundations of Modern
Physics: 50 Years of the Einstein-Podolsky-Rosen Experiment
(Joensuu, Finland, 1985)},
World Scientific, Singapore, 1985, pp.~51-72.

\item {\bf [Aspect 96]}:
A. Aspect,
``Wave-particle duality for single photon'',
in A. Mann, \& M. Revzen (eds.),
{\em The dilemma of Einstein, Podolsky and Rosen -- 60 years
later. An international symposium in honour of Nathan Rosen
(Haifa, Israel, 1995)},
{\em Ann. Phys. Soc. Israel} {\bf 12}, 182-201 (1996).

\item {\bf [Aspect 98]}:
A. Aspect,
``?'',
in {\em Waves, Information and Foundation of Physics},
Conf. Proc. Vol. 60, Italian Physical Society,
Bologna, 1998, pp.~?-?.

\item {\bf [Aspect 99]}:
A. Aspect,
``Bell's inequality test: More ideal than ever'',
{\em Nature} {\bf 398}, 6724, 189-190 (1999).
See {\bf [Weihs-Jennewein-Simon-(+2) 98]}.

\item {\bf [Aspect 00]}:
A. Aspect,
``Testing Bell's inequalities'',
in {\bf [Ellis-Amati 00]}, pp.~69-78.

\item {\bf [Aspect 02]}:
A. Aspect,
``Bell's theorem: The naive view of an experimentalist'',
in {\bf [Bertlmann-Zeilinger 02]}, pp.~119-154;
quant-ph/0402001.

\item {\bf [Aspelmeyer-Bohm-Gyatso-(+10) 03]}:
M. Aspelmeyer, H. R. Bohm, T. Gyatso,
T. Jennewein, R. Kaltenbaek, M. Lindenthal,
G. Molina-Terriza, A. Poppe, K. Resch,
M. Taraba, R. Ursin, P. Walther, \& A. Zeilinger,
``Long-distance free-space distribution of quantum entanglement'',
{\em Science} {\bf 301}, 5633, 621-623 (2003).

\item {\bf [Aspelmeyer-Jennewein-Pfennigbauer-(+2) 03]}:
M. Aspelmeyer, T. Jennewein, M. Pfennigbauer, W. Leeb, \& A. Zeilinger,
``Long-distance quantum communication with entangled photons using
satellites'',
submitted to {\em IEEE J. of Selected Topics in Quantum Electronics}
(special issue on {\em Quantum Internet technologies}),
quant-ph/0305105.

\item {\bf [Atat\"{u}re-Sergienko-Saleh-Teich 01]}:
M. Atat\"{u}re, A. V. Sergienko, B. E. A. Saleh, \& M. C. Teich,
``Entanglement in cascaded-crystal parametric down-conversion'',
{\em Phys. Rev. Lett.} {\bf 86}, 18, 4013-4016 (2001).

\item {\bf [Atat\"{u}re-Di Giuseppe-Shaw-(+3) 02 a]}:
M. Atat\"{u}re, G. Di Giuseppe, M. D. Shaw,
A. V. Sergienko, B. E. A. Saleh, \& M. C. Teich,
``Multi-parameter entanglement in femtosecond parametric down-conversion'',
{\em Phys. Rev. A} {\bf 65}, 2, 023808 (2002);
quant-ph/0110154.

\item {\bf [Atat\"{u}re-Di Giuseppe-Shaw-(+3) 02 b]}:
M. Atat\"{u}re, G. Di Giuseppe, M. D. Shaw,
A. V. Sergienko, B. E. A. Saleh, \& M. C. Teich,
``Multiparameter entanglement in quantum interferometry'',
{\em Phys. Rev. A} {\bf 66}, 2, 023822 (2002);
quant-ph/0111024.

\item {\bf [Atmanspacher-R\"{o}mer-Walach 02]}:
H. Atmanspacher, H. R\"{o}mer, \& H. Walach,
``Weak quantum theory:
Complementarity and entanglement in physics and beyond'',
{\em Found. Phys.} {\bf 32}, 3, 379-406 (2002);
quant-ph/0104109.

\item {\bf [Atmanspacher-Primas 03]}:
H. Atmanspacher, \& H. Primas,
``Epistemic and ontic quantum realities'',
in L. Castell, \& O. Ischebeck (eds.)
{\em Logic, quantum, and spacetime},
Springer, Berlin, 2003;
PITT-PHIL-SCI00000938.

\item {\bf [Atmanspacher-Ehm-Gneiting 03]}:
H. Atmanspacher, W. Ehm, \& T. Gneiting,
``Necessary and sufficient conditions for the quantum Zeno and anti-Zeno effect'',
{\em J. Phys. A} {\bf 36}, 38, 9899-9905 (2003).

\item {\bf [Auberson-Mahoux-Roy-Singh 02]}:
G. Auberson, G. Mahoux, S. M. Roy, \& V. Singh,
``Bell inequalities in phase space and their violation
in quantum mechanics'',
{\em Phys. Lett. A} {\bf 300}, 4-5, 327-333 (2002).

\item {\bf [Auberson-Mahoux-Roy-Singh 03]}:
G. Auberson, G. Mahoux, S. M. Roy, \& V. Singh,
``Bell inequalities in four dimensional phase space and the three marginal
theorem'',
{\em J. Math. Phys.} {\bf 44}, ?, 2729-? (2003).

\item {\bf [Audenaert-Verstraete-De Bie-De Moor 00]}:
K. M. R. Audenaert, F. Verstraete, T. De Bie \& B. De Moor,
``Negativity and concurrence of mixed $2 \times 2$ states'',
quant-ph/0012074.
See {\bf [Eisert-Plenio 99]}.

\item {\bf [Audenaert-Verstraete-De Moor 01]}:
K. M. R. Audenaert, F. Verstraete, \& B. De Moor,
``Variational characterizations of separability and entanglement of formation'',
{\em Phys. Rev. A} {\bf 64}, 5, 052304 (2001);
quant-ph/0006128.

\item {\bf [Audenaert-Eisert-Jan\'{e}-(+3) 01]}:
K. M. R. Audenaert, J. Eisert, E. Jan\'{e},
M. B. Plenio, S. Virmani, \& B. De Moor,
``Asymptotic relative entropy of entanglement'',
{\em Phys. Rev. Lett.} {\bf 87}, 21, 217902 (2001);
quant-ph/0103096.

\item {\bf [Audenaert-De Moor 02]}:
K. M. R. Audenaert, \& B. De Moor,
``Optimizing completely positive maps using semidefinite programming'',
{\em Phys. Rev. A} {\bf 65}, 3, 030302 (2002);
quant-ph/0109155.

\item {\bf [Audenaert-De Moor-Vollbrecht-Werner 02]}:
K. M. R. Audenaert, B. De Moor, K. G. H. Vollbrecht, \& R. F. Werner,
``Asymptotic relative entropy of entanglement for orthogonally invariant states'',
{\em Phys. Rev. A} {\bf 66}, 3, 032310 (2002);
quant-ph/0204143.
See {\bf [Ishizaka 04]}.

\item {\bf [Audenaert-Eisert-Plenio-Werner 02]}:
K. M. R. Audenaert, J. Eisert, M. B. Plenio, \& R. F. Werner,
``Entanglement properties of the harmonic chain'',
{\em Phys. Rev. A} {\bf 66}, 4, 042327 (2002).

\item {\bf [Audenaert-Plenio-Eisert 03]}:
K. M. R. Audenaert, M. B. Plenio, \& J. Eisert,
``Entanglement cost under positive-partial-transpose-preserving operations'',
{\em Phys. Rev. Lett.} {\bf 90}, 2, 027901 (2003).

\item {\bf [Audenaert-Braunstein 03]}:
K. M. R. Audenaert, \& S. L. Braunstein,
``On strong superadditivity of the entanglement of formation'',
quant-ph/0303045.

\item {\bf [Audenaert-Fuchs-King-Winter 03]}:
K. M. R. Audenaert, C. A. Fuchs, C. King, \& A. Winter,
``Multiplicativity of accessible fidelity and quantumness for sets of
quantum states'',
quant-ph/0308120.
See {\bf [Fuchs-Sasaki 03 a]}.

\item {\bf [Audenaert 04]}:
K. M. R. Audenaert,
``There, and back again: Quantum theory and global optimisation'',
quant-ph/0402076.

\item {\bf [Audretsch-Di\'{o}si-Konrad 02]}:
J. Audretsch, L. Di\'{o}si, \& T. Konrad,
``Evolution of a qubit under the influence of a succession of weak
measurements with unitary feedback'',
{\em Phys. Rev. A} {\bf 66}, 2, 022310 (2002);
quant-ph/0201078.

\item {\bf [Audretsch-Di\'{o}si-Konrad 03]}:
J. Audretsch, L. Di\'{o}si, \& T. Konrad,
``Estimating the postmeasurement state'',
{\em Phys. Rev. A} {\bf 68}, 3, 034302 (2003).

\item {\bf [Auffeves, P. Maioli, T. Meunier-(+5) 03]}:
A. Auffeves, P. Maioli, T. Meunier,
S. Gleyzes, G. Nogues, M. Brune,
J. M. Raimond, \& S. Haroche,
``Entanglement of a mesoscopic field with an atom induced by photon graininess in a cavity'',
{\em Phys. Rev. Lett.} {\bf 91}, 23, 230405 (2003).

\item {\bf [Augusiak-Horodecki 04]}:
R. Augusiak, \& P. Horodecki,
``Bound entanglement maximally violating Bell inequalities: Quantum
entanglement is not equivalent to quantum security'',
quant-ph/0405187.

\item {\bf [Auletta 00]}:
G. Auletta,
{\em Foundations and interpretation of quantum mechanics
in the light of a critical-historical analysis of the problems
and of a synthesis of the results},
World Scientific, Singapore, 2000.
Review: {\bf [Vedral 01]}, {\bf [Introzzi 02]}.

\item {\bf [Averin 99]}:
D. V. Averin,
``Solid-state qubits under control'',
{\em Nature} {\bf 398}, 6730, 748-749 (1999).
See {\bf [Nakamura-Pashkin-Tsai 99]}.

\item {\bf [Averin 00]}:
D. V. Averin,
``Quantum computing and quantum measurement with mesoscopic Josephson junctions'',
{\em Fortschr. Phys.} {\bf 48}, 9-11 (Special issue:
Experimental proposals for quantum computation), 1055-1074 (2000);
quant-ph/0008114.

\item {\bf [Averin 02]}:
D. V. Averin,
``Quantum nondemolition measurements of a qubit'',
{\em Phys. Rev. Lett.} {\bf 88}, 20, 207901 (2002);
cond-mat/0202082.

\item {\bf [Averin-Bruder 03]}:
D. V. Averin, \& C. Bruder,
``Variable electrostatic transformer: Controllable coupling of two charge
qubits'',
{\em Phys. Rev. Lett.} {\bf 91}, 5, 057003 (2003).

\item {\bf [Averin-Fazio 03]}:
D. V. Averin, \& R. Fazio,
``Active suppression of dephasing in Josephson-junction qubits'',
{\em JETP Lett.} {\bf 78}, ?, 664-668 (2003).

\item {\bf [Avis-Imai-Ito-Sasaki 04]}:
D. Avis, H. Imai, T. Ito, \& Y. Sasaki,
``Deriving tight Bell inequalities for 2 parties with many 2-valued
observables from facets of cut polytopes'',
quant-ph/0404014.

\item {\bf [Awschalom-Loss-Samarth 02]}:
D. D. Awschalom, D. Loss, \& N. Samarth,
{\em Semiconductor spintronics and quantum computation},
Springer-Verlag, New York, 2002.

\item {\bf [Ax-Kochen 99]}:
J. Ax, \& S. Kochen,
``Extension of quantum mechanics to individual systems'',
quant-ph/9905077.

\item {\bf [Azuma 98]}:
H. Azuma,
``Building partially entangled states defined with even collision functions'',
quant-ph/9810093.

\item {\bf [Azuma-Ban 01]}:
H. Azuma, \& M. Ban,
``A method of enciphering quantum states'',
{\em J. Phys. A} {\bf 34}, 13, 2723-2742 (2001);
quant-ph/0006124.

\item {\bf [Azuma-Bose-Vedral 01]}:
H. Azuma, S. Bose, \& V. Vedral,
``Entangling capacity of global phases and implications for the Deutsch-Jozsa
algorithm'',
{\em Phys. Rev. A} {\bf 64}, 6, 062308 (2001);
quant-ph/0102029.

\item {\bf [Azuma 02]}:
H. Azuma,
``Decoherence in Grover's quantum algorithm: Perturbative approach'',
{\em Phys. Rev. A} {\bf 65}, 4, 042311 (2002).
Erratum: {\em Phys. Rev. A} {\bf 66}, 1, 019903 (2002);
quant-ph/0110101.

\item {\bf [Azuma 03]}:
H. Azuma,
``Interaction-free generation of entanglement'',
{\em Phys. Rev. A} {\bf 68}, 2, 022320 (2003);
quant-ph/0304031.

\item {\bf [Azuma 04]}:
H. Azuma,
``Interaction-free quantum computation'',
{\em Phys. Rev. A} {\bf 70}, 1, 012318 (2004);
quant-ph/0403159.


\newpage

\subsection{}


\item {\bf [Babichev-Brezger-Lvovsky 03]}:
S. A. Babichev, B. Brezger, \& A. I. Lvovsky,
``Remote preparation of a single-mode photonic qubit by measuring field
quadrature noise'',
quant-ph/0308127.

\item {\bf [Babichev-Appel-Lvovsky 03]}:
S. A. Babichev, J. Appel, \& A. I. Lvovsky,
``Homodyne tomography characterization and nonlocality of a dual-mode
optical qubit'',
quant-ph/0312135.

\item {\bf [Babichev-Brezger-Lvovsky 04]}:
S. A. Babichev, B. Brezger, \& A. I. Lvovsky,
``Remote preparation of a single-mode photonic qubit by measuring field quadrature noise'',
{\em Phys. Rev. Lett.} {\bf 92}, 4, 047903 (2004).

\item {\bf [Bacciagaluppi 94]}:
G. Bacciagaluppi,
``Separation theorems and Bell inequalities in algabraic quantum mechanics'',
in P. Busch, P. J. Lahti, \& P. Mittelstaedt (eds.),
{\em Proc.\ Symp.\ on the Foundations of Modern Physics
(Cologne, Germany, 1993)},
World Scientific, Singapore, 1994, pp.~29-37.

\item {\bf [Bacciagaluppi 95]}:
G. Bacciagaluppi,
``Kochen-Specker theorem in the
modal interpretation of quantum mechanics'',
{\em Int. J. Theor. Phys.} {\bf 34}, 8, 1205-1216 (1995).

\item {\bf [Bacciagaluppi 96]}:
G. Bacciagaluppi,
``Topics in the modal interpretation of quantum mechanics'',
Ph.\ D. thesis, Cambridge University, 1996.

\item {\bf [Bacciagaluppi-Donald-Vermaas 95]}:
G. Bacciagaluppi, M. J. Donald, \& P. E. Vermaas,
``Continuity and discontinuity of definite properties in the modal
interpretation'',
{\em Helv. Phys. Acta} {\bf 68}, 7-8, 679-704 (1995).

\item {\bf [Bacciagaluppi-Hemmo 96]}:
G. Bacciagaluppi, \& M. Hemmo,
``Modal interpretations, decoherence and measurements'',
{\em Stud. Hist. Philos. Sci. Part B:
Stud. Hist. Philos. Mod. Phys.} {\bf 27}, 3, 239-277 (1996).

\item {\bf [Bacciagaluppi-Hemmo 98 a]}:
G. Bacciagaluppi, \& M. Hemmo,
``State preparation in the modal interpretation'',
in G. Hellman, \& R. A. Healey (eds.),
{\em Quantum measurement, decoherence, and modal interpretations
(Minnesota Studies in Philosophy of Science)}, 1998.

\item {\bf [Bacciagaluppi-Hemmo 98 b]}:
G. Bacciagaluppi, \& M. Hemmo,
``Modal interpretations of imperfect measurements'',
{\em Stud. Hist. Philos. Sci. Part B:
Stud. Hist. Philos. Mod. Phys.} {\bf ?}, ?, ?-? (1998).

\item {\bf [Bacciagaluppi 98 a]}:
G. Bacciagaluppi,
``A note on dynamics in the modal interpretation'',
{\em Int. J. Theor. Phys.} {\bf 37}, 1, 421-426 (1998).

\item {\bf [Bacciagaluppi 98 b]}:
G. Bacciagaluppi,
``Nelsonian mechanics revisited'',
quant-ph/9811040.

\item {\bf [Bacciagaluppi-Dickson 99]}:
G. Bacciagaluppi, \& W. M. Dickson,
``Dynamics for modal interpretations'',
{\em Found. Phys.} {\bf 29}, 8, 1165-1202 (1999).

\item {\bf [Bacciagaluppi 99]}:
G. Bacciagaluppi,
``Nelsonian mechanics revisited'',
{\em Found. Phys. Lett.} {\bf 12}, 1, 1-16 (1999).

\item {\bf [Bacciagaluppi 00]}:
G. Bacciagaluppi,
``Delocalized properties in the modal interpretation
of a continuous model of decoherence'',
{\em Found. Phys.} {\bf 30}, 9, 1431-1444 (2000).

\item {\bf [Bacciagaluppi 01 a]}:
G. Bacciagaluppi,
{\em The modal interpretation of quantum mechanics},
Cambridge University Press, Cambridge, 2001.

\item {\bf [Bacciagaluppi 01 b]}:
G. Bacciagaluppi,
``Remarks on space-time and locality in Everett's Interpretation'' (2001),
PITT-PHIL-SCI00000504.

\item {\bf [Bacciagaluppi 03]}:
G. Bacciagaluppi,
``Derivation of the symmetry postulates for identical particles from
pilot-wave theories'',
quant-ph/0302099,
PITT-PHIL-SCI00000993.

\item {\bf [Bacciagaluppi 04]}:
G. Bacciagaluppi,
``Classical extensions, classical representations and Bayesian updating in
quantum mechanics'',
{\em Int.\ Conf.\ on Quantum
Theory: Reconsideration of Foundations (V\"{a}xj\"{o}, Sweden 2003)};
quant-ph/0403055.

\item {\bf [Bacon-Lidar-Whaley 99]}:
D. Bacon, D. A. Lidar, \& K. B. Whaley,
``Robustness of decoherence-free subspaces for quantum computation'',
{\em Phys. Rev. A} {\bf 60}, 3, 1944-1955 (1999).

\item {\bf [Bacon-Kempe-Lidar-Whaley 00]}:
D. Bacon, J. Kempe, D. A. Lidar, \& K. B. Whaley,
``Universal fault-tolerant quantum computation
on decoherence-free subspaces'',
{\em Phys. Rev. Lett.} {\bf 85}, 8, 1758-1761 (2000);
quant-ph/9909058.
See {\bf [Osborne 00 d]}.

\item {\bf [Bacon-Childs-Chuang-(+3) 01]}:
D. Bacon, A. M. Childs, I. L. Chuang,
J. Kempe, D. W. Leung, \& X. Zhou,
``Universal simulation of Markovian quantum dynamics'',
{\em Phys. Rev. A} {\bf 64}, 6, 062302 (2001);
quant-ph/0008070.

\item {\bf [Bacon-Kempe-DiVincenzo-(+2) 01]}:
D. Bacon, J. Kempe, D. P. DiVincenzo,
D. A. Lidar, \& K. B. Whaley,
``Encoded universality in physical implementations of a quantum
computer'',
to appear in {\em Experimental Implementation of Quantum Computation
(Sydney, 2001)};
quant-ph/0102140.

\item {\bf [Bacon-Brown-Whaley 01]}:
D. Bacon, K. R. Brown, \& K. B. Whaley,
``Coherence-preserving quantum bits'',
{\em Phys. Rev. Lett.} {\bf 87}, 24, 247902 (2001).

\item {\bf [Bacon 01]}:
D. Bacon,
{\em Decoherence, control, and symmetry in quantum computers},
Ph.\ D. thesis, University of California, Berkeley, 2001;
quant-ph/0305025.

\item {\bf [Bacon-Toner 03]}:
D. Bacon, \& B. F. Toner,
``Bell inequalities with auxiliary communication'',
{\em Phys. Rev. Lett.} {\bf 90}, 15, 157904 (2003);
quant-ph/0208057.

\item {\bf [Bacon-Chuang-Harrow 04]}:
D. Bacon, I. L. Chuang, \& A. Harrow,
``Efficient quantum circuits for Schur and Clench-Gordon transforms'',
quant-ph/0407082.

\item {\bf [Badagnani 02]}:
D. O. Badagnani,
``A Bell telegraph'',
quant-ph/0204108.

\item {\bf [Badurek-Rauch-Summhammer 83]}:
G. Badurek, H. Rauch, \& J. Summhammer,
``Time-dependent superposition of spinors'',
{\em Phys. Rev. Lett.} {\bf 51}, 12, 1015-1018 (1983).
See {\bf [Summhammer-Badurek-Rauch-Kischko 82]},
{\bf [Badurek-Rauch-Tuppinger 86]}.

\item {\bf [Badurek-Rauch-Tuppinger 86]}:
G. Badurek, H. Rauch, \& D. Tuppinger,
``Neutron interferometric double-resonance experiment'',
{\em Phys. Rev. A} {\bf 34}, 4, 2600-2608 (1986).
See {\bf [Summhammer-Badurek-Rauch-Kischko 82]},
{\bf [Badurek-Rauch-Summhammer 83]}.

\item {\bf [Badzi\c{a}g-Horodecki-Horodecki.Horodecki 00]}:
P. Badzi\c{a}g, M. Horodecki, P. Horodecki, \& R. Horodecki,
``Local environment can enhance fidelity of quantum teleportation'',
{\em Phys. Rev. A} {\bf 62}, 1, 012311 (2000);
quant-ph/9912098.

\item {\bf [Badzi\c{a}g-Deuar-Horodecki-(+2) 01]}:
P. Badzi\c{a}g, P. Deuar, M. Horodecki,
P. Horodecki, \& R. Horodecki,
``Concurrence in arbitrary dimensions'',
{\em J. Mod. Opt.} {\bf 49} 8, 1289-1297 (2002);
quant-ph/0107147.

\item {\bf [Badzi\c{a}g-Horodecki-Sen De-Sen 03]}:
P. Badziag, M. Horodecki, A. Sen De, \& U. Sen,
``Locally accessible information: How much can the parties gain by
cooperating?'',
{\em Phys. Rev. Lett.} {\bf 91}, 11, 117901 (2003).

\item {\bf [Bae-Kwon 02]}:
J. Bae, \& Y. Kwon,
``Generalized quantum search Hamiltonian'',
{\em Phys. Rev. A} {\bf 66}, 1, 012314 (2002);
quant-ph/0110020.

\item {\bf [Bae-Kwon 03 a]}:
J. Bae, \& Y. Kwon,
``Maximum speedup in quantum search: $O(1)$ running time'',
{\em Int. J. Theor. Phys.} {\bf 42}, 9, 2069-2074 (2003);
quant-ph/0204087.

\item {\bf [Bae-Kwon 03 a]}:
J. Bae, \& Y. Kwon,
``Perturbations can enhance quantum search
{\em Int. J. Theor. Phys.} {\bf 42}, 9, 2075-2080 (2003);
quant-ph/0211186.

\item {\bf [De Baere 96 a]}:
W. de Baere,
``How to avoid Peres' `Incompatible results
of quantum measurements'\,'',
{\em Found. Phys. Lett.} {\bf 9}, 1, 67-78 (1996).
See {\bf [Peres 90 b]}.

\item {\bf [De Baere 96 b]}:
W. de Baere,
``Quantum nonreproducibility at the individual level as
a resolution of quantum paradoxes'',
in A. Mann, \& M. Revzen (eds.),
{\em The dilemma of Einstein, Podolsky and Rosen -- 60 years
later. An international symposium in honour of Nathan Rosen
(Haifa, Israel, 1995)},
{\em Ann. Phys. Soc. Israel} {\bf 12}, 95-108 (1996).

\item {\bf [De Baere 98]}:
W. de Baere,
``On some consequences of the breakdown of counterfactual
definiteness in the quantum world'',
{\em Fortschr. Phys.} {\bf 46}, 6-8, 843-847 (1998).

\item {\bf [De Baere-Mann-Revzen 99]}:
W. de Baere, A. Mann, \& M. Revzen,
``Locality and Bell's theorem'',
{\em Found. Phys.} {\bf 29}, 1, 67-78 (1999).

\item {\bf [Bagan-Baig-Brey-(+2) 00 a]}:
E. Bagan, M. Baig, A. Brey,
R. Mu\~{n}oz-Tapia, \& R. Tarrach,
``Optimal strategies for sending information
through a quantum channel'',
{\em Phys. Rev. Lett.} {\bf 85}, 24, 5230-5233 (2000);
quant-ph/0006075.

\item {\bf [Bagan-Baig-Brey-(+2) 00 b]}:
E. Bagan, M. Baig, A. Brey,
R. Mu\~{n}oz-Tapia, \& R. Tarrach,
``Optimal encoding and decoding of a spin direction'',
{\em Phys. Rev. A} {\bf 63}, 5, 052309 (2001);
quant-ph/0012006.

\item {\bf [Bagan-Baig-Mu\~{n}oz Tapia 01 a]}:
E. Bagan, M. Baig, \& R. Mu\~{n}oz-Tapia,
``Communication of spin directions with product
states and finite measurements'',
{\em Phys. Rev. A} {\bf 64}, 2, 022305 (2001);
quant-ph/0012045.

\item {\bf [Bagan-Baig-Mu\~{n}oz Tapia 01 b]}:
E. Bagan, M. Baig, \& R. Mu\~{n}oz-Tapia,
``Aligning reference frames with quantum states'',
{\em Phys. Rev. Lett.} {\bf 87}, 25, 257903 (2001);
quant-ph/0106014.

\item {\bf [Bagan-Baig-Mu\~{n}oz Tapia 01 c]}:
E. Bagan, M. Baig, \& R. Mu\~{n}oz-Tapia,
``Communicating a direction using spin states'',
contribution to the {\em ICQI (Rochester, New York, 2001)};
quant-ph/0106155.

\item {\bf [Bagan-Baig-Mu\~{n}oz Tapia 02]}:
E. Bagan, M. Baig, \& R. Mu\~{n}oz-Tapia,
``Optimal scheme for estimating a pure qubit state via local measurements'',
{\em Phys. Rev. Lett.} {\bf 89}, 27, 277904 (2002);
quant-ph/0205026.

\item {\bf [Bagan-Baig-Mu\~{n}oz Tapia 03 a]}:
E. Bagan, M. Baig, \& R. Mu\~{n}oz-Tapia,
``Minimal measurements of the gate fidelity of a qudit map'',
{\em Phys. Rev. A} {\bf 67}, 1, 014303 (2003);
quant-ph/0207152.

\item {\bf [Bagan-Baig-Mu\~{n}oz Tapia 03 b]}:
E. Bagan, M. Baig, \& R. Mu\~{n}oz-Tapia,
``Entanglement assisted alignment of reference frames using a dense
covariant coding'',
{\em Phys. Rev. A};
quant-ph/0303019.

\item {\bf [Bagan-Baig-Monras-Mu\~{n}oz Tapia 03]}:
E. Bagan, M. Baig, A. Monras, \& R. Mu\~{n}oz-Tapia,
``Optimal reconstruction of a pure qubit state with local measurements'',
{\em Proc. of the 8th ICSSUR (Puebla, Mexico, 2003)},
Rinton Press, Princeton, New Jersey, 2003;
quant-ph/0312167.

\item {\bf [Bagan-Baig-Mu\~{n}oz Tapia-Rodr\'{\i}guez 04]}:
E. Bagan, M. Baig, R. Mu\~{n}oz-Tapia, \& A. Rodr\'{\i}guez,
``Collective versus local measurements in a qubit mixed-state estimation'',
{\em Phys. Rev. A} {\bf 69}, 1, 010304 (2004);
quant-ph/0307199.

\item {\bf [Bagan-Baig-Mu\~{n}oz Tapia 04]}:
E. Bagan, M. Baig, \& R. Mu\~{n}oz-Tapia,
``Quantum reverse-engineering and reference frame alignment without
non-local correlations
quant-ph/0405082.

\item {\bf [Baggott 92]}:
J. Baggott,
{\em The meaning of quantum theory},
Oxford University Press, Oxford, 1992.

\item {\bf [Bagherinezhad-Karimipour 03]}:
S. Bagherinezhad, \& V. Karimipour,
``Quantum secret sharing based on reusable Greenberger-Horne-Zeilinger states
as secure carriers'',
{\em Phys. Rev. A} {\bf 67}, 4, 044302 (2003);
quant-ph/0204124.

\item {\bf [Balachandran-Roy 00]}:
A. P. Balachandran, \& S. M. Roy,
``Quantum anti-Zeno paradox'',
{\em Phys. Rev. Lett.} {\bf 84}, 18, 4019-4022 (2000);
quant-ph/9909056.
Comment: {\bf [Di\'{o}si 01]}.

\item {\bf [Balachandran-Roy 01]}:
A. P. Balachandran, \& S. M. Roy,
``Continuous time-dependent measurements:
Quantum anti-Zeno paradox with applications'',
quant-ph/0102019.

\item {\bf [Balduz 01]}:
J. L. Balduz, Jr.,
``Quantum theory of observers'',
quant-ph/0108079.

\item {\bf [Balian 89]}:
R. Balian,
``On the principles of quantum mechanics and the
reduction of the wave packet'',
{\em Am. J. Phys.} {\bf 57}, 11, 1019-1027 (1989).

\item {\bf [Ball-Banaszek 04]}:
J. L. Ball, \& K. Banaszek,
``Hybrid noiseless subsystems for quantum communication over optical
fibers'',
quant-ph/0410077.

\item {\bf [Ballentine 70]}:
L. E. Ballentine,
``The statistical interpretation of quantum mechanics'',
{\em Rev. Mod. Phys.} {\bf 42}, 4, 358-381 (1970).
See {\bf [Kunstatter-Trainor 84 b]}.

\item {\bf [Ballentine-Pearle-Walker-(+4) 71]}:
L. E. Ballentine, P. Pearle, E. H. Walker,
M. Sachs, T. Koga, J. Gerver, \& B. DeWitt,
``Quantum mechanics debate'',
{\em Phys. Today} {\bf 24}, 4, 36-44 (1971).

\item {\bf [Ballentine 72]}:
L. E. Ballentine,
``Einstein's interpretation of quantum mechanics'',
{\em Am. J. Phys.} {\bf 40}, 12, 1763-1771 (1972).

\item {\bf [Ballentine 73]}:
L. E. Ballentine,
``Can the statistical postulate of quantum theory be derived?---A
critique of the many-universes interpretation'',
{\em Found. Phys.} {\bf 3}, ?, 229-240 (1973).

\item {\bf [Ballentine 74]}:
L. E. Ballentine,
`Comments on Stapp's ``Copenhagen interpretation'' and the significance
of Bell's theorem',
{\em Am. J. Phys.} {\bf 42}, 1, 81-83 (1974).
Comment on {\bf [Stapp 72]}. Reply: {\bf [Stapp 74]}.

\item {\bf [Ballentine 86 a]}:
L. E. Ballentine,
``Probability in quantum mechanics'',
{\em Am. J. Phys.} {\bf 54}, 10, 883-889 (1986).

\item {\bf [Ballentine 86 b]}:
L. E. Ballentine,
``What is the point of the quantum theory of measurement?'',
in L. M. Roth, \& A. Inomata (eds.),
{\em Fundamental questions in quantum mechanics
(Albany, New York, 1984)},
Gordon \& Breach, New York, 1986, pp.~65-75.

\item {\bf [Ballentine-Jarrett 87]}:
L. E. Ballentine, \& J. P. Jarrett,
``Bell's theorem: Does quantum mechanics contradict relativity?'',
{\em Am. J. Phys.} {\bf 55}, 8, 696-701 (1987).
See {\bf [Jarrett 84]}.

\item {\bf [Ballentine 87 a]}:
L. E. Ballentine,
``Resource letter IQM-2: Foundations
of quantum mechanics since the Bell inequalities'',
{\em Am. J. Phys.} {\bf 55}, 9, 785-792 (1987).
See {\bf [DeWitt-Graham 71]} (I).

\item {\bf [Ballentine 87 b]}:
L. E. Ballentine,
``Realism and quantum flux tunneling'',
{\em Phys. Rev. Lett.} {\bf 59}, 14, 1493-1495 (1987).

\item {\bf [Ballentine 88 a]}:
L. E. Ballentine,
``What do we learn about quantum
mechanics from the theory of measurement?'',
{\em Int. J. Theor. Phys.} {\bf 27}, 2, 211-218 (1988).

\item {\bf [Ballentine 88 b]}:
L. E. Ballentine (ed.),
{\em Foundations of quantum
mechanics since the Bell inequalities, selected reprints},
Am. Asoc. of Physics Techers,
College Park, Maryland, 1988.

\item {\bf [Ballentine 90 a]}:
L. E. Ballentine,
{\em Quantum mechanics},
Prentice-Hall, Englewood Cliffs, New Jersey, 1990.
Reviews: {\bf [Greenberger 91]}, {\bf [Griffiths 91]}.
See {\bf [Ballentine 98]}.

\item {\bf [Ballentine 90 b]}:
L. E. Ballentine,
``Limitations of the projection postulate'',
{\em Found. Phys.} {\bf 20}, 11, 1329-1343 (1990).

\item {\bf [Ballentine 91]}:
L. E. Ballentine,
`Comment on ``Quantum Zeno effect''\,'
{\em Phys. Rev. A} {\bf 43}, 9, 5165-5167 (1991).
Comment on {\bf [Itano-Heinzen-Bollinger-Wineland 90]}.
Reply: {\bf [Itano-Heinzen-Bollinger-Wineland 91]}.

\item {\bf [Ballentine 95 a]}:
L. E. Ballentine,
``The emergence of classical
properties from quantum mechanics: New problems from old'',
in M. Ferrero, \& A. van
der Merwe (eds.),
{\em Fundamental problems in quantum physics.
Proc.\ of an international symposium (Oviedo, Spain, 1993)},
Kluwer Academic, Dordrecht, Holland, 1995, pp.~15-28.
See {\bf [Ballentine 96]}.

\item {\bf [Ballentine 95 b]}:
L. E. Ballentine,
``Quantum theory: Concepts and methods'',
{\em Am. J. Phys.} {\bf 63}, 3, 285-286 (1995).
Review of {\bf [Peres 93 a]}.

\item {\bf [Ballentine 96]}:
L. E. Ballentine,
``The emergence of classical properties from quantum mechanics'',
in R. K. Clifton (ed.), {\em Perspectives on quantum reality:
Non-relativistic, relativistic, and field-theoretic
(London, Western Ontario, Canada, 1994)},
Kluwer Academic, Dordrecht, Holland, 1996, pp.~?-?.
See {\bf [Ballentine 95 a]}.

\item {\bf [Ballentine 98]}:
L. E. Ballentine,
{\em Quantum mechanics: A modern development},
World Scientific, Singapore, 1998.
Enlarged edition of {\bf [Ballentine 90 a]}.

\item {\bf [Ballentine-McRae 98]}:
L. E. Ballentine, \& S. M. McRae
``Moment equations for probability distributions in classical and quantum
mechanics'',
{\em Phys. Rev. A} {\bf 58}, 3, 1799-1809 (1998).

\item {\bf [Ballentine 99]}:
L. E. Ballentine,
``Fractal quantum probability distributions and the classical limit'',
{\em Phys. Lett. A} {\bf 261}, 3-4, 145-149 (1999).

\item {\bf [Ballentine 03]}:
L. E. Ballentine,
``Quantum measurement of a single system'',
{\em Am. J. Phys.} {\bf 71}, 6, 639-640 (2003).
Review of {\bf [Alter-Yamamoto 01]}.

\item {\bf [Ballester 04 a]}:
M. A. Ballester,
``Estimation of unitary quantum operations'',
{\em Phys. Rev. A} {\bf 69}, 2, 022303 (2004).

\item {\bf [Ballester 04 b]}:
M. A. Ballester,
``Entanglement is not very useful for estimating multiple phases'',
quant-ph/0403190.

\item {\bf [Balzer-Huesmann-Neuhauser-Toschek 00]}:
C. Balzer, R. Huesmann, W. Neuhauser, \& P. E. Toschek,
``The quantum Zeno effect -- Evolution of an atom impeded by
measurement'',
{\em Opt. Comm.} {\bf 180}, 115-120 (2000);
quant-ph/0105004.

\item {\bf [Balzer-Hannemann-Rei\ss,\, 01]}:
C. Balzer, T. Hannemann, D. Rei\ss,
W. Neuhauser, P. E. Toschek, \& C. Wunderlich,
``From spectral relaxation to quantified decoherence'',
quant-ph/0109037.

\item {\bf [Balzer-Hannemann-Rei\ss-(+3) 02]}:
C. Balzer, T. Hannemann, D. Rei\ss,
C. Wunderlich, W. Neuhauser, \& P. E. Toschek,
``A relaxationless demonstration of the quantum Zeno paradox on an
individual atom'',
{\em Opt. Comm.} {\bf 211}, 235-241 (2002);
quant-ph/0406027.

\item {\bf [Ban-Yamazaki-Hirota 97]}:
M. Ban, K. Yamazaki, \& O. Hirota,
``Accessible information in combined and sequential
quantum measurements on a binary-state signal'',
{\em Phys. Rev. A} {\bf 55}, 1, 22-26 (1997).
See {\bf [Peres-Wootters 91]}.

\item {\bf [Ban 98]}:
M. Ban,
``Information-theoretical properties of a sequence of quantum
nondemolition measurements'',
{\em Phys. Lett. A} {\bf 249}, 3, 167-179 (1998).

\item {\bf [Ban 99 a]}:
M. Ban,
``State reduction, information and entropy
in quantum measurement processes'',
{\em J. Phys. A} {\bf 32}, 9, 1643-1665 (1999).

\item {\bf [Ban 99 b]}:
M. Ban,
``Information transmission in quantum measurement processes with pure
operation'',
{\em J. Phys. A} {\bf 32}, 37, 6527-6533 (1999).

\item {\bf [Ban 99 c]}:
M. Ban,
``Quantum dense coding via two-mode squeezed-vacuum state'',
{\em J. Opt. B: Quantum Semiclass. Opt.} {\bf 1}, 6, L9-L11 (1999).

\item {\bf [Ban 00 a]}:
M. Ban,
``Entanglement-assisted quantum communication system with coherent signals'',
{\em J. Opt. B: Quantum Semiclass. Opt.} {\bf 2}, 5, 576-580 (2000).

\item {\bf [Ban 00 b]}:
M. Ban,
``Information transmission via dense coding in a noisy quantum
channel'',
{\em Phys. Lett. A} {\bf 276}, 5-6, 213-220 (2000).

\item {\bf [Ban 00 c]}:
M. Ban,
``Quantum dense coding of continuous variables in a noisy quantum channel`'',
{\em J. Opt. B: Quantum Semiclass. Opt.} {\bf 2}, 6, 786-791 (2000).

\item {\bf [Ban 02]}:
M. Ban,
``Optimal signal detection in entanglement-assisted quantum communication systems'',
{\em J. Opt. B: Quantum Semiclass. Opt.} {\bf 4}, 2, 143-148 (2002).

\item {\bf [Ban-Sasaki-Takeoka 02]}:
M. Ban, M. Sasaki, \& M. Takeoka,
``Continuous variable teleportation as a generalized thermalizing quantum channel'',
quant-ph/0202172.

\item {\bf [Ban 03 a]}:
M. Ban,
``Partially reversible quantum operations and their information-theoretical properties'',
{\em J. Phys. A} {\bf 36}, 24, 6443-6461 (2003).

\item {\bf [Ban 03 b]}:
M. Ban,
``Transmission rate of classical information through the thermalizing quantum channel'',
{\em J. Phys. A} {\bf 36}, 26, 7397-7409 (2003).

\item {\bf [Ban 04]}:
M. Ban,
``Phase-space approach to continuous variable quantum teleportation'',
{\em Phys. Rev. A} {\bf 69}, 5, 054304 (2004).

\item {\bf [Banaszek-Knight 97]}:
K. Banaszek, \& P. L. Knight,
``Quantum interference in three-photon down-conversion'',
{\em Phys. Rev. A} {\bf 55}, 3, 2368-2375 (1997).

\item {\bf [Banaszek-W\'{o}dkiewicz 98]}:
K. Banaszek, \& K. W\'{o}dkiewicz,
``Nonlocality of the Einstein-Podolsky-Rosen state in the Wigner representation'',
{\em Phys. Rev. A} {\bf 58}, 6, 4345-4347 (1998).

\item {\bf [Banaszek-W\'{o}dkiewicz 99 a]}:
K. Banaszek, \& K. W\'{o}dkiewicz,
``Testing quantum nonlocality in phase space'',
{\em Phys. Rev. Lett.} {\bf 82}, 10, 2009-2013 (1999).

\item {\bf [Banaszek-W\'{o}dkiewicz 99 b]}:
K. Banaszek, \& K. W\'{o}dkiewicz,
``Nonlocality of the Einstein-Podolsky-Rosen state in the phase
space'',
{\em Acta Phys. Slov.} {\bf 49}, ?, 491-500 (1999);
quant-ph/9904071.

\item {\bf [Banaszek 99 a]}:
K. Banaszek,
``Quantum homodyne tomography with a priori constraints'',
{\em Phys. Rev. A} {\bf 59}, 6, 4797-4800 (1999);
quant-ph/9901064.

\item {\bf [Banaszek 99 b]}:
K. Banaszek,
``Optimal receiver for quantum cryptography with two coherent states'',
{\em Phys. Lett. A} {\bf 253}, 1-2, 12-15 (1999);
quant-ph/9901067.

\item {\bf [Banaszek 99 c]}:
K. Banaszek,
``Maximum-likelihood algorithm for quantum tomography'',
{\em Acta Phys. Slov.} {\bf 49}, ?, 633-638 (1999);
quant-ph/9904063.

\item {\bf [Banaszek-D'Ariano-Paris-Sacchi 00]}:
K. Banaszek, G. M. D'Ariano, M. G. A. Paris, \& M. F. Sacchi,
``Maximum-likelihood estimation of the density matrix'',
{\em Phys. Rev. A} {\bf 61}, 1, 010304(R) (2000);
quant-ph/9909052.

\item {\bf [Banaszek 00 a]}:
K. Banaszek,
``Optimal quantum teleportation with an arbitrary pure state'',
{\em Phys. Rev. A} {\bf 62}, 2, 024301 (2000);
quant-ph/0002088.

\item {\bf [Banaszek 00 b]}:
K. Banaszek,
``Information gain versus disturbance for a single qubit'',
quant-ph/0006062.

\item {\bf [Banaszek-Walmsley-W\'{o}dkiewicz 00]}:
K. Banaszek, I. A. Walmsley, \& K. W\'{o}dkiewicz,
``Comment on `Proposal for the measurement of
Bell-type correlations from continuous variables'\,'',
quant-ph/0012097.
Comment on {\bf [Ralph-Munro-Polkinghorne 00]}.
Reply: {\bf [Ralph-Munro 01]}.

\item {\bf [Banaszek 01]}:
K. Banaszek,
``Fidelity balance in quantum operations'',
{\em Phys. Rev. Lett.} {\bf 86}, 7, 1366-1369 (2001);
quant-ph/0003123.

\item {\bf [Banaszek-U'Ren-Walmsley 01]}:
K. Banaszek, A. B. U'Ren, \& I. A. Walmsley,
``Generation of correlated photons in controlled spatial
modes by down-conversion in nonlinear waveguides'',
quant-ph/0103026.

\item {\bf [Banaszek-Devetak 01]}:
K. Banaszek, \& I. Devetak,
``Fidelity trade-off for finite ensembles of identically prepared qubits'',
{\em Phys. Rev. A} {\bf 64}, 5, 052307 (2001);
quant-ph/0104008.

\item {\bf [Banaszek-Dragan-W\'{o}dkiewicz-Radzewicz 02]}:
K. Banaszek, A. Dragan, K. W\'{o}dkiewicz, \& C. Radzewicz,
``Direct measurement of optical quasidistribution functions:
Multimode theory and homodyne tests of Bell's inequalities'',
{\em Phys. Rev. A} {\bf 66}, 4, 043803 (2002);
quant-ph/0204103.

\item {\bf [Banaszek-Dragan-Wasilewski-Radzewicz 04]}:
K. Banaszek, A. Dragan, W. Wasilewski, \& C. Radzewicz,
``Experimental demonstration of entanglement-enhanced classical
communication over a quantum channel with correlated noise'',
{\em Phys. Rev. Lett.} {\bf 92}, 25, 257901 (2004);
quant-ph/0403024.

\item {\bf [Banerjee 01]}:
A. Banerjee,
``Comment on relationship between atomic squeezed and entangled states'',
quant-ph/0110032.

\item {\bf [Band-Park 70]}:
W. Band, \& J. L. Park,
``The empirical determination of quantum states'',
{\em Found. Phys.} {\bf 1}, 2, 133-144 (1970).

\item {\bf [Band-Park 71]}:
W. Band, \& J. L. Park,
``A general method of empirical
state determination in quantum physics: Part II'',
{\em Found. Phys.} {\bf 1}, 4, 339-357 (1971).
See {\bf [Park-Band 71]} (I).

\item {\bf [Band-Park 79]}:
W. Band, \& J. L. Park,
``Quantum state determination:
Quorum for a particle in one dimension'',
{\em Am. J. Phys.} {\bf 47}, 2, 188-191 (1979).

\item {\bf [Bandyopadhyay-Lakshminarayan 02]}:
J. N. Bandyopadhyay, \& A. Lakshminarayan,
``Testing statistical bounds on entanglement using quantum chaos'',
{\em Phys. Rev. Lett.} {\bf 89}, 6, 060402 (2002);
quant-ph/0203117.

\item {\bf [Bandyopadhyay-Lakshminarayan 03]}:
J. N. Bandyopadhyay, \& A, Lakshminarayan,
``Entanglement production in coupled chaotic systems: Case of the kicked tops'',
{\em Phys. Rev. E} {\bf 69}, 1, 016201 (2004);
quant-ph/0307134.

\item {\bf [Bandyopadhyay-Kar-Roy 99]}:
S. Bandyopadhyay, G. Kar, \& A. Roy,
``Disentanglement of pure bipartite quantum states by local
cloning'',
{\em Phys. Lett. A} {\bf 258}, 4-6, 205-209 (1999);
quant-ph/9903015.

\item {\bf [Bandyopadhyay-Kar 99]}:
S. Bandyopadhyay, \& G. Kar,
``Broadcasting of entanglement and universal quantum cloners'',
{\em Phys. Rev. A} {\bf 60}, 4, 3296-3299 (1999);
quant-ph/9902073.

\item {\bf [Bandyopadhyay 99]}:
S. Bandyopadhyay,
``Qubit assisted conclusive teleportation'',
quant-ph/9907010.

\item {\bf [Bandyopadhyay 00 a]}:
S. Bandyopadhyay,
``Self-assembled nanoelectronic quantum computer
based on the Rashba effect in quantum dots'',
{\em Phys. Rev. B}, {\bf 61}, 20, 13813-13820 (2000);
quant-ph/9910032.

\item {\bf [Bandyopadhyay 00 b]}:
S. Bandyopadhyay,
``Teleportation and secret sharing with pure entangled states'',
{\em Phys. Rev. A} {\bf 62}, 1, 012308 (2000);
quant-ph/0002032.

\item {\bf [Bandyopadhyay 00 c]}:
S. Bandyopadhyay,
``Qubit- and entanglement-assisted optimal
entanglement concentration'',
{\em Phys. Rev. A} {\bf 62}, 3, 032308 (2000);
quant-ph/9911013.

\item {\bf [Bandyopadhyay 00 d]}:
S. Bandyopadhyay,
``Welcher weg experiments, duality and the orthodox Bohr's
complementarity principle'',
{\em Phys. Lett. A} {\bf 276}, 5-6, 233-239 (2000);
quant-ph/0003073.

\item {\bf [Bandyopadhyay 00 e]}:
S. Bandyopadhyay,
``Local interactions of an entangled state with the environment via
amplitude damping channel'',
quant-ph/0003120.

\item {\bf [Bandyopadhyay 01]}:
S. Bandyopadhyay,
``Prospects for a quantum dynamic random access memory (Q-DRAM)'',
quant-ph/0101058.

\item {\bf [Bandyopadhyay-Majumdar-Home 01]}:
S. Bandyopadhyay, A. S. Majumdar, \& D. Home,
``Quantum superarrivals'',
quant-ph/0103001.

\item {\bf [Bandyopadhyay-Roychowdhury-Sen 01]}:
S. Bandyopadhyay, V. P. Roychowdhury, \& U. Sen,
``A classification of incomparable states'',
quant-ph/0103131.
Comment: {\bf [Leung-Smolin 01]}.

\item {\bf [Bandyopadhyay 02]}:
S. Bandyopadhyay,
``Origin of noisy states whose teleportation fidelity can be enhanced through
dissipation'',
{\em Phys. Rev. A} {\bf 65}, 2, 022302 (2002).

\item {\bf [Bandyopadhyay-Roychowdhury-Vatan 02]}:
S. Bandyopadhyay, V. Roychowdhury, \& F. Vatan,
``Partial recovery of entanglement in bipartite-entanglement transformations'',
{\em Phys. Rev. A} {\bf 65}, 4, 040303 (2002);
quant-ph/0105019.

\item {\bf [Bandyopadhyay-Roychowdhury 02]}:
S. Bandyopadhyay, \& V. Roychowdhury,
``Efficient entanglement-assisted transformation for bipartite pure states'',
{\em Phys. Rev. A} {\bf 65}, 4, 042306 (2002).

\item {\bf [Bandyopadhyay-Roychowdhury-Sen 02]}:
S. Bandyopadhyay, V. Roychowdhury, \& U. Sen,
``Classification of nonasymptotic bipartite pure-state entanglement
transformations'',
{\em Phys. Rev. A} {\bf 65}, 5, 052315 (2002).

\item {\bf [Bandyopadhyay-Roychowdhury 03]}:
S. Bandyopadhyay, \& V. Roychowdhury,
``Classes of $n$-copy undistillable quantum states with negative partial
transposition'',
{\em Phys. Rev. A} {\bf 68}, 2, 022319 (2003);
quant-ph/0302093.

\item {\bf [Bandyopadhyay-Roychowdhury 04]}:
S. Bandyopadhyay, \& V. Roychowdhury,
``Maximally-disordered distillable quantum states'',
{\em Phys. Rev. A} {\bf 69}, 4, 040302 (2004);
quant-ph/0302043.

\item {\bf [Bandyopadhyay-Lidar 04]}:
S. Bandyopadhyay, \& D. A. Lidar,
``Entangling capacities of noisy two-qubit Hamiltonians'',
{\em Phys. Rev. A} {\bf 70}, 1, 010301 (2004);
quant-ph/0308014.

\item {\bf [Bandyopadhyay-Ghosh-Roychowdhury 04]}:
S. Bandyopadhyay, S. Ghosh, \& V. Roychowdhury,
``Non-full rank bound entangled states satisfying the range criterion'',
{\em Phys. Rev. A};
quant-ph/0406023.

\item {\bf [Bandyopadhyay-Lidar 04]}:
S. Bandyopadhyay, \& D. A. Lidar,
``Robustness of many-qubit entanglement under general decoherence'',
quant-ph/0408174.

\item {\bf [Banerjee-Ghosh 00]}:
S. Banerjee \& R. Ghosh,
``Quantum theory of a Stern-Gerlach system in contact
with a linearly dissipative environment'',
{\em Phys. Rev. A} {\bf 62}, 4, 042105 (2000).

\item {\bf [Bar 00]}:
D. Bar,
``The Zeno effect in the EPR paradox, in the teleportation process,
and in Wheeler's delayed-choice experiment'',
{\em Found. Phys.} {\bf 30}, 6, 813-838 (2000).

\item {\bf [Bar-Horwitz 02]}:
D. Bar, \& L. P. Horwitz,
``Lax-Phillips evolution as an evolution of Gell-Mann-Hartle-Griffiths
histories and emergence of the Schr\"{o}dinger equation for a stable history'',
{\em Phys. Lett. A} {\bf 303}, 2-3, 135-139 (2002).

\item {\bf [Baracca-Bergia-Livi-Restignoli 76]}:
A. Baracca, S. Bergia, R. Livi, \& M. Restignoli,
``Reinterpretation and extension of Bell's inequality
for multivalued observables'',
{\em Int. J. Theor. Phys.} {\bf 15}, 7, 473-486 (1976).

\item {\bf [Baranov-Pechen-Volovich 02]}:
A. A. Baranov, A. N. Pechen, \& I. V. Volovich,
``Space dependence of entangled states and Franson-type EPR experiments'',
quant-ph/0203152.

\item {\bf [Barat-Kimball 01]}:
N. Barat, \& J. C. Kimball,
``Localization and causality for a free particle'',
quant-ph/0111060.

\item {\bf [Barbieri-De Martini-Di Nepi-Mataloni 03]}:
M. Barbieri, F. De Martini, G. Di Nepi, \& P. Mataloni,
``Violation of Bell inequalities and quantum tomography with pure-states,
Werner-states and maximally entangled mixed states created by a universal
quantum entangler'',
quant-ph/0303018.

\item {\bf [Barbieri-De Martini-Di Nepi-(+3) 03]}:
M. Barbieri, F. De Martini, G. Di Nepi,
P. Mataloni, G. M. D'Ariano, \& C. Macchiavello,
``Detection of entanglement with polarized photons:
Experimental realization of an entanglement witness'',
{\em Phys. Rev. Lett.} {\bf 91}, 22, 227901 (2003);
quant-ph/0307003.

\item {\bf [Barbieri-De Martini-Di Nepi-Mataloni 04]}:
M. Barbieri, F. De Martini, G. Di Nepi, \& P. Mataloni,
`Nonlocality tests of Bell inequalities and of Hardy's ``ladder theorem''
without ``supplementary assumptions''\,',
quant-ph/0406156.

\item {\bf [Barbosa 98]}:
G. A. Barbosa,
``Quantum communications: Tetrat coding'',
{\em Phys. Rev. A} {\bf 58}, 4, 3332-3335 (1998).

\item {\bf [Barbosa-Arnaut 02]}:
G. A. Barbosa, \& H. H. Arnaut,
``Twin photons with angular-momentum entanglement: Phase matching'',
{\em Phys. Rev. A} {\bf 65}, 5, 053801 (2002).

\item {\bf [Barbosa-Corndorf-Kumar-Yuen 03]}:
G. A. Barbosa, E. Corndorf, P. Kumar, \& H. P. Yuen,
``Secure communication using mesoscopic coherent states'',
{\em Phys. Rev. Lett.};
quant-ph/0212018.

\item {\bf [Barbosa-Corndorf-Kumar-Yuen 03]}:
G. A. Barbosa, E. Corndorf, P. Kumar, H. P. Yuen,
G. M. D'Ariano, M. G. A. Paris, \& P. Perinotti,
``Secure communication using coherent states'',
{\em Proc.\ Sixth Int.\ Conf.\ on Quantum Communication,
Measurement and Computing Proceedings (QCMC'02), 2002};
quant-ph/0210089.

\item {\bf [Barbosa-Pinto Neto 04]}:
G. D. Barbosa, \& N. Pinto-Neto,
``Noncommutative quantum mechanics and Bohm's ontological interpretation'',
{\em Phys. Rev. D} {\bf 69}, 6, 065014 (2004);
hep-th/0304105.

\item {\bf [Barenco 95]}:
A. Barenco,
``A universal two-bit gate for quantum computation'',
{\em Proc. R. Soc. Lond. A} {\bf 449}, 1937, 679-683 (1995).

\item {\bf [Barenco-Deutsch-Ekert-Jozsa 95]}:
A. Barenco, D. Deutsch, A. K. Ekert, \& R. Jozsa,
``Conditional quantum dynamics and logic gates'',
{\em Phys. Rev. Lett.} {\bf 74}, 20, 4083-4086 (1995).

\item {\bf [Barenco-Bennett-Cleve-(+6) 95]}:
A. Barenco, C. H. Bennett, R. Cleve, D. P. DiVincenzo, N. Margolus,
P. W. Shor, T. Sleator, J. A. Smolin, \& H. Weinfurter,
``Elementary gates for quantum computation'',
{\em Phys. Rev. A} {\bf 52}, 5, 3457-3467 (1995);
quant-ph/9503016.

\item {\bf [Barenco-Ekert 95]}:
A. Barenco, \& A. K. Ekert,
``Dense coding based on quantum entanglement'',
{\em J. Mod. Opt.} {\bf 42}, 6, 1253-1259 (1995).

\item {\bf [Barenco 96]}:
A. Barenco,
``Quantum physics and computers'',
{\em Contemp. Phys.} {\bf 37}, 5, 375-389 (1996).

\item {\bf [Barenco-Ekert-Macchiavello-Sanpera 96]}:
A. Barenco, A. K. Ekert, C. Macchiavello, \& A. Sanpera,
``Un saut d'\'{e}chelle pour les calculateurs'',
{\em La Recherche} {\bf 27}, 292, 52-58 (1996).
Spanish version: ``Un salto de escala para los calculadores'',
{\em Mundo Cient\'{\i}fico} {\bf 17}, 175, 44-50 (1997).

\item {\bf [Barenco-Brun-Schack-Spiller 97]}:
A. Barenco, T. A. Brun, R. Schack, \& T. P. Spiller,
``Effects of noise on quantum error correction algorithms'',
{\em Phys. Rev. A} {\bf 56}, 2, 1177-1188 (1997).

\item {\bf [Barenco-Fern\'{a}ndez Huelga-Ekert 97]}:
A. Barenco, S. G. Fern\'{a}ndez Huelga, \& A. K. Ekert,
``Quantum computation'',
in M. Ferrero, \& A. van der Merwe (eds.),
{\em New developments on fundamental problems in quantum
physics (Oviedo, Spain, 1996)},
Kluwer Academic, Dordrecht, Holland, 1997, pp.~39-54.

\item {\bf [Barenco-Berthiaume-Deutsch-(+3) 97]}:
A. Barenco, A. Berthiaume, D. Deutsch,
A. K. Ekert, R. Jozsa, \& C. Macchiavello,
``Stabilization of quantum computations by symmetrization'',
{\em SIAM J. Comput.} {\bf 26}, 5, 1541-1557 (1997).

\item {\bf [Barenco 98]}:
A. Barenco,
``Quantum computation: An introduction'',
in {\bf [Lo-Spiller-Popescu 98]}, pp.~143-183.

\item {\bf [Barndorff Nielsen-Gill-Jupp 03]}:
O. S. Barndorff-Nielsen, R. D. Gill, \& P. E. Jupp,
``On quantum statistical inference, II'',
{\em J. Roy. Statist. Soc. B};
quant-ph/0307191.

\item {\bf [Barnes-Warren 99]}:
J. P. Barnes, \& W. S. Warren,
``Decoherence and programmable quantum computation'',
{\em Phys. Rev. A} {\bf 60}, 6, 4363-4374 (1999);
quant-ph/9902084.

\item {\bf [Barnes-Warren 00]}:
J. P. Barnes, \& W. S. Warren,
``Automatic quantum error correction'',
{\em Phys. Rev. Lett.} {\bf 85}, 4, 856-859 (2000);
quant-ph/9912104.

\item {\bf [Barnes 98]}:
S. E. Barnes,
``Efficient quantum computing on low temperature spin ensembles'',
quant-ph/9804065.

\item {\bf [Barnes-Bj\"{o}rk-G\'{e}rard-(+4) 02]}:
W. L. Barnes, G. Bj\"{o}rk, J. M. G\'{e}rard,
P. Jonsson, J. A. E. Wasey, P. T. Worthing, \& V. Zwiller,
``Solid-state single photon sources: Light collection strategies'',
{\em Eur. Phys. J. D} {\bf 18}, 2 (Special issue:
{\em Quantum interference and cryptographic keys:
Novel physics and advancing technologies (QUICK) (Corsica, 2001)}, 197-210 (2002).

\item {\bf [Barnett-Knight 87]}:
S. M. Barnett, \& P. L. Knight,
``Squeezing in correlated quantum systems'',
{\em J. Mod. Opt.} {\bf 34}, 6-7, 841-853 (1987).

\item {\bf [Barnett-Phoenix 91]}:
S. M. Barnett, \& S. J. D. Phoenix,
``Information theory, squeezing, and quantum correlations'',
{\em Phys. Rev. A} {\bf 44}, 1, 535-545 (1991).

\item {\bf [Barnett-Phoenix 92]}:
S. M. Barnett, \& S. J. D. Phoenix,
``Bell's inequality and the Schmidt decomposition'',
{\em Phys. Lett. A} {\bf 167}, 3, 233-237 (1992).

\item {\bf [Barnett-Phoenix 93 a]}:
S. M. Barnett, \& S. J. D. Phoenix,
``Information-theoretic limits to quantum cryptography'',
{\em Phys. Rev. A} {\bf 48}, 1, R5-R8 (1993).

\item {\bf [Barnett-Phoenix 93 b]}:
S. M. Barnett, \& S. J. D. Phoenix,
``Bell's inequality and rejected-data protocols for quantum cryptography'',
{\em J. Mod. Opt.} {\bf 40}, 8, 1443-1448 (1993).

\item {\bf [Barnett-Huttner-Phoenix 93]}:
S. M. Barnett, B. Huttner, \& S. J. D. Phoenix,
``Eavesdropping strategies and rejected-data protocols
in quantum cryptography'',
{\em J. Mod. Opt.} {\bf 40}, 12, 2501-2513 (1993).

\item {\bf [Barnett-Ekert-Phoenix 94]}:
S. M. Barnett, A. K. Ekert, \& S. J. D. Phoenix,
``Special issue on quantum communication. Introduction'',
in S. M. Barnett, A. K. Ekert, \& S. J. D. Phoenix (eds.),
{\em J. Mod. Opt.} {\bf 41}, 12 (Special issue: Quantum
communication), 2239-2240 (1994).

\item {\bf [Barnett-Loudon-Pegg-Phoenix 94]}:
S. M. Barnett, R. Loudon, D. T. Pegg, \& S. J. D. Phoenix,
``Communication using quantum states'',
in S. M. Barnett, A. K. Ekert, \& S. J. D. Phoenix (eds.),
{\em J. Mod. Opt.} {\bf 41}, 12 (Special issue: Quantum
communication), 2351-2373 (1994).

\item {\bf [Barnett-Phoenix 94]}:
S. M. Barnett, \& S. J. D. Phoenix,
``?'',
patent application EP94302359.8, 1994.
See {\bf [Townsend-Smith 93]}, {\bf [Townsend-Blow 93]},
{\bf [Townsend-Phoenix-Blow-Barnett 94]},
{\bf [Phoenix-Barnett-Townsend-Blow 95]}.

\item {\bf [Barnett 97]}:
S. M. Barnett,
``Quantum information via novel measurements'',
in P. L. Knight, B. Stoicheff, \& D. Walls (eds.),
{\em Highlight in Quantum Optics},
{\em Philos. Trans. R. Soc. Lond. A} {\bf 355}, 1733, 2279-2290 (1997).

\item {\bf [Barnett-Imoto-Huttner 98]}:
S. M. Barnett, N. Imoto \& B. Huttner,
``Photonic de Broglie wave interferometers'',
{\em J. Mod. Opt.} {\bf 45}, 11, 2217-2232 (1998).

\item {\bf [Barnett 98]}:
S. M. Barnett,
``Reflections on the EPR paradox'',
{\em Contemp. Phys.} {\bf 39}, 1, 81-82 (1998).
Review of {\bf [Mann-Revzen 96]}.

\item {\bf [Barnett-Chefles 98]}:
S. M. Barnett, \& A. Chefles,
``Nonlocality without inequalities for all pure entangled states'',
quant-ph/9807090 (withdrawn).
See {\bf [Cabello 00 b]}.

\item {\bf [Barnett-Pegg 99]}:
S. M. Barnett, \& D. T. Pegg,
``Optical state truncation'',
{\em Phys. Rev. A} {\bf 60}, 6, 4965-4973 (1999).

\item {\bf [Barnett-Pegg-Jeffers-(+2) 00]}:
S. M. Barnett, D. T. Pegg, J. Jeffers,
O. Jedrkiewicz, \& R. Loudon,
``Retrodiction for quantum optical communications'',
{\em Phys. Rev. A} {\bf 62}, 2, 022313 (2000).

\item {\bf [Barnett 00]}:
S. M. Barnett,
``Scientific revolutions, paradoxes and paradigms'',
{\em Contemp. Phys.} {\bf 41}, 3, 167-169 (2000).

\item {\bf [Barnett-Pegg-Jeffers-Jedrkiewicz 01]}:
S. M. Barnett, D. T. Pegg, J. Jeffers, \& O. Jedrkiewicz,
``Master equation for retrodiction of quantum communication
signals'',
{\em Phys. Rev. Lett.} {\bf 86}, 11, 2455-2458 (2001);
quant-ph/0107139.

\item {\bf [Barnett-Pegg-Jeffers 00]}:
S. M. Barnett, D. T. Pegg, \& J. Jeffers,
`\,``Bayes' theorem and quantum retrodiction'',
{\em J. Mod. Opt.} {\bf 47}, ?, 1779-? (2000);
quant-ph/0106139.

\item {\bf [Barnett-Gilson-Sasaki 01]}:
S. M. Barnett, C. R. Gilson, \& M. Sasaki,
``Fidelity and the communication of quantum information'',
in S. Popescu, N. Linden, \& R. Jozsa (eds.),
{\em J. Phys. A} {\bf 34}, 35
(Special issue: Quantum information and computation), 6755-6766 (2001);
quant-ph/0107024.

\item {\bf [Barnett 01]}:
S. M. Barnett,
``Minimum-error discrimination between multiply symmetric states'',
{\em Phys. Rev. A} {\bf 64}, 3, 030303(R) (2001).

\item {\bf [Barnett-Pegg-Jeffers-Jedrkiewicz 01]}:
S. M. Barnett, D. T. Pegg, J. Jeffers, \& O. Jedrkiewicz,
``Atomic retrodiction'',
{\em J. Phys. B} {\bf 33}, ? 3047-? (2000);
quant-ph/0107019.

\item {\bf [Barnett-Kraemer 02]}:
S. M. Barnett, \& T. Kraemer,
``CP violation, EPR correlations and quantum state discrimination'',
{\em Phys. Lett. A} {\bf 293}, 5-6, 211-215 (2002).

\item {\bf [Barnett-Andersson 02]}:
S. M. Barnett, \& E. Andersson,
``Bound on measurement based on the no-signaling condition'',
{\em Phys. Rev. A} {\bf 65}, 4, 044307 (2002).

\item {\bf [Barnett-Chefles-Jex 03]}:
S. M. Barnett, A. Chefles, \& I. Jex,
``Comparison of two unknown quantum states'',
{\em Phys. Lett. A} {\bf 307}, 4, 189-195 (2003);
quant-ph/0202087.

\item {\bf [Barndorff Nielsen-Gill 98]}:
O. E. Barndorff-Nielsen, \& R. D. Gill,
``An example of non-attainability of expected quantum information'',
quant-ph/9808009.

\item {\bf [Bartlett-Rice-Sanders-(+2) 00]}:
S. D. Bartlett, D. A. Rice, B. C. Sanders,
J. Daboul, \& H. de Guise,
``Unitary transformations for testing Bell inequalities'',
quant-ph/0010049.

\item {\bf [Bartlett-Sanders-Braunstein-Nemoto 02]}:
S. D. Bartlett, B. C. Sanders, S. L. Braunstein, \& K. Nemoto,
``Efficient classical simulation of continuous variable quantum information
processes'',
{\em Phys. Rev. Lett.} {\bf 88}, 9, 097904 (2002);
quant-ph/0109047.

\item {\bf [Bartlett-de Guise-Sanders 02]}:
S. D. Bartlett, H. de Guise, \& B. C. Sanders,
``Quantum encodings in spin systems and harmonic oscillators'',
{\em Phys. Rev. A} {\bf 65}, 5, 052316 (2002).

\item {\bf [Bartlett-Sanders 02]}:
S. D. Bartlett, \& B. C. Sanders,
``Efficient classical simulation of optical quantum information circuits'',
{\em Phys. Rev. Lett.} {\bf 89}, 20, 207903 (2002);
quant-ph/0204065.

\item {\bf [Bartlett-Diamanti-Sanders-Yamamoto 02]}:
S. D. Bartlett, E. Diamanti, B. C. Sanders, \& Y. Yamamoto,
``Photon counting schemes and performance of non-deterministic nonlinear gates in linear optics'',
in J. C. Ricklin, \& D. G. Voelz (eds.),
{\em Proc.\ of Free-Space Laser Communication and Laser Imaging II,
SPIE Int.\ Symp.\ on Optical Science and Technology (2002)},
SPIE, Bellingham, Washington;
quant-ph/0204073.

\item {\bf [Bartlett-Rudolph-Spekkens 03]}:
S. D. Bartlett, T. Rudolph, \& R. W. Spekkens,
``Classical and quantum communication without a shared reference frame'',
{\em Phys. Rev. Lett.} {\bf 91}, 2, 027901 (2003);
quant-ph/0302111.

\item {\bf [Bartlett-Sanders 03]}:
S. D. Bartlett, \& B. C. Sanders,
``Requirement for quantum computation'',
{\em J. Mod. Opt.} {\bf 50}, ?, 2331-2340 (2003);
quant-ph/0302125.


\item {\bf [Bartlett-Rudolph-Spekkens 04]}:
S. D. Bartlett, T. Rudolph, \& R. W. Spekkens,
``Optimal measurements for relative quantum information'',
{\em Phys. Rev. A} {\bf 70}, 3, 032321 (2004);
quant-ph/0310009.

\item {\bf [Bartlett-Rudolph-Spekkens 04]}:
S. D. Bartlett, T. Rudolph, \& R. W. Spekkens,
``Decoherence-full subsystems and the cryptographic power of a private
shared reference frame'',
{\em Phys. Rev. A} {\bf 70}, 3, 032307 (2004);
quant-ph/0403161.

\item {\bf [Bartell 80 a]}:
L. S. Bartell,
``Complementarity in the double-slit experiment: On simple realizable
systems for observing intermediate particle-wave behavior'',
{\em Phys. Rev. D} {\bf 21}, 6, 1698-1699 (1980).
Reprinted in {\bf [Wheeler-Zurek 83]}, pp.~455-456.

\item {\bf [Bartell 80 b]}:
L. S. Bartell,
``Local realism and the Einstein-Podolsky-Rosen paradox.
On concrete new tests'',
{\em Phys. Rev. D} {\bf 22}, 6, 1352-1360 (1980).

\item {\bf [Bartlett-de Guise-Sanders 01]}:
S. D. Bartlett, H. de Guise, \& B. C. Sanders,
``Quantum computation with qudits in spin systems and harmonic oscillators'',
quant-ph/0109066.
See quant-ph/0011080.

\item {\bf [Bartlett-Sanders 02]}:
S. D. Bartlett, \& B. C. Sanders,
``Universal continuous-variable quantum computation: Requirement of optical
nonlinearity for photon counting'',
{\em Phys. Rev. A} {\bf 65}, 4, 042304 (2002);
quant-ph/0110039.

\item {\bf [Bartlett-Munro 03]}:
S. D. Bartlett, \& W. J. Munro,
``Quantum teleportation of optical quantum gates'',
{\em Phys. Rev. Lett.} {\bf 90}, 11, 117901 (2003);
quant-ph/0208022.

\item {\bf [Bartlett-Wiseman 03]}:
S. D. Bartlett, \& H. M. Wiseman,
``Entanglement constrained by superselection rules'',
{\em Phys. Rev. Lett.} {\bf 91}, 9, 097903 (2003);
quant-ph/0303140.

\item {\bf [Bartlett-Terno 04]}:
S. D. Bartlett, \& D. R. Terno,
``Relativistically invariant quantum information'',
quant-ph/0403014.

\item {\bf [Barnum-Caves-Fuchs-(+2) 96]}:
H. N. Barnum, C. M. Caves, C. A. Fuchs, R. Jozsa, \& B. W. Schumacher,
``Noncomuting mixed states cannot be broadcast'',
{\em Phys. Rev. Lett.} {\bf 76}, 15, 2818-2821 (1996).
Reprinted in {\bf [Macchiavello-Palma-Zeilinger 00]}, pp.~195-198.

\item {\bf [Barnum-Nielsen-Schumacher 97]}:
H. N. Barnum, M. A. Nielsen, \& B. W. Schumacher,
``Information transmission through a noisy quantum channel'',
quant-ph/9702049.

\item {\bf [Barnum-Smolin-Terhal 97]}:
H. N. Barnum, J. A. Smolin, \& B. M. Terhal,
``Results on quantum channel capacity'',
quant-ph/9711032.

\item {\bf [Barnum-Smolin-Terhal 98]}:
H. N. Barnum, J. A. Smolin, \& B. M. Terhal,
``Quantum capacity is properly defined without encodings'',
{\em Phys. Rev. A} {\bf 58}, 5, 3496-3501 (1998).

\item {\bf [Barnum-Bernstein-Spector 99]}:
H. N. Barnum, H. J. Bernstein, \& L. Spector,
``A quantum circuit for OR'',
quant-ph/9907056.

\item {\bf [Barnum 99]}:
H. N. Barnum,
``Quantum secure identification using entanglement and
catalysis'',
quant-ph/9910072.

\item {\bf [Barnum-Knill-Nielsen 00]}:
H. N. Barnum, E. Knill, \& M. A. Nielsen,
``On quantum fidelities and channel capacities'',
{\em IEEE Trans. Inf. Theory} {\bf 46}, 1317-1329 (2000);
quant-ph/9809010.

\item {\bf [Barnum-Caves-Finkelstein-(+2) 00]}:
H. N. Barnum, C. M. Caves, J. Finkelstein, C. A. Fuchs, \& R. Schack,
``Quantum probability from decision theory?'',
{\em Proc. R. Soc. Lond. A} {\bf 456}, 1997, 1175-1182 (2000);
quant-ph/9907024.
See: {\bf [Deutsch 99]}, {\bf [Finkelstein 99 c]},
{\bf [Polley 99]}, {\bf [Summhammer 99]}.

\item {\bf [Barnum 00]}:
H. N. Barnum,
``Quantum rate-distortion coding'',
{\em Phys. Rev. A} {\bf 62}, 4, 042309 (2000).

\item {\bf [Barnum-Bernstein-Spector 00]}:
H. N. Barnum, H. J. Bernstein, \& L. Spector,
``Quantum circuits for OR and AND of ORs'',
{\em J. Phys. A} {\bf 33}, 45, 8047-8057 (2000).

\item {\bf [Barnum-Caves-Fuchs-(+2) 01]}:
H. N. Barnum, C. M. Caves, C. A. Fuchs, R. Jozsa, \& B. W. Schumacher,
``On quantum coding for ensembles of mixed states'',
in S. Popescu, N. Linden, \& R. Jozsa (eds.),
{\em J. Phys. A} {\bf 34}, 35
(Special issue: Quantum information and computation), 6767-6786 (2001);
quant-ph/0008024.

\item {\bf [Barnum-Linden 01]}:
H. N. Barnum, \& N. Linden,
``Monotones and invariants for multi-particle quantum states'',
in S. Popescu, N. Linden, \& R. Jozsa (eds.),
{\em J. Phys. A} {\bf 34}, 35
(Special issue: Quantum information and computation), 6787-6806 (2001).

\item {\bf [Barnum-Hayden-Jozsa-Winter 01]}:
H. N. Barnum, P. Hayden, R. Jozsa, \& A. Winter,
``On the reversible extraction of classical
information from a quantum source'',
{\em Proc. R. Soc. Lond. A} {\bf 457}, 2012, 2019-2039 (2001);
quant-ph/0011072.

\item {\bf [Barnum 01]}:
H. N. Barnum,
``Quantum message authentication codes'',
quant-ph/0103123.

\item {\bf [Barnum-Linden 01]}:
H. N. Barnum, \& N. Linden,
``Monotones and invariants for multi-particle quantum states'',
quant-ph/0103155.

\item {\bf [Barnum-Knill 02]}:
H. N. Barnum, \& E. Knill,
``Reversing quantum dynamics with near-optimal quantum and classical fidelity'',
{\em J. Math. Phys.} {\bf 43}, 5, 2097-2106 (2002);
quant-ph/0004088.

\item {\bf [Barnum-Saks 02]}:
H. N. Barnum, \& M. Saks,
``A lower bound on the quantum query complexity of read-once functions'',
quant-ph/0201007.

\item {\bf [Barnum-Knill-Ortiz-Viola 03]}:
H. N. Barnum, E. Knill, G. Ortiz, \& L. Viola,
``Generalizations of entanglement based on coherent states and convex sets'',
{\em Phys. Rev. A} {\bf 68}, 3, 032308 (2003).

\item {\bf [Barnum 03]}:
H. N. Barnum,
``Quantum information processing, operational quantum logic, convexity,
and the foundations of physics'',
quant-ph/0304159.

\item {\bf [Barnum-Knill-Ortiz-(+2) 04]}:
H. N. Barnum, E. Knill, G. Ortiz, R. Somma, \& L. Viola,
``A subsystem-independent generalization of entanglement'',
{\em Phys. Rev. Lett.} {\bf 92}, 10, 107902 (2004);
quant-ph/0305023.

\item {\bf [Barranco-Potel-Mu\~{n}oz Ale\~{n}ar-Bienvenido 02]}:
F. Barranco, G. Potel, M. Mu\~{n}oz Ale\~{n}ar, \& J. F. Bienvenido,
``Estabilidad de las trayectorias y distribuci\'{o}on de probabilidad
en din\'{a}mica bohmiana'',
in C. Mataix, \& A. Rivadulla (eds.),
{\em F\'{\i}sica cu\'{a}ntica y realidad.
Quantum physics and reality (Madrid, 2000)},
Editorial Complutense, Madrid, 2002, pp.~357-370.
See {\bf [Potel-Mu\~{n}oz Ale\~{n}ar-Barranco-Vigezzi 02]}.

\item {\bf [Barrett 99 a]}:
J. A. Barrett,
{\em The quantum mechanics of minds and worlds},
Oxford University Press, Oxford, 1999.

\item {\bf [Barrett 99 b]}:
J. A. Barrett,
``Book review. Quantum chance and nonlocality'',
{\em Found. Phys.} {\bf 29}, 6, 1011-1018 (1999).
Review of {\bf [Dickson 98]}.

\item {\bf [Barrett 00]}:
J. A. Barrett,
``The persistence of memory: Surreal trajectories in Bohm's
theory'',
{\em UC Irvine Laguna Beach Physics Conf.\ (2000)};
quant-ph/0002046.

\item {\bf [Barrett 01 a]}:
J. A. Barrett,
{\em The quantum mechanics of minds and worlds};
Oxford University Press, Oxford, 2001.

\item {\bf [Barrett 01 b]}:
J. Barrett,
``Implications of teleportation for nonlocality'',
{\em Phys. Rev. A} {\bf 64}, 4, 042305 (2001);
quant-ph/0103105.

\item {\bf [Barrett 01 c]}:
J. Barrett,
``A local hidden variable model for positive operator valued
measurements on a class of entangled mixed states'',
quant-ph/0107045.

\item {\bf [Barrett 02]}:
J. Barrett,
``Nonsequential positive-operator-valued measurements on entangled mixed
states do not always violate a Bell inequality'',
{\em Phys. Rev. A} {\bf 65}, 4, 042302 (2002).

\item {\bf [Barrett-Collins-Hardy-(+2) 02]}:
J. Barrett, D. G. Collins, L. Hardy, A. Kent, \& S. Popescu,
``Quantum nonlocality, Bell inequalities, and the memory loophole'',
{\em Phys. Rev. A} {\bf 66}, 4, 042111 (2002).

\item {\bf [Barrett 03]}:
J. Barrett,
``Countering quantum noise with supplementary classical information'',
{\em Phys. Rev. A} {\bf 68}, 1, 012303 (2003);
quant-ph/0201073.

\item {\bf [Barrett-Massar 03]}:
J. Barrett, \& S. Massar,
``Quantum coin tossing in the presence of noise'',
{\em Phys. Rev. A} {\bf 69}, 2, 022322 (2004);
quant-ph/0303182.

\item {\bf [Barrett-Kent 04]}:
J. Barrett, \& A. Kent,
``Non-contextuality, finite precision measurement and the Kochen-Specker theorem'',
{\em Stud. Hist. Philos. Mod. Phys.} {\bf 35}, 2, 151-176 (2004);
quant-ph/0309017.

\item {\bf [Barrett-Massar 04]}:
J. Barrett, \& S. Massar,
``Quantum coin tossing and bit-string generation in the presence of noise'',
{\em Phys. Rev. A} {\bf 69}, 2, 022322 (2004).

\item {\bf [Barrett-Linden-Massar-(+3) 04]}:
J. Barrett, N. Linden, S. Massar, S. Pironio, S. Popescu, \& D. Roberts,
``Non-local correlations as an information theoretic resource'',
quant-ph/0404097.

\item {\bf [Barrett-Hardy-Kent 04]}:
J. Barrett, L. Hardy, \& A. Kent,
``No signalling and quantum key distribution''.
quant-ph/0405101.

\item {\bf [Barrett-Massar 04]}:
J. Barrett, \& S. Massar,
``Security of quantum bit-string generation'',
{\em Phys. Rev. A};
quant-ph/0408120.
See {\bf [Lamoureux-Brainis-Amans-(+2) 04]}.

\item {\bf [Barrett 90]}:
J. W. Barrett,
``Quantum mechanics and algorithmic complexity'',
in {\bf [Zurek 90]}, pp.~375-380.

\item {\bf [Barrett-DeMarco-Schaetz-(+9) 03]}:
M. D. Barrett, B. DeMarco, T. Schaetz,
D. Leibfried, J. Britton, J. Chiaverini,
W. M. Itano, B. Jelenkovic, J. D. Jost,
C. Langer, T. Rosenband, \& D. J. Wineland,
``Sympathetic cooling of $^9$Be$^+$ and $^{24}$Mg$^+$ for quantum
logic'',
{\em Phys. Rev. A} {\bf 68}, 4, 042302 (2003);
quant-ph/0307088.

\item {\bf [Barrett-Chiaverini-Schaetz-(+8) 04]}:
M. D. Barrett, J. Chiaverini, T. Schaetz,
J. Britton, W. M. Itano, J. D. Jost,
E. Knill, C. Langer, D. Leibfried,
R. Ozeri, \& D. J. Wineland,
``Deterministic quantum teleportation of atomic qubits'',
{\em Nature} {\bf 429}, ?, 737-739 (2004).
See {\bf [Riebe-H\"{a}ffner-Roos-(+8) 04]}.

\item {\bf [Barrett-Milburn 03]}:
S. D. Barrett, \& G. J. Milburn,
``Measuring the decoherence rate in a semiconductor charge qubit'',
{\em Phys. Rev. B} {\bf 68}, 15, 155307 (2003).

\item {\bf [Barrett-Kok 04]}:
S. D. Barrett, \& P. Kok,
``Efficient high-fidelity quantum computation using matter qubits and
linear optics'',
quant-ph/0408040.

\item {\bf [Barrett-Kok-Nemoto-(+3) 04]}:
S. D. Barrett, P. Kok, K. Nemoto,
R. G. Beausoleil, W. J. Munro, \& T. P. Spiller,
``A symmetry analyser for non-destructive Bell state detection using EIT'',
quant-ph/0408117.

\item {\bf [de Barros-Suppes 00 a]}:
J. A. de Barros, \& P. Suppes,
``Some conceptual issues involving probability in quantum mechanics'',
quant-ph/0001017.

\item {\bf [de Barros-Suppes 00 b]}:
J. A. de Barros, \& P. Suppes,
``Inequalities for dealing with detector inefficiencies in
Greenberger-Horne-Zeilinger-type experiments'',
{\em Phys. Rev. Lett.} {\bf 84}, 5, 793-797 (2000);
quant-ph/0001034.

\item {\bf [Barut-van der Merwe-Vigier 84]}:
A. O. Barut, A. van der Merwe, \& J.-P. Vigier,
{\em Quantum, space and time -- The quest continues.
Studies and essays in honour of Louis de Broglie,
Paul Dirac and Eugene Wigner},
Cambridge University Press, Cambridge, 1984.
Initially published in {\em Found. Phys.} 1982-1983.

\item {\bf [Barut-Meystre 84 a]}:
A. O. Barut, \& P. Meystre,
``A classical model of
EPR experiment with quantum mechanical correlations and Bell inequalities'',
{\em Phys. Lett. A} {\bf 105}, 9, 458-462 (1984).

\item {\bf [Barut-Meystre 84 b]}:
A. O. Barut, \& P. Meystre,
``Rotational invariance, locality, and Einstein-Podolsky-Rosen experiments'',
{\em Phys. Rev. Lett.} {\bf 53}, 10, 1021 (1984).
Comment on {\bf [Angelidis 83]}.

\item {\bf [Barut 88]}:
A. O. Barut,
``Explicit calculations with a hidden-variable spin model'',
in F. Selleri (ed.),
{\em Quantum mechanics versus local realism:
The Einstein-Podolsky-Rosen paradox},
Plenum Press, New York, 1988, pp.~433-446.

\item {\bf [Barut-Bo\v{z}i\'{c} 88]}:
A. O. Barut, \& M. Bo\v{z}i\'{c},
`Classical and quantum ``EPR''-spin correlations in the triplet state',
{\em Nuovo Cimento B} {\bf 101}, 5, 595-605 (1988).

\item {\bf [Barut 92]}:
O. A. Barut,
``How to avoid `quantum paradoxes'\,'',
{\em Found. Phys.} {\bf 22}, 1, 137-142 (1992).

\item {\bf [Basdevant 96]}:
J.-L. Basdevant,
{\em Probl\`{e}mes de m\'{e}canique quantique},
Elipse, Paris, 1996.

\item {\bf [Basdevant-Dalibard 00]}:
J.-L. Basdevant, \& J. Dalibard,
{\em The quantum mechanics solver},
Springer, Berlin, 2000.

\item {\bf [Basdevant-Dalibard 02]}:
J.-L. Basdevant, \& J. Dalibard,
{\em Quantum mechanics},
Springer, Berlin, 2002.

\item {\bf [Basharov 02 a]}:
A. M. Basharov,
``Entanglement of atomic states upon collective radiative decay'',
{\em JETP Lett.} {\bf 75}, ?, 123-? (2002).

\item {\bf [Basharov 02 b]}:
``Decoherence and entanglement in radiative decay of a diatomic system'',
A. M. Basharov,
{\em JETP} {\bf 94}, ?, 1070-? (2002).

\item {\bf [Basharov-Manykin 04]}:
A. M. Basharov, \& \'{E}. A. Manykin,
``On entanglement of quantum states of noninteracting subsystems'',
{\em Opt. and Spectroscopy} {\bf 96}, 81-85 (2004).

\item {\bf [Basharov-Bashkeev 04]}:
A. M. Basharov, \& A. A. Bashkeev,
``Quantum correlations in a system of two two-level atoms'',
{\em Opt. and Spectroscopy} {\bf 96}, 650-657 (2004).

\item {\bf [Basharov-Moreva-Manykin 04]}:
A. M. Basharov, E. V. Moreva, and \'{E}. A. Manykin,
``Entanglement of atoms by a resonance classical electromagnetic field'',
{\em Opt. and Spectroscopy} {\bf 96}, 658-664 (2004).

\item {\bf [Basini-Capozziello-Longo 03]}:
G. Basini, S. Capozziello, \& G. Longo,
``The general conservation principle. Absolute validity of conservation laws
and their role as source of entanglement, topology changes, and generation of masses'',
{\em Phys. Lett. A} {\bf 311}, 6, 465-473 (2003).

\item {\bf [Basoalto-Percival 01]}:
R. M. Basoalto, \& I. C. Percival,
``A general computer program for the Bell detection loophole'',
{\em Phys. Lett. A} {\bf 280}, 1-2, 1-6 (2001);
quant-ph/0012024.

\item {\bf [Basoalto-Percival 03]}:
R. M. Basoalto, \& I. C. Percival,
``BellTest and CHSH experiments with more than two settings'',
{\em J. Phys. A} {\bf 36}, 26, 7411–7423 (2003).

\item {\bf [Bassi-Ghirardi 99]}:
A. Bassi, \& G.-C. Ghirardi,
``Can the decoherent histories description of reality be considered satisfactory?'',
{\em Phys. Lett. A} {\bf 257}, 5-6, 247-263 (1999).
See {\bf [Griffiths 00 a, b]}, {\bf [Bassi-Ghirardi 00 b, c]}.

\item {\bf [Bassi-Ghirardi 00]}:
A. Bassi, \& G.-C. Ghirardi,
``More about dynamical reduction and the enumeration principle'',
{\em Brit. J. Philos. Sci.} {\bf 51}, ?-? (2000);
quant-ph/9907050.
See {\bf [Clifton-Monton 99, 00]}.

\item {\bf [Bassi-Ghirardi 00 b]}:
A. Bassi, \& G.-C. Ghirardi,
``About the notion of truth in the decoherent histories approach:
A reply to Griffiths'',
{\em Phys. Lett. A} {\bf 265}, 1-2, 153-155 (2000);
quant-ph/9912065.
Reply to {\bf [Griffiths 00 a]}.
See {\bf [Bassi-Ghirardi 00 a, d]}, {\bf [Griffiths 00 b]}.

\item {\bf [Bassi-Ghirardi 00 c]}:
A. Bassi, \& G.-C. Ghirardi,
``Decoherent histories and realism'',
{\em J. Stat. Phys.};
quant-ph/9912031.
See {\bf [Bassi-Ghirardi 00 a, c]}, {\bf [Griffiths 00 a, b]}.

\item {\bf [Bassi-Ghirardi 00 c]}:
A. Bassi, \& G.-C. Ghirardi,
``A general argument against the universal validity of the superposition
principle'',
{\em Phys. Lett. A} {\bf 275}, 5-6, 373-381 (2000);
quant-ph/0009020.
See {\bf [d'Espagnat 01]}.

\item {\bf [Bassi-Ghirardi 02]}:
A. Bassi, \& G.-C. Ghirardi,
``Dynamical reduction models with general Gaussian noises'',
quant-ph/0201122.

\item {\bf [Bassi-Ghirardi 03 a]}:
A. Bassi, \& G.-C. Ghirardi,
``A general scheme for ensemble purification'',
{\em Phys. Lett. A} {\bf 309}, 1-2, 24-28 (2003).

\item {\bf [Bassi-Ghirardi 03 b]}:
A. Bassi, \& G.-C. Ghirardi,
``Dynamical reduction models'',
quant-ph/0302164.

\item {\bf [Bassi-Ghirardi 03 b]}:
A. Bassi, \& G.-C. Ghirardi,
``Quantum vs classical computation: A proposal opening a new perspective'',
quant-ph/0304074.

\item {\bf [Bassi 03]}:
A. Bassi,
``Stochastic Schr\"{o}dinger equations with general complex Gaussian noises'',
{\em Phys. Rev. A};
quant-ph/0209170.

\item {\bf [Bassi 04]}:
A. Bassi,
``Collapse models: analysis of the free particle dynamics'',
quant-ph/0410222.

\item {\bf [Basu-Bandyopadhyay-Kar-Home 01]}:
S. Basu, S. Bandyopadhyay, G. Kar, \& D. Home,
``Bell's inequality for a single spin 1/2 particle and quantum contextuality'',
{\em Phys. Lett. A} {\bf 279}, 5-6, 281-286 (2001);
quant-ph/9907030.

\item {\bf [Batelaan-Gay-Schwendiman 97]}:
H. Batelaan, T. J. Gay, \& J. J. Schwendiman,
``Stern-Gerlach effect for electron beams'',
{\em Phys. Rev. Lett.} {\bf 79}, 23, 4517-4521 (1997).
Comment: {\bf [Rutherford-Grobe 98]}.
Reply: {\bf [Batelaan-Gay 98]}.

\item {\bf [Batelaan-Gay 98]}:
H. Batelaan, \& T. J. Gay,
``Batelaan and Gay reply'',
{\em Phys. Rev. Lett.} {\bf 81}, 21, 4773 (1998).
Reply to {\bf [Rutherford-Grobe 98]}.
See {\bf [Batelaan-Gay-Schwendiman 97]}.

\item {\bf [Batelaan 02]}:
H. Batelaan,
``Electrons, Stern-Gerlach magnets, and quantum mechanical propagation'',
{\em Am. J. Phys.} {\bf 70}, 3, 325-331 (2002).

\item {\bf [Batle-Casas-Plastino-Plastino 01]}:
J. Batle, M. Casas, A. R. Plastino, \& A. Plastino,
`On the ``fake'' inferred entanglement associated with the maximum
entropy inference of quantum states',
quant-ph/0107023.

\item {\bf [Batle-Casas-Plastino-Plastino 02 a]}:
J. Batle, M. Casas, A. R. Plastino, \& A. Plastino,
``Inference schemes and entanglement determination'',
{\em Phys. Rev. A} {\bf 65}, 2, 024304 (2002).
Publisher's note: {\em Phys. Rev. A} {\bf 65}, 4, 049902 (2002).

\item {\bf [Batle-Casas-Plastino-Plastino 02 b]}:
J. Batle, M. Casas, A. R. Plastino, \& A. Plastino,
``Entanglement, mixedness, and $q$-entropies'',
{\em Phys. Lett. A} {\bf 296}, 6, 251-258 (2002).

\item {\bf [Batle-Plastino-Casas-Plastino 02]}:
J. Batle, A. R. Plastino, M. Casas, \& A. Plastino,
``On the entanglement properties of two-rebits systems'',
{\em Phys. Lett. A} {\bf 298}, 5-6, 301-307 (2002);
quant-ph/0203011.

\item {\bf [Batle-Plastino-Casas-Plastino 02]}:
J. Batle, A. R. Plastino, M. Casas, \& A. Plastino,
``Conditional $q$-entropies and quantum separability:
a numerical exploration'',
{\em J. Phys. A} {\bf 35}, 48, 10311-10324 (2002).

\item {\bf [Batle-Plastino-Casas-Plastino 03 a]}:
J. Batle, A. R. Plastino, M. Casas, \& A. Plastino,
``On the distribution of entanglement changes produced by unitary operations'',
{\em Phys. Lett. A} {\bf 307}, 5-6, 253-261 (2003).

\item {\bf [Batle-Plastino-Casas-Plastino 03 b]}:
J. Batle, A. R. Plastino, M. Casas, \& A. Plastino,
``Understanding quantum entanglement:
Qubits, rebits and the quaternionic approach'',
{\em Opt. Spectrosc.} {\bf 94}, 700 (2003).

\item {\bf [Batle-Plastino-Casas-Plastino 03 c]}:
J. Batle, M. Casas, A. Plastino, \& A. R. Plastino,
``On the correlations between quantum entanglement and $q$-information
measures'',
{\em Phys. Lett. A} {\bf 318}, 6, 506-513 (2003);
quant-ph/0302143.

\item {\bf [Batle-Plastino-Casas-Plastino 03 d]}:
J. Batle, A. R. Plastino, M. Casas, \& A. Plastino,
``Some features of the conditional $q$-entropies of composite quantum
systems'',
quant-ph/0306188.

\item {\bf [Batle-Plastino-Casas-Plastino 04]}:
J. Batle, A. R. Plastino, M. Casas, \& A. Plastino,
``Inclusion relations among separability criteria'',
{\em J. Phys. A} {\bf 37}, 3, 895-907 (2003);
quant-ph/0307167.

\item {\bf [Batty-Braunstein-Duncan-Rees 03]}:
M. Batty, S. L. Braunstein, A. J. Duncan, \& S. Rees,
``Quantum algorithms in group theory'',
in {\em Interactions between logic, group theory and computer
science}, AMS Contemporary Math.;
quant-ph/0310133.

\item {\bf [Baute-Egusquiza-Muga 01]}:
A. D. Baute, I. L. Egusquiza, \& J. G. Muga,
``Quantum times of arrival for multiparticle states'',
quant-ph/0109133.

\item {\bf [BBN Technologies 02]}:
{\em BBN Technologies},
www.bbn.com.

\item {\bf [Beals-Buhrman-Cleve-(+2) 98]}:
R. Beals, H. Buhrman, R. Cleve, M. Mosca, \& R. de Wolf,
``Quantum lower bounds by polynomials'',
in {\em Proc.\ of the 39th Annual Symp.\ on the Foundations of
Computer Science},
IEEE Computer Society Press, Los Alamitos, California, 1998,
pp.~352-361.

\item {\bf [de Beaudrap 04]}:
J. N. de Beaudrap,
``One-qubit fingerprinting schemes'',
{\em Phys. Rev. A} {\bf 69}, 2, 022307 (2004);
quant-ph/0309036.

\item {\bf [Beauregard 03]}:
S. Beauregard,
``Circuit for Shor's algorithm using $2n+3$ qubits'',
{\em Quant. Inf. Comp.} {\bf 3}, 2, 175-185 (2003);
quant-ph/0205095.

\item {\bf [Beausoleil-Kent-Munro-Spiller 03]}:
R. G. Beausoleil, A. P. A. Kent, W. J. Munro, \& T. P. Spiller,
``Applications of coherent population transfer to classical and quantum
information processing'',
quant-ph/0302109.

\item {\bf [Beausoleil-Munro-Rodrigues-Spiller 04]}:
R. G. Beausoleil, W. J. Munro, D. A. Rodrigues, \& T. P. Spiller
``Applications of electromagnetically induced transparency to quantum
information processing'',
quant-ph/0403028.

\item {\bf [Bechmann Johansen 94]}:
H. Bechmann Johansen,
`Comment on
``Getting contextual and nonlocal elements-of-reality the easy way'',
by R. Clifton',
{\em Am. J. Phys.} {\bf 62}, 5, 471 (1994).
Comment on {\bf [Clifton 93]}.

\item {\bf [Bechmann Pasquinucci-Huttner-Gisin 97]}:
H. Bechmann-Pasquinucci (Johansen before), B. Huttner, \& N. Gisin,
``Nonlinear quantum state transformation of spin-$\frac{1}{2}$'',
{\em Phys. Lett. A} {\bf 242}, 4-5, 198-204 (1998);
quant-ph/9708040.

\item {\bf [Bechmann Pasquinucci-Gisin 99]}:
H. Bechmann-Pasquinucci (Johansen before), \& N. Gisin,
``Incoherent and coherent eavesdropping in the six-state protocol of quantum
cryptography'',
{\em Phys. Rev. A} {\bf 59}, 6, 4238-4248 (1999);
quant-ph/9807041.

\item {\bf [Bechmann Pasquinucci-Peres 00]}:
H. Bechmann-Pasquinucci (Johansen before), \& A. Peres,
``Quantum cryptography with 3-state systems'',
{\em Phys. Rev. Lett.} {\bf 85}, 15, 3313-3316 (2000);
quant-ph/0001083.

\item {\bf [Bechmann Pasquinucci-Tittel 00]}:
H. Bechmann-Pasquinucci (Johansen before), \& W. Tittel,
``Quantum cryptography using larger alphabets'',
{\em Phys. Rev. A} {\bf 61}, 6, 062308 (2000);
quant-ph/9910095.

\item {\bf [Bechmann Pasquinucci-Gisin 02]}:
H. Bechmann-Pasquinucci (Johansen before), \& N. Gisin,
``Bell inequality for quNits with binary measurements'',
quant-ph/0204122.

\item {\bf [Bechmann Pasquinucci-Gisin 03]}:
H. Bechmann-Pasquinucci (Johansen before), \& N. Gisin,
``Intermediate states in quantum cryptography and Bell inequalities'',
{\em Phys. Rev. A} {\bf 67}, 6, 062310 (2003).

\item {\bf [Bechmann Pasquinucci 03]}:
H. Bechmann-Pasquinucci (Johansen before),
``Quantum seals'',
{\em Int. J. Quant. Inf.};
quant-ph/0303173.

\item {\bf [Bechmann Pasquinucci 04]}:
H. Bechmann-Pasquinucci,
``From quantum state targeting to Bell inequalities'',
{\em Found. Phys.} (Festschrift in honor of Asher Peres);
quant-ph/0406106.

\item {\bf [Becker 98]}:
L. Becker,
``A new form of quantum interference restoring experiment'',
{\em Phys. Lett. A} {\bf 249}, 1-2, 19-24 (1998).

\item {\bf [Beckman-Gottesman-Nielsen-Preskill 01]}:
D. Beckman, D. Gottesman, M. A. Nielsen, \& J. Preskill,
``Causal and localizable quantum operations'',
{\em Phys. Rev. A} {\bf 64}, 5, 052309 (2001);
quant-ph/0102043.

\item {\bf [Bedard 97]}:
K. Bedard,
``Reply to: `Quantum tunneling times: A crucial
test for the causal program?'\,'',
{\em Found. Phys. Lett.} {\bf 10}, 2, 183-187 (1997).
Reply to {\bf [Cushing 95 a]}.

\item {\bf [Bedard 99]}:
K. Bedard,
``?'',
{\em Phil. Sci.} {\bf 66}, ?, 221-? (1999).
Comment: {\bf [Dickson 00]}.

\item {\bf [Bedford-Selleri 85]}:
D. Bedford, \& F. Selleri,
``On Popper's new EPR-experiment'',
{\em Lettere al Nuovo Cimento} {\bf 42}, 7, 325-328 (1985).
See {\bf [Collett-Loudon 87]}, {\bf [Combourieu 92]}.

\item {\bf [Bedford-Stapp 89]}:
D. Bedford, \& H. P. Stapp,
``Nonlocal character of the Rastall model'',
{\em Found. Phys.} {\bf 19}, 4, 397-406 (1989).
See {\bf [Rastall 85]}.

\item {\bf [Bednorz-Zieli\'{n}ski 03]}:
A. Bednorz, \& J. Zieli\'{n}ski,
``The necessity of randomness in tests of Bell inequalities'',
{\em Phys. Lett. A} {\bf 314}, 5-6, 362-366 (2003).

\item {\bf [Beenakker-Emary-Kindermann-van Velsen 03]}:
C. W. J. Beenakker, C. Emary, M. Kindermann, \& J. L. van Velsen,
``Proposal for production and detection of entangled electron-hole pairs in a
degenerate electron gas'',
{\em Phys. Rev. Lett.} {\bf 91}, 14, 147901 (2003).

\item {\bf [Beenakker-DiVincenzo-Emary-Kindermann 04]}:
C. W. J. Beenakker, D. P. DiVincenzo, C. Emary, \& M. Kindermann,
``Charge detection enables free-electron quantum computation'',
{\em Phys. Rev. Lett.} {\bf 93}, 2, 020501 (2004);
quant-ph/0401066.

\item {\bf [Beenakker-Emary-Kindermann 04]}:
C. W. J. Beenakker, C. Emary, \& M. Kindermann,
``Production and detection of three-qubit entanglement in the Fermi sea'',
{\em Phys. Rev. B} {\bf 69}, 115320 (2004).

\item {\bf [Beenakker-Kindermann 04]}:
C. W. J. Beenakker, \& M. Kindermann,
``Quantum teleportation by particle-hole annihilation in the Fermi sea'',
{\em Phys. Rev. Lett.} {\bf 92}, 5, 056801 (2004).

\item {\bf [Behrman-Chandrashekar-Wang-(+3) 02]}:
E. C. Behrman, V. Chandrashekar, Z. Wang,
C. K. Belur, J. E. Steck, \& S. R. Skinner,
``A quantum neural network computes entanglement'',
quant-ph/0202131.

\item {\bf [Beige-Hegerfeldt 96]}:
A. Beige, \& G. C. Hegerfeldt,
``Projection postulate and atomic quantum Zeno effect'',
{\em Phys. Rev. A} {\bf 53}, 1, 53-65 (1996);
quant-ph/9512012.

\item {\bf [Beige-Hegerfeldt-Sondermann 97]}:
A. Beige, G. C. Hegerfeldt, \& D. G. Sondermann,
``Atomic quantum Zeno effect for ensembles and single systems'',
{\em Found. Phys.} {\bf 27}, ?, 1671-1688 (1997);
quant-ph/9610003.

\item {\bf [Beige-Hegerfeldt 97]}:
A. Beige, \& G. C. Hegerfeldt,
``Quantum Zeno effect and light-dark periods for a single atom'',
{\em J. Phys. A} {\bf 30}, 4, 1323-1334 (1997);
quant-ph/9611020.

\item {\bf [Beige-Fern\'{a}ndez Huelga-Knight-(+2) 00]}:
A. Beige, S. G. Fern\'{a}ndez Huelga, P. L. Knight,
M. B. Plenio, \& R. C. Thompson,
``Coherent manipulation of two dipole-dipole interacting ions'',
in V. Bu\v{z}zek, \& D. P. DiVincenzo (eds.),
{\em J. Mod. Opt.} {\bf 47}, 2-3 (Special issue:
Physics of quantum information), 401-414 (2000);
quant-ph/9903059.

\item {\bf [Beige-Braun-Tregenna-Knight 00]}:
A. Beige, D. Braun, B. Tregenna, \& P. L. Knight,
``Quantum computing using dissipation to remain
in a decoherence-free subspace'',
{\em Phys. Rev. Lett.} {\bf 85}, 8, 1762-1765 (2000);
quant-ph/0004043.
See {\bf [Osborne 00 d]}.

\item {\bf [Beige-Munro-Knight 00]}:
A. Beige, W. J. Munro, \& P. L. Knight,
``Bell's inequality test with entangled atoms'',
{\em Phys. Rev. A} {\bf 62}, 5, 052102 (2000).
Erratum: {\em Phys. Rev. A} {\bf 64}, 4, 049901(E) (2001);
quant-ph/0006054.

\item {\bf [Beige-Bose-Braun-(+4) 00]}:
A. Beige, S. Bose, D. Braun,
S. G. Fern\'{a}ndez Huelga, P. L. Knight, M. B. Plenio, \& V. Vedral,
``Entangling atoms and ions in dissipative environments'',
{\em J. Mod. Opt.};
quant-ph/0007082.

\item {\bf [Beige-Englert-Kurtsiefer-Weinfurter 01 a]}:
A. Beige, B.-G. Englert, C. Kurtsiefer, \& H. Weinfurter,
``Cryptography with single-photon two-qubit states'',
quant-ph/0101066.

\item {\bf [Beige-Englert-Kurtsiefer-Weinfurter 01 b]}:
A. Beige, B.-G. Englert, C. Kurtsiefer, \& H. Weinfurter,
``Secure communication with a publicly known key'',
quant-ph/0111106.

\item {\bf [Beige-Schoen-Pachos 01]}:
A. Beige, C. Schoen, \& J. Pachos,
``Interference of spontaneously emitted photons'',
{\em 100 Years of Werner Heisenberg - Works and Impact (Bamberg, Germany, 2001)},
{\em Fortschr. Phys.};
quant-ph/0112078.

\item {\bf [Beige 03 a]}:
A. Beige,
``Dissipation-assisted quantum gates with cold trapped ions'',
{\em Phys. Rev. A} {\bf 67}, 2, 020301 (2003).

\item {\bf [Beige-Cable-Knight 03]}:
A. Beige, H. Cable, \& P. L. Knight,
``Dissipation-assisted quantum computation in atom-cavity systems'',
{\em Proc.\ SPIE Conf.\ Fluctuations and Noise (Santa Fe, New Mexico, 2003)}
quant-ph/0303151.

\item {\bf [Beige 03 b]}:
A. Beige,
``Ion-trap quantum computing in the presence of cooling'',
{\em Phys. Rev. A} {\bf 69}, 1, 012303 (2004);
quant-ph/0304168.

\item {\bf [Beige 03 c]}:
A. Beige,
``Quantum computing using dissipation'',
in {\em Proc.\ for the XXIV Int.\ Colloquium on Group
Theoretical Methods in Physics (Paris, 2002)};
quant-ph/0306125.

\item {\bf [Beige-Knight-Vitiello 04]}:
A. Beige, P. L. Knight, \& G. Vitiello,
``Cooling many particles at once'',
quant-ph/0404160.

\item {\bf [Beige-Cable-Marr-Knight 04]}:
A. Beige, H. Cable, C. Marr, \& P. L. Knight,
``Speeding up gate operations through dissipation'',
{\em Laser Phys.} (special issue in
honour of the 70th birthday of Herbert Walther),
quant-ph/0405186.

\item {\bf [Bekenstein 01]}:
J. D. Bekenstein,
``Limitations on quantum information from black hole physics'',
in M. Biesiada (ed.),
{\em XXV International School of Theoretical Physics (Ustron, Poland, 2001)},
{\em Acta Physica Polonica B};
quant-ph/0110005.

\item {\bf [Bekenstein 02]}:
J. D. Bekenstein,
``Quantum information and quantum black holes'',
to appear in V. de Sabbata (ed.),
{\em Advances in the interplay between quantum and gravity physics (Erice, 2001)},
Kluwer Academic, Dordrecht, Holland, 2002;
gr-qc/0107049.

\item {\bf [Bekenstein 04]}:
J. D. Bekenstein,
``How does the entropy/information bound work?'',
in C. Fuchs, \& A. van der Merwe (eds.),
{\em Found. Phys.} (Festschrift in honor of Asher Peres);
quant-ph/0404042.

\item {\bf [Belavkin-Ohya 98]}:
V. P. Belavkin, \& M. Ohya,
``Quantum entanglements and entangled mutual entropy'',
quant-ph/9812082.

\item {\bf [Belavkin-Ohya 00]}:
V. P. Belavkin, \& M. Ohya,
``Entanglements and compound states in quantum information
theory'',
quant-ph/0004069.

\item {\bf [Belavkin 00]}:
V. P. Belavkin,
``?'',
{\em Open Sys. Information Dyn.} {\bf 7}, ?, 101-? (2000).

\item {\bf [Belavkin 01]}:
V. P. Belavkin,
``On entangled information and quantum capacity'',
{\em Open Sys. Information Dyn.} {\bf 8}, 1, 1-18 (2001).

\item {\bf [Belavkin-Ohya 02]}:
V. P. Belavkin, \& M. Ohya,
``Entanglement, quantum entropy and mutual information'',
{\em Proc. R. Soc. Lond. A} {\bf 458}, 2017, 209-231 (2002).

\item {\bf [Belavkin-D'Ariano-Raginsky 04]}:
V. P. Belavkin, G. M. D'Ariano, \& M. Raginsky,
``Operational distance and fidelity for quantum channels'',
quant-ph/0408159.

\item {\bf [Belinfante 73]}:
F. J. Belinfante,
{\em A survey of hidden-variables theories},
Pergamon Press, New York, 1973.

\item {\bf [Belinfante 75]}:
F. J. Belinfante,
{\em Measurements and time reversal in objective quantum theory},
Pergamon Press, Oxford, 1975.

\item {\bf [Belinfante 78]}:
F. J. Belinfante,
``Can individual elementary particles have individual properties?'',
{\em Am. J. Phys.} {\bf 46}, 4, 329-336 (1978).

\item {\bf [Belinsky-Klyshko 93 a]}:
A. V. Belinsky, \& D. N. Klyshko,
``A modified $N$-particle Bell theorem, the corresponding optical
experiment and its classical model'',
{\em Phys. Lett. A} {\bf 176}, 6, 415-420 (1993).

\item {\bf [Belinsky-Klyshko 93 b]}:
A. V. Belinsky, \& D. N. Klyshko,
``?'',
{\em Usp. Fiz. Nauk} {\bf 163}, 1, 1-? (1993);
English version:
``?'',
{\em Phys. Usp.} {\bf 36}, ?, 653-? (1993).
See {\bf [Klyshko 93]}.

\item {\bf [Belinsky 94 a]}:
A. V. Belinsky,
``?'',
{\em Zh. Eksp. Teor. Fiz.} {\bf 105}, 4, 818-827 (1994).
English version:
``A generalized Bell's theorem'',
{\em J. Exp. Theor. Phys.} {\bf 78}, 4, 436-440 (1994).

\item {\bf [Belinsky 94 b]}:
A. V. Belinsky,
``?'',
{\em Usp. Fiz. Nauk} {\bf 164}, 4, 435-441 (1994).
English version:
``Bell's paradoxes without the
introduction of hidden variables'',
{\em Phys. Usp.} {\bf 37}, 4, 413-419 (1994).

\item {\bf [Belinsky 94 c]}:
A. V. Belinsky,
``?'',
{\em Pis'ma Zh. Eksp. Teor. Fiz.} {\bf 59}, 2, 137-140 (1994).
English version:
``Nonlocal Bell's paradoxes for an arbitrary number of observers'',
{\em JETP Lett.} {\bf 59}, 2, 147-151 (1994).

\item {\bf [Belinsky 04]}:
A. V. Belinsky,
``Bell's theorem with the inclusion of losses'',
{\em Opt. and Spectroscopy} {\bf 96}, 665-667 (2004).

\item {\bf [Bell 66]}:
J. S. Bell,
``On the problem of hidden variables in quantum mechanics'',
{\em Rev. Mod. Phys.} {\bf 38}, 3, 447-452 (1966).
Reprinted in {\bf [Wheeler-Zurek 83]}, pp.~397-402;
in {\bf [Bell 87 b]}, pp.~1-13 (pp.~25-40 in the Spanish version);
in {\bf [Bell 95]};
and in {\bf [Bell 01]}, pp.~1-6.

\item {\bf [Bell 64]}:
J. S. Bell,
``On the Einstein Podolsky Rosen paradox'',
{\em Physics} {\bf 1}, 3, 195-200 (1964).
Reprinted in {\bf [Wheeler-Zurek 83]}, pp.~403-408;
in {\bf [Bell 87 b]}, pp.~14-21 (pp.~41-50 in the Spanish version);
in {\bf [Bell 95]};
and in {\bf [Bell 01]}, pp.~7-12

\item {\bf [Bell 71]}:
J. S. Bell,
``Introduction to the hidden-variable question'',
in {\bf [d'Espagnat 71]}, pp.~171-181.
Reprinted in {\bf [Bell 87 b]}, pp.~29-39 (pp.~60-73 in the Spanish version);
in {\bf [Bell 95]};
and in {\bf [Bell 01]}.

\item {\bf [Bell 72]}:
J. S. Bell,
``?'',
{\em Science} {\bf 177}, ?, 880-881 (1972).

\item {\bf [Bell 76 a]}:
J. S. Bell,
``The theory of local beables'',
{\em Epistemological Lett.} no. 9 (March 1976), ?-?.
Presented at the 6th GIFT Seminar, Jaca, Spain, 1975.
Reprinted in {\bf [Bell-Shimony-Horne-Clauser 85]} and
in {\bf [Bell 87 b]}, pp.~52-62 (pp.~89-103 in the Spanish version);
in {\bf [Bell 95]};
and in {\bf [Bell 01]}.
See {\bf [Bell 76 b, 77]}, {\bf [Shimony-Horne-Clauser 76, 78]}.

\item {\bf [Bell 76 b]}:
J. S. Bell,
``Locality in quantum mechanics: Reply to critics'',
{\em Epistemological Lett.} no. 13 (Nov. 1976), 2-6.
Reprinted in {\bf [Bell-Shimony-Horne-Clauser 85]};
in {\bf [Bell 87 b]}, pp.~63-66 (pp.~104-108 in the Spanish version);
in {\bf [Bell 95]};
and in {\bf [Bell 01]}.
See {\bf [Bell 76 a, 77]}, {\bf [Shimony-Horne-Clauser 76, 78]}.

\item {\bf [Bell 77]}:
J. S. Bell,
``Free variables and and local causality'',
{\em Epistemological Lett.} no. 15 (Feb. 1977), ?-?.
Reprinted in {\bf [Bell-Shimony-Horne-Clauser 85]} and
in {\bf [Bell 87 b]}, pp.~100-103 (pp.~149-154 in the Spanish version);
in {\bf [Bell 95]};
and in {\bf [Bell 01]}.
See {\bf [Bell 76 a, b]}, {\bf [Shimony-Horne-Clauser 76, 78]}.

\item {\bf [Bell 80]}:
J. S. Bell,
``Atomic-cascade photons and quantum-mechanical nonlocality'',
{\em Com. At. Molec. Phys.} {\bf 9}, ?, 121-126 (1980).
Reprinted in {\bf [Bell 87 b]}, pp.~105-110 (pp.~155-162 in the Spanish version);
in {\bf [Bell 95]};
and in {\bf [Bell 01]}.

\item {\bf [Bell 81 a]}:
J. S. Bell,
``Quantum mechanics for cosmologists'',
in C. J. Isham, R. Penrose, \& D. W. Sciama (eds.),
{\em Quantum gravity 2},
Clarendon Press, Oxford, 1981, pp.~611-637.
Reprinted in {\bf [Bell 87 b]}, pp.~117-138 (pp.~170-196 in the Spanish version);
in {\bf [Bell 95]};
and in {\bf [Bell 01]}.

\item {\bf [Bell 81 b]}:
J. S. Bell,
``Bertlmann's socks and the nature of reality'',
{\em Journal de Physique}, Colloque {\em C2}, suppl. au numero 3, Tome 42, 41-61 (1981).
Reprinted in {\bf [Bell 87 b]}, pp.~139-158 (pp.~197-220 in the Spanish version);
in {\bf [Bell 95]};
and in {\bf [Bell 01]}.

\item {\bf [Bell 82]}:
J. S. Bell,
``On the impossible pilot wave'',
{\em Found. Phys.} {\bf 12}, 10, 989-999 (1982).
Reprinted in {\bf [Barut-van der Merwe-Vigier 84]}, pp.~66-76;
in {\bf [Bell 87 b]}, pp.~159-168 (pp.~221-233 in the Spanish version);
in {\bf [Bell 95]};
and in {\bf [Bell 01]}, pp.~148-158.

\item {\bf [Bell 84]}:
J. S. Bell,
``Beables for quanum field theory'', CERN-TH 4035/84.
Reprinted in {\bf [Bell 87 b]}, pp.~173-180 (pp.~238-248 in the Spanish version);
in {\bf [Bell 95]};
and in {\bf [Bell 01]}.

\item {\bf [Bell-Shimony-Horne-Clauser 85]}:
J. S. Bell, A. Shimony, M. A. Horne, \& J. F. Clauser,
``An exchange on local beables'',
{\em Dialectica} {\bf 39}, ?, 85-110 (1985).

\item {\bf [Bell 87 a]}:
J. S. Bell,
``Beables for quantum field theory'',
in {\bf [Hiley-Peat 87]}, pp.~227-234.

\item {\bf [Bell 87 b]}:
J. S. Bell,
{\em Speakable and unspeakable in quantum mechanics},
Cambridge University Press, Cambridge, 1987, 2003 (2nd edition with an introduction by A. Aspect).
Spanish version: {\em Lo decible y lo indecible en mec\'{a}nica cu\'{a}ntica},
Alianza, Madrid, 1990.
The same papers and two more can be found in {\bf [Bell 95]}.
The same papers and three more can be found in {\bf [Bell 01]}.

\item {\bf [Bell 89]}:
J. S. Bell,
``Towards an exact quantum mechanics'',
in S. Deser, \& R. J. Finkelstein (eds.),
{\em Themes in contemporary physics II. Essays in
honor of Julian Schwinger's 70th birthday},
World Scientific, Singapore, 1989, pp.~?-?.
Reprinted in {\bf [Bell 95]};
and in {\bf [Bell 01]}.

\item {\bf [Bell 90 a]}:
J. S. Bell,
``Against `measurement'\,'',
{\em Phys. World} {\bf 3}, 8, 33-40 (1990). Also in A. I. Miller (ed.),
{\em Sixty-two years of uncertainty:
Historical, philosophical and physical inquiries into the foundations of quantum
mechanics.
Proc.\ Int. School of History of Science (Erice, Italy, 1989)},
Plenum Press, New York, 1990, pp.~17-31.
Reprinted in {\bf [Bell 95]};
in {\bf [Bell 01]}; and in {\bf [Bell 87 b]} (2nd edition).
See {\bf [Peierls 91]}, {\bf [Lyre 97]}.

\item {\bf [Bell 90 b]}:
J. S. Bell,
``La nouvelle cuisine'',
in A. Sarlenijn \& P. Kroes (eds.),
{\em Between science and technology},
Elesevier, Amsterdam, 1990, pp.~97-115.
Reprinted in {\bf [Bell 01]}, pp.~216-234;
and in {\bf [Bell 87 b]} (2nd edition).

\item {\bf [Bell 95]}:
J. S. Bell,
{\em Quantum mechanics, high energy physics and accelerators:
Selected papers of John Bell},
M. Bell, K. Gottfried, \& M. Veltman (eds.),
World Scientific, River Edge, New Jersey, 1995.

\item {\bf [Bell 01]}:
J. S. Bell,
{\em John S. Bell on the foundations of quantum mechanics},
M. Bell, K. Gottfried, \& M. Veltman (eds.),
World Scientific, Singapore, 2001.

\item {\bf [Bell-Clifton 95]}:
J. L. Bell, \& R. K. Clifton,
``Quasiboolean algebras
and simultaneously definite properties in quantum mechanics'',
{\em Int. J. Theor. Phys.} {\bf 43}, 12, 2409-2421 (1995).

\item {\bf [Bell 96]}:
J. L. Bell,
``Logical reflections on the Kochen-Specker theorem'',
in R. K. Clifton (ed.),
{\em Perspectives on quantum reality: Non-relativistic, relativistic,
and field-theoretic (London, Western Ontario, Canada, 1994)},
Kluwer Academic, Dordrecht, Holland, 1996, pp.~227–235.

\item {\bf [Bell 02]}:
M. Bell,
``Some reminiscences'',
in {\bf [Bertlmann-Zeilinger 02]}, pp.~3-6.

\item {\bf [Bell-Sawyer-Volkas-Wong 02]}:
N. F. Bell, R. F. Sawyer, R. R. Volkas, \& Y. Y. Y. Wong,
``Generation of entangled states and error protection from adiabatic avoided
level crossings'',
{\em Phys. Rev. A} {\bf 65}, 4, 042328 (2002);
quant-ph/0109014.

\item {\bf [Bell-Sawyer-Volkas 02]}:
N. F. Bell, R. F. Sawyer, \& R. R. Volkas,
``Entanglement and quantal coherence: Study of two limiting cases of rapid
system-bath interactions'',
{\em Phys. Rev. A} {\bf 65}, 5, 052105 (2002);
quant-ph/0106082.

\item {\bf [Bell-Sawyer-Volkas-Wong 03]}:
N. F. Bell, R. F. Sawyer, R. R. Volkas, \& Y. Y. Y. Wong,
``State permutations from manipulation of near-level-crossings'',
{\em Phys. Rev. A} {\bf 68}, 3, 032307 (2003);
quant-ph/0204136.

\item {\bf [Beller-Fine 94]}:
M. Beller, \& A. I. Fine,
``Bohr's response to EPR'',
in {\bf [Faye-Folse 94]}, pp.~1-31.

\item {\bf [Beller 99]}:
M. Beller,
{\em Quantum dialogue: The making of a revolution},
University of Chicago Press, Chicago, 1999.
Reviews: {\bf [Omn\`{e}s 00]}, {\bf [Greenberger 00 b]},
{\em [Post 01]}.

\item {\bf [Bellini-Marin-Viciani-(+2) 03]}:
M. Bellini, F. Marin, S. Viciani,
A. Zavatta, \& F. T. Arecchi,
``Nonlocal pulse shaping with entangled photon pairs'',
{\em Phys. Rev. Lett.} {\bf 90}, 4, 043602 (2003).

\item {\bf [Belnap-Szab\'{o} 96]}:
N. Belnap, \& L. E. Szab\'{o},
``Branching space-time analysis of the GHZ theorem'',
{\em Found. Phys.} {\bf 26}, 8, 989-1002 (1996).

\item {\bf [Belokurov-Khrustalev-Sadovnichy-Timofeevskaya 01]}:
V. V. Belokurov, O. A. Khrustalev, V. A. Sadovnichy, \& O. D. Timofeevskaya,
``Systems and subsystems in quantum communication'',
quant-ph/0111164.

\item {\bf [Belousek 96]}:
D. W. Belousek,
``Einstein's 1927 unpublished hidden-variable theory: Its background,
context and significance'',
{\em Stud. Hist. Philos. Sci. Part B: Stud. Hist. Philos. Mod. Phys.}
{\bf 27}, 4, 437-461 (1996).

\item {\bf [Belousek 00 a]}:
D. W. Belousek,
``Statistics, symmetry, and the conventionality of
indistinguishability in quantum mechanics'',
{\bf Found. Phys.} {\bf 30}, 1, 1-34 (2000).

\item {\bf [Belousek 00 b]}:
D. W. Belousek,
``Statistics, symmetry, and
(in)distinguishability in Bohmian mechanics'',
{\bf Found. Phys.} {\bf 30}, 1, 153-164 (2000).

\item {\bf [Beltrametti-Cassinelli 79]}:
E. G. Beltrametti, \& G. Cassinelli,
``Properties of states in quantum logic'',
in G. Toraldo di Francia (ed.),
{\em Problems in the foundations of physics.
Proc.\ of the Int. School of Physics ``Enrico Fermi''.
Course LXXII: Problems in the Foundations of Physics (Varenna, Italy, 1977)},
North-Holland, Amsterdam, 1979, pp.~29-70.

\item {\bf [Beltrametti-Cassinelli 81]}:
E. G. Beltrametti, \& G. Cassinelli,
{\em The logic of quantum mechanics},
Addison-Wesley, Reading, Massachusetts, 1981.

\item {\bf [Beltrametti-van Fraassen 81]}:
E. G. Beltrametti, \& B. van Fraassen (eds.),
{\em Current issues in quantum logic},
Plenum Press, New York, 1981.

\item {\bf [Beltrametti 85]}:
E. G. Beltrametti,
``The non-unique decomposition of mixtures: Some remarks'',
in P. J. Lahti, \& P. Mittelstaedt (eds.),
{\em Proc.\ Symp.\ on the Foundations of Modern
Physics: 50 Years of the Einstein-Podolsky-Rosen Experiment
(Joensuu, Finland, 1985)},
World Scientific, Singapore, 1985, pp.~85-96.

\item {\bf [Beltrametti-Cassinelli-Lahti 90]}:
E. G. Beltrametti, G. Cassinelli, \& P. J. Lahti,
``Unitary measurements of discrete quantities in quantum mechanics'',
{\em J. Math. Phys.} {\bf 31}, 1, 91-98 (1990).

\item {\bf [Beltrametti-Maczy\'{n}ski 93 a]}:
E. G. Beltrametti, \& M. J. Maczy\'{n}ski,
``On some probabilistic inequalities related to the Bell inequality'',
{\em Rep. Math. Phys.} {\bf 33}, 1-2, 123-129 (1993).

\item {\bf [Beltrametti-Maczy\'{n}ski 93 b]}:
E. G. Beltrametti, \& M. J. Maczy\'{n}ski,
``On the characterization of probabilities:
A generalization of Bell's inequalities'',
{\em J. Math. Phys.} {\bf 34}, 11, 4919-4929 (1993).

\item {\bf [Beltrametti-Maczy\'{n}ski 94]}:
E. G. Beltrametti, \& M. J. Maczy\'{n}ski,
``On Bell-type inequalities'',
{\em Found. Phys.} {\bf 24}, 8, 1153-1159 (1994).

\item {\bf [Beltrametti-Bugajski 96]}:
E. G. Beltrametti, \& S. Bugajski,
``The Bell phenomenon in classical frameworks'',
{\em J. Phys. A} {\bf 29}, 2, 247-261 (1996).

\item {\bf [Beltrametti-Bugajski 00]}:
E. G. Beltrametti, \& S. Bugajski,
``Remarks on two-slit probabilities'',
{\em Found. Phys.} {\bf 30}, 9, 1415-1429 (2000).

\item {\bf [Beltrametti-Bugajski 03]}:
E. G. Beltrametti, \& S. Bugajski,
``Entanglement and classical correlations in the quantum frame'',
{\em Int. J. Theor. Phys.} {\bf 42}, 5, 969-982 (2003).

\item {\bf [Ben Aryeh-Postan 92 a]}:
Y. Ben-Aryeh, \& A. Postan,
``Possible solution for EPR paradox in atomic cascade experiments'',
{\em Phys. Lett. A} {\bf 163}, 3, 139-142 (1992).

\item {\bf [Ben Aryeh-Postan 92 b]}:
Y. Ben-Aryeh, \& A. Postan,
``Comment on `Does quantum mechanics violate the Bell inequalities?'\,'',
{\em Phys. Rev. Lett.} {\bf 68}, 17, 2701 (1992).
Comment on {\bf [Santos 91 b]}.

\item {\bf [Ben Aryeh 93]}:
Y. Ben-Aryeh,
``Localiy is not violated for the singlet spin state'',
{\em Found. Phys. Lett.} {\bf 6}, 4, 317-325 (1993).

\item {\bf [Ben Aryeh 94]}:
Y. Ben-Aryeh,
``The use of Bell inequalities in quantum optics'',
{\em Found. Phys. Lett.} {\bf 7}, 5, 459-466 (1994).

\item {\bf [Ben Aryeh-Mann-Sanders 99]}:
Y. Ben-Aryeh, A. Mann, \& B. C. Sanders,
``Empirical state determination of entangled two-level systems and
its relation to information theory'',
{\em Found. Phys.} {\bf 29}, 12, 1963-1976 (1999);
quant-ph/9906013.

\item {\bf [Ben Aryeh-Mann 04]}:
Y. Ben-Aryeh, \& A. Mann,
``The correlations of Greenberger-Horne-Zeilinger states described by
Hilbert-Schmidt decomposition'',
{\em Found. Phys. Lett.} {\bf 17}, 1, 97-101 (2004);
quant-ph/0402034.

\item {\bf [Ben Dov 90 a]}:
Y. Ben-Dov,
``No collapse versions of quantum mechanics'',
in M. Cini, \& J. M. L\'{e}vy-Leblond (eds.),
{\em Quantum theory without reduction},
Adam Hilger, Bristol, 1990, pp.~140-150.

\item {\bf [Ben Dov 90 b]}:
Y. Ben-Dov,
``Everett's theory and the `many-worlds' interpretation'',
{\em Am. J. Phys.} {\bf 58}, 9, 829-832 (1990).

\item {\bf [Ben Kish-DeMarco-Meyer-(+8) 03]}:
A. Ben-Kish, B. DeMarco, V. Meyer,
M. Rowe, J. Britton, W. M. Itano, B. M. Jelenkovi\'{c},
C. Langer, D. Leibfried, T. Rosenband, \& D. J. Wineland,
``Experimental demonstration of a technique to generate arbitrary quantum
superposition states of a harmonically bound spin-1/2 particle'',
{\em Phys. Rev. Lett.} {\bf 90}, 3, 037902 (2003).

\item {\bf [Ben Menahem 89]}:
S. Ben-Menahem,
``Spin-measurement retrodiction'',
{\em Phys. Rev. A} {\bf 39}, 4, 1621-1627 (1989).

\item {\bf [Ben Menahem 97]}:
Y. Ben-Menahem,
``Dummett vs Bell on quantum mechanics'',
{\em Stud. Hist. Philos. Sci. Part B: Stud. Hist. Philos. Mod. Phys.}
{\bf 28}, 2, 277-290 (1997).

\item {\bf [Ben Or-Horodecki-Leung-(+2) 04]}:
M. Ben-Or, M. Horodecki, D. W. Leung,
D. Mayers, \& J. Oppenheim,
``The universal composable security of quantum key distribution'',
quant-ph/0409078.

\item {\bf [Bena-Vishveshwara-Balents-Fisher 02]}:
C. Bena, S. Vishveshwara, L. Balents, \& M. P. A. Fisher,
``Quantum entanglement in carbon nanotubes'',
{\em Phys. Rev. Lett.} {\bf 89}, 3, 037901 (2002).

\item {\bf [Benatti-Narnhofer 00]}:
F. Benatti, \& H. Narnhofer,
``On the additivity of entanglement of formation'',
quant-ph/0005126.

\item {\bf [Benatti-Floreanini 02]}:
F. Benatti, \& R. Floreanini,
``Dissipation and decoherence in photon interferometry'',
quant-ph/0204094.

\item {\bf [Benatti-Narnhofer-Uhlmann 03]}:
F. Benatti, H. Narnhofer, \& A. Uhlmann,
``Broken symmetries in the entanglement of formation'',
{\em J. Math. Phys.} {\bf 44}, 2402 (2003).

\item {\bf [Benatti-Floreanini-Piani 03]}:
F. Benatti, R. Floreanini, \& M. Piani,
``Environment induced entanglement in Markovian dissipative dynamics'',
{\em Phys. Rev. Lett.} {\bf 91}, 7, 070402 (2003).

\item {\bf [Benatti-Marinatto 03]}:
F. Benatti, \& L. Marinatto,
``On deciding whether a Boolean function is constant or not'',
{\em Int. J. Quant. Inf.};
quant-ph/0304073.

\item {\bf [Benatti-Floreanini 04]}:
F. Benatti, \& R. Floreanini,
``Entanglement generation in uniformly accelerating atoms: Reexamination of the Unruh effect'',
{\em Phys. Rev. A} {\bf 70}, 1, 012112 (2004).

\item {\bf [Bene 97 a]}:
G. Bene,
``Explanation of the violation of Bell's inequality by
maintaining Einstein separability'',
quant-ph/9708044.

\item {\bf [Bene 97 b]}:
G. Bene,
``On the solution of the EPR paradox and the
explanation of the violation of Bell's inequality'',
quant-ph/9708045.

\item {\bf [Bene 00]}:
G. Bene,
``Quantum reference systems: Reconciling locality with quantum
mechanics'',
quant-ph/0008128.

\item {\bf [Bene-Borsanyi 00]}:
G. Bene, \& S. Borsanyi,
``Decoherence within a single atom'',
quant-ph/0008131.

\item {\bf [Bene 01 a]}:
G. Bene,
``Lowest threshold visibility for testing local realistic theories'',
quant-ph/0104110.

\item {\bf [Bene 01 b]}:
G. Bene,
``Relational modal interpretation for relativistic quantum field theories'',
quant-ph/0104111.

\item {\bf [Bene 01 c]}:
G. Bene,
``Quantum origin of classical properties within the modal interpretations'',
quant-ph/0104112.

\item {\bf [Bene-Dieks 02]}:
G. Bene, \& D. Dieks,
``A perspectival version of the
modal interpretation of quantum mechanics and the origin of
macroscopic behavior'',
{\em Found. Phys.} {\bf 32}, 5, 645-671 (2002);
quant-ph/0112134,
PITT-PHIL-SCI00000556.

\item {\bf [Benenti-Casati-Montangero-Shepelyansky 01]}:
G. Benenti, G. Casati, S. Montangero, \& D. L. Shepelyansky,
``Efficient quantum computing of complex dynamics'',
{\em Phys. Rev. Lett.} {\bf 87}, 22, 227901 (2001);
quant-ph/0107036.

\item {\bf [Benenti-Casati-Montangero-Shepelyansky 03]}:
G. Benenti, G. Casati, S. Montangero, \& D. L. Shepelyansky,
``Dynamical localization simulated on a few-qubit quantum computer'',
{\em Phys. Rev. A} {\bf 67}, 5, 052312 (2003).

\item {\bf [Benenti-Casati-Strini 04]}:
G. Benenti, G. Casati, \& G. Strini,
{\em Principles of quantum computation and information. Volume 1: Basic concepts},
World Scientific, Singapore, 2004.

\item {\bf [Bengtsson-Braennlund-\.{Z}yczkowski 01]}:
I. Bengtsson, J. Braennlund, \& K. \.{Z}yczkowski,
``$CP^n$, or, entanglement illustrated'',
quant-ph/0108064.

\item {\bf [Bengtsson-Ericsson 03]}:
I. Bengtsson, \& A. Ericsson,
``How to mix a density matrix'',
{\em Phys. Rev. A} {\bf 67}, 1, 012107 (2003).

\item {\bf [Beniaminov 01]}:
E. M. Beniaminov,
``A method for justification of the view of
observables in quantum mechanics and probability distributions in
phase space'',
quant-ph/0106112.

\item {\bf [Benioff 72 a]}:
P. A. Benioff,
``Operator valued measures in quantum mechanics: Finite
and infinite processes'',
{\em J. Math. Phys.} {\bf 13}, ?, 231-242 (1972).

\item {\bf [Benioff 72 b]}:
P. A. Benioff,
``Decision procedures in quantum mechanics'',
{\em J. Math. Phys.} {\bf 13}, ?, 908-915 (1972).

\item {\bf [Benioff 72 c]}:
P. A. Benioff,
``Procedures in quantum mechanics without von Neumann's
projection axiom'',
{\em J. Math. Phys.} {\bf 13}, ?, 1347-1355 (1972).

\item {\bf [Benioff 77]}:
P. A. Benioff,
``Finite and infinite measurement sequences in quantum
mechanics and randomness: The Everett interpretation'',
{\em J. Math. Phys.} {\bf 18}, 12, 2289-2295 (1977).

\item {\bf [Benioff 80]}:
P. A. Benioff,
``The computer as a physical system:
A microscopic quantum mechanical Hamiltonian model of computers
as represented by Turing machines'',
{\em J. Stat. Phys.} {\bf 22}, ?, 563-591 (1980).

\item {\bf [Benioff 81]}:
P. A. Benioff,
``Quantum mechanical models of discrete processes'',
{\em J. Math. Phys.} {\bf 22}, 3, 495-507 (1981).

\item {\bf [Benioff 82 a]}:
P. A. Benioff,
``Quantum mechanical models of Turing machines that dissipate no
energy'',
{\em Phys. Rev. Lett.} {\bf 48}, 23, 1581-1585 (1982).

\item {\bf [Benioff 82 b]}:
P. A. Benioff,
``Quantum mechanical Hamiltonian models of Turing machines'',
{\em J. Stat. Phys.} {\bf 29}, 3, 515-? (1982).

\item {\bf [Benioff 82 c]}:
P. A. Benioff,
``Quantum mechanical Hamiltonian models of discrete processes
that erase their own histories: Application to Turing machines'',
{\em Int. J. Theor. Phys.} {\bf 21}, ?, 177-201 (1982).

\item {\bf [Benioff 86]}:
P. A. Benioff,
``Quantum mehcanical Hamiltonian models of computers'',
{\em Ann. N. Y. Acad. Sci.} {\bf 480}, 475-? (1986).

\item {\bf [Benioff 95]}:
P. A. Benioff,
``Unitary dilation models of Turing machines in quantum
mechanics'',
{\em Phys. Rev. A} {\bf 51}, 5, 3513-3524 (1995).

\item {\bf [Benioff 96]}:
P. A. Benioff,
``Quantum ballistic evolution in quantum mechanics:
Application to quantum computers'',
{\em Phys. Rev. A} {\bf 54}, 2, 1106-1123 (1996).

\item {\bf [Benioff 97 a]}:
P. A. Benioff,
``Tight binding Hamiltonians and quantum Turing machines'',
{\em Phys. Rev. Lett.} {\bf 78}, 4, 590-593 (1997).

\item {\bf [Benioff 97 b]}:
P. A. Benioff,
``Transmission and spectral aspects of tight-binding
Hamiltonians for the counting quantum Turing machine'',
{\em Phys. Rev. B} {\bf 55}, 15, 9482-9494 (1997).

\item {\bf [Benioff 98 a]}:
P. A. Benioff,
``The Landauer resistance and band spectra for the
counting quantum Turing machine'',
{\em Physica D} {\bf 120}, ?, 12-29 (1998).

\item {\bf [Benioff 98 b]}:
P. A. Benioff,
``Quantum robots and environments'',
{\em Phys. Rev. A} {\bf 58}, 2, 893-904 (1998);
quant-ph/9807032.

\item {\bf [Benioff 98 c]}:
P. A. Benioff,
``Foundational aspects of quantum computers and quantum robots'',
{\em Superlattices and Microstructures} {\bf 23}, ?, 407-417
(1998).

\item {\bf [Benioff 98 d]}:
P. A. Benioff,
``Models of quantum Turing machines'',
{\em Fortschr. Phys.} {\bf 46}, 4-5, 423-441 (1998).

\item {\bf [Benioff 99]}:
P. A. Benioff,
``Simple example of definitions of truth, validity, consistency,
and completeness in quantum mechanics'',
{\em Phys. Rev. A} {\bf 59}, 6, 4223-4237 (1999);
quant-ph/9811055.

\item {\bf [Benioff 00]}:
P. A. Benioff,
``The representation of numbers by states in quantum mechanics'',
quant-ph/0009124.

\item {\bf [Benioff 01 a]}:
P. A. Benioff,
``Representation of natural numbers in quantum mechanics'',
{\em Phys. Rev. A} {\bf 63}, 3, 032305 (2001);
quant-ph/0003063.

\item {\bf [Benioff 01 b]}:
P. Benioff,
``Efficient implementation and the product-state representation of numbers'',
{\em Phys. Rev. A} {\bf 64}, 5, 052310 (2001);
quant-ph/0104061.

\item {\bf [Benioff 01 c]}:
P. A. Benioff,
``The representation of numbers in quantum mechanics'',
{\em Algorithmica} {\bf 34}, 4, 529-559 (2002);
quant-ph/0103078.

\item {\bf [Benioff 01 d]}:
P. A. Benioff,
``Use of mathematical logical concepts in
quantum mechanics: An example'',
quant-ph/0106153.

\item {\bf [Benioff 02 a]}:
P. A. Benioff,
``Towards a coherent theory of physics and mathematics'',
{\em Found. Phys.} {\bf 32}, 7, 989-1029 (2002).

\item {\bf [Benioff 02 b]}:
P. A. Benioff,
``Towards a coherent theory of physics and mathematics'',
based on talk given at the {\em 1st Brazilian Symposium on the Philosophy of Nature};
quant-ph/0201093.

\item {\bf [Benioff 02 c]}:
P. A. Benioff,
``Space searches with a quantum robot'',
in {\bf [Lomonaco-Brandt 02]} 1-12;
quant-ph/0003006.

\item {\bf [Benioff 04 a]}:
P. A. Benioff,
``Towards a coherent theory of physics and mathematics: The
theory-experiment connection'',
quant-ph/0403209.

\item {\bf [Benioff 04 b]}:
P. A. Benioff,
``Tightening the theory-experiment connection in physics: $R_{n}$ based
space and time'',
in {\em Proc.\ of Foundations of Quantum Information" (Camerino, Italy, 2004)},
{\em Int. J. Quantum Inf.};
quant-ph/0408074.

\item {\bf [Benjamin-Johnson 99]}:
S. C. Benjamin, \& N. F. Johnson,
``Cellular structures for computation in the quantum regime'',
{\em Phys. Rev. A} {\bf 60}, 6, 4334-4337 (1999);
cond-mat/9808243.

\item {\bf [Benjamin 00 a]}:
S. C. Benjamin,
``Schemes for parallel quantum computation
without local control of qubits'',
{\em Phys. Rev. A} {\bf 61}, 2, 020301(R) (2000);
quant-ph/9909007.

\item {\bf [Benjamin 00 b]}:
S. C. Benjamin,
`Quantum cryptography: Single photons ``on demand''\,'
{\em Science} {\bf 290}, 5500, 2273-2274 (2000).
See {\bf [Michler-Kiraz-Becher-(+5) 00]}.

\item {\bf [Benjamin 00 c]}:
S. C. Benjamin,
`Comment on: ``A quantum approach to static games of complete
information''\,',
quant-ph/0008127.
Comment on {\bf [Marinatto-Weber 00 a]}.
Reply: {\bf [Marinatto-Weber 00 c]}.

\item {\bf [Benjamin-Hayden 01 a]}:
S. C. Benjamin, \& P. M. Hayden,
``Multiplayer quantum games'',
{\em Phys. Rev. A} {\bf 64}, 3, 030301(R) (2001);
quant-ph/0007038.

\item {\bf [Benjamin-Hayden 01 b]}:
S. C. Benjamin, \& P. M. Hayden,
``Comment on `Quantum games and quantum strategies'\,'',
{\em Phys. Rev. Lett.} {\bf 87}, 6, 069801 (2001);
quant-ph/0003036.
Comment on {\bf [Eisert-Wilkens-Lewenstein 99]}.
Reply: {\bf [Eisert-Wilkens-Lewenstein 01]}.

\item {\bf [Benjamin 01 a]}:
S. C. Benjamin,
``Simple pulses for universal quantum computation with
a Heisenberg $ABAB$ chain'',
{\em Phys. Rev. A} {\bf 64}, 5, 054303 (2001).

\item {\bf [Benjamin 01 b]}:
S. C. Benjamin,
``Quantum computing with globally controlled exchange-type
interactions'',
quant-ph/0104117.

\item {\bf [Benjamin 02]}:
S. C. Benjamin,
``Quantum computing without local control of qubit-qubit interactions'',
{\em Phys. Rev. Lett.} {\bf 88}, 1, 017904 (2002).

\item {\bf [Benjamin-Bose 03]}:
S. C. Benjamin, \& S. Bose,
``Quantum computing with an always-on Heisenberg interaction'',
{\em Phys. Rev. Lett.} {\bf 90}, 24, 247901 (2003).

\item {\bf [Benjamin-Bose 04]}:
S. C. Benjamin, \& S. Bose,
``Quantum computing in arrays coupled by `always on' interactions'',
quant-ph/0401071.

\item {\bf [Benjamin 04]}:
S. C. Benjamin,
``Multi-qubit gates in arrays coupled by 'always on' interactions'',
quant-ph/0403077.

\item {\bf [Benjamin-Lovett-Reina 04]}:
S. C. Benjamin, B. W. Lovett, \& J. H. Reina,
``Optical quantum computation with perpetually coupled spins'',
quant-ph/0407063.

\item {\bf [Bennett-Brassard-Breidbart-Wiesner 83]}:
C. H. Bennett, G. Brassard, S. Breidbart, \& S. Wiesner,
``Quantum cryptography, or unforgeable subway tokens'', in
D. Chaum, R. L. Rivest, \& A. T. Sherman,
{\em Advances in Cryptology: Proc.\ of Crypto `82},
Plenum Press, New York, 1983, pp.~267-275.

\item {\bf [Bennett 84]}:
C. H. Bennett,
``Eavesdrop-detecting quantum communications channel'',
{\em IBM Technical Disclosure Bulletin} Jan. 1984, pp.~4363-4366.

\item {\bf [Bennett-Brassard 84]}:
C. H. Bennett, \& G. Brassard,
``Quantum key distribution and coin tossing'',
in {\em Proc.\ of IEEE Int.\ Conf.\ on Computers,
Systems, and Signal Processing (Bangalore, India, 1984)},
IEEE, New York, 1984, pp.~175-179.

\item {\bf [Bennett-Brassard-Breidbart-Wiesner 85]}:
C. H. Bennett, G. Brassard, S. Breidbart, \& S. Wiesner,
``Eavesdrop-detecting quantum communications channel",
{\em IBM Technical Disclosure Bulletin} {\bf 26}, 3153-3163 (1985).

\item {\bf [Bennett-Brassard 85 a]}:
C. H. Bennett, \& G. Brassard,
``An update on quantum cryptography'',
in G. R. Blakley \& D. Chaum (eds.),
{\em Advances in Cryptology: Proc.\ of Crypto 84},
{\em Lecture Notes in Computer Science} {\bf 196},
Springer-Verlag, Berlin, 1985, pp.~475-480.

\item {\bf [Bennett-Brassard 85 b]}:
C. H. Bennett, \& G. Brassard,
``Quantum public key distribution'',
{\em IBM Technical Disclosure Bulletin} {\bf 28}, 3153-3163 (1985).

\item {\bf [Bennett-Brassard 87]}:
C. H. Bennett, \& G. Brassard,
``Quantum cryptography reinvented'',
{\em SIGACT News} {\bf 18}, 51-53 (1987).

\item {\bf [Bennett-Brassard-Robert 88]}:
C. H. Bennett, G. Brassard, \& J.-M. Robert,
``Privacy amplification by public discussion'',
{\em Soc. Ind. Appl. Math. J. Comp.} {\bf 17}, 2, 210-229 (1988).

\item {\bf [Bennett-Brassard 88]}:
C. H. Bennett, \& G. Brassard,
``Chapter 6: Quantum cryptography'',
in G. Brassard
{\em Modern cryptology-- a tutorial},
{\em Lecture Notes in Computer Science} {\bf 325},
Springer-Verlag, New York, 1988, pp.~75-90.

\item {\bf [Bennett-Brassard 89]}:
C. H. Bennett, \& G. Brassard,
``The dawn of a new era for quantum cryptography:
The experimental prototype is working!'',
{\em Special Interest Group on Automata and Computability Theory
News} {\bf 20}, 78-82 (1989).

\item {\bf [Bennett 90 a]}:
C. H. Bennett,
``The emperor's new mind'',
{\em American Scientist} {\bf 78}, 473-474 (1990).
Review of {\bf [Penrose 89]}.

\item {\bf [Bennett 90 b]}:
C. H. Bennett,
``How to define complexity in physics, and why'',
in {\bf [Zurek 90]}, pp.~137-148.

\item {\bf [Bennett-Bessette-Brassard-(+2) 92]}:
C. H. Bennett, F. Bessette, G. Brassard, L. Salvail, \& J. A. Smolin,
``Experimental quantum cryptography'',
{\em J. Cryptology} {\bf 5}, 1, 3-28 (1992).
See {\bf [Bennett 94 b]}.

\item {\bf [Bennett-Brassard-Mermin 92]}:
C. H. Bennett, G. Brassard, \& N. D. Mermin,
``Quantum cryptography without Bell's theorem'',
{\em Phys. Rev. Lett.} {\bf 68}, 5, 557-559 (1992).

\item {\bf [Bennett-Brassard-Cr\'{e}peau-Skubiszewska 92]}:
C. H. Bennett, G. Brassard, C. Cr\'{e}peau, \& M.-H. Skubiszewska,
``Practical quantum oblivious transfer'',
in {\em Advances in Cryptology: Proc.\ of Crypto '91},
{\em Lecture Notes in Computer Science} {\bf 576},
Springer-Verlag, Berlin, 1992, pp.~351-366.

\item {\bf [Bennett-Brassard-Ekert 92]}:
C. H. Bennett, G. Brassard, \& A. K. Ekert,
``Quantum cryptography'',
{\em Sci. Am.} {\bf 267}, 4, 26-33 (1992).
Spanish version: ``Criptograf\'{\i}a cu\'{a}ntica'',
{\em Investigaci\'{o}n y Ciencia} 195, 14-22 (1992).
Reprinted in {\bf [Cabello 97 c]}, pp.~75-83.

\item {\bf [Bennett 92 a]}:
C. H. Bennett,
``Quantum cryptography using any two nonorthogonal states'',
{\em Phys. Rev. Lett.} {\bf 68}, 21, 3121-3124 (1992).
See {\bf [Ekert 92]}.

\item {\bf [Bennett 92 b]}:
C. H. Bennett,
``Quantum cryptography: Uncertainty in the service of privacy'',
{\em Science} {\bf 257}, 5071, 752-753 (1992).

\item {\bf [Bennett-Wiesner 92]}:
C. H. Bennett, \& S. J. Wiesner,
``Communication via one- and two-particle operators on
Einstein-Podolsky-Rosen states'',
{\em Phys. Rev. Lett.} {\bf 69}, 20, 2881-2884 (1992).

\item {\bf [Bennett 93]}:
C. H. Bennett,
``Certainty from uncertainty'',
{\em Nature} {\bf 362}, 6422, 694-695 (1993).

\item {\bf [Bennett-Brassard-Cr\'{e}peau-(+3) 93]}:
C. H. Bennett, G. Brassard, C. Cr\'{e}peau, R. Jozsa,
A. Peres, \& W. K. Wootters,
``Teleporting an unknown quantum state via dual
classical and Einstein-Podolsky-Rosen channels'',
{\em Phys. Rev. Lett.} {\bf 70}, 13, 1895-1899 (1993).
Reprinted in {\bf [Macchiavello-Palma-Zeilinger 00]}, pp.~35-38.

\item {\bf [Bennett-Brassard-Jozsa-(+4) 94]}:
C. H. Bennett, G. Brassard, R. Jozsa,
D. Mayers, A. Peres, B. W. Schumacher, \& W. K. Wootters,
``Reduction of quantum entropy by reversible extraction of classical
information'',
in S. M. Barnett, A. K. Ekert, \& S. J. D. Phoenix (eds.),
{\em J. Mod. Opt.} {\bf 41}, 12 (Special issue: Quantum
communication), 2307-2314 (1994).

\item {\bf [Bennett 94 a]}:
C. H. Bennett,
``Night thoughts, dark sight'',
{\em Nature} {\bf 371}, 6497, 479-480 (1994).

\item {\bf [Bennett 94 b]}:
C. H. Bennett,
``Interferometric quantum cryptographic key distribution system'',
patent US5307410, 1994.
See {\bf [Bennett-Bessette-Brassard-(+2) 92]}.

\item {\bf [Bennett 95 a]}:
C. H. Bennett,
``Quantum information and computation'',
{\em Phys. Today} {\bf 48}, 10, 24-30 (1995).

\item {\bf [Bennett 95 b]}:
C. H. Bennett,
``Quantum and classical information transmission acts and reducibilities'',
{\em Proc.\ of EPR60 conference (Haifa, Israel, 1995)}.

\item {\bf [Bennett-DiVincenzo 95]}:
C. H. Bennett, \& D. P. DiVincenzo,
``Quantum computing: Towards an engineering era?'',
{\em Nature} {\bf 377}, 6548, 389-390 (1995).

\item {\bf [Bennett-Brassard-Cr\'{e}peau-Maurer 95]}:
C. H. Bennett, G. Brassard, C. Cr\'{e}peau, \& U. M. Maurer,
``Generalized privacy amplification'',
{\em IEEE Trans. Inf. Theory} {\bf IT41}, 6,
1915-1923 (1995).

\item {\bf [Bennett-Bernstein-Popescu-Schumacher 95]}:
C. H. Bennett, H. J. Bernstein, S. Popescu, \& B. W. Schumacher,
``Concentrating partial entanglement by local operations'',
{\em Phys. Rev. A} {\bf 53}, 4, 2046-2052 (1996);
quant-ph/9511030.

\item {\bf [Bennett 96 a]}:
C. H. Bennett,
``Freely communicating'',
{\em Nature} {\bf 382}, 6593, 669-670 (1996).
See {\bf [Landauer 96]}.

\item {\bf [Bennett 96 b]}:
C. H. Bennett,
``Quantum and classical information transmission acts and
reducibilities'',
in {\bf [Mann-Revzen 96]}, pp.~258-?.

\item {\bf [Bennett 96 c]}:
C. H. Bennett,
``Quantum and classical information transmission acts and reducibilities'',
in A. Mann, \& M. Revzen (eds.),
{\em The dilemma of Einstein, Podolsky and Rosen -- 60 years
later. An international symposium in honour of Nathan Rosen
(Haifa, Israel, 1995)},
{\em Ann. Phys. Soc. Israel} {\bf 12}, 258-277 (1996).

\item {\bf [Bennett-Brassard-Popescu-(+3) 96]}:
C. H. Bennett, G. Brassard, S. Popescu,
B. W. Schumacher, J. A. Smolin, \& W. K. Wootters,
``Purification of noisy entanglement and faithful
teleportation via noisy channels'',
{\em Phys. Rev. Lett.} {\bf 76}, 5, 722-725 (1996).
Erratum: {\em Phys. Rev. Lett.} {\bf 78}, 10, 2031 (1997);
quant-ph/9511027.
Reprinted in {\bf [Macchiavello-Palma-Zeilinger 00]}, pp.~221-224.
See {\bf [Aravind 97 a]}.

\item {\bf [Bennett-Bernstein-Popescu-Schumacher 96]}:
C. H. Bennett, H. J. Bernstein, S. Popescu, \& B. W. Schumacher,
``Concentrating partial entanglement by local operations'',
{\em Phys. Rev. A} {\bf 53}, 4, 2046-2052 (1996).

\item {\bf [Bennett-Mor-Smolin 96]}:
C. H. Bennett, T. Mor, \& J. A. Smolin,
``The parity bit in quantum cryptography'',
{\em Phys. Rev. A} {\bf 54}, 4, 2675-2684 (1996);
quant-ph/9604040.

\item {\bf [Bennett-DiVincenzo-Smolin-Wootters 96]}:
C. H. Bennett, D. P. DiVincenzo, J. A. Smolin, \& W. K. Wootters,
``Mixed-state entanglement and quantum error correction'',
{\em Phys. Rev. A} {\bf 54}, 4, 3824-3851 (1996);
quant-ph/9604024.

\item {\bf [Bennett-Wiesner 96]}:
C. H. Bennett, \& S. J. Wiesner,
``Quantum key distribution using non-orthogonal macroscopic
signals'',
patent US5515438, 1996.

\item {\bf [Bennett 97]}:
C. H. Bennett,
``Classical and quantum information transmission and interactions'',
in O. Hirota, A. S. Holevo (Kholevo), \& C. M. Caves (eds.),
{\em Quantum communication, computing, and measurement},
Plenum Press, New York, 1997, pp.~25-40.

\item {\bf [Bennett-Fuchs-Smolin 97]}:
C. H. Bennett, C. A. Fuchs, \& J. A. Smolin,
``Entanglement-enhanced classical communication on a noisy quantum channel'',
in O. Hirota, A. S. Holevo (Kholevo), \& C. M. Caves (eds.),
{\em Quantum communication, computing, and measurement},
Plenum Press, New York, 1997, pp.~79-88;
quant-ph/9611006.

\item {\bf [Bennett-Bernstein-Brassard-Vazirani 97]}:
C. H. Bennett, E. Bernstein, G. Brassard, \& U. Vazirani,
``Strengths and weaknesses of quantum computing'',
{\em SIAM J. Comput.} {\bf 26}, 5, 1510-1523 (1997);
quant-ph/9701001.

\item {\bf [Bennett-DiVincenzo-Smolin 97]}:
C. H. Bennett, D. P. DiVincenzo, \& J. A. Smolin,
``Capacities of quantum erasure channels'',
{\em Phys. Rev. Lett.} {\bf 78}, 16, 3217-3220 (1997);
quant-ph/9701015.

\item {\bf [Bennett 98 a]}:
C. H. Bennett,
``Classical and quantum information: Similarities and differences'',
in S. C. Lim, R. Abd-Shukor, \& K. H. Kwek (eds.), in
{\em Frontiers in quantum physics},
Springer-Verlag, Singapore, 1998, pp.~24-37.

\item {\bf [Bennett 98 b]}:
C. H. Bennett,
``Quantum information'',
in E. B. Karlsson, \& E. Br\"{a}ndas (eds.),
{\em Proc.\ of the 104th Nobel Symp.\ ``Modern Studies of Basic Quantum Concepts and Phenomena'' (Gimo, Sweden, 1997)},
{\em Physica Scripta} {\bf T76}, 210-217 (1998).

\item {\bf [Bennett 98 c]}:
C. H. Bennett,
``Future directions for quantum information theory'',
in {\bf [Lo-Spiller-Popescu 98]}, pp.~340-348.

\item {\bf [Bennett-Shor 98]}:
C. H. Bennett, \& P. W. Shor,
``Quantum information theory'',
{\em IEEE Trans. Inf. Theory} {\bf 44}, ?, 2724-2742 (1998).

\item {\bf [Bennett-DiVincenzo-Fuchs-(+5) 99]}:
C. H. Bennett, D. P. DiVincenzo, C. A. Fuchs, T. Mor,
E. M. Rains, P. W. Shor, J. A. Smolin, \& W. K. Wootters,
``Quantum nonlocality without entanglement'',
{\em Phys. Rev. A} {\bf 59}, 2, 1070-1091 (1999);
quant-ph/9804053.
See {\bf [Horodecki-Horodecki-Horodecki 99 d]}.

\item {\bf [Bennett-DiVincenzo-Mor-(+3) 99]}:
C. H. Bennett, D. P. DiVincenzo, T. Mor,
P. W. Shor, J. A. Smolin, \& B. M. Terhal,
``Unextendible product bases and bound entanglement'',
{\em Phys. Rev. Lett.} {\bf 82}, 26, 5385-5388 (1999);
quant-ph/9808030.

\item {\bf [Bennett-Shor-Smolin-Thapliyal 99]}:
C. H. Bennett, P. W. Shor, J. A. Smolin, \& A. V. Thapliyal,
``Entanglement-assisted classical capacity of noisy quantum channels'',
{\em Phys. Rev. Lett.} {\bf 83}, 15, 3081-3084 (1999);
quant-ph/9904023.
See {\bf [Bennett-Shor-Smolin-Thapliyal 02]}.

\item {\bf [Bennett 99 a]}:
C. H. Bennett,
``Quantum information theory'',
in T. Hey (ed.),
{\em Feynman and computation},
Perseus, Reading, Massachusetts, 1999, pp.~177-190.

\item {\bf [Bennett 99 b]}:
C. H. Bennett,
``Explorations in quantum computing'',
{\em Phys. Today} {\bf 52}, 2, 11-? (1999).
Review of {\bf [Williams-Clearwater 98]}.

\item {\bf [Bennett-Shor 99]}:
C. H. Bennett, \& P. W. Shor,
``Quantum cryptography: Privacy in a quantum world'',
{\em Science} {\bf 284}, 5415, 747-748 (1999).

\item {\bf [Bennett-DiVincenzo 00]}:
C. H. Bennett, \& D. P. DiVincenzo,
``Quantum information and computation'',
{\em Nature} {\bf 404}, 6775, 247-255 (2000).

\item {\bf [Bennett-DiVincenzo-Smolin-(+2) 01]}:
C. H. Bennett, D. P. DiVincenzo, J. A. Smolin,
B. M. Terhal, \& W. K. Wootters,
``Remote state preparation'',
{\em Phys. Rev. Lett.} {\bf 87}, 7, 077902 (2001).
Erratum: {\em Phys. Rev. Lett.} {\bf 88}, 9, 099902 (2002);
quant-ph/0006044.

\item {\bf [Bennett-Popescu-Rohrlich-(+2) 01]}:
C. H. Bennett, S. Popescu, D. Rohrlich,
J. A. Smolin, \& A. V. Thapliyal,
``Exact and asymptotic measures of multipartite pure state
entanglement'',
{\em Phys. Rev. A} {\bf 63}, 1, 012308 (2001);
quant-ph/9908073.

\item {\bf [Bennett-Shor-Smolin-Thapliyal 02]}:
C. H. Bennett, P. W. Shor, J. A. Smolin, \& A. V. Thapliyal,
``Entanglement-assisted capacity of a quantum
channel and the reverse Shannon theorem'',
{\em IEEE Trans. Inf. Theor.} {\bf 48}, 10, 2637-2655 (2002);
quant-ph/0106052.
See {\bf [Bennett-Shor-Smolin-Thapliyal 99]}.

\item {\bf [Bennett-Cirac-Leifer-(+4) 02]}:
C. H. Bennett, J. I. Cirac, M. S. Leifer,
D. W. Leung, N. Linden, S. Popescu, \& G. Vidal,
``Optimal simulation of two-qubit Hamiltonians using general local operations'',
{\em Phys. Rev. A} {\bf 66}, 1, 012305 (2002);
quant-ph/0107035.

\item {\bf [Bennett-Hayden-Leung-(+2) 02]}:
C. H. Bennett, P. Hayden, D. W. Leung, P. W. Shor, \& A. Winter,
``Remote preparation of quantum states'',
quant-ph/0307100.

\item {\bf [Bennett-Harrow-Leung-Smolin 03]}:
C. H. Bennett, A. W. Harrow, D. W. Leung, \& J. A. Smolin,
``?´´,
{\em IEEE Trans. Inf. Theory} {\bf 49}, 8, 1895-? (2003).

\item {\bf [Bennett-Harrow-Lloyd 04]}:
C. H. Bennett, A. W. Harrow, \& S. Lloyd,
``Universal quantum data compression via gentle tomography'',
quant-ph/0403078.

\item {\bf [Bennett-Devetak-Shor-Smolin 04]}:
C. H. Bennett, I. Devetak, P. W. Shor, \& J. A. Smolin,
``Inequalities and separations among assisted capacities of quantum
channels'',
quant-ph/0406086.

\item {\bf [Bennink-Bentley-Boyd 02]}:
R. S. Bennink, S. J. Bentley, \& R. W. Boyd,
`\,``Two-photon'' coincidence imaging with a classical source',
{\em Phys. Rev. Lett.} {\bf 89}, 11, 113601 (2002).

\item {\bf [Berardi-Garuccio-Lepore 00]}:
V. Berardi, A. Garuccio, \& V. L. Lepore,
``Correlation functions in EPR-type experiments:
The low-detection-efficiency loophole'',
{\em J. Opt. B: Quantum Semiclass. Opt.} {\bf 2}, 4, 476-481 (2000).

\item {\bf [Berberian 66]}:
S. K. Berberian,
{\em Notes on spectral theory},
Van Nostrand, Princeton, New Jersey, 1966.

\item {\bf [Bergeron 03]}:
H. Bergeron,
``New derivation of quantum equations from classical
stochastic arguments'',
quant-ph/0303153.

\item {\bf [Berglund 00]}:
A. J. Berglund,
``Quantum coherence and control in one- and two-photon
optical systems'',
undergraduate thesis, A. B. Dartmouth College, 2000;
quant-ph/0010001.

\item {\bf [Bergmann-Merzbacher-Peres 96]}:
P. G. Bergmann, E. Merzbacher, \& A. Peres,
``Obituary: Nathan Rosen'',
{\em Phys. Today} {\bf 49}, 3, 120 (1996).
See {\bf [Peres 96 a, b]}.

\item {\bf [Bergou-Hillery 97]}:
J. A. Bergou, \& M. Hillery,
``Generation of
highly entangled field states in multiparticle micromaser cavities'',
{\em Phys. Rev. A} {\bf 55}, 6, 4585-4588 (1997).

\item {\bf [Bergou-Englert 98]}:
J. A. Bergou, \& B.-G. Englert,
``Heisenberg's dog and quantum computing'',
{\em J. Mod. Opt.} {\bf 45}, 4, 701-711 (1998).

\item {\bf [Bergou-Hillery-Sun 00]}:
J. A. Bergou, M. Hillery, \& Y. Sun,
``Non-unitary transformations in quantum mechanics:
An optical realization'',
in V. Bu\v{z}zek, \& D. P. DiVincenzo (eds.),
{\em J. Mod. Opt.} {\bf 47}, 2-3 (Special issue:
Physics of quantum information), 487-497 (2000).

\item {\bf [Bergou-Herzog-Hillery 03]}:
J. A. Bergou, U. Herzog, \& M. Hillery,
``Quantum filtering and discrimination between sets of boolean functions'',
{\em Phys. Rev. Lett.} {\bf 90}, 25, 257901 (2003).

\item {\bf [Berkley-Xu-Ramos-(+7) 03]}:
A. J. Berkley, H. Xu, R. C. Ramos,
M. A. Gubrud, F. W. Strauch, P. R. Johnson,
J. R. Anderson, A. J. Dragt, C. J. Lobb, \& F. C. Wellstood,
``Entangled macroscopic quantum states in two superconducting qubits'',
{\em Science} {\bf 300}, ?, 1548-? (2003).

\item {\bf [Berkley-Xu-Gubrud-(+4) 03]}:
A. J. Berkley, H. Xu, M. A. Gubrud,
R. C. Ramos, J. R. Anderson, C. J. Lobb, \& F. C. Wellstood,
``Decoherence in a Josephson-junction qubit'',
{\em Phys. Rev. B} {\bf 68}, 6, 060502 (2003).

\item {\bf [Berkovitz 98 a]}:
J. Berkovitz,
``Aspects of quantum nonlocality. I: Superluminal signalling,
action-at-a-distance, non-separability and holism'',
{\em Stud. Hist. Philos. Sci. Part B:
Stud. Hist. Philos. Mod. Phys.} {\bf 29}, 2, 183-222 (1998).
See {\bf [Berkovitz 98 b]} (II).

\item {\bf [Berkovitz 98 b]}:
J. Berkovitz,
``Aspects of quantum nonlocality. II: Superluminal
causation and relativity'',
{\em Stud. Hist. Philos. Sci. Part B:
Stud. Hist. Philos. Mod. Phys.} {\bf 29}, 4, 509-545 (1998).
See {\bf [Berkovitz 98 a]} (I).

\item {\bf [Berman-Doolen-Holm-Tsifrinovich 94]}:
G. P. Berman, G. D. Doolen, D. D. Holm, \& V. I. Tsifrinovich,
``Quantum computer on a class of one-dimensional Ising systems'',
{\em Phys. Lett. A} {\bf 193}, ?, 444-450 (1994).

\item {\bf [Berman-Doolen-Mainieri-Tsifrinovich 98]}:
G. P. Berman, G. D. Doolen, R. Mainieri, \& V. I. Tsifrinovich,
{\em Introduction to quantum computers},
World Scientific, Singapore, 1998.

\item {\bf [Berman-Doolen-Hammel-Tsifrinovich 00]}:
G. P. Berman, G. D. Doolen, P. C. Hammel, \& V. I. Tsifrinovich,
``Solid-state nuclear spin quantum computer based on
magnetic resonance force microscopy'',
quant-ph/9909033.

\item {\bf [Berman-Doolen-Tsifrinovich 00]}:
G. P. Berman, G. D. Doolen, \& V. I. Tsifrinovich,
``Influence of superpositional wave function oscillations
on Shor's quantum algorithm'',
{\em Phys. Rev. Lett.} {\bf 84}, 7, 1615-1618 (2000);
quant-ph/9906045.

\item {\bf [Berman-Doolen-L\'{o}pez-Tsifrinovich 00 a]}:
G. P. Berman, G. D. Doolen, G. V. L\'{o}pez, \& V. I. Tsifrinovich,
``Nonresonant effects in the implementation of the quantum Shor algorithm'',
{\em Phys. Rev. A} {\bf 61}, 4, 042307 (2000);
quant-ph/9909027.

\item {\bf [Berman-Doolen-L\'{o}pez-Tsifrinovich 00 b]}:
G. P. Berman, G. D. Doolen, G. V. L\'{o}pez, \& V. I. Tsifrinovich,
``Simulations of quantum-logic operations in a quantum computer with
a large number of qubits'',
{\em Phys. Rev. A} {\bf 61}, 6, 062305 (2000);
quant-ph/9909032.

\item {\bf [Berman-Doolen-Hammel-Tsifrinovich 01]}:
G. P. Berman, G. D. Doolen, P. C. Hammel, \& V. I. Tsifrinovich,
``Magnetic resonance force microscopy quantum computer
with tellurium donors in silicon'',
{\em Phys. Rev. Lett.} {\bf 86}, 13, 2894-2896 (2001).

\item {\bf [Berman-Brown-Hawley-Tsifrinovich 01]}:
G. P. Berman, G. W. Brown, M. E. Hawley, \& V. I. Tsifrinovich,
``Solid-state quantum computer based on scanning tunneling
microscopy'',
{\em Phys. Rev. Lett.} {\bf 87}, 9, 097902 (2001);
quant-ph/0103008.

\item {\bf [Berman-Borgonovi-Chapline-(+5) 01]}:
G. P. Berman, F. Borgonovi, G. Chapline,
S. A. Gurvitz, P. C. Hammel, D. V. Pelekhov,
A. Suter, \& V. I. Tsifrinovich,
``Formation and dynamics of a Schr\"{o}dinger-cat state
in continuous quantum measurement'',
quant-ph/0101035.

\item {\bf [Berman-Doolen-L\'{o}pez-Tsifrinovich 02]}:
G. P. Berman, G. D. Doolen, G. V. L\'{o}pez, \& V. I. Tsifrinovich,
``A quantum full adder for a scalable nuclear spin quantum computer'',
{\em Proc. of the Quantum Computation for Physical Modeling
Workshop 2000 (North Falmouth, Massachusetts)},
{\em Comput. Phys. Comm.} {\bf 146}, 3, 324-330 (2002);
quant-ph/0105133.

\item {\bf [Berman-Doolen-Kamenev-Tsifrinovich 02]}:
G. P. Berman, G. D. Doolen, D. I. Kamenev, \& V. I. Tsifrinovich,
``Perturbation theory for quantum computation with a large number of qubits'',
{\em Phys. Rev. A} {\bf 65}, 1, 012321 (2002).

\item {\bf [Berman-Borgonovi-Izrailev-Tsifrinovich 02]}:
G. P. Berman, F. Borgonovi, F. M. Izrailev, \& V. I. Tsifrinovich,
``Avoiding quantum chaos in quantum computation'',
{\em Phys. Rev. E} {\bf 65}, 1, 015204 (2002);
quant-ph/0012106.

\item {\bf [Berman-Doolen-Hammel-Tsifrinovich 02]}:
G. P. Berman, G. D. Doolen, P. C. Hammel, \& V. I. Tsifrinovich,
``Static Stern-Gerlach effect in magnetic force microscopy'',
{\em Phys. Rev. A} {\bf 65}, 3, 032311 (2002).

\item {\bf [Berman-Ezhov-Kamenev-Yepez 02]}:
G. P. Berman, A. A. Ezhov, D. I. Kamenev, \& J. Yepez,
``Simulation of the diffusion equation on a type-II quantum computer'',
{\em Phys. Rev. A} {\bf 66}, 1, 012310 (2002).

\item {\bf [Berman-Borgonovi-Chapline-(+2) 02]}:
G. P. Berman, F. Borgonovi, G. Chapline,
P. C. Hammel, \& V. I. Tsifrinovich,
``Magnetic-resonance force microscopy measurement of entangled spin states'',
{\em Phys. Rev. A} {\bf 66}, 3, 032106 (2002).

\item {\bf [Berman-Tsifrinovich-Allara-(+2) 02]}:
G. P. Berman, V. I. Tsifrinovich, \& D. L. Allara,
``Entangled spin states in self-assembled monolayer systems'',
{\em Phys. Rev. B} {\bf 66}, 19, 193406 (2002).

\item {\bf [Berman-L\'{o}pez, \& V. I. Tsifrinovich 02]}:
G. P. Berman, G. V. L\'{o}pez, \& V. I. Tsifrinovich,
``Teleportation in a nuclear spin quantum computer'',
{\em Phys. Rev. A} {\bf 66}, 4, 042312 (2002).

\item {\bf [Berman-Borgonovi-Celardo-(+2) 02]}:
G. P. Berman, F. Borgonovi, G. Celardo,
F. M. Izrailev, \& D. I. Kamenev,
``Dynamical fidelity of a solid-state quantum computation'',
{\em Phys. Rev. E} {\bf 66}, 5, 056206 (2002).

\item {\bf [Berman-Borgonovi-Goan-(+2) 03]}:
G. P. Berman, F. Borgonovi, H.-S. Goan,
S. A. Gurvitz, \& V. I. Tsifrinovich,
``Single-spin measurement and decoherence in magnetic-resonance force
microscopy'',
{\em Phys. Rev. B} {\bf 67}, 9, 094425 (2003).

\item {\bf [Berman-Borgonovi-Gorshkov-Tsifrinovich 04]}:
G. P. Berman, F. Borgonovi, V. N. Gorshkov, \& V. I. Tsifrinovich,
``Modeling and simulations of a single-spin measurement using MRFM'',
quant-ph/0410002.

\item {\bf [Bernab\'{e}u-Mavromatos-Papavassiliou 04]}:
J. Bernab\'{e}u, N. Mavromatos, \& J. Papavassiliou,
``Novel type of CPT violation for correlated Einstein-Podolsky-Rosen states of neutral mesons'',
{\em Phys. Rev. Lett.} {\bf 92}, 13, 131601 (2004).

\item {\bf [Berndl-Goldstein 94]}:
K. Berndl, \& S. Goldstein,
``Comment on `Quantum mechanics, local realistic theories,
and Lorentz-invariant realistic theories'\,'',
{\em Phys. Rev. Lett.} {\bf 72}, 5, 780 (1994).
Comment on {\bf [Hardy 92 a]}.

\item {\bf [Berndl-D\"{u}rr-Goldstein-Zangh\`{\i} 96]}:
K. Berndl, D. D\"{u}rr, S. Goldstein, \& N. Zangh\`{\i},
``Nonlocality, Lorentz invariance, and
Bohmian quantum theory'',
{\em Phys. Rev. A} {\bf 53}, 4, 2062-2073 (1996).

\item {\bf [Berndl-Teufel 97]}:
K. Berndl, \& S. Teufel,
``Hidden quantum
nonlocality revealed by local filters'',
{\em Phys. Lett. A} {\bf 224}, 4-5, 314-316 (1997).
Comment on {\bf [Gisin 96 a]}.
See {\bf [Gisin 97]}.

\item {\bf [Bernevig-Chen 03]}:
B. A. Bernevig, \& H.-D. Chen,
``Geometry of the three-qubit state, entanglement and division algebras'',
{\em J. Phys. A} {\bf 36}, 30, 8325-8339 (2003).

\item {\bf [Bernstein-Vazirani 93]}:
E. Bernstein, \& U. Vazirani,
``Quantum complexity theory'',
in {\em Proc.\ of the 25th Annual ACM Symp.\ on the Theory of Computing},
ACM Press, New York, 1993, pp.~11-20.

\item {\bf [Bernstein 89]}:
J. Bernstein,
{\em Quantum profiles}, Princeton
University Press, Princeton, New Jersey, 1989.
Spanish version: {\em Perfiles
cu\'{a}nticos. Un an\'{a}lisis de la f\'{\i}sica cu\'{a}ntica},
McGraw-Hill, Madrid, 1991.

\item {\bf [Bernstein 01]}:
J. Bernstein,
``The Genius of Science: A Portrait of 20th-Century
Physicists by Abraham Pais'',
{\em Am. J. Phys.} {\bf 69}, 6, 725-? (2001).

\item {\bf [Bernstein-Greenberger-Horne-Zeilinger 93]}:
H. J. Bernstein, D. M. Greenberger, M. A. Horne, \& A. Zeilinger,
``Bell theorem without inequalities for two spinless particles'',
{\em Phys. Rev. A} {\bf 47}, 1, 78-84 (1993).
See {\bf [Greenberger-Bernstein-Horne-Zeilinger 93]}.

\item {\bf [Bernstein 99]}:
H. J. Bernstein,
``Simple version of the Greenberger-Horne-Zeilinger (GHZ) argument
against local realism'',
{\em Found. Phys.} {\bf 29}, 4, 521-526 (1999).

\item {\bf [Berry-Wiseman 00]}:
D. W. Berry, \& H. M. Wiseman,
``Optimal states and almost optimal adaptive measurements for quantum
interferometry'',
{\em Phys. Rev. Lett.} {\bf 85}, ?, 5098-? (2000);
quant-ph/0009117.

\item {\bf [Berry-Sanders 01]}:
D. W. Berry, \& B. C. Sanders,
``Quantum teleportation and entanglement swapping for spin systems'',
quant-ph/0111079.

\item {\bf [Berry-Wiseman 02]}:
D. W. Berry, \& H. M. Wiseman,
``Adaptive quantum measurements of a continuously varying phase'',
{\em Phys. Rev. A} {\bf 65}, 4, 043803 (2002).

\item {\bf [Berry-Sanders 03 a]}:
D. W. Berry, \& B. C. Sanders,
``Optimal remote state preparation'',
{\em Phys. Rev. Lett.} {\bf 90}, 5, 057901 (2003).

\item {\bf [Berry-Sanders 03 b]}:
D. W. Berry, \& B. C. Sanders,
``Relations for classical communication capacity and entanglement capability
of two-qubit operations'',
{\em Phys. Rev. A} {\bf 67}, 4, 040302 (2003);
quant-ph/0207065.

\item {\bf [Berry-Sanders 03 c]}:
D. W. Berry, \& B. C. Sanders,
``Bounds on general entropy measures'',
{\em J. Phys. A} {\bf 36}, 49, 12255-12265 (2003).

\item {\bf [Berry-Sanders 03 c]}:
D. W. Berry, \& B. C. Sanders,
``Numerical analysis of the capacities for two-qubit unitary operations'',
quant-ph/0307058.

\item {\bf [Berry-Scheel-Sanders-Knight 04]}:
D. W. Berry, S. Scheel, B. C. Sanders, \& P. L. Knight,
``Improving single-photon sources via linear optics and photodetection'',
{\em Phys. Rev. A} {\bf 69}, 3, 031806 (2004);
quant-ph/0309140.

\item {\bf [Berry-Scheel-Myers-(+3) 04]}:
D. W. Berry, S. Scheel, C. R. Myers, B. C. Sanders,
P. L. Knight, \& R. Laflamme,
``Post-processing with linear optics for improving the quality of
single-photon sources'',
quant-ph/0402018.

\item {\bf [Berry 04]}:
D. W. Berry,
``Resources required for exact remote state preparation'',
quant-ph/0404004.
See {\bf [Ye-Zhang-Guo 04]}.

\item {\bf [Berry 96]}:
R. S. Berry,
``The picture book of quantum mechanics'',
{\em Phys. Today} {\bf 49}, 1, 65-66 (1996).
Review of {\bf [Brandt-Dahmen 94]}.

\item {\bf [Bertet-Osnaghi-Rauschenbeutel-(+4) 01]}:
P. Bertet, S. Osnaghi, A. Rauschenbeutel,
G. Nogues, A. Auffeves, M. Brune, J. M. Raimond, \& S. Haroche,
``A complementarity experiment with an interferometer
at the quantum-classical boundary'',
{\em Nature} {\bf 411}, 6834, 166-170 (2001),

\item {\bf [Berthiaume-Brassard 94]}:
A. Berthiaume, \& G. Brassard,
``Oracle quantum computing'',
in S. M. Barnett, A. K. Ekert, \& S. J. D. Phoenix (eds.),
{\em J. Mod. Opt.} {\bf 41}, 12 (Special issue: Quantum
communication), 2521-2536 (1994).

\item {\bf [Bertlmann 90]}:
R. A. Bertlmann,
``Bell's theorem and the nature of reality'',
{\em Found. Phys.} {\bf 20}, 10, 1191-1212 (1990).

\item {\bf [Bertlmann 93]}:
R. A. Bertlmann,
``Magic moments with John Bell'',
in A. van der Merwe, \& F. Selleri (eds.),
{\em Bell's theorem and the foundations of modern physics.
Proc.\ of an international
conference (Cesena, Italy, 1991)},
World Scientific, Singapore, 1993, pp.~31-44.

\item {\bf [Bertlmann-Hiesmayr 01]}:
R. A. Bertlmann, \& B. C. Hiesmayr,
``Bell inequalities for entangled kaons and their unitary time
evolution'',
{\em Phys. Rev. A} {\bf 63}, 6, 062112 (2001);
hep-ph/0101356.

\item {\bf [Bertlmann-Grimus-Hiesmayr 01]}:
R. A. Bertlmann, W. Grimus, \& B. C. Hiesmayr,
``Bell inequality and $CP$ violation in the neutral kaon system'',
{\em Phys. Lett. A} {\bf 289}, 1-2, 21-26 (2001);
quant-ph/0107022.

\item {\bf [Bertlmann-Zeilinger 02]}:
R. A. Bertlmann, \& A. Zeilinger (eds.),
{\em Quantum [un]speakables: From Bell to quantum information (Vienna, 2000)},
Springer-Verlag, Berlin, 2002.
Review: {\bf [Zanghi-Tumulka 03]}.

\item {\bf [Bertlmann 02]}:
R. A. Bertlmann,
``Magic moments: A collaboration with John Bell'',
in {\bf [Bertlmann-Zeilinger 02]}, pp.~29-50.

\item {\bf [Bertlmann-Grimus-Hiesmayr 02]}:
R. A. Bertlmann, W. Grimus, \& B. C. Hiesmayr,
``The EPR-paradox in massive systems or about strange particles'',
in {\bf [Bertlmann-Zeilinger 02]}, pp.~163-184;
quant-ph/0106166.

\item {\bf [Bertlmann-Narnhofer-Thirring 02]}:
R. A. Bertlmann, H. Narnhofer, \& W. Thirring,
``Geometric picture of entanglement and Bell inequalities'',
{\em Phys. Rev. A} {\bf 66}, 3, 032319 (2002);
quant-ph/0111116.

\item {\bf [Bertlmann-Durstberger-Hiesmayr 03]}:
R. A. Bertlmann, K. Durstberger, \& B. C. Hiesmayr,
``Decoherence of entangled kaons and its connection to entanglement measures'',
{\em Phys. Rev. A} {\bf 68}, 1, 012111 (2003);
quant-ph/0209017.

\item {\bf [Bertlmann-Durstberger-Hasegawa-Hiesmayr 04]}:
R. A. Bertlmann, K. Durstberger, Y. Hasegawa, \& B. C. Hiesmayr,
``Berry phase in entangled systems: A proposed experiment with single neutrons'',
{\em Phys. Rev. A} {\bf 69}, 3, 032112 (2004).

\item {\bf [Bertlmann-Bramon-Garbarino-Hiesmayr 04]}:
R. A. Bertlmann, A. Bramon, G. Garbarino, \& B. C. Hiesmayr,
``Violation of a Bell inequality in particle physics experimentally
verified?'',
{\em Phys. Lett. A} {\bf 332}, 5-6, 355-360 (2004);
quant-ph/0409051.

\item {\bf [Bertlmann 04]}:
R. A. Bertlmann,
``Entanglement, Bell inequalities and decoherence in particle physics'',
{\em Quantum Coherence in Matter: From Quarks to Solids (Schladming, Austria, 2004)},
Springer, Berlin, 2004;
quant-ph/0410028.

\item {\bf [Bertoni-Bordone-Brunetti-(+2) 00]}:
A. Bertoni, P. Bordone, R. Brunetti, C. Jacoboni, \& S. Reggiani,
``Quantum logic gates based on coherent electron transport in quantum
wires'',
{\em Phys. Rev. Lett.} {\bf 84}, 25, 5912-5915 (2000).

\item {\bf [Bes 04]}:
D. R. Bes,
{\em Quantum mechanics: A modern and concise introductory course},
Springer-Verlag, Berlin, 2004.

\item {\bf [Bester-Shumway-Zunger 04]}:
G. Bester, J. Shumway, \& A. Zunger,
``Theory of excitonic spectra and entanglement engineering in dot molecules'',
{\em Phys. Rev. Lett.} {\bf 93}, 4, 047401 (2004).

\item {\bf [Beth-Grassl 98]}:
T. Beth, \& M. Grassl,
``The quantum Hamming and hexacodes'',
{\em Fortschr. Phys.} {\bf 46}, 4-5, 459-491 (1998).

\item {\bf [Bethune-Navarro-Risk 01]}:
D. S. Bethune, M. Navarro, \& W. P. Risk,
``Enhanced autocompensating quantum cryptography system'',
submitted to {\em App. Opt.};
quant-ph/0104089.

\item {\bf [Bethune-Risk 02]}:
D. S. Bethune, \& W. P. Risk,
``Autocompensating quantum cryptography'',
{\em New J. Phys} {\bf 4}, 42.1-42.15 (2002);
quant-ph/0204144.

\item {\bf [Bethune-Risk-Pabst 03]}:
D. S. Bethune, W. P. Risk, \& G. W. Pabst,
``A high-performance integrated single-photon detector for telecom
wavelengths'',
quant-ph/0311112.

\item {\bf [Bettelli-Shepelyansky 03]}:
S. Bettelli, \& D. L. Shepelyansky,
``Entanglement versus relaxation and decoherence in a quantum algorithm for
quantum chaos'',
{\em Phys. Rev. A} {\bf 67}, 5, 054303 (2003);
quant-ph/0301086.

\item {\bf [Beveratos-K\"{u}hn-Brouri-(+3) 02]}:
A. Beveratos, S. K\"{u}hn, R. Brouri,
T. Gacoin, J.-P. Poizat, \& P. Grangier,
``Room temperature stable single-photon source'',
{\em Eur. Phys. J. D} {\bf 18}, 2 (Special issue:
{\em Quantum interference and cryptographic keys:
Novel physics and advancing technologies (QUICK) (Corsica, 2001)}, 191-196 (2002).

\item {\bf [Beveratos-Brouri-Gacoin-(+3) 02]}:
A. Beveratos, R. Brouri, T. Gacoin,
A. Villing, J.-P. Poizat, \& P. Grangier,
``Single photon quantum cryptography'',
{\em Phys. Rev. Lett.} {\bf 89}, 18, 187901 (2002).

\item {\bf [Bhattacharya-Habib-Jacobs 02]}:
T. Bhattacharya, S. Habib, \& K. Jacobs,
``The emergence of classical dynamics in a quantum world'',
{\em Los Alamos Science} {\bf 27}, 110-? (2002);
quant-ph/0407096.

\item {\bf [Bhagwat-Khandekar-Menon-(+2) 00]}:
K. V. Bhagwat, D. C. Khandekar, S. V. G. Menon,
R. R. Puri, \& D. C. Sahni,
``A critique on `Quantum no-deleting principle'\,'',
quant-ph/0005006.
Comment on {\bf [Pati-Braunstein 00 a]}.

\item {\bf [Bhattacharya-van Linden van den Heuvell-Spreeuw 02]}:
N. Bhattacharya, H. B. van Linden van den Heuvell, \& R. J. C. Spreeuw,
``Implementation of quantum search algorithm using classical Fourier optics'',
{\em Phys. Rev. Lett.} {\bf 88}, 13, 137901 (2002);
quant-ph/0110034.

\item {\bf [Bialynicki Birula-Mycielski 76]}:
I. Bialynicki-Birula, \& J. Mycielski,
``Nonlinear wave mechanics'',
{\em Ann. Phys.} {\bf 100}, 1-2, 62-93 (1976).

\item {\bf [Bianucci-Paz-Saraceno 01]}:
P. Bianucci, J. P. Paz, \& M. Saraceno,
``Decoherence for classically chaotic quantum maps'',
quant-ph/0110033.

\item {\bf [Bianucci-Miquel-Paz-Saraceno 02]}:
P. Bianucci, C. Miquel, J. P. Paz, \& M. Saraceno,
``Discrete Wigner functions and the phase space representation of
quantum computers'',
{\em Phys. Lett. A} {\bf 297}, 5-6, 353-358 (2002);
quant-ph/0106091.

\item {\bf [Bianucci-Muller-Shih-(+3) 04]}:
P. Bianucci, A. Muller, C. K. Shih,
Q. Q. Wang, Q. K. Xue, \& C. Piermarocchi,
``Experimental realization of the one qubit Deutsch-Jozsa algorithm in a quantum dot'',
{\em Phys. Rev. B} {\bf 69}, 16, 161303 (2004);
cond-mat/0401226.

\item {\bf [Biedenharn-van Dam 65]}:
L. C. Biedenharn, \& H. van Dam (eds.),
{\em Quantum theory of angular momentum},
Academic Press, New York, 1965.

\item {\bf [Bienert-Freyberger 01]}:
M. Bienert, \& M. Freyberger,
``Coherence via decoherence'',
{\em Europhys. Lett.} {\bf 56}, 5, 619-625 (2001).

\item {\bf [Bienfang-Gross-Mink-(+9) 04]}:
J. C. Bienfang, A. J. Gross, A. Mink,
B. J. Hershman, A. Nakassis, X. Tang,
R. Lu, D. H. Su, C. W. Clark,
C. J. Williams, E. W. Hagley, \& J. Wen,
``Quantum key distribution with $1.25$ Gbps clock synchronization'',
{\em Opt. Express} {\bf 12}, 2011-2015 (2004);
quant-ph/0405097.
Comment: {\bf [Hughes-Nordholt 04]}.
Reply: {\bf [Bienfang-Clark-Williams-(+2) 04]}.

\item {\bf [Bienfang-Clark-Williams-(+2) 04]}:
J. C. Bienfang, C. W. Clark, C. J. Williams,
E. W. Hagley, \& J. Wen,
``Advantages of high-speed technique for quantum key distribution; reply
to quant-ph/0407050'',
quant-ph/0407139.
Reply to {\bf [Hughes-Nordholt 04]}.
See {\bf [Bienfang-Gross-Mink-(+9) 04]}.

\item {\bf [Bigaj 03]}:
T. Bigaj,
``Counterfactuals and non-locality of quantum mechanics'' (2003);
PITT-PHIL-SCI00000961.

\item {\bf [Bigelow 01]}:
N. Bigelow,
``Quantum engineering: Squeezing entanglement'',
{\em Nature} {\bf 409}, 6816, 27-28 (2001).
See {\bf [S{\o}rensen-Duan-Cirac-Zoller 01]}.

\item {\bf [Bignami 00]}:
G. F. Bignami,
``The genius of science.
A portrait gallery of 20th century physicists'',
{\em Nature} {\bf 404}, 6781, 927-928 (2000).
Review of {\bf [Pais 00]}.

\item {\bf [Biham-Huttner-Mor 96]}:
E. Biham, B. Huttner, \& T. Mor,
``Quantum cryptographic network based on quantum memories'',
{\em Phys. Rev. A} {\bf 54}, 4, 2651-2658 (1996).

\item {\bf [Biham-Mor 97 a]}:
E. Biham, \& T. Mor,
``Security of quantum cryptography against collective attacks'',
{\em Phys. Rev. Lett.} {\bf 78}, 11, 2256-2259 (1997);
quant-ph/9605007.

\item {\bf [Biham-Mor 97 b]}:
E. Biham, \& T. Mor,
``Bounds on information and the security of quantum cryptography'',
{\em Phys. Rev. Lett.} {\bf 79}, 20, 4034-4037 (1997);
quant-ph/9605010.

\item {\bf [Biham-Boyer-Brassard-(+2) 98]}:
E. Biham, M. Boyer, G. Brassard, J. van de Graaf, \& T. Mor,
``Security of quantum key distribution against all collective attacks'',
quant-ph/9801022.

\item {\bf [Biham-Biham-Biron-(+2) 99]}:
E. Biham, O. Biham, D. Biron, M. Grassl, \& D. A. Lidar,
``Grover's quantum search algorithm for an arbitrary initial amplitude distribution'',
{\em Phys. Rev. A} {\bf 60}, 4, 2742-2745 (1999).

\item {\bf [Biham-Boyer-Boykin-(+2) 00]}:
E. Biham, M. Boyer, P. O. Boykin, T. Mor, \& V. Roychowdhury,
``A proof of the security of quantum key distribution'',
in {\em Proc.\ of the 32nd Annual ACM Symp.\ on Theory
of Computing}, ACM Press, New York, 2000, pp.~715-724;
quant-ph/9912053.

\item {\bf [Biham-Biham-Biron-(+3) 01]}:
E. Biham, O. Biham, D. Biron,
M. Grassl, D. A. Lidar, \& D. Shapira,
``Analysis of generalized Grover quantum search algorithms using recursion
equations'',
{\em Phys. Rev. A} {\bf 63}, 1, 012310 (2001);
quant-ph/0010077.

\item {\bf [Biham-Kenigsberg 02]}:
E. Biham, \& D. Kenigsberg,
``Grover's quantum search algorithm for an arbitrary initial mixed state'',
{\em Phys. Rev. A} {\bf 66}, 6, 062301 (2002);
quant-ph/0306183.

\item {\bf [Biham-Brassard-Kenigsberg-Mor 03]}:
E. Biham, G. Brassard, D. Kenigsberg, \& T. Mor,
``Quantum computing without entanglement'',
quant-ph/0306182.

\item {\bf [Biham-Nielsen-Osborne 02]}:
O. Biham, M. A. Nielsen, \& T. J. Osborne,
``Entanglement monotone derived from Grover's algorithm'',
{\em Phys. Rev. A} {\bf 65}, 6, 062312 (2002);
quant-ph/0112097.

\item {\bf [Biham-Shapira-Shimoni03]}:
O. Biham, D. Shapira, \& Y. Shimoni,
``Analysis of Grover's quantum search algorithm as a dynamical system'',
{\em Phys. Rev. A} {\bf 68}, 2, 022326 (2003).

\item {\bf [Bihary-Glenn-Lidar-Apkarian 01]}:
Z. Bihary, D. R. Glenn, D. A. Lidar, \& V. A. Apkarian,
``An implementation of the Deutsch-Jozsa algorithm on molecular
vibronic coherences through four-wave mixing: A theoretical study'',
quant-ph/0110041.

\item {\bf [Bilodeau 98]}:
D. Bilodeau,
``Why quantum mechanics is hard to understand'',
quant-ph/9812050.

\item {\bf [Biolatti-Iotti-Zanardi-Rossi 00]}:
E. Biolatti, R. C. Iotti, P. Zanardi, \& F. Rossi,
``Quantum-information processing with semiconductor macroatoms'',
{\em Phys. Rev. Lett.} {\bf 85}, 26, 5647-5650 (2000);
quant-ph/0007034.

\item {\bf [Bimonte-Musto 03 a]}:
G. Bimonte, \& R. Musto,
`Comment on ``Quantitative wave-particle duality in multibeam
interferometers''\,',
{\em Phys. Rev. A} {\bf 67}, 6, 066101 (2003).
Comment on {\bf [D\"{u}rr 01]}.

\item {\bf [Bimonte-Musto 03 b]}:
G. Bimonte, \& R. Musto,
``On interferometric duality in multibeam experiments'',
{\em J. Phys. A} {\bf 36}, 45, 11481-11502 (2003).

\item {\bf [Birkhoff-von Neumann 36]}:
G. Birkhoff, \& J. von Neumann,
``The logic of quantum mechanics'',
{\em Ann. Math.} {\bf 37}, ?, 823-843 (1936).
Reprinted in {\bf [von Neumann 61]}, vol. 4, pp.~105-125;
and in {\bf [Hooker 75]}, pp.~1-26.

\item {\bf [Biron-Biham-Biham-(+2) 98]}:
D. Biron, O. Biham, E.
Biham, M. Grassl, \& D. A. Lidar,
``Generalized Grover search algorithm for arbitrary initial amplitude distribution'',
quant-ph/9801066.

\item {\bf [Biswas-Choudhury-Datta 91]}:
S. N. Biswas, S. R. Choudhury, \& K. Datta,
``Einstein-Podolsky-Rosen paradox, statistical locality and the time-energy
uncertainty relation'',
{\em Phys. Lett. A} {\bf 153}, 6-7, 279-284 (1991).

\item {\bf [Biswas-Agarwal 03 a]}:
A. Biswas, \& G. S. Agarwal,
``Strong subadditivity inequality for quantum
entropies and four-particle entanglement'',
{\em Phys. Rev. A} {\bf 68}, 5, 054303 (2003).

\item {\bf [Biswas-Agarwal 03 b]}:
A. Biswas, \& G. S. Agarwal,
``Preparation of $W$, GHZ, and two-qutrit states using bimodal cavities'',
{\em J. Mod. Opt.};
quant-ph/0311137.

\item {\bf [Biswas-Agarwal 04]}:
A. Biswas, \& G. S. Agarwal,
``Quantum logic gates using Stark-shifted Raman transitions in a cavity'',
{\em Phys. Rev. A} {\bf 69}, 6, 062306 (2004).

\item {\bf [Bitbol 83]}:
M. Bitbol,
``An analysis of the Einstein-Podolsky-Rosen correlations in terms of events'',
{\em Phys. Lett. A} {\bf 96}, 2, 66-70 (1983).
See {\bf [Page 82]}.

\item {\bf [Bitbol 96]}:
M. Bitbol,
{\em M\'{e}canique quantique: Une introduction philosophique},
Flammarion, Paris, 1996, 1998 (paperback).
Review: {\bf [Mugur Sch\"{a}chter 98]}.

\item {\bf [Bitton-Grice-Moreau-Zhang 02]}:
G. Bitton, W. P. Grice, J. Moreau, \& L. Zhang,
``Cascaded ultrabright source of polarization-entangled photons'',
{\em Phys. Rev. A} {\bf 65}, 6, 063805 (2002);
quant-ph/0106122.

\item {\bf [Bj{\o}rk-Karlsson 98]}:
G. Bj{\o}rk, \& A. Karlsson,
``Complementarity and quantum erasure in welcher Weg experiments'',
{\em Phys. Rev. A} {\bf 58}, 5, 3477-3483 (1998).

\item {\bf [Bj{\o}rk-S{\o}derholm-Trifonov-(+2) 99]}:
G. Bj{\o}rk, J. S{\o}derholm, A. Trifonov, T. Tsegaye, \& A. Karlsson
``Complementarity and the uncertainty relations'',
{\em Phys. Rev. A} {\bf 60}, 3, 1874-1882 (1999).

\item {\bf [Bj{\o}rk-Jonsson-S\'{a}nchez Soto 01]}:
G. Bj{\o}rk, P. Jonsson, \& L. L. S\'{a}nchez-Soto,
``Single-particle nonlocality and entanglement with the vacuum'',
{\em Phys. Rev. A} {\bf 64}, 4, 042106 (2001);
quant-ph/0103074.

\item {\bf [Bj{\o}rk-S\'{a}nchez Soto-S{\o}derholm 01]}:
G. Bj{\o}rk, L. L. S\'{a}nchez-Soto, \& J. S{\o}derholm,
``Entangled-state lithography: Tailoring any pattern with a single state'',
{\em Phys. Rev. Lett.} {\bf 86}, 20, 4516-4519 (2001).

\item {\bf [Bj{\o}rk-Jonsson-Heydari-(+2) 03]}:
G. Bj{\o}rk, P. Jonsson, H. Heydari,
J. S{\o}derholm, \& B. Hessmo,
``Hilbert space factorization and partial measurements'',
{\em Opt. Spectrosc.} {\bf 94}, 695 (2003).

\item {\bf [Bj{\o}rk-Mana 03]}:
G. Bj{\o}rk, \& P. G. L. Mana,
``A size criterion for macroscopic superposition states'',
quant-ph/0310193.

\item {\bf [Bl\"{a}si-Hardy 95]}:
B. Bl\"{a}si, \& L. Hardy,
``Realism and time symmetry in quantum mechanics'',
{\em Phys. Lett. A} {\bf 207}, 3-4, 119-125 (1995);
quant-ph/9505017.

\item {\bf [Blais-Zagoskin 00]}:
A. Blais, \& A. M. Zagoskin,
``Operation of universal gates in a solid-state quantum computer
based on clean Josephson junctions between d-wave superconductors'',
{\em Phys. Rev. A} {\bf 61}, 4, 042308 (2000).

\item {\bf [Blais 01]}:
A. Blais,
``Quantum network optimization'',
{\em Phys. Rev. A} {\bf 64}, 2, 022312 (2001).

\item {\bf [Blais-Tremblay 03]}:
A. Blais, \& A.-M. S. Tremblay,
``Effect of noise on geometric logic gates for quantum computation'',
{\em Phys. Rev. A} {\bf 67}, 1, 012308 (2003);
quant-ph/0105006.

\item {\bf [Blais-van den Brink-Zagoskin 03]}:
A. Blais, A. M. van den Brink, \& A. M. Zagoskin,
``Tunable coupling of superconducting qubits'',
{\em Phys. Rev. Lett.} {\bf 90}, 12, 127901 (2003).

\item {\bf [Blais-Huang-Wallraff-(+2) 04]}:
A. Blais, R.-S. Huang, A. Wallraff,
S. M. Girvin, \& R. J. Schoelkopf,
``Cavity quantum electrodynamics for superconducting electrical circuits:
An architecture for quantum computation'',
{\em Phys. Rev. A} {\bf 69}, 6, 062320 (2004).
Publisher's note: {\em Phys. Rev. A} {\bf 70}, 1, 019901 (2004).

\item {\bf [Blanchard-Olkiewicz 00]}:
P. Blanchard, \& R. Olkiewicz,
``Effectively classical quantum states for open systems'',
{\em Phys. Lett. A} {\bf 273}, 4, 223-231 (2000).

\item {\bf [Blanchard-Jak\'{o}bczyk-Olkiewicz 01 a]}:
P. Blanchard, L. Jak\'{o}bczyk, \& R. Olkiewicz,
``Entangled versus classical quantum states'',
{\em Phys. Lett. A} {\bf 280}, 1-2, 7-16 (2001).

\item {\bf [Blanchard-Jak\'{o}bczyk-Olkiewicz 01 b]}:
P. Blanchard, L. Jak\'{o}bczyk, \& R. Olkiewicz,
``Measures of entanglement based on decoherence'',
{\em J. Phys. A} {\bf 34}, 41, 8501-8516 (2001).

\item {\bf [Blanchard-{\L}ugiewicz-Olkiewicz 03]}:
P. Blanchard, P. {\L}ugiewicz, \& R. Olkiewicz,
``From quantum to quantum via decoherence'',
{\em Phys. Lett. A} {\bf 314}, 1-2, 29-36 (2003).

\item {\bf [Blasi-Pascazio-Takagi 98]}:
R. Blasi, S. Pascazio, \& S. Takagi,
``Particle tracks and the mechanism of decoherence
in a model bubble chamber'',
{\em Phys. Lett. A} {\bf 250}, 4-6, 230-240 (1998).

\item {\bf [Blatt-Gill-Thompson 92]}:
R. Blatt, P. Gill, \& R. C. Thompson,
``Current perspectives on the physics of trapped ions'',
{\em J. Mod. Opt.} {\bf 39}, 2, 193-220 (1992).

\item {\bf [Blatt 00]}:
R. Blatt,
``Quantum engineering: Push-button entanglement'',
{\em Nature} {\bf 404}, 6775, 231-232 (2000).
See {\bf [Sackett-Kielpinski-King-(+8) 00]}.

\item {\bf [Blatt-Lange 00]}:
R. Blatt, \& W. Lange,
``Quantum computation with ion traps'',
{\bf [Macchiavello-Palma-Zeilinger 00]}, pp.~313-319.

\item {\bf [Blatt 01]}:
R. Blatt,
``Delicate information'',
{\em Nature} {\bf 412}, 6849, 773 (2001).

\item {\bf [Blaut-Kowalski Glikman 98]}:
A. Blaut, \& J. Kowalski-Glikman,
``The time evolution of quantum universes
in the quantum potential picture'',
{\em Phys. Lett. A} {\bf 245}, 3-4, 197-202 (1998).

\item {\bf [Blencowe-Vitelli 00]}:
M. P. Blencowe, \& V. Vitelli,
``Universal quantum limits on single-channel information,
entropy, and heat flow'',
{\em Phys. Rev. A} {\bf 62}, 5, 052104 (2000);
quant-ph/0001007.

\item {\bf [Blinov-Moehring-Duan-Monroe 04]}:
B. B. Blinov, D. L. Moehring, L.-M. Duan, \& C. Monroe,
``Observation of entanglement between a single trapped atom and a single photon'',
{\em Nature} {\bf 428}, ?, 153-157 (2004).
See {\bf [Moehring-Madsen-Blinov-Monroe 04]}.

\item {\bf [Bloch 67]}:
I. Bloch,
``Some relativistic oddities in the quantum theory of observation'',
{\em Phys. Rev.} {\bf 156}, 5, 1377-1384 (1967).

\item {\bf [Blow-Phoenix 93]}:
K. J. Blow, \& S. J. D. Phoenix,
``On a fundamental theorem of quantum cryptography'',
{\em J. Mod. Opt.} {\bf 40}, 1, 33-36 (1993).

\item {\bf [Blow 94]}:
K. J. Blow,
``System and method for quantum cryptography'',
patent EP717897B1, 1994.
See {\bf [Blow 98]}.

\item {\bf [Blow 98]}:
K. J. Blow,
``System and method for quantum cryptography'',
patent US5757912, 1998.
See {\bf [Blow 94]}.

\item {\bf [Blume Kohout-Zurek 03]}:
R. Blume-Kohout, \& W. H. Zurek,
`Decoherence from a chaotic environment: An upside-down ``oscillator'' as a
model',
{\em Phys. Rev. A} {\bf 68}, 3, 032104 (2003);
quant-ph/0212153.

\item {\bf [Blume Kohout-Zurek 04]}:
R. Blume-Kohout, \& W. H. Zurek,
`A simple example of ``Quantum darwinism'': Redundant information storage
in many-spin environments'',
{\em Found. Phys.} (Festschrift in honor of Asher Peres);
quant-ph/0408147.

\item {\bf [Blute-Ivanov-Panangaden 01]}:
R. F. Blute, I. T. Ivanov, \& P. Panangaden,
``Decoherent histories on graphs'',
gr-qc/0111020.

\item {\bf [Bocko-Onofrio 96]}:
M. F. Bocko, \& R. Onofrio,
``On the measurement of a weak classical force coupled to a harmonic
oscillator: Experimental progress'',
{\em Rev. Mod. Phys.} {\bf 68}, 3, 755-799 (1996).

\item {\bf [Bog\'{a}r-Bergou 96]}:
P. Bog\'{a}r, \& J. A. Bergou,
``Entanglement
of atomic beams: Tests of complementarity and other applications'',
{\em Phys. Rev. A} {\bf 53}, 1, 49-52 (1996).

\item {\bf [Bogdanski-Bj\"{o}rk-Karlsson 04]}:
J. Bogdanski, G. Bj\"{o}rk, \& A. Karlsson,
``Quantum and classical correlated imaging'',
quant-ph/0407127.

\item {\bf [Boghosian-Taylor 98 a]}:
B. M. Boghosian, \& W. Taylor,
``Simulating quantum mechanics on a quantum computer'',
{\em Physica D} {\bf 120}, 30-42 (1998).

\item {\bf [Boghosian-Taylor 98 b]}:
B. M. Boghosian, \& W. Taylor,
``Quantum lattice-gas model for the many-particle Schr\"{o}dinger equation in $d$ dimensions'',
{\em Phys. Rev. E} {\bf 57}, 1, 54–66 (1998).

\item {\bf [Bohm 79]}:
A. Bohm,
{\em Quantum mechanics. Foundations and applications},
Springer-Verlag, New York, 1979, 1986 (2nd edition), 1993 (3rd edition).

\item {\bf [Bohm 99]}:
A. Bohm,
``Time-asymmetric quantum physics'',
{\em Phys. Rev. A} {\bf 60}, 2, 861-876 (1999).

\item {\bf [Bohm 51]}:
D. Bohm,
{\em Quantum theory},
Prentice-Hall, Englewood Cliffs, New Jersey, 1951;
Constable, London, 1954; Dover, New York, 1989.

\item {\bf [Bohm 52]}:
D. Bohm,
`A suggested interpretation of the quantum theory
in terms of ``hidden'' variables. I \& II',
{\em Phys. Rev.} {\bf 85}, 2, 166-193 (1952).
Reprinted in {\bf [Wheeler-Zurek 83]}, pp.~369-396;
{\bf [Stroke 95]} (I), pp.~1226-1239.

\item {\bf [Bohm 53]}:
D. Bohm,
``Comments on an article of Takabayasi concerning
the formulation of quantum mechanics with classical pictures'',
{\em Prog. Theor. Phys.} {\bf 9}, ?, 273-287 (1953).

\item {\bf [Bohm-Schiller-Tiomno 53]}:
D. Bohm, R. Schiller, \& J. Tiomno,
``A classical interpretation of the Pauli equation'',
{\em Nuovo Cimento Suppl.} (Ser. 10) {\bf 1}, 1, 48-66, 67-91 (1953).

\item {\bf [Bohm-Vigier 54]}:
D. Bohm, \& J.-P. Vigier,
``Model of the causal
interpretation in terms of a fluid with irregular fluctuations'',
{\em Phys. Rev.} {\bf 96}, 1, 208-216 (1954).

\item {\bf [Bohm-Aharonov 57]}:
D. Bohm, \& Y. Aharonov,
``Discussion of
experimental proof for the paradox of Einstein, Rosen, and Podolsky'',
{\em Phys. Rev.} {\bf 108}, 4, 1070-1076 (1957).

\item {\bf [Bohm-Aharonov 60]}:
D. Bohm, \& Y. Aharonov,
``Further discussion of possible experimental tests for the paradox of
Einstein, Podolsky and Rosen'',
{\em Nuovo Cimento} {\bf 17}, 6, 964-976 (1960).
See {\bf [Peres-Singer 60]}.

\item {\bf [Bohm-Bub 66 a]}:
D. Bohm, \& J. Bub,
``A proposed solution of the
measurement problem in quantum mechanics by a hidden variable theory'',
{\em Rev. Mod. Phys.} {\bf 38}, 3, 453-469 (1966).

\item {\bf [Bohm-Bub 66 b]}:
D. Bohm, \& J. Bub,
``A refutation of the proof by
Jauch and Piron that hidden variables can be excluded in quantum mechanics'',
{\em Rev. Mod. Phys.} {\bf 38}, 3, 470-475 (1966).

\item {\bf [Bohm 71 a]}:
D. Bohm,
``On Bohr's view concerning the quantum theory'',
in E. Bastin (ed.),
{\em Quantum theory and beyond},
Cambridge University Press, Cambridge, 1971, pp.~33-40.

\item {\bf [Bohm 71 b]}:
D. Bohm,
``On the role of hidden variables in the fundamental structure of physics'',
in E. Bastin (ed.),
{\em Quantum theory and beyond},
Cambridge University Press, Cambridge, 1971, pp.~95-116.

\item {\bf [Bohm-Hiley 81]}:
D. Bohm, \& B. J. Hiley,
``Nonlocality in quantum theory understood in terms of Einstein's nonlinear
field approach'',
{\em Found. Phys.} {\bf 11}, 7-8, 529-546 (1981).

\item {\bf [Bohm-Hiley 82]}:
D. Bohm, \& B. J. Hiley,
``The de Broglie pilot wave theory and the further development
of new insights arising out of it'',
{\em Found. Phys.} {\bf 12}, 10, 1001-1016 (1982).
Reprinted in {\bf [Barut-van der Merwe-Vigier 84]}, pp.~77-92.

\item {\bf [Bohm-Hiley 85]}:
D. Bohm, \& B. J. Hiley,
``Unbroken quantum realism, from microscopic to macroscopic levels'',
{\em Phys. Rev. Lett.} {\bf 55}, 23, 2511-2514 (1985).

\item {\bf [Bohm 86]}:
D. Bohm,
``Interview with D. Home'',
{\em Sci. Today} {\bf ?}, 11, 25-? (1986).

\item {\bf [Bohm 87]}:
D. Bohm,
``Hidden variables and the implicate order'',
in {\bf [Hiley-Peat 87]}, pp.~33-45.

\item {\bf [Bohm-Hiley-Kaloyerou 87]}:
D. Bohm, B. J. Hiley, \& P. Kaloyerou,
``An ontological basis for the quantum theory'',
{\em Phys. Rep.} {\bf 144}, 6, 321-375 (1987).

\item {\bf [Bohm-Hiley 89]}:
D. Bohm, \& B. J. Hiley,
``Nonlocality and locality in
the stochastic interpretation of quantum mechanics'',
{\em Phys. Rep.} {\bf 172}, ?, 94-122 (1989).

\item {\bf [Bohm-Hiley 93]}:
D. Bohm, \& B. J. Hiley,
{\em The undivided universe. An ontological interpretation of quantum theory},
Routledge, London, 1993.
Reviews: {\bf [Goldstein 94 b]}, {\bf [Greenberger 94 a]}, {\bf [Stapp 94 c]},
{\bf [Cushing 95 b]}, {\bf [Dickson 96 c]}.

\item {\bf [B\"{o}hm-B\"{o}hm-Aspelmeyer-(+2) 04]}:
H. R. B\"{o}hm, P. S. B\"{o}hm, M. Aspelmeyer,
\v{C}. Brukner, \& A. Zeilinger,
``Exploiting the randomness of the measurement basis in quantum
cryptography: Secure quantum key growing without privacy amplification'',
quant-ph/0408179.

\item {\bf [Bohn 91]}:
J. Bohn,
``Observable characteristics of pure quantum states'',
{\em Phys. Rev. Lett.} {\bf 66}, 12, 1547-1550 (1991).
Erratum: {\em Phys. Rev. Lett.} {\bf 67}, 7, 932 (1991).

\item {\bf [Bohr-Ulfbeck 95]}:
A. Bohr, \& O. Ulfbeck,
``Primary manifestation of symmetry. Origin of quantal indeterminacy'',
{\em Rev. Mod. Phys.} {\bf 67}, 1, 1-35 (1995).

\item {\bf [Bohr-Mottelson-Ulfbeck 04]}:
A. Bohr, B. R. Mottelson, \& O. Ulfbeck,
``The principle underlying quantum mechanics'',
{\em Found. Phys.} {\bf 34}, 3, 405-417 (2004).
See {\bf [Ulfbeck-Bohr 01]}.

\item {\bf [Bohr 28]}:
N. H. D. Bohr,
``The quantum postulate and the recent development of atomic theory'',
{\em Nature} {\bf 121}, 3050, 580-590 (1928).
Reprinted in {\bf [Bohr 34]}, pp.~52-91 (pp.~97-132 in the Spanish version).
Reprinted in {\bf [Wheeler-Zurek 83]}, pp.~87-126.

\item {\bf [Bohr 34]}:
N. H. D. Bohr,
{\em Atomic theory and the description of nature},
Cambridge University Press, Cambridge, 1934.
Ox Bow Press, Woodbridge, Connecticut, 1987.
Spanish version:
{\em La teor\'{\i}a at\'{o}mica y la descripci\'{o}n de la naturaleza},
Alianza, Madrid, 1988.
Review: {\bf [Mermin 89 b]}.
See {\bf [Ferrero 88]}.

\item {\bf [Bohr 35 a]}:
N. H. D. Bohr,
``Quantum mechanics and physical reality'',
{\em Nature} {\bf 136}, ?, 65 (1935).

\item {\bf [Bohr 35 b]}:
N. H. D. Bohr,
``Can quantum-mechanical description of physical
reality be considered complete?'',
{\em Phys. Rev.} {\bf 48}, 8, 696-702 (1935).
Reprinted in {\bf [Wheeler-Zurek 83]}, pp.~145-151
(pages 148 and 149 are out of order);
{\bf [Stroke 95]}, pp.~1219-1225;
{\bf [Bohr 98]}, pp.~73-82.

\item {\bf [Bohr 39]}:
N. H. D. Bohr,
``The causality problem in atomic physics'',
in {\em New Theories in Physics (Warsaw, 1938)},
Int. Institute of Intellectual Cooperation, Paris, 1939.
Reprinted in {\bf [Bohr 98]}, pp.~94-121.

\item {\bf [Bohr 48]}:
N. H. D. Bohr,
``On the notions of causality and complementarity'',
{\em Dialectica} {\bf 2}, ?, 312-319 (1948).
Reprinted in {\bf [Bohr 98]}, pp.~141-148.

\item {\bf [Bohr 49]}:
N. H. D. Bohr,
``Discussion with Einstein on epistemological
problems in atomic physics'',
in P. A. Schilpp (ed.),
{\em Albert Einstein: Philosopher-scientist},
Library of Living Philosophers, Open Court, La Salle,
Illinois, 1949, vol. 1, pp.~201-241.
Also Evanston, Illinois, 1949;
Tudor, New York, 1949;
Harper and Row, New York, 1959.
Reprinted in {\bf [Bohr 58 a]}, pp.~32-66;
{\bf [Wheeler-Zurek 83]}, pp.~9-49.
Review: {\bf [Zeilinger 99 b]}.

\item {\bf [Bohr 58 a]}:
N. H. D. Bohr,
{\em Essays 1932-1957 on atomic physics and human knowledge},
Wiley, New York, 1958;
Ox Bow Press, Woodbridge, Connecticut, 1987.
Spanish version:
{\em F\'{\i}sica at\'{o}mica y conocimiento humano},
Aguilar, Madrid, 1964.
Catalan version:
{\em F\'{\i}sica at\'{o}mica i coneixement huma},
Ediciones 62, Barcelona, 1967.
Review: {\bf [Mermin 89 b]}.

\item {\bf [Bohr 58 b]}:
N. H. D. Bohr,
``Quantum physics and philosophy---causality and complementarity'',
in R. Klibansky (ed.),
{\em Philosophy in mid-century},
La Nuova Italia, Firenze, 1958.
Reprinted in {\bf [Bohr 63]}, pp.~1-7.

\item {\bf [Bohr 63]}:
N. H. D. Bohr,
{\em Essays 1958-1962 on atomic physics and human knowledge},
Interscience, New York, 1963;
Vintage, New York, 1966;
Ox Bow Press, Woodbridge, Connecticut, 1987.
Spanish version:
{\em Nuevos ensayos sobre f\'{\i}sica at\'{o}mica y conocimiento humano},
Aguilar, Madrid, 1970.
Review: {\bf [Mermin 89 b]}.

\item {\bf [Bohr 84]}:
N. H. D. Bohr
(edited by E. R\"{u}dinger, \& K. Stolzenburg),
{\em Niels Bohr collected works (vol. 5): The
emergence of quantum mechanics (mainly 1924-1926)},
North-Holland, Amsterdam, 1984.

\item {\bf [Bohr 85]}:
N. H. D. Bohr
(edited by E. R\"{u}dinger, \& J. Kalckar),
{\em Niels Bohr collected works (vol. 6):
Foundations of quantum physics I (1926-1932)},
North-Holland, Amsterdam, 1985.

\item {\bf [Bohr 96]}:
N. H. D. Bohr
(edited by E. R\"{u}dinger, F. Aaserud, \& J. Kalckar),
{\em Niels Bohr collected works (vol. 7):
Foundations of quantum physics II (1933-1958)},
North-Holland, Amsterdam, 1996.

\item {\bf [Bohr 98]}:
N. H. D. Bohr
(edited by J. Faye, \& H. J. Folse),
{\em The philosophical writings of
Niels Bohr. Volume IV. Causality and complementarity},
Ox Bow Press, Woodbridge, Connecticut, 1998.
See {\bf [Bohr 34]} (I),
{\bf [Bohr 58 a]} (II),
{\bf [Bohr 63]} (III).

\item {\bf [Bohr 99]}:
N. H. D. Bohr
(edited by F. Aaserud, \& D. Favrholdt),
{\em Niels Bohr collected works (vol. 10):
Complementarity beyond physics (1928-1962)},
North-Holland, Amsterdam, 1999.

\item {\bf [Boileau-Gottesman-Laflamme-(+2) 04]}:
J.-C. Boileau, D. Gottesman, R. Laflamme,
D. Poulin, \& R. W. Spekkens,
``Robust polarization-based quantum key distribution over a collective-noise
channel'',
{\em Phys. Rev. Lett.} {\bf 92}, 1, 017901 (2004);
quant-ph/0306199.

\item {\bf [Boileau-Laflamme-Laforest-Myers 04]}:
J.-C. Boileau, R. Laflamme, M. Laforest, \& C. R. Myers,
``Robust quantum communication using a polarization-entangled photon pair'',
quant-ph/0406118.

\item {\bf [Boileau-Tamaki-Batuwantudawe-Laflamme 04]}:
J.-C. Boileau, K. Tamaki, J. Batuwantudawe, \& R. Laflamme,
``Unconditional security of three state quantum key distribution protocols'',
quant-ph/0408085.

\item {\bf [Bollinger-Itano-Heinzen-Wineland 89]}:
J. J. Bollinger, W. M. Itano, D. J. Heinzen, \& D. J. Wineland,
``?'',
{\em Science} {\bf 243}, 4893, 888-? (1989) (?).
See {\bf [Itano-Heinzen-Bollinger-Wineland 90]}.

\item {\bf [Bollinger-Heinzen-Itano-(+2) 89]}:
J. J. Bollinger, D. J. Heinzen, W. M. Itano,
S. L. Gilbert, \& D. J. Wineland,
``Test of the linearity of quantum mechanics by rf
spectroscopy of the $^9$Be$^+$ ground state'',
{\em Phys. Rev. Lett.} {\bf 63}, 10, 1031-1034 (1989).

\item {\bf [Bona 03]}:
P. Bona,
`Comment on ``No-signaling condition and quantum dynamics''\,',
{\em Phys. Rev. Lett.} {\bf 90}, 20, 208901 (2003);
quant-ph/0201002.
Comment on {\bf [Simon-Bu\v{z}zek-Gisin 01]}.

\item {\bf [Bonadeo-Erland-Gammon-(+3) 98]}:
N. H. Bonadeo, J. Erland, D. Gammon,
D. Park, D. S. Katzer, \& D. G. Steel,
``Coherent optical control of the
quantum state of a single quantum dot'',
{\em Science} {\bf 282}, 5393, 1473-1476 (1998).

\item {\bf [Bonesteel 00]}:
N. E. Bonesteel,
``Chiral spin liquids and quantum error-correcting codes'',
{\em Phys. Rev. A} {\bf 62}, 6, 062310 (2000);
quant-ph/0006092.

\item {\bf [Bonesteel-Stepanenko-DiVincenzo 01]}:
N. E. Bonesteel, D. Stepanenko, \& D. P. DiVincenzo,
``Anisotropic spin exchange in pulsed quantum gates'',
{\em Phys. Rev. Lett.} {\bf 87}, 20, 207901 (2001);
quant-ph/0106161.

\item {\bf [Bonifacio-Olivares-Tombesi-Vitali 99]}:
R. Bonifacio, S. Olivares, P. Tombesi, \& D. Vitali,
``A model independent approach to non dissipative decoherence'',
quant-ph/9911100.

\item {\bf [Bonifacio-Olivares 02]}:
R. Bonifacio, \& S. Olivares,
``Puzzling aspects of Young interference and spontaneous intrinsic decoherence'',
in R. Bonifacio, \& D. Vitali (eds.),
{\em Mysteries, Puzzles and Paradoxes in Quantum Mechanics IV:
Quantum Interference Phenomena (Gargnano, Italy, 2001)},
{\em J. Opt. B: Quantum Semiclass. Opt.} {\bf 4}, 4, S253-S259 (2002).

\item {\bf [Booth-Atat\"{u}re-Di Giuseppe-(+3) 02]}:
M. C. Booth, M. Atat\"{u}re, G. Di Giuseppe, B. E. A. Saleh,
A. V. Sergienko, \& M. C. Teich,
``Counterpropagating entangled photons from a waveguide with periodic
nonlinearity'',
{\em Phys. Rev. A} {\bf 66}, 2, 023815 (2002).

\item {\bf [Booth-Di Giuseppe-Saleh-(+2) 04]}:
M. C. Booth, G. Di Giuseppe, B. E. A. Saleh,
A. V. Sergienko, \& M. C. Teich,
``Polarization-sensitive quantum-optical coherence tomography'',
{\em Phys. Rev. A};
quant-ph/0401176.

\item {\bf [Borcherds 03]}:
P. Borcherds,
``The odd quantum'',
{\em Eur. J. Phys.} {\bf 24}, 4, 495 (2003).
Review of {\bf [Treiman 99]}.

\item {\bf [Borelli-Vidiella Barranco 04]}:
L. F. M. Borelli, \& A. Vidiella-Barranco,
``Quantum key distribution using bright polarized coherent states'',
quant-ph/0403076.

\item {\bf [Borhani-Loss 04]}:
M. Borhani, \& D. Loss,
``Cluster states from Heisenberg interaction'',
quant-ph/0410145.

\item {\bf [Born 26 a]}:
M. Born,
``Zur Quantenmechanik der Stossvorg\"{a}nge'',
{\em Zeitschrift f\"{u}r Physik} {\bf 37}, 863-867 (1926).
English version: ``On the
quantum mechanics of collisions'',
in {\bf [Wheeler-Zurek 83]}, pp.~52-55.

\item {\bf [Born 26 b]}:
M. Born,
``Quantenmechanik der Stossvorg\"{a}nge'',
{\em Zeitschrift f\"{u}r Physik} {\bf 38}, 803-827 (1926).
Partial English version in {\bf [Ludwig 68 b]}.

\item {\bf [Born 33]}:
M. Born,
{\em Moderne Physik},
?, ?, 1933.
English version:
{\em Atomic physics},
Blackie \& Son, London, 1935;
Dover, New York, 1989 (8th ed.).

\item {\bf [Born 60]}:
M. Born,
{\em Vorlesungen \"{u}ber Atommechanik},
?, ?, ?.
English version:
{\em The mechanics of the atom},
F. Ungar Pub. Co., New York, 1960.

\item {\bf [Born 62]}:
M. Born,
{\em Zur statistischen Deutung der Quantentheorie},
E. Battenberg, Stuttgart, 1962
(published to honor the author on the occasion of
his 80th birthday).
Reprinted from
{\em Zeitschrift f\"{u}ur Physik} {\bf 36-38},
40 (1926), and
{\em Nachrichten der Gesellschaft der Wissenschaften
zu G\"{o}ottingen}, no. 146 (1926).

\item {\bf [Born 68]}:
M. Born,
{\em My life and views},
Scribner, New York, 1968.

\item {\bf [Born-Einstein 69]}:
M. Born, \& A. Einstein,
{\em Der Einstein-Born Briefwechsel 1916-1955},
Nymphenburger, M\"{u}nchen, 1969.
English version: {\em The Born-Einstein letters.
Correspondence between Albert Einstein and Max and Hedwig
Born from 1916 to 1955 with commentaries by Max Born},
MacMillan, London, 1971.
Spanish version: {\em A. Einstein y M. Born. Correspondencia (1916-1955)},
Siglo XXI, M\'{e}xico, 1973.
It does not include (at least) one letter from B. to E. of Nov. 1926;
see {\bf [Pais 91]}, p.~288, and {\bf [Pais 00]}.

\item {\bf [Born 75]}:
M. Born,
{\em Mein Leben: Die Erinnerungen des Nobelpreistr\"{a}gers},
Nymphenburger Verlagshandlung, M\"{u}nchen, 1975.
English version:
{\em My life: Recollections of a Nobel laureate},
Taylor \& Francis, London, 1978;
Charles Scriber's Sons, New York, 1978.
Review: {\bf [Gowing 79 a]}.

\item {\bf [Boschi-De Martini-Di Giuseppe 97]}:
D. Boschi, F. De Martini, \& G. Di Giuseppe,
``Test of the violation of local realism in quantum mechanics without Bell
inequalities'',
{\em Phys. Lett. A} {\bf 228}, 4-5, 208-214 (1997).
Almost the same as {\bf [Di Giuseppe-De Martini-Boschi 97]}.
See {\bf [Torgerson-Branning-Monchen-Mandel 95]}.

\item {\bf [Boschi-Branca-De Martini-Hardy 97]}:
D. Boschi, S. Branca, F. De Martini, \& L. Hardy,
``Ladder proof of nonlocality without inequalities:
Theoretical and experimental results'',
{\em Phys. Rev. Lett.} {\bf 79}, 15, 2755-2758 (1997).
See {\bf [Hardy 97 a]}.

\item {\bf [Boschi-Branca-De Martini-(+2) 98]}:
D. Boschi, S.
Branca, F. De Martini, L. Hardy, \& S. Popescu,
``Experimental realization of teleporting an unknown pure quantum state
via dual classical an Einstein-Podolsky-Rosen channels'',
{\em Phys. Rev. Lett.} {\bf 80}, 6, 1121-1125 (1998);
quant-ph/9710013.
See {\bf [Sudbery 97]}.

\item {\bf [Bose-Chattopadhyay 02]}:
I. Bose, \& E. Chattopadhyay,
``Macroscopic entanglement jumps in model spin systems'',
{\em Phys. Rev. A} {\bf 66}, 6, 062320 (2002).

\item {\bf [Bose-Jacobs-Knight 97]}:
S. Bose, K. Jacobs, \& P. L. Knight,
``Preparation of nonclassical states in cavities with a moving mirror'',
quant-ph/9708002.

\item {\bf [Bose-Knight-Murao-(+2) 97]}:
S. Bose, P. L. Knight, M. Murao, M. B. Plenio, \& V. Vedral,
``Implementations of quantum logic: Fundamental and experimental limits'',
{\em Philos. Trans. R. Soc. Lond., Proc.\ of
the Royal Society discussion meeting on quantum computation:
Theory and experiment (London, 1997)};
quant-ph/9712021.

\item {\bf [Bose-Vedral-Knight 98]}:
S. Bose, V. Vedral, \& P. L. Knight,
``Multiparticle generalization of entanglement swapping'',
{\em Phys. Rev. A} {\bf 57}, 2, 822-829 (1998);
quant-ph/9708004.

\item {\bf [Bose-Knight-Murao-(+2) 98]}:
S. Bose, P. L. Knight, M. Murao, M. B. Plenio, \& V. Vedral,
``Implementations of quantum logic: Fundamental and
 experimental limits'',
in A. K. Ekert, R. Jozsa, \& R. Penrose (eds.),
{\em Quantum Computation: Theory and Experiment.
Proceedings of a Discussion Meeting held at the Royal
Society of London on 5 and 6 November 1997},
{\em Philos. Trans. R. Soc. Lond. A} {\bf 356}, 1743, 1823-1840
(1998).

\item {\bf [Bose-Jacobs-Knight 99]}:
S. Bose, K. Jacobs, \& P. L. Knight,
``Scheme to probe the decoherence of a macroscopic object'',
{\em Phys. Rev. A} {\bf 59}, 5, 3204-3210 (1999);
quant-ph/9712017.

\item {\bf [Bose-Vedral-Knight 99]}:
S. Bose, V. Vedral, \& P. L. Knight,
``Purification via entanglement swapping and conserved entanglement'',
{\em Phys. Rev. A} {\bf 60}, 1, 194-197 (1999);
quant-ph/9812013.

\item {\bf [Bose-Knight-Plenio-Vedral 99]}:
S. Bose, P. L. Knight, M. B. Plenio, \& V. Vedral,
``Proposal for teleportation of an atomic state via cavity decay'',
{\em Phys. Rev. Lett.} {\bf 83}, 24, 5158-5161 (1999);
quant-ph/9908004.

\item {\bf [Bose-Vedral 00]}:
S. Bose, \& V. Vedral,
``Mixedness and teleportation'',
{\em Phys. Rev. A} {\bf 61}, 4, 040101(R) (2000);
quant-ph/9912033.

\item {\bf [Bose-Plenio-Vedral 00]}:
S. Bose, M. B. Plenio, \& V. Vedral,
``Mixed state dense coding and its relation to entanglement
measures'',
in V. Bu\v{z}zek, \& D. P. DiVincenzo (eds.),
{\em J. Mod. Opt.} {\bf 47}, 2-3 (Special issue:
Physics of quantum information), 291-310 (2000);
quant-ph/9810025.

\item {\bf [Bose-Rallan-Vedral 00]}:
S. Bose, L. Rallan, \& V. Vedral,
``Communication capacity of quantum computation'',
{\em Phys. Rev. Lett.} {\bf 85}, 25, 5448-5451 (2000);
quant-ph/0003072.

\item {\bf [Bose-Vedral 00]}:
S. Bose, \& V. Vedral,
``Fundamental bounds on quantum measurements
with a mixed apparatus'',
quant-ph/0004016.

\item {\bf [Bose-Fern\'{a}ndez Huelga-Plenio 01]}:
S. Bose, S. G. Fern\'{a}ndez Huelga, \& M. B. Plenio,
``Singlet-aided infinite resource reduction
in the comparison of distant fields'',
{\em Phys. Rev. A} {\bf 63}, 3, 032313 (2001);
quant-ph/0005020.

\item {\bf [Bose-Fuentes Guridi-Knight-Vedral 01]}:
S. Bose, I. Fuentes-Guridi, P. L. Knight, \& V. Vedral,
``Subsystem purity as an enforcer of entanglement'',
{\em Phys. Rev. Lett.} {\bf 87}, 5, 050401 (2001).
Erratum:
{\em Phys. Rev. Lett.} {\bf 87}, 27, 279901 (2001);
quant-ph/0103063.

\item {\bf [Bose-Home 02]}:
S. Bose, \& D. Home,
``Generic entanglement generation, quantum statistics, and complementarity'',
{\em Phys. Rev. Lett.} {\bf 88}, 5, 050401 (2002);
quant-ph/0101093.

\item {\bf [Bose 02]}:
S. Bose,
``Quantum communication through an unmodulated spin chain'',
quant-ph/0212041.

\item {\bf [Bose-Ekert-Omar-(+2) 02]}:
S. Bose, A. K. Ekert, Y. Omar, N. Paunkovi\'{c}, \& V. Vedral,
``Optimal state Discrimination Using Particle Statistics
quant-ph/0309090.

\item {\bf [Bostroem 00 a]}:
K. J. Bostroem,
``Concepts of a quantum information theory of many letters'',
quant-ph/0009052.

\item {\bf [Bostroem 00 b]}:
K. J. Bostroem,
``Lossless quantum coding in many-letter spaces'',
quant-ph/0009073.

\item {\bf [Bostroem-Felbinger 02 a]}:
K. J. Bostroem, \& T. Felbinger,
``Lossless quantum data compression and variable-length coding'',
{\em Phys. Rev. A} {\bf 65}, 3, 032313 (2002);
quant-ph/0105026.

\item {\bf [Bostroem 02]}:
K. J. Bostroem,
``Secure direct communication using entanglement'',
quant-ph/0203064.

\item {\bf [Bostroem-Felbinger 02 b]}:
K. Bostroem, \& T. Felbinger,
``Deterministic secure direct communication using entanglement'',
{\em Phys. Rev. Lett.} {\bf 89}, 18, 187902 (2002).

\item {\bf [Botero-Reznik 00]}:
A. Botero, \& B. Reznik,
``Quantum-communication protocol employing weak measurements'',
{\em Phys. Rev. A} {\bf 61}, 5, 050301(R) (2000);
quant-ph/9908006.

\item {\bf [Botero-Reznik 03]}:
A. Botero, \& B. Reznik,
``Modewise entanglement of Gaussian states'',
{\em Phys. Rev. A} {\bf 67}, 5, 052311 (2003).

\item {\bf [Botero-Reznik 04]}:
A. Botero, \& B. Reznik,
``BCS-like modewise entanglement of fermion Gaussian states'',
{\em Phys. Lett. A} {\bf 331}, 1-2, 39-44 (2004);
quant-ph/0404176.

\item {\bf [Boto-Kok-Abrams-(+3) 00]}:
A. N. Boto, P. Kok, D. S. Abrams,
S. L. Braunstein, C. P. Williams, \& J. P. Dowling,
``Quantum interferometric optical lithography: Exploiting entanglement
to beat the diffraction limit'',
{\em Phys. Rev. Lett.} {\bf 85}, 13, 2733-2736 (2000).
See {\bf [Kok-Boto-Abrams-(+3) 01]}.

\item {\bf [Bouda-Djama 01]}:
A. Bouda, \& T. Djama,
``Quantum trajectories'',
quant-ph/0108022.

\item {\bf [Bouda-Bu\v{z}zek 01]}:
J. Bouda, \& V. Bu\v{z}zek,
``Entanglement swapping between multi-qudit systems'',
{\em J. Phys. A} {\bf 34}, 20, 4301-4311 (2001);
quant-ph/0112022.

\item {\bf [Bouda 01]}:
J. Bouda,
``Probability current and trajectory representation'',
{\em Found. Phys. Lett.} {\bf 14}, 1, 17-35 (2001).

\item {\bf [Bouda-Bu\v{z}zek 02]}:
J. Bouda, \& V. Bu\v{z}zek,
``Purification and correlated measurements of bipartite mixed states'',
{\em Phys. Rev. A} {\bf 65}, 3, 034304 (2002);
quant-ph/0112015.

\item {\bf [Bouda-Bu\v{z}zek 02]}:
J. Bouda, \& V. Bu\v{z}zek,
``Security of the private quantum channel'',
in M. Ferrero (ed.),
{\em Proc. of Quantum Information: Conceptual Foundations,
Developments and Perspectives (Oviedo, Spain, 2002)},
{\em J. Mod. Opt.} {\bf 50}, 6-7, 1071-1077 (2003).

\item {\bf [Bondurant 04]}:
R. S. Bondurant,
``Simple model for the detection of a particle by a point detector'',
{\em Phys. Rev. A} {\bf 69}, 6, 062104 (2004).

\item {\bf [Boulant-Fortunato-Pravia-(+3) 02]}:
N. Boulant, E. M. Fortunato, M. A. Pravia,
G. Teklemariam, D. G. Cory, \& T. F. Havel,
``Entanglement transfer experiment in NMR quantum information processing'',
{\em Phys. Rev. A} {\bf 65}, 2, 024302 (2002).

\item {\bf [Boulant-Havel-Pravia-Cory 03]}:
N. Boulant, T. F. Havel, M. A. Pravia, \& D. G. Cory,
``Robust method for estimating the Lindblad operators of a dissipative quantum
process from measurements of the density operator at multiple time points'',
{\em Phys. Rev. A} {\bf 67}, 4, 042322 (2003).

\item {\bf [Boulant-Edmonds-Yang-(+2) 03]}:
N. Boulant, K. Edmonds, J. Yang,
M. A. Pravia, \& D. G. Cory,
``Experimental demonstration of an entanglement swapping operation and
improved control in NMR quantum-information processing'',
{\em Phys. Rev. A} {\bf 68}, 3, 032305 (2003).

\item {\bf [Boulant-Emerson-Havel-(+2) 04]}:
N. Boulant, J. Emerson, T. F. Havel,
D. G. Cory, \& S. Furuta,
``Incoherent noise and quantum information processing'',
{\em J. Chem. Phys.} {\bf 121}, 2955-2961 (2004).

\item {\bf [Boulant-Viola-Fortunato-Cory 04]}:
N. Boulant, L. Viola, E. M. Fortunato, \& D. G. Cory,
``Experimental implementation of a concatenated quantum error-correcting
code'',
quant-ph/0409193.

\item {\bf [Boulatov-Smelyanskiy 03]}:
A. Boulatov, \& V. N. Smelyanskiy,
``Quantum adiabatic algorithms and large spin tunnelling'',
{\em Phys. Rev. A} {\bf 68}, 6, 062321 (2003).

\item {\bf [Bourdon-Williams 04]}:
P. S. Bourdon, \& H. T. Williams,
``Unital quantum operations on the Bloch ball and Bloch region'',
{\em Phys. Rev. A} {\bf 69}, 2, 022314 (2004);
quant-ph/0308089.

\item {\bf [Bourennane-Gibson-Karlsson-(+5) 99]}:
M. Bourennane, F. Gibson, A. Karlsson,
A. Hening, P. Jonsson, T. Tsegaye,
D. Ljunggren, \& E. Sundberg,
``Experiments on long wavelength (1550 nm) `plug and play' quantum cryptography system'',
{\em Opt. Express} {\bf 4}, 383-387 (1999).

\item {\bf [Bourennane-Ljunggren-Karlsson-(+3) 00]}:
M. Bourennane, D. Ljunggren, A. Karlsson, P. Jonsson,
A. Hening, \& J. Pena Ciscar,
``Experimental long wavelenght quantum cryptography:
From single-photon transmission to key extraction protocols'',
in V. Bu\v{z}zek, \& D. P. DiVincenzo (eds.),
{\em J. Mod. Opt.} {\bf 47}, 2-3 (Special issue:
Physics of quantum information), ?-? (2000).

\item {\bf [Bourennane-Karlsson-Bj\"{o}rk 01]}:
M. Bourennane, A. Karlsson, \& G. Bj\"{o}rk,
``Quantum key distribution using multilevel encoding'',
{\em Phys. Rev. A} {\bf 64}, 1, 012306 (2001).

\item {\bf [Bourennane-Karlsson-Bj\"{o}rk-(+3) 01]}:
M. Bourennane, A. Karlsson, G. Bj\"{o}rk,
N. Gisin, \& N. Cerf,
``Quantum key distribution using multilevel
encoding: Security analysis'',
{\em J. Phys. A} {\bf 35}, 47, 10065-10076 (2002);
quant-ph/0106049.

\item {\bf [Bourennane-Eibl-Kurtsiefer-(+7) 04]}:
M. Bourennane, M. Eibl, C. Kurtsiefer, S. Gaertner,
H. Weinfurter, O. G\"{u}hne, P. Hyllus,
D. Bru\ss, M. Lewenstein, \& A. Sanpera,
``Experimental detection of multipartite entanglement using witness operators'',
{\em Phys. Rev. Lett.} {\bf 92}, 8, 087902 (2004).
quant-ph/0309043.

\item {\bf [Bourennane-Eibl-Gaertner-(+3) 04]}:
M. Bourennane, M. Eibl, S. Gaertner,
C. Kurtsiefer, A. Cabello, \& H. Weinfurter,
``Decoherence-free quantum information processing
with four-photon entangled states'',
{\em Phys. Rev. Lett.} {\bf 92}, 10, 107901 (2004);
quant-ph/0309041.

\item {\bf [Bouwmeester-Zeilinger 97]}:
D. Bouwmeester, \& A. Zeilinger,
``Atoms that agree to differ'',
{\em Nature} {\bf 388}, 6645, 827-829 (1997).
See {\bf [Hagley-Ma\^{\i}tre-Nogues-(+4) 97]}.

\item {\bf [Bouwmeester-Pan-Mattle-(+3) 97]}:
D. Bouwmeester, J.-W. Pan, K. Mattle, M. Eibl,
H. Weinfurter, \& A. Zeilinger,
``Experimental quantum teleportation'',
{\em Nature} {\bf 390}, 6660, 575-579 (1997).
Reprinted in {\bf [Macchiavello-Palma-Zeilinger 00]}, pp.~40-44.
See {\bf [Sudbery 97]}.
Comment: {\bf [Braunstein-Kimble 98 b]}.
Reply: {\bf [Bouwmeester-Pan-Daniell-(+3) 98]}.

\item {\bf [Bouwmeester-Pan-Daniell-(+3) 98]}:
D. Bouwmeester, J.-W. Pan, K. Mattle, M. Daniell, H. Weinfurter,
M. \.{Z}ukowski, \& A. Zeilinger,
``Bouwmeester {\em et al.} reply'',
{\em Nature} {\bf 394}, 6696, 841 (1998).
Reply to {\bf [Braunstein-Kimble 98 b]}.
See {\bf [Bouwmeester-Pan-Mattle-(+3) 97]}.

\item {\bf [Bouwmeester-Pan-Daniell-(+3) 98]}:
D. Bouwmeester, J.-W. Pan, K. Mattle, M. Eibl,
H. Weinfurter, \& A. Zeilinger,
``Experimental quantum teleportation'',
in A. K. Ekert, R. Jozsa, \& R. Penrose (eds.),
{\em Quantum Computation: Theory and Experiment.
Proceedings of a Discussion Meeting held at the Royal
Society of London on 5 and 6 November 1997},
{\em Philos. Trans. R. Soc. Lond. A} {\bf 356}, 1743, 1733-1737 (1998).

\item {\bf [Bouwmeester-Pan-Weinfurter-Zeilinger 99]}:
D. Bouwmeester, J.-W. Pan, H. Weinfurter, \& A. Zeilinger,
``Experimental quantum teleportation of qubits and entanglement swapping'',
in {\bf [Greenberger-Reiter-Zeilinger 99]}, pp.~127-140.

\item {\bf [Bouwmeester-Pan-Daniell-(+2) 99]}:
D. Bouwmeester, J.-W. Pan, M. Daniell, H. Weinfurter, \& A. Zeilinger,
``Observation of three-photon Greenberger-Horne-Zeilinger entanglement'',
{\em Phys. Rev. Lett.} {\bf 82}, 7, 1345-1349 (1999);
quant-ph/9810035.
Reprinted in {\bf [Macchiavello-Palma-Zeilinger 00]}, pp.~55-59.
See {\bf [\.{Z}ukowski 98]},{\bf [Pan-Bouwmeester-Daniell-(+2) 00]}.

\item {\bf [Bouwmeester-Pan-Weinfurter-Zeilinger 00]}:
D. Bouwmeester, J.-W. Pan, H. Weinfurter, \& A. Zeilinger,
``High-fidelity teleportation of independent qubits'',
in V. Bu\v{z}zek, \& D. P. DiVincenzo (eds.),
{\em J. Mod. Opt.} {\bf 47}, 2-3 (Special issue:
Physics of quantum information), 279-289 (2000);
quant-ph/9910043.

\item {\bf [Bouwmeester-Ekert-Zeilinger 00]}:
D. Bouwmeester, A. K. Ekert, \& A. Zeilinger (eds.),
{\em The physics of quantum information: Quantum cryptography,
quantum teleportation, quantum computation},
Springer-Verlag, Berlin, 2000.

\item {\bf [Bouwmeester 00 a]}:
D. Bouwmeester,
``Quantum entanglement manipulation'',
{\bf [Macchiavello-Palma-Zeilinger 00]}, pp.~31-34.

\item {\bf [Bouwmeester 00 b]}:
D. Bouwmeester,
``Error-free optical quantum communication'',
quant-ph/0006108.

\item {\bf [Bowden-Pethel 99]}:
C. M. Bowden, \& S. D. Pethel,
``Novel scheme for universal quantum computation'',
quant-ph/9912003.

\item {\bf [Bowden-Chen-Diao-Klappenecker 02]}:
C. M. Bowden, G. Chen, Z. Diao, \& A. Klappenecker,
``The universality of the quantum Fourier transform in
forming the basis of quantum computing algorithms'',
{\em J. Math. Anal. Appl.} {\bf 274}, 1, 69-80 (2002);
quant-ph/0007122.

\item {\bf [Bowdrey-Oi-Short-(+2) 02]}:
M. D. Bowdrey, D. K. L. Oi, A. J. Short,
K. Banaszek, \& J. A. Jones,
``Fidelity of single qubit maps'',
{\em Phys. Lett. A} {\bf 294}, 5-6, 258-260 (2002);
quant-ph/0201106.

\item {\bf [Bowen 01 a]}:
G. Bowen,
``Classical information capacity of superdense coding'',
{\em Phys. Rev. A} {\bf 63}, 2, 022302 (2001);
quant-ph/0101117.

\item {\bf [Bowen 01 b]}:
G. Bowen,
``Dense coding using entangled input states'',
{\em Phys. Lett. A} {\bf 285}, 3-4, 115-118 (2001).

\item {\bf [Bowen-Bose 01]}:
G. Bowen, \& S. Bose,
``Teleportation as a depolarizing quantum channel,
relative entropy and classical capacity'',
{\em Phys. Rev. Lett.} {\bf 87}, 26, 267901 (2001);
quant-ph/0107132.

\item {\bf [Bowen 02]}:
G. Bowen,
``Entanglement required in achieving entanglement-assisted channel capacities'',
{\em Phys. Rev. A} {\bf 66}, 5, 052313 (2002).

\item {\bf [Bowen-Devetak-Mancini 03]}:
G. Bowen, I. Devetak, \& S. Mancini,
``Classical information capacities for a class of quantum memory channels'',
quant-ph/0312216.

\item {\bf [Bowen-Mancini 04 a]}:
G. Bowen, \& S. Mancini,
``Noise enhancing the classical information capacity of a quantum channel'',
{\em Phys. Lett. A} {\bf 321}, 1, 1-5 (2004).

\item {\bf [Bowen-Mancini 04 b]}:
G. Bowen, \& S. Mancini,
``Quantum channels with a finite memory'',
{\em Phys. Rev. A} {\bf 69}, 1, 012306 (2004);
quant-ph/0305010.

\item {\bf [Bowen 04]}:
G. Bowen,
``Feedback in quantum communication'',
quant-ph/0410191.

\item {\bf [Bowen-Lam-Ralph 01]}:
W. P. Bowen, P. K. Lam, \& T. C. Ralph,
``Concentrating pure continuous variable entanglement'',
quant-ph/0104108.

\item {\bf [Bowen-Treps-Schnabel-Lam 01]}:
W. P. Bowen, N. Treps, R. Schnabel, \& P. K. Lam,
``Experimental demonstration of continuous variable polarization entanglement'',
{\em Phys. Rev. Lett.} {\bf 89}, 25, 253601 (2002).

\item {\bf [Bowen-Schnabel-Lam-Ralph 03]}:
W. P. Bowen, R. Schnabel, P. K. Lam, \& T. C. Ralph,
``Experimental investigation of criteria for continuous variable entanglement'',
{\em Phys. Rev. Lett.} {\bf 90}, 4, 043601 (2003).

\item {\bf [Bowen-Treps-Buchler-(+5) 03]}:
W. P. Bowen, N. Treps, B. C. Buchler,
R. Schnabel, T. C. Ralph, H.-A. Bachor, T. Symul, \& P. K. Lam,
``Experimental investigation of continuous-variable quantum teleportation'',
{\em Phys. Rev. A} {\bf 67}, 3, 032302 (2003).

\item {\bf [Bowen-Treps-Buchler-(+4) 03]}:
W. P. Bowen, N. Treps, B. C. Buchler,
R. Schnabel, T. C. Ralph, T. Symul, \& P. K. Lam,
``Unity gain and non-unity gain quantum teleportation'',
{\em IEEE J. of selected topics in Quantum Electronics} {\bf 9}, 1519-? (2003);
quant-ph/0303179.

\item {\bf [Bowen-Treps-Schnabel-(+2) 03]}:
W. P. Bowen, N. Treps, R. Schnabel, T. C. Ralph, \& P. K. Lam,
``Continuous variable polarization entanglement, experiment and analysis'',
quant-ph/0303180.

\item {\bf [Bowen-Schnabel-Lam-Ralph 04]}:
W. P. Bowen, R. Schnabel, P. K. Lam, \& T. C. Ralph,
``Experimental characterization of continuous-variable entanglement'',
{\em Phys. Rev. A} {\bf 69}, 1, 012304 (2004);
quant-ph/0309013.

\item {\bf [Bowen-Mancini 04]}:
G. Bowen, \& S. Mancini,
``Quantum channels with a finite memory'',
{\em Phys. Rev. A} {\bf 69}, 1, 012306 (2004).

\item {\bf [Bowman 02 a]}:
G. E. Bowman,
``Bohmian mechanics as a heuristic device:
Wave packets in the harmonic oscillator'',
{\em Am. J. Phys.} {\bf 70}, 3, 313-318 (2002).

\item {\bf [Bowman 02 b]}:
G. E. Bowman,
``Wave packets and Bohmian mechanics in the kicked rotator'',
{\em Phys. Lett. A} {\bf 298}, 1, 7-17 (2002).

\item {\bf [Bovino-De Martini-Mussi 99]}:
F. A. Bovino, F. De Martini, \& V. Mussi,
``Quantum superposition of parametrically amplified multiphoton pure
states within a decoherence-free Schr\"{o}dinger-cat structure'';
quant-ph/9905048.

\item {\bf [Bovino-Colla-Castagnoli-(+3) 03]}:
F. A. Bovino, A. M. Colla, G. Castagnoli,
S. Castelletto, I. P. Degiovanni, \& M. L. Rastello,
``Experimental eavesdropping attack against Ekert's protocol based on Wigner's
inequality'',
{\em Phys. Rev. A} {\bf 68}, 3, 034309 (2003);
quant-ph/0308030.

\item {\bf [Bovino-Castagnoli-Degiovanni-Castelletto 04]}:
F. A. Bovino, G. Castagnoli,
I. P. Degiovanni, \& S. Castelletto,
``Experimental evidence for bounds on quantum correlations'',
{\em Phys. Rev. Lett.} {\bf 92}, 6, 060404 (2004);
quant-ph/0310042.
See {\bf [Cabello 04 a]}.

\item {\bf [Boyer-Brassard-H\o{}yer-Tapp 96]}:
M. Boyer, G. Brassard, P. H\o{}yer, \& A. Tapp,
``?'',
in T. Toffoli, M. Biafore, \& J. Le\"{a}lo (eds.),
{\em PhysComp 96: Proc.\ 4th Workshop on Physics and Computation},
New England Complex Systems Institute,
Cambridge, Massachusetts, 1996, pp.~36-?.
See {\bf [Grover 97 b]}.

\item {\bf [Boyer-Brassard-H\o{}yer-Tapp 98]}:
M. Boyer, G. Brassard, P. H\o{}yer, \& A. Tapp,
``Tight bounds on quantum searching'',
{\em Fortschr. Phys.} {\bf 46}, 4-5, 493-505 (1998);
quant-ph/9605034.

\item {\bf [Boyer 04]}:
M. Boyer,
``An extended GHZ $n$-party game with classical probability of winning tending to 0'',
quant-ph/0408090.

\item {\bf [Boykin-Mor-Pulver-(+2) 99]}:
P. O. Boykin, T. Mor, M. Pulver, V. Roychowdhury, \& F. Vatan,
``On universal and fault-tolerant quantum computing'',
quant-ph/9906054.

\item {\bf [Boykin-Mor-Roychowdhury-Vatan 99]}:
P. O. Boykin, T. Mor, V. Roychowdhury, \& F. Vatan,
``Algorithms on ensemble quantum computers'',
quant-ph/9907067.

\item {\bf [Boykin-Mor-Roychowdhury-(+2) 01]}:
P. O. Boykin, T. Mor, V. Roychowdhury,
F. Vatan, \& R. Vrijen,
``Algorithmic cooling and scalable NMR quantum computers'',
quant-ph/0106093.

\item {\bf [Boykin-Roychowdhury 03]}:
P. O. Boykin, \& V. Roychowdhury,
``Optimal encryption of quantum bits'',
{\em Phys. Rev. A} {\bf 67}, 4, 042317 (2003);
quant-ph/0003059.

\item {\bf [Boyle-Schafir 01 a]}:
C. F. Boyle, \& R. L. Schafir,
``Remarks on noncontextual hidden
variables and physical measurements'',
quant-ph/0106040.
Comment on {\bf [Kent 99]}.

\item {\bf [Boyle-Schafir 01 b]}:
C. F. Boyle, \& R. L. Schafir,
``A delayed-choice thought-experiment with later-time
entanglement, and a dependency on future choice in quantum
mechanics'',
quant-ph/0107098.

\item {\bf [Boyle-Schafir 01 c]}:
C. F. Boyle, \& R. L. Schafir,
``The $N$-box paradox in orthodox quantum mechanics'',
quant-ph/0108113.

\item {\bf [Bo\v{z}i\'{c}-Mari\'{c} 98]}:
M. Bo\v{z}i\'{c}, \& Z. Mari\'{c},
``Quantum interference, quantum theory of measurement,
and (in)completeness of quantum mechanics'',
{\em Found. Phys.} {\bf 28}, 3, 415-428 (1998).

\item {\bf [Bracken-Ellinas-Tsohantjis 04]}:
A. J. Bracken, D. Ellinas, \& I. Tsohantjis,
``Pseudo memory effects, majorization and entropy in quantum random walks'',
{\em J. Phys. A} {\bf 37}, ?, L91-L97 (2004);
quant-ph/0402187.

\item {\bf [Bracken 04]}:
A. J. Bracken,
``Entangled subspaces and quantum symmetries'',
{\em Phys. Rev. A} {\bf 69}, 5, 052331 (2004).

\item {\bf [Braginsky-Vorontsov 74]}:
V. B. Braginsky, \& Y. I. Vorontsov,
``?'',
{\em Usp. Fiz. Nauk} {\bf 114}, ?, 41-53 (1974).
English version:
``Quantum-mechanical limitations in macroscopic experiments and
modern experimental technique'',
{\em Sov. Phys. Usp.} {\bf 17}, ?, 644-650 (1975).

\item {\bf [Braginsky-Vorontsov-Khalili 77]}:
V. B. Braginsky, Y. I. Vorontsov, \& F. Y. Khalili,
``?'',
{\em Zh. Eksp. Teor. Fiz.} {\bf 73}, 4, 1340-1343 (1977).
English version:
``Quantum singularities of a ponderomotive meter of
electromagnetic energy'',
{\em Sov. Phys. JETP} {\bf 46}, 4, 705-706 (1977).

\item {\bf [Braginsky-Vorontsov-Thorne 80]}:
V. B. Braginsky, Y. I. Vorontsov, \& K. S. Thorne,
``Quantum nondemolition measurements'',
{\em Science} {\bf 209}, 4456, 547-557 (1980).
Reprinted in {\bf [Wheeler-Zurek 83]}, pp.~749-768.

\item {\bf [Braginsky-Khalili 92]}:
V. B. Braginsky, \& F. Y. Khalili,
{\em Quantum measurement},
Cambridge University Press, Cambridge, 1992.
Review: {\bf [Scarl 94]}.

\item {\bf [Braginsky 95]}:
V. B. Braginsky,
``Unsolved problems in quantum optics (several short notes)'',
{\em Appl. Phys. B} {\bf ?}, 2-3, 239-241 (1995).

\item {\bf [Braginsky-Khalili 96]}:
V. B. Braginsky, \& F. Y. Khalili,
``Quantum nondemolition measurements:
The route from toys to tools'',
{\em Rev. Mod. Phys.} {\bf 68}, 1, 1-11 (1996).

\item {\bf [Braginsky-Khalili 99]}:
V. B. Braginsky, \& F. Y. Khalili,
``Low noise rigidity in quantum measurements'',
{\em Phys. Lett. A} {\bf 257}, 5-6, 241-246 (1999).

\item {\bf [Brainis-Lamoureux-Cerf-(+3) 03]}:
E. Brainis, L.-P. Lamoureux, N. J. Cerf,
P. Emplit, M. Haelterman, \& S. Massar,
``Fiber-optics implementation of the Deutsch-Jozsa and Bernstein-Vazirani
quantum algorithms with three qubits'',
{\em Phys. Rev. Lett.} {\bf 90}, 15, 157902 (2003).

\item {\bf [Brambilla-Gatti-Navez-Lugiato 00]}:
E. Brambilla, A. Gatti, P. Navez, \& L. A. Lugiato,
``Spatial entanglement of twin quantum images'',
quant-ph/0010108.

\item {\bf [Bramon-Nowakowski 99]}:
A. Bramon, \& M. Nowakowski,
``Bell inequalities for entangled pairs of neutral kaons'',
{\em Phys. Rev. Lett.} {\bf 83}, 1, 1-5 (1999);
hep-ph/9811406.

\item {\bf [Bramon-Garbarino 02 a]}:
A. Bramon, \& G. Garbarino,
``Novel Bell's inequalities for entangled $K^0-\bar{K^0}$ pairs'',
{\em Phys. Rev. Lett.} {\bf 88}, 4, 040403 (2002);
quant-ph/0108047.

\item {\bf [Bramon-Garbarino 02 b]}:
A. Bramon, \& G. Garbarino,
``Test of local realism with entangled kaon pairs and without
inequalities'',
{\em Phys. Rev. Lett.} {\bf 89}, 16, 160401 (2002);
quant-ph/0205112.

\item {\bf [Bramon-Garbarino-Hiesmayr 03 a]}:
A. Bramon, G. Garbarino, \& B. C. Hiesmayr,
``Quantitative duality and quantum erasure for neutral kaons'',
invited talk at the {\em Workshop on $e^+e^-$ in the 1,2\,GeV range (Alghero, Italy, 2003)};
hep-ph/0311232.

\item {\bf [Bramon-Garbarino-Hiesmayr 03 b]}:
A. Bramon, G. Garbarino, \& B. C. Hiesmayr,
``Quantitative complementarity in two-path interferometry'',
quant-ph/0311179.

\item {\bf [Bramon-Garbarino-Hiesmayr 03 c]}:
A. Bramon, G. Garbarino, \& B. C. Hiesmayr,
``Active and passive quantum erasers for neutral kaons'',
{\em Phys. Rev. A};
quant-ph/0402212.

\item {\bf [Bramon-Garbarino-Hiesmayr 04 a]}:
A. Bramon, G. Garbarino, \& B. C. Hiesmayr,
``Quantum marking and quantum erasure for neutral kaons'',
{\em Phys. Rev. Lett.} {\bf 92}, 2, 020405 (2004);
quant-ph/0306114.

\item {\bf [Bramon-Garbarino-Hiesmayr 04 b]}:
A. Bramon, G. Garbarino, \& B. C. Hiesmayr,
``Passive quantum erasure for neutral kaons'',
in V. Aldaya, \& J. M. Cervero (eds.)
{\em Symmetries in gravity and field theory},
Ediciones de la Universidad de Salamanca, Salamanca, Spain, 2004, p.~223;
quant-ph/0404086.

\item {\bf [Bramon-Escribano-Garbarino 04]}:
A. Bramon, R. Escribano, \& G. Garbarino,
``Bell's inequality tests: From photons to B-mesons'',
quant-ph/0410122.
See {\bf [Go 04]}.

\item {\bf [Brandt-Myers 96]}:
H. E. Brandt, \& J. M. Myers,
``Invention disclosure: POVM receiver for quantum cryptography'',
U.S. Army Research Laboratory, Adelphi, Maryland, 1996.
See {\bf [Brandt-Myers 99]}.

\item {\bf [Brandt-Myers-Lomonaco 97]}:
H. E. Brandt, J. M. Myers, \& S. J. Lomonaco, Jr.,
``Aspects of entangled translucent eavesdropping in quantum cryptography'',
{\em Phys. Rev. A} {\bf 56}, 6, 4456-4465 (1997).
Erratum: {\em Phys. Rev. A} {\bf 58}, 3, 2617 (1998).

\item {\bf [Brandt 98]}:
H. E. Brandt,
``Qubit devices and the issue of quantum decoherence'',
{\em Prog. Quant. Electron.} {\bf 22}, 257-370 (1998).

\item {\bf [Brandt 99 a]}:
H. E. Brandt,
``Eavesdropping optimization for quantum cryptography
using a positive operator-valued measure'',
{\em Phys. Rev. A} {\bf 59}, 4, 2665-2669 (1999).
See {\bf [Brandt-Myers 99]}.

\item {\bf [Brandt 99 b]}:
H. E. Brandt,
``Positive operator valued measure in quantum information processing'',
{\em Am. J. Phys.} {\bf 67}, 5, 434-439 (1999).

\item {\bf [Brandt-Myers 99]}:
H. E. Brandt, \& J. M. Myers,
``Positive-operator-valued-measure receiver for quantum
cryptography'',
patent US5999285, 1999.
See {\bf [Brandt-Myers 96]}, {\bf [Brandt 99]}.

\item {\bf [Brandt 00]}:
H. E. Brandt,
``Inconclusive rate as a disturbance measure in quantum cryptography'',
{\em Phys. Rev. A} {\bf 62}, 4, 042310 (2000).

\item {\bf [Brandt 01]}:
H. E. Brandt,
``Inconclusive rate in quantum key distribution'',
{\em Phys. Rev. A} {\bf 64}, 4, 042316 (2001).

\item {\bf [Brandt 02 a]}:
H. E. Brandt,
``Probe optimization in four-state protocol of quantum cryptography'',
{\em Phys. Rev. A} {\bf 66}, 3, 032303 (2002).
Erratum: {\em Phys. Rev. A} {\bf 68}, 2, 029904 (2003).

\item {\bf [Brandt 02 b]}:
H. E. Brandt,
``Secrecy capacity in the four-state protocol of quantum key distribution'',
{\em J. Math. Phys.} {\bf 43}, 9, 4526-4530 (2002).

\item {\bf [Brandt 02 c]}:
H. E. Brandt,
``Qubit devices'',
in {\bf [Lomonaco 02 a]}, pp.~67-139.

\item {\bf [Brandt 02 d]}:
H. E. Brandt,
``Inconclusive rate with a positive operator valued measure'',
in {\bf [Lomonaco-Brandt 02]} 43-52.

\item {\bf [Brandt 02 e]}:
H. E. Brandt,
``Deconstructing Wigner's density matrix concerning the mind-body question'',
{\em Found. Phys. Lett.} {\bf 15}, 3, 287-292 (2002).

\item {\bf [Brandt 03]}:
H. E. Brandt,
``Optimum probe parameters for entangling probe in quantum key
distribution'',
quant-ph/0302033.

\item {\bf [Brandt-Dahmen 94]}:
S. Brandt, \& H. D. Dahmen,
{\em The picture book of quantum mechanics},
Springer-Verlag, New York, 1994 (1995, 2nd edition; 2001, 3rd edition).
Review: {\bf [Berry 96]}.

\item {\bf [Branning 97]}:
D. Branning,
``Does nature violate local realism?'',
{\em American Scientist} {\bf 85}, ?, 160-167 (1997).

\item {\bf [Bransden-Joachain 89]}:
B. H. Bransden, \& C. J. Joachain,
{\em Quantum mechanics},
Longman Scientific and Technical, Harlow, England, 1989;
2nd edition: Prentice-Hall, Harlow, England, 2000.

\item {\bf [Brassard 88]}:
G. Brassard,
{\em Modern cryptology: A tutorial},
Springer, New York, 1988.
French version: {\em Cryptologie contemporaine},
Masson, Paris, 1992.

\item {\bf [Brassard 89]}:
G. Brassard,
``The dawn of a new era for quantum cryptography:
The experimental proptotype is working!'',
{\em SIGACT News}, {\bf 20}, 4, 78-82 (1989).

\item {\bf [Brassard-Cr\'{e}peau-Jozsa-Langlois 93]}:
G. Brassard, C. Cr\'{e}peau, R. Jozsa, \& D. Langlois,
``A quantum bit commitment scheme provably unbreakable by both parties'',
in {\em Proc.\ of the 34th Annual IEEE Symp.\ on the
Foundation of Computer Science (1993)},
IEEE Computer Science Society Press, Los Alamitos, California, 1993, pp.~2379-2382.

\item {\bf [Brassard 96 a]}:
G. Brassard,
``New trends in quantum computing'',
quant-ph/9602014.

\item {\bf [Brassard 96 b]}:
G. Brassard,
``Teleportation as a quantum computation'',
quant-ph/9605035.
See {\bf [Brassard-Braunstein-Cleve 98]}.

\item {\bf [Brassard-H\o{}yer 96]}:
G. Brassard, \& P. H\o{}yer,
``On the power of exact quantum polynomial time'',
quant-ph/9612017.

\item {\bf [Brassard 97]}:
G. Brassard,
``Searching a quantum phone book'',
{\em Science} {\bf 275}, 5300, 627-628 (1997).

\item {\bf [Brassard-H\o{}yer 97]}:
G. Brassard, \& P. H\o{}yer,
``An exact quantum polynomial-time algorithm for Simon's
problem'',
in {\em Proc.\ 5th Israeli Symp.\ on Theory of
Computing and Systems (ISTCS'97)};
quant-ph/9704027.

\item {\bf [Brassard-H\o{}yer-Tapp 97]}:
G. Brassard, P. H\o{}yer, \& A. Tapp,
``Quantum algorithm for the collision problem'',
quant-ph/9705002.

\item {\bf [Brassard-Cr\'{e}peau-Mayers-Salvail 97]}:
G. Brassard, C. Cr\'{e}peau, D. Mayers, \& L. Salvail,
``A brief review on the impossibility of quantum bit commitment'',
quant-ph/9712023.
See {\bf [Brassard-Cr\'{e}peau-Mayers-Salvail 98]}.

\item {\bf [Brassard-Braunstein-Cleve 98]}:
G. Brassard, S. L. Braunstein, \& R. Cleve,
``Teleportation as a quantum computation'',
{\em Physica D} {\bf 120}, ?, 43-47 (1998).
See also {\bf [Brassard 96]}.

\item {\bf [Brassard-H\o{}yer-Tapp 98]}:
G. Brassard, P. H\o{}yer, \& A. Tapp,
``Quantum counting'',
in {\em Proc.\ 25th ICALP} {\bf 1443},
{\em Lecture Notes in Computer Science} {\bf 80},
Springer-Verlag, New York, 1998;
quant-ph/9805082.

\item {\bf [Brassard-Cr\'{e}peau-Mayers-Salvail 98]}:
G. Brassard, C. Cr\'{e}peau, D. Mayers, \& L. Salvail,
``Defeating classical bit commitments with a quantum computer'',
quant-ph/9806031.
Supersedes {\bf [Brassard-Cr\'{e}peau-Mayers-Salvail 97]}.

\item {\bf [Brassard-Cleve-Tapp 99]}:
G. Brassard, R. Cleve, \& A. Tapp,
``Cost of exactly simulating quantum entanglement with
classical communication'',
{\em Phys. Rev. Lett.} {\bf 83}, 9, 1874-1877 (1999);
quant-ph/9901035.

\item {\bf [Brassard-Mor-Sanders 99]}:
G. Brassard, T. Mor, \& B. C. Sanders,
``Quantum cryptography via parametric downconversion'',
quant-ph/9906074.
See {\bf [Brassard-L\"{u}tkenhaus-Mor-Sanders 00]}.

\item {\bf [Brassard-L\"{u}tkenhaus-Mor-Sanders 00]}:
G. Brassard, N. L\"{u}tkenhaus, T. Mor, \& B. C. Sanders,
``Limitations on practical quantum cryptography'',
{\em Phys. Rev. Lett.} {\bf 85}, 6, 1330-1333 (2000).
See {\bf [Brassard-Mor-Sanders 99]};
quant-ph/9911054.

\item {\bf [Brassard-Mor 01]}:
G. Brassard, \& T. Mor,
``Multi-particle entanglement via two-party entanglement'',
in S. Popescu, N. Linden, \& R. Jozsa (eds.),
{\em J. Phys. A} {\bf 34}, 35
(Special issue: Quantum information and computation), 6807-6814 (2001).

\item {\bf [Brassard 01]}:
G. Brassard,
``Quantum communication complexity (A survey)'',
quant-ph/0101005.

\item {\bf [Brassard-H\o{}yer-Mosca-Tapp 00]}:
G. Brassard, P. H\o{}yer, M. Mosca, \& A. Tapp,
``Quantum amplitude amplification and estimation'',
in {\bf [Lomonaco-Brandt 02]} 53-74;
quant-ph/0005055.

\item {\bf [Brassard-Broadbent-Tapp 03]}:
G. Brassard, A. Broadbent, \& A. Tapp,
``Multi-party pseudo-telepathy'',
{\em WADS 2003 Proc.};
quant-ph/0306042.

\item {\bf [Brassard 03]}:
G. Brassard,
``Quantum communication complexity'',
{\em Found. Phys.} {\bf 33}, 11, 1593-1616 (2003).

\item {\bf [Brassard-Broadbent-Tapp 04 a]}:
G. Brassard, A. Broadbent, \& A. Tapp,
``Quantum pseudo-telepathy'',
quant-ph/0407221.

\item {\bf [Brassard-Broadbent-Tapp 04 a]}:
G. Brassard, A. Broadbent, \& A. Tapp,
``Recasting Mermin's multi-player game into the framework of
pseudo-telepathy'',
quant-ph/0408052.

\item {\bf [Brattke-Guth\"{o}hrlein-Keller-(+3) 03]}:
S. Brattke, G. R. Guth\"{o}hrlein, M. Keller,
W. Lange, B. Varcoe, \& H. Walther,
``Generation of photon number states on demand'',
in M. Ferrero (ed.),
{\em Proc. of Quantum Information: Conceptual Foundations,
Developments and Perspectives (Oviedo, Spain, 2002)},
{\em J. Mod. Opt.} {\bf 50}, 6-7, 1103-1113 (2003).

\item {\bf [Braun-Haake-Strunz 01]}:
D. Braun, F. Haake, \& W. T. Strunz,
``Universality of decoherence'',
{\em Phys. Rev. Lett.} {\bf 86}, 14, 2913-2917 (2001);
quant-ph/0006117.

\item {\bf [Braun 02 a]}:
D. Braun,
``Quantum chaos and quantum algorithms'',
{\em Phys. Rev. A} {\bf 65}, 4, 042317 (2002);
quant-ph/0110037.

\item {\bf [Braun 02 b]}:
D. Braun,
``Creation of entanglement by interaction with a common heat bath'',
{\em Phys. Rev. Lett.} {\bf 89}, 27, 277901 (2002);
quant-ph/0205019.

\item {\bf [Braunstein 99]}:
P. Braunstein,
{\em Quantum computing: Where do we want to go tomorrow},
Vch Publishers, ?, 1999.

\item {\bf [Braunstein-Caves 88]}:
S. L. Braunstein, \& C. M. Caves,
``Information-theoretic Bell inequalities'',
{\em Phys. Rev. Lett.} {\bf 61}, 6, 662-665 (1988).

\item {\bf [Braunstein-Caves 89]}:
S. L. Braunstein, \& C. M. Caves,
``Chained Bell inequalities'',
in M. Kafatos (ed.),
{\em Bell's theorem, quantum theory, and conceptions of the universe.
Proc.\ of a workshop (George Mason University, 1988)},
Kluwer Academic, Dordrecht, Holland, 1989, pp.~27-36.

\item {\bf [Braunstein-Caves 90]}:
S. L. Braunstein, \& C. M. Caves,
``Wringing out better Bell inequalities'',
{\em Ann. Phys.} {\bf 202}, 1, 22-56 (1990).
See {\bf [Braunstein-Caves 89]}.

\item {\bf [Braunstein-Mann-Revzen 92]}:
S. L. Braunstein, A. Mann, \& M. Revzen,
``Maximal violation of Bell inequalities for mixed states'',
{\em Phys. Rev. Lett.} {\bf 68}, 22, 3259-3261 (1992).

\item {\bf [Braunstein-Mann 93]}:
S. L. Braunstein, \& A. Mann,
``Noise in Mermin's $n$-particle Bell inequality'',
{\em Phys. Rev. A} {\bf 47}, 4, Part A, R2427-R2430 (1993).

\item {\bf [Braunstein-Mann 95]}:
S. L. Braunstein, \& A. Mann,
``Measurement of the Bell operator and quantum teleportation'',
{\em Phys. Rev. A} {\bf 51}, 3, R1727-R1730 (1995).
Erratum: {\em Phys. Rev. A} {\bf 53}, 1, 630 (1996).

\item {\bf [Braunstein 96 a]}:
S. L. Braunstein,
``Quantum teleportation without irreversible detection'',
{\em Phys. Rev. A} {\bf 53}, 3, 1900-1902 (1996).

\item {\bf [Braunstein 96 b]}:
S. L. Braunstein,
``Geometry of quantum inference'',
in A. Mann, \& M. Revzen (eds.),
{\em The dilemma of Einstein, Podolsky and Rosen -- 60 years
later. An international symposium in honour of Nathan Rosen
(Haifa, Israel, 1995)},
{\em Ann. Phys. Soc. Israel} {\bf 12}, 218-225 (1996).

\item {\bf [Braunstein-Smolin 97]}:
S. L. Braunstein, \& J. A. Smolin,
``Perfect quantum-error-correction coding in 24 laser pulses'',
{\em Phys. Rev. A} {\bf 55}, 2, 945-950 (1997).

\item {\bf [Braunstein-Kimble 98 a]}:
S. L. Braunstein, \& H. J. Kimble
``Teleportation of continuous quantum variables'',
{\em Phys. Rev. Lett.} {\bf 80}, 4, 869-872 (1998).
Reprinted in {\bf [Macchiavello-Palma-Zeilinger 00]}, pp.~45-48.

\item {\bf [Braunstein 98 a]}:
S. L. Braunstein,
``Error correction for continuous quantum variables'',
{\em Phys. Rev. Lett.} {\bf 80}, 18, 4084-4087 (1998);
quant-ph/9711049.

\item {\bf [Braunstein 98 b]}:
S. L. Braunstein,
``Quantum error correction for communication with
linear optics'',
{\em Nature} {\bf 394}, 6688, 47-49 (1998).

\item {\bf [Braunstein-Kimble 98 b]}:
S. L. Braunstein, \& H. J. Kimble
``{\em A posteriori} teleportation'',
{\em Nature} {\bf 394}, 6696, 840-841 (1998);
quant-ph/9810001.
Comment on {\bf [Bouwmeester-Pan-Mattle-(+3) 97]}.
Reply: {\bf [Bouwmeester-Pan-Daniell-(+3) 98]}.

\item {\bf [Braunstein-Caves-Jozsa-(+3) 99]}:
S. L. Braunstein,
C. M. Caves, R. Jozsa, N. Linden, S. Popescu, \& R. Schack,
``Separability of very noisy mixed states and implications for NMR quantum
computing'',
{\em Phys. Rev. Lett.} {\bf 83}, 5, 1054-1057 (1999);
quant-ph/9811018.

\item {\bf [Braunstein-Fuchs-Kimble 00]}:
S. L. Braunstein, C. A. Fuchs, \& H. J. Kimble,
``Criteria for continuous-variable quantum teleportation'',
in V. Bu\v{z}zek, \& D. P. DiVincenzo (eds.),
{\em J. Mod. Opt.} {\bf 47}, 2-3 (Special issue:
Physics of quantum information), 267-278 (2000);
quant-ph/9910030.

\item {\bf [Braunstein-Kimble 00]}:
S. L. Braunstein, \& H. J. Kimble,
``Dense coding for continuous variables'',
{\em Phys. Rev. A} {\bf 61}, 4, 042302 (2000);
quant-ph/9910010.

\item {\bf [Braunstein-D'Ariano-Milburn-Sacchi 00]}:
S. L. Braunstein, G. M. D'Ariano, G. J. Milburn, \& M. F. Sacchi,
``Universal teleportation with a twist'',
{\em Phys. Rev. Lett.} {\bf 84}, 15, 3486-3489 (2000);
quant-ph/9908036.

\item {\bf [Braunstein-Pati 00]}:
S. L. Braunstein, \& A. K. Pati,
``Speed-up and entanglement in quantum searching'',
quant-ph/0008018.

\item {\bf [Braunstein-Fuchs-Kimble-van Loock 01]}:
S. L. Braunstein, C. A. Fuchs, H. J. Kimble, \& P. van Loock,
``Quantum versus classical domains for
teleportation with continuous variables'',
{\em Phys. Rev. A} {\bf 64}, 2, 022321 (2001);
quant-ph/0012001.

\item {\bf [Braunstein-Bu\v{z}ek-Hillery 01]}:
S. L. Braunstein, V. Bu\v{z}ek, \& M. Hillery,
``Quantum information distributors: Quantum network for symmetric and
asymmetric cloning in arbitrary dimension and continuous limit'',
{\em Phys. Rev. A} {\bf 63}, 5, 052313 (2001);
quant-ph/0009076.

\item {\bf [Braunstein-Cerf-Iblisdir-(+2) 01]}:
S. L. Braunstein, N. J. Cerf, S. Iblisdir,
P. van Loock, \& S. Massar,
``Optimal cloning of coherent states with a
linear amplifier and beam splitters'',
{\em Phys. Rev. Lett.} {\bf 86}, 21, 4938-4941 (2001);
quant-ph/0012046.

\item {\bf [Braunstein-Lo 01]}:
S. L. Braunstein, \& H.-K. Lo (eds.),
{\em Scalable quantum computers: Paving the way to realization},
Willey-VCH, New York, 2001.

\item {\bf [Braunstein 02]}:
S. L. Braunstein,
``Quantum teleportation'',
{\em Proc.\ of the Symp.\ ``100 Years Werner Heisenberg---Works and Impact'' (Bamberg, Germany, 2001)},
{\em Fortschr. Phys.} {\bf 50}, 5-7, 608-613 (2002).

\item {\bf [Braunstein-Ghosh-Severini 04]}:
S. L. Braunstein, S. Ghosh, \& S. Severini,
``The laplacian of a graph as a density matrix: a basic combinatorial
approach to separability of mixed states'',
quant-ph/0406165.

\item {\bf [Braunstein-van Loock 04]}:
S. L. Braunstein, \& P. van Loock,
``Quantum information with continuous variables'',
{\em Rev. Mod. Phys.};
quant-ph/0410100.

\item {\bf [Bravyi-Kitaev 00]}:
S. Bravyi, \& A. Kitaev,
``Fermionic quantum computation'',
quant-ph/0003137.

\item {\bf [Bravyi 03]}:
S. Bravyi,
``Entanglement entropy of multipartite pure states'',
{\em Phys. Rev. A} {\bf 67}, 1, 012313 (2003).

\item {\bf [Bravyi-Kitaev 04]}:
S. Bravyi, \& A. Kitaev,
``Universal quantum computation based on a magic states distillation'',
quant-ph/0403025.

\item {\bf [Bray-Moore 82]}:
A. J. Bray, \& M. A. Moore,
``Influence of dissipation on quantum coherence'',
{\em Phys. Rev. Lett.} {\bf 49}, 21, 1545-1549 (1982).

\item {\bf [Brazier-Plenio 03]}:
A. Brazier, \& M. B. Plenio,
``Broken promises and quantum algorithms'',
quant-ph/0304017.

\item {\bf [Breitenberger 65]}:
E. Breitenberger,
``On the so-called paradox of Einstein, Podolsy, and Rosen'',
{\em Nuovo Cimento} {\bf 38}, 1, 356-360 (1965).

\item {\bf [Breguet-Muller-Gisin 94]}:
J. Breguet, A. Muller, \& N. Gisin,
``Quantum cryptography with polarized photons in optical fibres.
Experimental and practical limits'',
in S. M. Barnett, A. K. Ekert, \& S. J. D. Phoenix (eds.),
{\em J. Mod. Opt.} {\bf 41}, 12 (Special issue: Quantum
communication), 2405-2412 (1994).

\item {\bf [Bremner-Dawson-Dodd-(+5) 02]}:
M. J. Bremner, C. M. Dawson, J. L. Dodd,
A. Gilchrist, A. W. Harrow, D. Mortimer, M. A. Nielsen, \& T. J. Osborne,
``Practical scheme for quantum computation with any two-qubit entangling gate'',
{\em Phys. Rev. Lett.} {\bf 89}, 24, 247902 (2002).

\item {\bf [Bremner-Dodd-Nielsen-Bacon 04]}:
M. J. Bremner, J. L. Dodd, M. A. Nielsen, \& D. Bacon,
``Fungible dynamics: There are only two types of entangling multiple-qubit interactions'',
{\em Phys. Rev. A} {\bf 69}, 1, 012313 (2004);
quant-ph/0307148.

\item {\bf [Brendel-Gisin-Tittel-Zbinden 99]}:
J. Brendel, N. Gisin, W. Tittel, \& H. Zbinden,
``Pulsed energy-time entangled twin-photon source for quantum communication'',
{\em Phys. Rev. Lett.} {\bf 82}, 12, 2594-2597 (1999);
quant-ph/9809034.
Comment: {\bf [Durt 01 a]}.
Reply: {\bf [Gisin-Tittel-Zbinden 01]}.

\item {\bf [Brennen-Caves-Jessen-Deutsch 99]}:
G. K. Brennen, C. M. Caves, P. S. Jessen, \& I. H. Deutsch,
``Quantum logic gates in optical lattices'',
{\em Phys. Rev. Lett.} {\bf 82}, 5, 1060-1063 (1999);
quant-ph/9806021.

\item {\bf [Brennen-Deutsch-Jessen 00]}:
G. K. Brennen, I. H. Deutsch, \& P. S. Jessen,
``Entangling dipole-dipole interactions for quantum logic with neutral
atoms'',
{\em Phys. Rev. A} {\bf 61}, 6, 062309 (2000).

\item {\bf [Brennen-Deutsch-Williams 02]}:
G. K. Brennen, I. H. Deutsch, \& C. J. Williams,
``Quantum logic for trapped atoms via molecular hyperfine interactions'',
{\em Phys. Rev. A} {\bf 65}, 2, 022313 (2002);
quant-ph/0107136.

\item {\bf [Brennen-Song-Williams 03]}:
G. K. Brennen, D. Song, \& C. J. Williams,
``Quantum-computer architecture using nonlocal interactions'',
{\em Phys. Rev. A} {\bf 67}, 5, 050302 (2003);
quant-ph/0301012.

\item {\bf [Brennen 03]}:
G. K. Brennen,
``An observable measure of entanglement for pure states of multi-qubit
systems'',
{\em Quant. Inf. Comp.} {\bf 3}, 6, 619-626 (2003);
quant-ph/0305094.
See {\bf [Meyer-Wallach 02]}.

\item {\bf [Brennen-Williams 03]}:
G. K. Brennen, \& J. E. Williams,
``Entanglement dynamics in one-dimensional quantum cellular automata'',
{\em Phys. Rev. A} {\bf 68}, 4, 042311 (2003).

\item {\bf [Breuer-Petruccione 01]}:
H.-P. Breuer, \& F. Petruccione,
``Destruction of quantum coherence through emission of
bremsstrahlung'',
{\em Phys. Rev. A} {\bf 63}, 3, 032102 (2001).

\item {\bf [Breuer 02 a]}:
T. Breuer,
``A Kochen-Specker theorem for unsharp spin 1 observables'',
in T. Placek, \& J. Butterfield (eds.),
{\em Non-locality and modality},
Kluwer Academic, Dordrecht, Holland, 2002, pp.~195-203;
quant-ph/0203103.

\item {\bf [Breuer 02 b]}:
T. Breuer,
``Kochen-Specker theorem for finite precision spin-one measurements'',
{\em Phys. Rev. Lett.} {\bf 88}, 24, 240402 (2002);
quant-ph/0206035.

\item {\bf [Brezger-Hackerm\"{u}ller-Uttenthaler-(+3) 02]}:
B. Brezger, L. Hackerm\"{u}ller, S. Uttenthaler,
J. Petschinka, M. Arndt, \& A. Zeilinger,
``Matter-wave interferometer for large molecules'',
{\em Phys. Rev. Lett.} {\bf 88}, 10, 100404 (2002);
quant-ph/0202158.
Comment: {\bf [Stadler-Hofer 02]}.

\item {\bf [Briand-Luque-Thibon-Verstraete 03]}:
E. Briand, J.-G. Luque, J.-Y. Thibon, \& F. Verstraete,
``The moduli space of three qutrit states'',
quant-ph/0306122.

\item {\bf [Briand-Luque-Thibon 03]}:
E. Briand, J.-G. Luque, \& J.-Y. Thibon,
``A complete set of covariants of the four qubit system'',
{\em J. Phys. A} {\bf 36}, 38, 9915-9927 (2003).

\item {\bf [Brida-Genovese-Novero 99]}:
G. Brida, M. Genovese, \& C. Novero,
``An application of two photon entangled states to quantum
metrology'',
{\em J. Mod. Opt.};
quant-ph/9911032.

\item {\bf [Brida-Genovese-Novero-Predazzi 00]}:
G. Brida, M. Genovese, C. Novero, \& E. Predazzi,
``New experimental test of Bell inequalities by the
use of a non-maximally entangled photon state'',
{\em Phys. Lett. A} {\bf 268}, 1-2, 12-16 (2000);
quant-ph/0003012.
See {\bf [Brida-Genovese-Novero-Predazzi 00 b]}.

\item {\bf [Brida-Genovese-Novero-Predazzi 00]}:
G. Brida, M. Genovese, C. Novero, \& E. Predazzi,
``A new conception experimental test of Bell inequalities
using non-maximally entangled states'',
to be published in
{\em Proc.\ of Int. Workshop on Optics and Spectroscopy (Hanoi,
Vietnam)};
quant-ph/0004033.
See {\bf [Brida-Genovese-Novero-Predazzi 00 a]}.

\item {\bf [Brida-Genovese-Gramegna-(+2) 02]}:
G. Brida, M. Genovese, M. Gramegna, C. Novero, \& E. Predazzi,
``A first test of Wigner function local realistic model'',
{\em Phys. Lett. A} {\bf 299}, 2-3, 121-124 (2002);
quant-ph/0203048.

\item {\bf [Brida-Cagliero-Falzetta-(+3) 03]}:
G. Brida, E. Cagliero, G. Falzetta,
M. Genovese M. Gramegna, \& E. Predazzi,
``A biphotons double slit experiment'',
{\em Phys. Rev. A} {\bf 68}, 3, 033803 (2003).
quant-ph/0310020.

\item {\bf [Brida-Genovese-Gramegna-Predazzi 04]}:
G. Brida, M. Genovese, M. Gramegna, \& E. Predazzi,
`A conclusive experiment to throw more light on ``light''\',
{\em Phys. Lett. A} {\bf 328}, 4-5, 313-318 (2004).

\item {\bf [Briegel-D\"{u}r-van Enk-(+2) 98]}:
H.-J. Briegel, W. D\"{u}r, S. J. van Enk, J. I. Cirac, \& P. Zoller,
``Quantum communication and the
creation of maximally entangled pair of atoms over a noisy channel'',
in A. K. Ekert, R. Jozsa, \& R. Penrose (eds.),
{\em Quantum Computation: Theory and Experiment.
Proceedings of a Discussion Meeting held at the Royal
Society of London on 5 and 6 November 1997},
{\em Philos. Trans. R. Soc. Lond. A} {\bf 356}, 1743, 1841-1852 (1998);
quant-ph/9712027.

\item {\bf [Briegel-D\"{u}r-Cirac-Zoller 98]}:
H.-J. Briegel, W. D\"{u}r, J. I. Cirac, \& P. Zoller,
``Quantum repeaters: The role of imperfect local operations in communication'',
{\em Phys. Rev. Lett.} {\bf 81}, 26, 5932-5935 (1998);
quant-ph/9803056.
Reprinted in {\bf [Macchiavello-Palma-Zeilinger 00]}, pp.~217-220.

\item {\bf [Briegel-Cirac-D\"{u}r-(+2) 99]}:
H.-J. Briegel, J. I. Cirac, W. D\"{u}r, G. Giedke, \& P. Zoller,
``?'',
in {\bf [Greenberger-Reiter-Zeilinger 99]}, pp.~147-154.

\item {\bf [Briegel 00 a]}:
H.-J. Briegel,
``Long-distance quantum communication'',
{\bf [Macchiavello-Palma-Zeilinger 00]}, pp.~213-216.

\item {\bf [Briegel 00 b]}:
H.-J. Briegel,
``Quantum information in optical lattices'',
{\bf [Macchiavello-Palma-Zeilinger 00]}, pp.~401-403.

\item {\bf [Briegel-Calarco-Jaksch-(+2) 00]}:
H.-J. Briegel, T. Calarco, D. Jaksch,
J. I. Cirac, \& P. Zoller,
``Quantum computing with neutral atoms'',
in V. Bu\v{z}zek, \& D. P. DiVincenzo (eds.),
{\em J. Mod. Opt.} {\bf 47}, 2-3 (Special issue:
Physics of quantum information), 415-451 (2000).
Reprinted in {\bf [Macchiavello-Palma-Zeilinger 00]}, pp.~404-424.

\item {\bf [Briegel-Raussendorf 01]}:
H.-J. Briegel, \& R. Raussendorf,
``Persistent entanglement in arrays of interacting particles'',
{\em Phys. Rev. Lett.} {\bf 86}, 5, 910-913 (2001);
quant-ph/0004051.

\item {\bf [Brif-Mann-Revzen 98 a]}:
C. Brif, A. Mann, \& M. Revzen,
``Generalized coherent states are unique Bell states of quantum
systems with Lie-group symmetries'',
{\em Phys. Rev. A} {\bf 57}, 2, 742-745 (1998);
quant-ph/9707019.

\item {\bf [Brif-Mann-Revzen 98 b]}:
C. Brif, A. Mann, \& M. Revzen,
``Classical properties of generalized coherent states:
From phase-space dynamics to Bell's inequality'',
in H. Ezawa (ed.),
{\em 2nd Jagna Interational Workshop
``Mathematical Methods of Quantum Physics'' (Jagna, Philippines, 1998)};
quant-ph/9805007.

\item {\bf [Brif-Mann 00]}:
C. Brif, \& A. Mann,
``Testing Bell's inequality with two-level atoms via population
spectroscopy'',
{\em Europhys. Lett.} {\bf 49}, 1, 1-7 (2000).

\item {\bf [Bririd-Benjamin 03]}:
A. Bririd, \& S. C. Benjamin,
``Quantum error correction in globally controlled arrays'',
quant-ph/0308113.

\item {\bf [Brody-Meister 96]}:
D. C. Brody, \& B. Meister,
``Minimum decision cost for quantum ensembles'',
{\em Phys. Rev. Lett.} {\bf 76}, 1, 1-5 (1996).

\item {\bf [Brody-Hughston 00]}:
D. C. Brody, \& L. P. Hughston,
``Information content for quantum states'',
{\em J. Math. Phys.} {\bf 41}, 5, 2586-2592 (2000);
quant-ph/9906085.

\item {\bf [Brody-Hughston 01]}:
D. C. Brody, \& L. P. Hughston,
``Experimental tests for stochastic reduction models'',
quant-ph/0104032.

\item {\bf [Brody-Hughston 02]}:
D. C. Brody, \& L. P. Hughston,
``Efficient simulation of quantum state reduction'',
quant-ph/0203035.

\item {\bf [de Broglie 23 a]}:
L. de Broglie,
``Ondes et quanta'',
{\em C. R. Acad. Sci. Paris} {\bf 177}, 507-510 (1923).

\item {\bf [de Broglie 23 b]}:
L. de Broglie,
``Quanta de lumi\`{e}re, diffraction et ineterf\'{e}rences'',
{\em C. R. Acad. Sci. Paris} {\bf 177}, 548-550 (1923).

\item {\bf [de Broglie 23 c]}:
L. de Broglie,
``Les quanta, la th\'{e}orie cin\'{e}tique des gaz et le
principie de Fermat'',
{\em C. R. Acad. Sci. Paris} {\bf 177}, 630-632 (1923).

\item {\bf [de Broglie 24 a]}:
L. de Broglie,
``Recherche sur la th\'{e}orie des quanta'',
Ph.\ D. thesis, Universit\'{e} de Paris, 1924,
Masson et Cie., Paris, 1924 (reprinted in 1963);
{\em Ann. de Phys.} {\bf 3}, 22-128 (1925).

\item {\bf [de Broglie 24 b]}:
L. de Broglie,
``A tentative theory of light quanta'',
{\em Phil. Mag.} {\bf 47}, 446-458 (1924).

\item {\bf [de Broglie 24 c]}:
L. de Broglie,
``Sur la d\'{e}finition g\'{e}n\'{e}rale de la
correspondance entre onde et mouvement'',
{\em C. R. Acad. Sci. Paris} {\bf 179}, 39 (1924).

\item {\bf [de Broglie 24 d]}:
L. de Broglie,
``Sur un th\'{e}or\`{e}me de Bohr'',
{\em C. R. Acad. Sci. Paris} {\bf 179}, 676 (1924).

\item {\bf [de Broglie 24 e]}:
L. de Broglie,
``Sur la dynamique du quantum de lumi\`{e}re et les interf\'{e}rence'',
{\em C. R. Acad. Sci. Paris} {\bf 179}, 1029 (1924).

\item {\bf [de Broglie 26 a]}:
L. de Broglie,
``?'',
{\em C. R. Acad. Sci. Paris} {\bf 183}, 447-448 (1926).

\item {\bf [de Broglie 26 b]}:
L. de Broglie,
{\em Ondes et mouvements},
Gauthier-Villars, Paris, 1926.

\item {\bf [de Broglie 28]}:
L. de Broglie,
``La nouvelle dynamique des quanta'',
in {\bf [Solvay 28]}, pp.~?-?.

\item {\bf [de Broglie 30]}:
L. de Broglie,
{\em Introduction \`{a} l'\'{e}tude de la
m\'{e}canique ondulatoire},
Hermann, Paris, 1930.

\item {\bf [de Broglie 53 a]}:
L. de Broglie,
{\em La physique quantique restera-t-elle ind\'{e}terministe?},
Gauthier-Villars, Paris, 1953.

\item {\bf [de Broglie 53 b]}:
L. de Broglie,
{\em The revolution in physics},
Noonday Press, New York, 1953.

\item {\bf [de Broglie 56]}:
L. de Broglie,
{\em Une tentative d'interpr\'{e}tation causale et nonlin\'{e}aire
de la m\'{e}canique ondulatoire},
Gauthier-Villars, Paris, 1956.
English version:
{\em Non-linear wave mechanics: A causal interpretation},
Elsevier, Amsterdam, 1960.

\item {\bf [de Broglie 71]}:
L. de Broglie,
``L'interpr\'{e}tation de la m\'{e}canique ondulatoire par
la th\'{e}orie de la double solution'',
in {\bf [d'Espagnat 71]}, pp.~346-367.

\item {\bf [de Broglie 74]}:
L. de Broglie,
``?'',
{\em C. R. Acad. Sci. Paris B} {\bf 278}, ?, 721-? (1974).

\item {\bf [de Broglie 78]}:
L. de Broglie,
``R\'{e}futation du th\'{e}or\`{e}me de Bell'',
in L. de Broglie,
{\em Jalons pour une nouvelle microphysique},
Gauthier-Villars, Paris, 1978, pp.~147-153.

\item {\bf [de Broglie 83]}:
L. de Broglie,
``On the true ideas underlying wave mechanics'',
in A. van der Merwe (ed.),
{\em Old and new questions in physics, cosmology, philosophy, and theoretical biology.
Essays in honor of Wolfgang Yourgrau},
Plenum Press, New York, 1983.

\item {\bf [Brooks 99]}:
M. Brooks (ed.),
{\em Quantum computing and communications},
Springer-Verlag, Berlin, 1999.

\item {\bf [Brouard-Plata 03]}:
S. Brouard, \& J. Plata,
``Internal-state dephasing of trapped ions'',
{\em Phys. Rev. A} {\bf 68}, 1, 012311 (2003).

\item {\bf [Brouard-Plata 04]}:
S. Brouard, \& J. Plata,
``Decoherence of trapped-ion internal and vibrational modes: The effect of fluctuating magnetic fields'',
{\em Phys. Rev. A} {\bf 70}, 1, 013413 (2004).

\item {\bf [Brown-Redhead 81]}:
H. R. Brown, \& M. L. G. Redhead,
``A critique of the disturbance theory of indeterminacy in quantum mechanics'',
{\em Found. Phys.} {\bf 11}, 1-2, 1-20 (1981).

\item {\bf [Brown 86]}:
H. R. Brown,
``The insolubility proof of the quantum measurement problem'',
{\em Found. Phys.} {\bf 16}, 9, 857-870 (1986).

\item {\bf [Brown-Svetlichny 90]}:
H. R. Brown, \& G. Svetlichny,
``Nonlocality and Gleason's lemma. Part I. Deterministic theories'',
{\em Found. Phys.} {\bf 20}, 11, 1379-1387 (1990).
See {\bf [Elby 90 a]} (II).

\item {\bf [Brown 93]}:
H. R. Brown,
``Bell's other theorem and its connection with nonlocality. Part I'',
in A. van der Merwe, \& F. Selleri (eds.),
{\em Bell's theorem and the foundations of modern physics.
Proc.\ of an international conference (Cesena, Italy, 1991)},
World Scientific, Singapore, 1993, pp.~104-116.

\item {\bf [Brown 98]}:
H. R. Brown,
``Aspects of objectivity in quantum mechanics'' (1998),
PITT-PHIL-SCI00000223.

\item {\bf [Brown-Sj\"{o}qvist-Bacciagaluppi 99]}:
H. R. Brown, E. Sj\"{o}qvist, \& G. Bacciagaluppi,
``Remarks on identical particles in de Broglie-Bohm theory'',
{\em Phys. Lett. A} {\bf 251}, 4, 229-235 (1999);
quant-ph/9811054.

\item {\bf [Brown 94]}:
J. P. Brown,
``A quantum revolution for computing'',
{\em New Scientist} {\bf 143}, 1944, 21-24 (1994).

\item {\bf [Brown 01]}:
J. Brown,
{\em Quest for the quantum computer},
Simon \& Schuster, New York, 2001.

\item {\bf [Brown-Lidar-Whaley 02]}:
K. R. Brown, D. A. Lidar, \& K. B. Whaley,
``Quantum computing with quantum dots on quantum linear supports'',
{\em Phys. Rev. A} {\bf 65}, 1, 012307 (2002);
quant-ph/0105102.

\item {\bf [Brown-Vala-Whaley 03]}:
K. R. Brown, J. Vala, \& K. B. Whaley,
``Scalable ion trap quantum computation in decoherence-free subspaces with
pairwise interactions only'',
{\em Phys. Rev. A} {\bf 67}, 1, 012309 (2003).

\item {\bf [Brown-Dani-Stamper Kurn-Whaley 03]}:
K. R. Brown, K. M. Dani, D. M. Stamper-Kurn, \& K. B. Whaley,
``Deterministic optical Fock-state generation'',
{\em Phys. Rev. A} {\bf 67}, 4, 043818 (2003).

\item {\bf [Brown-Harrow-Chuang 04]}:
K. Brown, A. Harrow, \& I. L. Chuang,
``Arbitrarily accurate composite pulses'',
quant-ph/0407022.

\item {\bf [Brown-Hiley 00]}:
M. R. Brown, \& B. J. Hiley,
``Schrodinger revisited:
The role of Dirac's 'standard' ket in the algebraic approach'',
quant-ph/0005026.

\item {\bf [Brown-Hiley 04]}:
M. R. Brown, \& B. J. Hiley,
``Solving the measurement problem: de Broglie-Bohm loses out to
Everett'',
{\em Found. Phys.},
quant-ph/0403094.

\item {\bf [Brown 04]}:
M. S. Brown,
``Classical cryptosystems in a quantum setting'',
M.\ Sc. thesis, University of Waterloo, 2004;
quant-ph/0404061.

\item {\bf [Browne-Plenio 02]}:
D. E. Browne, \& M. B. Plenio,
``Symmetric qubits from cavity states'',
{\em Phys. Rev. A} {\bf 66}, 4, 042307 (2002);
quant-ph/0206141.

\item {\bf [Browne-Plenio 03]}:
D. E. Browne, \& M. B. Plenio,
``Robust generation of entanglement between two cavities mediated by short
interactions with an atom'',
{\em Phys. Rev. A} {\bf 67}, 1, 012325 (2003);
quant-ph/0209115.

\item {\bf [Browne-Plenio-Fern\'{a}ndez Huelga 03]}:
D. E. Browne, M. B. Plenio, \& S. G. Fern\'{a}ndez Huelga,
``Robust creation of entanglement between ions in spatially separate cavities'',
{\em Phys. Rev. Lett.} {\bf 91}, 6, 067901 (2003);
quant-ph/0302185.

\item {\bf [Browne-Eisert-Scheel-Plenio 03]}:
D. E. Browne, J. Eisert, S. Scheel, \& M. B. Plenio,
``Driving non-Gaussian to Gaussian states with linear optics'',
{\em Phys. Rev. A} {\bf 67}, 6, 062320 (2003);
quant-ph/0211173.

\item {\bf [Browne-Rudolph 04]}:
D. E. Browne, \& T. Rudolph,
``Efficient linear optical quantum computation'',
quant-ph/0405157.

\item {\bf [Bruce 01]}:
S. Bruce,
``Discrete time in quantum mechanics'',
{\em Phys. Rev. A} {\bf 64}, 1, 014103 (2001).

\item {\bf [de Bruijn-Erd\"{o}s 51]}:
N. G. de Bruijn, \& P. Erd\"{o}s,
``A color problem for infinite graphs and a problem in the theory of relations'',
{\em Proc.\ Nederl. Akad. Wetensh. A} {\bf 54}, 371-373 (1951).

\item {\bf [Brukner-Zeilinger 99 a]}:
\v{C}. Brukner, \& A. Zeilinger,
``Operationally invariant information in quantum measurements'',
{\em Phys. Rev. Lett.} {\bf 83}, 17, 3354-3357 (1999);
quant-ph/0005084.

\item {\bf [Brukner-Zeilinger 99 b]}:
\v{C}. Brukner, \& A. Zeilinger,
``Quantum complementarity and information invariance'',
in {\bf [Greenberger-Reiter-Zeilinger 99]}, pp.~231-234.

\item {\bf [Brukner-Zeilinger 01 a]}:
\v{C}. Brukner, \& A. Zeilinger,
``Conceptual inadequacy of Shannon information
in quantum measurements'',
{\em Phys. Rev. A} {\bf 63}, 2, 022113 (2001).
Erratum: {\em Phys. Rev. A} {\bf 67}, 4, 049901 (2003);
quant-ph/0006087.
Comment: {\bf [Hall 00 b]}.
Reply: {\bf [Brukner-Zeilinger 00 b]}.
See: {\bf [Timpson 00]}, {\bf [Mana 03]}.

\item {\bf [Brukner-Zeilinger 01 b]}:
\v{C}. Brukner, \& A. Zeilinger,
``Quantum measurement and Shannon information,
a reply to M. J. W. Hall'',
quant-ph/0008091.
Reply to {\bf [Hall 00 b]}.
See {\bf [Brukner-Zeilinger 00 a]}.

\item {\bf [Brukner-\.{Z}ukowski-Zeilinger 01]}:
\v{C}. Brukner, M. \.{Z}ukowski, \& A. Zeilinger,
``The essence of entanglement'',
quant-ph/0106119.

\item {\bf [Brukner-Pan-Simon-(+2) 01]}:
\v{C}. Brukner, J.-W. Pan, C. Simon, G. Weihs, \& A. Zeilinger,
``Probabilistic instantaneous quantum computation'',
{\em Phys. Rev. A} {\bf 67}, 3, 034304 (2003);
quant-ph/0109022.

\item {\bf [Brukner-Zeilinger 02 a]}:
\v{C}. Brukner, \& A. Zeilinger,
``Young's experiment and the finiteness of information'',
{\em Proc. Roy. Soc. Lond. A};
quant-ph/0201026.

\item {\bf [Brukner-\.{Z}ukowski-Zeilinger 02]}:
\v{C}. Brukner, M. \.{Z}ukowski, \& A. Zeilinger,
``Quantum communication complexity protocol with two entangled qutrits'',
{\em Phys. Rev. Lett.} {\bf 89}, 19, 197901 (2002).

\item {\bf [Brukner-Zeilinger 02 b]}:
\v{C}. Brukner, \& A. Zeilinger,
``Information and fundamental elements of the structure of quantum theory'',
quant-ph/0212084.

\item {\bf [Brukner-Kim-Pan-Zeilinger 03]}:
\v{C}. Brukner, M. S. Kim, J.-W. Pan, \& A. Zeilinger,
``Correspondence between continuous-variable and discrete quantum systems of arbitrary dimensions'',
{\em Phys. Rev. A} {\bf 68}, 6, 062105 (2003).

\item {\bf [Brukner-Laskowski-\.{Z}ukowski 03]}:
\v{C}. Brukner, W. Laskowski, \& M. \.{Z}ukowski,
``Multiparticle Bell's inequalities involving many measurement settings'',
quant-ph/0303187.
See {\bf [Laskowski-Paterek-\.{Z}ukowski-Brukner 04]}.

\item {\bf [Brukner-Taylor-Cheung-Vedral 04]}:
\v{C}. Brukner, S. Taylor, S. Cheung, \& V. Vedral,
``Quantum entanglement in time'',
quant-ph/0402127.

\item {\bf [Brukner-Aspelmeyer-Zeilinger 04]}:
\v{C}. Brukner, M. Aspelmeyer, \& A. Zeilinger,
`Complementarity and information in ``Delayed-choice for entanglement
swapping'',\',
{\em Found. Phys.} (Festschrift in honor of Asher Peres);
quant-ph/0405036.

\item {\bf [Brukner-Vedral 04]}:
\v{C}. Brukner, \& V. Vedral,
``Macroscopic thermodynamical witnesses of quantum entanglement'',
quant-ph/0406040.

\item {\bf [Brukner-\.{Z}ukowski-Pan-Zeilinger 04]}:
\v{C}. Brukner, M. \.{Z}ukowski, J. W. Pan, \& A. Zeilinger,
``Bell's inequalities and quantum communication complexity'',
{\em Phys. Rev. Lett.} {\bf 92}, 12, 127901 (2004);
quant-ph/0210114.

\item {\bf [Brukner-Vedral-Zeilinger 04]}:
\v{C}. Brukner, V. Vedral, \& A. Zeilinger,
``Crucial role of quantum entanglement in bulk properties of solids'',
quant-ph/0410138.

\item {\bf [Brumby-Joshi-Anderson 95]}:
S. P. Brumby, G. C. Joshi, \& R. Anderson,
``Multiparticle correlations in quaternionic quantum systems'',
{\em Phys. Rev. A} {\bf 51}, 2, 976-981 (1995).

\item {\bf [Brun 98]}:
T. A. Brun,
``Quantum jumps as decoherent histories'',
{\em Phys. Rev. Lett.} {\bf 78}, 10, 1833-1837 (1998).

\item {\bf [Brun-Barnett 98]}:
T. A. Brun, \& S. M. Barnett,
``Interference in dielectrics and pseudo-measurements'',
{\em J. Mod. Opt.} {\bf 45}, 4, 777-783 (1998).

\item {\bf [Brun-Schac 99]}:
T. A. Brun, \& R. Schack,
``Realizing the quantum baker's map on a NMR quantum computer'',
{\em Phys. Rev. A} {\bf 59}, 4, 2649-2658 (1999);
quant-ph/9807050.

\item {\bf [Brun-Wang 00]}:
T. A. Brun, \& H. Wang,
``Coupling nanocrystals to a high-$Q$ silica microsphere:
Entanglement in quantum dots via photon exchange'',
{\em Phys. Rev. A} {\bf 61}, 3, 032307 (2000).

\item {\bf [Brun 00]}:
T. A. Brun,
``Continuous measurements, quantum trajectories, and decoherent histories'',
{\em Phys. Rev. A} {\bf 61}, 4, 042107 (2000).

\item {\bf [Brun-Griffiths 00]}:
T. A. Brun, \& R. B. Griffiths,
``Quantum theory -- Interpretation, formulation, inspiration'',
{\em Phys. Today} {\bf 53}, 9, ? (2000).
Comment on {\bf [Fuchs-Peres 00 a]}.
Reply: {\bf [Fuchs-Peres 00 b]}.

\item {\bf [Brun-Cohen 01]}:
T. A. Brun, \& O. Cohen,
``Parametrization and distillability of three-qubit
entanglement'',
{\em Phys. Lett. A} {\bf 281}, 2-3, 88-100 (2001);
quant-ph/0005124.

\item {\bf [Brun-Caves-Schack 01]}:
T. A. Brun, C. M. Caves, \& R. Schack,
``Entanglement purification of unknown quantum states'',
{\em Phys. Rev. A} {\bf 63}, 4, 042309 (2001);
quant-ph/0010038.

\item {\bf [Brun 01]}:
T. A. Brun,
``A quantum web page'',
submitted to special issue of {\em Algorithmica},
quant-ph/0102046.

\item {\bf [Brun 02]}:
T. A. Brun,
``A simple model of quantum trajectories'',
{\em Am. J. Phys.} {\bf 70}, 7, 719-737 (2002);
quant-ph/0108132.

\item {\bf [Brun-Finkelstein-Mermin 02]}:
T. A. Brun, J. Finkelstein, \& N. D. Mermin,
``How much state assignments can differ'',
{\em Phys. Rev. A} {\bf 65}, 3, 032315 (2002);
quant-ph/0109041.

\item {\bf [Brun-Carteret-Ambainis 03 a]}:
T. A. Brun, H. A. Carteret, \& A. Ambainis,
``Quantum random walks with decoherent coins'',
{\em Phys. Rev. A} {\bf 67}, 3, 032304 (2003);
quant-ph/0210180.

\item {\bf [Brun-Carteret-Ambainis 03 b]}:
T. A. Brun, H. A. Carteret, \& A. Ambainis,
``Quantum walks driven by many coins'',
{\em Phys. Rev. A} {\bf 67}, 5, 052317 (2003).

\item {\bf [Brun 03 a]}:
T. A. Brun,
``Computers with closed timelike curves can solve hard problems efficiently'',
{\em Found. Phys. Lett.} {\bf 16}, 4, 245-253 (2003).

\item {\bf [Brun-Carteret-Ambainis 03 c]}:
T. A. Brun, H. A. Carteret, \& A. Ambainis,
``Quantum to classical transition for random walks'',
{\em Phys. Rev. Lett.} {\bf 91}, 13, 130602 (2003).

\item {\bf [Brun 03 b]}:
T. A. Brun,
``Probability in decoherent histories'',
in {\em Foundations of Quantum Mechanics and Probability (V\"{a}xj\"{o}, Sweden, 2002)};
quant-ph/0302034.

\item {\bf [Brun-Goan 03]}:
T. A. Brun, \& H.-S. Goan,
``Realistic simulations of single-spin nondemolition measurement by magnetic
resonance force microscopy'',
{\em Phys. Rev. A} {\bf 68}, 3, 032301 (2003);
quant-ph/0302178.

\item {\bf [Brun-Klauck-Nayak-Zalka 03]}:
T. A. Brun, H. Klauck, A. Nayak, \& C. Zalka,
`Comment on ``Probabilistic quantum memories''\,',
{\em Phys. Rev. Lett.} {\bf 91}, 2, 209801 (2003);
quant-ph/0303091.
Comment on {\bf [Trugenberger 00]}.

\item {\bf [Brun 04]}:
T. A. Brun,
``Measuring polynomial functions of states'',
quant-ph/0401067.

\item {\bf [Brune-Hagley-Dreyer-(+5) 96]}:
M. Brune, E. Hagley, J. Dreyer, X. Ma\^{\i}tre, A. Maali, C.
Wunderlich, J.-M. Raimond, \& S. Haroche,
`Observing the progressive decoherence of
the ``meter'' in a quantum measurement',
{\em Phys. Rev. Lett.} {\bf 77}, 24, 4887-4890 (1996).
Reprinted in {\bf [Macchiavello-Palma-Zeilinger 00]}, pp.~302-305.
See {\bf [Davidovich-Brune-Raimond-Haroche 96]}, {\bf [Zurek 97]}.

\item {\bf [Brunner-Ac\'{\i}n-Collins-(+2) 03]}:
N. Brunner, A. Ac\'{\i}n, D. Collins, N. Gisin, \& V. Scarani,
``Optical telecom networks as weak quantum measurements with post-selection'',
quant-ph/0306108.
Extended version: {\bf [Brunner 03]}.

\item {\bf [Brunner 03]}:
N. Brunner,
``Optical telecom networks as weak measurements with post-selection'',
diploma thesis,
quant-ph/0309055.
Extended version of {\bf [Brunner-Ac\'{\i}n-Collins-(+2) 03]}.

\item {\bf [Brunner-Scarani, M. Wegmuller-(+2) 04]}:
N. Brunner, V. Scarani, M. Wegmuller,
M. Legre, \& N. Gisin,
``Direct measurement of superluminal group velocity and of signal velocity
in an optical fiber'',
quant-ph/0407155.

\item {\bf [Bru\ss-Ekert-Fern\'{a}ndez Huelga-(+2) 97]}:
D. Bru\ss, A. K. Ekert, S. G. Fern\'{a}ndez Huelga, J.-W. Pan, \& A. Zeilinger,
``Quantum computing with controlled-NOT and few qubits'',
in P. L. Knight, B. Stoicheff, \& D. Walls (eds.),
{\em Highlight in Quantum Optics},
{\em Philos. Trans. R. Soc. Lond. A} {\bf 355}, 1733, 2259-2266 (1997).

\item {\bf [Bru\ss-DiVincenzo-Ekert-(+2) 98]}:
D. Bru\ss, D. P. DiVincenzo, A. K. Ekert, C. Macchiavello, \& J. A. Smolin,
``Optimal universal and state-dependent quantum cloning'',
{\em Phys. Rev. A} {\bf 57}, 4, 2368-2378 (1998);
quant-ph/9705038.

\item {\bf [Bru\ss-Ekert-Macchiavello 98]}:
D. Bru\ss, A. K. Ekert, \& C. Macchiavello,
``Optimal universal quantum cloning and state estimation'',
{\em Phys. Rev. Lett.} {\bf 81}, 12, 2598-2601 (1998);
quant-ph/9712019.

\item {\bf [Bru\ss\, 98]}:
D. Bru\ss,
``Optimal eavesdropping in quantum cryptography with six states'',
{\em Phys. Rev. Lett.} {\bf 81}, 14, 3018-3021 (1998);
quant-ph/9805019.

\item {\bf [Bru\ss\-L\"{u}tkenhaus ?]}:
D. Bru\ss, \& N. L\"{u}tkenhaus,
``Quantum key distribution: From principles to practicalities'',
in {\em Applicable algebra in engineering, communication and computation}, ?, ?.
Reprinted in {\bf [Macchiavello-Palma-Zeilinger 00]}, pp.~259-274.

\item {\bf [Bru\ss-Macchiavello 99]}:
D. Bru\ss, \& C. Macchiavello,
``Optimal state estimation for $d$-dimensional quantum systems'',
{\em Phys. Lett. A} {\bf 253}, 5-6, 249-251 (1999);
quant-ph/9812016.

\item {\bf [Bru\ss\, 99]}:
D. Bru\ss,
``Entanglement splitting of pure bipartite quantum states'',
{\em Phys. Rev. A} {\bf 60}, 6, 4344-4348 (1999).

\item {\bf [Bru\ss-Fauro-Macchiavello-Palma 00]}:
D. Bru\ss, L. Fauro, C. Macchiavello, \& G. M. Palma,
``Quantum entanglement and classical communication through
a depolarizing channel'',
in V. Bu\v{z}zek, \& D. P. DiVincenzo (eds.),
{\em J. Mod. Opt.} {\bf 47}, 2-3 (Special issue:
Physics of quantum information), 325-331 (2000).

\item {\bf [Bru\ss-Peres 00]}:
D. Bru\ss, \& A. Peres,
``Construction of quantum states with bound entanglement'',
{\em Phys. Rev. A} {\bf 61}, 3, 030301(R) (2000);
quant-ph/9911056.

\item {\bf [Bru\ss-Cinchetti-D'Ariano-Macchiavello 00]}:
D. Bru\ss, M. Cinchetti, G. M. D'Ariano, \& C. Macchiavello,
``Phase-covariant quantum cloning'',
{\em Phys. Rev. A} {\bf 62}, 1, 012302 (2000);
quant-ph/9909046.

\item {\bf [Bru\ss-D'Ariano-Macchiavello-Sacchi 00]}:
D. Bru\ss, G. M. D'Ariano, C. Macchiavello, \& M. F. Sacchi,
``Approximate quantum cloning and the impossibility of
superluminal information transfer'',
{\em Phys. Rev. A} {\bf 62}, 6, 062302 (2000);
quant-ph/0010070.

\item {\bf [Bru\ss-Calsamiglia-L\"{u}tkenhaus 00]}:
D. Bru\ss, J. Calsamiglia, \& N. L\"{u}tkenhaus,
``Quantum cloning and distributed measurements'',
quant-ph/0011073.

\item {\bf [Bru\ss-Macchiavello 01]}:
D. Bru\ss, \& C. Macchiavello,
``Optimal cloning for two pairs of orthogonal states'',
in S. Popescu, N. Linden, \& R. Jozsa (eds.),
{\em J. Phys. A} {\bf 34}, 35
(Special issue: Quantum information and computation), 6815-6820 (2001);
quant-ph/0110099.

\item {\bf [Bru\ss\, 01]}:
D. Bru\ss,
``Characterizing entanglement'',
{\em Proc. ICQI (Rochester, 2001)};
quant-ph/0110078.

\item {\bf [Bru\ss-Cirac-Horodecki 02]}:
D. Bru\ss, J. I. Cirac, P. Horodecki,
F. Hulpke, B. Kraus, M. Lewenstein, \& A. Sanpera,
``Reflections upon separability and distillability'',
{\em Proc.\ ESF QIT Conf.\ Quantum Information: Theory, Experiment and Perspectives
(Gdansk, Poland, 2001)}, {\em J. Mod. Opt.} {\bf 49}, 8, 1399-1418 (2002);
quant-ph/0110081.

\item {\bf [Bru\ss-Macchiavello 02]}:
D. Bru\ss, \& C. Macchiavello,
``Optimal eavesdropping in cryptography with three-dimensional quantum states'',
{\em Phys. Rev. Lett.} {\bf 88}, 12, 127901 (2002);
quant-ph/0106126.

\item {\bf [Bru\ss\, 02]}:
D. Bru\ss,
``Characterizing entanglement'',
{\em J. Math. Phys.} {\bf 43}, 9, 4237-4251 (2002).

\item {\bf [Bru\ss\-Christandl-Ekert-(+3) 02]}:
D. Bru\ss, M. Christandl, A. K. Ekert,
B.-G. Englert, D. Kaszlikowski, \& C. Macchiavello,
``Tomographic quantum cryptography: Equivalence of quantum and classical key
distillation'',
{\em Phys. Rev. Lett.} {\bf 91}, 9, 097901 (2003);
quant-ph/0303184.

\item {\bf [Bru\ss\-Macchiavello 03]}:
D. Bru\ss, \& C. Macchiavello,
``On the entanglement structure in quantum cloning'',
{\em Found. Phys.} {\bf 33}, 11, 1617-1628 (2003).

\item {\bf [Bru\ss\-D'Ariano-Lewenstein-(+3) 04]}:
D. Bru\ss, G. M. D'Ariano, M. Lewenstein,
C. Macchiavello, A. Sen De, \& U. Sen,
``Distributed quantum dense coding'',
quant-ph/0407037.

\item {\bf [Bru\ss\-Datta-Ekert-(+2) 04]}:
D. Bru\ss, N. Datta, A. K. Ekert,
L. C. Kwek, \& C. Macchiavello,
``Multipartite entanglement and quantum phase transitions'',
quant-ph/0411080.

\item {\bf [Br\"{u}schweiler 00]}:
R. Br\"{u}schweiler,
``Novel strategy for database searching in spin
Liouville space by NMR ensemble computing'',
{\em Phys. Rev. Lett.} {\bf 85}, 22, 4815-4818 (2000).

\item {\bf [Brylinski 00]}:
J.-L. Brylinski,
``Algebraic measures of entanglement'',
quant-ph/0008031.

\item {\bf [Brylinski-Brylinski 01]}:
J.-L. Brylinski, \& R. Brylinski,
``Universal quantum gates'',
quant-ph/0108062.

\item {\bf [Bschorr-Fischer-Freyberger 01]}:
T. C. Bschorr, D. G. Fischer, \& M. Freyberger,
``Channel estimation with noisy entanglement'',
{\em Phys. Lett. A} {\bf 292}, 1-2, 15-22 (2001);
quant-ph/0108036.

\item {\bf [Bub 68]}:
J. Bub,
``The Dainieri-Loinger-Prosperi quantum theory of measurement'',
{\em Nuovo Cimento B} {\bf 57}, 2, 503-520 (1968).

\item {\bf [Bub 69]}:
J. Bub,
``What ia a hidden variable theory of quantum phenomena?,
{\em Int. J. Theor. Phys.} {\bf 2}, 2, 101-123 (1969).

\item {\bf [Bub 73 a]}:
J. Bub,
``On the completeness of quantum mechanics'',
in C. A. Hooker (ed.),
{\em Contemporary
research in the foundations and philosophy of quantum theory
Proc. of a Conf. held at
the University of Western Ontario, London, Canada},
Reidel, Dordrecht, Holland, 1973, pp.~1-65.

\item {\bf [Bub 73 b]}:
J. Bub,
``?'',
{\em Brit. J. Philos. Sci.} {\bf 24}, ?, 78-? (1973).

\item {\bf [Bub 74]}:
J. Bub,
{\em The interpretation of quantum mechanics}, Reidel,
Boston, Massachusetts, 1974.

\item {\bf [Bub 75]}:
J. Bub,
``Popper's propensity interpretation of probability and quantum mechanics'',
in G. Maxwell, \& R. M. Anderson, Jr. (eds.),
{\em Induction, probability and confirmation.
Minnesota Studies in Philosophy of Science. Vol. 6},
University of Minnesota Press, Minneapolis, 1975, pp.~416-429.

\item {\bf [Bub 76]}:
J. Bub,
``Hidden variables and locality'',
{\em Found. Phys.}
{\bf 6}, 5, 511-525 (1976).

\item {\bf [Bub 77]}:
J. Bub,
``Von Neumann's projection postulate as a possibility
conditionalization rule in quantum mechanics'',
{\em J. Philos. Logic} {\bf 6}, ?, 381-390 (1977).

\item {\bf [Bub 79]}:
J. Bub,
``The measurement problem of quantum mechanics'',
{\em Problems in the Philosophy of Physics. 72d Corso},
Societ\`a Italiana di Fisica, Bologna, 1979.

\item {\bf [Bub 80]}:
J. Bub,
`Comment on ``Locality an the algebraic structure of
quantum mechanics'' by William Demopoulos',
in {\bf [Suppes 80]}, pp.~149-153.
Comment on {\bf [Demopoulos 80]}.
See {\bf [Humphreys 80]}.

\item {\bf [Bub 82]}:
J. Bub,
``Quantum logic, conditional probability, and inference'',
{\em Philos. Sci.} {\bf 49}, ?, 402-421 (1982).

\item {\bf [Bub 85]}:
J. Bub,
``On the non-locality of pre- and post-selected quantum ensembles'',
in P. J. Lahti, \& P. Mittelstaedt (eds.),
{\em Proc.\ Symp.\ on the Foundations of Modern
Physics: 50 Years of the Einstein-Podolsky-Rosen Experiment
(Joensuu, Finland, 1985)},
World Scientific, Singapore, 1985, pp.~333-341.

\item {\bf [Bub-Brown 86]}:
J. Bub, \& H. R. Brown,
``Curious properties of
quantum ensembles which have been both preselected and post-selected'',
{\em Phys. Rev. Lett.} {\bf 56}, 22, 2337-2340 (1986).
Comment: {\bf [Albert-Aharonov-D'Amato 86]}.
See {\bf [Albert-Aharonov-D'Amato 85]}, {\bf [Hu 90]}.

\item {\bf [Bub 89]}:
J. Bub,
``On Bohr's response to EPR: A quantum logical analysis'',
{\em Found. Phys.} {\bf 19}, 7, 793-805 (1989).
See {\bf [Bub 90]} (II).

\item {\bf [Bub 90]}:
J. Bub,
``On Bohr's response to EPR: II'',
{\em Found. Phys.} {\bf 20}, 8, 929-941 (1989).
See {\bf [Bub 89]} (I).

\item {\bf [Bub 92]}:
J. Bub,
``Quantum mechanics without the projection postulate'',
{\em Found. Phys.} {\bf 22}, 5, 737-754 (1992).

\item {\bf [Bub 93 a]}:
J. Bub,
``Measurement: It ain't over till it's over'',
{\em Found. Phys. Lett.} {\bf 6}, 1, 21-35 (1993).

\item {\bf [Bub 93 b]}:
J. Bub,
`Quantum mechanics as a theory of ``beables''\,',
in A. van der Merwe, \& F. Selleri (eds.),
{\em Bell's theorem and the foundations of modern physics.
Proc.\ of an international
conference (Cesena, Italy, 1991)},
World Scientific, Singapore, 1993, pp.~117-124.

\item {\bf [Bub 95]}:
J. Bub,
``Maximal structures of determinate propositions in
quantum mechanics'',
{\em Int. J. Theor. Phys.} {\bf 34}, 8, 1255-1264 (1995).

\item {\bf [Bub 96]}:
J. Bub,
``Sch\"{u}tte's tautology and the KS theorem'',
{\em Found. Phys.} {\bf 26}, 6, 787-806 (1996).

\item {\bf [Bub-Clifton 96]}:
J. Bub, \& R. K. Clifton,
`A uniqueness theorem for
``no collapse'' interpretations of quantum mechanics',
{\em Stud. Hist. Philos. Sci. Part B: Stud. Hist. Philos. Mod. Phys.}
{\bf 27}, 2, 181-219 (1996).
Reprinted in {\bf [Clifton 04]}.
Revised proof: {\bf [Bub-Clifton-Goldstein 00]}.

\item {\bf [Bub 97]}:
J. Bub,
{\em Interpreting the quantum world}, Cambridge
University Press, Cambridge, 1997, 1999 (with corrections).
Reviews: {\bf [Rae 97]}, {\bf [Healey 98 b]}, {\bf [Peres 98 c]},
{\bf [Greenberger 98 b]}, {\bf [van Fraassen 98]}, {\bf [Cabello 00 e]}.
See {\bf [Bub 00]}.

\item {\bf [Bub-Greenberger 97]}:
J. Bub, \& D. M. Greenberger,
``Quantum measurement theory'',
in W. Trigg (ed.),
{\em Encyclopedia of Applied Physics},
American Institute of Physics and VHC, Berlin, 1997.

\item {\bf [Bub-Clifton-Monton 98]}:
J. Bub, R. K. Clifton, \& B. Monton,
``The bare theory has no clothes'',
in G. Hellman, \& R. A. Healey (eds.),
{\em Quantum measurement, decoherence, and modal interpretations
(Minnesota Studies in Philosophy of Science)}, 1998.

\item {\bf [Bub 00]}:
J. Bub,
``Quantum mechanics as a principle theory'',
{\em Stud. Hist. Philos. Sci. Part B: Stud. Hist. Philos. Mod. Phys.}
{\bf 31}, 1, 75-94 (2000);
quant-ph/9910096.
See {\bf [Bub 97]}, {\bf [Peres 98 c]}, {\bf [Bub-Clifton-Goldstein 00]}.

\item {\bf [Bub-Clifton-Goldstein 00]}:
J. Bub, R. K. Clifton, \& S. Goldstein,
``Revised proof of the uniqueness theorem for `no collapse'
interpretations of quantum mechanics'',
{\em Stud. Hist. Philos. Sci. Part B: Stud. Hist. Philos. Mod. Phys.}
{\bf 31}, 1, 95-98 (2000);
quant-ph/9910097.
Reprinted in {\bf [Clifton 04]}.
See {\bf [Bub 97, 00]}.

\item {\bf [Bub 01 a]}:
J. Bub,
``Secure key distribution via pre- and postselected quantum states'',
{\em Phys. Rev. A} {\bf 63}, 3, 032309 (2001);
quant-ph/0006086,
PITT-PHIL-SCI00000229.

\item {\bf [Bub 01 b]}:
J. Bub,
``The quantum bit commitment theorem'',
{\em Found. Phys.} {\bf 31}, 5, 735-756 (2001);
quant-ph/0007090.

\item {\bf [Bub 01 c]}:
J. Bub,
``Von Neumann's theory of quantum measurement'',
in M. R\'{e}dei, \& M. St\"{o}ltzner (eds.),
{\em John von Neumann and the foundations of quantum physics},
Kluwer Academic, Dordrecht, Holland, 2001, pp.~?-?.

\item {\bf [Bub-Clifton-Halvorson 02]}:
J. Bub, R. K. Clifton, \& H. Halvorson,
``Characterizing quantum theory in terms of information-theoretic constraints'' (2002),
PITT-PHIL-SCI00000887.

\item {\bf [Bub 04 a]}:
J. Bub,
``Why the quantum?'',
{\em Stud. Hist. Philos. Sci. Part B: Stud. Hist. Philos. Mod. Phys.};
quant-ph/0402149.

\item {\bf [Bub 04 b]}:
J. Bub,
``Quantum mechanics is about quantum information'',
{\em Found. Phys.} James Cushing Festschrift issue;
quant-ph/0408020.

\item {\bf [Buch 66]}:
T. Buch,
``Quantum mechanics as a model'',
{\em Am. J. Phys.} {\bf 34}, ?, 653-? (1966).

\item {\bf [Buchkremer-Dumke-Volk-(+3) 01]}:
F. B. J. Buchkremer, R. Dumke, M. Volk,
T. Muether, G. Birkl, \& W. Ertmer,
``Quantum information processing with microfabricated optical elements'',
quant-ph/0110119.

\item {\bf [Buchler-Lam-Ralph 99]}:
B. C. Buchler, P. K. Lam, \& T. C. Ralph,
``Enhancement of quantum nondemolition measurements with
an electro-optic feed-forward amplifier'',
{\em Phys. Rev. A} {\bf 60}, 6, 4943-4950 (1999).

\item {\bf [Buchler-Lam-Bachor-(+2) 02]}:
B. C. Buchler, P. K. Lam, H.-A. Bachor, U. L. Andersen, \& T. C. Ralph,
``Squeezing more from a quantum nondemolition measurement'',
{\em Phys. Rev. A} {\bf 65}, 1, 011803 (2002).

\item {\bf [Buck-van Enk-Fuchs 00]}:
J. R. Buck, S. J. van Enk, \& C. A. Fuchs,
``Experimental proposal for achieving superadditive communication
capacities with a binary quantum alphabet'',
{\em Phys. Rev. A} {\bf 61}, 3, 032309 (2000).

\item {\bf [Bucksbaum-Ahn-Weinacht 00]}:
P. H. Bucksbaum, J. Ahn, \& T. C. Weinacht,
``Does Rydberg state manipulation equal quantum computation?'',
{\em Science} {\bf 289}, 5484, 1431a (2000).
Reply to: {\bf [Meyer 00 c]}, {\bf [Kwiat-Hughes 00]}.
See {\bf [Ahn-Weinacht-Bucksbaum 00]}.

\item {\bf [Buehler-Reilly-Brenner-(+3) 00]}:
T. M. Buehler, D. J. Reilly, R. Brenner,
A. R. Hamilton, A. S. Dzurak, R. G. Clark,
``Correlated charge detection for readout of a solid-state quantum computer'',
{\em Appl. Phys. Lett.} {\bf 82}, 577-? (2003).

\item {\bf [Buks-Schuster-Heiblum-(+2) 98]}:
E. Buks, R. Schuster, M. Heiblum, D. Mahalu, \& V. Umansky,
``Dephasing in electron interference by a `which-path' detector'',
{\em Nature} {\bf 391}, 6670, 871-874 (1998).

\item {\bf [Bugajska-Bugajski 72]}:
K. Bugajska, \& S. Bugajski,
``Hidden variables and 2-dimensional Hilbert space'',
{\em Ann. Inst. H. Poincar\'{e} A} {\bf 16}, 93-102 (1972).

\item {\bf [Buhrman-Cleve-Wigderson 98]}:
H. Buhrman, R. Cleve, \& A. Wigderson,
``Quantum vs. classical communication and computation'',
in {\em Proc.\ of the 30th Annual ACM Symp.\ on the Theory of
Computing (El Paso, Texas, 1998)}, ACM Press, New York, 1998, pp.~63-68;
quant-ph/9802040.

\item {\bf [Buhrman-de Wolf 98]}:
H. Buhrman, \& R. de Wolf,
``Lower bounds for quantum search and derandomization'',
quant-ph/9811046.

\item {\bf [Buhrman-van Dam-H\o{}yer-Tapp 99]}:
H. Buhrman, W. van Dam, P. H\o{}yer, \& A. Tapp,
``Multiparty quantum communication complexity'',
{\em Phys. Rev. A} {\bf 60}, 4, 2737-2741 (1999);
quant-ph/9710054.

\item {\bf [Buhrman-van Dam 99]}:
H. Buhrman, \& W. van Dam,
``Bounded quantum query complexity'',
in {\em Proc.\ of the 14th Annual IEEE Conference on Computational Complexity (Complexity'99)},
149-156;
quant-ph/9903035.

\item {\bf [Buhrman-Durr-Heiligman-(+4) 00]}:
H. Buhrman, C. Durr, M. Heiligman,
P. H\o{}yer, F. Magniez, M. Santha, \& R. de Wolf,
``Quantum algorithms for element distinctness'',
quant-ph/0007016.

\item {\bf [Buhrman-Cleve-van Dam 01]}:
H. Buhrman, R. Cleve, \& W. van Dam,
``Quantum entanglement and communication complexity'',
{\em SIAM J. Comput.} {\bf 30}, 1829-1841 (2001);
quant-ph/9705033.

\item {\bf [Buhrman-Tromp-Vit\'{a}nyi 01]}:
H. Buhrman, J. Tromp, \& P. Vit\'{a}nyi,
``Time and space bounds for reversible simulation'',
in S. Popescu, N. Linden, \& R. Jozsa (eds.),
{\em J. Phys. A} {\bf 34}, 35
(Special issue: Quantum information and computation), 6821-6830 (2001).

\item {\bf [Buhrman-Cleve-Watrous-de Wolf 01]}:
H. Buhrman, R. Cleve, J. Watrous, \& R. de Wolf,
``Quantum fingerprinting'',
{\em Phys. Rev. Lett.} {\bf 87}, 16, 167902 (2001);
quant-ph/0102001.

\item {\bf [Buhrman-Fortnow-Newman-Roehrig 02]}:
H. Buhrman, L. Fortnow, I. Newman, \& H. Roehrig,
``Quantum property testing'',
quant-ph/0201117.

\item {\bf [Buhrman-H\o{}yer-Massar-Roehrig 03]}:
H. Buhrman, P. H\o{}yer, S. Massar, \& H. Roehrig,
``Combinatorics and quantum nonlocality'',
{\em Phys. Rev. Lett.} {\bf 91}, 4, 047903 (2003);
quant-ph/0209052.

\item {\bf [Buhrman-de Wolf 03]}:
H. Buhrman, \& R. de Wolf,
``Quantum zero-error algorithms cannot be composed'',
{\em Inf. Processing Lett.} {\bf 87}, 2, 79-84 (2003);
quant-ph/0211029.

\item {\bf [Buhrman-Newman-Roehrig-de Wolf 03]}:
H. Buhrman, I. Newman, H. Roehrig, \& R. de Wolf,
``Robust quantum algorithms and polynomials'',
quant-ph/0309220.

\item {\bf [Buhrman-Spalek 04]}:
H. Buhrman, \& R. Spalek,
``Quantum verification of matrix products'',
quant-ph/0409035.

\item {\bf [Buhrman-Massar 04]}:
H. Buhrman, \& S. Massar,
``Causality and Cirel'son bounds'',
quant-ph/0409066.

\item {\bf [Buhrman-H\o{}yer-Massar-Roehrig 03]}:
H. Buhrman, P. H\o{}yer, S. Massar, \& H. Roehrig,
``Multipartite nonlocal quantum correlations resistant to imperfections''
quant-ph/0410139.

\item {\bf [Buisson-Balestro-Pekola-Hekking 03]}:
O. Buisson, F. Balestro, J. P. Pekola, \& F. W. J. Hekking,
``One-shot quantum measurement using a hysteretic dc SQUID'',
{\em Phys. Rev. Lett.} {\bf 90}, 23, 238304 (2003).

\item {\bf [Bullock-Brennen 04]}:
S. S. Bullock, \& G. K. Brennen,
``Canonical decompositions of $n$-qubit quantum computations and concurrence'',
{\em J. Math. Phys.} {\bf 45}, ?, 2447-2467 (2004).

\item {\bf [Bunge 67]}:
M. Bunge (ed.),
{\em Quantum theory and reality (Oberwolfach, 1966)},
Springer-Verlag, New York, 1967.

\item {\bf [Bunge-Kalnay 83]}:
M. Bunge, \& A. J. Kalnay,
``Solution of two paradoxes in the quantum theory of unstable systems'',
{\em Nuovo Cimento B} {\bf 77}, 1, 1-9 (1983).

\item {\bf [Burgarth-Bose 04]}:
D. Burgarth, \& S. Bose,
``Conclusive and arbitrarily perfect quantum state transfer using
amplitude delaying spin chain channels'',
quant-ph/0406112.

\item {\bf [Burgarth-Giovannetti-Bose 04]}:
D. Burgarth, V. Giovannetti, \& S. Bose,
``Efficient and perfect state transfer in quantum chains'',
quant-ph/0410175.

\item {\bf [Burgers 63]}:
J. M. Burgers,
``The measuring process in quantum theory'',
{\em Rev. Mod. Phys.} {\bf 35}, 1, 145-150 (1963).

\item {\bf [Burgos 84 a]}:
M. E. Burgos,
``?'',
{\em Found. Phys.} {\bf 14}, ?, 739-? (1984).

\item {\bf [Burgos 84 b]}:
M. E. Burgos,
``?'',
{\em Found. Phys.} {\bf 14}, ?, 753-? (1984).

\item {\bf [Burgos 87 a]}:
M. E. Burgos,
``?'',
{\em Phys. Lett. A} {\bf 123}, ?, 313-? (1987).

\item {\bf [Burgos 87 b]}:
M. E. Burgos,
``?'',
{\em Found. Phys.} {\bf 17}, ?, 809-? (1987).

\item {\bf [Burgos 98]}:
M. E. Burgos,
``Which natural processes have the special status of measurements?'',
{\em Found. Phys.} {\bf 28}, 8, 1323-1346 (1998).

\item {\bf [Burkard-Loss-DiVincenzo 99]}:
G. Burkard, D. Loss, \& D. P. DiVincenzo,
``Coupled quantum dots as quantum gates'',
{\em Phys. Rev. B} {\bf 59}, 3, 2070-2078 (1999);
quant-ph/9808026.
Reprinted in {\bf [Macchiavello-Palma-Zeilinger 00]}, pp.~440-448.

\item {\bf [Burkard-Loss-DiVincenzo-Smolin 99]}:
G. Burkard, D. Loss, D. P. DiVincenzo, \& J. A. Smolin,
``Physical optimization of quantum error correction circuits'',
{\em Phys. Rev. B} {\bf 60}, 16, 11404-11416 (1999);
cond-mat/9905230.

\item {\bf [Burkard-Engel-Loss 00]}:
G. Burkard, H.-A. Engel, \& D. Loss,
``Spintronics and quantum dots for quantum computing and quantum communication'',
{\em Fortschr. Phys.} {\bf 48}, 9-11 (Special issue: Experimental proposals for quantum computation), 965-986 (2000).

\item {\bf [Burkard-Loss 03]}:
G. Burkard, \& D. Loss,
``Lower bound for electron spin entanglement from beam splitter current
correlations'',
{\em Phys. Rev. Lett.} {\bf 91}, 8, 087903 (2003).

\item {\bf [Burkard-Koch-DiVincenzo 04]}:
G. Burkard, R. H. Koch, \& D. P. DiVincenzo,
``Multi-level quantum description of decoherence in superconducting qubits'',
{\em Phys. Rev. B} {\bf 69}, 6, 064503 (2004);
cond-mat/0308025.

\item {\bf [Burkard-Koch-DiVincenzo 04]}:
G. Burkard, R. H. Koch, \& D. P. DiVincenzo,
``Multilevel quantum description of decoherence in superconducting qubits'',
{\em Phys. Rev. B} {\bf 69}, 064503 (2004).

\item {\bf [Burkard-DiVincenzo-Bertet-(+2) 04]}:
G. Burkard, D. P. DiVincenzo, P. Bertet,
I. Chiorescu, \& J. E. Mooij,
``Asymmetry and decoherence in a double-layer persistent-current qubit'',
cond-mat/0405273.

\item {\bf [Burlakov-Chekhova-Karabutova-(+2) 99]}:
A. V. Burlakov, M. V. Chekhova, O. A. Karabutova, D. N. Klyshko, \& S. P. Kulik,
``Polarization state of a biphoton: Quantum ternary logic'',
{\em Phys. Rev. A} {\bf 60}, 6, R4209-R4212 (1999).

\item {\bf [Burlakov-Krivitskii-Kulik-(+2) 03]}:
A. V. Burlakov, L. A. Krivitskii, S. P. Kulik,
G. A. Maslennikov, \& M. V. Chekhova,
``Measurement of qutrits'',
{\em Opt. Spectrosc.} {\bf 94}, 684 (2003).

\item {\bf [Burnashev-Holevo 97]}:
M. V. Burnashev, \& A. S. Holevo (Kholevo),
``On reliability function of quantum communication channel'',
quant-ph/9703013.

\item {\bf [Burt-Ekstrom-Swanson 01]}:
E. A. Burt, C. R. Ekstrom, \& T. B. Swanson,
`Comment on ``Quantum clock synchronization based
on shared prior entanglement''\,',
{\em Phys. Rev. Lett.} {\bf 87}, 12, 129801 (2001);
quant-ph/0007030,
Comment on {\bf [Jozsa-Abrams-Dowling-Williams 00]}.
Reply: {\bf [Jozsa-Abrams-Dowling-Williams 01]}.
See {\bf [Chuang 00]}.

\item {\bf [Buscemi-D'Ariano-Sacchi 03 a]}:
F. Buscemi, G. M. D'Ariano, \& M. F. Sacchi,
``Unitary realizations of the ideal phase measurement'',
{\em Phys. Lett. A} {\bf 312}, 5-6, 315-318 (2003).

\item {\bf [Buscemi-D'Ariano-Perinotti-Sacchi 03]}:
F. Buscemi, G. M. D'Ariano, P. Perinotti, \& M. F. Sacchi,
``Optimal realization of the transposition maps'',
{\em Phys. Lett. A} {\bf 314}, 5-6, 374-379 (2003);
quant-ph/0304175.

\item {\bf [Buscemi-D'Ariano-Sacchi 03 b]}:
F. Buscemi, G. M. D'Ariano, \& M. F. Sacchi,
``Physical realizations of quantum operations'',
{\em Phys. Rev. A};
quant-ph/0305180.

\item {\bf [Buscemi-D'Ariano-Perinotti 04]}:
F. Buscemi, G. M. D'Ariano, \& P. Perinotti,
``There exist nonorthogonal quantum measurements that are perfectly repeatable'',
{\em Phys. Rev. Lett.} {\bf 92}, 7, 070403 (2004);
quant-ph/0310041.

\item {\bf [Buscemi-D'Ariano-Macchiavello 04]}:
F. Buscemi, G. M. D'Ariano, \& C. Macchiavello,
``Economical phase-covariant cloning of qudits'',
quant-ph/0407103.

\item {\bf [Busch-Lahti 84]}:
P. Busch, \& P. J. Lahti,
``On various joint
measurements of position and momentum observables'',
{\em Phys. Rev. D} {\bf 29}, 8, 1634-1646 (1984).

\item {\bf [Busch-Lahti 85]}:
P. Busch, \& P. J. Lahti,
``A note on quantum thory, complementarity and uncertainty'',
{\em Philos. Sci.} {\bf 52}, ?, 64-77 (1985).

\item {\bf [Busch 88]}:
P. Busch,
``Surprising features of unsharp quantum measurements'',
{\em Phys. Lett. A} {\bf 130}, 6-7, 323-329 (1988).
See {\bf [Aharonov-Albert-Casher-Vaidman 88]}.

\item {\bf [Busch-Cassinelli-Lahti 90]}:
P. Busch, G. Cassinelli, \& P. J. Lahti,
``On the quantum theory of sequential measurements'',
{\em Found. Phys.} {\bf 20}, 7, 757-775 (1990).

\item {\bf [Busch-Mittelstaedt 91]}:
P. Busch, \& P. Mittelstaedt,
``The problem of objectification in quantum mechanics'',
{\em Found. Phys.} {\bf 21}, 8, 889-904 (1991).

\item {\bf [Busch-Lahti-Mittelstaedt 91]}:
P. Busch, P. J. Lahti, \& P. Mittelstaedt,
{\em The quantum theory of measurement},
Springer-Verlag, London, 1991 (1st edition);
Springer-Verlag, Berlin, 1996 (2nd edition).

\item {\bf [Busch-Lahti-Mittelstaedt 92]}:
P. Busch, P. J. Lahti, \& P. Mittelstaedt,
``Weak objectification, joint probabilities,
and Bell inequalities in quantum mechanics'',
{\em Found. Phys.} {\bf 22}, 7, 949-962 (1992).

\item {\bf [Busch-Kienzler-Lahti-Mittelstaedt 93]}:
P. Busch, P. Kienzler, P. J. Lahti, \& P. Mittelstaedt,
``Testing quantum mechanics against a full set of Bell inequalities'',
{\em Phys. Rev. A} {\bf 47}, 6, 4627-4631 (1993).

\item {\bf [Busch-Grabowski-Lahti 95]}:
P. Busch, M. Grabowski, \& P. J. Lahti,
{\em Operational quantum physics},
Springer-Verlag, Berlin, 1995.
Review: {\bf [Fleming 00]}.

\item {\bf [Busch-Lahti 95]}:
P. Busch, \& P. J. Lahti,
``The complementarity of quantum observables: Theory and experiment'',
{\em Rivista del Nuovo Cimento} {\bf 18}, 4, 1-27 (1995);
quant-ph/0406132.

\item {\bf [Busch-Shimony 96]}:
P. Busch, \& A. Shimony,
``Insolubility of the
quantum measurement problem for unsharp observables'',
{\em Stud. Hist. Philos. Sci.
Part B: Stud. Hist. Philos. Mod. Phys.} {\bf 27}, 4, 397-404 (1996).

\item {\bf [Busch-Lahti 97]}:
P. Busch, \& P. J. Lahti,
``Remarks on separability of compound quantum systems and time reversal'',
{\em Found. Phys. Lett.} {\bf 10}, 2, 113-117 (1997).

\item {\bf [Busch 98 a]}:
P. Busch,
``Remarks on unsharp quantum observables,
objectification, and modal interpretations'',
quant-ph/9802006.

\item {\bf [Busch 98 b]}:
P. Busch,
`Can ``unsharp objectification'' solve the quantum
measurement problem?',
{\em Int. J. Theor. Phys.} {\bf 37}, 1, 241-248 (1998);
quant-ph/9802011.

\item {\bf [Busch-Singh 98]}:
P. Busch, \& J. Singh,
``L\"{a}ders theorem for unsharp quantum measurements'',
{\em Phys. Lett. A} {\bf 249}, 1-2, 10-12 (1998).

\item {\bf [Busch 99 a]}:
P. Busch,
``Unsharp localization and causality in relativistic quantum
theory'',
{\em J. Phys. A} {\bf 32}, 37, 6535-6546 (1999).

\item {\bf [Busch 99 b]}:
P. Busch,
``Resurrection of von Neumann's no-hidden-variables theorem'',
quant-ph/9909073.
See {\bf [Busch 03 a]}.

\item {\bf [Busch 00]}:
P. Busch,
``Classical versus quantum ontology'',
quant-ph/0010115.
Review of {\bf [Home 97]}.

\item {\bf [Busch-Cassinelli-De Vito-(+2) 01]}:
P. Busch, G. Cassinelli, E. De Vito,
P. J. Lahti, \& A. Levrero,
``Teleportation and measurement'',
{\em Phys. Lett. A} {\bf 284}, 4-5, 141-145 (2001);
quant-ph/0102121.

\item {\bf [Busch 01]}:
P. Busch,
``Just how final are today's quantum structures?'',
quant-ph/0103139.

\item {\bf [Busch 02 a]}:
P. Busch,
``EPR-Bell tests with unsharp observables and relativistic quantum measurement'',
in T. Placek, \& J. N. Butterfield (eds.),
{\em NATO Advanced Research Workshop ``Modality, Probability, and Bell's Theorem''
(Cracow, Poland, 2001)},
Kluwer Academic, Dordrecht, Holland, 2002;
quant-ph/0110023.

\item {\bf [Busch 02 b]}:
P. Busch,
``The time energy uncertainty relation'',
in J. G. Muga, R. Sala Mayato, \& I. L. Egusquiza (eds.),
{\em Time in quantum mechanics},
Springer-Verlag, Berlin (2002), pp.~69-98;
quant-ph/0105049.

\item {\bf [Busch 03 a]}:
P. Busch,
``Quantum states and generalized observables: A simple proof of Gleason's
theorem'',
{\em Phys. Rev. Lett.} {\bf 91}, 12, 120403 (2003).
See {\bf [Busch 99 b]}.

\item {\bf [Busch 03 b]}:
P. Busch,
``The role of entanglement in quantum measurement and information processing'',
{\em Int. J. Theor. Phys.} {\bf 42}, 5, 937-941 (2003);
quant-ph/0209090.

\item {\bf [Busch-Heinonen-Lahti 03]}:
P. Busch, T. Heinonen, \& P. J. Lahti,
``Noise and disturbance in quantum mechanics'',
{\em Phys. Lett. A} {\bf 320}, 4, 261-270 (2003);
quant-ph/0312006.

\item {\bf [Busch-Shilladay 03]}:
P. Busch, \& C. R. Shilladay,
``Uncertainty reconciles complementarity with joint measurability'',
{\em Phys. Rev. A} {\bf 68}, 3, 034102 (2003).

\item {\bf [Busch 04]}:
P. Busch,
``Less (precision) is more (information): Quantum information in fuzzy
probability theory'',
in {\em Int.\ Conf.\ on Quantum
Theory: Reconsideration of Foundations (V\"{a}xj\"{o}, Sweden 2003)};
quant-ph/0401027.

\item {\bf [Bussey 83]}:
P. J. Bussey 83,
``On replicating photons'',
{\em Nature} {\bf 304}, 5922, 188 (1983).
Comment on {\bf [Wootters-Zurek 82]}.
Reply: {\bf [Wootters-Zurek 83]}.

\item {\bf [Bussey 88]}:
P. J. Bussey,
`Comments on ``When is a quantum
measurement?'' [{\em Am. J. Phys.} {\bf 54}, 688 (1986)]',
{\em Am. J. Phys.} {\bf 56}, 6, 569-570 (1988).
Comment on {\bf [Peres 86 a]}.

\item {\bf [Butler-Hartel 98]}:
M. Butler, \& P. Hartel,
``Reasoning about Grover's quantum search algorithm using probabilistic wp'',
quant-ph/9810066.

\item {\bf [Butterfield 90]}:
J. N. Butterfield,
``A space-time approach to the Bell inequalities'',
in A. I. Miller (ed.),
{\em Sixty-two years of uncertainty: Historical, philosophical and
physical inquiries into the foundations of quantum mechanics.
Proc.\ Int. School of History of Science
(Erice, Italy, 1989)},
Plenum Press, New York, 1990, pp.~114-144.

\item {\bf [Butterfield 91]}:
J. N. Butterfield,
``Causal independence in EPR arguments'',
in A. I. Fine, M. Forbes, \& L. Wessels (eds.),
{\em Proc.\ of the 1990
Biennial Meeting of the Philosophy of Science Association},
East Lansing, Michigan, 1991, vol. 2, pp.~213-225.

\item {\bf [Butterfield 92 a]}:
J. N. Butterfield,
``Bell's theorem: What it takes'',
{\em Brit. J. Philos. Sci.} {\bf 43}, 1, 41-83 (1992).

\item {\bf [Butterfield 92 b]}:
J. N. Butterfield,
``David Lewis meets John Bell'',
{\em Philos. Sci.} {\bf 59}, ?, 26-42 (1992).

\item {\bf [Butterfield et al. 93 a]}:
J. N. Butterfield et al.,
``Parameter dependence
and outcome dependence in dynamical models for statevector reduction'',
{\em Found. Phys.} {\bf 23}, 3, 341-364 (1993).
See {\bf [Butterfield et al. 93 b]}.

\item {\bf [Butterfield et al. 93 b]}:
J. N. Butterfield et al.,
``Parameter dependence
and outcome dependence in dynamical models for statevector reduction'',
{\em Int. J. Theor. Phys.} {\bf 32}, ?, 2287-2304 (1993).
See {\bf [Butterfield et al. 93 a]}.

\item {\bf [Butterfield-Pagonis 00]}:
J. N. Butterfield, \& C. Pagonis (eds.),
{\em From physics to philosophy},
Cambridge University Press, Cambridge, 2000.
Review: {\bf [Stenholm 00 b]}.

\item {\bf [Butterfield 01]}:
J. N. Butterfield,
``Some worlds of quantum theory'',
in {\bf [Rusell-Clayton-Wegter McNelly-Polkinghorne 01]}, pp.~111-140;
quant-ph/0105052,
PITT-PHIL-SCI00000203.

\item {\bf [Butterfield-Isham 02]}:
J. N. Butterfield, \& C. J. Isham,
``A topos perspective on the Kochen-Specker theorem: IV. Interval valuations'',
{\em Int. J. Theor. Phys.} {\bf 41}, 4, 613-639 (2002);
quant-ph/0107123.
See {\bf [Isham-Butterfield 98]} (I),
{\bf [Isham-Butterfield 99]} (II),
{\bf [Hamilton-Isham-Butterfield 00]} (III).

\item {\bf [Butterfield 03]}:
J. N. Butterfield,
``On Hamilton-Jacobi theory as a classical root of quantum theory'',
PITTPHILSCI, 1193 (2003).

\item {\bf [Butterfield 04]}:
J. N. Butterfield,
``On Hamilton-Jacobi theory: Its geometry and relation to pilot-wave
theory'',
in A. C. Elitzur, S. Dolev, \& N. Kolenda (eds.),
{\em Quo vadis quantum mechanics? Possible
developments in quantum theory in the 21st century},
Springer, New York, 2004;
quant-ph/0210140.

\item {\bf [Buttler-Hughes-Kwiat-(+6) 98]}:
W. T. Buttler, R. J. Hughes, P. G. Kwiat, S. K. Lamoreaux, G. G. Luther,
G. L. Morgan, J. E. Nordholt, C. G. Peterson, \& C. M. Simmons,
``Practical free-space quantum key distribution over 1 km'',
{\em Phys. Rev. Lett.} {\bf 81}, 15, 3283-3286 (1998).

\item {\bf [Buttler-Hughes-Kwiat-(+5) 98]}:
W. T. Buttler, R. J. Hughes, P. G. Kwiat, G. G. Luther, G. L.
Morgan, J. E. Nordholt, C. G. Peterson, \& C. M. Simmons,
``Free-space quantum key distribution'',
{\em Phys. Rev. A} {\bf 57}, 4, 2379-2382 (1998);
quant-ph/9801006.

\item {\bf [Buttler-Hughes-Lamoreaux-(+3) 00]}:
W. T. Buttler, R. J. Hughes, S. K. Lamoreaux, G. L. Morgan,
J. E. Nordholt, \& C. G. Peterson,
``Daylight quantum key distribution over 1.6 km'',
{\em Phys. Rev. Lett.} {\bf 84}, 24, 5652-5655 (2000);
quant-ph/0001088.

\item {\bf [Buttler-Torgerson-Lamoreaux 02]}:
W. T. Buttler, J. R. Torgerson, \& S. K. Lamoreaux,
``New, efficient and robust, fiber-based quantum key distribution schemes'',
{\em Phys. Lett. A} {\bf 299}, 1, 38-42 (2002);
quant-ph/0203098.

\item {\bf [Buttler-Lamoreaux-Torgerson-(+2) 03]}:
W. T. Buttler, S. K. Lamoreaux, J. R. Torgerson,
G. H. Nickel, C. H. Donahue, \& C. G. Peterson,
``Fast, efficient error reconciliation for quantum cryptography'',
{\em Phys. Rev. A} {\bf 67}, 5, 052303 (2003);
quant-ph/0203096.

\item {\bf [Bu\v{z}ek-Hladk\'{y} 93]}:
V. Bu\v{z}ek, \& B. Hladk\'{y},
``Macroscopic superposition states of light via two-photon
resonant interaction of atoms with cavity field'',
{\em J. Mod. Opt.} {\bf 40}, 7, 1309-1324 (1993).

\item {\bf [Bu\v{z}ek-Gantsog-Kim 94]}:
V. Bu\v{z}ek, T. Gantsog, \& M. S. Kim,
``Production of macroscopic Schr\"{o}dinger cat states
from weak quantized fields interacting with atoms driven by
classical fields'',
{\em J. Mod. Opt.} {\bf 41}, 8, 1625-1635 (1994).

\item {\bf [Bu\v{z}ek-Hillery 96]}:
V. Bu\v{z}ek, \& M. Hillery,
``Quantum copying: Beyond the no-cloning theorem'',
{\em Phys. Rev. A} {\bf 54}, 3, 1844-1852 (1996);
quant-ph/9607018.

\item {\bf [Bu\v{z}ek-Drobn\'{y}-Adam-(+2) 97]}:
V. Bu\v{z}ek, G. Drobn\'{y}, G. Adam,
R. Derka, \& P. L. Knight,
``Reconstruction of quantum states of spin systems via
the Jaynes principle of maximum entropy'',
{\em J. Mod. Opt.} {\bf 44}, 11-12 (Special issue:
Quantum state preparation and measurement), 2607-2627 (1997);
quant-ph/9701038.

\item {\bf [Bu\v{z}ek-Vedral-Plenio-(+2) 97]}:
V. Bu\v{z}ek, V. Vedral, M. B. Plenio,
P. L. Knight, \& M. Hillery,
``Broadcasting of entanglement via local copying'',
{\em Phys. Rev. A} {\bf 55}, 5, 3327-3332 (1997);
quant-ph/9701028.

\item {\bf [Bu\v{z}ek-Braunstein-Hillery-Bru\ss\, 97]}:
V. Bu\v{z}ek, S. L. Braunstein, M. Hillery, \& D. Bru\ss,
``Quantum copying: A network'',
{\em Phys. Rev. A} {\bf 56}, 5, 3446-3452 (1997);
quant-ph/9703046.

\item {\bf [Bu\v{z}ek-Hillery 98 a]}:
V. Bu\v{z}ek, \& M. Hillery,
``Universal optimal cloning of arbitrary quantum
states: From qubits to quantum registers'',
{\em Phys. Rev. Lett.} {\bf 81}, 22, 5003-5006 (1998).

\item {\bf [Bu\v{z}ek-Hillery 98 b]}:
V. Bu\v{z}ek, \& M. Hillery,
``Universal optimal cloning of qubits and quantum registers'',
in C. P. Williams (ed.),
{\em 1st NASA Int.\ Conf.\ on Quantum Computing and Quantum Communications
(Palm Springs, California, 1998)},
{\em Lecture Notes in Computer Science} {\bf 1509},
Springer-Verlag, New York, 1999, pp.~?-?;
quant-ph/9801009.

\item {\bf [Bu\v{z}ek-Drobny-Derka-(+2) 98]}:
V. Bu\v{z}ek, G. Drobny, R. Derka,
G. Adam, \& H. Wiedemann,
``Quantum state recostruction from incomplete data'',
quant-ph/9805020.

\item {\bf [Bu\v{z}ek 98]}:
V. Bu\v{z}ek,
``Reconstruction of Liouvillian superoperators'',
{\em Phys. Rev. A} {\bf 58}, 3, 1723-1727 (1998);
quant-ph/9806093.

\item {\bf [Bu\v{z}ek-Hillery-Bednik 98]}:
V. Bu\v{z}ek, M. Hillery, \& R. Bednik,
``Controlling the flow of information in quantum cloners:
Asymmetric cloning'',
{\em Acta Phys. Slovaca}, {\bf 48}, ?, 177-184 (1998);
quant-ph/9807086.

\item {\bf [Bu\v{z}ek-Hillery-Knight 98]}:
V. Bu\v{z}ek, M. Hillery, \& P. L. Knight,
``Flocks of quantum clones: Multiple copying of qubits'',
{\em Fortschr. Phys.} {\bf 46}, 4-5, 521-533 (1998).

\item {\bf [Bu\v{z}ek-Derka-Massar 99]}:
V. Bu\v{z}ek, R. Derka, \& S. Massar,
``Optimal quantum clocks'',
{\em Phys. Rev. Lett.} {\bf 82}, 10, 2207-2210 (1999);
quant-ph/9808042.

\item {\bf [Bu\v{z}ek-Hillery-Werner 99]}:
V. Bu\v{z}ek, M. Hillery, \& R. F. Werner,
``Optimal manipulations with qubits: Universal-NOT gate'',
{\em Phys. Rev. A} {\bf 60}, 4, R2626-R2629 (1999);
quant-ph/9901053.
See {\bf [Bu\v{z}ek-Hillery-Werner 00]}.

\item {\bf [Bu\v{z}ek-Hillery-Werner 00]}:
V. Bu\v{z}ek, M. Hillery, \& R. F. Werner,
``Universal-NOT gate'',
in V. Bu\v{z}zek, \& D. P. DiVincenzo (eds.),
{\em J. Mod. Opt.} {\bf 47}, 2-3 (Special issue:
Physics of quantum information), 211-232 (2000).

\item {\bf [Bu\v{z}ek-Hillery 00]}:
V. Bu\v{z}ek, \& M. Hillery,
``Optimal manipulations with qubits: Universal quantum entanglers'',
{\em Phys. Rev. A} {\bf 62}, 2, 022303 (2000);
quant-ph/0006045.

\item {\bf [Bu\v{z}ek-Hillery 00]}:
V. Bu\v{z}ek, \& M. Hillery,
``Quantum disentanglers'',
{\em Phys. Rev. A} {\bf 62}, 5, 052303 (2000);
quant-ph/0006047.

\item {\bf [Bu\v{z}ek-Knight-Imoto 00]}:
V. Bu\v{z}ek, P. L. Knight, \& N. Imoto,
``Multiple observations of quantum clocks'',
{\em Phys. Rev. A} {\bf 62}, 6, 062309 (2000);
quant-ph/0006048.

\item {\bf [Byrd-Slater 01]}:
M. S. Byrd, \& P. B. Slater,
``Bures measures over the spaces of two- and three-dimensional density matrices'',
{\em Phys. Lett. A} {\bf 283}, 3-4, 152-156 (2001).

\item {\bf [Byrd-Lidar 02]}:
M. S. Byrd, \& D. A. Lidar,
``Comprehensive encoding and decoupling solution to problems of decoherence
and design in solid-state quantum computing'',
{\em Phys. Rev. Lett.} {\bf 89}, 4, 047901 (2002);
quant-ph/0112054.

\item {\bf [Byrd-Lidar 03]}:
M. S. Byrd, \& D. A. Lidar,
``Empirical determination of dynamical decoupling operations'',
{\em Phys. Rev. A} {\bf 67}, 1, 012324 (2003).

\item {\bf [Byrd-Wu-Lidar 04]}:
M. S. Byrd, L.-A. Wu, \& D. A. Lidar,
``Overview of quantum error prevention and leakage elimination'',
{\em Proc.\ of the Physics of Quantum Electronics}, 2004
quant-ph/0402098.

\item {\bf [Byrd-Khaneja 03]}:
M. S. Byrd, \& N. Khaneja,
``Characterization of the positivity of the density
matrix in terms of the coherence vector representation'',
{\em Phys. Rev. A} {\bf 68}, 6, 062322 (2003).

\item {\bf [Byrd-Lidar-Wu-Zanardi 04]}:
M. S. Byrd, D. A. Lidar, L.-A. Wu, \& P. Zanardi,
``Universal leakage elimination'',
quant-ph/0409049.


\newpage

\subsection{}


\item {\bf [Caban-Rembieli\'{n}ski 99]}:
P. Caban, \& J. Rembieli\'{n}ski,
``Lorentz-covariant quantum mechanics and preferred frame'',
{\em Phys. Rev. A} {\bf 59}, 6, 4187-4196 (1999).

\item {\bf [Caban-Rembieli\'{n}ski-Smolinski-Walczak 01]}:
P. Caban, J. Rembieli\'{n}ski, K. A. Smolinski, \& Z. Walczak,
``Destruction of states in quantum mechanics'',
quant-ph/0112092.

\item {\bf [Caban-Rembieli\'{n}ski-Smolinski-Walczak 03]}:
P. Caban, J. Rembieli\'{n}ski, K. A. Smolinski, \& Z. Walczak,
``Einstein-Podolsky-Rosen correlations and Galilean transformations'',
{\em Phys. Rev. A} {\bf 67}, 1, 012109 (2003);
quant-ph/0302026.

\item {\bf [Caban-Smolinski-Walczak 03 a]}:
P. Caban, K. A. Smolinski, \& Z. Walczak,
``Kraus representation of destruction of states for one qudit'',
{\em Phys. Rev. A} {\bf 68}, 3, 034308 (2003);
quant-ph/0307080.

\item {\bf [Caban-Rembieli\'{n}ski 03]}:
P. Caban, \& J. Rembieli\'{n}ski,
``Photon polarization and Wigner's little group'',
{\em Phys. Rev. A} {\bf 68}, 4, 042107 (2003).

\item {\bf [Caban-Smolinski-Walczak 03 b]}:
P. Caban, K. A. Smoliski, \& Z. Walczak,
``Galilean covariance of a reduced density matrix'',
{\em Phys. Rev. A} {\bf 68}, 4, 044101 (2003).

\item {\bf [Cabauy-Benioff 03]}:
P. Cabauy, \& P. Benioff,
``Cyclic networks of quantum gates'',
{\em Phys. Rev. A} {\bf 68}, 3, 032315 (2003);
quant-ph/0211175.

\item {\bf [Cabello 94]}:
A. Cabello,
``A simple proof of the Kochen-Specker theorem'',
{\em Eur. J. Phys.} {\bf 15}, 4, 179-183 (1994).

\item {\bf [Cabello 95]}:
A. Cabello,
``Kochen-Specker diagram of the Peres-Mermin example'',
in M. Ferrero, \& A. van der Merwe (eds.),
{\em Fundamental problems in quantum physics.
Proc.\ of an international symposium (Oviedo, Spain, 1993)},
Kluwer Academic, Dordrecht, Holland, 1995, pp.~43-46.

\item {\bf [Cabello-Garc\'{\i}a Alcaine 95 a]}:
A. Cabello, \& G. Garc\'{\i}a Alcaine,
``La sorprendente incompatibilidad de la idea de realidad einsteiniana
con la mec\'{a}nica cu\'{a}ntica (o de c\'{o}mo la mec\'{a}nica
cu\'{a}ntica es m\'{a}s extra\~{n}a de lo que usualmente se cree)'',
{\em Revista Espa\~{n}ola de F\'{\i}sica} {\bf 9}, 2, 11-17 (1995).

\item {\bf [Cabello-Garc\'{\i}a Alcaine 95 b]}:
A. Cabello, \& G. Garc\'{\i}a Alcaine,
``A hidden-variables versus quantum mechanics experiment'',
{\em J. Phys. A} {\bf 28}, 13, 3719-3724 (1995).

\item {\bf [Cabello-Garc\'{\i}a Alcaine 96 a]}:
A. Cabello, \& G. Garc\'{\i}a Alcaine,
``Bell-Kochen-Specker theorem for any finite dimensions $n \geq 3$'',
{\em J. Phys. A} {\bf 29}, 5, 1025-1036 (1996).

\item {\bf [Cabello-Garc\'{\i}a Alcaine 96 b]}:
A. Cabello, \& G. Garc\'{\i}a Alcaine,
``Elementos de realidad locales versus mec\'{a}nica cu\'{a}ntica'',
in M. Ferrero, A. Fern\'{a}ndez Ra\~{n}ada, J. L. S\'{a}nchez G\'{o}mez,
\& E. Santos (eds.),
{\em Fundamentos de la F\'{\i}sica Cu\'{a}ntica (San Lorenzo de El
Escorial, Spain, 1995)},
Editorial Complutense, Madrid, 1996, pp.~83-91.

\item {\bf [Cabello-Estebaranz-Garc\'{\i}a Alcaine 96 a]}:
A. Cabello, J. M. Estebaranz, \& G. Garc\'{\i}a Alcaine,
``Bell-Kochen-Specker theorem: A proof with 18 vectors'',
{\em Phys. Lett. A} {\bf 212}, 4, 183-187 (1996);
quant-ph/9706009.
See {\bf [Peres 91 a]}, {\bf [Kernaghan 94]}.

\item {\bf [Cabello-Santos 96]}:
A. Cabello, \& E. Santos,
``Comment on `Experimental demonstration of the violation of
local realism without Bell inequalities'\,'',
{\em Phys. Lett. A} {\bf 214}, 5-6, 316-318 (1996);
quant-ph/9709057.
Comment on {\bf [Torgerson-Branning-Monken-Mandel 95]}.
See {\bf [Garuccio 95 b]}.
Reply: {\bf [Torgerson-Branning-Monken-Mandel 96]}.

\item {\bf [Cabello-Estebaranz-Garc\'{\i}a Alcaine 96 b]}:
A. Cabello, J. M. Estebaranz, \& G. Garc\'{\i}a Alcaine,
``New variants of the Bell-Kochen-Specker theorem'',
{\em Phys. Lett. A} {\bf 218}, 3-6, 115-118 (1996);
quant-ph/9706010.

\item {\bf [Cabello-Estebaranz-Garc\'{\i}a Alcaine 96 c]}:
A. Cabello, J. M. Estebaranz, \& G. Garc\'{\i}a Alcaine,
``Recursive proof of the Bell-Kochen-Specker
theorem in dimension $n \geq 3$'', preprint, 1996.

\item {\bf [Cabello 96]}:
A. Cabello,
``Pruebas algebraicas de imposibilidad de
variables ocultas en mec\'{a}nica cu\'{a}ntica'',
Ph.\ D. thesis, Universidad Complutense de Madrid, 1996.
Published version: Universidad Complutense de Madrid, 2001.

\item {\bf [Cabello-Garc\'{\i}a Alcaine 97 a]}:
A. Cabello, \& G. Garc\'{\i}a Alcaine,
``Quantum mechanics and elements of reality inferred from joint
measurements'',
{\em J. Phys. A} {\bf 30}, 2, 725-732 (1997);
quant-ph/9709056.

\item {\bf [Cabello 97 a]}:
A. Cabello,
``No-hidden-variables proof for two spin-$\frac{1}{2}$ particles
preselected and postselected in unentangled states'',
{\em Phys. Rev. A} {\bf 55}, 6, 4109-4111 (1997);
quant-ph/9706016.

\item {\bf [Cabello 97 b]}:
A. Cabello,
``A proof with 18 vectors of the Bell-Kochen-Specker theorem'',
in M. Ferrero, \& A. van der Merwe (eds.),
{\em New developments
on fundamental problems in quantum physics (Oviedo, Spain, 1996)},
Kluwer Academic, Dordrecht, Holland, 1997, pp.~59-62.

\item {\bf [Cabello 97 c]}:
A. Cabello (comp.),
{\em Misterios de la f\'{\i}sica cu\'{a}ntica},
Temas de Investigaci\'{o}n y Ciencia n.\ 10, Prensa Cient\'{\i}fica,
Barcelona, 1997.

\item {\bf [Cabello 97 d]}:
A. Cabello,
``Introducci\'{o}n'',
in {\bf [Cabello 97 c]}, pp.~2-3.

\item {\bf [Cabello 97 e]}:
A. Cabello,
``Los experimentos no realizados no tienen resultados'',
in {\bf [Cabello 97 c]}, pp.~65-67.

\item {\bf [Cabello-Garc\'{\i}a Alcaine 98]}:
A. Cabello, \& G. Garc\'{\i}a Alcaine,
``Proposed experimental tests of the Bell-Kochen-Specker theorem'',
{\em Phys. Rev. Lett.} {\bf 80}, 9, 1797-1799 (1998);
quant-ph/9709047.

\item {\bf [Cabello 98]}:
A. Cabello,
``Ladder proof of nonlocality without inequalities and
without probabilities'',
{\em Phys. Rev. A} {\bf 58}, 3, 1687-1693 (1998);
quant-ph/9712055.

\item {\bf [Cabello-Cereceda-Garc\'{\i}a de Polavieja 98]}:
A. Cabello, J. L. Cereceda, \& G. Garc\'{\i}a de Polavieja,
``Sobre la transmisi\'{o}n de se\~{n}ales a velocidades superlum\'{\i}nicas
utilizando las correlaciones cu\'{a}nticas'',
{\em Revista Espa\~{n}ola de F\'{\i}sica} {\bf 12}, 3, 44-46 (1998).
See {\bf [Garc\'{\i}a Alcaine 98 b]}.

\item {\bf [Cabello 99 a]}:
A. Cabello,
``Quantum correlations are not local elements of reality'',
{\em Phys. Rev. A} {\bf 59}, 1, 113-115 (1999);
quant-ph/98012088.
See {\bf [Mermin 98 a, b, 99 a]},
{\bf [Cabello 99 c]}, {\bf [Jordan 99]},
{\bf [Fuchs 03 a]} (Chaps.~18, 33).

\item {\bf [Cabello 99 b]}:
A. Cabello,
``Mec\'{a}nica cu\'{a}ntica. Interpretaciones'',
{\em Investigaci\'{o}n y Ciencia} 269, 95 (1999).
Review of {\bf [Dickson 98]}.

\item {\bf [Cabello 99 c]}:
A. Cabello,
``Quantum correlations are not contained in the initial state'',
{\em Phys. Rev. A} {\bf 60}, 2, 877-880 (1999);
quant-ph/9905060.
See {\bf [Mermin 98 a, b, 99 a]},
{\bf [Cabello 99 a]}, {\bf [Jordan 99]},
{\bf [Fuchs 03 a]} (Chaps.~18, 33).

\item {\bf [Cabello 99 d]}:
A. Cabello,
``Comment on `Hidden variables are compatible with physical measurements'\,'',
quant-ph/9911024.
Comment on {\bf [Kent 99 b]}.
See {\bf [Meyer 99 b]}, {\bf [Clifton-Kent 00]},
{\bf [Havlicek-Krenn-Summhammer-Svozil 01]}, {\bf [Mermin 99 b]},
{\bf [Appleby 00, 01, 02]}, {\bf [Boyle-Schafir 01 a]}, {\bf [Cabello 02 c]}.

\item {\bf [Cabello 00 a]}:
A. Cabello,
``Kochen-Specker theorem and experimental test on hidden variables'',
{\em Int. J. Mod. Phys. A} {\bf 15}, 18, 2813-2820 (2000);
quant-ph/9911022.

\item {\bf [Cabello 00 b]}:
A. Cabello,
``Nonlocality without inequalities has not
been proved for maximally entangled states'',
{\em Phys. Rev. A} {\bf 61}, 2, 022119 (2000);
quant-ph/9911023.
See {\bf [Wu-Xie-Huang-Hsia 96]},
{\bf [Barnett-Chefles 98]}, {\bf [Cereceda 99 c]}.

\item {\bf [Cabello 00 c]}:
A. Cabello,
``Quantum key distribution without alternative measurements'',
{\em Phys. Rev. A} {\bf 61}, 5, 052312 (2000);
quant-ph/9911025.
Comment: {\bf [Zhang-Li-Guo 01 a]}.
Reply: {\bf [Cabello 01 b, e]}.

\item {\bf [Cabello 00 d]}:
A. Cabello,
``Introducci\'{o}n a la l\'{o}gica cu\'{a}ntica'',
{\em Arbor} {\bf 167}, 659-660, 489-507 (2000).

\item {\bf [Cabello 00 e]}:
A. Cabello,
``Mec\'{a}nica cu\'{a}ntica desde una perspectiva moderna'',
{\em Investigaci\'{o}n y Ciencia} 291, 86 (2000).
Review of {\bf [Bub 97]}.

\item {\bf [Cabello 00 f]}:
A. Cabello,
``Quantum key distribution in the Holevo limit'',
{\em Phys. Rev. Lett.} {\bf 85}, 26, 5635-5638 (2000);
quant-ph/0007064.

\item {\bf [Cabello 00 g]}:
A. Cabello,
``Procedimiento cu\'{a}ntico para distribuir claves criptogr\'{a}ficas
sin descartar datos'',
patent application,
Oficina Espa\~{n}ola de Patentes y Marcas, P200000713, 2000.

\item {\bf [Cabello 00 h]}:
A. Cabello,
``Bibliographic guide to the foundations of
quantum mechanics and quantum information'',
quant-ph/0012089.

\item {\bf [Cabello 00 i]}:
A. Cabello,
``Multiparty key distribution and secret sharing based
on entanglement swapping'';
quant-ph/0009025.
See {\bf [Lee-Lee-Kim-Oh 03]}.

\item {\bf [Cabello 01 a]}:
A. Cabello,
``Multiparty multilevel Greenberger-Horne-Zeilinger states'',
{\em Phys. Rev. A} {\bf 63}, 2, 022104 (2001);
quant-ph/0007065.
See {\bf [Savinien-Taron-Tarrach 00]}.

\item {\bf [Cabello 01 b]}:
A. Cabello,
`Reply to ``Comment on `Quantum key distribution without
alternative measurements'\,''
[Phys. Rev. A {\bf 63}, 036301 (2001)]',
{\em Phys. Rev. A} {\bf 63}, 3, 036302 (2001).
Reply to {\bf [Zhang-Li-Guo 01 a]}.
See {\bf [Cabello 00 c, 01 e]}.

\item {\bf [Cabello 01 c]}:
A. Cabello,
``Bell's theorem without inequalities and without
probabilities for two observers'',
{\em Phys. Rev. Lett.} {\bf 86}, 10, 1911-1914 (2001);
quant-ph/0008085.
Comment: {\bf [Marinatto 03]}.
Reply: {\bf [Cabello 03 f]}.

\item {\bf [Cabello 01 d]}:
A. Cabello,
`\,``All versus nothing'' inseparability for two observers',
{\em Phys. Rev. Lett.} {\bf 87}, 1, 010403 (2001);
quant-ph/0101108.
Comment: {\bf [Lvovsky 02]}.

\item {\bf [Cabello 01 e]}:
A. Cabello,
`Addendum to ``Quantum key distribution without
alternative measurements''\,',
{\em Phys. Rev. A} {\bf 64}, 2, 024301 (2001);
quant-ph/0009051.
Reply to {\bf [Zhang-Li-Guo 01 a]}.
See {\bf [Cabello 00 c, 01 b]}.

\item {\bf [Cabello 01 f]}:
A. Cabello,
``Mec\'{a}nica cu\'{a}ntica avanzada'',
{\em Investigaci\'{o}n y Ciencia} 300, 93 (2001).
Review of {\bf [Hecht 00]}.

\item {\bf [Cabello 01 g]}:
A. Cabello,
``Efficient quantum cryptography'',
in S. G. Pandalai (ed.),
{\em Recent Res. Devel. Phys.} {\bf 2}, Part II, 249-257 (2001).

\item {\bf [Cabello 02 a]}:
A. Cabello,
``Violating Bell's inequality beyond Cirel'son's bound'',
{\em Phys. Rev. Lett.} {\bf 88}, 6, 060403 (2002);
quant-ph/0108084.
See {\bf [Cabello 02 g]}.

\item {\bf [Cabello 02 b]}:
A. Cabello,
``Bell's theorem with and without inequalities for
the three-qubit Greenberger-Horne-Zeilinger and $W$ states'',
{\em Phys. Rev. A} {\bf 65}, 3, 032108 (2002);
quant-ph/0107146.

\item {\bf [Cabello 02 c]}:
A. Cabello,
``Finite-precision measurement does not nullify the Kochen-Specker theorem'',
{\em Phys. Rev. A} {\bf 65}, 5, 052101 (2002);
quant-ph/0104024.

\item {\bf [Cabello 02 d]}:
A. Cabello,
``Bell's inequality for $n$ spin-$s$ particles'',
{\em Phys. Rev. A} {\bf 65}, 6, 062105 (2002);
quant-ph/0202126.

\item {\bf [Cabello 02 e]}:
A. Cabello,
``$N$-particle $N$-level singlet states: Some properties and applications'',
{\em Phys. Rev. Lett.} {\bf 89}, 10, 100402 (2002);
quant-ph/0203119.

\item {\bf [Cabello 02 f]}:
A. Cabello,
``Mec\'{a}nica cu\'{a}ntica'',
{\em Investigaci\'{o}n y Ciencia} 312, 95 (2002).
Review of {\bf [Levin 02]}.

\item {\bf [Cabello 02 g]}:
A. Cabello,
``Two qubits of a $W$ state violate
Bell's inequality beyond Cirel'son's bound'',
{\em Phys. Rev. A} {\bf 66}, 4, 042114 (2002).
Erratum: {\em Phys. Rev. A} {\bf 67}, 2, 029901 (2003);
quant-ph/0205183.

\item {\bf [Cabello 02 h]}:
A. Cabello,
``Criptograf\'{\i}a cu\'{a}ntica eficiente'',
in C. Mataix, \& A. Rivadulla (eds.),
{\em F\'{\i}sica cu\'{a}ntica y realidad.
Quantum physics and reality (Madrid, 2000)},
Editorial Complutense, Madrid, 2002, pp.~333-344.

\item {\bf [Cabello 02 i]}:
A. Cabello,
``Quantum measurements and decoherence. Models and phenomenology'',
{\em Math. Rev.}, 2002.
Review of {\bf [Mensky 00]}.

\item {\bf [Cabello 02 j]}:
A. Cabello,
``The four-qubit singlet state and decoherence-free subspaces'',
quant-ph/0210080.

\item {\bf [Cabello 03 a]}:
A. Cabello (comp.),
{\em Fen\'{o}menos cu\'{a}nticos},
Temas de Investigaci\'{o}n y Ciencia n.\ 31, Prensa Cient\'{\i}fica,
Barcelona, 2003.

\item {\bf [Cabello 03 b]}:
A. Cabello,
``Rotationally invariant proof of Bell's theorem without inequalities'',
{\em Phys. Rev. A} {\bf 67}, 3, 032107 (2003);
quant-ph/0306073.

\item {\bf [Cabello 03 c]}:
A. Cabello,
``Kochen-Specker theorem for a single qubit using positive operator-valued measures'',
{\em Phys. Rev. Lett.} {\bf 90}, 19, 190401 (2003);
quant-ph/0210082.
See {\bf [Aravind 03 a]}.

\item {\bf [Cabello 03 d]}:
A. Cabello,
``Supersinglets'',
in M. Ferrero (ed.),
{\em Proc. of Quantum Information: Conceptual Foundations,
Developments and Perspectives (Oviedo, Spain, 2002)},
{\em J. Mod. Opt.} {\bf 50}, 6-7, 1049-1061 (2003);
quant-ph/0306074.

\item {\bf [Cabello 03 e]}:
A. Cabello,
``Bell's inequality without alternative settings'',
{\em Phys. Lett. A} {\bf 313}, 1-2, 1-7 (2003);
quant-ph/0210081.

\item {\bf [Cabello 03 f]}:
A. Cabello,
``Cabello replies'',
{\em Phys. Rev. Lett.} {\bf 90}, 25, 258902 (2003);
quant-ph/0306180.
Reply to {\bf [Marinatto 03]}.
See {\bf [Cabello 01 c]}.

\item {\bf [Cabello 03 g]}:
A. Cabello,
``Solving the liar detection problem using the four-qubit singlet state'',
{\em Phys. Rev. A} {\bf 68}, 1, 012304 (2003);
quant-ph/0210079.

\item {\bf [Cabello 03 h]}:
A. Cabello,
``Greenberger-Horne-Zeilinger-like proof of Bell's theorem
involving observers who do not share a reference frame'',
{\em Phys. Rev. A} {\bf 68}, 4, 042104 (2003);
quant-ph/0306075.

\item {\bf [Cabello 03 i]}:
A. Cabello,
``Bell's theorem without inequalities and without alignments'',
{\em Phys. Rev. Lett.} {\bf 91}, 23, 230403 (2003);
quant-ph/0303076.
Comment: {\bf [Marinatto 04]}.
Reply: {\bf [Cabello 04 b]}.

\item {\bf [Cabello 04 a]}:
A. Cabello,
``Proposed experiment to test the bounds of quantum correlations'',
{\em Phys. Rev. Lett.} {\bf 92}, 6, 060403 (2004);
quant-ph/0309172.
See {\bf [Bovino-Castagnoli-Degiovanni-Castelletto 04]}.

\item {\bf [Cabello 04 b]}:
A. Cabello,
``Cabello replies'',
{\em Phys. Rev. Lett.} {\bf 93}, 12, 128902 (2004);
quant-ph/0409189.
Reply to {\bf [Marinatto 04]}.
See {\bf [Cabello 03 i]}.

\item {\bf [Cabello 04 c]}:
A. Cabello,
``Bell's theorem without inequalities and without unspeakable information'',
{\em Found. Phys.} (Festschrift in honor of Asher Peres);
quant-ph/0409190.

\item {\bf [Cabello-L\'{o}pez Tarrida 04]}:
A. Cabello, \& A. J. L\'{o}pez Tarrida,
`Proposed experiment for the quantum ``{\em Guess My Number}'' protocol using four-qubit entangled states',
quant-ph/0409191.

\item {\bf [Cabello 04 d]}:
A. Cabello,
``How much larger quantum correlations are than classical ones'';
quant-ph/0409192.

\item {\bf [Cabello 04 e]}:
A. Cabello,
``Examples of mathematical beauty when comparing classical and quantum worlds'',
{\em Festschrift in honor of Alberto Galindo},
World Scientific, Singapore, 2004.

\item {\bf [Cabello-Calsamiglia 04]}:
A. Cabello, \& J. Calsamiglia,
``Quantum entanglement, indistinguishability, and the absent-minded driver's problem'',
preprint 2004.

\item {\bf [Cabrillo-Cirac-Garc\'{\i}a Fern\'{a}ndez-Zoller 98]}:
C. Cabrillo, J. I. Cirac, P. Garc\'{\i}a Fern\'{a}ndez, \& P. Zoller,
``Creation of entangled states of distant atoms by interference'',
{\em Phys. Rev. A} {\bf 59}, 2, 1025-1033 (1999);
quant-ph/9810013.

\item {\bf [Caetano-Souto Ribeiro 01]}:
D. P. Caetano, \& P. H. Souto Ribeiro,
``Quantum distillation of position entanglement
with the polarization degrees of freedom'',
submitted to {\em Opt. Comm.};
quant-ph/0111065.

\item {\bf [Caetano-Souto Ribeiro-Pardal-Khoury 03]}:
D. P. Caetano, P. H. Souto Ribeiro, J. T. C. Pardal, \& A. Z. Khoury,
``Quantum image control through polarization entanglement in parametric
down-conversion'',
{\em Phys. Rev. A} {\bf 68}, 2, 023805 (2003).

\item {\bf [Cai 03]}:
Q.-Y. Cai,
`The ``ping-pong'' protocol can be attacked without eavesdropping',
{\em Phys. Rev. Lett.} {\bf 91}, 10, 109801 (2003);
quant-ph/0402052.

\item {\bf [Cai-Kuang 02]}:
X.-H. Cai, \& L.-M. Kuang,
``Proposal for teleporting a superposition state of squeezed
vacuum states'',
{\em Phys. Lett. A} {\bf 300}, 2-3, 103-106 (2002).

\item {\bf [Cain-Ahmed-Williams 02]}:
P. A. Cain, H. Ahmed, \& D. A. Williams,
``Hole transport in coupled SiGe quantum dots for quantum computation'',
{\em J. Appl. Phys.} {\bf 92}, ?, 346-? (2002).

\item {\bf [Calarco-Cini-Onofrio 99]}:
T. Calarco, M. Cini, \& R. Onofrio,
``Are violations to temporal Bell inequalities there when somebody
looks?'',
{\em Europhys. Lett.} {\bf 47}, 4, 407-413 (1999);
quant-ph/9908030.

\item {\bf [Calarco-Hinds-Jaksch-(+3) 00]}:
T. Calarco, E. A. Hinds, D. Jaksch, J. Schmiedmayer,
J. I. Cirac, \& P. Zoller,
``Quantum gates with neutral atoms:
Controlling collisional interactions in time-dependent traps'',
{\em Phys. Rev. A} {\bf 61}, 2, 022304 (2000);
quant-ph/9905013.

\item {\bf [Calarco-Briegel-Jaksch-(+2) 00]}:
T. Calarco, H.-J. Briegel, D. Jaksch, J. I. Cirac, \& P. Zoller,
``Quantum computing with trapped particles in microscopic potentials'',
{\em Fortschr. Phys.} {\bf 48}, 9-11 (Special issue:
Experimental proposals for quantum computation), 945-955 (2000).

\item {\bf [Calarco-Cirac-Zoller 01]}:
T. Calarco, J. I. Cirac, \& P. Zoller,
``Entangling ions in arrays of microscopic traps'',
{\em Phys. Rev. A} {\bf 63}, 6, 062304 (2001);
quant-ph/0010105.

\item {\bf [Calarco-Datta-Fedichev-(+3) 03]}:
T. Calarco, A. Datta, P. Fedichev,
E. Pazy, \& P. Zoller,
``Spin-based all-optical quantum computation with quantum dots: Understanding
and suppressing decoherence'',
{\em Phys. Rev. A} {\bf 68}, 1, 012310 (2003);
quant-ph/0304044.

\item {\bf [Calarco-Dorner-Julienne-(+2) 04]}:
T. Calarco, U. Dorner, P. S. Julienne,
C. J. Williams, \& P. Zoller,
``Quantum computations with atoms in optical lattices:
 Marker qubits and molecular interactions'',
{\em Phys. Rev. A} {\bf 70}, 1, 012306 (2004).

\item {\bf [Calderbank-Rains-Shor-Sloane 96]}:
A. R. Calderbank, E. M. Rains, P. W. Shor, \& N. J. A. Sloane,
``Quantum error correction via codes over GF(4)'',
{\em IEEE Trans. Inf. Theory};
quant-ph/9608006.

\item {\bf [Calderbank-Shor 96]}:
A. R. Calderbank, \& P. W. Shor,
``Good quantum error-correcting codes exist'',
{\em Phys. Rev. A} {\bf 54}, 2, 1098-1105 (1996).

\item {\bf [Calderbank-Rains-Shor-Sloane 97]}:
A. R. Calderbank, E. M. Rains, P. W. Shor, \& N. J. A. Sloane,
``Quantum error correction and orthogonal geometry'',
{\em Phys. Rev. Lett.} {\bf 78}, 3, 405-408 (1997).

\item {\bf [Calderbank-Sloane 01]}:
A. R. Calderbank, \& N. J. A. Sloane,
``Obituary: Claude Shannon (1916-2001).
Inventor, mathematician and leader of the digital revolution'',
{\em Nature} {\bf 410}, 6830, 768 (2001).

\item {\bf [Caldeira-Leggett 81]}:
A. O. Caldeira, \& A. J. Leggett,
``Influence of dissipation on quantum tunneling in macroscopic
systems'',
{\em Phys. Rev. Lett.} {\bf 46}, 4, 211-214 (1981)

\item {\bf [Caldeira-Leggett 85]}:
A. O. Caldeira, \& A. J. Leggett,
``Influence of damping on quantum interference: An exactly soluble model'',
{\em Phys. Rev. A} {\bf 31}, 2, 1059-1066 (1985).

\item {\bf [Caldeira-Naddeo 04]}:
C. R. Calidonna, \& A. Naddeo,
``Towards a CA model for quantum computation with fully frustrated linear Josephson junction arrays'',
{\em Phys. Lett. A} {\bf 327}, 5-6, 409-415 (2004).

\item {\bf [Calixto 01]}:
M. Calixto,
``Computaci\'{o}n cu\'{a}ntica: Un reto tecnol\'{o}gico'',
{\em Revista Espa\~{n}ola de F\'{\i}sica} {\bf 15}, 2, 35-43 (2001).

\item {\bf [Calsamiglia-L\"{u}tkenhaus 01]}:
J. Calsamiglia, \& N. L\"{u}tkenhaus,
``Maximum efficiency of a linear-optical Bell-state analyzer'',
{\em Appl. Phys. B} {\bf 72}, 67-71 (2001);
quant-ph/0007058.

\item {\bf [Calsamiglia-Barnett-L\"{u}tkenhaus 02]}:
J. Calsamiglia, S. M. Barnett, \& N. L\"{u}tkenhaus,
``Conditional beam-splitting attack on quantum key distribution'',
{\em Phys. Rev. A} {\bf 65}, 1, 012312 (2002);
quant-ph/0107148.

\item {\bf [Calsamiglia 02]}:
J. Calsamiglia,
``Generalized measurements by linear elements'',
{\em Phys. Rev. A} {\bf 65}, 3, 030301 (2002);
quant-ph/0108108.

\item {\bf [Calude-Hertling-Svozil 98]}:
C. S. Calude, P. H. Hertling, \& K. Svozil,
``Kochen-Specker theorem: Two geometric proofs'',
{\em Tatra Mt. Math. Publ.} {\bf 15}, 133-142 (1998).

\item {\bf [Calude-Hertling-Svozil 99]}:
C. S. Calude, P. H. Hertling, \& K. Svozil,
``Embedding quantum universes into classical ones'',
{\em Found. Phys.} {\bf 29}, 3, 349-379 (1999).

\item {\bf [Camacho 00]}:
A. Camacho,
``Quantum nondemolition measurements in a Paul trap'',
{\em Phys. Lett. A} {\bf 277}, 1, 7-12 (2000);
quant-ph/0010037.

\item {\bf [Camacho 01]}:
A. Camacho,
``Decoherence-induced violations of Einstein equivalence principle'',
{\em Int. J. Mod. Phys. D};
gr-qc/0107028.

\item {\bf [Campos-Gerry 02]}:
R. A. Campos, \& C. C. Gerry,
``A single-photon test of Gleason's theorem'',
{\em Phys. Lett. A} {\bf 299}, 1, 15-18 (2002).

\item {\bf [Can-Klyachko-Shumovsky 02]}:
M. A. Can, A. A. Klyachko, \& A. S. Shumovsky,
``Easily monitored entangled state'',
{\em Appl. Phys. Lett.} {\bf 81}, ?, 5072-? (2002).

\item {\bf [Can-Cakir-Klyachko-Shumovsky 03]}:
M. A. Can, O. Cakir, A. Klyachko, \& A. Shumovsky,
``Robust entanglement in atomic systems via lambda-type processes'',
{\em Phys. Rev. A} {\bf 68}, 2, 022305 (2003).

\item {\bf [Canosa-Rossignoli 02]}:
N. Canosa, \& R. Rossignoli,
``Generalized nonadditive entropies and quantum entanglement'',
{\em Phys. Rev. Lett.} {\bf 88}, 17, 170401 (2002).

\item {\bf [Cantrell-Scully 78]}:
C. D. Cantrell, \& M. O. Scully,
``The EPR paradox revisited'',
{\em Phys. Rep.} {\bf 43}, 13, 499-508 (1978).
Comment: {\bf [Whitaker-Singh 81]}.
See {\bf [Whitaker-Singh 82]}.

\item {\bf [Cao-Yang-Guo 03]}:
Z.-L. Cao, M. Yang, \& G.-C. Guo,
``The scheme for realizing probabilistic teleportation of atomic
states and purifying the quantum channel on cavity QED'',
{\em Phys. Lett. A} {\bf 308}, 5-6, 349-354 (2003).

\item {\bf [Cao-Yang 03 a]}:
Z.-L. Cao, \& M. Yang,
``Entanglement distillation for three-particle $W$ class states'',
{\em J. Phys. B} {\bf 36}, 21, 4245-4253 (2003);
quant-ph/0310118.

\item {\bf [Cao-Yang 03 b]}:
Z.-L. Cao, \& M. Yang,
``Entanglement distillation for $W$ class states'',
quant-ph/0307115.

\item {\bf [Cao-Yang 03 c]}:
Z.-L. Cao, \& M. Yang,
``Scheme for preparation of $W$ state via cavity QED'',
quant-ph/0307173.

\item {\bf [Cao-Yang 03 d]}:
Z.-L. Cao, \& M. Yang,
``Entanglement distillation for atomic states via cavity QED'',
quant-ph/0307174.

\item {\bf [Cao-Song 04 a]}:
Z.-L. Cao, \& W. Song,
``Scheme for the implementation of optimal cloning of two pairs of
orthogonal states'',
{\em Phys. Lett. A} {\bf 325}, 5-6, 309-314 (2004);
quant-ph/0405043.

\item {\bf [Cao-Song 04 b]}:
Z.-L. Cao, \& W. Song,
``Teleportation of a two-particle entangled state via $W$ class states'',
quant-ph/0401054.

\item {\bf [Capasso-Fortunato-Selleri 70]}:
V. Capasso, D. Fortunato, \& F. Selleri,
``Von Neumann's theorem and hidden variable models'',
{\em Rivista del Nuovo Cimento} {\bf ?}, ?, 149-? (1970).

\item {\bf [Capasso-Fortunato-Selleri 73]}:
V. Capasso, D. Fortunato, \& F. Selleri,
``Sensitive observables of quantum mechanics'',
{\em Int. J. Theor. Phys.} {\bf 7}, 5, 319-326 (1973).
See {\bf [Fortunato-Selleri 76]}.

\item {\bf [Carazza 99]}:
B. Carazza,
``Decoherence within a simple model for the environment'',
{\em Found. Phys. Lett.} {\bf 12}, ?, 485-495 (1999);
quant-ph/9907051.

\item {\bf [Cardone-Mignani-Olkhovsky 01]}:
F. Cardone, R. Mignani, \& V. S. Olkhovsky,
``Are particle and photon tunneling and filling
in barriers local or non-local phenomena?'',
{\em Phys. Lett. A} {\bf 289}, 6, 279-286 (2001).

\item {\bf [Carlini-Hosoya 98]}:
A. Carlini, \& A. Hosoya,
``An alternative algorithm for the database search problem on
a quantum computer'',
quant-ph/9808028.

\item {\bf [Carlini-Hosoya 99]}:
A. Carlini, \& A. Hosoya,
``Carmichael numbers on a quantum computer'',
quant-ph/9908022.

\item {\bf [Carlini-Hosoya 00]}:
A. Carlini, \& A. Hosoya,
``Quantum probabilistic subroutines and problems in number theory'',
{\em Phys. Rev. A} {\bf 62}, 3, 032312 (2000);
quant-ph/9907020.

\item {\bf [Carlini-Hosoya 01]}:
A. Carlini, \& A. Hosoya,
``Quantum computers and unstructured search:
Finding and counting items with an arbitrarily entangled initial state'',
{\em Phys. Lett. A} {\bf 280}, 3, 114-120 (2001);
quant-ph/9909089.

\item {\bf [Carlo-Benenti-Casati 03]}:
G. G. Carlo, G. Benenti, \& G. Casati,
``Teleportation in a noisy environment:
A quantum trajectories approach'',
{\em Phys. Rev. Lett.} {\bf 91}, 25, 257903 (2003).

\item {\bf [Carlo-Benenti-Casati-Mej\'{\i}a Monasterio 04]}:
G. G. Carlo, G. Benenti, G. Casati, \& C. Mej\'{\i}a-Monasterio,
``Simulating noisy quantum protocols with quantum trajectories'',
{\em Phys. Rev. A} {\bf 69}, 6, 062317 (2004).

\item {\bf [Carmeli-Cassinelli-DeVito-(+2) 04]}:
C. Carmeli, G. Cassinelli, E. DeVito, A. Toigo, \& B. Vacchini,
``A complete characterization of phase space measurements'',
{\em J. Phys. A} {\bf 37}, ?, 5057-5066 (2004).
quant-ph/0405026.

\item {\bf [Caro-Salcedo 98]}:
J. Caro, \& L. L. Salcedo,
``Obstructions to mixing classical and quantum dynamics'',
quant-ph/9812046.

\item {\bf [De Caro-Garuccio 94]}:
L. de Caro, \& A. Garuccio,
``Reliability of Bell-inequality measurements using polarization
correlations in parametric-down-conversion photon sources'',
{\em Phys. Rev. A} {\bf 50}, 4, R2803-R2805 (1994).
Comment: {\bf [Kwiat 95]}.

\item {\bf [De Caro-Garuccio 96]}:
L. de Caro, \& A. Garuccio,
``Bell's inequality,
trichotomic observables, and supplementary assumptions'',
{\em Phys. Rev. A} {\bf 54}, 1, 174-181 (1996).

\item {\bf [Carollo-Palma-Simon-Zeilinger 01]}:
A. Carollo, G. M. Palma, C. Simon, \& A. Zeilinger,
``Linear optical implementation of nonlocal
product states and their indistinguishability'',
{\em Phys. Rev. A} {\bf 64}, 2, 022318 (2001);
quant-ph/0102124.

\item {\bf [Carollo-Palma 01]}:
A. Carollo, \& G. M. Palma,
``The role of auxiliary states in state
discrimination with linear optical devices'',
quant-ph/0106041.

\item {\bf [Carteret-Linden-Popescu-Sudbery 99]}:
H. A. Carteret, N. Linden, S. Popescu, \& A. Sudbery,
``Multiparticle entanglement'',
{\em Found. Phys.} {\bf 29}, 4, 527-552 (1999).

\item {\bf [Carteret-Sudbery 00]}:
H. A. Carteret, \& A. Sudbery,
``Local symmetry properties of pure three-qubit states'',
{\em J. Phys. A} {\bf 33}, 28, 4981-5002 (2000).

\item {\bf [Carteret-Higuchi-Sudbery 00]}:
H. A. Carteret, A. Higuchi \& A. Sudbery,
``Multipartite generalisation of the Schmidt decomposition'',
{\em J. Math. Phys.};
quant-ph/0006125.

\item {\bf [Carteret-Ismail-Richmond 03]}:
H. A. Carteret, M. E. H. Ismail, \& B. Richmond,
``Three routes to the exact asymptotics for the one-dimensional quantum
walk'',
{\em J. Phys. A} {\bf 36}, 33, 8775-8795 (2003);
quant-ph/0303105.

\item {\bf [Carteret 03 a]}:
H. A. Carteret,
``Noiseless circuits for the concurrence and residual 3-tangle'',
quant-ph/0309212.

\item {\bf [Carteret 03 b]}:
H. A. Carteret,
``Noiseless circuits for the Peres criterion'',
quant-ph/0309216.

\item {\bf [Cartwright 74]}:
N. Cartwright,
``Van Fraassen's modal model of quantum mechanics'',
{\em Philos. Sci.} {\bf 41}, ?, 199-202 (1974).

\item {\bf [Cartwright 77]}:
N. Cartwright,
``The sum rule has not been tested'',
{\em Philos. Sci.} {\bf 44}, ?, 107-112 (1977).

\item {\bf [Carvalho-Milman-de Matos Filho-Davidovich 01]}:
A. R. R. Carvalho, P. Milman,
R. L. de Matos Filho, \& L. Davidovich,
``Decoherence, pointer engineering, and quantum state protection'',
{\em Phys. Rev. Lett.} {\bf 86}, 22, 4988-4991 (2001);
quant-ph/0009024.
See {\bf [Paz 01]}.

\item {\bf [Carvalho-Mintert-Buchleitner 04]}:
A. R. R. Carvalho, F. Mintert, \& A. Buchleitner,
``Decoherence and multipartite entanglement'',
{\em Phys. Rev. Lett.};
quant-ph/0410208.

\item {\bf [Casado-Marshall-Santos 97]}:
A. Casado, T. W. Marshall, \& E. Santos,
``Parametric downconversion experiments in the Wigner representation'',
{\em J. Opt. Soc. Am. B} {\bf 14}, 3, 494-502 (1997).

\item {\bf [Casado-Fern\'{a}ndez Rueda-Marshall-(+2)
97 a]}:
A. Casado, A. Fern\'{a}ndez Rueda, T. W. Marshall,
R. Risco Delgado, \& E. Santos,
``Fourth-order interference in the Wigner representation for parametric
down-conversion experiments'',
{\em Phys. Rev. A} {\bf 55}, 5, 3879-3890 (1997).

\item {\bf [Casado-Fern\'{a}ndez Rueda-Marshall-(+2)
97 b]}:
A. Casado, A. Fern\'{a}ndez Rueda, T. W. Marshall,
R. Risco Delgado, \& E. Santos,
``Dispersion cancellation and quantum eraser experiments analyzed
in the Wigner function formalism'',
{\em Phys. Rev. A} {\bf 56}, 3, 2477-2480 (1997).
See {\bf [Kwiat-Steinberg-Chiao 92]}.

\item {\bf [Casado 98]}:
A. Casado,
``Estudio de los expe\-ri\-mentos de conversi\'{o}n
param\'{e}trica a la baja con el formalismo de la funci\'{o}n de Wigner'',
Ph.\ D. thesis, Universidad de Sevilla, 1998.

\item {\bf [Casado-Marshall-Santos 98]}:
A. Casado, T. W. Marshall, \& E. Santos,
``Type-II parametric down conversion in the Wigner-function formalism.
Entanglement and Bell's inequalities'',
{\em J. Opt. Soc. Am. B} {\bf 15}, 5, 1572-1577 (1998);
quant-ph/9711042.

\item {\bf [Casado-Fern\'{a}ndez Rueda-Marshall-(+2) 00]}:
A. Casado, A. Fern\'{a}ndez Rueda, T. W. Marshall,
R. Risco Delgado, \& E. Santos,
``Dependence on crystal parameters of the correlation time between
signal and idler beams in parametric down conversion calculated in
the Wigner representation'',
{\em Eur. Phys. J. D} {\bf 11}, 3, 465-472 (2000).
See {\bf [Casado-Marshall-Risco Delgado-Santos 01]}.

\item {\bf [Casado-Marshall-Risco Delgado-Santos 01]}:
A. Casado, T. W. Marshall,
R. Risco Delgado, \& E. Santos,
``Spectrum of the parametric down converted radiation calculated in
the Wigner representation'',
{\em Eur. Phys. J. D} {\bf 13}, 1, 109-119 (2001).
See {\bf [Casado-Fern\'{a}ndez Rueda-Marshall-(+2) 00]}.

\item {\bf [Casado-Risco Delgado-Santos 01]}:
A. Casado, R. Risco Delgado, \& E. Santos,
``Local realistic theory for PDC experiments based on the Wigner
formalism'',
in {\em Mysteries, Puzzles, and Paradoxes in Quantum Mechanics
(Gargnano, Italy, 2000)},
{\em Naturforsch.} {\bf 56A}, 178-181 (2001);
quant-ph/0103045.
See {\bf [Casado-Risco Delgado-Santos 02]}.

\item {\bf [Casado-Risco Delgado-Santos 02]}:
A. Casado, R. Risco Delgado, \& E. Santos,
``Modelo realista local para los experimentos de conversi\'{o}on param\'{e}trica
a la baja basado en la representaci\'{o}n de Wigner de \'{o}ptica cu\'{a}ntica'',
in C. Mataix, \& A. Rivadulla (eds.),
{\em F\'{\i}sica cu\'{a}ntica y realidad.
Quantum physics and reality (Madrid, 2000)},
Editorial Complutense, Madrid, 2002, pp.~285-296.
See {\bf [Casado-Risco Delgado-Santos 01]}.

\item {\bf [Casado-Marshall-Risco Delgado-Santos 02]}:
A. Casado, T. Marshall, R. Risco Delgado, \& E. Santos,
``A local hidden variables model for experiments involving
photon pairs produced in parametric down conversion'',
quant-ph/0202097.

\item {\bf [Casas-Lamberti-Plastino-Plastino 04]}:
M. Casas, P. W. Lamberti, A. Plastino, \& A. R. Plastino,
``Jensen-Shannon divergence, Fisher information, and Wootters' hypothesis'',
quant-ph/0407147.

\item {\bf [Casati-Chirikov 95]}:
G. Casati, \& B. V. Chirikov,
`Comment on ``Decoherence, chaos, and the second law''\,'
{\em Phys. Rev. Lett.} {\bf 75}, 2, 350 (1995).
Comment on {\bf [Zurek-Paz 94]}.
Reply: {\bf [Zurek-Paz 95 b]}.

\item {\bf [Caser 82]}:
S. Caser,
``On a possible role of the vector potential in the
Einstein-Podolsky-Rosen paradox'',
{\em Phys. Lett. A} {\bf 92}, 1, 13-16 (1982).

\item {\bf [Cassidy 92]}:
D. C. Cassidy,
{\em Uncertainty. The life and science of
Werner Heisenberg}, Freeman, New York, 1992.

\item {\bf [Cassinelli-Lahti 90]}:
G. Cassinelli, \& P. J. Lahti,
``Strong-correlation measurements in quantum mechanics'',
{\em Nuovo Cimento B} {\bf 105}, 11, 1223-1233 (1990).

\item {\bf [Cassinelli-Lahti 95]}:
G. Cassinelli, \& P. J. Lahti,
``Quantum theory of
measurement and the modal interpretations of quantum mechanics'',
{\em Int. J. Theor. Phys.} {\bf 34}, 8, 1271-1281 (1995).

\item {\bf [Cassinello 94]}:
A. Cassinello,
``La interpretaci\'{o}n de los muchos universos de la
mec\'{a}nica cu\'{a}ntica. Apuntes hist\'{o}ricos'',
{\em Arbor} {\bf 148}, ?, 47-? (1994).

\item {\bf [Cassinello-S\'{a}nchez G\'{o}mez 96]}:
A. Cassinello, \& Jos\'{e} Luis S\'{a}nchez G\'{o}mez,
``On the probabilistic postulate of quantum mechanics'',
{\em Found. Phys.} {\bf 26}, 10, 1357-1374 (1996).
See {\bf [Cassinello 96]}.

\item {\bf [Cassinello 96]}:
A. Cassinello,
``?'',
Ph.\ D. thesis, Universidad Aut\'{o}noma de Madrid, 1996.

\item {\bf [Castagnino-Gunzig 99]}:
M. Castagnino, \& E. Gunzig,
``Minimal irreversible quantum mechanics: An axiomatic formalism'',
{\em Int. J. Theor. Phys.} {\bf 38}, ?, 47-? (1999);
quant-ph/0005078.

\item {\bf [Castagnino-Laura 00 a]}:
M. Castagnino, \& R. Laura,
``Decoherence and the final pointer basis'',
in {\em Workshop on Quantum Mechanics (Goslar, 1998)};
quant-ph/0005098.

\item {\bf [Castagnino-Laura 00 b]}:
M. Castagnino, \& R. Laura,
``Functional approach to quantum decoherence
and the classical final limit'',
{\em Phys. Rev. A} {\bf 62}, 2, 022107 (2000);
quant-ph/0005099.
See {\bf [Castagnino-Laura-Betan 01]} (II).

\item {\bf [Castagnino-Lombardi 02]}:
M. Castagnino, \& O. Lombardi,
``Self-induced selection: A new approach to quantum decoherence'' (2002),
PITT-PHIL-SCI00000801.

\item {\bf [Castagnino-Laura-Betan 01]}:
M. Castagnino, R. Laura, \& R. I. Betan,
``Functional approach to quantum decoherence and the classical final
limit II: The pointer basis and the quantum measurements'',
quant-ph/0103107.
See {\bf [Castagnino-Laura 00 b]} (I).

\item {\bf [Castagnoli-Monti 97]}:
G. Castagnoli, \& D. Monti,
``A reductionist approach to quantum computation'',
quant-ph/9711045.

\item {\bf [Castagnoli-Ekert-Macchiavello 98]}:
G. Castagnoli, A. K. Ekert, \& C. Macchiavello,
``Quantum computation: From the sequential approach
to simulated annealing'',
{\em Int. J. Theor. Phys.} {\bf 32}, 1, 463-470 (1998).

\item {\bf [Castagnoli-Monti 98 a]}:
G. Castagnoli, \& D. Monti,
``Quantum computation based on particle statistics'',
quant-ph/9806010.
See {\bf [Castagnoli-Monti 98 b]}.

\item {\bf [Castagnoli-Monti 98 b]}:
G. Castagnoli, \& D. Monti,
``Exploiting particle statistics in quantum computation'',
submitted to 4h Biannual IQSA Meeting Quantum Structures '98;
quant-ph/9806086.
Further development of {\bf [Castagnoli-Monti 98 a]}.

\item {\bf [Castagnoli-Monti 98 c]}:
G. Castagnoli, \& D. Monti,
``The non-mechanistic character of quantum computation'',
{\em Int.\ Quantum Structures Association Conf.\
(Liptovsky, 1998)};
quant-ph/9811039.

\item {\bf [Castagnoli-Monti-Sergienko 99]}:
G. Castagnoli, D. Monti, \& A. Sergienko,
``Performing quantum measurement in suitably entangled states
originates the quantum computation speed up'',
quant-ph/9908015.

\item {\bf [Castagnoli 99]}:
G. Castagnoli,
``A quantum logic gate representation of quantum measurement:
Reversing and unifying the two steps of von Neumann's model'',
submitted to {\em Int. J. Theor. Phys.};
quant-ph/9912020.

\item {\bf [Castagnoli 00 a]}:
G. Castagnoli,
``Parallel quantum computation, the Library of Babel and quantum
measurement as the efficient librarian'',
quant-ph/0003003.

\item {\bf [Castagnoli 00 b]}:
G. Castagnoli,
``Performing quantum measurement in suitably entangled
states originates the quantum computation speed up'',
quant-ph/0005069.

\item {\bf [Castagnoli-Finkelstein 01 a]}:
G. Castagnoli, \& D. R. Finkelstein,
`` Theory of the quantum speed-up'',
{\em Proc. R. Soc. Lond. A} {\bf 457}, 2012, 1799-1806 (2001);
quant-ph/0010081.

\item {\bf [Castagnoli-Finkelstein 01 b]}:
G. Castagnoli, \& D. R. Finkelstein,
``Quantum-statistical computation'',
submitted to {\em Proc. R. Soc. Lond. A};
quant-ph/0111120.

\item {\bf [Castelletto-Degiovanni-Rastello 03 a]}:
S. Castelletto, I. P. Degiovanni, \& M. L. Rastello,
``Quantum and classical noise in practical quantum-cryptography systems based
on polarization-entangled photons'',
{\em Phys. Rev. A} {\bf 67}, 2, 022305 (2003);
quant-ph/0302196.

\item {\bf [Castelletto-Degiovanni-Rastello 03 b]}:
S. Castelletto, I. P. Degiovanni, \& M. L. Rastello,
``Modified Wigner inequality for secure quantum-key distribution'',
{\em Phys. Rev. A} {\bf 67}, 4, 044303 (2003).

\item {\bf [Castelli-Lugiato 97]}:
F. Castelli, \& L. A. Lugiato,
``Realization of the Einstein-Podolsky-Rosen paradox in the
far-field of the optical parametric oscillator above threshold'',
{\em J. Mod. Opt.} {\bf 44}, 4, 765-783 (1997).

\item {\bf [Caticha 98 a]}:
A. Caticha,
``Consistency, amplitudes, and probabilities in quantum theory'',
{\em Phys. Rev. A} {\bf 57}, 3, 1572-1582 (1998);
quant-ph/9804012.
Comment: {\bf [Finkelstein 99 b]}.
Reply: {\bf [Caticha 99]}.
See {\bf [Tikochinsky-Gull 00]}.

\item {\bf [Caticha 98 b]}:
A. Caticha,
``Consistency and linearity in quantum theory'',
{\em Phys. Lett. A} {\bf 244}, 1-3, 13-17 (1998);
quant-ph/9803086.

\item {\bf [Caticha 99]}:
A. Caticha,
``Reply to `Comment on ``Consistency, amplitudes, and probabilities in
quantum theory''\,'\,'',
{\em Phys. Rev. A} {\bf 60}, 2, 1725 (1999).
Reply to {\bf [Finkelstein 99 b]}.
See {\bf [Caticha 98 a]}.

\item {\bf [Caticha 00]}:
A. Caticha,
``Insufficient reason and entropy in quantum theory'',
{\em Found. Phys. Lett.} {\bf 30}, 2, 227-251 (2000).

\item {\bf [Cattaneo-Gudder 99]}:
G. Cattaneo, \& S. Gudder,
``Algebraic structures arising in axiomatic unsharp quantum physics'',
{\em Found. Phys.} {\bf 29}, 10, 1607-1638 (1999).

\item {\bf [Cavalcanti-Cioletti-Terra Cunha 04]}:
D. Cavalcanti, L. M. Cioletti, \& M. O. Terra Cunha,
``Tomographic characterization of three qubits pure states with only two
qubits detectors'',
quant-ph/0408022.

\item {\bf [Caves-Thorne-Drever-(+2) 80]}:
C. M. Caves, K. S. Thorne, R. W. P. Drever,
V. D. Sandberg, \& M. Zimmermann,
``On the measurement of a weak classical force coupled to a
quantum-mechanical oscillator. I. Issues of principles'',
{\em Rev. Mod. Phys.} {\bf 52}, 2, 341-392 (1980).

\item {\bf [Caves-Milburn 87]}:
C. M. Caves, \& G. J. Milburn,
``Quantum-mechanical model for continuous position measurements'',
{\em Phys. Rev. A} {\bf 36}, 12, 5543-5555 (1987).

\item {\bf [Caves 90]}:
C. M. Caves,
``Entropy and information: How much information is needed to assign a probability?'',
in {\bf [Zurek 90]}, pp.~91-115.

\item {\bf [Caves-Drummond 94]}:
C. M. Caves, \& P. D. Drummond,
``Quantum limits of bosonic communication rates'',
{\em Rev. Mod. Phys.} {\bf 66}, 2, 481-537 (1994).

\item {\bf [Caves 94]}:
C. M. Caves,
``Book review. Quantum theory: Concepts and methods'',
{\em Found. Phys.} {\bf 24}, 11, 1583-1585 (1994).
Review of {\bf [Peres 93 a]}.

\item {\bf [Caves-Fuchs 96]}:
C. M. Caves, \& C. A. Fuchs,
``Quantum information: How much information in a state vector?'',
in A. Mann, \& M. Revzen (eds.),
{\em The dilemma of Einstein, Podolsky and Rosen -- 60 years
later. An international symposium in honour of Nathan Rosen
(Haifa, Israel, 1995)},
{\em Ann. Phys. Soc. Israel} {\bf 12}, 226-257 (1996).

\item {\bf [Caves 98 a]}:
C. M. Caves,
``Quantum teleportation: Enhanced: A tale of two cities'',
{\em Science} {\bf 282}, 5389, 637-638 (1998).
See {\bf [Furusawa-S{\o}rensen-Braunstein-(+3) 98]}.

\item {\bf [Caves 98 b]}:
C. M. Caves,
``Quantum error correction and reversible operations'',
based on a talk presented at the {\em Int. Workshop on
Macroscopic Quantum Tunneling and Coherence (Naples, Italy, 1998)},
{\em Superconductivity};
quant-ph/9811082.

\item {\bf [Caves-Milburn 99]}:
C. M. Caves, \& G. J. Milburn,
``Qutrit entanglement'',
submitted to {\em Opt. Comm.};
quant-ph/9910001.

\item {\bf [Caves-Fuchs-Rungta 01]}:
C. M. Caves, C. A. Fuchs, \& P. Rungta,
``Entanglement of formation of an arbitrary state of two rebits'',
{\em Found. Phys. Lett.} {\bf 14}, 3, 199-212 (2001);
quant-ph/0009063.

\item {\bf [Caves-Fuchs-Schack 01]}:
C. M. Caves, C. A. Fuchs, \& R. Schack,
``Making good sense of quantum probabilities'',
quant-ph/0106133.

\item {\bf [Caves-Fuchs-Schack 02 a]}:
C. M. Caves, C. A. Fuchs, \& R. Schack,
``Quantum probabilities as Bayesian probabilities'',
{\em Phys. Rev. A} {\bf 65}, 2, 022305 (2002);
quant-ph/0106133.

\item {\bf [Caves-Fuchs-Schack 02 b]}:
C. M. Caves, C. A. Fuchs, \& R. Schack,
``Unknown quantum states: The quantum de Finetti representation'',
{\em J. Math. Phys.} {\bf 43}, 9, 4537-4559 (2002);
quant-ph/0104088.

\item {\bf [Caves-Fuchs-Schack 02 c]}:
C. M. Caves, C. A. Fuchs, \& R. Schack,
``Conditions for compatibility of quantum state assignments'',
quant-ph/0206110.

\item {\bf [Caves-Deutsch-Blume Kohout 03]}:
C. M. Caves, I. H. Deutsch, R. Blume-Kohout,
``Physical-resource demands for scalable quantum computation'',
{\em Proc.\ of the SPIE Conf.\ on Fluctuations and Noise in Photonics and Quantum
Optics (Santa Fe, New Mexico, 2003)};
quant-ph/0304083.

\item {\bf [Caves-Fuchs-Manne-Renes 04]}:
C. M. Caves, C. A. Fuchs, K. Manne, \& J. M. Renes,
``Gleason-type derivations of the quantum probability rule for generalized
measurements'',
{\em Found. Phys.} {\bf 34}, 2, 193-209 (2004);
quant-ph/0306179.

\item {\bf [Caves-W\'{o}dkiewicz 04 a]}:
C. M. Caves, \& K. W\'{o}dkiewicz,
``Classical phase-space descriptions of continuous-variable teleportation'',
{\em Phys. Rev. Lett.} {\bf 93}, 4, 040506 (2004);
quant-ph/0401149.

\item {\bf [Caves-W\'{o}dkiewicz 04 b]}:
C. M. Caves, \& K. W\'{o}dkiewicz,
``Fidelity of Gaussian channels'',
quant-ph/0409063.

\item {\bf [Caves-Schack 04]}:
C. M. Caves, \& R. Schack,
``Properties of the frequency operator do not imply the quantum
probability postulate'',
quant-ph/0409144.

\item {\bf [Cen-Li-Zhu 00]}:
L.-X. Cen, F.-L. Li, \& S.-Y. Zhu,
``Optimal entanglement manipulation for an arbitrary state
of two qubits'',
{\em Phys. Lett. A} {\bf 275}, 5-6, 368-372 (2000).

\item {\bf [Cen-Wang 00]}:
L.-X. Cen, \& S.-J. Wang,
``Distilling a Greenberger-Horne-Zeilinger state from
an arbitrary pure state of three qubits'',
quant-ph/0012022.

\item {\bf [Cen-Wu-Yang-An 02]}:
L.-X. Cen, N.-J. Wu, F.-H. Yang, \& J.-H. An,
``Local transformation of mixed states of two qubits to Bell diagonal states'',
{\em Phys. Rev. A} {\bf 65}, 5, 052318 (2002);
quant-ph/0203092.

\item {\bf [Cen-Li-Yan-(+2) 03]}:
L.-X. Cen, X.-Q. Li, Y.-J. Yan, H.-Z. Zheng, \& S.-J. Wang,
``Evaluation of holonomic quantum computation: Adiabatic versus nonadiabatic'',
{\em Phys. Rev. Lett.} {\bf 90}, 14, 147902 (2003);
quant-ph/0208120.

\item {\bf [Cen-Li-Yan 03]}:
L.-X. Cen, X.-Q. Li, Y.-J. Yan,
``Characterization of entanglement transformation via group representation theory'',
{\em J. Phys. A} {\bf 36}, 49, 12267-12273 (2003).

\item {\bf [Cen-Zanardi 04 a]}:
L.-X. Cen, \& P. Zanardi,
``Refocusing schemes for holonomic quantum computation in presence of
dissipation'',
quant-ph/0403143.

\item {\bf [Cen-Zanardi 04 b]}:
L.-X. Cen, \& P. Zanardi,
``Decoherence suppression for oscillator-assisted geometric quantum gates
via symmetrization'',
quant-ph/0410114.

\item {\bf [Cereceda 95]}:
J. L. Cereceda,
``The Kochen-Specker theorem and Bell's theorem: An algebraic approach'',
{\em Found. Phys.} {\bf 25}, 6, 925-949 (1995).

\item {\bf [Cereceda 96 a]}:
J. L. Cereceda,
``An apparent paradox at the heart of quantum mechanics'',
{\em Am. J. Phys.} {\bf 64}, 4, 459-466 (1996).
See {\bf [Kar-Roy-Ghosh-Sarkar 99]}.

\item {\bf [Cereceda 96 b]}:
J. L. Cereceda,
``Maximally entangled states and the Bell inequality'',
{\em Phys. Lett. A} {\bf 212}, 3, 123-129 (1996);
quant-ph/9812012.
See {\bf [Kar 95]}.

\item {\bf [Cereceda 97 a]}:
J. L. Cereceda,
``El teorema de Bell sin desigualdades: Estados GHZ'',
{\em Revista Espa\~{n}ola de F\'{\i}sica} {\bf 11}, 1, 27-31 (1997).

\item {\bf [Cereceda 97 b]}:
J. L. Cereceda,
``Comment on `Local-realism violations in two-particle interferometry'\,'',
{\em Phys. Rev. A} {\bf 55}, 5, 3968-3969 (1997).
Comment on {\bf [Wu-Xie-Huang-Hsia 96]}.

\item {\bf [Cereceda 97 c]}:
J. L. Cereceda,
``Two-particle entanglement as a property of three-particle entangled states'',
{\em Phys. Rev. A} {\bf 56}, 3, 1733-1738 (1997);
quant-ph/9905095.
See {\bf [Krenn-Zeilinger 96]}.

\item {\bf [Cereceda 98]}:
J. L. Cereceda,
``Generalized probability for Hardy's nonlocality contradiction'',
{\em Phys. Rev. A} {\bf 57}, 1, 659-662 (1998).

\item {\bf [Cereceda 99 a]}:
J. L. Cereceda,
``A simple proof of the converse of Hardy's theorem'',
quant-ph/9907094.

\item {\bf [Cereceda 99 b]}:
J. L. Cereceda,
``Quantum perfect correlations and Hardy's nonlocality theorem'',
{\em Found. Phys. Lett.} {\bf 12}, 3, 211-231 (1999).
Erratum: {\em Found. Phys. Lett.} {\bf 13}, 1, 105 (2000);
quant-ph/9908039.

\item {\bf [Cereceda 99 c]}:
J. L. Cereceda,
``Hardy-type experiment for the maximally entangled state:
Illustrating the problem of subensemble postselection'',
{\em Phys. Lett. A} {\bf 263}, 4-6, 232-244 (1999);
quant-ph/0001030.
See {\bf [Wu-Xie-Huang-Hsia 96]}, {\bf [Cabello 00 b]}.

\item {\bf [Cereceda 00 a]}:
J. L. Cereceda,
``Quantum mechanical probabilities and general probabilistic
constraints for Einstein-Podolsky-Rosen-Bohm experiments'',
{\em Found. Phys. Lett.} {\bf 13}, 5, 427-442 (2000);
quant-ph/0003026.

\item {\bf [Cereceda 00 c]}:
J. L. Cereceda,
``Local hidden-variable models and negative-probability
measures'',
quant-ph/0010091.

\item {\bf [Cereceda 01 a]}:
J. L. Cereceda,
``Mermin's $n$-particle Bell inequality and operators' noncommutativity'',
{\em Phys. Lett. A} {\bf 286}, 6, 376-382 (2001);
quant-ph/0007006.

\item {\bf [Cereceda 01 b]}:
J. L. Cereceda,
``Identification of all Hardy-type correlations for
two photons or particles with spin 1/2'',
{\em Found. Phys. Lett.} {\bf 14}, 5, 401-424 (2001);
quant-ph/0101143.

\item {\bf [Cereceda 01 c]}:
J. L. Cereceda,
``Quantum dense coding using three qubits'',
quant-ph/0105096.

\item {\bf [Cereceda 02 a]}:
J. L. Cereceda,
``A feasible quantum optical experiment capable of refuting
noncontextuality for single photons'',
{\em J. Opt. B: Quantum Semiclass. Opt.} {\bf 4}, 2, 87-90 (2002);
quant-ph/0107032.

\item {\bf [Cereceda 02 b]}:
J. L. Cereceda,
``Three-particle entanglement versus three-particle nonlocality'',
{\em Phys. Rev. A} {\bf 66}, 2, 024102 (2002);
quant-ph/0202139.

\item {\bf [Cereceda 02 c]}:
J. L. Cereceda,
``Ladder proof of nonlocality for two spin-half particles revisited'',
{\em J. Phys. A} {\bf 35}, 43, 9105-9111 (2002);
quant-ph/0206015.

\item {\bf [Cereceda 03 a]}:
J. L. Cereceda,
``On the equivalence of the CH and CHSH inequalities for two three-level
systems'',
{\em Int. J. Quant. Inf.} {\bf 1}, 1, 115-133 (2003);
quant-ph/0212117.

\item {\bf [Cereceda 03 b]}:
J. L. Cereceda,
``Degree of entanglement for two qutrits in a pure state'',
quant-ph/0305043.

\item {\bf [Cereceda 03 c]}:
J. L. Cereceda,
`Comment on ``Quantum nonlocality for a three-particle nonmaximally
entangled state without inequalities''\,',
quant-ph/0312046.
Comment on {\bf [Zheng 02 b]}.

\item {\bf [Cereceda 04 a]}:
J. L. Cereceda,
``Hardy's nonlocality for generalized $n$-particle GHZ states'',
{\em Phys. Lett. A} {\bf 327}, 5-6, 433-437 (2004);
quant-ph/0401122.

\item {\bf [Cereceda 04 b]}:
J. L. Cereceda,
`Comment on ``Experimental realization of a three-qubit entangled $W$ state''\,',
quant-ph/0402198.
Comment on {\bf [Eibl-Kiesel-Bourennane-(+3) 04]}.

\item {\bf [Cereceda 04 c]}:
J. L. Cereceda,
``Generalization of the Deutsch algorithm using two qudits'',
quant-ph/0407253.

\item {\bf [Cerf-Cleve 97]}:
N. J. Cerf, \& R. Cleve,
``Information-theoretic interpretation of quantum error-correcting codes'',
{\em Phys. Rev. A} {\bf 56}, 3, 1721-1732 (1997);
quant-ph/9702031.

\item {\bf [Cerf-Adami 97 a]}:
N. J. Cerf, \& C. Adami,
``Negative entropy in quantum information theory'',
in M. Ferrero, \& A. van der Merwe (eds.),
{\em New developments on fundamental problems in quantum physics
(Oviedo, Spain, 1996)},
Kluwer Academic, Dordrecht, Holland, 1997, pp.~77-84.
See {\bf [Cerf-Adami 97 b]}.

\item {\bf [Cerf-Adami 97 b]}:
N. J. Cerf, \& C. Adami,
``Negative entropy and information in quantum mechanics'',
{\em Phys. Rev. Lett.} {\bf 79}, 26, 5194-5197 (1997);
quant-ph/9512022.
See {\bf [Cerf-Adami 97 a]}.

\item {\bf [Cerf 98 a]}:
N. J. Cerf,
``Entropic bounds on coding for noisy quantum channels'',
{\em Phys. Rev. A} {\bf 57}, 5, 3330-3347 (1998);
quant-ph/9707023.

\item {\bf [Cerf-Adami-Kwiat 98]}:
N. J. Cerf, C. Adami, \& P. G. Kwiat,
``Optical simulation of quantum logic'',
{\em Phys. Rev. A} {\bf 57}, 3, R1477-R1480 (1998);
quant-ph/9706022.

\item {\bf [Cerf-Adami 98]}:
N. J. Cerf, \& C. Adami,
``Quantum information theory of entanglement and measurement'',
{\em Physica D} {\bf 120}, ?, 62-81 (1998).

\item {\bf [Cerf-Adami 99]}:
N. J. Cerf, \& C. Adami,
``Quantum extension of conditional probability'',
{\em Phys. Rev. A} {\bf 60}, 2, 893-897 (1999);
quant-ph/9710001.

\item {\bf [Cerf-Adami-Gingrich 99]}:
N. J. Cerf, C. Adami, \& R. M. Gingrich,
``Reduction criterion for separability'',
{\em Phys. Rev. A} {\bf 60}, 2, 898-909 (1999);
quant-ph/9710001.

\item {\bf [Cerf 00]}:
N. J. Cerf,
``Asymmetric quantum cloning machines in any dimension'',
J. Mod. Opt. {\bf 47}, ?, 187-? (2000);
quant-ph/9805024.

\item {\bf [Cerf-Gisin 00]}:
N. J. Cerf, \& N. Gisin,
``Les promesses de l'information quantique'',
{\em La Recherche} {\bf 31}, 1327, 46-53 (2000).
Spanish version: ``Las promesas de la informaci\'{o}n cu\'{a}ntica'',
{\em Mundo Cient\'{\i}fico} {\bf 20}, 210, 48-55 (2000).

\item {\bf [Cerf-Grover-Williams 00]}:
N. J. Cerf, L. K. Grover, \& C. P. Williams,
``Nested quantum search and structured problems'',
{\em Phys. Rev. A} {\bf 61}, 3, 032303 (2000);
quant-ph/980678.

\item {\bf [Cerf-Gisin-Massar 00]}:
N. J. Cerf, N. Gisin, \& S. Massar,
``Classical teleportation of a quantum bit'',
{\em Phys. Rev. Lett.} {\bf 84}, 11, 2521-2524 (2000);
quant-ph/9906105.

\item {\bf [Cerf 00 a]}:
N. J. Cerf,
``Pauli cloning of a quantum bit'',
{\em Phys. Rev. Lett.} {\bf 84}, 19, 4497-4500 (2000);
quant-ph/9803058.

\item {\bf [Cerf-Ipe-Rottenberg 00]}:
N. J. Cerf, A. Ipe, \& X. Rottenberg,
``Cloning of continuous quantum variables'',
{\em Phys. Rev. Lett.} {\bf 85}, 8, 1754-1757 (2000);
quant-ph/9909037.

\item {\bf [Cerf 00 b]}:
N. J. Cerf,
``Asymmetric quantum cloning in any dimension'',
in V. Bu\v{z}zek, \& D. P. DiVincenzo (eds.),
{\em J. Mod. Opt.} {\bf 47}, 2-3 (Special issue:
Physics of quantum information), 187-209 (2000).

\item {\bf [Cerf-Iblisdir 00]}:
N. J. Cerf, \& S. Iblisdir,
``Optimal $N$-to-$M$ cloning of conjugate quantum variables'',
{\em Phys. Rev. A} {\bf 62}, 4, 040301(R) (2000);
quant-ph/0005044.

\item {\bf [Cerf-L\'{e}vy-Van Assche 01]}:
N. J. Cerf, M. L\'{e}vy, \& G. Van Assche,
``Quantum distribution of Gaussian keys using squeezed states'',
{\em Phys. Rev. A} {\bf 63}, 5, 052311 (2001);
quant-ph/0008058.

\item {\bf [Cerf-Massar-Pironio 02]}:
N. J. Cerf, S. Massar, \& S. Pironio,
``Greenberger-Horne-Zeilinger paradoxes for many qudits'',
{\em Phys. Rev. Lett.} {\bf 89}, 8, 080402 (2002);
quant-ph/0107031.

\item {\bf [Cerf-Iblisdir-Van Assche 02]}:
N. J. Cerf, S. Iblisdir, \& G. Van Assche,
``Cloning and cryptography with quantum continuous variables'',
{\em Eur. Phys. J. D} {\bf 18}, 2 (Special issue:
{\em Quantum interference and cryptographic keys:
Novel physics and advancing technologies (QUICK) (Corsica, 2001)}, 211-218 (2002);
quant-ph/0107077.

\item {\bf [Cerf-Bourennane-Karlsson-Gisin 02]}:
N. J. Cerf, M. Bourennane, A. Karlsson, \& N. Gisin,
``Security of quantum key distribution using $d$-level systems'',
{\em Phys. Rev. Lett.} {\bf 88}, 12, 127902 (2002);
quant-ph/0107130.

\item {\bf [Cerf-Iblisdir 01]}:
N. J. Cerf, \& S. Iblisdir,
``Quantum cloning machines with phase-conjugate input modes'',
{\em Phys. Rev. Lett.} {\bf 87}, 24, 247903 (2001);
quant-ph/0102077.

\item {\bf [Cerf-Durt-Gisin 02]}:
N. J. Cerf, T. Durt, \& N. Gisin,
``Cloning a qutrit'',
{\em Proc.\ ESF QIT Conf.\ Quantum Information: Theory, Experiment and Perspectives
(Gdansk, Poland, 2001)}, {\em J. Mod. Opt.} {\bf 49}, 8, 1355-1373 (2002);
quant-ph/0110092.

\item {\bf [Cerf-Massar-Pironio 02]}:
N. J. Cerf, S. Massar, \& S. Pironio,
``Greenberger-Horne-Zeilinger paradoxes for many qudits'',
{\em Phys. Rev. Lett.} {\bf 89}, 8, 080402 (2002);
quant-ph/0107031.

\item {\bf [Cerf-Massar-Schneider 02]}:
N. J. Cerf, S. Massar, \& S. Schneider,
``Multipartite classical and quantum secrecy monotones'',
{\em Phys. Rev. A} {\bf 66}, 4, 042309 (2002);
quant-ph/0202103.

\item {\bf [Cerf-Gisin-Massar-Popescu 04]}:
N. J. Cerf, N. Gisin, S. Massar, \& S. Popescu,
``Quantum entanglement can be simulated without communication'',
quant-ph/0410027.

\item {\bf [Cerletti-Gywat-Loss 04]}:
V. Cerletti, O. Gywat, \& D. Loss,
``Entanglement transfer from electron spins to photons'',
cond-mat/0411235.

\item {\bf [Ceruzzi 93]}:
P. E. Ceruzzi,
``Unpopularized genius'',
{\em Science} {\bf 260}, ?, 1164-1165 (1993).
Review of {\bf [MacRae 91]}.

\item {\bf [Cetto-de la Pe\~{n}a Auerbach-Santos 85]}:
A. M. Cetto, L. de la Pe\~{n}a-Auerbach, \& E. Santos,
``A Bell inequality involving position, momentum and energy'',
{\em Phys. Lett. A} {\bf 113}, 6, 304-306 (1985).

\item {\bf [Chakrabarti-Vasan 03]}:
R. Chakrabarti, \& S. S. Vasan,
``Entanglement via Barut–Girardello coherent state
for suq(1,1) quantum algebra: Bipartite composite system'',
{\em Phys. Lett. A} {\bf 312}, 5-6, 287-295 (2003).

\item {\bf [Chan-Law-Eberly 02]}:
K. W. Chan, C. K. Law, \& J. H. Eberly,
``Localized single-photon wave functions in free space'',
{\em Phys. Rev. Lett.} {\bf 88}, 10, 100402 (2002).

\item {\bf [Chan-Law-Eberly 03]}:
K. W. Chan, C. K. Law, \& J. H. Eberly,
``Quantum entanglement in photon-atom scattering'',
{\em Phys. Rev. A} {\bf 68}, 2, 022110 (2003).

\item {\bf [Chandra-Ghosh 04]}:
N. Chandra, \& R. Ghosh,
``Entanglement in double photoionization of rotating linear molecules'',
{\em Phys. Rev. A} {\bf 69}, 1, 012315 (2004).

\item {\bf [Chang-Pal 92]}:
D. Chang, \& P. Pal,
``Kochen-Specker-Bell paradox for spins larger than 1'',
preprint OITS-491, 1992.

\item {\bf [Chang-Vandersypen-Steffen 00]}:
D. E. Chang, L. M. K. Vandersypen, \& M. Steffen,
``Implementation of a building block for scalable
NMR quantum computation'',
quant-ph/0011055.

\item {\bf [Chapline 02]}:
G. Chapline,
``Entangled states, holography, and quantum surfaces'',
{\em Chaos, Solitons and Fractals} {\bf 14}, 6, 809-816 (2002).

\item {\bf [Chapman-Hammond-Lenef-(+4) 95]}:
M. S. Chapman, T. D. Hammond, A. Lenef, J. Schiedmeyer, R. A. Rubinstein,
E. Smith, \& D. E. Pritchard,
``Photon scattering from atoms in an atom interferometer:
Coherence lost and regained'',
{\em Phys. Rev. Lett.} {\bf 75}, 21, 3783-3787 (1995).

\item {\bf [Charron-Tiesinga-Mies-Williams 02]}:
E. Charron, E. Tiesinga, F. Mies, \& C. Williams,
``Optimizing a phase gate using quantum interference'',
{\em Phys. Rev. Lett.} {\bf 88}, 7, 077901 (2002).

\item {\bf [Chattopadhyay-Das-Gupta Bhaya 00]}:
A. P. Chattopadhyay, B. Das, \& P. Gupta-Bhaya,
`Comment on ``DNA molecular cousin of Schr\"{o}dinger's cat:
A curious example of quantum measuremen''\,',
{\em Phys. Rev. Lett.} {\bf 76}, 1, 195 (2000).
Comment on {\bf [Home-Chattopadhyay 96]}.

\item {\bf [Chaturvedi 02]}:
S. Chaturvedi,
``Measure of strength of an unextendible product basis'',
{\em Phys. Rev. A} {\bf 65}, 4, 042322 (2002);
quant-ph/0105125.

\item {\bf [Chaturvedi 02]}:
S. Chaturvedi,
``Aspects of mutually unbiased bases in odd-prime-power dimensions'',
{\em Phys. Rev. A} {\bf 65}, 4, 044301 (2002).

\item {\bf [Chatzisavvas-Daskaloyannis-Panos 01]}:
K. C. Chatzisavvas, C. Daskaloyannis, \& C. P. Panos,
``Quantum machine language and quantum computation with Josephson junctions'',
quant-ph/0109098.

\item {\bf [Chau 97 a]}:
H. F. Chau,
``Correcting quantum errors in higher spin systems'',
{\em Phys. Rev. A} {\bf 55}, 2, R839-R841 (1997).

\item {\bf [Chau 97 b]}:
H. F. Chau,
``Five quantum register error correction code for higher spin
systems'',
{\em Phys. Rev. A} {\bf 56}, 1, R1-R4 (1997).

\item {\bf [Chau-Lo 98]}:
H. F. Chau, \& H.-K. Lo,
``Making an empty promise with a quantum computer'',
{\bf Fortschr. Phys.} {\bf 46}, 4-5, 507-519 (1998);
quant-ph/9709053.

\item {\bf [Chau 98]}:
H. F. Chau,
``Quantum convolutional error correction codes'',
{\em Phys. Rev. A} {\bf 58}, 2, 905-909 (1998);
quant-ph/9802009.

\item {\bf [Chau 99]}:
H. F. Chau,
``Good quantum-convolutional error-correction codes
and their decoding algorithm exist'',
{\em Phys. Rev. A} {\bf 60}, 3, 1966-1974 (1999);
quant-ph/9806032.

\item {\bf [Chau 00]}:
H. F. Chau,
``Quantum-classical complexity-security tradeoff in secure multiparty
computations'',
{\em Phys. Rev. A} {\bf 61}, 3, 032308 (2000).

\item {\bf [Chau 02]}:
H. F. Chau,
``Practical scheme to share a secret key through a quantum channel with a
27.6\% bit error rate'',
{\em Phys. Rev. A} {\bf 66}, 6, 060302 (2002).

\item {\bf [Chefles-Barnett 96]}:
A. Chefles, \& S. M. Barnett,
``Complementarity and Cirel'son's inequality'',
{\em J. Phys. A} {\bf 29}, 10, L237-L239 (1996).

\item {\bf [Chefles-Barnett 97 a]}:
A. Chefles, \& S. M. Barnett,
``Collective observables and enhanced violation of Bell's inequality'',
{\em Phys. Rev. A} {\bf 55}, 3, 1721-1731 (1997).

\item {\bf [Chefles-Barnett 97 b]}:
A. Chefles, \& S. M. Barnett,
``Diagonalisation of the Bell-CHSH operator'',
{\em Phys. Lett. A} {\bf 232}, 1-2, 4-8 (1997).

\item {\bf [Chefles-Barnett 97 c]}:
A. Chefles, \& S. M. Barnett,
``Entanglement and unanbiguous discrimination between non-orthogonal states'',
{\em Phys. Lett. A} {\bf 236}, 3, 177-179 (1997).

\item {\bf [Chefles 98]}:
A. Chefles,
``Unambiguous discrimination between linearly independent quantum states'',
{\em Phys. Lett. A} {\bf 239}, 6, 339-347 (1998);
quant-ph/9807022.

\item {\bf [Chefles-Barnett 98 a]}:
A. Chefles, \& S. M. Barnett,
``Optimum unambiguous discrimination between linearly independent symmetric states'',
{\em Phys. Lett. A} {\bf 250}, 4-6, 223-229 (1998).

\item {\bf [Chefles-Barnett 98 b]}:
A. Chefles, \& S. M. Barnett,
``Quantum state separation, unambiguous discrimination and exact cloning'',
{\em J. Phys. A} {\bf 31}, 50, 10097-10103 (1998);
quant-ph/9808018.

\item {\bf [Chefles-Barnett 99]}:
A. Chefles, \& S. M. Barnett,
``Strategies and networks for state-dependent quantum cloning'',
{\em Phys. Rev. A} {\bf 60}, 1, 136-144 (1999).

\item {\bf [Chefles 00 a]}:
A. Chefles,
``Deterministic quantum state transformations'',
{\em Phys. Lett. A} {\bf 270}, 1-2, 14-19 (2000);
quant-ph/9911086.

\item {\bf [Chefles-Gilson-Barnett 00]}:
A. Chefles, C. R. Gilson, \& S. M. Barnett,
``Entanglement and collective quantum operations'',
{\em Phys. Lett. A} {\bf 273}, 1-2, 10-14 (2000);
quant-ph/0003062.

\item {\bf [Chefles-Gilson-Barnett 01]}:
A. Chefles, C. R. Gilson, \& S. M. Barnett,
``Entanglement, information, and multiparticle quantum
operations'',
{\em Phys. Rev. A} {\bf 63}, 3, 032314 (2001);
quant-ph/0006106.

\item {\bf [Chefles 00 b]}:
A. Chefles,
``Quantum state discrimination'',
{\em Contemp. Phys.} {\bf 41}, 6, 401-424 (2000);
quant-ph/0010114.

\item {\bf [Chefles 01 a]}:
A. Chefles,
``Unambiguous discrimination between linearly dependent states with multiple
copies'',
{\em Phys. Rev. A} {\bf 64}, 6, 062305 (2001);
quant-ph/0105016.

\item {\bf [Chefles 01 b]}:
A. Chefles,
``Distinguishability measures and ensemble orderings'',
quant-ph/0109141.

\item {\bf [Chefles 02]}:
A. Chefles,
``Quantum operations, state transformations and probabilities'',
{\em Phys. Rev. A} {\bf 65}, 5, 052314 (2002);
quant-ph/0109060.

\item {\bf [Chefles-Sasaki 03]}:
A. Chefles, \& M. Sasaki,
``Retrodiction of generalized measurement outcomes'',
{\em Phys. Rev. A} {\bf 67}, 3, 032112 (2003).

\item {\bf [Chefles-Jozsa-Winter 03]}:
A. Chefles, R. Jozsa, \& A. Winter,
``On the existence of physical transformations between sets of quantum states'',
quant-ph/0307227.

\item {\bf [Chefles-Andresson-Jex 04]}:
A. Chefles, E. Andresson, \& I. Jex,
``Unambiguous comparison of the states of multiple quantum systems'',
{\em J. Phys. A};
quant-ph/0402125.

\item {\bf [Chefles 04]}:
A. Chefles,
``Condition for unambiguous state discrimination with local operations and
classical communication'',
{\em Phys. Rev. A} {\bf 69}, 5, 050307 (2004);
quant-ph/0302066.

\item {\bf [Chen-Ma-Long 04]}:
B. Chen, Y.-J. Ma, \& G. L. Long,
``Quantum game with restricted matrix strategies'',
{\em Commun. Theor. Phys.} {\bf 40} 655-658 (2003);
quant-ph/0301062.

\item {\bf [Chen-Fulling-Scully 99]}:
G. Chen, S. A. Fulling, \& M. O. Scully,
``Grover's algorithm for multiobject search in quantum
computing'',
in {\em Proc.\ of Jackson Hole Quantum Optics Workshop
in honor of Daniel Walls};
quant-ph/9909040.

\item {\bf [Chen-Bonadeo-Steel-(+4) 00]}:
G. Chen, N. H. Bonadeo, D. G. Steel, D. Gammon,
D. S. Katzer, D. Park, \& L. J. Sham,
``Optically induced entanglement of
excitons in a single quantum dot'',
{\em Science} {\bf 289}, 5486, 1906-1909 (2000).

\item {\bf [Chen-Fulling-Chen 00]}:
G. Chen, S. A. Fulling, \& J. Chen,
``Generalization of Grover's algorithm to multiobject search
in quantum computing, part I: Continuous time and discrete time'',
quant-ph/0007123.
See {\bf [Chen-Sun 00]} (II).

\item {\bf [Chen-Sun 00]}:
G. Chen, \& S. Sun,
``Generalization of Grover's algorithm to multiobject search
in quantum computing, part II: General unitary transformations'',
quant-ph/0007124.
See {\bf [Chen-Fulling-Chen 00]} (I).

\item {\bf [Chen-Diao 00 a]}:
G. Chen, \& Z. Diao,
``Quantum multi-object search algorithm with the availability
of partial information'',
quant-ph/0011019.

\item {\bf [Chen-Diao 00 b]}:
G. Chen, \& Z. Diao,
``An exponentially fast quantum search algorithm'',
quant-ph/0011109.
See {\bf [Tu-Long 01]}.

\item {\bf [Chen-Stievater-Batteh-(+6) 02]}:
G. Chen, T. H. Stievater, E. T. Batteh, X. Li, D. G. Steel, D.
Gammon, D. S. Katzer, D. Park, \& L. J. Sham,
``Biexciton quantum coherence in a single quantum dot'',
{\em Phys. Rev. Lett.} {\bf 88}, 11, 117901 (2002).

\item {\bf [Chen-Church-Englert-Zubairy 03]}:
G. Chen, D. A. Church, B.-G. Englert, \& M. S. Zubairy,
``Mathematical models of contemporary elementary quantum computing devices'',
in A. Bandrauk, M. C. Delfour, \& C. Le Bris (eds.),
{\em Quantum control: Mathematical and numerical challenges
(Centre de Recherches Math\'{e}matiques, Universit\'{e} de Montr\'{e}al)}
American Mathematical Society, Providence, Rhode Island, 2003;
quant-ph/0303163.

\item {\bf [Chen 01 a]}:
H. Chen,
``Some good error-correcting codes from algebraic-geometric codes'',
{\em IEEE Trans. Inf. Theory} {\bf 47}, 5, ?-? (2001);
quant-ph/0107102.

\item {\bf [Chen 01 b]}:
H. Chen,
``New invariants and separability criterion of the mixed states: Bipartite case'',
quant-ph/0107111.
See {\bf [Chen 01 c]}.

\item {\bf [Chen 01 c]}:
H. Chen,
``New invariants and separability criterion of the mixed states: Multipartite case'',
quant-ph/0107112.
See {\bf [Chen 01 b]}.

\item {\bf [Chen 01 c]}:
H. Chen,
``Schmidt number of pure states in bipartite quantum systems
as an algebraic-geometric invariant'',
quant-ph/0108034.

\item {\bf [Chen 01 d]}:
H. Chen,
``Quantum entanglement without eigenvalue spectra'',
quant-ph/0108093.
See {\bf [Chen 01 e]}.

\item {\bf [Chen 01 e]}:
H. Chen,
``Quantum entanglement without eigenvalue spectra: Multipartite case'',
quant-ph/0109056.
See {\bf [Chen 01 d]}.

\item {\bf [Chen 01 f]}:
H. Chen,
``Necessary conditions for the efficient simulation of
Hamiltonians using local unitary transformations'',
quant-ph/0109115.

\item {\bf [Chen 01 g]}:
H. Chen,
``Quantum entanglement and geometry of determinantal varieties'',
quant-ph/0110103.

\item {\bf [Chen 01 h]}:
H. Chen,
``Random low rank mixed states are highly entangled'',
quant-ph/0111004.

\item {\bf [Chen 03]}:
H. Chen,
``Schmidt numbers of low-rank bipartite mixed states'',
{\em Phys. Rev. A} {\bf 67}, 6, 062301 (2003).

\item {\bf [Chen-Wu 03]}:
K. Chen, \& L.-A. Wu,
``Detection of entanglement and Bell's inequality violation'',
quant-ph/0306137.

\item {\bf [Chen-Wu 04]}:
K. Chen, \& L.-A. Wu,
``Test for entanglement using physically observable witness
operators and positive maps'',
{\em Phys. Rev. A} {\em 69}, 2, 022312 (2004);
quant-ph/0306041.

\item {\bf [Chen-Kaszlikowski-Kwek-Oh 01]}:
J.-L. Chen, D. Kaszlikowski, L. C. Kwek, \& C. H. Oh,
``Searching a database under decoherence'',
quant-ph/0102033.

\item {\bf [Chen-Kaszlikowski-Kwek-(+2) 01]}:
J.-L. Chen, D. Kaszlikowski, L. C. Kwek, C. H. Oh, \& M. \.{Z}ukowski,
``Entangled three-state systems violate local realism
more strongly than qubits: An analytical proof'',
{\em Phys. Rev. A} {\bf 64}, 5, 052109 (2001);
quant-ph/0103099.
See {\bf [Kaszlikowski-Gnaci\'{n}ski-\.{Z}ukowski-(+2) 00]}.

\item {\bf [Chen-Fu-Ungar-Zhao 02]}:
J.-L. Chen, L. Fu, A. A. Ungar, \& X.-G. Zhao,
``Geometric observation for Bures fidelity between two states of a qubit'',
{\em Phys. Rev. A} {\bf 65}, 2, 024303 (2002).

\item {\bf [Chen-Kaszlikowski-Kwek-Oh 02]}:
J.-L. Chen, D. Kaszlikowski, L. C. Kwek, \& C.H. Oh,
``Clauser-Horne-Bell inequality for three three-dimensional systems'',
quant-ph/0202115.

\item {\bf [Chen-Fu-Ungar-Zhao 02]}:
J.-L. Chen, L. Fu, A. A. Ungar, \& X.-G. Zhao,
``Degree of entanglement for two qubits'',
{\em Phys. Rev. A} {\bf 65}, 4, 044303 (2002).

\item {\bf [Chen-Kwek-Oh 02]}:
J.-L. Chen, L. C. Kwek, \& C. H. Oh,
``Noisy quantum game'',
{\em Phys. Rev. A} {\bf 65}, 5, 052320 (2002).

\item {\bf [Chen-Fu-Ungar-Zhao 02]}:
J.-L. Chen, L. Fu, A. A. Ungar, \& X.-G. Zhao,
``Alternative fidelity measure between two states of an $N$-state quantum system'',
{\em Phys. Rev. A} {\bf 65}, 5, 054304 (2002).

\item {\bf [Chen-Kaszlikowski-Kwek-Oh 03]}:
J.-L. Chen, D. Kaszlikowski, L. C. Kwek, \& C.H. Oh,
``Searching a database under decoherence'',
{\em Phys. Lett. A} {\bf 306}, 5-6, 296-305 (2003).

\item {\bf [Chen-Kuang 04]}:
J.-L. Chen, \& L.-M. Kuang,
``Quantum dense coding in multiparticle entangled states via local
measurements'',
{\em Chin. Phys. Lett.} {\bf 21}, 12-14 (2004);
quant-ph/0402089.

\item {\bf [Chen-Wu-Kwek-Oh 04]}:
J.-L. Chen, C. Wu, L. C. Kwek, \& C. H. Oh,
``Gisin's theorem for three qubits'',
{\em Phys. Rev. Lett.} {\bf 93}, 14, 140407 (2004);
quant-ph/0311180.

\item {\bf [Chen-Wu 02]}:
K. Chen, \& L.-A. Wu,
``The generalized partial transposition criterion for separability of multipartite quantum states'',
{\em Phys. Lett. A} {\bf 306}, 1, 14-20 (2002).

\item {\bf [Chen-Lo 04]}:
K. Chen, \& H.-K. Lo,
``Conference key agreement and quantum sharing of classical secrets with
noisy GHZ states'',
quant-ph/0404133.

\item {\bf [Chen-Hogg-Beausoleil 03]}:
K.-Y. Chen, T. Hogg, \& R. Beausoleil,
``A practical quantum mechanism for the public goods game'',
quant-ph/0301013.

\item {\bf [Chen-Ang-Kiang-(+2) 03]}:
L. K. Chen, H. Ang, D. Kiang, L. C. Kwek, \& C. F. Lo,
``Quantum prisoner dilemma under decoherence'',
{\em Phys. Lett. A} {\bf 316}, 5, 317-323 (2003).

\item {\bf [Chen-Liang-Li-Huang 01 a]}:
P.-X. Chen, L.-M. Liang, C.-Z. Li, \& M.-Q. Huang,
``Necessary and sufficient condition of separability of any
system'',
{\em Phys. Rev. A} {\bf 63}, 5, 052306 (2001);
quant-ph/0102133.
Comment: {\bf [Eggeling-Vollbrecht-Wolf 01]}.
Reply: {\bf [Chen-Liang-Li-Huang 01 b]}.

\item {\bf [Chen-Liang-Li-Huang 01 b]}:
P.-X. Chen, L.-M. Liang, C.-Z. Li, \& M.-Q. Huang,
``Reply to comment'',
quant-ph/0103104.
Reply to {\bf [Eggeling-Vollbrecht-Wolf 01]}.
See {\bf [Chen-Liang-Li-Huang 01 a]}.

\item {\bf [Chen-Liang-Li-Huang 02 a]}:
P.-X. Chen, L.-M. Liang, C.-Z. Li, \& M.-Q. Huang,
``Impossibility criterion for obtaining pure entangled states from mixed
states by purifying protocols'',
{\em Phys. Rev. A} {\bf 65}, 1, 012317 (2002);
quant-ph/0110044.

\item {\bf [Chen-Li 02]}:
P.-X. Chen, \& C.-Z. Li,
``Distinguishing locally of quantum states and the distillation of entanglement'',
quant-ph/0202165.

\item {\bf [Chen-Liang-Li-Huang 02 b]}:
P.-X. Chen, L.-M. Liang, C.-Z. Li, \& M.-Q. Huang,
``A lower bound on entanglement of formation of $2n$ system'',
{\em Phys. Lett. A} {\bf 295}, 4, 175-177 (2002).

\item {\bf [Chen-Liang-Li-Huang 02]}:
P.-X. Chen, L.-M. Liang, C.-Z. Li, \& M.-Q. Huang,
``Necessary and sufficient condition for distillability with unit fidelity from
finite copies of a mixed state: The most efficient purification protocol'',
{\em Phys. Rev. A} {\bf 66}, 2, 022309 (2002);
quant-ph/0110045.

\item {\bf [Chen-Li 03 a]}:
P.-X. Chen, \& C.-Z. Li,
``Orthogonality and distinguishability:
Criterion for local distinguishability of arbitrary orthogonal states'',
{\em Phys. Rev. A} {\bf 68}, 6, 062107 (2003);
quant-ph/0209048.
See {\bf [Walgate-Short-Hardy-Vedral 00]}.

\item {\bf [Chen-Li 03 b]}:
P.-X. Chen, \& C.-Z. Li,
``Distilling multipartite pure states from a finite
number of copies of multipartite mixed states'',
{\em Phys. Rev. A} {\bf 69}, 1, 012308 (2004);
quant-ph/0311095.

\item {\bf [Chen-Li 04]}:
P.-X. Chen, \& C.-Z. Li,
``Distinguishing the elements of a full product basis set needs
only projective measurements and classical communication'',
{\em Phys. Rev. A} {\bf 70}, 2, 022306 (2004).

\item {\bf [Chen-Wang-Liu-Wang 04]}:
Q. Chen, Y. Wang, J.-T. Liu, \& K.-L. Wang,
``$N$-player quantum minority game'',
{\em Phys. Lett. A} {\bf 327}, 2-3, 98-102 (2004).

\item {\bf [Chen-Law-Leung 03]}:
T. W. Chen, C. K. Law, \& P. T. Leung,
``Generation of entangled states of two atoms inside a leaky cavity'',
{\em Phys. Rev. A} {\bf 68}, 5, 052312 (2003).

\item {\bf [Chen-Law-Leung 04]}:
T. W. Chen, C. K. Law, \& P. T. Leung,
``Single-photon scattering and quantum-state transformations in cavity QED'',
{\em Phys. Rev. A} {\bf 69}, 6, 063810 (2004).

\item {\bf [Chen-Zhou-Wu-Zeng 04]}:
X. Chen, C. Zhou, G. Wu, \& H. Zeng,
``Stable differential phase shift quantum key distribution
with a key creation efficiency of 3/4'',
{\em App. Phys. Lett.} {\bf 84}, 2691-2693 (2004).

\item {\bf [Chen-Qiu 03]}:
X.-Y. Chen, \& P.-L. Qiu,
``Bounds for entanglement of formation of two mode squeezed thermal states'',
{\em Phys. Lett. A} {\bf 314}, 3, 191-196 (2003).

\item {\bf [Chen-Zanardi-Wang-Zhang 04]}:
Y. Chen, P. Zanardi, Z. D. Wang, \& F. C. Zhang,
``Entanglement and quantum phase transition in low dimensional spin
 systems'',
quant-ph/0407228.

\item {\bf [Chen-Chuu-Brandes 03]}:
Y. N. Chen, D. S. Chuu, \& T. Brandes,
``Current detection of superradiance and induced entanglement of double
quantum dot excitons'',
{\em Phys. Rev. Lett.} {\bf 90}, 16, 166802 (2003).

\item {\bf [Chen-Yang 00]}:
Y.-X. Chen, \& D. Yang,
``Transmitting one qubit can increase one ebit
between two parties at most'',
quant-ph/0006051.

\item {\bf [Chen-Yang 01]}:
Y.-X. Chen, \& D. Yang,
``Optimal conclusive discrimination of two nonorthogonal pure product
multipartite states through local operations'',
{\em Phys. Rev. A} {\bf 64}, 6, 064303 (2001);
quant-ph/0103111.

\item {\bf [Chen-Yang 02 a]}:
Y.-X. Chen, \& D. Yang,
``Optimally conclusive discrimination of nonorthogonal entangled states by
local operations and classical communications'',
{\em Phys. Rev. A} {\bf 65}, 2, 022320 (2002);
quant-ph/0104068.

\item {\bf [Chen-Yang 02 b]}:
Y.-X. Chen, \& D. Yang,
``Distillable entanglement of multiple copies of Bell states'',
{\em Phys. Rev. A} {\bf 66}, 1, 014303 (2002);
quant-ph/0204004.

\item {\bf [Chen-Jin-Yang 03]}:
Y.-X. Chen, J.-S. Jin, \& D. Yang,
``Distillation of multiple copies of Bell states'',
{\em Phys. Rev. A} {\bf 67}, 1, 014302 (2003).

\item {\bf [Chen 03]}:
Z. Chen,
``Greenberger-Horne-Zeilinger theorem cannot be extended to a Bell state'',
{\em Phys. Rev. A} {\bf 68}, 5, 052106 (2003).

\item {\bf [Chen 04 a]}:
Z. Chen,
``Maximal violation of Mermin's inequalities'',
quant-ph/0407029.

\item {\bf [Chen 04 b]}:
Z. Chen,
``Maximal violation of the Ardehali's inequality of $n$ qubits'',
quant-ph/0407110.

\item {\bf [Chen-Bu-Zhu 04]}:
Z. Chen, T. Bu, \& H. Zhu,
``Statistically secure quantum oblivious transfer'',
quant-ph/0408108.

\item {\bf [Chen-Yepez-Cory 04]}:
Z. Chen, J. Yepez, \& D. G. Cory,
``Simulation of the Burgers equation by NMR quantum information processing'',
quant-ph/0410198.

\item {\bf [Chen-Zhang 01]}:
Z.-B. Chen, \& Y.-D. Zhang,
``Greenberger-Horne-Zeilinger nonlocality
for continuous quantum variables'',
quant-ph/0103082.

\item {\bf [Chen-Pan-Zhang 01]}:
Z.-B. Chen, J.-W. Pan, \& Y.-D. Zhang,
``Feasible linear-optics generation of polarization-entangled photons
assisted with single-photon quantum non-demolition measurement'',
quant-ph/0105100.

\item {\bf [Chen-Pan-Hou-Zhang 02]}:
Z.-B. Chen, J.-W. Pan, G. Hou, \& Y.-D. Zhang,
``Maximal violation of Bell's inequalities for continuous variable systems'',
{\em Phys. Rev. Lett.} {\bf 88}, 4, 040406 (2002).

\item {\bf [Chen-Hou-Zhang 02]}:
Z.-B. Chen, G. Hou, \& Y.-D. Zhang,
``Quantum nonlocality and applications in quantum-information processing of
hybrid entangled states'',
{\em Phys. Rev. A} {\bf 65}, 3, 032317 (2002);
quant-ph/0203038.

\item {\bf [Chen-Lu-Zhang 02]}:
Z.-B. Chen, H.-X. Lu, \& Y.-D. Zhang,
``Linear optics quantum communication over long distances'',
quant-ph/0202040.

\item {\bf [Chen-Pan-Zhang-(+2) 03]}:
Z.-B. Chen, J.-W. Pan, Y.-D. Zhang, \v{C}. Brukner, \& A. Zeilinger,
``All-versus-nothing violation of local realism for two entangled photons'',
{\em Phys. Rev. Lett.} {\bf 90}, 16, 160408 (2003);
quant-ph/0211075.

\item {\bf [Chen-Liu-Yu-Zhang 03]}:
Z.-B. Chen, N.-L. Liu, S. Yu, \& Y.-D. Zhang,
``Violations of locality beyond Bell's theorem'',
quant-ph/0307143.

\item {\bf [Chen-Yu-Zhang 03]}:
Z.-B. Chen, S. Yu, \& Y.-D. Zhang,
``Violation of locality without inequalities for multiparticle perfect correlations'',
quant-ph/0307144.

\item {\bf [Chen-Han 04]}:
J. Cheng, \& S. Han,
``Incoherent coincidence imaging and its applicability in X-ray
diffraction'',
{\em Phys. Rev. Lett.} {\bf 92}, 9, 093903 (2004)
quant-ph/0408135.

\item {\bf [Cheng-Silbey 04]}:
Y. C. Cheng, \& R. J. Silbey,
``Stochastic Liouville equation approach for the effect of noise in quantum computations'',
{\em Phys. Rev. A} {\bf 69}, 5, 052325 (2004).

\item {\bf [Cheon-Tsutsui-F\"{u}l\"{o}p 04]}:
T. Cheon, I. Tsutsui, \& T. F\"{u}l\"{o}p,
``Quantum abacus'',
{\em Phys. Lett. A} {\bf 330}, 5, 338-342 (2004).

\item {\bf [Cheong-Kim-Lee 04]}:
Y. W. Cheong, H. Kim, \& H.-W. Lee,
``Near-complete teleportation of a superposed coherent state'',
quant-ph/0404173.

\item {\bf [Chepelianskii-Shepelyansky 02 a]}:
A. D. Chepelianskii, \& D. L. Shepelyansky,
``Simulation of chaos-assisted tunneling in a semiclassical regime on existing
quantum computers'',
{\em Phys. Rev. A} {\bf 66}, 5, 054301 (2002).

\item {\bf [Chepelianskii-Shepelyansky 02 b]}:
A. D. Chepelianskii, \& D. L. Shepelyansky,
``Schr\"{o}dinger cat animated on a quantum computer'',
quant-ph/0202113.

\item {\bf [Cheung 01]}:
C.-Y. Cheung,
``Quantum bit commitment can be unconditionally secure'',
quant-ph/0112120.

\item {\bf [Chevalier 99]}:
G. Chevalier,
``Why do we find Bohr obscure?'',
in {\bf [Greenberger-Reiter-Zeilinger 99]}, pp.~59-73.

\item {\bf [Chevalier-Dvure\v{c}enskij-Svozil 00]}:
G. Chevalier, A. Dvure\v{c}enskij, \& K. Svozil,
``Piron's and Bell's geometric lemmas and Gleason's theorem'',
{\em Found. Phys.} {\bf 30}, 10, 1737-1755 (2000).

\item {\bf [Chi-Kim 97]}:
D. P. Chi, \& J. Kim,
``Quantum database searching by a single query'',
quant-ph/9708005.

\item {\bf [Chi-Kim-Lee 00 a]}:
D. P. Chi, J. Kim, \& S. Lee,
``Quantum algorithm for generalized Deutsch-Jozsa problem'',
quant-ph/0005059.

\item {\bf [Chi-Kim-Lee 00 b]}:
D. P. Chi, J. Kim, \& S. Lee,
``Function-dependent phase transform in quantum computing'',
quant-ph/0006039.

\item {\bf [Chi-Kim-Lee 01]}:
D. P. Chi, J. Kim, \& S. Lee,
``Initialization-free generalized Deutsch-Jozsa algorithm'',
{\em J. Phys. A} {\bf 34}, 25, 5251-5258 (2001).

\item {\bf [Chi-Lee 03]}:
D. P. Chi, \& S. Lee,
``Entanglement for a two-parameter class of states in $2 \otimes n$ quantum
system'',
{\em J. Phys. A} {\bf 47}, 45, 11503-11511 (2003);
quant-ph/0309073.

\item {\bf [Chiao-Kwiat-Steinberg 93]}:
R. Y. Chiao, P. G. Kwiat, \& A. M. Steinberg,
``Faster than light?'',
{\em Sci. Am.} {\bf 269}, 2, 38-46 (1993).
Spanish version: ``?'M\'{a}s veloz que la luz?'',
{\em Investigaci\'{o}n y Ciencia} 205, 14-23 (1993).
Reprinted in {\bf [Cabello 97 c]}, pp.~46-55.

\item {\bf [Chiao-Kwiat-Steinberg 95 a]}:
R. Y. Chiao, P. G. Kwiat, \& A. M. Steinberg,
``A non-polarization EPR experiment: Observation of high-visibility Franson
interference fringes'',
in J. S. Anandan, \& J. L. Safko (eds.),
{\em Quantum coherence and reality.
In celebration of the 60th birthday of Yakir Aharonov.
Int.\ Conf.\ on Fundamental Aspects of Quantum Theory (?, ?)},
World Scientific, Singapore, 1995, pp.~?-?.

\item {\bf [Chiao-Kwiat-Steinberg 95 b]}:
R. Y. Chiao, P. G. Kwiat, \& A. M. Steinberg,
``Quantum non-locality in two-photon experiments at Berkeley'',
{\em Quantum Semiclass. Opt.} {\bf 7}, 3, 259-278 (1995).
Presented in the Int.
Workshop on Laser and Quantum Optics (Nathiagali, Pakistan, 1994).

\item {\bf [Chiao-Garrison 98]}:
R. Y. Chiao, \& J. C. Garrison,
``Realism or locality: Which should we abandon?'',
{\em Found. Phys.} {\bf 29}, 4, 553-560 (1998);
quant-ph/9807042.

\item {\bf [Chiao 00]}:
R. Y. Chiao,
``Testing quantum mechanics on new ground by Partha Ghose'',
{\em Am. J. Phys.} {\bf 68}, 2, 201-202 (2000).
Review of {\bf [Ghose 99]}.

\item {\bf [Chiao 01]}:
R. Y. Chiao,
``Quantum nonlocalities: Experimental evidence'',
in {\bf [Rusell-Clayton-Wegter McNelly-Polkinghorne 01]}, pp.~17-39.

\item {\bf [Chiao-Kwiat 02]}:
R. Y. Chiao, \& P. G. Kwiat,
``Heisenberg's introduction of the `collapse
of the wavepacket' into quantum mechanics'',
{\em 100 Years of Werner Heisenberg--Works and Impact (Bamberg, Germany, 2001)},
{\em Fortschr. Phys.};
quant-ph/0201036.
Comment: {\bf [Nakhmanson 02]}.

\item {\bf [De Chiara-Fazio-Macchiavello-Palma 04]}:
G. De Chiara, R. Fazio, C. Macchiavello, \& G. M. Palma,
``Entanglement production by quantum error correction in the presence of
correlated environment'',
{\em Europhys. Lett.} {\bf 67}, ?, 714-? (2004);
quant-ph/0310173.

\item {\bf [De Chiara-Fazio-Macchiavello-(+2) 04 a]}:
G. De Chiara, R. Fazio, C. Macchiavello,
S. Montangero, \& G. M. Palma,
``Quantum cloning in spin networks'',
quant-ph/0402071.

\item {\bf [De Chiara-Fazio-Macchiavello-(+2) 04 b]}:
G. De Chiara, R. Fazio, C. Macchiavello,
S. Montangero, \& G. M. Palma,
``Quantum cloning without external control'',
quant-ph/0410211.

\item {\bf [Chibeni 01]}:
S. S. Chibeni,
``Indeterminacy, EPR and Bell'',
{\em Eur. J. Phys.} {\bf 22}, 1, 9-15 (2001).
See {\bf [Tartaglia 98]}.

\item {\bf [Chiorescu-Nakamura-Harmans-Mooij 03]}:
I. Chiorescu, Y. Nakamura, C. J. P. M. Harmans, \& J. E. Mooij,
``Coherent quantum dynamics of a superconducting flux qubit'',
{\em Science} {\bf 299}, ?, 1869-? (2003).

\item {\bf [Chiribella-D'Ariano-Perinotti-Sacchi 03]}:
G. Chiribella, G. M. D'Ariano, P. Perinotti, \& M. F. Sacchi,
``Covariant quantum measurements which maximize the likelihood'',
quant-ph/0403083.

\item {\bf [Chiribella-D'Ariano-Perinotti-Sacchi 04]}:
G. Chiribella, G. M. D'Ariano, P. Perinotti, \& M. F. Sacchi,
``Efficient use of quantum resources for the transmission of a reference
frame'',
{\em Phys. Rev. Lett.};
quant-ph/0405095.

\item {\bf [Chiribella-D'Ariano 04]}:
G. Chiribella, \& G. M. D'Ariano,
``Extremal covariant POVM's'',
quant-ph/0406237.

\item {\bf [Chisolm 00]}:
E. D. Chisolm,
``Generalizing the Heisenberg uncertainty relation'',
{\em Am. J. Phys.};
quant-ph/0011115.

\item {\bf [Chizhov-Schmidt-Knoll-Welsch 00]}:
A. V. Chizhov, E. Schmidt, L. Knoll, \& D.-G. Welsch,
``Propagation of entangled light pulses through
dispersive and absorbing channels'',
quant-ph/0003143.

\item {\bf [Chizhov-Kn\"{o}ll-Welsch 02]}:
A. V. Chizhov, L. Kn\"{o}ll, \& D.-G. Welsch,
``Continuous-variable quantum teleportation through lossy channels'',
{\em Phys. Rev. A} {\bf 65}, 2, 022310 (2002);
quant-ph/0106022.

\item {\bf [Childress-S{\o}rensen-Lukin 03]}:
L. I. Childress, A. S. S{\o}rensen, \& M. D. Lukin,
``Mesoscopic cavity quantum electrodynamics with quantum dots'',
quant-ph/0309106.

\item {\bf [Childress-Taylor-S{\o}rensen-Lukin 03]}:
L. I. Childress, J. M. Taylor, A. S. S{\o}rensen, \& M. D. Lukin,
``Fault-tolerant quantum communication based on solid-state photon
emitters'',
quant-ph/0410123.

\item {\bf [Childs-Preskill-Renes 00]}:
A. M. Childs, J. Preskill, \& J. Renes,
``Quantum information and precision measurement'',
in V. Bu\v{z}zek, \& D. P. DiVincenzo (eds.),
{\em J. Mod. Opt.} {\bf 47}, 2-3 (Special issue:
Physics of quantum information), 155-176 (2000).

\item {\bf [Childs-Chuang 01]}:
A. M. Childs, \& I. L. Chuang,
``Universal quantum computation with two-level trapped ions'',
{\em Phys. Rev. A} {\bf 63}, 1, 012306 (2001);
quant-ph/0008065.

\item {\bf [Childs-Chuang-Leung 01]}:
A. M. Childs, I. L. Chuang, \& D. W. Leung,
``Realization of quantum process tomography in NMR'',
{\em Phys. Rev. A} {\bf 64}, 1, 012314 (2001);
quant-ph/0012032.

\item {\bf [Childs 01]}:
A. M. Childs,
``Secure assisted quantum computation'',
quant-ph/0111046.

\item {\bf [Childs-Farhi-Preskill 02]}:
A. M. Childs, E. Farhi, \& J. Preskill,
``Robustness of adiabatic quantum computation'',
{\em Phys. Rev. A} {\bf 65}, 1, 012322 (2002);
quant-ph/0108048.

\item {\bf [Childs-Deotto-Farhi-(+3) 02]}:
A. M. Childs, E. Deotto, E. Farhi,
J. Goldstone, S. Gutmann, \& A. J. Landahl,
``Quantum search by measurement'',
{\em Phys. Rev. A} {\bf 66}, 3, 032314 (2002);
quant-ph/0204013.

\item {\bf [Childs-Leung-Vidal 03]}:
A. M. Childs, D. W. Leung, \& G. Vidal,
``Reversible simulation of bipartite product Hamiltonians'',
quant-ph/0303097.

\item {\bf [Childs-Goldstone 04]}:
A. M. Childs, \& J. Goldstone,
``Spatial search by quantum walk'',
{\em Phys. Rev. A} {\bf 70}, 2, 022314 (2004);
quant-ph/0306054.

\item {\bf [Childs-Haselgrove-Nielsen 03]}:
A. M. Childs, H. L. Haselgrove, \& M. A. Nielsen,
``Lower bounds on the complexity of simulating quantum gates'',
{\em Phys. Rev. A} {\bf 68}, 5, 052311 (2003);
quant-ph/0307190.

\item {\bf [Childs-Eisenberg 03]}:
A. M. Childs, \& J. M. Eisenberg,
``Quantum algorithms for subset finding'',
quant-ph/0311038.

\item {\bf [Childs-Leung-Nielsen 04]}:
A. M. Childs, D. W. Leung, \& M. A. Nielsen,
``Unified derivations of measurement-based schemes for quantum computation'',
quant-ph/0404132.

\item {\bf [Childs-Goldstone 04]}:
A. M. Childs, \& J. Goldstone,
``Spatial search and the Dirac equation'',
quant-ph/0405120.

\item {\bf [Chiu-Sudarshan-Misra 77]}:
C. B. Chiu, E. C. G. Sudarshan, \& B. Misra,
``Time evolution of unstable quantum states and a resolution of Zeno's paradox'',
{\em Phys. Rev. D} {\bf 16}, 2, 520-529 (1977).

\item {\bf [Cho 00]}:
A. Cho,
``Quantum mechanics:
Physicists unveil Schr\"{o}dinger's SQUID'',
{\em Science} {\bf 287}, 5462, 2395 (2000).

\item {\bf [Choi 91]}:
H. S. Choi,
``Measurement of photon polarization in the many-particle EPR experiment'',
{\em Phys. Lett. A} {\bf 153}, 6-7, 285-287 (1991).

\item {\bf [Choi 75]}:
M.-D. Choi,
``Completely positive linear maps on complex matrices'',
{\em Lin. Alg. Appl.} {\bf 10}, 285-289 (1975).
Reprinted in {\bf [Macchiavello-Palma-Zeilinger 00]}, pp.~174-178.

\item {\bf [Choi 01]}:
M.-S. Choi,
``Solid-state implementation of quantum teleportation and quantum dense
coding'',
{\em Phys. Rev. A} {\bf 64}, 5, 054301 (2001).

\item {\bf [Chong-Vourdas 01]}:
C. C. Chong, \& A. Vourdas,
``Quantum correlations and the extended phase space'',
{\em J. Phys. A} {\bf 34}, 9849-6980 (2001).

\item {\bf [Chou-Hsu 03]}:
C.-L. Chou, \& L. Y. Hsu,
``Minimum-error discrimination between symmetric mixed quantum states'',
{\em Phys. Rev. A} {\bf 68}, 4, 042305 (2003).

\item {\bf [Chou-Polyakov-Kuzmich-Kimble 04]}:
C. W. Chou, S. V. Polyakov, A. Kuzmich, \& H. J. Kimble,
C. W. Chou, S. V. Polyakov, A. Kuzmich, \& H. J. Kimble,
``Single-photon generation from stored excitation in an atomic ensemble'',
{\em Phys. Rev. Lett.} {\bf 92}, 21, 213601 (2004);
quant-ph/0401147.

\item {\bf [Chough-Nha-Kim-(+2) 01]}:
Y.-T. Chough, H. Nha, S. W. Kim,
S.-H. Youn, \& K. An,
``Decoherence-induced wave packet splitting'',
quant-ph/0108005.

\item {\bf [Chow-Cohen 00]}:
C.-K. Chow, \& T. D. Cohen,
``Quantum coins, dice, and children: Probability and quantum statistics''.
{\em Am. J. Phys.} {\bf 68}, 9, 829-834 (2000);
quant-ph/9911101.

\item {\bf [Choy-Ziegeler 99]}:
T. C. Choy, \& D. Ziegeler,
``The meaning of 'counterfactual' statements and
non-locality in quantum mechanics'',
quant-ph/9907027.

\item {\bf [Christian 99]}:
J. Christian,
``Potentiality, entanglement and passion-at-distance'',
{\em Stud. Hist. Philos. Mod. Phys.} {\bf 30}, 4, 561-567 (1999).
Review of {\bf [Cohen-Horne-Stachel 97 b]};
quant-ph/9901008.

\item {\bf [Christandl-Winter 04]}:
M. Christandl, \& A. Winter,
```Squashed entanglement'': An additive entanglement measure',
{\em J. Math. Phys.} {\bf 45}, 3, 829-840 (2004);
quant-ph/0308088.

\item {\bf [Christandl-Datta-Ekert-Landahl 03]}:
M. Christandl, N. Datta, A. K. Ekert, \& A. J. Landahl,
``Perfect state transfer in quantum spin networks'',
quant-ph/0309131.

\item {\bf [Christandl-Renner-Ekert 04]}:
M. Christandl, R. Renner, \& A. K. Ekert,
``A generic security proof for quantum key distribution'',
quant-ph/0402131.

\item {\bf [Christandl-Wehner 04]}:
M. Christandl, \& S. Wehner,
``Quantum anonymous transmissions'',
quant-ph/0409201.

\item {\bf [Christandl-Datta, T. C. Dorlas-(+3) 04]}:
M. Christandl, N. Datta, T. C. Dorlas,
A. K. Ekert, A. Kay, \& A. J. Landahl
``Perfect transfer of arbitrary states in quantum spin networks'',
quant-ph/0411020.

\item {\bf [Christiaens 03]}:
W. Christiaens,
``Non-spatiality and EPR-experiments according to the creation-discovery view'',
{\em Found. Phys. Lett.} {\bf 16}, 4, 379-387 (2003).

\item {\bf [Chtchelkatchev-Blatter-Lesovik-Martin 02]}:
N. M. Chtchelkatchev, G. Blatter, G. B. Lesovik, \& T. Martin,
``Bell inequalities and entanglement in solid-state devices'',
{\em Phys. Rev. B} {\bf 66}, 66, 161320 (2002).

\item {\bf [Chtchelkatchev-Nazarov 03]}:
N. M. Chtchelkatchev, \& Y. V. Nazarov,
``Andreev quantum dots for spin manipulation'',
{\em Phys. Rev. Lett.} {\bf 90}, 22, 226806 (2003).

\item {\bf [Chu 98]}:
S. Y. Chu,
``Is it possible to disentangle an entangled quantum state?'',
gr-qc/9802071 (1998).

\item {\bf [Chuang-Laflamme-Shor-Zurek 95]}:
I. L. Chuang, R. Laflamme, P. W. Shor, \& W. H. Zurek,
``Quantum computers, factoring, and decoherence'',
{\em Science} {\bf 270}, 5242, 1633-1635 (1995).

\item {\bf [Chuang-Yamamoto 95]}:
I. L. Chuang, \& Y. Yamamoto,
``Simple quantum computer'',
{\em Phys. Rev. A} {\bf 52}, 5, 3489-3496 (1995).

\item {\bf [Chuang-Yamamoto 96]}:
I. L. Chuang, \& Y. Yamamoto,
``Quantum bit regeneration'',
{\em Phys. Rev. Lett.} {\bf 76}, 22, 4281-4284 (1996).

\item {\bf [Chuang-Yamamoto 97]}:
I. L. Chuang, \& Y. Yamamoto,
``Creation of a persistent quantum bit using error correction'',
{\em Phys. Rev. A} {\bf 55}, 1, 114-127 (1997).

\item {\bf [Chuang-Leung-Yamamoto 97]}:
I. L. Chuang, D. W. Leung, \& Y. Yamamoto,
``Bosonic quantum codes for amplitude damping'',
{\em Phys. Rev. A} {\bf 56}, 2, 1114-1125 (1997).

\item {\bf [Chuang-Nielsen 97]}:
I. L. Chuang, \& M. A. Nielsen,
``Prescription for experimental determination of the dynamics of
a quantum black box'',
{\em J. Mod. Opt.} {\bf 44}, 11-12 (Special issue: Quantum
state preparation and measurement), 2455-2467 (1997).

\item {\bf [Chuang-Vandersypen-Zhou-(+2) 98]}:
I. L. Chuang, L. M.
K. Vandersypen, X. Zhou, D. W. Leung, \& S. Lloyd,
``Experimental realization of a quantum algorithm'',
{\em Nature} {\bf 393}, 6681, 143-146 (1998);
quant-ph/9801037.
See {\bf [DiVincenzo 98 a]}.

\item {\bf [Chuang-Gershenfeld-Kubinec 98]}:
I. L. Chuang, N. Gershenfeld, \& M. G. Kubinec,
``Experimental implementation of fast quantum searching'',
{\em Phys. Rev. Lett.} {\bf 80}, 15, 3408-3411 (1998).
Reprinted in {\bf [Macchiavello-Palma-Zeilinger 00]}, pp.~482-485.

\item {\bf [Chuang-Gershenfeld-Kubinec-Leung 98]}:
I. L. Chuang, N. Gershenfeld, M. G. Kubinec, \& D. W. Leung,
``Bulk quantum computation with nuclear magnetic resonance:
Theory and experiment'',
in D. P. DiVincenzo. E. Knill, R. Laflamme, \& W. H. Zurek (eds.),
{\em Quantum Coherence and Decoherence.
Proc.\ of the ITP Conf.\ (Santa Barbara, California, 1996)},
{\em Proc. R. Soc. Lond. A} {\bf 454}, 1969, 447-467 (1998).

\item {\bf [Chuang 98]}:
I. L. Chuang,
``Quantum computation with nuclear magnetic resonance'',
in {\bf [Lo-Spiller-Popescu 98]}, pp.~311-339.

\item {\bf [Chuang 00]}:
I. L. Chuang,
``Quantum algorithm for distributed clock synchronization'',
{\em Phys. Rev. Lett.} {\bf 85}, 9, 2006-2009 (2000);
quant-ph/0005092.
See {\bf [Jozsa-Abrams-Dowling-Williams 00]},
{\bf [Burt-Ekstrom-Swanson 00]}.

\item {\bf [Chuprikov 01]}:
N. L. Chuprikov,
`From the ``paradoxes'' of wave packet tunneling
to the definition of tunneling times for particles',
quant-ph/0106129.

\item {\bf [Chun-Lee 03]}:
Y.-J. Chun, \& H.-W. Lee,
``Measurement-induced decoherence and Gaussian smoothing of the Wigner
distribution function'',
{\em Ann. Phys.};
quant-ph/0305011.

\item {\bf [Ciaramicoli-Tombesi-Vitali 00]}:
G. Ciaramicoli, P. Tombesi, \& D. Vitali,
``Performance of a deterministic source of entangled photonic qubits'',
quant-ph/0009045.

\item {\bf [Ciaramicoli-Marzoli-Tombesi 01]}:
G. Ciaramicoli, I. Marzoli, \& P. Tombesi,
``Realization of a quantum algorithm using a trapped electron'',
{\em Phys. Rev. A} {\bf 63}, 5, 052307 (2001).

\item {\bf [Ciaramicoli-Marzoli-Tombesi 02]}:
G. Ciaramicoli, I. Marzoli, \& P. Tombesi,
``Three-qubit network with a single trapped electron'',
{\em Proc.\ ESF QIT Conf.\ Quantum Information: Theory, Experiment and Perspectives
(Gdansk, Poland, 2001)}, {\em J. Mod. Opt.} {\bf 49}, 8, 1307-1323 (2002).

\item {\bf [Ciaramicoli-Marzoli-Tombesi 03]}:
G. Ciaramicoli, I. Marzoli, \& P. Tombesi,
``Scalable quantum processor with trapped electrons'',
{\em Phys. Rev. Lett.} {\bf 91}, 1, 017901 (2003).

\item {\bf [Cinchetti-Twamley 01]}:
M. Cinchetti, \& J. Twamley,
``Entanglement distribution between $N$ distant users via a
center'',
{\em Phys. Rev. A} {\bf 63}, 5, 052310 (2001).

\item {\bf [Cini-de Maria-Mattioli-Nicol\'{o} 79]}:
M. Cini, M. de Maria, G. Mattioli, \& F. Nicol\'{o},
``Wave packet reduction in quantum mechanics:
A model of a measuring apparatus'',
{\em Found. Phys.} {\bf 9}, 7-8, 479-500 (1979).

\item {\bf [Cini 83]}:
M. Cini,
``Quantum theory of measurement without wave packet collapse'',
{\em Nuovo Cimento B} {\bf 73}, 1, 27-56 (1983).

\item {\bf [Cipra 96]}:
B. Cipra,
``Quantum mechanics:
Error-correcting code keeps quantum computers on track'',
{\em Science} {\bf 272}, 5259, 199-200 (1996).

\item {\bf [Cirac-Parkins 94]}:
J. I. Cirac, \& A. S. Parkins,
``Schemes for atomic-state teleportation'',
{\em Phys. Rev. A} {\bf 50}, 6, R4441-R4444 (1994).

\item {\bf [Cirac-Zoller 94]}:
J. I. Cirac, \& P. Zoller,
``Preparation of macroscopic superpositions in many-atom systems'',
{\em Phys. Rev. A} {\bf 50}, 4, R2799-R2802 (1994).

\item {\bf [Cirac-Zoller 95]}:
J. I. Cirac, \& P. Zoller,
``Quantum computations with cold trapped ions'',
{\em Phys. Rev. Lett.} {\bf 74}, 20, 4091-4094 (1995).
Reprinted in {\bf [Macchiavello-Palma-Zeilinger 00]}, pp.~330-333.

\item {\bf [Cirac-Pellizzari-Zoller 96]}:
J. I. Cirac, T. Pellizzari, \& P. Zoller,
``Enforcing coherent evolution in dissipative quantum dynamics'',
{\em Science} {\bf 273}, 5279, 1207-1210 (1996).

\item {\bf [Cirac-Zoller-Kimble-Mabuchi 97]}:
J. I. Cirac, P. Zoller, H. J. Kimble, \& H. Mabuchi,
``Quantum state transfer and entanglement distribution among
distant nodes in a quantum network'',
{\em Phys. Rev. Lett.} {\bf 78}, 16, 3221-3224 (1997).

\item {\bf [Cirac-Gisin 97]}:
J. I. Cirac, \& N. Gisin,
``Coherent eavesdropping
strategies for the four state quantum cryptography protocol'',
{\em Phys. Lett. A} {\bf 229}, 1, 1-7 (1997);
quant-ph/9702002.
See {\bf [Xiang Bin 01 b]}.

\item {\bf [Cirac-van Enk-Zoller-(+2) 98]}:
J. I. Cirac, S. J. van Enk, P. Zoller, H. J. Kimble, \& H. Mabuchi,
``Quantum communication in a quantum network'',
in E. B. Karlsson, \& E. Br\"{a}ndas (eds.),
{\em Proc.\ of the 104th Nobel Symp.\ ``Modern Studies of Basic Quantum Concepts and Phenomena'' (Gimo, Sweden, 1997)},
{\em Physica Scripta} {\bf T26}, 223-? (1998).

\item {\bf [Cirac-Ekert-Fern\'{a}ndez Huelga-Macchiavello 99]}:
J. I. Cirac, A. K. Ekert, S. G. Fern\'{a}ndez Huelga, \& C. Macchiavello,
``Distributed quantum computation over noisy channels'',
{\em Phys. Rev. A} {\bf 59}, 6, 4249-4254 (1999);
quant-ph/9803017.

\item {\bf [Cirac-Ekert-Macchiavello 99]}:
J. I. Cirac, A. K. Ekert, \& C. Macchiavello,
``Optimal purification of single qubits'',
{\em Phys. Rev. Lett.} {\bf 82}, 21, 4344-4347 (1999);
quant-ph/9812075.

\item {\bf [Cirac-Zoller 99]}:
J. I. Cirac, \& P. Zoller,
``Quantum engineering moves on'',
{\em Phys. World} {\bf 12}, 1, 22-23 (1999).

\item {\bf [Cirac 00 a]}:
J. I. Cirac,
``Quanta y computaci\'{o}n'',
{\em Revista Espa\~{n}ola de F\'{\i}sica} {\bf 14}, 1, 48-53 (2000).

\item {\bf [Cirac-Zoller 00]}:
J. I. Cirac, \& P. Zoller,
``A scalable quantum computer with ions in an array of
microtraps'',
{\em Nature} {\bf 404}, 6778, 579-581 (2000).

\item {\bf [Cirac 00 b]}:
J. I. Cirac,
``La revoluci\`{o}n cu\'{a}ntica'',
{\em ABC-El Cultural}, 20 Dec. 2000, pp.~62-63.

\item {\bf [Cirac 01]}:
J. I. Cirac,
``Quantum physics: Entangled atomic samples'',
{\em Nature} {\bf 413}, 6854, 375-377 (2001).
See {\bf [Julsgaard-Kozhekin-Polzik 01]}.

\item {\bf [Cirac-D\"{u}r-Kraus-Lewenstein 01]}:
J. I. Cirac, W. D\"{u}r, B. Kraus, \& M. Lewenstein,
``Entangling operations and their implementation using a small amount of
entanglement'',
{\em Phys. Rev. Lett.} {\bf 86}, 3, 544-547 (2001);
quant-ph/0007057.

\item {\bf [Cirac 02]}:
J. I. Cirac,
``Presente y futuro de la computaci\'{o}n cu\'{a}ntica'',
en J. M. S\'{a}nchez Ron (ed.),
{\em La ciencia y la tecnolog\'{\i}a ante el tercer milenio (I)},
Sociedad Estatal Espa\~{n}a Nuevo Milenio, Madrid, 2002, pp.~185-202.

\item {\bf [Cirac-Duan-Zoller 02]}:
J. I. Cirac, L.-M. Duan, \& P. Zoller,
in F. Di Martini, \& C. Monroe (eds.),
{\em Experimental quantum computation and information.
Proc.\ of the Int.\ School of Physics ``Enrico Fermi'', Course CXLVII)},
IOS Press, Amsterdam, 2002, pp.~263-?;
quant-ph/0405030.

\item {\bf [Cirac-Zoller 03]}:
J. I. Cirac, \& P. Zoller,
``How to manipulate cold atoms'',
{\em Science} {\bf 301}, 5630, 176-177 (2003).

\item {\bf [Cirac-Zoller 04]}:
J. I. Cirac, \& P. Zoller,
``New frontiers in quantum information with atoms and ions'',
{\em Phys. Today} {\bf 57}, ?, 38-? (2004).

\item {\bf [Cirel'son 80]}:
B. S. Cirel'son [Tsirelson],
``Quantum generalizations of Bell's inequality'',
{\em Lett. Math. Phys.} {\bf 4}, 2, 93-100 (1980).

\item {\bf [Cirel'son 85]}:
B. S. Cirel'son [Tsirelson],
``?'',
{\em Zapiski LOMI} {\bf 142}, 174-194 (1985).
English version:
``Quantum analogues of the Bell inequalities. The case of two spatially separated domains'',
{\em J. Soviet Math.} {\bf 36}, 4, 557-570 (1987).

\item {\bf [Cirel'son 93 a]}:
B. S. Cirel'son [Tsirelson],
``Some results and problems on quantum Bell-type inequalities'',
{\em Hadronic J. Supplement} {\bf 8}, 4, 329-345 (1993).

\item {\bf [Cirel'son 93 b]}:
B. S. Cirel'son [Tsirelson],
``A new framework for old Bell inequalities'',
{\em Helv. Phys. Acta} {\bf 66}, 7-8, 858-874 (1993).

\item {\bf [Cirel'son 96]}:
B. S. Cirel'son [Tsirelson],
``Fine structure of EPR state and universal quantum correlation'',
in A. Mann, \& M. Revzen (eds.),
{\em The dilemma of Einstein, Podolsky and Rosen -- 60 years
later. An international symposium in honour of Nathan Rosen
(Haifa, Israel, 1995)},
{\em Ann. Phys. Soc. Israel} {\bf 12}, 83-86 (1996).

\item {\bf [Cirel'son 01]}:
B. S. Cirel'son [Tsirelson],
``The quantum algorithm of Kieu does not solve the Hilbert's tenth problem'',
quant-ph/0111009.
Reply: {\bf [Kieu 01 b]}.
See {\bf [Kieu 01 a, c]}.

\item {\bf [Cirone-Delgado-Fischer-(+3) 01]}:
M. A. Cirone, A. Delgado, D. G. Fischer,
M. Freyberger, H. Mack, \& M. Mussinger,
``Estimation of quantum channels with finite resources'',
quant-ph/0108037.

\item {\bf [Cirone-Rzazewski-Schleich 01]}:
M. A. Cirone, K. Rzazewski, W. P. Schleich, F. Straub, \& J. A. Wheeler,
``Quantum anti-cetrifugal force'',
quant-ph/0108069.

\item {\bf [Cirone 01]}:
M. A. Cirone,
``Quantifying entanglement with probabilities'',
quant-ph/0110139.

\item {\bf [Cirone-Compagno-Palma-(+2) 04]}:
M. A. Cirone, G. Compagno, G. M. Palma,
R. Passante, \& F. S. Persico,
``Entanglement between a pair of spatially separated two-level atoms
induced by zero-point field fluctuations'',
quant-ph/0407032.

\item {\bf [Cisneros-Mart\'{\i}nez y
Romero-N\'{u}\~{n}ez Y\'{e}pez-Salas Brito 98]}:
C. Cisneros, R. P. Mart\'{\i}nez y Romero, H. N. N\'{u}\~{n}ez
Y\'{e}pez, \& A. L. Salas Brito,
``Limitations on the superposition principle:
Superselection rules in non-relativistic quantum mechanics'',
{\em Eur. J. Phys.} {\bf 19}, 3, 237-243 (1998);
quant-ph/9809059.

\item {\bf [Clark-Turner 68]}:
P. M. Clark, \& J. E. Turner,
``Experimental tests of quantum mechanics'',
{\em Phys. Lett. A} {\bf 26}, 10, 447 (1968).

\item {\bf [Clark 01]}:
R. Clark (ed.),
{\em Experimental implementation of quantum computation (IQC'01)},
Rinton Press, Princeton, New Jersey, 2001.

\item {\bf [Clark-Parkins 03]}:
S. G. Clark, \& A. S. Parkins,
``Entanglement and entropy engineering of atomic two-qubit states'',
{\em Phys. Rev. Lett.} {\bf 90}, 4, 047905 (2003);
quant-ph/0203137.

\item {\bf [Clark-Peng-Gu-Parkins 03]}:
S. G. Clark, A. Peng, M. Gu, \& A. S. Parkins,
``Unconditional preparation of entanglement between
atoms in cascaded optical cavities'',
{\em Phys. Rev. Lett.} {\bf 91}, 17, 177901 (2003).

\item {\bf [Clarke 74]}:
C. J. S. Clarke,
``Quantum theory and cosmology'',
{\em Philos. Sci.} {\bf 41}, ?, 317-332 (1974).

\item {\bf [Clarke 01 a]}:
C. J. S. Clarke,
``The histories interpretation of quantum theory and
the problem of human/divine action'',
in {\bf [Rusell-Clayton-Wegter McNelly-Polkinghorne 01]}, pp.~159-178.

\item {\bf [Clarke 01 b]}:
C. J. S. Clarke,
``The histories interpretation: Stability instead of
consistency?'',
{\em Found. Phys. Lett.} {\bf 14}, 2, 179-186 (2001);
quant-ph/0008060.

\item {\bf [Clarke 02]}:
C. J. S. Clarke,
``Entanglement and statistical independence for mixed quantum states'',
{\em Found. Phys. Lett.} {\bf 15}, 5, 495-500 (2002).

\item {\bf [Clarke-Cleland-Devoret-(+2) 88]}:
J. Clarke, A. N. Cleland, M. H. Devoret, D. Esteve, \& J. M. Martinis,
``Quantum mechanics of a macroscopic variable: The phase difference of a
Josephson junction'',
{\em Science} {\bf 239}, 4843, 992-997 (1988).

\item {\bf [Clarke-Chefles-Barnett-Riis 00]}:
R. B. M. Clarke, A. Chefles, S. M. Barnett, \& E. Riis,
``Experimental demonstration of optimal unambiguous state discrimination'',
{\em Phys. Rev. A};
quant-ph/0007063.

\item {\bf [Clarke-Kendon-Chefles-(+3) 01]}:
R. B. M. Clarke, V. M. Kendon, A. Chefles,
S. M. Barnett, E. Riis, \& M. Sasaki,
``Experimental realization of optimal
detection strategies for overcomplete states'',
{\em Phys. Rev. A} {\bf 64}, 1, 012303 (2001);
quant-ph/0008028.

\item {\bf [Clausen-Dakna-Knoll-Welsch 98]}:
J. Clausen, M. Dakna, L. Knoll, \& D.-G. Welsch,
``Conditional quantum-state transformation at a beam splitter'',
quant-ph/9811063.

\item {\bf [Clausen-Opatrn\'{y}-Welsch 00]}:
J. Clausen, T. Opatrn\'{y}, \& D.-G. Welsch,
``Conditional teleportation using optical
squeezers and photon counting'',
{\em Phys. Rev. A} {\bf 62}, 4, 042308 (2000);
quant-ph/0003142.

\item {\bf [Clausen-Kn\"{o}ll-Welsch 02]}:
J. Clausen, L. Kn\"{o}ll, \& D.-G. Welsch,
``Lossy purification and detection of entangled coherent states'',
{\em Phys. Rev. A} {\bf 66}, 6, 062303 (2002);
quant-ph/0203144.

\item {\bf [Clausen-Kn\"{o}ll-Welsch 03]}:
J. Clausen, L. Kn\"{o}ll, \& D.-G. Welsch,
``Entanglement purification of multi-mode quantum states'',
quant-ph/0302103.

\item {\bf [Clauser 69]}:
J. F. Clauser,
``?'',
{\em Bull. Am. Phys. Soc.} {\bf 14}, ?, 578-? (1969).

\item {\bf [Clauser-Horne-Shimony-Holt 69]}:
J. F. Clauser, M. A. Horne, A. Shimony, \& R. A. Holt,
``Proposed experiment to test local hidden-variable theories'',
{\em Phys. Rev. Lett.} {\bf 23}, 15, 880-884 (1969).
Reprinted in {\bf [Wheeler-Zurek 83]}, pp.~409-413.

\item {\bf [Clauser 71 a]}:
J. F. Clauser,
``Von Neumann's informal hidden-variable argument'',
{\em Am. J. Phys.} {\bf 39}, 9, 1095-1096 (1971).
See {\bf [Wigner 70, 71 a]}, {\bf [Clauser 71 b]}.

\item {\bf [Clauser 71 b]}:
J. F. Clauser,
``Reply to Dr. Wigner's objections'',
{\em Am. J. Phys.} {\bf 39}, 9, 1098-1099 (1971).
See {\bf [Wigner 70, 71 a]}, {\bf [Clauser 71 a]}.

\item {\bf [Clauser 71 b]}:
J. F. Clauser,
``Experimental limitations to the validity of
semiclassical radiation theories'',
{\em Phys. Rev. A} {\bf 6}, 1, 49-54 (1972).

\item {\bf [Clauser 74]}:
J. F. Clauser,
``Experimental distinction between the quantum and classical
field-theoretic predictions for the photoelectric effect'',
{\em Phys. Rev. D} {\bf 9}, 4, 853-860 (1974).

\item {\bf [Clauser-Horne 74]}:
J. F. Clauser, \& M. A. Horne,
``Experimental consequences of objective local theories'',
{\em Phys. Rev. D} {\bf 10}, 2, 526-535 (1974).

\item {\bf [Clauser 76 a]}:
J. F. Clauser,
``Experimental investigation of a polarization correlation anomaly'',
{\em Phys. Rev. Lett.} {\bf 36}, 21, 1223-1226 (1976).

\item {\bf [Clauser 76 b]}:
J. F. Clauser,
``Measurement of the circular-polarization
correlation in photons from an atomic cascade'',
{\em Nuovo Cimento B} {\bf 33}, 2, 740-746 (1976).

\item {\bf [Clauser-Shimony 78]}:
J. F. Clauser, \& A. Shimony,
``Bell's theorem: Experimental tests and implications'',
{\em Rep. Prog. Phys.} {\bf 41}, 12, 1881-1927 (1978).
Reprinted in {\bf [Stenholm 85]}.

\item {\bf [Clauser 92]}:
J. F. Clauser,
``Early history of Bell's theorem theory and experiment'',
in T. D. Black, M. Mart\'{\i}n Nieto, H. S. Pilloff,
M. O. Scully, \& R. M. Sinclair (eds.),
{\em Foundations of quantum mechanics. Workshop (Santa Fe, New
Mexico, 1991)},
World Scientific, Singapore, 1992, pp.~168-174.

\item {\bf [Clauser-Dowling 96]}:
J. F. Clauser, \& J. P. Dowling,
``Factoring integers with Young's $N$-slit interferometer'',
{\em Phys. Rev. A} {\bf 53}, 6, 4587-4590 (1996).
See {\bf [Summhammer 97]}.

\item {\bf [Clauser 97]}:
J. F. Clauser,
``De Broglie-wave interference of small rocks and live viruses'',
in {\bf [Cohen-Horne-Stachel 97 a]}, 1997.

\item {\bf [Clauser 02]}:
J. F. Clauser,
``Early history of Bell's theorem'',
in {\bf [Bertlmann-Zeilinger 02]}, pp.~61-98.

\item {\bf [Cleland-Geller 04]}:
A. N. Cleland, \& M. R. Geller,
``Superconducting qubit storage and entanglement with nanomechanical resonators'',
{\em Phys. Rev. Lett.} {\bf 93}, 7, 070501 (2004).

\item {\bf [Clemens-Siddiqui-Gea Banacloche 04]}:
J. P. Clemens, S. Siddiqui, \& J. Gea-Banacloche,
``Quantum error correction against correlated noise'',
{\em Phys. Rev. A} {\bf 69}, 6, 062313 (2004).

\item {\bf [Cleve 96]}:
R. Cleve,
``Quantum stabilizer codes and classical linear codes'',
quant-ph/9612048.

\item {\bf [Cleve 97]}:
R. Cleve,
``Quantum stabilizer codes and classical linear codes'',
{\em Phys. Rev. A} {\bf 55}, 6, 4054-4059 (1997).

\item {\bf [Cleve-Gottesman 97]}:
R. Cleve, \& D. Gottesman,
``Efficient computations of encodings for quantum error
correction'',
{\em Phys. Rev. A} {\bf 56}, 2, 76-82 (1997).

\item {\bf [Cleve-Buhrman 97]}:
R. Cleve, \& H. Buhrman,
``Substituting quantum entanglement for communication'',
{\em Phys. Rev. A} {\bf 56}, 2, 1201-1204 (1997);
quant-ph/9704026.

\item {\bf [Cleve-Ekert-Macchiavello-Mosca 98]}:
R. Cleve, A. K. Ekert, C. Macchiavello, \& M. Mosca,
``Quantum algorithms revisited'',
in D. P. DiVincenzo. E. Knill, R. Laflamme, \& W. H. Zurek (eds.),
{\em Quantum Coherence and Decoherence.
Proc.\ of the ITP Conf.\ (Santa Barbara, California, 1996)},
{\em Proc. R. Soc. Lond. A} {\bf 454}, 1969, 339-354 (1998);
quant-ph/9708016.

\item {\bf [Cleve-van Dam-Nielsen-Tapp 98]}:
R. Cleve, W. van Dam, M. A. Nielsen, \& A. Tapp,
``Quantum entanglement and the
communication complexity of the inner product function'',
in C.P. Williams (editor),
{\em Proc.\ of the 1st NASA International Conference, QCQC'98 (Palm Springs, California, 1998)},
61-74;
quant-ph/9708019.

\item {\bf [Cleve-Ekert-Henderson-Macchiavello-Mosca 98]}:
R. Cleve, A. K. Ekert, L. Henderson, C. Macchiavello, \& M. Mosca,
``On quantum algorithms'',
{\em Complexity} {\bf 4}, 1, 33-42 (1998);
quant-ph/9903061.
Reprinted in {\bf [Macchiavello-Palma-Zeilinger 00]}, pp.~86-100.

\item {\bf [Cleve-Gottesman-Lo 99]}:
R. Cleve, D. Gottesman, \& H.-K. Lo,
``How to share a quantum secret'',
{\em Phys. Rev. Lett.} {\bf 83}, 3, 648-651 (1999);
quant-ph/9901025.

\item {\bf [Cleve 99]}:
R. Cleve,
``An introduction to quantum complexity theory'',
in {\bf [Macchiavello-Palma-Zeilinger 00]}, pp.~103-128;
quant-ph/9906111.

\item {\bf [Cleve-Watrous 00]}:
R. Cleve, \& J. Watrous,
``Fast parallel circuits for the quantum Fourier transform'',
quant-ph/0006004.

\item {\bf [Cleve-H\o{}yer-Toner-Watrous 04]}:
R. Cleve, P. H\o{}yer, B. Toner, \& J. Watrous,
``Consequences and limits of nonlocal strategies'',
in {\em Proc.\ 19th IEEE Conference on Computational Complexity (2004)};
quant-ph/0404076.

\item {\bf [Clifton-Redhead 88]}:
R. K. Clifton, \& M. L. G. Redhead,
``The compatibility of correlated $CP$ violating systems with
statistical locality'',
{\em Phys. Lett. A} {\bf 126}, 5-6, 295-299 (1988).

\item {\bf [Clifton 89 a]}:
R. K. Clifton,
``Determinism, realism, and Stapp's 1985 proof of nonlocality'', Foundations of
{\em Found. Phys. Lett.} {\bf 2}, ?, 347-359 (1989).

\item {\bf [Clifton 89 b]}:
R. K. Clifton,
``The Rastall model and Stapp's proof of nonlocality'',
{\em Found. Phys. Lett.} {\bf 2}, ?, ?-? (1989).

\item {\bf [Clifton-Butterfield-Redhead 90]}:
R. K. Clifton, J. N. Butterfield, \& M. L. G. Redhead,
``Nonlocal influences and possible worlds---A Stapp in the wrong
direction'',
{\em Brit. J. Philos. Sci.} {\bf 41}, 1, 5-58 (1990).
Comment: {\bf [Stapp 90]}.

\item {\bf [Clifton 91 a]}:
R. K. Clifton,
``Noninvasive measurability, negative-result measurements, and watched-pots:
Another look at Leggett's arguments for the incompatibility between macro-realism and quantum mechanics'',
in P. J. Lahti, \& P. Mittelstaedt (eds.),
{\em Proc.\ Symp.\ on the Foundations of Modern Physics 1990.
Quantum Theory of Measurement and Related Philosophical Problems
(Joensuu, Finland, 1990)},
World Scientific, Singapore, 1990, pp.~77-88.

\item {\bf [Clifton-Redhead-Butterfield 91 a]}:
R. K. Clifton, M. L. G. Redhead, \& J. N. Butterfield,
``Generalization of the Greenberger-Horne-Zeilinger algebraic proof
of nonlocality'',
{\em Found. Phys.} {\bf 21}, 2, 149-184 (1991).
Comment: {\bf [Jones 91]}.
See {\bf [Clifton-Redhead-Butterfield 91 b]}.

\item {\bf [Clifton-Redhead-Butterfield 91 b]}:
R. K. Clifton, M. L. G. Redhead, \& J. N. Butterfield,
``A second look at a recent algebraic proof of nonlocality'',
{\em Found. Phys. Lett.} {\bf 4}, 4, 395-403 (1991).
Reply to {\bf [Jones 91]}.
See {\bf [Clifton-Redhead-Butterfield 91 a]}.

\item {\bf [Clifton 91]}:
R. K. Clifton,
``Nonlocality in quantum mechanics:
Signalling, counterfactuals, probability and causation'',
Ph.\ D. thesis, Cambridge University, 1991.

\item {\bf [Clifton-Niemann 92]}:
R. K. Clifton, \& P. Niemann,
``Locality, Lorentz invariance, and linear algebra: Hardy's theorem
for two entangled spin-$s$ particles'',
{\em Phys. Lett. A} {\bf 166}, 3-4, 177-184 (1992).
See {\bf [Ghosh-Kar 98]}.

\item {\bf [Clifton-Pagonis-Pitowsky 92]}:
R. K. Clifton, C. Pagonis, \& I. Pitowsky,
``Relativity, quantum mechanics, and EPR'',
in D. Hall, M. Forbes, \& K. Okruhlik (eds.),
{\em Proc.\ of the 1992 Biennial Meeting of the Philosophy of Science
Association}, East Lansing, Michigan, 1992, vol. 1, pp.~114-128.
See {\bf [Pagonis-Redhead-La Rivi\`{e}re 96]}.

\item {\bf [Clifton 93]}:
R. K. Clifton,
``Getting contextual and nonlocal elements-of-reality the easy way'',
{\em Am. J. Phys.} {\bf 61}, 5, 443-447 (1993).
Comments: {\bf [Bechmann Johansen 94]}, {\bf [Vermaas 94]}.

\item {\bf [Clifton 95 a]}:
R. K. Clifton,
``Book review. Quantum theory: Concepts and methods'',
{\em Found. Phys.} {\bf 25}, 1, 205-209 (1995).
Review of {\bf [Peres 93 a]}.

\item {\bf [Clifton 95 b]}:
R. K. Clifton,
``Independently motivating the Kochen-Dieks modal
interpretation of quantum mechanics'',
{\em Brit. J. Philos. Sci.} {\bf 46}, 1, 33-57 (1995).
Reprinted in {\bf [Clifton 04]}.

\item {\bf [Clifton 95 c]}:
R. K. Clifton,
`Making sense of the Kochen-Dieks ``no-collapse''
interpretation of quantum mechanics independent of the measurement problem',
in D. M. Greenberger, \& A. Zeilinger (eds.),
{\em Fundamental problems in
quantum theory: A conference held in honor of professor John A. Wheeler,
Ann. N. Y. Acad. Sci.} {\bf 755}, 570-578 (1995).

\item {\bf [Clifton 95 d]}:
R. K. Clifton,
``The triorthogonal uniqueness theorem and its irrelevance
to the modal interpretation of quantum mechanics'',
in K. V. Laurikainen et al. (eds.),
{\em Proc.\ Symp.\ on the Foundations of Modern Physics 1994:
70 Years of Matter Waves}
World Scientific, Singapore, 1995, pp.~45-60.

\item {\bf [Clifton 95 e]}:
R. K. Clifton,
``Why modal interpretations of quantum
mechanics must abandon classical reasoning about physical properties'',
{\em Int. J. Theor. Phys.} {\bf 34}, 8, 1303-1312 (1995).

\item {\bf [Clifton 96]}:
R. K. Clifton,
``The properties of modal interpretations of quantum mechanics'',
{\em Brit. J. Philos. Sci.} {\bf 47}, ?, 371-398 (1995).

\item {\bf [Clifton-Feldman-Redhead-Wilce 97]}:
R. K. Clifton, D. V. Feldman, M. L. G. Redhead, \& A. Wilce,
``Superentangled states'',
{\em Phys. Rev. A} {\bf 58}, 1, 135-145 (1998);
quant-ph/9711020.

\item {\bf [Clifton-Dickson 98]}:
R. K. Clifton, \& W. M. Dickson,
``Lorentz-invariance in modal interpretatons'',
in {\bf [Dieks-Vermaas 98]}, pp.~9-47.
Reprinted in {\bf [Clifton 04]}.

\item {\bf [Clifton 99]}:
R. K. Clifton,
``Beables in algebraic quantum mechanics'',
in J. N. Butterfield, \& C. Pagonis (eds.),
{\em From physics to philosophy: Essays in honour of Michael Redhead},
Cambridge University Press, Cambridge, 1999, pp.~12-43;
quant-ph/9711009.

\item {\bf [Clifton-Monton 99]}:
R. K. Clifton, \& B. Monton,
``Losing your marbles in wavefunction collapse theories'',
{\em Brit. J. Philos. Sci.} {\bf 50}, 697-717 (1999);
quant-ph/9905065.
See {\bf [Bassi-Ghirardi 99 b]}, {\bf [Clifton-Monton 00]};

\item {\bf [Clifton-Monton 00]}:
R. K. Clifton, \& B. Monton,
``Counting marbles with 'accessible' mass density:
A reply to Bassi and Ghirardi'',
{\em Brit. J. Philos. Sci.} {\bf 51} 155-164 (2000);
quant-ph/9909071.
See {\bf [Clifton-Monton 99]}, {\bf [Bassi-Ghirardi 99 b]}.

\item {\bf [Clifton-Kent 00]}:
R. K. Clifton, \& A. Kent,
``Simulating quantum mechanics by non-contextual hidden variables'',
{\em Proc. R. Soc. Lond. A} {\bf 456}, 45, 2001, 2101-2114 (2000);
quant-ph/9908031.
Reprinted in {\bf [Clifton 04]}.
See {\bf [Meyer 99 b]}, {\bf [Kent 99 b]}, {\bf [Cabello 99 d, 02 c]},
{\bf [Havlicek-Krenn-Summhammer-Svozil 01]},
{\bf [Appleby 00, 01, 02]}, {\bf [Boyle-Schafir 01 a]}.

\item {\bf [Clifton-Halvorson 00 a]}:
R. K. Clifton, \& H. Halvorson,
``Bipartite-mixed-states of infinite-dimensional systems are generically nonseparable'',
{\em Phys. Rev. A} {\bf 61}, 1, 012108 (2000);
quant-ph/9908028.

\item {\bf [Clifton-Halvorson-Kent 00]}:
R. K. Clifton, H. Halvorson, \& A. Kent,
``Nonlocal correlations are generic in infinite-dimensional bipartite systems'',
{\em Phys. Rev. A} {\bf 61}, 4, 042101 (2000);
quant-ph/9909016.
Reprinted in {\bf [Clifton 04]}.

\item {\bf [Clifton 00 a]}:
R. K. Clifton,
``Complementarity between position and momentum as
a consequence of Kochen-Specker arguments'',
{\em Phys. Lett. A} {\bf 271}, 1-2, 1-7 (2000);
quant-ph/9912108.
Reprinted in {\bf [Clifton 04]}.

\item {\bf [Clifton 00 b]}:
R. K. Clifton,
``The modal interpretation of algebraic quantum field theory'',
{\em Phys. Lett. A} {\bf 271}, 3, 167-177 (2000);
quant-ph/0003018.
Reprinted in {\bf [Clifton 04]}.

\item {\bf [Clifton-Halvorson 01 a]}:
R. K. Clifton, \& H. Halvorson,
``Entanglement and open systems in algebraic quantum field theory'',
{\em Stud. Hist. Philos. Sci. Part B: Stud. Hist. Philos. Mod. Phys.}
{\bf 32}, 1, 1-31 (2001).
Reprinted in {\bf [Clifton 04]}.

\item {\bf [Clifton-Halvorson 01 b]}:
R. K. Clifton, \& H. Halvorson,
``Are Rindler quanta real?
Inequivalent particle concepts in quantum field theory'',
{\em Brit. J. Philos. Sci.} {\bf 52}, 417-470 (2001);
quant-ph/0008030.
Reprinted in {\bf [Clifton 04]}.

\item {\bf [Clifton 01 a]}:
R. K. Clifton,
``The subtleties of entanglement and its role in quantum information theory'',
{\em PSA} 2000 Vol. II;
PITT-PHIL-SCI00000196.
Reprinted in {\bf [Clifton 04]}.

\item {\bf [Clifton-Pope 01]}:
R. K. Clifton, \& D. Pope,
``On the nonlocality of the quantum channel in the
standard teleportation protocol'',
{\em Phys. Lett. A} {\bf 292}, 1-2, 1-11 (2001);
quant-ph/0103075,
PITT-PHIL-SCI00000198.

\item {\bf [Clifton 01]}:
R. K. Clifton,
``Introductory notes on the mathematics needed for quantum theory'',
the beginning of a book that was abandoned (1996);
PITT-PHIL-SCI00000390.

\item {\bf [Clifton-Hepburn-W\"{u}thrich 02]}:
R. K. Clifton, B. Hepburn, \& C. W\"{u}thrich,
``Generic incomparability of infinite-dimensional entangled states'',
{\em Phys. Lett. A} {\bf 303}, 2-3, 121-124 (2002);
quant-ph/0205063.

\item {\bf [Clifton-Halvorson 02 a]}:
R. K. Clifton, \& H. Halvorson,
``A fresh perspective on Bohr's reply to EPR'',
{\em Cracow Nato Workshop}, 2002.

\item {\bf [Clifton-Halvorson 02 b]}:
R. K. Clifton, \& H. Halvorson,
{\em Suspended in language: Algebraic quantum theory and Bohr's Copenhagen interpretation},
in progress.

\item {\bf [Clifton-Bub-Halvorson 03]}:
R. K. Clifton, J. Bub, \& H. Halvorson,
``Characterizing quantum theory in terms of information-theoretic constraints'',
{\em Found. Phys.} {\bf 33}, 11, 1561-1591 (2003);
quant-ph/0211089.

\item {\bf [Clifton 04]}:
R. K. Clifton (edited by J. Butterfield, \& H. Halvorson),
{\em Quantum entanglements --- Selected papers};
Clarendon Press, Oxford, 2004.

\item {\bf [Clover 03]}:
M. Clover,
``The Innsbruck EPR experiment: A time-retarded local description of
space-like separated correlations'',
quant-ph/0304115.

\item {\bf [Clover 04]}:
M. Clover,
``Bell's theorem: A new derivation and a new experiment'',
quant-ph/0409058.

\item {\bf [Coates 02]}:
A. Coates,
A quantum measurement scenario which requires exponential classical communication for
simulation'',
quant-ph/0203112.

\item {\bf [Cockhott 97]}:
P. Cockhott,
``Quantum relational databases'',
quant-ph/9712025.

\item {\bf [Cochrane-Milburn-Munro 99]}:
P. T. Cochrane, G. J. Milburn, \& W. J. Munro,
``Macroscopically distinct quantum-superposition states
as a bosonic code for amplitude damping'',
{\em Phys. Rev. A} {\bf 59}, 4, 2631-2634 (1999).

\item {\bf [Cochrane-Milburn-Munro 00]}:
P. T. Cochrane, G. J. Milburn, \& W. J. Munro,
``Teleportation using coupled oscillator states'',
{\em Phys. Rev. A} {\bf 62}, 6, 062307 (2000);
quant-ph/0004048.

\item {\bf [Cochrane-Milburn 01]}:
P. T. Cochrane, \& G. J. Milburn,
``Teleportation with the entangled states of a beam splitter'',
{\em Phys. Rev. A} {\bf 64}, 6, 062312 (2001);
quant-ph/0103159.

\item {\bf [Cochrane-Ralph-Milburn 02]}:
P. T. Cochrane, T. C. Ralph, \& G. J. Milburn,
``Teleportation improvement by conditional measurements on the two-mode
squeezed vacuum'',
{\em Phys. Rev. A} {\bf 65}, 6, 062306 (2002);
quant-ph/0108051.

\item {\bf [Cochrane-Ralph 03]}:
P. T. Cochrane, \& T. C. Ralph,
``Tailoring teleportation to the quantum alphabet'',
{\em Phys. Rev. A} {\bf 67}, 2, 022313 (2003).

\item {\bf [Coecke 95]}:
B. Coecke,
``Representation of a spin-1 entity as a joint system of
two spin-1/2 entities on which we introduce correlations of the
second kind'',
{\em Helv. Phys. Acta} {\bf 68}, ?, 396-406 (1995).
See {\bf [Coecke 98 b]}.

\item {\bf [Coecke 98 a]}:
B. Coecke,
``A representation for compound quantum systems as individual entities:
Hard acts of creation and hidden correlations'',
{\em Found. Phys.} {\bf 28}, 7, 1109-1136 (1998);
quant-ph/0105093.

\item {\bf [Coecke 98 b]}:
B. Coecke,
``A representation for a spin-$S$ quantum entity as a compound system
in ${\cal R}^3$ consisting of $2S$ individual spin-1/2 entities'',
{\em Found. Phys.} {\bf 28}, 8, 1347-1365 (1998);
quant-ph/0105094.
See {\bf [Coecke 95]}.

\item {\bf [Coecke 00]}:
B. Coecke,
``Structural characterization of compoundness'',
{\em Int. J. Theor. Phys.} {\bf 39}, 3, 585-594 (2000);
quant-ph/0008054.

\item {\bf [Coecke-Smets 00]}:
B. Coecke, \& S. Smets,
``A logical description for perfect measurements'',
{\em Int. J. Theor. Phys.} {\bf 39}, 3, 595-604 (2000);
quant-ph/0008017.

\item {\bf [Coecke-Moore-Wilce 00]}:
B. Coecke, D. Moore, \& A. Wilce,
``Operational quantum logic: An overview'',
in {\em Current research in operational quantum logic:
Algebras, categories, languages},
Kluwer Academic, 2000;
quant-ph/0008019.

\item {\bf [Coecke 01]}:
B. Coecke,
``Forcing discretization and determination in quantum history theories'',
in L. Accardi (ed.),
in {\em Quantum probability and related topics (V\"{a}xj\"{o}, Sweden, 2000)} {\bf IX},
World Scientific, Singapore, to appear;
quant-ph/0105085.

\item {\bf [Coecke 04]}:
B. Coecke,
``The logic of entanglement'',
quant-ph/0402014.

\item {\bf [Coffey 02]}:
M. W. Coffey,
`Comment on ``Quantum search protocol for an atomic array''\,',
{\em Phys. Rev. A} {\bf 66}, 4, 046301 (2002).
Comment on {\bf [Scully-Zubairy 01]}.
Reply: {\bf [Scully-Zubairy 02 b]}.

\item {\bf [Coffman-Kundu-Wootters 00]}:
V. Coffman, J. Kundu, \& W. K. Wootters,
``Distributed entanglement'',
{\em Phys. Rev. A} {\bf 61}, 5, 052306 (2000);
quant-ph/9907047.

\item {\bf [Cohen Tannoudji-Diu-Lalo\"{e} 73]}:
C. Cohen Tannoudji, B. Diu, \& F. Lalo\"{e},
{\em M\'{e}canique quantique}, 2 Vols.,
Hermann, Paris, 1973 (1st edition), 1977 (2nd edition).
English version: {\em Quantum mechanics},
John Wiley \& Sons-Hermann, Paris, 1977.

\item {\bf [Cohen 03]}:
D. Cohen,
``Quantum pumping and dissipation: From closed to open systems'',
{\em Phys. Rev. B} {\bf 68}, 2, 0201303 (2003);
cond-mat/0304678.

\item {\bf [Cohen-Encheva-Litsyn 98]}:
G. Cohen, S. Encheva, \& S. Litsyn,
``On binary constructions of quantum codes'',
quant-ph/9812065.

\item {\bf [Cohen 95]}:
O. Cohen,
``Pre- and postselected quantum systems,
counterfactual measurements, and consistent histories'',
{\em Phys. Rev. A} {\bf 51}, 6, 4373-4380 (1995).
Erratum: {\em Phys. Rev. A} {\bf 56}, 6, 5191 (1997).

\item {\bf [Cohen-Hiley 95 a]}:
O. Cohen, \& B. J. Hiley,
``Reexamining the assumption that elements of reality can be Lorentz invariant'',
{\em Phys. Rev. A} {\bf 52}, 1, 76-81 (1995).
See {\bf [Vaidman 93, 97]}, {\bf [Cohen-Hiley 96]}.

\item {\bf [Cohen-Hiley 95 b]}:
O. Cohen, \& B. J. Hiley,
``Retrodiction in
quantum mechanics, preferred Lorentz frames, and nonlocal measurements'',
{\em Found. Phys.} {\bf 25}, 12, 1669-1698 (1995).

\item {\bf [Cohen-Hiley 96]}:
O. Cohen, \& B. J. Hiley,
``Elements of reality, Lorentz invariance, and the product rule'',
{\em Found. Phys.} {\bf 26}, 1, 1-15 (1996).
See {\bf [Vaidman 93]}, {\bf [Cohen-Hiley 95 a]}.

\item {\bf [Cohen 97 a]}:
O. Cohen,
``Nonlocality of the original Einstein-Podolsky-Rosen state'',
{\em Phys. Rev. A} {\bf 56}, 5, 3484-3492 (1997).

\item {\bf [Cohen 97 b]}:
O. Cohen,
``Quantum cryptography using nonlocal measurements'',
{\em Helv. Phys. Acta} {\bf 70}, 5, 710-726 (1997).

\item {\bf [Cohen 98 a]}:
O. Cohen,
``Reply to `Validity of the Aharonov-Bergman-Lebowitz rule'\,'',
{\em Phys. Rev. A} {\bf 57}, 3, 2254-2255 (1998).
Reply to {\bf [Vaidman 98 a]}.

\item {\bf [Cohen 98 b]}:
O. Cohen,
``Unlocking hidden entanglement with classical information'',
{\em Phys. Rev. Lett.} {\bf 80}, 11, 2493-2496 (1998).

\item {\bf [Cohen 99 a]}:
O. Cohen,
``Counterfactual entanglement and nonlocal correlations in separable states'',
{\em Phys. Rev. A} {\bf 60}, 1, 80-84 (1999);
quant-ph/9907109.
Comment: {\bf [Terno 01]}.
Reply: {\bf [Cohen 01]}.

\item {\bf [Cohen 99 b]}:
O. Cohen,
``Quantifying nonorthogonality'',
{\em Phys. Rev. A} {\bf 60}, 6, 4349-4353 (1999);
quant-ph/9907110.

\item {\bf [Cohen-Brun 00]}:
O. Cohen, \& T. A. Brun,
``Distillation of Greenberger-Horne-Zeilinger states by selective
information manipulation'',
{\em Phys. Rev. Lett.} {\bf 84}, 25, 5908-5911 (2000);
quant-ph/0001084.

\item {\bf [Cohen 01]}:
O. Cohen,
`Reply to ``Comment on `Counterfactual entanglement and
nonlocal correlations in separable states'\,''\,',
{\em Phys. Rev. A} {\bf 63}, 1, 016102 (2001);
quant-ph/0008080.
Reply to {\bf [Terno 01]}.
See {\bf [Cohen 99 a]}.

\item {\bf [Cohen 03]}:
O. Cohen,
``Classical teleportation of classical states'',
quant-ph/0310017.

\item {\bf [Cohen-Horne-Stachel 97 a]}:
R. S. Cohen, M. A. Horne, \& J. Stachel (eds.),
{\em Experimental metaphysics:
Quantum mechanical studies for Abner Shimony, volume one},
Kluwer Academic, Dordrecht, Holland, 1997.
Review: {\bf [Dieks 99]}.
See {\bf [Cohen-Horne-Stachel 97 b]} (II), {\bf [Christian 99]}.

\item {\bf [Cohen-Horne-Stachel 97 b]}:
R. S. Cohen, M. A. Horne, \& J. Stachel (eds.),
{\em Potentiality, entanglement and passion-at-distance:
Quantum mechanical studies for Abner Shimony, volume two},
Kluwer Academic, Dordrecht, Holland, 1997.
Review: {\bf [Christian 99]}.
See {\bf [Cohen-Horne-Stachel 97 a]} (I), {\bf [Dieks 99]}.

\item {\bf [Cola-Paris-Piovella 04]}:
M. M. Cola, M. G. A. Paris, \& N. Piovella,
``Robust generation of entanglement in Bose-Einstein condensates by
collective atomic recoil'',
quant-ph/0404109.

\item {\bf [Colijn-Vrscay 02]}:
C. Colijn, \& E. R. Vrscay,
``Spin-dependent Bohm trajectories for hydrogen eigenstates'',
{\em Phys. Lett. A} {\bf 300}, 4-5, 334-340 (2002).
Erratum: {\em Phys. Lett. A} {\bf 316}, 6, 424 (2003).

\item {\bf [Colijn-Vrscay 03 a]}:
C. Colijn, \& E. R. Vrscay,
``Spin-dependent Bohm trajectories associated with an electronic transition in hydrogen'',
{\em J. Phys. A} {\bf 36}, 16, 4689-4702 (2003).

\item {\bf [Colijn-Vrscay 03 b]}:
C. Colijn, \& E. R. Vrscay,
``Spin-dependent Bohm trajectories for hydrogen eigenstates'',
{\em Phys. Lett. A} {\bf 300}, ?, 334-340 (2002);
quant-ph/0308105.

\item {\bf [Colijn-Vrscay 03 c]}:
C. Colijn, \& E. R. Vrscay,
``Spin-dependent Bohm trajectories for Pauli and Dirac eigenstates of hydrogen'',
{\em Found. Phys. Lett.} {\bf 16}, 4, 303-323 (2003).

\item {\bf [Colin 03 a]}:
S. Colin,
``A deterministic Bell model'',
{\em Phys. Lett. A} {\bf 317}, 5-6, 349-358 (2003);
quant-ph/0310055.

\item {\bf [Colin 03 b]}:
S. Colin,
``Beables for quantum electrodynamics'',
{\em Proc.\ Peyresq Conf.\ on
Electromagnetism (?, 2002)}, {\em Ann. Fond. L. de Broglie};
quant-ph/0310056.

\item {\bf [Collett-Pearle 03]}:
B. Collett, \& P. Pearle,
``Wavefunction collapse and random walk'',
{\em Found. Phys.} {\bf 33}, 10, 1495-1541 (2003).

\item {\bf [Collett-Loudon 87]}:
M. J. Collett, \& R. Loudon,
``Analysis of a proposed crucial test of quantum mechanics'',
{\em Nature} {\bf 326}, 6114, 671-672 (1987).
See {\bf [Bedford-Selleri 85]}, {\bf [Combourieu 92]}.

\item {\bf [Collins-Kim-Holton 98]}:
D. G. Collins, K. W. Kim, \& W. C. Holton,
``Deutsch-Jozsa algorithm as a test of quantum computation'',
{\em Phys. Rev. A} {\bf 58}, 3, R1633-R1636 (1998).

\item {\bf [Collins-Kim-Holton-(+2) 00]}:
D. G. Collins, K. W. Kim, W. C. Holton,
H. Sierzputowska-Gracz, \& E. O. Stejskal,
``NMR quantum computation with indirectly coupled gates'',
{\em Phys. Rev. A} {\bf 62}, 2, 022304 (2000);
quant-ph/9910006.

\item {\bf [Collins-Linden-Popescu 01]}:
D. G. Collins, N. Linden, \& S. Popescu,
``Nonlocal content of quantum operations'',
{\em Phys. Rev. A} {\bf 64}, 3, 032302 (2001);
quant-ph/0005102.

\item {\bf [Collins-Kim-Holton-(+2) 01]}:
D. G. Collins, K. W. Kim, W. C. Holton,
H. Sierzputowska-Gracz, \& E. O. Stejskal,
``Orchestrating an NMR quantum computation:
The $N=3$ Deutsch-Jozsa algorithm'',
quant-ph/0105045.

\item {\bf [Collins-Gisin-Linden-(+2) 02]}:
D. G. Collins, N. Gisin, N. Linden,
S. Massar, \& S. Popescu,
``Bell inequalities for arbitrarily high-dimensional systems'',
{\em Phys. Rev. Lett.} {\bf 88}, 4, 040404 (2002);
quant-ph/0106024.

\item {\bf [Collins-Gisin-Popescu-(+2) 02]}:
D. G. Collins, N. Gisin, S. Popescu,
D. Roberts, \& V. Scarani,
``Bell-type inequalities to detect true $n$-body nonseparability'',
{\em Phys. Rev. Lett.} {\bf 88}, 17, 170405 (2002);
quant-ph/0201058.
See {\bf [Seevinck-Svetlichny 02]}.

\item {\bf [Collins 02 a]}:
D. G. Collins,
``Modified Grover's algorithm for an expectation-value quantum computer'',
{\em Phys. Rev. A} {\bf 65}, 5, 052321 (2002);
quant-ph/0111108.

\item {\bf [Collins 02 b]}:
D. G. Collins,
``Perspectives on quantum non-locality'',
Ph.\ D. thesis, University of Bristol, 2002.

\item {\bf [Collins-Gisin 03]}:
D. G. Collins, \& N. Gisin,
``A relevant two qubit Bell inequality inequivalent to the CHSH inequality'',
quant-ph/0306129.

\item {\bf [Collins-Gisin-de Riedmatten 03]}:
D. G. Collins, N. Gisin, \& H. de Riedmatten,
``Quantum relays for long distance quantum cryptography'',
quant-ph/0311101.

\item {\bf [Collins-Di\'{o}si-Gisin-(+2) 04]}:
D. Collins, L. Di\'{o}si, N. Gisin, S. Massar, \& S. Popescu,
``Quantum gloves'',
quant-ph/0409221.

\item {\bf [Collins 92]}:
G. P. Collins,
``Quantum cryptography defies eavesdropping'',
{\em Phys. Today} {\bf 45}, 11, 21-23 (1992).

\item {\bf [Collins 97]}:
G. P. Collins,
``Exhaustive searching is less tiring with a bit of quantum magic'',
{\em Phys. Today} {\bf 50}, 10, 19-21 (1997).
See {\bf [Grover 97 b]}.

\item {\bf [Collins 98]}:
G. P. Collins,
``Quantum teleportation channels opened in Rome and Innsbruck'',
{\em Phys. Today} {\bf 51}, 2, 18-21 (1998).

\item {\bf [Collins 99]}:
G. P. Collins,
``Qubit chip'',
{\em Sci. Am.} {\bf 281}, 2, 14-15 (1999).
Spanish version: ``Chip de cubits'',
{\em Investigaci\'{o}n y Ciencia} 275, 38 (1999).
See {\bf [Nakamura-Pashkin-Tsai 99]}.

\item {\bf [Collins-Popescu 01]}:
D. G. Collins, \& S. Popescu,
``Violations of local realism by two entangled quNits'',
in S. Popescu, N. Linden, \& R. Jozsa (eds.),
{\em J. Phys. A} {\bf 34}, 35
(Special issue: Quantum information and computation), 6831-6836 (2001);
quant-ph/0106156.

\item {\bf [Collins-Popescu 02]}:
D. G. Collins, \& S. Popescu,
``Classical analog of entanglement'',
{\em Phys. Rev. A} {\bf 65}, 3, 032321 (2002);
quant-ph/0107082.

\item {\bf [Collins-Popescu 04]}:
D. G. Collins, \& S. Popescu,
``Frames of reference and the intrinsic directional information of a
particle with spin'',
quant-ph/0401096.

\item {\bf [Combourieu 92]}:
M. C. Combourieu,
``Karl R. Popper, 1992: About the EPR controversy'',
{\em Found. Phys.} {\bf 22}, 10, 1303-1323 (1992).
See {\bf [Bedford-Selleri 85]}, {\bf [Collett-Loudon 87]}.

\item {\bf [Conway-Kochen 02]}:
J. H. Conway, \& S. Kochen,
``The geometry of the quantum paradoxes'',
in {\bf [Bertlmann-Zeilinger 02]}, pp.~257-269.

\item {\bf [Cooke-Keane-Moran 84]}:
R. Cooke, M. Keane, \& W. Moran,
``Gleason's theorem'',
{\em Delft Progress Rep.} {\bf 9}, 135-150 (1984).

\item {\bf [Cooke-Keane-Moran 85]}:
R. Cooke, M. Keane, \& W. Moran,
``An elementary proof of Gleason's theorem'',
{\em Math. Proc.\ Cambridge Philos. Soc.} {\bf 98}, 1, 117-128 (1985).
Reprinted as an appendix in {\bf [Hughes 89]}.
See {\bf [Gill-Keane 96]}.

\item {\bf [Cooper-Van Vechten 69]}:
L. N. Cooper, \& D. Van Vechten,
``On the interpretation of measurement within the quantum theory'',
{\em Am. J. Phys.} {\bf 37}, 12, 1212-1220 (1969).

\item {\bf [Cooper 99]}:
T. L. Cooper,
``Dynamic rules for decoherence'',
quant-ph/9911084.

\item {\bf [Coppersmith 02]}:
D. Coppersmith,
``An approximate Fourier transform useful in quantum factoring'',
quant-ph/0201067.

\item {\bf [Corato-Silvestrini-Stodolsky-Wosiek 03]}:
V. Corato, P. Silvestrini, L. Stodolsky, \& J. Wosiek,
``Design of adiabatic logic for a quantum CNOT gate'',
{\em Phys. Lett. A} {\bf 309}, 3-4, 206-210 (2003).

\item {\bf [Corbett-Home 00]}:
J. V. Corbett, \& D. Home,
``Quantum effects involving interplay between unitary
dynamics and kinematic entanglement'',
{\em Phys. Rev. A} {\bf 62}, 6, 062103 (2000).

\item {\bf [Corbett-Home 01]}:
J. V. Corbett, \& D. Home,
``{\em Ipso}-information-transfer'',
quant-ph/0103146.

\item {\bf [Cortese 04]}:
J. Cortese,
``Holevo-Schumacher-Westmoreland channel capacity for a class of qudit unital channels'',
{\em Phys. Rev. A} {\bf 69}, 2, 022302 (2004).

\item {\bf [Corwin 84]}:
T. M. Corwin,
``Quantum mechanics and separability'',
{\em Am. J. Phys.} {\bf 52}, 4, 371-372 (1984).

\item {\bf [Cory-Fahmy-Havel 96]}:
D. G. Cory, A. F. Fahmy, \& T. F. Havel,
``Nuclear magnetic resonance spectroscopy: An experimentally accessible paradigm
for quantum computing'',
in T. Toffoli, M. Biafore, \& J. Le\~{a}o (eds.),
{\em Proc.\ PhysComp '96},
New England Complex Systems Institute, 1996, pp.~87-91.
Reprinted in {\bf [Macchiavello-Palma-Zeilinger 00]}, pp.~471-475.

\item {\bf [Cory-Fahmy-Havel 97]}:
D. G. Cory, A. F. Fahmy, \& T. F. Havel,
``Ensemble quantum computing by nuclear magnetic resonance spectroscopy'',
{\em Proc.\ Natl. Acad. Sci. USA} {\bf 94}, 5, 1634-1639 (1997).

\item {\bf [Cory-Price-Havel 98]}:
D. G. Cory, M. D. Price, \& T. F. Havel,
``Nuclear magnetic resonance spectroscopy: An experimentally
accessible paradigm for quantum computing'',
{\em Physica D} {\bf 120}, ?, 82-101 (1998).

\item {\bf [Cory-Price-Maas-(+5) 98]}:
D. G. Cory, M. D. Price, W. Mass, E. Knill, R. Laflamme, W. H. Zurek,
T. F. Havel, \& S. S. Somaroo,
``Experimental quantum error correction'',
{\em Phys. Rev. Lett.} {\bf 81}, 10, 2152-2155 (1998);
quant-ph/9802018.

\item {\bf [Cory-Laflamme-Knill-(+13) 00]}:
D. G. Cory, R. Laflamme, E. Knill,
L. Viola, T. F. Havel, N. Boulant,
G. Boutis, E. Fortunato, S. Lloyd,
R. Mart\'{\i}nez, C. Negrevergne, M. Pravia,
Y. Sharf, G. Teklemariam, Y. S. Weinstein, \& W. H. Zurek,
``NMR based quantum information processing: Achievements and prospects'',
{\em Fortschr. Phys.} {\bf 48}, 9-11 (Special issue: Experimental proposals for quantum computation), 875-907 (2000);
quant-ph/0004104.

\item {\bf [Costa-Bose 01]}:
A. T. Costa, Jr., \& S. Bose,
``Impurity scattering induced entanglement of ballistic electrons'',
{\em Phys. Rev. Lett.} {\bf 87}, 27, 277901 (2001);
quant-ph/0109045.

\item {\bf [Costa de Beauregard 81]}:
O. Costa de Beauregard,
``Comments on a recent proposal by Garuccio and Vigier'',
{\em Found. Phys.} {\bf 11}, 11-12, 947-948 (1981).
Comment on {\bf [Garuccio-Vigier 80]}.

\item {\bf [Costa de Beauregard 83]}:
O. Costa de Beauregard,
``Lorentz and CPT
invariances and the Einstein-Podolsky-Rosen correlations'',
{\em Phys. Rev. Lett.} {\bf 50}, 12, 867-869 (1983).

\item {\bf [Costa de Beauregard 01]}:
O. Costa de Beauregard,
``Comments on the experimental disproof of multisimultaneity'',
quant-ph/0112150.
Comment on {\bf [Stefanov-Gisin-Suarez-Zbinden 01]}.

\item {\bf [Costa Dias-Prata 01]}:
N. Costa Dias, \& J. N. Prata,
``Causal interpretation and quantum phase space'',
{\em Phys. Lett. A} {\bf 291}, 6, 355-366 (2001);
quant-ph/0110062.

\item {\bf [Costi-McKenzie 03]}:
T. A. Costi, \& R. H. McKenzie,
``Entanglement between a qubit and the environment in the spin-boson model'',
{\em Phys. Rev. A} {\bf 68}, 3, 034301 (2003).

\item {\bf [Courty-Grassia-Reynaud 01]}:
J.-M. Courty, F. Grassia, \& S. Reynaud,
``Quantum non ideal measurements'',
quant-ph/0110022.

\item {\bf [Cramer 80]}:
J. G. Cramer,
``Generalized absorber theory and the Einstein-
Podolsky-Rosen paradox'',
{\em Phys. Rev. D} {\bf 22}, 2, 362-376 (1980).

\item {\bf [Cramer 86]}:
J. G. Cramer,
``The transactional interpretation of quantum mechanics'',
{\em Rev. Mod. Phys.} {\bf 58}, 3, 647-687 (1986).

\item {\bf [Cramer 88]}:
J. G. Cramer,
``An overview of the transactional interpretation of quantum
mechanics'',
{\em Int. J. Theor. Phys.} {\bf 27}, ? 227-236 (1988).

\item {\bf [Cr\'{e}peau 94]}:
C. Cr\'{e}peau,
``Quantum oblivious transfer'',
in S. M. Barnett, A. K. Ekert, \& S. J. D. Phoenix (eds.),
{\em J. Mod. Opt.} {\bf 41}, 12 (Special issue: Quantum
communication), 2445-2454 (1994).

\item {\bf [Croca-Ferrero-Garuccio-Lepore 97]}:
J. R. Croca, M. Ferrero, A. Garuccio, \& V. L. Lepore,
``An experiment to test the reality of de Broglie waves'',
{\em Found. Phys. Lett.} {\bf 10}, 5, 441-448 (1997).

\item {\bf [Croca 03]}:
J. R. Croca,
{\em Towards a nonlinear quantum physics},
World Scientific, Singapore, 2003.
Review: {\bf [Selleri 04]}.

\item {\bf [Csirik 02]}:
J. A. Csirik,
``Cost of exactly simulating a Bell pair using classical communication'',
{\em Phys. Rev. A} {\bf 66}, 1, 014302 (2002).

\item {\bf [Cubitt-Verstraete-D\"{u}r-Cirac 03]}:
T. S. Cubitt, F. Verstraete, W. D\"{u}r, \& J. I. Cirac,
``Separable states can be used to distribute entanglement'',
{\em Phys. Rev. Lett.} {\bf 91}, 3, 037902 (2003);
quant-ph/0302168.

\item {\bf [Cubitt-Verstraete-Cirac 04]}:
T. S. Cubitt, F. Verstraete, \& J. I. Cirac,
``Entanglement flow in multipartite systems'',
quant-ph/0404179.

\item {\bf [Cucchietti-Pastawski-Wisniacki 02]}:
F. M. Cucchietti, H. M. Pastawski, \& D. A. Wisniacki,
``Decoherence as decay of the Loschmidt echo in a Lorentz gas'',
{\em Phys. Rev. E} {\bf 65}, 4, 045206 (2002).

\item {\bf [Cucchietti-Dalvit-Paz-Zurek 03]}:
F. M. Cucchietti, D. A. R. Dalvit, J. P. Paz, \& W. H. Zurek,
``Decoherence and the Loschmidt echo'',
quant-ph/0306142.

\item {\bf [Cummins-Jones 00 a]}:
H. K. Cummins, \& J. A. Jones,
``Use of composite rotations to correct systematic
errors in NMR quantum computation'',
{\em New J. Phys.} {\bf 2}, 6.1-6.12 (2000);
quant-ph/9911072.

\item {\bf [Cummins-Jones 00 b]}:
H. K. Cummins, \& J. A. Jones,
``Resonance offset tailored pulses for NMR quantum computation'',
submitted to {\em J. Magnetic Resonance},
quant-ph/0008034.

\item {\bf [Cummins-Jones-Furze-(+3) 02]}:
H. K. Cummins, C. Jones, A. Furze,
N. F. Soffe, M. Mosca, J. M. Peach, \& J. A. Jones,
``Approximate quantum cloning with nuclear magnetic resonance'',
{\em Phys. Rev. Lett.} {\bf 88}, 18, 187901 (2002);
quant-ph/0111098.

\item {\bf [Cummins-Llewellyn-Jones 03]}:
H. K. Cummins, G. Llewellyn, \& J. A. Jones,
``Tackling systematic errors in quantum logic gates with composite rotations'',
{\em Phys. Rev. A} {\bf 67}, 4, 042308 (2003).

\item {\bf [Curty-Santos 01 a]}:
M. Curty, \& D. J. Santos,
``Quantum authentication of classical messages'',
{\em Phys. Rev. A} {\bf 64}, 6, 062309 (2001).

\item {\bf [Curty-Santos 01 b]}:
M. Curty, \& D. J. Santos,
``Quantum information processing: A linear systems perspective'',
quant-ph/0101060.

\item {\bf [Curty-Santos 01 c]}:
M. Curty, \& D. J. Santos,
``Quantum cryptography without a quantum channel'',
quant-ph/0101079.

\item {\bf [Curty-Santos 01 d]}:
M. Curty, \& D. J. Santos,
``Secure authentication of classical messages with a one-ebit quantum
key'',
quant-ph/0103122.

\item {\bf [Curty-Santos-P\'{e}rez-Garc\'{\i}a Fern\'{a}ndez 02]}:
M. Curty, D. J. Santos, E. P\'{e}rez, \& P. Garc\'{\i}a-Fern\'{a}ndez,
``Qubit authentication'',
{\em Phys. Rev. A} {\bf 66}, 2, 022301 (2002);
quant-ph/0108100.

\item {\bf [Curty-L\"{u}tkenhaus 03]}:
M. Curty, \& N. L\"{u}tkenhaus,
``Practical quantum key distribution: On the security evaluation with
inefficient single-photon detectors'',
quant-ph/0311066.

\item {\bf [Curty-Lewenstein-L\"{u}tkenhaus 04]}:
M. Curty, M. Lewenstein, \& N. L\"{u}tkenhaus,
``Entanglement as precondition for secure quantum key distribution'',
{\em Phys. Rev. Lett.} {\bf 92}, 21, 217903 (2004);
quant-ph/0307151.

\item {\bf [Curty 04]}:
M. Curty,
``Protocolos cu\'{a}nticos para la autenticaci\'{o}n de informaci\'{o}n en sistemas de telecomunicaci\'{o}n'',
Ph.\ D. thesis, Universidad de Vigo, 2004.

\item {\bf [Curty-G\"{u}hne-Lewenstein-L\"{u}tkenhaus 04]}:
M. Curty, O. G\"{u}hne, M. Lewenstein, \& N. L\"{u}tkenhaus,
``Detecting two-party quantum correlations in quantum key distribution
protocols'',
quant-ph/0409047.

\item {\bf [Curty-L\"{u}tkenhaus 04]}:
M. Curty, \& N. L\"{u}tkenhaus,
``Intercept-resend attacks in the Bennett-Brassard 1984 quantum key
distribution protocol with weak coherent pulses'',
quant-ph/0411041.

\item {\bf [Cushing 85]}:
J. T. Cushing,
``Comment on Angelidis's universality claim'',
{\em Phys. Rev. Lett.} {\bf 54}, 18, 2059 (1985).
Comment on {\bf [Angelidis 83]}.

\item {\bf [Cushing 93]}:
J. T. Cushing,
`What if Bell had come before ``Copenhagen''?',
in A. van der Merwe, \& F. Selleri (eds.),
{\em Bell's theorem and the foundations of modern physics.
Proc.\ of an international
conference (Cesena, Italy, 1991)},
World Scientific, Singapore, 1993, pp.~125-134.

\item {\bf [Cushing 94 a]}:
J. T. Cushing,
``Locality/separability: Is this necessarily a useful distinction?'',
in D. Hull, M. Forbes, \& R. Burian (eds.),
{\em Proc.\ of the 1994
Biennial Meeting of the Philosophy of Science Association},
East Lansing, Michigan, 1994, vol. 1, pp.~107-116.

\item {\bf [Cushing 94 b]}:
J. T. Cushing,
{\em Quantum mechanics. Historical
contingency and the Copenhagen hegemony},
University of Chicago Press, Chicago, 1994.
Review: {\bf [Hiley 97]}.

\item {\bf [Cushing 95 a]}:
J. T. Cushing,
``Quantum tunneling times: A crucial test for the causal program?'',
{\em Found. Phys.} {\bf 25}, 2, 269-280 (1995).
Reply: {\bf [Bedard 97]}.

\item {\bf [Cushing 95 b]}:
J. T. Cushing,
``Book review. The undivided universe:
An ontological interpretation of quantum theory'',
{\em Found. Phys.} {\bf 25}, 3, 507-510 (1995).
Review of {\bf [Bohm-Hiley 93]}.

\item {\bf [Cushing-Fine-Goldstein 96]}:
J. T. Cushing, A. I. Fine, \& S. Goldstein (eds.),
{\em Bohmian mechanics and quantum theory: An appraisal},
Kluwer Academic, Dordrecht, Holland, 1996.
Review: {\bf [Jaeger 00]}.

\item {\bf [Cushing 97]}:
J. T. Cushing,
``Infinite potential: The life and times of David Bohm'',
{\em Phys. Today} {\bf 50}, 3, 77-78 (1997).
Review of {\bf [Peat 97]}.

\item {\bf [Cushing 98]}:
J. T. Cushing,
``Conceptual foundation of quantum physics:
An overview from modern perspectives'',
{\em Phys. Today} {\bf 51}, 10, 79 (1998).
Review of {\bf [Home 97]}.

\item {\bf [Cushing 01]}:
J. T. Cushing,
`Determinism versus indeterminism in quantum mechanics:
A ``free'' choice',
in {\bf [Rusell-Clayton-Wegter McNelly-Polkinghorne 01]}, pp.~99-110.

\item {\bf [Czachor 92]}:
M. Czachor,
``On classical models of spin'',
{\em Found. Phys. Lett.} {\bf 5}, 3, 249-264 (1992).

\item {\bf [Czachor 93]}:
M. Czachor,
``Nonlocality in nonlinear quantum mechanics'',
in A. van der Merwe, \& F. Selleri (eds.),
{\em Bell's theorem and the foundations of modern physics.
Proc.\ of an international
conference (Cesena, Italy, 1991)},
World Scientific, Singapore, 1993, pp.~135-138.

\item {\bf [Czachor 94]}:
M. Czachor,
``Bell theorem without inequalities: A single-particle formulation'',
{\em Phys. Rev. A} {\bf 49}, 4, 2231-2240 (1994).

\item {\bf [Czachor 97]}:
M. Czachor,
``Einstein-Podolsky-Rosen-Bohm experiment
with relativistic massive particles'',
{\em Phys. Rev. A} {\bf 55}, 1, 72-77 (1997);
quant-ph/9609022.

\item {\bf [Czachor 99]}:
M. Czachor,
``Quantum cryptography with polarizing interferometers'',
{\em Phys. Lett. A} {\bf 257}, 3-4, 107-112 (1999);
quant-ph/9812030.

\item {\bf [Czachor-Doebner 02]}:
M. Czachor, \& H.-D. Doebner,
``Correlation experiments in nonlinear quantum mechanics'',
{\em Phys. Lett. A} {\bf 301}, 3-4, 139-152 (2002);
quant-ph/0110008.

\item {\bf [Czachor-Wilczewski 03]}:
M. Czachor, \& M. Wilczewski,
``Relativistic BB84, relativistic errors, and how to correct them'',
quant-ph/0303077.

\item {\bf [Czachor 03 a]}:
M. Czachor,
``States of light via reducible quantization'',
{\em Phys. Lett. A};
quant-ph/0211009.

\item {\bf [Czachor 03 b]}:
M. Czachor,
`Comment on ``Quantum entropy and special relativity'' [by A. Peres, P. F.
Scudo, and D. R. Terno, Phys. Rev. Lett. {\bf 88}, 230402 (2002)]',
quant-ph/0312040.
Comment on {\bf [Peres-Scudo-Terno 02]}.

\item {\bf [Czachor-Wilczewski 03]}:
M. Czachor, \& M. Wilczewski,
``Relativistic Bennett-Brassard cryptographic scheme, relativistic errors, and
how to correct them'',
{\em Phys. Rev. A} {\bf 68}, 1, 010302 (2003).


\newpage

\subsection{}


\item {\bf [Daboul-Wang-Sanders 03]}:
J. Daboul, X. Wang, \& B. C. Sanders,
``Quantum gates on hybrid qudits'',
{\em J. Phys. A} {\bf 36}, 10, 2525-2536 (2003).

\item {\bf [Daffer-W\'{o}dkiewicz-McIver 03 a]}:
S. Daffer, K. W\'{o}dkiewicz, \& J. K. McIver,
``Nonlocality of two-mode squeezing with internal noise'',
{\em Phys. Rev. A} {\bf 68}, 1, 012104 (2003).

\item {\bf [Daffer-W\'{o}dkiewicz-McIver 03 b]}:
S. Daffer, K. W\'{o}dkiewicz, \& J. K. McIver,
``?'',
{\em Phys. Rev. A} {\bf 67}, 6, 062312 (2003).

\item {\bf [Daffer-W\'{o}dkiewicz-Cresser-McIver 03 b]}:
S. Daffer, K. W\'{o}dkiewicz, J. D. Cresser, \& J. K. McIver,
``Depolarizing channel as a completely positive map with memory'',
{\em Phys. Rev. A} {\bf 70}, 1, 010304 (2004).

\item {\bf [Daffertshofer-Plastino-Plastino 02]}:
A. Daffertshofer, A. R. Plastino, \& A. Plastino,
``Classical no-cloning theorem'',
{\em Phys. Rev. Lett.} {\bf 88}, 21, 210601 (2002).

\item {\bf [Daftuar-Klimesh 01 a]}:
S. Daftuar, \& M. Klimesh,
``Mathematical structure of entanglement catalysis'',
{\em Phys. Rev. A} {\bf 64}, 4, 042314 (2001).

\item {\bf [Daftuar-Klimesh 01 b]}:
S. Daftuar, \& M. Klimesh,
``The trumping relation and the structure of the bipartite entangled
states'',
quant-ph/0104058.

\item {\bf [Daftuar-Hayden 04]}:
S. Daftuar, \& P. Hayden,
``Quantum state transformations and the Schubert calculus'',
{\em Ann. Phys.};
quant-ph/0410052.

\item {\bf [Dai-Zhang-Li 04]}:
H.-Y. Dai, M. Zhang, \& C.-Z. Li,
``Probabilistic teleportation of an unknown entangled state of two three-level
particles using a partially entangled state of three three-level particles'',
{\em Phys. Lett. A} {\bf 323}, 5-6, 360-364 (2004).

\item {\bf [Dai-Chen 04]}:
L. Dai, \& Q. Chen,
`Comment on: ``Quantum cooperative games'' [Phys. Lett. A 293 (2002) 103]'',
{\em Phys. Lett. A} {\bf 328}, 4-5, 414-415 (2004).
Comment on {\bf [Iqbal-Toor 02 b]}.

\item {\bf [Dahm-Goodkind-Karakurt-Pilla 01]}:
A. J. Dahm, J. M. Goodkind, I. Karakurt, \& S. Pilla,
``Using electrons on liquid helium for quantum computing'',
quant-ph/0111029.

\item {\bf [Dahm 03]}:
A. J. Dahm,
``Quantum computing with bits made of electrons on a helium surface'',
{\em Low Temp. Phys.} {\bf 29}, 489 (2003).

\item {\bf [Dakna-Kn\"{o}ll-Welsch 98]}:
M. Dakna, L. Kn\"{o}ll, \& D.-G. Welsch,
``Quantum state engineering using conditional measurement on a beam
splitter'',
{\em Eur. Phys. J. D} {\bf 3}, 3, 295-308 (1998).

\item {\bf [Dakna-Clausen-Kn\"{o}ll-Welsch 98 a]}:
M. Dakna, J. Clausen, L. Kn\"{o}ll, \& D.-G. Welsch,
``Generating and monitoring Schr\"{o}dinger cats in conditional
measurement on a beam splitter'',
quant-ph/9805048.

\item {\bf [Dakna-Clausen-Kn\"{o}ll-Welsch 98 b]}:
M. Dakna, J. Clausen, L. Kn\"{o}ll, \& D.-G. Welsch,
``Quantum state engineering by alternate state displacement and photon
adding'',
quant-ph/9807089.

\item {\bf [Daley-Fedichev-Zoller, 04]}:
A. J. Daley, P. O. Fedichev, \& P. Zoller,
``Single-atom cooling by superfluid immersion: A nondestructive method for qubits'',
{\em Phys. Rev. A} {\bf 69}, 2, 022306 (2004);
quant-ph/0308129.

\item {\bf [Daley-Kollath-Schollwoeck-Vidal 04]}:
A. J. Daley, C. Kollath, U. Schollwoeck, \& G. Vidal,
``Time-dependent density-matrix renormalization-group using adaptive
effective Hilbert spaces'',
cond-mat/0403313.

\item {\bf [Dalitz-Garbarino 00]}:
R. H. Dalitz, \& G. Garbarino,
``Local realistic theories and quantum mechanics
for the two-neutral-kaon system'',
{\em Nucl. Phys. B} {\bf 606}, 483-? (2001);
quant-ph/0011108.

\item {\bf [Dalla Chiara-Giuntini 01]}:
M. L. Dalla Chiara, \& R. Giuntini,
``Quantum logic'',
quant-ph/0101028.

\item {\bf [Dalton 01]}:
B. Dalton,
``Two-particle correlations via quasi-deterministic analyzer model'',
quant-ph/0101053.

\item {\bf [Dalton 03]}:
B. J. Dalton,
``Scaling of decoherence effects in quantum computers'',
in M. Ferrero (ed.),
{\em Proc. of Quantum Information: Conceptual Foundations,
Developments and Perspectives (Oviedo, Spain, 2002)},
{\em J. Mod. Opt.} {\bf 50}, 6-7, 951-966 (2003).

\item {\bf [Dalvit-Maia Neto 00]}:
D. A. R. Dalvit, \& P. A. Maia Neto,
``Decoherence via the dynamical Casimir effect'',
{\em Phys. Rev. Lett.} {\bf 84}, 5, 798-801 (2000).

\item {\bf [Dalvit-Dziarmaga-Zurek 01]}:
D. A. R. Dalvit, J. Dziarmaga, \& W. H. Zurek,
``Unconditional pointer states from conditional master
equations'',
{\em Phys. Rev. Lett.} {\bf 86}, 3, 373-376 (2001).

\item {\bf [Das-Mahesh-Kumar 02]}:
R. Das, T. S. Mahesh, \& A. Kumar,
``Implementation of conditional phase shift gate for quantum information
processing by NMR, using transition-selective pulses'',
{\em J. Magnetic Resonance} {\bf 159}, 46-54 (2002);
quant-ph/0311103.

\item {\bf [Das-Mahesh-Kumar 03]}:
R. Das, T. S. Mahesh, \& A. Kumar,
``Efficient quantum-state tomography for quantum-information processing using
a two-dimensional Fourier-transform technique'',
{\em Phys. Rev. A} {\bf 67}, 6, 062304 (2003).

\item {\bf [Das-Bhattacharyya-Kumar 04]}:
R. Das, R. Bhattacharyya, \& A. Kumar,
``Quantum information processing by NMR using a 5-qubit system formed by
dipolar coupled spins in an oriented molecule'',
{\em J. Magnetic Resonance} {\bf 170}, 2, 310-? (2004);
quant-ph/0409186.

\item {\bf [Das-Mahesh-hakraborty-Rukmani-Kumar 04]}:
R. Das, S. Chakraborty, K. Rukmani, \& A. Kumar,
``Search for optimum labeling schemes in qubit systems
for quantum-information processing by nuclear magnetic resonance'',
{\em Phys. Rev. A} {\bf 70}, 1, 012314 (2004).

\item {\bf [Das-Kobes-Kunstatter 02]}:
S. Das, R. Kobes, \& G. Kunstatter,
``Adiabatic quantum computation and Deutsch's algorithm'',
{\em Phys. Rev. A} {\bf 65}, 6, 062310 (2002);
quant-ph/0111032.

\item {\bf [Das-Kobes-Kunstatter 03]}:
S. Das, R. Kobes, \& G. Kunstatter,
``Energy and efficiency of adiabatic quantum search algorithms'',
{\em J. Phys. A} {\bf 36}, 11, 2839-2845 (2003).

\item {\bf [Das-Kumar 03]}:
R. Das, \& A. Kumar,
``Use of quadrupolar nuclei for quantum-information processing by nuclear
magnetic resonance: Implementation of a quantum algorithm'',
{\em Phys. Rev. A} {\bf 68}, 3, 032304 (2003).

\item {\bf [Dasgupta-Agarwal 01]}:
S. Dasgupta, \& G. S. Agarwal,
``Improving the fidelity of quantum cloning by field-induced
inhibition of the unwanted transition'',
{\em Phys. Rev. A} {\bf 64}, 2, 022315 (2001);
quant-ph/0010035.

\item {\bf [Dass-Qureshi 98]}:
N. D. H. Dass, \& T. Qureshi,
`Measurable and unmeasurable in ``protective'' measurements',
quant-ph/9805012.

\item {\bf [Dass-Qureshi 99]}:
N. D. H. Dass, \& T. Qureshi,
``Critique of protective measurements'',
{\em Phys. Rev. A} {\bf 59}, 4, 2590-2601 (1999).

\item {\bf [Dass-Joglekar 98]}:
T. Dass, \& Y. Joglekar,
``Symmetries and conservation laws in histories-based generalized
quantum mechanics'',
gr-qc/9812018.

\item {\bf [Datta-Home-Raychaudhuri 87]}:
A. Datta, D. Home, \& A. Raychaudhuri,
``A curious gedanken example of the Einstein-Podolsky-Rosen paradox
using CP nonconservation'',
{\em Phys. Lett. A} {\bf 123}, 1, 4-8 (1987).

\item {\bf [Datta-Ghosh-Majumdar-Nayak 04]}:
A. Datta, B. Ghosh, A. S. Majumdar, \& N. Nayak,
``Information transfer through a one-atom micromaser'',
{\em Europhys. Lett.} (2004);
quant-ph/0307207.

\item {\bf [Datta-Ghose-Samal 04]}:
A. Datta, P. Ghose, \& M. K. Samal,
``Bohmian picture of Rydberg atoms'',
{\em Phys. Lett. A} {\bf 322}, 5-6, 277-281 (2004).

\item {\bf [Datta-Holevo-Suhov 04]}:
N. Datta, A. S. Holevo, \& Y. M. Suhov,
``A quantum channel with additive minimum output entropy'',
quant-ph/0403072.

\item {\bf [Davidovich-Maali-Brune-(+2) 93]}:
L. Davidovich,
A. Maali, M. Brune, J.-M. Raimond, \& S. Haroche,
``Quantum switches and nonlocal microwave fields'',
{\em Phys. Rev. Lett.} {\bf 71}, 15, 2360-2363 (1993).

\item {\bf [Davidovich-Zagury-Brune-(+2) 94]}:
L. Davidovich, N. Zagury, M. Brune, J.-M. Raimond, \& S. Haroche,
``Teleportation of an atomic state
between two cavities using nonlocal microwave fields'',
{\em Phys. Rev. A} {\bf 50}, 2, Part A, R895-R898 (1994).

\item {\bf [Davidovich-Brune-Raimond-Haroche 96]}:
L. Davidovich, M. Brune, J.-M. Raimond, \& S. Haroche,
``Mesoscopic quantum coherences in cavity
QED: Preparation and decoherence monitoring schemes'',
{\em Phys. Rev. A} {\bf 53}, 3, 1295-1309 (1996).
See {\bf [Brune-Hagley-Dreyer-(+5) 96]},
{\bf [Zurek 97]}.

\item {\bf [Davidson 79]}:
M. Davidson,
``A generalization of the Fenyes-Nelson stochastic
model of quantum mechanics'',
{\em Lett. Math. Phys.} {\bf 3}, ?, 271-277 (1979).

\item {\bf [Davidson 01]}:
M. P. Davidson,
``Comments on the nonlinear Schr\"{o}dinger equation'',
quant-ph/0106124.

\item {\bf [Davidson 03]}:
M. P. Davidson,
``A proposed experiment to test the hydrodynamic interpretation
of quantum mechanics using bremsstrahlung'',
quant-ph/0302045.

\item {\bf [Davies-Lewis 70]}:
E. B. Davies, \& J. T. Lewis,
``An operational approach to quantum probability'',
{\em Comm. Math. Phys.} {\bf 17}, ?, 239-260 (1970).

\item {\bf [Davies 76]}:
E. B. Davies,
{\em Quantum theory of open systems},
Academic Press, New York, 1976.

\item {\bf [Davies 78]}:
E. B. Davies,
``Information and quantum measurement'',
{\em IEEE Trans. Inf. Theory} {\bf IT-24}, ?, 596-599 (1978).

\item {\bf [Davies-Brown 86]}:
P. C. W. Davies, \& J. R. Brown (eds.),
{\em The
ghost in the atom. A discussion of the mysteries of quantum physics},
Cambridge University Press, Cambridge, 1986.
Spanish version: {\em El esp\'{\i}ritu en el \'{a}tomo.
Una discusi\'{o}n sobre los misterios de la f\'{\i}sica cu\'{a}ntica},
Alianza, Madrid, 1989.

\item {\bf [Davies 04 a]}:
P. C. W. Davies,
``Quantum tunneling time'',
quant-ph/0403010.

\item {\bf [Davies 04 b]}:
P. C. W. Davies,
``Quantum fluctuations and life'',
quant-ph/0403017.

\item {\bf [Davies 04 c]}:
P. C. W. Davies,
``Quantum mechanics and the equivalence principle'',
quant-ph/0403027.

\item {\bf [Davies 04 d]}:
P. C. W. Davies,
``Transit time of a freely-falling quantum particle in a background
gravitational field'',
quant-ph/0407028.

\item {\bf [Davis-Delbourgo-Jarvis 00]}:
R. I. A. Davis, R. Delbourgo, \& P. D. Jarvis,
``Covariance, correlation and entanglement'',
{\em J. Phys. A} {\bf 33}, 9 1895-1914 (2000);
quant-ph/0001076.

\item {\bf [Dawson-Nielsen 04]}:
C. M. Dawson, \& M. A. Nielsen,
``Frustration, interaction strength and ground-state entanglement in
complex quantum systems'',
quant-ph/0401061.

\item {\bf [Dawson-Hines-McKenzie-Milburn 04]}:
C. M. Dawson, A. P. Hines, R. H. McKenzie, \& G. J. Milburn,
``Entanglement sharing and decoherence in the spin-bath'',
quant-ph/0407206.

\item {\bf [Dawson-Haselgrove-Hines-(+3) 04]}:
C. M. Dawson, H. L. Haselgrove, A. P. Hines,
D. Mortimer, M. A. Nielsen, \& T. J. Osborne,
``Quantum computing and polynomial equations over the finite field $Z_2$'',
quant-ph/0408129.

\item {\bf [Deb-Agarwal 02]}:
B. Deb, \& G. S. Agarwal,
``Tripartite entanglement in a Bose-Einstein condensate by stimulated Bragg
scattering'',
{\em Phys. Rev. A} {\bf 65}, 6, 063618 (2002).

\item {\bf [Deb-Agarwal 03]}:
B. Deb, \& G. S. Agarwal,
``Entangling two Bose-Einstein condensates by stimulated Bragg scattering'',
{\em Phys. Rev. A} {\bf 67}, 2, 023603 (2003).

\item {\bf [Debuisschert-Boucher 03 a]}:
T. Debuisschert, \& W. Boucher,
``Quantum key distribution based on time coding'',
{\em Proc. QELS Conf. (2002)}, QThG4, pp.~201-202;
quant-ph/0309138.

\item {\bf [Debuisschert-Boucher 03 b]}:
T. Debuisschert, \& W. Boucher,
``4 states protocol for time coding quantum key distribution'',
{\em Proc. QELS Conf. (2003)}, QTuB2;
quant-ph/0309139.

\item {\bf [Decker-Janzing-Beth 03]}:
T. Decker, D. Janzing, \& T. Beth,
``Quantum circuits for single-qubit measurements corresponding to platonic
solids'',
quant-ph/0308098.

\item {\bf [Decker-Janzing-R\"{o}tteler 04]}:
T. Decker, D. Janzing, \& M. R\"{o}tteler,
``Implementation of group-covariant POVMs by orthogonal measurements'',
quant-ph/0407054.

\item {\bf [Degiovanni-Ruo Berchera-Castelletto-(+4) 04 a]}:
I. P. Degiovanni, I. Ruo Berchera, S. Castelletto,
M. L. Rastello, F. A. Bovino, A. M. Colla, \& G. Castagnoli,
``Quantum dense key distribution'',
{\em Phys. Rev. A} {\bf 69}, 3, 032310 (2004);
quant-ph/0312128.
See {\bf [Degiovanni-Ruo Berchera-Castelletto-(+4) 04 b]}.

\item {\bf [Degiovanni-Ruo Berchera-Castelletto-(+4) 04 b]}:
I. P. Degiovanni, I. Ruo Berchera, S. Castelletto,
M. L. Rastello, F. A. Bovino, A. M. Colla, \& G. Castagnoli,
``Reply to `Comment on ``Quantum dense key distribution''\,'\,'',
{\em Phys. Rev. A};
quant-ph/0410221.
See {\bf [Degiovanni-Ruo Berchera-Castelletto-(+4) 04 a]}.

\item {\bf [Dehaene-Van den Nest-De Moor-Verstraete 03]}:
J. Dehaene, M. Van den Nest, B. De Moor, \& F. Verstraete,
``Local permutations of products of Bell states and entanglement distillation'',
{\em Phys. Rev. A} {\bf 67}, 2, 022310 (2003).

\item {\bf [Dehesa-Martinez Finkelshtein-Sorokin 02]}:
J. S. Dehesa, A. Martinez-Finkelshtein, \& V. N. Sorokin,
``Quantum-information entropies for highly excited states of single-particle
systems with power-type potentials'',
{\em Phys. Rev. A} {\bf 66}, 6, 062109 (2002).

\item {\bf [Dehlinger-Mitchell 02 a]}:
D. Dehlinger, \& M. W. Mitchell,
``Entangled photon apparatus for the undergraduate laboratory'',
{\em Am. J. Phys.} {\bf 70}, 9, 898-902 (2002).
See {\bf [Dehlinger-Mitchell 02 b]}.

\item {\bf [Dehlinger-Mitchell 02 b]}:
D. Dehlinger, \& M. W. Mitchell,
``Entangled photons, nonlocality, and Bell inequalities in the undergraduate
laboratory'',
{\em Am. J. Phys.} {\bf 70}, 9, 903-910 (2002).
See {\bf [Dehlinger-Mitchell 02 a]}.

\item {\bf [DelRe-Crosignani-Di Porto 00]}:
E. DelRe, B. Crosignani, \& P. Di Porto,
``Scheme for total quantum teleportation'',
{\em Phys. Rev. Lett.} {\bf 84}, 13, 2989-2992 (2000);
quant-ph/9909091.

\item {\bf [Deltete-Guy 90]}:
R. Deltete, \& R. Guy,
``Einstein's opposition to the quantum theory'',
{\em Am. J. Phys.} {\bf 58}, 7, 673-683 (1990).
See {\bf [Guy-Deltete 90]}.

\item {\bf [Deltete-Guy 91]}:
R. Deltete, \& R. Guy,
``Einstein and EPR'',
{\em Philos. Sci.} {\bf 58}, 3, 377-397 (1991).
See {\bf [Howard 85]}, {\bf [Fine 86]},
{\bf [Fine 89]} (Sec. 1), {\bf [H\'{a}jek-Bub 92]}, {\bf [Combourieu 92]}.

\item {\bf [DeMarco-Ben Kish-Leibfried 02]}:
B. DeMarco, A. Ben-Kish, D. Leibfried,
V. Meyer, M. Rowe, B. M. Jelenkovi\'{c},
W. M. Itano, J. Britton, C. Langer, T. Rosenband, \& D. J. Wineland,
``Experimental demonstration of a controlled-NOT wave-packet gate'',
{\em Phys. Rev. Lett.} {\bf 89}, 26, 267901 (2002).

\item {\bf [DeMille 02]}:
D. DeMille,
``Quantum computation with trapped polar molecules'',
{\em Phys. Rev. Lett.} {\bf 88}, 6, 067901 (2002);
quant-ph/0109083.

\item {\bf [Demkowicz Dobrzanski-Kus-W\'{o}dkiewicz 04]}:
R. Demkowicz-Dobrzanski, M. Kus, \& K. W\'{o}dkiewicz,
``Cloning of spin-coherent states'',
{\em Phys. Rev. A} {\bf 69}, 1, 012301 (2004);
quant-ph/0307061.

\item {\bf [Demopoulos 79]}:
W. Demopoulos,
``Boolean representations of physical magnitudes and locality'',
{\em Synthese} {\bf 42}, 1, 101-119 (1979).

\item {\bf [Demopoulos 80]}:
W. Demopoulos,
``Locality and the algebraic structure of quantum mechanics'',
in {\bf [Suppes 80]}, pp.~119-144.
Comments: {\bf [Humphreys 80]}, {\bf [Bub 80]}.

\item {\bf [Deng-Long 03]}:
F.-G. Deng, \& G. L. Long,
``Controlled order rearrangement encryption for quantum key distribution'',
{\em Phys. Rev. A} {\bf 68}, 4, 042315 (2003).

\item {\bf [Deng-Long-Liu 03]}:
F.-G. Deng, G. L. Long, \& X.-S. Liu,
``Two-step quantum direct communication protocol
using the Einstein-Podolsky-Rosen pair block'',
{\em Phys. Rev. A} {\bf 68}, 4, 042317 (2003);
quant-ph/0308173.

\item {\bf [Deng-Long 04 a]}:
F.-G. Deng, \& G. L. Long,
``Secure direct communication with a quantum one-time-pad'',
{\em Phys. Rev. A} {\bf 69}, 5, 052319 (2004);
quant-ph/0405177.
Comment: {\bf [Hoffmann-Bostroem-Felbinger 04]}.

\item {\bf [Deng-Long 04 b]}:
F.-G. Deng, \& G. L. Long,
``Bidirectional quantum key distribution protocol with practical faint laser pulses'',
{\em Phys. Rev. A} {\bf 70}, 1, 012311 (2004).

\item {\bf [Dennis 01]}:
E. Dennis,
``Toward fault-tolerant computation without concatenation'',
{\em Phys. Rev. A} {\bf 63}, 5, 052314 (2001);
quant-ph/9905027.

\item {\bf [Dennis 02]}:
E. Dennis,
`Comment on ``Experimental nonlocality proof of quantum
teleportation and entanglement swapping''\,',
quant-ph/0202025.
Comment on {\bf [Jennewein-Weihs-Pan-Zeilinger 02]}.

\item {\bf [Dennis-Kitaev-Landahl-Preskill 02]}:
E. Dennis, A. Kitaev, A. Landahl, \& J. Preskill,
``Topological quantum memory'',
{\em J. Math. Phys.} {\bf 43}, 9, 4452-4505 (2002);
quant-ph/0110143.

\item {\bf [Dennis-Norsen 04]}:
E. Dennis, \& T. Norsen,
``Quantum theory: Interpretation cannot be avoided'',
quant-ph/0408178.
See {\bf [Fuchs-Peres 00 a]}.

\item {\bf [Dennison-Wootters 02]}:
K. A. Dennison, \& W. K. Wootters,
``Entanglement sharing among quantum particles with more than two orthogonal
states'',
{\em Phys. Rev. A} {\bf 65}, 1, 010301 (2002);
quant-ph/0106058.

\item {\bf [Deotto-Ghirardi 98]}:
E. Deotto, \& G.-C. Ghirardi,
``Bohmian mechanics revisited'',
{\em Found. Phys.} {\bf 28}, 1, 1-30 (1998);
quant-ph/9704021.

\item {\bf [Derka-Bu\v{z}ek-Ekert 98]}:
R. Derka, V. Bu\v{z}ek, \& A. K. Ekert,
``Universal algorithm for optimal estimation of quantum states from
finite ensembles via realizable generalized measurement'',
{\em Phys. Rev. Lett.} {\bf 80}, 8, 1571-1575 (1998);
quant-ph/9707028.

\item {\bf [Derkacz-Jak\'{o}bczyk 04]}:
\L. Derkacz, \& L. Jak\'{o}bczyk,
``Bell inequalities versus entanglement and mixedness for a class of two-qubit states'',
{\em Phys. Lett. A} {\bf 328}, 1, 26-35 (2004);
quant-ph/0402112.

\item {\bf [Dersch 03]}:
H. Dersch,
``Observer created violation of Bell's inequalities'',
quant-ph/0301010.

\item {\bf [Detlefsen 98]}:
M. Detlefsen,
``Mind in the shadows'',
{\em Stud. Hist. Philos. Sci. Part B: Stud. Hist. Philos. Mod. Phys.}
{\bf 29}, 1, 123-136 (1998).
Review of {\bf [Penrose 89, 94 b, 97 a]}.

\item {\bf [Deuar-Munro 00 a]}:
P. Deuar, \& W. J. Munro,
``Improving detectors using entangling quantum copiers'',
{\em Phys. Rev. A} {\bf 61}, 1, 010306(R) (2000);
quant-ph/9911103.

\item {\bf [Deuar-Munro 00 b]}:
P. Deuar, \& W. J. Munro,
``Information transfer and fidelity in quantum copiers'',
{\em Phys. Rev. A} {\bf 61}, 6, 062304 (2000);
quant-ph/0003054.

\item {\bf [Deuar-Munro 00 c]}:
P. Deuar, \& W. J. Munro,
``Quantum copying can increase the practically available information'',
{\em Phys. Rev. A} {\bf 62}, 4, 042304 (2000);
quant-ph/0008032.

\item {\bf [Deuar-Munro-Nemoto 00]}:
P. Deuar, W. J. Munro, \& K. Nemoto,
``Upper bound on the region of separable states near
the maximally mixed state'',
{\em J. Opt. B: Quantum Semiclass. Opt.} {\bf 2}, 3, 225-229 (2000);
quant-ph/0002002.

\item {\bf [Deutsch 83]}:
D. Deutsch,
``Uncertainty in quantum measurements'',
{\em Phys. Rev. Lett.} {\bf 50}, 9, 631-633 (1983).

\item {\bf [Deutsch 85 a]}:
D. Deutsch,
``Quantum theory as a universal physical theory'',
{\em Int. J. Theor. Phys.} {\bf 24}, 1, 1-41 (1985).

\item {\bf [Deutsch 85 b]}:
D. Deutsch,
``Quantum theory, the Church-Turing principle
and the universal quantum computer'',
{\em Proc. R. Soc. Lond. A} {\bf 400}, 1818, 97-117 (1985).

\item {\bf [Deutsch 86]}:
D. Deutsch,
``Three connections between Everett's interpretation and experiment'',
in R. Penrose, \& C. Isham (eds.),
{\em Quantum concepts in space and time},
Clarendon Press, Oxford, 1986, pp.~?-?.

\item {\bf [Deutsch 88]}:
D. Deutsch,
`On Wheeler's notion of ``Law without law'' in physics',
in {\bf [Zurek-van der Merwe-Miller 88]}, pp.~583-590.

\item {\bf [Deutsch 89 a]}:
D. Deutsch,
``Quantum computational networks'',
{\em Proc. R. Soc. Lond. A} {\bf 425}, 1868, 73-90 (1989).

\item {\bf [Deutsch 89 b]}:
D. Deutsch,
``?'',
{\em New Scientist} {\bf 124}, 1694, 25 (1989).

\item {\bf [Deutsch 92]}:
D. Deutsch,
``Quantum computation'',
{\em Phys. World} {\bf 5}, 6, 57-61 (1993).

\item {\bf [Deutsch-Jozsa 92]}:
D. Deutsch, \& R. Jozsa,
``Rapid solution of problems by quantum computation'',
{\em Proc. R. Soc. Lond. A} {\bf 439}, 1907, 553-558 (1992).

\item {\bf [Deutsch-Ekert 93]}:
D. Deutsch, \& A. K. Ekert,
``Quantum communication moves into the unknown'',
{\em Phys. World} {\bf 6}, 6, 22-23 (1993).

\item {\bf [Deutsch-Lockwood 93]}:
D. Deutsch, \& M. Lockwood,
``The quantum physics of time travel'',
{\em Sci. Am.} {\bf 270}, 4, 68-? (1994).
Spanish version: ``F\'{\i}sica cu\'{a}ntica de los viajes a
trav\'{e}s del tiempo'',
{\em Investigaci\'{o}n y Ciencia} 212, 48-54 (1994).

\item {\bf [Deutsch-Barenco-Ekert 95]}:
D. Deutsch, A. Barenco, \& A. K. Ekert,
``Universality in quantum computation'',
{\em Proc. R. Soc. Lond. A} {\bf 449}, 1937, 669-677 (1995).

\item {\bf [Deutsch 96]}:
D. Deutsch,
``Comment on `The many minds interpretation of quantum mechanics'\,'',
{\em Brit. J. Philos. Sci.} {\bf 47}, 2, 222-? (1996).
Comment on {\bf [Lockwood 96]}.

\item {\bf [Deutsch-Ekert-Jozsa-(+3) 96]}:
D. Deutsch, A. K. Ekert, R. Jozsa, C. Macchiavello,
S. Popescu, \& A. Sanpera,
``Quantum privacy amplification and the security of quantum
cryptography over noisy channels'',
{\em Phys. Rev. Lett.} {\bf 77}, 13, 2818-2821 (1996).
Erratum: {\em Phys. Rev. Lett.} {\bf 80}, 9, 2022 (1998);
quant-ph/9604039.
Reprinted in {\bf [Macchiavello-Palma-Zeilinger 00]}, pp.~225-228.

\item {\bf [Deutsch 97]}:
D. Deutsch,
{\em The fabric of reality: The science of
parallel universes and its implications},
Allen Lane, The Penguin Press, London, 1997.
Spanish version: {\em La estructura de la realidad},
Anagrama, Barcelona, 1999.

\item {\bf [Deutsch-Ekert 98]}:
D. Deutsch, \& A. K. Ekert,
``Quantum computation'',
{\em Phys. World} {\bf 11}, 3, 53-57 (1998).

\item {\bf [Deutsch 99]}:
D. Deutsch,
``Quantum theory of probability and decisions'',
{\em Proc. R. Soc. Lond. A} {\bf 455}, 1988, 3129-3137 (1999);
quant-ph/9906015.
See {\bf [Finkelstein 99 c]},
{\bf [Polley 99]},
{\bf [Summhammer 99]},
{\bf [Barnum-Caves-Finkelstein-(+2) 00]}.

\item {\bf [Deutsch-Ekert-Lupacchini 00]}:
D. Deutsch, A. K. Ekert, \& R. Lupacchini,
``Machines, logic and quantum physics'',
{\em Bull. Symbolic Logic} {\bf 6}, 3, 265-283 (2000);
math-HO/9911150.

\item {\bf [Deutsch-Hayden 00]}:
D. Deutsch, \& P. Hayden,
``Information flow in entangled quantum systems'',
{\em Proc. R. Soc. Lond. A} {\bf 456}, 1999, 1759-1774 (2000);
quant-ph/9906007.
Comment: {\bf [Schafir 02 a]}.

\item {\bf [Deutsch 01]}:
D. Deutsch,
``The structure of the multiverse'',
quant-ph/0104033.

\item {\bf [Deutsch 04]}:
D. Deutsch,
``Qubit field theory'',
quant-ph/0401024.

\item {\bf [Deutsch-Brennen-Jessen 00]}:
I. H. Deutsch, G. K. Brennen, \& P. S. Jessen,
``Quantum computing with neutral atoms in an optical lattice'',
{\em Fortschr. Phys.} {\bf 48}, 9-11 (Special issue:
Experimental proposals for quantum computation), 925-943 (2000);
quant-ph/0003022.

\item {\bf [Devetak-Berger 01]}:
I. Devetak, \& T. Berger,
``Low-entanglement remote state preparation'',
{\em Phys. Rev. Lett.} {\bf 87}, 19, 197901 (2001);
quant-ph/0102123.

\item {\bf [Devetak-Winter 03 a]}:
I. Devetak, \& A. Winter,
``Classical data compression with quantum side information'',
{\em Phys. Rev. A} {\bf 68}, 4, 042301 (2003);
quant-ph/0209029.

\item {\bf [Devetak-Winter 03 b]}:
I. Devetak, \& A. Winter,
``Distilling common randomness from bipartite quantum states'',
quant-ph/0304196.

\item {\bf [Devetak-Winter 03 c]}:
I. Devetak, \& A. Winter,
``Distillation of secret key and entanglement from quantum states'',
quant-ph/0306078.

\item {\bf [Devetak-Winter 03 d]}:
I. Devetak, \& A. Winter,
``Relating quantum privacy and quantum coherence: An operational approach'',
quant-ph/0307053.

\item {\bf [Devetak-Winter 03 e]}:
I. Devetak, \& P. W. Shor,
``The capacity of a quantum channel for simultaneous transmission of
classical and quantum information'',
quant-ph/0311131.

\item {\bf [Devetak 03]}:
I. Devetak,
``The private classical capacity and quantum capacity of a quantum channel'',
quant-ph/0304127.

\item {\bf [Devetak 04]}:
I. Devetak,
``Distillation of local purity from quantum states'',
quant-ph/0406234.

\item {\bf [Devi-Sirsi-Ramachandran 97]}:
A. R. U. Devi, S. Sirsi, \& G. Ramachandran,
``Nonlocality in Einstein-Podolsky-Rosen spin correlations'',
{\em Int. J. Mod. Phys. A} {\bf 12}, 29, 5279-5288 (1997).

\item {\bf [Devi 00 a]}:
A. R. U. Devi,
``Normalized Braunstein-Caves inequalities'',
{\em J. Phys. A} {\bf 33}, 1, 227-232 (2000).

\item {\bf [Devi 00 b]}:
A. R. U. Devi,
``Spin distributions for bipartite quantum systems'',
quant-ph/0009092.

\item {\bf [DeVoe 98]}:
R. G. DeVoe,
``Elliptical ion traps and trap arrays for quantum computation'',
{\em Phys. Rev. A} {\bf 58}, 2, 910-914 (1998).

\item {\bf [Dewdney-Hiley 82]}:
C. Dewdney, \& B. J. Hiley,
``A quantum potential description of one-dimensional time-dependent
scattering from square barriers and square wells'',
{\em Found. Phys.} {\bf 12}, 1, 27-48 (1982).

\item {\bf [Dewdney-Holland-Kyprianidis 86]}:
C. Dewdney, P. R. Holland, \& A. Kyprianidis,
``What happens in a spin measurement?'',
{\em Phys. Lett. A} {\bf 119}, 6, 259-267 (1986).

\item {\bf [Dewdney 87]}:
C. Dewdney,
``Calculations in the causal interpretation of quantum mechanics'',
in W. M. Honig, D. W. Kraft, \& E. Panarella,
{\em Quantum uncertainties:
Recent and future experiments and interpretations},
Plenum Press, New York, 1987, pp.~19-40.

\item {\bf [Dewdney-Holland-Kyprianidis 87]}:
C. Dewdney, P. R. Holland, \& A. Kyprianidis,
``A causal account of non-local Einstein-Podolsky-Rosen spin correlations'',
{\em J. Phys. A} {\bf 20}, 14, 4717-4732 (1987).

\item {\bf [Dewdney 88]}:
C. Dewdney,
``The quantum potential approach to neutron interferometry experiments'',
{\em Physica B} {\bf 151}, ?, 160-170 (1988).

\item {\bf [Dewdney-Holland 88]}:
C. Dewdney, \& P. R. Holland,
``Particle trajectories and quantum correlations'',
in F. Selleri (ed.),
{\em Quantum mechanics
versus local realism: The Einstein-Podolsky-Rosen paradox},
Plenum Press, New York, 1988, pp.~301-325.

\item {\bf [Dewdney-Holland-Kyprianidis-Vigier 88]}:
C. Dewdney, P. R.
Holland, A. Kyprianidis, \& J.-P. Vigier,
``Spin and non-locality in quantum mechanics'',
{\em Nature} {\bf 336}, 6199, 536-544 (1988).

\item {\bf [Dewdney 92]}:
C. Dewdney,
``Constraints on quantum hidden-variables and the Bohm theory'',
{\em J. Phys. A} {\bf 25}, 12, 3615-3626 (1992).
See {\bf [Dewdney 93]}.

\item {\bf [Dewdney 93]}:
C. Dewdney,
``Constraints on quantum hidden variables and the Bohm theory'',
in A. van der Merwe, \& F. Selleri (eds.),
{\em Bell's theorem and the foundations of modern physics.
Proc.\ of an international conference (Cesena, Italy, 1991)},
World Scientific, Singapore, 1993, pp.~147-160.
See {\bf [Dewdney 92]}.

\item {\bf [Dewdney-Malik 93]}:
C. Dewdney, \& Z. Malik,
``Angular-momentum
measurement and nonlocality in Bohm's interpretation of quantum theory'',
{\em Phys. Rev. A} {\bf 48}, 5, 3513-3524 (1993).

\item {\bf [Dewdney-Hardy-Squires 93]}:
C. Dewdney, L. Hardy, \& E. J. Squires,
``How late measurements of quantum trajectories can fool a detector'',
{\em Phys. Lett. A} {\bf 184}, 1, 6-11 (1993).

\item {\bf [Dewdney-Horton 02]}:
C. Dewdney, \& G. Horton,
``Relativistically invariant extension of
the de Broglie-Bohm theory of quantum mechanics'',
quant-ph/0202104.

\item {\bf [DeWeerd 02]}:
A. J. DeWeerd,
``Interaction-free measurement'',
{\em Am. J. Phys.} {\bf 70}, 3, 272-275 (2002).

\item {\bf [DeWitt 68]}:
B. S. DeWitt,
``The Everett-Wheeler interpretation of quantum mechanics'',
in C. M. DeWitt, \& J. A. Wheeler (eds.),
{\em Battelle rencontres, 1967 lectures in mathematics and physics},
Benjamin, New York, 1968.

\item {\bf [DeWitt 70]}:
B. S. DeWitt,
``Quantum mechanics and reality'',
{\em Phys. Today} {\bf 23}, 9, 30-35 (1970).
Reprinted in {\bf [DeWitt-Graham 73]}, pp.~155-165.

\item {\bf [DeWitt 71 a]}:
B. S. DeWitt,
``?'',
{\em Phys. Today} {\bf 24}, 4, 36-? (1971).

\item {\bf [DeWitt 71 b]}:
B. S. DeWitt,
``The many-universes interpretation of quantum mechanics'',
in {\bf [d'Espagnat 71]}, pp.~211-262.
Reprinted in {\bf [DeWitt-Graham 73]}, pp.~167-218.

\item {\bf [DeWitt-Graham 71]}:
B. S. DeWitt, \& R. N. Graham,
``Resource letter IQM-1 on the interpretation of quantum mechanics'',
{\em Am. J. Phys.} {\bf 39}, 7, 724-738 (1971).
See {\bf [Ballentine 87 a]} (II).

\item {\bf [DeWitt-Graham 73]}:
B. S. DeWitt, \& R. N. Graham (eds.),
{\em The many-worlds interpretation of quantum mechanics},
Princeton University Press, Princeton, New Jersey, 1973.

\item {\bf [DeWitt 98]}:
B. S. DeWitt,
``The quantum mechanics of isolated systems'',
{\em Int. J. Mod. Phys. A} {\bf 13}, 12, 1881-1916 (1998).

\item {\bf [Di Giuseppe-De Martini-Boschi 97]}:
G. Di Giuseppe, F. De Martini, \& D. Boschi,
``Experimental test of the violation of local realism in quantum mechanics
without Bell inequalities'',
{\em Phys. Rev. A} {\bf 56}, 1, 176-181 (1997).
Almost the same as {\bf [Boschi-De Martini-Di Giuseppe 97]}.
See {\bf [Torgerson-Branning-Monchen-Mandel 95]}.

\item {\bf [Di Giuseppe-De Martini-Boschi-Branca 98]}:
G. Di Giuseppe, F. De Martini, D. Boschi, S. Branca,
``Entangled non-local quantum interferometry'',
{\em Fortschr. Phys.} {\bf 46}, 6-8, 643-661 (1998).

\item {\bf [Di Giuseppe-Atat\"{u}re-Shaw-(+3) 01]}:
G. Di Giuseppe, M. Atat\"{u}re, M. D. Shaw,
A. V. Sergienko, B. E. A. Saleh, \& M. C. Teich,
``Entangled-photon generation from parametric down-conversion
in media with inhomogeneous nonlinearity'',
quant-ph/0112140.

\item {\bf [Di Lisi-M{\o}lmer 02]}:
A. Di Lisi, \& K. M{\o}lmer,
``Entanglement of two atomic samples by quantum-nondemolition measurements'',
{\em Phys. Rev. A} {\bf 66}, 5, 052303 (2002).

\item {\bf [Di Lisi-De Siena-Illuminati-Vitali 04]}:
A. Di Lisi, S. De Siena, F. Illuminati, \& D. Vitali,
``Efficient and robust generation of maximally entangled states of two
atomic ensembles by adiabatic quantum feedback'',
quant-ph/0407254.

\item {\bf [Di Lisi-De Siena-Illuminati 04]}:
A. Di Lisi, S. De Siena, \& F. Illuminati,
``Dynamics of entanglement between two atomic samples with spontaneous scattering'',
{\em Phys. Rev. A} {\bf 70}, 1, 012301 (2004).

\item {\bf [Di Nepi-De Martini-Barbieri-Mataloni 03]}:
G. Di Nepi, F. De Martini, M. Barbieri, \& P. Mataloni,
``Universal ultrabright source of entangled photon states: Generation and
tomographic analysis of Werner states and of maximally entangled mixed states'',
quant-ph/0307204.

\item {\bf [Di Rienzi 01]}:
J. Di Rienzi,
``Quantum contradictions in the context of
special relativity'',
{\em Phys. Essays} {\bf 14}, 2, 149-153 (2001).

\item {\bf [Dicke 81]}:
R. H. Dicke,
``Interaction-free quantum measurements: A paradox?'',
{\em Am. J. Phys.} {\bf 49}, 10, 925-930 (1981).

\item {\bf [Dicke 86]}:
R. H. Dicke,
``On observing the absence of an atom'',
{\em Found. Phys.} {\bf 16}, 2, 107-113 (1986).

\item {\bf [Dickson 93]}:
W. M. Dickson,
``Stapp's theorem without counterfactual
commitments: Why it fails nonetheless'',
{\em Stud. Hist. Philos. Sci.} {\bf 24}, 5, 791-814 (1993).
See {\bf [Stapp 90]}.

\item {\bf [Dickson 94 a]}:
W. M. Dickson,
``Wavefunction tails in the modal interpretation'',
in D. Hull, M. Forbes, \& R. Burian (eds.),
{\em Proc.\ of the 1994
Biennial Meeting of the Philosophy of Science Association},
East Lansing, Michigan, 1994, vol. 1, pp.~366-376.

\item {\bf [Dickson 94 b]}:
W. M. Dickson,
``What is preferred about the preferred basis?'',
{\em Found. Phys.} {\bf 25}, 3, 423-441 (1994).

\item {\bf [Dickson-Clifton 94]}:
W. M. Dickson, \& R. K. Clifton,
``Stapp's algebraic argument for nonlocality'',
{\em Phys. Rev. A} {\bf 49}, 5, Part B, 4251-4256 (1994).
See {\bf [Stapp 94 b]}.

\item {\bf [Dickson 95 a]}:
W. M. Dickson,
``Is there really no projection postulate in the modal interpretation?'',
{\em Brit. J. Philos. Sci.} {\bf 46}, ?, 167-188 (1995).

\item {\bf [Dickson 95 b]}:
W. M. Dickson,
``Probability and nonlocality:
Determinism versus indeterminism in quantum mechanic'',
Ph.\ D. thesis, University of Notre Dame, Indiana, 1995.

\item {\bf [Dickson 96 a]}:
W. M. Dickson,
``Determinism and locality in quantum systems'',
{\em Synthese} {\bf 107}, ?, 52-82 (1996).

\item {\bf [Dickson 96 b]}:
W. M. Dickson,
``Logical foundations for modal interpretations'',
{\em Philos. Sci.} {\bf 63} (Supp.), ?, 322-329 (1996).

\item {\bf [Dickson 96 c]}:
W. M. Dickson,
``Antidote or theory?'',
{\em Stud. Hist. Philos. Sci. Part B: Stud. Hist. Philos. Mod. Phys.}
{\bf 27}, 2, 229-238 (1996).
Review of {\bf [Bohm-Hiley 93]} and {\bf [Holland 93]}.

\item {\bf [Dickson 96 d]}:
W. M. Dickson,
``Is the Bohm theory local?'',
in {\bf [Cushing-Fine-Goldstein 96]}, pp.~321-330 (1996).

\item {\bf [Dickson 98]}:
W. M. Dickson,
{\em Quantum chance and nonlocality:
Probability and nonlocality in the interpretation of quantum mechanics},
Cambridge University Press, Cambridge, 1998.
Reviews: {\bf [Cabello 99 b]}, {\bf [Healey 99]},
{\bf [Barrett 99 b]}, {\bf [Ghirardi 00 a]}.

\item {\bf [Dickson-Clifton 98]}:
W. M. Dickson, \& R. K. Clifton,
``Lorentz-invariance in modal interpretations'',
{\bf [Dieks-Vermaas 98]}, pp.~?-?.

\item {\bf [Dickson 00]}:
W. M. Dickson,
``Discussion: Are there material objects in Bohm's theory'',
quant-ph/0003102.
Reply to {\bf [Bedard 99]}.

\item {\bf [Dickson 01]}:
W. M. Dickson,
``The EPR experiment: A prelude to Bohr's reply to EPR'',
to appear in the {\em Selected Archive of HOPOS 2000},
quant-ph/0102053.

\item {\bf [Diddams-Udem-Bergquist-(+8) 01]}:
S. A. Diddams, T. Udem, J. C. Bergquist,
E. A. Curtis, R. E. Drullinger, L. Hollberg,
W. M. Itano, W. D. Lee, C. W. Oates,
K. R. Vogel, \& D. J. Wineland,
``An optical clock based on a single trapped ${}^{199}$Hg$^+$ ion'',
{\em Science} {\bf 293}, 5531, 825-828 (2001).

\item {\bf [Dieks 82]}:
D. Dieks,
``Communication by EPR devices'',
{\em Phys. Lett. A} {\bf 92}, 6, 271-272 (1982).

\item {\bf [Dieks 88]}:
D. Dieks,
``Overlap and distinguishability of quantum states'',
{\em Phys. Lett. A} {\bf 126}, 5-6, 303-306 (1987).
See {\bf [Ivanovic 87]}, {\bf [Peres 88 b]}.

\item {\bf [Dieks 89]}:
D. Dieks,
``Resolution of the measurement problem through
decoherence of the quantum state'',
{\em Phys. Lett. A} {\bf 142}, 8-9, 439-446 (1989).

\item {\bf [Dieks 91]}:
D. Dieks,
``On some alleged difficulties in interpretation of quantum mechanics'',
{\em Synthese} {\bf 86}, ?, 77-86 (1991).

\item {\bf [Dieks 94]}:
D. Dieks,
``Modal interpretation of quantum mechanics,
measurements, and macroscopic behavior'',
{\em Phys. Rev. A} {\bf 49}, 4, 2290-2300 (1994).
See {\bf [Elby 93 a]}.

\item {\bf [Dieks 95]}:
D. Dieks,
``Physical motivation of the modal interpretation of quantum mechanics'',
{\em Phys. Lett. A} {\bf 197}, 5-6, 367-371 (1995).

\item {\bf [Dieks-Vermaas 98]}:
D. Deks, \& P. Vermaas (eds.),
{\em The modal interpretation of quantum mechanics},
Kluwer Academic, Dordrecht, Holland, 1998.

\item {\bf [Dieks 99]}:
D. Dieks,
``Experimental metaphysics'',
{\em Stud. Hist. Philos. Mod. Phys.} {\bf 30}, 4, 555-560 (1999).
Review of {\bf [Cohen-Horne-Stachel 97 a]}.

\item {\bf [Dieks 00]}:
D. Dieks,
``Consistent histories and relativistic invariance
in the modal interpretation of quantum mechanics'',
{\em Phys. Lett. A} {\bf 265}, 5-6, 317-325 (2000);
quant-ph/9912075.

\item {\bf [Dieks 02 a]}:
D. Dieks,
``Events and covariance in the interpretation of quantum field theory'' (2002),
PITT-PHIL-SCI00000554.

\item {\bf [Dieks 02 b]}:
D. Dieks,
``Inequalities that test locality in quantum mechanics'',
{\em Phys. Rev. A} {\bf 66}, 6, 062104 (2002);
quant-ph/0206172.

\item {\bf [Dieks 02 b]}:
D. Dieks,
``Consistent quantum theory'',
{\em Math. Rev.} 2003:g, 81001, 1 (2003).
Review of {\bf [Griffiths 01]}.

\item {\bf [Dieks 03]}:
D. Dieks,
``Foundations of quantum mechanics, an empiricist approach'',
{\em Found. Phys.} {\bf 33}, 6, 1003-1006 (2003).
Report: {\bf [de Muynck 02]}.

\item {\bf [Ding-McDowell-Ye-(+6) 01]}:
S. Ding, C. A. McDowell, C. Ye,
M. Zhan, X. Zhu, K. Gao, X. Sun, X.-A. Mao, \& M. Liu,
``Quantum computation based on magic-angle-spinning solid
state nuclear magnetic resonance spectroscopy'',
{\em Eur. Phys. J. B};
quant-ph/0110014.

\item {\bf [Di\'{o}si 88]}:
L. Di\'{o}si,
``Quantum stochastic processes a models for state vector reduction'',
{\em J. Phys. A} {\bf 21}, 13, 2885-2898 (1988).

\item {\bf [Di\'{o}si-Gisin-Halliwell-Percival 95]}:
L. Di\'{o}si, N. Gisin, J. Halliwell, \& I. C. Percival,
``Decoherent histories and quantum state diffusion'',
{\em Phys. Rev. Lett.} {\bf 74}, 2, 203-207 (1995).

\item {\bf [Di\'{o}si-Halliwell 98]}:
L. Di\'{o}si, \& J. Halliwell,
``Coupling classical and quantum variables using
continuous quantum measurement theory'',
{\em Phys. Rev. Lett.} {\bf 81}, 14, 2846-2849 (1998).

\item {\bf [Di\'{o}si-Gisin-Strunz 00]}:
L. Di\'{o}si, N. Gisin, \& W. T. Strunz,
``Quantum approach to coupling classical and quantum dynamics'',
{\em Phys. Rev. A} {\bf 61}, 2, 022108 (2000).

\item {\bf [Di\'{o}si 00 a]}:
L. Di\'{o}si,
``Comment on `Nonclassical states: An observable criterion'\,'',
{\em Phys. Rev. Lett.} {\bf 85}, 13, 2841 (2000).
Comment on {\bf [Vogel 00 a]}.
Reply: {\bf [Vogel 00 b]}.

\item {\bf [Di\'{o}si-Kiefer 00 b]}:
L. Di\'{o}si, \& C. Kiefer,
``Robustness and diffusion of pointer states'',
{\em Phys. Rev. Lett.} {\bf 85}, 17, 3552-3555 (2000);
quant-ph/0005071.

\item {\bf [Di\'{o}si 01]}:
L. Di\'{o}si,
``Comment on `Quantum-anti-Zeno paradox'\,'',
quant-ph/0104029.
Comment on {\bf [Balachandran-Roy 00]}.

\item {\bf [Di\'{o}si 02 a]}:
L. Di\'{o}si,
`Comment on ``Stable quantum computation of unstable classical chaos''\,',
{\em Phys. Rev. Lett.} {\bf 88}, 21, 219801 (2002);
quant-ph/0110026.
Comment on {\bf [Georgeot-Shepelyansky 01 b]}.
Reply: {\bf [Georgeot-Shepelyansky 02]}.
See {\bf [Zalka 01]}.

\item {\bf [Di\'{o}si 02 b]}:
L. Di\'{o}si,
``Advancement of estimation fidelity in continuous quantum measurement'',
quant-ph/0202052.

\item {\bf [Di\'{o}si 02 c]}:
L. Di\'{o}si,
``Wigner Centennial: His function, and its environmental decoherence'',
talk at {\em Wigner Centennial Conf. (Pecs, Hungary, 2002)},
quant-ph/0212103.

\item {\bf [Di\'{o}si 03]}:
L. Di\'{o}si,
``Anomalies of weakened decoherence criteria for quantum histories'',
quant-ph/0310181.

\item {\bf [Di\'{o}si 04]}:
L. Di\'{o}si,
``Three-party pure quantum states are determined by two two-party reduced
states'',
{\em Phys. Rev. A} {\bf 70}, 1, 010302 (2004);
quant-ph/0403200.

\item {\bf [Dirac 30]}:
P. A. M. Dirac,
{\em The principles of quantum mechanics},
Oxford University Press, Oxford, 1930 (1st edition), 1958 (4th edition).
Spanish version: {\em Principios de mec\'{a}nica cu\'{a}ntica},
Ariel, Barcelona, 1968.

\item {\bf [Dirac 37]}:
P. A. M. Dirac,
``Physical science and philosophy'',
{\em Nature} {\bf 139}, ? (suppl.), 1001-1002 (1937).

\item {\bf [Dirac 39]}:
P. A. M. Dirac,
``A new notation for quantum mechanics'',
{\em Proc.\ Cambridge Philos. Soc.} {\bf 35} part 3, 416-418 (1939).
Reprinted in {\bf [Dirac 95]}, pp.~915-919.
See {\bf [Gieres 99]}.

\item {\bf [Dirac 41]}:
P. A. M. Dirac,
``The physical interpretation of quantum mechanics'',
{\em Proc. R. Soc. Lond. A} {\bf 180}, 1, 1-40 (1941).
Reprinted in {\bf [Dirac 95]}.

\item {\bf [Dirac 63]}:
P. A. M. Dirac,
``The evolution of the physicist's picture of nature'',
{\em Sci. Am.} {\bf 208}, 5, 45-53 (1963).
Spanish version: ``La concepci\'{o}n f\'{\i}sica de la naturaleza'',
in {\bf [Cabello 97 c]}, pp.~4-12.

\item {\bf [Dirac 64]}:
P. A. M. Dirac,
``Foundations of quantum mechanics'',
{\em Nature} {\bf 203}, 4941, 115-116 (1964).

\item {\bf [Dirac 71]}:
P. A. M. Dirac,
``The development of quantum theory'',
J. Robert Oppenheimer Memorial Prize acceptance speech,
Gordon and Breach, New York, 1971.

\item {\bf [Dirac 95]}:
P. A. M. Dirac,
{\em The collected works of P. A. M. Dirac, 1924-1948},
R. H. Dalitz (ed.),
Cambridge University Press, New York, 1995.
Review: {\bf [Hovis 96]}.

\item {\bf [Divakaran 97]}:
P. P. Divakaran,
``Quantum theory as the representation theory of symmetries'',
{\em Phys. Rev. Lett.} {\bf 79}, 12, 2159-2163 (1997).

\item {\bf [DiVincenzo 95 a]}:
D. P. DiVincenzo,
``Quantum computation'',
{\em Science} {\bf 270}, 5234, 255-261 (1995).

\item {\bf [DiVincenzo 95 b]}:
D. P. DiVincenzo,
``Two-bit gates are universal for quantum computation'',
{\em Phys. Rev. A} {\bf 51}, 2, 1015-1022 (1995);
cond-mat/9407022.

\item {\bf [DiVincenzo 96 a]}:
D. P. DiVincenzo,
``Quantum computation and spin physics'',
{\em Proc.\ of the Annual MMM Meeting (1996)},
to be published in {\em J. Appl. Phys.};
cond-mat/9612125.

\item {\bf [DiVincenzo 96 b]}:
D. P. DiVincenzo,
``Topics in quantum computers'',
to be published in
 L. Kowenhoven, G. Schoen, \& L. Sohn (eds.),
{\em Mesoscopic electron transport},
NATO ASI Series E,
Kluwer Academic, Dordrecht, Holland;
cond-mat/9612126.

\item {\bf [DiVincenzo-Shor 96]}:
D. P. DiVincenzo, \& P. W. Shor,
``Fault-tolerant error correction with efficient quantum codes'',
{\em Phys. Rev. Lett.} {\bf 77}, 15, 3260-3263 (1996).
Reprinted in {\bf [Macchiavello-Palma-Zeilinger 00]}, pp.~150-153.

\item {\bf [DiVincenzo-Peres 97]}:
D. P. DiVincenzo, \& A. Peres,
``Quantum codewords contradict local realism'',
{\em Phys. Rev. A} {\bf 55}, 6, 4089-4092 (1997);
quant-ph/9611011.

\item {\bf [DiVincenzo-Loss 98 a]}:
D. P. DiVincenzo, \& D. Loss,
``Quantum information is physical'',
{\em Superlatt. Microstruct.} {\bf 23}, ?, 419-432 (1998).
cond-mat/9710259.

\item {\bf [DiVincenzo-Shor-Smolin 98]}:
D. P. DiVincenzo, P. W. Shor, \& J. A. Smolin,
``Quantum-channel capacity of very noisy channels'',
{\em Phys. Rev. A} {\bf 57}, 2, 830-839 (1998).
Erratum: {\em Phys. Rev. A} {\bf 59}, 2, 1717 (1999).

\item {\bf [DiVincenzo-Fuchs-Mabuchi-(+3) 98]}:
D. P. DiVincenzo, C. A. Fuchs, H. Mabuchi, J. A. Smolin,
A. Thapliyal, \& A. Uhlmann,
``Entanglement of assistance'',
quant-ph/9803033.

\item {\bf [DiVincenzo-Terhal 98]}:
D. P. DiVincenzo, \& B. Terhal,
``Decoherence: The obstacle to quantum computation'',
{\em Phys. World} {\bf 11}, 3, 53-57 (1998).

\item {\bf [DiVincenzo 98 a]}:
D. P. DiVincenzo,
``Real and realistic quantum computers'',
{\em Nature} {\bf 393}, 6681, 113-114 (1998).

\item {\bf [DiVincenzo 98 b]}:
D. P. DiVincenzo,
``Quantum gates and circuits'',
in D. P. DiVincenzo. E. Knill, R. Laflamme, \& W. H. Zurek (eds.),
{\em Quantum Coherence and Decoherence.
Proc.\ of the ITP Conf.\ (Santa Barbara, California, 1996)},
{\em Proc. R. Soc. Lond. A} {\bf 454}, 1969, 261-276 (1998).

\item {\bf [DiVincenzo-Loss 99]}:
D. P. DiVincenzo, \& D. Loss,
``Quantum computers and quantum coherence'',
{\em J. Mag. Magn. Matl.} {\bf 200}, ?, 202-? (1999);
cond-mat/9901137.
Reprinted in {\bf [Macchiavello-Palma-Zeilinger 00]}, pp.~449-462.

\item {\bf [DiVincenzo 99 a]}:
D. P. DiVincenzo,
``Quantum computing and single-qubit measurements
using the spin filter effect'',
{\em J. Appl. Phys.} {\bf 85}, ?, 4785-? (1999);
cond-mat/9810295.

\item {\bf [DiVincenzo 99 b]}:
D. P. DiVincenzo,
``Quantum behaviour'',
{\em Nature} {\bf 399}, 6732, 119-120 (1999).
Review of {\bf [Lo-Spiller-Popescu 98]}.

\item {\bf [DiVincenzo-Mor-Shor-(+2) 99]}:
D. P. DiVincenzo, T. Mor, P. W. Shor,
J. A. Smolin, \& B. M. Terhal,
``Unextendible product bases, uncompletable product bases and bound
entanglement'',
{\em Comm. Math. Phys.};
quant-ph/9908070.

\item {\bf [DiVincenzo-Burkard-Loss-Sukhorukov 99]}:
D. P. DiVincenzo, G. Burkard, D. Loss, \& E. V. Sukhorukov,
``Quantum computation and spin electronics'',
to be published in I. O. Kulik, \& R. Ellialtioglu (eds.),
{\em Quantum Mesoscopic Phenomena and Mesoscopic
Devices in Microelectronics (Turkey, 1999)};
cond-mat/9911245.

\item {\bf [DiVincenzo-Shor-Smolin-(+2) 00]}:
D. P. DiVincenzo, P. W. Shor, J. A. Smolin,
B. M. Terhal, \& V. A. Thapliyal,
``Evidence for bound entangled states with negative partial
transpose'',
{\em Phys. Rev. A} {\bf 61}, 6, 062312 (2000);
quant-ph/9910026.

\item {\bf [DiVincenzo-Terhal-Thapliyal 00]}:
D. P. DiVincenzo, B. M. Terhal, \& V. A. Thapliyal,
``Optimal decompositions of barely separable states'',
in V. Bu\v{z}zek, \& D. P. DiVincenzo (eds.),
{\em J. Mod. Opt.} {\bf 47}, 2-3 (Special issue:
Physics of quantum information), 377-385 (2000).

\item {\bf [DiVincenzo 00 a]}:
D. P. DiVincenzo,
``The physical implementation of quantum computation'',
{\em Fortschr. Phys.} {\bf 48}, 9-11 (Special issue:
Experimental proposals for quantum computation), 771-783 (2000);
quant-ph/0002077.

\item {\bf [DiVincenzo 00 b]}:
D. P. DiVincenzo,
``?'',
{\em Quant. Inf. Theor.} {\bf 1}, 1, 1-? (2000).

\item {\bf [DiVincenzo 00 c]}:
D. P. DiVincenzo,
``Quantum error correction'',
{\bf [Macchiavello-Palma-Zeilinger 00]}, pp.~131-133.

\item {\bf [DiVincenzo-Bacon-Kempe-(+2) 00]}:
D. P. DiVincenzo, D. Bacon, J. Kempe, G. Burkard, \& K. B. Whaley,
``Universal quantum computation with the exchange interaction'',
{\em Nature} {\bf 408}, 6810, 339-342 (2000);
quant-ph/0005116.

\item {\bf [DiVincenzo-Terhal 00]}:
D. P. DiVincenzo, \& B. M. Terhal,
``Product bases in quantum information theory'',
submitted to
{\em Proc.\ of the XIII Int.\ Congress on Mathematical Physics};
quant-ph/0008055.

\item {\bf [DiVincenzo-Leung-Terhal 02]}:
D. P. DiVincenzo, D. W. Leung, \& B. M. Terhal,
``Quantum data hiding'',
{\em IEEE Trans. Inf. Theory} {\bf 48}, 3, 580-598 (2002);
quant-ph/0103098.
See {\bf [Terhal-DiVincenzo-Leung 01]}.

\item {\bf [DiVincenzo-Hayden-Terhal 03]}:
D. P. DiVincenzo, P. Hayden, \& B. M. Terhal,
``Hiding quantum data'',
{\em Found. Phys.} {\bf 33}, 11, 1629-1647 (2003).

\item {\bf [DiVincenzo-Horodecki-Leung-(+2) 04]}:
D. P. DiVincenzo, M. Horodecki, D. W. Leung,
J. A. Smolin, \& B. M. Terhal,
``Locking classical correlations in quantum states'',
{\em Phys. Rev. Lett.} {\bf 92}, 6, 067902 (2004);
quant-ph/0303088.

\item {\bf [DiVincenzo-Terhal 04]}:
D. P. DiVincenzo, \& B. M. Terhal,
``Fermionic linear optics revisited'',
{\em Found. Phys.},
quant-ph/0403031.

\item {\bf [DiVincenzo-Smolin-Terhal 04]}:
 D. P. DiVincenzo, J. A. Smolin, \& B. M. Terhal,
``Security trade-offs in ancilla-free quantum bit commitment in the
presence of selection rules'',
{\em New J. Phys.};
quant-ph/0405111.

\item {\bf [Dobrovitski-De Raedt 03]}:
V. V. Dobrovitski, \& H. A. De Raedt,
``Efficient scheme for numerical simulations of the spin-bath decoherence'',
{\em Phys. Rev. E} {\bf 67}, 5, 056702 (2003).

\item {\bf [Dodd-Nielsen-Bremner-Thew 02]}:
J. L. Dodd, M. A. Nielsen, M. J. Bremner, \& R. T. Thew,
``Universal quantum computation and simulation using any entangling
Hamiltonian and local unitaries'',
{\em Phys. Rev. A} {\bf 65}, 4, 040301 (2002);
quant-ph/0106064.

\item {\bf [Dodd-Nielsen 02]}:
J. L. Dodd, \& M. A. Nielsen,
``Simple operational interpretation of the fidelity of mixed states'',
{\em Phys. Rev. A} {\bf 66}, 4, 044301 (2002);
quant-ph/0111053.

\item {\bf [Dodd-Ralph-Milburn 03]}:
J. L. Dodd, T. C. Ralph, \& G. J. Milburn,
``Experimental requirements for Grover's algorithm in optical quantum
computation'',
quant-ph/0306081.

\item {\bf [Dodd-Halliwell 03 a]}:
P. J. Dodd, \& J. J. Halliwell,
``Decoherence and records for the case of a scattering environment'',
quant-ph/0301104.

\item {\bf [Dodd-Halliwell 03 b]}:
P. J. Dodd, \& J. J. Halliwell,
``Disentanglement and decoherence by open system dynamics'',
quant-ph/0312068.

\item {\bf [Dodonov-Man'ko-Rudenko 80]}:
V. V. Dodonov, V. I. Man'ko, \& V. N. Rudenko,
``?'',
{\em Zhurn. Exper. Teor. Fiz.} {\bf 78}, 3, 881-896 (1980).
English version:
``Nondemolition measurements in gravity wave experiments'',
{\em Sov. Phys. JETP} {\bf 51}, 3, 443-450 (1980).

\item {\bf [Dodonov-Man'ko-Rudenko 83]}:
V. V. Dodonov, V. I. Man'ko, \& V. N. Rudenko,
``Quantum integrals of motion and
gravity wave experiment: Measurements in pure quantum states'',
{\em Found. Phys.} {\bf 13}, 6, 607-628 (1983).

\item {\bf [Dodonov-Mizrahi 93]}:
V. V. Dodonov, \& S. S. Mizrahi,
``A new class of nonlinear generalisations of the Schr\"{o}dinger
equation'',
{\em J. Phys. A} {\bf 26}, 23, 7163-7168 (1993).

\item {\bf [Dodonov-Mizrahi 95 a]}:
V. V. Dodonov, \& S. S. Mizrahi,
``Uniform nonlinear evolution equations for pure and mixed quantum
states'',
{\em Ann. Phys.} {\bf 237}, 2, 226-268 (1995).

\item {\bf [Dodonov-Mizrahi 95 b]}:
V. V. Dodonov, \& S. S. Mizrahi,
``Generalized nonlinear Doebner-Goldin Schr\"{o}dinger equation
and the relaxation of quantum systems'',
{\em Physica A} {\bf 214}, 4, 619-628 (1995).

\item {\bf [Dodonov-Mizrahi 98]}:
V. V. Dodonov, \& S. S. Mizrahi,
``Non-Lorentzian line shape of a
forced two-level system described by a nonlinear Schr\"{o}dinger
equation'',
{\em Physica A} {\bf 260}, 3-4, 430-438 (1998).

\item {\bf [Dodonov-Mizrahi-Silva 00]}:
V. V. Dodonov, S. S. Mizrahi, \& A. L. de Souza Silva,
``Decoherence and thermalization dynamics of a quantum oscillator'',
{\em J. Opt. B: Quantum Semiclass. Opt.} {\bf 2}, 3, 271-281 (2000).

\item {\bf [Dodonov-de Castro-Mizrahi 02]}:
V. V. Dodonov, A. S. M. de Castro, \& S. S. Mizrahi,
``Covariance entanglement measure for two-mode continuous variable systems'',
{\em Phys. Lett. A} {\bf 296}, 2-3, 73-81 (2002).

\item {\bf [Dodonov 02]}:
V. V. Dodonov,
``Purity- and entropy-bounded uncertainty relations for mixed quantum states'',
in Y. S. Kim, M. A. Man'ko, \& A. Sergienko (eds.),
{\em Seventh International Conference on Squeezed States and Uncertainty Relations (Boston, 2001)},
{\em J. Opt. B: Quantum Semiclass. Opt.} {\bf 4}, 3, S98-S108 (2002).

\item {\bf [Dodonov-Ren\'{o} 03]}:
V. V. Dodonov, \& M. B. Ren\'{o},
``Classicality and anticlassicality measures of pure and mixed quantum states'',
{\em Phys. Lett. A} {\bf 308}, 4, 249-255 (2003);
quant-ph/0302019.

\item {\bf [Doering 04]}:
A. Doering,
``Kochen-Specker theorem for von Neumann algebras'',
quant-ph/0408106.

\item {\bf [Doherty-Tan-Parkins-Walls 99]}:
A. C. Doherty, S. M. Tan, A. S. Parkins, \& D. F. Walls,
``State determination in continuous measurement'',
{\em Phys. Rev. A} {\bf 60}, 3, 2380-2392 (1999).

\item {\bf [Doherty-Parrilo-Spedalieri 02]}:
A. C. Doherty, P. A. Parrilo, \& F. M. Spedalieri,
``Distinguishing separable and entangled states'',
{\em Phys. Rev. Lett.} {\bf 88}, 18, 187904 (2002);
quant-ph/0112007.

\item {\bf [Doherty-Parrilo-Spedalieri 04]}:
A. C. Doherty, P. A. Parrilo, \& F. M. Spedalieri,
``Complete family of separability criteria'',
{\em Phys. Rev. A} {\bf 69}, 2, 022308 (2004);
quant-ph/0308032.

\item {\bf [Dohrn-Guerra 78]}:
D. Dohrn, \& F. Guerra,
``Nelson's stochastic mechanics on Riemannian manifolds'',
{\em Lettere al Nuovo Cimento} {\bf 22}, ?, 121-127 (1978).

\item {\bf [Dohrn-Guerra-Ruggiero 79]}:
D. Dohrn, F. Guerra, \& P. Ruggiero,
``Spinning particles and relativistic particles in the framework of
Nelson's stochastic mechanics'',
in S. Albeverio (ed.),
{\em Feynman path integrals},
Spinger, New York, 1979.

\item {\bf [Dolev-Elitzur 01]}:
S. Dolev, \& A. C. Elitzur,
``Non-sequential behavior of the wave function'',
quant-ph/0102109.

\item {\bf [Dolinska-Buchler-Bowen-(+2) 03]}:
A. Doliska, B. C. Buchler, W. P. Bowen,
T. C. Ralph, \& P. K. Lam,
``Teleportation of continuous-variable polarization states'',
{\em Phys. Rev. A} {\bf 68}, 5, 052308 (2003);
quant-ph/0308090.

\item {\bf [Domingos-Nogeira-Caldeira-dos Aidos 96]}:
J. M. Domingos, F. Nogeira, M. H. Caldeira, \& F. D. dos Aidos,
``EPR: Copenhagen interpretation has got what it takes'',
{\em Eur. J. Phys.} {\bf 17}, 3, 125-130 (1996).
Comment: {\bf [Hnizdo 97]}.
Reply: {\bf [Domingos-Nogeira-Caldeira-dos Aidos 98]}.

\item {\bf [Domingos-Nogeira-Caldeira-dos Aidos 98]}:
J. M. Domingos, F. Nogeira, M. H. Caldeira, \& F. D. dos Aidos,
``A comment on `EPR and the Copenhagen interpretation'\,'',
{\em Eur. J. Phys.} {\bf 19}, 2, L197 (1998).
Reply to {\bf [Hnizdo 97]}.
See {\bf [Domingos-Nogeira-Caldeira-dos Aidos 98]}.

\item {\bf [Domokos-Raimond-Brune-Haroche 95]}:
P. Domokos, J.-M. Raimond, M. Brune, \& S. Haroche,
``Simple cavity-QED two-bit universal quantum
logic gate: The principle and expected performances'',
{\em Phys. Rev. A} {\bf 52}, 5, 3554-3559 (1995).

\item {\bf [Domokos-Brune-Raimond-Haroche 98]}:
P. Domokos, M. Brune, J.-M. Raimond, \& S. Haroche,
``Photon-number-state generation with a single two-level atom in a cavity:
A proposal'',
{\em Eur. Phys. J. D} {\bf 1}, 1, 1-4 (1998).

\item {\bf [Donald 99]}:
M. J. Donald,
``Decoherence and the appearance of a classical world in quantum theory'',
{\em Stud. Hist. Philos. Sci. Part B: Stud. Hist. Philos. Mod. Phys.}
{\bf 30}, 3, 437-441 (1999).
Review of {\bf [Giulini-Joos-Kiefer-(+3) 96]}.

\item {\bf [Donald-Horodecki 99]}:
M. J. Donald, \& M. Horodecki,
``Continuity of relative entropy of entanglement'',
{\em Phys. Lett. A} {\bf 264}, 4, 257-260 (1999);
quant-ph/9910002.

\item {\bf [Donald-Horodecki-Rudolph 02]}:
M. J. Donald, M. Horodecki, \& O. Rudolph,
``The uniqueness theorem for entanglement measures'',
{\em J. Math. Phys.} {\bf 43}, 9, 4252-4272 (2002);
quant-ph/0105017.
See {\bf [Rudolph 01 a]}.

\item {\bf [Donald 03]}:
M. J. Donald,
``On the work of Henry P. Stapp'',
quant-ph/0311158.

\item {\bf [Donath-Svozil 02]}:
N. Donath, \& K. Svozil,
``Finding a state among a complete set of orthogonal states'',
{\em Phys. Rev. A} {\bf 65}, 4, 044302 (2002);
quant-ph/0105046.

\item {\bf [Donley-Claussen-Thompson-Wieman 02]}:
E. A. Donley, N. R. Claussen, S. T. Thompson, \& C. E. Wieman,
``Atom-molecule coherence in a Bose-Einstein condensate'',
{\em Nature} {\bf 417}, 6888, 529-533 (2002).

\item {\bf [de Dood-Irvine-Bouwmeester 04]}:
M. J. A. de Dood, W. T. M. Irvine, \& D. Bouwmeester,
``Nonlinear photonic crystals as a source of entangled photons'',
{\em Phys. Rev. Lett.} {\bf 93}, 4, 040504 (2004);
quant-ph/0409171.

\item {\bf [Dorai-Arvind-Kumar 00]}:
K. Dorai, Arvind, \& A. Kumar,
``Implementing quantum-logic operations, pseudopure states, and
the Deutsch-Jozsa algorithm using noncommuting selective pulses in NMR'',
{\em Phys. Rev. A} {\bf 61}, 4, 042306 (2000);
quant-ph/9906027.

\item {\bf [Dorai-Arvind-Kumar 01]}:
K. Dorai, Arvind, \& A. Kumar,
``Implementation of a Deutsch-like quantum algorithm utilizing entanglement
at the two-qubit level on an NMR quantum-information processor'',
{\em Phys. Rev. A} {\bf 63}, 3, 034101 (2001);
quant-ph/0006103.

\item {\bf [Dorner-Fedichev-Jaksch-(+2) 03]}:
U. Dorner, P. Fedichev, D. Jaksch, M. Lewenstein, \& P. Zoller,
``Entangling strings of neutral atoms in 1D atomic pipeline structures'',
{\em Phys. Rev. Lett.} {\bf 91}, 7, 073601 (2003).

\item {\bf [Doronin 03]}:
S. I. Doronin,
``Multiple quantum spin dynamics of entanglement'',
{\em Phys. Rev. A} {\bf 68}, 5, 052306 (2003).

\item {\bf [Dotson 86]}:
A. C. Dotson,
``Bell's theorem and the features of physical properties'',
{\em Am. J. Phys.} {\bf 54}, 3, 218-221 (1986).

\item {\bf [Dotson 95]}:
A. C. Dotson,
``On creating values of physical properties nonlocally'',
{\em Found. Phys.} {\bf 25}, 9, 1359-1379 (1995).

\item {\bf [Dotson 98]}:
A. C. Dotson,
``Interpretive principles and the quantum mysteries'',
{\em Am. J. Phys.} {\bf 66}, 11, 967-969 (1998).

\item {\bf [Dou\c{c}ot-Ioffe-Vidal 04]}:
B. Dou\c{c}ot, L. B. Ioffe, \& J. Vidal,
``Discrete non-Abelian gauge theories in Josephson-junction arrays and quantum computation'',
{\em Phys. Rev. B} {\bf 69}, 21, 214501 (2004).

\item {\bf [Dove 95]}:
C. Dove,
``Quantum computers and possible wavefunction collapse'',
{\em Phys. Lett. A} {\bf 207}, 6, 315-319 (1995).

\item {\bf [Dowker-Kent 95]}:
F. Dowker, \& A. Kent,
``Properties of consistent histories'',
{\em Phys. Rev. Lett.} {\bf 75}, 17, 3038-3041 (1995).

\item {\bf [Dowker-Kent 96]}:
F. Dowker, \& A. Kent,
``On the consistent histories approach to quantum mechanics'',
{\em J. Stat. Phys.} {\bf 82}, 5-6, 1575-1646 (1996);
gr-qc/9412067.

\item {\bf [Dowker-Herbauts 04]}:
F. Dowker, \& I. Herbauts,
``Simulating causal collapse models'',
{\em Class. Quant. Grav.} {\bf 21} 2963-2980 (2004);
quant-ph/0401075.

\item {\bf [Dowling-Franson-Lee-Milburn 04]}:
J. P. Dowling, J. D. Franson, H. Lee, \& G. J. Milburn,
``Towards linear optical quantum computers'',
{\em Quant. Inf. Proc.};
quant-ph/0402090.

\item {\bf [Dowling-Doherty-Bartlett 04]}:
M. R. Dowling, A. C. Doherty, \& S. D. Bartlett,
``Energy as an entanglement witness for quantum many-body systems'',
quant-ph/0408086.

\item {\bf [Dragan-Banaszek 01]}:
A. Dragan, \& K. Banaszek,
``Homodyne Bell's inequalities for entangled mesoscopic superpositions'',
{\em Phys. Rev. A} {\bf 63}, 6, 062102 (2001);
quant-ph/0010062.

\item {\bf [Dragoman 01]}:
D. Dragoman,
``Proposal for a three-qubit teleportation experiment'',
{\em Phys. Lett. A} {\bf 288}, 3-4, 121-124 (2001).

\item {\bf [Draper-Kutin-Rains-Svore 04]}:
T. G. Draper, S. A. Kutin, E. M. Rains, \& K. M. Svore,
``A logarithmic-depth quantum carry-lookahead adder'',
quant-ph/0406142.

\item {\bf [Drisch 79]}:
T. Drisch,
``Generalization of Gleason's theorem'',
{\em Int. J. Theor. Phys.} {\bf 18}, 4, 239-243 (1979).

\item {\bf [Du-Shi-Zhou-Han 00]}:
J. F. Du, M. J. Shi, X. Y. Zhou, \& R. D. Han,
``Geometrical interpretation for robustness of entanglement'',
{\em Phys. Lett. A} {\bf 267}, 4, 244-250 (2000).

\item {\bf [Du-Shi-Wu-(+2) 00]}:
J. F. Du, M. J. Shi, J. H. Wu, X. Y. Zhou, \& R. D. Han,
``Implementing universal multi-qubit quantum logic gates
in three and four-spin systems at room temperature'',
quant-ph/0008110.

\item {\bf [Du-Xu-Li-(+2) 00]}:
J. F. Du, X. Xu, H. Li, X. Zhou, \& R. Han,
``Nash equilibrium in quantum games'',
quant-ph/0010050.

\item {\bf [Du-Li-Xu-(+3) 00]}:
J. F. Du, H. Li, X. Xu, M. Shi, X. Zhou, \& R. Han,
``Multi-player and multi-choice quantum game'',
quant-ph/0010092.

\item {\bf [Du-Xu-Li-(+3) 00]}:
J. F. Du, X. Xu, H. Li, M. Shi, X. Zhou, \& R. Han,
``Quantum strategy without entanglement'',
quant-ph/0011078.

\item {\bf [Du-Shi-Zhou-(+4) 01]}:
J. F. Du, M. Shi, X. Zhou,
Y. Fan, B. Ye, R. Han, \& J. Wu,
``Implementation of a quantum algorithm to solve the Bernstein-Vazirani
parity problem without entanglement on an ensemble quantum computer'',
{\em Phys. Rev. A} {\bf 64}, 4, 042306 (2001).

\item {\bf [Du-Xu-Li-(+2) 01]}:
J. F. Du, X. Xu, H. Li, X. Zhou, \& R. Han,
``Entanglement playing a dominating role in quantum games'',
{\em Phys. Lett. A} {\bf 289}, 1-2, 9-15 (2001).

\item {\bf [Du-Li-Xu-(+3) 01 a]}:
J. F. Du, H. Li, X. Xu, M. Shi, X. Zhou, \& R. Han,
``Remark on quantum battle of the sexes game'',
quant-ph/0103004.

\item {\bf [Du-Li-Xu-(+3) 01 v]}:
J. Du, H. Li, X. Xu, M. Shi, X. Zhou, \& R. Han,
``Entanglement correlated phase changes in quantum games'',
quant-ph/0111138.

\item {\bf [Du-Li-Xu-(+4) 02]}:
J. Du, H. Li, X. Xu,
M. Shi, J. Wu, X. Zhou, \& R. Han,
``Experimental realization of quantum games on a quantum computer'',
{\em Phys. Rev. Lett.} {\bf 88}, 13, 137902 (2002);
quant-ph/0104087.

\item {\bf [Du-Li-Xu-(+2) 02]}:
J. Du, H. Li, X. Xu, X. Zhou, \& R. Han,
``Entanglement enhanced multiplayer quantum games'',
{\em Phys. Lett. A} {\bf 302}, 5-6, 229-233 (2002);
quant-ph/0110122.

\item {\bf [Du-Xu-Li-(+2) 02]}:
J. Du, X. Xu, H. Li, X. Zhou, \& R. Han,
``Playing prisoner's dilemma with quantum rules'',
{\em Fluctuation and Noise Lett.}
({\em Special Issue on Game Theory and
Evolutionary Processes: Order from Disorder -- The Role of Noise in Creative
Processes})
{\bf 2}, 4, R189-? (2002);
quant-ph/0301042.

\item {\bf [Du-Li-Xu-(+4) 03]}:
J. Du, H. Li, X. Xu, M. Shi, J. Wu, X. Zhou, \& R. Han,
``Experimental implementation of the quantum random-walk algorithm'',
{\em Phys. Rev. A} {\bf 67}, 4, 042316 (2003);
quant-ph/0203120.

\item {\bf [Du-Li-Xu-(+2) 03]}:
J. Du, H. Li, X. Xu, X. Zhou, \& R. Han,
``Phase-transition-like behaviour of quantum games'',
{\em J. Phys. A} {\bf 36}, 23, 6551-6562 (2003).

\item {\bf [Du-Li-Ju 03]}:
J. Du, H. Li, \& C. Ju,
``Quantum games of asymmetric information'',
{\em Phys. Rev. E} {\bf 68}, 1, 016124 (2003);
quant-ph/0308034.

\item {\bf [Du-Durt-Zou-(+4) 03]}:
J. Du, T. Durt, P. Zou, L. C. Kwek,
C. H. Lai, C. H. Oh, \& A. K. Ekert,
``Experimental demonstration of an efficient quantum phase-covariant
cloning and its possible applications to simulating eavesdropping in quantum
cryptography'',
quant-ph/0311010.

\item {\bf [Du-Durt-Zou-(+5) 04]}:
J. Du, T. Durt, P. Zou,
H. Li, L. C. Kwek, C. H. Lai,
C. H. Oh, \& A. K. Ekert,
``Experimental quantum cloning with prior partial information'',
quant-ph/0405094.

\item {\bf [Duan-Guo 97 a]}:
L.-M. Duan, \& G.-C. Guo,
``Preserving coherence in quantum
computation by pairing quantum bits'',
{\em Phys. Rev. Lett.} {\bf 79}, 10, 1953-1956 (1997).
Comment: {\bf [Zanardi-Rasetti 97 c]}.

\item {\bf [Duan-Guo 97 b]}:
L.-M. Duan, \& G.-C. Guo,
``Influence of noise on the fidelity and the entanglement fidelity
of states'',
{\em Quantum Semiclass. Opt} {\bf 9}, 6, 953-959 (1997).

\item {\bf [Duan-Guo 98 a]}:
L.-M. Duan, \& G.-C. Guo,
``Reducing decoherence in quantum-computer memory with all quantum
bits coupling to the same environment'',
{\em Phys. Rev. A} {\bf 57}, 2, 737-741 (1998).

\item {\bf [Duan-Guo 98 b]}:
L.-M. Duan, \& G.-C. Guo,
``Prevention of dissipation with two particles'',
{\em Phys. Rev. A} {\bf 57}, 4, 2399-2402 (1998).

\item {\bf [Duan-Guo 98 c]}:
L.-M. Duan, \& G.-C. Guo,
``A probabilistic cloning machine for replicating two non-orthogonal states'',
{\em Phys. Lett. A} {\bf 243}, 5-6, 261-264 (1998).

\item {\bf [Duan-Guo 98 d]}:
L.-M. Duan, \& G.-C. Guo,
``Scheme for reducing collective decoherence in quantum memory'',
{\em Phys. Lett. A} {\bf 243}, 5-6, 265-269 (1998).

\item {\bf [Duan-Guo 98 e]}:
L.-M. Duan, \& G.-C. Guo,
``Optimal quantum codes for preventing collective amplitude damping'',
{\em Phys. Rev. A} {\bf 58}, 5, 3491-3495 (1998);
quant-ph/9804014.

\item {\bf [Duan-Guo 98 f]}:
L.-M. Duan, \& G.-C. Guo,
``Probabilistic cloning and
identification of linearly independent quantum states'',
{\em Phys. Rev. Lett.} {\bf 80}, 22, 4999-5002 (1998);
quant-ph/9804064.

\item {\bf [Duan-Guo 98 g]}:
L.-M. Duan, \& G.-C. Guo,
``Correlations in interference and diffraction'',
{\em J. Mod. Opt.} {\bf 45}, 1, 79-89 (1998).

\item {\bf [Duan-Guo 98 h]}:
L.-M. Duan, \& G.-C. Guo,
``Pulse controlled noise suppressed quantum computation'',
quant-ph/9807072.

\item {\bf [Duan-Guo 98 i]}:
L.-M. Duan, \& G.-C. Guo,
``Reducing spatially correlated noise and decoherence
with quantum error correcting codes'',
quant-ph/9811058.

\item {\bf [Duan-Guo 99 a]}:
L.-M. Duan, \& G.-C. Guo,
``Quantum error correction with spatially correlated decoherence'',
{\em Phys. Rev. A} {\bf 59}, 5, 4058-4060 (1999).

\item {\bf [Duan-Guo 99 b]}:
L.-M. Duan, \& G.-C. Guo,
``Quantum error avoiding codes verses quantum error
correcting codes'',
{\em Phys. Lett. A} {\bf 255}, 4-6, 209-212 (1999);
quant-ph/9809057.

\item {\bf [Duan-Guo 99 c]}:
L.-M. Duan, \& G.-C. Guo,
``Suppressing environmental noise in quantum computation through pulse control'',
{\em Phys. Lett. A} {\bf 261}, 3-4, 139-144 (1999).

\item {\bf [Duan-Giedke-Cirac-Zoller 00 a]}:
L.-M. Duan, G. Giedke, J. I. Cirac, \& P. Zoller,
``Inseparability criterion for continuous variable systems'',
{\em Phys. Rev. Lett.} {\bf 84}, 12, 2722-2725 (2000);
quant-ph/9908056.
See {\bf [Julsgaard-Kozhekin-Polzik 01]},
{\bf [Cirac 01]}.

\item {\bf [Duan-Giedke-Cirac-Zoller 00 b]}:
L.-M. Duan, G. Giedke, J. I. Cirac, \& P. Zoller,
``Entanglement purification of Gaussian continuous variable quantum
states'',
{\em Phys. Rev. Lett.} {\bf 84}, 17, 4002-4005 (2000);
quant-ph/9912017.

\item {\bf [Duan-Giedke-Cirac-Zoller 00 b]}:
L.-M. Duan, G. Giedke, J. I. Cirac, \& P. Zoller,
``Physical implementation for entanglement purification of
Gaussian continuous-variable quantum states'',
{\em Phys. Rev. A} {\bf 62}, 3, 032304 (2000);
quant-ph/0003116.

\item {\bf [Duan-S{\o}rensen-Cirac-Zoller 00]}:
L.-M. Duan, A. S. S{\o}rensen, J. I. Cirac, \& P. Zoller,
``Squeezing and entanglement of atomic beams'',
{\em Phys. Rev. Lett.} {\bf 85}, 19, 3991-3994 (2000);
quant-ph/0007048.

\item {\bf [Duan-Cirac-Zoller-Polzik 00]}:
L.-M. Duan, J. I. Cirac, P. Zoller, \& E. S. Polzik,
``Quantum communication between atomic ensembles using coherent
light'',
{\em Phys. Rev. Lett.} {\bf 85}, 26, 5643-5646 (2000);
quant-ph/0003111.

\item {\bf [Duan-Lukin-Cirac-Zoller 01]}:
L.-M. Duan, M. D. Lukin, J. I. Cirac, \& P. Zoller,
``Long-distance quantum communication with atomic ensembles and linear optics'',
{\em Nature} {\bf 414}, 6862, 413 (2001);
quant-ph/0105105.

\item {\bf [Duan-Cirac-Zoller 01]}:
L.-M. Duan, J. I. Cirac, \& P. Zoller,
``Geometric manipulation of trapped ions for quantum computation'',
{\em Science} {\bf 292}, ?, 1695-1697 (2001);
quant-ph/0111086.

\item {\bf [Duan-Cirac-Zoller 02]}:
L.-M. Duan, J. I. Cirac, \& P. Zoller,
``Quantum entanglement in spinor Bose-Einstein condensates'',
{\em Phys. Rev. A} {\bf 65}, 3, 033619 (2002);
quant-ph/0107055.

\item {\bf [Duan-Cirac-Zoller 02]}:
L.-M. Duan, J. I. Cirac, \& P. Zoller,
``Three-dimensional theory for interaction between atomic ensembles and
free-space light'',
{\em Phys. Rev. A} {\bf 66}, 2, 023818 (2002).

\item {\bf [Duan 02]}:
L.-M. Duan,
``Entangling many atomic ensembles through laser manipulation'',
{\em Phys. Rev. Lett.} {\bf 88}, 17, 170402 (2002);
quant-ph/0201128.

\item {\bf [Duan-Kuzmich-Kimble 03]}:
L.-M. Duan, A. Kuzmich, \& H. J. Kimble,
`Cavity QED and quantum-information processing with ``hot'' trapped atoms',
{\em Phys. Rev. A} {\bf 67}, 3, 032305 (2003).

\item {\bf [Duan-Kimble 03]}:
L.-M. Duan, \& H. J. Kimble,
``Efficient engineering of multiatom entanglement through single-photon
detections'',
{\em Phys. Rev. Lett.} {\bf 90}, 25, 253601 (2003);
quant-ph/0301164.

\item {\bf [Duan-Demler-Lukin 03]}:
L.-M. Duan, E. Demler, \& M. D. Lukin,
``Controlling spin exchange interactions of ultracold atoms in optical
lattices'',
{\em Phys. Rev. Lett.} {\bf 91}, 9, 090402 (2003);
cond-mat/0210564.

\item {\bf [Duan-Kimble 04]}:
L.-M. Duan, \& H. J. Kimble,
``Scalable photonic quantum computation through cavity-assisted interactions'',
{\em Phys. Rev. Lett.} {\bf 92}, 12, 127902 (2004);
quant-ph/0309187.

\item {\bf [Duan 04]}:
L.-M. Duan,
``Scaling ion trap quantum computation through fast quantum gates'',
quant-ph/0401185.

\item {\bf [Duan-Blinov-Moehring-Monroe 04]}:
L.-M. Duan, B. B. Blinov, D. L. Moehring, \& C. Monroe,
``Scalable trapped ion quantum computation with a probabilistic ion-photon
mapping'',
quant-ph/0401020.

\item {\bf [Duan-Feng-Ying 03]}:
R. Duan, Y. Feng, \& M. Ying,
``Increasing number of copies can help entanglement transformation'',
quant-ph/0312010.

\item {\bf [Duan-Ji-Feng-Ying 04]}:
R. Duan, Z. Ji, Y. Feng, \& M. Ying,
``Quantum operation, quantum Fourier transform and semi-definite programming'',
{\em Phys. Lett. A} {\bf 323}, 1-2, 48-56 (2004).

\item {\bf [Duatti 01]}:
A. Duatti,
``On the interpretation of the Namiki-Pascazio order parameter
$\epsilon_{ij}$'',
{\em Phys. Lett. A} {\bf 279}, 5-6, 291-293 (2001).

\item {\bf [Duch-Aerts 86]}:
W. Duch, \& Aerts,
``Microphysical reality'',
{\em Phys. Today} {\bf 39}, 6, 13-15 (1986).

\item {\bf [Duck-Stevenson-Sudarshan 89]}:
I. M. Duck, P. M. Stevenson, \& E.
C. G. Sudarshan,
`The sense in which a ``weak measurement'' of a spin-$\frac{1}{2}$ particle's
spin component yields a value 100',
{\em Phys. Rev. D} {\bf 40}, 6, 2112-2117 (1989).
See {\bf [Aharonov-Albert-Vaidman 88]}, {\bf [Leggett 89]}, {\bf [Peres 89 a]},
{\bf [Aharonov-Vaidman 89]}.

\item {\bf [Duck 03]}:
I. Duck,
``Discovering quantum mechanics once again'',
quant-ph/0307121.
See {\bf [Hardy 01, 02 a]}.

\item {\bf [Dudarev-Diener-Wu-Raizen-Niu 03]}:
A. M. Dudarev, R. B. Diener, B. Wu, M. G. Raizen, \& Q. Niu,
``Entanglement generation and multiparticle interferometry with neutral atoms'',
{\em Phys. Rev. Lett.} {\bf 91}, 1, 010402 (2003).

\item {\bf [Dugi\'c 98]}:
M. Dugi\'c,
``On the necessary conditions for the occurrence of
the `environment-induced superselection rules'\,'',
quant-ph/9802041.

\item {\bf [Dugi\'c 99]}:
M. Dugi\'c,
``Quantum mechanical modeling of the CNOT (XOR) gate'',
quant-ph/9907011.

\item {\bf [Dugi\'c 00]}:
M. Dugi\'c,
``New strategy for suppressing decoherence in quantum computation'',
quant-ph/0001009.

\item {\bf [Dugi\'c 01]}:
M. Dugi\'c,
`On the maximum speed of operation of a quantum ``black box''\,',
quant-ph/0103150.

\item {\bf [Dugi\'c-\'{C}irkovi\'{c} 02]}:
M. Dugi\'c, \& M. M. \'{C}irkovi\'{c},
``Quantum parallelism in quantum information processing'',
{\em Int. J. Theor. Phys.} {\bf 41}, 9, 1641-1649 (2002).

\item {\bf [Dukelsky-Roman-Sierra 03]}:
J. Dukelsky, J. M. Roman, \& G. Sierra,
`Comment on ``Polynomial-time simulation of pairing models on a quantum
computer''\,',
{\em Phys. Rev. Lett.} {\bf 90}, 24, 249803 (2003);
quant-ph/0305139.
Comment on {\bf [Wu-Byrd-Lidar 02 a]}.
Reply {\bf [Wu-Byrd-Lidar 03]}.

\item {\bf [Dumachev-Orlov 02]}:
V. N. Dumachev, \& S. V. Orlov,
``Cloning of qubits of a quantum computer'',
{\em JETP Lett.} {\bf 76}, 7, 481-? (2002);
quant-ph/0212029.

\item {\bf [Dumitru 00]}:
S. Dumitru,
``A genuine reinterpretation of the Heisenberg's
(`uncertainty') relations'',
quant-ph/0004013.

\item {\bf [Dumitru 01]}:
S. Dumitru,
``Reconsideration of quantum measurements'',
quant-ph/0111141.

\item {\bf [Dumke-Volk-Muether-(+3) 01]}:
R. Dumke, M. Volk, T. Muether,
F. B. J. Buchkremer, G. Birkl, \& W. Ertmer,
``A scalable configuration for quantum computation with atomic qubits:
Microoptical realization of arrays of selectively addressable dipole traps'',
quant-ph/0110140.

\item {\bf [Duncan-Kleinpoppen 88]}:
A. J. Duncan, \& H. Kleinpoppen,
``The experimental investigation of the Einstein-Podolsky-Rosen
question and Bell's inequality'',
in F. Selleri (ed.),
{\em Quantum mechanics versus local realism: The
Einstein-Podolsky-Rosen paradox},
Plenum Press, New York, 1988, pp.~175-218.

\item {\bf [Dung-Scheel-Welsch-Kn\"{o}ll 01]}:
H. T. Dung, S. Scheel, D.-G. Welsch, \& L. Kn\"{o}ll,
``Atomic entanglement near a realistic microsphere'',
quant-ph/0110013.

\item {\bf [Dunningham-Bose-Henderson-(+2) 02]}:
J. A. Dunningham, S. Bose, L. Henderson, V. Vedral, \& K. Burnett,
``Entanglement concentration in Bose-Einstein condensates'',
{\em Phys. Rev. A} {\bf 65}, 6, 064302 (2002).

\item {\bf [Durdevi\'{c}-Vujici\'{c}-Herbut 91]}:
M. Durdevi\'{c}, M. Vujici\'{c}, \& F. Herbut,
``Symplectic hidden-variables theories---The missing link
in algebraic contextual approaches'',
{\em J. Math. Phys.} {\bf 32}, 11, 3088-3093 (1991).

\item {\bf [Durdevic-Makaruk-Owczarek 00]}:
M. Durdevi\'{c}, H. E. Makaruk, \& R. Owczarek,
``Generalized noiseless quantum codes utilizing quantum enveloping
algebras'',
quant-ph/0003134.

\item {\bf [D\"{u}r-Briegel-Cirac-Zoller 99]}:
W. D\"{u}r, H.-J. Briegel, J. I. Cirac, \& P. Zoller,
``Quantum repeaters based on entanglement purification'',
{\em Phys. Rev. A} {\bf 59}, 1, 169-181 (1999).
Erratum: {\em Phys. Rev. A} {\bf 60}, 1, 725 (1999);
quant-ph/9808065.

\item {\bf [D\"{u}r-Cirac-Tarrach 99]}:
W. D\"{u}r, J. I. Cirac, \& R. Tarrach,
``Separability and distillability of multiparticle quantum
systems'',
{\em Phys. Rev. Lett.} {\bf 83}, 17, 3562-3565 (1999).

\item {\bf [D\"{u}r-Cirac 00 a]}:
W. D\"{u}r, \& J. I. Cirac,
``Classification of multiqubit mixed states:
Separability and distillability properties'',
{\em Phys. Rev. A} {\bf 61}, 4, 042314 (2000);
quant-ph/9911044.

\item {\bf [D\"{u}r-Cirac-Lewenstein-Bru\ss\, 00]}:
W. D\"{u}r, J. I. Cirac, M. Lewenstein, \& D. Bru\ss,
``Distillability and partial transposition in bipartite systems'',
{\em Phys. Rev. A} {\bf 61}, 6, 062313 (2000);
quant-ph/9910022.

\item {\bf [D\"{u}r-Cirac 00 b]}:
W. D\"{u}r, \& J. I. Cirac,
``Multiparty teleportation'',
in V. Bu\v{z}zek, \& D. P. DiVincenzo (eds.),
{\em J. Mod. Opt.} {\bf 47}, 2-3 (Special issue:
Physics of quantum information), 247-255 (2000).

\item {\bf [D\"{u}r-Cirac 00 c]}:
W. D\"{u}r, \& J. I. Cirac,
``Activating bound entanglement in multiparticle systems'',
{\em Phys. Rev. A} {\bf 62}, 2, 022302 (2000);
quant-ph/0002028.

\item {\bf [D\"{u}r-Vidal-Cirac 00]}:
W. D\"{u}r, G. Vidal, \& J. I. Cirac,
``Three qubits can be entangled in two inequivalent ways'',
{\em Phys. Rev. A} {\bf 62}, 6, 062314 (2000);
quant-ph/0005115.

\item {\bf [D\"{u}r-Vidal-Cirac-(+2) 00]}:
W. D\"{u}r, G. Vidal, J. I. Cirac, N. Linden, \& S. Popescu,
``Entanglement capabilities of non-local Hamiltonians'',
{\em Phys. Rev. Lett.} {\bf 87}, 13, 137901 (2001);
quant-ph/0006034.

\item {\bf [D\"{u}r 00]}:
W. D\"{u}r,
``Entanglement molecules'',
quant-ph/0006105.

\item {\bf [D\"{u}r-Cirac 01 d]}:
W. D\"{u}r, \& J. I. Cirac,
``Multiparticle entanglement and its experimental detection'',
in S. Popescu, N. Linden, \& R. Jozsa (eds.),
{\em J. Phys. A} {\bf 34}, 35
(Special issue: Quantum information and computation), 6837-6850 (2001);
quant-ph/0011025.

\item {\bf [D\"{u}r 01 a]}:
W. D\"{u}r,
``Multipartite entanglement that is robust against disposal of particles'',
{\em Phys. Rev. A} {\bf 63}, 2, 020303(R) (2001).

\item {\bf [D\"{u}r-Vidal-Cirac 01 a]}:
W. D\"{u}r, G. Vidal, \& J. I. Cirac,
``Visible compression of commuting mixed states'',
{\em Phys. Rev. A} {\bf 64}, 2, 022308 (2001);
quant-ph/0101111.

\item {\bf [D\"{u}r-Cirac 01]}:
W. D\"{u}r, \& J. I. Cirac,
``Non-local operations: Purification, storage, compression,
tomography, and probabilistic implementation'',
{\em Phys. Rev. A} {\bf 64}, 1, 012317 (2001);
quant-ph/0012148.

\item {\bf [D\"{u}r 01 b]}:
W. D\"{u}r,
``Multipartite bound entangled states that violate Bell's
inequality'',
{\em Phys. Rev. Lett.} {\bf 87}, 23, 230402 (2001);
quant-ph/0107050.

\item {\bf [D\"{u}r-Vidal-Cirac 02]}:
W. D\"{u}r, G. Vidal, \& J. I. Cirac,
``Optimal conversion of nonlocal unitary operations'',
{\em Phys. Rev. Lett.} {\bf 89}, 5, 057901 (2002);
quant-ph/0112124.

\item {\bf [D\"{u}r-Simon-Cirac 02]}:
W. D\"{u}r, C. Simon, \& J. I. Cirac,
``Effective size of certain macroscopic quantum superpositions'',
{\em Phys. Rev. Lett.} {\bf 89}, 21, 210402 (2002).

\item {\bf [D\"{u}r-Cirac 02]}:
W. D\"{u}r, \& J. I. Cirac,
``Equivalence classes of non-local unitary operations'',
submitted to {\em Quant. Inf. Comp.};
quant-ph/0201112.

\item {\bf [D\"{u}r-Raussendorf-Kendon-Briegel 02]}:
W. D\"{u}r, R. Raussendorf, V. M. Kendon, \& H.-J. Briegel,
``Quantum walks in optical lattices'',
{\em Phys. Rev. A} {\bf 66}, 5, 052319 (2002).

\item {\bf [D\"{u}r-Briegel 03 a]}:
W. D\"{u}r, \& H.-J. Briegel,
``Entanglement purification for quantum computation'',
{\em Phys. Rev. Lett.} {\bf 90}, 6, 067901 (2003).
quant-ph/0210069.

\item {\bf [D\"{u}r-Aschauer-Briegel 03]}:
W. D\"{u}ur, H. Aschauer, \& H.-J. Briegel,
``Multiparticle entanglement purification for graph states'',
{\em Phys. Rev. Lett.} {\bf 91}, 10, 107903 (2003);
quant-ph/0303087.

\item {\bf [D\"{u}r-Briegel 03 b]}:
W. D\"{u}r, \& H.-J. Briegel,
``Stability of macroscopic entanglement under decoherence'',
quant-ph/0307180.

\item {\bf [D\"{u}r-Cirac-Horodecki 04]}:
W. D\"{u}r, J. I. Cirac, \& P. Horodecki,
``Nonadditivity of quantum capacity for multiparty communication channels'',
{\em Phys. Rev. Lett.} {\bf 93}, 2, 020503 (2004);
quant-ph/0403068.

\item {\bf [D\"{u}r-Hartmann-Hein-Briegel 04]}:
W. D\"{u}r, L. Hartmann, M. Hein, \& H. J. Briegel,
``Entanglement in spin chains and lattices with long-range interactions'',
quant-ph/0407075.

\item {\bf [D\"{u}rr-Goldstein-Tumulka-Zanghi 04]}:
D. D\"{u}rr, S. Goldstein, R. Tumulka, \& N. Zanghi,
``Bell-type quantum field theories'',
quant-ph/0407116.

\item {\bf [Durkin-Simon-Bouwmeester 02]}:
G. A. Durkin, C. Simon, \& D. Bouwmeester,
``Multiphoton entanglement concentration and quantum cryptography'',
{\em Phys. Rev. Lett.} {\bf 88}, 18, 187902 (2002);
quant-ph/0109132.

\item {\bf [Durkin-Simon-Eisert-Bouwmeester 04]}:
G. A. Durkin, C. Simon, J. Eisert, \& D. Bouwmeester,
``Quantifying multi-photon entanglement'',
quant-ph/0402053.

\item {\bf [Durr-Heiligman-H\o{}yer-Mhalla 04]}:
C. Durr, \& P. H\o{}yer,
``Quantum query complexity of some graph problems'',
quant-ph/0401091

\item {\bf [D\"{u}rr-Goldstein-Zangh\`{\i} 92]}:
D. D\"{u}rr, S. Goldstein, \& N. Zangh\`{\i},
``Quantum equilibrium and the origin of absolute uncertainty'',
{\em J. Stat. Phys.} {\bf 67}, 843-907 (1992);
quant-ph/0308039.

\item {\bf [D\"{u}rr-Fusseder-Goldstein-Zangh\`{\i} 93]}:
D. D\"{u}rr, W. Fusseder, S. Goldstein, \& N. Zangh\`{\i},
``Comment on `Surrealistic Bohm trajectories'\,'',
{\em Zeitschrift f\"{u}r Naturforschung A} {\bf 48}, ?, 1261-1262 (1993).
Comment on {\bf [Englert-Scully-S\"{u}ssmann-Walther 93 a]}.
Reply: {\bf [Englert-Scully-S\"{u}ssmann-Walther 93 b]}.

\item {\bf [D\"{u}rr-Goldstein-Zangh\`{\i} 96]}:
D. D\"{u}rr, S. Goldstein, \& N. Zangh\`{\i},
``Bohmian mechanics as the foundation of quantum mechanics'',
in {\bf [Cushing-Fine-Goldstein 96]}, pp.~21-44.

\item {\bf [D\"{u}rr-Goldstein-M\"{u}nch Berndl-Zangh\`{\i} 99]}:
D. D\"{u}rr, S. Goldstein, K. M\"{u}nch-Berndl, \& N. Zangh\`{\i},
``Hypersurface Bohm-Dirac models'',
{\em Phys. Rev. A} {\bf 60}, 4, 2729-2736 (1999).

\item {\bf [D\"{u}rr 01]}:
D. D\"{u}rr,
{\em Bohmsche Mechanik als Grundlage der Quantenmechanik},
Springer, Berlin, 2001.

\item {\bf [D\"{u}rr-Goldstein-Tumulka-Zanghi 03]}:
D. D\"{u}rr, S. Goldstein, R. Tumulka, \& N. Zanghi,
``Quantum Hamiltonians and stochastic jumps'',
quant-ph/0303056.

\item {\bf [D\"{u}rr-Goldstein-Zanghi 03]}:
D. D\"{u}rr, S. Goldstein, \& N. Zanghi,
``Quantum equilibrium and the role of operators as observables in quantum
theory'',
quant-ph/0308038.

\item {\bf [D\"{u}rr-Goldstein-Tumulka-Zanghi 03]}:
D. D\"{u}rr, S. Goldstein, R. Tumulka, \& N. Zanghi,
``On the role of density matrices in Bohmian mechanics'',
quant-ph/0311127.

\item {\bf [D\"{u}rr-Goldstein-Tumulka-Zanghi 04]}:
D. D\"{u}rr, S. Goldstein, R. Tumulka, \& N. Zanghi,
``Bohmian mechanics and quantum field theory'',
{\em Phys. Rev. Lett.} {\bf 93}, 9, 090402 (2004);
quant-ph/0303156.

\item {\bf [D\"{u}rr-Nonn-Rempe 98 a]}:
S. D\"{u}rr, T. Nonn, \& G. Rempe,
`Origin of quantum-mechanical complementarity probed by a ``which way''
experiment in an atom interferometer',
{\em Nature} {\bf 395}, 6697, 33-37 (1998).
See {\bf [Knight 98]}, {\bf [Paul 98]}.

\item {\bf [D\"{u}rr-Nonn-Rempe 98 b]}:
S. D\"{u}rr, T. Nonn, \& G. Rempe,
``Fringe visibility and which-way information in an
atom interferometer'',
{\em Phys. Rev. Lett.} {\bf 81}, 26, 5705-5709 (1998).

\item {\bf [D\"{u}rr-Rempe 00]}:
S. D\"{u}rr, \& G. Rempe,
``Can wave-particle duality be based on the uncertainty relation?'',
{\em Am. J. Phys.} {\bf 68}, 11, 1021-1024 (2000).

\item {\bf [D\"{u}rr 01]}:
S. D\"{u}rr,
``Quantitative wave-particle duality in multibeam interferometers'',
{\em Phys. Rev. A} {\bf 64}, 4, 042113 (2001).
Comment: {\bf [Bimonte-Musto 03 a]}.

\item {\bf [Durt 97]}:
T. Durt,
``Three interpretations of the violation of Bell's inequalities'',
{\em Found. Phys.} {\bf 27}, 3, 415-434 (1997).

\item {\bf [Durt 99]}:
T. Durt,
`Comment on ``A local hidden variable model of quantum
correlations exploiting the detection loophole''\,',
quant-ph/9907061.
Comment on {\bf [Gisin-Gisin 99]}.

\item {\bf [Durt 01 a]}:
T. Durt,
`Comment on ``Pulsed energy-time entangled twin-photon
source for quantum communication''\,'
{\em Phys. Rev. Lett.} {\bf 86}, 7, 1392 (2001).
Comment on {\bf [Brendel-Gisin-Tittel-Zbinden 99]}.
Reply: {\bf [Gisin-Tittel-Zbinden 01]}.

\item {\bf [Durt-Kaszlikowski-\.{Z}ukowski 01]}:
T. Durt, D. Kaszlikowski, \& \.{Z}ukowski,
``Violations of local realism with quantum systems described
by $N$-dimensional Hilbert spaces up to $N = 16$'',
{\em Phys. Rev. A} {\bf 64}, 2, 024101 (2001);
quant-ph/0101084.

\item {\bf [Durt 01 b]}:
T. Durt,
``Characterisation of an entanglement-free evolution'',
quant-ph/0109112.

\item {\bf [Durt-Pierseaux 02]}:
T. Durt, \& Y. Pierseaux,
``Bohm's interpretation and maximally entangled states'',
{\em Phys. Rev. A} {\bf 66}, 5, 052109 (2002).

\item {\bf [Durt-Cerf-Gisin-\.{Z}ukowski 03]}:
T. Durt, N. J. Cerf, N. Gisin, \& \.{Z}ukowski,
``Security of quantum key distribution with entangled qutrits'',
{\em Phys. Rev. A} {\bf 67}, 1, 012311 (2003);
quant-ph/0207057.

\item {\bf [Durt-Nagler 03]}:
T. Durt, \& B. Nagler,
``Covariant cloning machines for four-level systems'',
{\em Phys. Rev. A} {\bf 68}, 4, 042323 (2003).

\item {\bf [Durt 03]}:
T. Durt,
``Security of quantum key distribution with entangled quNits'',
quant-ph/0302078.

\item {\bf [Durt-Kaszlikowski-Chen-Kwek 04]}:
T. Durt, D. Kaszlikowski, J.-L. Chen, \& L. C. Kwek,
``Security of quantum key distributions with entangled qudits'',
{\em Phys. Rev. A} {\bf 69}, 3, 032313 (2004).

\item {\bf [Durt-Du 04]}:
T. Durt, \& J. Du,
``Characterization of low-cost one-to-two qubit cloning'',
{\em Phys. Rev. A} {\bf 69}, 6, 062316 (2004);
quant-ph/0309072.

\item {\bf [Durt 04 a]}:
T. Durt,
``Bell states, mutually unbiased bases and the Mean King's problem'',
quant-ph/0401037.

\item {\bf [Durt 04 b]}:
T. Durt,
``If $1=2+3$, then $1=2.3$: Bell states, finite groups, and mutually unbiased
bases, a unifying approach'',
quant-ph/0401046.
See {\bf [Durt 04 d]}.

\item {\bf [Durt 04 c]}:
T. Durt,
``Quantum entanglement, interaction, and the classical limit'',
quant-ph/0401121.

\item {\bf [Durt 04 d]}:
T. Durt,
``A new expression for mutually unbiased bases in prime power dimensions'',
quant-ph/0409090.
See {\bf [Durt 04 b]}.

\item {\bf [Du\v{s}ek-Haderka-Hendrych-My\v{z}ka 99]}:
M. Du\v{s}ek, O. Haderka, M. Hendrych, \& R. My\v{z}ka,
``Quantum identification system'',
{\em Phys. Rev. A} {\bf 60}, 1, 149-156 (1999);
quant-ph/9809024.

\item {\bf [Du\v{s}ek-Jahma-L\"{u}tkenhaus 00]}:
M. Du\v{s}ek, M. Jahma, \& N. L\"{u}tkenhaus,
``Unambiguous state discrimination in quantum
cryptography with weak coherent states'',
{\em Phys. Rev. A} {\bf 62}, 2, 022306 (2000);
quant-ph/9910106.

\item {\bf [Du\v{s}ek 01]}:
M. Du\v{s}ek,
``Discrimination of the Bell states of qudits by means of linear optics'',
quant-ph/0107119.

\item {\bf [Du\v{s}ek-Br\'{a}dler 02]}:
M. Du\v{s}ek, \& K. Br\'{a}dler,
``The effect of multi-pair signal states in quantum
cryptography with entangled photons'',
{\em J. Opt. B: Quantum Semiclass. Opt.} {\bf 4}, 2, 109-113 (2002);
quant-ph/0011007.

\item {\bf [Du\v{s}ek-Bu\v{z}zek 02]}:
M. Du\v{s}ek, \& V. Bu\v{z}zek,
``Quantum-controlled measurement device for quantum-state discrimination'',
{\em Phys. Rev. A} {\bf 66}, 2, 022112 (2002);
quant-ph/0201097.

\item {\bf [Duty-Gunnarsson-Bladh-Delsing 04]}:
T. Duty, D. Gunnarsson, K. Bladh, \& P. Delsing,
``Coherent dynamics of a Josephson charge qubit'',
{\em Phys. Rev. B} {\bf 69}, 14, 140503 (2004).

\item {\bf [Duvenhage 02]}:
R. Duvenhage,
``The nature of information in quantum mechanics'',
{\em Found. Phys.} {\bf 32}, 9, 1399-1417 (2002);
quant-ph/0203070.

\item {\bf [Drummond 83]}:
P. D. Drummond,
``Violations of Bell's inequality in cooperative states'',
{\em Phys. Rev. Lett.} {\bf 50}, 19, 1407-1410 (1983).

\item {\bf [Dyakonov 01]}:
M. I. Dyakonov,
``Quantum computing: A view from the enemy camp'',
in S. Luryi, J. Xu, \& A. Zaslavsky (eds.),
{\em Future Trends in Microelectronics: The Nano Millenium},
Wiley, 2001;
cond-mat/0110326.

\item {\bf [Dyakonov 03]}:
M. I. Dyakonov,
``Quantum computing: A view from the enemy camp'',
{\em Opt. Spectrosc.} {\bf 95}, 261-? (2003).

\item {\bf [Dykman-Platzman 00]}:
M. I. Dykman, \& P. M. Platzman,
``Quantum computing using electrons floating on liquid helium'',
{\em Fortschr. Phys.} {\bf 48}, 9-11 (Special issue:
Experimental proposals for quantum computation), 1095-1108 (2000);
quant-ph/0007113.

\item {\bf [Dykman-Platzman 01]}:
M. I. Dykman, \& P. M. Platzman,
``A quantum computer based on electrons floating on liquid helium'',
in {\bf [Clark 01]};
quant-ph/0109030.

\item {\bf [Dykman-Platzman-Seddighrad 03]}:
M. I. Dykman, P. M. Platzman, \& P. Seddighrad,
``Qubits with electrons on liquid helium'',
{\em Phys. Rev. B} {\bf 67}, 15, 155402 (2003).

\item {\bf [Dykman-Izrailev-Santos-Shapiro 04]}:
M. I. Dykman, F. M. Izrailev, L. F. Santos, \& M. Shapiro,
``Many-particle localization by constructed disorder: Enabling quantum
computing with perpetually coupled qubits'',
cond-mat/0401201.
See {\bf [Santos-Dykman-Shapiro-Izrailev 04]}.

\item {\bf [Dyson 02]}:
F. Dyson,
``The conscience of physics. The life, work and dreams of Wolfgang Pauli'',
{\em Nature} {\bf 420}, 6916, 607-608 (2002).
Review of {\bf [Enz 02]}.

\item {\bf [Dziarmaga-Dalvit-Zurek 01 a]}:
J. Dziarmaga, D. A. R. Dalvit, \& W. H. Zurek,
``Conditional dynamics of open quantum systems:
The case of multiple observers'',
quant-ph/0106036.

\item {\bf [Dziarmaga-Dalvit-Zurek 01 b]}:
J. Dziarmaga, D. A. R. Dalvit, \& W. H. Zurek,
``Continuous quantum measurement with multiple observers'',
quant-ph/0107033.


\newpage

\subsection{}


\item {\bf [Eakins-Jaroszkiewicz 03]}:
J. Eakins, \& G. Jaroszkiewicz,
``Factorization and entanglement in quantum systems'',
{\em J. Phys. A} {\bf 36}, 2, 517-526 (2003).

\item {\bf [Eberhard 77]}:
P. H. Eberhard,
``Bell's theorem without hidden variables'',
{\em Nuovo Cimento B} {\bf 38}, 1, 75-80 (1977).

\item {\bf [Eberhard 78]}:
P. H. Eberhard,
``Bell's theorem and the different concepts of locality'',
{\em Nuovo Cimento B} {\bf 46}, 2, 392-419 (1978).

\item {\bf [Eberhard 82]}:
P. H. Eberhard,
``Constraints of determinism and of Bell's inequalities are not equivalent'',
{\em Phys. Rev. Lett.} {\bf 49}, 20, 1474-1477 (1982).
See {\bf [Fine 82 a, d]}.

\item {\bf [Eberhard 93]}:
P. H. Eberhard,
``Background level and counter
efficiences required for a loophole-free Einstein-Podolsky-Rosen experiment'',
{\em Phys. Rev. A} {\bf 47}, 2, R747-R750 (1993).

\item {\bf [Eberhard-Rosselet 95]}:
P. H. Eberhard, \& P. Rosselet,
``Bell's theorem based on a generalized EPR criterion of reality'',
{\em Found. Phys.} {\bf 25}, 1, 91-111 (1995).

\item {\bf [Eberly 02]}:
J. H. Eberly,
``Bell inequalities and quantum mechanics'',
{\em Am. J. Phys.} {\bf 70}, 3, 276-279 (2002).

\item {\bf [Echternach-Williams-Dultz-(+3) 01]}:
P. Echternach, C. P. Williams, S. C. Dultz,
P. Delsing, S. L. Braunstein, \& J. P. Dowling,
``Universal quantum gates for single Cooper pair box based quantum computing'',
{\em Quantum Information and Computation} {\bf 1}, special issue, 143-150 (2001);
quant-ph/0112025.

\item {\bf [Eckert-Schliemann-Bru\ss-Lewenstein 02]}:
K. Eckert, J. Schliemann, D. Bru\ss, \& M. Lewenstein,
``Quantum correlations in systems of indistinguishable particles'',
{\em Ann. Phys.} {\bf 299}, ?, 88-127 (2002);
quant-ph/0203060.

\item {\bf [Eckert-Mompart-Yi-(+4) 02]}:
K. Eckert, J. Mompart, X. X. Yi,
J. Schliemann, D. Bru\ss, G. Birkl, \& M. Lewenstein,
``Quantum computing in optical microtraps based on the motional states of
neutral atoms'',
{\em Phys. Rev. A} {\bf 66}, 4, 042317 (2002).

\item {\bf [Edalat 03]}:
A. Edalat,
``An extension of Gleason's theorem for quantum computation'',
quant-ph/0311070.

\item {\bf [Edamatsu-Shimizu-Itoh 01]}:
K. Edamatsu, R. Shimizu, \& T. Itoh,
``Measurement of the photonic de Broglie wavelength of parametric
down-converted photons using a Mach-Zehnder interferometer'',
quant-ph/0109005.

\item {\bf [Edamatsu-Shimizu-Itoh 02]}:
K. Edamatsu, R. Shimizu, \& T. Itoh,
``Measurement of the photonic de Broglie wavelength of entangled photon pairs
generated by spontaneous parametric down-conversion'',
{\em Phys. Rev. Lett.} {\bf 89}, 21, 213601 (2002).

\item {\bf [Edwards-Cheung-van Pham-(+3) 00]}:
P. J. Edwards, W. N. Cheung, H. van Pham,
G. Ganesharajah, P. Lynam, \& L. Barbopoulos,
``A note on quantum key channel efficiency and security using correlated
photon beam transmitters'',
quant-ph/0008013.

\item {\bf [Edwards-Pollard-Cheung 02]}:
P. J. Edwards, G. H. Pollard, \& W. N. Cheung,
``Quantum key distribution using quantum-correlated photon sources'',
{\em Eur. Phys. J. D} {\bf 18}, 2 (Special issue:
{\em Quantum interference and cryptographic keys:
Novel physics and advancing technologies (QUICK) (Corsica, 2001)}, 147-153 (2002).

\item {\bf [Eggeling-Werner 00]}:
T. Eggeling, \& R. F. Werner,
``Separability properties of tripartite states with
$U \times U \times U$-symmetry'',
quant-ph/0003008;
quant-ph/0010096.

\item {\bf [Eggeling-Vollbrecht-Wolf 01]}:
T. Eggeling, K. G. H. Vollbrecht, \& M. M. Wolf,
``Comment on: `Necessary and sufficient condition of separability of any
system'\,'',
quant-ph/0103003.
Comment on {\bf [Chen-Liang-Li-Huang 01]}.

\item {\bf [Eggeling-Vollbrecht-Werner-Wolf 01]}:
T. Eggeling, K. G. H. Vollbrecht, R. F. Werner, \& M. M. Wolf,
``Distillability via protocols respecting the positivity of partial transpose'',
{\em Phys. Rev. Lett.} {\bf 87}, 25, 257902 (2001);
quant-ph/0104095.

\item {\bf [Eggeling-Werner 02]}:
T. Eggeling, \& R. F. Werner,
``Hiding classical data in multipartite quantum states'',
{\em Phys. Rev. Lett.} {\bf 89}, 9, 097905 (2002);
quant-ph/0203004.

\item {\bf [Egusquiza-Muga 00]}:
I. L. Egusquiza, \& J. G. Muga,
``Consistent histories, the quantum Zeno effect, and time of arrival'',
{\em Phys. Rev. A} {\bf 62}, 3, 032103 (2000);
quant-ph/0003041.

\item {\bf [Eibl-Gaertner-Bourennane-(+3) 03]}:
M. Eibl, S. Gaertner, M. Bourennane,
C. Kurtsiefer, M. \.{Z}ukowski, \& H. Weinfurter,
``Experimental observation of four-photon entanglement from parametric
down-conversion'',
{\em Phys. Rev. Lett.} {\bf 90}, 20, 200403 (2003).

\item {\bf [Eibl-Kiesel-Bourennane-(+3) 04]}:
M. Eibl, N. Kiesel, M. Bourennane,
C. Kurtsiefer, \& H. Weinfurter,
``Experimental realization of a three-qubit entangled $W$ state'',
{\em Phys. Rev. Lett.} {\bf 92}, 7, 077901 (2004).
Comment: {\bf [Cereceda 04 b]}.

\item {\bf [Eichmann-Bergquist-Bollinger-(+4) 93]}:
U. Eichmann, J. C. Bergquist, J. J. Bollinger,
J. M. Gilligan, W. M. Itano, D. J. Wineland, \& M. G. Raizen,
``Young's interference experiment with light scattered from two
atoms'',
{\em Phys. Rev. Lett.} {\bf 70}, 16, 2359-2362 (1993).

\item {\bf [Eilers-Horst 75]}:
M. Eilers, \& E. Horst,
``The theorem of Gleason for nonseparable Hilbert spaces'',
{\em Int. J. Theor. Phys.} {\bf 13}, 6, 419-424 (1975).



\item {\bf [Einstein-Tolman-Podolsky 31]}:
A. Einstein, R. C. Tolman, \& B. Podolsky,
``Knowledge of past and future in quantum mechanics'',
{\em Phys. Rev.} {\bf 37}, 6, 780-781 (1935).
Reprinted in {\bf [Wheeler-Zurek 83]}, pp.~135-136.

\item {\bf [Einstein-Podolsky-Rosen 35]}:
A. Einstein, B. Podolsky, \& N. Rosen,
``Can quantum-mechanical description of physical reality be considered
complete?'',
{\em Phys. Rev.} {\bf 47}, 10, 777-780 (1935).
Reprinted in {\bf [Toulmin 70]}, pp.~122-142;
{\bf [Wheeler-Zurek 83]}, pp.~138-141;
{\bf [Stroke 95]}, pp.~1215-1218.

\item {\bf [Einstein 36]}:
A. Einstein,
``Physik und Realit\"{a}t'' (``Physics and reality''),
{\em J. Frankin Institute} {\bf 221}, ?, 313-347 (349-382) (1936).
Other English version in A. Einstein,
{\em Ideas and opinions}, Bonanza Books, New York, 1954, pp.~290-323.

\item {\bf [Einstein 48]}:
A. Einstein,
``Quanten-Mechanik und Wirklichkeit'',
{\em Dialectica} {\bf 2}, 3, 320-324 (1948).
Also in {\bf [Born-Einstein 69]}, in the April 5, 1948 letter
(pp.~168-173 in the English version).

\item {\bf [Einstein 49]}:
A. Einstein,
``Autobiographisches'' (even pages),
``Autobiographical notes'' (odd pages), in P. A. Schilpp (ed.),
{\em Albert Einstein:
Philosopher-scientist}, Library of Living Philosophers,
Open Court, La Salle, Illinois, 1949, vol. 1, pp.~2-94.
Also Evanston, Illinois, 1949; Tudor, New York, 1949;
Harper and Row, New York, 1959.
Spanish version: {\em Notas autobiogr\'{a}ficas},
Alianza, Madrid, 1984.
Review: {\bf [Zeilinger 99 b]}.


\item {\bf [Eisaman-Childress-Andre-(+3) 04]}:
M. D. Eisaman, L. Childress, A. Andre,
F. Massou, A. S. Zibrov, \& M. D. Lukin,
``Shaping quantum pulses of light via coherent atomic memory'',
quant-ph/0406093.

\item {\bf [Eisenberg-Khoury-Durkin-(+2) 04]}:
H. S. Eisenberg, G. Khoury, G. Durkin,
C. Simon, \& D. Bouwmeester,
``Quantum entanglement of a large number of photons'',
quant-ph/0408030.

\item {\bf [Eisenberg-Hodelin-Khoury-Bouwmeester 04]}:
H. S. Eisenberg, J. F. Hodelin, G. Khoury, \& D. Bouwmeester,
``Multiphoton path entanglement by non-local bunching'',
quant-ph/0410093.

\item {\bf [Eisert-Wilkens-Lewenstein 99]}:
J. Eisert, M. Wilkens, \& M. Lewenstein,
``Quantum games and quantum strategies'',
{\em Phys. Rev. Lett.} {\bf 83}, 15, 3077-3080 (1999);
quant-ph/9806088.
Comment: {\bf [Benjamin-Hayden 01]}.
Reply: {\bf [Eisert-Wilkens-Lewenstein 01]}.
See {\bf [Eisert-Wilkens 00 b]}.

\item {\bf [Eisert-Plenio 99]}:
J. Eisert, \& M. B. Plenio,
``A comparison of entanglement measures'',
{\em J. Mod. Opt.} {\bf 46}, 1, 145-154 (1999);
quant-ph/9807034.
See {\bf [Audenaert-Verstraete-De Bie-De Moor 00]}.

\item {\bf [Eisert-Felbinger-Papadopoulos-(+2) 00]}:
J. Eisert, T. Felbinger, P. Papadopoulos,
M. B. Plenio, \& M. Wilkens,
``Classical information and distillable entanglement'',
{\em Phys. Rev. Lett.} {\bf 84}, 7, 1611-1614 (2000);
quant-ph/9907021.

\item {\bf [Eisert-Wilkens 00 a]}:
J. Eisert, \& M. Wilkens,
``Catalysis of entanglement manipulation for mixed states'',
{\em Phys. Rev. Lett.} {\bf 85}, 2, 437-440 (2000);
quant-ph/9912080.

\item {\bf [Eisert-Wilkens 00 b]}:
J. Eisert, \& M. Wilkens,
``Quantum games'',
quant-ph/0004076.
See {\bf [Eisert-Wilkens-Lewenstein 99]}.

\item {\bf [Eisert-Jacobs-Papadopoulos-Plenio 00]}:
J. Eisert, K. Jacobs, P. Papadopoulos, \& M. B. Plenio,
``Optimal local implementation of nonlocal quantum gates'',
{\em Phys. Rev. A} {\bf 62}, 5, 052317 (2000);
quant-ph/0005101.

\item {\bf [Eisert-Wilkens-Lewenstein 01]}:
J. Eisert, M. Wilkens, \& M. Lewenstein,
``Reply: Eisert, Wilkens, and Lewenstein'',
{\em Phys. Rev. Lett.} {\bf 87}, 6, 069802 (2001).
Reply to {\bf [Benjamin-Hayden 01]}.
See {\bf [Eisert-Wilkens-Lewenstein 99]}.

\item {\bf [Eisert-Briegel 01]}:
J. Eisert, \& H.-J. Briegel,
``Schmidt measure as a tool for quantifying multiparticle entanglement'',
{\em Phys. Rev. A} {\bf 64}, 2, 022306 (2001);
quant-ph/0007081.

\item {\bf [Eisert-Simon-Plenio 01]}:
J. Eisert, C. Simon, \& M. B. Plenio,
``On the quantification of entanglement
in infinite-dimensional quantum systems'',
{\em J. Phys. A} {\bf 35}, 17, 3911-3923 (2002);
quant-ph/0112064.

\item {\bf [Eisert-Plenio 02 a]}:
J. Eisert, \& M. B. Plenio,
``Conditions for the local manipulation of Gaussian states'',
{\em Phys. Rev. Lett.} {\bf 89}, 9, 097901 (2002);
quant-ph/0109126.

\item {\bf [Eisert-Plenio 02 b]}:
J. Eisert, \& M. B. Plenio,
``Quantum and classical correlations in quantum Brownian motion'',
{\em Phys. Rev. Lett.} {\bf 89}, 13, 137902 (2002);
quant-ph/0111016.

\item {\bf [Eisert-Scheel-Plenio 02]}:
J. Eisert, S. Scheel, \& M. B. Plenio,
``Distilling Gaussian states with Gaussian operations is impossible'',
{\em Phys. Rev. Lett.} {\bf 89}, 13, 137903 (2002);
quant-ph/0204052.

\item {\bf [Eisert-Audenaert-Plenio 03]}:
J. Eisert, K. M. R. Audenaert, \& M. B. Plenio,
``Remarks on entanglement measures and non-local state distinguishability'',
{\em J. Phys. A} {\bf 36}, 20, 5605-5615 (2003);
quant-ph/0212007.

\item {\bf [Eisert-Plenio-Bose-Hartley 03]}:
J. Eisert, M. B. Plenio, S. Bose, \& J. Hartley,
``Towards mechanical entanglement in nano-electromechanical devices'',
quant-ph/0311113.

\item {\bf [Eisert-Plenio 03]}:
J. Eisert, \& M. B. Plenio,
``Introduction to the basics of entanglement theory in continuous-variable
systems'',
{\em EQIS' 03}, {\em Int. J. Quant. Inf.};
quant-ph/0312071.

\item {\bf [Eisert-Browne-Scheel-Plenio 04]}:
J. Eisert, D. Browne, S. Scheel, \& M. B. Plenio,
``Distillation of continuous-variable entanglement with optical means'',
{\em Ann. Phys.} {\bf 311}, 431-? (2004);
quant-ph/0307106.

\item {\bf [Eisert 04 a]}:
J. Eisert,
``Exact decoherence to pointer states in free open quantum systems is
universal'',
{\em Phys. Rev. Lett.} {\bf 92}, 21, 210401 (2004);
quant-ph/0311022.

\item {\bf [Eisert-Wolf 04]}:
J. Eisert, \& M. M. Wolf,
``Quantum computing'',
in G. J. Milburn, J. Dongarra, D. Bader,
R. Brent, M. Eshaghian-Wilner, \& F. Seredynski (eds.),
{\em Handbook innovative computing},
Springer, Berlin, 2004.
quant-ph/0401019.

\item {\bf [Eisert-Hyllus-G\"{u}hne-Curty 04]}:
J. Eisert, P. Hyllus, O. G\"{u}hne, \& M. Curty,
``Complete hierarchies of efficient approximations to problems in
entanglement theory'',
{\em Phys. Rev. A};
quant-ph/0407135.

\item {\bf [Eisert 04 b]}:
J. Eisert,
``Optimizing linear optics quantum gates'',
quant-ph/0409156.

\item {\bf [Ekert 91 a]}:
A. K. Ekert,
``Quantum cryptography based on Bell's theorem'',
{\em Phys. Rev. Lett.} {\bf 67}, 6, 661-663 (1991).

\item {\bf [Ekert 91 b]}:
A. K. Ekert,
``La m\'{e}canique quantique au secours des agents secrets'',
{\em La Recherche} {\bf 22}, 233, 790-791 (1991).
Spanish version:
``La mec\'{a}nica cu\'{a}ntica en auxilio de los agentes secretos'',
{\em Mundo Cient\'{\i}fico} {\bf 11}, 116, 886-887 (1991).

\item {\bf [Ekert-Phoenix 91]}:
A. K. Ekert, \& S. J. D. Phoenix,
``The evolution of correlations in a dissipative system'',
{\em J. Mod. Opt.} {\bf 38}, 1, 19-29 (1991).

\item {\bf [Ekert 92]}:
A. K. Ekert,
``Beating the code breakers'',
{\em Nature} {\bf 358}, 6381, 14-15 (1992).
See {\bf [Bennett 92]}.

\item {\bf [Ekert-Rarity-Tapster-Palma 92]}:
A. K. Ekert, J. G. Rarity, P. R. Tapster, \& G. M. Palma,
``Practical quantum cryptography based on two-photon interferometry'',
{\em Phys. Rev. Lett.} {\bf 69}, 9, 1293-1295 (1992).

\item {\bf [Ekert 93]}:
A. K. Ekert,
``Quantum keys for keeping secrets'',
{\em New Scientist}, 16 January 1993.

\item {\bf [Ekert 94]}:
A. K. Ekert,
``Shannon's theorem revisited'',
{\em Nature} {\bf 367}, ?, 513-514 (1994).

\item {\bf [Ekert-Huttner-Palma-Peres 94]}:
A. K. Ekert, B. Huttner, G. M. Palma, \& A. Peres,
``Eavesdropping on quantum-cryptographical systems'',
{\em Phys. Rev. A} {\bf 50}, 2, 1047-1056 (1994).

\item {\bf [Ekert-Huttner 94]}:
A. K. Ekert, \& B. Huttner,
``Eavesdropping techniques in quantum cryptosystems'',
{\em J. Mod. Opt.} {\bf 41}, 2455-2466 (2002).

\item {\bf [Ekert-Palma 94]}:
A. K. Ekert, \& G. M. Palma,
``Quantum cryptography with interferometric quantum entanglement'',
{\em J. Mod. Opt.} {\bf 41}, 12 (Special
issue: Quantum communication), 2413-2423 (1994).

\item {\bf [Ekert-Knight 95]}:
A. K. Ekert, \& P. L. Knight,
``Entangled quantum systems and the Schmidt decomposition'',
{\em Am. J. Phys.} {\bf 63}, 5, 415-423 (1995).

\item {\bf [Ekert-Jozsa 96]}:
A. K. Ekert, \& R. Jozsa,
``Quantum computation and Schor's factoring algorithm'',
{\em Rev. Mod. Phys.} {\bf 68}, 3, 733-753 (1996).

\item {\bf [Ekert-Macchiavello 96]}:
A. K. Ekert, \& C. Macchiavello,
``Quantum error correction for communication'',
{\em Phys. Rev. Lett.} {\bf 77}, 12, 2585-2588 (1996).

\item {\bf [Ekert-Jozsa 98]}:
A. K. Ekert, \& R. Jozsa,
``Quantum algorithms: Entanglement enhanced information processing'',
in A. K. Ekert, R. Jozsa, \& R. Penrose (eds.),
{\em Quantum Computation: Theory and Experiment.
Proceedings of a Discussion Meeting held at the Royal
Society of London on 5 and 6 November 1997},
{\em Philos. Trans. R. Soc. Lond. A} {\bf 356}, 1743, 1769-1782 (1998).

\item {\bf [Ekert 98 a]}:
A. K. Ekert,
``From quantum code-making to quantum code-breaking'',
in S. A. Huggett, L. J. Mason, K. P. Tod, S. T. Tsou, \& N. M. J. Woodhouse (eds.),
{\em The geometric universe: Science, geometry and the work of Roger Penrose},
Oxford University Press, Oxford, 1998, pp.~195-214.

\item {\bf [Ekert 98 b]}:
A. K. Ekert,
``Quantum interferometers as quantum computers'',
in E. B. Karlsson, \& E. Br\"{a}ndas (eds.),
{\em Proc.\ of the 104th Nobel Symp.\ ``Modern Studies of Basic Quantum Concepts and Phenomena'' (Gimo, Sweden, 1997)},
{\em Physica Scripta} {\bf T76}, 218-222 (1998);

\item {\bf [Ekert-Macchiavello 98]}:
A. K. Ekert, \& C. Macchiavello,
``An overview of quantum computing'',
in C. S. Calude, J. Casti, \& M. J. Dinneen (eds.),
{\em Unconventional models of computation},
Springer, Singapore, 1988, pp.~19-44.
Reprinted in {\bf [Macchiavello-Palma-Zeilinger 00]}, pp.~66-85.

\item {\bf [Ekert-Ericsson-Hayden-(+4) 00]}:
A. K. Ekert, M. Ericsson, P. Hayden, H. Inamori,
J. A. Jones, D. K. L. Oi, \& V. Vedral,
``Geometric quantum computation'',
{\em J. Mod. Opt.};
quant-ph/0004015.

\item {\bf [Ekert 00 a]}:
A. K. Ekert,
``Quantum algorithms'',
{\bf [Macchiavello-Palma-Zeilinger 00]}, pp.~63-65.

\item {\bf [Ekert 00 b]}:
A. K. Ekert,
``Coded secrets cracked open'',
{\em Phys. World} {\bf 13}, 2, 39-40 (2000).

\item {\bf [Ekert-Hayden-Inamori 00]}:
A. K. Ekert, P. Hayden, \& H. Inamori,
``Basic concepts in quantum computation'',
quant-ph/0011013.

\item {\bf [Ekert 01]}:
A. K. Ekert,
``Quantum entanglement and secrecy'',
{\em Int. J. Mod. Phys. A} {\bf 16}, 19, 3191-3202 (2001).

\item {\bf [Ekert-Hayden-Inamori-Oi 01]}:
A. K. Ekert, P. Hayden, H. Inamori, \& D. K. L. Oi,
``What is quantum computation?'',
{\em Int. J. Mod. Phys. A} {\bf 16}, 20, 3335-3363 (2001).

\item {\bf [Ekert 02]}:
A. K. Ekert,
``Secret sides of Bell's theorem'',
in {\bf [Bertlmann-Zeilinger 02]}, pp.~209-220.

\item {\bf [Ekert-Alves-Oi-(+3) 02]}:
A. K. Ekert, C. M. Alves, D. K. L. Oi,
M. Horodecki, P. Horodecki, \& L. C. Kwek,
``Direct estimations of linear and nonlinear functionals of a quantum state'',
{\em Phys. Rev. Lett.} {\bf 88}, 21, 217901 (2002);
quant-ph/0112073;
quant-ph/0203016.

\item {\bf [Ekstrand 85]}:
K. E. Ekstrand,
in ``Reality and the quantum theory'',
{\em Phys. Today} {\bf 38}, 11, 136 (1985).
Comment on {\bf [Mermin 85]}.

\item {\bf [Elbaz 98]}:
E. Elbaz,
{\em Quantique},
Ellipses, Paris, 1995.
English version:
{\em Quantum. The quantum theory of particles, fields, and cosmology},
Springer-Verlag, Berlin, 1998.

\item {\bf [Elby 90 a]}:
A. Elby,
``Nonlocality and Gleason's lemma. Part 2. Stochastic theories'',
{\em Found. Phys.} {\bf 20}, 11, 1389-1397 (1990).
See {\bf [Brown-Svetlichny 90]} (I).

\item {\bf [Elby 90 b]}:
A. Elby,
``On the physical interpretation of Heywood and
Redhead's algebraic impossibility theorem'',
{\em Found. Phys. Lett.} {\bf 3}, 3, 239-247 (1990).

\item {\bf [Elby 90 c]}:
A. Elby,
``Critique of Home and Sengupta's derivation of a Bell inequality'',
{\em Found. Phys. Lett.} {\bf 3}, 4, 317-324 (1990).
Comment on {\bf [Home-Sengupta 84]}.
Comment: {\bf [Home-Sengupta 91]}.
See {\bf [Elby 91]}.

\item {\bf [Elby 91]}:
A. Elby,
``Reply: How is Home and Sengupta's
noncontextuality condition related to locality'',
{\em Found. Phys. Lett.} {\bf 4}, 5, 455-457 (1991).
Reply to {\bf [Home-Sengupta 91]}.
See {\bf [Home-Sengupta 84]}, {\bf [Elby 90 c]}.

\item {\bf [Elby-Foster 92]}:
A. Elby, \& S. Foster,
``Why SQUID experiments can rule out non-invasive measurability'',
{\em Phys. Lett. A} {\bf 166}, 1, 17-23 (1992).

\item {\bf [Elby-Jones 92]}:
A. Elby, \& M. R. Jones,
``Weakening the locality conditions in algebraic nonlocality proofs'',
{\em Phys. Lett. A} {\bf 171}, 1-2, 11-16 (1992).

\item {\bf [Elby 93 a]}:
A. Elby,
`Why ``modal'' interpretations of quantum mechanics
don't solve the measurement problem',
{\em Found. Phys. Lett.} {\bf 6}, 1, 5-19 (1993).
See {\bf [Dieks 94]}.

\item {\bf [Elby 93 b]}:
A. Elby,
``Bell's other theorem and its connection with nonlocality, part 2'',
in A. van der Merwe, \& F. Selleri (eds.),
{\em Bell's theorem and
the foundations of modern physics.
Proc.\ of an international conference (Cesena, Italy, 1991)},
World Scientific, Singapore, 1993, pp.~184-193.

\item {\bf [Elby 93 c]}:
A. Elby,
``Why local realistic theories violate, nontrivially, the
quantum mechanical EPR perfect correlations'',
{\em Brit. J. Philos. Sci.} {\bf 44}, 2, 213-230 (1993).

\item {\bf [Elby-Brown-Foster 93]}:
A. Elby, H. R. Brown, \& S. Foster,
`What makes a theory physically ``complete''?',
{\em Found. Phys.} {\bf 23}, 7, 971-985 (1993).

\item {\bf [Elby 94 a]}:
A. Elby,
``Decoherence and Zurek's existential intrpretation of quantum mechanics'',
in P. Busch, P. J. Lahti, \& P. Mittelstaedt (eds.),
{\em Symp.\ on the Foundations of Modern Physics
(Cologne, Germany, 1993)},
World Scientific, Singapore, 1994, pp.~?-?.

\item {\bf [Elby 94 b]}:
A. Elby,
`The ``docoherence'' approach to the measurement
problem in quantum mechanics',
in D. Hull, M. Forbes, \& R. Burian (eds.),
{\em Proc.\ of the 1994 Biennial Meeting of
the Philosophy of Science Association},
East Lansing, Michigan, 1994, vol. 1, pp.~?-?.

\item {\bf [Elby-Bub 94]}:
A. Elby, \& J. Bub,
``Triorthogonal uniqueness theorem
and its relevance to the interpretation of quantum mechanics'',
{\em Phys. Rev. A} {\bf 49}, 5, 4213-4216 (1994).
See {\bf [Peres 95 c]}.

\item {\bf [Eldar-Forney 00]}:
Y. C. Eldar, \& G. D. Forney, Jr.,
``On quantum detection and the square-root measurement'',
{\em IEEE Trans. Inf. Theory},
quant-ph/0005132.

\item {\bf [Eldar 03]}:
Y. C. Eldar,
``Mixed-quantum-state detection with inconclusive results'',
{\em Phys. Rev. A} {\bf 67}, 4, 042309 (2003).

\item {\bf [Eldar-Stojnic-Hassibi 04]}:
Y. C. Eldar, M. Stojnic, \& B. Hassibi,
``Optimal quantum detectors for unambiguous detection of mixed states'',
{\em Phys. Rev. A} {\bf 69}, 6, 062318 (2004).

\item {\bf [Elitzur-Popescu-Rohrlich 92]}:
A. Elitzur, S. Popescu, \& D. Rohrlich,
``Quantum nonlocality for each pair in an ensemble'',
{\em Phys. Lett. A} {\bf 162}, 1, 25-28 (1992).

\item {\bf [Elitzur-Vaidman 93 a]}:
A. Elitzur, \& L. Vaidman,
``Is it possible to know about something without ever interacting with it?'',
{\em Vist. Astr.} {\bf 37}, 253-? (1993)

\item {\bf [Elitzur-Vaidman 93 b]}:
A. Elitzur, \& L. Vaidman,
``Quantum mechanical interaction-free measurements'',
{\em Found. Phys.} {\bf 23}, 7, 987-997 (1993);
hep-th/9305002.
See {\bf [Vaidman 94 b, c]}.

\item {\bf [Elitzur-Dolev 01]}:
A. C. Elitzur, \& S. Dolev,
``Nonlocal effects of partial measurements and quantum erasure'',
{\em Phys. Rev. A} {\bf 63}, 6, 062109 (2001);
quant-ph/0012091.

\item {\bf [Elizalde 00]}:
E. Elizalde,
``Quantum deletion is possible via a partial randomization
procedure'',
quant-ph/0007035.

\item {\bf [Ellinas-Floratos 99]}:
D. Ellinas, \& E. G. Floratos,
``Prime decomposition and correlation measure of
finite quantum systems'',
{\em J. Phys. A} {\bf 32}, 5, L63-L69 (1999).

\item {\bf [Ellinas-Pachos 01 a]}:
D. Ellinas, \& J. Pachos,
``Universal quantum computation by holonomic
and nonlocal gates with imperfections'',
{\em Phys. Rev. A} {\bf 64}, 2, 022310 (2001);
quant-ph/0009043.

\item {\bf [Ellinas-Konstadakis 01]}:
D. Ellinas, \& C. Konstadakis,
``Noisy Grover's search algorithm'',
{\em ICQI, 2001};
quant-ph/0110010.

\item {\bf [Ellinas-Pachos 01 b]}:
D. Ellinas, \& J. Pachos,
``Optical holonomic quantum computation'',
{\em ICQI, 2001};
quant-ph/0111075.

\item {\bf [Ellis-Amati 00]}:
J. Ellis, \& D. Amati (eds.),
{\em Quantum reflections},
Cambridge University Press, Cambridge, 2000.
Review: {\bf [Greenberger 00 a]}, {\bf [Pearle 02]}.

\item {\bf [Elliot 02]}:
C. Elliot,
``Building the quantum network'',
{\em New J. Phys} {\bf 4}, 46.1-46.12 (2002).

\item {\bf [Elliot-Pearson-Troxel 03]}:
C. Elliott, D. Pearson, \& G. Troxel,
``Quantum cryptography in practice'',
{\em SIGCOMM 2003};
quant-ph/0307049.

\item {\bf [Emary-Beenakker 04]}:
C. Emary, \& C. W. J. Beenakker,
``Relation between entanglement measures and Bell inequalities for three qubits'',
{\em Phys. Rev. A} {\bf 69}, 3, 032317 (2004);
quant-ph/0311105.

\item {\bf [Emerson-Ballentine 01]}:
J. Emerson, \& L. E. Ballentine,
``Characteristics of quantum-classical correspondence
for two interacting spins'',
{\em Phys. Rev. A} {\bf 63}, 5, 052103 (2001).

\item {\bf [Emerson-Weinstein-Lloyd-Cory 02]}:
J. Emerson, Y. S. Weinstein, S. Lloyd, \& D. G. Cory,
``Fidelity decay as an efficient indicator of quantum chaos'',
{\em Phys. Rev. Lett.} {\bf 89}, 28, 284102 (2002).

\item {\bf [Emerson-Weinstein-Saraceno-(+2) 03]}:
J. Emerson, Y. S. Weinstein, M. Saraceno,
S. Lloyd, \& D. G. Cory,
``Pseudo-random unitary operators for quantum information processing'',
{\em Science} {\bf 302}, ?, 2098-2101 (2003).
See {\bf [Emerson 04]}.

\item {\bf [Emerson-Lloyd-Poulin-Cory 03]}:
J. Emerson, S. Lloyd, D. Poulin, \& D. Cory,
``Estimation of the local density of states on a quantum computer'',
{\em Phys. Rev. A} {\bf 69}, 5, 050305 (2004);
quant-ph/0308164.

\item {\bf [Emerson 04]}:
J. Emerson,
``Random quantum circuits and pseudo-random operators: Theory and
applications'',
quant-ph/0410087.
See {\bf [Emerson-Weinstein-Saraceno-(+2) 03]}.

\item {\bf [Endo 03]}:
T. Endo,
``Verification of Born's rule by a quantum mechanical meter'',
{\em Phys. Lett. A} {\bf 308}, 4, 256-258 (2003).

\item {\bf [Englert-Schwinger-Scully 88]}:
B.-G. Englert, J. Schwinger, \& M. O. Scully,
``Is spin coherence like Humpty-Dumpty? I. Simplified treatment'',
{\em Found. Phys.} {\bf 18}, 10, 1045-56 (1988).
See {\bf [Schwinger-Scully-Englert 88]} (II),
{\bf [Scully-Englert-Schwinger 89]} (III).

\item {\bf [Englert-Walther-Scully 92]}:
B.-G. Englert, H. Walther, \& M. O. Scully,
``Quantum optical Ramsey fringes and complementarity'',
{\em Appl. Phys. B} {\bf 54}, 5, 366-368 (1992).

\item {\bf [Englert-Scully-S\"{u}ssmann-Walther 93 a]}:
B.-G. Englert, M. O. Scully, G. S\"{u}ssmann, \& H. Walther,
``Surrealistic Bohm trajectories'',
{\em Zeitschrift f\"{u}r Naturforschung A} {\bf 47}, ?, 1175-1186 (1993).
Comment: {\bf [D\"{u}rr-Fusseder-Goldstein-Zangh\`{\i} 93]}.
Reply: {\bf [Englert-Scully-S\"{u}ssmann-Walther 93 b]}.
See {\bf [Terra Cunha 98]}.

\item {\bf [Englert-Scully-S\"{u}ssmann-Walther 93 a]}:
B.-G. Englert, M. O. Scully, G. S\"{u}ssmann, \& H. Walther,
``Reply to comment on `Surrealistic Bohm trajectories'\,'',
{\em Zeitschrift f\"{u}r Naturforschung A} {\bf 46}, ?, 1263-1264 (1993).
Reply to {\bf [D\"{u}rr-Fusseder-Goldstein-Zangh\`{\i} 93]}.
See {\bf [Englert-Scully-S\"{u}ssmann-Walther 93 b]}.

\item {\bf [Englert-Fearn-Scully-Walther 94]}:
B.-G. Englert, H. Fearn, M. O. Scully, \& H. Walther,
``?'',
in F. De Martini, G. Denardo, \& A. Zeilinger (eds.),
{\em Quantum interferometry},
World Scientific, Singapore, 1994, pp.~103-119.
See {\bf [Storey-Tan-Collett-Walls 94 a, b, 95]},
{\bf [Englert-Scully-Walther 95]},
{\bf [Wiseman-Harrison 95]}.

\item {\bf [Englert-Scully-Walther 94]}:
B.-G. Englert, M. O. Scully, \& H. Walther,
``The duality in matter and light'',
{\em Sci. Am.} {\bf 269}, 6, 56-61 (1994).
Spanish version: ``La dualidad en la materia y en la luz'',
{\em Investigaci\'{o}n y Ciencia} 221, 46-52 (1995).
Reprinted in {\bf [Cabello 97 c]}, pp.~68-74.
See {\bf [Scully-Englert-Walther 91]}, {\bf [Mohrhoff 96]}.

\item {\bf [Englert-Scully-Walther 95]}:
B.-G. Englert, M. O. Scully, \& H. Walther,
``Complementarity and uncertainty'',
{\em Nature} {\bf 375}, 6530, 367-368 (1995).
See {\bf [Englert-Fearn-Scully-Walther 94]},
{\bf [Storey-Tan-Collett-Walls 94, 95]},
{\bf [Wiseman-Harrison 95]}.

\item {\bf [Englert 96 a]}:
B.-G. Englert,
``?'',
{\em Acta Phys. Slov.} {\bf 46}, ?, 249-? (1996).

\item {\bf [Englert 96 b]}:
B.-G. Englert,
``Fringe visibility and which-way information: An inequality'',
{\em Phys. Rev. Lett.} {\bf 77}, 11, 2154-2157 (1996).
See {\bf [Mart\'{\i}nez Linares-Harmin 98]}.

\item {\bf [Englert-L\"{o}ffler-Benson-(+3) 98]}:
B.-G. Englert, M. L\"{o}ffler, O. Benson,
B. Varcoe, M. Weidinger, \& H. Walther,
``Entangled atoms in micromaser physics'',
{\em Fortschr. Phys.} {\bf 46}, 6-8, 897-926 (1998).

\item {\bf [Englert-Scully-Walther 99 a]}:
B.-G. Englert, M. O. Scully, \& H. Walther,
``Quantum erasure in double-slit interferometers
with which-way detectors'',
{\em Am. J. Phys.} {\bf 67}, 4, 325-329 (1999).
See {\bf [Mohrhoff 99]}.

\item {\bf [Englert 98]}:
B.-G. Englert,
``Classical analogs of unitarily equivalent Hamilton operators'',
{\em Found. Phys.} {\bf 28}, 3, 375-384 (1998).

\item {\bf [Englert 99]}:
B.-G. Englert,
``Book review. Quantum optics'',
{\em Found. Phys.} {\bf 29}, 5, 829-831 (1999).
Review of {\bf [Scully-Zubairy 97]}.

\item {\bf [Englert-Scully-Walther 99 b]}:
B.-G. Englert, M. O. Scully, \& H. Walther,
``On mechanisms that enforce complementarity'',
quant-ph/9910037.

\item {\bf [Englert-Metwally 99]}:
B.-G. Englert, \& N. Metwally,
``Separability of entangled q-bit pairs'',
quant-ph/9912089.

\item {\bf [Englert-Metwally 00]}:
B.-G. Englert, \& N. Metwally,
``Remarks on 2-q-bit states'',
quant-ph/0007053.

\item {\bf [Englert-Kurtsiefer-Weinfurter 01]}:
B.-G. Englert, C. Kurtsiefer, \& H. Weinfurter,
``Universal unitary gate for single-photon two-qubit states'',
{\em Phys. Rev. A} {\bf 63}, 3, 032303 (2001);
quant-ph/0101064.

\item {\bf [Englert-Aharonov 01]}:
B.-G. Englert, \& Y. Aharonov,
``The mean king's problem: Prime degrees of freedom'',
{\em Phys. Lett. A} {\bf 284}, 1, 1-5 (2001);
quant-ph/0101134.
See {\bf [Aharonov-Englert 01]}.

\item {\bf [Englert-W\'{o}dkiewicz 02]}:
B.-G. Englert, \& K. W\'{o}dkiewicz,
``Separability of two-party Gaussian states'',
{\em Phys. Rev. A} {\bf 65}, 5, 054303 (2002);
quant-ph/0107131.

\item {\bf [Englert-W\'{o}dkiewicz 03]}:
B.-G. Englert, \& K. W\'{o}dkiewicz,
``Tutorial notes on one-party and two-party Gaussian states'',
{\em Int. J. Quant. Inf.} {\bf 1}, 2, 153-188 (2003);
quant-ph/0307196.

\item {\bf [Ensslin 00]}:
K. Ensslin,
``The spin degree of freedom in quantum dots'',
{\em Fortschr. Phys.} {\bf 48}, 9-11 (Special issue:
Experimental proposals for quantum computation), 999-1004 (2000).

\item {\bf [Enz 02]}:
C. P. Enz,
{\em No time to be brief: A scientific biography of Wolfgang Pauli},
Oxford University Press, Oxford, 2002.
Review: {\bf [Dyson 02]}, {\bf [Telegdi 03]}.

\item {\bf [Enzer-Hadley-Hughes-(+2) 02]}:
D. G. Enzer, P. G. Hadley, R. J. Hughes,
C. G. Peterson, \& P. G. Kwiat,
``Entangled-photon six-state quantum cryptography'',
{\em New J. Phys} {\bf 4}, 45.1-45.8 (2002).

\item {\bf [Epstein 45]}:
P. S. Epstein,
``?'',
{\em Am. J. Phys.} {\bf 13}, ?, 127-? (1945).

\item {\bf [Erez-Aharonov-Reznik-Vaidman 04]}:
N. Erez, Y. Aharonov, B. Reznik, \& L. Vaidman,
``Quantum error correction with the Zeno effect'',
{\em Phys. Rev. A} {\bf 69}, 6, 062315 (2004);
quant-ph/0309162.

\item {\bf [Ericsson 02]}:
\AA. Ericsson
``Separability and the stella octangula'',
{\em Phys. Lett. A} {\bf 295}, 5-6, 256-258 (2002);
quant-ph/0109099.

\item {\bf [Ericsson-Sj\"{o}qvist 02]}:
M. Ericsson, \& E. Sj\"{o}qvist,
``Quantum computation using the Aharonov-Casher set up'',
{\em Phys. Lett. A} {\bf 303}, 1, 7-10 (2002);
quant-ph/0209006.

\item {\bf [Ericsson-Pati-Sj\"{o}qvist-(+2) 03]}:
M. Ericsson, A. K. Pati, E. Sj\"{o}qvist,
J. Brannlund, \& D. K. L. Oi,
``Mixed state geometric phases, entangled systems, and local unitary
transformations'',
{\em Phys. Rev. Lett.} {\bf 91}, 9, 090405 (2003).

\item {\bf [Erlichson 72]}:
H. Erlichson,
``Bohr and the Einstein-Podolsky-Rosen paradox'',
{\em Am. J. Phys.} {\bf 40}, 4, 634-636 (1972).
Comment on {\bf [Hooker 70]}.

\item {\bf [Ermakov-Fung 02]}:
V. L. Ermakov, \& B. M. Fung,
``Experimental realization of a continuous version of the Grover algorithm'',
{\em Phys. Rev. A} {\bf 66}, 4, 042310 (2002).

\item {\bf [Ermakov-Fung 03]}:
V. L. Ermakov, \& B. M. Fung,
``Nuclear magnetic resonance implementation of the Deutsch-Jozsa algorithm
using different initial states'',
{\em J. Chem. Phys.};
quant-ph/0304058.

\item {\bf [Eschner-Raab-Schmidt Kaler-Blatt 01]}:
J. Eschner, C. Raab, F. Schmidt-Kaler, \& R. Blatt,
``Light interference from single atoms and their mirror images'',
{\em Nature} {\bf 413}, 6855, 495-498 (2001).

\item {\bf [Esfeld 99]}:
M. Esfeld,
``Wigner's view of physical reality'',
{\em Stud. Hist. Philos. Sci. Part B:
Stud. Hist. Philos. Mod. Phys.} {\bf 30}, 1, 145-154 (1999).
Review of {\bf [Wigner 97 b]}.

\item {\bf [d'Espagnat 71]}:
B. d'Espagnat (ed.),
{\em Foundations of quantum mechanics.
Proc.\ of the Int.
School of Physics ``Enrico Fermi''. Course IL:
Foundations of Quantum Mechanics (Varenna, Italy, 1970)},
Academic Press, New York, 1971.

\item {\bf [d'Espagnat 72]}:
B. d'Espagnat,
{\em Foundations of quantum mechanics},
Academic Press, New York, 1972.

\item {\bf [d'Espagnat 75]}:
B. d'Espagnat,
``Use of inequalities for the experimental
test of a general conception of the foundation of microphysics'',
{\em Phys. Rev. D} {\bf 11}, 6, 1424-1435 (1975).
See {\bf [d'Espagnat 78]} (II).

\item {\bf [d'Espagnat 76]}:
B. d'Espagnat,
{\em Conceptual foundations of quantum mechanics},
Benjamin, Reading, Massachusetts, 1976;
Addison-Wesley, Reading, Massachusetts, 1994.
Review: {\bf [Hiley 77]}.

\item {\bf [d'Espagnat 78]}:
B. d'Espagnat,
``Use of inequalities for the experimental
test of a general conception of the foundations of microphysics. II'',
{\em Phys. Rev. D} {\bf 18}, 2, 349-358 (1978).
See {\bf [d'Espagnat 75]} (I).

\item {\bf [d'Espagnat 79]}:
B. d'Espagnat,
``The quantum theory and reality'',
{\em Sci. Am.} {\bf 241}, 11, 158-181 (1979).
Spanish version: ``Teor\'{\i}a cu\'{a}ntica y realidad'',
{\em Investigaci\'{o}n y Ciencia} 40, 80-95 (1980).
Reprinted in {\bf [Cabello 97 c]}, pp.~13-27.

\item {\bf [d'Espagnat 81]}:
B. d'Espagnat,
``The concepts of influences and of attributes as seen in connection with
Bell's theorem'',
{\em Found. Phys.} {\bf 11}, 3-4, 205-234 (1981).

\item {\bf [d'Espagnat 84]}:
B. d'Espagnat,
``Nonseparability and the tentative descriptions of reality'',
{\em Phys. Rep.} {\bf 110}, 4, 201-264 (1984).

\item {\bf [d'Espagnat 85]}:
B. d'Espagnat,
{\em Une incertaine r\'{e}alit\'{e}},
Bordas, Paris, 1985.
English version: {\em Reality and the physicist: Knowledge,
duration and the quantum world},
Cambridge University Press, Cambridge, 1989.

\item {\bf [d'Espagnat 87]}:
B. d'Espagnat,
``Meaning and being in contemporary physics'',
in {\bf [Hiley-Peat 87]}, pp.~151-168.

\item {\bf [d'Espagnat 88]}:
B. d'Espagnat,
``Are the quantum rules exact? The case of imperfect measurements'',
in {\bf [Zurek-van der Merwe-Miller 88]}, pp.~413-423.

\item {\bf [d'Espagnat 89 a]}:
B. d'Espagnat,
{\em Reality and the physicist},
Cambridge University Press, Cambridge, 1989.

\item {\bf [d'Espagnat 89 b]}:
B. d'Espagnat,
``Nonseparability and the tentative descriptions of reality'',
in W. Schommers (ed.),
{\em Quantum theory and pictures of reality},
Springer-Verlag, Berlin, 1989, pp.~89-168.

\item {\bf [d'Espagnat 93]}:
B. d'Espagnat,
``One or two Bell theorems?'',
in A. van der Merwe, \& F. Selleri (eds.),
{\em Bell's theorem and the foundations of modern
physics. Proc.\ of an international conference (Cesena, Italy, 1991)},
World Scientific, Singapore, 1993, pp.~139-146.

\item {\bf [d'Espagnat 95]}:
B. d'Espagnat,
{\em Veiled reality, an analysis of present-day quantum mechanical concepts},
Addison-Wesley, Reading, Massachusetts, 1995.

\item {\bf [d'Espagnat 97]}:
B. d'Espagnat,
``Aiming at describing empirical reality'',
in {\bf [Cohen-Horne-Stachel 97 b]}.

\item {\bf [d'Espagnat 98 a]}:
B. d'Espagnat,
``Quantum theory: A pointer to an independent reality'',
quant-ph/9802046.

\item {\bf [d'Espagnat 98 b]}:
B. d'Espagnat,
``Reply to Aharonov and Anandan's `Meaning of the density matrix'\,'',
quant-ph/9804063.
See {\bf [Aharonov-Anandan 98]}.

\item {\bf [d'Espagnat 01]}:
B. d'Espagnat,
``A note on measurement'',
{\em Phys. Lett. A} {\bf 282}, 3, 133-137 (2001);
quant-ph/0101141.
See {\bf [Bassi-Ghirardi 00 c]}.
Comment: {\bf [Mohrhoff 01 d]}.

\item {\bf [d'Espagnat 01]}:
B. d'Espagnat,
``Reply to K. A. Kirkpatrick'',
quant-ph/0111081.

\item {\bf [d'Espagnat 02]}:
B. d'Espagnat,
``My Interaction with John Bell'',
in {\bf [Bertlmann-Zeilinger 02]}, pp.~21-28.

\item {\bf [d'Espagnat 03]}:
B. d'Espagnat,
``On the Unnikrishnan approach to the notion of locality'',
quant-ph/0302167.
Comment on {\bf [Unnikrishnan 02]}.

 \item {\bf [d'Espagnat 04]}:
``Consciousness and the Wigner's friend problem'',
quant-ph/0402121.

\item {\bf [Ettinger-H\o{}yer 99]}:
J. M. Ettinger, \& P. H\o{}yer,
``Quantum state detection via elimination'',
quant-ph/9905099.

\item {\bf [Ettinger-H\o{}yer-Knill 04]}:
J. M. Ettinger, P. H\o{}yer, \& E. Knill,
``The quantum query complexity of the hidden subgroup problem is
polynomial'',
{\em Inf. Processing Lett.};
quant-ph/0401083

\item {\bf [Everett 57 a]}:
H. Everett III,
``The theory of the universal wave function'',
Ph.\ D. thesis, Princeton University, 1957.
Reprinted in {\bf [DeWitt-Graham 73]}.

\item {\bf [Everett 57 b]}:
H. Everett III,
`\,``Relative state'' formulation of quantum mechanics',
{\em Rev. Mod. Phys.} {\bf 29}, 3, 454-462 (1957).
Reprinted in {\bf [DeWitt-Graham 73]}.
Reprinted in {\bf [Wheeler-Zurek 83]}, pp.~315-323.
See {\bf [Wheeler 57]}.

\item {\bf [Everett 63]}:
H. Everett III,
``The theory of the universal wave function'',
in {\bf [DeWitt-Graham 73]}, pp.~3-140.
See {\bf [DeWitt 63]}.

\item {\bf [Everitt-Clark-Stiffell-(+4) 04]}:
M. J. Everitt, T. D. Clark, P. B. Stiffell,
A. Vourdas, J. F. Ralph, R. J. Prance, \& H. Prance,
``Superconducting analogs of quantum optical phenomena:
Macroscopic quantum superpositions and squeezing in
a superconducting quantum-interference device ring'',
{\em Phys. Rev. A} {\bf 69}, 4, 043804 (2004).

\item {\bf [Everitt-Clark-Stiffell-(+3) 04]}:
M. J. Everitt, T. D. Clark, P. B. Stiffell,
J. F. Ralph, A. R. Bulsara, \& C. J. Harland,
``Persistent entanglement in the classical limit'',
quant-ph/0409112.


\newpage

\subsection{}


\item {\bf [Facchi-Pascazio 98]}:
P. Facchi, \& S. Pascazio,
``Temporal behavior and quantum Zeno time of an excited state of the hydrogen atom'',
{\em Phys. Lett. A} {\bf 241}, 3, 139-144 (1998).

\item {\bf [Facchi-Gorini-Marmo-(+2) 00]}:
P. Facchi, V. Gorini, G. Marmo,
S. Pascazio, \& E. C. G. Sudarshan,
``Quantum Zeno dynamics'',
{\em Phys. Lett. A} {\bf 275}, 1-2, 12-19 (2000);
quant-ph/0004040.

\item {\bf [Facchi-Nakazato-Pascazio 01]}:
P. Facchi, H. Nakazato, \& S. Pascazio,
``From the quantum Zeno to the inverse quantum Zeno effect'',
{\em Phys. Rev. Lett.} {\bf 86}, 13, 2699-2703 (2001);
quant-ph/0006094.

\item {\bf [Facchi-Mariano-Pascazio 01 a]}:
P. Facchi, A. Mariano, \& S. Pascazio,
``Decoherence versus entropy in neutron interferometry'',
{\em Phys. Rev. A} {\bf 63}, 5, 052108 (2001);
quant-ph/9906118.

\item {\bf [Facchi-Mariano-Pascazio 01 b]}:
P. Facchi, A. Mariano, \& S. Pascazio,
``Mesoscopic interference'',
quant-ph/0105110.

\item {\bf [Facchi-Pascazio 01]}:
P. Facchi, \& S. Pascazio,
``Quantum Zeno phenomena: Pulsed versus
continuous measurement'',
quant-ph/0106026.

\item {\bf [Facchi-Pascazio 02]}:
P. Facchi, \& S. Pascazio,
``Quantum Zeno subspaces'',
{\em Phys. Rev. Lett.} {\bf 89}, 8, 080401 (2002);
quant-ph/0201115.

\item {\bf [Facchi 02]}:
P. Facchi,
``Quantum Zeno effect, adiabaticity and dynamical superselection rules'',
quant-ph/0202174.

\item {\bf [Facchi-Pascazio 03]}:
P. Facchi, \& S. Pascazio,
``Three different manifestations of the quantum Zeno effect'',
quant-ph/0303161.

\item {\bf [Facchi-Lidar-Pascazio 04]}:
P. Facchi, D. A. Lidar, \& S. Pascazio,
``Unification of dynamical decoupling and the quantum Zeno effect'',
{\em Phys. Rev. A} {\bf 69}, 3, 032314 (2004).

\item {\bf [Facchi-Montangero-Fazio-Pascazio 04]}:
P. Facchi, S. Montangero, R. Fazio, \& S. Pascazio,
``Dynamical imperfections in quantum computers'',
quant-ph/0407098.

\item {\bf [Fahmi-Golshani 03]}:
A. Fahmi,
``Locality, Bell's inequality and the GHZ theorem'',
{\em Phys. Lett. A} {\bf 303}, 1, 1-6 (2002).
Erratum: {\em Phys. Lett. A} {\bf 306}, 5-6, 258 (2003).
Revised version:
A. Fahmi, \& M. Golshani,
``Locality, Bell's inequality and the GHZ theorem'',
{\em Phys. Lett. A} {\bf 306}, 5-6, 259-264 (2003).

\item {\bf [Facchi-Tasaki-Pascazio-(+3) 04]}:
P. Facchi, S. Tasaki, S. Pascazio,
H. Nakazato, A. Tokuse, \& D. A. Lidar,
``Control of decoherence: Analysis and comparison of three different
strategies'',
quant-ph/0403205.

\item {\bf [Fahmy-Jones 98]}:
A. F. Fahmy, \& J. A. Jones,
``Quantum computing'',
{\em Science} {\bf 281}, ?, 1961 (1998).

\item {\bf [Falci-Paladino-Fazio 04]}:
G. Falci, E. Paladino, \& R. Fazio,
``Decoherence in Josephson qubits'',
in B. Altshuler, \& V. Tognetti (eds.),
{\em Quantum Phenomena of Mesoscopic Systems},
{\em Proc. of the Int. School of Physics
``Enrico Fermi'', Course CLI (Varenna, Italy, 2002)},
IOS Press Amsterdam, 2004;
cond-mat/0312550.

\item {\bf [de Falco-Tamascelli 04]}:
D. de Falco, \& D. Tamascelli,
``Grover's algorithm on a Feynman computer'',
{\em J. Phys. A} {\bf 37}, 3, 909-930 (2003).

\item {\bf [Falsaperla-Fonte 03]}:
P. Falsaperla, \& G. Fonte,
``On the motion of a single particle near a nodal line in
the de Broglie–Bohm interpretation of quantum mechanics'',
{\em Phys. Lett. A} {\bf 316}, 6, 382-390 (2003).

\item {\bf [Fan-Zhang 98]}:
H. Fan, \& Y. Zhang,
``Common eigenkets of three-particle compatible observables'',
{\em Phys. Rev. A} {\bf 57}, 5, 3225-3228 (1998).

\item {\bf [Fan-Chen 00]}:
H. Fan, \& Z. Chen,
``Einstein-Podolsky-Rosen pair states and the charge-amplitude
representation for complex scalar fields'',
{\em J. Phys. A} {\bf 33}, 10, 2145-2150 (2000).

\item {\bf [Fan-Chen-Lin 00]}:
H. Fan, Z. Chen, \& J. Lin,
``Demonstration of Einstein-Podolsky-Rosen entanglement
for an electron in a uniform magnetic field'',
{\em Phys. Lett. A} {\bf 272}, 1-2, 20-25 (2000).

\item {\bf [Fan-Matsumoto-Wadati 01 a]}:
H. Fan, K. Matsumoto, \& M. Wadati,
``Quantum cloning machines of a $d$-level system'',
{\em Phys. Rev. A} {\bf 64}, 6, 064301 (2001);
quant-ph/0103053.

\item {\bf [Fan-Matsumoto-Wadati 01 b]}:
H. Fan, K. Matsumoto, \& M. Wadati,
``Optimal quantum cloning networks for equatorial qubits`'',
quant-ph/0101101.

\item {\bf [Fan-Matsumoto-Wang-(+2) 01]}:
H. Fan, K. Matsumoto, X.-B. Wang, H. Imai, \& M. Wadati,
``A universal cloner allowing the input to be arbitrary states in symmetric subspace'',
quant-ph/0107113.

\item {\bf [Fan-Matsumoto-Wang-Wadati 02]}:
H. Fan, K. Matsumoto, X.-B. Wang, \& M. Wadati,
``Quantum cloning machines for equatorial qubits'',
{\em Phys. Rev. A} {\bf 65}, 1, 012304 (2002);
quant-ph/0012033.

\item {\bf [Fan-Matsumoto-Wang-Imai 02]}:
H. Fan, K. Matsumoto, X.-B. Wang, \& H. Imai,
``Phase-covariant quantum cloning'',
{\em J. Phys. A} {\bf 35}, 34, 7415-7423 (2002).

\item {\bf [Fan-Cheng 02 a]}:
H. Fan, \& H. Cheng,
``Nonlinear entangled state representation in quantum mechanics'',
{\em Phys. Lett. A} {\bf 295}, 2-3, 65-73 (2002).

\item {\bf [Fan-Cheng 02 b]}:
H. Fan, \& H. Cheng,
``Bipartite entangled state $|\zeta\rangle$ and its applications in quantum communication'',
{\em J. Opt. B: Quantum Semiclass. Opt.} {\bf 4}, 3, 228-234 (2002).

\item {\bf [Fan 02]}:
H. Fan,
``Time evolution of the Wigner function in the entangled-state representation'',
{\em Phys. Rev. A} {\bf 65}, 6, 064102 (2002).

\item {\bf [Fan-Weihs-Matsumoto-Imai 02]}:
H. Fan, G. Weihs, K. Matsumoto, \& H. Imai,
``Cloning of symmetric $d$-level photonic states in physical systems'',
{\em Phys. Rev. A} {\bf 66}, 2, 024307 (2002);
quant-ph/0112094.

\item {\bf [Fan-Imai-Matsumoto-Wang 03]}:
H. Fan, H. Imai, K. Matsumoto, \& X.-B. Wang,
``Phase-covariant quantum cloning of qudits'',
{\em Phys. Rev. A} {\bf 67}, 2, 022317 (2003).

\item {\bf [Fan-Matsumoto-Imai 03]}:
H. Fan, K. Matsumoto, \& H. Imai,
``Quantify entanglement by concurrence hierarchy'',
{\em J. Phys. A} {\bf 36}, 14, 4151-4158 (2003);
quant-ph/0204041.

\item {\bf [Fan 03]}:
H. Fan,
``Remarks on entanglement assisted classical capacity'',
{\em Phys. Lett. A} {\bf 313}, 3, 182-187 (2003).

\item {\bf [Fan-Song 03]}:
H. Fan, \& T. Song,
``Multipartite entangled state of continuum variables generated by an optical network'',
{\em J. Phys. A} {\bf 36}, 28, 7803–7811 (2003).

\item {\bf [Fan-Lloyd 04]}:
H. Fan, \& S. Lloyd,
``Entanglement of eta-pairing state with off-diagonal long-range order'',
quant-ph/0405130.

\item {\bf [Fan-Chen 01]}:
H.-Y. Fan, \& Z.-B. Chen,
``Einstein-Podolsky-Rosen entanglement for self-interacting
complex scalar fields'',
{\em J. Phys. A} {\bf 34}, 9, 1853-1860 (2001).

\item {\bf [Fan 01 a]}:
H.-Y. Fan,
``Entanglement swapping transformation and swapping operator
for two pairs of EPR entangled states with continuous variables'',
{\em Phys. Lett. A} {\bf 286}, 2-3, 81-86 (2001).

\item {\bf [Fan 01 b]}:
H.-Y. Fan,
``Entanglement swapping for two pairs of EPR eigenstates
generated via Bell measurement'',
{\em Mod. Phys. Lett. B} {\bf 15}, 3, 119-126 (2001).

\item {\bf [Fan-Yu 01]}:
H.-Y. Fan, \& G.-C. Yu,
``Resolution of unity in differential form for EPR complex entangled
state representation and its application in nonlinear operator's normal
ordering expansion'',
{\em Mod. Phys. Lett. A} {\bf 16}, 32, 2067-2074 (2001).

\item {\bf [Fan 02]}:
H.-Y. Fan,
``Application of EPR entangled state representation in
quantum teleportation of continuous variables'',
{\em Phys. Lett. A} {\bf 294}, 5-6, 253-257 (2002).

\item {\bf [Fan-Chen 02]}:
H.-Y. Fan, \& J. Chen,
``EPR entangled state and generalized Bargmann transformation'',
{\em Phys. Lett. A} {\bf 303}, 5-6, 311-317 (2002).

\item {\bf [Fan-Fan-Song 02]}:
H.-Y. Fan, Y. Fan, \& T.-Q. Song,
``Quantum theory of mesoscopic electric circuits in entangled state representation'',
{\em Phys. Lett. A} {\bf 305}, 5, 222-230 (2002).

\item {\bf [Fan-Fan 02]}:
H.-Y. Fan, \& Y. Fan,
``Reducing projection calculation in quantum teleportation by virtue of
the IWOP technique and Schmidt decomposition of $\vert \eta\rangle$ state'',
{\em Commun. Theor. Phys. (Beijing)} {\bf 37}, 1, 35-38 (2002).

\item {\bf [Fan-Sun 02 a]}:
H.-Y. Fan, \& M.-Z. Fan,
``Entanglement of the common eigenvector of two particles' center-of-mass
coordinate and mass-weighted relative momentum'',
{\em Commun. Theor. Phys.} {\bf 37}, 5, 535-538 (2002).

\item {\bf [Fan-Sun 02 b]}:
H.-Y. Fan, \& M.-Z. Fan,
``Bipartite state $|\zeta\rangle$ and its applications in quantum communication'',
{\em J. Opt. B: Quantum Semiclass. Opt.} {\bf 4}, 3, 228-234 (2002).

\item {\bf [Fan-Yang-Li-(+2) 01]}:
X.-F. Fan, T. Yang, J. Li, C.-F. Li, \& G.-C. Guo,
``Generating three-particle entanglement states'',
{\em Phys. Lett. A} {\bf 284}, 2-3, 59-62 (2001).

\item {\bf [Fanchi 00]}:
J. R. Fanchi,
``Quantum potential in relativistic dynamics'',
{\em Found. Phys.} {\bf 30}, 8, 1161-1189 (2000).

\item {\bf [Fanchi 01]}:
J. R. Fanchi,
``Nonlocality in relativistic dynamics'',
{\em Found. Phys.} {\bf 31}, 9, 1267-1285 (2001).

\item {\bf [Fang-Chang-Tucker 02]}:
A. Fang, Y. C. Chang, \& J. R. Tucker,
``Effects of $J$-gate potential and uniform electric field on a coupled donor
pair in Si for quantum computing'',
{\em Phys. Rev. B} {\bf 66}, 15, 155331 (2002).

\item {\bf [Fang-Chang 03]}:
A. Fang, \& Y. C. Chang,
``Entanglement and correlation for identical particles in quantum computing'',
{\em Phys. Lett. A} {\bf 311}, 6, 443-458 (2003).

\item {\bf [Fang-Lin-Zhu-Chen 03]}:
J. Fang, Y. Lin, S. Zhu, \& X. Chen,
``Probabilistic teleportation of a three-particle state via three pairs of
entangled particles'',
{\em Phys. Rev. A} {\bf 67}, 1, 014305 (2003).

\item {\bf [Fang-Zhu-Feng-Mao-Du 00]}:
X. Fang, X. Zhu, M. Feng, X. Mao, \& F. Du,
``Experimental implementation of dense coding using
nuclear magnetic resonance'',
{\em Phys. Rev. A} {\bf 61}, 2, 022307 (2000);
quant-ph/9906041.

\item {\bf [Fano 57]}:
U. Fano,
``Description of states in quantum mechanics by
density matrix and operator techniques'',
{\em Rev. Mod. Phys.} {\bf 29}, 1, 74-93 (1957).

\item {\bf [Fano-Racah 59]}:
U. Fano, \& G. Racah,
{\em Irreducible tensorial sets},
Academic Press, New York, 1959.

\item {\bf [Faoro-Siewert-Fazio 03]}:
L. Faoro, J. Siewert, \& R. Fazio,
``Non-Abelian holonomies, charge pumping, and quantum computation with
Josephson junctions'',
{\em Phys. Rev. Lett.} {\bf 90}, 2, 028301 (2003).

\item {\bf [Faoro-Taddei-Fazio 04]}:
L. Faoro, F. Taddei, \& R. Fazio,
``Clauser-Horne inequality for electron-counting statistics in multiterminal mesoscopic conductors'',
{\em Phys. Rev. B} {\bf 69}, 12, 125326 (2004).

\item {\bf [Faoro-Viola 04]}:
L. Faoro, \& L. Viola,
``Dynamical suppression of $1/f$ noise processes in qubit systems'',
{\em Phys. Rev. Lett.} {\bf 92}, 11, 117905 (2004).

\item {\bf [Farhi-Gutmann 97]}:
E. Farhi, \& S. Gutmann,
``Quantum mechanical square root speedup in a structured search problem'',
quant-ph/9711035.

\item {\bf [Farhi-Gutmann 98 a]}:
E. Farhi, \& S. Gutmann,
``Analog analogue of a digital quantum computation'',
{\em Phys. Rev. A} {\bf 57}, 4, 2403-2406 (1998);
quant-ph/9612026.

\item {\bf [Farhi-Gutmann 98 b]}:
E. Farhi, \& S. Gutmann,
``Quantum computation and decision trees'',
{\em Phys. Rev. A} {\bf 58}, 2, 915-? (1998).

\item {\bf [Farhi-Goldstone-Gutmann-Sipser 98]}:
E. Farhi, J. Goldstone, S. Gutmann, \& M. Sipser,
``A limit on the speed of quantum computation for
insertion into an ordered list'',
quant-ph/9812057.

\item {\bf [Farhi-Goldstone-Gutmann-Sipser 99]}:
E. Farhi, J. Goldstone, S. Gutmann, \& M. Sipser,
``Bound on the number of functions that can be distinguished with $k$ quantum queries'',
{\em Phys. Rev. A} {\bf 60}, 6, 4331-4333 (1999);
quant-ph/9901012.

\item {\bf [de Farias Neto 04]}:
J. J. de Farias Neto,
``Quantum battle of the sexes revisited'',
quant-ph/0408019.

\item {\bf [Faris 99]}:
W. G. Faris,
``Book review'',
{\em Stud. Hist. Philos. Sci. Part B:
Stud. Hist. Philos. Mod. Phys.} {\bf 30}, 1, 141-143 (1999).
Review of {\bf [Wigner 97 a]}.

\item {\bf [Fasel-Gisin-Ribordy-(+2) 02]}:
S. Fasel, N. Gisin, G. Ribordy,
V. Scarani, \& H. Zbinden,
``Quantum cloning with an optical fiber amplifier'',
{\em Phys. Rev. Lett.} {\bf 89}, 10, 107901 (2002);
quant-ph/0203056.

\item {\bf [Fasel-Gisin-Ribordy-Zbinden 04]}:
S. Fasel, N. Gisin, G. Ribordy, \& H. Zbinden,
``Quantum key distribution over 30km of standard fiber using energy-time
entangled photon pairs: a comparison of two chromatic dispersion reduction
methods'',
quant-ph/0403144.

\item {\bf [Fasel-Alibart-Beveratos-(+4) 04]}:
S. Fasel, O. Alibart, A. Beveratos,
S. Tanzilli, H. Zbinden, P. Baldi, \& N. Gisin,
``High quality asynchronous heralded single photon source at telecom
wavelength'',
quant-ph/0408136.

\item {\bf [Fasel-Gisin-Zbinden-(+3) 04]}:
S. Fasel, N. Gisin, H. Zbinden,
D. Erni, E. Moreno, \& F. Robin,
``Energy-time entanglement preservation in plasmon-assisted light
transmission'',
quant-ph/0410064.

\item {\bf [Fattal-Inoue-Vukovi-(+3) 04]}:
D. Fattal, K. Inoue, J. Vukovi,
C. Santori, G. S. Solomon, \& Y. Yamamoto,
``Entanglement formation and violation of Bell's inequality with a semiconductor single photon source'',
{\em Phys. Rev. Lett.} {\bf 92}, 3, 037903 (2004);
quant-ph/0305048.

\item {\bf [Fattal-Diamanti-Inoue-Yamamoto 04]}:
D. Fattal, E. Diamanti, K. Inoue, \& Y. Yamamoto,
``Quantum teleportation with a quantum dot single photon source'',
{\em Phys. Rev. Lett.} {\bf 92}, 3, 037904 (2004);
quant-ph/0307105.

\item {\bf [Fattal-Cubitt-Yamamoto-(+2) 04]}:
D. Fattal, T. S. Cubitt, Y. Yamamoto,
S. Bravyi, \& I. L. Chuang,
``Entanglement in the stabilizer formalism'',
quant-ph/0406168.

\item {\bf [Favrholdt 94]}:
D. Favrholdt,
``Niels Bohr and realism'',
in {\bf [Faye-Folse 94]}, pp.~77-96.

\item {\bf [Faye 91]}:
J. Faye,
{\em Niels Bohr: His heritage and legacy.
An anti-realist view of quantum mechanics},
Kluwer Academic, Dordrecht, Holland, 1991.

\item {\bf [Faye 94]}:
J. Faye,
``Non-locality or non-separability? A defense of Bohr's
anti-realist approach to quantum mechanics'',
in {\bf [Faye-Folse 94]}, pp.~97-118.

\item {\bf [Faye-Folse 94]}:
J. Faye, \& H. J. Folse (eds.),
{\em Niels Bohr and contemporary philosophy},
Kluwer Academic, Dordrecht, Holland, 1994.

\item {\bf [Fazio-Sch\"{o}n 97]}:
R. Fazio, \& G. Sch\"{o}n,
``Mesoscopic effects in superconductivity'',
in {\em Mesoscopic electron transport},
{\em NATO ASI series E} {\bf 345}, 407-? (1997).
Reprinted in {\bf [Macchiavello-Palma-Zeilinger 00]}, pp.~368-390.

\item {\bf [Fazio-Massimo Palma-Siewert 99]}:
R. Fazio, G. Massimo Palma, \& J. Siewert,
``Fidelity and leakage of Josephson qubits'',
{\em Phys. Rev. Lett.} {\bf 83}, 25, 5385-5388 (1999).

\item {\bf [Fazio-Sch\"{o}n 00]}:
R. Fazio, \& G. Sch\"{o}n,
``Josephson junctions and quantum computation'',
{\bf [Macchiavello-Palma-Zeilinger 00]}, pp.~363-367.

\item {\bf [Feagin 04]}:
J. M. Feagin,
``Hardy nonlocality via few-body fragmentation imaging'',
{\em Phys. Rev. A} {\bf 69}, 6, 062103 (2004).

\item {\bf [Fearn-Lamb 95]}:
H. Fearn, \& W. E. Lamb, Jr.,
``Comments on the quantum Zeno effect'',
{\em Quantum Semiclass. Opt.} {\bf 7}, 3, 211-214 (1995).

\item {\bf [Fedichkin 00]}:
L. Fedichkin,
``Polynomial procedure of avoiding multiqubit errors
arising due to qubit-qubit interaction'',
quant-ph/0011005.

\item {\bf [Fedichkin-Fedorov 04]}:
L. Fedichkin, \& A. Fedorov,
``Error rate of a charge qubit coupled to an acoustic phonon reservoir'',
{\em Phys. Rev. A} {\bf 69}, 3, 032311 (2004).

\item {\bf [Fedichkin-Fedorov-Privman 04]}:
L. Fedichkin, Fedorov, \& A. V. Privman,
``Additivity of decoherence measures for multiqubit quantum systems'',
{\em Phys. Lett. A} {\bf 328}, 2-3, 87-93 (2004).

\item {\bf [Fedorov-Efremov-Kazakov-(+3) 04]}:
M. V. Fedorov, M. A. Efremov, A. E. Kazakov,
K. W. Chan, C. K. Law, \& J. H. Eberly,
``Packet narrowing and quantum entanglement in photoionization
and photodissociation'',
{\em Phys. Rev. A} {\bf 69}, 5, 052117 (2004).

\item {\bf [Fei-Jost-Popescu-(+2) 97]}:
H.-B. Fei, B. M. Jost, S. Popescu, B. E. A. Saleh, \& M. C. Teich,
``Entanglement induced two-photon transparency'',
{\em Phys. Rev. Lett.} {\bf 78}, 9, 1679-1682 (1997).

\item {\bf [Fei-Gao-Wang-(+2) 02]}:
S.-M. Fei, X.-H. Gao, X.-H. Wang, Z.-X. Wang, \& K. Wu,
``Separability of rank two quantum states on multiple
quantum spaces'',
{\em Phys. Lett. A} {\bf 300}, 6, 559-566 (2002).

\item {\bf [Fei-Jost-Li Jost-Wang 03]}:
S.-M. Fei, J. Jost, X. Li-Jost, \& G.-F. Wang,
``Entanglement of formation for a class of quantum states'',
{\em Phys. Lett. A} {\bf 310}, 5-6, 333-338 (2003);
quant-ph/0304095.

\item {\bf [Fei-Gao-Wang-(+2) 03 a]}:
S.-M. Fei, X.-H. Gao, X.-H. Wang, Z.-X. Wang, \& K. Wu,
``Separability and entanglement in $C^2 \otimes C^3 \otimes C^N$ composite quantum
systems'',
{\em Phys. Rev. A} {\bf 68}, 2, 022315 (2003);
quant-ph/0308159.

\item {\bf [Fei-Gao-Wang-(+2) 03 b]}:
S.-M. Fei, X.-H. Gao, X.-H. Wang, Z.-X. Wang, \& K. Wu,
``Separability of rank-two quantum states in $C^M \otimes C^N$ composite quantum
systems'',
{\em Commun. Theor. Phys.} {\bf 39}, 5, 525-528 (2003).

\item {\bf [Fei-Gao-Wang-(+2) 03 c]}:
S.-M. Fei, X.-H. Gao, X.-H. Wang, Z.-X. Wang, \& K. Wu,
``Separability of rank two quantum states on multiple quantum spaces with different dimensions'',
{\em Int. J. Quant. Inform.} {\bf 1}, ?, 37-49 (2003);
quant-ph/0305080.

\item {\bf [Fei-Gao-Wang-(+2) 03 d]}:
S.-M. Fei, X.-H. Gao, X.-H. Wang, Z.-X. Wang, \& K. Wu,
``Canonical form and separability of PPT states in $2 \times 2 \times \times 2 \times N$
composite quantum systems'',
{\em Commun. Theor. Phys.} {\bf 40}, 515-518 (2003);
quant-ph/0310146.

\item {\bf [Fei 04]}:
S.-M. Fei,
``Exactly solvable many-body systems and pseudo-Hermitian point
interactions'',
{\em Czech J. Phys.} {\bf 54}, 43-49 (2004);
quant-ph/0402185.

\item {\bf [Fei-Li Jost 04]}:
S.-M. Fei, \& X. Li-Jost,
``A class of special matrices and quantum entanglement'',
{\em Rep. Math. Phys.} {\bf 53}, ?, 195-210 (2004);
quant-ph/0405078.

\item {\bf [Fei-Wang-Zhao 04]}:
S.-M. Fei, Z.-X. Wang, \& H. Zhao,
``A note on entanglement of formation and generalized concurrence'',
{\em Phys. Lett. A} {\bf 329}, 6, 414-419 (2004);
quant-ph/0408131.

\item {\bf [Fei-Jihui-Mingjun-(+3) 02]}:
X. Fei, W. Jihui, S. Mingjun,
D. Jiangfeng, Z. Xianyi, \& H. Rongdian,
``Realization of the Fredkin gate by three transition
pulses in NMR quantum information processor'',
quant-ph/0202014.

\item {\bf [Feigel'man-Ioffe-Geshkenbein-(+2) 04]}:
M. V. Feigel'man, L. B. Ioffe, V. B. Geshkenbein,
P. Dayal, \& G. Blatter,
``Superconducting tetrahedral quantum bits'',
{\em Phys. Rev. Lett.} {\bf 92}, 9, 098301 (2004).

\item {\bf [Feingold-Moiseyev-Peres 84]}:
M. Feingold, N. Moiseyev, \& A. Peres,
``Ergodicity and mixing in quantum theory. II'',
{\em Phys. Rev. A} {\bf 30}, 1, 509-511 (1984).
See {\bf [Peres 84 b]} (I).

\item {\bf [Feingold-Peres 85]}:
S. J. Feingold, \& A. Peres,
in ``Reality and the quantum theory'',
{\em Phys. Today} {\bf 38}, 11, 15 (1985).
Comment on {\bf [Mermin 85]}.

\item {\bf [Fel'dman-Lacelle 01]}:
E. B. Fel'dman, \& S. Lacelle,
``Perspectives on a solid state NMR quantum computer'',
quant-ph/0108106.

\item {\bf [Feldman-Hillery 04]}:
E. Feldman, \& M. Hillery,
``Quantum walks on graphs and quantum scattering theory'',
{\em Proc. Conf. on Coding Theory and Quantum Computing}
quant-ph/0403066.

\item {\bf [Feldman-Zhou 02]}:
M. Feldmann, \& X. Zhou,
``Superconducting quantum computing without switches'',
in {\em Quantum computing and quantum bits in mesoscopic systems},
Kluwer, Dordrecht, Holland, 2003;
quant-ph/0211158.

\item {\bf [Feldmann 95]}:
M. Feldmann,
``New loophole for the Einstein-Podolsky-Rosen paradox'',
{\em Found. Phys. Lett.} {\bf 8}, 1, 41-53 (1995).

\item {\bf [Felber-Gahler-Golub-(+4) 99]}:
J. Felber, R. Gahler, R. Golub, P. Hank, V. Ignatovich,
T. Keller, \& U. Rauch,
``Neutron time interferometry'',
{\em Found. Phys.} {\bf 29}, 3, 381-398 (1999).

\item {\bf [Felicetti-Mancini-Tombesi 02]}:
A. Felicetti, S. Mancini, \& P. Tombesi,
``Quantum characterization of a Werner-like mixture'',
{\em Phys. Rev. A} {\bf 65}, 6, 062107 (2002);
quant-ph/0201071.

\item {\bf [Feligioni-Panella-Srivastava-Widom 02]}:
L. Feligioni, O. Panella, Y. N. Srivastava, \& A. Widom,
``Two-time correlation functions:
Bohm theory and conventional quantum mechanics'',
quant-ph/0202045.

\item {\bf [Felix-Gisin-Stefanov-Zbinden 012]}:
s. Felix, N. Gisin, A. Stefanov, \& H. Zbinden,
``Faint laser quantum key distribution:
Eavesdropping exploiting multiphoton pulses'',
{\em J. Mod. Opt.} {\bf 48}, ?, 2009-2021 (2001);
quant-ph/0102062.

\item {\bf [Fellows 88]}:
F. Fellows,
`Comment on ``Bell's theorem and the
foundations of quantum physics''[{\em Am. J. Phys.} {\bf 53}, 306 (1985)]',
{\em Am. J. Phys.} {\bf 56}, 6, 567-568 (1988).
Comment on {\bf [Stapp 85 a]}.
Reply: {\bf [Stapp 88 c, d]}.

\item {\bf [Feng-Gao-Zhan 01]}:
J. Feng, Y.-F. Gao, \& M.-S. Zhan,
``Disentanglement of multiparticle quantum systems'',
{\em Phys. Lett. A} {\bf 279}, 3-4, 110-116 (2001).

\item {\bf [Feng-Wang-Gao-Zhan 01]}:
J. Feng, J.-S. Wang, Y.-F. Gao, \& M.-S. Zhan,
``Sufficient conditions for exact disentanglement of a
set of triparticle entangled states'',
{\em Phys. Lett. A} {\bf 288}, 3-4, 125-131 (2001).

\item {\bf [Feng-Gao-Cao-(+2) 01]}:
J. Feng, Y.-F. Gao, J.-W. Cao, J.-S. Wang, \& M.-S. Zhan,
``Probabilistic exact deletion of copies of two non-orthogonal states'',
{\em Phys. Lett. A} {\bf 292}, 1-2, 12-14 (2001).

\item {\bf [Feng-Gao-Wang-Zhan 02]}:
J. Feng, Y.-F. Gao, J.-S. Wang, \& M.-S. Zhan,
``Probabilistic deletion of copies of linearly independent quantum states'',
{\em Phys. Rev. A} {\bf 65}, 5, 052311 (2002).

\item {\bf [Feng-Gao-Wang-Zhan 02]}:
J. Feng, Y.-F. Gao, J.-S. Wang, \& M.-S. Zhan,
``Quantum no-deletion theorem for entangled states'',
{\em Phys. Lett. A} {\bf 298}, 4, 225-228 (2002).

\item {\bf [Feng-Yu-Wang-(+2) 03]}:
J. Feng, Y.-F. Yu, J.-S. Wang, T.-K. Liu, \& M.-S. Zhan,
``Probabilistic comparing and sorting non-orthogonal quantum states'',
{\em Phys. Lett. A} {\bf 312}, 3-4, 131-135 (2003).

\item {\bf [Feng 01 a]}:
M. Feng,
``Grover search with pairs of trapped ions'',
{\em Phys. Rev. A} {\bf 63}, 5, 052308 (2001);
quant-ph/0102122.
Comment: {\bf [Hill-Goan 04 a]}:.

\item {\bf [Feng 01 b]}:
M. Feng,
``Simultaneous intraportation of many quantum states
within the quantum computing network'',
{\em Phys. Lett. A} {\bf 282}, 3, 138-144 (2001);
quant-ph/0103069.

\item {\bf [Feng-Wang 02 a]}:
M. Feng, \& X. Wang,
``Quantum computing with four-particle decoherence-free states in an ion trap'',
{\em Phys. Rev. A} {\bf 65}, 4, 044304 (2002).

\item {\bf [Feng-Wang 02 b]}:
M. Feng, \& X. Wang,
``Implementation of quantum gates and preparation of
entangled states in cavity QED with cold trapped ions'',
{\em J. Opt. B: Quantum Semiclass. Opt.} {\bf 4}, 5, 283-288 (2002);
quant-ph/0112031.

\item {\bf [Feng 02 a]}:
M. Feng,
``Quantum computing in cavity QED with cold trapped ions by bichromatic
radiation'',
{\em Phys. Rev. A} {\bf 65}, 6, 064301 (2002).

\item {\bf [Feng-Zhu-Gao-Fang 02]}:
M. Feng, X. Zhu, K. Gao, \& X. Fang,
``Testing wave-particle duality with the hot-ion quantum computation'',
{\em Phys. Lett. A} {\bf 294}, 5-6, 271-277 (2002).

\item {\bf [Feng 02 b]}:
M. Feng,
``Quantum computing with trapped ions in an optical cavity via Raman
transition'',
{\em Phys. Rev. A} {\bf 66}, 5, 054303 (2002).

\item {\bf [Feng 03]}:
M. Feng,
``A scheme of quantum computing with semiconductor quantum dots in optical cavity'',
{\em Phys. Lett.} {\bf 306}, 5-6, 353-357 (2003).

\item {\bf [Feng-D'Amico-Zanardi-Rossi 03]}:
M. Feng, I. D'Amico, P. Zanardi, \& F. Rossi,
``Spin-based quantum-information processing with semiconductor quantum dots
and cavity QED'',
{\em Phys. Rev. A} {\bf 67}, 1, 014306 (2003).

\item {\bf [Feng-Twamley 03]}:
M. Feng, \& J. Twamley,
``One-step implementation of two-qubit gates for spin-based fullerene
quantum information processing'',
quant-ph/0311178.

\item {\bf [Feng-Twamley 04]}:
M. Feng, \& J. Twamley,
``Readout scheme of the fullerene-based quantum computer by a single
electron transistor'',
quant-ph/0401157.

\item {\bf [Feng-Pfister 04]}:
S. Feng, \& O. Pfister,
``Quantum interference of ultrastable twin optical beams'',
{\em Phys. Rev. Lett.} {\bf 92}, 20, 203601 (2004);
quant-ph/0310002.

\item {\bf [Feng-Gong-Xu 00]}:
X.-L. Feng, S.-Q. Gong, \& Z.-Z. Xu,
``Entanglement purification via controlled-controlled-NOT
operations'',
{\em Phys. Lett. A} {\bf 271}, 1-2, 44-47 (2000).

\item {\bf [Feng-Wang-Xu 02]}:
X.-L. Feng, Z.-Y. Wang, \& Z.-Z. Xu,
``Mutual catalysis of entanglement transformations for pure entangled states'',
{\em Phys. Rev. A} {\bf 65}, 2, 022307 (2002).

\item {\bf [Feng-Zhang-Li-(+2) 03]}:
X.-L. Feng, Z.-M. Zhang, X.-D. Li,
S.-Q. Gong, \& Z.-Z. Xu,
``Entangling distant atoms by interference of polarized photons'',
{\em Phys. Rev. Lett.} {\bf 90}, 21, 217902 (2003).

\item {\bf [Feng-Zhang-Ying 02]}:
Y. Feng, S. Zhang, \& M. Ying,
``Probabilistic cloning and deleting of quantum states'',
{\em Phys. Rev. A} {\bf 65}, 4, 042324 (2002).

\item {\bf [Feng-Zhang-Sun-Ying 02]}:
Y. Feng, S. Zhang, X. Sun, \& M. Ying,
``Universal and original-preserving quantum copying is impossible'',
{\em Phys. Lett. A} {\bf 297}, 1-2, 1-3 (2002).

\item {\bf [Feng-Zhang-Duan-Ying 02]}:
Y. Feng, S. Zhang, R. Duan, \& M. Ying,
``Lower bound on inconclusive probability of unambiguous discrimination'',
{\em Phys. Rev. A} {\bf 66}, 6, 062313 (2002).

\item {\bf [Feng-Duan-Ying 04]}:
Y. Feng, R. Duan, \& M. Ying,
``When catalysis is useful for probabilistic entanglement transformation'',
{\em Phys. Rev. A} {\bf 69}, 6, 062310 (2004).

Y. Feng, R. Duan, \& M. Ying,
``Unambiguous discrimination between mixed quantum states'',
{\em Phys. Rev. A} {\bf 70}, 1, 012308 (2004).

\item {\bf [Fenner-Green-Homer-Pruim 98]}:
S. A. Fenner, F. Green, S. Homer, \& R. Pruim,
``Determining acceptance possibility for a quantum
computation is hard for the polynomial hierarchy'',
quant-ph/9812056.

\item {\bf [Fenner 00]}:
S. A. Fenner,
``An intuitive Hamiltonian for quantum search'',
quant-ph/0004091.

\item {\bf [Fenner-Zhang 01]}:
S. A. Fenner, \& Y. Zhang,
``Universal quantum computation with two- and
three-qubit projective measurements'',
quant-ph/0111077.

\item {\bf [Ferguson-Cain-Williams-Briggs 02]}:
A. J. Ferguson, P. A. Cain, D. A. Williams, \& G. A. D. Briggs,
``Ammonia-based quantum computer'',
{\em Phys. Rev. A} {\bf 65}, 3, 034303 (2002).

\item {\bf [Fermani-Mancini-Tombesi 03]}:
R. Fermani, S. Mancini, \& P. Tombesi,
``Sensitivity of a cavityless optomechanical system'',
quant-ph/0312166.

\item {\bf [Fern-Kempe-Simic-Sastry 04]}:
J. Fern, J. Kempe, S. Simic, \& S. Sastry,
``Fault-tolerant quantum computation -- a dynamical systems approach'',
quant-ph/0409084.

\item {\bf [Fern\'{a}ndez Huelga 93]}:
S. G. Fern\'{a}ndez Huelga,
``Encuesta sobre la f\'{\i}sica cu\'{a}ntica'',
{\em Revista Espa\~{n}ola de F\'{\i}sica} {\bf 7}, 4, 15-18 (1993).

\item {\bf [Fern\'{a}ndez Huelga-Ferrero-Santos 94]}:
S. G. Fern\'{a}ndez
Huelga, M. Ferrero, \& E. Santos,
``Atomic-cascade experiment with detection of the recoil atom'',
{\em Europhys. Lett.} {\bf 27}, 3, 181-186 (1994).

\item {\bf [Fern\'{a}ndez Huelga-Ferrero-Santos 95 a]}:
S. G. Fern\'{a}ndez
Huelga, M. Ferrero, \& E. Santos,
``Loophole-free test of the Bell inequality'',
{\em Phys. Rev. A} {\bf 51}, 6, 5008-5011 (1995).

\item {\bf [Fern\'{a}ndez Huelga-Ferrero-Santos 95 b]}:
S. G. Fern\'{a}ndez Huelga, M. Ferrero, \& E. Santos,
``Proposed new polarization correlation test of local realism'',
in D. M. Greenberger, \& A. Zeilinger (eds.),
{\em Fundamental problems in
quantum theory: A conference held in honor of professor John A. Wheeler,
Ann. N. Y. Acad. Sci.} {\bf 755}, 429-437 (1995).

\item {\bf [Fern\'{a}ndez Huelga-Marshall-Santos 95]}:
S. G. Fern\'{a}ndez Huelga, T. W. Marshall, \& E. Santos,
``Proposed test for realistic theories using
Rydberg atoms coupled to a high-$Q$ resonator'',
{\em Phys. Rev. A} {\bf 52}, 4, R2497-R2500 (1995).
See {\bf [Fern\'{a}ndez Huelga-Marshall-Santos 96]}.

\item {\bf [Fern\'{a}ndez Huelga-Marshall-Santos 96]}:
S. G. Fern\'{a}ndez Huelga, T. W. Marshall, \& E. Santos,
``Temporal Bell-type inequalities for two-level
Rydberg atoms coupled to a high-$Q$ resonator'',
{\em Phys. Rev. A} {\bf 54}, 3, 1798-1807 (1996).
See {\bf [Fern\'{a}ndez Huelga-Marshall-Santos 95]}.

\item {\bf [Fern\'{a}ndez Huelga-Macchiavello-Pellizzari-(+3) 97]}:
S. G. Fern\'{a}ndez Huelga, C. Macchiavello, T. Pellizzari, A. K. Ekert,
M. B. Plenio, \& J. I. Cirac,
``On the improvement of frequency standards with quantum entanglement'',
{\em Phys. Rev. Lett.} {\bf 79}, 20, 3865-3868 (1997);
quant-ph/9707014.

\item {\bf [Fern\'{a}ndez Huelga-Plenio 00]}:
S. G. Fern\'{a}ndez Huelga, \& M. B. Plenio,
``Quantum stochastic resonance in electron shelving'',
{\em Phys. Rev. A} {\bf 62}, 5, 052111 (2000);
quant-ph/0001102.

\item {\bf [Fern\'{a}ndez Huelga-Vaccaro-Chefles-Plenio 00]}:
S. G. Fern\'{a}ndez Huelga, J. A. Vaccaro, A. Chefles, \& M. B. Plenio,
``Quantum remote control: Teleportation of unitary operations'',
quant-ph/0005061.

\item {\bf [Fern\'{a}ndez Huelga-Plenio-Vaccaro 02]}:
S. Fern\'{a}ndez Huelga, M. B. Plenio, \& J. A. Vaccaro,
``Remote control of restricted sets of operations: Teleportation of angles'',
{\em Phys. Rev. A} {\bf 65}, 4, 042316 (2002);
quant-ph/0107110.

\item {\bf [Fern\'{a}ndez Ra\~{n}ada 04]}:
A. Fern\'{a}ndez Ra\~{n}ada,
{\em Ciencia, incertidumbre y conciencia. Heisenberg},
Nivola, Madrid, 2004.

\item {\bf [Ferrero 85]}:
M. Ferrero,
``What kind of realism'',
in E. Bitsakis, \& N. Tambakis (eds.),
{\em Determinism in physics},
Gutemberg, Athens, 1985.

\item {\bf [Ferrero 86]}:
M. Ferrero,
``La desigualdad de Bell y los experimentos de
cascada at\'{o}mica: ?'Es posible el realismo local frente a la mec\'{a}nica
cu\'{a}ntica?'',
Ph.\ D. thesis, Universidad de Cantabria, Spain, 1986.

\item {\bf [Ferrero 88]}:
M. Ferrero,
``Pr\'{o}logo'',
in N. H. D. Bohr,
{\em La teor\'{\i}a at\'{o}mica y la descripci\'{o}n de la naturaleza},
Alianza, Madrid, 1988, pp.~9-46.
Prologue to the Spanish version of {\bf [Bohr 34]}.

\item {\bf [Ferrero-Marshall-Santos 88]}:
M. Ferrero, T. W. Marshall, \& E. Santos,
``Symmetric and asymmetric models for atomic cascade experiments'',
in F. Selleri (ed.),
{\em Quantum mechanics versus local realism: The Einstein-Podolsky-Rosen paradox},
Plenum Press, New York, 1988, pp.~447-456.

\item {\bf [Ferrero-Marshall-Santos 90]}:
M. Ferrero, T. W. Marshall, \& E. Santos,
``Bell's theorem: Local realism versus quantum mechanics'',
{\em Am. J. Phys.} {\bf 58}, 7, 683-688 (1990).

\item {\bf [Ferrero-Marshall 91]}:
M. Ferrero, \& T. W. Marshall,
``The compatibility between quantum mechanics and local realistic theories in
atomic cascade experiments'',
{\em Found. Phys.} {\bf 21}, 4, 403-415 (1991).

\item {\bf [Ferrero-Getino 94]}:
M. Ferrero, \& J. M. Getino,
``Bounds to the difference between the predictions of stochastic and
quantum optics in the optical test of Bell inequalities'',
{\em Found. Phys. Lett.} {\bf 7}, 2, 201-208 (1994).

\item {\bf [Ferrero-Santos 96 a]}:
M. Ferrero, \& E. Santos,
``Realismo local y mec\'{a}nica cu\'{a}ntica'',
in M. Ferrero, A. Fern\'{a}ndez Ra\~{n}ada, J. L.
S\'{a}nchez G\'{o}mez, \& E. Santos (eds.),
{\em Fundamentos de la F\'{\i}sica
Cu\'{a}ntica (San Lorenzo de El Escorial, Spain, 1995)},
Editorial Complutense, Madrid, 1996, pp.~9-42.

\item {\bf [Ferrero-Santos 96 b]}:
M. Ferrero, \& E. Santos,
``Contraste de las desigualdades de Bell con fotones de cascadas at\'{o}micas'',
in M. Ferrero, A. Fern\'{a}ndez Ra\~{n}ada, J. L. S\'{a}nchez G\'{o}mez,
\& E. Santos (eds.),
{\em Fundamentos de la F\'{\i}sica Cu\'{a}ntica
(San Lorenzo de El Escorial, Spain, 1995)},
Editorial Complutense, Madrid, 1996, pp.~43-61.

\item {\bf [Ferrero-Santos 97]}:
M. Ferrero, \& E. Santos,
``Empirical consequences of the scientific construction: The program of local
hidden-variables theories in quantum mechanics'',
{\em Found. Phys.} {\bf 27}, 6, 765-800 (1997).

\item {\bf [Ferrero 00]}:
M. Ferrero,
``F\'{\i}sica cu\'{a}ntica y objetividad'',
{\em Arbor} {\bf 167}, 659-660, 459-473 (2000).

\item {\bf [Ferrero 02]}:
M. Ferrero,
``Quantum physics and philosophy'',
in R. Blanco, A. Ma\~{n}anes, S. Marcos, L. Pesquera, \& M. A. Rodr\'{\i}guez (eds.),
{\em Foundations of quantum physics},
?, Madrid, 2002.

\item {\bf [Ferrero-Salgado-S\'{a}nchez G\'{o}mez 03]}:
M. Ferrero, D. Salgado, \& J. L. S\'{a}nchez-G\'{o}mez,
``Some reflections upon the reduction postulate in the light of today's technology'',
in M. Ferrero (ed.),
{\em Proc. of Quantum Information: Conceptual Foundations,
Developments and Perspectives (Oviedo, Spain, 2002)},
{\em J. Mod. Opt.} {\bf 50}, 6-7, 967-974 (2003).

\item {\bf [Ferrero 03]}:
M. Ferrero,
``The information interpretation and the conceptual problems of quantum mechanics'',
{\em Found. Phys.} {\bf 33}, 4, 665-676 (2003).

\item {\bf [Ferry-Akis-Bird 04]}:
D. K. Ferry, R. Akis, \& J. P. Bird,
``Einselection in action: Decoherence and pointer states in open quantum dots'',
{\em Phys. Rev. Lett.} {\bf 93}, 2, 026803 (2004).

\item {\bf [Feynman 60]}:
R. P. Feynman,
``There's plenty of room at the bottom'',
{\em Engineering and Science} {\bf 23}, 22-36 (1960).

\item {\bf [Feynman-Vernon 63]}:
R. P. Feynman, \& F. Vernon,
``The theory of a general quantum system interacting with a
linear dissipative system'',
{\em Annals of Physics} {\bf 24}, 118-173 (1963).

\item {\bf [Feynman 65]}:
R. P. Feynman,
{\em The character of physical law},
M. I. T. Press, Cambridge, Massachusetts, 1965.
Spanish version: {\em El car\'{a}cter de la ley f\'{\i}sica},
Orbis, Barcelona, 1987.

\item {\bf [Feynman-Hibbs 65]}:
R. P. Feynman, \& A. R. Hibbs,
{\em Quantum mechanics and path integrals},
McGraw-Hill, New York, 1965.

\item {\bf [Feynman-Leighton-Sands 65]}:
R. P. Feynman, R. B. Leighton, \& M. Sands,
{\em The Feynman lectures on physics. Quantum mechanics},
Addison-Wesley, Reading, Massachusetts, 1965.
English-Spanish version: {\em Feynman. F\'{\i}sica.
Volumen III: Mec\'{a}nica cu\'{a}ntica},
Fondo Educativo Interamericano, Bogot\'{a}, Colombia, 1971.

\item {\bf [Feynman 82]}:
R. P. Feynman,
``Simulating physics with computers'',
{\em Int. J. Theor. Phys.} {\bf 21}, 6-7, 467-488 (1982).
Reprinted in {\bf [Feynman 00]}.

\item {\bf [Feynman 85]}:
R. P. Feynman,
``Quantum mechanical computers'',
{\em Opt. News} {\bf 11}, 2, 11-20 (1985).
Reprinted in {\bf [Feynman 96, 00]}.
Spanish version in the Spanish version of {\bf [Feynman 96]}, 205-231 (1985).

\item {\bf [Feynman 86]}:
R. P. Feynman,
``Quantum-mechanical computers'',
{\em Found. Phys.} {\bf 16}, 6, 507-531 (1986).
Reprinted in {\bf [Zurek-van der Merwe-Miller 88]}, pp.~523-547.

\item {\bf [Feynman 87]}:
R. P. Feynman,
``Negative probability'',
in {\bf [Hiley-Peat 87]}, pp.~235-248.

\item {\bf [Feynman 96]}:
R. P. Feynman (edited by A. J. G. Hey, \& R. W. Allen),
{\em Feynman lectures on computation},
Addison-Wesley, Reading, Massachusetts, 1996.
Spanish version: {\em Conferencias sobre computaci\'{o}n},
Cr\'{\i}tica, Barcelona, 2003.
See {\bf [Galindo 03]}.

\item {\bf [Feynman 00]}:
R. P. Feynman (edited by L. M. Brown),
{\em Selected papers of Richard Feynman},
World Scientific, Singapore, 2000.

\item {\bf [Fichtner-Ohya 99]}:
K.-H. Fichtner, \& M. Ohya,
``Quantum teleportation with entangled states given by beam
splitting'',
quant-ph/9912083.

\item {\bf [Ficek-Tanas 02]}:
Z. Ficek, \& R. Tanas,
``Entangled states and collective nonclassical effects in two-atom systems'',
{\em Phys. Rep.} {\bf 372}, 369-? (2002);
quant-ph/0302082.

\item {\bf [Fidani-Tombesi 02]}:
C. Fidani, \& P. Tombesi,
``Decreasing the error probability in optical transmission lines'',
{\em Phys. Rev. A} {\bf 65}, 3, 033815 (2002).

\item {\bf [Filip-Reh\'{a}cek-Du\v{s}ek 01]}:
R. Filip, J. Reh\'{a}cek, \& M. Du\v{s}ek,
``Entanglement of coherent states and decoherence'',
{\em J. Opt. B: Quantum Semiclass. Opt.} {\bf 3}, 5, 341-345 (2001);
quant-ph/0011006.

\item {\bf [Filip-Mista 02 a]}:
R. Filip, \& L. Mista, Jr.,
``Violation of Bell's inequalities for a two-mode squeezed vacuum state in
lossy transmission lines'',
{\em Phys. Rev. A} {\bf 66}, 4, 044309 (2002).

\item {\bf [Filip-Mista 02 b]}:
R. Filip, \& L. Mista Jr.,
``Squeezing concentration for Gaussian states with unknown parameter'',
quant-ph/0204105.

\item {\bf [Filip 01 a]}:
R. Filip,
``A device for feasible fidelity, purity, Hilbert-Schmidt
distance and entanglement witness measurements'',
quant-ph/0108119.

\item {\bf [Filip 01 b]}:
R. Filip,
``Relation between complementarity and nonlocality under decoherence'',
quant-ph/0108135.

\item {\bf [Filip 02 a]}:
R. Filip,
``Complementarity, entanglement and quantum erasing
in continuous-variable quantum nondemolition experiments'',
{\em J. Opt. B: Quantum Semiclass. Opt.} {\bf 4}, 3, 202-207 (2002).

\item {\bf [Filip 02 b]}:
R. Filip,
``Optimization of security costs in nested purification protocol'',
quant-ph/0202171.

\item {\bf [Filip-Du\v{s}ek-Fiur\'{a}\v{s}ek-Mista 02]}:
R. Filip, M. Du\v{s}ek, J. Fiur\'{a}\v{s}ek, \& L. Mista,
`Bell-inequality violation with ``thermal'' radiation',
{\em Phys. Rev. A} {\bf 65}, 4, 043802 (2002);
quant-ph/0105080.

\item {\bf [Filip 02]}:
R. Filip,
``Overlap and entanglement-witness measurements'',
{\em Phys. Rev. A} {\bf 65}, 6, 062320 (2002).

\item {\bf [Filip 03]}:
R. Filip,
``Continuous-variable quantum erasing'',
{\em Phys. Rev. A} {\bf 67}, 4, 042111 (2003).

\item {\bf [Filip-Fiur\'{a}\v{s}ek-Marek 04]}:
R. Filip, J. Fiur\'{a}\v{s}ek, \& P. Marek,
``Reversibility of continuous-variable quantum cloning'',
{\em Phys. Rev. A} {\bf 69}, 1, 012314 (2004);
quant-ph/0310123.

\item {\bf [Filip 04 a]}:
R. Filip,
``Conditional implementation of an asymmetrical universal quantum cloning machine'',
{\em Phys. Rev. A} {\bf 69}, 3, 032309 (2004);
quant-ph/0401036.

\item {\bf [Filip 04 b]}:
R. Filip,
``Quantum partial teleportation as optimal cloning at a distance'',
quant-ph/0401039.

\item {\bf [Filip 04 c]}:
R. Filip,
``Continuous-variable quantum non-demolishing interaction at a distance'',
quant-ph/0404010.

\item {\bf [Filipp-Svozil 01]}:
S. Filipp, \& K. Svozil,
``Boole-Bell-type inequalities in Mathematica'',
quant-ph/0105083.

\item {\bf [Filipp-Svozil 04 a]}:
S. Filipp, \& K. Svozil,
``Testing the bounds on quantum probabilities'',
{\em Phys. Rev. A} {\bf 69}, 3, 032101 (2004);
quant-ph/0306092.

\item {\bf [Filipp-Svozil 04 b]}:
S. Filipp, \& K. Svozil,
``Generalizing Tsirelson's bound on Bell inequalities using a min-max principle'',
{\em Phys. Rev. Lett.} {\bf 93}, 13, 130407 (2004);
quant-ph/0403175.

\item {\bf [Filipp-Svozil 04 c]}:
S. Filipp, \& K. Svozil,
``Tracing the bounds on Bell-type inequalities'',
in {\em Foundations of Probability and Physics-3 (V\"{a}xj\"{o}, Sweden, 2004)};
quant-ph/0407145.

\item {\bf [Filipp-Gavenda 04]}:
R. Filip, \& M. Gavenda,
``Knowledge excess duality and violation of Bell inequalities'',
quant-ph/0404145.

\item {\bf [De Filippo 00]}:
S. De Filippo,
``Quantum computation using decoherence-free states of
the physical operator algebra'',
{\em Phys. Rev. A} {\bf 62}, 5, 052307 (2000);
quant-ph/9910005.

\item {\bf [Fine 70]}:
A. I. Fine,
``Insolubility of the quantum measurement problem'',
{\em Phys. Rev. D} {\bf 2}, 12, 2783-2787 (1970).

\item {\bf [Fine 72 a]}:
A. I. Fine,
``There is a measurement problem: A comment'',
{\em Phys. Rev. D} {\bf 5}, 4, 1033 (1972).
Comment on {\bf [Moldauer 72]}.

\item {\bf [Fine 72 b]}:
A. I. Fine,
``Some conceptual problems of quantum theory'',
in R. G. Colodny (ed.),
{\em Paradigms and paradoxes. The philosophical challenge of the
quantum domain}, University of Pittsburgh Press, Pittsburgh, 1972, pp.~3-31.

\item {\bf [Fine 74]}:
A. I. Fine,
``On the completeness of quantum theory'',
{\em Synthese} {\bf 29}, 1-4, 257-289 (1974).
Reprinted in {\bf [Suppes 76]}, pp.~249-281.

\item {\bf [Fine 77]}:
A. I. Fine,
``Conservation, the sum rule and confirmation'',
{\em Philos. Sci.} {\bf 44}, ?, 95-106 (1977).

\item {\bf [Fine-Teller 78]}:
A. I. Fine, \& P. Teller,
``Algebraic constraints on hidden variables'',
{\em Found. Phys.} {\bf 8}, 7-8, 629-636 (1978).

\item {\bf [Fine 79]}:
A. I. Fine,
``How to count frequencies: A primer for quantum realists'',
{\em Synthese} {\bf 42}, 1, 145-154 (1979).

\item {\bf [Fine 82 a]}:
A. I. Fine,
``Hidden variables, joint probability, and the Bell inequalities'',
{\em Phys. Rev. Lett.} {\bf 48}, 5, 291-295 (1982).
Comment: {\bf [Garg-Mermin 82 a]}. Reply: {\bf [Fine 82 c]}.
See {\bf [Garg-Mermin 82 c]},
{\bf [Stapp 82 b]}, {\bf [Eberhard 82]}, {\bf [Fine 82 c, d]}.

\item {\bf [Fine 82 b]}:
A. I. Fine,
``Joint distributions, quantum correlations, and
commuting observables'',
{\em J. Math. Phys.} {\bf 23}, 7, 1306-1310 (1982).

\item {\bf [Fine 82 c]}:
A. I. Fine,
``Fine responds'',
{\em Phys. Rev. Lett.} {\bf 49}, 3, 243 (1982).
Reply to {\bf [Garg-Mermin 82 a]}.

\item {\bf [Fine 82 d]}:
A. I. Fine,
``Comments on the significance of Bell's theorem'',
{\em Phys. Rev. Lett.} {\bf 49}, 20, 1536 (1982).
See {\bf [Stapp 82 b]}, {\bf [Eberhard 82]}.

\item {\bf [Fine 86]}:
A. I. Fine,
{\em The shaky game: Einstein, realism, and the quantum theory},
University of Chicago Press, Chicago, 1986, 1996 (2nd edition).
Comment: {\bf [H\'{a}jek-Bub 92]}.
See {\bf [Howard 85]}, {\bf [Fine 89]} (Sec. 1), {\bf [Deltete-Guy 91]}.

\item {\bf [Fine 89]}:
A. I. Fine,
``Correlations and efficiency: Testing the Bell inequalities'',
{\em Found. Phys.} {\bf 19}, 5, 453-478 (1989).

\item {\bf [Fine 90 a]}:
A. I. Fine,
``Einstein and ensembles: Response'',
{\em Found. Phys.} {\bf 20}, 8, 967-989 (1990).
Reply to {\bf [Guy-Deltete 90]}.

\item {\bf [Fine 90 b]}:
A. I. Fine,
``Do correlations need to be explained?'',
in A. I. Miller (ed.),
{\em Sixty-two years of uncertainty: Historical, philosophical and physical
inquiries into the foundations of quantum mechanics.
Proc.\ from the Int. School
of History of Science (Erice, Italy, 1989)},
Plenum Press, New York, 1990, pp.~175-194.

\item {\bf [Fine 91]}:
A. I. Fine,
``Inequalities for nonideal correlation experiments'',
{\em Found. Phys.} {\bf 21}, 3, 365-378 (1991).

\item {\bf [Fine 97]}:
A. I. Fine,
``Contextualism, locality, and the no-go theorems'',
in M. Ferrero, \& A. van der Merwe (eds.),
{\em New developments on fundamental
problems in quantum physics (Oviedo, Spain, 1996)},
Kluwer Academic, Dordrecht,
Holland, 1997, pp.~125-132.

\item {\bf [Fine 99]}:
A. I. Fine,
``Locality and the Hardy theorem'',
in J. N. Butterfield, \& C. Pagonis (eds.),
{\em From physics to philosophy: Essays in honour of Michael Redhead},
Cambridge University Press, Cambridge, 1999, pp.~1-11.

\item {\bf [Fink-Leschke 00]}:
H. Fink, \& H. Leschke,
``Is the universe a quantum system?'',
{\em Found. Phys. Lett.} {\bf 13}, 4, 345-356 (2000);
quant-ph/0005095.

\item {\bf [Finkelstein 69]}:
D. Finkelstein,
``Matter, space and logic'', in
R. S. Cohen, \& M. W. Wartofsky (eds.),
{\em Boston studies in the philosophy of science. Vol. 5},
Reidel, Dordrecht, 1969, pp.~199-215.

\item {\bf [Finkelstein 72]}:
D. Finkelstein,
``The physics of logic'',
in R. G. Colodny (ed.),
{\em Paradigms and paradoxes. The philosophical challenge of the
quantum domain}, University of Pittsburgh Press, Pittsburgh, 1972, pp.~47-66.
Reprinted in {\bf [Hooker 79]}, pp.~141-160.

\item {\bf [Finkelstein 98 a]}:
J. Finkelstein,
``Does Schr\"{o}dinger's cat know
something that Schr\"{o}dinger does not know?'',
quant-ph/9801004.

\item {\bf [Finkelstein 98 b]}:
J. Finkelstein,
``Yet another comment on `Nonlocal character of quantum theory'\,'',
quant-ph/9801011.
Comment on {\bf [Stapp 97 a]}.

\item {\bf [Finkelstein 98]}:
J. Finkelstein,
``Space-time counterfactuals'',
quant-ph/9811057.

\item {\bf [Finkelstein 99 a]}:
J. Finkelstein,
``Ambiguities of arrival-time distributions in quantum theory'',
{\em Phys. Rev. A} {\bf 59}, 5, 3218-3222 (1999);
quant-ph/9809085.

\item {\bf [Finkelstein 99 b]}:
J. Finkelstein,
``Comment on
`Consistency, amplitudes, and probabilities in quantum theory'\,'',
{\em Phys. Rev. A} {\bf 60}, 2, 1723-1724 (1999);
quant-ph/9809017
Comment on {\bf [Caticha 98 a]}.
Reply: {\bf [Caticha 99]}.

\item {\bf [Finkelstein 99 c]}:
J. Finkelstein,
``Quantum probability from decision theory?'',
quant-ph/9907004.
See {\bf [Deutsch 99]},
{\bf [Polley 99]},
{\bf [Summhammer 99]},
{\bf [Barnum-Caves-Finkelstein-(+2) 00]}.

\item {\bf [Finkelstein 00]}:
J. Finkelstein,
``Property attribution and the projection postulate in
relativistic quantum theory'',
{\em Phys. Lett. A} {\bf 278}, 1-2, 19-24 (2000);
quant-ph/0007105.

\item {\bf [Finkelstein 02]}:
J. Finkelstein,
``Do macroscopic properties dictate microscopic probabilities?'',
quant-ph/0204006.

\item {\bf [Finkelstein 04]}:
J. Finkelstein,
``On PSI-complete and PSIR-complete measurements'',
quant-ph/0407078.
See {\bf [Flammia-Silberfarb-Caves 04]}.

\item {\bf [Fiorentino-Messin-Kuklewicz-(+2) 04]}:
M. Fiorentino, G. Messin, C. E. Kuklewicz,
F. N. C. Wong, \& J. H. Shapiro,
``Generation of ultrabright tunable polarization entanglement
without spatial, spectral, or temporal constraints'',
{\em Phys. Rev. A} {\bf 69}, 4, 041801 (2004);
quant-ph/0309071.

\item {\bf [Fiorentino-Wong 04]}:
M. Fiorentino, \& F. N. C. Wong,
``Deterministic controlled-NOT gate for single-photon two-qubit quantum logic'',
{\em Phys. Rev. Lett.} {\bf 93}, 7, 070502 (2004).

\item {\bf [Fischer-Kienle-Freyberger 00]}:
D. G. Fischer, S. H. Kienle, \& M. Freyberger,
``Quantum-state estimation by self-learning measurements'',
{\em Phys. Rev. A} {\bf 61}, 3, 032306 (2000).

\item {\bf [Fischer-Freyberger 00]}:
D. G. Fischer, \& M. Freyberger,
``Estimating mixed quantum states'',
{\em Phys. Lett. A} {\bf 273}, 5-6, 293-302 (2000);
quant-ph/0005090.

\item {\bf [Fischer-Mack-Freyberger 00]}:
D. G. Fischer, H. Mack, \& M. Freyberger,
``Transfer of quantum states using finite resources'',
accpeted in {\em Phys. Rev. A};
quant-ph/0008038.

\item {\bf [Fischer-Mack-Cirone-Freyberger 01]}:
D. G. Fischer, H. Mack, M. A. Cirone, \& M. Freyberger,
``Enhanced estimation of a noisy
quantum channel using entanglement'',
{\em Phys. Rev. A} {\bf 64}, 2, 022309 (2001);
quant-ph/0103160.

\item {\bf [Fischer-Guti\'{e}rrez Medina-Raizen 01]}:
M. C. Fischer, B. Guti\'{e}rrez-Medina, \& M. G. Raizen,
``Observation of the quantum Zeno and anti-Zeno effects in an unstable system'',
{\em Phys. Rev. Lett.} {\bf 87}, 4, 040402 (2001).

\item {\bf [Fitch-Franson 02]}:
M. J. Fitch, \& J. D. Franson,
``Dispersion cancellation and nonclassical noise reduction for
large-photon-number states'',
{\em Phys. Rev. A} {\bf 65}, 5, 053809 (2002);
quant-ph/0201075.

\item {\bf [Fitch-Jacobs-Pittman-Franson 03]}:
M. J. Fitch, B. C. Jacobs, T. B. Pittman, \& J. D. Franson,
``Photon-number resolution using time-multiplexed single-photon detectors'',
{\em Phys. Rev. A} {\bf 68}, 4, 043814 (2003);
quant-ph/0305193.

\item {\bf [Fitchard 79]}:
E. E. Fitchard,
``Proposed experimental test of wave packet reduction and the
uncertainty principle'',
{\em Found. Phys.} {\bf 9}, 7-8, 525-535 (1979).

\item {\bf [Fitzgerald 00]}:
R. Fitzgerald,
``What really gives a quantum computer its power?'',
{\em Phys. Today} {\bf 53}, 1, 20-22 (2000).

\item {\bf [Fitzi-Gisin-Maurer 01]}:
M. Fitzi, N. Gisin, \& U. Maurer,
``Quantum solution to the Byzantine agreement problem'',
{\em Phys. Rev. Lett.} {\bf 87}, 21, 217901 (2001);
quant-ph/0107127.

\item {\bf [Fitzi-Gisin-Maurer-von Ritz 01]}:
M. Fitzi, N. Gisin, U. Maurer, \& O. von Ritz,
``Unconditional Byzantine agreement and multi-party computation
secure against dishonest minorities from scratch'',
{\em Advances in Cryptology -- EUROCRYPT 2001},
{\em Lecture Notes in Computer Science}.

\item {\bf [Fitzi-Gottesman-Hirt-(+2) 02]}:
M. Fitzi, D. Gottesman, M. Hirt, T. Holenstein, \& A. Smith,
``Detectable Byzantine agreement secure against faulty majorities'',
{\em Proc.\ 21st ACM Symp.\ on Principles of Distributed Computing},
118-126 (2002).

\item {\bf [Fiur\'{a}\v{s}ek 01 a]}:
J. Fiur\'{a}\v{s}ek,
``Optical implementation of continuous-variable
quantum cloning machines'',
{\em Phys. Rev. Lett.} {\bf 86}, 21, 4942-4945 (2001);
quant-ph/0012048.

\item {\bf [Fiur\'{a}\v{s}ek 01 b]}:
J. Fiur\'{a}\v{s}ek,
``Maximum-likelihood estimation of quantum measurement'',
{\em Phys. Rev. A} {\bf 64}, 2, 024102 (2001).

\item {\bf [Fiur\'{a}\v{s}ek 01 c]}:
J. Fiur\'{a}\v{s}ek,
``Extremal equation for optimal completely positive maps'',
{\em Phys. Rev. A} {\bf 64}, 6, 062310 (2001);
quant-ph/0105124.

\item {\bf [Fiur\'{a}\v{s}ek 01 d]}:
J. Fiur\'{a}\v{s}ek,
``Structural physical approximations of unphysical maps
and generalized quantum measurements'',
quant-ph/0111107.

\item {\bf [Fiur\'{a}\v{s}ek-Iblisdir-Massar-Cerf 02]}:
J. Fiur\'{a}\v{s}ek, S. Iblisdir, S. Massar, \& N. J. Cerf,
``Quantum cloning of orthogonal qubits'',
{\em Phys. Rev. A} {\bf 65}, 4, 040302 (2002);
quant-ph/0110016.

\item {\bf [Fiur\'{a}\v{s}ek 02 a]}:
J. Fiur\'{a}\v{s}ek,
``Conditional generation of $N$-photon entangled states of light'',
{\em Phys. Rev. A} {\bf 65}, 5, 053818 (2002);
quant-ph/0110138.

\item {\bf [Fiur\'{a}\v{s}ek 02 b]}:
J. Fiur\'{a}\v{s}ek,
``Improving the fidelity of continuous-variable teleportation via local
operations'',
{\em Phys. Rev. A} {\bf 66}, 1, 012304 (2002);
quant-ph/0202102.

\item {\bf [Fiur\'{a}\v{s}ek 02 c]}:
J. Fiur\'{a}\v{s}ek,
``Encoding the quantum state of cavity mode into an atomic beam'',
{\em Phys. Rev. A} {\bf 66}, 1, 015801 (2002).

\item {\bf [Fiur\'{a}\v{s}ek 02 d]}:
J. Fiur\'{a}\v{s}ek,
``Gaussian transformations and distillation of entangled Gaussian states'',
{\em Phys. Rev. Lett.} {\bf 89}, 13, 137904 (2002).

\item {\bf [Fiur\'{a}\v{s}ek-Du\v{s}ek-Filip 02]}:
J. Fiur\'{a}\v{s}ek, M. Du\v{s}ek, \& R. Filip,
``Universal measurement apparatus controlled by quantum software'',
{\em Phys. Rev. Lett.} {\bf 89}, 19, 190401 (2002);
quant-ph/0202152.

\item {\bf [Fiur\'{a}\v{s}ek 02 e]}:
J. Fiur\'{a}\v{s}ek,
``Structural physical approximations of unphysical maps and generalized
quantum measurements'',
{\em Phys. Rev. A} {\bf 66}, 5, 052315 (2002).

\item {\bf [Fiur\'{a}\v{s}ek 02 f]}:
J. Fiur\'{a}\v{s}ek,
``Gaussian transformations and distillation of entangled Gaussian states'',
quant-ph/0204069.

\item {\bf [Fiur\'{a}\v{s}ek-Jezek 03]}:
J. Fiur\'{a}\v{s}ek, \& M. Jezek,
``Optimal discrimination of mixed quantum states involving inconclusive
results'',
{\em Phys. Rev. A} {\bf 67}, 1, 012321 (2003).

\item {\bf [Fiur\'{a}\v{s}ek-Mista-Filip 03]}:
J. Fiur\'{a}\v{s}ek, L. Mista, Jr., \& R. Filip,
``Entanglement concentration of continuous-variable quantum states'',
{\em Phys. Rev. A} {\bf 67}, 2, 022304 (2003).

\item {\bf [Fiur\'{a}\v{s}ek 03 a]}:
J. Fiur\'{a}\v{s}ek,
``Optical implementations of the optimal phase-covariant quantum cloning
machine'',
{\em Phys. Rev. A} {\bf 67}, 5, 052314 (2003).

\item {\bf [Fiur\'{a}\v{s}ek 03 b]}:
J. Fiur\'{a}\v{s}ek,
``Unitary-gate synthesis for continuous-variable systems'',
{\em Phys. Rev. A} {\bf 68}, 2, 022304 (2003).

\item {\bf [Fiur\'{a}\v{s}ek-Massar-Cerf 03]}:
J. Fiur\'{a}\v{s}ek, S. Massar, \& N. J. Cerf,
``Conditional generation of arbitrary multimode entangled states of light with linear optics'',
{\em Phys. Rev. A} {\bf 68}, 4, 042325 (2003).

\item {\bf [Fiur\'{a}\v{s}ek-Du\v{s}ek 04]}:
J. Fiur\'{a}\v{s}ek, \& M. Du\v{s}ek,
``Probabilistic quantum multimeters'',
{\em Phys. Rev. A} {\bf 69}, 3, 032302 (2004);
quant-ph/0308111.

\item {\bf [Fiur\'{a}\v{s}ek-Cerf 04]}:
J. Fiur\'{a}\v{s}ek, \& N. J. Cerf,
``How to measure squeezing and entanglement of Gaussian states without
homodyning'',
{\em Phys. Rev. Lett.} {\bf 93}, 6, 063601 (2004);
quant-ph/0311119.

\item {\bf [Fiur\'{a}\v{s}ek 04]}:
J. Fiur\'{a}\v{s}ek,
``Optimal probabilistic cloning and purification of quantum states'',
quant-ph/0403165.

\item {\bf [Fiur\'{a}\v{s}ek-Cerf-Polzik 04]}:
J. Fiur\'{a}\v{s}ek, N. J. Cerf, \& E. S. Polzik,
``Quantum cloning at the light-atoms interface: Copying a coherent light
state into two atomic quantum memories'',
quant-ph/0404054.

\item {\bf [Fivel 91]}:
D. I. Fivel,
``Geometry underlying no-hidden-variable theorems'',
{\em Phys. Rev. Lett.} {\bf 67}, 3, 285-289 (1991).

\item {\bf [Fivel 94]}:
D. I. Fivel,
``How interference effects in mixtures determine
the rules of quantum mechanics'',
{\em Phys. Rev. A} {\bf 50}, 3, 2108-2119 (1994).

\item {\bf [Fivel 95]}:
D. I. Fivel,
``Remarkable phase oscillations appearing in the
lattice dynamics of Einstein-Podolsky-Rosen states'',
{\em Phys. Rev. Lett.} {\bf 74}, 6, 835-838 (1995).

\item {\bf [Fivel 97]}:
D. I. Fivel,
``Dynamical reduction theory of Einstein-Podolsky-Rosen
correlations and a possible origin of CP violations'',
{\em Phys. Rev. A} {\bf 56}, 1, 146-156 (1997).

\item {\bf [Fivel 98]}:
D. I. Fivel,
``How to probe for dynamical structure in the
collapse of entangled states using nuclear magnetic resonance'',
{\em Phys. Lett. A} {\bf 248}, 2-4, 139-144 (1998).

\item {\bf [Fivel 99]}:
D. I. Fivel,
``Multiparticle entanglement'',
{\em Found. Phys.} {\bf 29}, 4, 527-552 (1999).

\item {\bf [Fivel 01]}:
D. I. Fivel,
``How a quantum theory based on generalized coherent
states resolves the EPR and measurement problems'',
quant-ph/0104123.

\item {\bf [Fivel 02]}:
D. I. Fivel,
``How entangled pairs act as measuring rods on manifolds of generalized
coherent states'',
{\em Phys. Rev. A} {\bf 66}, 5, 052108 (2002).

\item {\bf [Fivel 03]}:
D. I. Fivel,
``Disappearance of the measurement paradox in a metaplectic extension of
quantum dynamics'',
quant-ph/0311145.

\item {\bf [Flam 03]}:
F. Flam,
``Quantum cryptography's only certainty: Secrecy'',
{\bf Science} {\bf 253}, 5022, 858 (1991).

\item {\bf [Flammia-Silberfarb-Caves 04]}:
S. T. Flammia, A. Silberfarb, \& C. M. Caves,
``Minimal informationally complete measurements for pure states'',
quant-ph/0404137.
See {\bf [Finkelstein 04]}.

\item {\bf [Fleischhauer 00]}:
M. Fleischhauer,
``On quantum logic operations based on photon-exchange
interactions in an ensemble of non-interacting atoms'',
quant-ph/0006042.

\item {\bf [Fleischhauer-Gong 02]}:
M. Fleischhauer, \& S. Gong,
``Stationary source of nonclassical or entangled atoms'',
{\em Phys. Rev. Lett.} {\bf 88}, 7, 070404 (2002);
quant-ph/0110057.

\item {\bf [Fleming 95]}:
G. N. Fleming,
``A GHZ argument for a single spinless particle'',
in D. M. Greenberger, \& A. Zeilinger (eds.),
{\em Fundamental problems in
quantum theory: A conference held in honor of professor
John A. Wheeler, Ann. N. Y.
Acad. Sci.} {\bf 755}, 646-653 (1995).

\item {\bf [Fleming-Butterfield 93]}:
G. Fleming, \& J. N. Butterfield,
``Is there superluminal causation in quantum theory?'',
in A. van der Merwe, \& F. Selleri (eds.),
{\em Bell's theorem and the foundations of modern physics.
Proc.\ of an international
conference (Cesena, Italy, 1991)},
World Scientific, Singapore, 1993, pp.~203-205.

\item {\bf [Fleming 00]}:
G. N. Fleming,
``Operational quantum physics'',
{\em Stud. Hist. Philos. Sci. Part B: Stud. Hist. Philos. Mod. Phys.}
{\bf 31}, 1, 117-125 (2000).
Review of {\bf [Busch-Grabowski-Lahti 95]}.

\item {\bf [Flitney-Abbott 02 a]}:
A. P. Flitney, \& D. Abbott,
``Quantum version of the Monty Hall problem'',
{\em Phys. Rev. A} {\bf 65}, 6, 062318 (2002);
quant-ph/0109035.

\item {\bf [Flitney-Abbott 02 b]}:
A. P. Flitney, \& D. Abbott,
``An introduction to quantum game theory'',
{\em Fluctuation and Noise Lett.} {\bf 2}, 4, R175-R187 (2002).

\item {\bf [Flitney-Ng-Abbott 02]}:
A. P. Flitney, J. Ng, \& D. Abbott,
``Quantum Parrondo's games'',
{\em Physica A} {\bf 314}, special issue (Gene Stanley Conf., Sicily, 2001), 35-42 (2002);
quant-ph/0201037.

\item {\bf [Flitney-Abbott 03 a]}:
A. P. Flitney, \& D. Abbott,
``Advantage of a quantum player over a classical one in $2 \times 2$ quantum games'',
{\em Proc. R. Soc. Lond. A} {\bf 459} 2463-2474 (2003);
quant-ph/0209121.

\item {\bf [Flitney-Abbott-Johnson 03]}:
A. P. Flitney, D. Abbott, \& N. F. Johnson,
``Quantum random walks with history dependence'',
quant-ph/0311009.

\item {\bf [Flitney-Abbott 03 b]}:
A. P. Flitney, \& D. Abbott,
``Quantum two and three person duels'',
{\em SPIE Symposium on Fluctuations and Noise (Santa Fe, New Mexico 2003)};
quant-ph/0305058.

\item {\bf [Flitney-Abbott 04]}:
A. P. Flitney, \& D. Abbott,
``Quantum games with decoherence'',
quant-ph/0408070.

\item {\bf [Flores-Ramakrishna 01]}:
K. L. Flores, \& V. Ramakrishna,
``Control of switched networks via quantum methods'',
quant-ph/0110001.

\item {\bf [Floyd 99]}:
E. R. Floyd,
``Which causality? Differences between the trajectory and Copenhagen
analysis of an impulsive perturbation'',
{\em Int. J. Mod. Phys. A} {\bf 14}, ?, 1111-1124 (1999).
Erratum: {\em Int. J. Mod. Phys. A} {\bf 16}, ?, 2447 (2001);
quant-ph/9708026.

\item {\bf [Floyd 00 a]}:
E. R. Floyd,
``Classical limit of the trajectory representation of quantum mechanics,
loss of information and residual indeterminacy'',
{\em Int. J. Mod. Phys. A} {\bf 15}, 9, 1363-1378 (2000);
quant-ph/9907092.

\item {\bf [Floyd 00 b]}:
E. R. Floyd,
`Extended version of ``The philosophy of the trajectory representation
of quantum mechanics'',\',
quant-ph/0009070.

\item {\bf [Floyd 03]}:
E. R. Floyd,
``Differences between the trajectory representation and Copenhagen
regarding the past and present in quantum theory'',
quant-ph/0307090.

\item {\bf [Foadi-Selleri 00]}:
R. Foadi, \& F. Selleri,
``Quantum mechanics versus local realism for neutral kaon pairs'',
{\em Phys. Rev. A} {\bf 61}, 1, 012106 (2000).

\item {\bf [Fock 31]}:
V. A. Fock,
{\em Fundamentals of quantum mechanics}, 1931;
(English version of the 2nd edition) URSS Publishers, Moscow, 1976.

\item {\bf [Foldi-Czirjak-Benedict 01]}:
P. Foldi, A. Czirjak, \& M. G. Benedict,
``Rapid and slow decoherence in conjunction with dissipation
in a system of two-level atoms'',
{\em Phys. Rev. A} (2001);
quant-ph/0101072.

\item {\bf [Foldi-Benedict-Czirjak 02]}:
P. Foldi, M. G. Benedict, \& A. Czirjak,
``Preparation of decoherence-free, subradiant states in a cavity'',
{\em Phys. Rev. A} {\bf 65}, 2, 021802 (2002);
quant-ph/0106025.

\item {\bf [Folk-Potok-Marcus-Umansky 03]}:
J. A. Folk, R. M. Potok, C. M. Marcus, \& V. Umansky,
``A gate-controlled bidirectional spin filter using quantum coherence'',
{\em Science} {\bf 299}, ?, 679-? (2003).

\item {\bf [Folman-Schmiedmayer-Ritsch-Vitali 01]}:
R. Folman, J. Schmiedmayer, H. Ritsch, \& D. Vitali,
``On the observation of decoherence with a movable mirror'',
{\em Eur. Phys. J. D} {\bf 13}, 1, 93-107 (2001).

\item {\bf [Folman-Schmiedmayer 01]}:
R. Folman, \& J. Schmiedmayer,
``Bose-Einstein condensates: Mastering the language of atoms'',
{\em Nature} {\bf 413}, 6855, 466-467 (2001).
See {\bf [H\"{a}nsel-Hommelhoff-H\"{a}nsch-Reichel 01]}.

\item {\bf [Folse 85]}:
H. J. Folse,
{\em The philosophy of Niels Bohr},
North-Holland, Amsterdam, 1985.

\item {\bf [Folse 87]}:
H. J. Folse,
``Niels Bohr's concept of reality'',
in P. J. Lahti, \& P. Mittelstaedt (eds.),
{\em Symp.\ on the Foundations of Modern Physics 1987:
The Copenhagen Interpretation 60 Years after the Como Lecture},
World Scientific, Singapore, 1987, pp.~161-179.

\item {\bf [Folse 89]}:
H. J. Folse,
``Bohr on Bell'',
in J. T. Cushing, \& E. McMullin (eds.),
{\em Philosophical consequences of quantum mechanics},
University of Notre Dame Press, Notre Dame, Indiana, 1989, pp.~254-271.

\item {\bf [Folse 94]}:
H. J. Folse,
``Bohr's framework of complementarity and the realism debate'',
in {\bf [Faye-Folse 94]}, pp.~119-139.

\item {\bf [Folse 96]}:
H. J. Folse,
``The Bohr-Einstein debate and the philosophers' debate over
realism versus anti-realism'',
in R. S. Cohen, R. Hilpinen, \& Q. Renzong (eds.),
{\em Realism and anti-realism in the philosophy of science},
Kluwer Academic, Dordrecht, Holland, 1996, pp.~?-?.

\item {\bf [Fonseca-Souto Ribeiro-P\'{a}dua-Monken 99]}:
E. J. S. Fonseca, P. H. Souto Ribeiro, S. P\'{a}dua, \& C. H. Monken,
``Quantum interference by a nonlocal double slit'',
{\em Phys. Rev. A} {\bf 60}, 2, 1530-1533 (1999).

\item {\bf [Fonseca-Machado da Silva-Monken-P\'{a}dua 00]}:
E. J. S. Fonseca, J. C. Machado da Silva, C. H. Monken, \& S. P\'{a}dua,
``Controlling two-particle conditional interference'',
{\em Phys. Rev. A} {\bf 61}, 2, 023801 (2000).

\item {\bf [Fonseca Romero-Useche Laverde-Torres Ardila 03]}:
K. M. Fonseca Romero, G. Useche Laverde, \& F. Torres Ardila,
``Optimal control of one-qubit gates'',
{\em J. Phys. A} {\bf 36}, 3, 841-849 (2003);
quant-ph/0203100.

\item {\bf [Fonseca Romero-Mokarzel-Terra Cunha-Nemes 03]}:
K. M. Fonseca Romero, S. G. Mokarzel, M. O. Terra Cunha, \& M. C. Nemes,
``Realistic decoherence free subspaces'',
quant-ph/0304018.

\item {\bf [Ford-O'Connell 01]}:
G. W. Ford, \& R. F. O'Connell,
``Decoherence without dissipation'',
{\em Phys. Lett. A} {\bf 286}, ?, 87-? (2001);
quant-ph/0301054.

\item {\bf [Ford-Lewis-O'Connell 01]}:
G. W. Ford, J. T. Lewis, \& R. F. O'Connell,
``Quantum measurement and decoherence'',
{\em Phys. Rev. A} {\bf 64}, 3, 032101 (2001);
quant-ph/0301055.
Comment: {\bf [Gobert-von Delft-Ambegaokar 03]}.
Reply: {\bf [Ford-O'Connell 04]}.

\item {\bf [Ford-O'Connell 02]}:
G. W. Ford, \& R. F. O'Connell,
``Wave packet spreading: Temperature and squeezing
effects with applications to quantum measurement and decoherence'',
{\em Am. J. Phys.} {\bf 70}, 3, 319-324 (2002);
quant-ph/0301057.

\item {\bf [Ford-O'Connell 03]}:
G. W. Ford, \& R. F. O'Connell,
``Decoherence at zero temperature'',
{\em J. Opt. B: Quantum Semiclass. Opt.} {\bf 5}, S349-? (2003);
quant-ph/0311019.

\item {\bf [Ford-O'Connell 04 a]}:
G. W. Ford, \& R. F. O'Connell,
``Reply to `Comment on ``Quantum measurement and decoherence''\,'\,'',
{\em Phys. Rev. A} {\bf 70}, 2, 026102 (2004);
quant-ph/0409178.
Reply to {\bf [Gobert-von Delft-Ambegaokar 03]}.
See {\bf [Ford-Lewis-O'Connell 01]}.

\item {\bf [Ford-O'Connell 04 b]}:
G. W. Ford, \& R. F. O'Connell,
``Wigner distribution analysis of a Schr\"{o}dinger cat superposition of
displaced equilibrium coherent states'',
{\em Acta Phys. Hung. B} {\bf 20}, 91-? (2004);
quant-ph/0409188.

\item {\bf [Fortnow 02]}:
L. Fortnow,
``Theory of quantum computing and communication'',
quant-ph/0203074.

\item {\bf [Fortunato-Selleri 76]}:
D. Fortunato, \& F. Selleri,
``Sensitive observables on infinite-dimensional Hilbert spaces'',
{\em Int. J. Theor. Phys.} {\bf 15}, 5, 333-338 (1976).
See {\bf [Capasso-Fortunato-Selleri 73]}.

\item {\bf [Fortunato-Tombesi-Schleich 98]}:
M. Fortunato, P. Tombesi, \& W. P. Schleich,
``Quantum-non-demolition endoscopic tomography'',
{\em Proc.\ of the 7th Int. Seminar on Quantum Optics, Raubiki
(Minsk), Belarus, 1998};
quant-ph/9808052.

\item {\bf [Fortunato-Harel-Kurizki 98 a]}:
M. Fortunato, G. Harel, \& G. Kurizki,
``?'',
{\em Optics Communications} {\bf 147}, ?, 71-? (1998).
See {\bf [Fortunato-Harel-Kurizki 98 b]}.

\item {\bf [Fortunato-Harel-Kurizki 98 b]}:
M. Fortunato, G. Harel, \& G. Kurizki,
``Using conditional measurements to combat decoherence'',
{\em Proc.\ of the Workshop ``Macroscopic Quantum Tunneling and
Coherence'',
Naples (Italy), June 10-13 1998, J. Superconductivity, October 1998};
quant-ph/9808057.
See {\bf [Fortunato-Harel-Kurizki 98 a]}.

\item {\bf [Fortunato-Tombesi-Schleich 98]}:
M. Fortunato, P. Tombesi, \& W. P. Schleich,
``Endoscopic tomography and quantum-nondemolition'',
{\em Phys. Rev. A} {\bf 59}, 1, 718-727 (1999);
quant-ph/9812043.

\item {\bf [Fortunato-Harel-Kurizki 98]}:
M. Fortunato, G. Harel, \& G. Kurizki,
``Recovering coherence via conditional measurements'',
{\em Opt. Comm.} {\bf 147}, ?, 71-? (1998);
quant-ph/0003119.

\item {\bf [Fortunato-Tombesi-Vitali-Raimond 00]}:
M. Fortunato, P. Tombesi, D. Vitali, \& J.-M. Raimond,
``Quantum feedback for protection of Schr\"{o}dinger-cat states'',
in {\em Foundation of Quantum Mechanics (Lecce, Italy, 1998)},
World Scientific, Singapore, 2000;
quant-ph/0004025.

\item {\bf [Fortunato-Tombesi-Vitali-Raimond 00]}:
M. Fortunato, P. Tombesi, D. Vitali, \& J.-M. Raimond,
``Autofeedback scheme for Schr\"{o}dinger cat preservation in microwave
cavities'',
{\em Fortschr. Phys.} {\bf 48}, 5-7, 431-436 (2000).

\item {\bf [Fortunato-Viola-Hodges-(+2) 02]}:
E. M. Fortunato, L. Viola, J. Hodges,
G. Teklemariam, \& D. G. Cory,
``Implementation of universal control on a decoherence-free qubit'',
{\em New. J. Phys} {\bf 4}, 5.1-5.20 (2002);
quant-ph/0111166.

\item {\bf [Fortunato-Pravia-Boulant-(+3) 02]}:
E. M. Fortunato, M. A. Pravia, N. Boulant, G. Teklemariam,
T. F. Havel, \& D. G. Cory,
``Design of strongly modulating pulses to implement precise effective
Hamiltonians for quantum information processing'',
{\em J. Chem. Phys.} {\bf 116}, 7599 (2002);
quant-ph/0202065.

\item {\bf [Fortunato-Viola-Pravia-(+4) 03]}:
E. M. Fortunato, L. Viola, M. A. Pravia,
E. Knill, R. Laflamme, T. F. Havel, \& D. G. Cory,
``Exploring noiseless subsystems via nuclear magnetic resonance'',
{\em Phys. Rev. A} {\bf 67}, 6, 062303 (2003).

\item {\bf [Foster-Orozco-Castro Beltran-Carmichael 00]}:
G. T. Foster, L. A. Orozco, H. M. Castro-Beltran, \& H. J.
Carmichael,
``Quantum state reduction and conditional time evolution of wave-particle
correlations in cavity QED'',
{\em Phys. Rev. Lett.} {\bf 85}, 15, 3149-3152 (2000).
See {\bf [Osborne 00 e]}.

\item {\bf [Foster-Elby 91]}:
S. Foster, \& A. Elby,
``A SQUID no-go theorem
without macrorealism: What SQUID's really tell us about nature'',
{\em Found. Phys.}
{\bf 21}, 7, 773-785 (1991).

\item {\bf [Fowler-Wellard-Hollenberg 03]}:
A. G. Fowler, C. J. Wellard, \& L. C. L. Hollenberg,
``Error rate of the Kane quantum computer controlled-NOT gate in
the presence of dephasing'',
{\em Phys. Rev. A} {\bf 67}, 1, 012301 (2003).

\item {\bf [Fowler-Hill-Hollenberg 03]}:
A. G. Fowler, C. D. Hill, \& L. C. L. Hollenberg,
``Quantum error correction on linear nearest neighbor qubit arrays'',
quant-ph/0311116.

\item {\bf [Fowler-Devitt-Hollenberg 03]}:
A. G. Fowler, S. J. Devitt, \& L. C. L. Hollenberg,
``Implementation of Shor's algorithm on a linear nearest neighbour qubit
array'',
quant-ph/0402196.

\item {\bf [Franca-Marshall-Santos-Watson 92]}:
H. M. Franca, T. W. Marshall,
E. Santos, \& E. J. Watson,
``Possible interference in the Stern-Gerlach phenomenon'',
{\em Phys. Rev. A} {\bf 46}, 5, 2265-2270 (1992).

\item {\bf [Franke-Huget-Barnett 00]}:
S. Franke, G. Huget, \& S. M. Barnett,
``Hardy state correlations for two trapped ions'',
in V. Bu\v{z}zek, \& D. P. DiVincenzo (eds.),
{\em J. Mod. Opt.} {\bf 47}, 2-3
(Special issue: Physics of quantum information), 145-153 (2000).

\item {\bf [Franke Arnold-Andersson-Barnett-Stenholm 01]}:
S. Franke-Arnold, E. Andersson, S. M. Barnett, \& S. Stenholm,
``Generalized measurements of atomic qubits'',
{\em Phys. Rev. A} {\bf 63}, 5, 052301 (2001).

\item {\bf [Franson-Potocki 88]}:
J. D. Franson, \& K. A. Potocki,
``Single-photon interference over large distances'',
{\em Phys. Rev. A} {\bf 37}, 7, 2511-2515 (1988).

\item {\bf [Franson 89]}:
J. D. Franson,
``Bell inequality for position and time'',
{\em Phys. Rev. Lett.} {\bf 62}, 19, 2205-2208 (1989).
See {\bf [Larsson-Aerts-\.{Z}ukowski 98]}.

\item {\bf [Franson 91 a]}:
J. D. Franson,
``Violations of a simple inequality for classical fields'',
{\em Phys. Rev. Lett.} {\bf 67}, 3, 290-293 (1991).

\item {\bf [Franson 91 b]}:
J. D. Franson,
``Two photon interferometry over large distances'',
{\em Phys. Rev. A} {\bf 44}, 7, 4552-4555 (1991).

\item {\bf [Franson 92]}:
J. D. Franson,
``Photon entanglement in macroscopic systems'',
{\em Phys. Rev. A} {\bf 45}, 11, 8074-8077 (1992).

\item {\bf [Franson 93]}:
J. D. Franson,
``Apparatus and method for quantum mechanical encryption
for the transmission of secure communications'',
patent US5243649, 1993.

\item {\bf [Franson-Ilves 94 a]}:
J. D. Franson, \& H. Ilves,
``Quantum cryptgraphy using polarization feedback'',
in S. M. Barnett, A. K. Ekert, \& S. J. D. Phoenix (eds.),
{\em J. Mod. Opt.} {\bf 41}, 12 (Special issue: Quantum
communication), 2391-2396 (1994).

\item {\bf [Franson-Ilves 94 b]}:
J. D. Franson, \& H. Ilves,
``?'',
{\em Appl. Opt.} {\bf 30}, ?, 2949-? (1994).

\item {\bf [Franson-Jacobs 95]}:
J. D. Franson, \& B. C. Jacobs,
``Operational system for quantum cryptography'',
{\em Electron. Lett.} {\bf 31}, ?, 232-234 (1995).

\item {\bf [Franson 97 a]}:
J. D. Franson,
``Experimental observation of the splitting of
single photons by a beam splitter'',
{\em Phys. Rev. A} {\bf 56}, 3, 1800-1805 (1997).

\item {\bf [Franson 97 b]}:
J. D. Franson,
``Cooperative enhancement of optical quantum gates'',
{\em Phys. Rev. Lett.} {\bf 78}, 20, 3852-3855 (1997).

\item {\bf [Franson-Pittman 98]}:
J. D. Franson, \& T. B. Pittman,
``Nonlocality in quantum computing'',
{\em Fortschr. Phys.} {\bf 46}, 6-8, 697-705 (1998).

\item {\bf [Franson-Pittman 99]}:
J. D. Franson, \& T. B. Pittman,
``Quantum logic operations based on photon-exchange interactions'',
{\em Phys. Rev. A} {\bf 60}, 2, 917-936 (1999).


\item {\bf [Franson 00 a]}:
J. D. Franson,
``Inconsistency of local realistic descriptions
of two-photon interferometer experiments'',
{\em Phys. Rev. A} {\bf 61}, 1, 012105 (2000).

\item {\bf [Franson 00 b]}:
J. D. Franson,
``Reply to a review of photon-exchange interactions by Opatrn\'{y} and Kurizki'',
{\em Fortschr. Phys.} {\bf 48}, 9-11 (Special issue: Experimental proposals for quantum computation), 1133-1138 (2000).
See {\bf [Opatrn\'{y}-Kurizki 00]}.

\item {\bf [Franson-Donegan 01]}:
J. D. Franson, \& M. M. Donegan,
``Perturbation theory for quantum-mechanical observables'',
quant-ph/0108018.

\item {\bf [Franson-Donegan-Fitch-(+2) 02]}:
J. D. Franson, M. M. Donegan, M. J. Fitch,
B. C. Jacobs, \& T. B. Pittman,
``High-fidelity quantum logic operations using linear optical elements'',
{\em Phys. Rev. Lett.} {\bf 89}, 13, 137901 (2002);
quant-ph/0202160.

\item {\bf [Franson-Donegan-Jacobs 04]}:
J. D. Franson, M. M. Donegan, \& B. C. Jacobs,
``Generation of entangled ancilla states for use in linear optics quantum computing'',
{\em Phys. Rev. A} {\bf 69}, 5, 052328 (2004);
quant-ph/0303137.

\item {\bf [Franson-Jacobs-Pittman 04 a]}:
J. D. Franson, B. C. Jacobs, \& T. B. Pittman,
``Quantum computing using single photons and the Zeno effect'',
quant-ph/0401133.
Extended version: {\bf [Franson-Jacobs-Pittman 04 c]}.

\item {\bf [Franson-Jacobs-Pittman 04 b]}:
J. D. Franson, B. C. Jacobs, \& T. B. Pittman,
``Quantum logic using linear optics'',
quant-ph/0402097.

\item {\bf [Franson 04]}:
J. D. Franson,
``Comment on photon exchange interactions'',
quant-ph/0405141.

\item {\bf [Franson-Jacobs-Pittman 04 c]}:
J. D. Franson, B. C. Jacobs, \& T. B. Pittman,
``Quantum computing using single photons and the Zeno effect'',
quant-ph/0408097.
Extended version of {\bf [Franson-Jacobs-Pittman 04 a]}.

\item {\bf [Frasca 01]}:
M. Frasca,
``Decoherence as thermodynamic limit of the unitary
evolution in quantum mechanics'',
{\em Phys. Lett. A} {\bf 283}, 5-6, 271-275 (2001).
Erratum: {\em Phys. Lett. A} {\bf 306}, 2-3, 185 (2003).
See {\bf [Frasca 03]}.

\item {\bf [Frasca 02]}:
M. Frasca,
``Classical states and decoherence by unitary
evolution in the thermodynamic limit'',
in R. Bonifacio, \& D. Vitali (eds.),
{\em Mysteries, Puzzles and Paradoxes in Quantum Mechanics IV:
Quantum Interference Phenomena (Gargnano, Italy, 2001)},
{\em J. Opt. B: Quantum Semiclass. Opt.} {\bf 4}, 4, S443-S445 (2002);
quant-ph/0201040.

\item {\bf [Frasca 03]}:
M. Frasca,
``General theorems on decoherence in the thermodynamic limit'',
{\em Phys. Lett. A} {\bf 308}, 2-3, 135-139 (2003).
See {\bf [Frasca 01]}.

\item {\bf [Frasca 04 a]}:
M. Frasca,
``Stern-Gerlach experiment and Bohm limit'',
quant-ph/0402072.

\item {\bf [Frasca 04 b]}:
M. Frasca,
``Decoherence in the thermodynamic limit: A general result'',
quant-ph/0403111.

\item {\bf [Freedman-Meyer 98]}:
M. H. Freedman, \& D. Meyer,
``Projective plane and planar quantum codes'',
quant-ph/9810055.

\item {\bf [Freedman 02]}:
M. H. Freedman,
``Poly-locality in quantum computing'',
{\em Found. Comp. Math.} {\bf 2}, 2, 145-154 (2002);
quant-ph/0001077.

\item {\bf [Freedman-Clauser 72]}:
S. J. Freedman, \& J. F. Clauser,
``Experimental test of local hidden-variable theories'',
{\em Phys. Rev. Lett.} {\bf 28}, 14, 938-941 (1972).
Reprinted in {\bf [Wheeler-Zurek 83]}, pp.~414-417.

\item {\bf [Freedman 72]}:
S. J. Freedman,
``Experimental test of local hidden-variable theories'',
Ph.\ D. thesis, University of California at Berkeley, 1972.

\item {\bf [Freegarde-Segal 03]}:
T. Freegarde, \& D. Segal,
``Algorithmic cooling in a momentum state quantum computer'',
{\em Phys. Rev. Lett.} {\bf 91}, 3, 037904 (2003).

\item {\bf [Freyberger 95]}:
M. Freyberger,
``Simple example of nonlocality: Atoms
interacting with correlated quantized fields'',
{\em Phys. Rev. A} {\bf 51}, 4, 3347-3350 (1995).

\item {\bf [Freyberger-Aravind-Horne-Shimony 96]}:
M. Freyberger, P. K.
Aravind, M. A. Horne, \& A. Shimony,
``Proposed test of Bell's inequality without
detection loophole by using entangled Rydberg atoms'',
{\em Phys. Rev. A} {\bf 53}, 3, 1232-1244 (1996).

\item {\bf [Freyberger-Bardroff-Leichtle-(+2) 97]}:
M. Freyberger, P. Bardroff, C. Leichtle, G. Schrade, \& W. Schleich,
``The art of measuring quantum states'',
{\em Phys. World} {\bf 10}, 11, 41-45 (1997).

\item {\bf [Freyberger 98]}:
M. Freyberger,
``Quantum state measurements and complementarity'',
{\em Phys. Lett. A} {\bf 242}, 4-5, 193-197 (1998).

\item {\bf [Friberg-Hong-Mandel 85]}:
S. Friberg, C. K. Hong, \& L. Mandel,
``Measurement of time delays in the parametric production of photon pairs'',
{\em Phys. Rev. Lett.} {\bf 54}, 18, 2011-2013 (1985).

\item {\bf [Friberg 97]}:
S. R. Friberg,
``Physics: Quantum nondemolition:
Probing the mystery of quantum mechanics'',
{\em Science} {\bf 278}, 5340, 1088-1089 (1997).

\item {\bf [Frieden-Plastino 00]}:
B. R. Frieden, \& A. Plastino,
``Classical trajectories compatible with quantum mechanics'',
quant-ph/0006012.

\item {\bf [Friedenauer-Sj\"{o}qvist 03]}:
A. Friedenauer, \& E. Sj\"{o}qvist,
``Noncyclic geometric quantum computation'',
{\em Phys. Rev. A} {\bf 67}, 2, 024303 (2003).

\item {\bf [Friedman-Putnam 78]}:
M. Friedman, \& H. Putnam,
``Quantum logic, conditional probability and inference'',
{\em Dialectica} {\bf 32}, ?, 305-315 (1978).

\item {\bf [Friedman-Patel-Chen-(+2) 00]}:
J. R. Friedman, V. Patel, W. Chen, S. K. Tolpygo, \& J. E. Lukens,
``Quantum superposition of distinct macroscopic states'',
{\em Nature} {\bf 406}, 6791, 43-46 (2000).

\item {\bf [Friedrich-Herschbach 03]}:
B. Friedrich, \& D. Herschbach,
``Stern and Gerlach: How a bad cigar helped reorient atomic physics'',
{\em Phys. Today} {\bf 56}, 12, ?-?, (2003).

\item {\bf [Friesen-Joynt-Eriksson 02]}:
M. Friesen, R. Joynt, \& M. A. Eriksson,
``Pseudo-digital quantum bits'',
{\em Appl. Phys. Lett.} {\bf 81}, ?, 4619-? (2002).

\item {\bf [Friesen-Rugheimer-Savage-(+4) 03]}:
M. Friesen, P. Rugheimer, D. E. Savage,
M. G. Lagally, D. W. van der Weide, R. Joynt, \& M. A. Eriksson,
``Practical design and simulation of silicon-based quantum-dot qubits'',
{\em Phys. Rev. B} {\bf 67}, 121301 (2003).

\item {\bf [Frigg 02]}:
R. Frigg,
`On the property structure of realist collapse of quantum mechanics
and the so-called ``counting anomaly''\,' (2002),
PITT-PHIL-SCI00000851.

\item {\bf [Frisch 71]}:
O. R. Frisch,
``The conceptual problem of quantum theory from the experimentalist's point of view'',
in E. Bastin (ed.),
{\em Quantum theory and beyond},
Cambridge University Press, Cambridge, 1971, pp.~13-21.

\item {\bf [Frisch 79]}:
O. R. Frisch,
{\em What little I remember},
Cambridge University Press, Cambridge, 1979.
Spanish version: {\em De la fisi\'{o}n del
\'{a}tomo a la bomba de hidr\'{o}geno. Recuerdos de un f\'{\i}sico nuclear},
Alianza, Madrid, 1982.
Review: {\bf [Gowing 79 b]}.

\item {\bf [Frodl 93]}:
P. Frodl,
``Bell's theorem and the concept of counterfactual
definiteness'',
in A. van der Merwe, \& F. Selleri (eds.),
{\em Bell's theorem and the
foundations of modern physics.
Proc.\ of an international conference (Cesena, Italy, 1991)},
World Scientific, Singapore, 1993, pp.~215-219.

\item {\bf [Froissart 81]}:
M. Froissart,
``Constructive generalization of Bell's inequalities'',
{\em Nuovo Cimento B} {\bf 64}, 2, 241-251 (1981).

\item {\bf [Fry-Thompson 76]}:
E. S. Fry, \& R. C. Thompson,
``Experimental test
of local hidden-variable theories'',
{\em Phys. Rev. Lett.} {\bf 37}, 8, 465-468 (1976).
Reprinted in {\bf [Wheeler-Zurek 83]}, pp.~418-421.

\item {\bf [Fry-Li 92]}:
E. S. Fry, \& S. Li,
``Bell inequalities with nearly 100\%
efficient detectors'',
in T. D. Black, M. Mart\'{\i}n Nieto, H. S. Pilloff, M. O. Scully,
\& R. M. Sinclair (eds.),
{\em Foundations of quantum mechanics. Workshop (Santa Fe,
New Mexico, 1991)},
World Scientific, Singapore, 1992, pp.~175-180.

\item {\bf [Fry 93]}:
E. S. Fry,
``?'',
in C. P. Wang (ed.),
{\em Proc.\ of the
Int.\ Conf.\ on Lasers '93 (?, 1993)},
STS Press, Mc Lean, Virginia, 1993, pp.~?-?.

\item {\bf [Fry 94]}:
E. S. Fry,
``The strong Bell inequalities. A proposed experimental test'',
in D. Han, Y. S. Kim, N. H. Rubin, Y. H. Shih, \& W. W. Zachary (eds.),
{\em 3rd Int. Workshop
on Squeezed States and Uncertainty Relations
(Baltimore, Maryland, 1993)}, NASA, 1994, pp.~575-580.

\item {\bf [Fry 95]}:
E. S. Fry,
``Bell inequalities and two experimental tests with mercury'',
{\em Quantum Semiclass. Opt.} {\bf 7}, 3, 259-278 (1995).
Presented in
the Int. Workshop on Laser and Quantum Optics
(Nathiagali, Pakistan, 1994).

\item {\bf [Fry-Walther-Li 95]}:
E. S. Fry, T. Walther, \& S. Li,
``Proposal for a loophole-free test of the Bell inequalities'',
{\em Phys. Rev. A} {\bf 52}, 6, 4381-4395 (1995).

\item {\bf [Fry-Walther 02]}:
E. S. Fry, \& T. Walther,
``Atom based tests of the Bell inequalities --
The legacy of John Bell continues'',
in {\bf [Bertlmann-Zeilinger 02]}, pp.~103-118.

\item {\bf [Fu-Wang-Solomon 01]}:
H. Fu, X. Wang, \& A. I. Solomon,
``Maximal entanglement of nonorthogonal states: classification'',
quant-ph/0105099.

\item {\bf [Fu-Luo-Xiao-Zeng 99]}:
L. Fu, J. Luo, L. Xiao, \& X. Zeng,
``Experimental realization of discrete Fourier transformation on NMR
quantum computer'',
quant-ph/9905083.

\item {\bf [Fu-Chen-Zhao-Chen 01]}:
L.-B. Fu, J.-L. Chen, X.-G. Zhao, \& S.-G. Chen,
``Understanding quantum entanglement of bipartite system
based on Bell inequality'',
quant-ph/0112155.

\item {\bf [Fu-Chen-Zhao 03]}:
L.-B. Fu, J.-L. Chen, \& X.-G. Zhao,
``Maximal violation of the Clauser-Horne-Shimony-Holt inequality for two
qutrits'',
{\em Phys. Rev. A} {\bf 68}, 2, 022323 (2003);
quant-ph/0208150.

\item {\bf [Fu-Chen-Chen 04]}:
L.-B. Fu, J.-L. Chen, \& S.-G. Chen,
``Maximal violation of Clauser-Horne-Shimony-Holt inequality for
four-level systems'',
{\em Phys. Rev. A} {\bf 69}, 3, 034305 (2004);
quant-ph/0307011.

\item {\bf [Fu 04]}:
L.-B. Fu,
``General correlation functions of the Clauser-Horne-Shimony-Holt inequality for arbitrarily high-dimensional systems'',
{\em Phys. Rev. Lett.} {\bf 92}, 13, 130404 (2004).

\item {\bf [Fuchs-Caves 94]}:
C. A. Fuchs, \& C. M. Caves,
``Ensemble-dependent bounds for accesible information in quantum
mechanics'',
{\em Phys. Rev. Lett.} {\bf 73}, 23, 3047-3050 (1994).

\item {\bf [Fuchs-Peres 96]}:
C. A. Fuchs, \& A. Peres,
``Quantum-state disturbance versus information gain:
Uncertainty relations for quantum information'',
{\em Phys. Rev. A} {\bf 53}, 4, 2038-2045 (1996);
quant-ph/9512023.

\item {\bf [Fuchs-Gisin-Griffiths-(+2) 97]}:
C. A. Fuchs, N. Gisin, R. B.
Griffiths, C. Niu, \& A. Peres,
``Optimal eavesdropping in quantum cryptography. I.
Information bound and optimal strategy'',
{\em Phys. Rev. A} {\bf 56}, 2, 1163-1172 (1997);
quant-ph/9701039.
See {\bf [Gisin-Huttner 97]}, {\bf [Griffiths-Niu 97]} (II).

\item {\bf [Fuchs 97]}:
C. A. Fuchs,
``Nonorthogonal quantum states maximize classical information capacity'',
{\em Phys. Rev. Lett.} {\bf 79}, 6, 1162-1165 (1997);
quant-ph/9703043.
Reprinted in {\bf [Macchiavello-Palma-Zeilinger 00]}, pp.~207-210.

\item {\bf [Fuchs-van de Graaf 97]}:
C. A. Fuchs, \& J. van de Graaf,
``Cryptographic distinguishability measures for quantum mechanical states'',
quant-ph/9712042.

\item {\bf [Fuchs 98]}:
C. A. Fuchs,
``Information gain vs. state disturbance in quantum theory'',
{\em Fortschr. Phys.} {\bf 46}, 4-5, 535-565 (1998).
Reprinted in S. L. Braunstein (ed.),
{\em Quantum computation: Where do we want to go tomorrow?},
Wiley-VCH, Weinheim, 1999, pp.~229–259;
quant-ph/9611010.

\item {\bf [Fuchs 00]}:
C. A. Fuchs,
``Just two nonorthogonal quantum states'',
in P. Kumar, G. M. D'Ariano, and O. Hirota (eds.),
{\em Quantum communication, computing, and measurement}
Kluwer, Dordrecht, Holland, 2000, pp.~11–16.
quant-ph/9810032.

\item {\bf [Fuchs-Peres 00 a]}:
C. A. Fuchs, \& A. Peres,
``Quantum theory needs no `interpretation'\,'',
{\em Phys. Today} {\bf 53}, 3, 70-71 (2000).
Reprinted in {\bf [Fuchs 03 a]}, pp.~416-420 (293-295).
Comments: {\bf [Styer 00 b]}, {\bf [Sobottka 00]}, {\bf [Holladay 00]},
{\bf [Brun-Griffiths 00]}, {\bf [Harris 00]}.
Reply: {\bf [Fuchs-Peres 00 b]}.
See: {\bf [Fuchs 03 a]} (Chap.~23), {\bf [Dennis-Norsen 04]}.

\item {\bf [Fuchs-Peres 00 b]}:
C. A. Fuchs, \& A. Peres,
``Quantum theory -- Interpretation, formulation, inspiration:
Fuchs and Peres reply'',
{\em Phys. Today} {\bf 53}, 9, 14, 90 (2000).
Reprinted in {\bf [Fuchs 03 a]}, pp.~421-424 (295-298).
Reply to {\bf [Styer 00 b]}, {\bf [Sobottka 00]}, {\bf [Holladay 00]},
{\bf [Brun-Griffiths 00]}, {\bf [Harris 00]}.
See {\bf [Fuchs-Peres 00 a]}, {\bf [Fuchs 03 a]} (Chap.~23).

\item {\bf [Fuchs 00 b]}:
C. A. Fuchs,
``Quantum channels'',
{\bf [Macchiavello-Palma-Zeilinger 00]}, pp.~157-160.

\item {\bf [Fuchs-Jacobs 01]}:
C. A. Fuchs, \& K. Jacobs,
``Information-tradeoff relations for finite-strength quantum
measurements'',
{\em Phys. Rev. A} {\bf 63}, 6, 062305 (2001);
quant-ph/0009101.

\item {\bf [Fuchs 01]}:
C. A. Fuchs,
``Quantum foundations in the light of quantum
information'',
(a very slightly different version)
in A. Gonis (ed.), {\em Proc.\ NATO Advanced Research
Workshop on Decoherence and its Implications in Quantum
Computation and Information Transfer};
quant-ph/0106166.

\item {\bf [Fuchs 02 a]}:
C. A. Fuchs,
``The anti-V\"{a}xj\"{o} interpretation of quantum mechanics'',
quant-ph/0204146.

\item {\bf [Fuchs 02 b]}:
C. A. Fuchs,
``Quantum mechanics as quantum information (and only a little more)'',
quant-ph/0205039.
See {\bf [Fuchs 03]}.

\item {\bf [Fuchs 03 a]}:
C. A. Fuchs (with a foreword by N. D. Mermin),
{\em Notes on a Paulian idea. Fundational, historical, anecdotal and
forward-looking thoughts on the quantum. Selected correspondence,
1995-2001},
V\"{a}xj\"{o} University Press, V\"{a}xj\"{o}, Sweden, 2003;
quant-ph/0105039.

\item {\bf [Fuchs 03 b]}:
C. A. Fuchs,
``Quantum mechanics as quantum information, mostly'',
in M. Ferrero (ed.),
{\em Proc. of Quantum Information: Conceptual Foundations,
Developments and Perspectives (Oviedo, Spain, 2002)},
{\em J. Mod. Opt.} {\bf 50}, 6-7, 987-1023 (2003).
See {\bf [Fuchs 02 b]}.

\item {\bf [Fuchs-Sasaki 03 a]}:
C. A. Fuchs, \& M. Sasaki,
`Squeezing quantum information through a classical channel: Measuring the
``quantumness'' of a set of quantum states',
quant-ph/0302092.
See {\bf [Fuchs-Sasaki 03 b]}, {\bf [Audenaert-Fuchs-King-Winter 03]}.

\item {\bf [Fuchs-Sasaki 03 b]}:
C. A. Fuchs, \& M. Sasaki,
``The quantumness of a set of quantum states'',
quant-ph/0302108.
Introduction to {\bf [Fuchs-Sasaki 03 a]}.

\item {\bf [Fuchs-Schack-Scudo 04]}:
C. A. Fuchs, R. Schack, \& P. F. Scudo,
``De Finetti representation theorem for quantum-process tomography'',
{\em Phys. Rev. A} {\bf 69}, 6, 062305 (2004);
quant-ph/0307198.

\item {\bf [Fuchs 04]}:
C. A. Fuchs,
``On the quantumness of a Hilbert space'',
for the Holevo festschrift;
quant-ph/0404122.

\item {\bf [Fuchs-Schack 04]}:
C. A. Fuchs, \& R. Schack,
``Unknown quantum states and operations, a Bayesian view'',
in M. G. A. Paris, \& J. Rehacek (eds.),
{\em Quantum estimation theory},
Springer-Verlag, Berlin, 2004;
quant-ph/0404156.

\item {\bf [Fuentes Schuller-Mann 04]}:
I. Fuentes-Schuller (Guridi before), \& R. B. Mann,
``Alice falls into a black hole: Entanglement in non-inertial frames'',
quant-ph/0410172.

\item {\bf [Fuji 01]}:
A. Fuji,
``A lecture on quantum logic gates'',
quant-ph/0101054.

\item {\bf [Fujii 01 a]}:
K. Fujii,
``A relation between coherent states and
generalized Bell states'',
quant-ph/0105077.

\item {\bf [Fujii 01 b]}:
K. Fujii,
``Generalized Bell states and quantum teleportation'',
quant-ph/0106018.

\item {\bf [Fujii 01 c]}:
K. Fujii,
``From geometry to quantum computation'',
{\em 2nd Int.\ Symp.\ Quantum Theory and Symmetries (Krakow, Poland, 2001)};
quant-ph/0107128.

\item {\bf [Fujii 01 d]}:
K. Fujii,
``Introduction to coherent states and quantum information theory'',
quant-ph/0112090.

\item {\bf [Fujii 03]}:
M. Fujii,
``Continuous-variable quantum teleportation with a conventional laser'',
{\em Phys. Rev. A} {\bf 68}, 5, 050302 (2003).

\item {\bf [Fujisaki-Miyadera-Tanaka 03]}:
H. Fujisaki, T. Miyadera, \& A. Tanaka,
``Dynamical aspects of quantum entanglement for weakly coupled kicked tops'',
{\em Phys. Rev. E} {\bf 67}, 6, 066201 (2003).

\item {\bf [Fujisaki 04]}:
H. Fujisaki,
``Entanglement induced by nonadiabatic chaos'',
{\em Phys. Rev. A} {\bf 70}, 1, 012313 (2004).

\item {\bf [Fujiwara-Algoet 99]}:
A. Fujiwara, \& P. Algoet,
``One-to-one parametrization of quantum channels'',
{\em Phys. Rev. A} {\bf 59}, 5, 3290-3294 (1999).

\item {\bf [Fujiwara 02]}:
A. Fujiwara,
``Estimation of SU(2) operation and dense coding: An information geometric
approach'',
{\em Phys. Rev. A} {\bf 65}, 1, 012316 (2002).

\item {\bf [Fujiwara-Hashizum\'{e} 02]}:
A. Fujiwara, \& T. Hashizum\'{e},
``Additivity of the capacity of depolarizing channels'',
{\em Phys. Lett. A} {\bf 299}, 5-6, 469-475 (2002).

\item {\bf [Fujiwara-Imai 03]}:
A. Fujiwara, \& H. Imai,
``Quantum parameter estimation of a generalized Pauli channel'',
{\em J. Phys. A} {\bf 36}, 29, 8093–8103 (2003).

\item {\bf [Fujiwara 04]}:
A. Fujiwara,
``Estimation of a generalized amplitude-damping channel'',
{\em Phys. Rev. A} {\bf 70}, 1, 012317 (2004).

\item {\bf [Fujiwara-Takeoka-Mizuno-Sasaki 03]}:
M. Fujiwara, M. Takeoka, J. Mizuno, \& M. Sasaki,
``Exceeding the classical capacity limit in a quantum optical channel'',
{\em Phys. Rev. Lett.} {\bf 90}, 16, 167906 (2003);
quant-ph/0304037.

\item {\bf [Fung 01]}:
B. M. Fung,
``Use of pairs of pseudopure states for NMR quantum computing'',
{\em Phys. Rev. A} {\bf 63}, 2, 022304 (2001).

\item {\bf [Funk-Raymer 02]}:
A. C. Funk, \& M. G. Raymer,
``Quantum key distribution using nonclassical photon-number correlations in
macroscopic light pulses'',
{\em Phys. Rev. A} {\bf 65}, 4, 042307 (2002);
quant-ph/0109071.

\item {\bf [Furry 36 a]}:
W. H. Furry,
``Note on the quantum-mechanical theory of measurement'',
{\em Phys. Rev.} {\bf 49}, 5, 393-399 (1936).
See {\bf [Furry 36 b]}.

\item {\bf [Furry 36 b]}:
W. H. Furry,
``Remarks on measurements in quantum theory'',
{\em Phys. Rev.} {\bf 49}, 6, 476 (1936).
See {\bf [Furry 36 a]}.


\item {\bf [Furuichi-Abdel Aty 01]}:
S. Furuichi, \& M. Abdel-Aty,
``Entanglement in squeezed two-level atom'',
to be published in {\em J. Phys. A};
quant-ph/0106012.

\item {\bf [Furusawa-S{\o}rensen-Braunstein-(+3) 98]}:
A. Furusawa, J. S{\o}rensen, S. L. Braunstein, C. A. Fuchs,
H. J. Kimble, \& E. S. Polzik,
``Unconditional quantum teleportation'',
{\em Science} {\bf 282}, 5389, 706-709 (1998).
See {\bf [Caves 98 a]}.

\item {\bf [Furuta 01]}:
S. Furuta,
``Measurement-induced decoherence in a simple quantum system'',
{\em Phys. Rev. A} {\bf 64}, 4, 042110 (2001).


\newpage

\subsection{}


\item {\bf [Gadomsky-Voronov 02]}:
O. N. Gadomsky, \& Y. Y. Voronov,
``On a physical implementation of logical operators NOT and CNOT in a
two-qubit quantum computer controlled by ultrashort optical pulses'',
{\em JETP} {\bf 94}, ?, 882-? (2002).

\item {\bf [Gadella-G\'{o}mez 02]}:
M. Gadella, \& F. G\'{o}mez,
``A unified mathematical formalism for the Dirac
formulation of quantum mechanics'',
{\em Found. Phys.} {\bf 32}, 6, 815-869 (2002).

\item {\bf [G\"{a}hler-Klein-Zeilinger 81]}:
R. G\"{a}hler, A. G. Klein, \& A. Zeilinger,
``Neutron optical test of nonlinear wave mechanics'',
{\em Phys. Rev. A} {\bf 23}, 4, 1611-1617 (1981).
See {\bf [Shull-Atwood-Arthur-Horne 80]}.

\item {\bf [Gainutdinov 01]}:
R. K. Gainutdinov,
``Nonlocal interactions and quantum dynamics'',
{\em J. Phys. A} {\bf 32}, ?, 5657-5677 (1999);
quant-ph/0106110.

\item {\bf [Gaitan 03]}:
F. Gaitan,
``Temporal interferometry: A mechanism for controlling qubit transitions during
twisted rapid passage with possible application to quantum computing'',
{\em Phys. Rev. A} {\bf 68}, 5, 052314 (2003).

\item {\bf [Gale-Guth-Trammell 68]}:
W. Gale, E. Guth, \& G. T. Trammell,
``Determination of the quantum state by measurements'',
{\em Phys. Rev.} {\bf 165}, 5, 1434-1436 (1968).

\item {\bf [Galindo 74]}:
A. Galindo,
``Variables ocultas'',
{\em Revista de la Real Academia de Ciencias Exactas, F\'{\i}sicas y
Naturales, de Madrid} {\bf 68}, 2, 195-207 (1974).

\item {\bf [Galindo 75]}:
A. Galindo,
``Another proof of the Kochen-Specker paradox'',
in {\em Algunas cuestiones de F\'{\i}sica Te\'{o}rica},
G.\ I.\ F.\ T., Zaragoza, 1975, pp.~3-9.

\item {\bf [Galindo-Pascual 78]}:
A. Galindo, \& P. Pascual,
{\em Mec\'{a}nica cu\'{a}ntica},
Alhambra, Madrid, 1978 (1st edition);
Eudema, Madrid, 1989 (2nd edition, 2 Vols.).
English version: {\em Quantum mechanics}, 2 Vols.,
Springer-Verlag, Berlin, 1990 (vol. 1), 1991 (vol. 2).

\item {\bf [Galindo-Pascual 89]}:
A. Galindo, \& P. Pascual,
{\em Problemas de mec\'{a}nica cu\'{a}ntica},
Eudema, Madrid, 1989.

\item {\bf [Galindo 00]}:
A. Galindo,
``Quanta e informaci\'{o}n'',
{\em Revista Espa\~{n}ola de F\'{\i}sica} {\bf 14}, 1, 30-47 (2000).

\item {\bf [Galindo-Mart\'{\i}n Delgado 00]}:
A. Galindo, \& M. A. Mart\'{\i}n-Delgado,
``Family of Grover's quantum-searching algorithms'',
{\em Phys. Rev. A} {\bf 62}, 6, 062303 (2000);
quant-ph/0009086.

\item {\bf [Galindo-Mart\'{\i}n Delgado 02]}:
A. Galindo, \& M. A. Mart\'{\i}n-Delgado,
``Information and computation: Classical and quantum aspects'',
{\em Rev. Mod. Phys.} {\bf 74}, 2, 347-423 (2002);
quant-ph/0112105.

\item {\bf [Galindo 03]}:
A. Galindo,
``Pr\'{o}logo''.
Introduction to the Spanish version of {\bf [Feynman 96]}.

\item {\bf [Gallup-Batelaan-Gay 01]}:
G. A. Gallup, H. Batelaan, \& T. J. Gay,
``Quantum-mechanical analysis of a longitudinal Stern-Gerlach
effect'',
{\em Phys. Rev. Lett.} {\bf 62}, 20, 4508-4511 (2001).

\item {\bf [Galv\~{a}o-Hardy 00 a]}:
E. F. Galv\~{a}o, \& L. Hardy,
``Building multiparticle states with teleportation'',
{\em Phys. Rev. A} {\bf 62}, 1, 012309 (2000);
quant-ph/9906080.

\item {\bf [Galv\~{a}o-Hardy 00 b]}:
E. F. Galv\~{a}o, \& L. Hardy,
``Cloning and quantum computation'',
{\em Phys. Rev. A} {\bf 62}, 2, 022301 (2000);
quant-ph/0002053.

\item {\bf [Galv\~{a}o-Plenio-Virmani 00]}:
E. F. Galv\~{a}o, M. B. Plenio, \& S. Virmani,
``Tripartite entanglement and quantum relative entropy'',
{\em J. Phys. A} {\bf 33}, 48, 8809-8818 (2000);
quant-ph/0008089.

\item {\bf [Galv\~{a}o 00]}:
E. F. Galv\~{a}o,
``Experimental requirements for quantum communication complexity protocols'',
quant-ph/0009014.

\item {\bf [Galv\~{a}o 02]}:
E. F. Galv\~{a}o,
``Feasible quantum communication complexity protocol'',
{\em Phys. Rev. A} {\bf 65}, 1, 012318 (2002);
quant-ph/0106121.

\item {\bf [Galv\~{a}o 02]}:
E. F. Galv\~{a}o,
``Foundations of quantum theory and quantum information applications'',
Ph.\ D. thesis, Oxford University, 2002;
quant-ph/0212124.

\item {\bf [Galv\~{a}o-Hardy 03]}:
E. F. Galv\~{a}o, \& L. Hardy,
``Substituting a qubit for an arbitrarily large number of classical bits'',
{\em Phys. Rev. Lett.} {\bf 90}, 8, 087902 (2003);
quant-ph/0110166.

\item {\bf [Galv\~{a}o 04]}:
E. F. Galv\~{a}o,
``Discrete Wigner functions and quantum computational speed-up'',
quant-ph/0405070.

\item {\bf [Gambetta-Wiseman 03]}:
J. Gambetta, \& H. M. Wiseman,
``The interpretation of non-Markovian stochastic Schr\"{o}dinger equations
as a hidden variables theory'',
{\em Phys. Rev. A} {\bf 68}, 6, 062104 (2003);
quant-ph/0307078.

\item {\bf [Gambetta-Wiseman 04]}:
J. Gambetta, \& H. M. Wiseman,
``Modal dynamics for positive operator measures'',
{\em Found. Phys.} {\bf 34}, 3, 419-448 (2004);
quant-ph/0306145.

\item {\bf [Gao 99]}:
B. Gao,
``Breakdown of Bohr's correspondence principle'',
{\em Phys. Rev. Lett.} {\bf 83}, 21, 4225-4228 (1999).
Comment: {\bf [Tannous-Langlois 99]}.

\item {\bf [Gao-Rosenberry-Batelaan 03]}:
H. Gao, M. Rosenberry, \& H. Batelaan,
``Light storage with light of arbitrary polarization'',
{\em Phys. Rev. A} {\bf 67}, 5, 053807 (2003).

\item {\bf [Gao-Yan-Wang 03 a]}:
T. Gao, F. Yan, \& Z. Wang,
``Achievable efficiencies for probabilistically cloning the states'',
quant-ph/0307186.

\item {\bf [Gao-Yan-Wang 03 b]}:
T. Gao, F. Yan, \& Z. Wang,
``Quantum logic network for probabilistic cloning the quantum states'',
quant-ph/0308036.

\item {\bf [Gao 03]}:
T. Gao,
``Controlled and secure direct communication using GHZ state and
teleportation'',
quant-ph/0312004.

\item {\bf [Garbarino 02]}:
G. Garbarino,
``Local realism for $K^0 \bar{K^0}$ pairs'',
in C. Mataix, \& A. Rivadulla (eds.),
{\em F\'{\i}sica cu\'{a}ntica y realidad.
Quantum physics and reality (Madrid, 2000)},
Editorial Complutense, Madrid, 2002, pp.~297-312;
quant-ph/0104013.

\item {\bf [Garc\'{\i}a Alcaine-\'{A}lvarez 87]}:
G. Garc\'{\i}a Alcaine, \& G. \'{A}lvarez,
``Las mediciones cu\'{a}nticas no violan la causalidad relativista'',
{\em Revista Espa\~{n}ola de F\'{\i}sica} {\bf 1}, 2, 29-35 (1987).

\item {\bf [Garc\'{\i}a Alcaine 97]}:
G. Garc\'{\i}a Alcaine,
``A simple state-dependent proof of the BKS theorem'',
in M. Ferrero, \& A. van der Merwe (eds.),
{\em New developments on fundamental problems in
quantum physics (Oviedo, Spain, 1996)},
Kluwer Academic, Dordrecht, Holland, 1997, pp.~133-136.

\item {\bf [Garc\'{\i}a Alcaine 98 a]}:
G. Garc\'{\i}a Alcaine,
``Teleportaci\'{o}n: Realidad y ficci\'{o}n'',
{\em Revista Espa\~{n}ola de F\'{\i}sica} {\bf 12}, 1, 6-9 (1998).

\item {\bf [Garc\'{\i}a Alcaine 98 b]}:
G. Garc\'{\i}a Alcaine,
``Sobre la transmisi\'{o}n de se\~{n}ales a velocidades superlum\'{\i}nicas
utilizando las correlaciones cu\'{a}nticas'',
{\em Revista Espa\~{n}ola de F\'{\i}sica} {\bf 12}, 2, 62-64 (1998).
See {\bf [Cabello-Cereceda-Garc\'{\i}a de Polavieja 98]}.

\item {\bf [Garc\'{\i}a Alcaine 00 a]}:
G. Garc\'{\i}a Alcaine,
``Enredo cu\'{a}ntico'',
{\em Revista Espa\~{n}ola de F\'{\i}sica} {\bf 14}, 1, 17-29 (2000).

\item {\bf [Garc\'{\i}a Alcaine 00 b]}:
G. Garc\'{\i}a Alcaine,
``Cartas a la direcci\'{o}n'',
{\em Revista Espa\~{n}ola de F\'{\i}sica} {\bf 14}, 3, 65-66 (2000).

\item {\bf [Garc\'{\i}a Alcaine-\'{A}lvarez 01]}:
G. Garc\'{\i}a Alcaine, \& G. \'{A}lvarez,
``Adversus collapsum'',
{\em Revista Espa\~{n}ola de F\'{\i}sica} {\bf 15}, 2, 29-34 (2001).

\item {\bf [Garc\'{\i}a Patr\'{o}n-Fiur\'{a}\v{s}ek-Cerf 04]}:
R. Garc\'{\i}a-Patr\'{o}n, J. Fiur\'{a}\v{s}ek, \& N. J. Cerf,
``Loophole-free test of quantum non-locality using high-efficiency
homodyne detectors'',
quant-ph/0407181.
See {\bf [Garc\'{\i}a Patr\'{o}n-Fiur\'{a}\v{s}ek-Cerf-(+3) 04]}.

\item {\bf [Garc\'{\i}a de Polavieja 96 a]}:
G. Garc\'{\i}a de Polavieja,
``A causal quantum theory in phase space'',
{\em Phys. Lett. A} {\bf 220}, 3, 303-314 (1996).

\item {\bf [Garc\'{\i}a de Polavieja 96 b]}:
G. Garc\'{\i}a de Polavieja,
``Nonstatistical quantum-classical correspondence in phase space'',
{\em Found. Phys. Lett.} {\bf 9}, 5, 411-424 (1996).

\item {\bf [Garc\'{\i}a de Polavieja 97 a]}:
G. Garc\'{\i}a de Polavieja,
``Properties of a causal quantum theory in phase space'',
in M. Ferrero, \& A. van der
Merwe (eds.),
in M. Ferrero, \& A. van der Merwe (eds.),
{\em New developments on fundamental problems in quantum
physics (Oviedo, Spain, 1996)},
Kluwer Academic, Dordrecht, Holland, 1997, pp.~137-139.

\item {\bf [Garc\'{\i}a de Polavieja 97 b]}:
G. Garc\'{\i}a de Polavieja,
``Geometric phase in the causal quantum theories'',
{\em Phys. Lett. A} {\bf 236}, 4, 296-300 (1997).

\item {\bf [Garc\'{\i}a Fern\'{a}ndez-Fern\'{a}ndez Mart\'{\i}nez-P\'{e}rez-Santos 03]}:
P. Garc\'{\i}a-Fern\'{a}ndez, E. Fern\'{a}ndez-Mart\'{\i}nez, E. P\'{e}rez, \& D. J. Santos,
``New encoding schemes for quantum authentication'',
quant-ph/0306068.

\item {\bf [Garc\'{\i}a Gonz\'{a}lez-Alvarellos Bermejo-Garc\'{\i}a Sanz 03]}:
P. Garc\'{\i}a Gonz\'{a}lez,
J. E. Alvarellos Bermejo, \&
J. J. Garc\'{\i}a Sanz,
{\em Introducci\'{o}n al formalismo de la Mec\'{a}nica Cu\'{a}ntica},
Universidad Nacional de Educaci\'{o}n a Distancia, Madrid, 2001.

\item {\bf [Garc\'{\i}a Mata-Saraceno-Spina 03]}:
I. Garc\'{\i}a-Mata, M. Saraceno, \& M. E. Spina,
``Classical decays in decoherent quantum maps'',
{\em Phys. Rev. Lett.} {\bf 91}, 6, 064101 (2003).

\item {\bf [Garc\'{\i}a Patr\'{o}n-Fiur\'{a}\v{s}ek-Cerf-(+3) 04]}:
R. Garc\'{\i}a-Patr\'{o}n, J. Fiur\'{a}\v{s}ek, N. J. Cerf,
J. Wenger, R. Tualle-Brouri, \& P. Grangier,
``Proposal for a loophole-free Bell test using homodyne detection'',
quant-ph/0403191.
See {\bf [Garc\'{\i}a Patr\'{o}n-Fiur\'{a}\v{s}ek-Cerf 04]}.

\item {\bf [Garc\'{\i}a Ripoll-Cirac 03 a]}:
J. J. Garc\'{\i}a-Ripoll, \& J. I. Cirac,
``Quantum computation with unknown parameters'',
{\em Phys. Rev. Lett.} {\bf 90}, 12, 127902 (2003).

\item {\bf [Garc\'{\i}a Ripoll-Zoller-Cirac 03 a]}:
J. J. Garc\'{\i}a-Ripoll, P. Zoller, \& J. I. Cirac,
``Speed optimized two-qubit gates with laser coherent control techniques for
ion trap quantum computing'',
{\em Phys. Rev. Lett.} {\bf 91}, 15, 157901 (2003).

\item {\bf [Garc\'{\i}a Ripoll-Zoller-Cirac 03 b]}:
J. J. Garc\'{\i}a-Ripoll, P. Zoller, \& J. I. Cirac,
``Fast and robust two-qubit gates for scalable ion trap quantum computing'',
quant-ph/0306006.

\item {\bf [Garc\'{\i}a Ripoll-Cirac 03 b]}:
J. J. Garc\'{\i}a-Ripoll, \& J. I. Cirac,
``Quantum computation with cold bosonic atoms in an optical lattice'',
{\em Phil. Trans. R. Soc. Lond. A} {\bf 361}, ?, 1537-1548 (2003);
quant-ph/0406144.

\item {\bf [Gardiner-Cirac-Zoller 97]}:
S. A. Gardiner, J. I. Cirac, \& P. Zoller,
``Nonclassical states and measurement of general
motional observables of a trapped ion'',
{\em Phys. Rev. A} {\bf 55}, 3, 1683-1694 (1997).

\item {\bf [Gardner 85]}:
M. Gardner,
in ``Reality and the quantum theory'',
{\em Phys. Today} {\bf 38}, 11, 136 (1985).
Comment on {\bf [Mermin 85]}.

\item {\bf [Garg 82]}:
A. Garg,
``?'',
Ph.\ D. thesis, Cornell University, 1982.

\item {\bf [Garg-Mermin 82 a]}:
A. Garg, \& N. D. Mermin,
``Comment on
`Hidden variables, joint probability, and the Bell inequalities'\,'',
{\em Phys. Rev. Lett.} {\bf 49}, 3, 242 (1982).
Comment on {\bf [Fine 82 a]}. Reply: {\bf [Fine 82 c]}.
See {\bf [Garg-Mermin 82 c]}.

\item {\bf [Garg-Mermin 82 b]}:
A. Garg, \& N. D. Mermin,
``Bell inequalities with a range of violation that does not
diminish as the spin becomes arbitrarily large'',
{\em Phys. Rev. Lett.} {\bf 49}, 13, 901-904 (1982).
Erratum: {\em Phys. Rev. Lett.} {\bf 49}, 17, 1294 (1982).
See {\bf [Mermin-Schwarz 82]}.

\item {\bf [Garg-Mermin 82 c]}:
A. Garg, \& N. D. Mermin,
``Correlation inequalities and hidden variables'',
{\em Phys. Rev. Lett.} {\bf 49}, 17, 1220-1223 (1982).
See {\bf [Fine 82 a, c, d]}, {\bf [Garg-Mermin 82 a]},
{\bf [Mermin-Garg 96]}.
Comment: {\bf [Horodecki-Horodecki 96 c]}.

\item {\bf [Garg-Mermin 83]}:
A. Garg, \& N. D. Mermin,
``Local realism and
measured correlations in the spin-$s$ Einstein-Podolsky-Rosen experiment'',
{\em Phys. Rev. D} {\bf 27}, 2, 339-348 (1983).

\item {\bf [Garg 83]}:
A. Garg,
``Detector error and Einstein-Podolsky-Rosen
correlations'',
{\em Phys. Rev. D} {\bf 28}, 4, 785-790 (1983).
See {\bf [Mermin-Schwarz 82]}, {\bf [Garg-Mermin 87]}.

\item {\bf [Garg-Mermin 84]}:
A. Garg, \& N. D. Mermin,
``Farkas's lemma and
the nature of reality: Statistical implications of quantum correlations'',
{\em Found. Phys.} {\bf 14}, 1, 1-39 (1984).

\item {\bf [Garg-Leggett 84]}:
A. Garg, \& A. J. Leggett,
``Comment on `Bell's
theorem: Does the Clauser-Horne inequality hold for all local theories?'\,'',
{\em Phys. Rev. Lett.} {\bf 53}, 10, 1019-1020 (1984).
Comment on {\bf [Angelidis 83]}.

\item {\bf [Garg-Mermin 87]}:
A. Garg, \& N. D. Mermin,
``Detector inefficiences
in the Einstein-Podolsky-Rosen experiment'',
{\em Phys. Rev. D} {\bf 35}, 12, 3831-3835 (1987).
See {\bf [Mermin-Schwarz 82]}, {\bf [Garg 83]}.

\item {\bf [Garg 96]}:
A. Garg,
``Decoherence in ion trap quantum computers'',
{\em Phys. Rev. Lett.} {\bf 77}, 5, 964-967 (1996).
See {\bf [Garg 98]}.

\item {\bf [Garg 98]}:
A. Garg,
``Vibrational decoherence in ion trap quantum computers'',
{\em Fortschr. Phys.} {\bf 46}, 6-8, 749-757 (1998);
quant-ph/9803071.
See {\bf [Garg 96]}.

\item {\bf [Garini-Frigeiro 86]}:
V. Garini, \& A. Frigeiro (eds.),
{\em Fundamental aspects of quantum theory (Como, Italy, 1985)},
Plenum Press, New York, 1986.

\item {\bf [Garisto-Hardy 99]}:
R. Garisto, \& L. Hardy,
``Entanglement of projection and a new class of quantum erasers'',
{\em Phys. Rev. A} {\bf 60}, 2, 827-831 (1999);
quant-ph/9808007.

\item {\bf [Garisto 02]}:
R. Garisto,
``What is the speed of quantum information?'',
quant-ph/0212078.

\item {\bf [Garola-Rossi 95]}:
C. Garola, \& A. Rossi (eds.),
{\em The foundations of quantum mechanics --
Historical analysis and open questions (Lecce, Italy, 1993)},
Kluwer Academic, Dordrecht, Holland, 1995.

\item {\bf [Garola 95 a]}:
C. Garola,
``Criticizing Bell: Local realism and quantum
phyiscs reconciled'',
{\em Int. J. Theor. Phys.} {\bf 34}, 2, 253-263 (1995).

\item {\bf [Garola 95 b]}:
C. Garola,
``Pragmatic versus semantic contextuality in
quantum physics'',
{\em Int. J. Theor. Phys.} {\bf 34}, 8, 1383-1396 (1995).

\item {\bf [Garola-Solombrino 96]}:
C. Garola, \& L. Solombrino,
``Semantic realism versus EPR-like paradoxes:
The Furry, Bohm-Aharonov, and Bell paradoxes'',
{\em Found. Phys.} {\bf 26}, 10, 1329-1356 (1996).

\item {\bf [Garola 00]}:
C. Garola,
``Objectivity versus nonobjectivity in quantum mechanics'',
{\em Found. Phys.} {\bf 30}, 9, 1539-1565 (2000).

\item {\bf [Garola 02 a]}:
C. Garola,
``A simple model for an objective interpretation of quantum mechanics'',
{\em Found. Phys.} {\bf 32}, 10, 1597-1615 (2002).

\item {\bf [Garola 02 b]}:
C. Garola,
``MGP versus Kochen-Specker condition in hidden variables theories'',
quant-ph/0211049.

\item {\bf [Garola-Pykacz 04]}:
C. Garola, \& J. Pykacz,
``Locality and measurements within the SR model for an objective
interpretation of quantum mechanics'',
{\em Found. Phys.} {\bf 34}, 3, 449-475 (2004);
quant-ph/0304025.

\item {\bf [Garraway-Stenholm 99]}:
B. M. Garraway, \& S. Stenholm,
``Observing the spin of a free electron'',
{\em Phys. Rev. A} {\bf 60}, 1, 63-79 (1999).

\item {\bf [Garraway 99]}:
B. M. Garraway,
``Quantum state conclusion?'',
{\em Contemp. Phys.} {\bf 40}, 6, 433-435 (1999).

\item {\bf [Garretson-Wiseman-Pope-Pegg 03]}:
J. L. Garretson, H. M. Wiseman, D. T. Pope, \& D. T. Pegg,
`The uncertainty relation in ``which-way'' experiments: How to observe
directly the momentum transfer using weak values',
submitted to {\em J. Opt. B: Quantum Semiclass. Opt.};
quant-ph/0310081.

\item {\bf [Garrett 91]}:
A. Garrett,
``Ockham's razor'',
{\em Phys. World} {\bf 4}, 5, 39-42 (1991).

\item {\bf [Garuccio 78]}:
A. Garuccio,
``Generalized inequalities following from
Einstein locality'',
{\em Lettere al Nuovo Cimento} {\bf 23}, 15, 559-565 (1978).

\item {\bf [Garrison-Mitchell-Chiao-Bolda 98]}:
J. C. Garrison, M. W. Mitchell, R. Y. Chiao, \& E. L. Bolda,
``Superluminal signals: Causal loop paradoxes revisited'',
{\em Phys. Lett. A} {\bf 245}, 1-2, 19-25 (1998);
quant-ph/9810031.

\item {\bf [Garuccio-Selleri 80]}:
A. Garuccio, \& F. Selleri,
``Systematic derivation of all the inequalities of Einstein locality'',
{\em Found. Phys.} {\bf 10}, 3-4, 209-216 (1980).

\item {\bf [Garuccio-Vigier 80]}:
A. Garuccio, \& J.-P. Vigier,
``Possible experimental test of the causal stochastic interpretation of
quantum mechanics: Physical reality of the Broglie waves",
{\em Found. Phys.} {\bf 10}, 9-10, 797-801 (1980).
Comment: {\bf [Costa de Beauregard 81]}.

\item {\bf [Garuccio-Rapisarda 81]}:
A. Garuccio, \& V. Rapisarda,
``?'',
{\em Nouvo Cimento A} {\bf 65}, ?, 269-? (1981).

\item {\bf [Garuccio 95 a]}:
A. Garuccio,
``Can the experiments based on parametric-down
conversion disprove Einstein locality?'',
in M. Ferrero, \& A. van der Merwe (eds.),
{\em Fundamental problems in quantum physics.
Proc.\ of an international symposium (Oviedo, Spain, 1993)},
Kluwer Academic, Dordrecht, Holland, 1995, pp.~103-112.

\item {\bf [Garuccio 95 b]}:
A. Garuccio,
``Hardy's approach, Eberhard's inequality,
and supplementary assumptions'',
{\em Phys. Rev. A} {\bf 52}, 4, 2535-2537 (1995).
See {\bf [Torgerson-Branning-Monken-Mandel 95, 96]}, {\bf [Cabello-Santos 96]}.

\item {\bf [Garuccio 96]}:
A. Garuccio,
``'',
in F. de Martini, G. Denardo, \& Y. H. Shih (eds.),
{\em Quantum interferometry},
VCH Publishers, Weinheim, 1996, pp.~?-?.

\item {\bf [Garuccio 98]}:
A. Garuccio,
``Entangled and factorized states in parametric down-conversion sources'',
{\em Fortschr. Phys.} {\bf 46}, 6-8, 663-671 (1998).

\item {\bf [Garuccio 00]}:
A. Garuccio,
``Coherent entangled states, quantum mechanics and relativity'',
{\em Fortschr. Phys.} {\bf 48}, 5-7, 481-487 (2000).

\item {\bf [Garuccio-Risco Delgado-Selleri 00]}:
A. Garuccio, R. Risco Delgado, \& F. Selleri,
``Nonlocality of some factorable quantum mechanical state
vectors'',
{\em Found. Phys.} {\bf 30}, 2, 321-330 (2000).

\item {\bf [Garuccio-Berardi 03]}:
A. Garuccio, \& V. Berardi,
``On the validity of Clauser and Horne factorizability'',
{\em Found. Phys.} {\bf 33}, 4, 657-664 (2003).

\item {\bf [Gasenzer-Roberts-Burnett 02]}:
T. Gasenzer, D. C. Roberts, \& K. Burnett,
``Limitations of entanglement between photons and atoms coupled out from a
Bose-Einstein condensate'',
{\em Phys. Rev. A} {\bf 65}, 2, 021605 (2002);
quant-ph/0106005.

\item {\bf [Gasparoni-Pan-Walther-(+2) 04]}:
S. Gasparoni, J.-W. Pan, P. Walther, T. Rudolph, \& A. Zeilinger,
``Realization of a photonic controlled-NOT gate sufficient for quantum computation'',
{\em Phys. Rev. Lett.} {\bf 93}, 2, 020504 (2004);
quant-ph/0404107.

\item {\bf [Gatti-Brambilla-Lugiato 03]}:
A. Gatti, E. Brambilla, \& L. A. Lugiato,
``Entangled imaging and wave-particle duality: From the microscopic to the
macroscopic realm'',
{\em Phys. Rev. Lett.} {\bf 90}, 13, 133603 (2003).

\item {\bf [Gavinsky-Kempe-de Wolf 04]}:
D. Gavinsky, J. Kempe, \& R. de Wolf,
``Quantum communication cannot simulate a public coin'',
quant-ph/0411051.

\item {\bf [Gea Banacloche 98]}:
J. Gea-Banacloche,
``Qubit-qubit interaction in quantum computers'',
{\em Phys. Rev. A} {\bf 57}, 1, R1-R4 (1998).
See {\bf [Gea Banacloche 99]} (II).

\item {\bf [Gea Banacloche 99]}:
J. Gea-Banacloche,
``Qubit-qubit interaction in quantum computers. II.
Adder algorithm with diagonal and off-diagonal interactions'',
{\em Phys. Rev. A} {\bf 60}, 1, 185-193 (1999).
See {\bf [Gea Banacloche 98]} (I).

\item {\bf [Gea Banacloche 00 a]}:
J. Gea-Banacloche,
``Quantum codes and macroscopic superpositions'',
{\em Phys. Rev. A} {\bf 61}, 2, 022302 (2000).

\item {\bf [Gea Banacloche 00 b]}:
J. Gea-Banacloche,
``Error correction for mutually interacting qubits'',
{\em Phys. Rev. A} {\bf 62}, 6, 062313 (2000);
quant-ph/0008027.

\item {\bf [Gea Banacloche 02 a]}:
J. Gea-Banacloche,
``Some implications of the quantum nature of laser fields for quantum
computations'',
{\em Phys. Rev. A} {\bf 65}, 2, 022308 (2002).
Comment: {\bf [Itano 03]}.

\item {\bf [Gea Banacloche 02 b]}:
J. Gea-Banacloche,
``Splitting the wave function of a particle in a box'',
{\em Am. J. Phys.} {\bf 70}, 3, 307-312 (2002).
Comment: {\bf [Lakner-Peternelj 03]}.

\item {\bf [Gea Banacloche 02 c]}:
J. Gea-Banacloche,
``Hiding messages in quantum data'',
{\em J. Math. Phys.} {\bf 43}, 9, 4531-4536 (2002).

\item {\bf [Gea Banacloche 02 d]}:
J. Gea-Banacloche,
``Minimum energy requirements for quantum computation'',
{\em Phys. Rev. Lett.} {\bf 89}, 21, 217901 (2002).

\item {\bf [Gea Banacloche 03]}:
J. Gea-Banacloche,
`Reply II to ``Comment on `Some implications of the quantum nature of laser
fields for quantum computations'\,''\,',
{\em Phys. Rev. A} {\bf 68}, 4, 046303 (2003).
Reply to: {\bf [Itano 03]}.
See: {\bf [Gea Banacloche 02 a]}, {\bf [van Enk-Kimble 03]}.

\item {\bf [Gea Banacloche-Burt-Rice-Orozco 04]}:
J. Gea-Banacloche, T. C. Burt, P. R. Rice, \& L. Orozco,
``Entangled and disentangled evolution for a single atom in a driven
cavity'',
quant-ph/0409164.

\item {\bf [Geiger-Obermair-Helm 99]}:
H. Geiger, G. Obermair, \& C. Helm,
``Classical behaviour of many-body systems in Bohmian quantum
mechanics'',
quant-ph/9906082.

\item {\bf [Geislinger 96]}:
E. Geislinger,
``Quantum voodoo'',
{\em Science} {\bf 274}, ?, 1285-1289 (1996).

\item {\bf [Gell-Mann-Hartle 90 a]}:
M. Gell-Mann, \& J. B. Hartle,
``?'',
in K. K. Phua, \& Y. Yamaguchi (eds.),
{\em Proc.\ of the 25th Int.\ Conf.\ on High-Energy Physics},
World Scientific, Singapore, 1990, pp.~?-?.

\item {\bf [Gell-Mann-Hartle 90 b]}:
M. Gell-Mann, \& J. B. Hartle,
``?'',
in S. Kobayashi, H. Ezawa, Y. Murayama, \& S. Nomura (eds.),
{\em Proc.\ of the 3rd Int.\ Symp.\ on the Foundations
of Quantum Mechanics in the Light of the New Technology (Tokyo, 1989)},
Physical Society of Japan, Tokyo, 1990, pp.~?-?.

\item {\bf [Gell-Mann-Hartle 90 c]}:
M. Gell-Mann, \& J. B. Hartle,
``Quantum mechanics in the light of quantum cosmology'',
in {\bf [Zurek 90]}, pp.~425-458.

\item {\bf [Gell-Mann-Hartle 93]}:
M. Gell-Mann, \& J. B. Hartle,
``Classical equations for quantum systems'',
{\em Phys. Rev. D} {\bf 47}, 8, 3345-3382 (1993).

\item {\bf [Gell-Mann-Hartle 94]}:
M. Gell-Mann, \& J. B. Hartle,
``?'', in J. Halliwell, J. P\'{e}rez-Mercader, \& W. Zurek (eds.),
{\em Proc.\ of the NATO Workshop on the Physical Origins
of Time Asymmetry (Mazag\'{o}n, Spain, 1991)},
Cambridge University Press, Cambridge, 1994, pp.~?-?.

\item {\bf [Gell-Mann 94]}:
M. Gell-Mann,
{\em The quark and the jaguar.
Adventures in the simple and the complex},
Freeman, New York (Little Brown, London), 1994.
Spanish version:
{\em El quark y el jaguar. Aventuras en lo simple y lo complejo},
Tusquets, Barcelona, 1995.
Review: {\bf [Mermin 94 b]}.

\item {\bf [Gemmer-Mahler 01]}:
J. Gemmer, \& G. Mahler,
``Entanglement and the factorization-approximation'',
{\em Eur. Phys. J. D} {\bf 17}, 3, 385-393 (2001);
quant-ph/0109140.

\item {\bf [Gemmer-Mahler 02]}:
J. Gemmer, \& G. Mahler,
``Distribution of local entropy in the Hilbert space of
bi-partite quantum systems: Origin of Jaynes' principle'',
quant-ph/0201136.

\item {\bf [Geng 92]}:
T. Geng,
``A new generalized inequality to test the locality assumption'',
{\em Phys. Lett. A} {\bf 162}, 1, 29-31 (1992).

\item {\bf [Genovese-Brida-Novero-Predazzi 00]}:
M. Genovese, G. Brida, C. Novero, \& E. Predazzi,
``First experimental test of Bell inequalities performed
using a non-maximally entangled state'', in
{\em Proc.\ of Winter Institute on Foundations of Quantum Mechanics
and Quantum Optics (Calcutta, 2000)};
quant-ph/0002025.

\item {\bf [Genovese-Brida-Novero-Predazzi 00]}:
M. Genovese, G. Brida, C. Novero, \& E. Predazzi,
``Experimental test of local realism using non-maximally entangled
states'',
quant-ph/0009067.

\item {\bf [Genovese-Novero 00 a]}:
M. Genovese, \& C. Novero,
``On the generation and identification of
optical Schr\"{o}dinger cats'',
{\em Phys. Lett. A} {\bf 271}, 1-2, 48-53 (2000);
quant-ph/0005083.

\item {\bf [Genovese-Novero 00 b]}:
M. Genovese, \& C. Novero,
``New schemes for manipulating quantum states using a Kerr cell'',
quant-ph/0009028.

\item {\bf [Genovese-Novero 00 c]}:
M. Genovese, \& C. Novero,
``Quantum clock synchronisation based on entangled photon pairs
transmission'',
quant-ph/0009119.

\item {\bf [Genovese 00]}:
M. Genovese,
``Proposal of an experimental scheme for realizing a translucent
eavesdropping on a quantum cryptographic channel'',
quant-ph/0012095.

\item {\bf [Genovese-Novero-Predazzi 01]}:
M. Genovese, C. Novero, \& E. Predazzi,
``Can experimental tests of Bell inequalities performed
with pseudoscalar mesons be definitive?'',
hep-ph/0103298.

\item {\bf [Genovese-Novero 02]}:
M. Genovese, \& C. Novero,
``Double entanglement and quantum cryptography'',
{\em Eur. Phys. J. D} {\bf 21}, 1, 109-113 (2002);
quant-ph/0107118.

\item {\bf [Genovese-Novero-Predazzi 02]}:
M. Genovese, C. Novero, \& E. Predazzi,
``On the conclusive tests of local
realism and pseudoscalar mesons'',
{\em Found. Phys.} {\bf 32}, 4, 589-605 (2002);
quant-ph/0106013.

\item {\bf [Genovese 03 a]}:
M. Genovese,
``Review of studies about quantum communication and foundations of quantum
mechanics at IENGF'',
{\em SPIE Proc.\ on Quantum Communications and Quantum Imaging (2003)};
quant-ph/0309160.

\item {\bf [Genovese 03 b]}:
M. Genovese,
``On DNA molecules as quantum measurement devices'',
{\em Found. Phys. Lett.} {\bf 16}, 5, 505-511 (2003).

\item {\bf [Genovese 04]}:
M. Genovese,
``Entanglement properties of kaons and tests of hidden-variable models'',
{\em Phys. Rev. A} {\bf 69}, 2, 022103 (2004);
quant-ph/0305087.

\item {\bf [Genovese-Brida-Chekhova-(+4) 04]}:
M. Genovese, G. Brida, M. Chekhova,
M. Gramegna, L. Krivitsky, S. Kulik, \& M. L. Rastello,
``Experimental realization of a measurement conditional unitary operation
at single photon level and application to detector characterization'',
quant-ph/0409055.

\item {\bf [Georgeot-Shepelyansky 99]}:
B. Georgeot, \& D. L. Shepelyansky,
``Quantum chaos border for quantum computing'',
quant-ph/9909074.

\item {\bf [Georgeot-Shepelyansky 00]}:
B. Georgeot, \& D. L. Shepelyansky,
``Emergence of quantum chaos in the quantum computer core
and how to manage it'',
{\em Phys. Rev. E} {\bf 62}, 5, 6366-6375 (2000);
quant-ph/0005015.

\item {\bf [Georgeot-Shepelyansky 01 a]}:
B. Georgeot, \& D. L. Shepelyansky,
``Exponential gain in quantum computing of quantum chaos and
localization'',
{\em Phys. Rev. Lett.} {\bf 86}, 13, 2890-2893 (2001);
quant-ph/0010005.

\item {\bf [Georgeot-Shepelyansky 01 b]}:
B. Georgeot, \& D. L. Shepelyansky,
``Stable quantum computation of unstable classical chaos'',
{\em Phys. Rev. Lett.} {\bf 86}, 25, 5393-5396 (2001).
Comments: {\bf [Zalka 01]}, {\bf [Di\'{o}si 02 a]}.
Reply:

\item {\bf [Georgeot-Shepelyansky 01 c]}:
B. Georgeot, \& D. L. Shepelyansky,
``Quantum computing of classical chaos:
Smile of the Arnold-Schr\"{o}dinger cat'',
quant-ph/0101004.

\item {\bf [Georgeot-Shepelyansky 01 d]}:
B. Georgeot, \& D. L. Shepelyansky,
``Efficient quantum computing insensitive to phase errors'',
quant-ph/0102082.

\item {\bf [Georgeot-Shepelyansky 01 e]}:
B. Georgeot, \& D. L. Shepelyansky,
``Efficient quantum computation of high
harmonics of the Liouville density distribution'',
quant-ph/0110142.

\item {\bf [Georgeot-Shepelyansky 02]}:
B. Georgeot, \& D. L. Shepelyansky,
``Reply'',
{\em Phys. Rev. Lett.} {\bf 88}, 21, 219802 (2002).
Reply to {\bf [Di\'{o}si 02 a]}.
See: {\bf [Georgeot-Shepelyansky 01 b]}.

\item {\bf [Georgeot-Shepelyansky 03]}:
B. Georgeot, \& D. L. Shepelyansky,
``Les ordinateurs quantiques affrontent le chaos'',
quant-ph/0307103.

\item {\bf [Georgeot 04]}:
B. Georgeot,
``Quantum computing of Poincar\'{e} recurrences and periodic orbits'',
{\em Phys. Rev. A} {\bf 69}, 3, 032301 (2004);
quant-ph/0307233.

\item {\bf [Gerber 98]}:
H.-J. Gerber,
``Searching for evolutions of pure states into
mixed states in the two-state system $K^{0}K^{0}$'',
{\em Phys. Rev. Lett.} {\bf 80}, 14, 2969-2971 (1998).

\item {\bf [Gerjuoy 03]}:
E. Gerjuoy,
``Lower bound on entanglement of formation for the qubit-qudit system'',
{\em Phys. Rev. A} {\bf 67}, 5, 052308 (2003).

\item {\bf [Gerlach-Stern 21]}:
W. Gerlach, \& O. Stern,
``Der experimentelle Nachweis des magnetischen Moments des Silberatoms'',
{\em Zeitschrift f\"{u}r Physik} {\bf 8}, 110-111 (1921).

\item {\bf [Gerlach-Stern 22 a]}:
W. Gerlach, \& O. Stern,
``Der experimentelle Nachweis der Richtungsquantelung im Magnetfeld'',
{\em Zeitschrift f\"{u}r Physik} {\bf 9}, 349-352 (1922).

\item {\bf [Gerlach-Stern 22 b]}:
W. Gerlach, \& O. Stern,
``Das magnetische Moment des Silberatoms'',
{\em Zeitschrift f\"{u}r Physik} {\bf 9}, 353-355 (1922).

\item {\bf [Gerlach 48]}:
W. Gerlach, \& O. Stern,
{\em Die Quantentheorie.
Max Planck, sein Werk und seine Wirkung.
Mit einer Bibliographie der Werke Max Plancks},
Universit\"{a}ts-Verlag, Bonn, 1948.

\item {\bf [Geroch 84]}:
R. Geroch,
``The Everett interpretation'',
{\em No\^{u}s} {\bf 18}, ?, 617-633 (1984).

\item {\bf [Gerry 96 a]}:
C. C. Gerry,
``Complementarity and quantum erasure with
dispersive atom-field interaction'',
{\em Phys. Rev. A} {\bf 53}, 2, 1179-1182 (1996).

\item {\bf [Gerry 96 b]}:
C. C. Gerry,
``Preparation of multiatom entangled states
through dispersive atom-cavity-field interactions'',
{\em Phys. Rev. A} {\bf 53}, 4, 2857-2860 (1996).

\item {\bf [Gerry 96 c]}:
C. C. Gerry,
``Nonlocality of a single photon in cavity QED'',
{\em Phys. Rev. A} {\bf 53}, 6, 4583-4586 (1996).

\item {\bf [Gerry 96 d]}:
C. C. Gerry,
``Preparation of a four-atom Greenberger-Horne-Zeilinger state'',
{\em Phys. Rev. A} {\bf 53}, 6, 4591-4593 (1996).

\item {\bf [Gerry 96 e]}:
C. C. Gerry,
``Proposal for a mesoscopic cavity QED
realization of the Greenberger-Horne-Zeilinger state'',
{\em Phys. Rev. A} {\bf 54}, 4, R2529-R2532 (1996).

\item {\bf [Gerry-Knight 97]}:
C. C. Gerry, \& P. L. Knight,
``Quantum superpositions and Schr\"{o}dinger cat states in
quantum optics'',
{\em Am. J. Phys.} {\bf 65}, 10, 964-973 (1997).

\item {\bf [Gerry-Grobe 99]}:
C. C. Gerry, \& R. Grobe,
``Entanglement-induced population trapping by
Schr\"{o}dinger's cat'',
{\em J. Mod. Opt.} {\bf 46}, 7, 1053-1059 (1999).

\item {\bf [Gershenfeld-Chuang 97]}:
N. A. Gershenfeld, \& I. L. Chuang,
``Bulk spin-resonance quantum computation'',
{\em Science} {\bf 275}, 5298, 350-356 (1997).
See {\bf [Taubes 97]}.

\item {\bf [Gershenfeld-Chuang 98]}:
N. A. Gershenfeld, \& I. L. Chuang,
``Quantum computing with molecules'',
{\em Sci. Am.} {\bf 278}, 6, 50-55 (1998).
Spanish version:
``Computaci\'{o}n cu\'{a}ntica con mol\'{e}culas'',
{\em Investigaci\'{o}n y Ciencia} 263, 44-49 (1998).
Reprinted in {\bf [Cabello 03 a]}, pp.~38-43.

\item {\bf [Gershenfeld 01]}:
N. A. Gershenfeld,
``A quantum conversation'',
{\em Science} {\bf 293} 2035-2038 (2001).

\item {\bf [Geszti 98]}:
T. Geszti,
``Interaction-free measurement and forward scattering'',
{\em Phys. Rev. A} {\bf 58}, 5, 4206-4209 (1998);
quant-ph/9804058.

\item {\bf [Geszti 98]}:
T. Geszti,
``Gravitational self-localization in quantum measurement'',
{\em Phys. Rev. A};
quant-ph/0401086.

\item {\bf [Geurdes 01]}:
J. F. Geurdes,
``Bell inequalities and pseudo-functional densities'',
{\em Int. J. Theor. Phys.} {\bf 7}, 3, 51-56 (2001);
quant-ph/0110094.

\item {\bf [Gheorghiu Svirschevski 02]}:
S. Gheorghiu-Svirschevski,
``Suppression of decoherence in quantum registers by entanglement with a
nonequilibrium environment'',
{\em Phys. Rev. A} {\bf 66}, 3, 032101 (2002);
quant-ph/0112045.

\item {\bf [Gheri-Ellinger-Pellizzari-Zoller 98]}:
K. M. Gheri, K. Ellinger, T. Pellizzari, \& P. Zoller
``Photon-wavepackets as flying quantum bits'',
{\em Fortschr. Phys.} {\bf 46}, 4-5, 401-415 (1998).

\item {\bf [Gheri-Saavedra-T\"{o}rm\"{a}-(+2) 98]}:
K. M. Gheri, C. Saavedra, P. T\"{o}rm\"{a}, J. I. Cirac, \& P. Zoller,
``Enatanglement engineering of one-photon wavepackets using a
single-atom source'',
{\em Phys. Rev. A} {\bf 58}, 4, R2627-R2630 (1998);
quant-ph/9802060.

\item {\bf [Ghim-Zhang 01]}:
Z.-Y. Ghim, \& H. I. Zhang,
`Bell's inequality and a strict assessment of the concept of
``possession''\,',
quant-ph/0102080.

\item {\bf [Ghirardi-Rimini-Weber 86]}:
G.-C. Ghirardi, A. Rimini, \& T. Weber,
``Unified dynamics for microscopic and macroscopic systems'',
{\em Phys. Rev. D} {\bf 34}, 2, 470-491 (1986).
Comment: {\bf [Joos 87]}.
Reply: {\bf [Ghirardi-Rimini-Weber 87]}.

\item {\bf [Ghirardi-Rimini-Weber 87]}:
G.-C. Ghirardi, A. Rimini, \& T. Weber,
``Disentanglement of quantum wave functions: Answer to `Comment
on ``Unified dynamics for microscopic and macroscopic systems''\,'\,'',
{\em Phys. Rev. D} {\bf 36}, 10, 3287-3289 (1987).
Reply to {\bf [Joos 87]}.
See {\bf [Ghirardi-Rimini-Weber 86]}.

\item {\bf [Ghirardi-Rimini-Weber 88]}:
G.-C. Ghirardi, A. Rimini, \& T. Weber,
``?'',
{\em Found. Phys.} {\bf 18}, 1, 1-? (1988).

\item {\bf [Ghirardi-Pearle-Rimini 90]}:
G.-C. Ghirardi, P. Pearle, \& A. Rimini,
``Markov processes in Hilbert space and continuous
localization of systems of identical particles'',
{\em Phys. Rev. A} {\bf 42}, 1, 78-79 (1990).

\item {\bf [Ghirardi-Grassi-Pearle 90 a]}:
G.-C. Ghirardi, R. Grassi, \& P. Pearle,
``Continuous-spontaneous-reduction involving gravity'',
{\em Phys. Rev. A} {\bf 42}, 3, 1057-1065 (1990).

\item {\bf [Ghirardi-Grassi-Pearle 90 b]}:
G.-C. Ghirardi, R. Grassi, \& P. Pearle,
``Relativistic dynamical reduction models:
General framework and examples'',
{\em Found. Phys.} {\bf 20}, 11, 1271-1316 (1990).

\item {\bf [Ghirardi 91]}:
G.-C. Ghirardi,
``?'',
in A. I. Fine, M. Forbes, \& L. Wessels (eds.),
{\em Proc.\ of the 1990 Biennial Meeting of the Philosophy of Science
Association}, East Lansing, Michigan, 1991, vol. 2, pp.~19-?.

\item {\bf [Ghirardi-Pearle 91]}:
G.-C. Ghirardi, \& P. Pearle,
``?'',
in A. I. Fine, M. Forbes, \& L. Wessels (eds.),
{\em Proc.\ of the 1990 Biennial Meeting of the Philosophy of Science
Association}, East Lansing, Michigan, 1991, vol. 2, pp.~35-47.

\item {\bf [Ghirardi 93]}:
G.-C. Ghirardi,
`Bell's requirements for a ``serious theory''\,',
in A. van der Merwe, \& F. Selleri (eds.),
{\em Bell's theorem and the foundations of modern physics.
Proc.\ of an international
conference (Cesena, Italy, 1991)},
World Scientific, Singapore, 1993, pp.~228-243.

\item {\bf [Ghirardi 95]}:
G.-C. Ghirardi,
``Spontaneous wave packet reduction'',
in D. M. Greenberger, \& A. Zeilinger (eds.),
{\em Fundamental problems in quantum theory:
A conference held in honor of professor John
A. Wheeler, Ann. N. Y. Acad. Sci.} {\bf 755}, 506-523 (1995).

\item {\bf [Ghirardi-Weber 95]}:
G.-C. Ghirardi, \& T. Weber,
``An interpretation which is appropriate for dynamical reduction theories'',
in {\bf [Cohen-Horne-Stachel 97 b]}.

\item {\bf [Ghirardi 95]}:
G.-C. Ghirardi,
``Properties and events in a relativistic context:
Revisiting the dynamical reduction program'',
{\em Found. Phys. Lett.} {\bf 9}, 4, 313-355 (1996).

\item {\bf [Ghirardi 98]}:
G.-C. Ghirardi,
``Quantum superpositions and definite perceptions:
Envisaging new feasible experimental tests'',
quant-ph/9810028.

\item {\bf [Ghirardi-Bassi 99]}:
G.-C. Ghirardi, \& A. Bassi,
``Do dynamical reduction models imply that arithmetic does not apply to
ordinary macroscopic objects?'',
{\em Brit. J. Philos. Sci.} {\bf ?}, ?, ?-? (1999);
quant-ph/9810041.

\item {\bf [Ghirardi 99 a]}:
G.-C. Ghirardi,
``The dynamical reduction program:
An example of a quantum theory without observers'',
in {\bf [Greenberger-Reiter-Zeilinger 99]}, pp.~43-58.

\item {\bf [Ghirardi 99 b]}:
G.-C. Ghirardi,
``Quantum superpositions and definite perceptions:
Envisaging new feasible experimental tests'',
{\em Phys. Lett. A} {\bf 262}, 1, 1-14 (1999).
Erratum: {\em Phys. Lett. A} {\bf 263}, 4-6, 465 (1999).
Comment: {\bf [Thaheld 00]}.,

\item {\bf [Ghirardi 99 c]}:
G.-C. Ghirardi,
``Replies and a comment on the problem of realism in modern'',
{\em Statistica (Bologna)} {\bf 59} 3, 379-423 (1999).

\item {\bf [Ghirardi 00 a]}:
G.-C. Ghirardi,
``Quantum chance and nonlocality:
Probability and nonlocality in the interpretation of quantum mechanics,
by W. Michael Dickson'',
{\em Am. J. Phys.} {\bf 68}, 3, 295-296 (2000).
Review of {\bf [Dickson 98]}.

\item {\bf [Ghirardi 00 b]}:
G.-C. Ghirardi,
``Beyond conventional quantum mechanics'',
in {\bf [Ellis-Amati 00]}, pp.~79-116.

\item {\bf [Ghirardi 00 c]}:
G.-C. Ghirardi,
``Local measurements of nonlocal observables and the relativistic reduction
process'',
{\em Found. Phys.} {\bf 30} 9, 1337-1385 (2000);
quant-ph/0003149.

\item {\bf [Ghirardi-Marinatto-Weber 01]}:
G.-C. Ghirardi, L. Marinatto, \& T. Weber,
``Entanglement and properties of composite quantum systems:
A conceptual and mathematical analysis'',
submitted to {\em J. Stat. Phys.};
quant-ph/0109017.

\item {\bf [Ghirardi 02]}:
G.-C. Ghirardi,
``John Stewart Bell and the dynamical reduction program'',
in {\bf [Bertlmann-Zeilinger 02]}, pp.~287-306.

\item {\bf [Ghirardi-Marinatto 03]}:
G.-C. Ghirardi, \& L. Marinatto,
``Entanglement and properties'',
{\em Fortschr. Phys.} {\bf 51}, 379-387 (2003);
quant-ph/0206021.

\item {\bf [Ghirardi-Marinatto-Romano 03]}:
G.-C. Ghirardi, L. Marinatto, \& R. Romano,
``An optimal entropic uncertainty relation in a two-dimensional Hilbert
space'',
{\em Phys. Lett. A} {\bf 317}, 1-2, 32-36 (2003);
quant-ph/0310120.

\item {\bf [Ghirardi-Marinatto 04 a]}:
G.-C. Ghirardi, \& L. Marinatto,
``General criterion for the entanglement of two indistinguishable particles'',
{\em Phys. Rev. A} {\bf 70}, 1, 012109 (2004).;
quant-ph/0401065.
See {\bf [Ghirardi-Marinatto 04 b]}.

\item {\bf [Ghirardi-Marinatto 04 b]}:
G.-C. Ghirardi, \& L. Marinatto,
``Criteria for the entanglement of composite systems of identical
particles'',
{\em Fortschr. Phys.} {\bf 52}, 11-12, 1045-1051 (2004);
quant-ph/0410086.
See {\bf [Ghirardi-Marinatto 04 a]}.

\item {\bf [Ghiu-Bourennane-Karlsson 01]}:
I. Ghiu, M. Bourennane, \& A. Karlsson,
``Entanglement-assisted local transformations between inequivalent
classes of three-particle entangled states'',
{\em Phys. Lett. A} {\bf 287}, 1-2, 12-18 (2001).

\item {\bf [Ghiu 03]}:
I. Ghiu,
`Asymmetric quantum telecloning of $d$-level systems and
broadcasting of entanglement to different locations using the
``many-to-many'' communication protocol',
{\em Phys. Rev. A} {\bf 67}, 1, 012323 (2003);
quant-ph/0303060.

\item {\bf [Ghose-Home 93]}:
P. Ghose, \& D. Home,
``The EPR problem in the light of the Tomonaga-Schwinger formalism'',
in A. van der Merwe, \& F. Selleri (eds.),
{\em Bell's theorem and the foundations of modern physics.
Proc.\ of an international
conference (Cesena, Italy, 1991)},
World Scientific, Singapore, 1993, pp.~244-249.

\item {\bf [Ghose 99]}:
P. Ghose,
{\em Testing quantum mechanics on new ground},
Cambridge University Press, Cambridge, 1999.
Review: {\bf [Chiao 00]}.

\item {\bf [Ghose 00 a]}:
P. Ghose,
``The incompatibility of the de Broglie-Bohm theory with quantum mechanics'',
quant-ph/0001024.
See {\bf [Neumaier 00]}, {\bf [Marchildon 00]}, {\bf [Ghose 00 d]},
{\bf [Ghose 01 b]} (II).

\item {\bf [Ghose 00 b]}:
P. Ghose,
``Quantum mechanics as a limiting case of classical mechanics'',
quant-ph/0001025.

\item {\bf [Ghose 00 c]}:
P. Ghose,
``An experiment to distinguish between de Broglie-Bohm and standard quantum
mechanics'',
quant-ph/0003037.
See {\bf [Ghose 01 b]}.

\item {\bf [Ghose 00 d]}:
P. Ghose,
``Reply to `No Contradictions between Bohmian and quantum
mechanics'\,'',
quant-ph/0008007.
Reply to {\bf [Marchildon 00]}.
See {\bf [Ghose 00 a]}.

\item {\bf [Ghose-Majumdar-Guha-Sau 01]}:
P. Ghose, A. S. Majumdar, S. Guha, \& J. Sau,
``Bohmian trajectories for photons'',
{\em Phys. Lett. A} {\bf 290}, 5-6, 205-213 (2001);
quant-ph/0102071.

\item {\bf [Ghose 01 a]}:
P. Ghose,
`Comments on ``On Bohm trajectories in two-particle
interference devices'' by L. Marchildon',
quant-ph/0102131.
Comment on {\bf [Marchildon 01]}.

\item {\bf [Ghose 01 b]}:
P. Ghose,
``On the incompatibility of quantum mechanics and the
de Broglie-Bohm theory II'',
quant-ph/0103126.
See {\bf [Ghose 00 a]} (I).

\item {\bf [Ghose-Samal 01]}:
P. Ghose, \& M. K. Samal,
``EPR type nonlocality in classical electrodynamics!'',
quant-ph/0111119.

\item {\bf [Ghose 02]}:
P. Ghose,
``A continuous transition between quantum and
classical mechanics. I'',
{\em Found. Phys.} {\bf 32}, 6, 871-892 (2002);
quant-ph/0104104.
See {\bf [Ghose-Samal 02]} (II).

\item {\bf [Ghose-Samal 02]}:
P. Ghose, \& M. K. Samal,
``A continuous transition between quantum and
classical mechanics. II'',
{\em Found. Phys.} {\bf 32}, 6, 893-906 (2002);
quant-ph/0104105.
See {\bf [Ghose 02]} (I).

\item {\bf [Ghose-Samal-Datta 02]}:
P. Ghose, M. K. Samal, \& A. Datta,
``Bohmian picture of Rydberg atoms'',
quant-ph/0201081.

\item {\bf [Ghosh-Mandel 98]}:
R. Ghosh, \& L. Mandel,
``Observation of nonclassical effects in the interference of two photons'',
{\em Phys. Rev. Lett.} {\bf 59}, 17, 1903-1905 (1987).

\item {\bf [Ghosh-Kar 98]}:
S. Ghosh, \& G. Kar,
``Hardy's nonlocality for two spin-$s$ particles'',
{\em Phys. Lett. A} {\bf 240}, 4-5, 191-195 (1998).

\item {\bf [Ghosh-Kar-Sarkar 98]}:
S. Ghosh, G. Kar, \& D. Sarkar,
``Hardy's nonlocality for entangled states of three spin-$\frac{1}{2}$ particles'',
{\em Phys. Lett. A} {\bf 243}, 5-6, 249-255 (1998).

\item {\bf [Ghosh-Bandyopadhyay-Roy-(+2) 99]}:
S. Ghosh, S. Bandyopadhyay, A. Roy, D. Sarkar, \& G. Kar,
``Optimal universal disentangling machine for two qubit quantum states'',
quant-ph/9905036.

\item {\bf [Ghosh-Kar-Roy 99]}:
S. Ghosh, G. Kar, \& A. Roy,
``Optimal cloning and no signaling'',
{\em Phys. Lett. A} {\bf 261}, 1-2, 17-19 (1999);
quant-ph/9907001.

\item {\bf [Ghosh-Bandyopadhyay-Roy-(+2) 00]}:
S. Ghosh, S. Bandyopadhyay, A. Roy, D. Sarkar, \& G. Kar,
``Optimal universal disentangling machine for two-qubit quantum states'',
{\em Phys. Rev. A} {\bf 61}, 5, 052301 (2000).

\item {\bf [Ghosh-Kar-Roy-(+2) 00 a]}:
S. Ghosh, G. Kar, A. Roy, D. Sarkar, \& U. Sen,
``Entanglement teleportation through cat-like states'',
quant-ph/0012110.

\item {\bf [Ghosh-Kar-Roy-(+2) 00 b]}:
S. Ghosh, G. Kar, A. Roy, D. Sarkar, \& U. Sen,
``Teleporting noncommuting qubits require maximal entanglement'',
quant-ph/0012118.

\item {\bf [Ghosh-Roy-Sen 01]}:
S. Ghosh, A. Roy, \& U. Sen,
``Antiparallel spin does not always contain more information'',
{\em Phys. Rev. A} {\bf 63}, 1, 014301 (2001);
quant-ph/0004071.
See {\bf [Gisin-Popescu 99]}.

\item {\bf [Ghosh-Kar-Roy-(+2) 01 a]}:
S. Ghosh, G. Kar, A. Roy, D. Sarkar, \& U. Sen,
``Realization of optimal disentanglement by
teleportation via separable channels'',
{\em Phys. Rev. A} {\bf 64}, 4, 042114 (2001);
quant-ph/0102010.

\item {\bf [Ghosh-Kar-Sen De-Sen 01]}:
S. Ghosh, G. Kar, A. Sen De, \& U. Sen,
``Mixedness in the Bell violation versus entanglement of formation'',
{\em Phys. Rev. A} {\bf 64}, 4, 044301 (2001);
quant-ph/0104007.

\item {\bf [Ghosh-Kar-Roy-(+2) 01]}:
S. Ghosh, G. Kar, A. Roy, A. Sen De, \& U. Sen,
``Distinguishability of Bell states'',
{\em Phys. Rev. Lett.} {\bf 87}, 27, 277902 (2001);
quant-ph/0106148.
See {\bf [Horodecki-Sen De-Sen-Horodecki 03]}.

\item {\bf [Ghosh-Kar-Roy-Sen 02]}:
S. Ghosh, G. Kar, A. Roy, \& U. Sen,
``Entanglement versus noncommutativity in teleportation'',
{\em Phys. Rev. A} {\bf 65}, 3, 032309 (2002);
quant-ph/0010012.

\item {\bf [Ghosh-Kar-Roy-(+3) 02]}:
S. Ghosh, G. Kar, A. Roy, D. Sarkar, A. Sen(De),
\& U. Sen,
``Local indistinguishability of orthogonal pure states by using a bound on
distillable entanglement'',
{\em Phys. Rev. A} {\bf 65}, 6, 062307 (2002);
quant-ph/0111136.

\item {\bf [Ghosh-Kar-Roy-(+2) 02 a]}:
S. Ghosh, G. Kar, A. Roy, D. Sarkar, \& U. Sen,
``Entanglement teleportation via Bell mixtures'',
{\em Phys. Rev. A} {\bf 66}, 2, 024301 (2002);
quant-ph/0107056.

\item {\bf [Ghosh-Kar-Roy-(+2) 02 b]}:
S. Ghosh, G. Kar, A. Roy, D. Sarkar, \& U. Sen,
``Entanglement teleportation through GHZ-class states'',
{\em New J. Phys.} {\bf 4}, 48.1-48.9 (2002).

\item {\bf [Ghosh-Kar-Kunkri-Roy 03]}:
S. Ghosh, G. Kar, S. Kunkri, \& A. Roy,
``Probabilistic cloning and signalling'',
quant-ph/0312045.

\item {\bf [Ghosh-Rosenbaum-Aeppli-Coppersmith 03]}:
S. Ghosh, T. F. Rosenbaum, G. Aeppli, \& S. N. Coppersmith,
``Entangled quantum state of magnetic dipoles'',
{\em Nature} {\bf 425}, 6953, 48-51 (2003).
See {\bf [Vedral 03 b]}.

\item {\bf [Ghosh 04]}:
S. Ghosh,
``Entanglement in quantum mechanics: Its manifestations in complementarity
and nonlocality, and its classification'',
Ph.\ D. thesis, Jadavpur University, Calcutta, 2004.

\item {\bf [Ghosh-Kar-Roy 04]}:
S. Ghosh, G. Kar, \& A. Roy,
``Local cloning of Bell states and distillable entanglement'',
{\em Phys. Rev. A} {\bf 69}, 5, 052312 (2004);
quant-ph/0311062.

\item {\bf [Ghosh-Kar-Roy-Sarkar 04]}:
S. Ghosh, G. Kar, A. Roy, \& D. Sarkar,
``Distinguishability of maximally entangled states'',
{\em Phys. Rev. A} {\bf 70}, 2, 022304 (2004).

\item {\bf [Ghosh-Joag-Kar-(+2) 04]}:
S. Ghosh, P. Joag, G. Kar,
S. Kunkri, \& A. Roy,
``Locally accessible information and distillation of entanglement'',
quant-ph/0403134.

\item {\bf [Ghose-Roy 91]}:
P. Ghose, \& M. N. S. Roy,
``Confronting the complementarity principle in an interference experiment'',
{\em Phys. Lett. A} {\bf 161}, 1, 5-8 (1991).

\item {\bf [Ghose-Alsing-Deutsch-(+3) 03]}:
S. Ghose, P. M. Alsing, I. H. Deutsch,
T. Bhattacharya, S. Habib, \& K. Jacobs,
``Recovering classical dynamics from coupled quantum systems through
continuous measurement'',
{\em Phys. Rev. A} {\bf 67}, 5, 052102 (2003);
quant-ph/0208064.

\item {\bf [Giacomini-Sciarrino-Lombardi-De Martini 02]}:
S. Giacomini, F. Sciarrino, E. Lombardi, \& F. De Martini,
``Active teleportation of a quantum bit'',
{\em Phys. Rev. A} {\bf 66}, 3, 030302 (2002).

\item {\bf [Gibbons 96]}:
G. Gibbons,
``A good foundation'',
{\em Phys. World} {\bf 9}, 11, 48 (1996).
Review of {\bf [Isham 95]}.

\item {\bf [Gibbons-Hoffman-Wootters 04]}:
K. S. Gibbons, M. J. Hoffman, \& W. K. Wootters,
``Discrete phase space based on finite fields'',
quant-ph/0401155.

\item {\bf [Giedke-Briegel-Cirac-Zoller 99]}:
G. Giedke, H.-J. Briegel, J. I. Cirac, \& P. Zoller,
``Lower bounds for attainable fidelities in entanglement purification'',
{\em Phys. Rev. A} {\bf 59}, 4, 2641-2648 (1999);
quant-ph/9809043.

\item {\bf [Giedke-Duan-Cirac-Zoller 99]}:
G. Giedke, L.-M. Duan, J. I. Cirac, \& P. Zoller,
``All inseparable two-mode Gaussian continuous variable states are
distillable'',
quant-ph/0007061.
See {\bf [Giedke-Duan-Zoller-Cirac 01]}.

\item {\bf [Giedke-Kraus-Lewenstein-Cirac 01]}:
G. Giedke, B. Kraus, M. Lewenstein, \& J. I. Cirac,
``Separability properties of three-mode Gaussian states'',
{\em Phys. Rev. A} {\bf 64}, 5, 052303 (2001);
quant-ph/0103137.

\item {\bf [Giedke-Kraus-Lewenstein-Cirac 01]}:
G. Giedke, B. Kraus, M. Lewenstein, \& J. I. Cirac,
``Separability criterion for all bipartite Gaussian states'',
{\em Phys. Rev. Lett.} {\bf 87}, 16, 167904 (2001),
quant-ph/0104050.

\item {\bf [Giedke-Duan-Zoller-Cirac 01]}:
G. Giedke, L.-M. Duan, P. Zoller, \& J. I. Cirac,
``Distillability criterion for all bipartite Gaussian states'',
{\em Quant. Inf. Comp.} 1, ?, 79-? (2001);
quant-ph/0104072.
Subsumes {\bf [Giedke-Duan-Cirac-Zoller 99]}.

\item {\bf [Giedke-Cirac 02]}:
G. Giedke, \& J. I. Cirac,
``Characterization of Gaussian operations and distillation of Gaussian states'',
{\em Phys. Rev. A} {\bf 66}, 3, 032316 (2002);
quant-ph/0204085.

\item {\bf [Giedke-Eisert-Cirac-Plenio 03]}:
G. Giedke, J. Eisert, J. I. Cirac, \& M. B. Plenio,
``Entanglement transformations of pure Gaussian states'',
quant-ph/0301038.

\item {\bf [Giedke-Wolf-Krueger-(+3) 03]}:
G. Giedke, M. M. Wolf, O. Kruger,
R. F. Werner, \& J. I. Cirac,
``Entanglement of formation for symmetric Gaussian states'',
{\em Phys. Rev. Lett.} {\bf 91}, 10, 107901 (2003);
quant-ph/0304042.

\item {\bf [Gieres 99]}:
F. Gieres,
``Dirac's formalism and mathematical surprises in quantum
mechanics'',
quant-ph/9907069.
French version:
``Formalisme de Dirac et surprises mathematiques en mecanique
quantique'',
quant-ph/9907070.
See {\bf [Dirac 39]}.

\item {\bf [Gilbert-Hamrick 00]}:
G. Gilbert, \& M. Hamrick,
``Practical quantum cryptography:
A comprehensive analysis (part one)'',
quant-ph/0009027.

\item {\bf [Gilbert-Hamrick 01 a]}:
G. Gilbert, \& M. Hamrick,
``The secrecy capacity of practical quantum
cryptography'',
quant-ph/0106033.

\item {\bf [Gilbert-Hamrick 01 b]}:
G. Gilbert, \& M. Hamrick,
``Constraints on eavesdropping on the BB84
protocol'',
quant-ph/0106034.

\item {\bf [Gilbert-Hamrick 01 c]}:
G. Gilbert, \& M. Hamrick,
``Secrecy, computational loads and rates in
practical quantum cryptography'',
quant-ph/0106043.

\item {\bf [Gilbert-Hamrick-Thayer 01]}:
G. Gilbert, M. Hamrick, \& F. J. Thayer,
``Privacy amplification in quantum key distribution:
Pointwise bound versus average bound'',
quant-ph/0108013.

\item {\bf [Gilchrist-Deuar-Reid 98]}:
A. Gilchrist, P. Deuar, \& M. D. Reid,
``Contradiction of quantum mechanics with local
hidden variables for quadrature phase amplitude measurements'',
{\em Phys. Rev. Lett.} {\bf 80}, 15, 3169-3172 (1998);
quant-ph/0010024.

\item {\bf [Gilchrist-Deuar-Reid 99]}:
A. Gilchrist, P. Deuar, \& M. D. Reid,
``Contradiction of quantum mechanics with local hidden variables for quadrature phase
measurements on pair-coherent states and squeezed macroscopic superpositions of coherent states'',
{\em Phys. Rev. A} {\bf 60}, 6, 4259-4271 (1999).

\item {\bf [Gilchrist-White-Munro 02]}:
A. Gilchrist, A. G. White, \& W. J. Munro,
``Entanglement creation using quantum interrogation'',
{\em Phys. Rev. A} {\bf 66}, 1, 012106 (2002);
quant-ph/0112101.

\item {\bf [Gilchrist-Munro-White 03]}:
A. Gilchrist, W. J. Munro, \& A. G. White,
``Input states for quantum gates'',
{\em Phys. Rev. A} {\bf 67}, 4, 040304 (2003);
quant-ph/0301112.

\item {\bf [Gilchrist-Nemoto-Munro-(+4) 03]}:
A. Gilchrist, K. Nemoto, W. J. Munro,
T. C. Ralph, S. Glancy, S. L. Braunstein, \& G. J. Milburn,
``Schr\"{o}dinger cats and their power for quantum information processing'',
{\em J. Opt. B: Quantum Semiclass. Opt.};
quant-ph/0312194.

\item {\bf [Gilchrist-Langford-Nielsen 04]}:
A. Gilchrist, N. K. Langford, \& M. A. Nielsen,
``Distance measures to compare real and ideal quantum processes'',
quant-ph/0408063.

\item {\bf [Giles 70]}:
R. Giles,
``Foundations for quantum mechanics'',
{\em J. Math. Phys.} {\bf 11}, 7, 2139-2160 (1970).

\item {\bf [Gill-Keane 96]}:
R. D. Gill, \& M. Keane,
``A geometric proof of the Kochen-Specker no-go theorem'',
{\em J. Phys. A} {\bf 29}, 12, L289-L291 (1996);
quant-ph/0304013.
See {\bf [Cooke-Keane-Moran 85]}.

\item {\bf [Gill-Massar 00]}:
R. D. Gill, \& S. Massar,
``State estimation for large ensembles'',
{\em Phys. Rev. A} {\bf 61}, 4, 042312 (2000).

\item {\bf [Gill 01 a]}:
R. D. Gill (with appendix by J.-\AA. Larsson),
``Accardi contra Bell (cum mundi): The impossible coupling'',
quant-ph/0110137.

\item {\bf [Gill 01 b]}:
R. D. Gill,
``Teleportation into quantum statistics'',
{\em J. Korean Statistical Soc.} {\bf 30}, 291-325 (2001);
math.ST/0405572.

\item {\bf [Gill-Weihs-Zeilinger-\.{Z}ukowski 02 a]}:
R. D. Gill, G. Weihs, A. Zeilinger, \& M. \.{Z}ukowski,
`Comment on ``Exclusion of time in the theorem of Bell'' by K. Hess and W. Philipp',
{\em Eurphys. Lett.};
quant-ph/0204169.
Comment on {\bf [Hess-Philipp 01 a, b, c, 02 a]}.
See {\bf [Gill-Weihs-Zeilinger-\.{Z}ukowski 02 b]}.
See {\bf [Hess-Philipp 03 b]}.

\item {\bf [Gill-Weihs-Zeilinger-\.{Z}ukowski 02 b]}:
R. D. Gill, G. Weihs, A. Zeilinger, \& M. \.{Z}ukowski,
``No time loophole in Bell's theorem; the Hess-Philipp model is non-local'',
{\em Proc. Natl. Acad. Sci. USA} {\bf 99}, 23, 14632-14635 (2002);
quant-ph/0208187.
Comment on {\bf [Hess-Philipp 01 a, b, c, 02 a]}.
See {\bf [Gill-Weihs-Zeilinger-\.{Z}ukowski 02 a]}.
See {\bf [Hess-Philipp 03 b]}.

\item {\bf [Gill 03]}:
R. D. Gill,
``Time, finite statistics, and Bell's fifth posit'',
in {\em Foundations of Quantum Mechanics and Probability (V\"{a}xj\"{o}, Sweden, 2002)};
quant-ph/0301059.

\item {\bf [Gill 04 a]}:
R. D. Gill,
``On an argument of David Deutsch'',
in {\em Quantum Probability and Infinite Dimensional Analysis (Greifswald, 2003)},
World Scientific, Singapore, 2004;
quant-ph/0307188.

\item {\bf [Gill 04 b]}:
R. D. Gill,
``The chaotic chameleon'',
in {\em Quantum Probability and Infinite Dimensional Analysis (Greifswald, 2003)},
World Scientific, Singapore, 2004;
quant-ph/0307217.

\item {\bf [Gill-Guta 04]}:
R. D. Gill, \& M. Guta,
``An invitation to quantum tomography'',
quant-ph/0303020.
See {\bf [Artiles-Gill-Guta 04]} (II).

\item {\bf [Gilles-Garc\'{\i}a Fern\'{a}ndez 95]}:
L. Gilles, \& P. Garc\'{\i}a Fern\'{a}ndez,
``The frequancy non-degenerate parametric amplifier as
a device to analyse the Einstein-Podolsky-Rosen paradox'',
{\em Quantum Semiclass. Opt.} {\bf 7}, 4, 625-638 (1995).

\item {\bf [Gillespie 86]}:
D. T. Gillespie,
``Untenability of simple ensemble
interpretations of quantum measurement probabilities'',
{\em Am. J. Phys.} {\bf 54}, 10, 889-894 (1986).

\item {\bf [Gillespie 94]}:
D. T. Gillespie,
``Why quantum mechanics cannot be formulated as a Markov
process'',
{\em Phys. Rev. A} {\bf 49}, 3, 1607-1612 (1994).
Comment: {\bf [Hardy-Home-Squires-Whitaker 97]}.
Reply: {\bf [Gillespie 97]}.
See: {\bf [Hardy-Home-Squires-Whitaker 92]}.

\item {\bf [Gillespie 97]}:
D. T. Gillespie,
`Reply to ``Comment on `Why quantum mechanics cannot
be formulated as a Markov process'\,''\,',
{\em Phys. Rev. A} {\bf 56}, 4, 3304-3306 (1997).
Reply to: {\bf [Hardy-Home-Squires-Whitaker 97]}.
See: {\bf [Hardy-Home-Squires-Whitaker 92]},
{\bf [Gillespie 94]}.

\item {\bf [Gilmore-Park 79 a]}:
T. Gilmore, Jr., \& J. L. Park,
``Superselection rules in quantum theory: Part I.
A new proposal for state restriction violation'',
{\em Found. Phys.} {\bf 9}, 7-8, 537-556 (1979).
See {\bf [Gilmore-Park 79 b]} (II).

\item {\bf [Gilmore-Park 79 b]}:
T. Gilmore, Jr., \& J. L. Park,
``Superselection rules in quantum theory: Part II.
Subensemble selection'',
{\em Found. Phys.} {\bf 9}, 9-10, 739-749 (1979).
See {\bf [Gilmore-Park 79 a]} (I).

\item {\bf [Gingrich-Williams-Cerf 00]}:
R. M. Gingrich, C. P. Williams, \& N. J. Cerf,
``Generalized quantum search with parallelism'',
{\em Phys. Rev. A} {\bf 61}, 5, 052313 (2000);
quant-ph/9904049.

\item {\bf [Gingrich 02]}:
R. M. Gingrich,
``Properties of entanglement monotones for three-qubit pure states'',
{\em Phys. Rev. A} {\bf 65}, 5, 052302 (2002);
quant-ph/0106042.

\item {\bf [Gingrich-Adami 02]}:
R. M. Gingrich, \& C. Adami,
``Quantum entanglement of moving bodies'',
{\em Phys. Rev. Lett.} {\bf 89}, 27, 270402 (2002);
quant-ph/0205179.

\item {\bf [Gingrich-Bergou-Adami 03]}:
R. M. Gingrich, A. J. Bergou, \& C. Adami,
``Entangled light in moving frames'',
{\em Phys. Rev. A} {\bf 68}, 4, 042102 (2003);
quant-ph/0302095.

\item {\bf [Gingrich-Kok-Lee-(+2) 03]}:
R. M. Gingrich, P. Kok, H. Lee,
F. Vatan, \& J. P. Dowling,
``All linear optical quantum memory based on quantum error correction'',
{\em Phys. Rev. Lett.} {\bf 91}, 21, 217901 (2003);
quant-ph/0306098.

\item {\bf [Giorda-Zanardi-Lloyd 03]}:
P. Giorda, P. Zanardi, \& S. Lloyd,
``Universal quantum control in irreducible state-space sectors:
Application to bosonic and spin-boson systems'',
{\em Phys. Rev. A} {\bf 68}, 6, 062320 (2003).

\item {\bf [Giorda-Iorio-Sen-Sen 03]}:
P. Giorda, A. Iorio, S. Sen, \& S. Sen,
``Semiclassical Shor's algorithm'',
quant-ph/0303037.

\item {\bf [Giorda-Zanardi 03]}:
P. Giorda, \& P. Zanardi,
``Mode entanglement and entangling power in bosonic graphs'',
{\em Phys. Rev. A} {\bf 68}, 6, 062108 (2003);
quant-ph/0311058.

\item {\bf [Giorda-Iorio-Sen-Sen 04]}:
P. Giorda, A. Iorio, S. Sen, \& S. Sen,
``Semiclassical Shor's algorithm'',
{\em Phys. Rev. A};
quant-ph/0303037.

\item {\bf [Giorgi-Mataloni-De Martini 03]}:
G. Giorgi, P. Mataloni, \& F. De Martini,
``Frequency hopping in quantum interferometry: Efficient up-down conversion
for qubits and ebits'',
{\em Phys. Rev. Lett.} {\bf 90}, 2, 027902 (2003).

\item {\bf [Giovannetti-Vitali-Tombesi-Ekert 00]}:
V. Giovannetti, D. Vitali, P. Tombesi, \& A. K. Ekert,
``Scalable quantum computation with cavity QED systems'',
{\em Phys. Rev. A} {\bf 62}, 3, 032306 (2000);
quant-ph/0004107.

\item {\bf [Giovannetti-Lloyd-Maccone 01 a]}:
V. Giovannetti, S. Lloyd, \& L. Maccone,
``Quantum-enhanced positioning and clock synchronization'',
{\em Nature} {\bf 412}, 6845, 417-419 (2001);
quant-ph/0103006.

\item {\bf [Giovannetti-Mancini-Tombesi 01]}:
V. Giovannetti, S. Mancini, \& P. Tombesi,
``Radiation pressure induced Einstein-Podolsky-Rosen paradox'',
{\em Europhys. Lett.} {\bf 54}, 5, 559-565 (2001);
quant-ph/0005066.

\item {\bf [Giovannetti-Lloyd-Maccone-Wong 01]}:
V. Giovannetti, S. Lloyd, L. Maccone, \& F. N. C. Wong,
``Clock synchronization with dispersion cancellation'',
{\em Phys. Rev. Lett.} {\bf 87}, 11, 117902 (2001);
quant-ph/0105156.

\item {\bf [Giovannetti-Lloyd-Maccone 01 b]}:
V. Giovannetti, S. Lloyd, \& L. Maccone,
``Quantum cryptographic ranging'',
{\em IV Mysteries, Puzzles, and Paradoxes in Quantum Mechanics (Gargnano, Italy, 2001},
{\em J. Opt. B: Quantum Semiclass. Opt.};
quant-ph/0112079.

\item {\bf [Giovannetti-Lloyd-Maccone 02 a]}:
V. Giovannetti, S. Lloyd, \& L. Maccone,
``Positioning and clock synchronization through entanglement'',
{\em Phys. Rev. A} {\bf 65}, 2, 022309 (2002);
quant-ph/0107140.

\item {\bf [Giovannetti-Maccone-Shapiro-Wong 02]}:
V. Giovannetti, L. Maccone, J. H. Shapiro, \& F. N. C. Wong,
``Generating entangled two-photon states with coincident frequencies'',
{\em Phys. Rev. Lett.} {\bf 88}, 18, 183602 (2002).

\item {\bf [Giovannetti-Lloyd-Maccone-Shahriar 02]}:
V. Giovannetti, S. Lloyd, L. Maccone, \& M. S. Shahriar,
``Limits to clock synchronization induced by completely dephasing
communication channels'',
{\em Phys. Rev. A} {\bf 65}, 6, 062319 (2002);
quant-ph/0110156.

\item {\bf [Giovannetti-Lloyd-Maccone 02 b]}:
V. Giovannetti, S. Lloyd, \& L. Maccone,
``Quantum cryptographic ranging'',
in R. Bonifacio, \& D. Vitali (eds.),
{\em Mysteries, Puzzles and Paradoxes in Quantum Mechanics IV:
Quantum Interference Phenomena (Gargnano, Italy, 2001)},
{\em J. Opt. B: Quantum Semiclass. Opt.} {\bf 4}, 4, S413-S414 (2002).

\item {\bf [Giovannetti-Lloyd-Maccone-Wong 02]}:
V. Giovannetti, S. Lloyd, L. Maccone, \& F. N. C. Wong,
``Clock synchronization and dispersion'',
in R. Bonifacio, \& D. Vitali (eds.),
{\em Mysteries, Puzzles and Paradoxes in Quantum Mechanics IV:
Quantum Interference Phenomena (Gargnano, Italy, 2001)},
{\em J. Opt. B: Quantum Semiclass. Opt.} {\bf 4}, 4, S415-S417 (2002).

\item {\bf [Giovannetti-Maccone-Shapiro-Wong 02]}:
V. Giovannetti, L. Maccone, J. H. Shapiro, \& F. N. C. Wong,
``Extended phase-matching conditions for improved entanglement generation'',
{\em Phys. Rev. A} {\bf 66}, 4, 043813 (2002).

\item {\bf [Giovannetti-Lloyd-Maccone 02]}:
V. Giovannetti, S. Lloyd, \& L. Maccone,
``The role of entanglement in dynamical evolution'',
{\em Europhys. Lett.} {\bf 62}, ?, 615-? (2003);
quant-ph/0206001.

\item {\bf [Giovannetti-Mancini-Vitali-Tombesi 03]}:
V. Giovannetti, S. Mancini, D. Vitali, \& P. Tombesi,
``Characterizing the entanglement of bipartite quantum systems'',
{\em Phys. Rev. A} {\bf 67}, 2, 022320 (2003).
quant-ph/0210155.

\item {\bf [Giovannetti-Lloyd-Maccone 03 a]}:
V. Giovannetti, S. Lloyd, \& L. Maccone,
``The quantum speed limit'',
{\em Proc.\ SPIE Conf.\ Fluctuations and Noise (Santa Fe, New Mexico, 2003)},
quant-ph/0303085.

\item {\bf [Giovannetti-Lloyd-Maccone 03 b]}:
V. Giovannetti, S. Lloyd, \& L. Maccone,
``Quantum limits to dynamical evolution'',
{\em Phys. Rev. A} {\bf 67}, 5, 052109 (2003).

\item {\bf [Giovannetti-Lloyd-Maccone-Shor 03]}:
V. Giovannetti, S. Lloyd, L. Maccone, \& P. W. Shor,
``Entanglement assisted capacity of the broadband lossy channel'',
{\em Phys. Rev. Lett.} {\bf 91}, 4, 047901 (2003);
quant-ph/0304020.

\item {\bf [Giovannetti-Lloyd-Maccone-Shor 03 b]}:
V. Giovannetti, S. Lloyd, L. Maccone, \& P. W. Shor,
``Broadband channel capacities'',
{\em Phys. Rev. A} {\bf 68}, 6, 062323 (2003);
quant-ph/0307098.

\item {\bf [Giovannetti-Lloyd-Maccone-Shapiro 03]}:
V. Giovannetti, S. Lloyd, L. Maccone, \& J. H. Shapiro,
``Information rate of waveguide'',
{\em Phys. Rev. A} {\bf 69}, 5, 052310 (2004);
quant-ph/0307112.

\item {\bf [Giovannetti-Guha-Lloyd-(+3) 03]}:
V. Giovannetti, S. Guha, S. Lloyd,
L. Maccone, J. H. Shapiro, \& H. P. Yuen,
``Classical capacity of the lossy bosonic channel: The exact solution'',
{\em Phys. Rev. Lett.} {\bf 92}, 2, 027902 (2004);
quant-ph/0308012.

\item {\bf [Giovannetti-Lloyd 04]}:
V. Giovannetti, \& S. Lloyd,
``Additivity properties of Gaussian channels'',
{\em Phys. Rev. A} {\bf 69}, 6, 062307 (2004);
quant-ph/0403075.

\item {\bf [Giovannetti 04]}:
V. Giovannetti,
``Separability conditions from entropic uncertainty relations'',
{\em Phys. Rev. A} {\bf 70}, 1, 012102 (2004);
quant-ph/0307171.

\item {\bf [Giovannetti-Lloyd-Maccone 04]}:
V. Giovannetti, S. Lloyd, \& L. Maccone,
``Capacity of nonlinear bosonic systems'',
{\em Phys. Rev. A} {\bf 70}, 1, 012307 (2004).

\item {\bf [Giovannetti-Guha-Lloyd-(+2) 04]}:
V. Giovannetti, S. Guha, S. Lloyd,
L. Maccone, \& J. H. Shapiro,
``Minimum output entropy of bosonic channels: A conjecture'',
quant-ph/0404005.

\item {\bf [Giovannetti-Lloyd-Maccone-(+2) 04]}:
V. Giovannetti, S. Lloyd, L. Maccone,
J. H. Shapiro, \& B. J. Yen,
``Minimum Renyi and Wehrl entropies at the output of bosonic channels'',
{\em Phys. Rev. A};
quant -ph/0404037.

\item {\bf [Giovannetti-Fazio 04]}:
V. Giovannetti, \& R. Fazio,
``Information capacities of spin chains'',
quant-ph/0405110.

\item {\bf [Giovannetti-Lloyd-Maccone-(+2) 04]}:
V. Giovannetti, S. Lloyd, L. Maccone,
J. H. Shapiro, \& F. N. C. Wong
``Conveyor belt clock synchronization'',
quant-ph/0405154.

\item {\bf [Giovannetti 04]}:
V. Giovannetti,
``Inequalities for quantum channels assisted with limited resources'',
quant-ph/0406111.

\item {\bf [Giovannetti-Lloyd-Ruskai 04]}:
V. Giovannetti, S. Lloyd, \& M. B. Ruskai,
``Conditions for the multiplicativity of maximal $l_p$-norms of channels for
fixed integer $p$'',
quant-ph/0408103.

\item {\bf [Giovannetti-Mancini 04]}:
V. Giovannetti, \& S. Mancini,
``Bosonic memory channels'',
quant-ph/0410176.

\item {\bf [Giraldi-Grigolini 01]}:
F. Giraldi \& P. Grigolini,
``Quantum entanglement and entropy'',
{\em Phys. Rev. A} {\bf 64}, 3, 032310 (2001).

\item {\bf [Gisin 84 a]}:
N. Gisin,
``Quantum measurement and stochastic processes'',
{\em Phys. Rev. Lett.} {\bf 52}, 19, 1657-1660 (1984).
Comment: {\bf [Pearle 84]}.
Reply: {\bf [Gisin 84 b]}.

\item {\bf [Gisin 84 b]}:
N. Gisin,
``Gisin responds'',
{\em Phys. Rev. Lett.} {\bf 53}, 18, 1776 (1984).
Reply to {\bf [Pearle 84]}.
See {\bf [Gisin 84 a]}.

\item {\bf [Gisin 89]}:
N. Gisin,
``Stochastic quantum dynamics and relativity'',
{\em Helv. Phys. Acta} {\bf 62}, 4, 363-371 (1989).

\item {\bf [Gisin 90]}:
N. Gisin,
``Weinberg's non-linear quantum mechanics and supraluminal communication'',
{\em Phys. Lett. A} {\bf 143}, 1-2, 1-2 (1990).
See {\bf [Weinberg 89 a, b, c, d]},
{\bf [Polchinski 91]}, {\bf [Mielnik 00]}.

\item {\bf [Gisin 91 a]}:
N. Gisin,
``Bell's inequality holds for all non-product states'',
{\em Phys. Lett. A} {\bf 154}, 5-6, 201-202 (1991).
Correction of some typographical errors in {\bf [Barnett-Phoenix 92]}.
See {\bf [Kar 95]}.

\item {\bf [Gisin 91 b]}:
N. Gisin,
``Can quantum entangled states collapse spontaneously?'',
{\em Phys. Lett. A} {\bf 155}, 8-9, 445-449 (1991).

\item {\bf [Gisin-Peres 92]}:
N. Gisin, \& A. Peres,
``Maximal violation of Bell's inequality for arbitrarirly large spin'',
{\em Phys. Lett. A} {\bf 162}, 1, 15-17 (1992).

\item {\bf [Gisin-Percival 92]}:
N. Gisin, \& I. C. Percival,
``The quantum-state diffusion model applied to open systems'',
{\em J. Phys. A} {\bf 25}, 21, 5677-5691 (1992).

\item {\bf [Gisin-Percival 93 a]}:
N. Gisin, \& I. C. Percival,
``Quantum-state diffusion, localization and quantum dispersion entropy'',
{\em J. Phys. A} {\bf 26}, 9, 2233-2243 (1993).

\item {\bf [Gisin-Percival 93 b]}:
N. Gisin, \& I. C. Percival,
``The quantum-state diffusion picture of physical processes'',
{\em J. Phys. A} {\bf 26}, 9, 2245-2260 (1993).

\item {\bf [Gisin-Percival 93 c]}:
N. Gisin, \& I. C. Percival,
``Stochastic wave equations versus parallel world components'',
{\em Phys. Lett. A} {\bf 175}, ?, 144-145 (1993).

\item {\bf [Gisin-Rigo 95]}:
N. Gisin, \& M. Rigo,
``?'',
{\em J. Phys. A} {\bf 28}, ?, 7375-7390 (1995).

\item {\bf [Gisin 96 a]}:
N. Gisin,
``Hidden quantum nonlocality revealed by local filters'',
{\em Phys. Lett. A} {\bf 210}, 3, 151-156 (1996).
Comment: {\bf [Berndl-Teufel 97]}.
Reply: {\bf [Gisin 97]}.

\item {\bf [Gisin 96 b]}:
N. Gisin,
``Nonlocality criteria for quantum teleportation'',
{\em Phys. Lett. A} {\bf 210}, 3, 157-159 (1996).

\item {\bf [Gisin 97]}:
N. Gisin,
``Hidden quantum nonlocality revealed by local filters'',
{\em Phys. Lett. A} {\bf 224}, 4-5, 317 (1997).
Reply to {\bf [Berndl-Teufel 97]}.
See {\bf [Gisin 96 a]}.

\item {\bf [Gisin-Huttner 97]}:
N. Gisin, \& B. Huttner,
``Quantum cloning, eavesdropping and Bell's inequality'',
{\em Phys. Lett. A} {\bf 228}, 1-2, 13-21 (1997).
Erratum: {\em Phys. Lett. A} {\bf 232}, 6, 463 (1997).
See {\bf [Fuchs-Gisin-Griffiths-(+2) 97]},
{\bf [Griffiths-Niu 97]}.

\item {\bf [Gisin-Massar 97]}:
N. Gisin, \& S. Massar,
``Optimal quantum cloning machines'',
{\em Phys. Rev. Lett.} {\bf 79}, 11, 2153-2156 (1997);
quant-ph/9705046.

\item {\bf [Gisin-Brendel-Zbinden-(+2) 98]}:
H. Gisin, J. Brendel, H. Zbinden, A. Sergienko, \& A. Muller,
``Twin-photon techniques for fiber measurements'',
quant-ph/9807063.

\item {\bf [Gisin 98]}:
N. Gisin,
``Quantum cloning without signaling'',
{\em Phys. Lett. A} {\bf 242}, 1-2, 1-3 (1998);
quant-ph/9801005.

\item {\bf [Gisin-Bechmann Pasquinucci 98]}:
N. Gisin, \& H. Bechmann-Pasquinucci,
``Bell inequality, Bell states and maximally entangled states for $n$ qubits'',
{\em Phys. Lett. A} {\bf 246}, 1-2, 1-6 (1998);
quant-ph/9804045.

\item {\bf [Gisin-Huttner-Muller-(+2) 98]}:
N. Gisin, B. Huttner, A. Muller, H. Zbinden, \& B. Perny,
``Quantum cryptography device and method'',
patent WO9810560A1, 1998.
See {\bf [Gisin-Huttner-Muller-(+2) 99]}.

\item {\bf [Gisin-Brendel-Gautier-(+5) 99]}:
N. Gisin, J. Brendel, J-D. Gautier, B. Gisin, B. Huttner,
G. Ribordy, W. Tittel, \& H. Zbinden,
``Quantum cryptography and long distance Bell experiments:
How to control decoherence'', in
{\em Decoherence (Bielefeld, Germany, 1998)};
quant-ph/9901043.

\item {\bf [Gisin-Popescu 99]}:
N. Gisin, \& S. Popescu,
``Spin flips and quantum information for antiparallel spins'',
{\em Phys. Rev. Lett.} {\bf 83}, 2, 432-435 (1999);
quant-ph/9901072.
See {\bf [Ghosh-Roy-Sen 00]}.

\item {\bf [Gisin-Gisin 99]}:
N. Gisin, \& B. Gisin,
``A local hidden variable model of quantum correlation exploiting the
detection loophole'',
{\em Phys. Lett. A} {\bf 260}, 5, 323-327 (1999);
quant-ph/9905018.
Comment: {\bf [Durt 99]}.

\item {\bf [Gisin 99]}:
N. Gisin,
``Bell inequality for arbitrary many settings of the analyzers'',
{\em Phys. Lett. A} {\bf 260}, 1-2, 1-3 (1999);
quant-ph/9905062.

\item {\bf [Gisin-Zbinden 99]}:
N. Gisin, \& H. Zbinden,
``Bell inequality and the locality loophole: Active versus passive switches'',
{\em Phys. Lett. A} {\bf 264}, 2-3, 103-107 (1999);
quant-ph/9906049.

\item {\bf [Gisin-Wolf 99]}:
N. Gisin, \& S. Wolf,
``Quantum cryptography on noisy channels:
Quantum versus classical key-agreement protocols'',
{\em Phys. Rev. Lett.} {\bf 83}, 20, 4200-4203 (1999).

\item {\bf [Gisin-Huttner-Muller-(+2) 99]}:
N. Gisin, B. Huttner, A. Muller, H. Zbinden, \& B. Perny,
``Quantum cryptography device and method'',
patent EP923828A1, 1999.
See {\bf [Gisin-Huttner-Muller-(+2) 98]}.

\item {\bf [Gisin-Wolf 00]}:
N. Gisin, \& S. Wolf,
``Linking classical and quantum key agreement:
Is there `bound information'?'',
accepted for {\em Crypto 2000};
quant-ph/0005042.

\item {\bf [Gisin-Scarani-Tittel-Zbinden 00]}:
N. Gisin, V. Scarani, W. Tittel, \& H. Zbinden,
``Optical tests of quantum nonlocality: From EPR-Bell tests
towards experiments with moving observers'',
quant-ph/0009055.

\item {\bf [Gisin-Tittel-Zbinden 01]}:
N. Gisin, W. Tittel, \& H. Zbinden,
``Reply: Gisin, Tittel, and Zbinden'',
{\em Phys. Rev. Lett.} {\bf 86}, 7, 1393 (2001).
Reply to {\bf [Durt 01 a]}.
See {\bf [Brendel-Gisin-Tittel-Zbinden 99]}.

\item {\bf [Gisin-Go 01]}:
N. Gisin, \& A. Go,
``EPR test with photons and kaons: Analogies'',
{\em Am. J. Phys.} {\bf 69}, 3, 264-270 (2001);
quant-ph/0004063.

\item {\bf [Gisin-Renner-Wolf 01]}:
N. Gisin, R. Renner, \& S. Wolf,
``Bound information: the classical analog to bound quantum information'',
{\em European Congress of Mathematics (Barcelona, 2000)},
Birkh\"{a}user, Basel, 2001; Vol.~II, pp.~439-447.

\item {\bf [Gisin 02 a]}:
N. Gisin,
``Sundays in a quantum engineer's life'',
in {\bf [Bertlmann-Zeilinger 02]}, pp.~199-208;
quant-ph/0104140.

\item {\bf [Gisin-Ribordy-Tittel-Zbinden 02]}:
N. Gisin, G. Ribordy, W. Tittel, \& H. Zbinden,
``Quantum cryptography'',
{\em Rev. Mod. Phys.} {\bf 74}, 1, 145-195 (2002);
quant-ph/0101098.

\item {\bf [Gisin-Gisin 02]}:
N. Gisin, \& B. Gisin,
``A local variable model for
entanglement swapping exploiting the detection loophole'',
{\em Phys. Lett. A} {\bf 297}, 5-6, 279-284 (2002);
quant-ph/0201077.

\item {\bf [Gisin 02 b]}:
N. Gisin,
``NOT logic'',
{\em Nature} {\bf 419}, 6909, 797-798 (2002).
See {\bf [De Martini-Bu\v{z}zek-Sciarrino-Sias 02]}.

\item {\bf [Gisin-Marcikic-de Riedmatten-(+2) 03]}:
N. Gisin, I. Marcikic, H. de Riedmatten,
W. Tittel, \& H. Zbinden,
``Quantum communications with time-bin entangled photons: Long distance
quantum teleportation and quantum repeaters'',
in {\em Proc.\ of the 6th Int.\ Conf.\
on Quantum Communication, Measurement and Computing (QCMC 02)};
quant-ph/0301181.

\item {\bf [Gisin 03]}:
N. Gisin,
``Quantum cryptography with fewer random numbers'',
quant-ph/0303052.

\item {\bf [Gisin-Brunner 03]}:
N. Gisin, \& N. Brunner,
``Quantum cryptography with and without entanglement'',
{\em Proc.\ for Les Houches Summer School on
Quantum entanglement and information processing (2003)};
quant-ph/0312011.

\item {\bf [Gisin-Linden-Massar-Popescu 04]}:
N. Gisin, N. Linden, S. Massar, \& S. Popescu,
``Error filtration and entanglement purification for quantum communication'',
quant-ph/0407021.

\item {\bf [Gisin 04]}:
N. Gisin,
``Quantum gloves: Physics and information'',
quant-ph/0408095.

\item {\bf [Gisin-Ribordy-Zbinden-(+3) 04]}:
N. Gisin, G. Ribordy, H. Zbinden,
D. Stucki, N. Brunner, \& V. Scarani,
``Towards practical and fast quantum cryptography'',
quant-ph/0411022.

\item {\bf [Gittings-Fisher 02]}:
J. R. Gittings, \& A. J. Fisher,
``Describing mixed spin-space entanglement of pure states
of indistinguishable particles using an occupation number basis'',
{\em Phys. Rev. A} {\bf 66}, 3, 032305 (2002);
quant-ph/0202051.

\item {\bf [Gittings-Fisher 03]}:
J. R. Gittings, \& A. J. Fisher,
``An efficient numerical method for calculating the entanglement of
formation of arbitrary mixed quantum states of any dimension'',
quant-ph/0302018.

\item {\bf [Giulini-Kiefer-Zeh 95]}:
D. Giulini, C. Kiefer, \& H. D. Zeh,
``Symmetries, superselection rules and decoherence'',
{\em Phys. Lett. A} {\bf 199}, 5-6, 291-298 (1995).

\item {\bf [Giulini-Joos-Kiefer-(+3) 96]}:
D. Giulini, E. Joos, C. Kiefer, J. Kupsch, I. Stamatescu, \& H. D. Zeh,
{\em Decoherence and the appearance of a classical world in quantum theory},
Springer-Verlag, Berlin, 1996.
Review: {\bf [Donald 99]}.

\item {\bf [Giulini 99]}:
D. Giulini,
``States, symmetries and superselection'',
in P. Blanchard, D. Giulini, E. Joos,
C. Kiefer, \& I-O. Stamatescu (eds.),
{\em Decoherence: Theoretical, experimental, and conceptual problems},
Springer-Verlag, Berlin, 1999;
quant-ph/9906108.

\item {\bf [Giulini 00]}:
D. Giulini,
``Decoherence: A dynamical approach to superselection rules?'',
in H.-P. Breuer, \& F. Petruccione (eds.),
{\em Relativistic quantum measurement and decoherence},
{\em Lecture notes in physics, vol. 538},
Springer-Verlag, Berlin, 2000;
quant-ph/0010090.

\item {\bf [Giuntini-Laudisa 01]}:
R. Giuntini, \& F. Laudisa,
``The impossible causality: The no-hidden variable theorem of von Neumann'',
in M. R\'{e}dei, \& M. St\"{o}ltzner (eds.),
{\em John von Neumann and the foundations of quantum physics},
Kluwer Academic, Dordrecht, Holland, 2001, pp.~?-?.

\item {\bf [Glancy-LoSecco-Vasconcelos-Tanner 02]}:
S. Glancy, J. M. LoSecco, H. M. Vasconcelos, \& C. E. Tanner,
``Imperfect detectors in linear optical quantum computers'',
{\em Phys. Rev. A} {\bf 65}, 6, 062317 (2002);
quant-ph/0201047.

\item {\bf [Glancy-Vasconcelos-Ralph 04]}:
S. Glancy, H. M. Vasconcelos, \& T. C. Ralph,
``Transmission of optical coherent state qubits'',
{\em Phys. Rev. A} {\bf 70}, 2, 022317 (2004);
quant-ph/0311093.

\item {\bf [Glanz 95 a]}:
J. Glanz,
``A quantum leap for computers?'',
{\em Science} {\bf 269}, 5220, 28-29 (1995).

\item {\bf [Glanz 95 b]}:
J. Glanz,
``Physics: Measurements are the only reality, say quantum tests'',
{\em Science} {\bf 270}, 5241, 1439-1440 (1995).

\item {\bf [Glanz 98]}:
J. Glanz,
``Physics: A first step toward wiring up a quantum computer'',
{\em Science} {\bf 282}, 5393, 1405 (1998).

\item {\bf [Glaser-Buttner-Fehske 03]}:
U. Glaser, H. Buttner, \& H. Fehske,
``Entanglement and correlation in anisotropic quantum spin systems'',
{\em Phys. Rev. A} {\bf 68}, 3, 032318 (2003).

\item {\bf [Gleason 57]}:
A. M. Gleason,
``Measures on the closed subspaces of a Hilbert space'',
{\em J. Math. Mech.} {\bf 6}, 6, 885-893 (1957).
Reprinted in {\bf [Hooker 75]}, pp.~123-134.

\item {\bf [Gleason 66]}:
A. M. Gleason,
{\em Fundamentals of abstract analysis},
Addison-Wesley, Reading, Massachusetts, 1966. pp.~77, 245-248.

\item {\bf [Gl\"{o}ckl-Lorenz-Marquardt-(+7) 03]}:
O. Gl\"{o}ckl, S. Lorenz, C. Marquardt,
J. Heersink, M. Brownnutt, C. Silberhorn,
Q. Pan, P. van Loock, N. Korolkova, \& G. Leuchs,
``Experiment towards continuous-variable entanglement swapping: Highly
correlated four-partite quantum state'',
{\em Phys. Rev. A} {\bf 68}, 1, 012319 (2003);
quant-ph/0302068.

\item {\bf [Go 04]}:
A. Go,
``Observation of Bell inequality violation in B mesons'',
{\em J. Mod. Opt.} {\bf 51}, ?, 991-? (2004);
quant-ph/0310192.
See {\bf [Bramon-Escribano-Garbarino 04]}.

\item {\bf [Goan-Milburn 01]}:
H.-S. Goan, \& G. J. Milburn,
``Dynamics of a mesoscopic charge quantum bit under continuous quantum
measurement'',
{\em Phys. Rev. B} {\bf 64}, 23, 235307 (2001).

\item {\bf [Gobert-von Delft-Ambegaokar 03]}:
D. Gobert, J. von Delft, \& V. Ambegaokar,
``Decoherence without dissipation?'',
quant-ph/0306019.
Comment on {\bf [Ford-Lewis-O'Connell 01]}.
Reply: {\bf [Ford-O'Connell 04]}.

\item {\bf [Goddard 98]}:
P. Goddard (ed.),
{\em Paul Dirac, the man and his work},
Cambridge University Press, Cambridge, 1998.

\item {\bf [Godsil-Zaks 88]}:
C. D. Godsil, \& J. Zaks,
``Colouring the sphere'',
University of Waterloo research report CORR 88-12 (1988).

\item {\bf [Goldberg-Schey-Schwartz 67]}:
A. Goldberg, H. M. Schey, \& J. L. Schwartz,
``Computer-generated motion pictures of one-dimensional
quantum-mechanical transmission and reflection phenomena'',
{\em Am. J. Phys.} {\bf 35}, 3, 177-186 (1967).

\item {\bf [Goldberger-Watson 64]}:
M. L. Goldberger, \& K. M. Watson,
``Measurement of time correlations for quantum-mechanical
systems'',
{\em Phys. Rev.} {\bf 134}, 4B, B919-B928 (1964).

\item {\bf [Goldenberg-Vaidman 95 a]}:
L. Goldenberg, \& L. Vaidman,
``Quantum cryptography based on orthogonal states'',
{\em Phys. Rev. Lett.} {\bf 75}, 7, 1239-1243 (1995);
quant-ph/9502021.
Comment: {\bf [Peres 96 f]}.
Reply: {\bf [Goldenberg-Vaidman 96]}.

\item {\bf [Goldenberg-Vaidman 95 b]}:
L. Goldenberg, \& L. Vaidman,
``Applications of a simple formula'',
quant-ph/9506030.

\item {\bf [Goldenberg-Vaidman 96]}:
L. Goldenberg, \& L. Vaidman,
`Goldenberg and Vaidman reply
[to ``Quantum cryptography with orthogonal states?'']',
{\em Phys. Rev. Lett.} {\bf 77}, 15, 3265 (1996);
quant-ph/9604029.
Reply to {\bf [Peres 96 f]}.
See {\bf [Goldenberg-Vaidman 95 a]}.

\item {\bf [Goldenberg-Vaidman-Wiesner 99]}:
L. Goldenberg, L. Vaidman, \& S. Wiesner,
``Quantum gambling'',
{\em Phys. Rev. Lett.} {\bf 82}, 16, 3356-3359 (1999);
quant-ph/9808001.

\item {\bf [Goldhaber 99]}:
M. Goldhaber,
``Geons, black holes and quantum foam'',
{\em Phys. Today} {\bf 52}, 5, 63-64 (1999).
Review of {\bf [Wheeler-Ford 98]}.

\item {\bf [Goldstein 94 a]}:
S. Goldstein,
``Nonlocality without inequalities for almost all
entangled states for two particles'',
{\em Phys. Rev. Lett.} {\bf 72}, 13, 1951 (1994).
See {\bf [Hardy 93]}.

\item {\bf [Goldstein 94 b]}:
S. Goldstein,
``The undivided universe: An ontological interpretation of quantum theory'',
{\em Phys. Today} {\bf 47}, 9, 90 (1994).
Review of {\bf [Bohm-Hiley 93]}.

\item {\bf [Goldstein 94 c]}:
S. Goldstein,
``The quantum theory of motion: An
account of the de Broglie-Bohm causal interpretation of quantum mechanics'',
{\em Science} {\bf 263}, 5144, 254-255 (1994).
Review of {\bf [Holland 93]}.

\item {\bf [Goldstein-Page 95]}:
S. Goldstein, \& D. Page,
``Linearly positive histories: Probabilities for a robust family of
sequences of quantum events'',
{\em Phys. Rev. Lett.} {\bf 74}, 19, 3715-3719 (1995).

\item {\bf [Goldstein 96]}:
S. Goldstein,
``Bohmian mechanics and the quantum revolution'',
{\em Synthese} {\bf 107}, ?, 145-165 (1996).

\item {\bf [Goldstein 98 a]}:
S. Goldstein,
``Quantum theory without observers---Part one'',
{\em Phys. Today} {\bf 51}, 3, 42-46 (1998).
See {\bf [Goldstein 98 b]} (II).
Comment: {\bf [Zeh 99 a]}.

\item {\bf [Goldstein 98 b]}:
S. Goldstein,
``Quantum theory without observers---Part two'',
{\em Phys. Today} {\bf 51}, 4, 38-42 (1998).
See {\bf [Goldstein 98 a]} (I).

\item {\bf [Goldstein 99]}:
S. Goldstein,
``Absence of chaos in Bohmian dynamics'',
quant-ph/9901005.

\item {\bf [Goldstein-Tumulka 01]}:
S. Goldstein, \& R. Tumulka,
``Opposite arrows of time can reconcile relativity and
nonlocality'',
quant-ph/0105040.

\item {\bf [Goldstein-Taylor-Tumulka-Zanghi 04 a]}:
S. Goldstein, J. Taylor, R. Tumulka, \& N. Zanghi,
``Are all particles real?'',
quant-ph/0404134.

\item {\bf [Goldstein-Taylor-Tumulka-Zanghi 04 b]}:
S. Goldstein, J. Taylor, R. Tumulka, \& N. Zanghi,
``Are all particles identical?'',
quant-ph/0405039.

\item {\bf [Golshani-Akhavan 00]}:
M. Golshani, \& O. Akhavan,
``A two-slit experiment which distinguishes between the
standard and Bohmian quantum mechanics'',
quant-ph/0009040.

\item {\bf [Golshani-Akhavan 01 a]}:
M. Golshani, \& O. Akhavan,
``Bohmian prediction about a two double-slit experiment
and its disagreement with standard quantum mechanics'',
{\em J. Phys. A} {\bf 34}, 25, 5259-5268 (2001);
quant-ph/0103101.
Comment: {\bf [Struyve-De Baere-De Neve-De Weirdt 03]}.

\item {\bf [Golshani-Akhavan 01 b]}:
M. Golshani, \& O. Akhavan,
``Experiment can decide between the standard and
Bohmian quantum mechanics'',
quant-ph/0103100.

\item {\bf [Golshani-Akhavan 01 c]}:
M. Golshani, \& O. Akhavan,
``On the experimental incompatibility between
standard and Bohmian quantum mechanics'',
quant-ph/0110123.

\item {\bf [Golub-G\"{a}hler 89]}:
R. Golub, \& R. G\"{a}hler,
`Measurement of $\left\langle {\sigma _z} \right\rangle \approx 100$
for a spin-$\frac{1}{2}$ particle or ``polarization amplification''
of $\left\langle \sigma \right\rangle \approx 1$?',
{\em Phys. Lett. A} {\bf 136}, 4-5, 178-182 (1989).

\item {\bf [G\'{o}mez Sal-Moreno-Santos 00]}:
J. C. G\'{o}mez Sal, M. Moreno, \& E. Santos,
``Entrevista con Emilio Santos Corchero (Medalla de la RSEF
1998)'',
{\em Revista Espa\~{n}ola de F\'{\i}sica} {\bf 14}, 2, 2-5 (2000).

\item {\bf [Gong-Brumer 02]}:
J. Gong, \& P. Brumer,
``When is quantum decoherence dynamics classical?'',
{\em Phys. Rev. Lett.};
quant-ph/0212106.

\item {\bf [Gong-Shapiro-Brumer 03]}:
J. Gong, M. Shapiro, \& P. Brumer,
``Entanglement-assisted coherent control in nonreactive diatom--diatom
scattering'',
{\em J. Chem. Phys.} {\bf 118}, ?, 2626 (2003).

\item {\bf [Gong-Brumer 03]}:
J. Gong, \& P. Brumer,
``Intrinsic decoherence dynamics in smooth Hamiltonian systems:
Quantum-classical correspondence'',
{\em Phys. Rev. A} {\bf 68}, 2, 022101 (2003);
quant-ph/0306002.

\item {\bf [Gontis-Kaulakys 98]}:
V. Gontis, \& B. Kaulakys,
``Quantum Zeno and quantum anti-Zeno efffects'',
{\em Lithuanian J. Phys.} {\bf 38}, 1, 118-121 (1998);
quant-ph/9806015.

\item {\bf [Gonz\'{a}lez Robles03]}:
V. M. Gonz\'{a}lez-Robles,
``About a generalization of Bell's inequality'',
{\em Found. Phys.} {\bf 33}, 5, 839-853 (2003).

\item {\bf [Gorbachev-Trubilko 99 a]}:
V. N. Gorbachev, \& A. I. Trubilko,
``Quantum teleportation of EPR pair by three-particle
entanglement'',
quant-ph/9906110.

\item {\bf [Gorbachev-Trubilko 99 b]}:
V. N. Gorbachev, \& A. I. Trubilko,
``Teleportation of entanglement for continuous variables'',
quant-ph/9912061.

\item {\bf [Gorbachev-Zhiliba-Trubilko-Yakovleva 00]}:
V. N. Gorbachev, A. I. Zhiliba, A. I. Trubilko, \& E. S. Yakovleva,
``Teleportation of entangled states and dense coding
using a multiparticle quantum channel'',
quant-ph/0011124.

\item {\bf [Gorbachev-Zhiliba-Trubilko 01 a]}:
V. N. Gorbachev, A. I. Zhiliba, \& A. I. Trubilko,
``Teleportation of entangled states'',
{\em J. Opt. B: Quantum Semiclass. Opt.} {\bf 3}, 1, S25-S29 (2001).

\item {\bf [Gorbachev-Zhiliba-Trubilko 01 b]}:
V. N. Gorbachev, A. I. Zhiliba, \& A. I. Trubilko,
``Continuous-variables teleportation of a two-particle entangled state'',
{\em Opt. Comm.} {\bf 187}, 4-6, 379-383 (2001).

\item {\bf [Gorbachev-Rodichkina-Trubilko-Zhiliba 03]}:
V. N. Gorbachev, A. A. Rodichkina, A. I. Trubilko, \& A. I. Zhiliba,
``On preparation of the entangled $W$-states from atomic ensembles'',
{\em Phys. Lett. A} {\bf 310}, 5-6, 339-343 (2003);
quant-ph/0211087.

\item {\bf [Gorbachev-Trubilko-Rodichkina 03]}:
V. N. Gorbachev, A. I. Trubilko, \& A. A. Rodichkina,
``Teleportation through $W$-class states'',
{\em Opt. Spectrosc.} {\bf 94}, 706 (2003).

\item {\bf [Gorbachev-Trubilko 03]}:
V. N. Gorbachev, \& A. I. Trubilko,
``Amplification of two-mode squeezed light in the Einstein-Podolsky-Rosen
state',
{\em JETP Lett.} {\bf 77}, 469 (2003).

\item {\bf [Gorbachev-Trubilko-Rodichkina-Zhiliba 03]}:
V. N. Gorbachev, A. I. Trubilko, A. A. Rodichkina, \& A. I. Zhiliba,
``Can the states of the $W$-class be suitable for teleportation?'',
{\em Phys. Lett. A} {\bf 314}, 4, 267-271 (2003);
quant-ph/0203028.

\item {\bf [Gorbachev-Trubilko-Rodichkina 04]}:
V. N. Gorbachev, A. I. Trubilko, \& A. A. Rodichkina,
``Origination of entangled states upon interaction of atoms with narrow-band light'',
{\em Opt. Spectroscopy} {\bf 97}, ?, 130-135 (2004).

\item {\bf [Gordon 64]}:
J. P. Gordon,
``?'',
in P. A. Milles (ed.),
{\em Proc.\ of the Int. School of Physics ``Enrico Fermi''.
Course XXXI: Quantum Electronics and Coherent Light (?, ?, ?)},
Academic Press, New York, 1964, pp.~156-181.

\item {\bf [Gorman-Tapster-Rarity 01]}:
P. M. Gorman, P. R. Tapster, \& J. G. Rarity,
``Secure free-space key exchange to 1.9 km and beyond'',
{\em J. Mod. Opt.} {\bf 48}, ?, 1887-1901 (2001).

\item {\bf [de Gosson 04]}:
M. de Gosson,
``The optimal pure Gaussian state canonically associated to a Gaussian quantum state'',
{\em Phys. Lett. A} {\bf 330}, 3-4, 161-167 (2004).

\item {\bf [Goswami 02]}:
D. Goswami,
``Laser phase modulation approaches towards ensemble quantum computing'',
{\em Phys. Rev. Lett.} {\bf 88}, 17, 177901 (2002).
Erratum: {\em Phys. Rev. Lett.} {\bf 89}, 27, 279901 (2002).

\item {\bf [Goto-Shimizu-Hashi-(+2) 03]}:
A. Goto, T. Shimizu, K. Hashi, H. Kitazawa, \& S. Ohki,
``Decoupling-free NMR quantum computer on a quantum spin chain'',
{\em Phys. Rev. A} {\bf 67}, 2, 022312 (2003).

\item {\bf [Goto-Ichimura 04]}:
H. Goto \& K. Ichimura,
``Multiqubit controlled unitary gate by adiabatic passage with an optical cavity'',
{\em Phys. Rev. A} {\bf 70}, 1, 012305 (2004).

\item {\bf [Gottesman 96]}:
D. Gottesman,
``Class of quantum error-correcting codes saturating the quantum Hamming bound'',
{\em Phys. Rev. A} {\bf 54}, 3, 1862-1868 (1998).
Reprinted in {\bf [Macchiavello-Palma-Zeilinger 00]}, pp.~143-149.

\item {\bf [Gottesman 98 a]}:
D. Gottesman,
``Theory of fault-tolerant quantum computation'',
{\em Phys. Rev. A} {\bf 57}, 1, 127-137 (1998).

\item {\bf [Gottesman 98 b]}:
D. Gottesman,
``Fault-tolerant quantum computation with
higher-dimensional systems'',
in C. P. Williams (ed.),
{\em 1st NASA Int.\ Conf.\ on Quantum Computing and Quantum Communications
(Palm Springs, California, 1998)},
{\em Lecture Notes in Computer Science} {\bf 1509},
Springer-Verlag, New York, 1999, pp.~?-?;
quant-ph/9802007.

\item {\bf [Gottesman-Chuang 99 a]}:
D. Gottesman, \& I. L. Chuang,
``Demonstrating the viability of universal quantum computation
using teleportation and single-qubit operations'',
{\em Nature} {\bf 402}, 6760, 390-393 (1999).
See {\bf [Preskill 99 b]}.

\item {\bf [Gottesman-Chuang 99 b]}:
D. Gottesman, \& I. L. Chuang,
``Quantum teleportation is a universal computational primitive'',
quant-ph/9908010.

\item {\bf [Gottesman 00 a]}:
D. Gottesman,
``Theory of quantum secret sharing'',
{\em Phys. Rev. A} {\bf 61}, 4, 042311 (2000);
quant-ph/9910067.

\item {\bf [Gottesman 00 b]}:
D. Gottesman,
``Fault-tolerant quantum computation with local gates'',
in V. Bu\v{z}zek, \& D. P. DiVincenzo (eds.),
{\em J. Mod. Opt.} {\bf 47}, 2-3 (Special issue:
Physics of quantum information), 333-345 (2000).

\item {\bf [Gottesman-Lo 00]}:
D. Gottesman, \& H.-K. Lo,
``From quantum cheating to quantum security'',
{\em Phys. Today} {\bf 53}, 11, 22-? (2000);
quant-ph/0111100.

\item {\bf [Gottesman-Preskill 01]}:
D. Gottesman, \& J. Preskill,
``Secure quantum key distribution using squeezed states'',
{\em Phys. Rev. A} {\bf 63}, 2, 022309 (2001);
quant-ph/0008046.

\item {\bf [Gottesman-Kitaev-Preskill 01]}:
D. Gottesman, A. Kitaev, \& J. Preskill,
``Encoding a qudit in an oscillator'',
{\em Phys. Rev. A} {\bf 64}, 1, 012310 (2001);
quant-ph/0008040.

\item {\bf [Gottesman-Chuang 01]}:
D. Gottesman, \& I. L. Chuang,
``Quantum digital signatures'',
quant-ph/0105032.

\item {\bf [Gottesman-Lo 02]}:
D. Gottesman, \& H.-K. Lo,
``Proof of security of quantum key
distribution with two-way classical communications'',
{\em IEEE Trans. Inf. Theory}, 2002;
quant-ph/0105121.

\item {\bf [Gottesman 02]}:
D. Gottesman,
``An introduction to quantum error correction'',
in {\bf [Lomonaco 02 a]}, pp.~221-235;
quant-ph/0004072.

\item {\bf [Gottesman-Lo-L\"{u}tkenhaus-Preskill 02]}:
D. Gottesman, H.-K. Lo, N. L\"{u}tkenhaus, \& J. Preskill,
``Security of quantum key distribution with imperfect devices'',
quant-ph/0212066.

\item {\bf [Gottesman 03]}:
D. Gottesman,
``Uncloneable encryption'',
{\em Quant. Inf. Comp.} {\bf 3}, ?, 581-602 (2003);
quant-ph/0210062.

\item {\bf [Gottfried 91]}:
K. Gottfried,
``Does quantum mechanics carry the seeds
of its own destruction?'',
{\em Phys. World} {\bf 4}, 10, 34-40 (1991).
Reprinted in {\bf [Ellis-Amati 00]}, pp.~165-185.

\item {\bf [Gottfried 99]}:
K. Gottfried,
``Is the statistical interpretation of quantum mechanics
implied by the correspondence principle?'',
in {\bf [Greenberger-Reiter-Zeilinger 99]}, pp.~?-?;
quant-ph/9812042.

\item {\bf [Gottfried 00 a]}:
K. Gottfried,
``Two-particle interferometry'',
{\em Am. J. Phys.} {\bf 68}, 2, 143-147 (2000).

\item {\bf [Gottfried 00 b]}:
K. Gottfried,
``Inferring the statistical interpretation of quantum mechanics from
the classical limit'',
{\em Nature} {\bf 405}, 6786, 533-536 (2000).

\item {\bf [Gottfried-Mermin 00]}:
K. Gottfried, \& N. D. Mermin,
``John Bell and the moral aspect of quantum mechanics'',
in {\bf [Ellis-Amati 00]}, pp.~186-192.

\item {\bf [Gottlieb-Janson-Scudo 04]}:
A. D. Gottlieb, S. Janson, \% P. F. Scudo,
``Convergence of coined quantum walks on $d$-dimensional Euclidean space'',
quant-ph/0406072.

\item {\bf [Gour-Khanna-Mann-Revzen 04]}:
G. Gour, F. C. Khanna, A. Mann, \& M. Revzen,
``Optimization of Bell's inequality violation for continuous variable
systems'',
{\em Phys. Lett. A} {\bf 325}, 5-6, 415-419 (2004);
quant-ph/0308063.

\item {\bf [Gour 04]}:
G. Gour,
``A family of concurrence monotones and its applications'',
{\em Phys. Rev. A};
quant-ph/0410148.

\item {\bf [Gour-Sanders 04]}:
G. Gour, \& B. C. Sanders,
``Remote preparation and distribution of bipartite entangled states'',
quant-ph/0410016.

\item {\bf [Gourlay-Snowdon 00]}:
I. Gourlay, \& J. F. Snowdon,
``Concatenated coding in the presence of dephasing'',
{\em Phys. Rev. A} {\bf 62}, 2, 022308 (2000).

\item {\bf [Gowing 79 a]}:
M. Gowing,
``My life: Recollections of a Nobel laureate'',
{\em Contemp. Phys.} {\bf 20}, 4, 485-486 (1979).
Review of {\bf [Born 75]}.

\item {\bf [Gowing 79 b]}:
M. Gowing,
``\dots I have always lived very much in the present\dots'',
{\em Contemp. Phys.} {\bf 20}, 6, 655-661 (1979).
Review of {\bf [Frisch 79]}.

\item {\bf [Gracia Bond\'{\i}a-Marshall-Santos 93]}:
J. M. Gracia-Bond\'{\i}a, T. W. Marshall, \& E. Santos,
``A phase-space description of the Stern-Gerlach phenomenon'',
{\em Phys. Lett. A} {\bf 183}, 1, 19-23 (1993).

\item {\bf [de Graaf-de Wolf 01]}:
M. de Graaf, \& R. de Wolf,
``On quantum versions of the Yao principle'',
quant-ph/0109070.

\item {\bf [Graham 71]}:
R. N. Graham,
``The Everett interpretation of quantum mechanics'',
Ph.\ D. thesis, University of North Carolina,
Chapel Hill, North Carolina, 1971.

\item {\bf [Gram\ss\, 98]}:
T. Gram\ss\, (ed.),
{\em Non-standard computation: Molecular computation,
cellular automata, evolutionary algorithms, quantum computers},
John Wiley \& Sons, New York, 1998.

\item {\bf [Grangier-Roger-Aspect 86]}:
P. Grangier, G. Roger, \& A. Aspect,
``Experimental evidence for a photon anticorrelation effect
on a beam splitter: A new light on single photon interferences'',
{\em Europhys. Lett.} {\bf 1}, 4, 173-179 (1986).
Comment: {\bf [Marshall-Santos 87]}.

\item {\bf [Grangier-Potasek-Yurke 88]}:
P. Grangier, M. J. Potasek, \& B. Yurke,
``Probing the phase coherence of parametrically
generated photon pairs: A new test of Bell's inequalities'',
{\em Phys. Rev. A} {\bf 38}, 6, 3132-3135 (1988).

\item {\bf [Grangier 98 a]}:
P. Grangier,
``Lighting up the dark'',
{\em Nature} {\bf 392}, 6677, 672 (1998).
Review of {\bf [Greenstein-Zajonc 98]}.

\item {\bf [Grangier 98 b]}:
P. Grangier,
``Atomic physics: Enhanced: Under control'',
{\em Science} {\bf 281}, 5373, 56-57 (1998).

\item {\bf [Grangier-Levenson-Poizat 98]}:
P. Grangier, J. A. Levenson, \& J-P. Poizat,
``Quantum non-demolition measurements in optics'',
{\em Nature} {\bf 396}, 6711, 537-542 (1998).

\item {\bf [Grangier 99]}:
P. Grangier,
``A box for a single photon'',
{\em Nature} {\bf 400}, 6741, 215-217 (1999).
See {\bf [Nogues-Rauschenbeutel-Osnaghi-(+3) 99]}.

\item {\bf [Grangier-Reymond-Schlosser 00]}:
P. Grangier, G. Reymond, \& N. Schlosser,
``Implementations of quantum computing using cavity quantum electrodynamics schemes'',
{\em Fortschr. Phys.} {\bf 48}, 9-11 (Special issue: Experimental proposals for quantum computation), 859-874 (2000).

\item {\bf [Grangier-Grosshans 00 a]}:
P. Grangier, \& F. Grosshans,
``Quantum teleportation criteria for continuous variables'',
quant-ph/0009079.

\item {\bf [Grangier-Grosshans 00 b]}:
P. Grangier, \& F. Grosshans,
``Evaluating quantum teleportation of coherent states'',
quant-ph/0010107.

\item {\bf [Grangier 01 a]}:
P. Grangier,
``Quantum physics: Count them all'',
{\em Nature} {\bf 409}, 6822, 774-775 (2001).
See {\bf [Rowe-Kielpinski-Meyer-(+4) 01]}.

\item {\bf [Grangier 01 b]}:
P. Grangier,
`Reconstructing the formalism of quantum mechanics
in the ``contextual objectivity'' point of view',
quant-ph/0111154.

\item {\bf [Grangier-Rarity-Karlsson 02]}:
P. Grangier, J. Rarity, \& A. Karlsson,
``Quantum interference and cryptographic keys:
Novel physics and advancing technologies (QUICK)'',
{\em Eur. Phys. J. D} {\bf 18}, 2 (Special issue:
{\em Quantum interference and cryptographic keys:
Novel physics and advancing technologies (QUICK) (Corsica, 2001)}, 139 (2002).

\item {\bf [Grangier 02 a]}:
P. Grangier,
``Contextual objectivity:
A realistic interpretation of quantum mechanics'',
{\em Eur. J. Phys.} {\bf 23}, 3, 331-337 (2002);
quant-ph/0012122.

\item {\bf [Grangier 02 b]}:
P. Grangier,
`FAQ about the ``contextual objectivity'' point of view',
quant-ph/0203131.
See {\bf [Grangier 03]}.

\item {\bf [Grangier 03]}:
P. Grangier,
``Contextual objectivity and quantum holism'',
quant-ph/0301001.
See {\bf [Grangier 02 b]}.

\item {\bf [Grangier 04]}:
P. Grangier,
``Contextual objectivity and the quantum formalism'',
in {\em Proc.\ of Foundations of Quantum Information" (Camerino, Italy, 2004)},
{\em Int. J. Quantum Inf.};
quant-ph/0407025.

\item {\bf [Granik-Caulfield 96]}:
A. Granik, \& H. J. Caulfield,
``Fuzziness in quantum mechanics'',
{\em Phys. Essays} {\bf 9}, ?, 496-? (1996);
quant-ph/0107054.

\item {\bf [Granik-Chapline 03]}:
A. Granik, \& G. Chapline,
``Energy dissipation in quantum computers'',
quant-ph/0310074.

\item {\bf [Grassl-Beth-Pellizzari 97]}:
M. Grassl, T. Beth, \& T. Pellizzari,
``Codes for the quantum erasure channel'',
{\em Phys. Rev. A} {\bf 56}, 1, 33-38 (1997).

\item {\bf [Grassl-Beth 99]}:
M. Grassl, \& T. Beth,
``Quantum BCH codes'',
in {\em Proc.\ 10th Int.\ Symp.\ on Theoretical
Electrical Engineering, Magdeburg, 1999}, pp.~207-212;
quant-ph/9910060.

\item {\bf [Grassl-Beth 00]}:
M. Grassl, \& T. Beth,
``Cyclic quantum error-correcting codes and
quantum shift registers'',
{\em Proc. R. Soc. Lond. A} {\bf 456}, 2003, 2689-2706 (2000);
quant-ph/9910061.

\item {\bf [Grassl-Beth-R\"{o}tteler 03]}:
M. Grassl, T. Beth, \& M. R\"{o}tteler,
``On optimal quantum codes'',
{\em Int. J. Quant. Inf.};
quant-ph/0312164.

\item {\bf [Gravier-Jorrand-Mhalla-Payan 03]}:
S. Gravier, P. Jorrand, M. Mhalla, \& C. Payan,
``Let us play with qubits'',
quant-ph/0311018.

\item {\bf [Greechie 78]}:
R. J. Greechie,
``Another nonstandard quantum logic (and how I found it)'',
in A. R. Marlow (ed.),
{\em Mathematical Foundations of Quantum Theory
(Loyola University, New Orleans, 1977)},
Academic Press, New York, 1978, 71-85.

\item {\bf [Green 98]}:
H. S. Green,
``Quantal information theory'',
{\em Int. J. Theor. Phys.} {\bf 37}, 11, 2735-2750 (1998).

\item {\bf [Green 00]}:
H. S. Green,
{\em Information theory and quantum physics},
Springer-Verlag, Berlin, 2000.

\item {\bf [Greenberger 83]}:
D. M. Greenberger,
``The neutron interferometer as a
device for illustrating the strange behavior of quantum systems'',
{\em Rev. Mod. Phys.} {\bf 55}, 4, 875-904 (1983).

\item {\bf [Greenberger 86]}:
D. M. Greenberger,
``The neutron interferometer and the quantum mechanical superposition principle'',
in L. M. Roth, \& I. Inomata (eds.),
{\em Fundamental questions in quantum mechanics},
Gordon \& Beach, New York, 1986, pp.~?-?.

\item {\bf [Greenberger-Yasin 88 a]}:
D. M. Greenberger, \& A. Yasin,
``The haunted measurement: A new twist on the quantum measurement problem'',
in {\em Urbino Conference on Quantum Measurement Theory},
Kluwer Academic, Dordrecht, Holland, 1988, pp.~?-?.

\item {\bf [Greenberger-Yasin 88 b]}:
D. M. Greenberger, \& A. Yasin,
``Simultaneous wave and particle knowledge in a neutron interferometer'',
{\em Phys. Lett. A} {\bf 128}, 8, 391-394 (1988).

\item {\bf [Greenberger 88]}:
D. M. Greenberger,
``A new non-local effect in quantum theory'',
{\em Physica B} {\bf 151}, ?, 374-? (1988).

\item {\bf [Greenberger-Yasin 89}:
D. M. Greenberger, \& A. Yasin,
``The haunted measurement theory'',
{\em Found. Phys.} {\bf 19}, ?, 679-704 (1989).

\item {\bf [Greenberger-Horne-Zeilinger 89]}:
D. M. Greenberger, M. A. Horne, \& A. Zeilinger,
``Going beyond Bell's theorem'',
in M. Kafatos (ed.),
{\em Bell's theorem, quantum theory, and conceptions of
the universe. Proc.\ of a workshop (George Mason University, 1988)},
Kluwer Academic, Dordrecht, Holland, 1989, pp.~69-72.
Reprinted in {\bf [Macchiavello-Palma-Zeilinger 00]}, pp.~49-52.
See {\bf [Greenberger-Horne-Shimony-Zeilinger 90]}.

\item {\bf [Greenberger-Horne-Zeilinger 90]}:
D. M. Greenberger, M. A. Horne, \& A. Zeilinger,
``Bell's theorem without inequalities'',
in A. I. Miller (ed.),
{\em Sixty-two years of uncertainty:
Historical, philosophical and physical inquiries into the foundations of quantum
mechanics.
Proc.\ Int. School of History of Science (Erice, Italy, 1989)},
Plenum Press, New York, 1990, pp.~?-?.

\item {\bf [Greenberger-Horne-Shimony-Zeilinger 90]}:
D. M. Greenberger, M. A. Horne, A. Shimony, \& A. Zeilinger,
``Bell's theorem without inequalities'',
{\em Am. J. Phys.} {\bf 58}, 12, 1131-1143 (1990).
See {\bf [Greenberger-Horne-Zeilinger 89]}.

\item {\bf [Greenberger 91]}:
D. M. Greenberger,
``Book review. Quantum mechanics'',
{\em Found. Phys.} {\bf 21}, 6, 751-752 (1991).
Review of {\bf [Ballentine 90 a]}.

\item {\bf [Greenberger-Zeilinger-et al. 91]}:
D. M. Greenberger, A. Zeilinger, et al.,
``Quantum reality and higher order correlations: Two remarks on entanglement'',
in P. J. Lahti, \& P. Mittelstaedt (eds.),
{\em Symp.\ on the Foundations of Modern Physics 1990.
Quantum Theory of Measurement and Related Philosophical Problems
(Joensuu, Finland, 1990)},
World Scientific, Singapore, 1991, pp.~?-?.

\item {\bf [Greenberger-Zeilinger-et al. 92]}:
D. M. Greenberger, A. Zeilinger, et al.,
``Higher order quantum entanglements'',
D. Han, (ed.)
in {\em Workshop on Squeezed States and Uncertainty Relations
(College Park, Maryland, 1991)},
NASA, ?, 1992, pp.~?-?.

\item {\bf [Greenberger-Horne-et al. 92]}:
D. M. Greenberger, M. A. Horne, et al.,
``Use of entanglement in quantum optics'',
D. Han, (ed.)
in {\em Workshop on Squeezed States and Uncertainty Relations
(College Park, Maryland, 1991)},
NASA, ?, 1992, pp.~?-?.

\item {\bf [Greenberger 92]}:
D. M. Greenberger,
``The breakdown of causality in modern physics'',
{\em Thesis Quarterly}, CUNY, New York, 1992.

\item {\bf [Greenberger-Horne-et al. 93]}:
D. M. Greenberger, M. A. Horne, et al.,
``Controlling entanglement in quantum optics'',
in H. Ezawa, \& Y. Muruyama (eds.),
{\em Quantum control and measurement},
Elsevier, Amsterdam, 1993, pp.~?-?.

\item {\bf [Greenberger-Greenberg-Greenbergest 93]}:
D. M. Greenberger, O. W. Greenberg, \& T. V. Greenbergest,
``On the superposition principle'',
in J. Anandan et al. (eds.),
{\em Symp.\ in Honor of Y. Aharonov},
World Scientific, Singapore, 1993, pp.~?-?.

\item {\bf [Greenberger-Bernstein-Horne-Zeilinger 93]}:
D. M. Greenberger, H. J. Bernstein, M. A. Horne, \& A. Zeilinger,
``A GHZ theorem for two spinless particles'',
in A. van der Merwe, \& F. Selleri (eds.),
{\em Bell's theorem and the foundations of modern physics.
Proc.\ of an international conference (Cesena, Italy, 1991)},
World Scientific, Singapore, 1993, pp.~250-262.
See {\bf [Bernstein-Greenberger-Horne-Zeilinger 93]}.

\item {\bf [Greenberger-Horne-Zeilinger 93]}:
D. M. Greenberger, M. A. Horne, \& A. Zeilinger,
``Multiparticle interferometry and the superposition principle'',
{\em Phys. Today} {\bf 46}, 8, 22-29 (1993).
Reprinted in {\bf [Macchiavello-Palma-Zeilinger 00]}, pp.~4-11.

\item {\bf [Greenberger-Horne-Zeilinger 94]}:
D. M. Greenberger, M. A. Horne, \& A. Zeilinger,
``Two-particle fringes dependent on the sum of the coordinates'',
in F. De Martini, \& A. Zeilinger (eds.),
{\em Quantum interferometry},
World scientific, Singapore, 1994, pp.~?-?.

\item {\bf [Greenberger 94 a]}:
D. M. Greenberger,
``Interpretations. The undivided
universe. An ontological interpretation of quantum theory'',
{\em Science} {\bf 266}, 5182, 147-148 (1994).
Review of {\bf [Bohm-Hiley 93]}.

\item {\bf [Greenberger 94 b]}:
D. M. Greenberger,
``Mind, matter, and quantum mechanics'',
{\em Phys. Today} {\bf 47}, 11, 88-90 (1994).
Review of {\bf [Stapp 93 b]}.

\item {\bf [Greenberger-Horne-Zeilinger 95]}:
D. M. Greenberger, M. A. Horne, \& A. Zeilinger,
``Nonlocality of a single photon?'',
{\em Phys. Rev. Lett.} {\bf 75}, 10, 2064 (1995).
Comment on {\bf [Hardy 94]}.
Reply: {\bf [Hardy 95 b]}.

\item {\bf [Greenberger 95]}:
D. M. Greenberger,
``Two-particle versus three-particle EPR experiments'',
in D. M. Greenberger, \& A. Zeilinger (eds.),
{\em Fundamental problems in quantum theory:
A conference held in honor of professor John
A. Wheeler, Ann. N. Y. Acad. Sci.} {\bf 755}, 585-599 (1995).

\item {\bf [Greenberger-Zeilinger 95]}:
D. M. Greenberger, \& A. Zeilinger,
``Quantum theory: Still crazy after all these years'',
{\em Phys. World} {\bf 8}, 9, 33-38 (1995).

\item {\bf [Greenberger 96]}:
D. M. Greenberger,
``Entangled states in Fock space'',
in F. De Martini, G. Denardo, \& Y. Shih (eds.),
{\em Proc. of a Workshop Coherence in Quantum Optics (Trieste, 1996)}
VCH, Weinheim, 1996, pp.~?-?.

\item {\bf [Greenberger 97]}:
D. M. Greenberger,
``A more proper role for proper time in physics?'',
in {\bf [Cohen-Horne-Stachel 97 a]}.

\item {\bf [Greenberger 98 a]}:
D. M. Greenberger,
``If one could build a macroscopic Schr\"{o}dinger cat state, one could communicate superluminally'',
in E. B. Karlsson, \& E. Br\"{a}ndas (eds.),
{\em Proc.\ of the 104th Nobel Symp.\ ``Modern Studies of Basic Quantum Concepts and Phenomena'' (Gimo, Sweden, 1997)},
{\em Physica Scripta} {\bf T76}, 57-60 (1998).

\item {\bf [Greenberger 98 b]}:
D. M. Greenberger,
``Interpreting the quantum world by Jeffrey Bub'',
{\em Am. J. Phys.} {\bf 66}, 11, 1031 (1998).
Review of {\bf [Bub 97]}.

\item {\bf [Greenberger 99]}:
D. M. Greenberger,
``Quantum mechanics'',
{\em Phys. Today} {\bf 52}, 5, 64-66 (1999).
Review of the 3rd edition (1998) of {\bf [Merzbacher 61]}.

\item {\bf [Greenberger-Reiter-Zeilinger 99]}:
D. M. Greenberger, W. L. Reiter, \& A. Zeilinger (eds.),
{\em Epistemological and experimental perspectives on quantum physics},
Kluwer Academic, Dordrecht, Holland, 1999.

\item {\bf [Greenberger-Horne-Zeilinger 00]}:
D. M. Greenberger, M. A. Horne, \& A. Zeilinger,
``Similarities and differences between two-particle and
three-particle interference'',
{\em Fortschr. Phys.} {\bf 48}, 4, 243-252 (2000).

\item {\bf [Greenberger-Hillary 00]}:
D. M. Greenberger, \& M. Hillary,
``The impossibility of keyless communication in quantum
cryptography'',
{\em Fortschr. Phys.} {\bf 48}, 5-7, 523-530 (2000).

\item {\bf [Greenberger 00 a]}:
D. M. Greenberger,
``An uncertain principal'',
{\em Nature} {\bf 408}, 6813, 644-645 (2000).
Review of {\bf [Ellis-Amati 00]}.

\item {\bf [Greenberger 00 b]}:
D. M. Greenberger,
``Physics: Bohr the innovator? Or Bohr the intimidator?'',
{\em Science} {\bf 287}, 5461, 2166 (2000).
Review of {\bf [Beller 99]}.

\item {\bf [Greenberger 01 a]}:
D. M. Greenberger,
``Quantum philosophy.
Understanding and interpreting contemporary science'',
{\em Am. J. Phys.} {\bf 69}, 1, 94 (2001).
Review of {\bf [Omn\`{e}s 94 b]}.

\item {\bf [Greenberger 01 b]}:
D. M. Greenberger,
``Inadequacy of the usual Galilean transformation in quantum mechanics'',
{\em Phys. Rev. Lett.} {\bf 87}, 10, 100405 (2001).

\item {\bf [Greenberger 01 c]}:
D. M. Greenberger,
``Book review. Conceptual foundations of quantum physics:
An overview from modern perspectives'',
{\em Found. Phys.} {\bf 31}, 5, 855-857 (2001).
Review of {\bf [Home 97]}.

\item {\bf [Greenberger 02]}:
D. M. Greenberger,
``The history of the GHZ paper'',
in {\bf [Bertlmann-Zeilinger 02]}, pp.~281-286.

\item {\bf [Greenberger-Shimony 03]}:
D. M. Greenberger, \& A. Shimony,
``The presence of David Mermin'',
{\em Found. Phys.} {\bf 33}, 10, 1419-1422 (2003).

\item {\bf [Greenberger 03]}:
D. M. Greenberger,
``Quantum mechanics: Symbolism of atomic measurements'',
{\em Am. J. Phys.} {\bf 71}, 9, 598 (2003).
Review of {\bf [Schwinger 01]}.

\item {\bf [Greenstein-Zajonc 98]}:
G. Greenstein, \& A. G. Zajonc,
{\em The quantum challenge: Modern research on the foundations
of quantum mechanics}, Jones and Bartlett, Sudbury, Massachusetts, 1998.
Review: {\bf [Grangier 98 a]}, {\bf [Peres 98 e]}.

\item {\bf [Greentree-Schirmer-Green-(+3) 04]}:
A. D. Greentree, S. G. Schirmer, F. Green,
L. C. L. Hollenberg, A. R. Hamilton, \& R. G. Clark,
``Maximizing the Hilbert space for a finite number of distinguishable quantum states'',
{\em Phys. Rev. Lett.} {\bf 92}, 9, 097901 (2004).

\item {\bf [Greentree-Hamilton-Green 04]}:
A. D. Greentree, A. R. Hamilton, \& F. Green,
``Charge shelving and bias spectroscopy for the readout
of a charge qubit on the basis of superposition states'',
{\em Phys. Rev. B} {\bf 70}, 4, 041305 (2004).

\item {\bf [Gregoratti-Werner 03]}:
M. Gregoratti, \& R. F. Werner,
``Quantum lost and found'',
in M. Ferrero (ed.),
{\em Proc. of Quantum Information: Conceptual Foundations,
Developments and Perspectives (Oviedo, Spain, 2002)},
{\em J. Mod. Opt.} {\bf 50}, 6-7, 915-933 (2003).

\item {\bf [Gregoratti-Werner 04]}:
M. Gregoratti, \& R. F. Werner,
``On quantum error-correction by classical feedback in discrete time'',
{\em J. Math. Phys.} {\bf 45}, ?, 2600-2612 (2004);
quant-ph/0403092.

\item {\bf [Greiner 89]}:
W. Greiner,
{\em Quantenmechanik Teil 1. Eine Einf\"{u}hrung},
Verlag Harri Deutsch, Thun, 1989.
English version:
{\em Quantum mechanics. An introduction},
Springer-Verlag, Berlin, 1989 (1st edition), 1993 (2nd edition).

\item {\bf [Gribbin 95]}:
J. Gribbin,
{\em Schr\"{o}dinger's kittens and the search for reality:
Solving the quantum mysteries},
Weidenfeld \& Nicolson, London, 1995.
Review: {\bf [Sudbery 95]}, {\bf [Knight 96]}.

\item {\bf [Grice-U'Ren-Walmsley 01]}:
W. P. Grice, A. B. U'Ren, \& I. A. Walmsley,
``Eliminating frequency and space-time correlations in multiphoton states'',
{\em Phys. Rev. A} {\bf 64}, 6, 063815 (2001).

\item {\bf [Griessner-Jaksch-Zoller 03]}:
A. Griessner, D. Jaksch, \& P. Zoller,
``Cavity assisted nondestructive laser cooling of atomic qubits'',
quant-ph/0311054.

\item {\bf [Griffiths 91]}:
D. J. Griffiths,
``Quantum mechanics. Leslie E. Ballentine'',
{\em Am. J. Phys.} {\bf 59}, 12, 1153-1154 (1991).
Review of {\bf [Ballentine 90 a]}.

\item {\bf [Griffiths 95]}:
D. J. Griffiths,
{\em Introduction to quantum mechanics},
Prentice-Hall, New York, 1995.
Review: {\bf [Taylor 01]}.

\item {\bf [Griffiths 84]}:
R. B. Griffiths,
``Consistent histories and the interpretation of quantum mechanics'',
{\em J. Stat. Phys.} {\bf 36}, 1-2, 219-272 (1984).

\item {\bf [Griffiths 86 a]}:
R. B. Griffiths,
``Correlations in separated quantum systems:
A consistent history analysis of the EPR problem'',
{\em Am. J. Phys.} {\bf 55}, 1, 11-17 (1986).

\item {\bf [Griffiths 86 b]}:
R. B. Griffiths,
``Quantum interpretation using consistent histories'',
in L. M. Roth, \& A. Inomata (eds.),
{\em Fundamental questions in quantum mechanics
(Albany, New York, 1984)},
Gordon \& Breach, New York, 1986, pp.~211-223.

\item {\bf [Griffiths 86 c]}:
R. B. Griffiths,
``Making consistent inferences from quantum measurements'',
in D. M. Greenberger (ed.),
{\em New techniques and ideas in quantum measurement theory.
Proc.\ of an international conference (New York, 1986),
Ann. N. Y. Acad. Sci.} {\bf 480}, 512-517 (1986).

\item {\bf [Griffiths 87]}:
R. B. Griffiths,
``Correlations in separated quantum
systems: A consistent history analysis of the EPR problem'',
{\em Am. J. Phys.} {\bf 55}, 1, 11-17 (1987).

\item {\bf [Griffiths 93 a]}:
R. B. Griffiths,
``Consistent interpretation of quantum
mechanics using quantum trajectories'',
{\em Phys. Rev. Lett.} {\bf 70}, 15, 2201-2204 (1993).

\item {\bf [Griffiths 93 b]}:
R. B. Griffiths,
``The consistency of consistent histories:
A reply to d'Espagnat'',
{\em Found. Phys.} {\bf 23}, 12, 1601-1610 (1993).

\item {\bf [Griffiths 95]}:
R. B. Griffiths,
``Book review. The interpretation of
quantum mechanics'',
{\em Found. Phys.} {\bf 25}, 8, 1231-1236 (1995).
Review of {\bf [Omn\`{e}s 94 a]}.

\item {\bf [Griffiths 96]}:
R. B. Griffiths,
``Consistent histories and quantum reasoning'',
{\em Phys. Rev. A} {\bf 54}, 4, 2759-2774 (1996).
Comment: {\bf [Schafir 98 b]}.
Reply: {\bf [Griffiths 99 a]}.

\item {\bf [Griffiths-Niu 96]}:
R. B. Griffiths, \& C. Niu,
``Semiclassical Fourier transform for quantum computation'',
{\em Phys. Rev. Lett.} {\bf 76}, 17, 3228-3231 (1996).

\item {\bf [Griffiths-Niu 97]}:
R. B. Griffiths, \& C. Niu,
``Optimal eavesdropping in quantum cryptography. II.
A quantum circuit'',
{\em Phys. Rev. A} {\bf 56}, 2, 1173-1176 (1997).
See {\bf [Fuchs-Gisin-Griffiths-(+2) 97]} (I).

\item {\bf [Griffiths-Hartle 98]}:
R. B. Griffiths, \& J. B. Hartle,
``Comment on `Consistent sets yield contrary inferences in quantum theory'\,'',
{\em Phys. Rev. Lett.} {\bf 81}, 9, 1981 (1998);
quant-ph/9710025.
Comment on {\bf [Kent 97 a]}.
Reply: {\bf [Kent 98 b]}.

\item {\bf [Griffiths 98 a]}:
R. B. Griffiths,
``Choice of consistent family, and quantum incompatibility'',
{\em Phys. Rev. A} {\bf 57}, 3, 1604-1618 (1998);
quant-ph/9708028.

\item {\bf [Griffiths 98 b]}:
R. B. Griffiths,
``Reply to `Comment on ``Consistent histories and quantum reasoning''\,'\,'',
{\em Phys. Rev. A} {\bf 58}, 4, 3356-3357 (1998).
Reply to {\bf [Schafir 98 b]}.
See {\bf [Griffiths 96]}.

\item {\bf [Griffiths 98 c]}:
R. B. Griffiths,
``Consistent histories and quantum delayed choice'',
{\em Fortschr. Phys.} {\bf 46}, 6-8, 741-748 (1998);
quant-ph/9810016.

\item {\bf [Griffiths 99 a]}:
R. B. Griffiths,
``Consistent quantum counterfactuals'',
{\em Phys. Rev. A} {\bf 60}, 1, R5-R8 (1999);
quant-ph/9805056.
See {\bf [Schafir 98 a]}.

\item {\bf [Griffiths 99 b]}:
R. B. Griffiths,
``Bohmian mechanics and consistent histories'',
{\em Phys. Lett. A} {\bf 261}, 5-6, 227-234 (1999).

\item {\bf [Griffiths-Omn\`{e}s 99]}:
R. B. Griffiths, \& R. Omn\`{e}s,
``Consistent histories and quantum measurements'',
{\em Phys. Today} {\bf 52}, 8, Part 1, 26-31 (1999).

\item {\bf [Griffiths 00 a]}:
R. B. Griffiths,
``Consistent histories, quantum truth functionals, and hidden
variables'',
{\em Phys. Lett. A} {\bf 265}, 1-2, 12-19 (2000);
quant-ph/9909049.

\item {\bf [Griffiths 00 b]}:
R. B. Griffiths,
``Consistent quantum realism: A reply to Bassi and Ghirardi'',
quant-ph/0001093.
See {\bf [Bassi-Ghirardi 99, 00 c]}, {\bf [Griffiths 00 a]}.

\item {\bf [Griffiths 01]}:
R. B. Griffiths,
{\em Consistent quantum theory},
Cambridge University Press, Cambridge, 2001.
Review: {\bf [Dieks 03]}.

\item {\bf [Griffiths 02]}:
R. B. Griffiths,
``Nature and location of quantum information'',
{\em Phys. Rev. A} {\bf 66}, 1, 012311 (2002);
quant-ph/0203058.

\item {\bf [Griffiths 03]}:
R. B. Griffiths,
``Probabilities and quantum reality: Are there correlata?'',
{\em Found. Phys.} {\bf 33}, 10, 1423-1459 (2003).

\item {\bf [Griffiths 04]}:
R. B. Griffiths,
``Channel kets, entangled states, and the location of quantum information'',
quant-ph/0409106.

\item {\bf [Grigorescu-Baylis 02]}:
M. Grigorescu, \& W. E. Baylis,
``Quasiclassical and statistical properties of fermion systems'',
{\em Phys. Rev. B} {\bf 66}, 1, 014530 (2002).

\item {\bf [Grimmett-Janson-Scudo 03]}:
G. Grimmett, S. Janson, \& P. Scudo,
``Weak limits for quantum random walks'',
quant-ph/0309135.

\item {\bf [Grishanin-Zadkov 98]}:
B. A. Grishanin, \& V. N. Zadkov,
``Natural capacity of a system of two two-level atoms
as a quantum information channel'',
{\em Laser Phys.} {\bf 8}, ?, 1074-1080 (1998);
quant-ph/9906069

\item {\bf [Grishanin-Zadkov 00]}:
B. A. Grishanin, \& V. N. Zadkov,
``Coherent-information analysis of quantum
channels in simple quantum systems'',
{\em Phys. Rev. A} {\bf 62}, 3, 032303 (2000);
quant-ph/9912113.

\item {\bf [Grishanin-Zadkov 01]}:
B. A. Grishanin, \& V. N. Zadkov,
``Coherent and compatible information: A basis to information analysis of quantum systems'',
{\em Proc. 17th Int.\ Conf.\ on Coherent and Nonlinear Optics (Minsk, 2001)};
quant-ph/0108035.

\item {\bf [Grishanin-Zadkov 03]}:
B. A. Grishanin, \& V. N. Zadkov,
``Entangling quantum measurements and their properties'',
{\em Phys. Rev. A} {\bf 68}, 2, 022309 (2003);
quant-ph/0306120.

\item {\bf [Grishanin-Zadkov 04]}:
B. A. Grishanin \& V. N. Zadkov,
``Entangling quantum measurements'',
{\em Opt. and Spectroscopy} {\bf 96}, 683-690 (2004).

\item {\bf [Groessing 02]}:
G. Groessing,
``Quantum cybernetics: A new perspective for Nelson's stochastic theory,
nonlocality, and the Klein-Gordon equation'',
{\em Phys. Lett. A};
quant-ph/0201035.

\item {\bf [Groisman-Vaidman 01]}:
B. Groisman, \& L. Vaidman,
``Nonlocal variables with product-state eigenstates'',
in S. Popescu, N. Linden, \& R. Jozsa (eds.),
{\em J. Phys. A} {\bf 34}, 35
(Special issue: Quantum information and computation), 6881-6889 (2001);
quant-ph/0103084.

\item {\bf [Groisman-Reznik 02]}:
B. Groisman, \& B. Reznik,
``Measurements of semilocal and nonmaximally entangled states'',
{\em Phys. Rev. A} {\bf 66}, 2, 022110 (2002).
Erratum: {\em Phys. Rev. A} {\bf 66}, 4, 049901 (2002);
quant-ph/0111012.

\item {\bf [Groisman-Reznik-Vaidman 01]}:
B. Groisman, B. Reznik, \& L. Vaidman,
``Instantaneous measurements of nonlocal variables'',
in M. Ferrero (ed.),
{\em Proc. of Quantum Information: Conceptual Foundations,
Developments and Perspectives (Oviedo, Spain, 2002)},
{\em J. Mod. Opt.} {\bf 50}, 6-7, 943-949 (2003).

\item {\bf [Groisman-Popescu-Winter 04]}:
B. Groisman, S. Popescu, \& A. Winter,
``On the quantum, classical and total amount of correlations in a quantum
state'',
quant-ph/0410091.

\item {\bf [Groisman-Reznik 04]}:
B. Groisman, \& B. Reznik,
``On the efficiency of nonlocal gates generation'',
quant-ph/0410170.

\item {\bf [Grometstein 99]}:
A. A. Grometstein,
{\em The roots of things. Topics in quantum mechanics},
Kluwer Academic/Plenum Publishers, New York, 1999.

\item {\bf [Grondalski-Etlinger-James 02]}:
J. Grondalski, D. M. Etlinger, \& D. F. V. James,
``The fully entangled fraction as an inclusive measure of entanglement applications'',
{\em Phys. Lett. A} {\bf 300}, 6, 573-580 (2002).

\item {\bf [Grosshans-Grangier 00]}:
F. Grosshans, \& P. Grangier,
``Quantum cloning and teleportation criteria for continuous
quantum variables'',
{\em Phys. Rev. A} {\bf 62}, 1, 010301(R) (2001);
quant-ph/0012121.

\item {\bf [Grosshans-Grangier 02 a]}:
F. Grosshans, \& P. Grangier,
``Continuous variable quantum cryptography using coherent states'',
{\em Phys. Rev. Lett.} {\bf 88}, 5, 057902 (2002);
quant-ph/0109084.

\item {\bf [Grosshans-Grangier 02 b]}:
F. Grosshans, \& P. Grangier,
``Reverse reconciliation protocols for quantum cryptography with continuous variables'',
quant-ph/0204127.

\item {\bf [Grosshans-Van Assche-Wenger-(+3) 03]}:
F. Grosshans, G. Van Assche, J. Wenger, R. Brouri, N.
J. Cerf, \& P. Grangier,
``Quantum key distribution using Gaussian-modulated coherent states'',
{\em Nature} {\bf 421}, 6920, 238-241 (2003);
quant-ph/0312016.

\item {\bf [Grosshans-Cerf-Wenger-(+2) 03]}:
F. Grosshans, N. J. Cerf, J. Wenger,
R. Tualle-Brouri, \& P. Grangier,
``Virtual entanglement and reconciliation protocols for quantum
cryptography with continuous variables'',
quant-ph/0306141.

\item {\bf [Grosshans-Cerf 04]}:
F. Grosshans, \& N. J. Cerf,
``Continuous-variable quantum cryptography is secure against non-Gaussian attacks'',
{\em Phys. Rev. Lett.} {\bf 92}, 4, 047905 (2004);
quant-ph/0311006.

\item {\bf [Grossman 96]}:
J. Grossman,
``Realizing generalized quantum measurements on the polarization
of photons'',
Williams College senior thesis, 1996.
See appendix A of {\bf [Brandt 99 b]}.

\item {\bf [Grover 96 a]}:
L. K. Grover,
``A fast quantum mechanical algorithm for estimating the median'',
quant-ph/9607024.

\item {\bf [Grover 96 b]}:
L. K. Grover,
``A fast quantum mechanical algorithm for database search'',
{\em Proc.\ of the 28th Annual ACM Symp.\ on Theory of Computing
(Philadelphia, Pennsylvania)}, ACM Press, New York, 1996, pp.~212-218.

\item {\bf [Grover 97 a]}:
L. K. Grover,
``Quantum telecomputation'',
quant-ph/9704012.

\item {\bf [Grover 97 b]}:
L. K. Grover,
``Quantum mechanics helps in searching for a needle in a haystack'',
{\em Phys. Rev. Lett.} {\bf 79}, 2, 325-328 (1997).
See {\bf [Boyer-Brassard-H\o{}yer-Tapp 96]}, {\bf [Collins 97]}.
Improved version: {\bf [Grover 97 c]}.

\item {\bf [Grover 97 c]}:
L. K. Grover,
``Quantum computers can search arbitrarily
large databases by a single query'',
{\em Phys. Rev. A} {\bf 79}, 23, 4709-4712 (1997);
quant-ph/9706005.
Improved version of {\bf [Grover 97 b]}.

\item {\bf [Grover 97 d]}:
L. K. Grover,
``Quantum computers can search rapidly by
using almost any transformation'',
quant-ph/9712011.

\item {\bf [Grover 98 a]}:
L. K. Grover,
``Quantum computing:
The advantages of superposition'',
{\em Science} {\bf 280}, 5361, 228 (1998).

\item {\bf [Grover 98 b]}:
L. K. Grover,
``Quantum computing:
Beyond factorization and search'',
{\em Science} {\bf 281}, 5378, 792-794 (1998).

\item {\bf [Grover 98 c]}:
L. K. Grover,
``Quantum search on structured problems'',
in C. P. Williams (ed.),
{\em 1st NASA Int.\ Conf.\ on Quantum Computing and Quantum Communications
(Palm Springs, California, 1998)},
{\em Lecture Notes in Computer Science} {\bf 1509},
Springer-Verlag, New York, 1999, pp.~?-?;
quant-ph/9802035.

\item {\bf [Grover 98 d]}:
L. K. Grover,
``How fast can a quantum computer search?'',
quant-ph/9809029.

\item {\bf [Grover 00 a]}:
L. K. Grover,
``Synthesis of quantum superpositions by quantum computation'',
{\em Phys. Rev. Lett.} {\bf 85}, 6, 1334-1337 (2000).

\item {\bf [Grover 00 b]}:
L. K. Grover,
``Rapid sampling through quantum computing'',
in
{\em Proc.\ of the 32nd Annual ACM
Symp.\ on Theory of Computing (2000)};
quant-ph/9912001.

\item {\bf [Grover 00 c]}:
L. K. Grover,
``Searching with quantum computers'',
quant-ph/0011118.

\item {\bf [Grover 01]}:
L. K. Grover,
``From Schr\"{o}dinger's equation to the quantum search algorithm'',
{\em Am. J. Phys.} {\bf 69}, 7, 769-777 (2001);
quant-ph/0109116.

\item {\bf [Grover-Sengupta 01]}:
L. K. Grover, \& A. Sengupta,
``From coupled pendulums to quantum search'',
condensed version of a chapter in {\em The mathematics of quantum computation}
CRC Press, ?, 2001;
quant-ph/0109123.

\item {\bf [Grover-Sengupta 02]}:
L. K. Grover, \& A. M. Sengupta,
``Classical analog of quantum search'',
{\em Phys. Rev. A} {\bf 65}, 3, 032319 (2002).

\item {\bf [Grover 02 a]}:
L. K. Grover,
``Quantum computation and quantum information,
{\em Am. J. Phys.} {\bf 70}, 5, 558-559 (2002).
Review of {\bf [Nielsen-Chuang 00]}.

\item {\bf [Grover 02 b]}:
L. K. Grover,
``Trade-offs in the quantum search algorithm'',
{\em Phys. Rev. A} {\bf 66}, 5, 052314 (2002);
quant-ph/0201152.

\item {\bf [Grover 02 c]}:
L. K. Grover,
``An improved quantum scheduling algorithm'',
quant-ph/0202033.

\item {\bf [Grover-Rudolph 03]}:
L. K. Grover, \& T. Rudolph,
``How significant are the known collision and element distinctness quantum
algorithms?'',
quant-ph/0309123.

\item {\bf [Grover-Radhakrishnan 04 a]}:
L. K. Grover, \& J. Radhakrishnan,
``Is partial quantum search of a database any easier?'',
quant-ph/0407122.

\item {\bf [Grover-Radhakrishnan 04 b]}:
L. K. Grover, \& J. Radhakrishnan,
``Quantum search for multiple items using parallel queries'',
quant-ph/0407217.

\item {\bf [Gr\"{o}ssing 00]}:
G. Gr\"{o}ssing,
{\em Quantum cybernetics. Toward a unification of relativity
and quantum theory via circularly causal modeling},
Springer-Verlag, New York, 2000.

\item {\bf [Gr\"{o}ssing 02]}:
G. Gr\"{o}ssing,
``Quantum cybernetics: A new perspective for Nelson's stochastic
theory, nonlocality, and the Klein-Gordon equation'',
{\em Phys. Lett. A} {\bf 296}, 1, 1-8 (2002).

\item {\bf [Gr\"{u}bl-Moser-Rheinberger 01]}:
G. Gr\"{u}bl, R. Moser, \& K. Rheinberger,
``Bohmian trajectories and Klein's paradox'',
{\em J. Phys. A} {\bf 34}, 13, 2753-2764 (2001);
quant-ph/0202098.

\item {\bf [Gr\"{u}bl-Rheinberger 02]}:
G. Gr\"{u}bl, \& K. Rheinberger,
``Time of arrival from Bohmian flow'',
{\em J. Phys. A};
quant-ph/0202084.

\item {\bf [Gr\"{u}bl 03]}:
G. Gr\"{u}bl,
``The quantum measurement problem enhanced'',
{\em Phys. Lett. A} {\bf 316}, 3-4, 153-158 (2003);
quant-ph/0202101.

\item {\bf [Grudka-W\'{o}jcik 02 a]}:
A. Grudka, \& A. W\'{o}jcik,
``Symmetric scheme for superdense coding between multiparties'',
{\em Phys. Rev. A} {\bf 66}, 1, 014301 (2002);
quant-ph/0202159.

\item {\bf [Grudka-W\'{o}jcik 02 b]}:
A. Grudka, \& A. W\'{o}jcik,
``Projective measurement of the two-photon polarization state:
Linear optics approach'',
{\em Phys. Rev. A} {\bf 66}, 6, 064303 (2002).

\item {\bf [Grudka-W\'{o}jcik 03]}:
A. Grudka, \& A. W\'{o}jcik,
``How to encode the states of two non-entangled qubits in one qutrit'',
{\em Phys. Lett. A} {\bf 314}, 5-6, 350-353 (2003);
quant-ph/0303168.

\item {\bf [Grudka 03]}:
A. Grudka,
``Quantum teleportation between multiparties'',
quant-ph/0303112.

\item {\bf [Grunhaus-Popescu-Rohrlich 96]}:
J. Grunhaus, S. Popescu, \& D. Rohrlich,
``Jamming non-local quantum correlations'',
{\em Phys. Rev. A} {\bf 53}, 6, 3781-3784 (1996);
quant-ph/9508001.

\item {\bf [Gruska 99]}:
J. Gruska,
{\em Quantum computing},
McGraw-Hill, New York, 1999.

\item {\bf [Gruska-Imai 01]}:
J. Gruska, \& H. Imai,
``Power, puzzles and properties of entanglement'',
in M. Margenstern, \& Y. Rogozhin (eds.),
{\em Machines, computations, and universality (Chi\c{s}n\v{a}u, 2001)},
Springer-Verlag, Berlin, 2001, pp.~25-68.

\item {\bf [Gu-Lin-Li 03]}:
S.-J. Gu, H.-Q. Lin, \& Y.-Q. Li,
``Entanglement, quantum phase transition, and scaling in the $XXZ$ chain'',
{\em Phys. Rev. A} {\bf 68}, 4, 042330 (2003);
quant-ph/0307131.

\item {\bf [Gu-Zheng-Guo 02]}:
Y.-J. Gu, Y.-Z. Zheng, \& G.-C. Guo,
``Conclusive teleportation and entanglement concentration'',
{\em Phys. Lett. A} {\bf 296}, 4-5, 157-160 (2002).

\item {\bf [Guay-Marchildon 03]}:
E. Guay, \& L. Marchildon,
``Two-particle interference in standard and Bohmian quantum mechanics'',
{\em J. Phys. A} {\bf 36}, 20, 5617-5624 (2003).

\item {\bf [Gudder 68]}:
S. P. Gudder,
``Hidden variables in quantum mechanics reconsidered'',
{\em Rev. Mod. Phys.} {\bf 40}, 1, 229-231 (1968).

\item {\bf [Gudder 70]}:
S. P. Gudder,
``On hidden-variable theories'',
{\em J. Math. Phys.} {\bf 11}, 2, 431-436 (1970).

\item {\bf [Gudder 80]}:
S. P. Gudder,
``Proposed test for a hidden variables theory'',
{\em Int. J. Theor. Phys.} {\bf 19}, 2, 163-168 (1980).

\item {\bf [Gudder 88]}:
S. P. Gudder,
{\em Quantum probability},
Academic Press, Boston, Massachusetts, 1988.

\item {\bf [Gudder 93]}:
S. P. Gudder,
``EPR, Bell and quantum probability'',
in A. van der Merwe, \& F. Selleri (eds.),
{\em Bell's theorem and the foundations of modern physics.
Proc.\ of an international conference (Cesena, Italy, 1991)},
World Scientific, Singapore, 1993, pp.~263-275.

\item {\bf [Gudder 98]}:
S. P. Gudder,
``Book review. Quantum logic in algebraic approach'',
{\em Found. Phys.} {\bf 28}, 11, 1729-1730 (1998).
Review of {\bf [R\'{e}dei 98]}.

\item {\bf [Gudder 99 a]}:
S. P. Gudder,
``Observables and statistical maps'',
{\em Found. Phys.} {\bf 29}, 6, 877-898 (1999).

\item {\bf [Gudder 99 b]}:
S. P. Gudder,
``Quantum automata: An overview'',
{\em Int. J. Theor. Phys.} {\bf 38}, 9, 2261-2282 (1999).

\item {\bf [Gudder 99 c]}:
S. P. Gudder,
``Book review. Quantum computing and quantum communications'',
{\em Found. Phys.} {\bf 29}, 10, 1639-1642 (1999).
Review of {\bf [Williams 99 a]}.

\item {\bf [Gudder 00 a]}:
S. P. Gudder,
``Basic properties of quantum automata'',
{\em Found. Phys.} {\bf 30}, 2, 301-319 (2000).

\item {\bf [Gudder 00 b]}:
S. P. Gudder,
``Book review. Ultimate zero and one: Computing at the quantum frontier'',
{\em Found. Phys.} {\bf 30}, 4, 607-609 (2000).
Review of {\bf [Williams 99 b]}.

\item {\bf [Gudder 01 a]}:
S. P. Gudder,
``Book review. Quantum computation and quantum information'',
{\em Found. Phys.} {\bf 31}, 11, 1665-1667 (2001).
Review of {\bf [Nielsen-Chuang 00]}.

\item {\bf [Gudder 01 b]}:
S. P. Gudder,
``Book review. Einstein and religion'',
{\em Found. Phys. Lett.} {\bf 14}, 2, 195-198 (2001).
Review of {\bf [Jammer 99]}.

\item {\bf [Gudder 03]}:
S. P. Gudder,
``Quantum computation'',
{\em Amer. Math. Monthly} {\bf 110}, 3, 181-201 (2003).

\item {\bf [Gu\'{e}ret-Vigier 84]}:
P. Gu\'{e}ret, \& J.-P. Vigier,
``de Broglie's wave-particle duality in the stochastic interpretation
of quantum mechanics: A testable physical assumption'',
in {\bf [Barut-van der Merwe-Vigier 84]}, pp.~129-155.

\item {\bf [Guerin-Unanyan-Yatsenko-Jauslin 02]}:
S. Guerin, R. G. Unanyan, L. P. Yatsenko, \& H. R. Jauslin,
``Adiabatic creation of entangled states by a bichromatic field designed from
the topology of the dressed eigenenergies'',
{\em Phys. Rev. A} {\bf 66}, 3, 032311 (2002).

\item {\bf [Guerra-Retamal 99]}:
E. S. Guerra, \& J. C. Retamal,
``Realization of atomic Greenberger-Horne-Zeilinger states
via cavity quantum electrodynamics'',
{\em J. Mod. Opt.} {\bf 46}, 2, 295-302 (1999).

\item {\bf [Guerra 04 a]}:
E. S. Guerra,
``On the complementarity principle and the uncertainty principle'',
{\em Opt. Comm.} {\bf 234}, ?, 295-? (2004);
quant-ph/0409172.

\item {\bf [Guerra 04 b]}:
E. S. Guerra,
``Teleportation of atomic states via cavity quantum electrodynamics
{\em Opt. Comm.};
quant-ph/0409194.

\item {\bf [Guerra 04 c]}:
E. S. Guerra,
``Teleportation of atomic states for atoms in a lambda configuration'',
quant-ph/0409195.

\item {\bf [Guerra 04 d]}:
E. S. Guerra,
``Realization of GHZ states and the GHZ test via cavity QED'',
quant-ph/0409196.

\item {\bf [Guerra 04 e]}:
E. S. Guerra,
``Realization of GHZ States and the GHZ test via cavity QED for a cavity
prepared in a superposition of zero and one Fock states'',
quant-ph/0410234.

\item {\bf [Guerra 04 f]}:
E. S. Guerra,
``Teleportation of atomic states via cavity QED for a cavity prepared in a
superposition of zero and one Fock states'',
quant-ph/0410239.

\item {\bf [Guerra 04 g]}:
E. S. Guerra,
``Teleportation of cavity field states via cavity QED'',
quant-ph/0410240.

\item {\bf [G\"{u}hne-Hyllus-Bru\ss-(+4) 02]}:
O. G\"{u}hne, P. Hyllus, D. Bru\ss,
A. K. Ekert, M. Lewenstein, C. Macchiavello, \& A. Sanpera,
``Detection of entanglement with few local measurements'',
{\em Phys. Rev. A} {\bf 66}, 6, 062305 (2002).

\item {\bf [G\"{u}hne-Hyllus-Bru\ss-(+4) 03]}:
O. G\"{u}hne, P. Hyllus, D. Bru\ss,
A. K. Ekert, M. Lewenstein, C. Macchiavello, \& A. Sanpera,
``Experimental detection of entanglement via witness operators and local measurements'',
in M. Ferrero (ed.),
{\em Proc. of Quantum Information: Conceptual Foundations,
Developments and Perspectives (Oviedo, Spain, 2002)},
{\em J. Mod. Opt.} {\bf 50}, 6-7, 1079-1102 (2003);
quant-ph/0210134.

\item {\bf [G\"{u}hne 04]}:
O. G\"{u}hne,
``Characterizing entanglement via uncertainty relations'',
{\em Phys. Rev. Lett.} {\bf 92}, 11, 117903 (2004);
quant-ph/0306194.

\item {\bf [G\"{u}hne-Lewenstein 04 a]}:
O. G\"{u}hne, \& M. Lewenstein,
``Entropic uncertainty relations and entanglement'',
{\em Phys. Rev. A} {\bf 70}, 2, 022316 (2004);
quant-ph/0403219.

\item {\bf [G\"{u}hne-Lewenstein 04 b]}:
O. G\"{u}hne, \& M. Lewenstein,
``Separability criteria from uncertainty relations'',
{\em Proc.\ of QCMC 2004 (Glasgow)};
quant-ph/0409140.

\item {\bf [G\"{u}hne-Toth-Hyllus-Briegel 04]}:
O. G\"{u}hne, G. Toth, P. Hyllus, \& H. J. Briegel,
``Bell inequalities for graph states'',
quant-ph/0410059.

\item {\bf [Guinea-Mart\'{\i}n Delgado 03]}:
F. Guinea, \& M. A. Mart\'{\i}n-Delgado,
``Quantum chinos game: Winning strategies through quantum fluctuations'',
{\em J. Phys. A} {\bf 36}, 13, L197-L204 (2003);
quant-ph/0201140.

\item {\bf [Gulde-Riebe-Lancaster-(+6) 03]}:
S. Gulde, M. Riebe, G. P. T. Lancaster,
C. Becher, J. Eschner, H. H\"{a}ffner,
F. Schmidt-Kaler, I. L. Chuang, \& R. Blatt,
``Implementation of the Deutsch-Jozsa algorithm on an ion-trap quantum computer'',
{\em Nature} {\bf 421}, 6918, 48-50 (2003).
See {\bf [Jones 03 a]}.

\item {\bf [Gulley-White-James 01]}:
M. S. Gulley, A. G. White, \& D. F. V. James,
``A Raman approach to quantum logic in calcium-like ions'',
submitted to {\em J. Quantum Information and Computation};
quant-ph/0112117.

\item {\bf [Gunlycke-Kendon-Vedral-Bose 01]}:
D. Gunlycke, V. M. Kendon, V. Vedral, \& S. Bose,
``Thermal concurrence mixing in a one-dimensional Ising model'',
{\em Phys. Rev. A} {\bf 64}, 4, 042302 (2001).

\item {\bf [Guo-Shi 99]}:
G.-C. Guo, \& B.-S. Shi,
``Quantum cryptography based on interaction-free measurement'',
{\em Phys. Lett. A} {\bf 256}, 2-3, 109-112 (1999).

\item {\bf [Guo-Li-Shi-(+2) 01]}:
G.-P. Guo, C.-F. Li, B.-S. Shi, J. Li, \& G.-C. Guo,
``Quantum key distribution scheme with orthogonal product states'',
{\em Phys. Rev. A} {\bf 64}, 4, 042301 (2001);
quant-ph/0102060.

\item {\bf [Guo-Li-Guo 01]}:
G.-P. Guo, C.-F. Li, \& G.-C. Guo,
``Quantum non-demolition measurement of nonlocal variables
and its application in quantum authentication'',
{\em Phys. Lett. A} {\bf 286}, 6, 401-404 (2001);
quant-ph/0103088.

\item {\bf [Guo-Li-Li-Guo 02]}:
G.-P. Guo, C.-F. Li, J. Li, \& G.-C. Guo,
``Scheme for the preparation of multiparticle entanglement in cavity QED'',
{\em Phys. Rev. A} {\bf 65}, 4, 042102 (2002);
quant-ph/0105123.

\item {\bf [Guo-Guo 03 a]}:
G.-P. Guo, \& G.-C. Guo,
``Quantum secret sharing without entanglement'',
{\em Phys. Lett. A} {\bf 310}, 4, 247-251 (2001);

\item {\bf [Guo-Zhang 02]}:
G.-C. Guo, \& Y.-S. Zhang,
``Scheme for preparation of the W state via cavity quantum electrodynamics'',
{\em Phys. Rev. A} {\bf 65}, 5, 054302 (2002).

\item {\bf [Guo-Guo 03 b]}:
G.-P. Guo, \& G.-C. Guo,
``Entanglement of individual photon and atomic ensembles'',
{\em Quant. Inf. Comp.} {\bf 3}, 6, 627-634 (2003);
quant-ph/0206041.

\item {\bf [Guo-Guo 03 b]}:
G.-C. Guo, \& G.-P. Guo,
``Quantum data hiding with spontaneous parameter down-conversion'',
{\em Phys. Rev. A} {\bf 68}, 4, 044303 (2003);
quant-ph/0301085.

\item {\bf [Guo-Guo 03 c]}:
G.-P. Guo, \& G.-C. Guo,
``Quantum memory for individual polarized photons'',
{\em Phys. Lett. A} {\bf 318}, 4-5, 337-341 (2003).

\item {\bf [Guo-Long-Sun 00]}:
H. Guo, G.-L. Long, \& Y. Sun,
``Effects of imperfect gate operations in Shor's
prime factorization algorithm'',
quant-ph/0012088.

\item {\bf [Gupta-Zia 02]}:
S. Gupta, \& R. K. P. Zia,
``Quantum neural networks'',
{\em J. Comp. Syst. Sci.}
quant-ph/0201144.

\item {\bf [Gurvits-Barnum 02]}:
L. Gurvits, \& H. N. Barnum,
``Largest separable balls around the maximally mixed bipartite quantum state'',
{\em Phys. Rev. A} {\bf 66}, 6, 062311 (2002);
quant-ph/0204159.

\item {\bf [Gurvits-Barnum 03]}:
L. Gurvits, \& H. N. Barnum,
``Separable balls around the maximally mixed multipartite quantum states'',
quant-ph/0302102.

\item {\bf [Gurvits-Barnum 03]}:
L. Gurvits, \& H. N. Barnum,
``Further results on the multipartite separable ball'',
quant-ph/0409095.

\item {\bf [Gurvits 03]}:
S. A. Gurvitz,
``Negative result measurements in mesoscopic systems'',
{\em Phys. Lett. A} {\bf 311}, 4-5, 292-296 (2003).

\item {\bf [Gurvitz-Fedichkin-Mozyrsky-Berman 03]}:
S. A. Gurvitz, L. Fedichkin, D. Mozyrsky, \& G. P. Berman,
``Relaxation and the Zeno effect in qubit measurements'',
{\em Phys. Rev. Lett.} {\bf 91}, 6, 066801 (2003).

\item {\bf [Gurvitz-Barnum 03]}:
L. Gurvits, \& H. Barnum,
``Separable balls around the maximally mixed multipartite quantum states'',
{\em Phys. Rev. A} {\bf 68}, 4, 042312 (2003).

\item {\bf [Guy-Deltete 88]}:
R. Guy, \& R. Deltete,
`Note on ``Bell's theorem and
the foundations of quantum physics''
[{\em Am. J. Phys.} {\bf 53}, 306 (1985)]',
{\em Am. J. Phys.} {\bf 56}, 6, 565-566 (1988).
Comment on {\bf [Stapp 85 a]}.
Reply: {\bf [Stapp 88 c, d]}.

\item {\bf [Guy-Deltete 90]}:
R. Guy, \& R. Deltete,
``Fine, Einstein, and ensembles'',
{\em Found. Phys.} {\bf 20}, 8, 943-965 (1990).
See {\bf [Deltete-Guy 90]}, {\bf [Fine 90 a]}.

\item {\bf [Guzm\'{a}n-Retamal-Romero-Saavedra 04]}:
R. Guzm\'{a}n, J. C. Retamal, J. L. Romero, \& C. Saavedra,
``Entanglement rate in qubits'',
{\em Phys. Lett. A} {\bf 323}, 5-6, 382-388 (2004).


\newpage

\subsection{}


\item {\bf [Ha-Kye-Park 03]}:
K.-C. Ha, S.-H. Kye, \& Y. S. Park,
``Entangled states with positive partial transposes arising from indecomposable positive linear maps'',
{\em Phys. Lett. A} {\bf 313}, 3, 163-174 (2003);
quant-ph/0310109.

\item {\bf [Ha-Kye 04]}:
K.-C. Ha, \& S.-H. Kye,
``Construction of entangled states with positive partial transposes based on indecomposable positive linear maps'',
{\em Phys. Lett. A} {\bf 325}, 5-6, 315-323 (2004).

\item {\bf [Haag 92]}:
R. Haag,
{\em Local quantum physics: Fields, particles, algebras},
Springer-Verlag, Berlin, 1992.

\item {\bf [Haba 98]}:
Z. Haba,
``Time dependence of the decoherence rate'',
{\em Phys. Rev. A} {\bf 57}, 5, 4034-4037 (1998).

\item {\bf [Habib-Shizume-Zurek 98]}:
S. Habib, K. Shizume, \& W. H. Zurek,
``Decoherence, chaos, and the correspondence principle'',
quant-ph/9803042.

\item {\bf [Hackenbroich-Rosenow-Weidenm\"{u}ller 98]}:
G. Hackenbroich, B. Rosenow, \& H. A. Weidenm\"{u}ller,
``A mesoscopic quantum eraser'',
{\em Europhys. Lett.} {\bf 44}, 6, 693-699 (1998).

\item {\bf [Hackermueller-Hornberger-Brezger-(+2) 03]}:
L. Hackermueller, K. Hornberger, B. Brezger,
A. Zeilinger, \& M. Arndt,
``Decoherence in a Talbot Lau interferometer: The influence of molecular
scattering'',
quant-ph/0307238.

\item {\bf [Hackermueller-Hornberger-Brezger-(+2) 04]}:
L. Hackermueller, K. Hornberger, B. Brezger,
A. Zeilinger, \& M. Arndt,
``Decoherence of matter waves by thermal emission of radiation'',
{\em Nature} {\bf 427}, ?, 711-714 (2004);
quant-ph/0402146.

\item {\bf [Hacyan 97 a]}:
S. Hacyan,
``Test of Bell's inequalities with harmonic oscillators'',
{\em Phys. Rev. A} {\bf 55}, 4, R2492-R2494 (1997).

\item {\bf [Hacyan 97 b]}:
S. Hacyan,
``Bell's inequality violation with four entangled fermions'',
{\em Phys. Rev. A} {\bf 56}, 4, R2489-R2490 (1997).

\item {\bf [Hacyan 01]}:
S. Hacyan,
``Relativistic invariance of Bell's inequality'',
{\em Phys. Lett. A} {\bf 288}, 2, 59-61 (2001).

\item {\bf [Hacyan 03]}:
S. Hacyan,
``Relativistic quantum correlations between spin-1/2 particles'',
{\em Found. Phys. Lett.} {\bf 16}, 3, 287-291 (2003).

\item {\bf [Haffner-Gulde-Riebe-(+5) 03]}:
H. Haffner, S. Gulde, M. Riebe,
G. Lancaster, C. Becher, J. Eschner, F. Schmidt-Kaler, \& R. Blatt,
``Precision measurement and compensation of optical Stark shifts for an
ion-trap quantum processor'',
{\em Phys. Rev. Lett.} {\bf 90}, 14, 143602 (2003).

\item {\bf [Hafner-Summhammer 97]}:
M. Hafner, \& J. Summhammer,
``Experiment on interaction-free measurement in neutron interferometry'',
{\em Phys. Lett. A} {\bf 235}, 6, 563-568 (1997);
quant-ph/9708048.

\item {\bf [Hagan-Hameroff-Tuszynski 02]}:
S. Hagan, S. R. Hameroff, \& J. A. Tuszynski,
``Quantum computation in brain microtubules: Decoherence and biological
feasibility'',
{\em Phys. Rev. E} {\bf 65}, 6, 061901 (2002);
quant-ph/0005025.

\item {\bf [Hagley-Ma\^{\i}tre-Nogues-(+4) 97]}:
E. Hagley, X. Ma\^{\i}tre, G. Nogues, C. Wunderlich,
M. Brune, J.-M. Raimond, \& S. Haroche,
``Generation of Einstein-Podolsky-Rosen pairs of atoms'',
{\em Phys. Rev. Lett.} {\bf 79}, 1, 1-5 (1997).
See {\bf [Bouwmeester-Zeilinger 97]}.

\item {\bf [Hahne 03]}:
G. E. Hahne,
``Time as an observable in nonrelativistic quantum mechanics'',
{\em J. Phys. A} {\bf 36}, 25, 7149–7172 (2003).

\item {\bf [H\'{a}jek-Bub 92]}:
A. H\'{a}jek, \& J. Bub,
``EPR'',
{\em Found. Phys.} {\bf 22}, 3, 313-332 (1992).
Comment on {\bf [Fine 86]}.
See {\bf [Howard 85]}, {\bf [Fine 89]} (Sec. 1), {\bf [Deltete-Guy 91]},
{\bf [Combourieu 92]}.

\item {\bf [Halder-Tanzilli-de Riedmatten-(+3) 04]}:
M. Halder, S. Tanzilli, H. de Riedmatten,
A. Beveratos, H. Zbinden, \& N. Gisin,
``Photon-bunching measurement after $2 \times 25$\,km of standard optical fibers'',
quant-ph/0408092.

\item {\bf [Hales-Straus 82]}:
A. W. Hales, \& E. G. Straus,
``Projetive colorings'',
{\em Pacific. J. Math.} {\bf 99}, ?, 31-43 (1982).

\item {\bf [Hales-Hallgren 98]}:
L. Hales, \& S. Hallgren,
``Sampling Fourier transforms on different domains'',
quant-ph/9812060.

\item {\bf [Hall-O'Rourke 93]}:
M. J. W. Hall, \& M. J. O'Rourke,
``?'',
{\em Quantum Opt.} {\bf 5}, ?, 161-? (1993).

\item {\bf [Hall 97]}:
M. J. W. Hall,
``Quantum information and correlation bounds'',
{\em Phys. Rev. A} {\bf 55}, 1, 100-113 (1997).

\item {\bf [Hall 00 a]}:
M. J. W. Hall,
``Quantum properties of classical Fisher information'',
{\em Phys. Rev. A} {\bf 62}, 1, 012107 (2000).

\item {\bf [Hall 00 b]}:
M. J. W. Hall,
`Comment on ``Conceptual inadequacy of Shannon information\dots''
by \v{C}. Brukner, and A. Zeilinger',
quant-ph/0007116.
Comment on {\bf [Brukner-Zeilinger 01 a]}.
Reply: {\bf [Brukner-Zeilinger 01 b]}.

\item {\bf [Hall 01 a]}:
M. J. W. Hall,
``Exact uncertainty relations'',
quant-ph/0103072.
Shortened version: {\bf [Hall 01 b]}.

\item {\bf [Hall 01 b]}:
M. J. W. Hall,
``Exact uncertainty relations: Physical significance'',
{\em Phys. Rev. A} {\bf 64}, 5, 052103 (2001);
quant-ph/0107149.
Shortened version of {\bf [Hall 01 a]}.

\item {\bf [Hall-Reginatto 02]}:
M. J. W. Hall, \& M. Reginatto,
``Quantum mechanics from a Heisenberg-type equality'',
{\em 100 Years of Werner Heisenberg--Works and Impact (Bamberg, Germany, 2001)},
{\em Fortschr. Phys.};
quant-ph/0201084.

\item {\bf [Hall 03]}:
M. J. W. Hall,
``Algebra for generalised quantum observables'',
quant-ph/0302007.

\item {\bf [Hall 04]}:
M. J. W. Hall,
``Superselection from canonical constraints'',
{\em J. Phys. A};
quant-ph/0404123.

\item {\bf [Hall-Blodwell 85]}:
P. J. Hall, \& J. F. Blodwell,
``Mixed states in the EPR experiment'',
{\em Phys. Lett. A} {\bf 108}, 7, 325-328 (1985).

\item {\bf [Hall 65]}:
R. Hall,
``The philosophical basis of Bohr's interpretation of quantum
mechanics'',
{\em Am. J. Phys.} {\bf 33}, ?, 624-? (1965).

\item {\bf [Halliwell 90]}:
J. J. Halliwell,
``Information dissipation in quantum cosmology and the emergence of classical spacetime'',
in {\bf [Zurek 90]}, pp.~459-469.

\item {\bf [Halliwell 95]}:
J. J. Halliwell,
``A review of the decoherent histories approach to quantum mechanics'',
in D. M. Greenberger, \& A. Zeilinger (eds.),
{\em Fundamental problems in quantum theory:
A conference held in honor of professor John
A. Wheeler, Ann. N. Y. Acad. Sci.} {\bf 755}, 726-740 (1995).

\item {\bf [Halliwell 98]}:
J. J. Halliwell,
``Decoherent histories and hydrodynamic equations'',
quant-ph/9805062.

\item {\bf [Halliwell 99 a]}:
J. J. Halliwell,
``Histories of local densities don't interfere'',
quant-ph/9905094.

\item {\bf [Halliwell 99 b]}:
J. J. Halliwell,
``Decoherent histories and the emergent classicality of local
densities'',
{\em Phys. Rev. Lett.} {\bf 83}, 13, 2481-2485 (1999).

\item {\bf [Halliwell 00]}:
J. J. Halliwell,
``Approximate decoherence of histories
and 't Hooft's deterministic quantum theory'',
quant-ph/0011103.

\item {\bf [Halliwell 01]}:
J. J. Halliwell,
``Decoherent Histories for Spacetime Domains'',
in J. G. Muga, R. Sala Mayato, \& I. L.Egususquiza (eds.),
{\em Time in quantum mechanics},
Springer-Verlag, Berlin, 2001;
quant-ph/0101099.

\item {\bf [Halliwell 03 a]}:
J. J. Halliwell,
``Some recent developments in the decoherent histories approach to quantum
theory'',
in T. Elze (ed.),
{\em Decoherence, Information, Complexity, Entropy (Piombino, Italy, 2002)};
quant-ph/0301117.

\item {\bf [Halliwell 03 b]}:
J. J. Halliwell,
``Decoherence of histories and hydrodynamic equations for a linear
oscillator chain'',
quant-ph/0305084.

\item {\bf [Hallgren-Russell-Ta Shma 03]}:
S. Hallgren, A. Russell, \& A. Ta-Shma,
``The hidden subgroup problem and quantum computation using group
representations'',
{\em Siam J. Comput.} {\bf 32}, 916-? (2003).

\item {\bf [Halvorson-Clifton 99]}:
H. Halvorson, \& R. K. Clifton,
``Maximal beable subalgebras of quantum-mechanical observables'',
{\em Int. J. Theor. Phys.} {\bf 38}, 10, 2441-2484 (1999);
quant-ph/9905042,
PITT-PHIL-SCI00000065.

\item {\bf [Halvorson-Clifton 00]}:
H. Halvorson, \& R. K. Clifton,
``Generic Bell correlation between arbitrary local
algebras in quantum field theory'',
{\em J. Math. Phys.} {\bf 41}, 4, 1711-1717 (2000);
math-ph/9909013.
Reprinted in {\bf [Clifton 04]}.

\item {\bf [Halvorson 00 a]}:
H. Halvorson,
``On the nature of continuous physical quantities
in classical and quantum mechanics'' (2000),
quant-ph/0003074,
PITT-PHIL-SCI00000097.

\item {\bf [Halvorson 00 b]}:
H. Halvorson,
``The Einstein-Podolsky-Rosen state maximally violates Bell's
inequalities'',
quant-ph/0009007.

\item {\bf [Halvorson 01]}:
H. Halvorson,
``Complementarity of representations in quantum mechanics'' (2001),
quant-ph/0110102,
PITT-PHIL-SCI00000444.

\item {\bf [Halvorson-Clifton 02 b]}:
H. Halvorson, \& R. K. Clifton,
``No place for particles in relativistic quantum theories?'',
{\em Phil. Sci.} {\bf 69}, 1-28 (2002);
quant-ph/0103041.
Reprinted in {\bf [Clifton 04]}.

\item {\bf [Halvorson-Clifton 02 b]}:
H. Halvorson, \& R. K. Clifton,
``Reconsidering Bohr's reply to EPR'',
in T. Placek, \& J. N. Butterfield (eds.),
{\em NATO Advanced Research Workshop ``Modality, Probability, and Bell's Theorem''
(Cracow, Poland, 2001)},
Kluwer Academic, Dordrecht, Holland, 2002;
quant-ph/0110107,
PITT-PHIL-SCI00000445.
Reprinted in {\bf [Clifton 04]}.

\item {\bf [Halvorson 03 a]}:
H. Halvorson,
``Generalization of the Hughston-Jozsa-Wootters theorem to hyperfinite von
Neumann algebras'',
quant-ph/0310001.

\item {\bf [Halvorson 03 b]}:
H. Halvorson,
``A note on information theoretic characterizations of physical theories'',
quant-ph/0310101.

\item {\bf [Halvorson-Bub 03]}:
H. Halvorson, \& J. Bub,
``Can quantum cryptography imply quantum mechanics? Reply to Smolin'',
quant-ph/0311065.
See {\bf [Smolin 03]}.

\item {\bf [Hamada 01]}:
M. Hamada,
``Lower bounds on the quantum capacity and error exponent of general memoryless channels'',
quant-ph/0112103.

\item {\bf [Hamada 02 a]}:
M. Hamada,
``Exponential lower bound on the highest fidelity achievable by quantum
error-correcting codes'',
{\em Phys. Rev. A} {\bf 65}, 5, 052305 (2002);
quant-ph/0109114.

\item {\bf [Hamada 02 b]}:
M. Hamada,
``A lower bound on the quantum capacity of channels with correlated errors'',
{\em J. Math. Phys.} {\bf 43}, 9, 4382-4390 (2002);
quant-ph/0201056.

\item {\bf [Hamada 03 a]}:
M. Hamada,
``Teleportation and entanglement distillation in the presence of correlation
among bipartite mixed states'',
{\em Phys. Rev. A} {\bf 68}, 1, 012301 (2003);
quant-ph/0302054.

\item {\bf [Hamada 03 b]}:
M. Hamada,
``Notes on the fidelity of symplectic quantum error-correcting codes'',
quant-ph/0311003.

\item {\bf [Hameroff 98]}:
S. Hameroff,
``Quantum computation in brain microtubules? The
Penrose-Hameroff `Orch OR' model of consciousness'',
in A. K. Ekert, R. Jozsa, \& R. Penrose (eds.),
{\em Quantum Computation: Theory and Experiment.
Proceedings of a Discussion Meeting held at the Royal
Society of London on 5 and 6 November 1997},
{\em Philos. Trans. R. Soc. Lond. A} {\bf 356}, 1743, 1869-1896 (1998).

\item {\bf [Hamieh-Qi-Siminovitch-Ali 03]}:
S. Hamieh, J. Qi, D. Siminovitch, \& M. K. Ali,
``Extracting classical correlations from a bipartite quantum system'',
{\em Phys. Rev. A} {\bf 67}, 1, 014301 (2003).

\item {\bf [Hamieh-Zaraket 03]}:
S. Hamieh, \& H. Zaraket,
``Distillable entanglement in $d \otimes d$ dimensions'',
{\em J. Phys. A} {\bf 36}, 24, L387-L391 (2003).

\item {\bf [Hamilton-Isham-Butterfield 00]}:
J. Hamilton, C. J. Isham, \& J. N. Butterfield,
``A topos perspective on the Kochen-Specker theorem:
III. Von Neumann algebras as the base category'',
{\em Int. J. Theor. Phys.} {\bf 39}, 6, 1413-1436 (2000);
quant-ph/9911020.
See {\bf [Isham-Butterfield 98]} (I),
{\bf [Isham-Butterfield 99]} (II),
{\bf [Butterfield-Isham 02]} (IV).

\item {\bf [Hamilton 00]}:
J. Hamilton,
``An obstruction-based approach to the Kochen-Specker theorem'',
{\em J. Phys. A} {\bf 33}, 20, 3783-3794 (2000);
quant-ph/9912018.

\item {\bf [Hamma-Zanardi 04]}:
A. Hamma, \& P. Zanardi,
``Quantum entangling power of adiabatically connected Hamiltonians'',
{\em Phys. Rev. A} {\bf 69}, 6, 062319 (2004);
quant-ph/0308131.

\item {\bf [Hamma-Ionicioiu-Zanardi 04 a]}:
A. Hamma, R. Ionicioiu, \& P. Zanardi,
``Ground state entanglement and geometric entropy in the Kitaev's model'',
quant-ph/0406202.

\item {\bf [Hamma-Ionicioiu-Zanardi 04 a]}:
A. Hamma, R. Ionicioiu, \& P. Zanardi,
``Bipartite entanglement and entropic boundary law in lattice spin systems'',
quant-ph/0409073.

\item {\bf [Hammerer-Vidal-Cirac 02]}:
K. Hammerer, G. Vidal, \& J. I. Cirac,
``Characterization of nonlocal gates'',
{\em Phys. Rev. A} {\bf 66}, 6, 062321 (2002);
quant-ph/0205100.

\item {\bf [Hammerer-M{\o}lmer-Polzik-Cirac 04]}:
K. Hammerer, K. M{\o}lmer, E. S. Polzik, \& J. I. Cirac,
``Light-matter quantum interface'',
quant-ph/0312156.

\item {\bf [Hammerer-Wolf-Polzik-Cirac 04]}:
K. Hammerer, M. M. Wolf, E. S. Polzik, \& J. I. Cirac,
``Quantum benchmark for storage and transmission of coherent states'',
quant-ph/0409109.

\item {\bf [Han-Kim-Noz 00]}:
D. Han, Y. S. Kim, \& M. E. Noz,
``Interferometers and decoherence matrices'',
{\em Phys. Rev. E} {\bf 61}, 5, 5907-5913 (2000);
quant-ph/0003044.

\item {\bf [Han-Hwang-Koh 96]}:
Y. D. Han, W.-Y. Hwang, \& I. G. Koh,
``Explicit solutions for negative-probability measures for all entangled states'',
{\em Phys. Lett. A} {\bf 221}, 5, 283-286 (1996).

\item {\bf [Han-Zhang-Guo 02 a]}:
Y.-J. Han, Y.-S. Zhang, \& G.-C. Guo,
``W state and Greenberger-Horne-Zeilinger state in quantum three-person
prisoner's dilemma'',
{\em Phys. Lett. A} {\bf 295}, 2-3, 61-64 (2002).

\item {\bf [Han-Zhang-Guo 02 b]}:
Y.-J. Han, Y.-S. Zhang, \& G.-C. Guo,
``Bounds for state-dependent quantum cloning'',
{\em Phys. Rev. A} {\bf 66}, 5, 052301 (2002).

\item {\bf [Han-Mo-Gui-Guo 04]}:
Z.-F. Han, X.-F. Mo, Y.-Z. Gui, \& G.-C. Guo,
``Stability of phase-modulated quantum key distribution system'',
quant-ph/0408031.

\item {\bf [Hannabuss 97]}:
K. Hannabuss,
{\em An introduction to quantum theory},
Oxford University Press, Oxford, 1997.
Review: {\bf [Rae 97]}.

\item {\bf [Hannemann-Reiss-Balzer-(+3) 02]}:
T. Hannemann, D. Reiss, C. Balzer,
W. Neuhauser, P. E. Toschek, \& C. Wunderlich,
``Self-learning estimation of quantum states'',
{\em Phys. Rev. A} {\bf 65}, 5, 050303 (2002);
quant-ph/0110068.

\item {\bf [Hannout-Hoyt-Kryowonos-Widom 98]}:
M. Hannout, S. Hoyt, A. Kryowonos, \& A. Widom,
``Quantum measurement theory and the Stern-Gerlach experiment'',
{\em Am. J. Phys.} {\bf 66}, 5, 377-379 (1998).

\item {\bf [H\"{a}nsel-Hommelhoff-H\"{a}nsch-Reichel 01]}:
W. H\"{a}nsel, P. Hommelhoff, T. W. H\"{a}nsch, \& J. Reichel,
``Bose-Einstein condensation on a microelectronic chip'',
{\em Nature} {\bf 413}, 6855, 498-501 (2001).
See {\bf [Folman-Schmiedmayer 01]}.

\item {\bf [H\"{a}nsel-Reichel-Hommelhoff-H\"{a}nsch 01]}:
W. H\"{a}nsel, J. Reichel, P. Hommelhoff, \& T. W. H\"{a}nsch,
``Trapped-atom-interferometer in a magnetic microtrap'',
quant-ph/0106162.

\item {\bf [Hanson 59]}:
N. R. Hanson,
``Copenhagen interpretation of quantum theory'',
{\em Am. J. Phys.} {\bf 27}, 1, 1-15 (1959).

\item {\bf [Hansson 00 a]}:
J. Hansson,
`Nonlinear gauge interactions - A solution to the ``measurement problem'' in
quantum mechanics?',
quant-ph/0003083.

\item {\bf [Hansson 00 b]}:
J. Hansson,
``A possible experimental test to decide if quantum mechanical
randomness is due to deterministic chaos in the underlying
dynamics'',
quant-ph/0006079.

\item {\bf [Hanson 03]}:
R. D. Hanson,
``When worlds collide: Quantum probability from observer selection?'',
{\em Found. Phys.} {\bf 33}, 7, 1129-1150 (2003);
quant-ph/0108070.

\item {\bf [Hansteen-Kocbach 00]}:
J. M. Hansteen, \& L. Kocbach,
``Restoring entanglement in atomic collisions: A gedanken
experiment'',
{\em Eur. Phys. J. D} {\bf 11}, 3, 323-326 (2000).

\item {\bf [Hao-Li-Guo 00]}:
J.-C. Hao, C.-F. Li, \& G.-C. Guo,
``Probabilistic dense coding and teleportation'',
{\em Phys. Lett. A} {\bf 278}, 3, 113-117 (2000).

\item {\bf [Hao-Li-Guo 01]}:
J.-C. Hao, C.-F. Li, \& G.-C. Guo,
``Controlled dense coding using the Greenberger-Horne-Zeilinger
state'',
{\em Phys. Rev. A} {\bf 63}, 5, 054301 (2001).

\item {\bf [Hao-Hou-Xi-Yue 02]}:
S.-R. Hao, B.-Y. Hou, X.-Q. Xi, \& R.-H. Yue,
``Accessible information for equally-distant partially-entangled alphabet state
resource'',
{\em Commun. Theor. Phys. (Beijing)} {\bf 37}, 2, 149-154 (2002).

\item {\bf [Hardy 91 a]}:
L. Hardy,
``$N$-measurement Bell inequalities, $N$-atom
entangled states, and the nonlocality of one photon'',
{\em Phys. Lett. A} {\bf 160}, 1, 1-7 (1991).

\item {\bf [Hardy 91 b]}:
L. Hardy,
``A new way to obtain Bell inequalities'',
{\em Phys. Lett. A} {\bf 161}, 1, 21-25 (1991).

\item {\bf [Hardy 91 c]}:
L. Hardy,
``Can classical wave theory explain the photon
anticorrelation effect on a beam splitter?'',
{\em Europhys. Lett.} {\bf 15}, 6, 591-595 (1991).

\item {\bf [Hardy 92 a]}:
L. Hardy,
``Quantum mechanics, local realistic theories, and
Lorentz-invariant realistic theories'',
{\em Phys. Rev. Lett.} {\bf 68}, 20, 2981-2984 (1992).
Comments: {\bf [Berndl-Goldstein 94]}, {\bf [Schauer 94 a]}.
See {\bf [Clifton-Niemann 92]}, {\bf [Vaidman 93]},
{\bf [Cohen-Hiley 95 a, 96]}, {\bf [Irvine-Hodelin-Simon-Bouwmeester 04]}.

\item {\bf [Hardy 92 b]}:
L. Hardy,
``Source of photons with correlated polarisations
and correlated directions'',
{\em Phys. Lett. A} {\bf 161}, 4, 326-328 (1992).

\item {\bf [Hardy 92 c]}:
L. Hardy,
``On the existence of empty waves in quantum theory'',
{\em Phys. Lett. A} {\bf 167}, 1, 11-16 (1992).
Comment: {\bf [Pagonis 92]}.
See {\bf [Hardy 92 e]}.

\item {\bf [Hardy 92 d]}:
L. Hardy,
``A quantum optical experiment to test local realism'',
{\em Phys. Lett. A} {\bf 167}, 1, 17-23 (1992).

\item {\bf [Hardy 92 e]}:
L. Hardy,
``Reply to `Empty waves: No necessarily effective'\,'',
{\em Phys. Lett. A} {\bf 169}, 3, 222-223 (1992).
Reply to {\bf [Pagonis 92]}.

\item {\bf [Hardy 92 f]}:
L. Hardy,
``Nonlocality, violation of Lorentz invariance,
and wave-particle duality in quantum theory'',
Ph.\ D. thesis, University of Durham, 1992.

\item {\bf [Hardy-Home-Squires-Whitaker 92]}:
L. Hardy, D. Home, E. J. Squires, \& M. A. B. Whitaker,
``Realism and the quantum-mechanical two-state oscillator'',
{\em Phys. Rev. A} {\bf 45}, 7, 4267-4270 (1992).
See: {\bf [Gillespie 94, 97]},
{\bf [Hardy-Home-Squires-Whitaker 97]}.

\item {\bf [Hardy-Squires 92]}:
L. Hardy, \& E. J. Squires,
``On the violation of Lorentz-invariance in deterministic
hidden-variable interpretations of quantum theory'',
{\em Phys. Lett. A} {\bf 168}, 3, 169-173 (1992).

\item {\bf [Hardy 93]}:
L. Hardy,
``Nonlocality for two particles without inequalities
for almost all entangled states'',
{\em Phys. Rev. Lett.} {\bf 71}, 11, 1665-1668 (1993).
See {\bf [Goldstein 94 a]}.

\item {\bf [Hardy 94]}:
L. Hardy,
``Nonlocality of a single photon revisited'',
{\em Phys. Rev. Lett.} {\bf 73}, 17, 2279-2283 (1994).
See {\bf [Peres 95 b]}.
Comments: {\bf [Vaidman 95 a]}, {\bf [Greenberger-Horne-Zeilinger 95]}.

\item {\bf [Hardy 95 a]}:
L. Hardy,
``The EPR argument and nonlocality without
inequalities for a single photon'',
in D. M. Greenberger, \& A. Zeilinger (eds.),
{\em Fundamental problems in quantum theory:
A conference held in honor of professor John
A. Wheeler, Ann. N. Y. Acad. Sci.} {\bf 755}, 600-615 (1995).

\item {\bf [Hardy 95 b]}:
L. Hardy,
`Hardy replies [to comments on ``Nonlocality of
a single photon revisited'']',
{\em Phys. Rev. Lett.} {\bf 75}, 10, 2065-2066 (1995).
Reply to {\bf [Vaidman 95 a]}, {\bf [Greenberger-Horne-Zeilinger 95]}.
See {\bf [Hardy 94]}.

\item {\bf [Hardy 96]}:
L. Hardy,
``Contextuality in Bohmian mechanics'',
in {\bf [Cushing-Fine-Goldstein 96]}, pp.~67-76.

\item {\bf [Hardy-Home-Squires-Whitaker 97]}:
L. Hardy, D. Home, E. J. Squires, \& M. A. B. Whitaker,
'Comment on ``Why quantum mechanics cannot be formulated as a Markov
process''\,',
{\em Phys. Rev. A} {\bf 56}, 4, 3301-3303 (1997).
Comment on {\bf [Gillespie 94]}.
Reply: {\bf [Gillespie 97]}.
See: {\bf [Hardy-Home-Squires-Whitaker 92]}.

\item {\bf [Hardy 97 a]}:
L. Hardy,
``A bigger contradiction between quantum theory
and locality for two particles without inequalities'',
in M. Ferrero, \& A. van der Merwe (eds.),
{\em New developments on fundamental problems in quantum physics
(Oviedo, Spain, 1996)},
Kluwer Academic, Dordrecht, Holland, 1997, pp.~163-170.
See {\bf [Boschi-Branca-De Martini-Hardy 97]}.

\item {\bf [Hardy 97 b]}:
L. Hardy,
``Einstein-Podolsky-Rosen reasoning in nonlocality theorems'',
in {\bf [Cohen-Horne-Stachel 97 b]}.

\item {\bf [Hardy 98]}:
L. Hardy,
``Spooky action at a distance in quantum mechanics'',
{\em Contemp. Phys.} {\bf 39}, 6, 419-429 (1998).

\item {\bf [Hardy-van Dam 99]}:
L. Hardy, \& W. van Dam,
``Quantum communication using a nonlocal Zeno effect'',
{\em Phys. Rev. A} {\bf 59}, 4, 2635-2640 (1999);
quant-ph/9805037.

\item {\bf [Hardy 99 a]}:
L. Hardy,
``Method of areas for manipulating the entanglement properties
of one copy of a two-particle pure entangled state'',
{\em Phys. Rev. A} {\bf 60}, 3, 1912-1923 (1999).

\item {\bf [Hardy-Song 99]}:
L. Hardy, \& D. D. Song,
``No signalling and probabilistic quantum cloning'',
{\em Phys. Lett. A} {\bf 259}, 5, 331-333 (1999);
quant-ph/9905024.

\item {\bf [Hardy 99 b]}:
L. Hardy,
``Disentangling nonlocality and teleportation'',
quant-ph/9906123.

\item {\bf [Hardy-Song 00]}:
L. Hardy, \& D. D. Song,
``Entanglement-swapping chains for general pure states'',
{\em Phys. Rev. A} {\bf 62}, 5, 052315 (2000);
quant-ph/0006132.

\item {\bf [Hardy-Song 01 a]}:
L. Hardy, \& D. D. Song,
``Universal manipulation of a single qubit'',
{\em Phys. Rev. A} {\bf 63}, 3, 032304 (2001);
quant-ph/0008011.

\item {\bf [Hardy 00]}:
L. Hardy,
``Can we obtain quantum theory from reasonable axioms?'',
quant-ph/0010083.

\item {\bf [Hardy 01]}:
L. Hardy,
``Quantum theory from five reasonable axioms'',
quant-ph/0101012.
See {\bf [Kirkpatrick 03 c]}, {\bf [Duck 03]}, {\bf [Schack 03]}.

\item {\bf [Hardy-Song 01]}:
L. Hardy, \& D. D. Song,
``Nonlinear qubit transformations'',
{\em Phys. Rev. A} {\bf 64}, 3, 032301 (2001);
quant-ph/0102100.

\item {\bf [Hardy 02 a]}:
L. Hardy,
``Why quantum theory?'',
in T. Placek, \& J. N. Butterfield (eds.),
{\em NATO Advanced Research Workshop ``Modality, Probability, and Bell's Theorem''
(Cracow, Poland, 2001)},
IOS Press, Holland, 2002;
quant-ph/0111068.
See {\bf [Kirkpatrick 03 c]}, {\bf [Duck 03]}, {\bf [Schack 03]}.

\item {\bf [Hardy 02 b]}:
L. Hardy,
``Manipulating the entanglement of one copy of a two-particle pure entangled state'',
in {\bf [Lomonaco-Brandt 02]} 75-80.

\item {\bf [Hardy-Steeb 01]}:
Y. Hardy, \& W.-H. Steeb,
{\em Classical and quantum computing}
Birkhauser, Berlin, 2001.

\item {\bf [Hardy-Kent 04]}:
L. Hardy, \& A. Kent,
``Cheat sensitive quantum bit commitment'',
{\em Phys. Rev. Lett.};
quant-ph/9911043.

\item {\bf [Hariharan-Samuel-Sinha 99]}:
P. Hariharan, J. Samuel, \& S. Sinha,
``Four-photon interference: A realizable experiment to demonstrate
violation of EPR postulates for perfect correlations'',
{\em J. Opt. B: Quantum Semiclass. Opt.} {\bf 1}, 2, 199-205
(1999).

\item {\bf [Hariharan-Sanders 00]}:
P. Hariharan, \& B. C. Sanders,
``Cavity-enhanced parametric down-conversion as a source of
correlated photons'',
{\em J. Mod. opt.} {\bf 47}, 10, 1739-1744 (2000).

\item {\bf [Hariharan-Sanders 02]}:
P. Hariharan, \& B. C. Sanders,
``Four-photon interferometry for secure quantum key distribution'',
{\em Opt. Express} {\bf 10}, 1222-? (2002).

\item {\bf [Harneit 02]}:
W. Harneit,
``Fullerene-based electron-spin quantum computer'',
{\em Phys. Rev. A} {\bf 65}, 3, 032322 (2002).

\item {\bf [Haroche-Brune-Raimond 92]}:
S. Haroche, M. Brune, \& J.-M. Raimond,
``Manipulation of optical fields by atomic interferometry:
Quantum variations of a theme by Young'',
{\em Appl. Phys. B} {\bf 54}, 5, 355-365 (1992).

\item {\bf [Haroche-Raimond 93]}:
S. Haroche, \& J.-M. Raimond,
``Cavity quantum electrodynamics'',
{\em Sci. Am.} {\bf 269}, 4, 26-33 (1993).
Spanish version:
``Electrodin\'{a}mica cu\'{a}ntica en cavidades'',
{\em Investigaci\'{o}n y Ciencia} 201, 51-59 (1993).
Reprinted in {\bf [Macchiavello-Palma-Zeilinger 00]}, pp.~282-289.

\item {\bf [Haroche 95]}:
S. Haroche,
``Atoms and photons in high-Q cavities: New
tests of quantum theory'',
in D. M. Greenberger, \& A. Zeilinger (eds.),
{\em Fundamental problems in quantum theory:
A conference held in honor of professor John
A. Wheeler, Ann. N. Y. Acad. Sci.} {\bf 755}, 73-86 (1995).

\item {\bf [Haroche-Raimond 96]}:
S. Haroche, \& J.-M. Raimond,
``Quantum computing: Dream or nightmare?'',
{\em Phys. Today} {\bf 49}, 8, 51-52 (1996).
French version (adapted):
``L'ordinateur quantique: R\^{e}ve ou cauchemar?'',
{\em La Recherche} {\bf 27}, 292, 58-60 (1996).
Spanish version: ``El ordenador cu\'{a}ntico:
?`Sue\~{n}o o pesadilla?'',
{\em Mundo Cient\'{\i}fico} {\bf 17}, 175, 50-51 (1997).

\item {\bf [Haroche-Brune-Raimond 97]}:
S. Haroche, M. Brune, \& J.-M. Raimond,
``Experiments with single atoms in a cavity:
Entanglement, Schr\"{o}dinger's cats and decoherence'',
in P. L. Knight, B. Stoicheff, \& D. Walls (eds.),
{\em Highlight in Quantum Optics},
{\em Philos. Trans. R. Soc. Lond. A} {\bf 355}, 2733, 2367-2380 (1997).

\item {\bf [Haroche-Raimond-Brune 97]}:
S. Haroche, J.-M. Raimond, \& M. Brune,
``Le chat de Schr\"{o}dinger se pr\`{e}te \`{a} l'exp\'{e}rience'',
{\em La Recherche} {\bf 28}, 301, 50-55 (1997).
Spanish version: ``Experimento con el gato de Schr\"{o}dinger'',
{\em Mundo Cient\'{\i}fico} {\bf 17}, 185, 1030-1035 (1997).

\item {\bf [Haroche 98]}:
S. Haroche,
``Entanglement, decoherence and the quantum/classical boundary'',
{\em Phys. Today} {\bf 51}, 7, 36-42 (1998).

\item {\bf [Harrelson-Kerenidis 01]}:
C. Harrelson, \& I. Kerenidis,
``Quantum clock synchronization with one qubit'',
cs.CC/0103021.

\item {\bf [Harrington-Preskill 01]}:
J. Harrington, \& Preskill,
``Achievable rates for the Gaussian quantum channel'',
{\em Phys. Rev. A} {\bf 64}, 6, 062301 (2001);
quant-ph/0105058.

\item {\bf [Harris-Akis-Ferry 01]}:
J. Harris, R. Akis, \& D. K. Ferry,
``Magnetically switched quantum waveguide qubit'',
{\em Appl. Phys. Lett.} {\bf 79}, ?, 2214-? (2001).

\item {\bf [Harris 00]}:
P. Harris,
``Quantum theory -- Interpretation, formulation, inspiration'',
{\em Phys. Today} {\bf 53}, 9, ? (2000).
Comment on {\bf [Fuchs-Peres 00 a]}.
Reply: {\bf [Fuchs-Peres 00 b]}.

\item {\bf [Harrow-Hayden-Leung 03]}:
A. Harrow, P. Hayden, \& D. Leung,
``Superdense coding of quantum states'',
quant-ph/0307221.

\item {\bf [Harrow 04]}:
A. Harrow,
``Coherent communication of classical messages'',
{\em Phys. Rev. Lett.} {\bf 92}, 9, 097902 (2004).

\item {\bf [Harrow-Recht-Chuang 02]}:
A. W. Harrow, B. Recht, \& I. L. Chuang,
``Efficient discrete approximations of quantum gates'',
{\em J. Math. Phys.} {\bf 43}, 9, 4445-4451 (2002);
quant-ph/0111031.

\item {\bf [Harrow-Lo 02]}:
A. W. Harrow, \& H.-K. Lo,
``A tight lower bound on the classical communication cost of entanglement dilution'',
quant-ph/0204096.

\item {\bf [Harrow-Nielsen 03]}:
A. W. Harrow, \& M. A. Nielsen,
``Robustness of quantum gates in the presence of noise'',
{\em Phys. Rev. A} {\bf 68}, 1, 012308 (2003);
quant-ph/0301108.

\item {\bf [Hartle 68]}:
J. B. Hartle,
``Quantum mechanics of individual systems'',
{\em Am. J. Phys.} {\bf 36}, 8, 704-712 (1968).

\item {\bf [Hartle-Hawking 83]}:
J. B. Hartle, \& S. W. Hawking,
``Wave function of the universe'',
{\em Phys. Rev. D} {\bf 28}, 12, 2960-2975 (1983).
See {\bf [Hawking 87]}.

\item {\bf [Hartle 91]}:
J. B. Hartle,
``Excess baggage'',
in J. H. Schwarz (ed.),
{\em Elementary particles and the universe},
?, ?, 1991.

\item {\bf [Hartle 95]}:
J. B. Hartle,
``Spacetime information'',
{\em Phys. Rev. D} {\bf 51}, 4, 1800-1817 (1995).
Comment: {\bf [Kent 97 c]}.

\item {\bf [Hartle 03]}:
J. B. Hartle,
``What connects different interpretations of quantum mechanics?'',
quant-ph/0305089.

\item {\bf [Hartle 04]}:
J. B. Hartle,
``Linear positivity and virtual probability'',
quant-ph/0401108.

\item {\bf [Hasegawa-Loidl-Badurek-(+2) 03 a]}:
Y. Hasegawa, R. Loidl, G. Badurek,
M. Baron, \& H. Rauch,
``Violation of a Bell-like inequality in single-neutron interferometry'',
{\em Nature} {\bf 425}, 6953, 45-48 (2003).

\item {\bf [Hasegawa-Loidl-Badurek-(+2) 03 b]}:
Y. Hasegawa, R. Loidl, G. Badurek,
M. Baron, \& H. Rauch,
``Violation of a Bell-like inequality in neutron optical experiments:
quantum contextuality'',
{\em Proc.\ of CEWQO10 Workshop};
quant-ph/0311121.

\item {\bf [Haselgrove-Nielsen-Osborne 03 a]}:
H. L. Haselgrove, M. A. Nielsen, \& T. J. Osborne,
``Quantum states far from the energy eigenstates of any local Hamiltonian'',
quant-ph/0303022.

\item {\bf [Haselgrove-Nielsen-Osborne 03 b]}:
H. L. Haselgrove, M. A. Nielsen, \& T. J. Osborne,
``Practicality of time-optimal two-qubit Hamiltonian simulation'',
{\em Phys. Rev. A} {\bf 68}, 4, 042303 (2003);
quant-ph/0303070.
See {\bf [Vidal-Hammerer-Cirac 02]}.

\item {\bf [Haselgrove-Nielsen-Osborne 04]}:
H. L. Haselgrove, M. A. Nielsen, \& T. J. Osborne,
``Entanglement, correlations, and the energy gap in many-body quantum systems'',
{\em Phys. Rev. A} {\bf 69}, 3, 032303 (2004);
quant-ph/0308083.

\item {\bf [Hasselmann 97]}:
K. Hasselmann,
``The metron model: Elements of a unified deterministic theory of
fields and particles. Part 3: Quantum phenomena'',
{\em Phys. Essays} {\bf 10}, 1, 64-86 (1997).

\item {\bf [Hastings 96]}:
J. Hastings,
``Optics: X-ray lenses near reality'',
{\em Nature} {\bf 384}, 6604, 22 (1996).
See {\bf [Snigirev-Kohn-Snigireva-Lengeler 96]}.

\item {\bf [Haug-Freyberger-W\'{o}dkiewicz 04]}:
F. Haug, M. Freyberger, \& K. W\'{o}dkiewicz,
``Nonlocality of a free atomic wave packet'',
{\em Phys. Lett. A} {\bf 321}, 1, 6-13 (2004);
quant-ph/0312020.

\item {\bf [Hauge-Stovneng 89]}:
E. H. Hauge, \& J. A. Stovneng,
``Tunneling times: A critical review'',
{\em Rev. Mod. Phys.} {\bf 61}, 4, 917-936 (1989).

\item {\bf [Hausladen-Wootters 94]}:
P. Hausladen, \& W. K. Wootters,
`A ``pretty good'' measurement for distinguishing quantum states',
in S. M. Barnett, A. K. Ekert, \& S. J. D. Phoenix (eds.),
{\em J. Mod. Opt.} {\bf 41}, 12 (Special issue: Quantum
communication), 2385-2390 (1994).

\item {\bf [Hausladen-Schumacher-Westmoreland-Wootters 95]}:
P. Hausladen, B. W. Schumacher, M. Westmoreland, \& W. K. Wootters,
``Sending classical bits via quantum its'',
in D. M. Greenberger, \& A. Zeilinger (eds.),
{\em Fundamental problems in quantum theory:
A conference held in honor of professor John A. Wheeler,
Ann. N. Y. Acad. Sci.} {\bf 755}, ?-? (1995).

\item {\bf [Hausladen-Jozsa-Schumacher-(+2) 96]}:
P. Hausladen, R. Jozsa, B. W. Schumacher,
M. Westmoreland, \& W. K. Wootters,
``Classical information capacity of a quantum channel'',
{\em Phys. Rev. A} {\bf 54}, 3, 1869-1876 (1996).
See {\bf [Schumacher-Westmoreland 97]}.

\item {\bf [Havel-Somaroo-Tseng-Cory 98]}:
T. F. Havel, S. S. Somaroo, C.-H. Tseng, \& D. G. Cory,
``Principles and demonstrations of quantum information
processing by NMR spectroscopy'',
quant-ph/9812086.

\item {\bf [Havel-Doran 01]}:
T. F. Havel, \& C. J. L. Doran,
``Interaction and entanglement in the multiparticle spacetime algebra'',
to appear in {\em Proc.\ AGACSE 2001},
Birkh\"{a}uer, Boston, Massachusetts, 2001;
quant-ph/0106063.

\item {\bf [Havel-Cory-Lloyd-(+7) 02]}:
T. F. Havel, D. G. Cory, S. Lloyd,
N. Boulant, E. M. Fortunato, M. A. Pravia, G. Teklemariam,
Y. S. Weinstein, A. Bhattacharyya, \& J. Hou,
``Quantum information processing by nuclear magnetic resonance spectroscopy'',
{\em Am. J. Phys.} {\bf 70}, 3, 345-362 (2002).

\item {\bf [Havel-Doran 02]}:
T. F. Havel, \& C. J. L. Doran,
``Geometric algebra in quantum information processing'',
in {\bf [Lomonaco-Brandt 02]} 81-100;
quant-ph/0004031.

\item {\bf [Havel 03]}:
T. F. Havel,
``The real density matrix'',
{\em Quant. Inf. Proc.} {\em (Feynman Festival 2002)};
quant-ph/0302176.

\item {\bf [Havel-Doran 04]}:
T. F. Havel, \& C. J. L. Doran,
``A Bloch-sphere-type model for two qubits in the geometric algebra of a
6-D Euclidean vector space'',
{\em Proc.\ SPIE Conference on Defense and Security};
quant-ph/0403136.

\item {\bf [Havlicek-Svozil 96]}:
H. Havlicek, \& K. Svozil,
``Density conditions for quantum propositions'',
{\em J. Math. Phys.} {\bf 37}, 11, 5337-5341 (1996).

\item {\bf [Havlicek-Krenn-Summhammer-Svozil 01]}:
H. Havlicek, G. Krenn, J. Summhammer, \& K. Svozil,
``Colouring the rational quantum sphere and the
Kochen-Specker theorem'',
{\em J. Phys. A} {\bf 34}, 14, 3071-3077 (2001);
quant-ph/9911040.
See {\bf [Meyer 99 b]}, {\bf [Kent 99 b]},
{\bf [Mermin 99 b]},
{\bf [Cabello 99 d, 02 c]},
{\bf [Appleby 00, 01, 02]}, {\bf [Boyle-Schafir 01 a]}.

\item {\bf [Haw 02]}:
M. Haw,
``Altered states'',
{\em Nature} {\bf 417}, 6892, 892-893 (2002).

\item {\bf [Hawking 87]}:
S. W. Hawking,
``Quantum cosmology'',
in S. W. Hawking, \& W. Israel (eds.),
{\em Three hundred years of gravitation (Cambridge, 1987)},
Cambridge University Press, Cambridge, 1987, pp.~17-49.
See {\bf [Hartle-Hawking 83]}.

\item {\bf [Hawton-Baylis 01]}:
M. Hawton, \& W. E. Baylis,
``Photon position operators and localized bases'',
{\em Phys. Rev. A} {\bf 64}, 1, 012101 (2001).

\item {\bf [Hay-Peres 97]}:
O. Hay, \& A. Peres,
``Quantum and classical descriptions of a measuring apparatus'',
{\em Phys. Rev. A} {\bf 58}, 1, 116-122 (1998);
quant-ph/9712044.

\item {\bf [Hay-Peres 00]}:
O. Hay, \& A. Peres,
``Dual classical and quantum descriptions of a measuring
apparatus'',
in P. Kumar, G. M. D'Ariano, \& O. Hirota (eds.),
{\em Quantum communication, computing and measurement II},
Kluwer/Plenum Press, ?, 2000, pp.~117-124.

\item {\bf [Hayashi 01]}:
M. Hayashi,
``Optimal sequence of POVMs in the sense of
Stein's lemma in quantum hypothesis testing'',
quant-ph/0107004.

\item {\bf [Hayashi-Matsumoto 01]}:
M. Hayashi, \& K. Matsumoto,
``Variable length universal entanglement concentration by local operations
and its application to teleportation and dense coding'',
quant-ph/0109028.

\item {\bf [Hayashi 02 a]}:
M. Hayashi,
``Exponents of quantum fixed-length pure-state source coding'',
{\em Phys. Rev. A} {\bf 66}, 3, 032321 (2002).
Erratum: {\em Phys. Rev. A} {\bf 66}, 6, 069901 (2002).
quant-ph/0202002.

\item {\bf [Hayashi 02 b]}:
M. Hayashi,
``Two quantum analogues of Fisher information from a large
deviation viewpoint of quantum estimation'',
quant-ph/0202003.

\item {\bf [Hayashi-Matsumoto 02]}:
M. Hayashi, \& K. Matsumoto,
``Quantum universal variable-length source coding'',
{\em Phys. Rev. A} {\bf 66}, 2, 022311 (2002);
quant-ph/0202001.

\item {\bf [Hayashi-Koashi-Matsumoto 03]}:
M. Hayashi, M. Koashi, K. Matsumoto, F. Morikoshi, \& A. Winter,
``Error exponents for entanglement concentration'',
{\em J. Phys. A} {\bf 36}, 2, 527-553 (2003);
quant-ph/0206097.

\item {\bf [Hayashi-Hashimoto-Horibe 03]}:
A. Hayashi, T. Hashimoto, \& M. Horibe,
``Remote state preparation without oblivious conditions'',
{\em Phys. Rev. A} {\bf 67}, 5, 052302 (2003).

\item {\bf [Hayashi-Imai-Matsumoto-(+2) 04]}:
M. Hayashi, H. Imai, K. Matsumoto,
M. B. Ruskai, \& T. Shimono,
``Qubit channels which require four inputs to achieve capacity:
Implications for additivity conjectures'',
quant-ph/0403176.

\item {\bf [Hayden-Terhal-Uhlmann 00]}:
P. M. Hayden, B. M. Terhal, \& A. Uhlmann,
``On the LOCC classification of bipartite density matrices'',
quant-ph/0011095.

\item {\bf [Hayden-Horodecki-Terhal 01]}:
P. M. Hayden, M. Horodecki, \& B. M. Terhal,
``The asymptotic entanglement cost of preparing a quantum state'',
in S. Popescu, N. Linden, \& R. Jozsa (eds.),
{\em J. Phys. A} {\bf 34}, 35
(Special issue: Quantum information and computation), 6891-6898 (2001),
quant-ph/0008134.

\item {\bf [Hayden-Jozsa-Winter 02]}:
P. Hayden, R. Jozsa, \& A. Winter,
``Trading quantum for classical resources in quantum data compression'',
{\em J. Math. Phys.} {\bf 43}, 9, 4404-4444 (2002);
quant-ph/0204038.

\item {\bf [Hayden-Winter 03]}:
P. Hayden, \& A. Winter,
``Communication cost of entanglement transformations'',
{\em Phys. Rev. A} {\bf 67}, 1, 012326 (2003);
quant-ph/0204092.

\item {\bf [Hayden-Jozsa-Winter 02]}:
P. Hayden, R. Jozsa, D. Petz, \& A. Winter,
``Structure of states which satisfy strong subadditivity of quantum
entropy with equality'',
quant-ph/0304007.

\item {\bf [Hayden-Leung-Shor-Winter 03]}:
P. Hayden, D. Leung, P. W. Shor, \& A. Winter,
``Randomizing quantum states: Constructions and applications'',
quant-ph/0307104.

\item {\bf [Hayden-Leung-Winter 04]}:
P. Hayden, D. Leung, \& A. Winter,
``Aspects of generic entanglement'',
quant-ph/0407049.

\item {\bf [Hayden-Leung-Smith 04]}:
P. Hayden, D. Leung, \& G. Smith,
``Multiparty data hiding of quantum information'',
quant-ph/0407152.

\item {\bf [Hayden-King 04]}:
P. Hayden, \& C. King,
``Correcting quantum channels by measuring the environment'',
quant-ph/0409026.

\item {\bf [Hayden 04]}:
P. Hayden,
``Entanglement in random subspaces'',
{\em Proc.\ QCMC 2004 (Glasgow)};
quant-ph/0409157.

\item {\bf [Hayes-Gilchrist-Myers-Ralph 04]}:
A. J. F. Hayes, A. Gilchrist, C. R. Myers, \& T. C. Ralph,
``Utilizing encoding in scalable linear optics quantum computing'',
quant-ph/0408098.

\item {\bf [He-Zhu-Wang-Li 03]}:
G.-P. He, S.-L. Zhu, Z. D. Wang, \& H.-Z. Li,
``Testing Bell's inequality and measuring the entanglement using
superconducting nanocircuits'',
{\em Phys. Rev. A} {\bf 68}, 1, 012315 (2003);
quant-ph/0304156.

\item {\bf [He-Wang 03]}:
G.-P. He, \& Z. D. Wang,
``Unconditionally secure quantum oblivious transfer'',
quant-ph/0312161.

\item {\bf [Healey 79]}:
R. A. Healey,
``Quantum realism: Na\"{\i}vet\'{e} is no excuse'',
{\em Synthese} {\bf 42}, 1, 121-144 (1979).

\item {\bf [Healey 84]}:
R. A. Healey,
``How many worlds?'',
{\em No\^{u}s} {\bf 18}, ?, 591-616 (1984).

\item {\bf [Healey 89]}:
R. A. Healey,
{\em The philosophy of quantum mechanics:
An interactive interpretation},
Cambridge University Press, Cambridge, 1989.
See {\bf [Reeder-Clifton 95]}.

\item {\bf [Healey 93]}:
R. A. Healey,
``Measurement and quantum indeterminateness'',
{\em Found. Phys. Lett.} {\bf 6}, 4, 307-316 (1993).
See {\bf [Albert-Loewer 93]}.

\item {\bf [Healey 94]}:
R. A. Healey,
``Nonseparable processes and causal explanation'',
{\em Stud. Hist. Philos. Sci.}
{\bf 25}, 3, 337-374 (1994).

\item {\bf [Healey 98 a]}:
R. A. Healey,
`\,``Modal'' interpretations, decoherence and the
quantum measurement problem',
in G. Hellman, \& R. A. Healey (eds.),
{\em Quantum measurement, decoherence, and modal interpretations
(Minnesota Studies in Philosophy of Science)}, 1998.

\item {\bf [Healey 98 b]}:
R. A. Healey,
``Interpreting the quantum world'',
{\em Phys. Today} {\bf 51}, 8, part I, 63-64 (1998).

\item {\bf [Healey 99]}:
R. A. Healey,
``Mining for metaphysics'',
{\em Stud. Hist. Philos. Sci. Part B: Stud. Hist. Philos. Mod. Phys.}
{\bf 30}, 3, 443-452 (1999).
Review of {\bf [Dickson 98]}.

\item {\bf [Hecht 00]}:
K. T. Hecht,
{\em Quantum mechanics},
Springer-Verlag, New York, 2002.
Review: {\bf [Cabello 01 f]}.

\item {\bf [Heelan 65]}:
P. A. Heelan,
{\em Quantum mechanics and objectivity:
A study of the physical philosophy of Werner Heisenberg},
Martinus Nijhof, The Hague, 1965.

\item {\bf [Heelan 75]}:
P. A. Heelan,
``Heisenberg and radical theoretic change'',
{\em Zeitschrift f\"{u}r allgemeine Wissenschaftstheorie} {\bf 6},
?, 113-138 (1975).
Reply: {\bf [Heisenberg 75]}.

\item {\bf [Hegerfeldt 98 a]}:
G. C. Hegerfeldt,
``Ensemble versus individual system in quantum optics'',
{\em Fortschr. Phys.} {\bf 46}, 6-8, 595-604 (1998).

\item {\bf [Hegerfeldt 98 b]}:
G. C. Hegerfeldt,
``Instantaneous spreading and Einstein causality in quantum theory'',
{\em Ann. Phys.} {\bf 7}, 7-8, 716-725 (1998);
quant-ph/9809030.

\item {\bf [Hegerfeldt 01]}:
G. C. Hegerfeldt,
``Particle localization and the notion of Einstein causality'',
in A. Horzela, \& E. Kapuscik (eds.),
{\em Extensions of quantum theory},
Apeiron, Montreal, 2001, pp.~9-16;
quant-ph/0109044.

\item {\bf [Heilbron 86]}:
J. L. Heilbron,
{\em The dilemmas of an upright man:
Max Planck as spokesman for German science},
University of California Press, Los Angeles, 1986, 1992 (paperback).
Revised version with a new afterword:
{\em The dilemmas of an upright man:
Max Planck and the fortunes of German science},
Harvard University Press, Cambridge, Massachusetts, 2000.

\item {\bf [Heiligman 00]}:
M. Heiligman,
``Finding matches between two databases on a quantum computer'',
quant-ph/0006136.

\item {\bf [Heims 80]}:
S. J. Heims,
{\em John von Neumann and Norbert Wiener.
From mathematics to the technologies of life and death},
M.\ I.\ T.\ Press, Cambridge, Massachusetts, 1980.
Spanish version:
{\em J. von Neumann y N. Wiener},
Salvat, Barcelona, 1986.
See {\bf [MacRae 91]}.

\item {\bf [Hein-Eisert-Briegel 04]}:
M. Hein, J. Eisert, \& H. J. Briegel,
``Multiparty entanglement in graph states'',
{\em Phys. Rev. A} {\bf 69}, 6, 062311 (2004);
quant-ph/0307130.

\item {\bf [Hein-D\"{u}r-Briegel 04]}:
M. Hein, W. D\"{u}r, \& H.-J. Briegel,
``Entanglement properties of multipartite entangled states under the
influence of decoherence'',
quant-ph/0408165.

\item {\bf [Heinrich-Novak 01]}:
S. Heinrich, \& E. Novak,
``Optimal summation and integration by
deterministic, randomized, and quantum algorithms'',
contribution to
{\em 4th Int.\ Conf.\ on Monte Carlo and Quasi-Monte
Carlo Methods (Hong-Kong, 2000)};
quant-ph/0105114.

\item {\bf [Heinrich 01]}:
S. Heinrich,
``Quantum summation with an application to integration'',
submitted to {\em J. of Complexity};
quant-ph/0105116.

\item {\bf [Heisenberg 27]}:
W. Heisenberg,
``\"{U}ber den anschaulichen Inhalt der quantentheorrtischen
Kinematik un Mechanik'',
{\em Zeitschrift f\"{u}r Physik} {\bf 43}, 172-198 (1927).
Reprinted in
{\em Dokumente der Naturwissenschaft} {\bf 4}, 9-35 (1963).
English version:
``The physical content of quantum kinematics and mechanics'',
in {\bf [Wheeler-Zurek 83]}, pp.~62-84.

\item {\bf [Heisenberg 30]}:
W. Heisenberg,
{\em Die physikalischen Prinzipien der Quantentheorie},
Hirzel, Leipzig, 1930.
English version: {\em The physical principles of
the quantum theory}, University of Chicago Press, Chicago, 1930;
(reprinted) Dover, New York.

\item {\bf [Heisenberg 44]}:
W. Heisenberg,
{\em Physik und Philosophie},
Hirzel, Leipzig, 1944.
English version:
{\em Physics and philosophy---The revolution in modern science},
Harper and Row, New York, 1958.

\item {\bf [Heisenberg 55 a]}:
W. Heisenberg,
``The development and interpretation of the quantum theory'',
in W. Pauli (ed.),
with the assistance of L. Rosenfeld, \& W. F. Weisskopf,
{\em Niels Bohr and the development of physics},
McGraw-Hill, New York, 1955, pp.~12-?.

\item {\bf [Heisenberg 55 b]}:
W. Heisenberg,
{\em Das Naturbild der heutigen Physik},
Rowholt Verlag, Hamburg, 1955.
English version: {\em The physicist's conception of nature},
Hutchinson, London, 1958.
Spanish version: {\em La imagen de la naturaleza en
la f\'{\i}sica actual},
Ariel 1976; Orbis 1986.

\item {\bf [Heisenberg 58]}:
W. Heisenberg,
``?'',
{\em Daedalus} {\bf 87}, ?, 95-? (1958).

\item {\bf [Heisenberg 75]}:
W. Heisenberg,
``?'',
{\em Zeitschrift f\"{u}r allgemeine Wissenschaftstheorie}
{\bf ?}, ?, ?-? (1975).
Reply to {\bf [Heelan 65]}.

\item {\bf [Heisenberg 95]}:
W. Heisenberg,
{\em Philosophical problems of quantum physics},
Ox Bow Press, Woodbridge, Connecticut, 1995.

\item {\bf [Held 94]}:
C. Held,
``The meaning of complementarity'',
{\em Stud. Hist. Philos. Sci.} {\bf 25}, ?, 871-893 (1994).

\item {\bf [Heller-Tomsovic 93]}:
E. J. Heller, \& S. Tomsovic,
``Postmodern quantum mechanics'',
{\em Phys. Today} {\bf 46}, 7, 38-46 (1993).

\item {\bf [Hellman 82 a]}:
G. Hellman,
``Einstein and Bell: Tightening the case for microphysical randomness'',
{\em Synthese} {\bf 53}, ?, 445-460 (1982).

\item {\bf [Hellman 82 b]}:
G. Hellman,
``Stochastic Einstein locality and the Bell theorems'',
{\em Synthese} {\bf 53}, ?, 461-504 (1982).

\item {\bf [Hellman 92]}:
G. Hellman,
``Bell-type inequalities in the nonideal case: Proof of a conjecture of Bell'',
{\em Found. Phys.} {\bf 22}, 6, 807-817 (1992).

\item {\bf [Hellman 93]}:
G. Hellman,
``?'',
{\em J. Pilos. Logic} {\bf 22}, ?, 193-203 (1993).

\item {\bf [D'Helon-Milburn 96]}:
C. D'Helon, \& G. J. Milburn,
``Measurements on trapped laser-cooled ions using quantum computations'',
{\em Phys. Rev. A} {\bf 54}, 6 5141-5146 (1996).
See {\bf [D'Helon-Milburn 98]},
{\bf [Schneider-Wiseman-Munro-Milburn 98]}.

\item {\bf [D'Helon-Milburn 98]}:
C. D'Helon, \& G. J. Milburn,
``Measurements on trapped laser-cooled ions using quantum computations'',
{\em Fortschr. Phys.} {\bf 46}, 6-8, 707-712 (1998).
See {\bf [D'Helon-Milburn 96]},
{\bf [Schneider-Wiseman-Munro-Milburn 98]}.

\item {\bf [D'Helon-Protopopescu 02]}:
C. D'Helon, \& V. Protopopescu,
``New summing algorithm using ensemble quantum computing'',
submitted to {\em Quant. Inf. Proc.};
quant-ph/0202142.

\item {\bf [D'Helon-Protopopescu-Perez 03]}:
C. D'Helon, V. Protopopescu, \& R. Perez,
``Targeting qubit states using open-loop control'',
{\em J. Phys. A} {\bf 36}, 25, 7129–7148 (2003).

\item {\bf [Helstrom 76]}:
C. W. Helstrom,
{\em Quantum detection and estimation theory},
Academic Press, New York, 1976.

\item {\bf [Hellwig-Kraus 70]}:
K.-E. Hellwig, \& K. Kraus,
``Formal description of
measurements in local quantum theory'',
{\em Phys. Rev. D} {\bf 1}, 2, 566-571 (1970).

\item {\bf [Hellwig 95]}:
K.-E. Hellwig,
``General scheme of measurement processes'',
{\em Int. J. Theor. Phys.} {\bf 34}, 8, 1467-1479 (1995).
Reprinted in {\bf [Macchiavello-Palma-Zeilinger 00]}, pp.~161-172.

\item {\bf [Hemaspaandra-Hemaspaandra-Zimand 99]}:
E. Hemaspaandra, L. A. Hemaspaandra, \& M. Zimand,
``Almost-everywhere superiority for quantum computing'',
quant-ph/9910033.

\item {\bf [Hemmerich 99]}:
A. Hemmerich,
``Quantum entanglement in dilute optical lattices'',
{\em Phys. Rev. A} {\bf 60}, 2, 943-946 (1999).

\item {\bf [Hemmo-Pitowsky 01]}:
M. Hemmo, \& I. Pitowsky,
``Probability and nonlocality in many minds interpretations of quantum mechanics'' (2001),
quant-ph/0112077,
PITT-PHIL-SCI00000765.

\item {\bf [Henderson-Vedral 00]}:
L. Henderson, \& V. Vedral,
``Information, relative entropy of entanglement, and irreversibility'',
{\em Phys. Rev. Lett.} {\bf 84}, 10, 2263-2266 (2000);
quant-ph/9909011.

\item {\bf [Henderson-Hardy-Vedral 00]}:
L. Henderson, L. Hardy, \& V. Vedral,
``Two-state teleportation'',
{\em Phys. Rev. A} {\bf 61}, 6, 062306 (2000);
quant-ph/9910028.

\item {\bf [Henderson-Linden-Popescu 01]}:
L. Henderson, N. Linden, \& S. Popescu,
``Are all noisy quantum states obtained from pure ones?'',
{\em Phys. Rev. Lett.} {\bf 87}, 23, 237901 (2001);
quant-ph/0104065.

\item {\bf [Henderson-Vedral 00]}:
L. Henderson, \& V. Vedral,
``Classical, quantum and total correlations'',
quant-ph/0105028.

\item {\bf [Henderson-Vedral 01]}:
L. Henderson, \& V. Vedral,
``Classical, quantum and total correlations'',
in S. Popescu, N. Linden, \& R. Jozsa (eds.),
{\em J. Phys. A} {\bf 34}, 35
(Special issue: Quantum information and computation), 6899-6906 (2001).

\item {\bf [Hendry 84]}:
J. Hendry,
{\em The creation of quantum mechanics and the Bohr-Pauli
dialogue},
Kluwer Academic, Dordrecht, Holland, 1984.

\item {\bf [Hendrych-Du\v{s}ek-Filip-Fiur\'{a}\v{s}ek 03]}:
M. Hendrych, M. Du\v{s}ek, R. Filip, \& J. Fiur\'{a}\v{s}ek,
``Simple optical measurement of the overlap and fidelity of quantum states'',
{\em Phys. Lett. A} {\bf 310}, 2-3, 95-100 (2003).

\item {\bf [Henkel-Steane-Kaiser-Dalibard 04]}:
C. Henkel, A. M. Steane, R. Kaiser, \& J. Dalibard,
``A modulated mirror for atomic interferometry'',
quant-ph/0408156.

\item {\bf [Hennrich-Legero-Kuhn-Rempe 04]}:
M. Hennrich, T. Legero, A. Kuhn, \& G. Rempe,
``Photon statistics of a non-stationary periodically driven single-photon
source'',
{\em New J. Phys.};
quant-ph/0406034.

\item {\bf [Herbert 75]}:
N. Herbert,
``Cryptographic approach to hidden variables'',
{\em Am. J. Phys.} {\bf 43}, 4, 315-316 (1975).
See {\bf [Stapp 85 a]}, {\bf [Mermin 89 a]}.

\item {\bf [Herbert 82]}:
N. Herbert,
``FLASH---A superluminal communicator based upon a new kind of measurement'',
{\em Found. Phys.} {\bf 12}, ?, 1171-? (1982).

\item {\bf [Herbert 85]}:
N. Herbert,
{\em Quantum reality: Beyond the new physics},
Rider, ?, 1985; Anchor Books/Doubleday, New York, 1985.

\item {\bf [Herbert 89]}:
N. Herbert,
{\em Faster than light: Superluminal loopholes in physics},
Plume Books, ?, 1989.

\item {\bf [Herbut-Vuji\v{c}i\'{c} 76]}:
F. Herbut, \& M. Vuji\v{c}i\'{c},
``?'',
{\em Ann. Phys.} {\bf 96}, ?, 382-? (1976).

\item {\bf [Herbut 92]}:
F. Herbut,
``On Pitowsky's `The relativity of quantum
predictions'\,'',
{\em Phys. Lett. A} {\bf 163}, 1-2, 5-6 (1992).
Comment on {\bf [Pitowsky 91 b]}.
Reply: {\bf [Pitowsky 92]}.

\item {\bf [Herbut 96]}:
F. Herbut,
``On retroactive occurrence and twin events in
quantum mechanics'',
{\em Found. Phys. Lett.} {\bf 9}, 5, 437-446 (1996).

\item {\bf [Herbut-Vuji\v{c}i\'{c} 97]}:
F. Herbut, \& M. Vuji\v{c}i\'{c},
``First-quantization quantum-mechanical insight into the
Hong-Ou-Mandel two-photon interferometer with
polarizers and its role as a quantum eraser'',
{\em Phys. Rev. A} {\bf 56}, 1, 931-935 (1997).
See {\bf [Hong-Ou-Mandel 87]}.

\item {\bf [Herbut-Damnjanovi\'{c} 00]}:
F. Herbut, \& M. Damnjanovi\'{c},
``Mixed-state twin observables'',
{\em J. Phys. A} {\bf 33}, ?, 6023-?;
quant-ph/0004085.

\item {\bf [Herbut 01 a]}:
F. Herbut,
``How to distinguish identical particles'',
{\em Am. J. Phys.} {\bf 69}, 2, 207-217 (2001).

\item {\bf [Herbut 01 b]}:
F. Herbut,
``On twin observables in entangled mixed states'',
quant-ph/0106038.
Superseded by {\bf [Herbut 02 a]}.

\item {\bf [Herbut 01 c]}:
F. Herbut,
``Strong twin events in mixed-state entanglement'',
quant-ph/0106101.
Superseded by {\bf [Herbut 02 a]}.

\item {\bf [Herbut 01 d]}:
F. Herbut,
``Is quantum decoherence reality or appearance?'',
quant-ph/0107049.

\item {\bf [Herbut 01 e]}:
F. Herbut,
``A theory of quantum preparation and the corresponding advantage
of the relative-collapse interpretation of quantum mechanics as
compared to the conventional one'',
quant-ph/0107064.

\item {\bf [Herbut 02 a]}:
F. Herbut,
``Hermitian Schmidt decomposition and twin observables of bipartite
mixed states'',
{\em J. Phys. A} {\bf 35}, 7, 1691-1708 (2002);
quant-ph/0305181.

\item {\bf [Herbut 02 b]}:
F. Herbut,
``Chains of quasiclassical information for bipartite correlations and the role
of twin observables'',
{\em Phys. Rev. A} {\bf 66}, 5, 052321 (2002);
quant-ph/0305187.

\item {\bf [Herbut 02 c]}:
F. Herbut,
``Necessary and sufficient range-dimension conditions for bipartite quantum
correlations'',
{\em J. Phys. A} {\bf 35}, ?, 7607-7611 (2002);
quant-ph/0305173.

\item {\bf [Herbut 03 a]}:
F. Herbut,
``The role of coherence entropy of physical twin observables in
entanglement'',
{\em J. Phys. A} {\bf 36}, 31, 8479-8495 (2003);
quant-ph/0309181.

\item {\bf [Herbut 03 b]}:
F. Herbut,
``Mixing property of quantum relative entropy'',
quant-ph/0309211.

\item {\bf [Herbut 03 c]}:
F. Herbut,
``On the meaning of entanglement in quantum measurement'',
quant-ph/0311192.

\item {\bf [Herbut 03 d]}:
F. Herbut,
``On mutual information in multipartite quantum states and equality in
strong subadditivity of entropy'',
quant-ph/0311193.

\item {\bf [Herbut 04]}:
F. Herbut,
``On compatibility and improvement of different quantum state assignments'',
{\em J. Phys. A} {\bf 37}, 1, 1-8 (2004);
quant-ph/0405135.

\item {\bf [Herbut 04 b]}:
F. Herbut,
``Distinguishing quantum measurements of observables in terms of
state transformers'',
quant-ph/0403101.

\item {\bf [Hermann 71]}:
A. Hermann,
{\em Fr\"{u}hgeschichte der Quantentheorie (1899-1913)},
Physik Verlag, Mosbach/Baden, 1971.
English version:
{\em The genesis of quantum theory (1899-1913)},
MIT Press, Cambridge, Massachusetts, 1971.

\item {\bf [Hermann 35]}:
G. Hermann,
``Die naturphilosophischen Grundlagen der
Quantenmechanik (Anzug)'',
{\em Abhandlungen de Freis'schen Schule} {\bf 6}, 75-152 (1935).

\item {\bf [Herzog-Rarity-Weinfurter-Zeilinger 94 a]}:
T. J. Herzog, J. G. Rarity
H. Weinfurter, \& A. Zeilinger,
``Frustrated two-photon creation via interference'',
{\em Phys. Rev. Lett.} {\bf 72}, 5, 629-632 (1994).
Erratum: {\em Phys. Rev. Lett.} {\bf 73}, 22, 3041 (1994).
Comment: {\bf [Senitzky 94]}.
Reply: {\bf [Herzog-Rarity-Weinfurter-Zeilinger 94 b]}.

\item {\bf [Herzog-Rarity-Weinfurter-Zeilinger 94 b]}:
T. J. Herzog, J. G. Rarity
H. Weinfurter, \& A. Zeilinger,
``Herzog {\em et al.} reply'',
{\em Phys. Rev. Lett.} {\bf 73}, 22, 3041.
Reply to {\bf [Senitzky 94]}.
See {\bf [Herzog-Rarity-Weinfurter-Zeilinger 94 a]}.

\item {\bf [Herzog-Kwiat-Weinfurter-Zeilinger 95]}:
T. J. Herzog, P. G. Kwiat, H. Weinfurter, \& A. Zeilinger,
``Complementarity and the quantum eraser'',
{\em Phys. Rev. Lett.} {\bf 75}, 17, 3034-3037 (1995).

\item {\bf [Herzog 01]}:
U. Herzog,
``Discrimination between nonorthogonal
two-photon polarization states'',
submitted to {\em Proc.\ of the 8th
Central-European Workshop on Quantum Optics, 2001};
quant-ph/0105139.

\item {\bf [Herzog-Bergou 02]}:
U. Herzog, \& J. A. Bergou,
``Minimum-error discrimination between subsets of linearly dependent quantum
states'',
{\em Phys. Rev. A} {\bf 65}, 5, 050305 (2002);
quant-ph/0112171.

\item {\bf [Herzog-Bergou 04]}:
U. Herzog, \& J. A. Bergou,
``Distinguishing mixed quantum states: Minimum-error discrimination versus optimum unambiguous discrimination'',
{\em Phys. Rev. A} {\bf 70}, 2, 022302 (2004);
quant-ph/0403124.

\item {\bf [Hess-Philipp 01 a]}:
K. Hess, \& W. Philipp,
``Einstein-separability, time related hidden parameters
for correlated spins, and the theorem of Bell'',
quant-ph/0103028.
Comments: {\bf [Gill-Weihs-Zeilinger-\.{Z}ukowski 02 a, b]}, {\bf [Mermin 02 e]},
{\bf [Myrvold 02 b]}, {\bf [Appleby 03 a]}.

\item {\bf [Hess-Philipp 01 b]}:
K. Hess, \& W. Philipp,
``?'',
{\em Proc. Nat. Acad. Sci.} {\bf 98}, ?, 14224-14227 (2001).
See {\bf [Hess-Philipp 01 c]} (II), {\bf [Philipp-Hess 02]}.
Comments: {\bf [Gill-Weihs-Zeilinger-\.{Z}ukowski 02 a, b]}, {\bf [Mermin 02 e]},
{\bf [Myrvold 02 b]}, {\bf [Appleby 03 a]}.

\item {\bf [Hess-Philipp 01 c]}:
K. Hess, \& W. Philipp,
``?'',
{\em Proc. Nat. Acad. Sci.} {\bf 98}, ?, 14227-14234 (2001).
See {\bf [Hess-Philipp 01 b]} (I), {\bf [Philipp-Hess 02]}.
Comments: {\bf [Gill-Weihs-Zeilinger-\.{Z}ukowski 02 a, b]}, {\bf [Mermin 02 e]},
{\bf [Myrvold 02 b]}, {\bf [Appleby 03 a]}.

\item {\bf [Hess-Philipp 02 a]}:
K. Hess, \& W. Philipp,
``Exclusion of time in the theorem of Bell'',
{\em Europhys. Lett.} {\bf 57}, ?, 775-781 (2002).
See {\bf [Philipp-Hess 02]}.
Comments: {\bf [Gill-Weihs-Zeilinger-\.{Z}ukowski 02 a, b]}, {\bf [Mermin 02 e]},
{\bf [Myrvold 02 b]}, {\bf [Appleby 03 a]}.

\item {\bf [Hess-Philipp 02 b]}:
K. Hess, \& W. Philipp,
``Logical inconsistencies in proofs of the theorem of Bell'',
quant-ph/0206046.

\item {\bf [Hess-Philipp 02 c]}:
K. Hess, \& W. Philipp,
``Comment on Mermin's recent proof of the theorem of Bell'',
quant-ph/0207110.
See {\bf [Mermin 02 e, f, 03 b]}.

\item {\bf [Hess-Philipp 02 d]}:
K. Hess, \& W. Philipp,
``Classical information and Mermin's non-technical proof of the theorem of Bell'',
quant-ph/0208086.
See {\bf [Mermin 02 e, f, 03 b]}.

\item {\bf [Hess-Philipp 02 e]}:
K. Hess, \& W. Philipp,
``On questions of non-locality in our EPR model'',
quant-ph/0209057.

\item {\bf [Hess-Philipp 02 f]}:
K. Hess, \& W. Philipp,
``Time and setting dependent instrument parameters and proofs of Bell-type inequalities'',
quant-ph/0211117.

\item {\bf [Hess-Philipp 02 g]}:
K. Hess, \& W. Philipp,
``Response to comment by Myrvold and Appleby'',
quant-ph/0211119.
Reply to {\bf [Myrvold 02 b]}, {\bf [Appleby 03 a]}.
See {\bf [Hess-Philipp 01 a, b, c, 02 a]}.

\item {\bf [Hess-Philipp 03 a]}:
K. Hess, \& W. Philipp,
``Exclusion of time in Mermin's proof of Bell-type inequalities'',
quant-ph/0305037.
See {\bf [Hess-Philipp 02 c, d]}, {\bf [Mermin 02 e, f, 03 b]}.

\item {\bf [Hess-Philipp 03 b]}:
K. Hess, \& W. Philipp,
``Comment on Papers by Gill and Gill, Weihs, Zeilinger and \.{Z}ukowski'',
quant-ph/0307092.
See {\bf [Gill-Weihs-Zeilinger-\.{Z}ukowski 02 a, b]}.

\item {\bf [Hess-Philipp 04 a]}:
K. Hess, \& W. Philipp,
``Breakdown of Bell's theorem for certain objective local parameter spaces'',
{\em Proc. Nat. Acad. Sci.} {\bf 101}, 1799-1805 (2004).

\item {\bf [Hess-Philipp 04 b]}:
K. Hess, \& W. Philipp,
``Bell's theorem: Critique of proofs with and without inequalities'',
quant-ph/0410015.

\item {\bf [Hessmo-Usachev-Heydari-Bj\"{o}rk 03]}:
B. Hessmo, P. Usachev, H. Heydari, \& G. Bj\"{o}rk,
``An experimental demonstration of single photon nonlocality'',
quant-ph/0311144.

\item {\bf [Hettich et al. 02]}:
C. Hettich et al.,
``?'',
{\em Science} {\bf 298}, ?, 385-? (2002).
See {\bf [Orrit 02]}.

\item {\bf [Hewitt-Horsman 03]}:
C. Hewitt-Horsman,
``The many worlds of uncertainty'',
quant-ph/0310014.

\item {\bf [Hey 99]}:
A. Hey (ed.),
{\em Feynman and computation: Exploring the limits of computers},
Perseus, Reading, Massachusetts, 1999.
Review: {\bf [Steane 99 c]}.

\item {\bf [Heydari-Bj\"{o}rk-S\'{a}nchez Soto 03]}:
H. Heydari, G. Bj\"{o}rk, \& L. L. S\'{a}nchez-Soto,
``Measurable entanglement criterion for two qubits'',
{\em Phys. Rev. A} {\bf 68}, 6, 062314 (2003);
quant-ph/0308091.

\item {\bf [Heydari-Bj\"{o}rk 04 a]}:
H. Heydari, \& G. Bj\"{o}rk,
``Entanglement criterion for pure $M\otimes N$ bipartite quantum states'',
quant-ph/0401128.

\item {\bf [Heydari-Bj\"{o}rk 04 b]}:
H. Heydari, \& G. Bj\"{o}rk,
``Entanglement measure for general multipartite quantum states'',
quant-ph/0401129.

\item {\bf [Heydari-Bj\"{o}rk 04 c]}:
H. Heydari, \& G. Bj\"{o}rk,
``Entanglement tensor for a general pure multipartite quantum state'',
quant-ph/0410124.

\item {\bf [Heywood-Redhead 83]}:
P. Heywood, \& M. L. G. Redhead,
``Nonlocality and the Kochen-Specker paradox'',
{\em Found. Phys.} {\bf 13}, 5, 481-499 (1983).

\item {\bf [Hida-Sait 99]}:
T. Hida, \& K. Sait (eds.),
{\em Quantum information:
Proc.\ of the First Int.\ Conf.\ (Meijo University,
Japan, 1997)},
World Scientific, Singapore, 1999.

\item {\bf [Hiesmayr 01]}:
B. C. Hiesmayr,
``A generalized Bell inequality and decoherence for
the $K^0 \bar{K^0}$ system'',
{\em Found. Phys. Lett.} {\bf 14}, 3, 231-245 (2001);
hep-ph/0010108.

\item {\bf [Higuchi-Sudbery 00]}:
A. Higuchi, \& A. Sudbery,
``How entangled can two couples get?'',
{\em Phys. Lett. A} {\bf 273}, 4, 213-217 (2000);
quant-ph/0005013.

\item {\bf [Higuchi-Sudbery-Szulc 03]}:
A. Higuchi, A. Sudbery, \& J. Szulc,
``One-qubit reduced states of a pure many-qubit state: Polygon
inequalities'',
{\em Phys. Rev. Lett.} {\bf 90}, 10, 107902 (2003);
quant-ph/0209085.

\item {\bf [Hiley 77]}:
B. J. Hiley,
``Foundations of quantum mechanics'',
{\em Contemp. Phys.} {\bf 18}, 4, 411-414 (1977).
Review of {\bf [d'Espagnat 76]}.

\item {\bf [Hiley-Peat 87]}:
B. J. Hiley, \& F. D. Peat (eds.),
{\em Quantum implications. Essays in honour of David Bohm},
Routledge \& Kegan Paul, London, 1987.

\item {\bf [Hiley 97]}:
B. J. Hiley,
``Quantum mechanics. Historical contingency and
the Copenhagen hegemony by James T. Cushing'',
{\em Stud. Hist. Philos. Sci. Part B:
Stud. Hist. Philos. Mod. Phys.} {\bf 28}, 2, 299-305 (1997).
See {\bf [Cushing 94 b]}.

\item {\bf [Hiley 99]}:
B. J. Hiley,
``Active information and teleportation'',
in {\bf [Greenberger-Reiter-Zeilinger 99]}, pp.~113-126.

\item {\bf [Hiley-Maroney 00]}:
B. J. Hiley, \& O. J. E. Maroney,
``Consistent histories and the Bohm approach'',
quant-ph/0009056.

\item {\bf [Hiley-Callaghan-Maroney 00]}:
B. J. Hiley, R. E. Callaghan, \& O. Maroney,
``Quantum trajectories, real, surreal or an
approximation to a deeper process?'',
quant-ph/0010020.

\item {\bf [Hilgevoord 02]}:
J. Hilgevoord,
``Time in quantum mechanics'',
{\em Am. J. Phys.} {\bf 70}, 3, 301-306 (2002).

\item {\bf [Hill-Goan 03]}:
C. D. Hill, \& H.-S. Goan,
``Fast nonadiabatic two-qubit gates for the Kane quantum computer'',
{\em Phys. Rev. A} {\bf 68}, 1, 012321 (2003).

\item {\bf [Hill-Goan 04 a]}::
C. D. Hill, \& H.-S. Goan,
`Comment on ``Grover search with pairs of trapped ions''\,',
{\em Phys. Rev. A} {\bf 69}, 5, 056301 (2004).
Comment on {\bf [Feng 01 a]}.

\item {\bf [Hill-Goan 04 b]}:
C. D. Hill, \& H.-S. Goan,
``Gates for the Kane quantum computer in the presence of dephasing'',
{\em Phys. Rev. A} {\bf 70}, 2, 022310 (2004).

\item {\bf [Hill-Wootters 97]}:
S. Hill, \& W. K. Wootters,
``Entanglement of a pair of quantum bits'',
{\em Phys. Rev. Lett.} {\bf 78}, 26, 5022-5025 (1997).

\item {\bf [Hillery-Scully 83]}:
M. Hillery, \& M. O. Scully,
``?'',
in P. Meystre, \& M. O. Scully (eds.),
{\em Quantum optics, experimental gravity and measurement
theory},
Plenum Press, New York, 1983, pp.~65-?.

\item {\bf [Hillery-Yurke 95]}:
M. Hillery, \& B. Yurke,
``Bell's theorem and beyond'',
{\em Quantum Semiclass. Opt.} {\bf 7}, 3, 215-227 (1995). Presented in the
Int. Workshop on Laser and Quantum Optics (Nathiagali, Pakistan, 1994).

\item {\bf [Hillery-Bu\v{z}ek 97]}:
M. Hillery, \& V. Bu\v{z}ek,
``Quantum copying: Fundamental inequalities'',
{\em Phys. Rev. A} {\bf 56}, 2, 1212-1216 (1997);
quant-ph/9701034.

\item {\bf [Hillery 98]}:
M. Hillery,
``Null test of quantum mechanics using two-photon down-conversion'',
{\em Phys. Rev. A} {\bf 57}, 5, 3285-3290 (1998).

\item {\bf [Hillery-Bu\v{z}ek-Berthiaume 99]}:
M. Hillery, V. Bu\v{z}ek, \& A. Berthiaume,
``Quantum secret sharing'',
{\em Phys. Rev. A} {\bf 59}, 3, 1829-1834 (1999);
quant-ph/9806063.

\item {\bf [Hillery 00]}:
M. Hillery,
``Quantum cryptography with squeezed states'',
{\em Phys. Rev. A} {\bf 61}, 2, 022309 (2000);
quant-ph/9909006.

\item {\bf [Hillery-Bu\v{z}ek 01]}:
M. Hillery, \& V. Bu\v{z}ek,
``Singlet states and the estimation of eigenstates
and eigenvalues of an unknown controlled-$U$ gate'',
{\em Phys. Rev. A} {\bf 64}, 4, 042303 (2001);
quant-ph/0104107.

\item {\bf [Hillery-Yurke 01]}:
M. Hillery, \& B. Yurke,
``Upper and lower bounds on maximal violation of local realism
in a Hardy-type test using continuous variables'',
{\em Phys. Rev. A} {\bf 63}, 6, 062111 (2001).

\item {\bf [Hillery-Bu\v{z}zek-Ziman 02]}:
M. Hillery, V. Bu\v{z}zek, \& M. Ziman,
``Probabilistic implementation of universal quantum processors'',
{\em Phys. Rev. A} {\bf 65}, 2, 022301 (2002);
quant-ph/0106088.

\item {\bf [Hillery-Ziman-Bu\v{z}ek 02]}:
M. Hillery, M. Ziman, \& V. Bu\v{z}zek,
``Implementation of quantum maps by programmable quantum processors'',
{\em Phys. Rev. A} {\bf 66}, 4, 042302 (2002);
quant-ph/0205050.

\item {\bf [Hillery-Mimih 03]}:
M. Hillery, \& J. Mimih,
``Distinguishing two-qubit states using local measurements and restricted
classical communication'',
{\em Phys. Rev. A} {\bf 67}, 4, 042304 (2003).

\item {\bf [Hillery-Bergou-Feldman 03]}:
M. Hillery, J. Bergou, \& E. Feldman,
``Quantum walks based on an interferometric analogy'',
{\em Phys. Rev. A} {\bf 68}, 3, 032314 (2003);
quant-ph/0302161.

\item {\bf [Hillery-Ziman-Bu\v{z}ek 03]}:
M. Hillery, M. Ziman, \& V. Bu\v{z}zek,
``Improving performance of probabilistic programmable quantum processors'',
quant-ph/0311170.

\item {\bf [Hillery-Feldman 03]}:
M. Hillery, \& E. Feldman,
``Scattering theory and quantum walks'',
quant-ph/0312062.

\item {\bf [Hines-McKenzie-Milburn 03]}:
A. P. Hines, R. H. McKenzie, \& G. J. Milburn,
``Entanglement of two-mode Bose-Einstein condensates'',
{\em Phys. Rev. A} {\bf 67}, 1, 013609 (2003).

\item {\bf [Hines-Dawson-McKenzie-Milburn 04]}:
A. P. Hines, C. M. Dawson, R. H. McKenzie, \& G. J. Milburn,
``Entanglement and bifurcations in Jahn-Teller models'',
{\em Phys. Rev. A} {\bf 70}, 2, 022303 (2004).

\item {\bf [Hirano-Konishi-Namiki 00]}:
T. Hirano, T. Konishi, \& R. Namiki,
``Quantum cryptography using balanced homodyne detection'',
quant-ph/0008037.

\item {\bf [Hirano-Yamanaka-Ashikaga-(+2) 03]}:
T. Hirano, H. Yamanaka, M. Ashikaga,
T. Konishi, \& R. Namiki,
``Quantum cryptography using pulsed homodyne detection'',
{\em Phys. Rev. A} {\bf 68}, 4, 042331 (2003).

\item {\bf [Hiroshima-Ishizaka 00]}:
T. Hiroshima, \& Satoshi Ishizaka,
``Local and nonlocal properties of Werner states'',
{\em Phys. Rev. A} {\bf 62}, 4, 044302 (2000);
quant-ph/0003058.

\item {\bf [Hiroshima 01 a]}:
T. Hiroshima,
``Decoherence and entanglement in two-mode squeezed vacuum states'',
{\em Phys. Rev. A} {\bf 63}, 2, 022305 (2001).

\item {\bf [Hiroshima 01 b]}:
T. Hiroshima,
``Optimal dense coding with mixed state entanglement'',
in S. Popescu, N. Linden, \& R. Jozsa (eds.),
{\em J. Phys. A} {\bf 34}, 35
(Special issue: Quantum information and computation), 6907-6912 (2001);
quant-ph/0009048.

\item {\bf [Hiroshima 02]}:
T. Hiroshima,
``An entanglement measure based on the capacity of dense coding'',
{\em Phys. Lett. A} {\bf 301}, 3-4, 263-268 (2002);
quant-ph/0204086.

\item {\bf [Hiroshima 03]}:
T. Hiroshima,
``Majorization criterion for distillability of a bipartite quantum state'',
{\em Phys. Rev. Lett.} {\bf 91}, 5, 057902 (2003);
quant-ph/0303057.

\item {\bf [Hirota-Sasaki 01]}:
O. Hirota, \& M. Sasaki,
``Entangled state based on nonorthogonal state'',
quant-ph/0101018.

\item {\bf [Hirota-van Enk-Nakamura-(+2) 01]}:
O. Hirota, S. J. van Enk, K. Nakamura,
M. Sohma, \& K. Kato,
``Entangled nonorthogonal states and their decoherence
properties'',
quant-ph/0101096.

\item {\bf [Hirota 03]}:
O. Hirota,
``Reply to comment on Y-00 quantum secure communication protocol'',
quant-ph/0312029.

\item {\bf [Hirvensalo 01]}:
M. Hirvensalo,
{\em Quantum computing},
Springer-Verlag, New York, 2001.
Report: {\bf [Moore 03]}.

\item {\bf [Hladk\'{y}-Drobn\'{y}-Bu\v{z}ek 00]}:
B. Hladk\'{y}, G. Drobn\'{y}, \& V. Bu\v{z}ek,
``Quantum synthesis of arbitrary unitary operators'',
{\em Phys. Rev. A} {\bf 61}, 2, 022102 (2000);
quant-ph/9905049.

\item {\bf [Hnilo 93]}:
A. A. Hnilo,
``On the convenience of using Greenberger-Horne-Zeilinger (GHZ) states for
testing quantum mechanics (QM) vs. objective local theories (OLT)'',
in A. van der Merwe, \& F. Selleri (eds.),
{\em Bell's theorem and the foundations of modern physics.
Proc.\ of an international conference (Cesena, Italy, 1991)},
World Scientific, Singapore, 1993, pp.~285-292.
Short version of {\bf [Hnilo 94]}.

\item {\bf [Hnilo 94]}:
A. A. Hnilo,
``On testing objective local theories by using
Greenberger-Horne-Zeilinger states'',
{\em Found. Phys.} {\bf 24}, 1, 139-162 (1994).
See {\bf [Hnilo 93]}.

\item {\bf [Hnilo-Peuriot-Santiago 02]}:
A. A. Hnilo, A. Peuriot, \& G. Santiago,
``Local realistic models tested by the EPRB experiment with random variable analyzers'',
{\em Found. Phys. Lett.} {\bf 15}, 4, 359-371 (2002).

\item {\bf [Hnizdo 97]}:
V. Hnizdo,
``EPR and the Copenhagen interpretation'',
{\em Eur. J. Phys.} {\bf 18}, 5, 404-406 (1997).
Comment on {\bf [Domingos-Nogueira-Caldeira-dos Aidos 96]}.
Reply: {\bf [Domingos-Nogueira-Caldeira-dos Aidos 98]}.

\item {\bf [Hnizdo 98]}:
V. Hnizdo,
``Common misrepresentation of the Einstein-Podolsky-Rosen
argument'',
{\em Found. Phys. Lett.} {\bf 11}, 4, 359-369 (1998);
quant-ph/0005131.

\item {\bf [Hnizdo 02]}:
V. Hnizdo,
``On Bohr's response to the clock-in-the-box thought experiment of
Einstein'',
{\em Eur. J. Phys.} {\bf 23}, 4, L9-L13 (2002);
quant-ph/0107028.
Comment on {\bf [de la Torre-Daleo-Garc\'{\i}a Mata 00]}.
Reply: {\bf [de la Torre-Daleo-Garc\'{\i}a Mata 02]}.

\item {\bf [Hockney 78]}:
D. Hockney,
``The significance of a hidden variable proof
and the logical interpretation of quantum mechanics'',
{\em Int. J. Theor. Phys.} {\bf 17}, 9, 658-707 (1978).

\item {\bf [Hockney-Kok-Dowling 03]}:
G. M. Hockney, P. Kok, \& J. P. Dowling,
``Suitability versus fidelity for rating single-photon guns'',
{\em Phys. Rev. A} {\bf 67}, 3, 032306 (2003).

\item {\bf [Hod 99]}:
S. Hod,
``Gravitation, the quantum, and Bohr's correspondence principle'',
{\em Gen. Rel. Grav.} {\bf 31}, ?, 1639-? (1999);
gr-qc/0002002.

\item {\bf [Hodgson 00]}:
P. E. Hodgson,
``Quantum philosophy'',
{\em Contemp. Phys.} {\bf 41}, 2, 105-108 (2000).

\item {\bf [Hofer 00 a]}:
W. A. Hofer,
``Measurements in quantum physics:
Towards a physical picture of relevant processes'',
talk given at {\em 6th Wigner Symp.\ (Istanbul, 1999)};
quant-ph/0003061.

\item {\bf [Hofer 00 b]}:
W. A. Hofer,
``Information transfer via the phase:
A local model of Einstein-Podolksy-Rosen experiments'',
quant-ph/0006005.

\item {\bf [Hofer 01 a]}:
W. A. Hofer,
``Simulation of Einstein-Podolsky-Rosen experiments with
limited efficiency and coherence'',
quant-ph/0103014.

\item {\bf [Hofer 01 b]}:
W. A. Hofer,
``Visibility, efficiency, and Bell violations in
real Einstein-Podolsky-Rosen experiments'',
quant-ph/0106120.

\item {\bf [Hofer 01 c]}:
W. A. Hofer,
``Numerical simulation of Einstein-Podolsky-Rosen
experiments in a local hidden variables model'',
quant-ph/0108141.

\item {\bf [Hofer 01 d]}:
W. A. Hofer,
``Numerical simulation of interference experiments
in a local hidden variables model'',
quant-ph/0111131.

\item {\bf [Hoffmann-Bostroem-Felbinger 04]}:
H. Hoffmann, K. Bostroem, \& T. Felbinger,
`Comment on ``Secure direct communication with a quantum one-time pad''\,',
quant-ph/0406115.
Comment on {\bf [Deng-Long 04]}.

\item {\bf [Hofmann 00 a]}:
H. F. Hofmann,
``Nonclassical correlations of phase noise and photon number
in quantum nondemolition measurements'',
{\em Phys. Rev. A} {\bf 61}, 033815 (2000);
quant-ph/9909013.

\item {\bf [Hofmann 00 b]}:
H. F. Hofmann,
``Information and noise in quantum measurement'',
{\em Phys. Rev. A} {\bf 62}, 2, 022103 (2000);
quant-ph/9907018.

\item {\bf [Hofmann-Ide-Kobayashi-Furusawa 00]}:
H. F. Hofmann, T. Ide, T. Kobayashi, \& A. Furusawa,
``Fidelity and information in the quantum teleportation
of continuous variables'',
{\em Phys. Rev. A} {\bf 62}, 6, 062304 (2000);
quant-ph/0003053.

\item {\bf [Hofmann-Kobayashi-Furusawa 00]}:
H. F. Hofmann, T. Kobayashi, \& A. Furusawa,
``Correlations of measurement information and noise in
quantum measurements with finite resolution'',
quant-ph/0003067.

\item {\bf [Hofmann 00 c]}:
H. F. Hofmann,
``Finite resolution measurement of the non-classical
polarization statistics of entangled photon pairs'',
quant-ph/0008122.

\item {\bf [Hofmann 00 d]}:
H. F. Hofmann,
``Information and noise in photon entanglement'',
quant-ph/0010022.

\item {\bf [Hofmann-Ide-Kobayashi-Furusawa 01 a]}:
H. F. Hofmann, T. Ide, T. Kobayashi, \& A. Furusawa,
``Information losses in continuous-variable quantum teleportation'',
{\em Phys. Rev. A} {\bf 64}, 4, 040301(R) (2001);
quant-ph/0102097.

\item {\bf [Hofmann-Ide-Kobayashi-Furusawa 01 b]}:
H. F. Hofmann, T. Ide, T. Kobayashi, \& A. Furusawa,
``Information extraction and quantum state distortions
in continuous variable quantum teleportation'',
{\em ISQM'01 (Tokyo, 2001)};
quant-ph/0110127.

\item {\bf [Hofmann-Tsujino-Takeuchi 01]}:
H. F. Hofmann, K. Tsujino, \& S. Takeuchi,
``Effects of noisy entangled state transmission on quantum teleportation'',
{\em Proc.\ Quantum Information Technology Conf.\ QIT5 (Tokyo, 2000)};
quant-ph/0111109.

\item {\bf [Hofmann-Takeuchi 02 a]}:
H. F. Hofmann, \& S. Takeuchi,
``Quantum filter for nonlocal polarization properties of photonic qubits'',
{\em Phys. Rev. Lett.} {\bf 88}, 14, 147901 (2002).

\item {\bf [Hofmann-Takeuchi 02 b]}:
H. F. Hofmann, \& S. Takeuchi,
``Quantum phase gate for photonic qubits using only beam splitters and
postselection'',
{\em Phys. Rev. A} {\bf 66}, 2, 024308 (2002).

\item {\bf [Hofmann 02]}:
H. F. Hofmann,
``Causality in quantum teleportation:
Information extraction and noise effects in entanglement distribution'',
{\em Phys. Rev. A} {\bf 66}, 3, 032317 (2002);
quant-ph/0203128.

\item {\bf [Hofmann-Takeuchi 02 c]}:
H. F. Hofmann, \& S. Takeuchi,
``Realization of quantum operations on photonic qubits by linear optics and post-selection'',
{\em Proc.\ of QIT6 (Kyoto, 2002)};
quant-ph/0204045.

\item {\bf [Hofmann 03 a]}:
H. F. Hofmann,
``Uncertainty characteristics of generalized quantum measurements'',
{\em Phys. Rev. A} {\bf 67}, 2, 022106 (2003).

\item {\bf [Hofmann 03 b]}:
H. F. Hofmann,
``Bound entangled states violate a nonsymmetric local uncertainty relation'',
{\em Phys. Rev. A} {\bf 68}, 3, 034307 (2003);
quant-ph/0305003.

\item {\bf [Hofmann-Takeuchi 03]}:
H. F. Hofmann, \& S. Takeuchi,
``Violation of local uncertainty relations as a signature of entanglement'',
{\em Phys. Rev. A} {\bf 68}, 3, 032103 (2003).

\item {\bf [Hofmann-Kojima-Takeuchi-Sasaki 03]}:
H. F. Hofmann, K. Kojima, S. Takeuchi, \& K. Sasaki,
``Entanglement and four wave mixing effects in the dissipation free
nonlinear interaction of two photons at a single atom'',
quant-ph/0301146.

\item {\bf [Hofmann-Takeuchi 04]}:
H. F. Hofmann, \& S. Takeuchi,
``Quantum-state tomography for spin-$1$ systems'',
{\em Phys. Rev. A};
quant-ph/0310003.

\item {\bf [Hofmann 04 a]}:
H. F. Hofmann,
``Complementary classical fidelities as an efficient criterion for the
evaluation of experimentally realized quantum operations'',
quant-ph/0409083.

\item {\bf [Hofmann 04 b]}:
H. F. Hofmann,
``Efficient tests for experimental quantum gates'',
quant-ph/0411011.

\item {\bf [Hofstadter 81]}:
D. R. Hofstadter,
``?'',
{\em Sci. Am.} {\bf 245}, 7 (?), ?-? (1981).
Spanish version:
``Temas metam\'{a}gicos: Falacias del principio de incertidumbre y
paradojas de la mec\'{a}nica cu\'{a}ntica'',
{\em Investigaci\'{o}n y Ciencia} 60, 108-115 (1981).

\item {\bf [Hogg 98 a]}:
T. Hogg,
``Highly structured searches with quantum computers'',
{\em Phys. Rev. Lett.} {\bf 80}, 11, 2473-2476 (1998).

\item {\bf [Hogg 98 b]}:
T. Hogg,
``A framework for structured quantum search'',
{\em Physica D} {\bf 120} 102-116 (1998).

\item {\bf [Hogg 98 c]}:
T. Hogg,
``Single-step quantum search using problem structure'',
quant-ph/9812049.

\item {\bf [Hogg-Yanik 98]}:
T. Hogg, \& M. Yanik,
``Local search for quantum computers'',
quant-ph/9802043.

\item {\bf [Hogg-Mochon-Polak-Rieffel 98]}:
T. Hogg, C. Mochon, W. Polak, \& E. Rieffel,
``Tools for quantum algorithms'',
quant-ph/9811073.

\item {\bf [Hogg 00]}:
T. Hogg,
``Quantum search heuristics'',
{\em Phys. Rev. A} {\bf 61}, 5, 052311 (2000).

\item {\bf [Hogg-Portnov 00]}:
T. Hogg, \& D. Portnov,
``Quantum optimization'',
quant-ph/0006090.

\item {\bf [Hogg 00]}:
T. Hogg,
``Solving random satisfiability problems with quantum computers'',
quant-ph/0104048.

\item {\bf [Hogg 03]}:
T. Hogg,
``Adiabatic quantum computing for random satisfiability problems'',
{\em Phys. Rev. A} {\bf 67}, 2, 022314 (2003).

\item {\bf [Hohenester 02]}:
U. Hohenester,
``Massive creation of entangled exciton states in semiconductor quantum dots'',
{\em Phys. Rev. B} {\bf 66}, 24, 245323 (2002).

\item {\bf [Hohenester-Sifel-Koskinen 03]}:
U. Hohenester, C. Sifel, \& P. Koskinen,
``Single scatterings in single artificial atoms:
Quantum coherence and entanglement'',
{\em Phys. Rev. B} {\bf 68}, 24, 245304 (2003).

\item {\bf [Holbrook-Kribs-Laflamme 04]}:
J. A. Holbrook, D. W. Kribs, \& R. Laflamme,
``Noiseless subsystems and the structure of the commutant in quantum error
correction'',
{\em Quant. Inf. Proc.};
quant-ph/0402056.

\item {\bf [Holbrow-Galvez-Parks 02]}:
C. H. Holbrow, E. Galvez, \& M. E. Parks,
``Photon quantum mechanics and beam splitters'',
{\em Am. J. Phys.} {\bf 70}, 3, 260-265 (2002).

\item {\bf [Holdsworth-Hooke 83]}:
D. G. Holdsworth, \& C. A. Hooke,
``A critical survey of quantum logic'',
{\em Scientia} {\bf ?}, ?
(Special issue: Logic in the 20th Century),
127-246 (1983).

\item {\bf [Holevo 72]}:
A. S. Holevo (Kholevo),
``An analogue of the theory of statistical decisions in noncommutative probability theory'',
{\em Trans. Moscow Math. Soc.} {\bf 26}, 133-149 (1972).

\item {\bf [Holevo 73 a]}:
A. S. Holevo (Kholevo),
``?'',
{\em Probl. Peredachi. Inf.} {\bf 9}, 3, 3-11 (1973).
English version:
``Some estimates of the information transmitted by a quantum communication channel'',
{\em Probl. Inf. Transm.} {\bf 9}, ?, 177-183 (1973).

\item {\bf [Holevo 73 b]}:
A. S. Holevo (Kholevo),
``?'',
{\em Probl. Peredachi. Inf.} {\bf 9}, ?, 110-? (1973).
English version:
``Information theoretical aspects of quantum measurement'',
{\em Probl. Inf. Transm.} {\bf 9}, 2, 31-42 (1973).

\item {\bf [Holevo 73 c]}:
A. S. Holevo (Kholevo),
``Statistical decision theory for quantum systems'',
{\em J. Multivar. Anal.} {\bf 3}, ?, 337-394 (1973).

\item {\bf [Holevo 77]}:
A. S. Holevo (Kholevo),
``Problems in the mathematical theory of quantum communication
channels'',
{\em Rep. Math. Phys.} {\bf 12}, ?, 273-278 (1977).

\item {\bf [Holevo 79]}:
A. S. Holevo (Kholevo),
``On the capacity of a quantum communications channel'',
{\em Probl. Peredachi. Inf.} {\bf 15}, ?, 3-11 (1979).
English version in {\em Probl. Inf. Transm.} {\bf 15}, ?, 247-253 (1979).

\item {\bf [Holevo 82]}:
A. S. Holevo (Kholevo),
{\em Probabilistic and statistical aspects of quantum theory},
North-Holland, Amsterdam, 1982.

\item {\bf [Holevo 97 a]}:
A. S. Holevo (Kholevo),
``On quantum communication channels with constrained inputs'',
quant-ph/9705054.

\item {\bf [Holevo 97 b]}:
A. S. Holevo (Kholevo),
``Coding theorems for quantum communication channels'',
quant-ph/9708046.

\item {\bf [Holevo 98 a]}:
A. S. Holevo (Kholevo),
``The capacity of quantum channel with general signal states'',
{\em IEEE Trans. Inf. Theory} {\bf 44}, ?, 269-273 (1998);
quant-ph/9611023.

\item {\bf [Holevo 98 b]}:
A. S. Holevo (Kholevo),
``Sending quantum information with Gaussian states'', in
{\em Proc.\ of the 4th Int.\ Conf.\ on
Quantum Communication, Measurement and Computing (Evanston, Illinois, 1998)};
quant-ph/9809022.

\item {\bf [Holevo 98 c]}:
A. S. Holevo (Kholevo),
``Coding theorems for quantum channels'',
extended version of report appeared in
{\em Tamagawa University Research Review} no. 4, 1998;
quant-ph/9809023.

\item {\bf [Holevo-Sohma-Hirota 99]}:
A. S. Holevo (Kholevo), M. Sohma, \& O. Hirota,
``Capacity of quantum Gaussian channels'',
{\em Phys. Rev. A} {\bf 59}, 3, 1820-1828 (1999).

\item {\bf [Holevo 99]}:
A. S. Holevo (Kholevo),
``Reliability function of general classical-quantum channel'',
{\em IEEE Trans. Inf. Theory};
quant-ph/9907087.

\item {\bf [Holevo-Werner 01]}:
A. S. Holevo (Kholevo), \& R. F. Werner,
``Evaluating capacities of bosonic Gaussian channels'',
{\em Phys. Rev. A} {\bf 63}, 3, 032312 (2001);
quant-ph/9912067.

\item {\bf [Holevo 02 a]}:
A. S. Holevo (Kholevo),
``On entanglement-assisted classical capacity'',
{\em J. Math. Phys.} {\bf 43}, 9, 4326-4333 (2002);
quant-ph/0106075.

\item {\bf [Holevo 02 b]}:
A. S. Holevo (Kholevo),
``On states, channels and purification'',
quant-ph/0204077.

\item {\bf [Holevo 02 c]}:
A. S. Holevo (Kholevo),
``Entanglement-assisted capacity of constrained quantum channel'',
extended version of paper presented at {\em Quantum Informatics Symposium
(Zvenigorod, 2002)};
quant-ph/0211170.

\item {\bf [Holevo 02 d]}:
A. S. Holevo (Kholevo),
``Remarks on the classical capacity of quantum channel'',
quant-ph/0212025.

\item {\bf [Holevo-Shirokov 03]}:
A. S. Holevo (Kholevo), \& M. E. Shirokov,
``On Shor's channel extension and constrained channels'',
quant-ph/0306196.

\item {\bf [Holevo 03]}:
A. S. Holevo (Kholevo),
``Asymptotic estimation of shift parameter of a quantum state'',
quant-ph/0307225.

\item {\bf [Holevo-Shirokov 04]}:
A. S. Holevo (Kholevo), \& M. E. Shirokov,
``Continuous ensembles and the $\chi$-capacity of infinite-dimensional
channels'',
quant-ph/0408176.

\item {\bf [Holladay 98]}:
W. G. Holladay,
``The nature of particle-wave complementarity'',
{\em Am. J. Phys.} {\bf 66}, 1, 27-33 (1998).

\item {\bf [Holladay 00]}:
W. G. Holladay,
``Quantum theory -- Interpretation, formulation, inspiration'',
{\em Phys. Today} {\bf 53}, 9, ? (2000).
Comment on {\bf [Fuchs-Peres 00 a]}.
Reply: {\bf [Fuchs-Peres 00 b]}.

\item {\bf [Holland-Walls-Zller 91]}:
M. J. Holland, D. F. Walls, \& P. Zoller,
``Quantum nondemolition measurements of photon
number by atomic-beam deflection'',
{\em Phys. Rev. Lett.} {\bf 67}, 13, 1716-1719 (1991).

\item {\bf [Holland 88]}:
P. R. Holland,
``Causal interpretation of a system of two spin-$\frac{1}{2}$ particles'',
{\em Phys. Rep.} {\bf 169}, 5, 293-327 (1988).

\item {\bf [Holland-Vigier 91]}:
P. R. Holland, \& J.-P. Vigier,
``Comment on `Experimental test of the de Broglie guided-wave theory
for photons'\,'',
{\em Phys. Rev. Lett.} {\bf 67}, 3, 402 (1991).
Comment on {\bf [Wang-Zou-Mandel 91 b]}.
Reply: {\bf [Wang-Zou-Mandel 91 b]}.

\item {\bf [Holland 92]}:
P. R. Holland,
``The Dirac equation in the de Broglie-Bohm theory of motion'',
{\em Found. Phys.} {\bf 22}, 10, 1287-1301 (1992).

\item {\bf [Holland 93]}:
P. R. Holland,
{\em The quantum theory of motion. An
account of the de Broglie-Bohm causal interpretation of quantum mechanics},
Cambridge University Press, Cambridge, 1993.
Reviews: {\bf [Pagonis 93]}, {\bf [Goldstein 94 c]},
{\bf [Knight 94 a]}, {\bf [Dickson 96 c]}.

\item {\bf [Holland 98]}:
P. R. Holland,
``New trajectory interpretation of quantum mechanics'',
{\em Found. Phys.} {\bf 28}, 6, 881-912 (1998).

\item {\bf [Holland 99]}:
P. R. Holland,
``Uniqueness of paths in quantum mechanics'',
{\em Phys. Rev. A} {\bf 60}, 6, 4326-4330 (1999).

\item {\bf [Holland 00]}:
P. R. Holland,
``Quantum weirdness'',
{\em Nature} {\bf 406}, 6792, 123 (2000).
Review of {\bf [Styer 00 a]}.

\item {\bf [Holland 04 a]}:
P. R. Holland,
``What's wrong with Einstein's 1927 hidden-variable interpretation of
quantum mechanics?'',
quant-ph/0401017.

\item {\bf [Holland 04 b]}:
P. R. Holland,
``Computing the wavefunction from trajectories: particle and wave pictures
in quantum mechanics and their relation'',
quant-ph/0405145.

\item {\bf [Hollenberg 00]}:
L. C. L. Hollenberg,
``Fast quantum search algorithms in protein sequence
comparison---quantum biocomputing'',
quant-ph/0002076.

\item {\bf [Hollenberg-Salgueiro-Nemes 01]}:
L. C. L. Hollenberg, A. N. Salgueiro, \& M. C. Nemes,
``Cavity QED Deutsch quantum computer'',
{\em Phys. Rev. A} {\bf 64}, 4, 042309 (2001).

\item {\bf [Holstein 01]}:
B. Holstein,
``The odd quantum by Sam B. Treiman'',
{\em Am. J. Phys.} {\bf 69}, 3, 395-396 (2001).
Review of {\bf [Treiman 99]}.

\item {\bf [Holt 73]}:
R. A. Holt,
``Atomic cascade experiments'',
Ph.\ D. thesis,
Harvard University, Cambridge, Massachusetts, 1973.

\item {\bf [Holt-Pipkin 74]}:
R. A. Holt, \& F. M. Pipkin,
``Quantum mechanics vs. hidden variables: Polarization correlation
measurement on an atomic mercury cascade'',
unpublished,
Harvard University, Cambridge, Massachusetts, 1974.

\item {\bf [Holton 80]}:
G. Holton (ed.),
{\em Certain selected ``classic'' papers in quantum physics.
Vol. I: Back and Goudsmit to Debye.
Vol. II: Dennison to Pauli.
Vol. III: Planck to Wilson},
133 informally collected papers duplicated in entirely by the
Xerox Flow-camera process or microfilmed from the original publication
(or the collected works) of each author, arranged alphabetically by first
author in 3 volumes (and microfilms),
Department of Physics, Harvard University, Cambridge, Massachusetts, 1980 (?).

\item {\bf [Holton 00]}:
G. Holton,
``Werner Heisenberg and Albert Einstein'',
{\em Phys. Today} {\bf 53}, 7, 38-42 (2000).

\item {\bf [Home-Sengupta 84]}:
D. Home, \& S. Sengupta,
``Bell's inequality and
non-contextual dispersion-free states'',
{\em Phys. Lett. A} {\bf 102}, 4, 159-162 (1984).
See {\bf [Elby 90 c]}, {\bf [Home-Sengupta 91]}, {\bf [Elby 91]}.

\item {\bf [Home-Marshall 85]}:
D. Home, \& T. W. Marshall,
``A stochastic local realist model for the EPR atomic-cascade experiment
which reproduces the quantum-mechanical coincidence rates'',
{\em Phys. Lett. A} {\bf 113}, 4, 183-186 (1985).

\item {\bf [Home-Sengupta 91]}:
D. Home, \& S. Sengupta,
``A comment on
`Critique of Home and Sengupta's derivation of a Bell inequality'\,'',
{\em Found. Phys. Lett.} {\bf 4}, 5, 451-454 (1991).
Comment on {\bf [Elby 90 c]}.
Reply: {\bf [Elby 91]}.
See {\bf [Home-Sengupta 84]}.

\item {\bf [Home-Selleri 91 a]}:
D. Home, \& F. Selleri,
``Neutral kaon physics from the point of view of realism'',
{\em J. Phys. A} {\bf 24}, 18, L1073-L1078 (1991).

\item {\bf [Home-Selleri 91 b]}:
D. Home, \& F. Selleri,
``Bell's theorem and the EPR paradox'',
{\em Rivista del Nuovo Cimento} {\bf 14}, 9, 1-95 (1991).

\item {\bf [Home-Nair 94]}:
D. Home, \& R. Nair,
``Wave function collapse as a nonlocal quantum effect'',
{\em Phys. Lett. A} {\bf 187}, 3, 224-226 (1994).

\item {\bf [Home 94]}:
D. Home,
``Position and contextuality in Bohm's causal
completion of quantum mechanics'',
{\em Phys. Lett. A} {\bf 190}, 5-6, 353-356 (1994).

\item {\bf [Home-Agarwal 95]}:
D. Home, \& G. S. Agarwal,
``Quantum nonlocality of a single photon states'',
{\em Phys. Lett. A} {\bf 209}, 1-2, 1-5 (1995).

\item {\bf [Home-Majumdar 95]}:
D. Home, \& A. S. Majumdar,
`Incompatibility between quantum mechanics and classical realism in the
``strong'' macroscopic limit',
{\em Phys. Rev. A} {\bf 52}, 6, 4959-4962 (1995).

\item {\bf [Home 97]}:
D. Home,
{\em Conceptual foundations of quantum physics:
An overview from modern perspectives},
Plenum Press, New York, 1997.
Reviews: {\bf [Cushing 98]}, {\bf [Busch 00]}, {\bf [Greenberger 01 c]}.

\item {\bf [Home-Whitaker 86]}:
D. Home, \& M. A. B. Whitaker,
``Reflections on the quantum Zeno paradox'',
{\em J. Phys. A} {\bf 19}, 10, 1847-1854 (1986).

\item {\bf [Home-Whitaker 87]}:
D. Home, \& M. A. B. Whitaker,
``The many-worlds and relative states interpretations of quantum mechanics
and the quantum Zeno paradox'',
{\em J. Phys. A} {\bf 20}, 11, 3339-3345 (1987).

\item {\bf [Home-Whitaker 92 a]}:
D. Home, \& M. A. B. Whitaker,
``A critical re-examination of the quantum Zeno paradox'',
{\em J. Phys. A} {\bf 25}, 3, 657-664 (1992).

\item {\bf [Home-Whitaker 92 b]}:
D. Home, \& M. A. B. Whitaker,
``Negative-result experiments, and the requirement of wavefunction collapse'',
{\em J. Phys. A} {\bf 25}, 8, 2387-2394 (1992).

\item {\bf [Home-Whitaker 92 c]}:
D. Home, \& M. A. B. Whitaker,
``Ensemble interpretations of quantum mechanica: A modern perspective'',
{\em Phys. Rep.} {\bf 210}, ?, 223-317 (1992).

\item {\bf [Home-Whitaker 93]}:
D. Home, \& M. A. B. Whitaker,
``A unified framework for quantum Zeno processes'',
{\em Phys. Lett. A} {\bf 173}, 4-5, 327-331 (1993).

\item {\bf [Home-Whitaker 94]}:
D. Home, \& M. A. B. Whitaker,
``Parameter dependence in the EPR-Bohm type experiment'',
{\em Phys. Lett. A} {\bf 187}, 3, 227-230 (1994).

\item {\bf [Home-Chattopadhyay 96]}:
D. Home, \& R. Chattopadhyay,
``DNA molecular cousin of Schr\"{o}dinger's cat:
A curious example of quantum measurement'',
{\em Phys. Rev. Lett.} {\bf 76}, 16, 2836-2839 (1996).
Comment: {\bf [Pearle-Squires 98]}.
Reply: {\bf [Home-Chattopadhyay 98]}.
Comment: {\bf [Chattopadhyay-Das-Gupta Bhaya 00]}.

\item {\bf [Home-Chattopadhyay 98]}:
D. Home, \& R. Chattopadhyay,
``Home and Chattopadhyaya reply'',
{\em Phys. Rev. Lett.}, 80, 6, 1349 (1996).
Reply to {\bf [Pearle-Squires 98]}.
See {\bf [Home-Chattopadhyay 96]}.

\item {\bf [Home-Whitaker 98]}:
D. Home, \& M. A. B. Whitaker,
``Quantum Zeno effect: Relevance for local realism, macroscopic realism,
and non-invasive measurability at the macroscopic level'',
{\em Phys. Lett. A} {\bf 239}, 1-2, 6-12 (1998).

\item {\bf [Home-Majumdar 99]}:
D. Home, \& A. S. Majumdar,
``On the importance of the Bohmian approach
for interpreting CP-violation experiments'',
{\em Found. Phys.} {\bf 29}, 5, 721-728 (1999);
quant-ph/9901065.

\item {\bf [Home-Samal 00]}:
D. Home, \& M. K. Samal,
``Does nonorthogonality necessarily imply
nonmaximal entanglement?'',
quant-ph/0012064.

\item {\bf [Home-Whitaker 03]}:
D. Home, \& M. A. B. Whitaker,
``Information flow and quantum cryptography using statistical fluctuations'',
{\em Phys. Rev. A} {\bf 67}, 2, 022306 (2003).
See {\bf [Larsson 03 b]}.

\item {\bf [Hong-Ou-Mandel 87]}:
C. K. Hong, Z. Y. Ou, \& L. Mandel,
``Measurements of subpicosecond time intervals between two photons by interference'',
{\em Phys. Rev. Lett.} {\bf 59}, 18, 2044-2046 (1987).
See {\bf [Herbut-Vuji\v{c}i\'{c} 97]}.

\item {\bf [Hong-Lee 02]}:
J. Hong, \& H.-W. Lee,
``Quasideterministic generation of entangled atoms in a cavity'',
{\em Phys. Rev. Lett.} {\bf 89}, 23, 237901 (2002).

\item {\bf [Hong-Jack-Yamashita 04]}:
T. Hong, M. W. Jack, \& M. Yamashita,
``On-demand single-photon state generation via nonlinear absorption'',
{\em Phys. Rev. A} {\bf 70}, 1, 013814 (2004);
quant-ph/0405131.

\item {\bf [Hong yi-Klauder 94]}:
F. Hong-yi, \& J. R. Klauder,
``Eigenvectors of two particles' relative position and total momentum'',
{\em Phys. Rev. A} {\bf 49}, 2, 704-707 (1994).

\item {\bf [Hong yi-Xiong 95]}:
F. Hong-yi, \& Y. Xiong,
``Common eigenstates of two particles' center-of-mass coordinates
and mass-weighted relative momentum'',
{\em Phys. Rev. A} {\bf 51}, 4, 3343-3346 (1995).

\item {\bf [Honner 87]}:
J. Honner,
{\em The description of nature: Niels Bohr and the philosophy of
quantum physics},
Clarendon Press, Oxford, 1987.

\item {\bf [Hood-Chapman-Lynn-Kimble 98]}:
C. J. Hood, M. S. Chapman, T. W. Lynn, \& H. J. Kimble,
``Real-time cavity QED with single atoms'',
{\em Phys. Rev. Lett.} {\bf 80}, 19, 4157-4160 (1998).
Reprinted in {\bf [Macchiavello-Palma-Zeilinger 00]}, pp.~294-297.

\item {\bf ['t Hooft 99]}:
G. 't Hooft,
``Quantum gravity as a dissipative deterministic system'',
{\em Classical and Quantum Gravity} {\bf 16}, 10, 3263-3279 (1999).

\item {\bf ['t Hooft 02 a]}:
G. 't Hooft,
``How does God play dice?
(Pre-)determinism at the Planck scale'',
in {\bf [Bertlmann-Zeilinger 02]}, pp.~307-318.

\item {\bf ['t Hooft 02 b]}:
G. 't Hooft,
``Determinism beneath quantum mechanics'',
in
{\em Quo Vadis Quantum Mechanics (Philadelphia, Pennsylvania, 2002)},
quant-ph/0212095.

\item {\bf [Hooker 70]}:
C. A. Hooker,
``Concerning Einstein's, Podolsky's, and
Rosen's objection to quantum theory'',
{\em Am. J. Phys.} {\bf 38}, 7, 851-857 (1970).
Comments: {\bf [Schlegel 71]}, {\bf [Erlichson 72]}.
See {\bf [Hooker 72]}.

\item {\bf [Hooker 71]}:
C. A. Hooker,
``?'',
{\em Philos. Sci.} {\bf 38}, ?, 244,-?; 418-? (1971).

\item {\bf [Hooker 72 a]}:
C. A. Hooker,
``Concerning measurements in quantum
theory: A critique of a recent proposal'',
{\em Int. J. Theor. Phys.} {\bf 5}, 4, 231-250 (1972).

\item {\bf [Hooker 72 b]}:
C. A. Hooker,
``The nature of quantum mechanical reality:
Einstein versus Bohr'',
in R. G. Colodny (ed.),
{\em Paradigms and paradoxes. The
philosophical challenge of the quantum domain},
University of Pittsburgh Press,
Pittsburgh, Pennsylvania, 1972, pp.~67-302.

\item {\bf [Hooker 73]}:
C. A. Hooker (ed.),
{\em Contemporary research in the
foundations and philosophy of quantum theory},
Reidel, Dordrecht, Holland, 1973.

\item {\bf [Hooker 75]}:
C. A. Hooker (ed.),
{\em The logico-algebraic approach to
quantum mechanics. Volume I: Historical evolution},
Reidel, Dordrecht, Holland, 1975.

\item {\bf [Hooker 79]}:
C. A. Hooker (ed.),
{\em The logico-algebraic approach to
quantum mechanics. Volume II. Contemporary consolidation},
Reidel, Dordrecht, Holland, 1979.

\item {\bf [Hooker 94]}:
C. A. Hooker,
``Bohr and the crisis of empirical intelligibility: An essay
on the depth of Bohr's thought and our philosophical ignorance'',
in {\bf [Faye-Folse 94]}, pp.~155-199.

\item {\bf [Horak 03]}:
P. Horak,
``The role of squeezing in quantum key distribution based on homodyne
detection and post-selection'',
quant-ph/0306138.

\item {\bf [Horgan 92]}:
J. Horgan,
``Quantum philosophy'',
{\em Sci. Am.} {\bf 267}, 1, 72-80 (1992).
Spanish version: ``Filosof\'{\i}a cu\'{a}ntica'',
{\em Investigaci\'{o}n y Ciencia} 192, 70-79 (1992).
Reprinted in {\bf [Cabello 97 c]}, pp.~36-45.

\item {\bf [Horgan 93]}:
J. Horgan,
``The artist, the physicist and the waterfall'',
{\em Sci. Am.} {\bf 268}, 2, 12 (1993).
Spanish version: ``Escher y Penrose junto a la cascada'',
{\em Investigaci\'{o}n y Ciencia} 198, 36-37 (1993).

\item {\bf [Horn-Babichev-Marzlin-(+2) 04]}:
R. T. Horn, S. A. Babichev, K.-P. Marzlin,
A. I. Lvovsky, \& B. C. Sanders,
``Single-qubit optical quantum fingerprinting'',
quant-ph/0410232.

\item {\bf [Horne 70]}:
M. A. Horne,
``Experimental consequences of local hidden
variables theories'',
Ph.\ D. thesis, Boston University, 1970.

\item {\bf [Horne-Shimony 84]}:
M. A. Horne, \& A. Shimony,
``Comment on
`Bell's theorem: Does the Clauser-Horne inequality hold for all local theories?'\,'',
{\em Phys. Rev. Lett.} {\bf 53}, 13, 1296 (1984).
Comment on {\bf [Angelidis 83]}.896.

\item {\bf [Horne-Zeilinger 85]}:
M. A. Horne, \& A. Zeilinger,
``A Bell-type EPR experiment using linear momenta'',
in P. J. Lahti, \& P. Mittelstaedt (eds.),
{\em Symp.\ on the Foundations of Modern Physics:
50 Years of the Einstein-Podolsky-Rosen gedanken Experiment
(Joensuu, Finland, 1985)},
World Scientific, Singapore, 1985, pp.~435-439.

\item {\bf [Horne-Zeilinger 86]}:
M. A. Horne, \& A. Zeilinger,
``Einstein-Podolsky-Rosen interferometry'',
in D. M. Greenberger (ed.),
{\em New techniques and ideas in quantum measurement theory.
Proc.\ of an international conference (New York, 1986),
Ann. N. Y. Acad. Sci.} {\bf 480}, 469-474 (1986).

\item {\bf [Horne-Zeilinger 88]}:
M. A. Horne, \& A. Zeilinger,
``A possible spin-less experimental test of Bell's inequality'',
in A. van der Merwe, F. Selleri, \& G.
Tarozzi (eds.),
{\em Microphysical reality and quantum formalism.
Proc.\ of an
international conference (Urbino, Italy, 1985)},
Kluwer Academic, Dordrecht, Holland, 1988, vol. 2, pp.~401-411.

\item {\bf [Horne-Shimony-Zeilinger 89]}:
M. A. Horne, A. Shimony, \& A. Zeilinger,
``Two particle interferometry'',
{\em Phys. Rev. Lett.} {\bf 62}, 19, 2209-2212 (1989).

\item {\bf [Horne-Shimony-Zeilinger 90 a]}:
M. A. Horne, A. Shimony, \& A. Zeilinger,
``Two-particle interferometry'',
{\em Nature} {\bf 347}, 6292, 429-430 (1990).

\item {\bf [Horne-Shimony-Zeilinger 90 b]}:
M. A. Horne, A. Shimony, \& A. Zeilinger,
``Introduction to two-particle interferometry'',
in A. I. Miller (ed.),
{\em Sixty-two years of uncertainty: Historical, philosophical and
physical inquiries into the foundations of quantum mechanics.
Proc.\ Int. School of History of
Science (Erice, Italy, 1989)},
Plenum Press, New York, 1990, pp.~?-?.

\item {\bf [Horne 98]}:
M. A. Horne,
``Fringe visibility for a three-particle beam-entanglement'',
{\em Fortschr. Phys.} {\bf 46}, 6-8, 683-688 (1998).

\item {\bf [Horne 99]}:
M. A. Horne,
``Schr\"{o}dinger interferometry in a gravity field as seen from free fall'',
{\em Found. Phys.} {\bf 29}, 3, 423-433 (1999).

\item {\bf [Horne 02]}:
M. A. Horne,
``On four decades of interaction with John Bell'',
in {\bf [Bertlmann-Zeilinger 02]}, pp.~99-102.

\item {\bf [Horodecki-Horodecki-Horodecki 95]}:
R. Horodecki, P. Horodecki, \& M. Horodecki,
``Violating Bell inequality by mixed spin-$\frac{1}{2}$ states:
necessary and sufficient condition'',
{\em Phys. Lett. A} {\bf 200}, 5, 340-344 (1995).
See {\bf [Aravind 95]}.

\item {\bf [Horodecki 96]}:
R. Horodecki,
``Two spin-$\frac{1}{2}$ mixtures and Bell's inequalities'',
{\em Phys. Lett. A} {\bf 210}, 4-5, 223-226 (1996).

\item {\bf [Horodecki-Horodecki 96 a]}:
R. Horodecki, \& P. Horodecki,
``Perfect
correlations in the Einstein-Podolsky-Rosen experiment and Bell's inequalities'',
{\em Phys. Lett. A} {\bf 210}, 4-5, 227-231 (1996).

\item {\bf [Horodecki-Horodecki-Horodecki 96 a]}:
R. Horodecki, P.
Horodecki, \& M. Horodecki,
``Quantum $\alpha$-entropy inequalities: Independent condition
for local realism'',
{\em Phys. Lett. A} {\bf 210}, 6, 377-381 (1996).

\item {\bf [Horodecki-Horodecki 96 b]}:
R. Horodecki, \& M. Horodecki,
``Information-theoretic aspects of inseparability'',
{\em Phys. Rev. A} {\bf 53}, 3, 1838-1843 (1996).

\item {\bf [Horodecki-Horodecki 96 c]}:
P. Horodecki, \& R. Horodecki,
``Comment on `Correlation inequalities and hidden variables'\,'',
{\em Phys. Rev. Lett.} {\bf 76}, 12, 2196-2197 (1996).
Comment on {\bf [Garg-Mermin 82 c]}.
Reply: {\bf [Mermin-Garg 96]}.

\item {\bf [Horodecki-Horodecki-Horodecki 96 b]}:
R. Horodecki, M. Horodecki, \& P. Horodecki,
``Teleportation, Bell's inequalities and inseparability'',
{\em Phys. Lett. A} {\bf 222}, 1-2, 21-25 (1996).

\item {\bf [Horodecki-Horodecki-Horodecki 96 c]}:
M. Horodecki, P. Horodecki, \& R. Horodecki,
``Separability of mixed states: Necessary and sufficient conditions'',
{\em Phys. Lett. A} {\bf 223}, 1, 1-8 (1996);
quant-ph/9605038.
See {\bf [Peres 96 d]}, {\bf [Horodecki 97 a]}.

\item {\bf [Horodecki-Horodecki-Horodecki 97 a]}:
M. Horodecki, P. Horodecki, \& R. Horodecki,
``Inseparable two spin-$\frac{1}{2}$ density matrices can
be distilled to a singlet form'',
{\em Phys. Rev. Lett.} {\bf 78}, 4, 574-577 (1997).

\item {\bf [Horodecki 97]}:
P. Horodecki,
``Separability criterion and inseparable
mixed states with positive partial transposition'',
{\em Phys. Lett. A} {\bf 232}, 5, 333-339 (1997);
quant-ph/9703004.
See {\bf [Peres 96 d]},
{\bf [Horodecki-Horodecki-Horodecki 96 c]}.

\item {\bf [Horodecki-Horodecki 97]}:
M. Horodecki, \& P. Horodecki,
``Positive maps and limits for a class of
protocols of entanglement distillation'',
quant-ph/9708015.

\item {\bf [Horodecki-Horodecki-Horodecki 97 b]}:
M. Horodecki, P.
Horodecki, \& R. Horodecki,
``Jaynes principle versus entanglement'',
quant-ph/9709010.

\item {\bf [Horodecki-Horodecki-Horodecki 98 a]}:
M. Horodecki, R. Horodecki, \& P. Horodecki,
``Optimal compression of quantum information for one-qubit source
at incomplete data: A new aspect of Jaynes principle'',
quant-ph/9803080.

\item {\bf [Horodecki 98]}:
M. Horodecki,
``Limits for compression of quantum
information carried by ensembles of mixed states'',
{\em Phys. Rev. A} {\bf 57}, 5, 3364-3363 (1998);
quant-ph/9712035.

\item {\bf [Horodecki-Horodecki-Horodecki 98 b]}:
M. Horodecki, P. Horodecki, \& R. Horodecki,
`Mixed-state entanglement and distillation: Is there a
``bound'' entanglement in nature?',
{\em Phys. Rev. Lett.} {\bf 80}, 24, 5239-5242 (1998).

\item {\bf [Horodecki-Horodecki 98]}:
M. Horodecki, \& R. Horodecki,
``Are there basic laws of quantum information processing?'',
{\em Phys. Lett. A} {\bf 244}, 6, 473-481 (1998).

\item {\bf [Horodecki-Horodecki-Horodecki 99 a]}:
P. Horodecki, M. Horodecki, \& R. Horodecki,
``Bound entanglement can be activated'',
{\em Phys. Rev. Lett.} {\bf 82}, 5, 1056-1059 (1999);
quant-ph/9806058.

\item {\bf [Horodecki-Horodecki-Horodecki 99 b]}:
R. Horodecki, M. Horodecki, \& P. Horodecki,
``Entanglement processing and statistical inference:
The Jaynes principle can produce fake entanglement'',
{\em Phys. Rev. A} {\bf 59}, 3, 1799-1803 (1999).

\item {\bf [Horodecki-Horodecki 99]}:
M. Horodecki, \& P. Horodecki,
``Reduction criterion of separability and limits for a
class of distillation protocols'',
{\em Phys. Rev. A} {\bf 59}, 6, 4206-4216 (1999).

\item {\bf [Horodecki-Horodecki-Horodecki 99 c]}:
M. Horodecki, P. Horodecki, \& R. Horodecki,
``General teleportation channel, singlet fraction, and quasidistillation'',
{\em Phys. Rev. A} {\bf 60}, 3, 1888-1898 (1999);
quant-ph/9807091.

\item {\bf [Horodecki-Horodecki-Horodecki 99 d]}:
R. Horodecki, M. Horodecki, \& P. Horodecki,
``Einstein-Podolsky-Rosen paradox without entanglement'',
{\em Phys. Rev. A} {\bf 60}, 5, 4144-4145 (1999);
quant-ph/9811004.
See {\bf [Bennett-DiVincenzo-Fuchs-(+5) 98]}.

\item {\bf [Horodecki-Smolin-Terhal-Thapliyal 99]}:
P. Horodecki, J. A. Smolin, B. M. Terhal, \& A. V. Thapliyal,
``Rank two bipartite bound entangled states do not exist'',
quant-ph/9910122.

\item {\bf [Horodecki-Lewenstein 00]}:
P. Horodecki, \& M. Lewenstein,
``Bound entanglement and continuous variables'',
{\em Phys. Rev. Lett.} {\bf 85}, 13, 2657-2660 (2000);
quant-ph/0001035.

\item {\bf [Horodecki-Horodecki-Horodecki 00 a]}:
M. Horodecki, P. Horodecki, \& R. Horodecki,
``Limits for entanglement measures'',
{\em Phys. Rev. Lett.} {\bf 84}, 9, 2014-2017 (2000);
quant-ph/9908065.

\item {\bf [Horodecki-Horodecki-Horodecki 00 b]}:
M. Horodecki, P. Horodecki, \& R. Horodecki,
``Asymptotic manipulations of entanglement can exhibit genuine
irreversibility'',
{\em Phys. Rev. Lett.} {\bf 84}, 19, 4260-4263 (2000).
Erratum: {\em Phys. Rev. Lett.} {\bf 86}, 25, 5844 (2001).
quant-ph/9912076.

\item {\bf [Horodecki 00]}:
M. Horodecki,
``Optimal compression for mixed signal states'',
{\em Phys. Rev. A} {\bf 61}, 5, 052309 (2000);
quant-ph/9905058.

\item {\bf [Horodecki-Horodecki-Horodecki 00 a]}:
M. Horodecki, P. Horodecki, \& R. Horodecki,
``Unified approach to quantum capacities:
Towards quantum noisy coding theorem'',
{\em Phys. Rev. Lett.} {\bf 85}, 2, 433-436 (2000);
quant-ph/0003040.

\item {\bf [Horodecki-Horodecki-Horodecki 00 b]}:
P. Horodecki, M. Horodecki, \& R. Horodecki,
``Binding entanglement channels'',
in V. Bu\v{z}zek, \& D. P. DiVincenzo (eds.),
{\em J. Mod. Opt.} {\bf 47}, 2-3 (Special issue:
Physics of quantum information), 347-354 (2000).

\item {\bf [Horodecki-Lewenstein-Vidal-Cirac 00]}:
P. Horodecki, M. Lewenstein, G. Vidal, \& J. I. Cirac,
``Operational criterion and constructive checks for the
separability of low-rank density matrices'',
{\em Phys. Rev. A} {\bf 62}, 3, 032310 (2000);
quant-ph/0002089.

\item {\bf [Horodecki-Horodecki-Horodecki 00 c]}:
P. Horodecki, M. Horodecki, \& R. Horodecki,
``Zero knowledge convincing protocol on quantum bit is
impossible'',
quant-ph/0010048.

\item {\bf [Horodecki 01 a]}:
P. Horodecki,
`\,``Interaction-free'' interaction:
Entangling evolution coming from the possibility of detection',
{\em Phys. Rev. A} {\bf 63}, 2, 022108 (2001);
quant-ph/9807030.

\item {\bf [Horodecki-Horodecki-Horodecki 01 a]}:
R. Horodecki, M. Horodecki, \& P. Horodecki,
``Balance of information in bipartite quantum-communication systems:
Entanglement-energy analogy'',
{\em Phys. Rev. A} {\bf 63}, 2, 022310 (2001);
quant-ph/0002021.

\item {\bf [Horodecki-Horodecki-Horodecki 01 b]}:
M. Horodecki, P. Horodecki, \& R. Horodecki,
``Separability of $n$-particle mixed states:
Necessary and sufficient conditions in terms of linear maps'',
{\em Phys. Lett. A} {\bf 283}, 1-2, 1-7 (2001);
quant-ph/0006071.

\item {\bf [Horodecki-Lewenstein 01]}:
P. Horodecki, \& M. Lewenstein,
``Is bound entanglement for continuous variables a rare
phenomenon?'',
quant-ph/0103076.

\item {\bf [Horodecki-Horodecki-Horodecki-(+2) 01]}:
M. Horodecki, P. Horodecki, R. Horodecki,
D. Leung, \& B. Terhal,
``Classical capacity of a noiseless quantum
channel assisted by noisy entanglement'',
submitted to {\em Quant. Inf. Comp.};
quant-ph/0106080.

\item {\bf [Horodecki-Horodecki-Horodecki 01 c]}:
M. Horodecki, P. Horodecki, \& R. Horodecki,
``Mixed-state entanglement and quantum communication'',
chapter in {\bf [Alber-Beth-Horodecki-(+6) 01]};
quant-ph/0109124.

\item {\bf [Horodecki 01 b]}:
P. Horodecki,
``From limits of quantum nonlinear operations to multicopy
entanglement witnesses and state spectrum estimation'',
quant-ph/0111036.

\item {\bf [Horodecki-Ekert 01]}:
P. Horodecki, \& A. K. Ekert,
``Direct detection of quantum entanglement'',
quant-ph/0111064.

\item {\bf [Horodecki 01 c]}:
P. Horodecki,
``How to measure amount of entanglement contained in unknown state'',
quant-ph/0111082.

\item {\bf [Horodecki-Ekert 02]}:
P. Horodecki, \& A. K. Ekert,
``Method for direct detection of quantum entanglement'',
{\em Phys. Rev. Lett.} {\bf 89}, 12, 127902 (2002).

\item {\bf [Horodecki-Oppenheim-Horodecki 02]}:
M. Horodecki, J. Oppenheim, \& R. Horodecki,
``Are the laws of entanglement theory thermodynamical?'',
{\em Phys. Rev. Lett.} {\bf 89}, 24, 240403 (2002);
quant-ph/0207177.

\item {\bf [Horodecki-Sen De-Sen-Horodecki 02]}:
M. Horodecki, A. Sen De, U. Sen, \& K. Horodecki,
``Local indistinguishability and LOCC monotones'',
quant-ph/0204116.
Partially supersedes by {\bf [Horodecki-Sen De-Sen-Horodecki 03]}.

\item {\bf [Horodecki-Sen De-Sen 02]}:
M. Horodecki, A. Sen De, \& U. Sen,
``The rates of asymptotic entanglement transformations for bipartite mixed
states: Maximally entangled states are not special'',
quant-ph/0207031.

\item {\bf [Horodecki-Sen De-Sen-Horodecki 03]}:
M. Horodecki, A. Sen De, U. Sen, \& K. Horodecki,
``Local indistinguishability: More nonlocality with less entanglement'',
{\em Phys. Rev. Lett.} {\bf 90}, 4, 047902 (2003);
quant-ph/0301106.
See {\bf [Ghosh-Kar-Roy-(+2) 01]}.
Partially supersedes {\bf [Horodecki-Sen De-Sen-Horodecki 02]}.

\item {\bf [Horodecki-Horodecki-Horodecki-(+4) 03]}:
M. Horodecki, K. Horodecki, P. Horodecki, R. Horodecki,
J. Oppenheim, A. Sen De, \& U. Sen,
``Local information as a resource in distributed quantum systems'',
{\em Phys. Rev. Lett.} {\bf 90}, 10, 100402 (2003);
quant-ph/0207168.

\item {\bf [Horodecki-Horodecki-Oppenheim 03]}:
M. Horodecki, P. Horodecki, \& J. Oppenheim,
``Reversible transformations from pure to mixed states and the unique measure
of information'',
{\em Phys. Rev. A} {\bf 67}, 6, 062104 (2003);
quant-ph/0212019.

\item {\bf [Horodecki-Shor-Ruskai 03]}:
M. Horodecki, P. W. Shor, \& M. B. Ruskai,
``General entanglement breaking channels'',
{\em Rev. Math. Phys.};
quant-ph/0302031.
See {\bf [Ruskai 02 d, 03]}.

\item {\bf [Horodecki-Horodecki-Oppenheim 03]}:
M. Horodecki, P. Horodecki, \& J. Oppenheim,
``Compressing compound states'',
{\em Int. J. Theor. Phys.};
quant-ph/0302139.

\item {\bf [Horodecki-Horodecki-Horodecki 03]}:
R. Horodecki, M. Horodecki, \& P. Horodecki,
``Quantum information isomorphism: Beyond the dilemma of Scylla of
ontology and Charybdis of instrumentalism'',
{\em IBM J. Res. Dev.};
quant-ph/0305024.

\item {\bf [Horodecki 03 a]}:
P. Horodecki,
``Measuring quantum entanglement without prior state reconstruction'',
{\em Phys. Rev. Lett.} {\bf 90}, 16, 167901 (2003).

\item {\bf [Horodecki-Horodecki-Sen De-Sen 03]}:
P. Horodecki, M. Horodecki, A. Sen De, \& U. Sen,
``Locally accessible information: How much can the parties gain by
cooperating?'',
{\em Phys. Rev. Lett.} {\bf 91}, 11, 117901 (2003);
quant-ph/0304040.

\item {\bf [Horodecki 03 b]}:
P. Horodecki,
``Direct estimation of elements of quantum states algebra
and entanglement detection {\em via} linear contractions'',
{\em Phys. Lett. A} {\bf 219}, 1-2, 1-7 (2003).

\item {\bf [Horodecki 03 c]}:
P. Horodecki,
``From limits of quantum operations to multicopy
entanglement witnesses and state-spectrum estimation'',
{\em Phys. Rev. A} {\bf 68}, 5, 052101 (2003).

\item {\bf [Horodecki-Sen De-Sen 03 a]}:
M. Horodecki, A. Sen De, \& U. Sen,
``Rates of asymptotic entanglement transformations for bipartite mixed states:
Maximally entangled states are not special'',
{\em Phys. Rev. A} {\bf 67}, 6, 062314 (2003).

\item {\bf [Horodecki-Horodecki-Sen De-Sen 03]}:
M. Horodecki, R. Horodecki, A. Sen De, \& U. Sen,
``No-deleting and no-cloning principles as consequences of conservation of
quantum information'',
quant-ph/0306044.

\item {\bf [Horodecki-Horodecki-Horodecki-Oppenheim 03]}:
K. Horodecki, M. Horodecki, P. Horodecki, \& J. Oppenheim,
``Secure key from bound entanglement'',
quant-ph/0309110.

\item {\bf [Horodecki-Sen De-Sen 03]}:
M. Horodecki, A. Sen De, \& U. Sen,
``Quantification of quantum correlation of ensemble of states'',
quant-ph/0310100.

\item {\bf [Horodecki-Sen De-Sen 03 b]}:
M. Horodecki, A. Sen De, \& U. Sen,
``Cloning of orthogonal mixed states entails irreversibility'',
quant-ph/0310142.

\item {\bf [Horodecki-Sen De-Sen 03 b]}:
M. Horodecki, A. Sen De, \& U. Sen,
``Dual entanglement measures based on no local cloning and no local
deleting'',
quant-ph/0403169.

\item {\bf [Horodecki-Horodecki-Horodecki-Oppenheim 04]}:
K. Horodecki, M. Horodecki, P. Horodecki, \& J. Oppenheim,
``Locking entanglement measures with a single qubit'',
quant-ph/0404096.

\item {\bf [Horodecki-Oppenheim-Sen De-Sen 04]}:
M. Horodecki, J. Oppenheim, A. Sen De, \& U. Sen,
``Distillation protocols: Output entanglement and local mutual information'',
quant-ph/0405185.

\item {\bf [Horodecki-Horodecki-Sen De-Sen 04]}:
M. Horodecki, R. Horodecki, A. Sen De, \& U. Sen,
``Common origin of no-cloning and no-deleting principles -- Conservation of
information'',
quant-ph/0407038.

\item {\bf [Horodecki-Horodecki-Horodecki-(+4) 04]}:
M. Horodecki, P. Horodecki, R. Horodecki,
J. Oppenheim, A. Sen De, U. Sen, \& B. Synak,
``Local versus non-local information in quantum information theory:
Formalism and phenomena'',
quant-ph/0410090.

\item {\bf [Horoshko-Kilin 00]}:
D. B. Horoshko, \& S. Y. Kilin,
``Quantum teleportation using quantum nondemolition technique'',
{\em Phys. Rev. A} {\bf 61}, 3, 032304 (2000);
quant-ph/9908049.

\item {\bf [Horoshko-Kilin 03]}:
D. B. Horoshko, \& S. Y. Kilin,
``Optimal dimensionality for quantum cryptography'',
{\em Opt. Spectrosc.} {\bf 94}, 691 (2003);
quant-ph/0203095.

\item {\bf [Horton-Dewdney 01 a]}:
G. Horton, \& C. Dewdney,
``A non-local, Lorentz-invariant, hidden-variable interpretation
of relativistic quantum mechanics based on particle trajectories'',
{\em J. Phys. A} {\bf 34}, 46, 9871-9878 (2001).

\item {\bf [Horton-Dewdney 01 b]}:
G. Horton, \& C. Dewdney,
``De Broglie's pilot wave theory for the Klein-Gordon equation'',
quant-ph/0109059.

\item {\bf [Horton-Dewdney 01 c]}:
G. Horton, \& C. Dewdney,
``A non-local, Lorentz-invariant, hidden-variable interpretation
of relativistic quantum mechanics based on particle trajectories'',
quant-ph/0110007.

\item {\bf [Horton-Dewdney-Ne'eman 02]}:
G. Horton, C. Dewdney, \& U. Ne'eman,
``De Broglie's pilot-wave theory for
the Klein-Gordon equation and its space-time pathologies'',
{\em Found. Phys.} {\bf 32}, 3, 463-476 (2002).

\item {\bf [Horton-Dewdney 04]}:
G. Horton, \& C. Dewdney,
``A relativistically covariant version of Bohm's quantum field theory for
the scalar field'',
{\em J. Phys. A};
quant-ph/0407089.

\item {\bf [Horvath-Thompson-Knight 97]}:
G. Z. K. Horvath, R. C. Thompson, \& P. L. Knight,
``Fundamental physics with trapped ions'',
{\em Contemp. Phys.} {\bf 38}, 1, 25-48 (1997).

\item {\bf [Hostens-Dehaene-De Moor 04]}:
E. Hostens, J. Dehaene, \& B. De Moor,
``Stabilizer states and Clifford operations for systems of arbitrary
dimensions, and modular arithmetic'',
quant-ph/0408190.

\item {\bf [Hotaling 98]}:
S. P. Hotaling,
``Radix-$R>2$ quantum computation'',
{\em Int. J. Theor. Phys.} {\bf 37}, 1, 481-485 (1998).

\item {\bf [Hotta-Morikawa 04]}:
M. Hotta, \& M. Morikawa,
``Impossibility of distant indirect measurement of the quantum Zeno effect'',
{\em Phys. Rev. A} {\bf 69}, 5, 052114 (2004);
quant-ph/0310090.

\item {\bf [Hotta-Ozawa 04]}:
M. Hotta, \& M.Ozawa,
``Quantum estimation by local observables'',
{\em Phys. Rev. A};
quant-ph/0401187.

\item {\bf [Hovis 96]}:
R. C. Hovis,
``The collected works of P. A. M. Dirac, 1924-1948'',
{\em Phys. Today} {\bf 49}, 9, 84-88 (1996).
Review of {\bf [Dirac 95]}.

\item {\bf [Howard 85]}:
D. Howard,
``Einstein on locality and separability'',
{\em Stud. Hist. Philos. Sci.} {\bf 16}, 3, 171-201 (1985).
See {\bf [Fine 86]}, {\bf [Fine 89]} (Sec. 1), {\bf [Deltete-Guy 91]},
{\bf [H\'{a}jek-Bub 92]}, {\bf [Combourieu 92]}.

\item {\bf [Howard 89]}:
D. Howard,
``Holism, separability, and
the metaphysical implications of the Bell experiments'',
in J. Cushing, \& E. McMullin (eds.),
{\em Philosophical consequences of quantum theory: Reflections on Bell's theorem},
University of Notre Dame Press, Notre Dame, Indiana, 1989, pp.~224-253.

\item {\bf [Howard 93]}:
D. Howard,
``Locality, separability, and
the physical implications of the Bell experiments'',
in A. van der Merwe, \& F. Selleri (eds.),
{\em Bell's theorem and the foundations of modern physics.
Proc.\ of an international
conference (Cesena, Italy, 1991)},
World Scientific, Singapore, 1993, pp.~306-314.

\item {\bf [Howard 94]}:
D. Howard,
``What makes a classical concept classical? Toward a
reconstruction of Niels Bohr's philosophy of physics'',
in {\bf [Faye-Folse 94]}, pp.~201-229.

\item {\bf [Howard 97]}:
D. Howard,
``Spacetime and separability:
Problems of identity and individuation in fundamental physics'',
in {\bf [Cohen-Horne-Stachel 97 b]}.

\item {\bf [Howell-Yeazell 00 a]}:
J. C. Howell, \& J. A. Yeazell,
``Quantum computation through entangling single photons
in multipath interferometers'',
{\em Phys. Rev. Lett.} {\bf 85}, 1, 198-201 (2000).

\item {\bf [Howell-Yeazell 00 b]}:
J. C. Howell, \& J. A. Yeazell,
``Entangling macroscopic quantum states'',
{\em Phys. Rev. A} {\bf 62}, 1, 012102 (2000).

\item {\bf [Howell-Yeazell 00 c]}:
J. C. Howell, \& J. A. Yeazell,
``Nondestructive single-photon trigger'',
{\em Phys. Rev. A} {\bf 62}, 3, 032311 (2000).

\item {\bf [Howell-Yeazell-Ventura 00]}:
J. C. Howell, J. A. Yeazell, \& D. Ventura,
``Optically simulating a quantum associative memory'',
{\em Phys. Rev. A} {\bf 62}, 4, 042303 (2000).

\item {\bf [Howell-Lamas Linares-Bouwmeester 02]}:
J. C. Howell, A. Lamas-Linares, \& D. Bouwmeester,
``Experimental violation of a spin-1 Bell inequality using maximally entangled
four-photon states'',
{\em Phys. Rev. Lett.} {\bf 88}, 3, 030401 (2002);
quant-ph/0105132.


\item {\bf [Howell-Bennink-Bentley-Boyd 04]}:
J. C. Howell, R. S. Bennink, S. J. Bentley, \& R. W. Boyd,
``Realization of the Einstein-Podolsky-Rosen paradox using momentum-
and position-entangled photons from spontaneous parametric down conversion'',
{\em Phys. Rev. Lett.} {\bf 92}, 21, 210403 (2004);
quant-ph/0309122.

\item {\bf [H\o{}yer 99]}:
P. H\o{}yer,
``Conjugated operators in quantum algorithms'',
{\em Phys. Rev. A} {\bf 59}, 5, 3280-3289 (1999).

\item {\bf [H\o{}yer 00]}:
P. H\o{}yer,
``Arbitrary phases in quantum amplitude amplification'',
{\em Phys. Rev. A} {\bf 62}, 5, 052304 (2000).

\item {\bf [H\o{}yer-Neerbek 00]}:
P. H\o{}yer, \& J. Neerbek,
``Bounds on quantum ordered searching'',
quant-ph/0009032.
See {\bf [H\o{}yer-Neerbek-Shi 01]}.

\item {\bf [H\o{}yer-Neerbek-Shi 01]}:
P. H\o{}yer, J. Neerbek, \& Y. Shi,
``Quantum bounds for ordered searching and sorting'',
a combined version of {\bf [H\o{}yer-Neerbek 00]}
and {\bf [Shi 00]};
quant-ph/0102078.

\item {\bf [H\o{}yer-de Wolf 01]}:
P. H\o{}yer, \& R. de Wolf,
``Improved quantum communication complexity bounds for disjointness and equality'',
quant-ph/0109068.

\item {\bf [H\o{}yer-Mosca-de Wolf 03]}:
P. H\o{}yer, M. Mosca, \& R. de Wolf,
``Quantum search on bounded-error inputs'',
{\em Proc.\ ICALP, 2003};
quant-ph/0304052.

\item {\bf [Hoyt 68]}:
G. D. Hoyt,
``On explaining the meaning of the wave function'',
{\em Am. J. Phys.} {\bf 36}, 4, 368-369 (1968).

\item {\bf [Hradil 97]}:
Z. Hradil,
``Quantum state estimation'',
{\em Phys. Rev. A} {\bf 55}, 3, R1561-R1564 (1997).

\item {\bf [Hradil-Summhammer-Rauch 99]}:
Z. Hradil, J. Summhammer, \& H. Rauch,
``Quantum tomography as normalization of incompatible observations'',
{\em Phys. Lett. A} {\bf 261}, 1-2, 20-24 (1999);
quant-ph/9806014.

\item {\bf [Hradil-Summhammer-Badurek-Rauch 00]}:
Z. Hradil, J. Summhammer, G. Badurek, \& H. Rauch,
``Reconstruction of the spin state'',
{\em Phys. Rev. A} {\bf 62}, 1, 014101 (2000);
quant-ph/9911068.

\item {\bf [Hradil-Du\v{s}ek 99]}:
Z. Hradil, \& M. Du\v{s}ek,
``Analogy between optimal spin estimation and interferometry'',
quant-ph/9911071.

\item {\bf [Hradil-Summhammer 00]}:
Z. Hradil, \& J. Summhammer,
``Quantum theory of incompatible observations'',
{\em J. Phys. A} {\bf 33}, 42, 7607-7612 (2000);
quant-ph/9911067.

\item {\bf [Hrushovski-Pitowsky 03]}:
E. Hrushovski, \& I. Pitowsky,
``Generalizations of Kochen and Specker's theorem and the effectiveness of
Gleason's theorem'',
{\em Robert Clifton Memorial Conf.};
quant-ph/0307139,
PITT-PHIL-SCI00001263.

\item {\bf [Hsieh-Li 02]}:
J.-Y. Hsieh, \& C.-M. Li,
``General $SU(2)$ formulation for quantum searching with certainty'',
{\em Phys. Rev. A} {\bf 65}, 5, 052322 (2002);
quant-ph/0112035.

\item {\bf [Hsieh-Li-Chuu 04]}:
J.-Y. Hsieh, C.-M. Li, \& D.-S. Chuu,
``A simplification of entanglement purification'',
{\em Phys. Lett. A} {\bf 328}, 2-3, 94-101 (2004).

\item {\bf [Hsieh-Kempe-Myrgren-Whaley 03]}:
M. Hsieh, J. Kempe, S. Myrgren, \& K. B. Whaley,
``An explicit universal gate-set for exchange-only quantum computation'',
quant-ph/0309002.

\item {\bf [Hsu-Chen 68]}:
L.-Y. Hsu, \& Y.-Y. Chen,
``Faster database search in quantum computing'',
quant-ph/0102068.

\item {\bf [Hsu 02 a]}:
L.-Y. Hsu,
``Nonlocality swapping'',
{\em Phys. Rev. A} {\bf 65}, 6, 062302 (2002).

\item {\bf [Hsu 02 b]}:
L.-Y. Hsu,
``Optimal entanglement purification via entanglement
swapping with least classical communication'',
{\em Phys. Lett. A} {\bf 297}, 3-4, 126-128 (2002).

\item {\bf [Hsu 02 c]}:
L.-Y. Hsu,
``Optimal information extraction in probabilistic teleportation'',
{\em Phys. Rev. A} {\bf 66}, 1, 012308 (2002).

\item {\bf [Hsu 03 a]}:
L.-Y. Hsu,
``Optimal probabilistic teleportation of an unknown $N$-level qudit via information extraction'',
{\em Phys. Lett. A} {\bf 311}, 6, 459-464 (2003).

\item {\bf [Hsu 03 b]}:
L.-Y. Hsu,
``Quantum secret-sharing protocol based on Grover's algorithm'',
{\em Phys. Rev. A} {\bf 68}, 2, 022306 (2003).

\item {\bf [Hu 02 a]}:
B. L. Hu,
``Decoherence of two-level systems can be very different from Brownian particles'',
talk given at the {\em Workshop on Mechanisms of Decoherence (Austin, 2001)};
quant-ph/0203001.

\item {\bf [Hu 02 b]}:
C.-R. Hu,
``Family of sure-success quantum algorithms for solving a generalized Grover
search problem'',
{\em Phys. Rev. A} {\bf 66}, 4, 042301 (2002);
quant-ph/0201049.

\item {\bf [Hu-Li-Ju 03]}:
J. Du, H. Li, \& C. Ju,
``Quantum games of asymmetric information'',
{\em Phys. Rev. E} {\bf 68}, 1, 016124 (2003).

\item {\bf [Hu 90]}:
W. Hu,
``The curious quantum mechanics of pre- and post-selected ensembles'',
{\em Found. Phys.} {\bf 20}, 4, 447-458 (1990).
See {\bf [Zachar-Alter 91]}.

\item {\bf [Hu-Das Sarma 00]}:
X. Hu, \& S. Das Sarma,
``Hilbert-space structure of a solid-state quantum computer:
Two-electron states of a double-quantum-dot artificial molecule'',
{\em Phys. Rev. A} {\bf 61}, 6, 062301 (2000).

\item {\bf [Hu-Das Sarma 01]}:
X. Hu, \& S. Das Sarma,
``Spin-based quantum computation in multielectron quantum dots'',
{\em Phys. Rev. A} {\bf 64}, 4, 042312 (2001);
cond-mat/0101102.

\item {\bf [Hu-de Sousa-Das Sarma 01]}:
X. Hu, R. de Sousa, \& S. Das Sarma,
``Interplay between Zeeman coupling
and swap action in spin-based quantum computer models: Error correction
in inhomogeneous magnetic fields'',
{\em Phys. Rev. Lett.} 86, ?, 918-921 (2001).

\item {\bf [Hu-Das Sarma 02]}:
X. Hu, \& S. Das Sarma,
``Gate errors in solid-state quantum-computer architectures'',
{\em Phys. Rev. A} {\bf 66}, 1, 012312 (2002).

\item {\bf [Hu-Das Sarma 03]}:
X. Hu, \& S. Das Sarma,
``Spin-swap gate in the presence of qubit inhomogeneity in a double quantum dot'',
{\em Phys. Rev. A} {\bf 68}, 5, 052310 (2003).

\item {\bf [Hu-Das Sarma 04]}:
X. Hu, \& S. Das Sarma,
``Double quantum dot turnstile as an electron spin entangler'',
{\em Phys. Rev. B} {\bf 69}, 11, 115312 (2004).

\item {\bf [Huang-Zhang-Hou 00]}:
M. Huang, Y. Zhang, \& G. Hou,
``Classical capacity of a quantum multiple-access channel'',
{\em Phys. Rev. A} {\bf 62}, 5, 052106 (2000);
quant-ph/9911120.

\item {\bf [Huang-Li-Li-(+3) 01]}:
Y.-F. Huang, W.-L. Li, C.-F. Li,
Y.-S. Zhang, Y.-K. Jiang, \& G.-C. Guo,
``Optical realization of universal quantum cloning'',
{\em Phys. Rev. A} {\bf 64}, 1, 012315 (2001).

\item {\bf [Huang-Li-Zhang-Guo 01]}:
Y.-F. Huang, C.-F. Li, Y.-S. Zhang, \& G.-C. Guo,
``Experimental test of CHSH inequality for non-maximally entangled states'',
{\em Phys. Lett. A} {\bf 287}, 5-6, 317-320 (2001).

\item {\bf [Huang-Li-Zhang-(+2) 03]}:
Y.-F. Huang, C.-F. Li, Y.-S. Zhang,
J.-W. Pan, \& G.-C. Guo,
``Experimental test of the Kochen-Specker theorem with single photons'',
{\em Phys. Rev. Lett.} {\bf 90}, 25, 250401 (2003);
quant-ph/0209038.
See {\bf [Simon-\.{Z}ukowski-Weinfurter-Zeilinger 00]}.

\item {\bf [Huang-Ren-Zhang-(+2) 04]}:
Y.-F. Huang, X.-F. Ren, Y.-S. Zhang, L.-M. Duan, \& G.-C. Guo,
``Experimental teleportation of a quantum controlled-NOT gate'',
quant-ph/0408007.

\item {\bf [Huang-Zhan 04]}:
Y.-X. Huang, \& M.-S. Zhan,
``Remote preparation of multipartite pure state'',
{\em Phys. Lett. A} {\bf 327}, 5-6, 404-408 (2004).

\item {\bf [Huang-Osenda-Kais 04]}:
Z. Huang, O. Osenda, \& S. Kais,
``Entanglement of formation for one-dimensional magnetic systems with defects'',
{\em Phys. Lett. A} {\bf 322}, 3-4, 137-145 (2004).

\item {\bf [Huberman-Hogg 03]}:
B. A. Huberman, \& T. Hogg,
``Quantum solution of coordination problems'',
quant-ph/0306112.

\item {\bf [Huertas Rosero 04]}:
A. F. Huertas-Rosero,
``Classification of quantum symmetric non-zero sum $2 \times 2$ games in the Eisert
scheme'',
M.\ Sc. thesis, Universidad de Los Andes, Colombia;
quant-ph/0402117.

\item {\bf [Huesmann-Balzer-Courteille-(+2) 99]}:
R. Huesmann, C. Balzer, P. Courteille, W. Neuhauser, \& P. E.
Toschek,
``Single-atom interferometry'',
{\em Phys. Rev. Lett.} {\bf 82}, 8, 1611-1615 (1999).

\item {\bf [Hughes 81]}:
R. I. G. Hughes,
``Quantum logic'',
{\em Sci. Am.} {\bf 245}, 4, 146-157 (1981).
Spanish version: ``L\'{o}gica cu\'{a}ntica'',
{\em Investigaci\'{o}n y Ciencia} 63, 80-94 (1981).

\item {\bf [Hughes-van Fraassen 88]}:
R. I. G. Hughes, \& B. C. van Fraassen,
``Can the measurement problem be solved by superselection rules?'',
{\em ?} {\bf ?}, ?, ?-? (1988).

\item {\bf [Hughes 89]}:
R. I. G. Hughes,
{\em The structure and interpretation of quantum mechanics},
Harvard University Press, Cambridge, Massachusetts, 1989.

\item {\bf [Hughes-Alde-Dyer-(+3) 95]}:
R. J. Hughes, D. M. Alde, P. Dyer, G. G. Luther, G. L. Morgan, \& M. Schauer,
``Quantum cryptography'',
{\em Contemp. Phys.} {\bf 36}, 3, 149-163 (1995);
quant-ph/9504002.

\item {\bf [Hughes-James-Knill-(+2) 96]}:
R. J. Hughes, D. F. V. James, E. H. Knill, R. Laflamme, \& A. G. Petschek,
``Decoherence bounds on quantum computation with trapped ions'',
{\em Phys. Rev. Lett.} {\bf 77}, 15, 3240-3243 (1996).

\item {\bf [Hughes-Luther-Morgan-(+2) 96]}:
R. J. Hughes, G. Luther, G. L. Morgan, C. G. Peterson, \& C. Simmons,
``Quantum cryptography over underground optical fibers'',
in {\em Lecture notes in Computer Science} {\bf 1109}, 329-342
(1996).

\item {\bf [Hughes 97}:
R. J. Hughes,
``Quantum security is spookily certain'',
{\em Nature} {\bf 385}, 6611, 17-18 (1997).

\item {\bf [Hughes-James-G\'{o}mez-(+12) 98]}:
R. J. Hughes, D. F. V. James, J. J. G\'{o}mez, M. S. Gulley,
M. H. Holzscheiter, P. G. Kwiat, S. K. Lamoreaux, C. G. Peterson,
V. D. Sandberg, M. M. Schauer, C. M. Simmons, C. E. Thorburn,
D. Tupa, P. Z. Wang, \& A. G. White,
``The Los Alamos trapped ion quantum computer experiment'',
{\em Fortschr. Phys.} {\bf 46}, 4-5, 329-361 (1998);
quant-ph/9708050.

\item {\bf [Hughes-James 98]}:
R. J. Hughes, \& D. F. V. James,
``Prospects for quantum computation with trapped ions'',
{\em Fortschr. Phys.} {\bf 46}, 6-8, 759-769 (1998).

\item {\bf [Hughes 98]}:
R. J. Hughes,
``Cryptography, quantum computation and trapped ions'',
in A. K. Ekert, R. Jozsa, \& R. Penrose (eds.),
{\em Quantum Computation: Theory and Experiment.
Proceedings of a Discussion Meeting held at the Royal
Society of London on 5 and 6 November 1997},
{\em Philos. Trans. R. Soc. Lond. A} {\bf 356}, 1743, 1853-1866 (1998);
quant-ph/9712054.

\item {\bf [Hughes-Nordholt 99]}:
R. J. Hughes, \& J. E. Nordholt,
``Quantum cryptography takes to the air'',
{\em Phys. World} {\bf 12}, 5, 31-35 (1999).

\item {\bf [Hughes-Buttler-Kwiat-(+4) 99]}:
R. J. Hughes, W. T. Buttler, P. G. Kwiat, S. K. Lamoreaux, G. L. Morgan,
J. E. Nordholt, \& C. G. Peterson,
``Practical quantum cryptography for secure free-space communications'',
quant-ph/9905009.
See {\bf [Hughes-Nordholt 99]}, {\bf [Hughes-Buttler-Kwiat-(+5) 99]},
{\bf [Hughes-Buttler-Kwiat-(+4) 00]}.

\item {\bf [Hughes-Buttler-Kwiat-(+5) 99]}:
R. J. Hughes, W. T. Buttler, P. G. Kwiat, G. G. Luther, G. L. Morgan,
J. E. Nordholt, C. G. Peterson, \& C. M. Simmons,
``Secure communications with low-orbit spacecraft using quantum cryptography'',
patent US5966224, 1999.
See {\bf [Hughes-Nordholt 99]},
{\bf [Hughes-Buttler-Kwiat-(+4) 99, 00]}.

\item {\bf [Hughes-Morgan-Peterson 00]}:
R. J. Hughes, G. L. Morgan, \& C. G. Peterson,
``Quantum key distribution over a 48 km optical fibre network'',
in V. Bu\v{z}zek, \& D. P. DiVincenzo (eds.),
{\em J. Mod. Opt.} {\bf 47}, 2-3 (Special issue:
Physics of quantum information), 533-547 (2000);
quant-ph/9904038.

\item {\bf [Hughes-Buttler-Kwiat-(+4) 00]}:
R. J. Hughes, W. T. Buttler, P. G. Kwiat, S. K. Lamoreaux, G. L. Morgan,
J. E. Nordholt, \& C. G. Peterson,
``Free-space quantum key distribution in daylight'',
in V. Bu\v{z}zek, \& D. P. DiVincenzo (eds.),
{\em J. Mod. Opt.} {\bf 47}, 2-3 (Special issue:
Physics of quantum information), 549-562 (2000).
See {\bf [Hughes-Nordholt 99]}, {\bf [Hughes-Buttler-Kwiat-(+4) 99]},
{\bf [Hughes-Buttler-Kwiat-(+5) 99]}.

\item {\bf [Hughes-Nordholt-Derkacs-Peterson 02]}:
R. J. Hughes, J. E. Nordholt, D. Derkacs, \& C. G. Peterson,
``Practical free-space quantum key distribution over 10 km in daylight and at night'',
{\em New J. Phys} {\bf 4}, 43.1-43.14 (2002).

\item {\bf [Hughes-Nordholt 04]}:
R. J. Hughes, \& J. E. Nordholt,
`Comment on ``Quantum key distribution with $1.25$ Gbps clock
synchronization'' by J. C. Bienfang et al., quant-ph/0405097',
quant-ph/0407050.
Comment on {\bf [Bienfang-Gross-Mink-(+9) 04]}.
Reply: {\bf [Bienfang-Clark-Williams-(+2) 04]}.

\item {\bf [Hughston-Jozsa-Wootters 93]}:
L. P. Hughston, R. Jozsa, \& W. K. Wootters,
``A complete classification of quantum ensembles having a given density matrix'',
{\em Phys. Lett. A} {\bf 183}, 1, 14-18 (1993).
See {\bf [Kirkpatrick 03 c]}.

\item {\bf [Hultgren-Shimony 77]}:
B. O. Hultgren III, \& A. Shimony,
``The lattice of verifiable propositions of the spin-1 system'',
{\em J. Math. Phys.} {\bf 18}, 3, 381-394 (1977).

\item {\bf [Hulpke-Bru\ss-Lewenstein-Sanpera 04]}:
F. Hulpke, D. Bru\ss, M. Lewenstein, \& A. Sanpera,
``Simplifying Schmidt number witnesses via higher-dimensional embeddings'',
quant-ph/0401118.


\item {\bf [Hulpke-Bru\ss 04]}:
F. Hulpke, \& D. Bru\ss,
``A two-way algorithm for the entanglement problem'',
quant-ph/0407179.

\item {\bf [Humphreys 80]}:
P. Humphreys,
``A note on Demopoulos' paper
`Locality and the algebraic structure of quantum mechanics'\,'',
in {\bf [Suppes 80]}, pp.~145-147.

\item {\bf [Hunter-Andersson-Gilson-Barnett 03]}:
K. Hunter, E. Andersson, C. R. Gilson, \& S. M. Barnett,
``Maximum fidelity for a mirror symmetric set of qubit states'',
{\em J. Phys. A} {\bf 36}, 14, 4159-4172 (2003).

\item {\bf [Hunter 03]}:
K. Hunter,
``Measurement does not always aid state discrimination'',
{\em Phys. Rev. A} {\bf 68}, 1, 012306 (2003).

\item {\bf [Huntington-Ralph 02]}:
E. H. Huntington, \& T. C. Ralph,
``Continuous-variable Bell-type correlations from two bright squeezed beams'',
{\em Phys. Rev. A} {\bf 65}, 1, 012306 (2002).

\item {\bf [Huntington-Ralph 03]}:
E. H. Huntington, \& T. C. Ralph,
``Components for optical qubits in the radio frequency basis'',
quant-ph/0311073.

\item {\bf [Hunziker-Meyer-Park-(+3) 02]}:
M. Hunziker, D. A. Meyer, J. Park,
J. Pommersheim, \& M. Rothstein,
``The geometry of quantum learning'',
quant-ph/0309059.

\item {\bf [Hupkes 04]}:
H. J. Hupkes,
``Unconditional security of practical quantum key distribution'',
quant-ph/0402170.

\item {\bf [Hutchinson-Milburn 04]}:
G. D. Hutchinson, \& G. J. Milburn,
``Nonlinear quantum optical computing via measurement'',
{\em J. Mod. Opt.} {\bf 20}, 51, 8, 1211-1222 (2004);
quant-ph/0409198.

\item {\bf [Huttner-Peres 94]}:
B. Huttner, \& A. Peres,
``Quantum cryptography with photon pairs'',
in S. M. Barnett, A. K. Ekert, \& S. J. D. Phoenix (eds.),
{\em J. Mod. Opt.} {\bf 41}, 12 (Special issue: Quantum
communication), 2397-2404 (1994).

\item {\bf [Huttner-Ekert 94]}:
B. Huttner, \& A. K. Ekert,
``Information gain in quantum eavesdropping'',
in S. M. Barnett, A. K. Ekert, \& S. J. D. Phoenix (eds.),
{\em J. Mod. Opt.} {\bf 41}, 12 (Special issue: Quantum
communication), 2455-2466 (1994).

\item {\bf [Huttner-Imoto-Gisin-Mor 95]}:
B. Huttner, N. Imoto, N. Gisin, \& T. Mor,
``Quantum cryptography with coherent states'',
{\em Phys. Rev. A} {\bf 51}, 3, 1863-1869 (1995);
quant-ph/9502020.

\item {\bf [Huttner 96]}:
B. Huttner,
``Quantum communication arrives in the laboratory'',
{\em Phys. World} {\bf 9}, 10, 20-21 (1996).

\item {\bf [Huttner-Muller-Gautier-(+2) 96]}:
B. Huttner, A. Muller, J. D. Gautier, H. Zbinden, \& N. Gisin,
``Unambiguous quantum measurement of nonorthogonal states'',
{\em Phys. Rev. A} {\bf 54}, 5, 3783-3789 (1996).

\item {\bf [Huttner-Imoto-Barnett 96]}:
B. Huttner, N. Imoto, \& S. M. Barnett,
``Short distance applications of quantum cryptography'',
{\em J. Nonlinear Opt. Phys. Mater.} {\bf 5}, 823-832 (1996).

\item {\bf [Hutton-Bose 02 a]}:
A. Hutton, \& S. Bose,
``Resource reduction via repeaters in entanglement distribution'',
{\em Phys. Rev. A} {\bf 65}, 2, 022321 (2002);
quant-ph/0103023.

\item {\bf [Hutton-Bose 02 b]}:
A. Hutton, \& S. Bose,
``Comparison of star and ring topologies for entanglement distribution'',
{\em Phys. Rev. A} {\bf 66}, 3, 032320 (2002);
quant-ph/0202116.

\item {\bf [Hutton-Bose 04]}:
A. Hutton, \& S. Bose,
``Ground state entanglement in a combination of star and ring geometries
of interacting spins'',
quant-ph/0408077.

\item {\bf [Hwang-Koh-Han 98]}:
W.-Y. Hwang, I. G. Koh, \& Y. D. Han,
``Quantum cryptography without public announcement of bases'',
{\em Phys. Lett. A} {\bf 244}, 6, 489-494 (1998);
quant-ph/9702009.

\item {\bf [Hwang-Lee-Ahn-Hwang 00]}:
W.-Y. Hwang, H. Lee, D. Ahn, \& S. W. Hwang,
``Efficient schemes for reducing imperfect collective decoherences'',
{\em Phys. Rev. A} {\bf 62}, 6, 062305 (2000);
quant-ph/0008050.

\item {\bf [Hwang-Ahn-Hwang-Lee 00]}:
W.-Y. Hwang, D. (D.) Ahn, S. W. Hwang, \& H. Lee,
``Relativity of entanglement'',
quant-ph/0011010.

\item {\bf [Hwang-Ahn-Hwang-Han 00]}:
W.-Y. Hwang, D. (D.) Ahn, S. W. Hwang, \& Y. D. Han,
``Entangled quantum clocks for measuring proper-time difference'',
quant-ph/0012003.

\item {\bf [Hwang-Ahn-Hwang 01 a]}:
W.-Y. Hwang, D. (D.) Ahn, \& S. W. Hwang,
``Correlated errors in quantum-error corrections'',
{\em Phys. Rev. A} {\bf 63}, 2, 022303 (2001);
quant-ph/0008049.

\item {\bf [Hwang-Ahn-Hwang 01 a]}:
W.-Y. Hwang, D. (D.) Ahn, \& S. W. Hwang,
``Eavesdropper's optimal information in variations of
Bennett-Brassard 1984 quantum key distribution in the coherent
attacks'',
{\em Phys. Lett. A} {\bf 279}, 3-4, 133-138 (2001);
quant-ph/0009006.

\item {\bf [Hwang-Ahn-Hwang 01]}:
W.-Y. Hwang, D. (D.) Ahn, \& S. W. Hwang,
``Quantum gambling using two nonorthogonal states'',
{\em Phys. Rev. A} {\bf 64}, 6, 064302 (2001);
quant-ph/0010103.

\item {\bf [Hwang-Kim-Lee 01]}:
W.-Y. Hwang, J. Kim, \& H.-W. Lee,
``Shor-Preskill type security-proofs for concatenated
six-state quantum key distribution scheme'',
quant-ph/0110079.

\item {\bf [Hwang-Matsumoto 02 a]}:
W.-Y. Hwang, \& K. Matsumoto,
``Entanglement measures with weak-monotonicity as lower (upper)
bound for the entanglement of cost (distillation)'',
{\em Phys. Lett. A} {\bf 300}, 6, 581-585 (2002);
quant-ph/0110080.

\item {\bf [Hwang-Matsumoto 02 b]}:
W.-Y. Hwang, \& K. Matsumoto,
``Quantum gambling using three nonorthogonal states'',
{\em Phys. Rev. A} {\bf 66}, 5, 052311 (2002);
quant-ph/0203148.

\item {\bf [Hwang-Matsumoto 02 c]}:
W.-Y. Hwang, \& K. Matsumoto,
``Irreversibility of entanglement manipulations:
Vagueness of the entanglement of cost and distillation'',
quant-ph/0203116.

\item {\bf [Hwang-Yuen 02]}:
W.-Y. Hwang, \& H. P. Yuen,
``Quantum key distribution with high loss: Toward global secure
communication'',
quant-ph/0211153.

\item {\bf [Hwang-Wang-Matsumoto-(+2) 03 a]}:
W.-Y. Hwang, X. Wang, K. Matsumoto, J. Kim, \& H.-W. Lee,
``Shor-Preskill type security-proof for the
quantum key distribution without public announcement of bases'',
{\em Phys. Rev. A} {\bf 67}, 1, 012302 (2003);
quant-ph/0201053.

\item {\bf [Hwang-Wang-Matsumoto-(+2) 03 a]}:
W.-Y. Hwang, X. Wang, K. Matsumoto, J. Kim, \& H.-W. Lee,
``Shor-Preskill-type security proof for concatenated Bennett-Brassard 1984
quantum-key-distribution protocol'',
{\em Phys. Rev. A} {\bf 67}, 2, 024302 (2003).

\item {\bf [Hwang-Matsumoto 03]}:
W.-Y. Hwang, \& K. Matsumoto,
``Irreversibility of entanglement manipulations: Vagueness of the entanglement
of cost and entanglement of distillation'',
{\em Phys. Lett. A} {\bf 310}, 2-3, 119-122 (2003).

\item {\bf [Hwang 03]}:
W.-Y. Hwang,
``Quantum key distribution with high loss: Toward global secure communication'',
{\em Phys. Rev. Lett.} {\bf 91}, 5, 057901 (2003).

\item {\bf [Hyafil-Mozley-Perrin-(+5) 04]}:
P. Hyafil, J. Mozley, A. Perrin,
J. Tailleur, G. Nogues, M. Brune,
J.-M. Raimond, \& S. Haroche,
``Coherence-preserving trap architecture for long-term control of giant
Rydberg atoms'',
{\em Phys. Rev. Lett.};
quant-ph/0407202.

\item {\bf [Hyllus-Alves-Bru\ss-Macchiavello 04]}:
P. Hyllus, C. M. Alves, D. D. Bru\ss, \& C. Macchiavello,
``Generation and detection of bound entanglement'',
quant-ph/0405164.


\newpage

\subsection{}


\item {\bf [Iannuzzi-Merlo 01]}:
M. Iannuzzi, \& V. Merlo,
``Interpretation of the experiments on Bell's inequality'',
{\em Found. Phys. Lett.} {\bf 14}, 4, 355-365 (2000).

\item {\bf [Iblisdir-Van Assche-Cerf 03]}:
S. Iblisdir, G. Van Assche, \& N. J. Cerf,
``Security of quantum key distribution with
coherent states and homodyne detection'',
quant-ph/0312018.

\item {\bf [Iblisdir-Gisin 04]}:
S. Iblisdir, \& N. Gisin,
``Byzantine agreement with two quantum key distribution setups'',
quant-ph/0405167.

\item {\bf [Iblisdir-Roland 04]}:
S. Iblisdir, \& J. Roland,
``Optimal finite measurements and Gauss quadratures'',
quant-ph/0410237.

\item {\bf [Icaza 91]}:
J. J. Icaza,
{\em La construcci\'{o}n de la mec\'{a}nica cu\'{a}ntica},
Universidad del Pa\'{\i}s Vasco, Bilbao, 1991.

\item {\bf [ID Quantique 01]}:
{\em ID Quantique},
www.idquantique.com.

\item {\bf [Ide-Hofmann-Kobayashi-Furusawa 01]}:
T. Ide, H. F. Hofmann, T. Kobayashi, \& A. Furusawa,
``Continuous variable teleportation of single photon states'',
{\em ISQM (Tokyo, 2001)};
quant-ph/0112018.

\item {\bf [Ide-Hofmann-Kobayashi-Furusawa 02]}:
T. Ide, H. F. Hofmann, T. Kobayashi, \& A. Furusawa,
``Continuous-variable teleportation of single-photon states'',
{\em Phys. Rev. A} {\bf 65}, 1, 012313 (2002);
quant-ph/0104014.

\item {\bf [Ide-Hofmann-Furusawa-Kobayashi 02]}:
T. Ide, H. F. Hofmann, A. Furusawa, \& T. Kobayashi,
``Gain tuning and fidelity in continuous-variable quantum teleportation'',
{\em Phys. Rev. A} {\bf 65}, 6, 062303 (2002);
quant-ph/0111127.

\item {\bf [Ikram-Zhu-Zubairy 00]}:
M. Ikram, S.-Y. Zhu, \& M. S. Zubairy,
``Quantum teleportation of an entangled state'',
{\em Phys. Rev. A} {\bf 62}, 2, 022307 (2000).

\item {\bf [Ikram-Zubairy 02]}:
M. Ikram, \& M. S. Zubairy,
``Reconstruction of an entangled state in a cavity via Autler-Townes
spectroscopy'',
{\em Phys. Rev. A} {\bf 65}, 4, 044305 (2002).

\item {\bf [Ikram-Saif 02]}:
M. Ikram, \& F. Saif,
``Engineering entanglement between two cavity modes'',
{\em Phys. Rev. A} {\bf 66}, 1, 014304 (2002).
Erratum: {\em Phys. Rev. A} {\bf 67}, 6, 069901 (2003);
quant-ph/0201044.

\item {\bf [Il'ichev 03]}:
L. V. Il'ichev,
``Extremal unraveling of a quantum operation'',
{\em JETP} {\bf 96}, 982 (2003).

\item {\bf [Imamo\={g}lu-Awschalom-Burkard-(+4) 99]}:
A. Imamo\={g}lu, D. D. Awschalom, G. Burkard, D. P. DiVincenzo,
D. Loss, M. Sherwin, \& A. Small,
``Quantum information processing using quantum dot spins and cavity QED'',
{\em Phys. Rev. Lett.} {\bf 83}, 20, 4204-4207 (1999);
quant-ph/9904096.

\item {\bf [Imamo\={g}lu 00]}:
A. Imamo\={g}lu,
``Quantum computation using quantum dot spins and microcavities'',
{\em Fortschr. Phys.} {\bf 48}, 9-11 (Special issue: Experimental proposals for quantum computation), 987-997 (2000).

\item {\bf [Imamo\={g}lu 02]}:
A. Imamo\={g}lu,
``High efficiency photon counting using stored light'',
{\em Phys. Rev. Lett.} {\bf 89}, 16, 163602 (2002).

\item {\bf [Inagaki-Namiki-Tajiri 92]}:
S. Inagaki, M. Namiki, \& T. Tajiri,
``Possible observation of the quantum Zeno effect by means of
neutron spin-flipping'',
{\em Phys. Lett. A} {\bf 166}, 1, 5-12 (1992).

\item {\bf [Inamori 00 a]}:
H. Inamori,
``Security of EPR-based quantum key distribution'',
quant-ph/0008064.

\item {\bf [Inamori 00 b]}:
H. Inamori,
``Security of EPR-based quantum key distribution using three
bases'',
quant-ph/0008076.

\item {\bf [Inamori-Rallan-Vedral 01]}:
H. Inamori, L. Rallan, \& V. Vedral,
``Security of EPR-based quantum cryptography against incoherent
symmetric attacks'',
in S. Popescu, N. Linden, \& R. Jozsa (eds.),
{\em J. Phys. A} {\bf 34}, 35
(Special issue: Quantum information and computation), 6913-6918 (2001);
quant-ph/0103058.

\item {\bf [Inamori-L\"{u}tkenhaus-Mayers 01]}:
H. Inamori, N. L\"{u}tkenhaus, \& D. Mayers,
``Unconditional security of practical quantum key distribution'',
quant-ph/0107017.

\item {\bf [Inglis 61]}:
D. R. Inglis,
``Completeness of quantum mechanics and
charge-conjugation correlations of theta particles'',
{\em Rev. Mod. Phys.} {\bf 33}, 1, 1-7 (1961).

\item {\bf [Ingraham 94]}:
R. L. Ingraham,
``Quantum nonlocality in a delayed-choice experiment
with partial, controllable memory erasing'',
{\em Phys. Rev. A} {\bf 50}, 6, 4502-4505 (1994).
Comment: {\bf [Aharonov-Popescu-Vaidman 95]}.

\item {\bf [Inoue-Waks-Yamamoto 02]}:
K. Inoue, E. Waks, \& Y. Yamamoto,
``Differential phase shift quantum key distribution'',
{\em Phys. Rev. Lett.} {\bf 89}, 3, 037902 (2002).

\item {\bf [Inoue-Santori-Waks-Yamamoto 03]}:
K. Inoue, C. Santori, E. Waks, \& Y. Yamamoto,
``Entanglement-based quantum key distribution without an entangled-photon
source'',
{\em Phys. Rev. A} {\bf 67}, 6, 062319 (2003).

\item {\bf [Inoue-Waks-Yamamoto 03]}:
K. Inoue, E. Waks, \& Y. Yamamoto,
``Differential-phase-shift quantum key distribution using coherent light'',
{\em Phys. Rev. A} {\bf 68}, 2, 022317 (2003).

\item {\bf [Introzzi 02]}:
G. Introzzi,
``Foundations and interpretation of quantum mechanics'',
{\em Found. Phys.} {\bf 32}, 10, 1635-1637 (2002).

\item {\bf [Inui-Konishi-Konno 04]}:
N. Inui, Y. Konishi, \& N. Konno,
``Localization of two-dimensional quantum walks'',
{\em Phys. Rev. A} {\bf 69}, 5, 052323 (2004).

\item {\bf [Ioannou-Travaglione-Cheung-Ekert 04]}:
L. M. Ioannou, B. C. Travaglione, D. C. Cheung, \& A. K. Ekert,
``An improved algorithm for quantum separability and entanglement
detection'',
quant-ph/0403041.

\item {\bf [Ioffe-Feigel'man-Ioselevich-(+3) 02]}:
L. B. Ioffe, M. V. Feigel'man, A. Ioselevich,
D. Ivanov, M. Troyer, \& G. Blatter,
``Topologically protected quantum bits using Josephson junction arrays'',
{\em Nature} {\bf 415}, 6871, 503-506 (2002).

\item {\bf [Ioffe-Feigel'man 02]}:
L. B. Ioffe, \& M. V. Feigel'man,
``Possible realization of an ideal quantum computer in Josephson junction
array'',
{\em Phys. Rev. B} {\bf 66}, 22, 224503 (2002).

\item {\bf [Ioffe-Geshkenbein-Helm-Blatter 04]}:
L. B. Ioffe, V. B. Geshkenbein, C. Helm, \& G. Blatter,
``Decoherence in superconducting quantum bits by phonon radiation'',
{\em Phys. Rev. Lett.} {\bf 93}, 5, 057001 (2004).

\item {\bf [Ionicioiu-Zanardi-Rossi 00]}:
R. Ionicioiu, P. Zanardi, \& F. Rossi,
``Testing Bell's inequality with ballistic electrons in semiconductors'',
{\em Phys. Rev. A} {\bf 63}, 5, 050101(R) (2001);
quant-ph/0009026.

\item {\bf [Ionicioiu-Amaratunga-Udrea 00]}:
R. Ionicioiu, G. Amaratunga, \& F. Udrea,
``Quantum computation with ballistic electrons'',
quant-ph/0011051.

\item {\bf [Ionicioiu-Zanardi 02]}:
R. Ionicioiu, \& P. Zanardi,
``Quantum-information processing in bosonic lattices'',
{\em Phys. Rev. A} {\bf 66}, 5, 050301 (2002);
quant-ph/0204118.

\item {\bf [Ionicioiu-D'Amico 03]}:
R. Ionicioiu, \& I. D'Amico,
``Mesoscopic Stern-Gerlach device to polarize spin currents'',
{\em Phys. Rev. B} {\bf 67}, 4, 041307 (2003).

\item {\bf [Ionicioiu 03]}:
R. Ionicioiu,
``Quantum gates with topological phases'',
{\em Phys. Rev. A} {\bf 68}, 3, 034305 (2003).

\item {\bf [Ionicioiu-Popescu 03]}:
R. Ionicioiu, \& A. E. Popescu,
``Single spin measurement using spin-orbital entanglement'',
quant-ph/0310047.

\item {\bf [Iqbal-Toor 01 a]}:
A. Iqbal, \& A. H. Toor,
``Evolutionarily stable strategies in quantum games'',
{\em Phys. Lett. A} {\bf 280}, 5-6, 249-256 (2001);
quant-ph/0007100.

\item {\bf [Iqbal-Toor 01 b]}:
A. Iqbal, \& A. H. Toor,
``Entanglement and dynamic stability of Nash
equilibrium in a symmetric quantum game'',
{\em Phys. Lett. A} {\bf 286}, 4, 245-250 (2001);
quant-ph/0101106.

\item {\bf [Iqbal-Toor 01 c]}:
A. Iqbal, \& A. H. Toor,
``Equilibria of replicator dynamic in quantum games'',
quant-ph/0106135.

\item {\bf [Iqbal-Toor 02 a]}:
A. Iqbal, \& A. H. Toor,
``Quantum mechanics gives stability to a Nash equilibrium'',
{\em Phys. Rev. A} {\bf 65}, 2, 022306 (2002);
quant-ph/0104091.

\item {\bf [Iqbal-Toor 02 b]}:
A. Iqbal, \& A. H. Toor,
``Quantum cooperative games'',
{\em Phys. Lett. A} {\bf 293}, 3-4, 103-108 (2002).
quant-ph/0108091.
Comment: {\bf [Dai-Chen 04]}.

\item {\bf [Iqbal-Toor 02 c]}:
A. Iqbal, \& A. H. Toor,
``Backwards-induction outcome in a quantum game'',
{\em Phys. Rev. A} {\bf 65}, 5, 052328 (2002);
quant-ph/0111090.

\item {\bf [Iqbal-Toor 02 d]}:
A. Iqbal, \& A. H. Toor,
``Darwinism in quantum systems?'',
{\em Phys. Lett. A} {\bf 294}, 5-6, 261-270 (2002);
quant-ph/0103085.

\item {\bf [Iqbal-Toor 02 e]}:
A. Iqbal, \& A. H. Toor,
``Quantum repeated games'',
{\em Phys. Lett. A} {\bf 300}, 6, 541-546 (2002);
quant-ph/0203044.

\item {\bf [Iqbal 02]}:
A. Iqbal,
``Quantum games with a multi-slit electron diffraction setup'',
quant-ph/0207078.

\item {\bf [Iqbal-Wegert 03]}:
A. Iqbal, \& S. Weigert,
``Quantum correlation games'',
{\em J. Phys. A};
quant-ph/0306176.

\item {\bf [Iqbal 04]}:
A. Iqbal,
``Quantum correlations and Nash equilibria of a bi-matrix game'',
{\em J. Phys. A} {\bf 37}, ?, L353-L359 (2004);
quant-ph/0406107.

\item {\bf [Iqbal-Toor 04]}:
A. Iqbal, \& A. H. Toor,
``Stability of mixed Nash equilibria in symmetric quantum games'',
{\em Comm. Theor. Phys.} (2004);
quant-ph/0106056.

\item {\bf [Irby 03]}:
V. D. Irby,
``Physical limitations on quantum nonlocality in the detection of photons
emitted from positron-electron annihilation'',
{\em Phys. Rev. A} {\bf 67}, 3, 034102 (2003).

\item {\bf [Irvine-Lamas Linares-de Dood-Bouwmeester 04]}:
W. T. M. Irvine, A. Lamas-Linares, M. J. A. de Dood, \& D. Bouwmeester,
``Optimal quantum cloning on a beam splitter'',
{\em Phys. Rev. Lett.} {\bf 92}, 4, 047902 (2004).

\item {\bf [Irvine-Hodelin-Simon-Bouwmeester 04]}:
W. T. M. Irvine, J. F. Hodelin, C. Simon, \& D. Bouwmeester,
``Realisation of Hardy's thought experiment'',
quant-ph/0410160.
See {\bf [Hardy 92 a]}.

\item {\bf [Ischi 01]}:
B. Ischi,
``Generalized quantum mechanics for
separated systems'',
{\em Found. Phys. Lett.} {\bf 14}, 6, 501-518 (2001).

\item {\bf [Isham 95]}:
C. J. Isham,
{\em Lectures on quantum theory. Mathematical
and structural foundations},
Imperial College Press (World Scientific), London, 1995.
Reviews: {\bf [Mayer 96]}, {\bf [Gibbons 96]}.

\item {\bf [Isham-Linden 97]}:
C. J. Isham, \& N. Linden,
``Information entropy and the space of decoherence
functions in generalized quantum theory'',
{\em Phys. Rev. A} {\bf 55}, 6, 4030-4040 (1997).

\item {\bf [Isham-Linden-Savvidou-Schreckenberg 97]}:
C. J. Isham, N. Linden, K. Savvidou, \& S. Schreckenberg,
``Continuous time and consistent histories'',
quant-ph/9711031.

\item {\bf [Isham-Butterfield 98]}:
C. J. Isham, \& J. N. Butterfield,
``A topos perspective on the Kochen-Specker theorem:
I. Mathematical development'',
{\em Int. J. Theor. Phys.} {\bf 37}, 11, 2669-2733 (1998);
quant-ph/9803055.
See {\bf [Isham-Butterfield 99]} (II),
{\bf [Hamilton-Isham-Butterfield 00]} (III),
{\bf [Hamilton 00]},
{\bf [Butterfield-Isham 02]} (IV).

\item {\bf [Isham-Butterfield 99]}:
C. J. Isham, \& J. N. Butterfield,
``A topos perspective on the Kochen-Specker theorem: II. Conceptual
aspects, and classical analogues'',
{\em Int. J. Theor. Phys.} {\bf 38}, 3, 827-859 (1999).
quant-ph/9808067.
See {\bf [Isham-Butterfield 98]} (I),
{\bf [Hamilton-Isham-Butterfield 00]} (III),
{\bf [Hamilton 00]},
{\bf [Butterfield-Isham 02]} (IV).

\item {\bf [Isham-Butterfield 00]}:
C. J. Isham, \& J. N. Butterfield,
``Some possible roles for topos theory in
quantum theory and quantum gravity'',
{\em Found. Phys.} {\bf 30}, 10, 1707-1735 (2000).

\item {\bf [Ishizaka-Hiroshima 00]}:
S. Ishizaka, \& T. Hiroshima,
``Maximally entangled mixed states under nonlocal
unitary operations in two qubits'',
{\em Phys. Rev. A} {\bf 62}, 2, 022310 (2000);
quant-ph/0003023.

\item {\bf [Ishizaka 01]}:
S. Ishizaka,
``Quantum channel locally interacting with environment'',
{\em Phys. Rev. A} {\bf 63}, 3, 034301 (2001);
quant-ph/0011069.

\item {\bf [Ishizaka 03]}:
S. Ishizaka,
``Analytical formula connecting entangled states and the closest disentangled
state'',
{\em Phys. Rev. A} {\bf 67}, 6, 060301 (2003);
quant-ph/0203084.

\item {\bf [Ishizaka 04]}:
S. Ishizaka,
``Binegativity and geometry of entangled states in two qubits'',
{\em Phys. Rev. A} {\bf 69}, 2, 020301 (2004);
quant-ph/0308056.
See {\bf [Audenaert-De Moor-Vollbrecht-Werner 02]}.

\item {\bf [Israel-Gasca 95]}:
G. Israel, \& A. Mill\'{a}n Gasca,
{\em Il mondo come gioco matematico. John von Neumann, scienziato del Novecento},
La Nuova Italia Scientifica, Roma 1995.
Spanish version:
{\em El mundo como un juego matem\'{a}tico. John von Neumann, un cient\'{\i}fico del siglo XX},
Nivola, Madrid, 2001.

\item {\bf [Israelit 96]}:
M. Israelit,
``Nathan Rosen: 1909-1995'',
{\em Found. Phys. Lett.} {\bf 9}, 2, 105-108 (1996).

\item {\bf [Itano-Heinzen-Bollinger-Wineland 90]}:
W. M. Itano, D. J. Heinzen, J. J. Bollinger, \& D. J. Wineland,
``Quantum Zeno effect'',
{\em Phys. Rev. A} {\bf 41}, 5, 2295-2300 (1990).
Comment: {\bf [Ballentine 91]}.
Reply: {\bf [Itano-Heinzen-Bollinger-Wineland 91]}.
See {\bf [Bollinger-Itano-Heinzen-Wineland 89]}.

\item {\bf [Itano-Heinzen-Bollinger-Wineland 91]}:
W. M. Itano, D. J. Heinzen, J. J. Bollinger, \& D. J. Wineland,
`Reply to ``Comment on `Quantum Zeno effect''\,',
{\em Phys. Rev. A} {\bf 43}, 9, 5168-5169 (1991).
Reply to {\bf [Ballentine 91]}.

\item {\bf [Itano 03]}:
W. M. Itano,
`Comment on ``Some implications of the quantum nature of laser fields for
quantum computations''\,',
{\em Phys. Rev. A} {\bf 68}, 4, 046301 (2003).
Comment on {\bf [Gea Banacloche 02 a]}.
Replies: {\bf [van Enk-Kimble 03]}, {\bf [Gea Banacloche 03]}.

\item {\bf [Ivanovic 81]}:
I. D. Ivanovic,
``Geometrical description of quantal state determination'',
{\em J. Phys. A} {\bf 14}, 3241-3245 (1981).

\item {\bf [Ivanovic 83]}:
I. D. Ivanovic,
``Formal state determination'',
{\em J. Math. Phys.} {\bf 24}, 5, 1199-1205 (1983).

\item {\bf [Ivanovic 87]}:
I. D. Ivanovic,
``How to differenciate between non-orthogonal states'',
{\em Phys. Lett. A} {\bf 123}, 6, 257-259 (1987).
See {\bf [Dieks 88]}, {\bf [Peres 88 b]}.

\item {\bf [Ivanovic 93]}:
I. D. Ivanovic,
``Determination of pure spin state from three measurements'',
{\em J. Phys. A} {\bf 26}, 13, L579-L582 (1993).

\item {\bf [Ivanyos-Magniez-Santha 01]}:
G. Ivanyos, F. Magniez, \& M. Santha,
``Efficient quantum algorithms for some instances
of the non-Abelian hidden subgroup problem'',
quant-ph/0102014.

\item {\bf [Izmalkov-Grajcar-Il'ichev-(+6) 04]}:
A. Izmalkov, M. Grajcar, E. Il'ichev,
T. Wagner, H.-G. Meyer, A. Y. Smirnov,
M. H. S. Amin, A. M. van den Brink, \& A. M. Zagoskin,
``Evidence for entangled states of two coupled flux qubits'',
{\em Phys. Rev. Lett.} {\bf 93}, 3, 037003 (2004).
Publisher's Note: {\em Phys. Rev. Lett.} {\bf 93}, 4, 049902 (2004).


\newpage

\subsection{}


\item {\bf [Jabs 92]}:
A. Jabs,
``An interpretation of the formalism of quantum
mechanics in terms of epistemological realism'',
{\em Brit. J. Philos. Sci.} {\bf 43}, ?, 405-421 (1992).

\item {\bf [Jabs 96]}:
A. Jabs,
``Quantum mechanics in terms of realism'',
{\em Phys. Essays} {\bf 9}, 36-95, 354 (1996);
quant-ph/9606017.

\item {\bf [Jabs 98]}:
A. Jabs,
``Comment on testing Bell's inequality with Rydberg atoms entangled
through cavity photons'',
quant-ph/9811042.

\item {\bf [Jack 95]}:
C. Jack,
``Sherlock Holmes investigates the EPR paradox'',
{\em Phys. World} {\bf 8}, 4, 39-42 (1995).
Comments: {\em Phys. World} {\bf 8}, 5, 20-21 (1995).

\item {\bf [Jack 96]}:
C. Jack,
``Probing the limits of quantum theory'',
{\em Phys. World} {\bf 9}, 7, 53 (1996).
Review of {\bf [Wick 95]}.

\item {\bf [Jackiw-Shimony 02]}:
R. Jackiw, \& A. Shimony,
``The depth and breadth of John Bell's physics'',
{\em Phys. in Perspective} {\bf 4}, 1, 78-116 (2002);
physics/0105046.

\item {\bf [Jakobczyk 04]}:
 L. Jakobczyk,
`Bell inequalities and linear entropy. Comment on the paper of E. Santos
``Entropy inequalities and Bell inequalities for two-qubit systems''\,',
quant-ph/0409087.
Comment on {\bf [Santos 04 a]}.

\item {\bf [Jacobs-Franson 96]}:
B. C. Jacobs, \& J. D. Franson,
``Quantum cryptography in free space'',
{\em Opt. Lett.} {\bf 21}, ?, 1854-1856 (1996).

\item {\bf [Jacobs-Pittman-Franson 02]}:
B. C. Jacobs, T. B. Pittman, \& J. D. Franson,
``Quantum relays and noise suppression using linear optics'',
{\em Phys. Rev. A} {\bf 66}, 5, 052307 (2002);
quant-ph/0204097.

\item {\bf [Jacobs-Knight-Vedral 97]}:
K. Jacobs, P. L. Knight, \& V. Vedral,
``Determining the state of a single cavity mode from photon
statistics'',
{\em J. Mod. Opt.} {\bf 44}, 11-12 (Special issue:
Quantum state preparation and measurement), 2427-2439 (1997).

\item {\bf [Jacobs 98]}:
K. Jacobs,
``Topics in quantum measurement and quantum noise'',
Ph.\ D. thesis, Imperial College, London, 1998;
quant-ph/9810015.

\item {\bf [Jacobs 02]}:
K. Jacobs,
``How do two observers pool their knowledge about a quantum system?'',
quant-ph/0201096.

\item {\bf [Jacobs 03 a]}:
K. Jacobs,
``How to project qubits faster using quantum feedback'',
{\em Phys. Rev. A} {\bf 67}, 3, 030301 (2003).

\item {\bf [Jacobs 03 a]}:
K. Jacobs,
``Bounds on the ability of a measurement to extract information'',
quant-ph/0306039.

\item {\bf [Jacobs 04]}:
K. Jacobs,
``Optimal feedback control for the rapid preparation of a single qubit'',
quant-ph/0410018.

\item {\bf [Jacquod 04]}:
P. Jacquod,
``Semiclassical time evolution of the reduced density matrix and
dynamically assisted generation of entanglement for bipartite quantum systems'',
{\em Phys. Rev. Lett.} {\bf 92}, 15, 150403 (2004).

\item {\bf [Jaeger-Horne-Shimony 93]}:
G. Jaeger, M. A. Horne, \& A. Shimony,
``Complementarity of one-particle and two-particle interference'',
{\em Phys. Rev. A} {\bf 48}, 2, 1023-1027 (1993).

\item {\bf [Jaeger-Shimony 95]}:
G. Jaeger, \& A. Shimony,
``Optimal distinction between two non-orthogonal quantum states'',
{\em Phys. Lett. A} {\bf 197}, 2, 83-87 (1995).

\item {\bf [Jaeger-Viger-Sarkar 96]}:
G. Jaeger, C. Viger, \& S. Sarkar,
``Bell-type equalities for SQUIDs on the assumptions of macroscopic
realism and non-invasive measurability'',
{\em Phys. Lett. A} {\bf 210}, 1-2, 5-10 (1996).

\item {\bf [Jaeger-Shimony 95]}:
G. Jaeger, \& A. Shimony,
``An extremum principle for a neutron diffraction experiment'',
{\em Found. Phys.} {\bf 29}, 3, 435-444 (1999).

\item {\bf [Jaeger 00]}:
G. Jaeger,
``Bohmian mechanics and quantum theory'',
{\em Stud. Hist. Philos. Sci. Part B: Stud. Hist. Philos. Mod. Phys.}
{\bf 31}, 1, 105-108 (2000);
quant-ph/9906096.
Review of {\bf [Cushing-Fine-Goldstein 96]}.

\item {\bf [Jaeger-Teodorescu Frumosu-Sergienko-(+2) 03 a]}:
G. Jaeger, M. Teodorescu-Frumosu, A. Sergienko,
B. E. A. Saleh, \& M. C. Teich,
``Invariants of multiple-qubit systems under stochastic local operations'',
in {\em Foundations of Quantum Mechanics and Probability (V\"{a}xj\"{o}, Sweden, 2002)};
quant-ph/0301174.

\item {\bf [Jaeger-Teodorescu Frumosu-Sergienko-(+2) 03 b]}:
G. Jaeger, M. Teodorescu-Frumosu, A. Sergienko,
B. E. A. Saleh, \& M. C. Teich,
``Multiphoton Stokes-parameter invariant for entangled states'',
{\em Phys. Rev. A} {\bf 67}, 3, 032307 (2003).

\item {\bf [Jaeger-Sergienko-Saleh-Teich 03]}:
G. Jaeger, A. V. Sergienko, B. E. A. Saleh, \& M. C. Teich,
``Entanglement, mixedness, and spin-flip symmetry in multiple-qubit systems'',
{\em Phys. Rev. A} {\bf 68}, 2, 022318 (2003);
quant-ph/0307124.

\item {\bf [Jaeger 04]}:
G. Jaeger,
``Bell gems: The Bell basis generalized'',
{\em Phys. Lett. A} {\bf 329}, 6, 425-429 (2004);
quant-ph/0407251.

\item {\bf [Jaekel 01]}:
C. Jaekel,
``Some comments on entanglement and local thermofield theory'',
{\em Found. Phys. Lett.} {\bf 14}, 1, 1-16 (2001);
quant-ph/0405024.

\item {\em [Jagannathan 02]}:
K. Jagannathan,
``The strange world of quantum mechanics'',
{\em Am. J. Phys.} {\bf 70}, 12, 1271-1272 (2002).
Review of {\bf [Styer 00 a]}.

\item {\bf [Jakob-Abranyos-Bergou 01]}:
M. Jakob, Y. Abranyos, \& J. A. Bergou,
``Quantum measurement apparatus with a squeezed reservoir: Control of
decoherence and nonlocality in phase space'',
{\em Phys. Rev. A} {\bf 64}, 6, 062102 (2001).

\item {\bf [Jakob-Abranyos-Bergou 02]}:
M. Jakob, Y. Abranyos, \& J. A. Bergou,
``Degree of entanglement in a quantum-measurement process'',
{\em Phys. Rev. A} {\bf 66}, 2, 022113 (2002);
quant-ph/0204054.

\item {\bf [Jakob-Bergou 02]}:
M. Jakob, \& J. Bergou,
``Quantitative conditional quantum erasure in two-atom resonance fluorescence'',
{\em Phys. Rev. A} {\bf 66}, 6, 062107 (2002).

\item {\bf [Jak\'{o}bczyk-Siennicki 01]}:
L. Jak\'{o}bczyk, \& M. Siennicki,
``Geometry of Bloch vectors in two-qubit system'',
{\em Phys. Lett. A} {\bf 286}, 6, 383-390 (2001).

\item {\bf [Jak\'{o}bczyk 02]}:
L. Jak\'{o}bczyk,
``Entangling two qubits by dissipation'',
{\em J. Phys. A} {\bf 35}, 30, 6383-6392 (2003).
Erratum: {\em J. Phys. A} {\bf 36}, 5, 1537 (2003);
quant-ph/0204140.

\item {\bf [Jak\'{o}bczyk-Jamr\'{o}z 03]}:
L. Jak\'{o}bczyk, \& A. Jamr\'{o}z,
``Entanglement and nonlocality versus spontaneous emission in two-atom systems'',
{\em Phys. Lett. A} {\bf 318}, 4-5, 318-326 (2003).

\item {\bf [Jaksch-Briegel-Cirac-(+2) 99]}:
D. Jaksch, H.-J. Briegel, J. I. Cirac, C. W. Gardiner, \& P. Zoller,
``Entanglement of atoms via cold controlled collisions'',
{\em Phys. Rev. Lett.} {\bf 82}, 9, 1975-1978 (1999);
quant-ph/9810087.

\item {\bf [Jaksch-Cirac-Zoller-(+3) 00]}:
D. Jaksch, J. I. Cirac, P. Zoller,
S. L. Rolston, R. C\^{o}t\'{e}, \& M. D. Lukin,
``Fast quantum gates for neutral atoms'',
{\em Phys. Rev. Lett.} {\bf 85}, 10, 2208-2211 (2000);
quant-ph/0004038.

\item {\bf [Jaksch-Papageorgiou 03]}:
P. Jaksch, \& A. Papageorgiou,
``Eigenvector approximation leading to exponential speedup
of quantum eigenvalue calculation'',
{\em Phys. Rev. Lett.} {\bf 91}, 25, 257902 (2003);
quant-ph/0308016.

\item {\bf [Jaksch 04]}:
D. Jaksch,
``Optical lattices, ultracold atoms and quantum information processing'',
{\em Contemp. Phys.};
quant-ph/0407048.

\item {\bf [James 98]}:
D. F. V. James,
``Quantum dynamics of cold trapped ions with application to quantum
computation'',
{\em Appl. Phys. B - Lasers Opt.} {\bf 66}, 181-190 (1998).
Reprinted in {\bf [Macchiavello-Palma-Zeilinger 00]}, pp.~345-354.

\item {\bf [James-Gulley-Holzscheiter-(+10) 98]}:
D. F. V. James, M. S. Gulley, M. H. Holzscheiter, R. J. Hughes, P. G.
Kwiat, S. K. Lamoreaux, C. G. Peterson, V. D. Sandberg, M. M. Schauer,
C. M. Simmons, D. Tupa, P. Z. Wang, \& A. G. White,
``Trapped ion quantum computer research at Los Alamos'',
quant-ph/9807071.

\item {\bf [James 00]}:
D. F. V. James,
``Quantum computation with hot and cold ions:
An assessment of proposed schemes'',
{\em Fortschr. Phys.} {\bf 48}, 9-11 (Special issue: Experimental proposals for quantum computation), 823-837 (2000);
quant-ph/0003122.

\item {\bf [James-Kwiat-Munro-White 01]}:
D. F. V. James, P. G. Kwiat, W. J. Munro, \& A. G. White,
``Measurement of qubits'',
{\em Phys. Rev. A} {\bf 64}, 5, 052312 (2001);
quant-ph/0103121.

\item {\bf [James 01]}:
D. F. V. James,
``Quantum computation and quantum information'',
{\em Phys. Today} {\bf 54}, 11, 60-? (2001).
Review of {\bf [Nielsen-Chuang 00]}.

\item {\bf [James-Kwiat 02]}:
D. F. V. James, \& P. G. Kwiat,
``Atomic-vapor-based high efficiency optical detectors with photon
number resolution'',
{\em Phys. Rev. Lett.} {\bf 89}, 18, 183601 (2002).

\item {\bf [Jammer 66]}:
M. Jammer,
{\em The conceptual development of quantum mechanics},
McGraw-Hill, New York, 1966. 2nd edition: Tomash Publishers, American
Institute of Physics, New York, 1989.

\item {\bf [Jammer 74]}:
M. Jammer,
{\em The philosophy of quantum mechanics:
The interpretations of quantum mechanics in historical perspective},
Wiley, New York, 1974.

\item {\bf [Jammer 80]}:
M. Jammer,
``Le paradoxe d'Einstein-Podolsky-Rosen'',
{\em La Recherche} {\bf 11}, 111, 510-519 (1980).

\item {\bf [Jammer 85]}:
M. Jammer,
``The EPR problem in its historical development'',
in P. J. Lahti, \& P. Mittelstaedt (eds.),
{\em Symp.\ on the Foundations of Modern
Physics: 50 Years of the Einstein-Podolsky-Rosen Experiment
(Joensuu, Finland, 1985)},
World Scientific, Singapore, 1985, pp.~129-149.

\item {\bf [Jammer 88]}:
M. Jammer,
``David Bohm and his work---On the
occasion of his seventieth birthday'',
{\em Found. Phys.} {\bf 18}, 7, 691-699 (1988).

\item {\bf [Jammer 90]}:
M. Jammer,
``John Stewart Bell and his work---On the
occasion of his sixtieth birthday'',
{\em Found. Phys.} {\bf 20}, 10, 1139-1145 (1990).

\item {\bf [Jammer 93]}:
M. Jammer,
``John Stewart Bell and the debate on the
significance of his contributions to the foundations of quantum mechanics'',
in A. van der Merwe, \& F. Selleri (eds.),
{\em Bell's theorem and the foundations of modern physics.
Proc.\ of an international conference (Cesena, Italy, 1991)},
World Scientific, Singapore, 1993, pp.~1-23.

\item {\bf [Jammer 99]}:
M. Jammer,
{\em Einstein and religion. Physics and theology},
Princeton University Press, Princeton, New Jersey, 1999.
Review: {\bf [Gudder 01 b]}.

\item {\bf [Jan\'{e}-Plenio-Jonathan 02]}:
E. Jan\'{e}, M. B. Plenio, \& D. Jonathan,
``Quantum-information processing in strongly detuned optical cavities'',
{\em Phys. Rev. A} {\bf 65}, 5, 050302 (2002).

\item {\bf [Jan\'{e} 02 a]}:
E. Jan\'{e},
``Correlacions qu\`{a}ntiques de sistemes bipartits i tripartits.
Quantum correlations of bipartite and tripartite systems'',
Ph.\ D. thesis, Universitat de Barcelona, 2002.

\item {\bf [Jan\'{e} 02 b]}:
E. Jan\'{e},
``Purification of two-qubit mixed states'',
{\em Quant. Inf. Comp.} {\bf 2}, 5, 348-354 (2002).

\item {\bf [Jan\'{e}-Vidal-D\"{u}r-(+2) 03]}:
E. Jan\'{e}, G. Vidal, W. D\"{u}r, P. Zoller, \& J. I. Cirac,
``Simulation of quantum dynamics with quantum optical systems'',
{\em Quant. Inf. Comp.} {\bf 3}, 1, 15-37 (2003);
quant-ph/0207011.

\item {\bf [Jang-Lee-Kim-Park 01]}:
J. Jang, J. Lee, M. S. Kim, \& Y.-J. Park,
``Probabilistic nonlocal gate operation via imperfect entanglement'',
quant-ph/0101107.

\item {\bf [Janszky-Koniorczyk-G\'{a}bris 01]}:
J. Janszky, M. Koniorczyk, \& A. G\'{a}bris,
``One-complex-plane representation approach to continuous
variable quantum teleportation'',
{\em Phys. Rev. A} {\bf 64}, 3, 034302 (2001).

\item {\bf [Janzing-Beth 00]}:
D. Janzing, \& T. Beth,
``Fragility of a class of highly entangled states with $n$ qubits'',
{\em Phys. Rev. A} {\bf 61}, 5, 052308 (2000);
quant-ph/9907042.

\item {\bf [Janzing-Beth 01 a]}:
D. Janzing, \& T. Beth,
``Complexity measure for continuous-time quantum algorithms'',
{\em Phys. Rev. A} {\bf 64}, 2, 022301 (2001).

\item {\bf [Janzing 01]}:
D. Janzing,
``Quantum algorithm for measuring the energy of $n$
qubits with unknown pair-interactions'',
quant-ph/0108052.

\item {\bf [Janzing-Beth 01 b]}:
D. Janzing, \& T. Beth,
``Quantum algorithm for finding periodicities in the spectrum of a
black-box Hamiltonian or unitary transformation'',
quant-ph/0108053.

\item {\bf [Janzing-Beth 01 c]}:
D. Janzing, \& T. Beth,
``Quasi-order of clocks and synchronism and
quantum bounds for copying timing information'',
quant-ph/0112138.

\item {\bf [Janzing-Armknecht-Zeier-Beth 02]}:
D. Janzing, F. Armknecht, R. Zeier, \& T. Beth,
``Quantum control without access to the controlling interaction'',
{\em Phys. Rev. A} {\bf 65}, 2, 022104 (2002).

\item {\bf [Janzing-Beth 02 a]}:
D. Janzing, \& T. Beth,
``Distinguishing $n$ Hamiltonians on $C^n$ by a single measurement'',
{\em Phys. Rev. A} {\bf 65}, 2, 022303 (2002).

\item {\bf [Janzing-Wocjan-Beth 02 a]}:
D. Janzing, P. Wocjan, \& T. Beth,
``Complexity of decoupling and time reversal for $n$ spins with pair
interactions: Arrow of time in quantum control'',
{\em Phys. Rev. A} {\bf 66}, 4, 042311 (2002).

\item {\bf [Janzing-Beth 02 b]}:
D. Janzing, \& T. Beth,
``Are there quantum bounds on the recyclability of
clock signals in low power computers?'',
quant-ph/0202059.

\item {\bf [Janzing-Decker-Beth 03]}:
D. Janzing, T. Decker, \& T. Beth,
`Performing joint measurements and transformations on several qubits by
operating on a single ``control'' qubit',
{\em Phys. Rev. A} {\bf 67}, 4, 042320 (2003).

\item {\bf [Japha-Kurizki 99]}:
Y. Japha, \& G. Kurizki,
``Faraday quantum clock and nonlocal photon pair correlations'',
{\em Phys. Rev. A} {\bf 60}, 3, 1811-1816 (1999).

\item {\bf [Jarrett 83]}:
J. P. Jarrett,
``Bell's theorem, quantum mechanics and local realism'',
Ph.\ D. thesis, University of Chicago, 1983.

\item {\bf [Jarrett 84]}:
J. P. Jarrett,
``On the physical significance of the locality
condition in the Bell arguments'',
{\em No\^{u}s} {\bf 18}, ?, 569-589 (1984).
See {\bf [Ballentine-Jarrett 87]}.

\item {\bf [Jarrett 89]}:
J. P. Jarrett,
``Bell's theorem: A guide to the implications'',
in J. T. Cushing, \& E. McMullin (eds.),
{\em Philosophical consequences of quantum
theory: Reflections on Bell's theorem},
University of Notre Dame Press,
Notre Dame, Indiana, 1989, pp.~60-79.

\item {\bf [Jauch-Piron 63]}:
J. M. Jauch, \& C. Piron,
``Can hidden variables be excluded in quantum mechanics?'',
{\em Helv. Phys. Acta} {\bf 36}, 7, 827-837 (1963).

\item {\bf [Jauch-Piron 67]}:
J. M. Jauch, \& C. Piron,
``Generalized localizability'',
{\em Helv. Phys. Acta} {\bf 40}, ?, 559-570 (1967).

\item {\bf [Jauch-Wigner-Yanase 67]}:
J. M. Jauch, E. P. Wigner, \& M. M. Yanase,
``Some comments concerning measurements in quantum mechanics'',
{\em Nuovo Cimento B} {\bf 48}, 1, 144-151 (1967).

\item {\bf [Jauch 68]}:
J. M. Jauch,
{\em Foundations of quantum mechanics},
Addison-Wesley, Reading, Massachusetts, 1968.

\item {\bf [Jauch-Piron 68]}:
J. M. Jauch, \& C. Piron,
``Hidden variables revisited'',
{\em Rev. Mod. Phys.} {\bf 40}, 1, 228-229 (1968).

\item {\bf [Jauch 71]}:
J. M. Jauch,
``Foundations of quantum mechanics'',
in {\bf [d'Espagnat 71]}, pp.~20-55.

\item {\bf [Jaynes 80]}:
E. Jaynes,
``?'',
in A. O. Barut (ed.),
{\em Foundations of
radiation theory and quantum electronics},
Plenum Press, New York, 1980, pp.~37-?.

\item {\bf [Jaynes 90]}:
E. T. Jaynes,
``Probability in quantum theory'',
in {\bf [Zurek 90]}, pp.~381-403.

\item {\bf [Jefferson-Fearn-Tipton-Spiller 02]}:
J. H. Jefferson, M. Fearn, D. L. J. Tipton, \& T. P. Spiller,
``Two-electron quantum dots as scalable qubits'',
{\em Phys. Rev. A} {\bf 66}, 4, 042328 (2002).

\item {\bf [Jelezko-Popa-Gruber-(+4) 02]}:
F. Jelezko, I. Popa, A. Gruber,
C. Tietz, J. Wrachtrup, A. Nizovtsev, \& S. Y. Kilin,
``Single spin states in a defect center resolved by optical spectroscopy'',
{\em Appl. Phys. Lett.} {\bf 81}, ?, 2160-? (2002).

\item {\bf [Jencov\'{a} 02]}:
A. Jencov\'{a},
``Quantum information geometry and standard purification'',
{\em J. Math. Phys.} {\bf 43}, 5, 2187-2201 (2002).

\item {\bf [Jennewein-Simon-Weihs-(+2) 00]}:
T. Jennewein, C. Simon, G. Weihs, H. Weinfurter, \& A. Zeilinger,
``Quantum cryptography with entangled photons'',
{\em Phys. Rev. Lett.} {\bf 84}, 20, 4729-4732 (2000);
quant-ph/9912117.

\item {\bf [Jennewein-Achleitner-Weihs-(+2) 00]}:
T. Jennewein, U. Achleitner, G. Weihs, H. Weinfurter, \& A. Zeilinger,
``A fast and compact quantum random number generator'',
{\em Rev. Sci. Instrum.} {\bf 71}, 1675-1680 (2000);
quant-ph/9912118.

\item {\bf [Jennewein-Weihs-Pan-Zeilinger 02]}:
T. Jennewein, G. Weihs, J.-W. Pan, \& A. Zeilinger,
``Experimental nonlocality proof of quantum teleportation and entanglement
swapping'',
{\em Phys. Rev. Lett.} {\bf 88}, 1, 017903 (2002);
quant-ph/0201134.
Comments: {\bf [Dennis 02]}, {\bf [Ryff 03]}.
See {\bf [Jennewein-Weihs-Pan-Zeilinger 03]}.

\item {\bf [Jennewein-Weihs-Pan-Zeilinger 03]}:
T. Jennewein, G. Weihs, J.-W. Pan, \& A. Zeilinger,
`Reply to Ryff's comment on `Experimental nonlocality proof of quantum
teleportation and entanglement swapping''\,',
quant-ph/0303104.
Reply to {\bf [Ryff 03]}.
See {\bf [Jennewein-Weihs-Pan-Zeilinger 02]}.

\item {\bf [Jennewein-Ursin-Aspelmeyer-Zeilinger 04]}:
T. Jennewein, R. Ursin, M. Aspelmeyer, \& A. Zeilinger,
``Experimental teleportation of quantum entanglement with an optimal
linear optics Bell-state analyzer'',
quant-ph/0409008.

\item {\bf [Jensen-Schack 00]}:
J. G. Jensen, \& R. Schack,
``Quantum authentication and key distribution using catalysis'',
quant-ph/0003104.

\item {\bf [Jensen-Schack 01]}:
J. G. Jensen, \& R. Schack,
``Simple algorithm for local conversion of pure states'',
{\em Phys. Rev. A} {\bf 63}, 6, 062303 (2001);
quant-ph/0006049.

\item {\bf [Jeong-Kim-Lee 01]}:
H. Jeong, M. S. Kim, \& J. Lee,
``Quantum-information processing for a coherent superposition
state via a mixed entangled coherent channel'',
{\em Phys. Rev. A} {\bf 64}, 5, 052308 (2001);
quant-ph/0104090.

\item {\bf [Jeong-Kim 01]}:
H. Jeong, \& M. S. Kim,
``Entanglement purification for entangled coherent states'',
quant-ph/0111015.

\item {\bf [Jeong-Kim 02]}:
H. Jeong, \& M. S. Kim,
``Efficient quantum computation using coherent states'',
{\em Phys. Rev. A} {\bf 65}, 4, 042305 (2002);
quant-ph/0109077.

\item {\bf [Jeong-Son-Kim-(+2) 03]}:
H. Jeong, W. Son, M. S. Kim, D. Ahn, \& \v{C}. Brukner,
``Quantum nonlocality test for continuous-variable states with dichotomic
observables'',
{\em Phys. Rev. A} {\bf 67}, 1, 012106 (2003).

\item {\bf [Jeong-Paternostro-Kim 04]}:
H. Jeong, M. Paternostro, \& M. S. Kim,
``Simulation of quantum random walks using the interference of a classical field'',
{\em Phys. Rev. A} {\bf 69}, 1, 012310 (2004);
quant-ph/0305008.

\item {\bf [Jeong-Kim-Ralph-Ham 04]}:
H. Jeong, M. S. Kim, T. C. Ralph, \& B. S. Ham,
``Generation of macroscopic superposition states with small nonlinearity'',
{\em Phys. Rev. A};
quant-ph/0405041.

\item {\bf [Jeong-Ralph-Bowen 04]}:
H. Jeong, T. C. Ralph, \& W. P. Bowen,
``Quantum and classical fidelities for Gaussian states'',
quant-ph/0409101.

\item {\bf [Jeong-Lund-Ralph 04]}:
H. Jeong, A. P. Lund, \& T. C. Ralph,
``Production of superpositions of coherent states in traveling optical
fields with inefficient photon detection'',
quant-ph/0410022.

\item {\bf [Jeong-Ralph 04]}:
H. Jeong, \& T. C. Ralph,
``Schrodinger's cat paradox revisited: Transfer of nonclassical properties
from microscopic superpositions to macroscopic thermal states'',
quant-ph/0410210.

\item {\bf [Jex-Andersson-Chefles 03]}:
I. Jex, E. Andersson, \& A. Chefles,
``Comparing the states of many quantum systems'',
{\em J. Mod. Opt.};
quant-ph/0305120.

\item {\bf [Je\v{z}ek-\v{R}eh\'{a}\v{c}ek-Fiur\'{a}\v{s}ek 02]}:
M. Je\v{z}ek, J. \v{R}eh\'{a}\v{c}ek, \& J. Fiur\'{a}\v{s}ek,
``Finding optimal strategies for minimum-error quantum-state discrimination'',
{\em Phys. Rev. A} {\bf 65}, 6, 060301 (2002);
quant-ph/0201109.

\item {\bf [Je\v{z}ek 02]}:
M. Je\v{z}ek,
``Discrimination between non-perfectly known states'',
{\em Phys. Lett. A} {\bf 299}, 5-6, 433-440 (2002);
quant-ph/0202085.

\item {\bf [Je\v{z}ek-Fiur\'{a}\v{s}ek-Hradil 03]}:
M. Je\v{z}ek, J. Fiur\'{a}\v{s}ek, \& Z. Hradil,
``Quantum inference of states and processes'',
{\em Phys. Rev. A} {\bf 68}, 1, 012305 (2003).

\item {\bf [Ji-Duan-Ying 04]}:
Z. Ji, R. Duan, \& M. Ying,
``Comparability of multipartite entanglement'',
{\em Phys. Lett. A} {\bf 330}, 6, 418-423 (2004).

\item {\bf [Jiang-Han-Xue-(+3) 03]}:
W. Jiang, C. Han, P. Xue, L.-M. Duan, \& G.-C. Guo,
``Nonclassical photon pairs generated from a room-temperature atomic
ensemble'',
quant-ph/0309175.

\item {\bf [Jin-Korepin 04]}:
B.-Q. Jin, \& V. E. Korepin,
``Localizable entanglement in antiferromagnetic spin chains'',
{\em Phys. Rev. A} {\bf 69}, 6, 062314 (2004).

\item {\bf [Jin-Li-Feng-Zheng 04]}:
G.-S. Jin, S.-S. Li, S.-L. Feng, \& H.-Z. Zheng,
``Method for generating maximally entangled states of multiple three-level atoms in cavity QED'',
{\em Phys. Rev. A} {\bf 69}, 3, 034302 (2004).

\item {\bf [Jing-Pan-Xie-Peng 02]}:
J. Jing, Q. Pan, C. Xie, \& K. Peng,
``Quantum cryptography using Einstein-Podolsky-Rosen correlations of continuous variables'',
quant-ph/0204111.

\item {\bf [Jing-Zhang-Yan-(+3) 03]}:
J. Jing, J. Zhang, Y. Yan,
F. Zhao, C. Xie, \& K. Peng,
``Experimental demonstration of tripartite entanglement and controlled dense
coding for continuous variables'',
{\em Phys. Rev. Lett.} {\bf 90}, 16, 167903 (2003).

\item {\bf [Johansen 96]}:
L. M. Johansen,
``Bell's inequality for the Mach-Zehnder
interferometer'',
{\em Phys. Lett. A} {\bf 219}, 1-2, 15-18 (1996);
quant-ph/9606002.

\item {\bf [Johansen 97 a]}:
L. M. Johansen,
``Equivalence between Bell's inequality and a constraint on
stochastic field theories for Einstein-Podolsky-Rosen states'',
{\em Phys. Rev. A} {\bf 56}, 1, 100-107 (1997);
quant-ph/9704032.

\item {\bf [Johansen 97 b]}:
L. M. Johansen,
``EPR correlations and EPW
distributions revisited'',
{\em Phys. Lett. A} {\bf 236}, 3, 173-176 (1997);
quant-ph/9709007.

\item {\bf [Johansen 98]}:
L. M. Johansen,
``Hydrodynamical quantum state reconstruction'',
{\em Phys. Rev. Lett.} {\bf 80}, 25, 5461-5465 (1998);
quant-ph/9805046.

\item {\bf [Johansen 03]}:
L. M. Johansen,
``What is the value of an observable between pre- and postselection?'',
{\em Phys. Lett. A};
quant-ph/0308137.

\item {\bf [Johansen 04 a]}:
L. M. Johansen,
``What is the value of an observable between pre- and postselection?'',
{\em Phys. Lett. A} {\bf 322}, 5-6, 298-300 (2004).

\item {\bf [Johansen 04 a]}:
L. M. Johansen,
``Nonclassical properties of coherent states'',
{\em Phys. Lett. A} {\bf 329}, 3, 184-187 (2004).

\item {\bf [Johansen 04 b]}:
L. M. Johansen,
``Weak measurements with arbitrary pointer states'',
quant-ph/0402050.

\item {\bf [Johansen 04 c]}:
L. M. Johansen,
``Nonclassicality of thermal radiation'',
quant-ph/0402105.

\item {\bf [Johansen-Luis 04]}:
L. M. Johansen, \& A. Luis,
``Nonclassicality in weak measurements'',
quant-ph/0408038.

\item {\bf [John 01]}:
M. V. John,
``Modified de Broglie-Bohm approach closer to
classical Hamilton-Jacobi theory'',
quant-ph/0102087.

\item {\bf [John 02]}:
M. V. John,
``Modified de Broglie-Bohm approach to quantum mechanics'',
{\em Found. Phys. Lett.} {\bf 15}, 4, 329-343 (2002);
quant-ph/0109093.

\item {\bf [Johnson 03]}:
G. Johnson,
{\em A shortcut through time: The path to a quantum computer},
Knopf, ?, 2003.

\item {\bf [Johnson 01]}:
N. F. Johnson,
``Playing a quantum game with a corrupted source'',
{\em Phys. Rev. A} {\bf 63}, 2, 020302(R) (2001);
quant-ph/0009050.

\item {\bf [Johnson-Bartlett-Sanders 02]}:
T. J. Johnson, S. D. Bartlett, \& B. C. Sanders,
``Continuous-variable quantum teleportation of entanglement'',
{\em Phys. Rev. A} {\bf 66}, 4, 042326 (2002);
quant-ph/0204011.

\item {\bf [Jona Lasinio-Presilla-Toninelli 01 a]}:
G. Jona-Lasinio, C. Presilla, \& C. Toninelli,
``Environment induced localization and superselection
rules in a gas of pyramidal molecules'',
cond-mat/0107341.

\item {\bf [Jona Lasinio-Presilla-Toninelli 01 b]}:
G. Jona-Lasinio, C. Presilla, \& C. Toninelli,
``Interaction induced localization in a gas of pyramidal molecules'',
cond-mat/0107342.

\item {\bf [Jonathan-Plenio 99 a]}:
D. Jonathan, \& M. B. Plenio,
``Minimal conditions for local pure-state entanglement
manipulation'',
{\em Phys. Rev. Lett.} {\bf 83}, 7, 1455-1458 (1999).
Erratum: {\em Phys. Rev. Lett.} {\bf 84}, 20, 4781 (2000).

\item {\bf [Jonathan-Plenio 99 b]}:
D. Jonathan, \& M. B. Plenio,
``Entanglement-assisted local manipulation of pure quantum states'',
{\em Phys. Rev. Lett.} {\bf 83}, 17, 3566-3569 (1999);
quant-ph/9905071.

\item {\bf [Jonathan-Plenio-Knight 00]}:
D. Jonathan, M. B. Plenio, \& P. L. Knight,
``Fast quantum gates for cold trapped ions'',
{\em Phys. Rev. A} {\bf 62}, 4, 042307 (2000).

\item {\bf [Jonathan-Plenio 01]}:
D. Jonathan, \& M. B. Plenio,
``Light-shift-induced quantum gates for ions in thermal motion'',
quant-ph/0103140.

\item {\bf [Jones-Pearle-Ring 04]}:
G. Jones, P. Pearle, \& J. Ring,
``Consequence for wavefunction collapse model of the Sudbury neutrino
observatory experiment'',
{\em Found. Phys.} {\bf 34}, 10, 1467-1474 (2004);
quant-ph/0411019.

\item {\bf [Jones-Mosca 98]}:
J. A. Jones, \& M. Mosca,
``Implementation of a quantum algorithm to solve Deutsch's problem
on a nuclear magnetic resonance quantum computer'',
{\em J. Chem. Phys.} {\bf 109}, 5, 1648-1653 (1998);
quant-ph/9801027.
Reprinted in {\bf [Macchiavello-Palma-Zeilinger 00]}, pp.~476-481.

\item {\bf [Jones-Mosca-Hansen 98]}:
J. A. Jones, M. Mosca, \& R. H. Hansen,
``Implementation of a quantum search algorithm on a quantum computer'',
{\em Nature} {\bf 393}, 6683, 344-346 (1998).

\item {\bf [Jones 98]}:
J. A. Jones,
``Quantum computing: Fast searches with nuclear magnetic resonance computers'',
{\em Science} {\bf 280}, 5361, 229 (1998).

\item {\bf [Jones-Knill 99]}:
J. A. Jones, \& E. Knill,
``Efficient refocussing of one spin and two spin interactions for
NMR quantum computation'',
{\em Journal of Magnetic Resonance}, 1999;
quant-ph/9905008.

\item {\bf[Jones-Mosca 99]}:
J. A. Jones, \& M. Mosca,
``Approximate quantum counting on an NMR ensemble quantum computer'',
{\em Phys. Rev. Lett.} {\bf 83}, 5, 1050-1053 (1999);
quant-ph/9808056.

\item {\bf [Jones-Vedral-Ekert-Castagnoli 00]}:
J. A. Jones, V. Vedral, A. K. Ekert, G. Castagnoli,
``Geometric quantum computation using nuclear magnetic resonance'',
{\em Nature} {\bf 403}, 6772, 869-871 (2000);
quant-ph/9910052.

\item {\bf [Jones 00 a]}:
J. A. Jones,
``NMR quantum computation: A critical evaluation'',
{\em Fortschr. Phys.} {\bf 48}, 9-11 (Special issue: Experimental proposals for quantum computation), 909-924 (2000);
quant-ph/0002085.

\item {\bf [Jones 00 b]}:
J. A. Jones,
``Quantum computing with NMR'',
{\bf [Macchiavello-Palma-Zeilinger 00]}, pp.~465-470.

\item {\bf [Jones 00 c]}:
J. A. Jones,
``NMR quantum computation'',
quant-ph/0009002.

\item {\bf [Jones 01]}:
J. A. Jones,
``Quantum computing and nuclear magnetic resonance'',
{\em Phys. Chem. Comm.} 11 (2001);
quant-ph/0106069.

\item {\bf [Jones 03 a]}:
J. A. Jones,
``Quantum computing: Putting it into practice'',
{\em Nature} {\bf 421}, 6918, 28-29 (2003).
See {\bf [Gulde-Riebe-Lancaster-(+6) 03]}.

\item {\bf [Jones 03 b]}:
J. A. Jones,
``Robust Ising gates for practical quantum computation'',
{\em Phys. Rev. A} {\bf 67}, 1, 012317 (2003).

\item {\bf [Jones 03 c]}:
J. A. Jones,
``Robust quantum information processing with techniques from liquid state NMR'',
submitted for Proc.\ of the Discussion Meeting on Practical Realisations of Quantum
Information Processing (Royal Society, London, 2002);
quant-ph/0301019.

\item {\bf [Jones 03 d]}:
J. A. Jones,
``Suppressing weak Ising couplings: Tailored gates for quantum computation'',
{\em Phys. Lett. A} {\bf 316}, 1-2, 24-28 (2003);
quant-ph/0307050.

\item {\bf [Jones 91]}:
M. R. Jones,
``Some difficulties for Clifton, Redhead, and
Butterfield's recent proof of nonlocality'',
{\em Found. Phys. Lett.} {\bf 4}, 4, 385-394 (1991).
Comment on {\bf [Clifton-Redhead-Butterfield 91 a]}.
See {\bf [Clifton-Redhead-Butterfield 91 b]}.

\item {\bf [Jones-Linden-Massar 04]}:
N. S. Jones, N. Linden, \& S. Massar,
``The extent of multi-particle quantum non-locality'',
quant-ph/0407018.

\item {\bf [Jones-Linden 04]}:
N. S. Jones, \& N. Linden,
``Parts of quantum states'',
quant-ph/0407117.

\item {\bf [Jones-Adelberger 94]}:
R. T. Jones, \& E. G. Adelberger,
``Quantum mechanics and Bell's inequalities'',
{\em Phys. Rev. Lett.} {\bf 72}, 17, 2675-2677 (1994).
Comment: {\bf [Santos 95 c]}.
See {\bf [Santos 91 b]}.

\item {\bf [Jongejan 01]}:
B. Jongejan,
``On Bell's paradox'',
quant-ph/0105020.

\item {\bf [Joo-Park 02]}:
J. Joo, \& Y.-J. Park,
``Comment on `Teleportation of an unknown state
by $W$ state'\,'',
{\em Phys. Lett. A} {\bf 300}, 2-3, 324-326 (2002).
Comment on {\bf [Shi-Tomita 02 a]}.
Reply: {\bf [Shi-Tomita 02 b]}.

\item {\bf [Joo-Lee-Jang-Park 02]}:
J. Joo, J. Lee, J. Jang, \& Y.-J. Park,
``Quantum secure communication with $W$ states'',
quant-ph/0204003.

\item {\bf [Joo-Park-Oh-Kim 02]}:
J. Joo, Y.-J. Park, S. Oh, \& J. Kim,
``Quantum teleportation via a $W$ state'',
quant-ph/0306175.

\item {\bf [Joos 84]}:
E. Joos,
``Continuous measurement: Watchdog effect versus golden rule'',
{\em Phys. Rev. D} {\bf 29}, 8, 1626-1633 (1984).

\item {\bf [Joos-Zeh 85]}:
E. Joos, \& H. Zeh,
``The emergence of classical properties through interaction with the environment'',
{\em Zeitschrift f\"{u}r Physik B} {\bf 59}, ?, 223-243 (1985).

\item {\bf [Joos 87]}:
E. Joos,
``Comment on `Unified dynamics for microscopic and macroscopic systems'\,'',
{\em Phys. Rev. D} {\bf 36}, 10, 3285-3286 (1987).
Comment on {\bf [Ghirardi-Rimini-Weber 86]}.
Reply: {\bf [Ghirardi-Rimini-Weber 87]}.

\item {\bf [Joos 99]}:
E. Joos,
``Elements of environmental decoherence'',
in P. Blanchard, D. Giulini, E. Joos,
C. Kiefer, \& I.-O. Stamatescu (eds.),
{\em Decoherence: Theoretical, Experimental, and
Conceptual Problems (Bielefeld, Germany, 1998)},
Springer-Verlag, Berlin, 1999;
quant-ph/9908008.

\item {\bf [Jordan-B\"{u}ttiker 04]}:
A. N. Jordan, \& M. B\"{u}ttiker,
``Entanglement energetics at zero temperature'',
{\em Phys. Rev. Lett.} {\bf 92}, 24, 247901 (2004).

\item {\bf [Jordan 27]}:
P. Jordan,
``Philosophical foundations of quantum theory'',
{\em Nature} {\bf 119}, ?, 566-569 (1927).

\item {\bf [Jordan 04 a]}:
S. P. Jordan,
``Efficient quantum algorithm for numerical gradient estimation'',
quant-ph/0405146.

\item {\bf [Jordan 85]}:
T. M. Jordan,
``Reality and the quantum theory'',
{\em Phys. Today} {\bf 38}, 11, 11 (1985).
Comment on {\bf [Mermin 85]}.

\item {\bf [Jordan-Sudarshan 91]}:
T. F. Jordan, \& E. C. G. Sudarshan,
``Simply no hidden variables'',
{\em Am. J. Phys.} {\bf 59}, 8, 698-700 (1991).
See {\bf [Mermin 92 a]}.

\item {\bf [Jordan 93 a]}:
T. F. Jordan,
``Disappearance and reapperance of macroscopic quantum interference'',
{\em Phys. Rev. A} {\bf 48}, 3, 2449-2450 (1993).

\item {\bf [Jordan 93 b]}:
T. F. Jordan,
``Reconstructing a nonlinear dynamical framework for testing
quantum mechanics'',
{\em Ann. Phys.} {\bf 225}, 83-113 (1993).

\item {\bf [Jordan 94 a]}:
T. F. Jordan,
``Testing Einstein-Podolsky-Rosen assumptions without inequalities
with two photons or particles with spin $\frac{1}{2}$'',
{\em Phys. Rev. A} {\bf 50}, 1, 62-66 (1994).

\item {\bf [Jordan 94 b]}:
T. F. Jordan,
``Quantum mysteries explored'',
{\em Am. J. Phys.} {\bf 62}, 10, 874-880 (1994).

\item {\bf [Jordan 99]}:
T. F. Jordan,
``Quantum correlations violate Einstein-Podolsky-Rosen assumptions'',
{\em Phys. Rev. A} {\bf 60}, 4, 2726-2728 (1999).
See {\bf [Mermin 98 a, b, 99 a]},
{\bf [Cabello 99 a, c]}.

\item {\bf [Jordan-Sariyianni 99]}:
T. F. Jordan, \& Z.-E. Sariyianni,
``Signals faster than light in nonlinear quantum dynamics'',
{\em Phys. Lett. A} {\bf 263}, 4-6, 263-267 (1999).

\item {\bf [Jordan 01]}:
T. F. Jordan,
``Choosing and rechoosing to have or have not interference'',
{\em Am. J. Phys.} {\bf 69}, 2, 155-157 (2001).

\item {\bf [Jordan 04 b]}:
T. F. Jordan,
``Equivalence of linear and affine maps for density matrices'',
quant-ph/0407203.

\item {\bf [Josse-Dantan-Bramati-(+2) 04]}:
V. Josse, A. Dantan, A. Bramati,
M. Pinard, \& E. Giacobino,
``Continuous variable entanglement using cold atoms'',
{\em Phys. Rev. Lett.} {\bf 92}, 12, 123601 (2004).

\item {\bf [Jost 76]}:
R. Jost,
``Measures on the finite dimensional subspaces of a
Hilbert space: Remarks on a theorem by A. M. Gleason'',
in E. H. Lieb, B. Simon, \& S. Wightman (eds.),
{\em Studies in mathematical physics: Essays in honour of
Valentine Bargmann},
Princeton University Press, Princeton, New Jersey, 1976, pp.~221-45.

\item {\bf [Jozsa 91]}:
R. Jozsa,
``Characterizing classes of functions computable by quantum parallelism'',
{\em Proc. R. Soc. Lond. A} {\bf 435}, 1895, 563-574 (1991).

\item {\bf [Jozsa-Robb-Wootters 94]}:
R. Jozsa, D. Robb, \& W. K. Wootters,
``A lower bound for accessible information in quantum mechanics'',
{\em Phys. Rev. A} {\bf 49}, 2, 668-677 (1994).

\item {\bf [Jozsa 94 a]}:
R. Jozsa,
``?'',
{\em J. Mod. Opt.} {\bf 41}, 12-? (1994).

\item {\bf [Jozsa 94 b]}:
R. Jozsa,
``Fidelity for mixed quantum states'',
in S. M. Barnett, A. K. Ekert, \& S. J. D. Phoenix (eds.),
{\em J. Mod. Opt.} {\bf 41}, 12 (Special issue: Quantum
communication), 2315-2323 (1994).

\item {\bf [Jozsa-Schumacher 94]}:
R. Jozsa, \& B. W. Schumacher,
``A new proof of the quantum noiseless coding theorem'',
{\em J. Mod. Opt.} {\bf 41}, 12 (Special
issue: Quantum communication), 2343-2349 (1994).

\item {\bf [Jozsa 97 a]}:
R. Jozsa,
``Entanglement and quantum computation'',
in S. Huggett, L. Mason, K. P. Tod, S. T. Tsou, \& N. M. J. Woodhouse (eds.),
{\em Geometric issues in the foundations of science},
Oxford University Press, Oxford, 1997,
?-?;
quant-ph/9707034.

\item {\bf [Jozsa 98 a]}:
R. Jozsa,
``Quantum information and its properties'',
in {\bf [Lo-Spiller-Popescu 98]}, pp.~49-75.

\item {\bf [Jozsa 98 b]}:
R. Jozsa,
``Quantum algorithms and the Fourier transform'',
in D. P. DiVincenzo. E. Knill, R. Laflamme, \& W. H. Zurek (eds.),
{\em Quantum Coherence and Decoherence.
Proc.\ of the ITP Conf.\ (Santa Barbara, California, 1996)},
{\em Proc. R. Soc. Lond. A} {\bf 454}, 1969, 323-337 (1998);
quant-ph/9707033.

\item {\bf [Jozsa-Horodecki-Horodecki-Horodecki 98]}:
R. Jozsa, M. Horodecki, P. Horodecki, \& R. Horodecki,
``Universal quantum information compression'',
{\em Phys. Rev. Lett.} {\bf 81}, 8, 1714-1717 (1998);
quant-ph/9805017.

\item {\bf [Jozsa 99]}:
R. Jozsa,
``Searching in Grover's algorithm'',
quant-ph/9901021.

\item {\bf [Jozsa-Abrams-Dowling-Williams 00]}:
R. Jozsa, D. S. Abrams, J. P. Dowling, \& C. P. Williams,
``Quantum clock synchronization based on shared prior
entanglement'',
{\em Phys. Rev. Lett.} {\bf 85}, 9, 2010-2013 (2000);
quant-ph/0004105.
Comment: {\bf [Burt-Ekstrom-Swanson 01]}.
Reply: {\bf [Jozsa-Abrams-Dowling-Williams 01]}.
See {\bf [Chuang 00]}.

\item {\bf [Jozsa-Schlienz 00]}:
R. Jozsa, \& J. Schlienz,
``Distinguishability of states and von Neumann entropy'',
{\em Phys. Rev. A} {\bf 62}, 1, 012301 (2000);
quant-ph/9911009.

\item {\bf [Jozsa 00]}:
R. Jozsa,
``Quantum factoring, discrete logarithms and the hidden subgroup
problem'',
{\em IEEE Computing in Science and Engineering};
quant-ph/0012084.

\item {\bf [Jozsa-Abrams-Dowling-Williams 01]}:
R. Jozsa, D. S. Abrams, J. P. Dowling, \& C. P. Williams,
``Reply: Jozsa {\em et al.}'',
{\em Phys. Rev. Lett.} {\bf 87}, 12, 129802 (2001);
Reply to {\bf [Burt-Ekstrom-Swanson 01]}.
See {\bf [Jozsa-Abrams-Dowling-Williams 00]}.
See {\bf [Chuang 00]}.

\item {\bf [Jozsa-Linden 02]}:
R. Jozsa, \& N. Linden,
``On the role of entanglement in quantum computational speed-up'',
quant-ph/0201143.

\item {\bf [Jozsa 03 a]}:
R. Jozsa,
``Notes on Hallgren's efficient quantum algorithm for solving Pell's equation'',
quant-ph/0302134.

\item {\bf [Jozsa-Koashi-Linden-(+4) 03]}:
R. Jozsa, M. Koashi, N. Linden, S. Popescu, S.
Presnell, D. Shepherd, \& A. Winter,
``Entanglement cost of generalised measurements'',
quant-ph/0303167.

\item {\bf [Jozsa 03 b]}:
R. Jozsa,
``Illustrating the concept of quantum information'',
{\em IBM J. Res. Dev.};
quant-ph/0305114.

\item {\bf [Julsgaard-Kozhekin-Polzik 01]}:
B. Julsgaard, A. Kozhekin, \& E. S. Polzik,
``Experimental long-lived entanglement
of two macroscopic objects'',
{\em Nature} {\bf 413}, 6854, 400-403 (2001);
quant-ph/0106057.
See {\bf [Duan-Giedke-Cirac-Zoller 00 a]},
{\bf [Cirac 01]}.

\item {\bf [Julsgaard-Sherson, J. I. Cirac-(+2) 04]}:
B. Julsgaard, J. Sherson, J. I. Cirac,
J. Fiur\'{a}\v{s}ek, \& E. S. Polzik,
``Experimental demonstration of quantum memory for light'',
{\em Nature};
quant-ph/0410072.


\newpage

\subsection{}


\item {\bf [Kadomtsev 96]}:
B. B. Kadomtsev,
``Quantum telegraph: Is it possible?,
{\em Phys. Lett. A} {\bf 210}, 6, 371-376 (1996).
Comment: {\bf [Wu 99]}.

\item {\bf [Kadomtsev 99]}:
M. B. Kadomtsev,
`Reply to the Comment by Wu on ``Quantum telegraph: Is
it possible?'' by B. B. Kadomtsev' (Phys. Lett. A 210
(1996) 371)',
{\em Phys. Lett. A} {\bf 255}, 3, 122 (1999).
Reply to {\bf [Wu 99]}.

\item {\bf [Kalamidas 04 a]}:
D. Kalamidas,
``Feasible quantum error detection with linear optics'',
{\em Phys. Lett. A} {\bf 332}, 2, 87-93 (2004).

\item {\bf [Kalamidas 04 b]}:
D. Kalamidas,
``Hardy-type nonlocality proof for two maximally entangled particles'',
{\em Phys. Lett. A} {\bf 332}, 3-4, 183-186 (2004);
quant-ph/0410060.

\item {\bf [Kaltenbaek-Aspelmeyer-Jennewein-(+4) 03]}:
R. Kaltenbaek, M. Aspelmeyer, T. Jennewein,
\v{C}. Brukner, M. Pfennigbauer, W. R. Leeb, \& A. Zeilinger,
``Proof-of-concept experiments for quantum physics in space'',
{\em SPIE Proc.\ on Quantum Communications and Quantum Imaging (2003)};
quant-ph/0308174.

\item {\bf [Kak 98 a]}:
S. C. Kak,
``Quantum information in a distributed apparatus'',
{\em Found. Phys.} {\bf 28}, 6, 1005-1012 (1998);
quant-ph/9804047.

\item {\bf [Kak 98 b]}:
S. C. Kak,
``A note on quantum errors and their correction'',
quant-ph/9811005.

\item {\bf [Kak 99 a]}:
S. C. Kak,
``The initialization problem in quantum computing'',
{\em Found. Phys.} {\bf 29}, 2, 267-280 (1999);
quant-ph/9805002.

\item {\bf [Kak 99 b]}:
S. C. Kak,
``Quantum key distribution using three basis states'',
{\em Pramana} {\bf 54}, ?, 709-713 (2000);
quant-ph/9902038.

\item {\bf [Kak 00 a]}:
S. C. Kak,
``On rotating a qubit'',
{\em Information Sciences} {\bf 128}, 149-154 (2000);
quant-ph/9910107.

\item {\bf [Kak 00 b]}:
S. C. Kak,
``From many to one: On starting a quantum computation'',
quant-ph/0010109.

\item {\bf [Kak 01]}:
S. Kak,
``Are quantum computing models realistic?'',
quant-ph/0110040.

\item {\bf [Kak 01]}:
S. Kak,
``Teleportation protocol requiring only one classical bit'',
quant-ph/0305085.

\item {\bf [Kamalov 01 a]}:
T. F. Kamalov,
``Quantum computer and its quasiclassical model'',
quant-ph/0109152.

\item {\bf [Kamalov 01 b]}:
T. F. Kamalov,
``Bell's inequalities in 4-dimension Rieman's space'',
quant-ph/0109153.

\item {\bf [Kamber 65]}:
F. Kamber,
``Zweiwertige Wahrscheinlichkeitsfunktionen auf orthokomplement\"{a}ren
Verb\"{a}nden'',
{\em Mathematische Annalen} {\bf 158}, 158-196 (1965).

\item {\bf [Kaminsky-Lloyd 02]}:
W. M. Kaminsky, \& S. Lloyd,
``Scalable architecture for adiabatic quantum computing of NP-hard
problems'', in
{\em Quantum computing and quantum bits in mesoscopic systems}
Kluwer, Dordrecht, Holland, 2003;
quant-ph/0211152.

\item {\bf [Kaminsky-Lloyd-Orlando 04]}:
W. M. Kaminsky, S. Lloyd, \& T. P. Orlando,
``Scalable superconducting architecture for adiabatic quantum computation'',
quant-ph/0403090.

\item {\bf [Kamta-Starace 02]}:
G. L. Kamta, \& A. F. Starace,
``Anisotropy and magnetic field effects on the entanglement of a two qubit
Heisenberg $XY$ chain'',
{\em Phys. Rev. Lett.} {\bf 88}, 10, 107901 (2002).

\item {\bf [Kane 98]}:
B. E. Kane,
``A silicon-based nuclear spin quantum computer'',
{\em Nature} {\bf 393}, 6681, 133-137 (1998).
See {\bf [DiVincenzo 98 a]}.

\item {\bf [Kane 00]}:
B. E. Kane,
``Silicon-based quantum computation'',
{\em Fortschr. Phys.} {\bf 48}, 9-11 (Special issue:
Experimental proposals for quantum computation), 1023-1041 (2000);
quant-ph/0003031.

\item {\bf [Kangro 72]}:
H. Kangro (ed.),
{\em Original papers in quantum physics},
Taylor and Francis, London, 1972.

\item {\bf [Kaniadakis 02]}:
G. Kaniadakis,
``Statistical origin of quantum mechanics'',
{\em Physica A} {\bf 307}, ?, 172-? (2002);
quant-ph/0112049.

\item {\bf [Kaniadakis 03 a]}:
G. Kaniadakis,
``Nonrelativistic quantum mechanics with spin in the framework
of a classical subquantum kinetics'',
{\em Found. Phys. Lett} {\bf 16}, 2, 99-? (2003);
quant-ph/0209033.

\item {\bf [Kaniadakis 03 b]}:
G. Kaniadakis,
``BBGKY hierarchy underlying many particle quantum mechanics'',
{\em Phys. Lett. A};
quant-ph/0303159.

\item {\bf [Kanter-Saad 99]}:
I. Kanter, \& D. Saad,
``Error-correcting codes that nearly saturate Shannon's bound'',
{\em Phys. Rev. Lett.} {\bf 83}, 13, 2660-2663 (1999).

\item {\bf [Kantor 00]}:
P. B. Kantor,
``Quantum message disruption: A two-state model'',
quant-ph/0007054.

\item {\bf [Kar-Roy 95]}:
G. Kar, \& S. Roy,
``Unsharp spin-$\frac{1}{2}$
observables and CHSH inequalities'',
{\em Phys. Lett. A} {\bf 199}, 1-2, 12-14 (1995).

\item {\bf [Kar 95]}:
G. Kar,
``Noncommuting spin-$\frac{1}{2}$ observables and
the CHSH inequality'',
{\em Phys. Lett. A} {\bf 204}, 2, 99-101 (1995).
See {\bf [Cereceda 96 b]}.

\item {\bf [Kar 96]}:
G. Kar,
``A simple proof of the converse of Hardy's nonlocality
theorem'', preprint, 1996.

\item {\bf [Kar 97 a]}:
G. Kar,
``Hardy's nonlocality for mixed states'',
{\em Phys. Lett. A} {\bf 228}, 3, 119-120 (1997).

\item {\bf [Kar 97 b]}:
G. Kar,
``Uniqueness of Hardy's state for fixed choice of observables'',
{\em J. Phys. A} {\bf 30}, 8, L217-L219 (1997).

\item {\bf [Kar 97 c]}:
G. Kar,
``Testing Hardy's nonlocality with $n$ spin-$\frac{1}{2}$ particles'',
{\em Phys. Rev. A} {\bf 56}, 1, 1023-1024 (1997).

\item {\bf [Kar-Roy 97]}:
G. Kar, \& S. Roy,
``Unsharp observables and quantum
objectification problem in quantum theory'', preprint 1997.

\item {\bf [Kar 98]}:
G. Kar,
``?'',
{\em Rivista del Nuovo Cimento} {\bf 22}, ?, 1-? (1999).

\item {\bf [Kar-Roy-Ghosh-Sarkar 99]}:
G. Kar, A. Roy, S. Ghosh, \& D. Sarkar,
``Noncomplementary wave-particle phenomena revisited'',
quant-ph/9901026.
See {\bf [Cereceda 96 a]}.

\item {\bf [Karafyllidis 04]}:
I. G. Karafyllidis,
``Cellular quantum computer architecture'',
{\em Phys. Lett. A} {\bf 320}, 1, 35-38 (2003).

\item {\bf [Karelin-Lazaruk 00]}:
M. U. Karelin, \& A. M. Lazaruk,
``Structure of the density matrix providing the minimum of
generalized uncertainty relation for mixed states'',
quant-ph/0006055.

\item {\bf [Kargin 04]}:
V. Kargin,
``Coordination games with quantum information'',
quant-ph/0409091.

\item {\bf [Karimipour-Bahraminasab-Bagherinezhad 02 a]}:
V. Karimipour, A. Bahraminasab, \& S. Bagherinezhad,
``Entanglement swapping of generalized cat states and secret sharing'',
{\em Phys. Rev. A} {\bf 65}, 4, 042320 (2002);
quant-ph/0112050.

\item {\bf [Karimipour-Bahraminasab-Bagherinezhad 02 b]}:
V. Karimipour, A. Bahraminasab, \& S. Bagherinezhad,
``Quantum key distribution for $d$-level systems with generalized Bell states'',
{\em Phys. Rev. A} {\bf 65}, 5, 052331 (2002);
quant-ph/0111091.

\item {\bf [Karimipour-Rezakhani 02]}:
V. Karimipour, \& A. T. Rezakhani,
``Generation of phase-covariant quantum cloning'',
{\em Phys. Rev. A} {\bf 66}, 5, 052111 (2002).

\item {\bf [Karimipour-Majd 04]}:
V. Karimipour, \& N. Majd,
``Exact solutions for universal holonomic quantum gates'',
{\em Phys. Rev. A} {\bf 70}, 1, 012320 (2004);
quant-ph/0404101.

\item {\bf [Karlsson-Bj{\o}rk-Fosberg 97]}:
A. Karlsson, G. Bj{\o}rk, \& E. Fosberg,
``Interaction-free measurements, atom localisation and complementarity'',
quant-ph/9705006.

\item {\bf [Karlsson-Bj{\o}rk-Fosberg 98]}:
A. Karlsson, G. Bj{\o}rk, \& E. Fosberg,
``Interaction (energy exchange) free and quantum nondemolition measurements'',
{\em Phys. Rev. Lett.} {\bf 80}, 6, 1198-1201 (1998).

\item {\bf [Karlsson-Bourennane 98]}:
A. Karlsson, \& M. Bourennane,
``Quantum teleportation using three-particle entanglement'',
{\em Phys. Rev. A} {\bf 58}, 6, 4394-4400 (1998).

\item {\bf [Karlsson-Koashi-Imoto 99]}:
A. Karlsson, M. Koashi, \& N. Imoto,
``Quantum entanglement for secret sharing and secret splitting'',
{\em Phys. Rev. A} {\bf 59}, 1, 162-168 (1999).

\item {\bf [Karnas-Lewenstein 01]}:
S. Karnas, \& M. Lewenstein,
``Separable approximations of density matrices of composite quantum systems'',
in S. Popescu, N. Linden, \& R. Jozsa (eds.),
{\em J. Phys. A} {\bf 34}, 35
(Special issue: Quantum information and computation), 6919-6938 (2001);
quant-ph/0011066.

\item {\bf [Karnas-Lewenstein 01]}:
S. Karnas, \& M. Lewenstein,
``Separability and entanglement
in $C^2 \otimes C^2 \otimes C^N$ composite quantum systems'',
{\em Phys. Rev. A} {\bf 64}, 4, 042313 (2001);
quant-ph/0102115.

\item {\bf [Kasday-Ullman-Wu 70]}:
L. R. Kasday, J. D. Ullman, \& C. S. Wu,
``The Einstein-Podolsky-Rosen argument: Positron annihilation experiment'',
{\em Bull. Am. Phys. Soc.} {\bf 15}, ?, 586 (1970).

\item {\bf [Kasday 71]}:
L. R. Kasday,
``Experimental test of quantum predictions for widely separated photons'',
in {\bf [d'Espagnat 71]}, pp.~?-?.

\item {\bf [Kasday-Ullman-Wu 75]}:
L. R. Kasday, J. D. Ullman, \& C. S. Wu,
``Angular correlation of Compton-scattered annihilation photons of hidden variables'',
{\em Nuovo Cimmento B} {\bf 25}, 633-661 (1975).

\item {\bf [Kashefi-Nishimura-Vedral 01]}:
E. Kashefi, H. Nishimura, \& V. Vedral,
``A note on quantum one-way permutations'',
quant-ph/0109157.

\item {\bf [Kashefi-Kent-Vedral-Banaszek 02]}:
E. Kashefi, A. Kent, V. Vedral, \& K. Banaszek,
``Comparison of quantum oracles'',
{\em Phys. Rev. A} {\bf 65}, 5, 050304 (2002);
quant-ph/0109104.

\item {\bf [Kassman-Berman-Tsifrinovich-L\'{o}pez 02]}:
R. B. Kassman, G. P. Berman, V. I. Tsifrinovich, G. V. L\'{o}pez,
``Qubitless quantum logic'',
quant-ph/0203059.

\item {\bf [Kastner 98 a]}:
R. E. Kastner,
``Closing a loophole in the case against the
counterfactual usage of the ABL rule'',
quant-ph/9807015.
See {\bf [Kastner 98 b, 99 b, c]}, {\bf [Vaidman 99 a, c, d]}.

\item {\bf [Kastner 98 b]}:
R. E. Kastner,
``Resolving the three-box paradox'',
quant-ph/9807037.
See {\bf [Kastner 98 a, 99 b, c]}, {\bf [Vaidman 99 a, c, d]}.

\item {\bf [Kastner 98 c]}:
R. E. Kastner,
``Time-symmetrized quantum theory and counterfactuals'',
{\em Fortschr. Phys.} {\bf 46}, 6-8, 881-887 (1998).

\item {\bf [Kastner 99 a]}:
R. E. Kastner,
``Time-symmetrized quantum theory, counterfactuals and `advanced action'\,'',
{\em Stud. Hist. Philos. Sci. Part B: Stud. Hist. Philos. Mod. Phys.}
{\bf 30}, 2, 237-259 (1999);
quant-ph/9806002.
See {\bf [Vaidman 98 c]}.

\item {\bf [Kastner 99 b]}:
R. E. Kastner,
``TSQT `elements of possibility'?'',
{\em Stud. Hist. Philos. Sci. Part B: Stud. Hist. Philos. Mod. Phys.}
{\bf 30}, 3, 399-402 (1999);
quant-ph/9812024.
See {\bf [Kastner 98 a, b, 99 c]}, {\bf [Vaidman 99 a, c]}.

\item {\bf [Kastner 99 c]}:
R. E. Kastner,
``The three-box `paradox' and other reasons to reject the counterfactual
usage of the ABL rule'',
{\em Found. Phys.} {\bf 29}, 6, 851-865 (1999).
See {\bf [Kastner 98 a, b, 99 b]}, {\bf [Vaidman 99 a, c]}.

\item {\bf [Kastner 01]}:
R. E. Kastner,
`Comment on ``What quantum mechanics is trying to tell us,'' by
Ulrich Mohrhoff [Am. J. Phys. {\bf 68} (8), 728-745 (2000)]',
{\em Am. J. Phys.} {\bf 69}, 8, 860-863 (2001);
quant-ph/0003098.
Comment on {\bf [Mohrhoff 00 a]}.
Reply: {\bf [Mohrhoff 01 a]}.

\item {\bf [Kastner 02]}:
R. E. Kastner,
``The nature of the controversy over time-symmetric quantum counterfactuals'',
{\em Philos. Sci.} (2002);
PITT-PHIL-SCI00000868.

\item {\bf [Kastner 03]}:
R. E. Kastner,
``Weak values and consistent histories in quantum theory'',
{\em Stud. Hist. Philos. Sci. Part B: Stud. Hist. Philos. Mod. Phys.};
quant-ph/0207182.

\item {\bf [Kastner 04]}:
R. E. Kastner,
``Cramer's transactional interpretation and causal loop problems'',
{\em Synthese};
quant-ph/0408109.

\item {\bf [Kaszlikowski-\.{Z}ukowski 00]}:
D. Kaszlikowski, \& M. \.{Z}ukowski,
``Bell theorem involving all possible local measurements'',
{\em Phys. Rev. A} {\bf 61}, 2, 022114 (2000);
quant-ph/9908009.

\item {\bf [Kaszlikowski-Gnaci\'{n}ski-\.{Z}ukowski-(+2) 00]}:
D. Kaszlikowski, P. Gnaci\'{n}ski, M. \.{Z}ukowski,
W. Miklaszewski, \& A. Zeilinger,
``Violations of local realism by two entangled $N$-dimensional systems
are stronger than for two qubits'',
{\em Phys. Rev. Lett.} {\bf 85}, 21, 4418-4421 (2000);
quant-ph/0005028.
See {\bf [Chen-Kaszlikowski-Kwek-(+2) 01]}.

\item {\bf [Kaszlikowski 00]}:
D. Kaszlikowski,
``Nonclassical phenomena in multiphoton interferometry:
More stringent tests against local realism'',
Ph.\ D. thesis;
quant-ph/0008086.

\item {\bf [Kaszlikowski-Kwek-Chen-(+2) 02]}:
D. Kaszlikowski, L. C. Kwek, J.-L. Chen, M. \.{Z}ukowski, \& C. H. Oh,
``Clauser-Horne inequality for three-state systems'',
{\em Phys. Rev. A} {\bf 65}, 3, 032118 (2002);
quant-ph/0106010.

\item {\bf [Kaszlikowski-\.{Z}ukowski-Gnaci\'{n}ski 02]}:
D. Kaszlikowski, M. \.{Z}ukowski, \& P. Gnaci\'{n}ski,
``Bound entanglement and local realism'',
{\em Phys. Rev. A} {\bf 65}, 3, 032107 (2002);
quant-ph/0107154.

\item {\bf [Kaszlikowski-Gosal-Ling-(+3) 02]}:
D. Kaszlikowski, D. Gosal, E. J. Ling,
L. C. Kwek, M. \.{Z}ukowski, \& C. H. Oh,
``Three-qutrit correlations violate local realism more strongly than those of
three qubits'',
{\em Phys. Rev. A} {\bf 66}, 3, 032103 (2002);
quant-ph/0202019.

\item {\bf [Kaszlikowski-\.{Z}ukowski 02]}:
D. Kaszlikowski, \& M. \.{Z}ukowski,
``Greenberger-Horne-Zeilinger paradoxes for $N$ $N$-dimensional systems'',
{\em Phys. Rev. A} {\bf 66}, 4, 042107 (2002);
quant-ph/0108097.

\item {\bf [Kaszlikowski-Kwek-Chen-Oh 02]}:
D. Kaszlikowski, L. C. Kwek, J. Chen, \& C. H. Oh,
``Multipartite bound entanglement and three-setting Bell inequalities'',
{\em Phys. Rev. A} {\bf 66}, 5, 052309 (2002).

\item {\bf [Kaszlikowski-Kwek-\.{Z}ukowski-Englert 03]}:
D. Kaszlikowski, L. C. Kwek, M. \.{Z}ukowski, \& B.-G. Englert,
``Information-theoretic approach to single-particle and two-particle
interference in multipath interferometers'',
{\em Phys. Rev. Lett.} {\bf 91}, 3, 037901 (2003);
quant-ph/0302140.

\item {\bf [Kaszlikowski-\.{Z}ukowski 03]}:
D. Kaszlikowski, \& M. \.{Z}ukowski,
``Three qubit GHZ correlations and generalized Bell experiments'',
quant-ph/0302165.

\item {\bf [Kaszlikowski-Oi-Christandl-(+4) 03]}:
D. Kaszlikowski, D. K. L. Oi, M. Christandl, K. Chang,
A. K. Ekert, L. C. Kwek, \& C. H. Oh,
``Quantum cryptography based on qutrit Bell inequalities'',
{\em Phys. Rev. A} {\bf 67}, 1, 012310 (2003).

\item {\bf [Kaszlikowski-Gopinathan-Liang-(+2) 03]}:
D. Kaszlikowski, A. Gopinathan, Y. C. Liang,
L. C. Kwek, B.-G. Englert,
``How well can you know the edge of a quantum pyramid?'',
quant-ph/0307086.

\item {\bf [Kaszlikowski-Yang-Kwek-Englert 03]}:
D. Kaszlikowski, L. J. Yang, L. C. Kwek, \& B.-G. Englert,
``Quantum and classical advantage distillation are not equivalent'',
quant-ph/0310156.
See {\bf [Kaszlikowski-Lim-Kwek-Englert 03]}.

\item {\bf [Kaszlikowski-Lim-Kwek-Englert 03]}:
D. Kaszlikowski, J. Y. Lim, L. C. Kwek, \& B.-G. Englert,
``Coherent eavesdropping attacks in quantum cryptography: Nonequivalence
of quantum and classical key distillation'',
quant-ph/0312172.
See {\bf [Kaszlikowski-Yang-Kwek-Englert 03]}.

\item {\bf [Kaszlikowski-Gopinathan-Liang-(+2) 04]}:
D. Kaszlikowski, A. Gopinathan, Y. C. Liang,
L. C. Kwek, \& B.-G. Englert,
``Quantum cryptography: Security criteria reexamined'',
{\em Phys. Rev. A} {\bf 70}, 3, 032306 (2004);
quant-ph/0310144.

\item {\bf [Kato-Osaki-Hirota 99]}:
K. Kato, M. Osaki, \& O. Hirota,
``Derivation of classical capacity of a quantum channel
for a discrete information source'',
{\em Phys. Lett. A} {\bf 251}, 3, 157-163 (1999).

\item {\bf [Katsnelson-Dobrovitski-Harmon 00]}:
M. I. Katsnelson, V. V. Dobrovitski, \& B. N. Harmon,
``Propagation of local decohering action
in distributed quantum systems'',
{\em Phys. Rev. A} {\bf 62}, 2, 022118 (2000);
quant-ph/9909028.

\item {\bf [Kauffman-Lomonaco 02]}:
L. H. Kauffman, \& S. J. Lomonaco, Jr.,
``Quantum entanglement and topological entanglement'',
{\em New J. Phys.} {\bf 4}, 73.1-73.18 (2002).

\item {\bf [Kauffman 02]}:
L. H. Kauffman,
``Quantum topology and quantum computing'',
in {\bf [Lomonaco 02 a]}, pp.~273-303.

\item {\bf [Kauffman-Lomonaco 03]}:
L. H. Kauffman, \& S. J. Lomonaco, Jr.,
``Entanglement criteria -- Quantum and topological'',
{\em Prcp.\ of SPIE Conf.\ (Orlando, Florida, 2003)},
quant-ph/0304091.

\item {\bf [Kauffman-Lomonaco 04 a]}:
L. H. Kauffman, \& S. J. Lomonaco, Jr.,
``Braiding operators are universal quantum gates'',
quant-ph/0401090.

\item {\bf [Kauffman-Lomonaco 04 b]}:
L. H. Kauffman, \& S. J. Lomonaco, Jr.,
``Quantum knots'',
quant-ph/0403228.

\item {\bf [Kaulakys-Gontis 97]}:
B. Kaulakys, \& V. Gontis,
``Quantum anti-Zeno effect'',
{\em Phys. Rev. A} {\bf 56}, 2, 1131-1137 (1997).

\item {\bf [Kawabata 00]}:
S. Kawabata,
``Test of Bell's inequality using the spin filter effect
in ferromagnetic semiconductor micro-structures'',
cond-mat/0012475.

\item {\bf [Kawabata 03]}:
S. Kawabata,
``Information-theoretical approach to control of quantum-mechanical systems'',
{\em Phys. Rev. A} {\bf 68}, 6, 064302 (2003).

\item {\bf [Kay-Pachos 04]}:
A. Kay, \& J. K. Pachos,
``Quantum computation in optical lattices via global addressing'',
quant-ph/0406073.

\item {\bf [Kay-Lee-Pachos-(+3) 04]}:
A. Kay, D. K. K. Lee, J. K. Pachos,
M. B. Plenio, M. E. Reuter, \& E. Rico,
``Quantum information and triangular optical lattices'',
quant-ph/0407121.

\item {\bf [Kay 98]}:
B. S. Kay,
``Decoherence of macroscopic closed systems within newtonian quantum gravity'',
{\em Classical and Quantum Gravity} {\bf 15}, 12, L89-L98 (1998);
hep-th/9810077.

\item {\bf [Kay-Johnson-Benjamin 01]}:
R. Kay, N. F. Johnson, \& S. C. Benjamin,
``Evolutionary quantum game'',
{\em J. Phys. A} {\bf 34}, 41, L547-L552 (2001);
quant-ph/0102008.

\item {\bf [Kaye-Mosca 01 a]}:
P. Kaye, \& M. Mosca,
``Quantum networks for concentrating entanglement'',
in S. Popescu, N. Linden, \& R. Jozsa (eds.),
{\em J. Phys. A} {\bf 34}, 35
(Special issue: Quantum information and computation), 6939-6948 (2001);
quant-ph/0101009.

\item {\bf [Kaye-Mosca 01 b]}:
P. Kaye, \& M. Mosca,
``Quantum networks for generating arbitrary quantum states'',
{\em Quantum Networks for Generating Arbitrary Quantum States (Rochester, New York, 2001)}
quant-ph/0407102.

\item {\bf [Kaye-Zalka 04]}:
P. Kaye, \& C. Zalka,
``Optimized quantum implementation of elliptic curve arithmetic over
binary fields'',
quant-ph/0407095.

\item {\bf [Kazakov 03]}:
A. Y. Kazakov,
``Modified Jaynes–Cummings systems and a quantum
algorithm for the Knapsack problem'',
{\em J. Exp. Theor. Phys.} {\bf 97}, 1131-1136 (2003).

\item {\bf [Kemble 37]}:
E. C. Kemble,
{\em The fundamental principles of quantum
mechanics},
McGraw-Hill, New York, 1937; (reprinted) Dover, New York.

\item {\bf [Kempe 99]}:
J. Kempe,
``Multiparticle entanglement and its applications to cryptography'',
{\em Phys. Rev. A} {\bf 60}, 2, 910-916 (1999).

\item {\bf [Kempe-Simon-Weihs 00]}:
J. Kempe, C. Simon, \& G. Weihs,
``Optimal photon cloning'',
{\em Phys. Rev. A} {\bf 62}, 3, 032302 (2000);
quant-ph/0003025.

\item {\bf [Kempe-Bacon-Lidar-Whaley 01]}:
J. Kempe, D. Bacon, D. A. Lidar, \& K. B. Whaley,
``Theory of decoherence-free fault-tolerant universal quantum
computation'',
{\em Phys. Rev. A} {\bf 63}, 4, 042307 (2001);
quant-ph/0004064.

\item {\bf [Kempe-Bacon-DiVincenzo-Whaley 01]}:
J. Kempe, D. Bacon, D. P. DiVincenzo, \& K. B. Whaley,
``Encoded universality from a single physical interaction'',
{\em Quant. Inf. Comp.} {\bf 1}, {\em suppl.}, 33-55 (2001);
quant-ph/0112013.

\item {\bf [Kempe-Whaley 02]}:
J. Kempe, \& K. B. Whaley,
``Exact gate sequences for universal quantum computation using the $XY$
interaction alone'',
{\em Phys. Rev. A} {\bf 65}, 5, 052330 (2002);
quant-ph/0112014.

\item {\bf [Kempe-Regev 03]}:
J. Kempe, \& O. Regev,
``3-Local Hamiltonian is QMA-complete'',
{\em Quant. Comp. Inf.} {\bf 3}, 3, 258-264 (2003);
quant-ph/0302079.

\item {\bf [Kempe 03]}:
J. Kempe,
``Quantum random walks -- An introductory overview'',
{\em Contemp. Phys.} {\bf 44}, 4, 307-327 (2003);
quant-ph/0303081.

\item {\bf [Kempe-Shalev 04]}:
J. Kempe, \& A. Shalev,
``The hidden subgroup problem and permutation group theory'',
quant-ph/0406046.

\item {\bf [Kempe-Kitaev-Regev 04]}:
J. Kempe, A. Kitaev, \& O. Regev,
``The complexity of the local Hamiltonian problem'',
quant-ph/0406180.

\item {\bf [Keller-Mahler 94]}:
M. Keller, \& G. Mahler,
``Nanostructures, entanglement and the physics of quantum control'',
in S. M. Barnett, A. K. Ekert, \& S. J. D. Phoenix (eds.),
{\em J. Mod. Opt.} {\bf 41}, 12 (Special issue: Quantum
communication), 2537-2556 (1994).

\item {\bf [Keller-Rubin-Shih-Wu 98]}:
T. E. Keller, M. H. Rubin, Y. H. Shih, \& L.-A. Wu,
``Theory of the three-photon entangled state'',
{\em Phys. Rev. A} {\bf 57}, 3, 2076-2079 (1998).

\item {\bf [Keller-Rubin-Shih 98 a]}:
T. E. Keller, M. H. Rubin, \& Y. H. Shih,
``Two-photon interference from separate pulses'',
{\em Phys. Lett. A} {\bf 244}, 6, 507-511 (1998).

\item {\bf [Keller-Rubin-Shih 98 b]}:
T. E. Keller, M. H. Rubin, \& Y. H. Shih,
``Three-photon entangled state'',
{\em Fortschr. Phys.} {\bf 46}, 6-8, 673-682 (1998).

\item {\bf [Kells-Twamley-Heffernan 03]}:
G. A. Kells, J. Twamley, \& D. M. Heffernan,
``Dynamical properties of the delta kicked harmonic oscillator'',
quant-ph/0307213.

\item {\bf [Kendon-Nemoto-Munro 02]}:
V. M. Kendon, K. Nemoto, \& W. J. Munro,
``Typical entanglement in multiple-qubit systems'',
{\em Proc.\ ESF QIT Conf.\ Quantum Information: Theory, Experiment and Perspectives
(Gdansk, Poland, 2001)}, {\em J. Mod. Opt.} {\bf 49}, 8, 1709-1716 (2002);
quant-ph/0106023.

\item {\bf [Kendon-\.{Z}yczkowski-Munro 02]}:
V. M. Kendon, K. \.{Z}yczkowski, \& W. J. Munro,
``Bounds on entanglement in qudit subsystems'',
{\em Phys. Rev. A} {\bf 66}, 6, 062310 (2002);
quant-ph/0203037.

\item {\bf [Kendon-Tregenna 03]}:
V. Kendon, \& B. Tregenna,
``Decoherence can be useful in quantum walks'',
{\em Phys. Rev. A} {\bf 67}, 4, 042315 (2003).

\item {\bf [Kendon 03]}:
V. Kendon,
``Quantum walks on general graphs'',
quant-ph/0306140.

\item {\bf [Kendon-Sanders 04]}:
V. Kendon, \& B. C. Sanders,
``Complementarity and quantum walks'',
quant-ph/0404043.

\item {\bf [Kenfack-\.{Z}yczkowski 04]}:
A. Kenfack, \& K. \.{Z}yczkowski,
``Negativity of the Wigner function as an indicator of nonclassicality'',
quant-ph/0406015.

\item {\bf [Kennedy-Zhou 01]}:
T. A. B. Kennedy, \& P. Zhou,
``Atomic dark states and motional entanglement in cavity QED'',
{\em Phys. Rev. A} {\bf 64}, 6, 063805 (2001).

\item {\bf [Kent 90]}:
A. Kent,
``Against many-worlds interpretations'',
{\em Int. J. Mod. Phys.} {\bf 5}, 9, 1745-1762 (1990);
gr-qc/9703089.

\item {\bf [Kent 95]}:
A. Kent,
``A note on Schmidt states and consistency'',
{\em Phys. Lett. A} {\bf 196}, 5-6, 313-317 (1995).

\item {\bf [Kent 96 a]}:
A. Kent,
``Quasiclassical dynamics in a closed quantum system'',
{\em Phys. Rev. A} {\bf 54}, 6, 4670-4675 (1996);
gr-qc/9512023.

\item {\bf [Kent 96 b]}:
A. Kent,
``Remarks on consistent histories and Bohmian mechanics'',
in {\bf [Cushing-Fine-Goldstein 96]}, pp.~343-352;
quant-ph/9511032.

\item {\bf [Kent-McElwaine 97]}:
A. Kent, \& J. McElwaine,
``Quantum prediction algorithms'',
{\em Phys. Rev. A} {\bf 55}, 3, 1703-1720 (1997);
gr-qc/9610028.

\item {\bf [Kent 97 a]}:
A. Kent,
``Consistent sets yield contrary inferences in quantum theory'',
{\em Phys. Rev. Lett.} {\bf 78}, 15, 2874-2877 (1997);
gr-qc/9604012.
See {\bf [Griffiths-Hartle 97]}.

\item {\bf [Kent 97 b]}:
A. Kent,
``Permanently secure quantum bit commitment from
a temporary computation bound'';
quant-ph/9712002.

\item {\bf [Kent 97 c]}:
A. Kent,
`Comment on ``Spacetime information''\,',
{\em Phys. Rev. D} {\bf 56}, 4, 2469-2472 (1997);
gr-qc/9610075.
Comment on {\bf [Hartle 95]}.

\item {\bf [Kent 98 a]}:
A. Kent,
``Entangled mixed states and local purification'',
{\em Phys. Rev. Lett.} {\bf 81}, 14, 2839-2841 (1998);
quant-ph/9805088.

\item {\bf [Kent 98 b]}:
A. Kent,
``Consistent sets and contrary inferences: Reply to Griffiths and Hartle'',
{\em Phys. Rev. Lett.} {\bf 81}, 9, 1982 (1998);
gr-qc/9808016.
Reply to {\bf [Griffiths-Hartle 98]}.
See {\bf [Kent 97 a]}.

\item {\bf [Kent 98 c]}:
A. Kent,
``Quantum histories'',
in E. B. Karlsson, \& E. Br\"{a}ndas (eds.),
{\em Proc.\ of the 104th Nobel Symp.\ ``Modern Studies of Basic Quantum Concepts and Phenomena'' (Gimo, Sweden, 1997)},
{\em Physica Scripta} {\bf T76}, 57-60 (1998);
gr-qc/9809026.

\item {\bf [Kent 99 a]}:
A. Kent,
``Unconditionally secure bit commitment'',
{\em Phys. Rev. Lett.} {\bf 83}, 7, 1447-1450 (1999);
quant-ph/9810068.

\item {\bf [Kent-Linden-Massar 99]}:
A. Kent, N. Linden, \& S. Massar,
``Optimal entanglement enhancement for mixed states'',
{\em Phys. Rev. Lett.} {\bf 83}, 13, 2656-2659 (1999);
quant-ph/9902022.

\item {\bf [Kent 99 b]}:
A. Kent,
``Hidden variables are compatible with physical measurements'',
{\em Phys. Rev. Lett.} {\bf 83}, 19, 3755-3757 (1999);
quant-ph/9906006.
See {\bf [Meyer 99 b]}, {\bf [Clifton-Kent 00]}, {\bf [Cabello 99 d, 02 c]},
{\bf [Havlicek-Krenn-Summhammer-Svozil 01]}, {\bf [Mermin 99 b]},
{\bf [Appleby 00, 01, 02]}, {\bf [Boyle-Schafir 01 a]}, {\bf [Wilce 04]}.

\item {\bf [Kent 99 c]}:
A. Kent,
``Secure classical bit commitment using fixed capacity communication
channels'',
quant-ph/9906103.

\item {\bf [Kent 99 d]}:
A. Kent,
``Coin tossing is strictly weaker than bit commitment'',
{\em Phys. Rev. Lett.} {\bf 83}, 25, 5382-5384 (1999);
quant-ph/9810067.

\item {\bf [Kent 99 e]}:
A. Kent,
``Causality in time-neutral cosmologies'',
{\em Phys. Rev. D} {\bf 59}, 4, 043505 (1999).
gr-qc/9703041.

\item {\bf [Kent 00 a]}:
A. Kent,
``Impossibility of unconditionally secure
commitment of a certified classical bit'',
{\em Phys. Rev. A} {\bf 61}, 4, 042301 (2000);
quant-ph/9910087.

\item {\bf [Kent 00 b]}:
A. Kent,
``Quantum histories and their implications'',
in F. Petruccione (ed.),
{\em Relativistic quantum measurement and decoherence},
{\em Lecture Notes in Physics} {\bf 559},
Springer-Verlag, New York, 2000, pp.~93-115;
gr-qc/9607073.
See {\bf [Peruzzi-Rimini 98]}.

\item {\bf [Kent-Wallace 01]}:
A. Kent, \& D. Wallace,
``Quantum interrogation and the safer X-ray'',
quant-ph/0102118.

\item {\bf [Kent 01 a]}:
A. Kent,
``A proposal for founding mistrustful quantum cryptography on coin tossing'',
quant-ph/0111097.

\item {\bf [Kent 01 b]}:
A. Kent,
``Quantum bit string commitment'',
quant-ph/0111099.

\item {\bf [Kent 02 a]}:
A. Kent,
``Locality and causality revisited'',
in T. Placek, \& J. Butterfield (eds.),
{\em Modality, probability, and Bell's theorems (Cracow, Poland, 2001)},
Kluwer Academic, Dordrecht, Holland, 2002;
quant-ph/0202064.

\item {\bf [Kent 02 b]}:
A. Kent,
``Causal quantum theory and the collapse locality loophole'',
quant-ph/0204104.

\item {\bf [Kent 02 c]}:
A. Kent,
``Nonlinearity without superluminality'',
quant-ph/0204106.

\item {\bf [Kent 02 d]}:
A. Kent,
``Large $N$ quantum cryptography'',
{\em Proc.\ QCMC02};
quant-ph/0212043.

\item {\bf [Kent 03 a]}:
A. Kent,
``Quantum bit string commitment'',
{\em Phys. Rev. Lett.} {\bf 90}, 23, 237901 (2003).

\item {\bf [Kent 03 b]}:
A. Kent,
``Proposal for founding mistrustful quantum cryptography on coin tossing'',
{\em Phys. Rev. A} {\bf 68}, 1, 012312 (2003).

\item {\bf [Kent 04]}:
A. Kent,
``Promising the impossible: Classical certification in a quantum world'',
quant-ph/0409029.

\item {\bf [Kerenidis-de Wolf 03]}:
I. Kerenidis, \& R. de Wolf,
``Quantum symmetrically-private information retrieval'',
quant-ph/0307076.

\item {\bf [Kernaghan 94]}:
M. Kernaghan,
``Bell-Kochen-Specker theorem for 20 vectors'',
{\em J. Phys. A} {\bf 27}, 21, L829-L830 (1994).
See {\bf [Peres 91 a]},
{\bf [Cabello-Estebaranz-Garc\'{\i}a Alcaine 96 a]}.

\item {\bf [Kernaghan-Peres 95]}:
M. Kernaghan, \& A. Peres,
``Kochen-Specker theorem for eight-dimensional space'',
{\em Phys. Lett. A} {\bf 198}, 1, 1-5 (1995);
quant-ph/9412006.

\item {\bf [Kernaghan 95]}:
M. Kernaghan,
``The Kochen-Specker paradox. A
doctoral dissertation in the History and Philosophy of Science'',
Ph.\ D. thesis, University of Western Ontario, London, Canada, 1995.

\item {\bf [Kessel-Ermakov 99]}:
A. R. Kessel, \& V. L. Ermakov,
``?'',
{\em Pis'ma Zh. Eksp. Teor. Fiz.} {\bf 70}, 1, 59-63 (1999).
English version:
``Multiqubit spin'',
{\em JETP Lett.} {\bf 70}, 1, 61-65 (1999);
quant-ph/9912047.

\item {\bf [Kessel-Ermakov 00 a]}:
A. R. Kessel, \& V. L. Ermakov,
``Three-qubit gate realization using single quantum particle'',
quant-ph/0002016.

\item {\bf [Kessel-Ermakov 00 b]}:
A. R. Kessel, \& V. L. Ermakov,
``Delocalized qubits as a computational basis in
the system of interacting spins'',
quant-ph/0006082.

\item {\bf [Kessel-Ermakov 00 c]}:
A. R. Kessel, \& V. L. Ermakov,
``Quantum entanglement and classical separability in NMR
computing'',
quant-ph/0011002.

\item {\bf [Kessel-Yakovleva 02]}:
A. R. Kessel, \& N. M. Yakovleva,
``Implementation schemes in NMR of quantum processors and the Deutsch-Jozsa
algorithm by using virtual spin representation'',
{\em Phys. Rev. A} {\bf 66}, 6, 062322 (2002).

\item {\bf [Keyl-Werner 98]}:
M. Keyl, \& R. F. Werner,
``Optimal cloning of pure states, judging single clones'',
quant-ph/9807010.

\item {\bf [Keyl-Werner 99]}:
M. Keyl, \& R. F. Werner,
``The rate of optimal purification procedures'',
quant-ph/9910124.

\item {\bf [Keyl-Werner 01]}:
M. Keyl, \& R. F. Werner,
``Estimating the spectrum of a density operator'',
{\em Phys. Rev. A} {\bf 64}, 5, 052311 (2001);
quant-ph/0102027.

\item {\bf [Keyl 02]}:
M. Keyl,
``Fundamentals of quantum information theory'',
{\em Phys. Rep.} {\bf 369}, 5, 431-548 (2002);
quant-ph/0202122.

\item {\bf [Keyl-Schlingemann-Werner 02]}:
M. Keyl, D. Schlingemann, \& R. F. Werner,
``Infinitely entangled states'',
quant-ph/0212014.

\item {\bf [Khalfin-Cirel'son 85]}:
L. A. Khalfin, \& B. S. Cirel'son [Tsirelson],
``Quantum and quasi-classical analogs of Bell inequalities'',
in P. J. Lahti, \& P. Mittelstaedt (eds.),
{\em Symp.\ on the Foundations of Modern
Physics: 50 Years of the Einstein-Podolsky-Rosen Experiment
(Joensuu, Finland, 1985)},
World Scientific, Singapore, 1985, pp.~441-460.

\item {\bf [Khalfin-Cirel'son 87]}:
L. A. Khalfin, \& B. S. Cirel'son [Tsirelson],
``A quantitative criterion of the applicability
of the classical description within the quantum theory'',
in P. J. Lahti, \& P. Mittelstaedt (eds.),
{\em Symp.\ on the Foundations of Modern Physics 1987},
World Scientific, Singapore, 1985, pp.~369-401.

\item {\bf [Khalfin 82]}:
L. A. Khalfin,
``?'',
{\em Phys. Lett. B} {\bf 112}, ?, 223-? (1982).

\item {\bf [Khalfin 83]}:
L. A. Khalfin,
``Bell's inequalities, Tsirelson inequalities and
$K^0-\bar{K^0}$, $D^0-\bar{D^0}$, $B^0-\bar{B^0}$ mesons'',
report on the scientific session of the Nuclear Division of the Academy of Sciences USSR,
1983; unpublished.

\item {\bf [Khalfin 90]}:
L. A. Khalfin,
``The quantum-classical correspondence in
light of classical Bell's and quantum Tsirelson's inequalities'',
in {\bf [Zurek 90]}, pp.~477-493.

\item {\bf [Khalfin-Cirel'son 92]}:
L. A. Khalfin, \& B. S. Cirel'son [Tsirelson],
``Quantum/classical correspondence in the light of Bell's inequalities'',
{\em Found. Phys.} {\bf 22}, 7, 879-948 (1992).

\item {\bf [Khalique-Saif 03]}:
A. Khalique, \& F. Saif,
``Engineering entanglement between external degrees of freedom of atoms via Bragg scattering'',
{\em Phys. Lett. A} {\bf 314}, 1-2, 37-43 (2003).

\item {\bf [Khan-Zubairy 98]}:
A. A. Khan, \& M. S. Zubairy,
``Quantum logic gate operating on atomic scattering
by standing wave field in Bragg regime'',
{\em Fortschr. Phys.} {\bf 46}, 4-5, 417-422 (1998).

\item {\bf [Khanna-Mann-Revzen-Roy 02]}:
F. Khanna, A. Mann, M. Revzen, \& S. Roy,
``Bell's inequality violation and symmetry'',
{\em Phys. Lett. A} {\bf 294}, 1, 1-5 (2002).

\item {\bf [Khaneja-Glaser-Brockett 02]}:
N. Khaneja, S. J. Glaser, \& R. Brockett,
``Sub-Riemannian geometry and time optimal control of three spin systems:
Quantum gates and coherence transfer'',
{\em Phys. Rev. A} {\bf 65}, 3, 032301 (2002).

\item {\bf [Khaneja-Glaser 02]}:
N. Khaneja, \& S. J. Glaser,
``Efficient transfer of coherence through Ising spin chains'',
{\em Phys. Rev. A} {\bf 66}, 6, 060301 (2002).

\item {\bf [Khitrin-Sun-Fung 01]}:
A. Khitrin, H. Sun, \& B. M. Fung,
``Method of multifrequency excitation for creating pseudopure
states for NMR quantum computing'',
{\em Phys. Rev. A} {\bf 63}, 2, 020301(R) (2001).

\item {\bf [Khitrin-Ermakov-Fung 02]}:
A. K. Khitrin, V. L. Ermakov, \& B. M. Fung,
``NMR implementation of a parallel search algorithm'',
{\em Phys. Rev. Lett.} {\bf 89}, 27, 277902 (2002).

\item {\bf [Khitun-Ostroumov-Wang 01]}:
A. Khitun, R. Ostroumov, \& K. L. Wang,
``Spin-wave utilization in a quantum computer'',
{\em Phys. Rev. A} {\bf 64}, 6, 062304 (2001).

\item {\bf [Khlebnikov-Sadiek 02]}:
S. Khlebnikov, \& G. Sadiek,
``Decoherence by a nonlinear environment: Canonical versus microcanonical case'',
{\em Phys. Rev. A} {\bf 66}, 3, 032312 (2002);
quant-ph/0202125.

\item {\bf [Khodjasteh-Lidar 02]}:
K. Khodjasteh, \& D. A. Lidar,
``Universal fault-tolerant quantum computation in the presence of spontaneous
emission and collective dephasing'',
{\em Phys. Rev. Lett.} {\bf 89}, 19, 197904 (2002).

\item {\bf [Khodjasteh-Lidar 03]}:
K. Khodjasteh, \& D. A. Lidar,
``Quantum computing in the presence of spontaneous emission by a combined
dynamical decoupling and quantum-error-correction strategy'',
{\em Phys. Rev. A} {\bf 68}, 2, 022322 (2003);
quant-ph/0301105.

\item {\bf [Khodjasteh-Lidar 04]}:
K. Khodjasteh, \& D. A. Lidar,
``Concatenated dynamical decoupling'',
quant-ph/0408128.

\item {\bf [Khrennikov 00 a]}:
A. Khrennikov,
``Einstein and Bell, von Mises and Kolmogorov:
Reality and locality, frequency and probability'',
quant-ph/0006016.

\item {\bf [Khrennikov 00 b]}:
A. Khrennikov,
``Kolmogorov and von Mises viewpoints to
the Greenburger-Horne-Zeilinger paradox'',
quant-ph/0006017.

\item {\bf [Khrennikov 01 a]}:
A. Khrennikov,
``Contextualist viewpoint to Greenberger-Horne-Zeilinger
paradox'',
{\em Phys. Lett. A} {\bf 278}, 6, 307-314 (2001);
quant-ph/0309065.

\item {\bf [Khrennikov 01 b]}:
A. Khrennikov,
`\,``Quantum probabilities' as context depending
probabilities'',
quant-ph/0106073.

\item {\bf [Khrennikov-Volovich 01]}:
A. Y. Khrennikov, \& Y. I. Volovich,
``Numerical experiment on interference for macroscopic particles'',
quant-ph/0111159.

\item {\bf [Khrennikov 02 a]}:
A. Khrennikov,
``Frequency analysis of the EPR-Bell argumentation'',
{\em Found. Phys.} {\bf 32}, 7, 1159-1174 (2002).

\item {\bf [Khrennikov 02 b]}:
A. Khrennikov,
``Fundamental principle for quantum theory'',
in A. Khrennikov (ed.),
{\em Quantum Theory: Reconsideration of Foundations (V\"{a}xj\"{o}, Sweden, 2001)},
V\"{a}xj\"{o} University Press, V\"{a}xj\"{o}, Sweden, 2002, pp.~188-196;
quant-ph/0204008.

\item {\bf [Khrennikov 02 c]}:
A. Khrennikov,
``V\"{a}xj\"{o} interpretation of quantum mechanics'',
quant-ph/0202107.

\item {\bf [Khrennikov 03 a]}:
A. Khrennikov,
``Bell's inequality for conditional probabilities and its violation by one
particle quantum system'',
quant-ph/0308078.

\item {\bf [Khrennikov 03 b]}:
A. Khrennikov,
``Probabilistic foundations of quantum mechanics and quantum information'',
quant-ph/0309066.

\item {\bf [Khrennikov 03 b]}:
A. Khrennikov,
``How can one find nonlocality in Bohmian mechanics?'',
quant-ph/0311108.

\item {\bf [Khrennikov 04 a]}:
A. Khrennikov,
``V\"{a}xj\"{o} interpretation-2003: Realism of contexts'',
quant-ph/0401072.

\item {\bf [Khrennikov 04 b]}:
A. Khrennikov,
``EPR-Bohm experiment and interference of probabilities'',
quant-ph/0402068.

\item {\bf [Khrennikov 04 c]}:
A. Khrennikov,
``Bell's inequality for conditional probabilities as a test for
quantum-like behaviour of mind'',
quant-ph/0402169.

\item {\bf [Khrennikov 04 d]}:
A. Khrennikov,
``On the notion of a macroscopic quantum system'',
quant-ph/0408164.

\item {\bf [Khrennikov 04 e]}:
A. Khrennikov,
``Principle of supplementarity: Contextual probabilistic viewpoint to
interference, complementarity and incompatibility'',
quant-ph/0408187.

\item {\bf [Khrennikov-Loubenets 02]}:
A. Y. Khrennikov, \& E. Loubenets,
``On relations between probabilities under quantum and classical measurements'',
quant-ph/0204001.

\item {\bf [Khveshchenko 03]}:
D. V. Khveshchenko,
``Entanglement and decoherence in near-critical qubit chains'',
{\em Phys. Rev. B} {\bf 68}, 19, 193307 (2003).

\item {\bf [Kiefer-Joos 98]}:
C. Kiefer, \& E. Joos,
``Decoherence: Concepts and
examples'',
quant-ph/9803052.

\item {\bf [Kielpinski-King-Myatt-(+6) 00]}:
D. Kielpinski, B. E. King, C. J. Myatt, C. A. Sackett, Q. A. Turchette,
W. M. Itano, C. Monroe, D. J. Wineland, \& W. H. Zurek,
``Sympathetic cooling of trapped ions for quantum logic'',
{\em Phys. Rev. A} {\bf 61}, 3, 032310 (2000);
quant-ph/9909035.

\item {\bf [Kielpinski-Meyer-Rowe-(+4) 01]}:
D. Kielpinski, V. Meyer, M. A. Rowe, C. A. Sackett,
W. M. Itano, C. Monroe, \& D. J. Wineland,
``A decoherence-free quantum memory using trapped ions'',
{\em Science} {\bf 291}, ?, 1013-1015 (2001).

\item {\bf [Kielpinski-Ben Kish-Britton-(+6) 01]}:
D. Kielpinski, A. Ben-Kish, J. Britton,
V. Meyer, M. A. Rowe, C. A. Sackett,
W. M. Itano, C. Monroe, \& D. J. Wineland,
``Recent results in trapped-ion quantum computing'',
{\em Experimental Implementation of Quantum Computation (Sydney, 2001)};
quant-ph/0102086.

\item {\bf [Kielpinski-Monroe-Wineland 02]}:
D. Kielpinski, C. Monroe, \& D. J. Wineland,
``Architecture for a large-scale ion-trap quantum computer'',
{\em Nature} {\bf 417}, 6890, 709-711 (2002).

\item {\bf [Kiesel-Bourennane-Kurtsiefer-(+4) 03]}:
N. Kiesel, M. Bourennane, C. Kurtsiefer,
H. Weinfurter, D. Kaszlikowski, W. Laskowski, \& M. \.{Z}ukowski,
``Three-photon $W$-state'',
in M. Ferrero (ed.),
{\em Proc. of Quantum Information: Conceptual Foundations,
Developments and Perspectives (Oviedo, Spain, 2002)},
{\em J. Mod. Opt.} {\bf 50}, 6-7, 1131-1138 (2003).

\item {\bf [Kiess-Shih-Sergienko-Alley 93]}:
T. E. Kiess, Y. H. Shih, A. V.
Sergienko, \& C. O. Alley,
``Einstein-Podolsky-Rosen-Bohm experiment using pairs of
light quanta produced by type-II parametric down-conversion'',
{\em Phys. Rev. Lett.} {\bf 71}, 24, 3893-3897 (1993).

\item {\bf [Kiess-Shih-Sergienko-Alley 95]}:
T. E. Kiess, Y. H. Shih, A. V.
Sergienko, \& C. O. Alley,
``Tunable Bell-inequality violations by non-maximally-violating states
in type-II parametric down-conversion'',
{\em Phys. Rev. A} {\bf 52}, 4, 3344-3347 (1995).

\item {\bf [Kieu-Danos 98]}:
T. D. Kieu, \& M. Danos,
``The halting problem for universal quantum computers'',
quant-ph/9811001.

\item {\bf [Kieu 01 a]}:
T. D. Kieu,
``Quantum algorithm for the Hilbert's tenth problem'',
{\em Int. J. Theor. Phys.} {\bf 42}, 1451-1468 (2003);
quant-ph/0110136.
See {\bf [Cirel'son 01]}, {\bf [Kieu 01 b, c]}.

\item {\bf [Kieu 01 b]}:
T. D. Kieu,
`Reply to ``The quantum algorithm of Kieu does
not solve the Hilbert's tenth problem''\,'
quant-ph/0111020.
Reply to {\bf [Cirel'son 01]}.
See {\bf [Kieu 01 a, c]}.

\item {\bf [Kieu 01 c]}:
T. D. Kieu,
``A reformulation of the Hilbert's tenth problem through quantum mechanics'',
quant-ph/0111063.
See {\bf [Kieu 01 a, b]}, {\bf [Cirel'son 01]}.

\item {\bf [Kieu 03]}:
T. D. Kieu,
``Computing the noncomputable'',
{\em Contemp. Phys.} {\bf 44}, 51-71 (2003).

\item {\bf [Kieu 04]}:
T. D. Kieu,
``An anatomy of a quantum adiabatic algorithm that transcends the Turing
computability'',
in {\em Proc.\ of Foundations of Quantum Information" (Camerino, Italy, 2004)},
{\em Int. J. Quantum Inf.};
quant-ph/0407090.

\item {\bf [Kilin 03]}:
S. Y. Kilin,
``Entangled states and nanoobjects in quantum optics'',
{\em Opt. Spectrosc.} {\bf 94}, 649 (2003).

\item {\bf [Kim-Ko-Kim 03 a]}:
H. Kim, J. Ko, \& T. Kim,
``Two-particle interference experiment with frequency-entangled photon pairs'',
{\em J. Opt. Soc. Am. B} {\bf 20}, ?, 760-? (2003).

\item {\bf [Kim-Ko-Kim 03 b]}:
H. Kim, J. Ko, \& T. Kim,
``Quantum-eraser experiment with frequency-entangled photon pairs'',
{\em Phys. Rev. A} {\bf 67}, 5, 054102 (2003).

\item {\bf [Kim-Cheong-Lee 04]}:
H. Kim, Y. W. Cheong, \& H.-W. Lee,
``Generalized measurement and conclusive teleportation with nonmaximal entanglement'',
{\em Phys. Rev. A} {\bf 70}, 1, 012309 (2004);
quant-ph/0404172.

\item {\bf [Kim-Mahler 99]}:
I. Kim, \& G. Mahler,
``Pattern formation in quantum Turing machines'',
{\em Phys. Rev. A} {\bf 60}, 1, 692-695 (1999).

\item {\bf [Kim-Mahler 00 a]}:
I. Kim, \& Mahler,
``Uncertainty rescued:
Bohr's complementarity for composite systems'',
{\em Phys. Lett. A} {\bf 269}, 5-6, 287-292 (2000);
quant-ph/0004056.

\item {\bf [Kim-Mahler 00 b]}:
I. Kim, \& Mahler,
``Quantum network architecture of tight-binding models
with substitution sequences'',
quant-ph/0001092.

\item {\bf [Kim-Mahler 01]}:
I. Kim, \& G. Mahler,
``Delayed-choice measurement and temporal nonlocality'',
{\em Zeitschrift f\"{u}r Naturforschung A} {\bf 56}, 202-? (2001);
quant-ph/0105136.

\item {\bf [Kim-Lee-Lee 00]}:
J. Kim, J.-S. Lee, \& S. Lee,
``Implementing unitary operators in quantum computation'',
{\em Phys. Rev. A} {\bf 61}, 3, 032312 (2000);
quant-ph/9908052.

\item {\bf [Kim-Lee-Lee-Cheong 00]}:
J. Kim, J.-S. Lee, S. Lee, \& C. Cheong,
``Implementation of the refined Deutsch-Jozsa algorithm
on a three-bit NMR quantum computer'',
{\em Phys. Rev. A} {\bf 62}, 2, 022312 (2000);
quant-ph/9910015.

\item {\bf [Kim-Cheong-Lee-Lee 02]}:
J. Kim, Y. Cheong, J.-S. Lee, \& S. Lee,
``Storing unitary operators in quantum states'',
{\em Phys. Rev. A} {\bf 65}, 1, 012302 (2002);
quant-ph/0109097.

\item {\bf [Kim-Lee-Lee 02]}:
J. Kim, J.-S. Lee, \& S. Lee,
``Experimental realization of a target-accepting quantum search by NMR'',
{\em Phys. Rev. A} {\bf 65}, 5, 054301 (2002).

\item {\bf [Kim 00]}:
M. Kim,
``?'',
{\em J. Kor. Phys.} {\bf 37}, ?, 490-? (2000).

\item {\bf [Kim-Bu\v{z}ek 92]}:
M. S. Kim, \& V. Bu\v{z}ek,
``Decay of quantum coherences under the influence of a thermal
bath. Schr\"{o}dinger cat states at finite temperature'',
{\em J. Mod. Opt.} {\bf 39}, 8, 1609-1614 (1992).

\item {\bf [Kim-Jo-Choi-(+3) 99]}:
M. S. Kim, S. G. Jo, S. D. Choi, J. H. Kim,
H. M. Jeon, \& H. I. Kim,
``The contextuality of the possessed values'',
{\em J. Phys. A} {\bf 32}, 17, 3117-3125 (1999).

\item {\bf [Kim-Lee 00]}:
M. S. Kim, \& J. Lee,
``Test of quantum nonlocality for cavity fields'',
{\em Phys. Rev. A} {\bf 61}, 4, 042102 (2000);
quant-ph/0002074.

\item {\bf [Kim-Lee 01]}:
M. S. Kim, \& J. Lee,
``Asymmetric quantum channel for quantum teleportation'',
{\em Phys. Rev. A} {\bf 64}, 1, 012309 (2001);
quant-ph/0005022.

\item {\bf [Kim-Son-Bu\v{z}zek-Knight 02]}:
M. S. Kim, W. Son, V. Bu\v{z}zek, \& P. L. Knight,
``Entanglement by a beam splitter: Nonclassicality as a prerequisite for
entanglement'',
{\em Phys. Rev. A} {\bf 65}, 3, 032323 (2002);
quant-ph/0106136.

\item {\bf [Kim-Lee-Ahn-Knight 02]}:
M. S. Kim, J. Lee, D. Ahn, \& P. L. Knight,
``Entanglement induced by a single-mode heat environment'',
{\em Phys. Rev. A} {\bf 65}, 4, 040101 (2002);
quant-ph/0109052.

\item {\bf [Kim-Lee-Munro 02]}:
M. S. Kim, J. Lee, \& W. J. Munro,
``Experimentally realizable characterizations of continuous-variable Gaussian states'',
{\em Phys. Rev. A} {\bf 66}, 3, 030301 (2002);
quant-ph/0203151.

\item {\bf [Kim-Park-Knight-Jeong 04]}:
M. S. Kim, E. Park, P. L. Knight, \& H. Jeong,
``Nonclassicality of a photon-subtracted Gaussian field'',
quant-ph/0409218.

\item {\bf [Kim-Yu-Kulik-(+2) 00]}:
Y.-H. Kim, R. Yu, S. P. Kulik, Y. H. Shih, \& M. O. Scully,
``Delayed `choice' quantum eraser'',
{\em Phys. Rev. Lett.} {\bf 84}, 1, 1-5 (2000);
quantum-ph/9903047.

\item {\bf [Kim-Chekhova-Kulik-Shih 99]}:
Y.-H. Kim, M. V. Chekhova, S. P. Kulik, \& Y. H. Shih,
``Quantum interference by two temporally distinguishable pulses'',
{\em Phys. Rev. A} {\bf 60}, 1, R37-R40 (1999);
quant-ph/9903048.

\item {\bf [Kim-Chekhova-Kulik-(+2) 99]}:
Y.-H. Kim, M. V. Chekhova, S. P. Kulik,
Y. H. Shih, \& M. H. Rubin,
``First-order interference of nonclassical light emitted
spontaneously at different times'',
{\em Phys. Rev. A} {\bf 61}, 5, 051803(R).
quant-ph/9911014.

\item {\bf [Kim-Shih 99]}:
Y.-H. Kim, \& Y. Shih,
``Experimental realization of Popper's experiment:
Violation of the uncertainty principle?'',
{\em Found. Phys.} {\bf 29}, 12, 1849-1862 (1999).
Comment: {\bf [Short 00]}.

\item {\bf [Kim-Chekhova-Kulik-(+3) 00]}:
Y.-H. Kim, M. V. Chekhova, S. P. Kulik, Y. H. Shih,
T. E. Keller, \& M. H. Rubin,
``Quantum interference by two temporally distinguishable pulses'',
{\em Fortschr. Phys.} {\bf 48}, 5-7, 505-510 (2000).

\item {\bf [Kim-Kulik-Shih 01 a]}:
Y.-H. Kim, S. P. Kulik, \& Y. H. Shih,
``Quantum teleportation of a polarization state with a complete
Bell state measurement'',
{\em Phys. Rev. Lett.} {\bf 86}, 7, 1370-1373 (2001);
quant-ph/0010046.

\item {\bf [Kim-Kulik-Shih 01 b]}:
Y.-H. Kim, S. P. Kulik, \& Y. H. Shih,
``Bell-state preparation using pulsed nondegenerate
two-photon entanglement'',
{\em Phys. Rev. A} {\bf 63}, 6, 060301(R) (2001);
quant-ph/0007067.

\item {\bf [Kim-Chekhova-Kulik-(+2) 01]}:
Y.-H. Kim, M. V. Chekhova, S. P. Kulik,
M. H. Rubin, \& Y. Shih,
``Interferometric Bell-state preparation using
femtosecond-pulse-pumped spontaneous parametric down-conversion'',
{\em Phys. Rev. A} {\bf 63}, 6, 062301 (2001);
quant-ph/0103168.

\item {\bf [Kim-Kulik-Shih 02]}:
Y.-H. Kim, S. P. Kulik, \& Y. Shih,
``Quantum teleportation with a complete Bell state measurement'',
{\em J. Mod. Opt.} {\bf 49}, 1-2, 221-236 (2002).

\item {\bf [Kim-Kulik-Chekhova-(+2) 03]}:
Y.-H. Kim, S. P. Kulik, M. V. Chekhova,
W. P. Grice, \& Y. Shih,
``Experimental entanglement concentration and universal Bell-state synthesizer'',
{\em Phys. Rev. A} {\bf 67}, 1, 010301 (2003);
quant-ph/0209041.

\item {\bf [Kim 03 a]}:
Y.-H. Kim,
``Single-photon two-qubit entangled states: Preparation and measurement'',
{\em Phys. Rev. A} {\bf 67}, 4, 040301 (2003);
quant-ph/0303125.

\item {\bf [Kim 03 b]}:
Y.-H. Kim,
``Quantum interference with beamlike type-II spontaneous parametric
down-conversion'',
{\em Phys. Rev. A} {\bf 68}, 1, 013804 (2003).

\item {\bf [Kim-Grice 03 a]}:
Y.-H. Kim, \& W. P. Grice,
``Observation of correlated-photon statistics using a single detector'',
{\em Phys. Rev. A} {\bf 67}, 6, 065802 (2003).

\item {\bf [Kim 03 c]}:
Y.-H. Kim,
``Measurement of one-photon and two-photon wave packets in spontaneous
parametric downconversion'',
{\em J. Opt. Soc. Am. B} {\bf 20}, 1959-? (2003).

\item {\bf [Kim-Grice 03 b]}:
Y.-H. Kim, \& W. P. Grice,
``Reliability of the beam-splitter–based Bell-state measurement'',
{\em Phys. Rev. A} {\bf 68}, 6, 062305 (2003).

\item {\bf [Kimble-Ou-Pereira 94]}:
H. J. Kimble, Z. Y. Ou, \& S. E. Pereira,
``Method and apparatus for quantum communication employing
nonclassical correlations of quadrature-phase amplitudes'',
patent US5339182, 1994.

\item {\bf [Kimble 99]}:
H. J. Kimble,
``Quantum teleportation'',
in {\bf [Greenberger-Reiter-Zeilinger 99]}, pp.~141-146.

\item {\bf [Kimble-Wolf 01]}:
H. J. Kimble, \& E. Wolf,
``Leonard Mandel'',
{\em Phys. Today} {\bf 54}, 8, ?-? (2001).

\item {\bf [Kimble 03]}:
H. J. Kimble,
Comment on ``Deterministic single-photon source for distributed quantum
networking'',
{\em Phys. Rev. Lett.} {\bf 90}, 24, 249801 (2003).
Comment on {\bf [Kuhn-Hennrich-Rempe 02]}.
Reply: {\bf [Kuhn-Hennrich-Rempe 03]}.

\item {\bf [Kimm-Kwon 02]}:
K. Kimm, \& H.-H. Kwon,
``Decoherence of the quantum gate in Milburn's model of decoherence'',
{\em Phys. Rev. A} {\bf 65}, 2, 022311 (2002).

\item {\bf [Kimura 03]}:
G. Kimura,
``The Bloch vector for $N$-level systems'',
{\em Phys. Lett. A} {\bf 314}, 5-6, 339-349 (2003);
quant-ph/0301152.

\item {\bf [Kimura-Nambu-Hatanaka-(+3) 04]}:
T. Kimura, Y. Nambu, T. Hatanaka,
A. Tomita, H. Kosaka, \& K. Nakamura,
``Single-photon interference over 150 km transmission using silica-based
integrated-optic interferometers for quantum cryptography'',
{\em Jpn. J. Appl. Phys.} {\bf 43}, 9, L1217-L1219 (2004);
quant-ph/0403104.

\item {\bf [Kindermann-Cory 04]}:
M. Kindermann, \& D. G. Cory,
``Read out of a nuclear spin qubit'',
quant-ph/0411038.

\item {\bf [King-Wood-Myatt-(+5) 98]}:
B. E. King, C. S. Wood, C. J. Myatt, Q. A. Turchette,
D. Leibfried, W. M. Itano, C. Monroe, \& D. J. Wineland,
``Initializing the collective motion of trapped ions for quantum logic'',
quant-ph/9803023.

\item {\bf [King-Ruskai 99]}:
C. King, \& M. B. Ruskai,
``Minimal entropy of states emerging from noisy quantum
channels'',
quant-ph/9911079.

\item {\bf [King-Ruskai 01]}:
C. King, \& M. B. Ruskai,
``Capacity of quantum channels using product measurements'',
{\em J. Math. Phys.} {\bf 42}, 1, 87-98 (2001);
quant-ph/0004062.

\item {\bf [King 02 a]}:
C. King,
``Maximization of capacity and $l_p$ norms for some product channels'',
{\em J. Math. Phys.} {\bf 43}, 3, 1247-1260 (2002).

\item {\bf [King 02 b]}:
C. King,
``Additivity for a class of unital qubit channels'',
{\em J. Math. Phys.} {\bf 43}, 10, 4641-4653 (2002).
quant-ph/0103156.

\item {\bf [King-Nathanson-Ruskai 02]}:
C. King, M. Nathanson, \& M. B. Ruskai,
``Qubit channels can require more than two inputs to achieve capacity'',
{\em Phys. Rev. Lett.} {\bf 88}, 5, 057901 (2002);
quant-ph/0109079.

\item {\bf [King-Ruskai 04]}:
C. King, \& M. B. Ruskai,
``Comments on multiplicativity of maximal p-norms when $p = 2$
for the Festschrift in honor of A. Holevo's 60th birthday;
quant-ph/0401026.

\item {\bf [King-Nathanson-Ruskai 04]}:
C. King, M. Nathanson, \& M. B. Ruskai,
``Multiplicativity properties of entrywise positive maps on matrix
algebras'',
quant-ph/0409181.

\item {\bf [Kinnunen-Torma-Pekola 03]}:
J. Kinnunen, P. Torma, \& J. P. Pekola,
``Measuring charge-based quantum bits by a superconducting single-electron
transistor'',
{\em Phys. Rev. B} {\bf 68}, 2, 020506 (2003).

\item {\bf [Kiraz-Atat\"{u}re-Imamolu 04]}:
A. Kiraz, M. Atat\"{u}re, \& A. Imamolu,
``Quantum-dot single-photon sources:
Prospects for applications in linear optics quantum-information processing'',
{\em Phys. Rev. A} {\bf 69}, 3, 032305 (2004);
quant-ph/0308117.

\item {\bf [Kirkpatrick 01 a]}:
K. A. Kirkpatrick,
``Indistinguishability and improper mixtures'',
quant-ph/0109146.
See {\bf [Kirkpatrick 04]}.

\item {\bf [Kirkpatrick 01 b]}:
K. A. Kirkpatrick,
``Formalization of the distinguishability heuristics'',
quant-ph/0110052.
Replaced by {\bf [Kirkpatrick 03 d]}.

\item {\bf [Kirkpatrick 03 a]}:
K. A. Kirkpatrick,
``Classical three-box `paradox'\,'',
{\em J. Phys. A} {\bf 36}, 17, 4891-4900 (2003).
See {\bf [Aharonov-Vaidman 91]}.

\item {\bf [Kirkpatrick 03 b]}:
K. A. Kirkpatrick,
`\,``Quantal'' behavior in classical probability',
{\em Found. Phys. Lett.} {\bf 14}, 3, 199-224 (2003).

\item {\bf [Kirkpatrick 03 c]}:
K. A. Kirkpatrick,
``Hardy's second axiom is insufficiently general'',
quant-ph/0302158.
See {\bf [Hardy 01, 02 a]}.

\item {\bf [Kirkpatrick 03 d]}:
K. A. Kirkpatrick,
``The Schr\"{o}dinger-HJW theorem'',
quant-ph/0305068.
See {\bf [Hughston-Jozsa-Wootters 93]}.

\item {\bf [Kirkpatrick 03 e]}:
K. A. Kirkpatrick,
``Indistinguishability and the external correlation of mixtures'',
quant-ph/0308160.
Replaces {\bf [Kirkpatrick 01 b]}.

\item {\bf [Kirkpatrick 04]}:
K. A. Kirkpatrick,
`Error in an argument regarding ``improper'' mixtures',
quant-ph/0405058.
See {\bf [Kirkpatrick 01 a]}.

\item {\bf [Kis-Stenholm 01]}:
Z. Kis, \& S. Stenholm,
``Measuring the density matrix by local addressing'',
{\em Phys. Rev. A} {\bf 64}, 6, 065401 (2001);
quant-ph/0309067.

\item {\bf [Kis-Renzoni 02]}:
Z. Kis, \& F. Renzoni,
``Qubit rotation by stimulated Raman adiabatic passage'',
{\em Phys. Rev. A} {\bf 65}, 3, 032318 (2002).

\item {\bf [Kis-Paspalakis 04]}:
Z. Kis, \& E. Paspalakis,
``Arbitrary rotation and entanglement of flux SQUID qubits'',
{\em Phys. Rev. B} {\bf 69}, 024510 (2004).

\item {\bf [Kisil 01]}:
V. V. Kisil,
``Two slits interference is compatible with particles' trajectories'',
quant-ph/0111094.

\item {\bf [Kitaev 95]}:
A. Y. Kitaev,
``Quantum measurements and the Abelian stabilizer problem'',
quant-ph/9511026.

\item {\bf [Kitaev 97]}:
A. Y. Kitaev,
``Quantum computations: Algorithms and error correction'',
{\em Russ. Math. Surveys} {\bf 52}, ?, 1191-1249 (1997).

\item {\bf [Kitaev-Shen-Vyalyi 02]}:
A. Y. Kitaev, A. H. Shen, \& M. N. Vyalyi,
{\em Classical and quantum computation},
American Mathematical Society (hardback), Providence, Rhode Island, 2002;
Oxford University Press (paperback), Oxford, 2002.

\item {\bf [Kitaev 02]}:
A. Y. Kitaev,
``Topological quantum codes and anyons'',
in {\bf [Lomonaco 02 a]}, pp.~267-272.

\item {\bf [Kitaev-Mayers-Preskill 04]}:
A. Kitaev, D. Mayers, \& J. Preskill,
``Superselection rules and quantum protocols'',
{\em Phys. Rev. A} {\bf 69}, 5, 052326 (2004);
quant-ph/0310088.

\item {\bf [Kitagawa-Yamamoto 02]}:
A. Kitagawa, \& K. Yamamoto,
``Maximal entanglement of squeezed vacuum states via swapping with
number-phase measurement'',
{\em Phys. Rev. A} {\bf 66}, 5, 052312 (2002).

\item {\bf [Kitano 97]}:
M. Kitano,
``Quantum Zeno effect and adiabatic change'',
{\em Phys. Rev. A} {\bf 56}, 2, 1138-1141 (1997).

\item {\bf [Kitagawa-Yamamoto 03]}:
A. Kitagawa, \& K. Yamamoto,
``Teleportation-based number-state
manipulation with number-sum measurement'',
{\em Phys. Rev. A} {\bf 68}, 4, 042324 (2003).

\item {\bf [Klappenecker-R\"{o}tteler 01 a]}:
A. Klappenecker, \& M. R\"{o}tteler,
``Discrete cosine transforms on quantum computers'',
{\em IEEE ISPA01 (Pula, Croatia, 2001)}.
quant-ph/0111038.

\item {\bf [Klappenecker-R\"{o}tteler 01 b]}:
A. Klappenecker, \& M. R\"{o}tteler,
``On the irresistible efficiency of signal processing methods in quantum computing'',
{\em SPECLOG 2000 (Tampere, Finland, 2000)};
quant-ph/0111039.

\item {\bf [Klappenecker-R\"{o}tteler 03 a]}:
A. Klappenecker, \& M. R\"{o}tteler,
``Engineering functional quantum algorithms'',
{\em Phys. Rev. A} {\bf 67}, 1, 010302 (2003).

\item {\bf [Klappenecker-R\"{o}tteler 03 b]}:
A. Klappenecker, \& M. R\"{o}tteler,
``Unitary error bases: Constructions, equivalence, and applications'',
in M. Frossorier, T. Hoeholdt, \& A. Poli (eds.),
{\em AAECC 2003, LNCS 2643},
Springer-Verlag, Berlin, 2003, pp.~139-149.

\item {\bf [Klappenecker-R\"{o}tteler 03 c]}:
A. Klappenecker, \& M. R\"{o}tteler,
``Constructions of mutually unbiased bases'',
quant-ph/0309120.

\item {\bf [Klappenecker-R\"{o}tteler 03 d]}:
A. Klappenecker, \& M. R\"{o}tteler,
``Quantum software reusability'',
quant-ph/0309121.

\item {\bf [Klarreich 01]}:
E. Klarreich,
``Playing by quantum rules'',
{\em Nature} {\bf 414}, 6861, 244-245 (2001).

\item {\bf [Klarreich 02]}:
E. Klarreich,
``Quantum cryptography: Can you keep a secret?'',
{\em Nature} {\bf 418}, 6895, 270-272 (2002).

\item {\bf [Klauck 00 a]}:
H. Klauck,
``On rounds in quantum communication'',
quant-ph/0004100.

\item {\bf [Klauck 00 b]}:
H. Klauck,
``Quantum communication complexity'',
to be published in {\em Workshop Proc.\ of the 27th Int.\
Colloquium on Automata, Languages and Programming 2000,
Workshop on Boolean Functions and Applications},
quant-ph/0005032.

\item {\bf [Klauck 01 a]}:
H. Klauck,
``Lower bounds for quantum communication complexity'',
quant-ph/0106160.

\item {\bf [Klauck 01 b]}:
H. Klauck,
``On quantum and approximate privacy'',
quant-ph/0110038.

\item {\bf [Kleckner-Ron 01]}:
M. Kleckner, \& A. Ron,
``Decoherence of a pointer by a gas reservoir'',
{\em Phys. Rev. A} {\bf 63}, 2, 022110 (2001);
quant-ph/0011087.

\item {\bf [Klein-Nystrand 03]}:
S. R. Klein, \& J. Nystrand,
``Does particle decay cause wave function collapse: an experimental test'',
{\em Phys. Lett. A} {\bf 308}, 5-6, 323-328 (2003).

\item {\bf [Klembovsky-Gorodetsky-Becker-Walther 04]}:
M. P. Klembovsky, M. L. Gorodetsky, T. Becker, \& H. Walther,
``Quantum bit detector'',
{\em JETP Lett.} {\bf 79}, 9, 441-444 (2004);
quant-ph/0404155.

\item {\bf [Kleppner-Jackiw 00]}:
D. Kleppner, \& R. Jackiw,
``One hundred years of quantum physics'',
{\em Science} {\bf 289}, 5481, 893-898 (2000).
Corrections and clarifications:
{\em Science} {\bf 289}, ?, 2052 (2000).

\item {\bf [Klimov-Guzman-Retamal-Saavedra 03]}:
A. B. Klimov, R. Guzman, J. C. Retamal, \& C. Saavedra,
``Qutrit quantum computer with trapped ions'',
{\em Phys. Rev. A} {\bf 67}, 6, 062313 (2003).

\item {\bf [Klimov-de Guise-S\'{a}nchez Soto 03]}:
A. B. Klimov, H. de Guise, \& L. L. S\'{a}nchez-Soto,
``Inequivalent classes of closed three-level systems'',
quant-ph/0310122.

\item {\bf [Klimov-S\'{a}nchez Soto-de Guise-Bj{\o}rk 03]}:
A. B. Klimov, L. L. S\'{a}nchez-Soto, H. de Guise, \& G. Bj{\o}rk,
``Quantum phases of a qutrit'',
quant-ph/0312127.

\item {\bf [Klimovitch 04]}:
G. V. Klimovitch,
``How quantum entanglement helps to coordinate non-communicating players'',
quant-ph/0405093.

\item {\bf [Klose-Smith-Jessen 01]}:
G. Klose, G. Smith, \& P. S. Jessen,
``Measuring the quantum state of a large angular momentum'',
quant-ph/0101017.

\item {\bf [Klyachko-Shumovsky 02]}:
A. A. Klyachko, \& A. S. Shumovsky,
``Entangled states and local measurements'',
quant-ph/0203099.

\item {\bf [Klyachko-Shumovsky 03]}:
A. A. Klyachko, \& A. S. Shumovsky,
``Entanglement, local measurements, and symmetry'',
{\em J. Opt. B: Quantum Semiclass. Opt.};
quant-ph/0302008.

\item {\bf [Klyshko 93]}:
D. N. Klyshko,
``The Bell and GHZ theorems: A possible
three-photon interference experiment and the question of nonlocality'',
{\em Phys. Lett. A} {\bf 172}, 6, 399-403 (1993).
See {\bf [Belinsky-Klyshko 93 b]}.

\item {\bf [Klyshko 96]}:
D. N. Klyshko,
``The Bell theorem and the problem of moments'',
{\em Phys. Lett. A} {\bf 218}, 3-6, 119-127 (1996).

\item {\bf [Klyshko 97]}:
D. N. Klyshko,
``Quantum cryptography using
multicolored or multidirectional photons'',
{\em Phys. Lett. A} {\bf 227}, 1-2, 1-4 (1997).

\item {\bf [Klyshko 98 a]}:
D. N. Klyshko,
``Reduction of the wave function: An operational approach'',
{\em Phys. Lett. A} {\bf 243}, 4, 179-186 (1998).

\item {\bf [Klyshko 98 b]}:
D. N. Klyshko,
``Reduction of the wave-function: An operational approach'',
{\em Fortschr. Phys.} {\bf 46}, 6-8, 605-614 (1998).

\item {\bf [Klyshko 98 c]}:
D. N. Klyshko,
``On the realization and interpretation of `quantum teleportation'\,'',
{\em Phys. Lett. A} {\bf 247}, 4-5, 261-266 (1998).

\item {\bf [Knight-Vaidman 90]}:
J. M. Knight, \& L. Vaidman,
``Weak measurement of photon polarization'',
{\em Phys. Lett. A} {\bf 143}, 8, 357-361 (1990).

\item {\bf [Knight 94 a]}:
P. L. Knight,
``The quantum theory of motion'',
{\em J. Mod. Opt.} {\bf 41}, 1, 168-169 (1994).
Review of {\bf [Holland 93]}.

\item {\bf [Knight 94 b]}:
P. L. Knight,
``?'',
{\em Contemp. Phys.} {\bf 35}, 3, ?-? (1994).
Review of {\bf [Peres 93 a]}.

\item {\bf [Knight 96]}:
P. L. Knight,
``Schr\"{o}dinger's kittens transmission and the search for reality'',
{\em J. Mod. Opt.} {\bf 43}, 4, 865 (1996).
Review of {\bf [Gribbin 95]}.

\item {\bf [Knight 97]}:
P. L. Knight,
``Einstein, Bohr and the quantum dilemma'',
{\em J. Mod. Opt.} {\bf 44}, 8, 1596-1597 (1997).
Review of {\bf [Whitaker 95]}.

\item {\bf [Knight-Plenio-Vedral 97]}:
P. L. Knight, M. B. Plenio, \& V. Vedral,
``Decoherence and quantum error correction'',
in P. L. Knight, B. Stoicheff, \& D. Walls (eds.),
{\em Highlight in Quantum Optics},
{\em Philos. Trans. R. Soc. Lond. A} {\bf 355}, 1733, 2381-2385 (1997).

\item {\bf [Knight 98]}:
P. L. Knight,
``Quantum mechanics: Where the weirdness comes from'',
{\em Nature} {\bf 395}, 6697, 12-13 (1998).
See {\bf [D\"{u}rr-Nonn-Rempe 98]}.

\item {\bf [Knight-Murao-Plenio-Vedral 99]}:
P. L. Knight, M. Murao, M. B. Plenio, \& V. Vedral,
``Ion trap quantum gates, decoherence and error correction'',
{\em Chaos Solitons \& Fractals} {\bf 10}, 10, 1621-1635 (1999).

\item {\bf [Knight 00]}:
P. L. Knight,
``Quantum computing:
Enhanced: Quantum information processing without entanglement'',
{\em Science} {\bf 287}, 5452, 441-442 (2000).
See {\bf [Ahn-Weinacht-Bucksbaum 00]}.

\item {\bf [Knight-Roldan-Sipe 03]}:
P. L. Knight, E. Roldan, \& J. E. Sipe,
``Optical cavity implementations of the quantum walk'',
quant-ph/0305165.

\item {\bf [Knill-Laflamme 97]}:
E. Knill, \& R. Laflamme,
``Theory of quantum error-correcting codes'',
{\em Phys. Rev. A} {\bf 55}, 2, 900-911 (1997).

\item {\bf [Knill-Laflamme-Zurek 98 a]}:
E. Knill, R. Laflamme, \& W. H. Zurek,
``Resilient quantum computation: Error models and thresholds'',
in D. P. DiVincenzo. E. Knill, R. Laflamme, \& W. H. Zurek (eds.),
{\em Quantum Coherence and Decoherence.
Proc.\ of the ITP Conf.\ (Santa Barbara, California, 1996)},
{\em Proc. R. Soc. Lond. A} {\bf 454}, 1969, 365-384 (1998).

\item {\bf [Knill-Laflamme-Zurek 98 b]}:
E. Knill, R. Laflamme, \& W. H. Zurek,
``Resilient quantum computation'',
{\em Science} {\bf 279}, 5349, 342-345 (1998).

\item {\bf [Knill-Chuang-Laflamme 98]}:
E. Knill, I. L. Chuang, \& R. Laflamme,
``Effective pure states for bulk quantum computation'',
{\em Phys. Rev. A} {\bf 57}, 5, 3348-3363 (1998).

\item {\bf [Knill-Laflamme 98]}:
E. Knill, \& R. Laflamme,
``On the power of one bit of quantum information'',
quant-ph/9802037.

\item {\bf [Knill-Laflamme-Viola 00]}:
E. Knill, R. Laflamme, \& L. Viola,
``Theory of quantum error correction for general noise'',
{\em Phys. Rev. Lett.} {\bf 84}, 11, 2525-2528 (2000);
quant-ph/9908066.

\item {\bf [Knill-Laflamme 99]}:
E. Knill, \& R. Laflamme,
``Quantum computation and quadratically signed weight
enumerators'',
quant-ph/9909094.

\item {\bf [Knill-Laflamme-Martinez-Tseng 00]}:
E. Knill, R. Laflamme, R. Martinez, \& C.-H. Tseng,
``An algorithmic benchmark for quantum information processing'',
{\em Nature} {\bf 404}, 6776, 368-370 (2000);
quant-ph/9908051.

\item {\bf [Knill-Laflamme-Milburn 00]}:
E. Knill, R. Laflamme, \& G. Milburn,
``Thresholds for linear optics quantum computation'',
quant-ph/0006120.

\item {\bf [Knill-Nielsen 01 a]}:
E. H. Knill, \& M. A. Nielsen,
``Theory of quantum computation'',
in {\em Encyclopaedia of Mathematics. Suplement III}, 2001;
quant-ph/0010057.

\item {\bf [Knill-Nielsen 01 b]}:
E. H. Knill, \& M. A. Nielsen,
``Quantum information processing'',
in {\em Encyclopaedia of Mathematics. Suplement III}, 2001;
quant-ph/0010058.

\item {\bf [Knill-Laflamme-Milburn 01]}:
E. Knill, R. Laflamme, \& G. J. Milburn,
``A scheme for efficient quantum computation with linear optics'',
{\em Nature} {\bf 409}, 6816, 46-52 (2001);
quant-ph/0006088.

\item {\bf [Knill-Laflamme-Martinez-Negrevergne 01]}:
E. Knill, R. Laflamme, R. Martinez, \& C. Negrevergne,
``Benchmarking quantum computers:
The five-qubit error correcting code'',
{\em Phys. Rev. Lett.} {\bf 86}, 25, 5811-5814 (2001);
quant-ph/0101034.

\item {\bf [Knill 02]}:
E. Knill,
``Quantum gates using linear optics and postselection'',
{\em Phys. Rev. A} {\bf 66}, 5, 052306 (2002);
quant-ph/0110144.

\item {\bf [Knill 03 a]}:
E. Knill,
``Bounds on the probability of success of postselected
nonlinear sign shifts implemented with linear optics'',
{\em Phys. Rev. A} {\bf 68}, 6, 064303 (2003).

\item {\bf [Knill 03 b]}:
E. Knill,
``Scalable quantum computation in the presence of large detected-error
rates'',
quant-ph/0312190.

\item {\bf [Knill 04 a]}:
E. Knill,
``Fault-tolerant postselected quantum computation: Schemes'',
quant-ph/0402171.

\item {\bf [Knill 04 b]}:
E. Knill,
``Quantum computing with very noisy devices'',
quant-ph/0410199.

\item {\bf [Koashi-Imoto 96]}:
M. Koashi, \& N. Imoto,
``Quantum cryptography based on two mixed states'',
{\em Phys. Rev. Lett.} {\bf 77}, 10, 2137-2140 (1996).

\item {\bf [Koashi-Imoto 97]}:
M. Koashi, \& N. Imoto,
``Quantum cryptography
based on split transmission of one-bit information in two steps'',
{\em Phys. Rev. Lett.} {\bf 79}, 12, 2383-2386 (1997).

\item {\bf [Koashi-Imoto 98 a]}:
M. Koashi, \& N. Imoto,
``No-cloning theorem of entangled states'',
{\em Phys. Rev. Lett.} {\bf 81}, 19, 4264-4266 (1998).

\item {\bf [Koashi-Imoto 98 b]}:
M. Koashi, \& N. Imoto,
``Maximum amount of information obtainable from
a single quantum query of a database'',
{\em Phys. Rev. Lett.} {\bf 81}, 23, 5233-5236 (1998).

\item {\bf [Koashi-Ueda 99]}:
M. Koashi, \& M. Ueda,
``Reversing measurement and probabilistic quantum error
correction'',
{\em Phys. Rev. Lett.} {\bf 82}, 12, 2598-2601 (1999).

\item {\bf [Koashi-Bu\v{z}ek-Imoto 00]}:
M. Koashi, V. Bu\v{z}ek, \& N. Imoto,
``Entangled webs: Tight bound for symmetric sharing of
entanglement'',
{\em Phys. Rev. A} {\bf 62}, 5, 050302(R) (2000);
quant-ph/0007086.

\item {\bf [Koashi-Yamamoto-Imoto 01]}:
M. Koashi, T. Yamamoto, \& N. Imoto,
``Probabilistic manipulation of entangled photons'',
{\em Phys. Rev. A} {\bf 63}, 3, 030301(R) (2001);
quant-ph/0102141.

\item {\bf [Koashi-Imoto 01 a]}:
M. Koashi, \& N. Imoto,
``Compressibility of quantum mixed-state signals'',
{\em Phys. Rev. Lett.} {\bf 87}, 1, 017902 (2001);
quant-ph/0103128.

\item {\bf [Koashi-Imoto 01 b]}:
M. Koashi, \& N. Imoto,
``Teleportation cost and hybrid compression of quantum signals'',
quant-ph/0104001.

\item {\bf [Koashi-Imoto 02 a]}:
M. Koashi, \& N. Imoto,
``Quantum information is incompressible without errors'',
{\em Phys. Rev. Lett.} {\bf 89}, 9, 097904 (2002);
quant-ph/0203045.

\item {\bf [Koashi-Imoto 02 b]}:
M. Koashi, \& N. Imoto,
``Operations that do not disturb partially known quantum states'',
{\em Phys. Rev. A} {\bf 66}, 2, 022318 (2002),
quant-ph/0101144.
Comment: {\bf [Tucci 03]}.

\item {\bf [Koashi-Preskill 03]}:
M. Koashi, \& J. Preskill,
``Secure quantum key distribution with an uncharacterized source'',
{\em Phys. Rev. Lett.} {\bf 90}, 5, 057902 (2003).

\item {\bf [Koashi-Winter 04]}:
M. Koashi, \& A. Winter,
``Monogamy of quantum entanglement and other correlations'',
{\em Phys. Rev. A} {\bf 69}, 2, 022309 (2004);
quant-ph/0310037.

\item {\bf [Koashi 04]}:
M. Koashi,
``Unconditional security of coherent-state quantum key distribution with
strong phase-reference pulse'',
quant-ph/0403131.

\item {\bf [Kobayashi-Matsumoto-Yamakami 01]}:
H. Kobayashi, K. Matsumoto, \& T. Yamakami,
``Quantum certificate verification:
Single versus multiple quantum certificates'',
quant-ph/0110006.
Extended version: {\bf [Kobayashi-Matsumoto-Yamakami 03]}.

\item {\bf [Kobayashi-Matsumoto-Yamakami 03]}:
H. Kobayashi, K. Matsumoto, \& T. Yamakami,
``Quantum Merlin-Arthur proof systems: Are multiple Merlins more helpful to Arthur?'',
quant-ph/0306051.
See: {\bf [Kobayashi-Matsumoto-Yamakami 01]}.

\item {\bf [Kobayashi-Khanna-Mann-(+2) 04]}:
M. Kobayashi, F. Khanna, A. Mann, M. Revzen, \& A. Santana,
``Maximal Bell's inequality violation for non-maximal entanglement'',
{\em Phys. Lett. A} {\bf 323}, 3-4, 190-193 (2004);
quant-ph/0311125.

\item {\bf [Ko\c{c} 80]}:
Y. Ko\c{c},
``A non-completeness argument for quantum
mechanics (analysis of the EPR paper)'',
{\em Phys. Lett. A} {\bf 79}, 1, 9-12 (1980).
See {\bf [Ko\c{c} 81, 82]}.

\item {\bf [Ko\c{c} 81]}:
Y. Ko\c{c},
``A critical analysis of N. Bohr's reply to the EPR argument'',
{\em Phys. Lett. A} {\bf 81}, 8, 436-440 (1981).
Comment: {\bf [Madsen 81]}.
See {\bf [Ko\c{c} 80, 82]}.

\item {\bf [Ko\c{c} 82]}:
Y. Ko\c{c},
`Some remarks on the ``no-interaction''
assumption in the Einstein-Podolsky-Rosen argument',
{\em Phys. Lett. A} {\bf 90}, 9, 451-454 (1982).
See {\bf [Ko\c{c} 80, 81]}.

\item {\bf [Ko\c{c} 93]}:
Y. Ko\c{c},
``Wigner's inequality, quantum-mechanical
probability functions and hidden-variable theories'',
{\em Nuovo Cimento B} {\bf 108}, 10, 1115-1126 (1993).

\item {\bf [Kochen-Specker 65 a]}:
S. Kochen, \& E. P. Specker,
``Logical structures arising in quantum theory'',
in J. W. Addison et al. (eds.),
{\em Symp.\ on the Theory of Models,
Proc.\ of the 1963 Int.\ Symp.\ at Berkeley},
North-Holland, Amsterdam, Holland, 1965, pp.~177-189.
Reprinted in {\bf [Hooker 75]}, pp.~263-276.
Reprinted in {\bf [Specker 90]}, pp.~209-221.

\item {\bf [Kochen-Specker 65 b]}:
S. Kochen, \& E. P. Specker,
``The calculus of partial propositional functions'',
in Y. Bar-Hillel (ed.),
{\em Proc.\ of the 1964 Int.\ Congress
for Logic, Methodology and
Philosophy of Science, Jerusalem}, 1965, pp.~45-57.
Reprinted in {\bf [Hooker 75]}, pp.~277-292.
Reprinted in {\bf [Specker 90]}, pp.~222-234.

\item {\bf [Kochen-Specker 67]}:
S. Kochen, \& E. P. Specker,
``The problem of hidden variables in quantum mechanics'',
{\em J. Math. Mech.} {\bf 17}, 1, 59-87 (1967).
Reprinted in {\bf [Hooker 75]}, pp.~293-328.
Reprinted in {\bf [Specker 90]}, pp.~235-263.

\item {\bf [Kochen 78]}:
S. Kochen,
``The interpretation of quantum mechanics'',
address to the
{\em Biennial Conf.\ of the Philosophy of Science Association, 1978},
mimeo,
Princeton University, Princeton, New Jersey, 1978.

\item {\bf [Kochen 85]}:
S. Kochen,
``A new interpretation of quantum mechanics'',
in P. J. Lahti, \& P. Mittelstaedt (eds.),
{\em Symp.\ on the Foundations of Modern
Physics: 50 Years of the Einstein-Podolsky-Rosen Experiment
(Joensuu, Finland, 1985)},
World Scientific, Singapore, 1985, pp.~151-169.

\item {\bf [Kochen 96]}:
S. Kochen,
``Construction of quantum mechanics via commutative operations'',
in R. K. Clifton (ed.), {\em Perspectives on quantum reality:
Non-relativistic, relativistic, and field-theoretic
(London, Western Ontario, Canada, 1994)},
Kluwer Academic, Dordrecht, Holland, 1996, pp.~237-243.

\item {\bf [Kocher-Commins 67]}:
C. A. Kocher, \& E. D. Commins,
``Polarization correlation of photons emitted in an atomic cascade'',
{\em Phys. Rev. Lett.} {\bf 18}, 15, 575-577 (1967).

\item {\bf [Koenig 00]}:
R. Koenig,
``European science: Teleportation guru stakes out new ground'',
{\em Science} {\bf 288}, 5470, 1327 (2000).

\item {\bf [Kofman-Kurizki-Opatrn\'{y} 00]}:
A. G. Kofman, G. Kurizki, \& T. Opatrn\'{y},
``Zeno and anti-Zeno effects for photon polarization dephasing'',
quant-ph/0011077.

\item {\bf [Kofman-Kurizki 01]}:
A. G. Kofman, \& G. Kurizki,
``Frequent observations accelerate decay: The anti-Zeno effect'',
submitted to {\em Zeitschrift f\"{u}r Naturforschung A};
quant-ph/0102002.

\item {\bf [Kofman 01]}:
A. G. Kofman,
``Relaxation of a two-level system strongly
coupled to a reservoir: Anomalously slow decoherence'',
{\em Phys. Rev. A};
quant-ph/0106015.

\item {\bf [Koiller-Hu-Das Sarma 02 a]}:
B. Koiller, X. Hu, \& S. Das Sarma,
``Exchange in silicon based quantum computer architecture'',
{\em Phys. Rev. Lett.} {\bf 88}, 2, 027903 (2002).

\item {\bf [Koiller-Hu-Das Sarma 02 b]}:
B. Koiller, X. Hu, \& S. Das Sarma,
``Strain effects on silicon donor exchange: Quantum computer architecture
considerations'',
{\em Phys. Rev. B} {\bf 66}, 11, 115201 (2002).

\item {\bf [Koiller-Hu-Drew-Das Sarma 03]}:
B. Koiller, X. Hu, H. D. Drew, \& S. Das Sarma,
``Disentangling the exchange coupling of entangled donors in the Si quantum
computer architecture'',
{\em Phys. Rev. Lett.} {\bf 90}, 6, 067401 (2003).

\item {\bf [Kojima-Hofmann-Takeuchi-Sasaki 04]}:
K. Kojima, H. F. Hofmann, S. Takeuchi, \& K. Sasaki,
``Efficiencies for the single-mode operation of a quantum optical nonlinear shift gate'',
{\em Phys. Rev. A} {\bf 70}, 1, 013810 (2004).

\item {\bf [Kok-Braunstein 99 a]}:
P. Kok, \& S. L. Braunstein,
``Comment on `Experimental entanglement swapping:
Entangling photons that never interacted'\,'',
quant-ph/9907016.
Comment on {\bf [Pan-Bouwmeester-Weinfurter-Zeilinger 98]}.

\item {\bf [Kok-Braunstein 99 b]}:
P. Kok, \& S. L. Braunstein,
``How useful is detector cascading?'',
quant-ph/9910084.

\item {\bf [Kok-Braunstein 00 a]}:
P. Kok, \& S. L. Braunstein,
``Postselected versus nonpostselected quantum teleportation
using parametric down-conversion'',
{\em Phys. Rev. A} {\bf 61}, 4, 042304 (2000).

\item {\bf [Kok-Braunstein 00 b]}:
P. Kok, \& S. L. Braunstein,
``Entanglement swapping as event-ready entanglement preparation'',
{\em Fortschr. Phys.} {\bf 48}, 5-7, 553-557 (2000).

\item {\bf [Kok-Braunstein 00 c]}:
P. Kok, \& S. L. Braunstein,
``Limitations on the creation of maximal entanglement'',
{\em Phys. Rev. A} {\bf 62}, 6, 064301 (2000);
quant-ph/0001080.

\item {\bf [Kok 01]}:
P. Kok,
``State preparation in quantum optics'',
Ph.\ D. thesis;
quant-ph/0102070.

\item {\bf [Kok-Boto-Abrams-(+3) 01]}:
P. Kok, A. N. Boto, D. S. Abrams,
C. P. Williams, S. L. Braunstein, \& J. P. Dowling,
``Quantum-interferometric optical lithography:
Towards arbitrary two-dimensional patterns'',
{\em Phys. Rev. A} {\bf 63}, 6, 063407 (2001).
See {\bf [Boto-Kok-Abrams-(+3) 00]}.

\item {\bf [Kok-Lee-Dowling 02]}:
P. Kok, H. Lee, \& J. P. Dowling,
``Creation of large-photon-number path entanglement conditioned on
photodetection'',
{\em Phys. Rev. A} {\bf 65}, 5, 052104 (2002);
quant-ph/0112002.

\item {\bf [Kok-Lee-Dowling 02 a]}:
P. Kok, H. Lee, \& J. P. Dowling,
``Implementing a single-photon quantum nondemolition
device with only linear optics and projective measurements'',
quant-ph/0202046.

\item {\bf [Kok-Lee-Dowling 02 b]}:
P. Kok, H. Lee, \& J. P. Dowling,
``Single-photon quantum-nondemolition detectors constructed with linear optics
and projective measurements'',
{\em Phys. Rev. A} {\bf 66}, 6, 063814 (2002);
quant-ph/0202046.

\item {\bf [Kok-Williams-Dowling 03]}:
P. Kok, C. P. Williams, \& J. P. Dowling,
``Construction of a quantum repeater with linear optics'',
{\em Phys. Rev. A} {\bf 68}, 2, 022301 (2003);
quant-ph/0203134.

\item {\bf [Kok-Braunstein-Dowling 04]}:
P. Kok, S. L. Braunstein, \& J. P. Dowling,
``Quantum lithography, entanglement and Heisenberg-limited parameter
estimation'',
{\em J. Opt. B};
quant-ph/0402083.

\item {\bf [Kok-Munro 04]}:
P. Kok, \& W. J. Munro,
`Comment on ``Measuring a photonic qubit without destroying it''\,',
quant-ph/0406120.
Comment on {\bf [Pryde-O'Brien-White-(+2) 04]}.

\item {\bf [Kok-Ralph-Milburn 04]}:
P. Kok, T. C. Ralph, \& G. J. Milburn,
``Communicating continuous quantum variables between different Lorentz
frames'',
quant-ph/0406219.

\item {\bf [Kok-Braunstein 00 c]}:
P. Kok, \& S. L. Braunstein,
``Relativistic quantum communication with bosonic and fermionic
interferometers'',
quant-ph/0407259.

\item {\bf [Kokin 00]}:
A. A. Kokin,
``A model for ensemble NMR quantum computer using
antiferromagnetic structure'',
quant-ph/0002034.

\item {\bf [Kokin-Valiev 02]}:
A. A. Kokin, \& K. A. Valiev,
``Problems in realization of large-scale ensemble
silicon-based NMR quantum computers'',
quant-ph/0201083.

\item {\bf [Kokorowski-Cronin-Roberts-Pritchard 01]}:
D. A. Kokorowski, A. D. Cronin, A. D. Roberts, \& D. E. Pritchard,
``From single- to multiple-photon decoherence in an atom
interferometer'',
{\em Phys. Rev. Lett.} {\bf 86}, 11, 2191-2195 (2001).

\item {\bf [Kolar-Opatrny-Kurizki-(+2) 04]}:
M. Kolar, T. Opatrny, G. Kurizki, N. Bar-Gill, \& G. Rempe,
``Wave-particle duality of interfering internally-translationally
entangled particles'',
quant-ph/0409149.

\item {\bf [Komar 62]}:
A. Komar,
``Indeterminate character of the reduction of
the wave packet in quantum theory'',
{\em Phys. Rev.} {\bf 126}, 1, 365-369 (1962).

\item {\bf [Konno 02]}:
N. Konno,
``Quantum random walks in one dimension'',
{\em Quant. Inf. Processing} {\bf 1}, 5, 345-354 (2002);
quant-ph/0206053.

\item {\bf [Konno-Mistuda-Soshi-Yoo 04]}:
N. Konno, K. Mistuda, T. Soshi, \& H. J. Yoo,
``Quantum walks and reversible cellular automata'',
{\em Phys. Lett. A} {\bf 330}, 6, 408-417 (2004).

\item {\bf [Konopka-Bu\v{z}ek 99]}:
M. Konopka, \& V. Bu\v{z}ek,
``Entangling atoms in photonic crystals'',
quant-ph/9901069.

\item {\bf [Koniorczyk-Janszky-Kis 99]}:
M. Koniorczyk, J. Janszky, \& Z. Kis,
``Photon number teleportation'',
{\em Phys. Lett. A} {\bf 256}, 5-6, 334-338 (1999).

\item {\bf [Koniorczyk-Kiss-Janszky 01]}:
M. Koniorczyk, T. Kiss, \& J. Janszky,
``Teleportation: From probability distributions to quantum states'',
in S. Popescu, N. Linden, \& R. Jozsa (eds.),
{\em J. Phys. A} {\bf 34}, 35
(Special issue: Quantum information and computation), 6949-6956 (2001);
quant-ph/0011083.

\item {\bf [Koniorczyk-Bu\v{z}zek-Janszky 01]}:
M. Koniorczyk, V. Bu\v{z}zek, \& J. Janszky,
``Wigner-function description of quantum teleportation
in arbitrary dimensions and a continuous limit'',
{\em Phys. Rev. A} {\bf 64}, 3, 034301 (2001);
quant-ph/0106109.

\item {\bf [Korbicz-Cirac-Wehr-Lewenstein 04]}:
J. Korbicz, J. I. Cirac, J. Wehr, \& M. Lewenstein,
``Hilbert's 17th problem and the quantumness of states'',
quant-ph/0408029.

\item {\bf [Korepin-Terilla 02]}:
V. Korepin, \& J. Terilla,
``Quantum error correction by means of thermodynamics'',
quant-ph/0202054.

\item {\bf [Korolkova-Silberhorn-Gl\"{o}ckl-(+3) 02]}:
N. Korolkova, C. Silberhorn, O. Gl\"{o}ckl,
S. Lorenz, C. Marquardt, \& G. Leuchs,
``Direct experimental test of non-separability and other
quantum techniques using continuous variables of light'',
{\em Eur. Phys. J. D} {\bf 18}, 2
(Special issue: {\em Quantum interference and cryptographic keys:
Novel physics and advancing technologies (QUICK) (Corsica, 2001)},
229-235 (2002);
quant-ph/0109011.

\item {\bf [Korolkova-Leuchs-Loudon-(+2) 02]}:
N. Korolkova, G. Leuchs, R. Loudon, T. C. Ralph, \& C. Silberhorn,
``Polarization squeezing and continuous-variable polarization entanglement'',
{\em Phys. Rev. A} {\bf 65}, 5, 052306 (2002).

\item {\bf [Korotkov 98]}:
A. Korotkov,
``Continuous quantum measurement with observer: Pure wavefunction
evolution instead of decoherence'',
quant-ph/9807051.

\item {\bf [Korotkov 00]}:
A. N. Korotkov,
``Selective quantum evolution of a qubit state due to continuous
measurement'',
cond-mat/0008461.

\item {\bf [Korotkov 01]}:
A. N. Korotkov,
``Correlated quantum measurement of a solid-state qubit'',
{\em Phys. Rev. B} {\bf 64}, 19, 193407 (2001);
cond-mat/0008003.

\item {\bf [Korotkov 02]}:
A. N. Korotkov,
``Continuous measurement of entangled qubits'',
{\em Phys. Rev. A} {\bf 65}, 5, 052304 (2002).

\item {\bf [Korotkov 03]}:
A. N. Korotkov,
``Nonideal quantum detectors in Bayesian formalism'',
{\em Phys. Rev. B} {\bf 67}, 23, 235408 (2003).

\item {\bf [Ko\v{s}\'{\i}k-Bu\v{z}ek 04]}:
J. Ko\v{s}\'{\i}k, \& V. Bu\v{z}ek,
``Scattering model for quantum random walk on the hypercube'',
{\em Phys. Rev. A};
quant-ph/0410154.

\item {\bf [Kostenko-Kuznetsov-Miller-(+2) 99]}:
B. F. Kostenko, V. D. Kuznetsov, M. B. Miller,
A. V. Sermyagin, \& D. V. Kamanin,
``Nuclear teleportation'',
quant-ph/9912119.

\item {\bf [Kostenko-Kuznetsov-Miller-(+2) 00]}:
B. F. Kostenko, V. D. Kuznetsov, M. B. Miller,
A. V. Sermyagin, \& D. V. Kamanin,
``Possibility for teleportation of nuclei'',
quant-ph/0012133.

\item {\bf [Kowalski-Plastino-Proto 02]}:
A. M. Kowalski, A. Plastino, \& A. N. Proto,
``Classical limits'',
{\em Phys. Lett. A} {\bf 297}, 3-4, 162-172 (2002).

\item {\bf [Kox 02]}:
A. J. Kox,
``The genius of science: A portrait gallery of twentieth-century physicists'',
{\em Eur. J. Phys.} {\bf 23}, 2, 233 (2002).
Review of {\bf [Pais 00]}.

\item {\bf [Koziel-Majewski 01]}:
S. Koziel, \& W. A. Majewski,
``Evolution of entanglement for spin-flip dynamics'',
quant-ph/0101033.

\item {\bf [Kracklauer 98 a]}:
A. F. Kracklauer,
``Skirting hidden-variable no-go theorems'',
quant-ph/9804059.

\item {\bf [Kracklauer 98 b]}:
A. F. Kracklauer,
``The error in Bell's theorem'',
quant-ph/9810081.

\item {\bf [Kracklauer 98 c]}:
A. F. Kracklauer,
``EPR correlations as an angular
Hanbury-Brown--twiss effect'',
quant-ph/9812072.

\item {\bf [Kracklauer 00 a]}:
A. F. Kracklauer,
``La `th\'{e}orie' de Bell, est-elle la plus grande m\'{e}prise de l'histoire de la physique?''
[``Is Bell's `theory' the biggest mistake in the history of physics?''],
{\em Ann. Fond. Louis de Broglie} {\bf 25}, 2, 193-207 (2000).

\item {\bf [Kracklauer 00 b]}:
A. F. Kracklauer,
``Bell's inequalities III. Logical loophole in their
formulation'',
quant-ph/0005001.

\item {\bf [Kracklauer-Kracklauer 00]}:
A. F. Kracklauer, \& N. A. Kracklauer,
`\,``Quantum mysteries for anyone'', or classical verities for
everyone?',
quant-ph/0007101.

\item {\bf [Kracklauer 01]}:
A. F. Kracklauer,
``Four-photon entanglement as stochastic-signal correlation'',
quant-ph/0105034.

\item {\bf [Kracklauer 02 a]}:
A. F. Kracklauer,
`Is ``entanglement'' always entangled?',
in Y. S. Kim, M. A. Man'ko, \& A. Sergienko (eds.),
{\em Seventh International Conference on Squeezed States and Uncertainty Relations (Boston, 2001)},
{\em J. Opt. B: Quantum Semiclass. Opt.} {\bf 4}, 3, S121-S126 (2002);
quant-ph/0108057.

\item {\bf [Kracklauer 02 b]}:
A. F. Kracklauer,
``One less quantum mystery'',
in R. Bonifacio, \& D. Vitali (eds.),
{\em Mysteries, Puzzles and Paradoxes in Quantum Mechanics IV:
Quantum Interference Phenomena (Gargnano, Italy, 2001)},
{\em J. Opt. B: Quantum Semiclass. Opt.} {\bf 4}, 4, S469-S472 (2002).

\item {\bf [Kracklauer 03]}:
A. F. Kracklauer,
``Betting on Bell'',
quant-ph/0302113.

\item {\bf [Kragh 90]}:
H. S. Kragh,
{\em Dirac: A scientific biography},
Cambridge University Press, New York, 1990.

\item {\bf [Kragh 02]}:
H. S. Kragh,
``Book review. The historical development of quantum theory,
volume 6: The completion of quantum mechanics 1926-1941'',
{\em Found. Phys.} {\bf 32}, 1, 187-189 (2002).
Review of {\bf [Mehra-Rechenberg 00 a, b]}.

\item {\bf [Kraichnan 85]}:
R. H. Kraichnan,
``Comment on `Quantum mechanics
versus macroscopic realism: Is the flux there when nobody looks?'\,'',
{\em Phys. Rev. Lett.} {\bf 54}, 25, 2723 (1985).
Comment on {\bf [Leggett-Garg 85 a]}.
Reply: {\bf [Leggett-Garg 85 b]}.

\item {\bf [Kramer-Savari 01]}:
G. Kramer, \& S. A. Savari,
``Quantum data compression of ensembles of mixed states
with commuting density operators'',
quant-ph/0101119.

\item {\bf [Krasnoholovets 01]}:
V. Krasnoholovets,
``Submicroscopic deterministic quantum mechanics'',
{\em 5th Int.\ Conf.\ on Computing Anticipatory Systems (Liege, Belgium, 2001)};
quant-ph/0109012.

\item {\bf [Kraus-Cirac-Karnas-Lewenstein 00]}:
B. Kraus, J. I. Cirac, S. Karnas, \& M. Lewenstein,
``Separability in $2 \times N$ composite quantum systems'',
{\em Phys. Rev. A} {\bf 61}, 6, 062302 (2000);
quant-ph/9912010.

\item {\bf [Kraus-Cirac 01]}:
B. Kraus, \& J. I. Cirac,
``Optimal creation of entanglement using a two-qubit gate'',
{\em Phys. Rev. A} {\bf 63}, 6, 062309 (2001);
quant-ph/0011050.

\item {\bf [Kraus-Lewenstein-Cirac 02]}:
B. Kraus, M. Lewenstein, \& J. I. Cirac,
``Characterization of distillable and activatable states using entanglement
witnesses'',
{\em Phys. Rev. A} {\bf 65}, 4, 042327 (2002);
quant-ph/0110174.

\item {\bf [Kraus-D\"{u}r-Vidal 02]}:
B. Kraus, W. D\"{u}r, G. Vidal,
J. I. Cirac, M. Lewenstein, N. Linden, \& S. Popescu,
``Entanglement capability of two-qubit unitary operations'',
{\em Fortschr. Phys.}.

\item {\bf [Kraus-Hammerer-Giedke-Cirac 03]}:
B. Kraus, K. Hammerer, G. Giedke, \& J. I. Cirac,
``Entanglement generation and Hamiltonian simulation in continuous-variable
systems'',
{\em Phys. Rev. A} {\bf 67}, 4, 042314 (2003);
quant-ph/0210136.

\item {\bf [Kraus-Cirac 03]}:
B. Kraus, \& J. I. Cirac,
``Discrete entanglement distribution with squeezed light'',
{\em Phys. Rev. Lett.} {\bf 92}, 1, 013602 (2004);
quant-ph/0307158.

\item {\bf [Kraus-Gisin-Renner 04]}:
B. Kraus, N. Gisin, \& R. Renner,
``Lower and upper bounds on the secret key rate for QKD protocols using
one--way classical communication'',
quant-ph/0410215.

\item {\bf [Kraus 83]}:
K. Kraus,
{\em States, effects and operations},
Springer-Verlag, Berlin, 1983.

\item {\bf [Kraus 89]}:
K. Kraus,
``Quantum theory does not require action at a distance'',
{\em Found. Phys. Lett.} {\bf 2}, 1, 1-6 (1989).
Comment: {\bf [Stapp 89 b]}.

\item {\bf [Kr\v{c}o-Paul 02]}:
M. Kr\v{c}o, \& P. Paul,
``Quantum clock synchronization: Multiparty protocol'',
{\em Phys. Rev. A} {\bf 66}, 2, 024305 (2002);
quant-ph/0112111.

\item {\bf [Kreidl-Gr\"{u}bl-Embacher 03]}:
S. Kreidl, G. Gr\"{u}bl, \& H. G. Embacher,
``Bohmian arrival time without trajectories'',
{\em J. Phys. A} {\bf 36}, 33, 8851-8865 (2003);
quant-ph/0305163.

\item {\bf [Krenn-Summhammer-Svozil 96]}:
G. Krenn, J. Summhammer, \& K. Svozil,
``Interaction-free preparation'',
{\em Phys. Rev. A} {\bf 53}, 3, 1228-1231 (1996).

\item {\bf [Krenn-Zeilinger 95]}:
G. Krenn, \& A. Zeilinger,
``Entangled entanglement'',
in D. M. Greenberger, \& A. Zeilinger (eds.),
{\em Fundamental problems in quantum theory:
A conference held in honor of professor John A. Wheeler,
Ann. N. Y. Acad. Sci.} {\bf 755}, 873-876 (1995).
Almost the same as {\bf [Krenn-Zeilinger 96]}.

\item {\bf [Krenn-Zeilinger 96]}:
G. Krenn, \& A. Zeilinger,
``Entangled entanglement'',
{\em Phys. Rev. A} {\bf 54}, 3, 1793-1797 (1996). Almost the same
as {\bf [Krenn-Zeilinger 95]}.
Erratum: {\em Phys. Rev. A} {\bf 55}, 5, 3970 (1997).
See {\bf [Cereceda 97 c]}, {\bf [Schafir 97]}.

\item {\bf [Krenn-Zeilinger 97]}:
G. Krenn, \& A. Zeilinger,
``Reply to `Comment on ``Entangled entanglement''\,'\,'',
{\em Phys. Rev. A} {\bf 56}, 5, 4336 (1997).
Reply to {\bf [Schafir 97]}.
See {\bf [Krenn-Zeilinger 95, 96]},
{\bf [Cereceda 97 c]}.

\item {\bf [Krenn-Svozil 98]}:
G. Krenn, \& K. Svozil,
``Stronger-than-quantum correlations'',
{\em Found. Phys.} {\bf 28}, 6, 971-984 (1998).

\item {\bf [Krenn-Summhammer-Svozil 00]}:
G. Krenn, J. Summhammer, \& K. Svozil,
``Interferometric information gain versus interaction-free measurement'',
{\em Phys. Rev. A} {\bf 61}, 5, 052102 (2000).

\item {\bf [Kretschmann-Werner 03]}:
D. Kretschmann, \& R. F. Werner,
``Tema con variazioni: Quantum channel capacity'',
quant-ph/0311037.

\item {\bf [Krips 71]}:
H. P. Krips,
``Defence of a measurement theory'',
{\em Nuovo Cimento B} {\bf 1}, 1, 23-33 (1971).

\item {\bf [Krips 74]}:
H. Krips,
``Foundations of quantum theory. Part I'',
{\em Found. Phys.} {\bf 4}, 2, 181-193 (1974).

\item {\bf [Krips 77]}:
H. Krips,
``Quantum theory and measures on Hilbert space'',
{\em J. Math. Phys.} {\bf 18}, 5, 1015-1021 (1977).

\item {\bf [Krips 87]}:
H. Krips,
{\em The metaphysics of quantum theory},
Clarendon Press, Oxford, 1987.

\item {\bf [Krips 94]}:
H. Krips,
``A critique of local realism'',
in {\bf [Faye-Folse 94]}, pp.~269-277.

\item {\bf [Krishnan 01]}:
V. V. Krishnan,
``Estimating the efficiency of ensemble quantum computing'',
{\em Phys. Lett. A} {\bf 291}, 1, 27-33 (2001).

\item {\bf [Kronz 01 a]}:
F. M. Kronz,
``Range of violations of Bell's inequality by entangled photon
pairs'',
{\em Phys. Lett. A} {\bf 279}, 5-6, 287-290 (2001);
PITT-PHIL-SCI00000264.

\item {\bf [Kronz 01 b]}:
F. M. Kronz,
``Bohm's ontological interpretation and its relations
to three formulations of quantum mechanics'' (2001),
PITT-PHIL-SCI00000265.

\item {\bf [Kr\"{u}ger-Werner-Wolf 04]}:
O. Kr\"{u}ger, R. F. Werner, \& M. M. Wolf,
``Optimal cloning of coherent states'',
quant-ph/0410058.

\item {\bf [Kr\"{u}ger-Luo-Klein-(+7) 03]}:
P. Kr\"{u}ger, X. Luo, M. W. Klein,
K. Brugger, A. Haase, S. Wildermuth,
S. Groth, I. Bar-Joseph, R. Folman, \& J. Schmiedmayer,
``Trapping and manipulating neutral atoms with electrostatic fields'',
{\em Phys. Rev. Lett.} {\bf 91}, 23, 233201 (2003).

\item {\bf [Kr\"{u}ger 00]}:
T. Kr\"{u}ger,
``Towards a deeper understanding of the Einstein-Podolsky-Rosen problem'',
{\em Found. Phys.} {\bf 30}, 11, 1869-1890 (2000).

\item {\bf [Kryuchkyan-Manukyan 04]}:
G. Y. Kryuchkyan, \& L. A. Manukyan,
``Entangled light in transition through the generation threshold'',
{\em Phys. Rev. A} {\bf 69}, 1, 013813 (2004).

\item {\bf [Kryukov 03]}:
A. A. Kryukov,
``Coordinate Formalism on Abstract Hilbert Space: Kinematics of a Quantum Measurement'',
{\em Found. Phys.} {\bf 33}, 3, 407-443 (2003).

\item {\bf [Kuang-Zhou 03 a]}:
L.-M. Kuang, \& L. Zhou,
``Generation of atom-photon entangled states in atomic Bose-Einstein
condensate via electromagnetically induced transparency'',
{\em Phys. Rev. A} {\bf 68}, 4, 043606 (2003).

\item {\bf [Kuang-Zhou 03 b]}:
L.-M. Kuang, A.-H. Zeng, \& Z.-H. Kuang,
``Generation of entangled squeezed states in atomic Bose-Einstein
condensates'',
{\em Phys. Lett. A} {\bf 319}, ?, 24-31 (2003);
quant-ph/0402058.

\item {\bf [Kudaka-Matsumoto 99 a]}:
S. Kudaka, \& S. Matsumoto,
``Uncertainty principle for proper time and mass'',
{\em J. Math. Phys.};
quant-ph/9901019.
See {\bf [Kudaka-Matsumoto 99 b]} (II).

\item {\bf [Kudaka-Matsumoto 99 b]}:
S. Kudaka, \& S. Matsumoto,
``Uncertainty principle for proper time and mass II'',
quant-ph/9901031.
See {\bf [Kudaka-Matsumoto 99 a]} (I).

\item {\bf [Kudryavtev-Knight 93]}:
I. K. Kudryavtev, \& P. L. Knight,
``Atomic entanglement and Bell's inequality violation'',
{\em J. Mod. Opt.} {\bf 40}, 9, 1673-1679 (1993).

\item {\bf [Kuhn-Hennrich-Rempe 02]}:
A. Kuhn, M. Hennrich, \& G. Rempe,
``Deterministic single-photon source for distributed quantum networking'',
{\em Phys. Rev. Lett.} {\bf 89}, 6, 067901 (2002);
quant-ph/0204147.
Comment: {\bf [Kimble 03]}.
Reply: {\bf [Kuhn-Hennrich-Rempe 03]}.

\item {\bf [Kuhn-Hennrich-Rempe 03]}:
A. Kuhn, M. Hennrich, \& G. Rempe,
``Kuhn, Hennrich, and Rempe reply'',
{\em Phys. Rev. Lett.} {\bf 90}, 24, 249802 (2003).
Reply to {\bf [Kimble 03]}.
See {\bf [Kuhn-Hennrich-Rempe 02]}.

\item {\bf [Kuhn 03]}:
D. R. Kuhn,
``A hybrid authentication protocol using quantum entanglement and
symmetric cryptography'',
quant-ph/0301150.

\item {\bf [Kuhn-Heilbron-Forman-Allen 67]}:
T. S. Kuhn, J. L. Heilbron, P. L. Forman, \& L. Allen,
{\em Sources for history of quantum physics: An inventory and report},
Philadelphia, Pennsylvania, 1967.

\item {\bf [Kuhn 77]}:
T. S. Kuhn,
{\em The essential tension. Selected studies in
scientific tradition and change},
University of Chicago Press, Chicago, 1977.
Spanish version: {\em La tensi\'{o}n esencial.
Estudios selectos sobre la tradici\'{o}n y el cambio
en el \'{a}mbito de la ciencia},
Fondo de Cultura Econ\'{o}mica, M\'{e}xico, 1982.

\item {\bf [Kuklewicz-Fiorentino-Messin-(+2) 04]}:
C. E. Kuklewicz, M. Fiorentino, G. Messin,
F. N. C. Wong, \& J. H. Shapiro,
``High-flux source of polarization-entangled photons from a periodically poled KTiOPO4 parametric down-converter'',
{\em Phys. Rev. A} {\bf 69}, 1, 013807 (2004).

\item {\bf [Kumar-Pandey 03]}:
D. Kumar, \& P. N. Pandey,
``Effect of noise on quantum teleportation'',
{\em Phys. Rev. A} {\bf 68}, 1, 012317 (2003).

\item {\bf [Kummer 01]}:
H. J. Kummer,
``Theory of pair of quantum bits'',
{\em Int. J. Theor. Phys.} {\bf 40}, 6, 1071-1112 (2001).

\item {\bf [Kuna 98]}:
M. Kuna,
``Entanglement and pseudomixtures'',
{\em Acta Phys. Slov.} june 1998
(Special issue on quantum optics and quantum information);
quant-ph/9806011.

\item {\bf [Kunstatter-Trainor 84 a]}:
G. Kunstatter, \& L. E. H. Trainor,
``Do intelligent observers exist in quantum mechanics?'',
{\em Phys. Lett. A} {\bf 103}, 1-2, 32-34 (1984).
Comment on {\bf [Page 82]}.
See {\bf [Todorov 83]}.

\item {\bf [Kunstatter-Trainor 84 b]}:
G. Kunstatter, \& L. E. H. Trainor,
``For whom the bell tolls'',
{\em Am. J. Phys.} {\bf 52}, 7, 598-602 (1984).

\item {\bf [Kupriyanov-Sokolov-Slavgorodskii 03]}:
D. V. Kupriyanov, I. M. Sokolov, \& A. V. Slavgorodskii,
``Entangled states in spin subsystems of polyatomic ensembles'',
{\em Opt. Spectrosc.} {\bf 94}, 717 (2003).

\item {\bf [Kupriyanov-Mishina-Sokolov-(+2) 04]}:
D. V. Kupriyanov, O. S. Mishina, I. M. Sokolov,
B. Julsgaard, \& E. S. Polzik,
``Multimode entanglement of light and atomic ensembles via off-resonant
coherent forward scattering'',
quant-ph/0411083.

\item {\bf [Kupsch 98]}:
J. Kupsch,
``Exactly soluble models of decoherence'',
quant-ph/9811010.

\item {\bf [Kurochkin-Ryabtsev-Neizvestniy 04]}:
V. L. Kurochkin, I. I. Ryabtsev, \& I. G. Neizvestniy,
``Quantum key generation based on coding of polarization states of photons'',
{\em Opt. and Spectroscopy} {\bf 96}, 703-706 (2004).

\item {\bf [Kurtsiefer-Mayer-Zarda-Weinfurter 00]}:
C. Kurtsiefer, S. Mayer, P. Zarda, \& H. Weinfurter,
``Stable solid-state source of single photons'',
{\em Phys. Rev. Lett.} {\bf 85}, 2, 290-293 (2000).
See {\bf [Osborne 00 c]}.

\item {\bf [Kurtsiefer-Oberparleiter-Weinfurter 01]}:
C. Kurtsiefer, M. Oberparleiter, \& H. Weinfurter,
``High efficiency entangled photon pair collection
in type II parametric fluorescence'',
quant-ph/0101074.

\item {\bf [Kurtsiefer-Zarda-Mayer-Weinfurter 01]}:
C. Kurtsiefer, P. Zarda, S. Mayer, \& H. Weinfurter,
``The breakdown flash of silicon avalance photodiodes --
backdoor for eavesdropper attacks?'',
submitted to {\em J. Mod. Opt.};
quant-ph/0104103.

\item {\bf [Kurtsiefer-Zarda-Halder-(+4) 02]}:
C. Kurtsiefer, P. Zarda, M. Halder,
H. Weinfurter, P. M. Gorman, P. R. Tapster, \& J. G. Rarity,
``Quantum cryptography: A step towards global key distribution'',
{\em Nature} {\bf 419}, 6906, 450 (2002).

\item {\bf [Kurucz-Koniorczyk-Janszky 01]}:
Z. Kurucz, M. Koniorczyk, \& J. Janszky,
``Teleportation with partially entangled states'',
{\em Fortschr. Phys.} {\bf 49}, 1019-1025 (2001);
quant-ph/0308020.

\item {\bf [Kurzeja-Parkins 02]}:
S. I. J. Kurzeja, \& A. S. Parkins,
``Continuous variable quantum teleportation
with a finite-basis entanglement resource'',
quant-ph/0201094.

\item {\bf [Ku\'{s}-\.{Z}yczkowski 01]}:
M. Ku\'{s}, \& K. \.{Z}yczkowski,
``Geometry of entangled states'',
{\em Phys. Rev. A} {\bf 63}, 3, 032307 (2001);
quant-ph/0006068.

\item {\bf [Kuvshinov-Kuzmin 03]}:
V. I. Kuvshinov, \& A. V. Kuzmin,
``Stability of holonomic quantum computations'',
{\em Phys. Lett. A} {\bf 316}, 6, 391-394 (2003);
quant-ph/0310069.

\item {\bf [Kuzmenko 00]}:
M. V. Kuzmenko,
``Two-particle model with noncommuting operators of coordinates and
momenta'',
quant-ph/0009036.

\item {\bf [Kuzmich-Branning-Mandel 98]}:
A. Kuzmich, D. Branning, \& L. Mandel,
``?'',
{\em J. Mod. Opt.} {\bf 45}, 11, 2233-2243 (1998).

\item {\bf [Kuzmich-Walmsley-Mandel 00]}:
A. Kuzmich, I. A. Walmsley, \& L. Mandel,
``Violation of Bell's inequality by a generalized
Einstein-Podolsky-Rosen state using homodyne detection'',
{\em Phys. Rev. Lett.} {\bf 85}, 7, 1349-1353 (2000).

\item {\bf [Kuzmich-Polzik 00]}:
A. Kuzmich, \& E. S. Polzik,
``Atomic quantum state teleportation and swapping'',
{\em Phys. Rev. Lett.} {\bf 85}, 26, 5639-5642 (2000);
quant-ph/0003015

\item {\bf [Kuzmich-Walmsley-Mandel 01]}:
A. Kuzmich, I. A. Walmsley, \& L. Mandel,
``Violation of a Bell-type inequality in the homodyne measurement of light in
an Einstein-Podolsky-Rosen state'',
{\em Phys. Rev. A} {\bf 64}, 6, 063804 (2001).

\item {\bf [Kuzmich-Bowen-Boozer-(+4) 01]}:
A. Kuzmich, W. P. Bowen, A. D. Boozer,
A. Boca, C. W. Chou, L.-M. Duan, \& H. J. Kimble,
``Generation of nonclassical photon pairs for scalable quantum communication
with atomic ensembles'',
{\em Nature} {\bf 423}, 6941, 731-734 (2003).

\item {\bf [Kuzmich-Kennedy 04]}:
A. Kuzmich, \& T. A. B. Kennedy,
``Nonsymmetric entanglement of atomic ensembles'',
{\em Phys. Rev. Lett.} {\bf 92}, 3, 030407 (2004);
quant-ph/0401143.

\item {\bf [Kuzovlev 03]}:
Y. E. Kuzovlev,
``Quantum interactions in a stochastic representation and two-level systems'',
{\em JETP Lett.} {\bf 78}, 92-? (2003).

\item {\bf [Kwek-Oh-Singh-Wang 00]}:
L. C. Kwek, C. H. Oh, K. Singh, \& X.-B. Wang,
``Diagonalization of diffusion matrix in Grover's algorithm'',
{\em Phys. Lett. A} {\bf 267}, 1, 24-30 (2000).

\item {\bf [Kwek-Oh-Wang-Yeo 00]}:
L. C. Kwek, C. H. Oh, X.-B. Wang, \& Y. Yeo,
``Bu\v{z}ek-Hillery cloning revisited
using the bures metric and trace norm'',
{\em Phys. Rev. A} {\bf 62}, 5, 052313 (2000).

\item {\bf [Kwiat-Vareka-Hong-(+2) 90]}:
P. G. Kwiat, W. A. Vareka, C. K. Hong, H. Nathel, \& R. Y. Chiao,
``Correlated two-photon interference in a dual-beam Michelson interferometer'',
{\em Phys. Rev. A} {\bf 41}, 5, 2910-2913 (1990).

\item {\bf [Kwiat-Steinberg-Chiao 92]}:
P. G. Kwiat, A. M. Steinberg, \& R. Y. Chiao,
`Observation of a ``quantum eraser'': A revival of coherence in a two-photon
interference experiment',
{\em Phys. Rev. A} {\bf 45}, 11, 7729-7739 (1992).
See {\bf [Casado-Fern\'{a}ndez Rueda-Marshall-(+2) 97]}.

\item {\bf [Kwiat 93]}:
P. G. Kwiat,
``?'',
Ph.\ D. thesis, University of California at Berkeley, 1993.

\item {\bf [Kwiat-Steinberg-Chiao 94]}:
P. G. Kwiat, A. M. Steinberg, \& R. Y. Chiao,
``Three proposed `quantum erasers'\,'',
{\em Phys. Rev. A} {\bf 49}, 1, 61-68 (1994).

\item {\bf [Kwiat-Eberhard-Steinberg-Chiao 94]}:
P. G. Kwiat, P. H.
Eberhard, A. M. Steinberg, \& R. Y. Chiao,
``Proporsal for a loophole-free Bell
inequality experiment'',
{\em Phys. Rev. A} {\bf 49}, 5, Part A, 3209-3220 (1994).

\item {\bf [Kwiat 95]}:
P. G. Kwiat,
``Comment on `Reliability of Bell-inequality measurements using
polarization correlations in parametric-down-conversion photon sources'\,'',
{\em Phys. Rev. A} {\bf 52}, 4, 3380-3381 (1995).
Comment on {\bf [De Caro-Garuccio 95]}.

\item {\bf [Kwiat-Weinfurter-Herzog-(+2) 95 a]}:
P. G. Kwiat,
H. Weinfurter, T. Herzog, A. Zeilinger, \& M. A. Kasevich,
``Experimental realization of interaction-free measurements'',
in D. M. Greenberger, \& A. Zeilinger (eds.),
{\em Fundamental problems in quantum theory:
A conference held in honor of professor John
A. Wheeler, Ann. N. Y. Acad. Sci.} {\bf 755}, 383-393 (1995).
See {\bf [Kwiat-Weinfurter-Herzog-(+2) 95 b]}.

\item {\bf [Kwiat-Weinfurter-Herzog-(+2) 95 b]}:
P. G. Kwiat, H. Weinfurter, T. Herzog, A. Zeilinger, \& M. A. Kasevich,
``Interaction-free measurement'',
{\em Phys. Rev. Lett.} {\bf 74}, 24, 4763-4766 (1995).
See {\bf [Kwiat-Weinfurter-Herzog-(+2) 95 a]}.

\item {\bf [Kwiat-Mattle-Weinfurter-(+3) 95]}:
P. G. Kwiat, K. Mattle, H. Weinfurter, A. Zeilinger,
A. V. Sergienko, \& Y. H. Shih,
``New high-intensity source of polarization-entangled photon pairs'',
{\em Phys. Rev. Lett.} {\bf 75}, 24, 4337-4341 (1995).

\item {\bf [Kwiat-Weinfurter-Zeilinger 96 a]}:
P. G. Kwiat, H. Weinfurter, \& A. Zeilinger,
``Quantum seeing in the dark'',
{\em Sci. Am.} {\bf 275}, 5, 72-78.
Spanish version: ``Visi\'{o}n cu\'{a}ntica de la obscuridad'',
{\em Investigaci\'{o}n y Ciencia} 244, 54-60 (1997).
Reprinted in {\bf [Cabello 97 c]}, pp.~91-97.

\item {\bf [Kwiat-Weinfurter-Zeilinger 96 b]}:
P. G. Kwiat, H. Weinfurter, \& A. Zeilinger,
``Interaction-free measurements of a quantum object: On the breeding of
`Schr\"{o}dinger's cats'\,'',
in ? Eberly et al. (eds.),
{\em Coherence and quantum optics VII},
Plenum Press, New York, 1996, pp.~673-674.

\item {\bf [Kwiat 97 a]}:
P. G. Kwiat,
``Production and uses of hyper-entangled states'',
in M. Ferrero, \& A. van der Merwe (eds.),
{\em New developments on
fundamental problems in quantum physics (Oviedo, Spain, 1996)},
Kluwer Academic,
Dordrecht, Holland, 1997, pp.~189-202.
See {\bf [Kwiat-Mattle-Weinfurter-(+3) 95]}.

\item {\bf [Kwiat 97 b]}:
P. G. Kwiat,
``Hyper-entangled states'',
{\em J. Mod. Opt.} {\bf 44}, 11-12 (Special issue: Quantum
state preparation and measurement), 2173-2184 (1997).

\item {\bf [Kwiat 98]}:
P. G. Kwiat,
``Experimental and theoretical progress in
interaction-free measurements'',
in E. B. Karlsson, \& E. Br\"{a}ndas (eds.),
{\em Proc.\ of the 104th Nobel Symp.\ ``Modern Studies of Basic Quantum Concepts and Phenomena'' (Gimo, Sweden, 1997)},
{\em Physica Scripta} {\bf T76}, 115-121 (1998).

\item {\bf [Kwiat-Weinfurter 98]}:
P. G. Kwiat, \& H. Weinfurter,
``Embedded Bell-state analysis'',
{\em Phys. Rev. A} {\bf 58}, 4, R2623-R2626 (1998).

\item {\bf [Kwiat-Schwindt-Englert 99]}:
P. G. Kwiat, P. D. D. Schwindt, \& B.-G. Englert,
``What does a quantum eraser really erase?'',
in R. Bonifacio (ed.),
{\em Mysteries, Puzzles, and Paradoxes in Quantum Mechanics (Gargnano, Italy, 1998)},
American Institute of Physics, Woodbury, New York, 1999, pp.~69-80.

\item {\bf [Kwiat-Waks-White-(+2) 99]}:
P. G. Kwiat, E. Waks, A. G. White, I. Appelbaum, \& P. H. Eberhard,
``Ultrabright source of polarization-entangled photons'',
{\em Phys. Rev. A} {\bf 60}, 2, R773-R776 (1999);
quant-ph/9810003.

\item {\bf [Kwiat-Mitchell-Schwindt-White 99]}:
P. G. Kwiat, J. R. Mitchell, P. D. D. Schwindt, \& A. G. White,
``Grover's search algorithm: An optical approach'',
{\em J. Mod. Opt.};
quant-ph/9905086.

\item {\bf [Kwiat-White-Mitchell-(+4) 99]}:
P. G. Kwiat, A. G. White, J. R. Mitchell,
O. Nairz, G. Weihs, H. Weinfurter, \& A. Zeilinger,
``High-efficiency quantum interrogation measurements
via the quantum Zeno effect'',
{\em Phys. Rev. Lett.} {\bf 83}, 23, 4725-4728 (1999);
quant-ph/9909083.

\item {\bf [Kwiat-Hardy 00]}:
P. G. Kwiat, \& L. Hardy,
``The mystery of the quantum cakes'',
{\em Am. J. Phys.} {\bf 68}, 1, 33-36 (2000).

\item {\bf [Kwiat-Mitchel-Schwindt-White 00]}:
P. G. Kwiat, J. R. Mitchel, P. D. D. Schwindt, \& A. G. White,
``?'',
{\em J. Mod. Opt.} {\bf 47}, ?, 257-? (2000).

\item {\bf [Kwiat 00]}:
P. G. Kwiat,
``Grover's search algorithm: An optical approach'',
in V. Bu\v{z}zek, \& D. P. DiVincenzo (eds.),
{\em J. Mod. Opt.} {\bf 47}, 2-3 (Special issue:
Physics of quantum information), 257-266 (2000).

\item {\bf [Kwiat-Hughes 00]}:
P. G. Kwiat, \& R. J. Hughes,
``Does Rydberg state manipulation equal quantum computation?'',
{\em Science} {\bf 289}, 5484, 1431a (2000).
Comment on {\bf [Ahn-Weinacht-Bucksbaum 00]}.
Reply: {\bf [Bucksbaum-Ahn-Weinacht 00]}.
See {\bf [Meyer 00 c]}.

\item {\bf [Kwiat-Berglund-Altepeter-White 00]}:
P. G. Kwiat, A. J. Berglund, J. B. Altepeter, \& A. G. White,
``Experimental verification of decoherence-free subspaces'',
{\em Science} {\bf 290}, 5491, 498-501 (2000).

\item {\bf [Kwiat-Barraza L\'{o}pez-Stefanov-Gisin 01]}:
P. G. Kwiat, S. Barraza L\'{o}pez, A. Stefanov, \& N. Gisin,
``Experimental entanglement distillation and `hidden'
non-locality'',
{\em Nature} {\bf 409}, 6823, 1014-1017 (2001).

\item {\bf [Kwiat 01]}:
P. G. Kwiat,
``Quantum optics: Photons yield to peer pressure'',
{\em Nature} {\bf 412}, 6850, 866-867 (2001).
See {\bf [Lamas Linares-Howell-Bouwmeester 01]}.


\newpage

\subsection{}


\item {\bf [La Mura 03]}:
P. La Mura,
``Correlated equilibria of classical strategic games with quantum signals'',
quant-ph/0309033.

\item {\bf [Ladd-Goldman-Dana-(+2) 00]}:
T. D. Ladd, J. R. Goldman, A. Dana,
F. Yamaguchi, \& Y. Yamamoto,
``Quantum computation in a one-dimensional crystal
lattice with NMR force microscopy'';
quant-ph/0009122.

\item {\bf [Ladd-Goldman-Yamaguchi-(+3) 02]}:
T. D. Ladd, J. R. Goldman, F. Yamaguchi,
Y. Yamamoto, E. Abe, \& K. M. Itoh,
``All-silicon quantum computer'',
{\em Phys. Rev. Lett.} {\bf 89}, 1, 017901 (2002);
quant-ph/0109039.

\item {\bf [Lakner-Peternelj 03]}:
M. Lakner, \& J. Peternelj,
`Comment on ``Splitting the wavefunction of a particle in a box,''
by Julio Gea-Banacloche [Am. J. Phys. {\bf 70} (3), 307–312 (2002)]',
{\em Am. J. Phys.} {\bf 71}, 6, 519 (2003).
Comment on {\bf [Gea Banacloche 02 b]}.

\item {\bf [Lakshminarayan-Subrahmanyam 03]}:
A. Lakshminarayan, \& V. Subrahmanyam,
A. Lakshminarayan, \& V. Subrahmanyam,
``Entanglement sharing in one-particle states'',
{\em Phys. Rev. A} {\bf 67}, 5, 052304 (2003);
quant-ph/0212049.

\item {\bf [Laudisa 00]}:
F. Laudisa,
``On time asymmetry and history in an Everett quantum world'',
{\em Found. Phys.} {\bf 30}, 9, 1525-1538 (2000).

\item {\bf [Laudisa 01]}:
F. Laudisa,
``The EPR argument in a relational interpretation
of quantum mechanics'',
{\em Found. Phys. Lett.} {\bf 14}, 2, 119-132 (2001);
quant-ph/0011016.

\item {\bf [Laudisa 02]}:
F. Laudisa,
``Non-locality and theories of causation'',
in T. Placek, \& J. N. Butterfield (eds.),
{\em NATO Advanced Research Workshop ``Modality, Probability, and Bell's Theorem''
(Cracow, Poland, 2001)},
Kluwer Academic, Dordrecht, Holland, 2002;
quant-ph/0111028.

\item {\bf [Laskowski-Paterek-\.{Z}ukowski 03]}:
W. Laskowski, T. Paterek, M. \.{Z}ukowski,
``Extending the class of quantum states that do not allow local realistic
description'',
quant-ph/0311182.
See {\bf [Laskowski-Paterek-\.{Z}ukowski-Brukner 04]}.

\item {\bf [Laskowski-Paterek-\.{Z}ukowski-Brukner 04]}:
W. Laskowski, T. Paterek, M. \.{Z}ukowski, \& \v{C}. Brukner,
``Tight multipartite Bell's inequalities involving many measurement settings'',
{\em Phys. Rev. Lett.} {\bf 93}, 200401 (2004);
quant-ph/0411066.

\item {\bf [Law-Eberly 04]}:
C. K. Law, \& J. H. Eberly,
``Analysis and interpretation of high transverse entanglement in optical parametric down conversion'',
{\em Phys. Rev. Lett.} {\bf 92}, 12, 127903 (2004).

\item {\bf [Lawrence-Brukner-Zeilinger 02]}:
J. Lawrence, \v{C}. Brukner, \& A. Zeilinger,
``Mutually unbiased binary observable sets on $N$ qubits'',
{\em Phys. Rev. A} {\bf 65}, 3, 032320 (2002);
quant-ph/0104012.

\item {\bf [Lawrence 04]}:
J. Lawrence,
``Mutually unbiased bases and trinary operator sets for $N$ qutrits'',
{\em Phys. Rev. A} {\bf 70}, 1, 012302 (2004);
quant-ph/0403095.

\item {\bf [Laflamme-Miquel-Paz-Zurek 96]}:
R. Laflamme, C. Miquel, J. P. Paz, \& W. H. Zurek,
``Perfect quantum error correcting code'',
{\em Phys. Rev. Lett.} {\bf 77}, 1, 198-201 (1996).

\item {\bf [Laflamme-Knill-Zurek-(+2) 98]}:
R. Laflamme, E. Knill, W. H. Zurek, P. Catasti, \& S. V. S. Mariappan,
``NMR Greenberger-Horne-Zeilinger states'',
in A. K. Ekert, R. Jozsa, \& R. Penrose (eds.),
{\em Quantum Computation: Theory and Experiment.
Proceedings of a Discussion Meeting held at the Royal
Society of London on 5 and 6 November 1997},
{\em Philos. Trans. R. Soc. Lond. A} {\bf 356}, 1743, 1941-1948
(1998);
quant-ph/9709025.

\item {\bf [Laflamme-Cory-Negrevergne-Viola 01]}:
R. Laflamme, D. G. Cory, C. Negrevergne, \& L. Viola,
``NMR quantum information processing and entanglement'',
quant-ph/0110029.

\item {\bf [Lahera 04]}:
J. Lahera,
{\em De la teor\'{\i}a at\'{o}mica a la f\'{\i}sica cu\'{a}ntica. Bohr},
Nivola, Madrid, 2004.

\item {\bf [Lahti 90]}:
P. J. Lahti,
``Quantum theory of measurement and the polar
decomposition of an interaction'',
{\em Int. J. Theor. Phys.} {\bf 29}, 4, 339-350 (1990).

\item {\bf [Lahti-Pellonpaa-Ylinen 04]}:
P. J. Lahti, J.-P. Pellonpaa, \& K. Ylinen,
``Two questions on quantum statistics'',
quant-ph/0408186.

\item {\bf [Laidlaw 01]}:
A. Laidlaw,
``Some advantages of a local realist, 3D wave soliton approach to EPR'',
quant-ph/0110160.

\item {\bf [Laiho-Molotkov-Nazin 00 a]}:
R. Laiho, S. N. Molotkov, \& S. S. Nazin,
``Teleportation of the relativistic quantum field'',
{\em Phys. Lett. A} {\bf 275}, 1-2, 36-47 (2000).

\item {\bf [Laiho-Molotkov-Nazin 00 b]}:
R. Laiho, S. N. Molotkov, \& S. S. Nazin,
``On teleportation of a completely unknown state of relativistic
photon'',
{\em Phys. Lett. A} {\bf 278}, 1-2, 9-18 (2000).

\item {\bf [Laiho-Molotkov-Nazin 00 c]}:
R. Laiho, S. N. Molotkov, \& S. S. Nazin,
``On the distinguishability of relativistic
quantum states in quantum cryptography'',
quant-ph/0006010.

\item {\bf [Lalo\"{e} 95]}:
F. Lalo\"{e},
``Correlating more than two particles in quantum-mechanics'',
{\em Current. Sci.} {\bf 68}, 10, 1026-1035 (1995).

\item {\bf [Lalo\"{e} 01]}:
F. Lalo\"{e},
``Do we really understand quantum mechanics? Strange
correlations, paradoxes and theorems'',
{\em Am. J. Phys.} {\bf 69}, 6, 655-701 (2001).
Erratum: {\em Am. J. Phys.} {\bf 70}, 5, 556 (2002).

\item {\bf [Lalo\"{e} 04]}:
F. Lalo\"{e},
``The hidden phase of Fock states; quantum non-local effects'',
quant-ph/0409097.

\item {\bf [Lamas Linares-Mikkelsen-Howell-Bouwmeester 01]}:
A. Lamas-Linares, C. Mikkelsen, J. C. Howell, \& D. Bouwmeester,
``Interference enhanced polarization entanglement and
the concept of an entangled-photon laser'',
quant-ph/0103056.
Preliminary version of {\bf [Lamas Linares-Howell-Bouwmeester 01]}.

\item {\bf [Lamas Linares-Howell-Bouwmeester 01]}:
A. Lamas-Linares, J. C. Howell, \& D. Bouwmeester,
``Stimulated emission of polarization-entangled photons'',
{\em Nature} {\bf 412}, 6850, 887-890 (2001);
quant-ph/0110048.
See {\bf [Kwiat 01]}.
Preliminary version: {\bf [Lamas Linares-Mikkelsen-Howell-Bouwmeester 01]}.

\item {\bf [Lamas Linares-Simon-Howell-Bouwmeester 02]}:
A. Lamas-Linares, C. Simon, J. C. Howell, \& D. Bouwmeester,
``Experimental quantum cloning of single photons'',
{\em Science} {\bf 296}, ?, 712-714 (2002).

\item {\bf [Lamata-Leon 03]}:
L. Lamata, \& J. Leon,
``Time of arrival with resonances: Beyond scattering states'',
{\em Time asymmetric quantum theory: The theory of resonances (Lisbon, 2003)};
quant-ph/0312034.

\item {\bf [Lamata-Leon 04]}:
L. Lamata, \& J. Leon,
``Dealing with entanglement of continuous variables: Schmidt decomposition
with discrete sets of orthogonal functions'',
quant-ph/0410167.

\item {\bf [Lamb 69]}:
W. E. Lamb, Jr.,
``An operational interpretation of nonrelativistic quantum mechanics'',
{\em Phys. Today} {\bf 22}, 4, 23-28 (1969).

\item {\bf [Lamb 01 a]}:
W. E. Lamb, Jr.,
``Super classical quantum mechanics:
The best interpretation of nonrelativistic quantum mechanics'',
{\em Am. J. Phys.} {\bf 69}, 4, 413-422 (2001).
Comment: {\bf [Luzuriaga 02]}.

\item {\bf [Lamb 01 b]}:
W. E. Lamb, Jr.,
{\em The interpretation of quantum mechanics},
J. Mehra (ed. and annotations),
Rinton Press, Princeton, New Jersey, 2001.

\item {\bf [Lambert-Emary-Brandes 04]}:
N. Lambert, C. Emary, \& T. Brandes,
``Entanglement and the phase transition in single-mode superradiance'',
{\em Phys. Rev. Lett.} {\bf 92}, 7, 073602 (2004).

\item {\bf [Lamehi Rachti-Mittig 76]}:
M. Lamehi Rachti, \& W. Mittig,
``Quantum mechanics and hidden variables: A test of Bell's inequality by the
measurement of the spin correlation in low-energy proton-proton scattering'',
{\em Phys. Rev. D} {\bf 14}, 10, 2543-2555 (1976).
Reprinted in {\bf [Wheeler-Zurek 83]}, pp.~422-434.

\item {\bf [Lamoureux-Navez-Fiur\'{a}\v{s}ek-Cerf 03]}:
L.-P. Lamoureux, P. Navez, J. Fiur\'{a}\v{s}ek, \& N. J. Cerf,
``Cloning the entanglement of a pair of quantum bits'',
{\em Phys. Rev. A} {\bf 69}, 4, 040301 (2004);
quant-ph/0302173.

\item {\bf [Lamoureux-Brainis-Cerf-(+3) 04]}:
L.-P. Lamoureux, E. Brainis, N. Cerf,
P. Emplit, M. Haelterman, \& S. Massar,
``Experimental quantum key distribution over highly noisy channels'',
quant-ph/0407031.

\item {\bf [Lamoureux-Brainis-Amans-(+2) 04]}:
L.-P. Lamoureux, E. Brainis, D. Amans,
J. Barrett, \& S. Massar,
``Provably secure experimental quantum bit-string generation'',
quant-ph/0408121.
See {\bf [Barrett-Massar 04]}.

\item {\bf [Lamoureux-Cerf 04]}:
L.-P. Lamoureux, \& N. J. Cerf,
``Asymmetric phase-covariant $d$-dimensional cloning'',
quant-ph/0410054.

\item {\bf [Lan 02]}:
B. L. Lan,
``A realist view of the canonical Einstein-Podolsky-Rosen-Bohm experiment
based on quantum theory and its consequences'',
in R. Bonifacio, \& D. Vitali (eds.),
{\em Mysteries, Puzzles and Paradoxes in Quantum Mechanics IV:
Quantum Interference Phenomena (Gargnano, Italy, 2001)},
{\em J. Opt. B: Quantum Semiclass. Opt.} {\bf 4}, 4, S384-S385 (2002);
quant-ph/0202151.

\item {\bf [Lance-Symul-Bowen-(+2) 02]}:
A. M. Lance, T. Symul, W. P. Bowen,
B. C. Sanders, \& P. K. Lam,
``Tripartite quantum state sharing'',
{\em Phys. Rev. Lett.} {\bf 92}, 17, 177903 (2004);
quant-ph/0311015.

\item {\bf [Landau-Lifshitz 48]}:
L. D. Landau, \& E. M. Lifshitz,
{\em Kvantovaja mekhanika}, G.\ I.\ T.\ T.\ L., Moscow, 1948.
English version: {\em Quantum mechanics},
Pergamon Press, Oxford, 1958;
Addison-Wesley, Reading, Massachusetts, 1958;
Pergamon Press, New York, 1977 (3rd edition).
Spanish version: {\em Mec\'{a}nica cu\'{a}ntica no-relativista},
Revert\'{e}, Barcelona, 1983 (2nd edition).

\item {\bf [Landau 87]}:
L. J. Landau,
``On the violation of Bell's inequality in quantum theory'',
{\em Phys. Lett. A} {\bf 120}, 2, 54-56 (1987).

\item {\bf [Landau 88]}:
L. J. Landau,
``Empirical two-point correlation functions'',
{\em Found. Phys.} {\bf 18}, 4, 449-460 (1988).

\item {\bf [Landauer 91]}:
R. Landauer,
``Information is physical'',
{\em Phys. Today} {\bf 44}, 5, 23-29 (1991).

\item {\bf [Landauer-Martin 94]}:
R. Landauer, \& T. Martin,
``Barrier interaction time in tunneling'',
{\em Rev. Mod. Phys.} {\bf 66}, 1, 217-228 (1994).

\item {\bf [Landauer 96]}:
R. Landauer,
``?'',
{\em Science} {\bf 272}, ?, 1841-2012 (1996).
See {\bf [Bennett 96 a]}.

\item {\bf [Landauer 98]}:
R. Landauer,
``Energy needed to send a bit'',
in D. P. DiVincenzo. E. Knill, R. Laflamme, \& W. H. Zurek (eds.),
{\em Quantum Coherence and Decoherence.
Proc.\ of the ITP Conf.\ (Santa Barbara, California, 1996)},
{\em Proc. R. Soc. Lond. A} {\bf 454}, 1969, 305-311 (1998).

\item {\bf [Landauer 99]}:
R. Landauer,
``Information is inevitably physical'',
in T. Hey (ed.),
{\em Feynman and computation},
Perseus, Reading, Massachusetts, 1999, pp.~?-?.

\item {\bf [Land\'{e} 60]}:
A. Land\'{e},
{\em From dualism to unity in quantum physics},
Cambridge University Press, London, 1960.

\item {\bf [Landsburg 04]}:
S. E. Landsburg,
``Quantum game theory'',
{\em Notices Am. Math. Soc.} {\bf 51}, 4, 394-399 (2004).

\item {\bf [Landsman 99]}:
N. P. Landsman,
``Quantum mechanics on phase space'',
{\em Stud. Hist. Philos. Sci. Part B: Stud. Hist. Philos. Mod. Phys.}
{\bf 30}, 2, 287-305 (1999).
Review of {\bf [Schroeck 96]}.

\item {\bf [Langford 02]}:
J. Langford,
``Generic quantum block compression'',
{\em Phys. Rev. A} {\bf 65}, 5, 052312 (2002);
quant-ph/0109074.

\item {\bf [Langford-Dalton-Harvey-(+5) 04]}:
N. K. Langford, R. B. Dalton, M. D. Harvey,
J. L. O'Brien, G. J. Pryde, A. Gilchrist,
S. D. Bartlett, \& A. G. White,
``Measuring entangled qutrits and their use for quantum bit commitment'',
{\em Phys. Rev. Lett.} {\bf 93}, 5, 053601 (2004);
quant-ph/0312072.

\item {\bf [Larsen 98]}:
U. Larsen,
``Measuring the entangled Bell and GHZ aspects
using a single-qubit shuttle'',
quant-ph/9812085.

\item {\bf [Larson-Garraway 04]}:
J. Larson, \& B. M. Garraway,
``Dynamics of a Raman coupled model: Entanglement and quantum computation'',
{\em J. Mod. Opt.} {\bf 51}, 1691 (2004);
quant-ph/0411090.

\item {\bf [Larsson 98 a]}:
J.-\AA. Larsson,
``Necessary and sufficient detector-efficiency conditions for the
Greenberger-Horne-Zeilinger paradox'',
{\em Phys. Rev. A} {\bf 57}, 5, R3145-R3149 (1998).

\item {\bf [Larsson 98 b]}:
J.-\AA. Larsson,
``Bell's inequality and detector inefficiency'',
{\em Phys. Rev. A} {\bf 57}, 5, 3304-3308 (1998).

\item {\bf [Larsson-Aerts-\.{Z}ukowski 98]}:
J.-\AA. Larsson, S. Aerts, \& M. \.{Z}ukowski,
``Two-photon Franson-type interference experiments are
not tests of local realism'',
quant-ph/9812053.
See {\bf [Franson 89]}.

\item {\bf [Larsson 99 a]}:
J.-\AA. Larsson,
``Modeling the singlet state with local variables'',
{\em Phys. Lett. A} {\bf 256}, 4, 245-252 (1999);
quant-ph/9901074.

\item {\bf [Larsson 99 b]}:
J.-\AA. Larsson,
``Detector efficiency in the Greenberger-Horne-Zeilinger paradox:
Independent errors'',
{\em Phys. Rev. A} {\bf 59}, 6, 4801-4804 (1999).

\item {\bf [Larsson 00 a]}:
J.-\AA. Larsson,
``A possible unification of the Copenhagen and the Bohm
interpretations using local realism'',
{\em Found. Phys. Lett.} {\bf 13}, 5, 477-486 (2000).

\item {\bf [Larsson 00 b]}:
J.-\AA. Larsson,
``Quantum paradoxes, probability theory, and change of ensemble'',
Ph.\ D. thesis, Link\"{o}pings Universitet, Sweden, 2000.

\item {\bf [Larsson-Semitecolos 01]}:
J.-\AA. Larsson, \& J. Semitecolos,
``Strict detector-efficiency bounds for $n$-site Clauser-Horne inequalities'',
{\em Phys. Rev. A} {\bf 63}, 2, 022117 (2001);
quant-ph/0006022.

\item {\bf [Larsson 02 a]}:
J.-\AA. Larsson,
``A Kochen-Specker inequality'',
{\em Europhys. Lett.} {\bf 58}, 6, 799-805 (2002);
quant-ph/0006134.
See {\bf [Simon-Brukner-Zeilinger 01]}.

\item {\bf [Larsson 02 b]}:
J.-\AA. Larsson,
``A practical Trojan horse for Bell-inequality-based quantum cryptography'',
{\em Quant. Inf. Comp.} {\bf 2}, ?, 434-? (2002);
quant-ph/0111073.

\item {\bf [Larsson 03 a]}:
J.-\AA. Larsson,
``Qubits from number states and Bell inequalities for number measurements'',
{\em Phys. Rev. A} {\bf 67}, 2, 022108 (2003);
quant-ph/0208125.

\item {\bf [Larsson 03 b]}:
J.-\AA. Larsson,
``No information flow using statistical fluctuations, and quantum
cryptography'',
quant-ph/0308107.
See {\bf [Home-Whitaker 03]}.

\item {\bf [Larsson 04]}:
J.-\AA. Larsson,
``Bell inequalities for position measurements'',
{\em Phys. Rev. A} {\bf 70}, 2, 022102 (2004);
quant-ph/0310140.

\item {\bf [Larsson-Gill 03]}:
J.-\AA. Larsson, \& R. D. Gill,
``Bell's inequality and the coincidence-time loophole'',
quant-ph/0312035.

\item {\bf [Latorre-Pascual-Tarrach 98]}:
J. I. Latorre, P. Pascual, \& R. Tarrach,
``Minimal optimal generalized quantum measurements'',
{\em Phys. Lett. A} {\bf 81}, 7, 1351-1354 (1998);
quant-ph/9803066.

\item {\bf [Latorre-Mart\'{\i}n Delgado 02]}:
J. I. Latorre, \& M. A. Mart\'{\i}n-Delgado,
``Majorization arrow in quantum-algorithm design'',
{\em Phys. Rev. A} {\bf 66}, 2, 022305 (2002);
quant-ph/0111146.

\item {\bf [Latorre-Orus 04]}:
J. I. Latorre, \& R. Orus,
``Adiabatic quantum computation and quantum phase transitions'',
{\em Phys. Rev. A} {\bf 69}, 6, 062302 (2004);
quant-ph/0308042.

\item {\bf [Latorre-Rico-Vidal 04]}:
J. I. Latorre, E. Rico, \& G. Vidal,
``Ground state entanglement in quantum spin chains'',
{\em Quant. Inf. Comp.} {\bf 4}, 1, 48-92 (2004);
quant-ph/0304098.

\item {\bf [Latorre-Lutken-Rico-Vidal 04]}:
J. I. Latorre, C. A. Lutken, E. Rico, \& G. Vidal,
``Fine-grained entanglement loss along renormalization group flows'',
quant-ph/0404120.

\item {\bf [Latorre-Orus-Rico-Vidal 04]}:
J. I. Latorre, R. Orus, E. Rico, \& J. Vidal,
``Entanglement entropy in the Lipkin-Meshkov-Glick model'',
cond-mat/0409611.

\item {\bf [Laurat-Coudreau-Treps-(+2) 03]}:
J. Laurat, T. Coudreau, N. Treps, A. Maitre, \& C. Fabre,
``Conditional preparation of a quantum state in the continuous variable
regime: Generation of a sub-Poissonian state from twin beams'',
quant-ph/0304111.

\item {\bf [Laurat-Coudreau-Keller-(+2) 04]}:
J. Laurat, T. Coudreau, G. Keller,
N. Treps, \& C. Fabre,
``Compact source of EPR entanglement and squeezing at very low noise
frequencies'',
quant-ph/0403224.

\item {\bf [Laurikainen 85]}:
K. V. Laurikainen,
``Wolfgang Pauli and the Copenhagen philosophy'',
in P. J. Lahti, \& P. Mittelstaedt (eds.),
{\em Symp.\ on the Foundations of Modern
Physics: 50 Years of the Einstein-Podolsky-Rosen Experiment
(Joensuu, Finland, 1985)},
World Scientific, Singapore, 1985, pp.~273-287.

\item {\bf [Laurikainen 88]}:
K. V. Laurikainen,
{\em Beyond the atom: The philosophical thought of Wolfgang Pauli},
Springer-Verlag, Berlin, 1988.

\item {\bf [Laurikainen 97]}:
K. V. Laurikainen,
{\em The message of the atoms.
Essays on Wolfgang Pauli and the unspeakable},
Springer-Verlag, Berlin, 1997.

\item {\bf [Laustsen-M{\o}lmer 02]}:
T. Laustsen, \& K. M{\o}lmer,
``Approximate quantum data storage and teleportation'',
{\em Phys. Rev. A} {\bf 65}, 6, 062313 (2002);
quant-ph/0201023.

\item {\bf [Lavor-Manssur-Portugal 03]}:
C. Lavor, L.R.U. Manssur, \& R. Portugal,
``Shor's algorithm for factoring large integers'',
quant-ph/0303175.

\item {\bf [Lea-Frayne-Mukharsky 00]}:
M. J. Lea, P. G. Frayne, \& Y. Mukharsky,
``Could we quantum compute with electrons on helium?'',
{\em Fortschr. Phys.} {\bf 48}, 9-11 (Special issue:
Experimental proposals for quantum computation), 1109-1124 (2000).

\item {\bf [Leach-Padgett-Barnett-(+2) 02]}:
J. Leach, M. J. Padgett, S. M. Barnett, S. Franke-Arnold,
\& J. Courtial,
``Measuring the orbital angular momentum of a single photon'',
{\em Phys. Rev. Lett.} {\bf 88}, 25, 257901 (2002).

\item {\bf [Leach-Courtial-Skeldon-(+3) 04]}:
J. Leach, J. Courtial, K. Skeldon,
S. M. Barnett, S. Franke-Arnold, \& M. J. Padgett,
``Interferometric methods to measure orbital and spin,
or the total angular momentum of a single photon'',
{\em Phys. Rev. Lett.} {\bf 92}, 1, 013601 (2004).

\item {\bf [Leavens 93]}:
C. R. Leavens,
``Arrival time distributions'',
{\em Phys. Let. A} {\bf 178}, ?, 27-32 (1993).

\item {\bf [Leavens-Aers 93]}:
C. R. Leavens, \& G. C. Aers,
``Bohm trajectories and the tunneling time problem'',
in R. Wiesendanger, \& H. J. Guntherrodt (eds.),
{\em Scanning Tunneling Microscopy III},
Springer-Verlag, New York, 1993, 105-140.

\item {\bf [Leavens 95]}:
C. R. Leavens,
``Bohm trajectory and Feynman path
approaches to the `tuneling time problem'\,'',
{\em Found. Phys.} {\bf 25}, 2, 229-268 (1995).

\item {\bf [Leavens-Iannaccone-McKinnon 95]}:
C. R. Leavens, G. Iannaccone, \& W. R. McKinnon,
``On the approach to the stationary-state scattering
limit within Bohmian mechanics'',
{\em Phys. Let. A} {\bf 208}, ?, 17-24 (1995).

\item {\bf [Leavens 96]}:
C. R. Leavens,
``Bohmian mechanics and the tunneling time
problem for electrons'',
in D. Mugnai, A. Rafani, \& L.S. Schulman (eds.),
{\em Proc.\ of the Adriatico Research Conf.\ on
Tunneling and its Implications}, ?, ?, 100-120, 1996.

\item {\bf [Leavens 98]}:
C. R. Leavens,
``Time of arrival in quantum and Bohmian mechanics'',
{\em Phys. Rev. A} {\bf 58}, 2, 840-847 (1998).

\item {\bf [Leavens-Sala Mayato 99]}:
C. R. Leavens, \& R. Sala Mayato,
``Superluminal systematic particle velocity in
relativistic stochastic Bohmian mechanics'',
{\em Phys. Lett. A} {\bf 263}, 1-2, 1-8 (1999).

\item {\bf [Leavens-Sala Mayato 99]}:
C. R. Leavens, \& R. Sala Mayato,
``On constructing the wave function of a quantum particle
from its Wigner phase-space distribution'',
{\em Phys. Lett. A} {\bf 280}, 4, 163-172 (2001).

\item {\bf [Lebedev-Blatter-Beenakker-Lesovik 04]}:
A. V. Lebedev, G. Blatter, C. W. J. Beenakker, \& G. B. Lesovik,
``Entanglement in mesoscopic structures: Role of projection'',
{\em Phys. Rev. B} {\bf 69}, 23, 235312 (2004);
cond-mat/0311649.

\item {\bf [Lee-Johnson 02 a]}:
C. F. Lee, \& N. F. Johnson,
``Exploiting randomness in quantum information processing'',
{\em Phys. Lett. A} {\bf 301}, 5-6, 343-349 (2002).

\item {\bf [Lee-Johnson 02 b]}:
C. F. Lee, \& N. F. Johnson,
``Let the quantum games begin'',
{\em Phys. World} {\bf 15}, 10, 25-29 (2002).

\item {\bf [Lee-Johnson 02 c]}:
C. F. Lee, \& N. F. Johnson,
``Parrondo games and quantum algorithms'',
quant-ph/0203043.

\item {\bf [Lee-Johnson 03 a]}:
C. F. Lee, \& N. F. Johnson,
``Efficiency and formalism of quantum games'',
{\em Phys. Rev. A} {\bf 67}, 2, 022311 (2003).

\item {\bf [Lee-Johnson 03 b]}:
C. F. Lee, \& N. F. Johnson,
``Game-theoretic discussion of quantum state estimation and cloning'',
{\em Phys. Lett. A} {\bf 319}, 5-6, 429-433 (2003).

\item {\bf [Lee-Johnson 04]}:
C. F. Lee, \& N. F. Johnson,
``Efficient quantum computation within a disordered Heisenberg spin-chain'',
{\em Phys. Rev. A};
quant-ph/0404163.

\item {\bf [Lee-Chang Young 04]}:
D. Lee, \& E. Chang-Young,
``Quantum entanglement under Lorentz boost'',
{\em New J. Phys.}
quant-ph/0308156.

\item {\bf [Lee-Kok-Cerf-Dowling 02]}:
H. Lee, P. Kok, N. J. Cerf, \& J. P. Dowling,
Linear optics and projective measurements alone suffice to create
large-photon-number path entanglement'',
{\em Phys. Rev. A} {\bf 65}, 3, 030101 (2002);
quant-ph/0109080.

\item {\bf [Lee-Kok-Dowling 02 a]}:
H. Lee, P. Kok, \& J. P. Dowling,
``A quantum Rosetta stone for interferometry'',
quant-ph/0202133.

\item {\bf [Lee-Kok-Dowling 02 b]}:
H. Lee, P. Kok, \& J. P. Dowling,
``Quantum imaging and metrology'',
in J. H. Shapiro \& O. Hirota (eds.),
{\em Proc.\ of the Sixth Int.\ Conf.\
on Quantum Communication, Measurement and Computing},
Rinton Press, Princeton, New Jersey, 2002;
quant-ph/0306113.

\item {\bf [Lee-Yurtsever-Kok-(+4) 03]}:
H. Lee, U. H. Yurtsever, P. Kok,
G. M. Hockney, C. Adami, S. L. Braunstein, \& J. P. Dowling,
``Towards photostatistics from photon-number discriminating detectors'',
quant-ph/0310161.

\item {\bf [Lee-Kok-Williams-Dowling 03]}:
H. Lee, P. Kok, C. P. Williams, \& J. P. Dowling,
``From linear optical quantum computing to Heisenberg-limited
interferometry'',
{\em J. Opt. B: Quantum Semiclass. Opt.};
quant-ph/0312169.

\item {\bf [Lee-Hong-Kim-(+2) 04]}:
H. Lee, C. Hong, H. Kim,
J. Lim, \& H. J. Yang,
``Arbitrated quantum signature scheme with message recovery'',
{\em Phys. Lett. A} {\bf 321}, 5-6, 295-300 (2004).

\item {\bf [Lee-Ahn-Hwang 02]}:
H.-J. Lee, D. Ahn, \& S. W. Hwang,
``Dense coding in entangled states'',
{\em Phys. Rev. A} {\bf 66}, 2, 024304 (2002).
Comments: {\bf [Wojcik-Grudka 03]}, {\bf [Akhavan-Rezakhani 03]}.

\item {\bf [Lee-Kim 01]}:
H.-W. Lee, \& J. Kim,
``Quantum teleportation and Bell's inequality
using single-particle entanglement'',
{\em Phys. Rev. A} {\bf 63}, 1, 012305 (2001);
quant-ph/0007106.

\item {\bf [Lee 01]}:
H.-W. Lee,
``Total teleportation of an entangled state'',
{\em Phys. Rev. A} {\bf 64}, 1, 014302 (2001);
quant-ph/0104097.

\item {\bf [Lee-Kim 00]}:
J. Lee, \& M. S. Kim,
``Entanglement teleportation via Werner states'',
{\em Phys. Rev. Lett.} {\bf 84}, 18, 4236-4239 (2000);
quant-ph/9907041.

\item {\bf [Lee-Kim-Jeong 00]}:
J. Lee, M. S. Kim, \& H. Jeong,
``Transfer of nonclassical features in quantum teleportation
via a mixed quantum channel'',
{\em Phys. Rev. A} {\bf 62}, 3, 032305 (2000);
quant-ph/0003078.

\item {\bf [Lee-Kim-Park-Lee 00]}:
J. Lee, M. S. Kim, Y. J. Park, \& S. Lee,
``Partial teleportation of entanglement in the noisy
environment'',
quant-ph/0003060.

\item {\bf [Lee-Min-Oh 02]}:
J. Lee, H. Min, \& S. D. Oh,
``Multipartite entanglement for entanglement teleportation'',
{\em Phys. Rev. A} {\bf 66}, 5, 052318 (2002);
quant-ph/0201069.

\item {\bf [Lee-Jang-Park 02]}:
J. Lee, J. Jang, \& Y.-J. Park,
``Quantum secure communication with $W$ states'',
quant-ph/0204003.

\item {\bf [Lee-Jang-Brukner 03]}:
J. Lee, M. S. Kim, \& \v{C}. Brukner,
``Operationally invariant measure of the distance between quantum states
by complementary measurements'',
{\em Phys. Rev. Lett.} {\bf 91}, 8, 087902 (2003);
quant-ph/0303111.

\item {\bf [Lee-Lee-Kim-Oh 03]}:
J. Lee, S. Lee, J. Kim, \& S. D. Oh,
``Entanglement swapping secures multiparty quantum communication'',
quant-ph/0309185.
See {\bf [Cabello 00 i]}.

\item {\bf [Lee-Kim-Ahn-(+2) 04]}:
J. Lee, I. Kim, D. Ahn,
H. McAneney, \& M. S. Kim,
``Completely positive non-Markovian decoherence'',
{\em Phys. Rev. A} {\bf 70}, 2, 024301 (2004).

\item {\bf [Lee-Lee-Kim 04]}:
J. Lee, S.-W. Lee, \& M. S. Kim,
``Greenberger-Horne-Zeilinger paradox in tripartite systems of arbitrary
dimension'',
quant-ph/0408072.

\item {\bf [Lee-Chung-Kim-Lee 99]}:
J.-S. Lee, Y. Chung, J. Kim, \& S. Lee,
``A practical method of constructing quantum
combinational logic circuits'',
quant-ph/9911053.

\item {\bf [Lee-Kim-Cheong-Lee 02]}:
J.-S. Lee, J. Kim, Y. Cheong, \& S. Lee,
``Implementation of phase estimation and quantum counting algorithms on an NMR
quantum-information processor'',
{\em Phys. Rev. A} {\bf 66}, 4, 042316 (2002).

\item {\bf [Lee-Lee-Chung-(+2) 03]}:
J.-W. Lee, E. K. Lee, Y. W. Chung, H.-W. Lee, \& J. Kim,
``Quantum cryptography using single-particle entanglement'',
{\em Phys. Rev. A} {\bf 68}, 1, 012324 (2003).

\item {\bf [Lee-Thomas 02]}:
K. F. Lee, \& J. E. Thomas,
``Experimental simulation of two-particle quantum entanglement using classical
fields'',
{\em Phys. Rev. Lett.} {\bf 88}, 9, 097902 (2002).

\item {\bf [Lee-Chi-Oh-Kim 03]}:
S. Lee, D. P. Chi, S. D. Oh, \& J. Kim,
``Convex-roof extended negativity as an entanglement measure for bipartite quantum systems'',
{\em Phys. Rev. A} {\bf 68}, 6, 062304 (2003);
quant-ph/0310027.

\item {\bf [Legero-Wilk-Hennrich-(+2) 04]}:
T. Legero, T. Wilk, M. Hennrich,
G. Rempe, \& A. Kuhn,
``Quantum beat of two single photons'',
{\em Phys. Rev. Lett.} {\bf 93}, 7, 070503 (2004);
quant-ph/0406096.

\item {\bf [Leggett 80]}:
A. J. Leggett,
``Macroscopic quantum systems and the quantum theory of measurement'',
{\em Prog. Theor. Phys. Supp.} {\bf 69}, 80-100 (1980).

\item {\bf [Leggett 84 a]}:
A. J. Leggett,
``Quantum tunneling in the presence of an arbitrary linear
dissipation mechanism'',
{\em Phys. Rev. B} {\bf 30}, 3 1208-1218 (1984).

\item {\bf [Leggett 84 b]}:
A. J. Leggett,
``Schr\"{o}dinger's cat and her laboratory cousins'',
{\em Contemp. Phys.} {\bf 25}, 6, 583-598 (1984).

\item {\bf [Leggett-Garg 85 a]}:
A. J. Leggett, \& A. Garg,
``Quantum mechanics
versus macroscopic realism: Is the flux there when nobody looks?'',
{\em Phys. Rev. Lett.} {\bf 54}, 9, 857-860 (1985).
Comment: {\bf [Kraichnan 85]}.
Reply: {\bf [Leggett-Garg 85 b]}.

\item {\bf [Leggett-Garg 85 b]}:
A. J. Leggett, \& A. Garg,
``Leggett and Garg respond'',
{\em Phys. Rev. Lett.} {\bf 54}, 25, 2724 (1985).
Reply to {\bf [Kraichnan 85]}.
See {\bf [Leggett-Garg 85 a]}.

\item {\bf [Leggett 86 a]}:
A. J. Leggett,
``Quantum mechanics at the macroscopic level'',
in G. Grinstein, \& G. Mezenko (eds.),
{\em Directions in condensed matter
physics: Memorial volume in honor of Shang-keng Ma},
World Scientific, Singapore, 1986, pp.~185-248.

\item {\bf [Leggett 86 b]}:
A. J. Leggett,
``Quantum mechanics at the macroscopic level'',
in J. de Boer, E. Dal, \& O. Ulfbect (eds.),
{\em The lesson of quantum theory:
Niels Bohr centenary symposium},
Elsevier, Amsterdam, 1986, pp.~35-58.

\item {\bf [Leggett 87 a]}:
A. J. Leggett,
``Reflections on the quantum measurement paradox'',
in {\bf [Hiley-Peat 87]}, pp.~85-104.

\item {\bf [Leggett 87 b]}:
A. J. Leggett,
``Macroscopic quantum tunnelling and related matters'',
{\em Jap. J. App. Phys.} {\bf 26}, supplement 3, 1986-1993 (1987).

\item {\bf [Leggett 88]}:
A. J. Leggett,
``Experimental approaches to the quantum measurement paradox'',
{\em Found. Phys.} {\bf 18}, ?, 939-952 (1988).

\item {\bf [Leggett 89]}:
A. J. Leggett,
``Comment on `How the result of a measurement of a component of the
spin of a spin-$\frac{1}{2}$ particle can turn out to be 100'\,'',
{\em Phys. Rev. Lett.} {\bf 62}, 19, 2325 (1989).
Comment on {\bf [Aharonov-Albert-Vaidman 88]}.
Reply: {\bf [Aharonov-Vaidman 89]}.
See {\bf [Peres 89 a]}, {\bf [Duck-Stevenson-Sudarshan 89]}.

\item {\bf [Leggett 99 a]}:
A. J. Leggett,
``Some thought-experiments involving macrosystems as illustrations
of various interpretations of quantum mechanics'',
{\em Found. Phys.} {\bf 29}, 3, 445-456 (1999).

\item {\bf [Leggett 99 b]}:
A. J. Leggett,
``The Feynman processor: Quantum entanglement and the computing
revolution'',
{\em Phys. Today} {\bf 52}, 7, 51-52 (1999).
Review of {\bf [Milburn 98]}.

\item {\bf [Leggett 99 c]}:
A. J. Leggett,
``Quantum theory: Weird and wonderful'',
{\em Phys. World} {\bf 12}, 73-77 (1999).

\item {\bf [Leggett 03]}:
A. J. Leggett,
``Nonlocal hidden-variable theories and quantum mechanics: An incompatibility theorem'',
{\em Found. Phys.} {\bf 33}, 10, 1469-1493 (2003).

\item {\bf [Legre-Wegmuller-Gisin 03]}:
M. Legre, M. Wegmuller, \& N. Gisin,
``Quantum measurement of the degree of polarization of a light beam'',
quant-ph/0305019.

\item {\bf [Leibfried-Meekhof-Monroe-(+2) 97]}:
D. Leibfried, D. M. Meekhof, C. Monroe, B. E. King,
W. M. Itano, \& D. J. Wineland,
``Experimental preparation and measurement of quantum states of
motion of a trapped atom'',
{\em J. Mod. Opt.} {\bf 44}, 11-12 (Special issue: Quantum
state preparation and measurement), 2485-2505 (1997).

\item {\bf [Leibfried-Pfau-Monroe 98]}:
D. Leibfried, T. Pfau, \& C. Monroe,
``Shadows and mirrors: Reconstructing quantum states of atom motion'',
{\em Phys. Today} {\bf 51}, 3, 22-28 (1998).

\item {\bf [Leibfried 99]}:
D. Leibfried,
``Individual addressing and state readout of trapped ions
utilizing rf micromotion'',
{\em Phys. Rev. A} {\bf 60}, 5, R3335-R3338 (1999).

\item {\bf [Leibfried-DeMarco-Meyer-(+8) 02]}:
D. Leibfried, B. DeMarco, V. Meyer,
M. Rowe, A. Ben-Kish, J. Britton, W. M. Itano,
B. Jelenkovi\'{c}, C. Langer, T. Rosenband, \& D. J. Wineland,
``Trapped-ion quantum simulator: Experimental application to nonlinear
interferometers'',
{\em Phys. Rev. Lett.} {\bf 89}, 24, 247901 (2002).

\item {\bf [Leibfried-DeMarco-Meyer-(+8) 03]}:
D. Leibfried, B. DeMarco, V. Meyer,
D. Lucas, M. Barrett, J. Britton, W. M. Itano,
B. Jelenkovi\'{c}, C. Langer, T. Rosenband, \& D. J. Wineland,
``Experimental demonstration of a robust, high-fidelity geometric two
ion-qubit phase gate'',
{\em Nature} {\bf 422}, 6930, 412-415 (2003).

\item {\bf [Leibfried-DeMarco-Meyer-(+11) 03]}:
D. Leibfried, B. Demarco, V. Meyer,
M. Rowe, A. Ben-Kish, M. Barrett,
J. Britton, J. Hughes, W. M. Itano,
B. M. Jelenkovi\'{c}, C. Langer, D. Lucas,
T. Rosenband, \& D. J. Wineland,
``Quantum information with trapped ions at NIST'',
in M. Ferrero (ed.),
{\em Proc. of Quantum Information: Conceptual Foundations,
Developments and Perspectives (Oviedo, Spain, 2002)},
{\em J. Mod. Opt.} {\bf 50}, 6-7, 1115-1129 (2003).

\item {\bf [Leibfried-Barrett-Schaetz-(+6) 04]}:
D. Leibfried, M. D. Barrett, T. Schaetz,
J. Britton, J. Chiaverini, W. M. Itano,
J. D. Jost, C. Langer, \& D. J. Wineland,
``Toward Heisenberg-limited spectroscopy with multiparticle entangled states'',
{\em Science} {\bf 304}, ?, 1476-1478 (2004).

\item {\bf [Leifer-Henderson-Linden 03]}:
M. S. Leifer, L. Henderson, \& N. Linden,
``Optimal entanglement generation from quantum operations'',
{\em Phys. Rev. A} {\bf 67}, 1, 012306 (2003).

\item {\bf [Leifer-Linden-Winter 03]}:
M. S. Leifer, N. Linden, \& A. Winter,
``Measuring polynomial invariants of multi-party quantum states'',
quant-ph/0308008.

\item {\bf [Lenard 74]}:
A. Lenard,
``A remark on the Kochen-Specker theorem'',
in C. P. Enz, \& J. Mehra (eds.),
{\em Physical reality and mathematical description},
Reidel, Dordrecht, Holland, 1974, pp.~226-233.

\item {\bf [Le\'{o}n 03]}:
J. Le\'{o}n,
``A position and a time for the photon'',
quant-ph/0309049.

\item {\bf [Leonhardt-Vaccaro 95]}:
U. Leonhardt, \& J. A. Vaccaro,
``Bell correlations in phase space: Application
to quantum optics'',
{\em J. Mod. Opt.} {\bf 42}, 5, 939-943 (1995).

\item {\bf [Leonhardt-Neumaier 03]}:
U. Leonhardt, \& A. Neumaier,
``Explicit effective Hamiltonians for general linear quantum-optical
networks'',
{\em J. Opt. B: Quantum Semiclass. Opt.} {\bf 6}, L1-? (2003);
quant-ph/0306123.

\item {\bf [Lepore 89]}:
V. L. Lepore,
``New inequalities from local realism'',
{\em Found. Phys. Lett.} {\bf 2}, 1, 15-26 (1989).

\item {\bf [Lepore-Selleri 90]}:
V. L. Lepore, \& F. Selleri,
``Do performed optical tests disprove local realism?'',
{\em Found. Phys. Lett.} {\bf 3}, 3, 203-220 (1990).

\item {\bf [Lepore 93]}:
V. L. Lepore,
``How many Bell's inequalities?'',
in A. van der Merwe, \& F. Selleri (eds.),
{\em Bell's theorem and the foundations of modern physics.
Proc.\ of an international conference (Cesena, Italy, 1991)},
World Scientific, Singapore, 1993, pp.~338-341.

\item {\bf [Lerner 91]}:
P. B. Lerner,
``Comment on `Proposed neutron interferometry test of
Einstein's ``einweg'' assumption in the Bohr-Einstein controversy'\,'',
{\em Phys. Lett. A} {\bf 157}, 4-5, 309-310 (1991).
Comment on {\bf [Rauch-Vigier 90]}.
Reply: {\bf [Rauch-Vigier 91]}.

\item {\bf [Leskowitz-Ghaderi-Olsen-Mueller 03]}:
G. M. Leskowitz, N. Ghaderi, R. A. Olsen, \& L. J. Mueller,
``Three-qubit nuclear magnetic resonance quantum information processing with a
single-crystal solid'',
{\em J. Chem. Phys.} {\bf 119}, 1643-? (2003).

\item {\bf [Leslau 01]}:
B. Leslau,
``Attacks on symmetric quantum coin-tossing protocols'',
quant-ph/0104075.
Comment on {\bf [Mayers-Salvail-Chiba Kohno 99]}.

\item {\bf [Leung-Nielsen-Chuang-Yamamoto 97]}:
D. W. Leung, M. A. Nielsen, I. L. Chuang, \& Y. Yamamoto,
``Approximate quantum error correction can lead to better codes'',
{\em Phys. Rev. A} {\bf 56}, 4, 2567-2573 (1997);
quant-ph/9704002.

\item {\bf [Leung-Vandersypen-Zhou-(+4) 99]}:
D. W. Leung, L. Vandersypen, X. Zhou, M. Sherwood,
C. Yannoni, M. Kubinec, \& I. L. Chuang,
``Experimental realization of a two-bit phase damping quantum code'',
{\em Phys. Rev. A} {\bf 60}, 3, 1924-1943 (1999);
quant-ph/9811068.

\item {\bf [Leung-Chuang-Yamaguchi-Yamamoto 00]}:
D. W. Leung, I. L. Chuang, F. Yamaguchi, \& Y. Yamamoto,
``Efficient implementation of coupled logic gates for quantum computation'',
{\em Phys. Rev. A} {\bf 61}, 4, 042310 (2000).

\item {\bf [Leung 00 a]}:
D. W. Leung,
``Towards robust quantum computation'',
Ph.\ D. thesis, Stanford University, 2000;
cs.CC/0012017.

\item {\bf [Leung 00 b]}:
D. W. Leung,
``Quantum Vernam cipher'',
quant-ph/0012077.

\item {\bf [Leung-Smolin 01]}:
D. W. Leung, \& J. A. Smolin,
``More is not necessarily easier'',
quant-ph/0103158.
Comment on {\bf [Bandyopadhyay-Roychowdhury-Sen 01]}.

\item {\bf [Leung 01 a]}:
D. W. Leung,
``Simulation and reversal of $n$-qubit Hamiltonians using Hadamard
matrices'',
quant-ph/0107041.

\item {\bf [Leung 01 b]}:
D. W. Leung,
``Two-qubit projective measurements are universal for quantum computation'',
quant-ph/0111122.

\item {\bf [Leung 02]}:
D. W. Leung,
``Choi's proof and quantum process tomography'',
submitted to special issue of JMP on QIT;
quant-ph/0201119.

\item {\bf [Leung-Shor 03]}:
D. W. Leung, \& P. W. Shor,
``Oblivious remote state preparation'',
{\em Phys. Rev. Lett.} {\bf 90}, 12, 127905 (2003);
quant-ph/0201008.

\item {\bf [Leung 03]}:
D. W. Leung,
``Quantum computation by measurements'',
quant-ph/0310189.

\item {\bf [Leuenberger-Loss 01]}:
M. N. Leuenberger, \& D. Loss,
``Quantum computing in molecular magnets'',
{\em Nature} {\bf 410}, 6830, 789-793 (2001).

\item {\bf [Leuenberger-Loss-Poggio-Awschalom 02]}:
M. N. Leuenberger, D. Loss, M. Poggio, \& D. D. Awschalom,
``Quantum information processing with large nuclear spins in GaAs
semiconductors'',
{\em Phys. Rev. Lett.} {\bf 89}, 20, 207601 (2002).

\item {\bf [Lev 01]}:
F. M. Lev,
``An integral version of Shor's factoring algorithm'',
quant-ph/0109103.

\item {\bf [Levenson-Abram-Rivera-(+3) 93]}:
J. A. Levenson, I. Abram, T. Rivera, P. Fayolle,
J. C. Garreau, \& P. Grangier,
``Quantum optical cloning amplifier'',
{\em Phys. Rev. Lett.} {\bf 70}, 3, 267-270 (1993).

\item {\bf [Levi-Georgeot-Shepelyansky 03]}:
B. Levi, B. Georgeot, \& D. L. Shepelyansky,
``Quantum computing of quantum chaos in the kicked rotator model'',
{\em Phys. Rev. E} {\bf 67}, 4, 046220 (2003).

\item {\bf [Levi-Georgeot 04]}:
B. Levi, \& B. Georgeot,
``Quantum computation of a complex system: The kicked Harper model
quant-ph/0409028.

\item {\bf [Levin 02]}:
F. S. Levin,
{\em An introduction to quantum theory},
Cambridge University Press, Cambridge, 2002.
Review: {\bf [Cabello 02 g]}.

\item {\bf [Levin-Peleg-Peres 93]}:
O. Levin, Y. Peleg, \& A. Peres,
``Bell's detector of vacuum fluctuations'',
in A. van der Merwe, F. Selleri, \& G. Tarozzi (eds.),
{\em Bell's theorem and the foundations of modern physics},
World Scientific, Singapore, 1993, pp.~342-346.

\item {\bf [Levine-Muthukumar 04]}:
G. Levine, \& V. N. Muthukumar,
``Entanglement of a qubit with a single oscillator mode'',
{\em Phys. Rev. B} {\bf 69}, 11, 113203 (2004).

\item {\bf [Levitin 69]}:
L. B. Levitin,
``On quantum measure of information'',
in {\em Proc.\ of the 4th All-Union Conf.\ on Information
and Coding Theory (Tashkent, 1969)}, pp.~111-115.
English version: ``?'',
in R. Hudson, V. P. Belavkin, \& O. Hirota (eds.),
{\em Quantum communication and measurement},
Plenum Press, New York, 1996, pp.~?-?.

\item {\bf [Levitin 87]}:
L. B. Levitin,
``?'',
in A. Blaqui\`{e}re, S. Diner, \& G. Lochak (eds.),
{\em Information, complexity, and control in quantum physics},
Springer-Verlag, Vienna, 1987, pp.~111-115.

\item {\bf [Levitin 93]}:
L. B. Levitin,
``?'',
in D. Matzke (ed.),
{\em Workshop on Physics and Computation: PhysComp '92},
IEEE Computer Science Society Press,
Los Alamitos, California, 1993, pp.~?-?.

\item {\bf [Levitin 98]}:
L. B. Levitin,
``Energy requirements in quantum communication'',
{\em Int. J. Theor. Phys.} {\bf 37}, 1, 487-494 (1998).

\item {\bf [Levitin-Toffoli-Walton 01]}:
L. B. Levitin, T. Toffoli, \& Z. D. Walton,
``Information and distinguishability of ensembles of identical quantum states'',
{\em IQSA 2001};
quant-ph/0112075.

\item {\bf [Levitin-Toffoli 03]}:
L. B. Levitin, \& T. Toffoli,
``Information between quantum systems via POVMs'',
quant-ph/0306058.

\item {\bf [Levy 01 a]}:
J. Levy,
``Quantum-information processing with ferroelectrically coupled quantum dots'',
{\em Phys. Rev. A} {\bf 64}, 5, 052306 (2001).

\item {\bf [Levy 01 b]}:
J. Levy,
``Universal quantum computation with spin-1/2 pairs and Heisenberg
exchange'',
{\em Phys. Rev. Lett.} {\bf 89}, 14, 147902 (2002);
quant-ph/0101057.

\item {\bf [L\'{e}vy Leblond 86]}:
J. M. L\'{e}vy-Leblond,
``Th\'{e}orie quantique: un d\'{e}bat toujours actuel'',
{\em La Recherche} {\bf 17}, 175, 394-396 (1986).
Spanish version: ``La teor\'{\i}a cu\'{a}ntica: un debate siempre actual'',
{\em Mundo Cient\'{\i}fico} {\bf 6}, 58, 560-562 (1986).

\item {\bf [Lewenstein 94]}:
M. Lewenstein,
``Quantum perceptrons'',
in S. M. Barnett, A. K. Ekert, \& S. J. D. Phoenix (eds.)
{\em J. Mod. Opt.} {\bf 41}, 12
(Special issue: Quantum communication), 2491-2501 (1994).

\item {\bf [Lewenstein-Sanpera 98]}:
M. Lewenstein, \& A. Sanpera,
``Separability and entanglement of composite quantum systems'',
{\em Phys. Rev. Lett.} {\bf 80}, 11, 2261-2264 (1998);
quant-ph/9707043.
See {\bf [Shi-Du 01 a]}.

\item {\bf [Lewenstein-Rzazewski 00]}:
M. Lewenstein, \& K. Rzazewski,
``Quantum anti-Zeno effect'',
{\em Phys. Rev. A} {\bf 61}, 2, 022105 (2000);
quant-ph/9901060.

\item {\bf [Lewenstein-Kraus-Cirac-Horodecki 00]}:
M. Lewenstein, B. Kraus, J. I. Cirac, \& P. Horodecki,
``Optimization of entanglement witnesses'',
{\em Phys. Rev. A} {\bf 62}, 5, 052310 (2000);
quant-ph/0005014.

\item {\bf [Lewenstein-Kraus-Horodecki-Cirac 00]}:
M. Lewenstein, B. Kraus, P. Horodecki, \& J. I. Cirac,
``Characterization of separable states and entanglement
witnesses'',
quant-ph/0005112.

\item {\bf [Lewenstein-Bruss-Cirac-(+5) 00]}:
M. Lewenstein, D. Bru\ss, J. I. Cirac, B. Kraus, M. Ku\'{s},
J. Samsonowicz, A. Sanpera, \& R. Tarrach,
``Separability and distillability in composite quantum
systems --a primer--'',
{\em J. Mod. Opt.} {\bf 77}, ?, 2481-? (2000);
quant-ph/0006064.

\item {\bf [Lewis 73]}:
D. Lewis,
{\em Counterfactuals}, Blackwell, Oxford, 1973.

\item {\bf [Li-Song-Luo 02]}:
C. Li, H.-S. Song, \& Y.-X. Luo,
``Criterion for general quantum teleportation'',
{\em Phys. Lett. A} {\bf 297}, 3-4, 121-125 (2002).
Erratum: {\em Phys. Lett. A} {\bf 300}, 6, 697 (2002);
quant-ph/0111048.

\item {\bf [Li-Song-Zhou-Wu 03]}:
C. Li, H.-S. Song, L. Zhou, \& C.-F. Wu,
``A random quantum key distribution by using Bell states'',
{\em J. Opt. B: Quantum Semiclass. Opt.};
quant-ph/0302193.

\item {\bf [Li 03]}:
C. Li,
``Four-photon, five-dimensional entanglement for quantum communication'',
{\em Phys. Lett. A} {\bf 313}, 5-6, 389-392 (2003).

\item {\bf [Li-Zhang-Huang-Guo 00]}:
C.-F. Li, Y.-S. Zhang, Y.-F. Huang, \& G.-C. Guo,
``Quantum Monty Hall problem'',
quant-ph/0007120.

\item {\bf [Li-Zhang-Huang-Guo 01]}:
C.-F. Li, Y.-S. Zhang, Y.-F. Huang, \& G.-C. Guo,
``Quantum strategies of quantum measurements'',
{\em Phys. Lett. A} {\bf 280}, 5-6, 257-260 (2001).

\item {\bf [Li-Guo 01]}:
C.-F. Li, \& G.-C. Guo,
``The existence of unconditionally secure
quantum bit commitment'',
quant-ph/0106002.

\item {\bf [Li-Hwang-Hsieh-Wang 02]}:
C.-M. Li, C.-C. Hwang, J.-Y. Hsieh, \& K.-S. Wang,
``General phase-matching condition for a quantum searching algorithm'',
{\em Phys. Rev. A} {\bf 65}, 3, 034305 (2002).

\item {\bf [Li-Li 01]}:
D. Li \& X. Li,
``More general quantum search algorithm
$Q=-I_{\gamma}VI_{\tau}U$ and the precise formula for the amplitude
and the non-symmetric effects of different rotating angles'',
{\em Phys. Lett. A} {\bf 287} 5-6, 304-316 (2001).

\item {\bf [Li-Li-Li 02]}:
D. Li, X. Li, H. Huang, \& X. Li,
``Invariants of Grover's algorithm and the rotation in space'',
{\em Phys. Rev. A} {\bf 66}, 4, 044304 (2002).

\item {\bf [Li-Li-Zhang-Zhu 02]}:
F.-L. Li, H.-R. Li, J.-X. Zhang, \& S.-Y. Zhu,
``Nonclassical properties of teleported optical fields in quantum
teleportation of continuous variables'',
{\em Phys. Rev. A} {\bf 66}, 2, 024302 (2002).

\item {\bf [Li-Yang-Allaart-Lenstra 04]}:
G.-X. Li, Y.-P. Yang, K. Allaart, \& D. Lenstra,
``Entanglement for excitons in two quantum dots in a cavity injected with squeezed vacuum'',
{\em Phys. Rev. A} {\bf 69}, 1, 014301 (2004).

\item {\bf [Li-Allaart-Lenstra 04]}:
G.-X. Li, K. Allaart, \& D. Lenstra,
``Entanglement between two atoms in an overdamped cavity injected
with squeezed vacuum'',
{\em Phys. Rev. A} {\bf 69}, 5, 055802 (2004).

\item {\bf [Li-Du-Massar 02]}:
H. Li, J. Du, \& S. Massar,
``Continuous-variable quantum games'',
{\em Phys. Lett. A} {\bf 306}, 2-3, 73-78 (2002);
quant-ph/0212122.

\item {\bf [Li-Du 03]}:
H. Li, \& J. Du,
``Relativistic invariant quantum entanglement between the spins of moving
bodies'',
{\em Phys. Rev. A} {\bf 68}, 2, 022108 (2003).

\item {\bf [Li-Du 04]}:
H. Li, \& J. Du,
``Spatial localization and relativistic transformation of quantum spins'',
{\em Phys. Rev. A} {\bf 70}, 1, 012111 (2004).

\item {\bf [Li-Guo 03]}:
J. Li, \& G.-C. Guo,
``Nonlocality proof without inequalities for $N$-qubit $W$ and
Greenberger–Horne–Zeilinger states'',
{\em Phys. Lett. A} {\bf 313}, 5-6, 397-400 (2003).

\item {\bf [Li-Guo 99]}:
L. Li, \& G.-C. Guo,
``Quantum logic gate operation between different ions in a trap'',
{\em Phys. Rev. A} {\bf 60}, 1, 696-699 (1999).

\item {\bf [Li-Xu 03 a]}:
S.-B. Li, \& J.-B. Xu,
``Quantum probabilistic teleportation via entangled coherent states'',
{\em Phys. Lett. A} {\bf 309}, 5-6, 321-328 (2003);
quant-ph/0312100.

\item {\bf [Li-Xu 03 b]}:
S.-B. Li, \& J.-B. Xu,
``Entanglement in quantum Heisenberg $XY$ chain with phase decoherence'',
{\em Phys. Lett. A} {\bf 311}, 4-5, 313-318 (2003).

\item {\bf [Li-Xu 03]}:
S.-B. Li, \& J.-B. Xu,
``Stationary state entanglement of interacting system of the qubit and thermal field with phase decoherence'',
{\em Phys. Lett. A} {\bf 313}, 3, 175-181 (2003).

\item {\bf [Li-Wu-Wang-Xu 04]}:
S.-B. Li, R.-K. Wu, Q.-M. Wang, \& J.-B. Xu,
``Entanglement of pair cat states and teleportation'',
{\em Phys. Lett. A} {\bf 325}, 3-4, 206-217 (2004).

\item {\bf [Li-Li-Guo 00]}:
W.-L. Li, C.-F. Li, \& G.-C. Guo,
``Probabilistic teleportation and entanglement matching'',
{\em Phys. Rev. A} {\bf 61}, 3, 034301 (2000);
quant-ph/9910021.

\item {\bf [Li-Li-Huang-(+3) 00]}:
W.-L. Li, C.-F. Li, Y.-F. Huang,
Y.-S. Zhang, Y. Jiang, \& G.-C. Guo,
``Optical realization of universal quantum cloning'',
quant-ph/0006032.

\item {\bf [Li-Pan-Jing-(+3) 02]}:
X. Li, Q. Pan, J. Jing,
J. Zhang, C. Xie, \& K. Peng,
``Quantum dense coding exploiting a bright Einstein-Podolsky-Rosen beam'',
{\em Phys. Rev. Lett.} {\bf 88}, 4, 047904 (2002);
quant-ph/0107068.

\item {\bf [Li-Wu-Steel-(+6) 02]}:
X. Li, Y. Wu, D. Steel,
D. Gammon, T. H. Stievater, D. S. Katzer,
D. Park, C. Piermarocchi, \& L. J. Sham,
``An all-optical quantum gate in a semiconductor quantum dot'',
{\em Science} {\bf 301}, 809-811 (2003).

\item {\bf [Li-Arakawa 01]}:
X.-Q. Li, \& Arakawa,
``Single qubit from two coupled quantum dots:
An approach to semiconductor quantum computations'',
{\em Phys. Rev. A} {\bf 63}, 1, 012302 (2001).

\item {\bf [Li-Yan 02 a]}:
X.-Q. Li, \& Y. Yan,
``Quantum computation with coupled quantum dots embedded in optical
microcavities'',
{\em Phys. Rev. B} {\bf 65}, 20, 205301 (2002);
quant-ph/0203145.

\item {\bf [Li-Yan 02 b]}:
X.-Q. Li, \& Y. Yan,
``Highly coherent solid-state quantum bit from a pair of quantum dots'',
{\em Appl. Phys. Lett.} {\bf 81}, ?, 168-? (2002);
quant-ph/0204027.

\item {\bf [Li-Cen-Huang-(+2) 02]}:
X.-Q. Li, L.-X. Cen, G.-X. Huang, L. Ma, \& Y. Yan,
``Nonadiabatic geometric quantum computation with trapped ions'',
{\em Phys. Rev. A} {\bf 66}, 4, 042320 (2002);
quant-ph/0204028.

\item {\bf [Li-Voss-Sharping-Kumar 04]}:
X. Li, P. L. Voss, J. E. Sharping, \& P. Kumar,
``Optical-fiber source of polarization-entangled photon pairs in the
1550\,nm telecom band'',
quant-ph/0402191.

\item {\bf [Li-Kobayashi 04 a]}:
Y. Li, \& T. Kobayashi,
``Four-photon entanglement from two-crystal geometry'',
{\em Phys. Rev. A} {\bf 69}, 2, 020302 (2004).

\item {\bf [Li-Kobayashi 04 b]}:
Y. Li, \& T. Kobayashi,
``Four-photon $W$ state using two-crystal geometry parametric down-conversion'',
{\em Phys. Rev. A} {\bf 70}, 1, 014301 (2004).

\item {\bf [Li-Zeng-Liu-Long 01]}:
Y. S. Li, B. Zeng, X. S. Liu, \& G. L. Long,
``Entanglement in a two-identical-particle system'',
{\em Phys. Rev. A} {\bf 64}, 5, 054302 (2001);
quant-ph/0104101.

\item {\bf [Li-Zhang-Peng 04]}:
Y. Li, K. Zhang, \& K. Peng,
``Multiparty secret sharing of quantum information based on entanglement swapping'',
{\em Phys. Lett. A} {\bf 324}, 5-6, 420-424 (2004).

\item {\bf [Li 01]}:
Z.-Y. Li,
``Local realistic model for two-particle Einstein-Podolsky-Rosen pairs'',
quant-ph/0108134.

\item {\bf [Li 01]}:
Z.-Y. Li,
``Atom interferometers: Beyond complementarity principles'',
quant-ph/0109023.

\item {\bf [Liang-Chen-Li-Huang 01]}:
L.-M. Liang, P.-X. Chen, C.-Z. Li, \& M.-Q. Huang,
``Purification of multipartite entangled mixed states under local manipulation'',
{\em Phys. Lett. A} {\bf 294}, 2, 71-73 (2002).

\item {\bf [Liang-Li 03 a]}:
L.-M. Liang, \& C.-Z. Li,
``Distillation of a $m$-GHZ state from an arbitrary pure states of
$m$ qubits under one successful branch protocol'',
{\em Phys. Lett. A} {\bf 308}, 5-6, 343-348 (2003).

\item {\bf [Liang-Li 03 b]}:
L.-M. Liang, \& C.-Z. Li,
``Bell's theorems without inequalities for continuous variable systems and even-dimensional systems'',
{\em Phys. Lett. A} {\bf 317}, 3-4, 206-209 (2003).

\item {\bf [Liang-Li 03 c]}:
L.-M. Liang, \& C.-Z. Li,
``Bell's theorem without inequalities for one class of two-qubit mixed states'',
{\em Phys. Lett. A} {\bf 318}, 4-5, 300-302 (2003).

\item {\bf [Liang-Kaszlikowski-Englert-(+3) 03]}:
Y. C. Liang, D. Kaszlikowski, B.-G. Englert,
L. C. Kwek, \& C. H. Oh,
``Tomographic quantum cryptography'',
{\em Phys. Rev. A} {\bf 68}, 2, 022324 (2003);
quant-ph/0305018.

\item {\bf [Liang-Xiong 04]}:
X.-T. Liang, Y.-J. Xiong,
``Short-time decoherence of Josephson charge qubit nonlinearly coupling
with its environment'',
{\em Phys. Lett. A} {\bf 332}, ?, 8-? (2004);
quant-ph/0409102.

\item {\bf [Lidar-Biham 97]}:
D. A. Lidar, \& O. Biham,
``Simulating Ising spin glasses on a quantum computer'',
{\em Phys. Rev. E} {\bf 56}, 3, 3661-3681 (1997);
quant-ph/9611038.

\item {\bf [Lidar-Wang 98]}:
D. A. Lidar, \& H. Wang,
``Calculating the thermal rate constant with exponential speed-up on
a quantum computer'',
{\em Phys. Rev. E} {\bf 59}, ?, 2429-? (1999);
quant-ph/9807009.

\item {\bf [Lidar-Chuang-Whaley 98]}:
D. A. Lidar, I. L. Chuang, \& K. B. Whaley,
``Decoherence-free subspaces for quantum computation'',
{\em Phys. Rev. Lett.} {\bf 81}, 12, 2594-2597 (1998).

\item {\bf [Lidar-Bacon-Whaley 99]}:
D. A. Lidar, D. Bacon, \& K. B. Whaley,
``Concatenating decoherence-free subspaces with quantum error correcting codes'',
{\em Phys. Rev. Lett.} {\bf 82}, 22, 4556-4559 (1999);
quant-ph/9809081.

\item {\bf [Lidar-Bacon-Kempe-Whaley 00]}:
D. A. Lidar, D. Bacon, J. Kempe, \& K. B. Whaley,
``Protecting quantum information encoded in
decoherence-free states against exchange errors'',
{\em Phys. Rev. A} {\bf 61}, 5, 052307 (2000).

\item {\bf [Lidar-Bacon-Kempe-Whaley 01 a]}:
D. A. Lidar, D. Bacon, J. Kempe, \& K. B. Whaley,
``Decoherence-free subspaces for multiple-qubit errors.
I. Characterization'',
{\em Phys. Rev. A} {\bf 63}, 2, 022306 (2001);
quant-ph/9908064.
See {\bf [Lidar-Bacon-Kempe-Whaley 01 b]} (II).

\item {\bf [Lidar-Bacon-Kempe-Whaley 01 b]}:
D. A. Lidar, D. Bacon, J. Kempe, \& K. B. Whaley,
``Decoherence-free subspaces for multiple-qubit errors.
II. Universal, fault-tolerant quantum computation'',
{\em Phys. Rev. A} {\bf 63}, 2, 022307 (2001);
quant-ph/0007013.
See {\bf [Lidar-Bacon-Kempe-Whaley 01 a]} (I).

\item {\bf [Lidar-Bihary-Whaley 01]}:
D. A. Lidar, Z. Bihary, \& K. B. Whaley,
``From completely positive maps to the quantum Markovian semigroup
master equation'',
{\em Chem. Phys.} {\bf 268}, ?, 35-? (2001);
cond-mat/0011204.

\item {\bf [Lidar-Wu 02 a]}:
D. A. Lidar, \& L.-A. Wu,
``Reducing constraints on quantum computer
design by encoded selective recoupling'',
{\em Phys. Rev. Lett.} {\bf 88}, 1, 017905 (2002);
quant-ph/0109021.

\item {\bf [Lidar 02]}:
D. A. Lidar,
`Comment on ``Quantum waveguide array generator for performing Fourier
transforms: Alternate route to quantum computing''
[Appl. Phys. Lett. {\bf 79}, 2823 (2001)]',
{\em Appl. Phys. Lett.} {\bf 80}, ?, 2419-? (2002);
quant-ph/0308124.
Comment to {\bf [Akis-Ferry 01]}.
Reply: {\bf [Akis-Ferry 02]}.

\item {\bf [Lidar-Wu 02 b]}:
D. A. Lidar, \& L.-A. Wu,
``Encoded recoupling and decoupling: An alternative to quantum error
correcting codes, applied to trapped ion quantum computation'',
quant-ph/0211088.

\item {\bf [Lidar-Whaley 03]}:
D. A. Lidar, \& K. B. Whaley,
``Decoherence-free subspaces and subsystems'',
to be published as a book chapter;
quant-ph/0301032.

\item {\bf [Lidar-Wu 03 a]}:
D. A. Lidar, \& L.-A. Wu,
``Encoded recoupling and decoupling: An alternative to quantum
error-correcting codes applied to trapped-ion quantum computation'',
{\em Phys. Rev. A} {\bf 67}, 3, 032313 (2003).

\item {\bf [Lidar-Wu 03 b]}:
D. A. Lidar, \& L.-A. Wu,
``Quantum computers and decoherence: Exorcising the demon from the machine'',
{\em Proc.\ SPIE Conf.\ Fluctuations and Noise (Santa Fe, New Mexico, 2003)};
quant-ph/0302198.

\item {\bf [Lidar 03]}:
D. A. Lidar,
`Comment on ``Conservative quantum computing''\,',
{\em Phys. Rev. Lett.} {\bf 91}, 8, 089801 (2003);
quant-ph/0308123.
Comment on {\bf [Ozawa 02 c]}.
Reply: {\bf [Ozawa 03 b]}.

\item {\bf [Lidar-Thywissen 04]}:
D. A. Lidar, \& J. H. Thywissen,
``Exponentially localized magnetic fields for single-spin quantum logic gates'',
{\em J. App. Phys.} {\bf 96}, ?, 754-758 (2004);
cond-mat/0310352.

\item {\bf [Lidar-Schneider 04]}:
D. A. Lidar, \& S. Schneider,
``Stabilizing qubit coherence via tracking-control'',
quant-ph/0410048.

\item {\bf [Lim-Abd Shukor-Kwek 98]}:
S. C. Lim, R. Abd-Shukor, \& K. W. Kwek (eds.),
{\em Frontiers in quantum physics (Kuala Lumpur, Malaysia, 1997)},
Springer-Verlag, Singapore, 1998.

\item {\bf [Lim-Beige 03]}:
Y. L. Lim, \& A. Beige,
``Photon polarisation entanglement from distant dipole sources'',
{\em J. Phys. A};
quant-ph/0308095.

\item {\bf [Lim-Beige 04 a]}:
Y. L. Lim, \& A. Beige,
``Push button generation of multiphoton entanglement'',
{\em Quantum Information and Computation II (Orlando, Florida, 2004)}'',
quant-ph/0403125.

\item {\bf [Lim-Beige 04 b]}:
Y. L. Lim, \& A. Beige,
``An efficient quantum filter for multiphoton states'',
{\em J. Mod. Opt.};
quant-ph/0406008.

\item {\bf [Lim-Beige 04 c]}:
Y. L. Lim, \& A. Beige,
``Postselected multiphoton entanglement through Bell-multiport beam
splitters'',
quant-ph/0406047.

\item {\bf [Lim-Beige-Kwek 04]}:
Y. L. Lim, A. Beige, \& L. C. Kwek,
``Quantum computing with distant single photon sources with insurance'',
quant-ph/0408043.

\item {\bf [LiMing-Tang-Liao 04]}:
W. LiMing, Z. L. Tang, \& C. J. Liao,
``Representation of the SO(3) group by a maximally entangled state'',
{\em Phys. Rev. A} {\bf 69}, 6, 064301 (2004).

\item {\bf [Lin-Zhou-Xue-(+2) 03]}:
X.-M. Lin, Z.-W. Zhou, P. Xue, Y.-J. Gu, \& G.-C. Guo,
``Scheme for implementing quantum dense coding via cavity QED'',
{\em Phys. Lett. A} {\bf 313}, 5-6, 351-355 (2003).

\item {\bf [Linden-Popescu 98 a]}:
N. Linden, \& S. Popescu,
``On multi-particle entanglement'',
{\em Fortschr. Phys.} {\bf 46}, 4-5, 567-578 (1998);
quant-ph/9711016.

\item {\bf [Linden 98]}:
N. Linden,
``Information-entropy and purity of decoherence functions'',
{\em Int. J. Theor. Phys.} {\bf 37}, 2, 685-690 (1998).

\item {\bf [Linden-Massar-Popescu 98]}:
N. Linden, S. Massar, \& S. Popescu,
``Purifying noisy entanglement requires collective measurements'',
{\em Phys. Rev. Lett.} {\bf 81}, 15, 3279-3282 (1998);
quant-ph/9805001.

\item {\bf [Linden-Popescu 98 b]}:
N. Linden, \& S. Popescu,
``The halting problem for quantum computers'',
quant-ph/9806054.

\item {\bf [Linden-Barjat-Fremann 98]}:
N. Linden, H. Barjat, \& R. Freeman,
``An implementation of the Deutsch-Jozsa algorithm on a three-qubit
NMR quantum computer'',
{\em Chem. Phys. Lett.} {\bf 296}, 61-67 (1998);
quant-ph/9808039.
Reprinted in {\bf [Macchiavello-Palma-Zeilinger 00]}, pp.~486-492.

\item {\bf [Linden-Barjat-Carbajo-Freeman 98]}:
N. Linden, H. Barjat, R. J. Carbajo, \& R. Freeman,
``Pulse sequences for NMR quantum computers: How to manipulate nuclear
spins while freezing the motion of coupled neighbours'',
quant-ph/9811043.

\item {\bf [Linden-Popescu 99 a]}:
N. Linden, \& S. Popescu,
``Bound entanglement and teleportation'',
{\em Phys. Rev. A} {\bf 59}, 1, 137-140 (1999);
quant-ph/9807069.

\item {\bf [Linden-Popescu-Sudbery 99]}:
N. Linden, S. Popescu, \& A. Sudbery,
``Nonlocal parameters for multiparticle density matrices'',
{\em Phys. Rev. Lett.} {\bf 83}, 2, 243-247 (1999);
quant-ph/9801076.

\item {\bf [Linden-Kupce-Freeman 99]}:
N. Linden, E. Kupce, \& R. Freeman,
``NMR quantum logic gates for homonuclear spin systems'',
quant-ph/9907003.

\item {\bf [Linden-Popescu-Schumacher-Westmoreland 99]}:
N. Linden, S. Popescu, B. W. Schumacher, \& M. Westmoreland,
``Reversibility of local transformations of multiparticle
entanglement'',
quant-ph/9912039.

\item {\bf [Linden-Popescu 01]}:
N. Linden, \& S. Popescu,
``Good dynamics versus bad kinematics:
Is entanglement needed for quantum computation?'',
{\em Phys. Rev. Lett.} {\bf 87}, 4, 047901 (2001);
quant-ph/9906008.

\item {\bf [Linden-Popescu-Wootters 02]}:
N. Linden, S. Popescu, \& W. K. Wootters,
``Almost every pure state of three qubits is completely determined by its
two-particle reduced density matrices'',
{\em Phys. Rev. Lett.} {\bf 89}, 20, 207901 (2002);
quant-ph/0207109.

\item {\bf [Linden-Wootters 02]}:
N. Linden, \& W. K. Wootters,
``The parts determine the whole in a generic pure quantum state'',
{\em Phys. Rev. Lett.} {\bf 89}, 27, 277906 (2002);
quant-ph/0208093.

\item {\bf [Lindley 96]}:
D. Lindley,
{\em Whore does the weirdness go??
Why quantum mechanics is strange, but not as strange as you think},
Basic Books, New York, 1996.

\item {\bf [Lindner-Peres-Terno 03]}:
N. H. Lindner, A. Peres, \& D. R. Terno,
``Elliptic Rydberg states as direction indicators'',
{\em Phys. Rev. A} {\bf 68}, 4, 042308 (2003);
quant-ph/0305171.

\item {\bf [Lindner-Terno 04]}:
N. H. Lindner, \& D. R. Terno,
``The effect of focusing on polarization qubits'',
quant-ph/0403029.

\item {\bf [Lipkin 68]}:
H. J. Lipkin,
``$CP$ violation and coherent decays of kaon pairs'',
{\em Phys. Rev.} {\bf 176}, 5, 1715-1718 (1968).

\item {\bf [Liu-Lu-Draayer 04]}:
D. Liu, G. Lu, \& J. P. Draayer,
``A simple entanglement measure for multipartite pure states'',
{\em Int. J. Theor. Phys.},
quant-ph/0405133.

\item {\bf [Liu-Wang 03]}:
J.-M. Liu, \& Y.-Z. Wang,
``Remote preparation of a two-particle entangled state'',
{\em Phys. Lett. A} {\bf 316}, 3-4, 159-167 (2003).

\item {\bf [Liu-Chen 03]}:
R.-F. Liu, \& C.-C. Chen,
``Generation and evolution of spin entanglement in nonrelativistic QED'',
{\em Phys. Rev. A} {\bf 68}, 4, 044302 (2003);
quant-ph/0305014.

\item {\bf [Liu-Sun 02]}:
X. F. Liu, \& C. P. Sun,
``Generalized quantum games with Nash equilibrium'',
quant-ph/0212045.

\item {\bf [Liu-Jing-Zhou-Ge 04]}:
X.-J. Liu, H. Jing, X.-T. Zhou, \& M.-L. Ge,
``Technique of quantum-state transfer for a double- atomic beam'',
{\em Phys. Rev. A} {\bf 70}, 1, 015603 (2004).

\item {\bf [Liu-Long-Tong-Li 02]}:
X. S. Liu, G. L. Long, D. M. Tong, \& F. Li,
``General scheme for superdense coding between multiparties'',
{\em Phys. Rev. A} {\bf 65}, 2, 022304 (2002);
quant-ph/0110112.

\item {\bf [Liu-\"{O}zdemir-Koashi-Imoto 02]}:
Y.-X. Liu, S. K. \"{O}zdemir, M. Koashi, \& N. Imoto,
``Dynamics of entanglement for coherent excitonic states in a system of two
coupled quantum dots and cavity QED'',
{\em Phys. Rev. A} {\bf 65}, 4, 042326 (2002).
Erratum: {\em Phys. Rev. A} {\bf 67}, 1, 019902 (2003).

\item {\bf [Liu-Miranowicz-Koashi-Imoto 02]}:
Y.-X. Liu, A. Miranowicz, M. Koashi, \& N. Imoto,
``Realization of symmetric sharing of entanglement in semiconductor
microcrystallites coupled by a cavity field'',
{\em Phys. Rev. A} {\bf 66}, 6, 062309 (2002);
quant-ph/0212150.

\item {\bf [Liu-Miranowicz-\"{O}zdemir-(+2) 02]}:
Y.-X. Liu, A. Miranowicz, S. K. \"{O}zdemir, M. Koashi, \& N. Imoto,
``Size-dependent decoherence of excitonic states in semiconductor
microcrystallites'',
{\em Phys. Rev. A} {\bf 67}, 3, 034303 (2003);
quant-ph/0304021.

\item {\bf [Liu-\"{O}zdemir-Miranowicz-(+2) 04]}:
Y.-X. Liu, S. K. \"{O}zdemir, A. Miranowicz,
M. Koashi, \& N. Imoto,
``Exciton entanglement in two coupled semiconductor microcrystallites'',
quant-ph/0401182.

\item {\bf [Ljunggren-Bourennane-Karlsson 00]}:
D. Ljunggren, M. Bourennane, \& A. Karlsson,
``Authority-based user authentication in quantum key distribution'',
{\em Phys. Rev. A} {\bf 62}, 2, 022305 (2000).

\item {\bf [Lloyd 90]}:
S. Lloyd,
``Valuable information'',
in {\bf [Zurek 90]}, pp.~193-198.

\item {\bf [Lloyd 92]}:
S. Lloyd,
``Any nonlinear gate, with linear gats, suffices for computation'',
{\em Phys. Lett. A} {\bf 167}, 3, 255-260 (1992).

\item {\bf [Lloyd 93]}:
S. Lloyd,
``A potentially realizable quantum computer'',
{\em Science} {\bf 261}, 5128, 1569-1571 (1993).
Extended version: quant-ph/9912086.

\item {\bf [Lloyd 94 a]}:
S. Lloyd,
``Envisioning a quantum supercomputer'',
{\em Science} {\bf 263}, 5147, 695 (1994).

\item {\bf [Lloyd 94 b]}:
S. Lloyd,
``Necessary and sufficient conditions for quantum computation'',
in S. M. Barnett, A. K. Ekert, \& S. J. D. Phoenix (eds.),
{\em J. Mod. Opt.} {\bf 41}, 12 (Special issue: Quantum
communication), 2503-2520 (1994).

\item {\bf [Lloyd 95 a]}:
S. Lloyd,
``Almost any quantum logic gate is universal'',
{\em Phys. Rev. Lett.} {\bf 75}, 2, 346-349 (1995).

\item {\bf [Lloyd 95 b]}:
S. Lloyd,
``Quantum-mechanical computers'',
{\em Sci. Am.} {\bf 273}, 4, 44-50.
Spanish version: ``Computaci\'{o}n mec\'{a}nico-cu\'{a}ntica'',
{\em Investigaci\'{o}n y Ciencia} 231, 20-26 (1995).
Reprinted in {\bf [Cabello 97 c]}, pp.~98-104.

\item {\bf [Lloyd 96]}:
S. Lloyd,
``Universal quantum simulators'',
{\em Science} {\bf 273}, 5278, 1073-1078 (1996).
Correction: {\em Science} {\bf 279}, 5354, 1113-1117 (1998).

\item {\bf [Lloyd 97 a]}:
S. Lloyd,
``Capacity of the noisy quantum channel'',
{\em Phys. Rev. A} {\bf 55}, 3, 1613-1622 (1997);
quant-ph/9604015,

\item {\bf [Lloyd 97 b]}:
S. Lloyd,
``A Greenberger-Horne-Zeilinger experiment for mixed states'',
quant-ph/9704013.

\item {\bf [Lloyd 97 c]}:
S. Lloyd,
``Universe as quantum computer'',
{\em Complexity} {\bf 3}, 1, 32-35 (1997);
quant-ph/9912088.

\item {\bf [Lloyd 98 a]}:
S. Lloyd,
``Microscopic analogs of the Greenberger-Horne-Zeilinger experiment'',
{\em Phys. Rev. A} {\bf 57}, 3, R1473-R1476 (1998).

\item {\bf [Lloyd 98 b]}:
S. Lloyd,
``Unconventional quantum computing devices'',
in C. S. Calude, J. Casti, \& M. J. Dinneen (eds.),
{\em Unconventional models of computation},
Springer-Verlag, Singapore, 1998;
quant-ph/0003151.

\item {\bf [Lloyd-Slotine 98]}:
S. Lloyd, \& J.-J. E. Slotine,
``Analog quantum error correction'',
{\em Phys. Rev. Lett.} {\bf 80}, 18, 4088-4091 (1998);
quant-ph/9711021.

\item {\bf [Lloyd-Braunstein 99]}:
S. Lloyd, \& S. L. Braunstein,
``Quantum computation over continuous variables'',
{\em Phys. Rev. Lett.} {\bf 82}, 8, 1784-1787 (1999);
quant-ph/9810082.

\item {\bf [Lloyd 99]}:
S. Lloyd,
``Ultimate physical limits to computation'',
quant-ph/9908043.

\item {\bf [Lloyd-Rahn-Ahn 99]}:
S. Lloyd, B. Rahn, \& C. Ahn,
``Robust quantum computation by simulation'',
quant-ph/9912040.

\item {\bf [Lloyd-Slotine 00]}:
S. Lloyd, \& J.-J. E. Slotine,
``Quantum feedback with weak measurements'',
{\em Phys. Rev. A} {\bf 62}, 1, 012307 (2000);
quant-ph/9905064.

\item {\bf [Lloyd 00 a]}:
S. Lloyd,
``Quantum search without entanglement'',
{\em Phys. Rev. A} {\bf 61}, 1, 010301(R) (2000).

\item {\bf [Lloyd 00 b]}:
S. Lloyd,
``Coherent quantum feedback'',
{\em Phys. Rev. A} {\bf 62}, 2, 022108 (2000).

\item {\bf [Lloyd 00 c]}:
S. Lloyd,
``Quantum computation with abelian anyons'',
quant-ph/0004010.

\item {\bf [Lloyd-Shahriar-Hemmer 00]}:
S. Lloyd, M. S. Shahriar, \& P. R. Hemmer,
``Teleportation and the quantum internet'',
quant-ph/0003147.

\item {\bf [Lloyd 00 d]}:
S. Lloyd,
``Hybrid quantum computing'',
quant-ph/0008057.

\item {\bf [Lloyd-Viola 00]}:
S. Lloyd, \& L. Viola,
``Control of open quantum systems dynamics'',
quant-ph/0008101.

\item {\bf [Lloyd-Shahriar-Shapiro-Hemmer 01]}:
S. Lloyd, M. S. Shahriar, J. H. Shapiro, \& P. R. Hemmer,
``Long distance, unconditional teleportation of atomic states
via complete Bell state measurements'',
{\em Phys. Rev. Lett.} {\bf 87}, 16, 167903 (2001).

\item {\bf [Lloyd-Shapiro-Wong 02]}:
S. Lloyd, J. H. Shapiro, \& F. N. C. Wong,
``Quantum magic bullets by means of entanglement'',
{\em J. Opt. Soc. Amer. B Opt. Phys.} {\bf 19}, 2, 312-318 (2002);
quant-ph/0009115.

\item {\bf [Lloyd-Viola 02]}:
S. Lloyd, \& L. Viola,
``Engineering quantum dynamics'',
{\em Phys. Rev. A} {\bf 65}, 1, 010101 (2002).

\item {\bf [Lloyd 02 a]}:
S. Lloyd,
``Quantum computation and quantum information'',
{\em Nature} {\bf 416}, 6876, 18-19 (2002).
Review of {\bf [Nielsen-Chuang 00]}.

\item {\bf [Lloyd 02 b]}:
S. Lloyd,
``Computational capacity of the universe'',
{\em Phys. Rev. Lett.} {\bf 88}, 23, 237901 (2002);
quant-ph/0110141.

\item {\bf [Lloyd-Landahl-Slotine 03]}:
S. Lloyd, A, J. Landahl, \& J.-J. E. Slotine,
``Universal quantum interfaces'',
{\em Phys. Rev. A} {\bf 69}, 1, 012305 (2004);
quant-ph/0303048.

\item {\bf [Lloyd 03]}:
S. Lloyd,
``Power of entanglement in quantum communication'',
{\em Phys. Rev. Lett.} {\bf 90}, 16, 167902 (2003);
quant-ph/0112034.

\item {\bf [Lloyd 04]}:
S. Lloyd,
``Almost certain escape from black holes'',
quant-ph/0406205.

\item {\bf [Lo-Shimony 81]}:
T. K. Lo, \& A. Shimony,
``Proposed molecular test of local hidden-variables theories'',
{\em Phys. Rev. A} {\bf 23}, 6, 3003-3012 (1981).
Comment: {\bf [Santos 84 b]}.
See {\bf [Shimony 84 b]}.

\item {\bf [Lo 95]}:
H.-K. Lo,
``Quantum coding theorem for mixed states'',
quant-ph/9504004.

\item {\bf [Lo-Chau 95]}:
H.-K. Lo, \& H. F. Chau,
``Quantum cryptography in noisy channels'',
quant-ph/9511025.

\item {\bf [Lo-Chau 96]}:
H.-K. Lo, \& H. F. Chau,
``Why quantum bit commitment
and ideal quantum coin tossing are impossible'',
quant-ph/9605026.
Enlarged version: {\bf [Lo-Chau 98 a]}.

\item {\bf [Lo 97]}:
H.-K. Lo,
``Insecurity of quantum secure computations'',
{\em Phys. Rev. A} {\bf 56}, 2, 1154-1162 (1997).

\item {\bf [Lo-Chau 97]}:
H.-K. Lo, \& H. F. Chau,
``Is quantum bit commitment really possible?'',
{\em Phys. Rev. Lett.} {\bf 78}, 17, 3410-3413 (1997).

\item {\bf [Lo-Chau 98 a]}:
H.-K. Lo, \& H. F. Chau,
``Why quantum bit commitment and ideal quantum coin tossing are impossible'',
{\em Physica D} {\bf 120}, 177-187 (1998);
quant-ph/9711065. Enlarged version of {\bf [Lo-Chau 96]}.

\item {\bf [Lo 98]}:
H.-K. Lo,
``Quantum cryptology'', in
{\bf [Lo-Spiller-Popescu 98]}, pp.~76-119.

\item {\bf [Lo-Spiller-Popescu 98]}:
H.-K. Lo, S. Popescu, \& T. Spiller (eds.),
{\em Introduction to quantum computation and information},
World Scientific, Singapore, 1998.
Review: {\bf [DiVincenzo 99 b]}.

\item {\bf [Lo-Chau 98 b]}:
H.-K. Lo, \& H. F. Chau,
``Security of quantum key distribution'',
quant-ph/9803006.
Extended version of {\bf [Lo-Chau 99]}.

\item {\bf [Lo-Chau 98 c]}:
H.-K. Lo, \& H. F. Chau,
``Quantum cryptographic system with reduced data loss'',
patent US5732139, 1998.
See {\bf [Ardehali-Chau-Lo 98]}, {\bf [Lo-Chau-Ardehali 00]}.

\item {\bf [Lo-Chau 98 d]}:
H.-K. Lo, \& H. F. Chau,
``Why quantum bit commitment and ideal quantum coin
tossing are impossible'',
{\em Physica D} {\bf 120}, ?, 177-? (1998).

\item {\bf [Lo-Chau 99]}:
H.-K. Lo, \& H. F. Chau,
``Unconditional security of quantum key distribution
over arbitrarily long distances'',
{\em Science} {\bf 283}, 5410, 2050-2056 (1999).
See {\bf [Lo-Chau 98 a]}.

\item {\bf [Lo-Popescu 99]}:
H.-K. Lo, \& S. Popescu,
``Classical communication cost of entanglement manipulation:
Is entanglement an interconvertible resource?'',
{\em Phys. Rev. Lett.} {\bf 83}, 7, 1459-1462 (1999).

\item {\bf [Lo 00 a]}:
H.-K. Lo,
``Classical-communication cost in distributed quantum-information
processing: A generalization of quantum-communication complexity'',
{\em Phys. Rev. A} {\bf 62}, 1, 012313 (2000);
quant-ph/9912009.

\item {\bf [Lo 00 b]}:
H.-K. Lo,
``Will quantum cryptography ever become a successful
technology in the marketplace?'',
{\em Phys. World} {\bf 13}, 6, 17-? (2000).
Extended version: quant-ph/9912011.

\item {\bf [Lo-Chau-Ardehali 00]}:
H.-K. Lo, H. F. Chau, \& M. Ardehali,
``Efficient quantum key distribution scheme and
proof of its unconditional security'',
quant-ph/0011056.
See {\bf [Lo-Chau 98 b]}, {\bf [Ardehali-Chau-Lo 98]}.

\item {\bf [Lo-Popescu 01]}:
H.-K. Lo, \& S. Popescu,
``Concentrating entanglement by local actions: Beyond mean values'',
{\em Phys. Rev. A} {\bf 63}, 2, 022301 (2001);
quant-ph/9707038.

\item {\bf [Lo 01 a]}:
H.-K. Lo,
``A simple proof of the unconditional security of quantum key distribution'',
in S. Popescu, N. Linden, \& R. Jozsa (eds.),
{\em J. Phys. A} {\bf 34}, 35
(Special issue: Quantum information and computation), 6957-6968 (2001).

\item {\bf [Lo 01 b]}:
H.-K. Lo,
``Proof of unconditional security of six-state quantum
key distribution scheme'',
quant-ph/0102138.

\item {\bf [Lo 02]}:
H.-K. Lo,
``Method for decoupling error correction from privacy amplification'',
quant-ph/0201030.

\item {\bf [Lo-Ko 03]}:
H.-K. Lo, \& T.-M. Ko,
``Some attacks on quantum-based cryptographic protocols'',
quant-ph/0309127.

\item {\bf [Lo 03]}:
H.-K. Lo,
``Why quantum information processing'',
quant-ph/0309128.

\item {\bf [Lo-Ma-Chen 04]}:
H.-K. Lo, X. Ma, \& K. Chen,
``Decoy state quantum key distribution'',
quant-ph/0411004.

\item {\bf [Lochak 84]}:
G. Lochak,
``The evolution of the ideas of Louis de Broglie on the interpretation of
wave mechanics'',
in {\bf [Barut-van der Merwe-Vigier 84]}, pp.~11-33.
See {\bf [van der Merwe 84]}.

\item {\bf [Lockhart 99]}:
R. B. Lockhart,
``Optimal ensemble length of mixed separable states'',
quant-ph/9908050.

\item {\bf [Lockhart-Steiner-Gerlach 00]}:
R. B. Lockhart, M. Steiner, \& K. Gerlach,
``Geometry and product states'',
quant-ph/0010013.
See {\bf [Sanpera-Tarrach-Vidal 98 a]}.

\item {\bf [Lockhart 01]}:
R. B. Lockhart,
``Non-maximal rank separable states are a set of measure 0
within the set of non-maximal rank states'',
quant-ph/0111022.

\item {\bf [Lockhart-Steiner 02]}:
R. B. Lockhart, \& M. J. Steiner,
``Preserving entanglement under perturbation and sandwiching all separable
states'',
{\em Phys. Rev. A} {\bf 65}, 2, 022107 (2002);
quant-ph/0009090.

\item {\bf [Lockhart 02]}:
R. Lockhart,
``Low-rank separable states are a set of measure zero within the set of
low-rank states'',
{\em Phys. Rev. A} {\bf 65}, 6, 064304 (2002);
quant-ph/0111051.

\item {\bf [Lockwood 96]}:
M. Lockwood,
`\,``Many minds'' interpretations of quantum mechanics',
{\em Brit. J. Philos. Sci.} {\bf 47}, 2, 159-188 (1996).
Comment: {\bf [Deutsch 96]}.

\item {\bf [Lodahl 03]}:
P. Lodahl,
``Einstein-Podolsky-Rosen correlations in second-harmonic generation'',
{\em Phys. Rev. A} {\bf 68}, 2, 023806 (2003).

\item {\bf [Loeber 99]}:
P. Loeber,
``Quantum channels and simultaneous ID coding'',
quant-ph/9907019.

\item {\bf [Lokajicek 98 a]}:
M. V. Lokajicek,
``Are quantum teleportation and cryptography predicted by quantum
mechanics?'',
quant-ph/9808019.

\item {\bf [Lokajicek 98 b]}:
M. V. Lokajicek,
``Realistic theory of microscopic phenomena; a new solution of
hidden-variable problem'',
quant-ph/9811030.

\item {\bf [Lombardi-Sciarrino-Popescu-De Martini 02]}:
E. Lombardi, F. Sciarrino, S. Popescu, \& F. De Martini,
``Teleportation of entangled states of a vacuum-one photon qubit'',
{\em Phys. Rev. Lett.} {\bf 88}, 7, 070402 (2002);
quant-ph/0109160.

\item {\bf [Lomonaco 98]}:
S. J. Lomonaco, Jr.,
``A quick glance at quantum cryptography'',
{\em Cryptologia},
quant-ph/9811056.

\item {\bf [Lomonaco 02 a]}:
S. J. Lomonaco, Jr. (ed.),
{\em Quantum computation:
A grand mathematical challenge for the twenty-first century and the millennium},
American Mathematical Society,
Providence, Rhode Island, 2002.

\item {\bf [Lomonaco 02 b]}:
S. J. Lomonaco, Jr.,
``A Rosetta stone for quantum mechanics with an
introduction to quantum computation'',
in {\bf [Lomonaco 02 a]}, pp.~3-65;
quant-ph/0007045.

\item {\bf [Lomonaco 02 c]}:
S. J. Lomonaco, Jr.,
``Shor's quantum factoring algorithm'',
in {\bf [Lomonaco 02 a]}, pp.~161-179;
quant-ph/0010034.

\item {\bf [Lomonaco 02 d]}:
S. J. Lomonaco, Jr.,
``Grover's quantum search algorithm'',
in {\bf [Lomonaco 02 a]}, pp.~181-192.

\item {\bf [Lomonaco 02 e]}:
S. J. Lomonaco, Jr.,
``A talk on quantum cryptography: Or how Alice outwits Eve'',
revised version of a paper published in D. Joyner (ed.),
{\em Coding theory, and cryptography: From
Geheimscheimschreiber and enigma to quantum theory},
Springer-Verlag, New York, 1999, pp.~144-174,
in {\bf [Lomonaco 02 a]}, pp.~237-264;
quant-ph/0102016.

\item {\bf [Lomonaco 02 f]}:
S. J. Lomonaco, Jr.,
``An entangled tale of quantum entanglement'',
in {\bf [Lomonaco 02 a]}, pp.~305-349;
quant-ph/0101120.

\item {\bf [Lomonaco-Brandt 02]}:
S. J. Lomonaco, Jr., \& H. E. Brandt (eds.),
{\em Quantum computation and information},
American Mathematical Society,
Providence, Rhode Island, 2002.

\item {\bf [Lomonaco-Kauffman 03]}:
S. J. Lomonaco, Jr., \& L. H. Kauffman,
``A continuous variable Shor algorithm'',
quant-ph/0210141.

\item {\bf [Lomonaco-Kauffman 04]}:
S. J. Lomonaco, Jr., \& L. H. Kauffman,
``Countinuous quantum hidden subgroup algorithms'',
quant-ph/0304084.

\item {\bf [Lomonaco-Kauffman 04]}:
S. J. Lomonaco, Jr., \& L. H. Kauffman,
``Quantum hidden subgroup algorithms: The devil is in the details'',
{\em Proc.\ of SPIE on Quantum Information and Computation};
quant-ph/0403229.

\item {\bf [Londero-Dorrer-Anderson-(+3) 04]}:
P. Londero, C. Dorrer, M. Anderson,
S. Wallentowitz, K. Banaszek, \& I. A. Walmsley,
``Efficient optical implementation of the Bernstein-Vazirani algorithm'',
{\em Phys. Rev. A} {\bf 69}, 1, 010302 (2004).

\item {\bf [Long-Li-Zhang-Niu 99]}:
G. L. Long, Y. S. Li, W. L. Zhang, \& L. Niu,
``Phase matching in quantum searching'',
{\em Phys. Lett. A} {\bf 262}, 1, 27-34 (1999);
quant-ph/9906020.

\item {\bf [Long-Li-Zhang-Tu 99]}:
G. L. Long, Y. S. Li, W. L. Zhang, \& C. C. Tu,
``An intrinsic limitation on the size of quantum database'',
quant-ph/9910076.

\item {\bf [Long-Tu-Li-Zhang-(+2) 99]}:
G. L. Long, C. C. Tu, Y. S. Li, W. L. Zhang, \& H. Y. Yan,
``A novel $SO(3)$ picture for quantum searching'',
quant-ph/9911004.

\item {\bf [Long-Li-Zhang-Tu 00]}:
G. L. Long, Y. S. Li, W. L. Zhang, \& C. C. Tu,
``Dominant gate imperfection in Grover's quantum search algorithm'',
{\em Phys. Rev. A} {\bf 61}, 4, 042305 (2000).

\item {\bf [Long-Yan-Li-(+6) 00]}:
G. L. Long, H. Y. Yan, Y. S. Li,
C. C. Tu, S. J. Zhu, D. Ruan, Y. Sun, J. X. Tao, \& H. M. Chen,
``On the quantum mechanical nature in liquid NMR quantum computing'',
quant-ph/0007077.

\item {\bf [Long-Liu 02]}:
G. L. Long, \& X. S. Liu,
``Theoretically efficient high-capacity quantum-key-distribution scheme'',
{\em Phys. Rev. A} {\bf 65}, 3, 032302 (2002);
quant-ph/0012056.

\item {\bf [Long-Sun 01]}:
G. L. Long, \& Y. Sun,
``Efficient scheme for initializing a quantum
register with an arbitrary superposed state'',
{\em Phys. Rev. A} {\bf 64}, 1, 014303 (2001);
quant-ph/0104030.

\item {\bf [Long 01]}:
G. L. Long,
``Grover algorithm with zero theoretical failure rate'',
{\em Phys. Rev. A} {\bf 64}, 2, 022307 (2001);
quant-ph/0106071.

\item {\bf [Long-Yan-Li-(+8) 01]}:
G. L. Long, H. Y. Yan, Y. S. Li,
C. C. Tu, J. X. Tao, H. M. Chen, M. L. Liu,
X. Zhang, J. Luo, L. Xiao, \& X. Z. Zheng,
``Experimental NMR realization of a generalized quantum search algorithm'',
{\em Phys. Lett. A} {\bf 286}, 2-3, 121-126 (2001);
quant-ph/0009059.

\item {\bf [Long-Yan-Sun 01]}:
G. L. Long, H. Y. Yan, \& Y. Sun,
``Analysis of density matrix reconstruction in NMR quantum computing'',
{\em J. Opt. B: Quantum Semiclass. Opt.} {\bf 3}, 6, 376-381 (2001);
quant-ph/0012047.

\item {\bf [Long-Li-Sun 01]}:
G. L. Long, X. Li, \& Y. Sun,
``Phase matching condition for quantum search with a generalized initial state'',
{\em Phys. Lett. A} {\bf 294}, 3-4, 143-152 (2002);
quant-ph/0107013.

\item {\bf [Long-Xiao 03]}:
G. L. Long, \& L. Xiao,
``Experimental realization of a fetching algorithm in a 7-qubit NMR spin
Liouville space computer'',
{\em J. Chem. Phys.} {\bf 119}, 8473-? (2003).

\item {\bf [Longdell-Sellars 04]}:
J. J. Longdell, \& M. J. Sellars,
``Experimental demonstration of quantum-state tomography and qubit-qubit
interactions for rare-earth-metal-ion-based solid-state qubits'',
{\em Phys. Rev. A} {\bf 69}, 3, 032307 (2004).

\item {\bf [L\'{o}pez-Paz 03]}:
C. C. L\'{o}pez, \& J. P. Paz,
``Phase-space approach to the study of decoherence in quantum walks'',
{\em Phys. Rev. A} {\bf 68}, 5, 052305 (2003);
quant-ph/0308104.

\item {\bf [Lorenz-Silberhorn-Korolkova-(+2) 01]}:
S. Lorenz, C. Silberhorn, N. Korolkova,
R. S. Windeler, \& G. Leuchs,
``Squeezed light from microstructured fibres:
Towards free space quantum cryptography'',
quant-ph/0109018.

\item {\bf [Loss-DiVincenzo 98]}:
D. Loss, \& D. P. DiVincenzo,
``Quantum computation with quantum dots'',
{\em Phys. Rev. A} {\bf 57}, 1, 120-126 (1998);
cond-mat/9701055.
Reprinted in {\bf [Macchiavello-Palma-Zeilinger 00]}, pp.~433-439.

\item {\bf [Loss-Burkard-Sukhorukov 99]}:
D. Loss, G. Burkard, \& E. V. Sukhorukov,
``Quantum computing and quantum communication with
electrons in nanostructures'',
to be published in the {\em Proc.\ of the XXXIVth Rencontres
de Moriond ``Quantum Physics at Mesoscopic Scale''
(Les Arcs, Savoie, France, 1999)};
cond-mat/9907133.

\item {\bf [Loss-Sukhorukov 00]}:
D. Loss, \& E. V. Sukhorukov,
``Probing entanglement and nonlocality of electrons
in a double-dot via transport and noise'',
{\em Phys. Rev. Lett.} {\bf 84}, 5, 1035-1038 (2000);
cond-mat/9907129.

\item {\bf [Loss 00]}:
D. Loss,
``Quantum computation and quantum communication with electrons'',
{\bf [Macchiavello-Palma-Zeilinger 00]}, pp.~427-432.

\item {\bf [Loubenets 01]}:
E. R. Loubenets,
``Quantum stochastic approach to the description of quantum measurements'',
{\em J. Phys. A} {\bf 34}, 37, 7639-7675 (2001);
quant-ph/0109094.

\item {\bf [Loubenets 03]}:
E. R. Loubenets,
`\,``Local realism'', Bell's theorem and quantum ``locally realistic''
inequalities',
quant-ph/0309111.

\item {\bf [Loubenets 04 a]}:
E. R. Loubenets,
``Separability of quantum states and the violation of Bell-type inequalities'',
{\em Phys. Rev. A} {\bf 69}, 4, 042102 (2004).
Comment: {\bf [Simon 04]}.

\item {\bf [Loubenets 04 b]}:
E. R. Loubenets,
``Quantum states satisfying classical probability constraints'',
quant-ph/0406139.

\item {\bf [Loubenets 04 c]}:
E. R. Loubenets,
``On validity of the original Bell inequality for the Werner nonseparable
state'',
quant-ph/0407097.

\item {\bf [Lougovski-Solano-Walther 03]}:
P. Lougovski, E. Solano, \& H. Walther,
``Generation and purification of maximally-entangled atomic states in
optical cavities'',
quant-ph/0308059.

\item {\bf [Lovett-Reina-Nazir-Briggs 03]}:
B. W. Lovett, J. H. Reina, A. Nazir, \& G. A. D. Briggs,
``Optical schemes for quantum computation in quantum dot molecules'',
{\em Phys. Rev. B} {\bf 68}, 20, 205319 (2003);
quant-ph/0307021.

\item {\bf [Lu-Guo 00]}:
H. Lu, \& G.-C. Guo,
``Teleportation of a two-particle entangled state via entanglement
swapping'',
{\em Phys. Lett. A} {\bf 276}, 5-6, 209-212 (2000).

\item {\bf [Lu-Chen-Pan-Zhan 02]}:
H.-X. Lu, Z.-B. Chen, J.-W. Pan, \& Y.-D. Zhang,
``Calculation of entanglement for continious variable states'',
quant-ph/0204098.

\item {\bf [Lu-Zhou-Kuang 04]}:
J. Lu, L. Zhou, \& L.-M. Kuang,
``Linear optics implementation for quantum game with two players'',
{\em Phys. Lett. A} {\bf 330}, 1-2, 48-53 (2004).

\item {\bf [Lu-Campbell-Ou 03]}:
Y. J. Lu, R. L. Campbell, \& Z. Y. Ou,
``Mode-locked two-photon states'',
{\em Phys. Rev. Lett.} {\bf 91}, 16, 163602 (2003).

\item {\bf [Lucamarini-Paganelli-Mancini 04]}:
M. Lucamarini, S. Paganelli, \& S. Mancini,
``Two qubits entanglement dynamics in a symmetry-broken environment'',
{\em Phys. Rev. A} {\bf 69}, 6, 062308 (2004).
quant-ph/0402073.

\item {\bf [Lucamarini-Di Giuseppe 04]}:
M. Lucamarini, \& G. Di Giuseppe,
``Deterministic plug-and-play for quantum communication'',
quant-ph/0407256.

\item {\bf [Lucarelli 02 a]}:
D. Lucarelli,
``Chow's theorem and universal holonomic quantum computation'',
{\em J. Phys. A} {\bf 35}, 24, 5107-5114 (2002);
quant-ph/0111078.
See {\bf [Lucarelli 02 b]}.

\item {\bf [Lucarelli 02 b]}:
D. Lucarelli,
``Control algebra for holonomic quantum computation
with squeezed coherent states'',
quant-ph/0202055.
Some overlap with {\bf [Lucarelli 02 a]}.

\item {\bf [L\"{u}ders 51]}:
L\"{u}ders,
``\"{U}ber die Zustands\"{a}nderung durch den messprozss'',
{\em Annalen der Physik} {\bf 8}, ?, 322-328 (1951).

\item {\bf [Ludwig 64]}:
G. Ludwig,
``Versuch einer axiomatischen Grundlegung der
Quantenmechanik und allgemeinerer physicalichen Theorien'',
{\em Z. Physik} {\bf 181}, 233-260 (1964).
See {\bf [Ludwig 67 c]} (II), {\bf [Ludwig 68 a]} (III).

\item {\bf [Ludwig 67 a]}:
G. Ludwig,
``Haupts\"{a}tze \"{u}ber das Messen als Grundlage der
Hilbert-Raumstruktur der Quantenmechanik'',
{\em Zeitschrift f\"{u}r Naturforschung A} {\bf 22}, 1303-1323 (1967).

\item {\bf [Ludwig 67 b]}:
G. Ludwig,
``Ein weiterer Hauptsatz \"{u}ber das Messen als Grundlage
der Hilbert-Raumstruktur der Quantenmechanik'',
{\em Zeitschrift f\"{u}r Naturforschung A} {\bf 22}, 1324-1327 (1967).

\item {\bf [Ludwig 67 c]}:
G. Ludwig,
``Attempt for an axiomatic foundation of quantum mechanics and
more general theories. II'',
{\em Comm. Math. Phys.} {\bf 4}, ?, 331-348 (1967).
See {\bf [Ludwig 64]} (I), {\bf [Ludwig 68 a]} (III).

\item {\bf [Ludwig 68 a]}:
G. Ludwig,
``Attempt for an axiomatic foundation of quantum mechanics and
more general theories. III'',
{\em Comm. Math. Phys.} {\bf 9}, 1, 1-12 (1967).
See {\bf [Ludwig 64]} (I), {\bf [Ludwig 67 c]} (II).

\item {\bf [Ludwig 68 b]}:
G. Ludwig,
{\em Wave mechanics},
Oxford University Press, Oxford, 1968.

\item {\bf [Ludwig 76]}:
G. Ludwig,
{\em Einf\"{u}hrung in die Grundlagen der Theoretischen Physik},
Vieweg, Braunschweig, 1976.

\item {\bf [Ludwig 83]}:
G. Ludwig,
{\em Foundations of quantum mechanics},
Springer-Verlag, New York, 1983.

\item {\bf [Ludwin-Ben Aryeh 01]}:
D. Ludwin, \& Y. Ben-Aryeh,
``The role of momentum transfer in
welcher-Weg experiments'',
{\em Found. Phys. Lett.} {\bf 14}, 6, 519-528 (2001).

\item {\bf [Luecke 97]}:
W. Luecke,
``Gisin nonlocality of the Doebner-Goldin 2-particle equation'',
quant-ph/9710033.

\item {\bf [Luis-S\'{a}nchez Soto 98 a]}:
A. Luis, \& L. L. S\'{a}nchez-Soto,
``Anti-Zeno effect in parametric down-conversion'',
{\em Phys. Rev. A} {\bf 57}, 2, 781-787 (1998).

\item {\bf [Luis-S\'{a}nchez Soto 98 b]}:
A. Luis, \& L. L. S\'{a}nchez-Soto,
``Dynamical analysis of seemingly interaction-free measurements'',
{\em Phys. Rev. A} {\bf 58}, 2, 836-839 (1998).

\item {\bf [Luis-S\'{a}nchez Soto 98 c]}:
A. Luis, \& L. L. S\'{a}nchez-Soto,
``Complementarity enforced by random classical phase kicks'',
{\em Phys. Rev. Lett.} {\bf 81}, 19, 4031-4035 (1998).

\item {\bf [Luis-S\'{a}nchez Soto 99 a]}:
A. Luis, \& L. L. S\'{a}nchez-Soto,
``Dynamics of a two-level atom observed via an interaction-free
measurement'',
{\em Phys. Rev. A} {\bf 60}, 1, 56-62 (1999).

\item {\bf [Luis-S\'{a}nchez Soto 99 b]}:
A. Luis, \& L. L. S\'{a}nchez-Soto,
``?'',
{\em Phys. Lett. A} {\bf 252}, ?, 130-? (1999).

\item {\bf [Luis-S\'{a}nchez Soto 99 c]}:
A. Luis, \& L. L. S\'{a}nchez-Soto,
``Measuring quantum input-output processes:
Phase-space representation of transformations'',
{\em Phys. Lett. A} {\bf 261}, 1-2, 12-16 (1999).

\item {\bf [Luis-S\'{a}nchez Soto 99 d]}:
A. Luis, \& L. L. S\'{a}nchez-Soto,
``Complete characterization of arbitrary quantum measurement
processes'',
{\em Phys. Rev. Lett.} {\bf 83}, 18, 3573-3576 (1999).

\item {\bf [Luis 00]}:
A. Luis,
``Quantum tomography of input-output processes'',
{\em Phys. Rev. A} {\bf 62}, 5, 054302 (2000).

\item {\bf [Luis 01 a]}:
A. Luis,
``Quantum-state preparation and control via the Zeno effect'',
{\em Phys. Rev. A} {\bf 63}, 5, 052112 (2001).

\item {\bf [Luis 01 b]}:
A. Luis,
``Complementarity and certainty relations for two-dimensional
systems'',
{\em Phys. Rev. A} {\bf 64}, 1, 012103 (2001).

\item {\bf [Luis 01 c]}:
A. Luis,
``Complementarity in multiple beam interference'',
{\em J. Phys. A} {\bf 34}, 41, 8597-8600 (2001).

\item {\bf [Luis 01 d]}:
A. Luis,
``Equivalence between macroscopic quantum superpositions and maximally
entangled states: Application to phase-shift detection'',
{\em Phys. Rev. A} {\bf 64}, 5, 054102 (2001).

\item {\bf [Luis 02]}:
A. Luis,
``Generation of maximally entangled states via dispersive interactions'',
{\em Phys. Rev. A} {\bf 65}, 3, 034102 (2002).

\item {\bf [Luis 03]}:
A. Luis,
``Visibility for multi-particle interference'',
{\em Phys. Lett. A} {\bf 314}, 3, 197-202 (2003).

\item {\bf [Lukin-Matsko-Fleischhauer-Scully 98]}:
M. D. Lukin, A. B. Matsko, M. Fleischhauer, \& M. O. Scully,
``Quantum noise and correlations in resonantly enhanced wave
mixing based on atomic coherence'',
quant-ph/9811028.

\item {\bf [Lukin-Hemmer 99]}:
M. D. Lukin, \& P. R. Hemmer,
``Coherent control of atom-atom interactions and entanglement using optical fields'',
quant-ph/9905025.

\item {\bf [Lukin-Imamo\u{g}lu 00]}:
M. D. Lukin, \& A. Imamo\u{g}lu,
``Nonlinear optics and quantum entanglement of ultraslow single
photons'',
{\em Phys. Rev. Lett.} {\bf 84}, 7, 1419-1422 (2000).

\item {\bf [Lukin-Yelin-Fleischhauer 00]}:
M. D. Lukin, S. F. Yelin, \& M. Fleischhauer,
``Entanglement of atomic ensembles by trapping correlated photon
states'',
{\em Phys. Rev. Lett.} {\bf 84}, 18, 4232-4235 (2000).

\item {\bf [Lukin-Fleischhauer-Cote-(+4) 01]}:
M. D. Lukin, M. Fleischhauer, R. Cote,
L. M. Duan, D. Jaksch, J. I. Cirac, \& P. Zoller,
``Dipole blockade and quantum information processing
in mesoscopic atomic ensembles'',
{\em Phys. Rev. Lett.} {\bf 87}, 3, 037901 (2001);
quant-ph/0011028.

\item {\bf [Lund-Ralph 02]}:
A. P. Lund, \& T. C. Ralph,
``Nondeterministic gates for photonic single-rail quantum logic'',
{\em Phys. Rev. A} {\bf 66}, 3, 032307 (2002).

\item {\bf [Lund-Bell-Ralph 03]}:
A. P. Lund, T. B. Bell, \& T. C. Ralph,
``Comparison of linear optics quantum-computation control-sign gates with
ancilla inefficiency and an improvement to functionality under these
conditions'',
{\em Phys. Rev. A} {\bf 68}, 2, 022313 (2003);
quant-ph/0308071.

\item {\bf [Lund-Jeong-Ralph-Kim 04]}:
A. P. Lund, H. Jeong, T. C. Ralph, \& M. S. Kim,
``Conditional production of superpositions of coherent states with
inefficient photon detection'',
quant-ph/0401001.

\item {\bf [Lund-Ralph 04]}:
A. P. Lund, \& T. C. Ralph,
``Coherent state LOQC gates using simplified diagonal superposition
resource states'',
quant-ph/0410204.

\item {\bf [Lunkes-Brukner-Vedral 04]}:
C. Lunkes, \v{C}. Brukner, \& V. Vedral,
``Equation of state for entanglement in a Fermi gas'',
quant-ph/0410166.

\item {\bf [Luo-Zeng 98]}:
J. Luo, \& X. Zeng,
``NMR quantum computation with a hyperpolarized nuclear spin bulk'',
quant-ph/9811044.

\item {\bf [Luo-Zhang 03]}:
S. Luo, \& Z. Zhang,
``Entanglement and interference'',
{\em Phys. Lett. A} {\bf 315}, 3-4, 189-193 (2003).

\item {\bf [Luo-Zhang 04]}:
S. Luo, \& Q. Zhang,
``Informational distance on quantum-state space'',
{\em Phys. Rev. A} {\bf 69}, 3, 032106 (2004).

\item {\bf [Luo-Kr\"{u}ger-Brugger-(+7) 03]}:
X. Luo, P. Kr\"{u}ger, K. Brugger,
S. Wildermuth, H. Gimpel, M. W. Klein,
S. Groth, R. Folman, I. Bar-Joseph, \& J. Schmiedmayer,
``An atom fiber for guiding cold neutral atoms'',
quant-ph/0311174.

\item {\bf [Lupascu-Verwijs-Schouten-(+2) 04]}:
A. Lupascu, C. J. M. Verwijs, R. N. Schouten,
C. J. P. M. Harmans, \& J. E. Mooij,
``Nondestructive readout for a superconducting flux qubit'',
{\em Phys. Rev. Lett.} {\bf 93}, 17, 177006 (2004);
cond-mat/0311510.

\item {\bf [Luque-Thibon 03]}:
J.-G. Luque, \& J.-Y. Thibon,
``Polynomial invariants of four qubits'',
{\em Phys. Rev. A} {\bf 67}, 4, 042303 (2003);
quant-ph/0212069.

\item {\bf [L\"{u}tkenhaus 96]}:
N. L\"{u}tkenhaus,
``Security against eavesdropping in quantum cryptography'',
{\em Phys. Rev. A} {\bf 54}, 1, 97-111 (1996).

\item {\bf [L\"{u}tkenhaus-Barnett 97]}:
N. L\"{u}tkenhaus, \& S. M. Barnett,
``Security against eavesdropping in quantum cryptography'',
{\em Int. workshop on quantum communication, computing,
and measurement (Shizuoka, Japan, 1996)},
Plenum Press, New York (1997),
quant-ph/9711033.

\item {\bf [L\"{u}tkenhaus-Cirac-Zoller 98]}:
N. L\"{u}tkenhaus, J. I. Cirac, \& P. Zoller,
``Mimicking a squeezed-bath interaction:
Quantum-reservoir engineering with atoms'',
{\em Phys. Rev. A} {\bf 57}, 11, 548-558 (1998).

\item {\bf [L\"{u}tkenhaus-Calsamiglia-Suominen 99]}:
N. L\"{u}tkenhaus, J. Calsamiglia, \& K.-A. Suominen,
``Bell measurements for teleportation'',
{\em Phys. Rev. A} {\bf 59}, 5, 3295-3300 (1999);
quant-ph/9809063.
See {\bf [Vaidman-Yoran 99]}.

\item {\bf [L\"{u}tkenhaus 99]}:
N. L\"{u}tkenhaus,
``Estimates for practical quantum cryptography'',
{\em Phys. Rev. A} {\bf 59}, 5, 3301-3319 (1999);
quant-ph/9806008.

\item {\bf [L\"{u}tkenhaus 00]}:
N. L\"{u}tkenhaus,
``Security against individual attacks for realistic quantum key distribution'',
{\em Phys. Rev. A} {\bf 61}, 5, 052304 (2000);
quant-ph/9910093.

\item {\bf [L\"{u}tkenhaus-Jahma 02]}:
N. L\"{u}tkenhaus, \& M. Jahma,
``Quantum key distribution with realistic states:
Photon-number statistics in the photon-number splitting attack'',
{\em New J. Phys} {\bf 4}, 44.1-44.9 (2002);
quant-ph/0112147.

\item {\bf [Luzuriaga 02]}:
J. Luzuriaga,
`Comment on ``Super classical quantum mechanics: The best
interpretation of nonrelativistic quantum mechanics,'' by W. E.
Lamb [{\em Am. J. Phys.} {\bf 69}, 413-422 (2001)]',
{\em Am. J. Phys.} {\bf 70}, 1, 10 (2002).
Comment on {\bf [Lamb 01 a]}.

\item {\bf [Lvovsky-Shapiro 01]}:
A. I. Lvovsky, \& J. H. Shapiro,
Submitted to {\em Phys. Rev. A};
``Nonclassical character of statistical mixtures
of the single-photon and vacuum optical states'',
quant-ph/0109057.

\item {\bf [Lvovsky 02]}:
A. I. Lvovsky,
``Cabello's nonlocality and linear optics'',
{\em Phys. Rev. Lett.} {\bf 88}, 9, 098901 (2002).
Comment to {\bf [Cabello 01 d]}.

\item {\bf [Lvovsky-Babichev 02]}:
A. I. Lvovsky, \& S. A. Babichev,
``Synthesis and tomographic characterization
of the displaced Fock state of light'',
quant-ph/0202163.

\item {\bf [Lvovsky-Mlynek 02]}:
A. I. Lvovsky, \& J. Mlynek,
``Quantum-optical catalysis: Generating nonclassical states of light by means
of linear optics'',
{\em Phys. Rev. Lett.} {\bf 88}, 25, 250401 (2002);
quant-ph/0202164.

\item {\bf [Lvovsky 03]}:
A. I. Lvovsky,
``Iterative maximum-likelihood reconstruction in quantum homodyne
tomography'',
quant-ph/0311097.

\item {\bf [Lyubomirsky-Shirasaki-K\"{a}rtner-Haus 93]}:
I. Lyubomirsky, M. Shirasaki, F. X. K\"{a}rtner, \& H. A. Haus,
``Test of Bell's inequality with
squeezed light from a Sagnac fibre ring'',
{\em Quantum Opt.} {\bf 5}, 4, 241-250 (1993).

\item {\bf [Lyre 97]}:
H. Lyre,
``Against measurement?---On the concept of information'',
in {\em 10th Max Born Symp.\ ``Quantum Future'' (Wroclaw, Poland, 1997)};
quant-ph/9709059.
See {\bf [Bell 90]}.


\newpage

\subsection{}


\item {\bf [Ma-Long-Deng-(+2) 02]}:
Y. J. Ma, G. L. Long, F. G. Deng, F. Li, \& S.-X. Zhang,
``Cooperative three- and four-player quantum games'',
{\em Phys. Lett. A} {\bf 301}, 3-4, 117-124 (2002).

\item {\bf [Maalas-Fleischhauer 04]}:
M. Maalas \& M. Fleischhauer,
``Scattering of dark-state polaritons in optical lattices and quantum phase gate for photons'',
{\em Phys. Rev. A} {\bf 69}, 6, 061801 (2004).

\item {\bf [Maassen-Uffink 88]}:
H. Maassen, \& J. Uffink,
``Generalized entropic uncertainty relations'',
{\em Phys. Rev. Lett.} {\bf 60}, 12, 1103-1106 (1988).

\item {\bf [Mabuchi-Zoller 96]}:
H. Mabuchi, \& P. Zoller,
``Inversion of quantum jumps in quantum optical systems under
continuous observation'',
{\em Phys. Rev. Lett.} {\bf 76}, 17, 3108-3111 (1996).
Reprinted in {\bf [Macchiavello-Palma-Zeilinger 00]}, pp.~306-309.

\item {\bf [Mabuchi 00]}:
H. Mabuchi,
``Cavity quantum electrodynamics'',
{\bf [Macchiavello-Palma-Zeilinger 00]}, pp.~277-281.

\item {\bf [Mabuchi-Doherty 02]}:
H. Mabuchi, \& A. C. Doherty,
``Cavity quantum electrodynamics: Coherence in context'',
{\em Science} {\bf 298}, ?, 1372-? (2002).

\item {\bf [Macchiavello-Palma-Zeilinger 00]}:
C. Macchiavello, G. M. Palma, \& A. Zeilinger (eds.),
{\em Quantum computation and quantum information theory:
Collected papers and notes (Villa Gualino, Torino, Italy, 1999)},
World Scientific, Singapore, 2000.

\item {\bf [Macchiavello 00 a]}:
C. Macchiavello,
``Universal transformations for finite dimensional quantum
systems'',
{\em Fortschr. Phys.} {\bf 48}, 5-7, 545-552 (2000).

\item {\bf [Macchiavello 00 b]}:
C. Macchiavello,
``Bounds on the efficiency of cloning for two-state quantum systems'',
{\em J. Opt. B: Quantum Semiclass. Opt.} {\bf 2}, 2, 144-148 (2000).

\item {\bf [Macchiavello-Palma 02]}:
C. Macchiavello, \& G. M. Palma,
``Entanglement-enhanced information transmission over a quantum channel with
correlated noise'',
{\em Phys. Rev. A} {\bf 65}, 5, 050301 (2002);
quant-ph/0107052.

\item {\bf [Macchiavello-Bru\ss 02]}:
C. Macchiavello, \& D. Bru\ss,
``Security aspects of quantum cryptography with $d$-dimensional systems'',
in M. Ferrero (ed.),
{\em Proc. of Quantum Information: Conceptual Foundations,
Developments and Perspectives (Oviedo, Spain, 2002)},
{\em J. Mod. Opt.} {\bf 50}, 6-7, 1025-1033 (2003).

\item {\bf [Macchiavello 03]}:
C. Macchiavello,
``Optimal estimation of multiple phases'',
{\em Phys. Rev. A} {\bf 67}, 6, 062302 (2003);
quant-ph/0304126.

\item {\bf [Macchiavello-Palma-Virmani 04]}:
C. Macchiavello, G. M. Palma, \& S. Virmani,
``Transition behavior in the channel capacity of two-quibit channels with memory'',
{\em Phys. Rev. A} {\bf 69}, 1, 010303 (2004);
quant-ph/0307016.

\item {\bf [Macdonald 82]}:
A. L. Macdonald,
``Comment on `Resolution of the
Einstein-Podolsky-Rosen and Bell paradoxes'\,'',
{\em Phys. Rev. Lett.} {\bf 49}, 16, 1215 (1982).
Comment on {\bf [Pitowsky 82 a]}.
Reply: {\bf [Pitowsky 82 b]}.

\item {\bf [Macdonald 01]}:
A. Macdonald,
``Joint measurement and state reduction'',
quant-ph/0108129.

\item {\bf [Machida-Nakazato-Pascazio-(+2) 99]}:
K. Machida, H. Nakazato, S. Pascazio, H. Rauch, \& S. Yu,
``Reflection and transmission in a neutron-spin test of the quantum Zeno effect'',
{\em Phys. Rev. A} {\bf 60}, 5, 3448-3460 (1999).

\item {\bf [Mack-Dietmar-Fischer-Freyberger 00]}:
H. Mack, D. G. Fischer, \& M. Freyberger,
``Enhanced quantum estimation via purification'',
{\em Phys. Rev. A} {\bf 62}, 4, 042301 (2000);
quant-ph/0004096.

\item {\bf [Mack-Bienert-Haug 02]}:
H. Mack, M. Bienert, F. Haug,
F. S. Strau\ss, M. Freyberger, \& W. P. Schleich,
``Wave packet dynamics and factorization of numbers'',
quant-ph/0204040.

\item {\bf [Mack-Freyberger 02]}:
H. Mack, \& M. Freyberger,
``Dynamics of entanglement between two trapped atoms'',
{\em Phys. Rev. A} {\bf 66}, 4, 042113 (2002).

\item {\bf [Mackey 57]}:
G. W. Mackey,
``Quantum mechanics and Hilbert space'',
{\em Am. Math. Monthly} {\bf 64}, 10, part II, 45-57 (1957).

\item {\bf [Mackey 60]}:
G. W. Mackey,
{\em Lecture notes on the mathematical foundations of quantum mechanics},
mimeograph, Harvard University, Cambridge, Massachusetts, 1960.
Origin of {\bf [Mackey 63]}.

\item {\bf [Mackey 63]}:
G. W. Mackey,
{\em The mathematical foundations of quantum mechanics},
Benjamin, New York, 1963.

\item {\bf [Mackman-Squires 95]}:
S. Mackman, \& E. Squires,
``Lorentz invariance and the retarded Bohm model'',
{\em Found. Phys.} {\bf 25}, 3, 391-397 (1995).

\item {\bf [MacRae 91]}:
N. MacRae,
{\em John von Neumann: The scientific genius who pioneered the
modern computer, game theory, nuclear deterrence, and much more},
Pantheon, New York, 1991;
American Mathematical Society, New York, 1999.
Review: {\bf [Ceruzzi 93]}.
See {\bf [Heims 80]}.

\item {\bf [Maczy\'{n}ski 67]}:
M. J. Maczy\'{n}ski,
``A remark on Mackey's axiom system for quantum mechanics'',
{\em Bulletin de L'Acad\'emie Polonaise des Sciences,
Serie des Sciences Mathematiques, Astronomiques et Physiques}
{\bf 15}, ?, 583-587 (1967).

\item {\bf [Maczy\'{n}ski 71 a]}:
M. J. Maczy\'{n}ski,
``Boolean properties of observables
in axiomatic quantum mechanics'',
{\em Rep. Math. Phys.} {\bf 2}, 2, 135-150 (1971).

\item {\bf [Maczy\'{n}ski 71 b]}:
M. J. Maczy\'{n}ski,
``On representing observables in
axiomatic quantum mechanics by point mappings'',
{\em Bull. Acad. Pol. Sci. Ser. Sci. Math. Aston. Phys.}
{\bf 19}, 4, 335-339 (1971).

\item {\bf [Madsen 81]}:
J. Madsen,
``Reply to `A critical analysis of N. Bohr's reply
to the EPR argument'\,'',
{\em Phys. Lett. A} {\bf 83}, 3, 103-104 (1981).
Comment on {\bf [Ko\c{c} 81]}.

\item {\bf [Maeda 89]}:
S. Maeda,
``Probability measures on projections in von Neumann algebras'',
{\em Rev. Math. Phys.} {\bf 1}, 2-3, 235-290 (1989).

\item {\bf [MagiQ Technologies 02]}:
{\em MagiQ Technologies},
www.magiqtech.com.

\item {\bf [Maguire 01]}:
Y. Maguire,
``?'',
{\em La Recherche} {\bf ?}, ?, ?-? (2001).
Spanish version: ``Unas gotas de ordenador'',
{\em Mundo Cient\'{\i}fico} {\bf 22}, 230, 28-31 (2002).

\item {\bf [Mahesh-Dorai-Arvind-Kumar 00]}:
T. S. Mahesh, K. Dorai, Arvind, \& A. Kumar,
``Quantum computing by two-dimensional NMR using spin- and
transition-selective pulses'',
quant-ph/0006123.

\item {\bf [Mahesh-Sinha-Ramanathan-Kumar 02]}:
T. S. Mahesh, N. Sinha, K. V. Ramanathan, \& A. Kumar,
``Ensemble quantum-information processing by NMR: Implementation of gates and
the creation of pseudopure states using dipolar coupled spins as qubits'',
{\em Phys. Rev. A} {\bf 65}, 2, 022312 (2002).

\item {\bf [Mahesh-Sinha-Ghosh-(+5) 03]}:
T. S. Mahesh, N. Sinha, A. Ghosh,
R. Das, N. Suryaprakash, M. H. Levitt,
K. V. Ramanathan, \& A. Kumar,
``Quantum information processing by NMR using strongly coupled spins'',
{\em Current Science} {\bf 85}, 7, 932-? (2003);
quant-ph/0212123.

\item {\bf [Mahler-Kim 98]}:
G. Mahler, \& I. Kim,
``Correlation between correlations: Process and time in quantum networks'',
in C. P. Williams (ed.),
{\em 1st NASA Int.\ Conf.\ on Quantum Computing and Quantum Communications
(Palm Springs, California, 1998)},
{\em Lecture Notes in Computer Science} {\bf 1509},
Springer-Verlag, New York, 1999, pp.~?-?;
quant-ph/9803008.

\item {\bf [Mahesh-Kumar 01]}:
T. S. Mahesh, \& A. Kumar,
``Ensemble quantum-information processing by NMR: Spatially averaged
logical labeling technique for creating pseudopure states'',
{\em Phys. Rev. A} {\bf 64}, 1, 012307 (2001).

\item {\bf [Mahler-Keller 96]}:
G. Mahler, \& M. Keller,
``Temporal Bell inequalities in quantum networks'',
in A. Mann, \& M. Revzen (eds.),
{\em The dilemma of Einstein, Podolsky and Rosen -- 60 years
later. An international symposium in honour of Nathan Rosen
(Haifa, Israel, 1995)},
{\em Ann. Phys. Soc. Israel} {\bf 12}, 132-137 (1996).

\item {\bf [Maia Neto-Dalvit 00]}:
P. A. Maia Neto, \& D. A. R. Dalvit,
``Radiation pressure as a source of decoherence'',
{\em Phys. Rev. A} {\bf 62}, 4, 042103 (2000);
quant-ph/0004057.

\item {\bf [Maierle-Lidar-Harris 98]}:
C. S. Maierle, D. A. Lidar, \& R. A. Harris,
``How to teleport superpositions of chirial amplitudes'',
{\em Phys. Rev. Lett.} {\bf 81}, 26, 5928-5931 (1998);
quant-ph/9807020.

\item {\bf [Mair-Zeilinger 99]}:
A. Mair, \& A. Zeilinger,
``Entangled states of orbital angular momentum of photons'',
in {\bf [Greenberger-Reiter-Zeilinger 99]}, pp.~249-252.

\item {\bf [Mair-Vaziri-Weihs-Zeilinger 01]}:
A. Mair, A. Vaziri, G. Weihs, \& A. Zeilinger,
``Entanglement of the orbital angular momentum states of photons'',
{\em Nature} {\bf 412}, 6844, 313-316 (2001);
quant-ph/0104070.

\item {\bf [Mair-Hager-Phillips-(+2) 02]}:
A. Mair, J. Hager, D. F. Phillips, R. L. Walsworth, \& M. D. Lukin,
``Phase coherence and control of stored photonic information'',
{\em Phys. Rev. A} {\bf 65}, 3, 031802 (2002).

\item {\bf [Maitra-Chakrabarti-Ghosh-(+2) 04]}:
S. Maitra, D. Chakrabarti, S. Ghosh,
P. Mukhopadhyay, \& S. L. Braunstein,
``Quantum algorithm to distinguish Boolean functions of different weights'',
quant-ph/0410043.

\item {\bf [Ma\^{\i}tre-Hagley-Nogues-(+5) 97]}:
X. Ma\^{\i}tre, E. Hagley, G. Nogues, C. Wunderlich, P. Goy,
M. Brune, J.-M. Raimond, \& S. Haroche,
``Quantum memory with a single photon in a cavity'',
{\em Phys. Rev. Lett.} {\bf 79}, 4, 769-772 (1997).
Reprinted in {\bf [Macchiavello-Palma-Zeilinger 00]}, pp.~298-301.

\item {\bf [Ma\^{\i}tre-Hagley-Dreyer-(+5) 97]}:
X. Ma\^{\i}tre, E. Hagley, J. Dreyer, A. Maali, C. Wunderlich,
M. Brune, J.-M. Raimond, \& S. Haroche,
``?'',
{\em J. Mod. Opt.} {\bf 44}, 11-12 (Special issue: Quantum
state preparation and measurement), 2023-2032 (1997).

\item {\bf [Majewski 97]}:
A. W. Majewski,
``Separable and entangled states of
composite quantum systems; Rigorous description'',
quant-ph/9711051.

\item {\bf [Majewski 00]}:
A. W. Majewski,
``Some remarks on separability of states'',
quant-ph/0003007.

\item {\bf [Majewski 01]}:
A. W. Majewski,
``On the measure of entanglement'',
quant-ph/0101030.

\item {\bf [Majewski 02]}:
A. W. Majewski,
``On entanglement of formation'',
{\em J. Phys. A} {\bf 35}, 1, 123-134 (2002).

\item {\bf [Majewski 04]}:
A. W. Majewski,
``Positive maps, states, entanglement and all that; some old and new
problems'',
quant-ph/0411043.

\item {\bf [Majorana 32]}:
E. Majorana,
``Atomi orientati in campo magnetico variabile'',
{\em Nuovo Cimento} {\bf 9}, 43-50 (1932).

\item {\bf [Majtey-Lamberti-Martin-Plastino 04]}:
A. P. Majtey, P. W. Lamberti, M. T. Martin, \& A. Plastino,
``Wootters' distance revisited: A new distinguishability criterium'',
quant-ph/0408082.

\item {\bf [Majumdar-Home 02]}:
A. S. Majumdar, \& D. Home,
``Interpreting the measurement of
time of decay: phenomenological significance of the Bohm model'',
{\em Phys. Lett. A} {\bf 296}, 4-5, 176-180 (2002);
quant-ph/0009019.

\item {\bf [Makhlin-Sch\"{o}n-Shnirman 99]}:
Y. Makhlin, G. Sch\"{o}n, \& A. Shnirman,
``Josephson-junction qubits with controlled couplings'',
{\em Nature} {\bf 398}, 6725, 305-307 (1999).
Reprinted in {\bf [Macchiavello-Palma-Zeilinger 00]}, pp.~395-397.

\item {\bf [Makhlin-Sch\"{o}n-Shnirman 00]}:
Y. Makhlin, G. Sch\"{o}n, \& A. Shnirman,
``Josephson-junction qubits'',
{\em Fortschr. Phys.} {\bf 48}, 9-11 (Special issue:
Experimental proposals for quantum computation), 1043-1054 (2000).

\item {\bf [Makhlin 00]}:
Y. Makhlin,
``Nonlocal properties of two-qubit gates and mixed states
and optimization of quantum computations'',
quant-ph/0002045.

\item {\bf [Makkaveev-Pomozov-Molotkov 04]}:
A. P. Makkaveev, D. I. Pomozov, \& S. N. Molotkov,
 ``Simple estimate of the critical length of a quantum communication channel
 with attenuation for coherent-state quantum cryptography'',
{\em JETP Lett.} {\bf 79}, ?, 510-514 (2004).

\item {\bf [Malley 98]}:
J. D. Malley,
``Quantum conditional probability and hidden-variables models'',
{\em Phys. Rev. A} {\bf 58}, 2, 812-820 (1998).

\item {\bf [Malley 04]}:
J. D. Malley,
``All quantum observables in a hidden-variables model must commute
simultaneously'',
{\em Phys. Rev. A};
quant-ph/0402126.

\item {\bf [Mana 03]}:
P. G. L. Mana,
``The properties of the Shannon entropy are not violated in quantum
measurements'',
quant-ph/0302049.
See {\bf [Brukner-Zeilinger 01 a]}.

\item {\bf [Mana 04]}:
P. G. L. Mana,
``Consistency of the Shannon entropy in quantum experiments'',
{\em Phys. Rev. A} {\bf 69}, 6, 062108 (2004);
quant-ph/0302049.

\item {\bf [Manabu-Imai 04]}:
H. Manabu, \& H. Imai,
``Non stabilizer Clifford codes with qupit'',
quant-ph/0402060.

\item {\bf [Mancini-Martins-Tombesi 00]}:
S. Mancini, A. M. Martins, \& P. Tombesi,
``Quantum logic with a single trapped electron'',
{\em Phys. Rev. A} {\bf 61}, 1, 012303 (2000);
quant-ph/9912008.

\item {\bf [Mancini-Bonifacio 01 a]}:
S. Mancini, \& R. Bonifacio,
``Nondissipative decoherence bounds on quantum computation'',
{\em Phys. Rev. A} {\bf 63}, 3, 032310 (2001);
quant-ph/0012057.

\item {\bf [Mancini 01]}:
S. Mancini,
``Ponderomotive entanglement purification'',
{\em Phys. Lett. A} {\bf 279}, 1-2, 1-6 (2001);
quant-ph/0012058.

\item {\bf [Mancini-Bose 01]}:
S. Mancini, \& S. Bose,
``Ponderomotive entangling of atomic motions'',
{\em Phys. Rev. A} {\bf 64}, 3, 032308 (2001);
quant-ph/0105003.

\item {\bf [Mancini-Bonifacio 01 b]}:
S. Mancini, \& R. Bonifacio,
``Quantum Zeno-like effect due to competing decoherence mechanisms'',
{\em Phys. Rev. A} {\bf 64}, 4, 042111 (2001);
quant-ph/0106137.

\item {\bf [Mancini-Man'ko 01]}:
S. Mancini, \& V. I. Man'ko,
``The survival of quantum coherence in deformed states
superposition'',
to appear in {\em Europhys. Lett.};
quant-ph/0103133.

\item {\bf [Mancini-Vitali-Bonifacio-Tombesi 01]}:
S. Mancini, D. Vitali, R. Bonifacio, \& P. Tombesi,
``Stochastic control of quantum coherence'',
quant-ph/0108011.

\item {\bf [Mancini-Man'ko 01]}:
S. Mancini, \& V. I. Man'ko,
``Deformed versus undeformed cat states encoding qubit'',
{\em J. Opt. B: Quantum Semiclass. Opt.};
quant-ph/0111128.

\item {\bf [Mancini-Bonifacio 01]}:
S. Mancini, \& R. Bonifacio,
``Temporal imperfections building up correcting codes'',
{\em J. Mod. Opt.};
quant-ph/0111129.

\item {\bf [Mancini-Giovannetti-Vitali-Tombesi 02]}:
S. Mancini, V. Giovannetti, D. Vitali, \& P. Tombesi,
``Entangling macroscopic oscillators exploiting radiation pressure'',
{\em Phys. Rev. Lett.} {\bf 88}, 12, 120401 (2002);
quant-ph/0108044.

\item {\bf [Mancini-Vitali-Tombesi-Bonifacio 02]}:
S. Mancini, D. Vitali, P. Tombesi, \& R. Bonifacio,
``Preserving quantum coherence via random modulation'',
in R. Bonifacio, \& D. Vitali (eds.),
{\em Mysteries, Puzzles and Paradoxes in Quantum Mechanics IV:
Quantum Interference Phenomena (Gargnano, Italy, 2001)},
{\em J. Opt. B: Quantum Semiclass. Opt.} {\bf 4}, 4, S300-S306 (2002).

\item {\bf [Mancini-Tombesi 03]}:
S. Mancini, \& P. Tombesi,
``High-sensitivity force measurement using entangled probes'',
{\em Europhys. Lett.} {\bf 61}, ?, 8-? (2003);
quant-ph/0109138.

\item {\bf [Mancini-Man'ko-Shchukin-Tombesi 03]}:
S. Mancini, V. I. Man'ko, E. V. Shchukin, \& P. Tombesi,
``A tomographic approach to quantum nonlocality'',
{\em J. Opt. B: Quantum Semiclass. Opt.};
quant-ph/0302089.

\item {\bf [Mancini-Vitali-Tombesi 03]}:
S. Mancini, D. Vitali, \& P. Tombesi,
``Scheme for teleportation of quantum states onto a mechanical resonator'',
{\em Phys. Rev. Lett.} {\bf 90}, 13, 137901 (2003);
quant-ph/0303011.

\item {\bf [Mancini-Giovannetti-Vitali-Tombesi 03]}:
S. Mancini, V. Giovannetti, D. Vitali, \& P. Tombesi,
``Entanglement from ponderomotive interaction'',
{\em Opt. Spectrosc.} {\bf 94}, 711 (2003).

\item {\bf [Mancini-Bose 04]}:
S. Mancini, \& S. Bose,
``Engineering an interaction and entanglement between distant atoms'',
{\em Phys. Rev. A} {\bf 70}, 2, 022307 (2004);
quant-ph/0111055.

\item {\bf [Mancini-Wang 04]}:
S. Mancini, \& J. Wang,
``Towards feedback control of entanglement'',
quant-ph/0410212.

\item {\bf [Mandel 83]}:
L. Mandel,
``Is a photon amplifier always polarization dependent?'',
{\em Nature} {\bf 304}, 5922, 188 (1983).
See {\bf [Wootters-Zurek 82, 83]}.

\item {\bf [Mandel 91]}:
L. Mandel,
``?'',
{\em Opt. Lett.} {\bf 16}, ?, 1882-? (1991).

\item {\bf [Mandel 93]}:
L. Mandel,
``Interference, indistinguishability, and nonlocality'',
in A. van der Merwe, \& F. Selleri (eds.),
{\em Bell's theorem and the foundations of modern physics.
Proc.\ of an international
conference (Cesena, Italy, 1991)},
World Scientific, Singapore, 1993, pp.~347-361.

\item {\bf [Mandel 97]}:
L. Mandel,
``Evidence for the failure of local realism based
on the Hardy-Jordan approach'',
in {\bf [Cohen-Horne-Stachel 97 a]}.

\item {\bf [Mandel-Greiner-Widera-(+3) 03]}:
O. Mandel, M. Greiner, A. Widera,
T. Rom, T. W. H\"{a}nsch, \& I. Bloch,
``Controlled collisions for multi-particle
entanglement of optically trapped atoms'',
{\em Nature} {\bf 425}, 6961, 937-940 (2003).

\item {\bf [Maneva-Smolin 00]}:
E. N. Maneva, \& J. A. Smolin,
``Improved two-party and multi-party purification protocols'',
to be published in S. J. Lomonaco, Jr., (ed.),
{\em Quantum computation and quantum information science},
AMS ``Contemporary Math'' series;
quant-ph/0003099.

\item {\bf [Mang-Xiwen-Kelin-Lei 00]}:
F. Mang, Z. Xiwen, G. Kelin, \& S. Lei,
``Quantum computing by pairing trapped ultracold ions'',
quant-ph/0011001.

\item {\bf [Manin 99]}:
Y. I. Manin,
``Classical computing, quantum computing, and Shor's factoring
algorithm'',
quant-ph/9903008.

\item {\bf [Maniscalco-Messina-Napoli-Vitali 01]}:
S. Maniscalco, A. Messina, A. Napoli, \& D. Vitali,
``Decoherence and robustness of parity-dependent entanglement
in the dynamics of a trapped ion'',
{\em J. Opt. B: Quantum Semiclass. Opt.} {\bf 3}, 5, 308-313 (2001).

\item {\bf [Man'ko-Marmo-Sudarshan-Zaccaria 00]}:
V. I. Man'ko, G. Marmo, E. C. G. Sudarshan, \& F. Zaccaria,
``Inner composition law of pure states as a purification of impure
states'',
{\em Phys. Lett. A} {\bf 273}, 1-2, 31-36 (2000);
quant-ph/9910080.

\item {\bf [Man'ko-Man'ko-Vilela Mendes 01]}:
M. A. Man'ko, V. I. Man'ko, \& R. Vilela Mendes,
``Quantum computation by quantumlike systems'',
{\em Phys. Lett. A} {\bf 288}, 3-4, 132-138 (2001).

\item {\bf [Man'ko-Vilela Mendes 02]}:
V. I. Man'ko, \& R. Vilela Mendes,
``Quantum sensitive dependence'',
{\em Phys. Lett. A} {\bf 300}, 4-5, 353-360 (2002).

\item {\bf [Man'ko-Rosa-Vitale 03]}:
V. I. Man'ko, L. Rosa, \& P. Vitale,
``Identical particle states in the probability representation of quantum mechanics'',
{\em J. Phys. A} {\bf 36}, 1, 255-265 (2003).

\item {\bf [Man'ko-Sharapov-Shchukin 03]}:
V. I. Man'ko, V. A. Sharapov, \& E. V. Shchukin,
``Tomographic representation of evolution equation for density matrix of open system'',
{\em Phys. Lett. A} {\bf 309}, 3-4, 176-182 (2003).

\item {\bf [Man'ko-Marmo-Sharapov-Shchukin-Zaccaria 04]}:
V. I. Man'ko, G. Marmo, E. C. G. Sudarshan, \& F. Zaccaria,
``Positive maps of density matrix and a tomographic criterion of entanglement'',
{\em Phys. Lett. A} {\bf 327}, 5-6, 353-364 (2004).

\item {\bf [Mann-Nakamura-Revzen 92]}:
A. Mann, K. Nakamura, \& M. Revzen,
``Bell's inequality for mixed states'',
{\em J. Phys. A} {\bf 25}, 13, L851-L854 (1992).

\item {\bf [Mann-Sanders-Munro 95]}:
A. Mann, B. C. Sanders, \& W. J. Munro,
``Bell's inequality for an entanglement of nonorthogonal states'',
{\em Phys. Rev. A} {\bf 51}, 2, 989-991 (1995).

\item {\bf [Mann-Revzen 96]}:
A. Mann, \& M. Revzen,
{\em The dilemma of Einstein, Podolsky and Rosen---60 years
later},
{\em Annals of the Israel Physical Society, vol. 12}, 1996.
Review: {\bf [Barnett 98]}.

\item {\bf [Mann-Revzen-Santos 98]}:
A. Mann, M. Revzen, \& E. Santos,
``Entanglement, information theory and Bell's inequality'',
{\em Phys. Lett. A} {\bf 238}, 2-3, 85-89 (1998).

\item {\bf [Mao-Averin-Ruskov-Korotkov 04]}:
W. Mao, D. V. Averin, R. Ruskov, \& A. N. Korotkov,
``Mesoscopic quadratic quantum measurements'',
{\em Phys. Rev. Lett.} {\bf 93}, 5, 056803 (2004).

\item {\bf [Marage-Wallenborn 95]}:
P. Marage, \& G. Wallenborn (eds.),
{\em Les conseils Solvay et les d\'{e}buts de la physique moderne},
Universit\'{e} Libre de Bruxelles, 1995.
English version:
{\em The Solvay councils and the birth of modern physics},
Birkh\"{a}user, Basel, 1998.

\item {\bf [Marand-Townsend 95]}:
C. Marand, \& P. D. Townsend,
``Quantum key distribution over distances as long as 30 km'',
{\em Opt. Lett.} {\bf 20}, 10, 1695-1697 (1995).

\item {\bf [March 00]}:
R. H. March.
``The odd quantum'',
{\em Phys. Today} {\bf 53}, 4, 66 (2000).
Review of {\bf [Treiman 99]}.

\item {\bf [du Marchie van Voorthuysen 96]}:
E. H. du Marchie van Voorthuysen,
``Realization of an interaction-free measurement of the presence of
an object in a light beam'',
{\em Am. J. Phys.} {\bf 64}, 12, 1504-1507 (1996).

\item {\bf [Marchildon 00]}:
L. Marchildon,
``No contradictions between Bohmian and quantum mechanics'',
quant-ph/0007068.
See {\bf [Neumaier 00]}, {\bf [Ghose 00 a]}.
Reply: {\bf [Ghose 00 d]}.

\item {\bf [Marchildon 01]}:
L. Marchildon,
``On Bohmian trajectories in two-particle interference devices'',
quant-ph/0101132.
Comment: {\bf [Ghose 01]}.

\item {\bf [Marchildon 03]}:
L. Marchildon,
``Two-particle interference devices and
compatibility of Bohmian and standard quantum mechanics'',
in M. Ferrero (ed.),
{\em Proc. of Quantum Information: Conceptual Foundations,
Developments and Perspectives (Oviedo, Spain, 2002)},
{\em J. Mod. Opt.} {\bf 50}, 6-7, 873-879 (2003).

\item {\bf [Marchildon 04 a]}:
L. Marchildon,
``Remarks on Mohrhoff's interpretation of quantum mechanics'',
{\em Found. Phys.} {\bf 34}, 1, 59-73 (2004);
quant-ph/0303170.

\item {\bf [Marchildon 04 b]}:
L. Marchildon,
``The counterfactual meaning of the ABL rule'',
in A. Khrennikov (ed.),
{\em Quantum Theory: Reconsideration of Foundations - 2 (V\"{a}xj\"{o}, Sweden 2003)},
V\"{a}xj\"{o} University Press, V\"{a}xj\"{o}, Sweden, 2004, pp.~403-412;
quant-ph/0307082.

\item {\bf [Marchildon 04 c]}:
L. Marchildon,
``Why should we interpret quantum mechanics?'',
{\em Found. Phys.} {\bf 34}, 10, 1453-1466 (2004);
quant-ph/0405126.

\item {\bf [Marchiolli-Missori-Roversi 03]}:
M. A. Marchiolli, R. J. Missori, \& J. A. Roversi,
``Qualitative aspects of entanglement in the Jaynes–Cummings model with an external quantum field'',
{\em J. Phys. A} {\bf 36}, 49, 12275-12292 (2003);
quant-ph/0404008.

\item {\bf [Marcikic-de Riedmatten-Tittel-(+3) 02]}:
I. Marcikic, H. de Riedmatten, W. Tittel,
V. Scarani, H. Zbinden, \& N. Gisin,
``Time-bin entangled qubits for quantum communication created by femtosecond
pulses'',
{\em Phys. Rev. A} {\bf 66}, 6, 062308 (2002).

\item {\bf [Marcikic-de Riedmatten-Tittel-(+2) 03]}:
I. Marcikic, H. de Riedmatten, W. Tittel,
H. Zbinden, \& N. Gisin,
``Long-distance teleportation of qubits
at telecommunication wavelengths'',
{\em Nature} {\bf 421}, 6922, 509-514 (2003);
quant-ph/0301178.

\item {\bf [Marcikic-de Riedmatten-Tittel-(+3) 04]}:
I. Marcikic, H. de Riedmatten, W. Tittel,
H. Zbinden, M. Legre, \& N. Gisin,
``Distribution of time-bin qubits over 50 km of optical fiber'',
quant-ph/0404124.

\item {\bf [Mardari 03]}:
G. N. Mardari,
``Inferential which-path determination as a test of locality'',
quant-ph/0312026.

\item {\bf [Marek-Filip 04]}:
P. Marek, \& R. Filip,
``Improved storage of coherent and squeezed states in an imperfect ring cavity'',
{\em Phys. Rev. A} {\bf 70}, 2, 022305 (2004).

\item {\bf [Marecki 04]}:
P. Marecki,
``Application of quantum inequalities to quantum optics'',
{\em Phys. Rev. A} {\bf 66}, 5, 053801 (2002);
quant-ph/0203027.

\item {\bf [Margenau 36]}:
H. Margenau,
``Quantum-mechanical description'',
{\em Phys. Rev.} {\bf 49}, 3, 240-242 (1936).

\item {\bf [Margenau 49]}:
H. Margenau,
``Reality in quantum mechanics'',
{\em Philos. Sci.} {\bf 16}, ?, 287-302 (1949).

\item {\bf [Margenau 50]}:
H. Margenau,
{\em The nature of physical reality},
McGraw-Hill, New York, 1950.

\item {\bf [Margenau 54]}:
H. Margenau,
``Advantages and disadvantages of various interpretations
of the quantum theory'',
{\em Phys. Today} {\bf 7}, 10, 6-13 (1954).

\item {\bf [Margenau 58]}:
H. Margenau,
``Philosophical problems concerning the meaning of
measurement in physics'',
{\em Philos. Sci.} {\bf 25}, ?, 23-33 (1958).

\item {\bf [Margenau 63 a]}:
H. Margenau,
``Measurement and quantum states. Part I'',
{\em Philos. Sci.} {\bf 11}, 1, 1-16 (1963).

\item {\bf [Margenau 63 b]}:
H. Margenau,
``Measurement and quantum states. Part II'',
{\em Philos. Sci.} {\bf 11}, ?, 138-157 (1963).

\item {\bf [Margenau 63 c]}:
H. Margenau,
``Measurements in quantum mechanics'',
{\em Ann. Phys.} {\bf 23}, ?, 469-485 (1963).

\item {\bf [Margenau-Park 73]}:
H. Margenau, \& J. L. Park,
``The physics and the semantics of quantum measurement'',
{\em Found. Phys.} {\bf 3}, 1, 19-28 (1973).

\item {\bf [Margolin-Strazhev-Tregubovich 01]}:
A. E. Margolin, V. I. Strazhev, \& A. Y. Tregubovich,
``Remarks on the Halting problem, unitarity and
reversibility in quantum theory of computations'',
quant-ph/0106108.

\item {\bf [Margolin-Strazhev-Tregubovich 01]}:
A. E. Margolin, V. I. Strazhev, \& A. Y. Tregubovich,
``On non-holonomic quantum computer'',
quant-ph/0110015.

\item {\bf [Margolin-Strazhev-Tregubovich 02]}:
A. E. Margolin, V. I. Strazhev, \& A. Y. Tregubovich,
``Geometric phases and quantum computations'',
{\em Phys. Lett. A} {\bf 303}, 2-3, 131-134 (2002);
quant-ph/0102030.

\item {\bf [Margolin-Strazhev-Tregubovich 03]}:
A. E. Margolin, V. I. Strazhev, \& A. Y. Tregubovich,
``On non-adiabatic holonomic quantum computer'',
{\em Phys. Lett. A} {\bf 312}, 5-6, 296-300 (2003).

\item {\bf [Marian-Marian-Scutaru 03]}:
P. Marian, T. A. Marian, \& H. Scutaru,
``Bures distance as a measure of entanglement for two-mode squeezed thermal states'',
{\em Phys. Rev. A} {\bf 68}, 6, 062309 (2003).

\item {\bf [Mariano-Facchi-Pascazio 01]}:
A. Mariano, P. Facchi, \& S. Pascazio,
``Decoherence and fluctuations in quantum
interference experiments'',
quant-ph/0106027.

\item {\bf [Marinatto-Weber 00 a]}:
L. Marinatto, \& T. Weber,
``A quantum approach to static games of complete information'',
{\em Phys. Lett. A} {\bf 272}, 5-6, 291-303 (2000);
quant-ph/0004081.
Comment: {\bf [Benjamin 00 c]}.
Reply: {\bf [Marinatto-Weber 00 c]}.

\item {\bf [Marinatto-Weber 00 b]}:
L. Marinatto, \& T. Weber,
``Which kind of two-particle states can be teleported
through a three-particle quantum channel?'',
{\em Found. Phys. Lett.} {\bf 13}, 2, 119-132 (2000);
quant-ph/0004054.

\item {\bf [Marinatto-Weber 00 c]}:
L. Marinatto, \& T. Weber,
`Reply to ``Comment on: A quantum approach
to static games of complete information''\,',
quant-ph/0009103.
Reply to {\bf [Benjamin 00 c]}.
See {\bf [Marinatto-Weber 00 a]}.

\item {\bf [Marinatto-Weber 01]}:
L. Marinatto, \& T. Weber,
``Teleportation with indistinguishable particles'',
{\em Phys. Lett. A} {\bf 287}, 1-2, 1-6 (2001);
quant-ph/0102025.

\item {\bf [Marinatto 03]}:
L. Marinatto,
`Comment on ``Bell's theorem without inequalities and without
probabilities for two observers''\,',
{\em Phys. Rev. Lett.} {\bf 90}, 25, 258901 (2003);
quant-ph/0306171.
Comment on {\bf [Cabello 01 c]}.
Reply: {\bf [Cabello 03 f]}.

\item {\bf [Marinatto 04]}:
L. Marinatto,
`Comment on ``Bell's
theorem without inequalities and without alignments''\,',
{\em Phys. Rev. Lett.} {\bf 93}, 12, 128901 (2004);
quant-ph/0409129.
Comment on {\bf [Cabello 03 i]}.
Reply: {\bf [Cabello 04 b]}.

\item {\bf [Markham-Vedral 03]}:
D. Markham, \& V. Vedral,
``Classicality of spin-coherent states via entanglement and distinguishability'',
{\em Phys. Rev. A} {\bf 67}, 4, 042113 (2003).

\item {\bf [Markham-Murao-Vedral 03]}:
D. Markham, M. Murao, \& V. Vedral,
``Entanglement generation from thermal spin states via unitary beam
splitters'',
quant-ph/0307147.

\item {\bf [Marmet 93]}:
P. Marmet,
``Quantum mechanics and its paradox: A realistic
solution to Mermin's EPR apparatus'',
{\em Phys. Essays} {\bf 6}, 3, 436-439 (1993).
See {\bf [Mermin 81 a, b, 85]}.

\item {\bf [Maroney-Hiley 99]}:
O. Maroney, \& B. J. Hiley,
``Quantum state teleportation understood through the Bohm
interpretation'',
{\em Found. Phys.} {\bf 29}, 9, 1403-1416 (1999).

\item {\bf [Marr-Beige-Rempe 03]}:
C. Marr, A. Beige, \& G. Rempe,
``Entangled-state preparation via dissipation-assisted adiabatic passages'',
{\em Phys. Rev. A} {\bf 68}, 3, 033817 (2003);
quant-ph/0305116.

\item {\bf [Marsh 01]}:
J. S. Marsh,
``Explicit measurement theory for quantum mechanics'',
{\em Phys. Rev. A} {\bf 64}, 4, 042109 (2001).

\item {\bf [Marshall-Santos-Selleri 83]}:
T. W. Marshall, E. Santos, \& F. Selleri,
``Local realism has not been refuted by atomic cascade experiments'',
{\em Phys. Lett. A} {\bf 98}, 1-2, 5-9 (1983).

\item {\bf [Marshall-Santos 85]}:
T. W. Marshall, \& E. Santos,
in ``Reality and the quantum theory'',
{\em Phys. Today} {\bf 38}, 11, 9-11 (1985).
Comment on {\bf [Mermin 85]}.

\item {\bf [Marshall-Santos 87]}:
T. W. Marshall, \& E. Santos,
``Comment on `Experimental evidence for a photon anticorrelation effect
on a beam splitter: A new light on single-photon interferences'\,'',
{\em Europhys. Lett.} {\bf 3}, 3, 293-296 (1987).
Comment on {\bf [Grangier-Roger-Aspect 86]}.

\item {\bf [Marshall-Santos 89]}:
T. W. Marshall, \& E. Santos,
``Stochastic optics: A local realistic analysis of optical
tests of Bell inequalities'',
{\em Phys. Rev. A} {\bf 39}, 12, 6271-6283 (1989).

\item {\bf [Marshall 91]}:
T. W. Marshall,
``What does noise do to the Bell inequalities?'',
{\em Found. Phys.} {\bf 21}, 2, 209-219 (1991).

\item {\bf [Marshall-Santos 97]}:
T. W. Marshall, \& E. Santos,
``The myth of the photon'',
in S. Jeffers, S. Roy, \& J.-P. Vigier (eds.),
{\em The present status of the quantum theory of light:
Proc.\ of a symposium in honour of Jean-Pierre Vigier},
Kluwer Academic, Dordrecht, Holland, 1997, pp.~67-77;
quant-ph/9711046.

\item {\bf [Marshall 97 a]}:
T. W. Marshall,
``Do we need photons in parametric down conversion?'',
quant-ph/9711029.
See {\bf [Marshall 97 c, d]}.

\item {\bf [Marshall 97 b]}:
T. W. Marshall,
``A local realistic theory of parametric down conversion'',
quant-ph/9711030.

\item {\bf [Marshall 97 c]}:
T. W. Marshall,
``The myth of the down converted photon'',
quant-ph/9712001.
See {\bf [Marshall 97 a, d]}.

\item {\bf [Marshall 97 d]}:
T. W. Marshall,
``The zeropoint field---no longer a ghost'',
quant-ph/9712050.
See {\bf [Marshall 97 a, c]}.

\item {\bf [Marshall 98]}:
T. W. Marshall,
``Parametric up conversion of the vacuum'',
quant-ph/9803054.

\item {\bf [Marshall 02]}:
T. W. Marshall,
``Nonlocality - The party may be over'',
quant-ph/0203042.

\item {\bf [Marshall 03]}:
T. W. Marshall,
``Are atoms waves or particles?'',
quant-ph/0409203.

\item {\bf [Martens-de Muynck 93]}:
H. Martens, \& W. M. de Muynck,
``Single and joint spin measurements with a Stern-Gerlach device'',
{\em J. Phys. A} {\bf 26}, 8, 2001-2010 (1993).
See {\bf [Martens-de Muynck 94]}.

\item {\bf [Martens-de Muynck 94]}:
H. Martens, \& W. M. de Muynck,
``On the possibility of measuring the electron spin in an
inhomogeneous magnetic field'',
{\em Found. Phys. Lett.} {\bf 7}, 4, 343-352 (1994).
See {\bf [Martens-de Muynck 93]}.

\item {\bf [Mart\'{\i}n Delgado-Navascues 03 a]}:
M. A. Mart\'{\i}n-Delgado, \& M. Navascues,
``Distillation protocols for mixed states of multilevel qubits and the
quantum renormalization group'',
quant-ph/0301099.

\item {\bf [Mart\'{\i}n Delgado-Navascues 03 b]}:
M. A. Mart\'{\i}n-Delgado, \& M. Navascues,
``Single-step distillation protocol with generalized beam splitters'',
{\em Phys. Rev. A} {\bf 68}, 1, 012322 (2003);
quant-ph/0305138.

\item {\bf [Mart\'{\i}nez Linares-Vargas Medina 03]}:
J. Mart\'{\i}nez Linares, \& J. Vargas-Medina,
``The quantum self-eraser'',
quant-ph/0309005.

\item {\bf [Mart\'{\i}nez Linares-Harmin 04]}:
J. Mart\'{\i}nez Linares, \& D. A. Harmin,
```Quality'' of a which-way detector',
{\em Phys. Rev. A} {\bf 69}, 6, 062109 (2004).
See {\bf [Englert 96 b]}.

\item {\bf [De Martini 98 a]}:
F. De Martini,
``Teleportation: Who was first?'',
{\em Phys. World} {\bf 11}, 3, 23-24 (1998).
See {\bf [Zeilinger 98 a]}.

\item {\bf [De Martini 98 b]}:
F. De Martini,
``Quantum superposition of parametrically amplified
multiphoton pure states'',
{\em Phys. Lett. A} {\bf 250}, 1-3, 15-19 (1998).

\item {\bf [De Martini 98 c]}:
F. De Martini,
``Amplification of quantum entanglement'',
{\em Phys. Rev. Lett.} {\bf 81}, 14, 2842-2845 (1998).

\item {\bf [De Martini-Mussi 00]}:
F. De Martini, \& V. Mussi,
``Universal quantum cloning and macroscopic superposition in parametric
amplification'',
{\em Fortschr. Phys.} {\bf 48}, 5-7, 413-422 (2000).

\item {\bf [De Martini-Mussi-Bovino 00]}:
F. De Martini, V. Mussi, \& F. Bovino,
``Schr\"{o}dinger cat states and optimum universal quantum cloning
by entangled parametric amplification'',
{\em Opt. Comm.} {\bf 179}, ?, 581-589 (2000).

\item {\bf [De Martini-Di Giuseppe-P\'{a}dua 01]}:
F. De Martini, G. Di Giuseppe, S. P\'{a}dua,
``Multiparticle quantum superposition and stimulated entanglement
by parity selective amplification of entangled states'',
{\em Phys. Rev. Lett.} {\bf 87}, 15, 150401 (2001);
quant-ph/0011081.

\item {\bf [De Martini-Bu\v{z}zek-Sciarrino-Sias 02]}:
F. De Martini, V. Bu\v{z}zek, F. Sciarrino, \& C. Sias,
``Experimental realization of the quantum universal NOT gate'',
{\em Nature} {\bf 419}, 6909, 815-818 (2002).
See {\bf [Gisin 02 b]}.

\item {\bf [De Martini-Mazzei-Ricci-D'Ariano 03]}:
F. De Martini, A. Mazzei, M. Ricci, \& G. M. D'Ariano,
``Exploiting quantum parallelism of entanglement for a complete experimental
quantum characterization of a single-qubit device'',
{\em Phys. Rev. A} {\bf 67}, 6, 062307 (2003).

\item {\bf [De Martini-Sciarrino-Secondi 03]}:
F. De Martini, F. Sciarrino, \& V. Secondi,
``A qubit on the Faraday mirror'',
quant-ph/0312113.

\item {\bf [De Martini-Pelliccia-Sciarrino 04]}:
F. De Martini, D. Pelliccia, \& F. Sciarrino,
``Contextual, optimal and universal realization of the quantum cloning
machine and of the NOT gate'',
{\em Phys. Rev. Lett.} {\bf 92}, 6, 067901 (2004);
quant-ph/0312124.

\item {\bf [De Martini-Sciarrino 04]}:
F. De Martini, \& F. Sciarrino,
``Realization of a classically distinguishable multi-photon quantum
superposition'',
quant-ph/0410225.

\item {\bf [Martinis-Devoret-Clarke 00]}:
J. M. Martinis, M. H. Devoret, \& J. Clarke,
``Experimental tests of the quantum behavior of
a macroscopic degree of freedom: The phase difference
across a Josephson junction'',
{\em Phys. Rev. B} {\bf 35}, 10, 4682-4698 (1987).

\item {\bf [Martinis-Nam-Aumentado-(+2) 03]}:
J. M. Martinis, S. Nam, J. Aumentado, K. M. Lang, \& C. Urbina,
``Decoherence of a superconducting qubit due to bias noise'',
{\em Phys. Rev. B} {\bf 67}, 9, 094510 (2003).

\item {\bf [Martins 02]}:
A. M. Martins,
``Proposed demonstration of the Einstein-Poldosky-Rosen paradox using trapped
electrons'',
{\em Phys. Rev. A} {\bf 65}, 5, 052114 (2002).

\item {\bf [Marto-Croca 02]}:
J. Marto, \& J. R. Croca,
``Non-local interferometry: A causal explanation by
means of local wavelet analysis'',
{\em Found. Phys.} {\bf 32}, 7, 1091-1109 (2002).

\item {\bf [Martonak-Santoro-Tosatti 04]}:
R. Martonak, G. E. Santoro, \& E. Tosatti,
``Quantum annealing of the traveling salesman problem'',
cond-mat/0402330.

\item {\bf [Maruyama-Knight 03 a]}:
K. Maruyama, \& P. L. Knight,
``Upper bounds for the number of quantum clones under decoherence'',
{\em Phys. Rev. A} {\bf 67}, 3, 032303 (2003).

\item {\bf [Maruyama-Knight 03 b]}:
K. Maruyama, \& P. L. Knight,
``Quantum state restoration by quantum cloning and measurement'',
quant-ph/0309167.

\item {\bf [Maruyama-Morikoshi-Vedral 03]}:
K. Maruyama, F. Morikoshi, \& V. Vedral,
``Entangled Maxwell's demons violating thermodynamical separability
criteria'',
quant-ph/0311083.

\item {\bf [Maruyama-Brukner-Vedral 04]}:
K. Maruyama, \v{C}. Brukner, \& V. Vedral,
``Thermodynamical cost of accessing quantum information'',
quant-ph/0407151.

\item {\bf [Marx-Fahmy-Myers-(+2) 99]}:
R. Marx, A. F. Fahmy, J. M. Myers, W. Bermel, \& S. J. Glaser,
``Realization of a 5-bit NMR quantum computer using a new molecular architecture'',
quant-ph/9905087.

\item {\bf [Marx-Fahmy-Myers-(+2) 00]}:
R. Marx, A. F. Fahmy, J. M. Myers,
W. Bermel, \& S. J. Glaser,
``Approaching five-bit NMR quantum computing'',
{\em Phys. Rev. A} {\bf 62}, 1, 012310 (2000).

\item {\bf [Marzlin-Bartlett-Sanders 03]}:
K.-P. Marzlin, S. D. Bartlett, \& B. C. Sanders,
``Entanglement gauge and the non-Abelian geometric phase with two photonic
qubits'',
{\em Phys. Rev. A} {\bf 67}, 2, 022316 (2003).

\item {\bf [Masanes-Vidal-Latorre 02]}:
L. Masanes, G. Vidal, \& J. I. Latorre,
``Time-optimal Hamiltonian simulation and gate
synthesis using homogeneous local unitaries'',
quant-ph/0202042.

\item {\bf [Masanes 03 a]}:
L. Masanes,
``Tight Bell inequality for $d$-outcome measurements correlations'',
{\em Quant. Inf. Comp.} {\bf 3}, 4, 345-358 (2003).

\item {\bf [Masanes 03 b]}:
L. Masanes,
``Necessary and sufficient condition for quantum-generated correlations'',
quant-ph/0309137.

\item {\bf [Masato-Nobuyuki 98]}:
K. Masato, \& I. Nobuyuki,
``Configuration method for quantum cryptography'',
patent JP10322329A, 1998.

\item {\bf [Mashkevich 98 a]}:
V. S. Mashkevich,
``On Stapp-Unruh-Mermin
dscussion on quantum nonlocality: Quantum jumps and relativity'',
quant-ph/9801032.
See {\bf [Stapp 97 a]}.

\item {\bf [Mashkevich 98 b]}:
V. S. Mashkevich,
``Comment on `Quantum
entanglement and the nonexistence of superluminal signals'\,'',
quant-ph/9801062.
Comment on {\bf [Westmoreland-Schumacher 98]}.
See {\bf [Mashkevich 98 b]}.

\item {\bf [Mashkevich 98 c]}:
V. S. Mashkevich,
``On quantum nonlocality: Using
prediction of a distant measurement outcome'',
quant-ph/9802071.

\item {\bf [Mashkevich 98 d]}:
V. S. Mashkevich,
``Another comment on `Quantum
entanglement and the nonexistence of superluminal signals'\,'',
quant-ph/9803027.
Comment on {\bf [Westmoreland-Schumacher 98]}.
See {\bf [Mashkevich 98 b]}.

\item {\bf [Masiak-Knight 98]}:
M. Masiak, \& P. L. Knight,
``Copying of entangled states and the degradation of correlations'',
quant-ph/9808043.

\item {\bf [Masiak 02]}:
P. Masiak,
``Quantum-entanglement production in a micromaser'',
{\em Phys. Rev. A} {\bf 66}, 2, 023804 (2002).

\item {\bf [Masiak 03]}:
P. Masiak,
``Entanglement preservation in quantum cloning'',
{\em J. Mod. Opt.} {\bf 50}, 12, 1873-1881 (2003);
quant-ph/0309019.

\item {\bf [Maslov-Dueck 03]}:
D. Maslov, \& G. W. Dueck,
``Improved quantum cost for $n$-bit Toffoli gates'',
{\em IEE Electronics Lett.} {\bf 39}, 25, 1790-1791 (2003);
quant-ph/0403053.

\item {\bf [Maslennikov-Zhukov-Chekhova-Kulik 03]}:
G. A. Maslennikov, A. A. Zhukov, M. V. Chekhova, \& S. P. Kulik,
``Practical realization of the quantum cryptography protocol exploiting
polarization encoding in the qutrits'',
quant-ph/0305115.

\item {\bf [Massad 98]}:
J. E. Massad,
``The Penrose dodecahedron'',
Project report, Worcester Polytechnic Institute.
See {\bf [Zimba-Penrose 93]}, {\bf [Massad-Aravind 99]}.

\item {\bf [Massad-Aravind 99]}:
J. E. Massad, \& P. K. Aravind,
``The Penrose dodecahedron revisited'',
{\em Am. J. Phys.} {\bf 67}, 7, 631-638 (1999).
See {\bf [Zimba-Penrose 93]}, {\bf [Massad 98]}.

\item {\bf [Massar-Popescu 95]}:
S. Massar, \& S. Popescu,
``Optimal extraction from finite quantum ensambles'',
{\em Phys. Rev. Lett.} {\bf 74}, 8, 1259-1263 (1995).

\item {\bf [Massar-Popescu 00]}:
S. Massar, \& S. Popescu,
``Amount of information obtained by a quantum measurement'',
{\em Phys. Rev. A} {\bf 61}, 6, 062303 (2000);
quant-ph/9907066.

\item {\bf [Massar 00]}:
S. Massar,
``Collective versus local measurements on
two parallel or antiparallel spins'',
{\em Phys. Rev. A} {\bf 62}, 4, 040101(R) (2000);
quant-ph/0004035.

\item {\bf [Massar-Bacon-Cerf-Cleve 01]}:
S. Massar, D. Bacon, N. Cerf, \& R. Cleve,
``Classical simulation of quantum entanglement without local hidden
variables'',
{\em Phys. Rev. A} {\bf 63}, 5, 052305 (2001);
quant-ph/0009088.

\item {\bf [Massar-Pironio 01]}:
S. Massar, \& S. Pironio,
``Greenberger-Horne-Zeilinger paradox for continuous variables'',
{\em Phys. Rev. A} {\bf 64}, 6, 062108 (2001);
quant-ph/0103048.

\item {\bf [Massar-Pironio-Roland-Gisin 02]}:
S. Massar, S. Pironio, J. Roland, \& B. Gisin,
``Bell inequalities resistant to detector inefficiency'',
{\em Phys. Rev. A} {\bf 66}, 5, 052112 (2002).

\item {\bf [Massar-Mitchison-Pironio 01]}:
S. Massar, G. Mitchison, \& S. Pironio,
``Minimal absorption measurements'',
{\em Phys. Rev. A} {\bf 64}, 6, 062303 (2001);
quant-ph/0102116.

\item {\bf [Massar 02]}:
S. Massar,
``Nonlocality, closing the detection loophole, and communication complexity'',
{\em Phys. Rev. A} {\bf 65}, 3, 032121 (2002);
quant-ph/0109008.

\item {\bf [Massar-Polzik 03]}:
S. Massar, \& E. S. Polzik,
``Generating a superposition of spin states in an atomic ensemble'',
{\em Phys. Rev. Lett.} {\bf 91}, 6, 060401 (2003);
quant-ph/0306121.

\item {\bf [Massar-Pironio 03]}:
S. Massar, \& S. Pironio,
``Violation of local realism versus detection efficiency'',
{\em Phys. Rev. A} {\bf 68}, 6, 062109 (2003).

\item {\bf [Massar 03]}:
S. Massar,
``Quantum fingerprinting with a single particle'',
quant-ph/0305112.

\item {\bf [Massen-Moustakidis-Panos 02]}:
S. E. Massen, C. C. Moustakidis, \& C. P. Panos,
``Comparison of the information entropy in fermionic and bosonic
systems'',
{\em Phys. Lett. A} {\bf 299}, 2-3, 131-136 (2002).

\item {\bf [Massini-Fortunato-Mancini-(+2) 00]}:
M. Massini, M. Fortunato, S. Mancini,
P. Tombesi, \& D. Vitali,
``Schr\"{o}dinger-cat entangled state reconstruction
in the Penning trap'',
{\em New J. Phys.} {\bf 2}, 20.1-20.14 (2002);
quant-ph/0007076.

\item {\bf [Massini-Fortunato-Mancini-Tombesi 00]}:
M. Massini, M. Fortunato, S. Mancini, \& P. Tombesi,
``Synthesis and characterization of entangled
mesoscopic superpositions for a trapped electron'',
{\bf Phys. Rev. A} {\bf 62}, 4, 041401 (2000);
quant-ph/0105131.

\item {\bf [Mataloni-Giorgi-De Martini 02]}:
P. Mataloni, G. Giorgi, \& F. De Martini,
``Frequency hopping in quantum interferometry:
Efficient up-down conversion for qubits and ebits'',
quant-ph/0201116.

\item {\bf [Matteucci-Beeli 98]}:
G. Matteucci, \& C. Beeli,
``An experiment on electron wave-particle duality including a
Planck constant measurement'',
{\em Am. J. Phys.} {\bf 66}, 12, 1055-1059 (1998).

\item {\bf [Mattle-Michler-Weinfurter-(+2) 95]}:
K. Mattle, M. Michler, H. Weinfurter, A. Zeilinger, \& M. \.{Z}ukowski,
``Nonclassical statistics at multiport beamsplitters'',
{\em Appl. Phys. B} {\bf 60}, 2-3, S111-S117 (1995).

\item {\bf [Mattle-Weinfurter-Kwiat-Zeilinger 96]}:
K. Mattle, H. Weinfurter, P. G. Kwiat, \& A. Zeilinger,
``Dense coding in experimental quantum communication'',
{\em Phys. Rev. Lett.} {\bf 76}, 25, 4656-4659 (1996).

\item {\bf [Matsukevich, \& A. Kuzmich 04]}:
D. N. Matsukevich, \& A. Kuzmich,
``Quantum state transfer between matter and light'',
{\em Science};
quant-ph/0410092.

\item {\bf [Matsumoto-Uyematsu 00]}:
R. Matsumoto, \& T. Uyematsu,
``Constructing quantum error-correcting codes for $p^m$-state
systems from classical error-correcting codes'',
{\em IEICE Trans. on Fundamentals of Electronics, Communications and Computer Sciences}
{\bf E83-A}, 10, 1878-1883 (2000);
quant-ph/9911011.

\item {\bf [Matsumoto 01 a]}:
R. Matsumoto,
``Fidelity of a $t$-error-correcting quantum code with more than $t$ errors'',
{\em Phys. Rev. A} {\bf 64}, 2, 022314 (2001).
Erratum: {\em Phys. Rev. A} {\bf 66}, 5, 059901 (2002).

\item {\bf [Matsumoto 01 b]}:
R. Matsumoto,
``Algebraic geometric construction of a quantum stabilizer code'',
quant-ph/0107128.

\item {\bf [Matsumoto-Uyematsu 02]}:
R. Matsumoto, \& T. Uyematsu,
``Lower bound for the quantum capacity of a discrete memoryless quantum
channel'',
{\em J. Math. Phys.} {\bf 43}, 9, 4391-4403 (2002);
quant-ph/0105151.

\item {\bf [Matsumoto 03]}:
R. Matsumoto,
``Conversion of a general quantum stabilizer code
to an entanglement distillation protocol'',
{\em J. Phys. A} {\bf 36}, 29, 8113–8127 (2003).

\item {\bf [Matsumoto-Shimono-Winter 02]}:
K. Matsumoto, T. Shimono, \& A. Winter,
``Remarks on additivity of the Holevo channel capacity and of the
entanglement of formation'',
quant-ph/0206148.

\item {\bf [Matsuoka-Hirano 03]}:
M. Matsuoka, \& T. Hirano,
``Quantum key distribution with a single photon from a squeezed coherent state'',
{\em Phys. Rev. A} {\bf 67}, 4, 042307 (2003).

\item {\bf [Matzkin 02]}:
A. Matzkin,
``Realism and the wavefunction'',
{\em Eur. J. Phys.} {\bf 23}, 3, 285-294 (2002).

\item {\bf [Maudlin 94]}:
T. Maudlin,
{\em Quantum non-locality and relativity:
Metaphysical intimations of modern physics},
Blackwell, Oxford, 1994.

\item {\bf [Maunz-Puppe-Schuster-(+3) 04]}:
P. Maunz, T. Puppe, I. Schuster,
N. Syassen, P. W. H. Pinkse, \& G. Rempe,
``Cavity cooling of a single atom'',
{\em Nature} {\bf 428}, 6978, 50-52 (2004);
quant-ph/0403033.

\item {\bf [Maurer-Hogg-Huberman 01]}:
S. M. Maurer, T. Hogg, \& B. A. Huberman,
``Portfolios of quantum algorithms'',
{\em Phys. Rev. Lett.} {\bf 87}, 25, 257901 (2001);
quant-ph/0105071.

\item {\bf [Mayburov 99]}:
S. Mayburov,
``Quantum information, irreversibility and state collapse
in some microscopic models of measurement'',
quant-ph/9911105.

\item {\bf [Mayburov 00]}:
S. Mayburov,
``Relational quantum measurements,
information and state collapse'',
quant-ph/0006104.

\item {\bf [Mayburov 00]}:
S. Mayburov,
``Quantum measurements, information, and dual states
representations'',
{\em Proc.\ of 4th Quantum Communications, Computers and Measurements Conf.},
Kluwer Academic, Dordrecht, Holland, 2001;
quant-ph/0103161.

\item {\bf [Mayer 94]}:
M. E. Mayer,
``Quantum theory: Concepts and methods'',
{\em Phys. Today} {\bf 47}, 12, 65-66 (1994).
Review of {\bf [Peres 93 a]}.

\item {\bf [Mayer 96]}:
M. E. Mayer,
``Book review. Lectures on quantum theory.
Mathematical and structural foundations'',
{\em Phys. Today} {\bf 49}, 8, 66 (1996).
Review of {\bf [Isham 95]}.

\item {\bf [Mayers 96 a]}:
D. Mayers,
``Unconditionally secure quantum bit commitment is impossible'',
in {\em 4h Workshop on Physics and Computation---PhysCom '96,
(Boston, 1996)}.

\item {\bf [Mayers 96 b]}:
D. Mayers,
``?'',
in N. Koblitz (ed.),
{\em Advances in Cryptology---Proc.\ of Crypto '96},
Springer-Verlag, New York, 1996, pp.~343-357.
Preliminary version of {\bf [Mayers 01]}.

\item {\bf [Mayers 96 c]}:
D. Mayers,
``The trouble with quantum bit commitment'',
quant-ph/9603015.

\item {\bf [Mayers 97]}:
D. Mayers,
``Unconditionally secure quantum bit commitment is impossible'',
{\em Phys. Rev. Lett.} {\bf 78}, 17, 3414-3417 (1997).

\item {\bf [Mayers-Yao 98]}:
D. Mayers, \& A. Yao,
``Quantum cryptography with imperfect apparatus'',
{\em Proc.\ of the 39th Annual Symposium on Foundations of Computer Science},
IEEE Computer Society, Los Alamitos, California, p.~503;
quant-ph/9809039.

\item {\bf [Mayers-Salvail-Chiba Kohno 99]}:
D. Mayers, L. Salvail, \& Y. Chiba-Kohno,
``Unconditionally secure quantum coin tossing'',
quant-ph/9904078.
Comment: {\bf [Leslau 01]}.

\item {\bf [Mayers 01]}:
D. Mayers,
``Unconditional security in quantum cryptography'',
{\em J. ACM} {\bf 48}, 3, 351-406 (2001);
quant-ph/9802025.
Preliminary version: {\bf [Mayers 96 b]}.

\item {\bf [Mayers 02 a]}:
D. Mayers,
``Shor and Preskill's and Mayers's security proof for the
BB84 quantum key distribution protocol'',
{\em Eur. Phys. J. D} {\bf 18}, 2 (Special issue:
{\em Quantum interference and cryptographic keys:
Novel physics and advancing technologies (QUICK) (Corsica, 2001)}, 161-170 (2002).

\item {\bf [Mayers 02 b]}:
D. Mayers,
``Superselection rules in quantum cryptography'',
quant-ph/0212159.

\item {\bf [Mayers-Yao 04]}:
D. Mayers, \& A. Yao,
``Self testing quantum apparatus'',
{\em Quant. Inf. Comp.} {\bf 4}, 4, 273-286 (2004);
quant-ph/0307205.

\item {\bf [Mazourenko-Merolla-Goedgebuer 98]}:
Y. Mazourenko, J.-M. Merolla, \& J.-P. Goedgebuer,
``Method and apparatus for quantum key distribution'',
patent EP877508A1, 1998.

\item {\bf [Mazzitelli-Paz-Villanueva 03]}:
F. D. Mazzitelli, J. P. Paz, \& A. Villanueva,
``Decoherence and recoherence from vacuum fluctuations near a conducting
plate'',
quant-ph/0307004.

\item {\bf [McAnally 01]}:
D. McAnally,
``A refinement of Shor's algorithm'',
quant-ph/0112055.

\item {\bf [McAneney-Paternostro-Kim 04]}:
H. McAneney, M. Paternostro, \& M. S. Kim,
``Multi-splitter interaction for entanglement distribution'',
quant-ph/0407161.

\item {\bf [McCall 01]}:
S. McCall,
``The Ithaca interpretation of quantum mechanics,
and objective probabilities'',
{\em Found. Phys. Lett.} {\bf 14}, 1, 95-101 (2001).

\item {\bf [McCrea 90]}:
W. McCrea,
``Origin of wave mechanics'',
{\em Contemp. Phys.} {\bf 31}, 1, 43-48 (1990).
Review of {\bf [Mehra-Rechenberg 87 a, b]}.

\item {\bf [McDermid 04]}:
K. McDermid,
``Bell inequality violations by partially entangled spin-1/2 pairs'',
{\em Phys. Lett. A} {\bf 331}, 3-4, 157-163 (2004).

\item {\bf [McElwaine 97]}:
J. McElwaine,
``Maximum information and quantum prediction algorithms'',
{\em Phys. Rev. A} {\bf 56}, 3, 1756-1766 (1997).

\item {\bf [McEvoy-Z\'{a}rate 96]}:
J. P. McEvoy, \& O. Z\'{a}rate,
{\em Quantum theory for beginners},
Icon Books, Cambridge, 1996.
Spanish version:
{\em Teor\'{\i}a cu\'{a}ntica para principiantes},
Era Naciente, SLR, Buenos Aires, Argentina, 1999.

\item {\bf [McGrath 78]}:
J. H. McGrath,
``A formal statement of the Einstein-Podolsky-Rosen argument'',
{\em Int. J. Theor. Phys.} {\bf 17}, 7, 557-571 (1978).

\item {\bf [McGuire-Fry 73]}:
J. H. McGuire, \& E. S. Fry,
``Restrictions on nonlocal hidden-variable theory'',
{\em Phys. Rev. D} {\bf 7}, 2, 555-557 (1973).

\item {\bf [Mc Hugh-Twamley 03]}:
D. Mc Hugh, \& J. Twamley,
``A quantum computer using a trapped-ion spin molecule and microwave
radiation'',
quant-ph/0310015.

\item {\bf [Mc Hugh-Twamley 04]}:
D. Mc Hugh, \& J. Twamley,
``6th order robust pulses for quantum control'',
quant-ph/0404127.

\item {\bf [McKay-Megill-Pavi\v{c}i\'{c} 00]}:
B. D. McKay, N. D. Megill, \& M. Pavi\v{c}i\'{c},
``Algorithms for Greechie diagrams'',
{\em Int. J. Theor. Phys.} {\bf 39}, ?, 2393-2417 (2000);
quant-ph/0009039.

\item {\bf [McKeever-Buck-Boozer-(+4) 03]}:
J. McKeever, J. R. Buck, A. D. Boozer,
A. Kuzmich, H.-C. Nagerl, D. M. Stamper-Kurn, \& H. J. Kimble,
``State-insensitive cooling and trapping of single atoms in an optical cavity'',
{\em Phys. Rev. Lett.} {\bf 90}, 13, 133602 (2003).

\item {\bf [McKeever-Boca-Boozer-(+4) 04]}:
J. McKeever, A. Boca, A. D. Boozer,
R. Miller, J. R. Buck, A. Kuzmich, \& H. J. Kimble,
``Deterministic generation of single photons from one atom trapped in a cavity'',
{\em Science} {\bf 303}, ?, 1992-1995 (2004).

\item {\bf [McKeever-Buck-Boozer-Kimble 04]}:
J. McKeever, J. R. Buck, A. D. Boozer, \& H. J. Kimble,
`Cavity QED ``by the numbers''\,',
quant-ph/0403121.

\item {\bf [McKinnon 94]}:
E. McKinnon,
``Bohr and the realism debates'',
in {\bf [Faye-Folse 94]}, pp.~279-302.

\item {\bf [McKinnon-Leavens 95]}:
W. R. McKinnon, \& C. R. Leavens,
``Distribution of delay times and transmission in
Bohm's causal interpretation of quantum mechanics'',
{\em Phys. Rev. A} {\bf 51}, 4, 2748-2757 (1995).

\item {\bf [Megill-Pavi\v{c}i\'{c} 00]}:
N. D. Megill, \& M. Pavi\v{c}i\'{c},
``Equations, states, and lattices of infinite-dimensional Hilbert spaces'',
{\em Int. J. Theor. Phys.} {\bf 39}, ?, 2349-2391 (2000);
quant-ph/0009038.

\item {\bf [Megill-Pavi\v{c}i\'{c} 01]}:
N. D. Megill, \& M. Pavi\v{c}i\'{c},
``Orthomodular lattices and a quantum algebra'',
{\em Int. J. Theor. Phys.} {\bf 40}, ?, 1387-1410 (2001);
quant-ph/0103135.

\item {\bf [Megill-Pavi\v{c}i\'{c} 02]}:
N. D. Megill, \& M. Pavi\v{c}i\'{c},
``Deduction, ordering, and operations in quantum logic'',
{\em Found. Phys.} {\bf 32}, 3, 357-378 (2002);
quant-ph/0108074.

\item {\bf [Megill-Pavi\v{c}i\'{c} 03 a]}:
N. D. Megill, \& M. Pavi\v{c}i\'{c},
``Quantum implication algebras'',
{\em Int. J. Theor. Phys.} {\bf 42}, 12, 2807-2822 (2003);
quant-ph/0310062.

\item {\bf [Megill-Pavi\v{c}i\'{c} 03 b]}:
N. D. Megill, \& M. Pavi\v{c}i\'{c},
``Equivalences, identities, symmetric differences, and congruences in
orthomodular lattices'',
{\em Int. J. Theor. Phys.} {\bf 42}, 12, 2797-2805 (2003);
quant-ph/0310063.

\item {\bf [Mehra-Rechenberg 82 a]}:
J. Mehra, \& H. Rechenberg,
{\em The historical development of quantum theory.
Vol. 1, Part 1:
The quantum theory of Planck, Einstein, Bohr and Sommerfeld:
Its foundations and the rise of its difficulties, 1900-1925},
Springer-Verlag, New York, 1982.

\item {\bf [Mehra-Rechenberg 82 b]}:
J. Mehra, \& H. Rechenberg,
{\em The historical development of quantum theory.
Vol. 1, Part 2:
The quantum theory of Planck, Einstein, Bohr and Sommerfeld:
Its foundations and the rise of its difficulties, 1900-1925},
Springer-Verlag, New York, 1982.

\item {\bf [Mehra-Rechenberg 82 c]}:
J. Mehra, \& H. Rechenberg,
{\em The historical development of quantum theory.
Vol. 2: The discovery of quantum mechanics, 1925},
Springer-Verlag, New York, 1982.

\item {\bf [Mehra-Rechenberg 82 d]}:
J. Mehra, \& H. Rechenberg,
{\em The historical development of quantum theory.
Vol. 3:
The formulation of matrix mechanics
and its modifications, 1925-1926},
Springer-Verlag, New York, 1982.

\item {\bf [Mehra-Rechenberg 82 e]}:
J. Mehra, \& H. Rechenberg,
{\em The historical development of quantum theory.
Vol. 4, Part 1:
The fundamental equations of quantum mechanics, 1925-1926},
Springer-Verlag, New York, 1982.

\item {\bf [Mehra-Rechenberg 82 f]}:
J. Mehra, \& H. Rechenberg,
{\em The historical development of quantum theory.
Vol. 4, Part 2:
The reception of the new quantum mechanics, 1925-1926},
Springer-Verlag, New York, 1982.

\item {\bf [Mehra-Rechenberg 87 a]}:
J. Mehra, \& H. Rechenberg,
{\em The historical development of quantum theory.
Vol. 5:
Erwin Schr\"{o}dinger and the rise of wave mechanics.
Part 1:
Schr\"{o}dinger in Vienna and Zurich, 1887-1925},
Springer-Verlag, New York, 1987.
Review: {\bf [McCrea 90]}.

\item {\bf [Mehra-Rechenberg 87 b]}:
J. Mehra, \& H. Rechenberg,
{\em The historical development of quantum theory.
Vol. 5:
Erwin Schr\"{o}dinger and the rise of wave mechanics.
Part 2:
The creation of wave mechanics:
Early response and applications, 1925-1926},
Springer-Verlag, New York, 1987.
Review: {\bf [McCrea 90]}.

\item {\bf [Mehra 99]}:
J. Mehra,
{\em Einstein, physics and reality},
World Sicentific, Singapore, 1999.

\item {\bf [Mehra-Rechenberg 00 a]}:
J. Mehra, \& H. Rechenberg,
{\em The historical development of quantum theory.
Vol. 6:
The completion of quantum mechanics, 1926-1941.
Part 1:
The probability interpretation,
the statistical transformation theory,
the physical interpretation and
the empirical and mathematical foundations
of quantum mechanics 1926-1932},
Springer-Verlag, New York, 2000.
Review: {\bf [Schweber 01]}, {\bf [Kragh 02]}.

\item {\bf [Mehra-Rechenberg 00 b]}:
J. Mehra, \& H. Rechenberg,
{\em The historical development of quantum theory.
Vol. 6:
The completion of quantum mechanics, 1926-1941.
Part 2: The conceptual completion and the extensions
of quantum mechanics, 1932-1941.
Epilogue: Aspects of the further
development of quantum theory, 1942-1999},
Springer-Verlag, New York, 2000.
Review: {\bf [Schweber 01]}, {\bf [Kragh 02]}.

\item {\bf [Mehring-Mende-Scherer 03]}:
M. Mehring, J. Mende, \& W. Scherer,
``Entanglement between an electron and a nuclear spin (1/2)'',
{\em Phys. Rev. Lett.} {\bf 90}, 15, 153001 (2003).

\item {\bf [Mei-Weitz 01]}:
M. Mei, \& M. Weitz,
``Controlled decoherence in multiple beam Ramsey interference'',
{\em Phys. Rev. Lett.} {\bf 86}, 4, 559-563 (2001).

\item {\bf [Meier-Levy-Loss 03 a]}:
F. Meier, J. Levy, \& D. Loss,
``Quantum computing with spin cluster qubits'',
{\em Phys. Rev. Lett.} {\bf 90}, 4, 047901 (2003).

\item {\bf [Meier-Levy-Loss 03 b]}:
F. Meier, J. Levy, \& D. Loss,
``Quantum computing with antiferromagnetic spin clusters'',
{\em Phys. Rev. B} {\bf 68}, 13, 134417 (2003).

\item {\bf [Mel\'{e}ndez-Sols 04]}:
J. Mel\'{e}ndez, \& F. Sols,
``Entrevista a Anthony J. Leggett'',
{\em Revista Espa\~{n}ola de F\'{\i}sica} {\bf 18}, 1, 4-9 (2004).

\item {\bf [Melloy 02]}:
G. F. Melloy,
``The generalized representation of
particle localization in quantum mechanics'',
{\em Found. Phys.} {\bf 32}, 4, 503-530 (2002).

\item {\bf [Menicucci-Caves 02]}:
N. C. Menicucci, \& C. M. Caves,
``Local realistic model for the dynamics of bulk-ensemble NMR information
processing'',
{\em Phys. Rev. Lett.} {\bf 88}, 16, 167901 (2002);
quant-ph/0111152.

\item {\bf [Mensky 96]}:
M. B. Mensky,
``Note on reversibility of quantum jumps'',
{\em Phys. Lett. A} {\bf 222}, ?, 137-140 (1996);
quant-ph/0007095.

\item {\bf [Mensky 98 a]}:
M. B. Mensky,
``?'',
{\em Usp. Fiz. Nauk} {\bf 168}, ?, 1017-1035.
English version:
``Effect of decoherence and the theory of continuous quantum measurements'',
{\em Phys. Usp.} {\bf 41}, ?, 923-? (1998);
quant-ph/9812017.
Comment: {\bf [Onofrio-Presilla 00]}.

\item {\bf [Mensky 98 b]}:
M. B. Mensky,
``Decoherence and continuous measurements:
Phenomenology and models'',
in P. Blanchard, D. Giulini, E. Joos,
C. Kiefer, \& I.-O. Stamatescu (eds.),
{\em Decoherence: Theoretical, Experimental, and
Conceptual Problems (Bielefeld, Germany, 1998)},
Springer-Verlag, Berlin, 1999;
quant-ph/9812078.

\item {\bf [Mensky 99]}:
M. B. Mensky,
``Quantum Zeno effect in the decay onto an unstable level'',
{\em Phys. Lett. A} {\bf 257}, 5-6, 227-231 (1999);
quant-ph/9908062.

\item {\bf [Mensky 00]}:
M. B. Mensky,
{\em Quantum measurements and decoherence: Models and phenomenology}
Kluwer Academic, Dordrecht, Holland, 2000.
Reviews: {\bf [von Borzeszkowski 00]}, {\bf [Cabello 02 i]}.

\item {\bf [Mensky 01]}:
M. B. Mensky,
``Once more about an interferometer with entangled atoms'',
{\em Phys. Lett. A} {\bf 290}, 5-6, 322-324 (2001).

\item {\bf [Mensky 02]}:
M. B. Mensky,
``Quantum continuous measurements, dynamical role of information and
restricted path integrals'',
reported at {\em Int.\ Conf.\ on Theoretical Physics (Paris, 2002)}.
quant-ph/0212112.

\item {\bf [Mensky 03]}:
M. B. Mensky,
``Evolution of an open system as a continuous measurement of this system by its environment'',
{\em Phys. Lett. A} {\bf 307}, 2-3, 85-92 (2003).

\item {\bf [Mensky-Stenholm 03]}:
M. B. Mensky, \& S. Stenholm,
``Quantum dissipative systems from theory of continuous measurements'',
{\em Phys. Lett. A} {\bf 208}, 4, 243-248 (2003).

\item {\bf [Mermin 80]}:
N. D. Mermin,
``Quantum mechanics vs local realism near
the classical limit: A Bell inequality for spin $s$'',
{\em Phys. Rev. D} {\bf 22}, 2, 356-361 (1980).

\item {\bf [Mermin 81 a]}:
N. D. Mermin,
``Bringing home the atomic world: Quantum mysteries for anyone'',
{\em Am. J. Phys.} {\bf 49}, 10, 940-943 (1981).
See {\bf [Mermin 81 b, 85]}, {\bf [Allen 92]},
{\bf [Marmet 93]}.

\item {\bf [Mermin 81 b]}:
N. D. Mermin,
``Quantum mysteries for anyone'',
{\em J. Philos.} {\bf 78}, ?, 397-408 (1981).
Reprinted in {\bf [Mermin 90 e]}, pp.~81-94.

\item {\bf [Mermin 82]}:
N. D. Mermin,
``Comment on `Resolution of the Einstein-Podolsky-Rosen and Bell paradoxes'\,'',
{\em Phys. Rev. Lett.} {\bf 49}, 16, 1214 (1982).
Comment on {\bf [Pitowsky 82 a]}.
Reply: {\bf [Pitowsky 82 b]}.

\item {\bf [Mermin-Schwarz 82]}:
N. D. Mermin, \& G. M. Schwarz,
``Joint distributions and local realism in the higher-spin
Einstein-Podolsky-Rosen experiment'',
{\em Found. Phys.} {\bf 12}, 2, 101-135 (1982).
See {\bf [Garg-Mermin 82 b, 83, 87]}, {\bf [Garg 83]}.

\item {\bf [Mermin 83]}:
N. D. Mermin,
``Pair distributions and conditional independence:
Some hints about the structure of strange quantum correlations'',
{\em Philos. Sci.} {\bf 50}, 3, 359-373 (1983).

\item {\bf [Mermin 85]}:
N. D. Mermin,
``Is the moon there when nobody looks?
Reality and the quantum theory'',
{\em Phys. Today} {\bf 38}, 4, 38-47 (1985).
Comments: {\bf [Aspect 85]}, {\bf [Marshall-Santos 85]},
{\bf [Rohrlich 85]}, {\bf [Feingold-Peres 85]}, {\bf [Mirman 85]},
{\bf [Gardner 85]}, {\bf [Ekstrand 85]}.
See {\bf [Mermin 81 a, b]}, {\bf [Allen 92]},
{\bf [Marmet 93]}.

\item {\bf [Mermin 86 a]}:
N. D. Mermin,
``Generalizations of Bell's theorem to
higher spins and higher correlations'',
in L. M. Roth, \& A. Inomata (eds.),
{\em Fundamental questions in quantum mechanics
(Albany, New York, 1984)},
Gordon \& Breach, London, 1986, pp.~7-20.

\item {\bf [Mermin 86 b]}:
N. D. Mermin,
``The EPR experiment---Thoughts about the `loophole'\,'',
in D. M. Greenberger (ed.),
{\em New techniques and ideas in quantum measurement theory.
Proc.\ of an international conference (New York, 1986), Ann. N. Y.
Acad. Sci.} {\bf 480}, 422-427 (1986).

\item {\bf [Mermin 88]}:
N. D. Mermin,
``A new representation for the quantum theoretic rotation matrix that
reveals the classical limit of Einstein-Podolsky-Rosen correlations'',
in A. van der Merwe, F. Selleri, \& G. Tarozzi (eds.),
{\em Microphysical reality and quantum formalism.
Proc.\ of an international conference (Urbino, Italy, 1985)},
Kluwer Academic, Dordrecht, Holland, 1988, vol. 2, pp.~339-344.

\item {\bf [Mermin 89 a]}:
N. D. Mermin,
``Can you help your team tonight by
watching on TV? More experimental methaphysics from
Einstein, Podolsky, and Rosen'',
in J. T. Cushing, \& E. McMullin (eds.),
{\em Philosophical consequences of
quantum theory: Reflections on Bell's theorem},
University of Notre Dame Press, Notre
Dame, Indiana, 1989, pp.~?-?.
Reprinted in {\bf [Mermin 90 e]}, pp.~95-109.
See {\bf [Herbert 75]}, {\bf [Stapp 85 a]}.

\item {\bf [Mermin 89 b]}:
N. D. Mermin,
``The philosophical writings of Niels Bohr'',
{\em Phys. Today} {\bf 42}, 2, 105 (1989).
Reprinted in {\bf [Mermin 90 e]}, pp.~186-189.
Review of {\bf [Bohr 34, 58 a, 63]}.

\item {\bf [Mermin 90 a]}:
N. D. Mermin,
``What's wrong with these elements of reality?'',
{\em Phys. Today} {\bf 43}, 6, 9-11 (1990).
Reprinted in {\bf [Macchiavello-Palma-Zeilinger 00]}, pp.~53-54.
Comments: {\bf [Sawicki 90]}, {\bf [Santos 90]}.

\item {\bf [Mermin 90 b]}:
N. D. Mermin,
``Quantum mysteries revisited'',
{\em Am. J. Phys.} {\bf 58}, 8, 731-734 (1990).
Comment: {\bf [Sen 91]}.

\item {\bf [Mermin 90 c]}:
N. D. Mermin,
``Extreme quantum entanglement in a
superposition of macroscopically distinct states'',
{\em Phys. Rev. Lett.} {\bf 65}, 15, 1838-1840 (1990).
See {\bf [Roy-Singh 91]}.

\item {\bf [Mermin 90 d]}:
N. D. Mermin,
``Simple unified form for the major no-hidden-variables theorems'',
{\em Phys. Rev. Lett.} {\bf 65}, 27, 3373-3376 (1990).

\item {\bf [Mermin 90 e]}:
N. D. Mermin,
{\em Boojums all the way through:
Communicating science in a prosaic age},
Cambridge University Press, Cambridge, 1990.

\item {\bf [Mermin 91]}:
N. D. Mermin,
``Can phase transition make quantum mechanics less embarrassing'',
{\em Physica A} {\bf 177}, ?, 561-? (1991).

\item {\bf [Mermin 92 a]}:
N. D. Mermin,
``Not quite so simply no hidden variables'',
{\em Am. J. Phys.} {\bf 60}, 1, 25-27 (1992).
See {\bf [Jordan-Sudarshan 91]}.

\item {\bf [Mermin 92 b]}:
N. D. Mermin,
``The (non)world (non)view of quantum mechanics'',
{\em New Literary History} {\bf 23}, ?, 855-875 (1992).
Erratum: {\em New Literary History} {\bf 24}, ?, 947 (1993).

\item {\bf [Mermin 93 a]}:
N. D. Mermin,
``Some simple unified versions of the two theorems of John Bell'',
in M. A. del Olmo, M. Santander, \& J. Mateos Guilarte (eds.),
{\em Group theoretical methods in physics.
Proc.\ of the XIX Int.\ Colloquium (Salamanca, Spain, 1992)},
Ciemat-Real Sociedad Espa\~{n}ola de F\'{\i}sica, Madrid, 1993,
vol. 2, pp.~3-17. Almost the same as {\bf [Mermin 93 b]}.

\item {\bf [Mermin 93 b]}:
N. D. Mermin,
``Hidden variables and the two theorems of John Bell'',
{\em Rev. Mod. Phys.} {\bf 65}, 3, 803-815 (1993).
Almost the same as {\bf [Mermin 93 a]}.

\item {\bf [Mermin 93 c]}:
N. D. Mermin,
``Two lectures on the wave-particle duality'',
{\em Phys. Today} {\bf 46}, 1, 9-11 (1993).

\item {\bf [Mermin 94 a]}:
N. D. Mermin,
``What's wrong with this temptation?'',
{\em Phys. Today} {\bf 47}, 6, 9-11 (1994).
Erratum: {\em Phys. Today} {\bf 47}, 11, 119 (1994).

\item {\bf [Mermin 94 b]}:
N. D. Mermin,
`A ``virtuosically adaptive'' system as seen by a
``marginally adaptive'' one',
{\em Phys. Today} {\bf 47}, 9, 89-90 (1994).
Review of {\bf [Gell-Mann 94]}.

\item {\bf [Mermin 94 c]}:
N. D. Mermin,
``Quantum mysteries refined'',
{\em Am. J. Phys.} {\bf 62}, 10, 880-887 (1994).

\item {\bf [Mermin 95 a]}:
N. D. Mermin,
``The best version of Bell's theorem'',
in D. M. Greenberger, \& A. Zeilinger (eds.),
{\em Fundamental problems in quantum theory: A conference held in
honor of professor John A. Wheeler, Ann. N. Y. Acad. Sci.}
{\bf 755}, 616-623 (1995).

\item {\bf [Mermin 95 b]}:
N. D. Mermin,
``Limits to quantum mechanics as a source
of magic tricks: Retrodiction and the Bell-Kochen-Specker theorem'',
{\em Phys. Rev. Lett.} {\bf 74}, 6, 831-834 (1995).

\item {\bf [Mermin 96]}:
N. D. Mermin,
``Hidden quantum non-locality'',
in R. K. Clifton (ed.),
{\em Perspectives on quantum reality: Non-relativistic, relativistic,
and field-theoretic (London, Western Ontario, Canada, 1994)},
Kluwer Academic, Dordrecht, Holland, 1996, pp.~57-71.

\item {\bf [Mermin-Garg 96]}:
N. D. Mermin, \& A. Garg,
``Reply to a comment on
`Correlation inequalities and hidden-variables'\,'',
{\em Phys. Rev. Lett.} {\bf 76}, 12, 2197 (1996).
Reply to {\bf [Horodecki-Horodecki 96 c]}.
See {\bf [Garg-Mermin 82 c]}.

\item {\bf [Mermin 97 a]}:
N. D. Mermin,
``Quantum theory: Concepts and methods'',
{\em Stud. Hist. Philos. Sci. Part B: Stud. Hist. Philos. Mod. Phys.}
{\bf 28}, 1, 131-136 (1997).
Review of {\bf [Peres 93 a]}.

\item {\bf [Mermin 97 b]}:
N. D. Mermin,
``Infinite potential: The life and times of David Bohm'',
{\em Am. J. Phys.} {\bf 65}, 10, 1027-1028 (1997).
Review of {\bf [Peat 97]}.

\item {\bf [Mermin 97 c]}:
N. D. Mermin,
``Nonlocality and Bohr's reply to EPR'',
quant-ph/9712003.

\item {\bf [Mermin 97 d]}:
N. D. Mermin,
``How to ascertain the values of every member of a set of
observables that cannot all have values'',
in {\bf [Cohen-Horne-Stachel 97 b]}, pp.~149-157.
Comment on {\bf [Stapp 97 b]}.
See {\bf [Stapp 97 a, 98 c]}, {\bf [Mermin 98 b]}.

\item {\bf [Mermin 98 a]}:
N. D. Mermin,
``The Ithaca interpretation of quantum mechanics'', in
{\em Golden Jubilee Workshop on Foundations of Quantum Theory
(Bombay, 1996)}, Pramana {\bf 51}, 5, 549-565 (1998);
quant-ph/9609013.
See {\bf [Mermin 98 b, 99 a]},
{\bf [Cabello 99 a, c]}, {\bf [Jordan 99]}.

\item {\bf [Mermin 98 b]}:
N. D. Mermin,
``What is quantum mechanics trying to tell us?'',
{\em Am. J. Phys.} {\bf 66}, 9, 753-767 (1998);
quant-ph/9801057.
See {\bf [Mermin 98 a, c]},
{\bf [Cabello 99 a, c]}, {\bf [Jordan 99]}.

\item {\bf [Mermin 98 c]}:
N. D. Mermin,
``Nonlocal character of quantum theory?'',
{\em Am. J. Phys.} {\bf 66}, 11, 920-923 (1998);
quant-ph/9711052.
Comment on {\bf [Stapp 97 a]}.
Reply: {\bf [Stapp 98 c]}.

\item {\bf [Mermin 99 a]}:
N. D. Mermin,
``What do these correlations know about reality?
Nonlocality and the absurd'',
{\em Found. Phys.} {\bf 29}, 4, 571-587 (1999);
quant-ph/9807055.
See {\bf [Mermin 98 a, b]},
{\bf [Cabello 99 a, c]}, {\bf [Jordan 99]}.

\item {\bf [Mermin 99 b]}:
N. D. Mermin,
``A Kochen-Specker theorem for imprecisely specified measurements'',
quant-ph/9912081.
See {\bf [Meyer 99 b]}.

\item {\bf [Mermin 99 c]}:
N. D. Mermin,
``Notes for physicists on the theory of quantum computation'',
informal notes for three lectures at LASSP Autumn School of Quantum
Computation, Cornell, September 20, 22, and 24, 1999,
www.lassp.cornell.edu/lassp\_data/NMermin.html.
See {\bf [Mermin 00 b]}.

\item {\bf [Mermin 00 a]}:
N. D. Mermin,
``The contemplation of quantum computation'',
{\em Phys. Today} {\bf 53}, 7, 11-12 (2000).

\item {\bf [Mermin 00 b]}:
N. D. Mermin,
``Lecture notes on quantum computation and quantum information
theory'',
Cornell University, Physics 481-681, CS-483 (2000),
www.lassp.cornell.edu/lassp\_data/NMermin.html.
See {\bf [Mermin 99 c]}.

\item {\bf [Mermin 02 a]}:
N. D. Mermin,
``Whose knowledge?'',
in {\bf [Bertlmann-Zeilinger 02]}, pp.~271-280;
quant-ph/0107151.

\item {\bf [Mermin 02 b]}:
N. D. Mermin,
``From classical state swapping to quantum teleportation'',
{\em Phys. Rev. A} {\bf 65}, 1, 012320 (2002);
quant-ph/0105117.

\item {\bf [Mermin 02 c]}:
N. D. Mermin,
``Compatibility of state assignments'',
{\em J. Math. Phys.} {\bf 43}, 9, 4560-4566 (2002).

\item {\bf [Mermin 02 d]}:
N. D. Mermin,
``Deconstructing dense coding'',
{\em Phys. Rev. A} {\bf 66}, 3, 032308 (2002).
Publisher's note: {\em Phys. Rev. A} {\bf 68}, 2, 029901 (2003);
quant-ph/0204107.

\item {\bf [Mermin 02 e]}:
N. D. Mermin,
`Shedding (red and green) light on ``time related hidden parameters''\,',
quant-ph/0206118.
See {\bf [Hess-Philipp 02 c, d, 03]}, {\bf [Mermin 02 f, 03 b]}.

\item {\bf [Mermin 02 f]}:
N. D. Mermin,
``Bell's theorem in the presence of classical communication'',
quant-ph/0207140.
See {\bf [Hess-Philipp 02 c, d, 03]}, {\bf [Mermin 02 e, 03 b]}.

\item {\bf [Mermin 03 a]}:
N. D. Mermin,
``From Cbits to Qbits: Teaching computer scientists quantum mechanics'',
{\em Am. J. Phys.} {\bf 71}, 1, 23-30 (2003);
quant-ph/0207118.

\item {\bf [Mermin 03 b]}:
N. D. Mermin,
``Inclusion of time in the theorem of Bell'',
{\em Europhys. Lett.} {\bf 61}, 2, 143-147 (2003).
See {\bf [Hess-Philipp 02 c, d, 03]}, {\bf [Mermin 02 e, f]}.

\item {\bf [Mermin 03 c]}:
N. D. Mermin,
``Copenhagen computation: How I learned to stop worrying and love Bohr'',
{\em IBM J. Res. Dev.};
quant-ph/0305088.

\item {\bf [Mermin 03 d]}:
N. D. Mermin,
``Writing physics'',
{\em Am. J. Phys.} {\bf 71}, 4, 296-301 (2003).

\item {\bf [Mermin 03 e]}:
N. D. Mermin,
``Publications of N. David Mermin'',
{\em Found. Phys.} {\bf 33}, 12, 1797-1809 (2003).

\item {\bf [M\'{e}rolla-Mazurenko-Goedgebuer-(+3) 99]}:
J.-M. M\'{e}rolla, Y. Mazurenko, J.-P. Goedgebuer, L. Duraffourg,
H. Porte, \& W. T. Rhodes,
``Quantum cryptographic device using single-photon phase modulation'',
{\em Phys. Rev. A} {\bf 60}, 3, 1899-1905 (1999).

\item {\bf [M\'{e}rolla-Duraffourg-Goedgebuer-(+6) 02]}:
J.-M. M\'{e}rolla, L. Duraffourg, J.-P. Goedgebuer,
A. Soujaeff, F. Patois, \& W. T. Rhodes,
``Integrated quantum key distribution system using single sideband detection'',
{\em Eur. Phys. J. D} {\bf 18}, 2 (Special issue:
{\em Quantum interference and cryptographic keys:
Novel physics and advancing technologies (QUICK) (Corsica, 2001)}, 141-146 (2002).

\item {\bf [Merzbacher 61]}:
E. Merzbacher,
{\em Quantum mechanics},
Wiley, New York, 1961, 1970 (2nd edition), 1998 (3rd edition).
Review: {\bf [Greenberger 99]}.

\item {\bf [Messiah 58]}:
A. M. L. Messiah,
{\em M\'{e}canique quantique}, 2 Vols.,
Dunod, Paris, 1958.
English version: {\em Quantum mechanics}, 2 Vols.,
John Wiley \& Sons, New York, 1958;
North-Holland, Amsterdam, 1961;
Dover, New York, 1999 (the two volumes bound as one).
Spanish version: {\em Mec\'{a}nica cu\'{a}ntica}, 2 Vols.,
Tecnos, Madrid, 1965.

\item {\bf [Messiah-Greenberg 64]}:
A. M. L. Messiah, \& O. W. Greenberg,
``Symmetrization postulate and its experimental foundation'',
{\em Phys. Rev.} {\bf 136}, 1B, B248-B267 (1964).

\item {\bf [Messikh-Ficek-Wahiddin 03]}:
A. Messikh, Z. Ficek, \& M. R. B. Wahiddin,
``Spin squeezing as a measure of entanglement in a two-qubit system'',
{\em Phys. Rev. A} {\bf 68}, 6, 064301 (2003).

\item {\bf [Messina 02]}:
A. Messina,
``A single atom-based generation of Bell states of two cavities'',
{\em Eur. Phys. J. D} {\bf 18}, 3, 379-383 (2002);
quant-ph/0203140.

\item {\bf [Methot-Wicker 01]}:
A. A. Methot, \& K. Wicker,
``Interaction-free measurement applied to quantum computation:
A new CNOT gate'',
quant-ph/0109105.

\item {\bf [Metwally 01]}:
N. Metwally,
``A more efficient variant of the Oxford protocol'',
quant-ph/0109051.

\item {\bf [Metwally 02]}:
N. Metwally,
``More efficient entanglement purification'',
{\em Phys. Rev. A} {\bf 66}, 5, 054302 (2002).

\item {\bf [Metzger 00]}:
S. Metzger,
``Spin measurement retrodiction revisited'',
quant-ph/0006115.

\item {\bf [Meyer 96 a]}:
D. A. Meyer,
``From quantum cellular automata to quantum lattice gases'',
{\em J. Stat. Phys.} {\bf 85}, 551-574 (1996).

\item {\bf [Meyer 96 b]}:
D. A. Meyer,
``On the absence of homogeneous scalar unitary cellular automata'',
{\em Phys. Lett. A} {\bf 223}, 5, 337–340 (1996).

\item {\bf [Meyer 97]}:
D. A. Meyer,
``Quantum mechanics of lattice gas automata: One-particle plane waves and potentials'',
{\em Phys. Rev. E} {\bf 55}, 5, 5261–5269 (1997).

\item {\bf [Meyer 99 a]}:
D. A. Meyer,
``Quantum strategies'',
{\em Phys. Rev. Lett.} {\bf 82}, 5, 1052-1055 (1999);
quant-ph/9804010.
Comment: {\bf [van Enk 00]}.
Reply: {\bf [Meyer 00 a]}.

\item {\bf [Meyer 99 b]}:
D. A. Meyer,
``Finite precision measurement nullifies the Kochen-Specker theorem'',
{\em Phys. Rev. Lett.} {\bf 83}, 19, 3751-3754 (1999);
quant-ph/9905080.
See {\bf [Kent 99 b]}, {\bf [Clifton-Kent 00]}, {\bf [Cabello 99 d, 02 c]},
{\bf [Havlicek-Krenn-Summhammer-Svozil 01]}, {\bf [Mermin 99 b]},
{\bf [Appleby 00, 01, 02]}, {\bf [Boyle-Schafir 01 a]}, {\bf [Peres 03 d]}, {\bf [Wilce 04]}.

\item {\bf [Meyer 00 a]}:
D. A. Meyer,
``Meyer replies'',
{\em Phys. Rev. Lett.} {\bf 84}, 4, 790 (2000).
Reply to {\bf [van Enk 00]}.
See {\bf [Meyer 99 a]}.

\item {\bf [Meyer 00 b]}:
D. A. Meyer,
``Sophisticated quantum search without entanglement'',
{\em Phys. Rev. Lett.} {\bf 85}, 9, 2014-2017 (2000);
quant-ph/0007070.

\item {\bf [Meyer 00 c]}:
D. A. Meyer,
``Does Rydberg state manipulation equal quantum computation?'',
{\em Science} {\bf 289}, 5484, 1431a (2000).
Comment on {\bf [Ahn-Weinacht-Bucksbaum 00]}.
Reply: {\bf [Bucksbaum-Ahn-Weinacht 00]}.
See {\bf [Kwiat-Hughes 00]}.

\item {\bf [Meyer 01]}:
D. A. Meyer,
``Quantum computing classical physics'',
quant-ph/0111069.

\item {\bf [Meyer 02]}:
D. A. Meyer,
``Physical quantum algorithms'',
{\em Proc. of the Quantum Computation for Physical Modeling
Workshop 2000 (North Falmouth, Massachusetts)},
{\em Comput. Phys. Comm.} {\bf 146}, 3, 295-301 (2002).

\item {\bf [Meyer-Wallach 02]}:
D. A. Meyer, \& N. R. Wallach,
``Global entanglement in multiparticle systems'',
{\em J. Math. Phys.} {\bf 43}, 9, 4273-4278 (2002);
quant-ph/0108104.
See {\bf [Brennen 03]}.

\item {\bf [Meyer 02]}:
D. A. Meyer,
``Quantum games and quantum algorithms'',
in {\bf [Lomonaco-Brandt 02]} 213-220;
quant-ph/0004092.

\item {\bf [Meyer-Poulsen-Eckert-(+2) 03]}:
T. Meyer, U. V. Poulsen, K. Eckert,
M. Lewenstein, \& D. Bru\ss,
``Finite size effects in entangled rings of qubits'',
quant-ph/0308082.

\item {\bf [Miao 00 a]}:
X. Miao,
``Universal construction of quantum computational networks
in superconducting Josephson junctions'',
quant-ph/0003113.

\item {\bf [Miao 00 b]}:
X. Miao,
``A convenient method to prepare the effective pure state in a quantum
ensemble'',
quant-ph/0008094.

\item {\bf [Miao 01 a]}:
X. Miao,
``Universal construction for the unsorted quantum search
algorithms'',
quant-ph/0101126.

\item {\bf [Miao 01 b]}:
X. Miao,
``A polynomial-time solution to the parity problem on an NMR quantum computer'',
quant-ph/0108116.

\item {\bf [Michler-Mattle-Weinfurter-Zeilinger 96]}:
M. Michler, K. Mattle, H. Weinfurter, \& A. Zeilinger,
``Interferometric Bell-state analysis'',
{\em Phys. Rev. A} {\bf 53}, 3, R1209-R1212 (1996).

\item {\bf [Michler-Weinfurter-\.{Z}ukowski 00]}:
M. Michler, H. Weinfurter, \& M. \.{Z}ukowski,
``Experiments towards falsification of noncontextual hidden variable
theories'',
{\em Phys. Rev. Lett.} {\bf 84}, 24, 5457-5461 (2000);
quant-ph/0009061.

\item {\bf [Michler-Kiraz-Becher-(+5) 00]}:
P. Michler, A. Kiraz, C. Becher,
W. V. Schoenfeld, P. M. Petroff, L. Zhang, E. Hu, \& A. Imamoglu,
``A quantum dot single-photon turnstile device'',
{\em Science} {\bf 290}, 5500, 2282-2285 (2000).
See {\bf [Benjamin 00 b]}.

\item {\bf [Mielnik 68]}:
B. Mielnik,
``Geometry of quantum states'',
{\em Com. Math. Phys.} {\bf 9}, 55-80 (1968).

\item {\bf [Mielnik 00]}:
B. Mielnik,
`Comments on: ``Weinberg's nonlinear quantum mechanics and
Einstein-Podolsky-Rosen paradox'', by Joseph Polchinski',
quant-ph/0012041.
Comment on {\bf [Polchinski 91]}.

\item {\bf [Mielnik 01]}:
B. Mielnik,
``Nonlinear quantum mechanics: A conflict with the Ptolomean structure?'',
{\em Phys. Lett. A} {\bf 289}, 1-2, 1-8 (2001).

\item {\bf [Migdall 99]}:
A. Migdall,
``Correlated-photon metrology without absolute standards'',
{\em Phys. Today} {\bf 52}, 1, 41-? (1999).

\item {\bf [Migdall-Branning-Castelletto 02]}:
A. L. Migdall, D. Branning, \& S. Castelletto,
``Tailoring single-photon and multiphoton probabilities of a single-photon
on-demand source'',
{\em Phys. Rev. A} {\bf 66}, 5, 053805 (2002).

\item {\bf [Migliore-Messina 03]}:
R. Migliore, \& A. Messina,
``Quantum superpositions of clockwise and counterclockwise supercurrent states
in the dynamics of an rf-SQUID exposed to a quantized electromagnetic field'',
{\em Phys. Rev. B} {\bf 67}, 13, 134505 (2003).

\item {\bf [Migliore-Konstadopoulou-Vourdas-(+2) 03]}:
R. Migliore, A. Konstadopoulou, A. Vourdas, T. P. Spiller, \& A. Messina,
``Entangling gates using Josephson circuits coupled through non-classical microwaves'',
{\em Phys. Lett. A} {\bf 319}, 1-2, 67-72 (2003).

\item {\bf [Mihara 02]}:
T. Mihara,
``Quantum identification schemes with entanglements'',
{\em Phys. Rev. A} {\bf 65}, 5, 052326 (2002).
Comment: {\bf [van Dam 03 b]}.

\item {\bf [Milburn 97]}:
G. J. Milburn,
{\em Schr\"{o}dinger's machines: The
quantum technology reshaping everyday life}, Freeman, New York, 1997.
Review: {\bf [Rae 97]}, {\bf [Zeilinger 97 a]}.

\item {\bf [Milburn 98]}:
G. J. Milburn,
{\em The Feynman processor: Quantum entanglement and the computing revolution},
Perseus, Reading, Massachusetts, 1998.
Review: {\bf [Preskill 99 a]},
{\bf [Leggett 99 b]}.

\item {\bf [Milburn-Braunstein 99]}:
G. J. Milburn, \& S. L. Braunstein,
``Quantum teleportation with squeezed vacuum states'',
{\em Phys. Rev. A} {\bf 60}, 2, 937-942 (1999);
quant-ph/9812018.

\item {\bf [Milburn-Schneider-James 00]}:
G. J. Milburn, S. Schneider, \& D. F. V. James,
``Ion trap quantum computing with warm ions'',
{\em Fortschr. Phys.} {\bf 48}, 9-11 (Special issue: Experimental proposals for quantum computation), 801-810 (2000).

\item {\bf [Milburn 00]}:
G. J. Milburn,
``Quantum computing using a neutral atom optical lattice: An appraisal'',
{\em Fortschr. Phys.} {\bf 48}, 9-11 (Special issue: Experimental proposals for quantum computation), 957-964 (2000).

\item {\bf [Milburn-Laflamme-Sanders-Knill 02]}:
G. J. Milburn, R. Laflamme, B. C. Sanders, \& E. Knill,
``Quantum dynamics of two coupled qubits'',
{\em Phys. Rev. A} {\bf 65}, 3, 032316 (2002);
quant-ph/0112001.

\item {\bf [Militello-Messina-Napoli 01]}:
B. Militello, A. Messina, \& A. Napoli,
``Quantum Zeno effect in trapped ions'',
quant-ph/0102129.

\item {\bf [Militello-Messina-Napoli 03]}:
B. Militello, A. Messina, \& A. Napoli,
``Measuring entanglement-induced quantum correlations in the oscillatory
motion of a trapped ion'',
{\em Opt. Spectrosc.} {\bf 94}, 813 (2003).

\item {\bf [Militello-Nakazato-Messina 04]}:
B. Militello, H. Nakazato, \& A. Messina,
`Steering distillation by ``pulsed'' and ``continuous'' measurements',
quant-ph/0404058.

\item {\bf [Miller 96]}:
D. J. Miller,
``Realism and time symmetry in quantum mechanics'',
{\em Phys. Lett. A} {\bf 222}, 1-2, 31-36 (1996).

\item {\bf [Miller-Wheeler 84]}:
W. A. Miller, \& J. A. Wheeler,
``Delayed-choice experiments and Bohr's elementary quantum phenomenon'',
in S. Kamefuchi et al. (eds.),
Foundations of quantum mechanics in the light of new technology.
Proc.\ of a conference (Kokubunji, Tokyo, 1983),
Physical Society of Japan, Tokyo, 1984, pp.~140-151.

\item {\bf [Milman-Ollivier-Raimond 03]}:
P. Milman, H. Ollivier, \& J. M. Raimond,
``Universal quantum cloning in cavity QED'',
{\em Phys. Rev. A} {\bf 67}, 1, 012314 (2003).

\item {\bf [Milman-Ollivier-Yamaguchi-(+3) 03]}:
P. Milman, H. Ollivier, F. Yamaguchi,
M. Brune, J. M. Raimond, \& S. Haroche,
``Simple quantum information algorithms in cavity QED'',
in M. Ferrero (ed.),
{\em Proc. of Quantum Information: Conceptual Foundations,
Developments and Perspectives (Oviedo, Spain, 2002)},
{\em J. Mod. Opt.} {\bf 50}, 6-7, 901-913 (2003).

\item {\bf [Milman-Mosseri 03]}:
P. Milman, \& R. Mosseri,
``Topological phase for entangled two-qubit states'',
{\em Phys. Rev. Lett.} {\bf 90}, 23, 230403 (2003).

\item {\bf [Milonni 98]}:
P. W. Milonni,
``Quantum optics'',
{\em Phys. Today} {\bf 51}, 10, 90-92 (1998).
Review of {\bf [Scully-Zubairy 97]}.

\item {\bf [Mintert-Wunderlich 01]}:
F. Mintert, \& C. Wunderlich,
``Ion-trap quantum logic using long-wavelength radiation'',
{\em Phys. Rev. Lett.} {\bf 87}, 25, 257904 (2001).

\item {\bf [Mintert-\.{Z}yczkowski 04]}:
F. Mintert, \& K. \.{Z}yczkowski,
``Wehrl entropy, Lieb conjecture, and entanglement monotones'',
{\em Phys. Rev. A} {\bf 69}, 2, 022317 (2004).

\item {\bf [Mintert-Kus-Buchleitner 04]}:
F. Mintert, M. Kus, \& A. Buchleitner,
``Concurrence of mixed bipartite quantum states in arbitrary dimensions'',
quant-ph/0403063.

\item {\bf [Miquel-Paz-Perazzo 96]}:
C. Miquel, J. P. Paz, \& R. Perazzo,
``Factoring in a dissipative quantum computer'',
quant-ph/9601021.

\item {\bf [Miquel-Paz-Zurek 97]}:
C. Miquel, J. P. Paz, \& W. H. Zurek,
``Quantum computation with phase drift errors'',
{\em Phys. Rev. Lett.} {\bf 78}, 20, 3971-3974 (1997);
quant-ph/9704003.

\item {\bf [Miquel-Paz-Saraceno 02]}:
C. Miquel, J. P. Paz, \& M. Saraceno,
``Quantum computers in phase space'',
{\em Phys. Rev. A} {\bf 65}, 6, 062309 (2002);
quant-ph/0204149.

\item {\bf [Miquel-Paz-Saraceno-(+3) 02]}:
C. Miquel, J. P. Paz, M. Saraceno,
E. Knill, R. Laflamme, \& C. Negrevergne,
``Interpretation of tomography and spectroscopy
as dual forms of quantum computation'',
{\em Nature} {\bf 417}, 6893, 59-62 (2002);
quant-ph/0109072.

\item {\bf [Miranowicz-Leonski-Imoto 01]}:
A. Miranowicz, W. Leonski, \& N. Imoto,
``Quantum-optical states in finite-dimensional Hilbert space. I. General formalism'',
in M. W. Evans (ed.),
{\em Modern Nonlinear Optics},
{\em Advances in Chemical Physics} {\bf 119} (I), Wiley, New York, 2001, pp.~155-193;
quant-ph/0108080.

\item {\bf [Miranowicz-\"{O}zdemir-Liu-(+3) 02]}:
A. Miranowicz, S. K. \"{O}zdemir, Y.-X. Liu,
M. Koashi, N. Imoto, Y. Hirayama,
``Generation of maximum spin entanglement induced by a cavity field in
quantum-dot systems'',
{\em Phys. Rev. A} {\bf 65}, 6, 062321 (2002).

\item {\bf [Miranowicz 04 a]}:
A. Miranowicz,
``Violation of Bell inequality and entanglement of decaying Werner states'',
{\em Phys. Lett. A} {\bf 327}, 4, 272-283 (2004);
quant-ph/0402023.

\item {\bf [Miranowicz-Grudka 04 a]}:
A. Miranowicz, \& A. Grudka,
``Ordering two-qubit states with concurrence and negativity'',
{\em Phys. Rev. A};
quant-ph/0404053.

\item {\bf [Miranowicz 04 b]}:
A. Miranowicz,
``Decoherence of two maximally entangled qubits in a lossy nonlinear
cavity'',
{\em J. Phys. A};
quant-ph/0402025.

\item {\bf [Miranowicz-Grudka 04 b]}:
A. Miranowicz, \& A. Grudka,
``A comparative study of relative entropy of entanglement, concurrence and
negativity'',
{\em J. Opt. B: Quantum Semiclass. Opt.} {\bf 6}, 542-548 (2004);
quant-ph/0409153.

\item {\bf [Mirell 94]}:
S. Mirell,
``Correlated photon asymmetry in local realism'',
{\em Phys. Rev. A} {\bf 50}, 1, 839-842 (1994).

\item {\bf [Mirell-Mirell 99]}:
S. Mirell, \& D. Mirell,
``High efficiency interaction-free measurement
from continuous wave multi-beam interference'',
quant-ph/9911076.

\item {\bf [Mirman 79]}:
R. Mirman,
``Nonexistence of superselection rules: Definition of term frame of reference'',
{\em Found. Phys.} {\bf 9}, 3-4, 283-299 (1979).

\item {\bf [Mirman 85]}:
R. Mirman,
in ``Reality and the quantum theory'',
{\em Phys. Today} {\bf 38}, 11, 15 and 136 (1985).
Comment on {\bf [Mermin 85]}.

\item {\bf [Misra 67]}:
B. Misra,
``When can hidden variables be excluded in quantum mechanics?'',
{\em Nuovo Cimento A} {\bf 47}, 4, 841-859 (1967).

\item {\bf [Misra-Sudarshan 77]}:
B. Misra, \& E. C. G. Sudarshan,
``The Zeno's paradox in quantum theory'',
{\em J. Math. Phys.} {\bf 18}, 4, 756-763 (1977).

\item {\bf [Mis\v{s}ta-Filip-Fiur\'{a}\v{s}ek 02]}:
L. Mi\v{s}ta, Jr., R. Filip, \& J. Fiur\'{a}\v{s}ek,
``Continuous-variable Werner state: Separability, nonlocality, squeezing, and
teleportation'',
{\em Phys. Rev. A} {\bf 65}, 6, 062315 (2002);
quant-ph/0112062.

\item {\bf [Mis\v{s}ta-Filip 04 a]}:
L. Mi\v{s}ta, Jr., \& R. Filip,
``Partial deterministic non-demolition Bell measurement'',
quant-ph/0407017.

\item {\bf [Mis\v{s}ta-Filip 04 b]}:
L. Mi\v{s}ta, Jr., \& R. Filip,
``Adapting continuous-variable teleportation'',
{\em Latsis Symposium on Quantum Optics for Communication and Computing
(2004, Lausanne, Switzerland)};
quant-ph/0410185.

\item {\bf [Mitchell-Adami-Lue-Williams 04]}:
D. R. Mitchell, C. Adami, W. Lue, \& C. P. Williams,
``A random matrix model of adiabatic quantum computing'',
quant-ph/0409088.

\item {\bf [Mitchell-Ellenor-Schneider-Steinberg 03]}:
M. W. Mitchell, C. W. Ellenor, S. Schneider, \& A. M. Steinberg,
``Diagnosis, prescription, and prognosis of a Bell-state filter by quantum
process tomography'',
{\em Phys. Rev. Lett.} {\bf 91}, 12, 120402 (2003);
quant-ph/0305001.

\item {\bf [Mitchell-Lundeen-Steinberg 03]}:
M. W. Mitchell, J. S. Lundeen, \& A. M. Steinberg,
``Super-resolving phase measurements with a multi-photon entangled state'',
quant-ph/0312186.

\item {\bf [Mitchell-Popescu-Roberts 02]}:
P. Mitchell, S. Popescu, \& D. Roberts,
``Conditions for the confirmation of three-particle non-locality'',
quant-ph/0202009.

\item {\bf [Mitchison-Massar 01]}:
G. Mitchison, \& S. Massar,
`\,``Absorption-free discrimination between semitransparent objects'',
{\em Phys. Rev. A} {\bf 63}, 3, 032105 (2001);
quant-ph/0003140.

\item {\bf [Mitchison-Jozsa 01]}:
G. Mitchison, \& R. Jozsa,
``Counterfactual computations'',
{\em Proc. R. Soc. Lond. A} {\bf 457}, 2009, 1175-1193 (2001);
quant-ph/9907007.

\item {\bf [Mitchison-Massar-Pironio 01]}:
G. Mitchison, S. Massar, \& S. Pironio,
``How many photons are needed to distinguish two
transparencies?'',
quant-ph/0105098.

\item {\bf [Mitchison-Jozsa 04]}:
G. Mitchison, \& R. Jozsa,
``Towards a geometrical interpretation of quantum-information compression'',
{\em Phys. Rev. A} {\bf 69}, 3, 032304 (2004);
quant-ph/0309177.

\item {\bf [Mitra 98]}:
A. Mitra,
``Alternative quantum cryptography'',
quant-ph/9812087.

\item {\bf [Mitra 99]}:
A. Mitra,
``Noised based cipher system'',
quant-ph/9912072.

\item {\bf [Mitsumori-Vaccaro-Barnett-(+4) 03]}:
Y. Mitsumori, J. A. Vaccaro, S. M. Barnett,
E. Andersson, A. Hasegawa, M. Takeoka, \& M. Sasaki,
``Experimental demonstration of quantum source coding'',
{\em Phys. Rev. Lett.} {\bf 91}, 21, 217902 (2003);
quant-ph/0304036.

\item {\bf [Mittelstaedt 78]}:
P. Mittelstaedt,
{\em Quantum logic},
Reidel, Dordrecht, 1978.

\item {\bf [Mittelstaedt-Stachow 83]}:
P. Mittelstaedt, \& E. W. Stachow,
``Analysis of the Einstein-Podolsky-Rosen experiment by
relativistic quantum logic'',
{\em Int. J. Theor. Phys.} {\bf 22}, 6, 517-540 (1983).

\item {\bf [Mittelstaedt 97]}:
P. Mittelstaedt,
{\em The interpretation of quantum mechanics and the measurement process},
Cambridge University Press, Cambridge, 1997.

\item {\bf [Mittelstaedt 98]}:
P. Mittelstaedt,
``Can EPR-correlations be used for the transmission of superluminal
signals?'',
{\em Ann. Phys.} {\bf 7}, 7-8, 710-715 (1998).

\item {\bf [Mittelstaedt 01]}:
P. Mittelstaedt,
``Quantum mechanics without probabilities'',
in M. R\'{e}dei, \& M. St\"{o}ltzner (eds.),
{\em John von Neumann and the foundations of quantum physics},
Kluwer Academic, Dordrecht, Holland, 2001, pp.~?-?.

\item {\bf [Mittelstaedt 02]}:
P. Mittelstaedt,
``Universality and consistency in quantum mechanics --New problems of
an old theory--'',
in C. Mataix, \& A. Rivadulla (eds.),
{\em F\'{\i}sica cu\'{a}ntica y realidad.
Quantum physics and reality (Madrid, 2000)},
Editorial Complutense, Madrid, 2002, pp.~197-213.

\item {\bf [Mittelstaedt 04]}:
P. Mittelstaedt,
{\em The interpretation of quantum mechanics and the measurement process},
Cambridge University Press, Cambridge, 2004.

\item {\bf [Miyake-Wadati 01]}:
A. Miyake, \& M. Wadati,
``Geometric strategy for the optimal quantum search'',
{\em Phys. Rev. A} {\bf 64}, 4, 042317 (2001);
quant-ph/0109109.

\item {\bf [Miyake 03]}:
A. Miyake,
``Classification of multipartite entangled states by multidimensional
determinants'',
{\em Phys. Rev. A} {\bf 67}, 1, 012108 (2003).

\item {\bf [Miyake-Verstraete 04]}:
A. Miyake \& F. Verstraete,
``Multipartite entanglement in $2 \times 2 \times n$ quantum systems'',
{\em Phys. Rev. A} {\bf 69}, 1, 012101 (2004);
quant-ph/0307067.

\item {\bf [Miyake 04]}:
A. Miyake,
``Multipartite entanglement under stochastic local operations and
classical communication'',
{\em EQIS'03},
{\em Int. J. Quant. Inf.};
quant-ph/0401023.

\item {\bf [Miyauchi 01]}:
A. Miyauchi,
{\em ERATO Workshop on Quantum Information Science (Tokyo, 2001)},
``Linear order matrix inversion method with help from quantum searching algorithm'',
quant-ph/0109131.

\item {\bf [Mizel-Mitchell-Cohen 99]}:
A. Mizel, M. W. Mitchell, \& M. L. Cohen,
``Ground state quantum computation'',
quant-ph/9908035.

\item {\bf [Mizel-Mitchell-Cohen 02]}:
A. Mizel, M. W. Mitchell, \& M. L. Cohen,
``Scaling considerations in ground-state quantum computation'',
{\em Phys. Rev. A} {\bf 65}, 2, 022315 (2002);
quant-ph/0007001.

\item {\bf [Mizel-Lidar 03]}:
A. Mizel, \& D. A. Lidar,
``Exchange interaction between three and four coupled quantum dots:
Theory and applications to quantum computing'',
cond-mat/0302018.
See {\bf [Mizel-Lidar 04]}.

\item {\bf [Mizel-Lidar 04]}:
A. Mizel, \& D. A. Lidar,
``Three- and four-body interactions in spin-based quantum computers'',
{\em Phys. Rev. Lett.} {\bf 92}, 7, 077903 (2004);
quant-ph/0401081.
See {\bf [Mizel-Lidar 03]}.

\item {\bf [Mizel 04]}:
A. Mizel,
``Mimicking time evolution within a quantum ground state:
Ground-state quantum computation, cloning, and teleportation'',
{\em Phys. Rev. A} {\bf 70}, 1, 012304 (2004).

\item {\bf [Mizuno-Fujiwara-Akiba-(+3) 02]}:
J. Mizuno, M. Fujiwara, M. Akiba,
T. Kawanishi, S. M. Barnett, \& M. Sasaki,
``Optimum detection for extracting maximum information from symmetric qubit
sets'',
{\em Phys. Rev. A} {\bf 65}, 1, 012315 (2002);
quant-ph/0106164.

\item {\bf [Mizuno-Wakui-Furusawa-Sasaki 04]}:
J. Mizuno, K. Wakui, A. Furusawa, \& M. Sasaki,
``Experimental demonstration of quantum dense coding using entanglement of
a two-mode squeezed vacuum state'',
quant-ph/0402040.

\item {\bf [Mo-Li-Guo 02]}:
Y.-N. Mo, C.-F. Li, \& G.-C. Guo,
``Distillation of the Greenberger-Horne-Zeilinger state from arbitrary
tripartite states'',
{\em Phys. Rev. A} {\bf 65}, 2, 024301 (2002).

\item {\bf [Mochon 04]}:
C. Mochon,
``Anyon computers with smaller groups'',
{\em Phys. Rev. A} {\bf 69}, 3, 032306 (2004).

\item {\bf [Moehring-Madsen-Blinov-Monroe 04]}:
D. L. Moehring, M. J. Madsen, B. B. Blinov, \& C. Monroe,
``Experimental Bell inequality violation with an atom and a photon'',
quant-ph/0406048.
See {\bf [Blinov-Moehring-Duan-Monroe 04]}.

\item {\bf [Mokarzel-Salgueiro-Nemes 01]}:
S. G. Mokarzel, A. N. Salgueiro, \& M. C. Nemes,
``Modelling the recoherence of mesoscopic
superpositions in dissipative environments'',
quant-ph/0106044.

\item {\bf [Moldauer 72]}:
P. A. Moldauer,
``Is there a quantum measurement problem?'',
{\em Phys. Rev. D} {\bf 5}, 4, 1028-1032 (1972).
Comment: {\bf [Fine 72 a]}.

\item {\bf [Moldauer 74]}:
P. A. Moldauer,
``Reexamination of the arguments of
Einstein, Podolsky, and Rosen'',
{\em Found. Phys.} {\bf 4}, 2, 195-205 (1974).

\item {\bf [Molina Terriza 04]}:
G. Molina-Terriza,
``Experimento en el Danubio. Fotones entrelazados'',
{\em Investigaci\'{o}n y Ciencia} 335, 40-41 (2004).

\item {\bf [Molina Terriza-Vaziri-Rehacek-(+2) 04]}:
G. Molina-Terriza, A. Vaziri, J. Rehacek,
Z. Hradil, \& A. Zeilinger,
``Triggered qutrits for quantum communication protocols'',
quant-ph/0401183.

\item {\bf [Molina Terriza-Vaziri-Ursin-Zeilinger 04]}:
G. Molina-Terriza, A. Vaziri, R. Ursin, \& A. Zeilinger,
``Experimental quantum coin tossing'',
quant-ph/0404027.

\item {\bf [M{\o}lmer 97]}:
K. M{\o}lmer,
``Quantum entanglement and classical behaviour'',
{\em J. Mod. Opt.} {\bf 44}, 10 (Special issue:
Fundamentals of quantum optics IV), 1937-1956 (1997).

\item {\bf [M{\o}lmer-S{\o}rensen 99]}:
K. M{\o}lmer, \& A. S. S{\o}rensen,
``Multi-particle entanglement of hot trapped ions'',
{\em Phys. Rev. Lett.} {\bf 82}, 9, 1835-1838 (1999);
quant-ph/9810040.

\item {\bf [M{\o}lmer-S{\o}rensen 00]}:
K. M{\o}lmer, \& A. S. S{\o}rensen,
``RISQ - reduced instruction set quantum computers'',
{\em J. Mod. Opt.};
quant-ph/0004014.

\item {\bf [M{\o}lmer 01]}:
K. M{\o}lmer,
``Counterfactual statements and weak measurements: An experimental proposal'',
{\em Phys. Lett. A} {\bf 292}, 3, 151-155 (2001);
quant-ph/0109042.

\item {\bf [Moln\'{a}r-Timmermann 03]}:
L. Moln\'{a}r, \& W. Timmermann,
``Isometries of quantum states'',
{\em J. Phys. A} {\bf 36}, 1, 267-273 (2003).

\item {\bf [Molotkov-Nazin 96]}:
S. N. Molotkov, \& S. S.Nazin,
``Quantum cryptography based on the
time--energy uncertainty relation'',
{\em JETP Lett.} {\bf 63}, ?, 924-? (1996).
quant-ph/9612013.

\item {\bf [Molotkov 98 a]}:
S. N. Molotkov,
``Quantum teleportation of a single-photon wave packet'',
{\em Phys. Lett. A} {\bf 245}, 5, 339-344 (1998);
quant-ph/9805045.

\item {\bf [Molotkov 98 b]}:
S. N. Molotkov,
``?'',
{\em JETP Lett.} {\bf 68}, ?, 263-? (1998).

\item {\bf [Molotkov 98 c]}:
S. N. Molotkov,
`Quantum cryptography based on photon ``frequency'' states:
Example of a possible realization',
{\em Sov. Phys. JETP} {\bf 87}, 288-293 (1998);
quant-ph/9811038.

\item {\bf [Molotkov-Nazin 99 a]}:
S. N. Molotkov, \& S. S. Nazin,
``Photon frequency entanglement swapping'',
{\em Phys. Lett. A} {\bf 252}, 1-2, 1-4 (1999);
quant-ph/9810076.

\item {\bf [Molotkov-Nazin 99 b]}:
S. N. Molotkov, \& S. S. Nazin,
``?'',
{\em Zh. Eksp. Teor. Fiz.} {\bf 116}, 777-792 (1999).
English version: ``On the teleportation of continuous variable'',
{\em JETP} {\bf 89}, 413-420 (1999);
quant-ph/9906018.

\item {\bf [Molotkov-Nazin 99 c]}:
S. N. Molotkov, \& S. S. Nazin,
``Unconditionally secure relativistic quantum bit commitment'',
quant-ph/9911055.

\item {\bf [Molotkov-Nazin 01 a]}:
S. N. Molotkov, \& S. S. Nazin,
``A simple proof of the unconditional security of
relativistic quantum cryptography'',
{\em J. Exper. and Theor. Phys.} {\bf 92}, 5, 871-878 (2001);
quant-ph/0008008.

\item {\bf [Molotkov-Nazin 01 b]}:
S. N. Molotkov, \& S. S. Nazin,
``Relativistic quantum protocols: Bit commitment and coin tossing'',
{\em JETP} {\bf 93}, ?, 876-? (2001);
quant-ph/0012075.

\item {\bf [Molotkov-Nazin 01 c]}:
S. N. Molotkov, \& S. S. Nazin,
``Quantum bit commitment in a noisy channel'',
quant-ph/0101105.

\item {\bf [Molotkov-Nazin 01 d]}:
S. N. Molotkov, \& S. S. Nazin,
``Relativistic restrictions on the distinguishability of
orthogonal quantum states'',
quant-ph/0103019.

\item {\bf [Molotkov-Nazin 01 e]}:
S. N. Molotkov, \& S. S. Nazin,
``The role of causality in ensuring the ultimate security
of relativistic quantum cryptography'',
{\em JETP Lett.} {\bf 73}, 12, 682-687 (2001);
quant-ph/0106046.

\item {\bf [Molotkov 02 a]}:
S. N. Molotkov,
``The efficiency of repeaters based on the Einstein--Podolsky--Rosen effect
for quantum cryptography in a damping channel'',
{\em JETP Lett.} {\bf 74}, ?, 517-? (2002).
Erratum: {\em JETP Lett.} {\bf 75}, ?, 210 (2002).

\item {\bf [Molotkov 02 b]}:
S. N. Molotkov,
``Limiting rate of secret-key generation in quantum cryptography in spacetime'',
{\em JETP Lett.} {\bf 75}, ?, 521-? (2002).

\item {\bf [Molotkov 02 c]}:
S. N. Molotkov,
`Relativistic quantum cryptography on ``arrested'' photons',
{\em JETP Lett.} {\bf 76}, ?, 71-? (2002).

\item {\bf [Molotkov 02 d]}:
S. N. Molotkov,
``Transmission capability of a sequential relativistic quantum communication
channel with limited observation time'',
{\em JETP Lett.} {\bf 76}, 584-? (2002).

\item {\bf [Molotkov 03 a]}:
S. N. Molotkov,
``Real-time coding in a parallel quantum communication channel'',
{\em JETP Lett.} {\bf 77}, ?, 47-? (2003).

\item {\bf [Molotkov 03 b]}:
S. N. Molotkov,
``Simple delay scheme for quantum cryptography based on a Mach-Zehnder optic
fiber interferometer'',
{\em JETP Lett.} {\bf 78}, ?, 162-? (2003).

\item {\bf [Molotkov 03 c]}:
S. N. Molotkov,
``Experimental scheme of quantum cryptography on the nonorthogonal states
with time shift and a minimum number of optical components'',
{\em JETP Lett.} {\bf 78}, ?, 659-663 (2003).

\item {\bf [Molotkov 03 d]}:
S. N. Molotkov,
``On coding of a quantum source of states with a finite frequency band:
Quantum analogue of the Kotel'nikov theorem on sampling'',
{\em JETP Lett.} {\bf 78}, ?, 597-601 (2003).

\item {\bf [Molotkov 04 a]}:
S. N. Molotkov,
``Multiplex quantum cryptography based on time coding without interferometers'',
{\em JETP Lett.} {\bf 79}, ?, 445-449 (2004).

\item {\bf [Molotkov 04 b]}:
S. N. Molotkov,
``Integration of quantum cryptography into fiber-optic telecommunication systems'',
{\em JETP Lett.} {\bf 79}, ?, 559-570 (2004).

\item {\bf [Mompart-Eckert-Ertmer-(+2) 03]}:
J. Mompart, K. Eckert, W. Ertmer,
G. Birkl, \& M. Lewenstein,
``Quantum computing with spatially delocalized qubits'',
{\em Phys. Rev. Lett.} {\bf 90}, 14, 147901 (2003).

\item {\bf [Monken-Souto Ribeiro-P\'{a}dua 98]}:
C. H. Monken, P. H. Souto Ribeiro, \& S. P\'{a}dua,
``Optimizing the photon pair collection efficiency:
A step toward a loophole-free Bell's inequalities experiment'',
{\em Phys. Rev. A} {\bf 57}, 4, R2267-R2269 (1998).

\item {\bf [Monroe-Meekhof-King-(+2) 95]}:
C. Monroe, D. M. Meekhof, B. E. King,
W. M. Itano, \& D. J. Wineland,
``Demonstration of a fundamental quantum logic gate'',
{\em Phys. Rev. Lett.} {\bf 75}, 25, 4714-4717 (1995).
Reprinted in {\bf [Macchiavello-Palma-Zeilinger 00]}, pp.~355-358.

\item {\bf [Monroe-Meekhof-King-Wineland 96]}:
C. Monroe, D. M. Meekhof, B. E. King, \& D. J. Wineland,
`A ``Schr\"{o}dinger cat'' superposition state of an atom'',
{\em Science} {\bf 272}, 5265, 1131-1136 (1996).

\item {\bf [Monroe-Leibfried-King-(+3) 97]}:
C. Monroe, D. Leibfried, B. E. King,
D. M. Meekhof, W. M. Itano, \& D. J. Wineland,
``Simplified quantum logic with trapped ions'',
{\em Phys. Rev. A} {\bf 55}, 4, R2489-R2491 (1997).

\item {\bf [Monroe 02]}:
C. Monroe,
``Quantum information processing with atoms and photons'',
{\em Nature} {\bf 416}, 6877, 238-246 (2002).

\item {\bf [Montangero-Benenti-Fazio 03]}:
S. Montangero, G. Benenti, \& R. Fazio,
``Dynamics of entanglement in quantum computers with imperfections'',
{\em Phys. Rev. Lett.} {\bf 91}, 18, 187901 (2003).

\item {\bf [Monteoliva-Paz 01]}:
D. Monteoliva, \& J. P. Paz,
``Decoherence in a classically chaotic quantum
system: Entropy production and quantum--classical correspondence'',
quant-ph/0106090.

\item {\bf [Montina-Arecchi 98]}:
A. Montina, \& F. T. Arecchi,
``Toward an optical evidence of quantum interference between
macroscopically distinct states'',
{\em Phys. Rev. A} {\bf 58}, 5, 3472-3476 (1998).

\item {\bf [Monton 03]}:
B. Monton,
``The problem of ontology for spontaneous collapse theories'',
PITT-PHIL-SCI00001258.

\item {\bf [Moore-Nilsson 98]}:
C. Moore, \& M. Nilsson,
``Some notes on parallel quantum computation'',
quant-ph/9804034.

\item {\bf [Moore-Nilsson 02]}:
C. Moore, \& M. Nilsson,
``Parallel quantum computation and quantum codes'',
{\em Siam J. Comput.} {\bf 31}, ?, 799-? (2002).

\item {\bf [Moore 03]}:
C. D. Moore,
``Quantum computing'',
{\em Math. Rev.} 2003h, 81031, 6 (2003).
Report on {\bf [Hirvensalo 01]}.

\item {\bf [Moore 89]}:
W. Moore,
{\em Schr\"{o}dinger. Life and thought},
Cambridge University Press, Cambridge, 1989. Compressed version:
{\em A life of Erwin Schr\"{o}dinger},
Cambridge University Press, Cambridge, 1994.
Spanish version: {\em Erwin Schr\"{o}dinger: Una vida},
Cambridge University Press, Cambridge, 1996.

\item {\bf [Mohan-Luo-Kr\"{o}ll-Mair 98]}:
R. K. Mohan, B. Luo, S. Kr\"{o}ll, \& A. Mair,
``Delayed single-photon self-interference'',
{\em Phys. Rev. A} {\bf 58}, 6, 4348-4358 (1998).

\item {\bf [Mohrhoff 96]}:
U. Mohrhoff,
``Restoration of interference and the fallacy of delayed choice:
Concerning an experiment proposed by Englert, Scully, and Walther'',
{\em Am. J. Phys.} {\bf 64}, 12, 1468-1475.
See {\bf [Scully-Englert-Walther 91]}, {\bf [Englert-Scully-Walther 94]}.

\item {\bf [Mohrhoff 99]}:
U. Mohrhoff,
``Objectivity, retrocausation, and the experiment of
Englert, Scully, and Walther'',
{\em Am. J. Phys.} {\bf 67}, 4, 330-335 (1999).
See {\bf [Englert-Scully-Walther 99]}.

\item {\bf [Mohrhoff 00 a]}:
U. Mohrhoff,
``What quantum mechanics is trying to tell us'',
{\em Am. J. Phys.} {\bf 68}, 8, 728-745 (2000).
Comment: {\bf [Kastner 01]}.

\item {\bf [Mohrhoff 00 b]}:
U. Mohrhoff,
``The one, the many, and the quantum'',
quant-ph/0005110.

\item {\bf [Mohrhoff 01 a]}:
U. Mohrhoff,
``Objective probabilities, quantum counterfactuals, and the ABL
rule--A response to R. E. Kastner'',
{\em Am. J. Phys.} {\bf 69}, 8, 864-873 (2001);
quant-ph/0006116.
Reply to {\bf [Kastner 01]}.

\item {\bf [Mohrhoff 01 b]}:
U. Mohrhoff,
``Quantum mechanics and consciousness: Fact and fiction'',
quant-ph/0102047.

\item {\bf [Mohrhoff 01 c]}:
U. Mohrhoff,
``Unveiled reality: comment on d'Espagnat's note on measurement'',
quant-ph/0102103.
Comment on {\bf [d'Espagnat 01]}.

\item {\bf [Mohrhoff 01 d]}:
U. Mohrhoff,
``A space for the quantum world'',
quant-ph/0107005.

\item {\bf [Mohrhoff 01 e]}:
U. Mohrhoff,
``Two theories of decoherence'',
quant-ph/0108002.

\item {\bf [Mohrhoff 01 f]}:
U. Mohrhoff,
`Against ``knowledge''\',
quant-ph/0109150.

\item {\bf [Mohrhoff 02 a]}:
U. Mohrhoff,
``The world according to quantum mechanics (or,
the 18 errors of Henry P. Stapp)'',
{\em Found. Phys.} {\bf 32}, 2, 217-254 (2002);
quant-ph/0105097.
Comment on {\bf [Stapp 01 c]}.
Reply: {\bf [Stapp 02]}.

\item {\bf [Mohrhoff 02 b]}:
U. Mohrhoff,
``Making sense of a world of clicks'',
{\em Found. Phys.} {\bf 32}, 8, 1295-1311 (2002);
quant-ph/0202148.

\item {\bf [Mohrhoff 02 c]}:
U. Mohrhoff,
``Why the laws of physics are just so'',
{\em Found. Phys.} {\bf 32}, 8, 1313-1324 (2002);
quant-ph/0202149.

\item {\bf [Mohrhoff 03]}:
U. Mohrhoff,
``Is the end in sight for theoretical pseudophysics?'',
in V. Krasnoholovets,
{\em Progress in quantum physics research},
Nova Science, ?, 2003;
quant-ph/0305095.

\item {\bf [Mohrhoff 04 a]}:
U. Mohrhoff,
``This elusive objective existence'',
{\em Int. J. Quant. Inf.};
quant-ph/0401179.

\item {\bf [Mohrhoff 04 b]}:
U. Mohrhoff,
``Do quantum states evolve? Apropos of Marchildon's remarks'',
{\em Found. Phys.} {\bf 34}, 1, 75-97 (2004);
quant-ph/0307113.

\item {\bf [Mohrhoff 04 c]}:
U. Mohrhoff,
``Probabilities from envariance?'',
quant-ph/0401180.
See {\bf [Zurek 03 b]}.

\item {\bf [Mohseni-Lidar 04]}:
M. Mohseni, \& D. A. Lidar,
``Fault-tolerant quantum computation via exchange interactions'',
quant-ph/0406198.

\item {\bf [Mor 96]}:
T. Mor,
``Reducing quantum errors and improving large scale quantum
cryptography'',
quant-ph/9608025.

\item {\bf [Mor 97]}:
T. Mor,
``Quantum memory in quantum cryptography'',
Ph.\ D. thesis, Technion, Haifa, 1997;
quant-ph/9906073.

\item {\bf [Mor 98 a]}:
T. Mor,
``No cloning of orthogonal states in composite systems'',
{\em Phys. Rev. Lett.} {\bf 80}, 14, 3137-3140 (1998);
quant-ph/9802036.

\item {\bf [Mor 99]}:
T. Mor,
``Disentangling quantum states while preserving all local properties'',
{\em Phys. Rev. Lett.} {\bf 83}, 7, 1451-1454 (1999);
quant-ph/9812020.

\item {\bf [Mor-Terno 99]}:
T. Mor, \& D. R. Terno,
``Sufficient conditions for a disentanglement'',
{\em Phys. Rev. A} {\bf 60}, 6, 4341-4343 (1999);
quant-ph/9907036.

\item {\bf [Mor-Horodecki 99]}:
T. Mor, \& P. Horodecki,
``Teleportation via generalized measurements, and conclusive
teleportation'',
quant-ph/9906039.

\item {\bf [Moreno-Garc\'{\i}a Vidal-Erni-(+2) 04]}:
E. Moreno, F. J. Garc\'{\i}a-Vidal, D. Erni,
J. I. Cirac, \& L. Mart\'{\i}n-Moreno,
``Theory of plasmon-assisted transmission of entangled photons'',
{\em Phys. Rev. Lett.} {\bf 92}, 23, 236801 (2004).

\item {\bf [Moret Bailly 01]}:
J. Moret-Bailly,
``Pointlessness and dangerousness of the
postulates of quantum mechanics'',
quant-ph/0107011.

\item {\bf [Morgan 00]}:
P. Morgan,
``The derivation of Bell inequalities for beables'',
quant-ph/0009087.

\item {\bf [Morgan 03 a]}:
P. Morgan,
``A relativistic variant of the Wigner function'',
{\em Phys. Lett. A};
quant-ph/0304171.

\item {\bf [Morgan 03 b]}:
P. Morgan,
``Bell inequalities and incompatible measurements'',
quant-ph/0307123.

\item {\bf [Morigi-Walther 01]}:
G. Morigi, \& H. Walther,
``Two-species Coulomb chains for quantum information'',
{\em Eur. Phys. J. D} {\bf 13}, 3, 261-269 (2001).

\item {\bf [Morikoshi 99]}:
F. Morikoshi,
``Entanglement measure and distance'',
quant-ph/9905022.

\item {\bf [Morikoshi 00]}:
F. Morikoshi,
``Recovery of entanglement lost in entanglement manipulation'',
{\em Phys. Rev. Lett.} {\bf 84}, 14, 3189-3192 (2000);
quant-ph/9911019.

\item {\bf [Morikoshi-Koashi 01]}:
F. Morikoshi, \& M. Koashi,
``Deterministic entanglement concentration'',
{\em Phys. Rev. A} {\bf 64}, 2, 022316 (2001);
quant-ph/0107120.

\item {\bf [Morikoshi-Santos-Vedral 04]}:
F. Morikoshi, M. F. Santos, \& V. Vedral,
``Accessibility of physical states and non-uniqueness of entanglement
measure'',
{\em J. Phys. A} {\bf 37}, ?, 5887-? (2004);
quant-ph/0306032.

\item {\bf [Mosca-Ekert 99]}:
M. Mosca, \& A. K. Ekert,
``The hidden subgroup problem and eigenvalue estimation on a quantum computer'',
in C. P. Williams (ed.),
{\em 1st NASA Int.\ Conf.\ on Quantum Computing and Quantum Communications
(Palm Springs, California, 1998)},
{\em Lecture Notes in Computer Science} {\bf 1509},
Springer-Verlag, New York, 1999, pp.~174-188;
quant-ph/9903071.

\item {\bf [Mosca-Tapp-de Wolf 00]}:
M. Mosca, A. Tapp, \& R. de Wolf,
``Private quantum channels and the cost of randomizing quantum
information'',
quant-ph/0003101.

\item {\bf [Mosna-Vaz 03]}:
R. A. Mosna, \& J. Vaz, Jr.,
``Quantum tomography for Dirac spinors'',
{\em Phys. Lett. A};
quant-ph/0303072.

\item {\bf [Motoyoshi 00]}:
A. Motoyoshi,
``Teleportation without resorting to Bell measurement'',
{\em Phys. Lett. A} {\bf 270}, 6, 293-295 (2000).

\item {\bf [Mould 98]}:
R. A. Mould,
``Consciousness and quantum mechanics'',
{\em Found. Phys.} {\bf 28}, 11, 1703-1718 (1998).

\item {\bf [Mould 01]}:
R. A. Mould,
``Consciousness and endogenous state reduction: Two
experiments'',
{\em Found. Phys. Lett.} {\bf 14}, 4, 377-386 (2001);
quant-ph/0106103.

\item {\bf [Mould 03 a]}:
R. A. Mould,
``Quantum brain states'',
{\em Found. Phys.} {\bf 33}, 4, 571-592 (2003);
quant-ph/0303064.

\item {\bf [Mould 03 b]}:
R. A. Mould,
``Objective vs observer measurements'',
quant-ph/0303065.

\item {\bf [Mould 03 c]}:
R. A. Mould,
``Timing of the Penrose R-process'',
quant-ph/0303067.

\item {\bf [Mould 03 d]}:
R. A. Mould,
``Rule (4) and continuous observation'',
quant-ph/0309029.

\item {\bf [Moura Alves-Horodecki-Oi-(+2) 03]}:
C. Moura Alves, P. Horodecki, D. K. L. Oi,
L. C. Kwek, \& A. K. Ekert,
``Direct estimation of functionals of density operators by local operations
and classical communication'',
{\em Phys. Rev. A} {\bf 68}, 3, 032306 (2003);
quant-ph/0304123.

\item {\bf [Moussa 97 a]}:
M. H. Y. Moussa,
``Teleportation with identity interchange of quantum states'',
{\em Phys. Rev. A} {\bf 55}, 5, R3287-R3290 (1997).

\item {\bf [Moussa 97 b]}:
M. H. Y. Moussa,
``Preparation of the Bell operator basis
through a restricted multiplying device'',
{\em Phys. Lett. A} {\bf 231}, 1, 23-28 (1997).

\item {\bf [Moussa-Baseia 98]}:
M. H. Y. Moussa, \& B. Baseia,
``Nonlocality of a single particle: From the Fock space to cavity QED'',
{\em Phys. Lett. A} {\bf 245}, 5, 335-338 (1998).

\item {\bf [Moya Cessa-Wallentowitz-Vogel 99]}:
H. Moya-Cessa, S. Wallentowitz, \& W. Vogel,
``Quantum-state engineering of a trapped ion
by coherent-state superpositions'',
{\em Phys. Rev. A} {\bf 59}, 4, 2920-2925 (1999).

\item {\bf [Moya Cessa-Dutra-Roversi-Vidiella Barranco 99]}:
H. Moya-Cessa, S. M. Dutra, J. A. Roversi, \& A. Vidiella-Barranco,
``Quantum state reconstruction in the presence of dissipation'',
{\em J. Mod. Opt.} {\bf 46}, 4, 555-558 (1999);
quant-ph/9907046.

\item {\bf [Moya Cessa-Roversi-Dutra-Vidiella Barranco 99]}:
H. Moya-Cessa, J. A. Roversi, S. M. Dutra, \& A. Vidiella-Barranco,
``Recovering coherence from decoherence:
A method of quantum-state reconstruction'',
{\em Phys. Rev. A} {\bf 60}, 5, 4029-4033 (1999);
quant-ph/9908034.

\item {\bf [Moyal 49]}:
J. E. Moyal,
``Quantum mechanics as a statistical theory'',
{\em Proc.\ Cambridge Philos. Soc.} {\bf 45}, 99-124 (1949).

\item {\bf [Mozes-Reznik-Oppenheim 04]}:
S. Mozes, B. Reznik, \& J. Oppenheim,
``Not so superDense coding - Deterministic dense coding with partially
entangled states'',
quant-ph/0403189.

\item {\bf [Mozyrsky-Privman-Hotaling 97]}:
D. Mozyrsky, V. Privman, \& S. P. Hotaling,
``Design of gates for quantum computation: the NOT gate'',
{\em Int. J. Mod. Phys. B} {\bf 11}, ?, 2207-2215 (1997);
quant-ph/9608029.

\item {\bf [Mozyrsky-Privman-Hillery 97]}:
D. Mozyrsky, V. Privman, \& M. Hillery,
``A Hamiltonian for quantum copying'',
{\em Phys. Lett. A} {\bf 226}, 5, 253-256 (1997).

\item {\bf [Mozyrsky-Privman 97]}:
D. Mozyrsky, \& V. Privman,
``Adiabatic decoherence'',
quant-ph/9709020.

\item {\bf [Mozyrsky-Martin 02]}:
D. Mozyrsky, \& I. Martin,
``Quantum-classical transition induced by electrical measurement'',
{\em Phys. Rev. Lett.} {\bf 89}, 1, 018301 (2002).

\item {\bf [Mozyrsky-Kogan-Gorshkov-Berman 02]}:
D. Mozyrsky, S. Kogan, V. N. Gorshkov, \& G. P. Berman,
``Time scales of phonon-induced decoherence of semiconductor spin qubits'',
{\em Phys. Rev. B} {\bf 65}, 24, 245213 (2002).

\item {\bf [M\"{u}ckenheim 82]}:
W. M\"{u}ckenheim,
``A resolution of the Einstein-Podolsky-Rosen paradox'',
{\em Lettere al Nuovo Cimento} {\bf 35}, 9, 300-304 (1982).

\item {\bf [M\"{u}ckenheim 86]}:
W. M\"{u}ckenheim,
``A review of extended probabilities'',
{\em Phys. Rep.} {\bf 133}, 6, 339-? (1986).

\item {\bf [M\"{u}ckenheim 88]}:
W. M\"{u}ckenheim,
``An extended-probability
response to the Einstein-Podolsky-Rosen argument'',
in F. Selleri (ed.),
{\em Quantum mechanics versus local realism:
The Einstein-Podolsky-Rosen paradox},
Plenum Press,
New York, 1988, pp.~345-364.

\item {\bf [M\"{u}ckenheim 93]}:
W. M\"{u}ckenheim,
``On quasi-realistic local
spin models and extended probabilities'',
{\em Phys. Lett. A} {\bf 175}, 165-168 (1993).

\item {\bf [Mueller Quade-Imai 00 a]}:
J. Mueller-Quade, \& H. Imai,
``Quantum cryptographic three party protocols'',
quant-ph/0010111.

\item {\bf [Mueller Quade-Imai 00 b]}:
J. Mueller-Quade, \& H. Imai,
``Temporary assumptions for quantum multi party protocols'',
quant-ph/0010112.

\item {\bf [Mugur Sch\"{a}chter 98]}:
M. Mugur-Sch\"{a}chter,
``Book review. M\'{e}canique quantique: Une introduction philosophique'',
{\em Found. Phys.} {\bf 28}, 12, 1817-1824 (1998).
Review of {\bf [Bitbol 96]}.

\item {\bf [Muller-Breguet-Gisin 93]}:
A. Muller, J. Breguet, \& N. Gisin,
``Experimental demonstration of quantum cryptography using
polarized photons in optical fibre over more than 1 km.'',
{\em Europhys. Lett.} {\bf 23}, 6, 383-388 (1993).

\item {\bf [Muller-Zbinden-Gisin 95]}:
A. Muller, H. Zbinden, \& N. Gisin,
``Underwater quantum coding'',
{\em Nature} {\bf 378}, 6556, 449 (1995).

\item {\bf [Muller-Zbinden-Gisin 96]}:
A. Muller, H. Zbinden, \& N. Gisin,
``Quantum cryptography over 23 km in installed under-lake telecom fibre'',
{\em Europhys. Lett.} {\bf 33}, 5, 335-339 (1996).

\item {\bf [Muller-Herzog-Huttner-(+3) 97]}:
A. Muller, T. Herzog, B. Huttner,
W. Tittel, H. Zbinden, \& N. Gisin,
``\,`Plug and play' systems for quantum cryptography'',
{\em Appl. Phys. Lett.} {\bf 70}, ?, 793-795 (1997);
quant-ph/9611042.

\item {\bf [Mundt-Kreuter-Becher-(+4) 02]}:
A. B. Mundt, A. Kreuter, C. Becher,
D. Leibfried, J. Eschner, F. Schmidt-Kaler, \& R. Blatt,
``Coupling a single atomic quantum bit to a high finesse optical cavity'',
{\em Phys. Rev. Lett.} {\bf 89}, 10, 103001 (2002);
quant-ph/0202112.

\item {\bf [Munro-Reid 93]}:
W. J. Munro, \& M. D. Reid,
``Violation of Bell's inequality by macroscopic states
generated via parametric down-conversion'',
{\em Phys. Rev. A} {\bf 47}, 4, 4412-4421 (1993).

\item {\bf [Munro-Reid 94 a]}:
W. J. Munro, \& M. D. Reid,
``Multiparticle and higher-spin tests of quantum mechanics using
parametric down-conversion'',
{\em Phys. Rev. A} {\bf 50}, 5, 3661-3681 (1994).

\item {\bf [Munro-Reid 94 b]}:
W. J. Munro, \& M. D. Reid,
``?'',
{\em Quantum Opt. Lett.} {\bf 6}, 1, 1-? (1994).

\item {\bf [Munro-Milburn 98]}:
W. J. Munro, \& G. J. Milburn,
``Characterizing Greenberger-Horne-Zeilinger correlations in
nondegenerate parametric oscillation via phase measurements'',
{\em Phys. Rev. Lett.} {\bf 81}, 20, 4285-4288 (1998);
quant-ph/9810075.

\item {\bf [Munro 99 a]}:
W. J. Munro,
``Optimal states for Bell-inequality violations using
quadrature phase homodyne measurements'',
{\em Phys. Rev. A} {\bf 59}, 6, 4197-4201 (1999);
quant-ph/0002001.

\item {\bf [Munro 99 b]}:
W. J. Munro,
``Discrete phase measurements and the Bell inequality'',
{\em J. Opt. B: Quantum Semiclass. Opt.} {\bf 1}, 6, 655-661 (1999);
quant-ph/0002005.

\item {\bf [Munro-Milburn-Sanders 00]}:
W. J. Munro, G. J. Milburn, \& B. C. Sanders,
``Entangled coherent-state qubits in an ion trap'',
{\em Phys. Rev. A} {\bf 62}, 5, 052108 (2000);
quant-ph/9910057.

\item {\bf [Munro-Nemoto-White 01]}:
W. J. Munro, K. Nemoto, \& A. G. White,
``The Bell inequality: A measure of entanglement?'',
{\em J. Mod. Opt.} {\bf 48}, 7, 1239-1246 (2001);
quant-ph/0102119.

\item {\bf [Munro-James-White-Kwiat 01]}:
W. J. Munro, D. F. V. James, A. G. White, \& P. G. Kwiat,
``Maximizing the entanglement of two mixed qubits'',
{\em Phys. Rev. A} {\bf 64}, 3, 030302(R) (2001);
quant-ph/0103113.

\item {\bf [Munro-Nemoto-Milburn-Braunstein 02]}:
W. J. Munro, K. Nemoto, G. J. Milburn, \& S. L. Braunstein,
``Weak-force detection with superposed coherent states'',
{\em Phys. Rev. A} {\bf 66}, 2, 023819 (2002).

\item {\bf [Munro-Nemoto-Beausoleil-Spiller 03]}:
W. J. Munro, K Nemoto, R. G. Beausoleil, \& T. P. Spiller,
``A high-efficiency quantum non-demolition single photon number resolving
detector'',
quant-ph/0310066.

\item {\bf [Murakami-Iinuma-Takahashi-(+2) 03]}:
T. Murakami, M. Iinuma, T. Takahashi,
Y. Kadoya, \& M. Yamanishi,
``Analysis of mutual communication between qubits by capacitive coupling'',
{\em Phys. Rev. A} {\bf 67}, 5, 050301 (2003).

\item {\bf [Murali-Sinha-Mahesh-(+3) 02]}:
K. V. R. M. Murali, N. Sinha, T. S. Mahesh, M. H. Levitt, K. V.
Ramanathan, \& A. Kumar,
``Quantum-information processing by nuclear magnetic resonance: Experimental
implementation of half-adder and subtractor operations using an oriented
spin-7/2 system'',
{\em Phys. Rev. A} {\bf 66}, 2, 022313 (2002).

\item {\bf [Murao-Plenio-Popescu-(+2) 98]}:
M. Murao, M. B. Plenio, S. Popescu, V. Vedral, \& P. L. Knight,
``Multiparticle entanglement purification protocols'',
{\em Phys. Rev. A} {\bf 57}, 6, R4075-R4078 (1998);
quant-ph/9712045.

\item {\bf [Murao-Jonathan-Plenio-Vedral 99]}:
M. Murao, D. Jonathan, M. B. Plenio, \& V. Vedral,
``Quantum telecloning and multiparticle entanglement'',
{\em Phys. Rev. A} {\bf 59}, 1, 156-161 (1999);
quant-ph/9806084.

\item {\bf [Murao-Plenio-Vedral 00]}:
M. Murao, M. B. Plenio, \& V. Vedral,
``Quantum-information distribution via entanglement'',
{\em Phys. Rev. A} {\bf 61}, 3, 032311 (2000);
quant-ph/9909031.

\item {\bf [Murao-Vedral 01]}:
M. Murao, \& V. Vedral,
``Remote information concentration using a bound entangled
state'',
{\em Phys. Rev. Lett.} {\bf 86}, 2, 352-355 (2001);
quant-ph/0008078.

\item {\bf [Murdoch 87]}:
D. Murdoch,
{\em Niels Bohr's philosophy of physics},
Cambridge University Press, Cambridge, 1987.

\item {\bf [Murdoch 94]}:
D. Murdoch,
``The Bohr-Einstein dispute'',
in {\bf [Faye-Folse 94]}, pp.~303-324.

\item {\bf [Murg-Cirac 03]}:
V. Murg, \& J. I. Cirac,
``Adiabatic time evolution in spin-systems'',
quant-ph/0309026.

\item {\bf [Murphy 01]}:
B. Murphy,
``A novel approach to quantum heuristics for
structured database search'',
quant-ph/0106152.

\item {\bf [Mushtari 01]}:
D. K. Mushtari,
``Some remarks on measures on orthogonal rational projections and the rational sphere'',
{\em Lobachevskii J. Math.} {\bf 9}, 47-53 (2001).

\item {\bf [Musser 04]}:
G. Musser,
``?'',
{\em Sci. Am.} {\bf ?}, 13, ?-? (2004);
Spanish version:
``?`Estaba Einstein en lo cierto?'',
{\em Investigaci\'{o}n y Ciencia} 338, 68-71 (2004).

\item {\bf [Mussinger-Delgado-Alber 00]}:
M. Mussinger, A. Delgado, \& G. Alber,
``Error avoiding quantum codes and dynamical
stabilization of Grover's algorithm'',
{\em New J. Phys.} {\bf 2}, 19.1-19.16 (2000);
quant-ph/0003141.

\item {\bf [Muthukrishnan-Stroud 00]}:
A. Muthukrishnan, \& C. R. Stroud, Jr.,
``Multivalued logic gates for quantum computation'',
{\em Phys. Rev. A} {\bf 62}, 5, 052309 (2000);
quant-ph/0002033.

\item {\bf [Muthukrishnan-Stroud 01 a]}:
A. Muthukrishnan, \& C. R. Stroud, Jr.,
``Atomic wave packet basis for quantum information'',
quant-ph/0106165.

\item {\bf [Muthukrishnan-Stroud 01 b]}:
A. Muthukrishnan, \& C. R. Stroud, Jr.,
``Quantum fast Fourier transform using multilevel atoms'',
{\em J. Mod. Opt.};
quant-ph/0112017.

\item {\bf [Muthukrishnan-Stroud 02]}:
A. Muthukrishnan, \& C. R. Stroud, Jr.,
``Entanglement of internal and external
angular momenta of a single atom'',
{\em J. Opt. B: Quantum Semiclass. Opt.} {\bf 4}, 2, S73-S77 (2002);
quant-ph/0111058.

\item {\bf [de Muynck-Janssen-Santman 79]}:
W. M. de Muynck, P. A. E. M. Janssen, \& A. Santman,
``Simultaneous measurement and joint probability distributions in
quantum mechanics'',
{\em Found. Phys.} {\bf 9}, 1-2, 71-122 (1979).

\item {\bf [de Muynck 86]}:
W. M. de Muynck,
``On the relation between the Einstein-Podolsky-Rosen paradox
and the problem of nonlocality in quantum mechanics'',
{\em Found. Phys.} {\bf 16}, 10, 973-1002 (1986).

\item {\bf [de Muynck-Martens 90]}:
W. M. de Muynck, \& H. Martens,
``Neutron interferometry and the joint measurement
of incompatible observables'',
{\em Phys. Rev. A} {\bf 42}, 9, 5079-5085 (1990).

\item {\bf [de Muynck-de Baere-Martens 94]}:
W. M. de Muynck, W. de Baere, \& H. Martens,
``Interpretations of quantum mechanics, joint measurements of
incompatible observables, and counterfactual definitness'',
{\em Found. Phys.} {\bf 24}, 12, 1589-1664 (1994).
Comment: {\bf [Stapp 94 d]}.

\item {\bf [de Muynck-de Baere 96]}:
W. M. de Muynck, \& W. de Baere,
``Quantum nonlocality and Bell's inequalities'',
in A. Mann, \& M. Revzen (eds.),
{\em The dilemma of Einstein, Podolsky and Rosen -- 60 years
later. An international symposium in honour of Nathan Rosen
(Haifa, Israel, 1995)},
{\em Ann. Phys. Soc. Israel} {\bf 12}, 109-126 (1996).

\item {\bf [de Muynck 00]}:
W. M. de Muynck,
``Preparation and measurement: Two independent
sources of uncertainty in quantum mechanics'',
{\em Found. Phys.} {\bf 30}, 2, 205-225 (2000);
quant-ph/9901010.

\item {\bf [de Muynck-Hendrikx 00]}:
W. M. de Muynck, \& A.J.A. Hendrikx,
``The Haroche-Ramsey experiment as a generalized measurement'',
quant-ph/0003043.

\item {\bf [de Muynck 01]}:
W. M. de Muynck,
``Interpretations of quantum mechanics, and interpretations of
violation of Bell's inequality'',
in {\em Foundations of Probability and Physics (V\"{a}xj\"{o}, Sweden, 2000)};
quant-ph/0102066.

\item {\bf [de Muynck 02]}:
W. M. de Muynck,
{\em Foundations of quantum mechanics, an empiricist approach},
Kluwer, Dordrecht, Holland, 2002.
See {\bf [Dieks 03]}.

\item {\bf [de Muynck 04]}:
W. M. de Muynck,
``Towards a neo-Copenhagen interpretation of quantum mechanics'',
{\em Found. Phys.} {\bf 34}, 5, 717-770 (2004);
quant-ph/0307235.

\item {\bf [Myatt-King-Turchette-(+5) 00]}:
C. J. Myatt, B. E. King, Q. A. Turchette, C. A. Sackett,
D. Kielpinski, W. M. Itano, C. Monroe, \& D. J. Wineland,
``Decoherence of quantum superpositions through coupling
to engineered reservoirs'',
{\em Nature} {\bf 403}, 6767, 269-273 (2000).
See {\bf [Schleich 00]}.

\item {\bf [Myers-Ericsson-Laflamme 04]}:
C. R. Myers, M. Ericsson, \& R. Laflamme,
``A single photon source with linear optics and squeezed states'',
quant-ph/0408194.

\item {\bf [Myers-Brandt 97]}:
J. M. Myers, \& H. E. Brandt,
``Converting a positive operator-valued measure to a design
for a measuring instrument on the laboratory bench'',
{\em Meas. Sci. Technol.} {\bf 8}, ?, 1222-1227 (1997).

\item {\bf [Myers 97]}:
J. M. Myers,
``Can a universal quantum computer be fully quantum?'',
{\em Phys. Rev. Lett.} {\bf 78}, 9, 1823-1824 (1997).

\item {\bf [Myers-Fahmy-Glaser-Marx 01]}:
J. M. Myers, A. F. Fahmy, S. J. Glaser, \& R. Marx,
``Rapid solution of problems by
nuclear-magnetic-resonance quantum computation'',
{\em Phys. Rev. A} {\bf 63}, 3, 032302 (2001);
quant-ph/0007043.

\item {\bf [Myers-Madjid 02 a]}:
J. M. Myers, \& F. H. Madjid,
``Gaps between equations and experiments in quantum cryptography'',
in Y. S. Kim, M. A. Man'ko, \& A. Sergienko (eds.),
{\em Seventh International Conference on Squeezed States and Uncertainty Relations (Boston, 2001)},
{\em J. Opt. B: Quantum Semiclass. Opt.} {\bf 4}, 3, S109-S116 (2002).

\item {\bf [Myers-Madjid 02 b]}:
J. M. Myers, \& F. H. Madjid,
``A proof that measured data and equations of quantum mechanics
can be linked only by guesswork'',
in {\bf [Lomonaco-Brandt 02]} 221-244.

\item {\bf [Myhr 04]}:
G. O. Myhr,
``Measures of entanglement in quantum mechanics'',
M.\ Sc. thesis;
quant-ph/0408094.

\item {\bf [Myrgren-Whaley 03]}:
E. S. Myrgren, \& K. B. Whaley,
``Implementing a quantum algorithm with exchange-coupled quantum dots: A
feasibility study'',
quant-ph/0309051.

\item {\bf [Myrvold 02 a]}:
W. C. Myrvold,
``Kochen-Specker obstruction for position and momentum
using a single degree of freedom'',
{\em Phys. Lett. A} {\bf 299}, 1, 8-14 (2002);
quant-ph/0201066,
PITT-PHIL-SCI00000533.

\item {\bf [Myrvold 02 b]}:
W. C. Myrvold,
``A loophole in Bell's theorem? Parameter dependence in the Hess-Philipp model'',
quant-ph/0205032.
Comment on {\bf [Hess-Philipp 01 a, b, c, 02 a]}.
Reply: {\bf [Hess-Philipp 02 f]}.
See {\bf [Appleby 03 a]}.

\item {\bf [Myrvold 02 c]}:
W. C. Myrvold,
``Modal interpretations and relativity'',
{\em Found. Phys. Lett.};
quant-ph/0209109,
PITT-PHIL-SCI00000811.


\newpage

\subsection{}


\item {\bf [Na-Wyatt 02]}:
K. Na, \& R. E. Wyatt,
``Decoherence demystified: The hydrodynamic viewpoint'',
quant-ph/0201108.

\item {\bf [Nagaj-Stelmachovic-Bu\v{z}zek-Kim 02]}:
D. Nagaj, P. Stelmachovic, V. Bu\v{z}zek, \& M. Kim,
``Quantum homogenization for continuous variables: Realization with linear
optical elements'',
{\em Phys. Rev. A} {\bf 66}, 6, 062307 (2002).

\item {\bf [Nagarajan-Gay 02]}:
R. Nagarajan, \& S. Gay,
``Formal verification of quantum protocols'',
quant-ph/0203086.

\item {\bf [Nagata-Koashi-Imoto 02 a]}:
K. Nagata, M. Koashi, \& N. Imoto,
``Observables suitable for restricting the fidelity to multipartite maximally
entangled states'',
{\em Phys. Rev. A} {\bf 65}, 4, 042314 (2002).

\item {\bf [Nagata 02]}:
K. Nagata,
``Classification of mixed high-dimensional multiparticle systems'',
{\em Phys. Rev. A} {\bf 66}, 6, 064101 (2002).

\item {\bf [Nagata-Koashi-Imoto 02 a]}:
K. Nagata, M. Koashi, \& N. Imoto,
``Configuration of separability and tests for multipartite entanglement in
Bell-type experiments'',
{\em Phys. Rev. Lett.} {\bf 89}, 26, 260401 (2002).

\item {\bf [Nagata 03 a]}:
K. Nagata,
``Incompatibility between two meanings of locality in quantum mechanics'',
quant-ph/0302090.

\item {\bf [Nagata 03 b]}:
K. Nagata,
``Inequalities as tests for Kochen-Specker theorem for multipartite states'',
quant-ph/0312052.

\item {\bf [Nagata-Laskowski-\.{Z}ukowski 04]}:
K. Nagata, W. Laskowski, \& M. \.{Z}ukowski,
``Rotational invariance as an additional constraint on local realism'',
quant-ph/0407232.

\item {\bf [N\"{a}gerl-Bechter-Eschner-(+2) 98]}:
H. C. N\"{a}gerl, W. Bechter, J. Eschner, F. Schmidt-Kaler, \& R. Blatt,
``Ion strings for quantum gates'',
{\em Appl. Phys. B - Lasers Opt.} {\bf 66}, 603-608 (1998).
Reprinted in {\bf [Macchiavello-Palma-Zeilinger 00]}, pp.~339-344.

\item {\bf [N\"{a}gerl-Leibfried-Rohde-(+4) 99]}:
H. C. N\"{a}gerl, D. Leibfried, H. Rohde,
G. Thalhammer, J. Eschner, F. Schmidt-Kaler, \& R. Blatt,
``Laser addressing of individual ions in a linear ion trap'',
{\em Phys. Rev. A} {\bf 60}, 1, 145-148 (1999).

\item {\bf [Nagy 90]}:
B. S. Nagy,
``Extensions of linear transformations in Hilbert space
which extend beyond this space'',
appendix in F. Riesz, \& B. S. Nagy,
{\em Functional analysis},
Dover, New York, 1990.

\item {\bf [Naik-Peterson-White-(+2) 00]}:
D. S. Naik, C. G. Peterson, A. G. White,
A. J. Berglund, \& P. G. Kwiat,
``Entangled state quantum cryptography:
Eavesdropping on the Ekert protocol'',
{\em Phys. Rev. Lett.} {\bf 84}, 20, 4733-4736 (2000);
quant-ph/9912105.

\item {\bf [Nairz-Arndt-Zeilinger 01]}:
O. Nairz, M. Arndt, \& A. Zeilinger,
``Experimental verification of the Heisenberg
uncertainty principle for hot fullerene molecules'',
quant-ph/0105061.

\item {\bf [Nairz-Brezger-Arndt-Zeilinger 01]}:
O. Nairz, B. Brezger, M. Arndt, \& A. Zeilinger,
``Diffraction of complex molecules by structures made of light'',
{\em Phys. Rev. Lett.} {\bf 87}, 16, 160401 (2001);
quant-ph/0110012.

\item {\bf [Nairz-Arndt-Zeilinger 03]}:
O. Nairz, M. Arndt, \& A. Zeilinger,
``Quantum interference experiments with large molecules`'',
{\em Am. J. Phys.} {\bf 71}, 4, 319–325 (2003),
Erratum: {\em Am. J. Phys.} {\bf 71}, 10, 1084 (2003).

\item {\bf [Nakamura-Pashkin-Tsai 99]}:
Y. Nakamura, Y. A. Pashkin, \& J. S. Tsai,
``Coherent control of macroscopic quantum states in a single-Cooper-pair box'',
{\em Nature} {\bf 398}, 6730, 786-788 (1999).
See {\bf [Averin 99]}, {\bf [Collins 99]}.

\item {\bf [Nakazato-Unoki-Yuasa 04]}:
H. Nakazato, M. Unoki, \& K. Yuasa,
``Preparation and entanglement purification of qubits through Zeno-like
measurements'',
{\em Phys. Rev. A} {\bf 70}, 1, 012303 (2004);
quant-ph/0402182.

\item {\bf [Nakhmanson 00 a]}:
R. Nakhmanson,
``Informational interpretation of quantum mechanics'',
physics/0004047.

\item {\bf [Nakhmanson 00 b]}:
R. Nakhmanson,
``Bell's theorem, entanglement, and common sense'',
physics/0005042.

\item {\bf [Nakhmanson 00 c]}:
R. Nakhmanson,
``Very promising hole in Bell's theorem'',
physics/0011022.

\item {\bf [Nakhmanson 02]}:
R. Nakhmanson,
``Wavepacket and its collapse'',
quant-ph/0204068.
Comment on {\bf [Chiao-Kwiat 02]}.

\item {\bf [Nambu-Tomita-Chiba Kohno-Nakamura 00]}:
Y. Nambu, A. Tomita, Y. Chiba-Kohno, \& K. Nakamura,
``Quantum key distribution using two coherent states
of light and their superposition'',
{\em Phys. Rev. A} {\bf 62}, 1, 012312 (2000);
quant-ph/9911035.

\item {\bf [Nambu-Chiba Kohno 00]}:
Y. Nambu, \& Y. Chiba-Kohno,
``Information-theoretic description of no-go theorem of a bit
commitment'',
quant-ph/0011068.

\item {\bf [Nambu-Usami-Tsuda-(+2) 02]}:
Y. Nambu, K. Usami, Y. Tsuda,
K. Matsumoto, \& K. Nakamura,
``Generation of polarization-entangled photon pairs in a cascade of two type-I
crystals pumped by femtosecond pulses'',
{\em Phys. Rev. A} {\bf 66}, 3, 033816 (2002).

\item {\bf [Nambu-Hatanaka-Nakamura 04]}:
Y. Nambu, T. Hatanaka, \& K. Nakamura,
``BB84 quantum key distribution system based on silica-based planar
lightwave circuits'',
{\em Jpn. J. Appl. Phys.} {\bf 43} 8B, L1109-L1110 (2004);
quant-ph/0404015.

\item {\bf [Nakamiki-Pascazio 93]}:
M. Nakamiki, \& S. Pascazio,
``Quantum theory of measurement based on the many-Hilbert-space approach'',
{\em Phys. Rep.} {\bf 232}, 6, 301-414 (1993).

\item {\bf [Nakamiki-Pascazio-Nakazato 97]}:
M. Nakamiki, S. Pascazio, \& H. Nakazato,
{\em Decoherence and quantum measurements},
World Scientific, Singapore, 1997.
Review: {\bf [Schulman 00]}.

\item {\bf [Nakamiki 99]}:
M. Nakamiki,
``Decoherence and wavefunction collapse in quantum measurements'',
{\em Found. Phys.} {\bf 29}, 3, 457-464 (1999).

\item {\bf [Namiki-Hirano 03]}:
R. Namiki, \& T. Hirano,
``Security of quantum cryptography using balanced homodyne detection'',
{\em Phys. Rev. A} {\bf 67}, 2, 022308 (2003);
quant-ph/0205191.

\item {\bf [Namiki-Hirano 04]}:
R. Namiki, \& T. Hirano,
``Practical limitation for continuous-variable quantum cryptography using coherent states'',
{\em Phys. Rev. Lett.} {\bf 92}, 11, 117901 (2004).

\item {\bf [Namiot-Chernavskii 03]}:
V. A. Namiot, \& D. S. Chernavskii,
``Efficient computing procedures and impossibility to solve the
problem of exact prediction of events in the quantum world'',
{\em Phys. Lett. A} {\bf 320}, 1, 1-4 (2003).

\item {\bf [Narnhofer-Thirring 02]}:
H. Narnhofer, \& W. Thirring,
``Entanglement of mesoscopic systems'',
{\em Phys. Rev. A} {\bf 66}, 5, 052304 (2002).

\item {\bf [Narnhofer 03]}:
H. Narnhofer,
``The structure of state space with respect to imbedding'',
{\em Int. J. Theor. Phys.} {\bf 42}, 5, 955-967 (2003).

\item {\bf [Nascimento-Mueller Quade-Imai 01]}:
A. C. A. Nascimento, J. Mueller-Quade, \& H. Imai,
``Improving quantum secret-sharing schemes'',
{\em Phys. Rev. A} {\bf 64}, 4, 042311 (2001).

\item {\bf [Nathanson 03]}:
M. Nathanson,
``Quantum guessing via Deutsch-Jozsa'',
quant-ph/0301025.

\item {\bf [Nayak-Ta Shma-Zuckerman 00]}:
A. Nayak, A. Ta-Shma, \& D. Zuckerman,
``Interaction in quantum communication complexity'',
quant-ph/0005106.

\item {\bf [Nayak-Vishwanath 00]}:
A. Nayak, \& A. Vishwanath,
``Quantum walk on a line'',
quant-ph/0010117.

\item {\bf [Nayak-Shor 03]}:
A. Nayak, \& P. Shor,
``Bit-commitment-based quantum coin flipping'',
{\em Phys. Rev. A} {\bf 67}, 1, 012304 (2003).

\item {\bf [Navascues-Bae-Cirac-(+3) 04]}:
M. Navascues, J. Bae, J. I. Cirac,
M. Lewestein, A. Sanpera, \& A. Ac\'{\i}n,
``Key distillation from Gaussian states by Gaussian operations'',
quant-ph/0405047.

\item {\bf [Navascues-Ac\'{\i}n 04]}:
M. Navascues, \& A. Ac\'{\i}n,
``Security bounds for continuous variables quantum key distribution'',
quant-ph/0407149.

\item {\bf [Navarro-Parra 81]}:
L. Navarro, \& J. M. Parra,
``Relevancia del problema E. P. R. en la fundamentaci\'{o}n de la
mec\'{a}nica cu\'{a}ntica'',
in {\em Libro homenaje a la memoria del profesor Ricardo Marqu\'{e}s
Fern\'{a}ndez}, Facultad de F\'{\i}sica de la Universidad de Barcelona,
Barcelona, 1981, pp.~343-358.

\item {\bf [Navarro 89]}:
L. Navarro,
``Heterod\`{o}xia i ci\`{e}ncia'',
{\em Quaderns Fundaci\'{o} Caixa de Pensions}, 42, 60-69 (1989).

\item {\bf [Navez-Gatti-Lugiato 01]}:
P. Navez, A. Gatti, \& L. A. Lugiato,
`A ``quantum public key'' based cryptographic scheme
for continuous variables',
quant-ph/0101113.

\item {\bf [Navez-Brambilla-Gatti-Lugiato 02]}:
P. Navez, E. Brambilla, A. Gatti, \& L. A. Lugiato,
``Spatial entanglement of twin quantum images'',
{\em Phys. Rev. A} {\bf 65}, 1, 013813 (2002).

\item {\bf [Navez-Gatti-Lugiato 02]}:
P. Navez, A. Gatti, \& L. A. Lugiato,
``Invisible transmission in quantum cryptography using continuous variables: A
proof of Eve's vulnerability'',
{\em Phys. Rev. A} {\bf 65}, 3, 032307 (2002).

\item {\bf [Navez 02]}:
P. Navez,
``Statistical confidentiality tests for a quantum transmission using continuous variables'',
{\em Eur. Phys. J. D} {\bf 18}, 2 (Special issue:
{\em Quantum interference and cryptographic keys:
Novel physics and advancing technologies (QUICK) (Corsica, 2001)}, 219-228 (2002).

\item {\bf [Navez-Cerf 03]}:
P. Navez, \& N. J. Cerf,
``Cloning a real $d$-dimensional quantum state on the edge of the no-signaling
condition'',
{\em Phys. Rev. A} {\bf 68}, 3, 032313 (2003);
quant-ph/0302172.

\item {\bf [Nawaz-Toor 01 a]}:
A. Nawaz, \& A. H. Toor,
``Evolutionarily stable strategies in quantum Hawk-Dove game'',
quant-ph/0108075.

\item {\bf [Nawaz-Toor 01 b]}:
A. Nawaz, \& A. H. Toor,
``Worst-case payoffs in quantum battle of sexes game'',
quant-ph/0110096.

\item {\bf [Nazarkin-Netz-Sauerbrey 04]}:
A. Nazarkin, R. Netz, \& R. Sauerbrey,
``Electromagnetically induced quantum memory'',
{\em Phys. Rev. Lett.} {\bf 92}, 4, 043002 (2004).

\item {\bf [Nazir-Spiller-Munro 02]}:
A. Nazir, T. P. Spiller, \& W. J. Munro,
``Decoherence of geometric phase gates'',
{\em Phys. Rev. A} {\bf 65}, 4, 042303 (2002).

\item {\bf [Negrevergne-Somma-Ortiz-(+2) 04]}:
C. Negrevergne, R. Somma, G. Ortiz,
E. Knill, \& R. Laflamme,
``Liquid state NMR simulations of quantum many-body problems'',
quant-ph/0410106.

\item {\bf [Nelson 66]}:
E. Nelson,
``Derivation of the Schr\"{o}dinger equation from Newtonian mechanics'',
{\em Phys. Rev.} {\bf 150}, 4, 1079-1085 (1966).

\item {\bf [Nelson 67]}:
E. Nelson,
{\em Dynamical theories of Brownian motion},
Princeton University Press, Princeton, New Jersey, 1967.

\item {\bf [Nelson 85]}:
E. Nelson,
{\em Quantum fluctuations},
Princeton University Press, Princeton, New Jersey, 1985.

\item {\bf [Nelson-Cory-Lloyd 00]}:
R. J. Nelson, D. G. Cory, \& S. Lloyd,
``Experimental demonstration of Greenberger-Horne-Zeilinger
correlations using nuclear magnetic resonance'',
{\em Phys. Rev. A} {\bf 61}, 2, 022106 (2000);
quant-ph/9905028.

\item {\bf [Nelson-Weinstein-Cory-Lloyd 00]}:
R. J. Nelson, Y. Weinstein, D. G. Cory, \& S. Lloyd,
``Experimental demonstration of fully coherent quantum feedback'',
{\em Phys. Rev. Lett.} {\bf 85}, 14, 3045-3048 (2000).

\item {\bf [Nemoto-Braunstein 02]}:
K. Nemoto, \& S. L. Braunstein,
``Equivalent efficiency of a simulated photon-number detector'',
{\em Phys. Rev. A} {\bf 66}, 3, 032306 (2002).

\item {\bf [Nemoto-Braunstein 03 a]}:
K. Nemoto, \& S. L. Braunstein,
``Quantum coherence in the presence of unobservable quantities'',
quant-ph/0306084.

\item {\bf [Nemoto-Munro-Milburn-Braunstein 03]}:
K. Nemoto, W. J. Munro, G. J. Milburn, \& S. L. Braunstein,
``Quantum metrology: Detection of weak forces using Schr\"{o}dinger cat
resources'',
in J. H. Shapiro and O. Hirota (eds.),
{\em Proc.\ of the Sixth Int.\ Conf.\ on Quantum
Communication, Measurement and Computing},
Rinton Press, Princeton, New Jersey, 2003, p.~333-?;
quant-ph/0312063.

\item {\bf [Nemoto-Braunstein 03 a]}:
K. Nemoto, \& S. L. Braunstein,
``Quantum coherence: Myth or fact?'',
quant-ph/0312108.

\item {\bf [Nemoto-Munro 04]}:
K. Nemoto, \& W. J. Munro,
``A near deterministic linear optical CNOT gate'',
quant-ph/0408118.

\item {\bf [Neumaier 99]}:
A. Neumaier,
``On a realistic interpretation of quantum mechanics'',
quant-ph/9908071.

\item {\bf [Neumaier 00]}:
A. Neumaier,
``Bohmian mechanics contradicts quantum mechanics'',
quant-ph/0001011.
See {\bf [Ghose 00 a]}, {\bf [Marchildon 00]}.

\item {\bf [Neumark 43]}:
M. A. Neumark,
``On a representation of additive operator set functions'',
{\em Dokl. Acad. Sci. URSS} {\bf 41}, ?, 359-361 (1943).

\item {\bf [Neumark 54]}:
M. A. Neumark,
``Operatorenalgebren im Hilbertschen Raum'',
in {\em Sowjetische Arbeiten zur Funktionalanalysis}, Verlag Kultur und
Fortschritt, Berlin, 1954. The theorem of von Neumann-Neumark is in p.~227;
see {\bf [von Neumann 31]}, {\bf [Park-Margenau 68]}, {\bf [Peres 90 a]}.

\item {\bf [Neves-P\'{a}dua-C. Saavedra 04]}:
L. Neves, S. P\'{a}dua, \& C. Saavedra,
``Controlled generation of maximally entangled qudits using twin photons'',
{\em Phys. Rev. A} {\bf 69}, 4, 042305 (2004).

\item {\bf [Neves-Lima-Aguirre G\'{o}mez-(+3) 04]}:
L. Neves, G. Lima, J. G. Aguirre G\'{o}mez,
C. H. Monken, C. Saavedra, \& S. P\'{a}dua,
``Generation of maximally entangled states of qudits using twin photons'',
quant-ph/0411054.

\item {\bf [Newton-Young 68]}:
R. G. Newton, \& B. Young,
``Measurability of the spin density matrix'',
{\em Ann. Phys.} {\bf 49}, 3, 393-402 (1968).

\item {\bf [Newton 80]}:
R. G. Newton,
``Probability interpretation of quantum mechanics'',
{\em Am. J. Phys.} {\bf 48}, 12, 1029-1034 (1980).

\item {\bf [Newton 04]}:
R. G. Newton,
``What is a state in quantum mechanics?'',
{\em Am. J. Phys.} {\bf 72}, 3, 348–350 (2004).
Erratum: {\em Am. J. Phys.} {\bf 72}, 9, 1261 (2004).

\item {\bf [Nguyen 04]}:
B. A. Nguyen,
``Quantum dialogue'',
{\em Phys. Lett. A} {\bf 328}, 1, 6-10 (2004).

\item {\bf [Nguyen-Van Assche-Cerf 04]}:
K.-C. Nguyen, G. Van Assche, \& N. J. Cerf,
``Side-information coding with turbo codes and its application to quantum
key distribution'',
submitted to {\em ISITA 2004};
cs.IT/0406001.

\item {\bf [Nha-Carmichael 04 a]}:
H. Nha, \& H. J. Carmichael,
``Proposed test of quantum nonlocality for continuous variables'',
{\em Phys. Rev. Lett.} {\bf 93}, 2, 020401 (2004);
quant-ph/0406101.

\item {\bf [Nha-Carmichael 04 b]}:
H. Nha, \& H. J. Carmichael,
``Entanglement within the quantum trajectory description of open quantum
systems'',
{\em Phys. Rev. Lett.} {\bf 93}, 12, 120408 (2004);
quant-ph/0408060.

\item {\bf [Nha-Carmichael 04 c]}:
H. Nha, \& H. J. Carmichael,
``Decoherence of a two-state atom driven by coherent light'',
{\em Phys. Rev. A};
quant-ph/0411007.

\item {\bf [Ni 98]}:
G. Ni,
``Einstein-Pauli-Yukawa paradox---What is the physical reality?'',
quant-ph/9803001.

\item {\bf [Ni-Guan 99]}:
G. Ni, \& H. Guan,
``Einstein-Podolsky-Rosen paradox and antiparticle'',
quant-ph/9901046.

\item {\bf [Ni-Guan-Zhou-Yan 00]}:
G. Ni, H. Guan, W. Zhou, \& J. Yan,
``Antiparticle in light of Einstein-Podolsky-Rosen paradox
and Klein paradox'',
{\em Chinese Phys. Lett.};
quant-ph/0001016.

\item {\bf [Ni 01]}:
G. Ni,
``What Schr\"{o}dinger's cat is telling'',
quant-ph/0103064.

\item {\bf [Nichols-Wootters 03]}:
S. R. Nichols, \& W. K. Wootters,
``Between entropy and subentropy'',
{\em Quant. Inf. Comp.} {\bf 3}, 1, 1-? (2003).

\item {\bf [Niederberger-Scarani-Gisin 04]}:
A. Niederberger, V. Scarani, \& N. Gisin,
``Photon-number-splitting versus cloning attacks in practical quantum
cryptography'',
quant-ph/0408122.

\item {\bf [Nielsen 96 a]}:
M. A. Nielsen,
``Properties of quantum trajectories for counting measurements'',
{\em J. Opt. B: Quantum Semiclass. Opt.} {\bf 8}, ?, 237-? (1996).

\item {\bf [Nielsen 96 b]}:
M. A. Nielsen,
``The entanglement fidelity and quantum error correction'',
quant-ph/9606012.

\item {\bf [Nielsen-Caves 97]}:
M. A. Nielsen, \& C. M. Caves,
``Reversible
quantum operations and their application to teleportation'',
{\em Phys. Rev. A} {\bf 55}, 4, 2547-2556 (1997);
quant-ph/9608001.

\item {\bf [Nielsen-Chuang 97]}:
M. A. Nielsen, \& I. L. Chuang,
``Programmable quantum gate arrays'',
{\em Phys. Rev. Lett.} {\bf 79}, 2, 321-324 (1997);
quant-ph/9703032.

\item {\bf [Nielsen 97]}:
M. A. Nielsen,
``Computable functions, quantum measurements, and quantum dynamics'',
{\em Phys. Rev. Lett.} {\bf 79}, 15, 2915-2918 (1997);
quant-ph/9706006.

\item {\bf [Nielsen-Caves-Schumacher-Barnum 98]}:
M. A. Nielsen, C. M. Caves, B. W. Schumacher, \& H. N. Barnum,
``Information-theoretic approach to quantum
error correction and reversible measurement'',
in D. P. DiVincenzo. E. Knill, R. Laflamme, \& W. H. Zurek (eds.),
{\em Quantum Coherence and Decoherence.
Proc.\ of the ITP Conf.\ (Santa Barbara, California, 1996)},
{\em Proc. R. Soc. Lond. A} {\bf 454}, 1969, 277-304 (1998);
quant-ph/9706064.

\item {\bf [Nielsen-Knill-Laflamme 98]}:
M. A. Nielsen, E. Knill, \& R. Laflamme,
``Complete quantum teleportation using nuclear magnetic resonance'',
{\em Nature} {\bf 396}, 6706, 52-55 (1998);
quant-ph/9811020.

\item {\bf [Nielsen 98 a]}:
M. A. Nielsen,
``Quantum information theory'',
Ph.\ D. thesis, University of New Mexico, 1998;
quant-ph/0011036.

\item {\bf [Nielsen 98 b]}:
M. A. Nielsen,
``A partial order on the entangled states'',
quant-ph/9811053.

\item {\bf [Nielsen 99]}:
M. A. Nielsen,
``Conditions for a class of entanglement transformations'',
{\em Phys. Rev. Lett.} {\bf 83}, 2, 436-439 (1999);
quant-ph/9811053.

\item {\bf [Nielsen 00 a]}:
M. A. Nielsen,
``Continuity bounds for entanglement'',
{\em Phys. Rev. A} {\bf 61}, 6, 064301 (2000);
quant-ph/9908086.

\item {\bf [Nielsen 00 b]}:
M. A. Nielsen,
``Probability distributions consistent with a mixed state'',
{\em Phys. Rev. A} {\bf 62}, 5, 052308 (2000);
quant-ph/9909020.

\item {\bf [Nielsen 00 c]}:
M. A. Nielsen,
``Introduction to quantum information theory'',
quant-ph/0011064.

\item {\bf [Nielsen-Chuang 00]}:
M. A. Nielsen, \& I. L. Chuang,
{\em Quantum computation and quantum information},
Cambridge University Press, Cambridge, 2000.
Review: {\bf [Gudder 01 a]}, {\bf [James 01]}, {\bf [Lloyd 02 a]},
{\bf [Grover 02 a]}.

\item {\bf [Nielsen 01 a]}:
M. A. Nielsen,
``Characterizing mixing and measurement in quantum mechanics'',
{\em Phys. Rev. A} {\bf 63}, 2, 022113 (2001);
quant-ph/0008073.

\item {\bf [Nielsen-Kempe 01]}:
M. A. Nielsen, \& J. Kempe,
``Separable states are more disordered globally than locally'',
{\em Phys. Rev. Lett.} {\bf 86}, 22, 5184-5187 (2001);
quant-ph/0011117.

\item {\bf [Nielsen-Vidal 01]}:
M. A. Nielsen, \& G. Vidal,
``Majorization and the interconversion of bipartite states'',
{\em Quantum Information and Computation} {\bf 1}, 1, 76-93 (2001).

\item {\bf [Nielsen 01 b]}:
M. A. Nielsen,
``On the units of bipartite entanglement:
Is sixteen ounces of entanglement always equal to one pound?'',
in S. Popescu, N. Linden, \& R. Jozsa (eds.),
{\em J. Phys. A} {\bf 34}, 35
(Special issue: Quantum information and computation), 6987-6996 (2001);
quant-ph/0011063.

\item {\bf [Nielsen-Bremner-Dodd-(+2) 02]}:
M. A. Nielsen, M. J. Bremner, J. L. Dodd,
A. M. Childs, \& C. M. Dawson,
``Universal simulation of Hamiltonian dynamics for quantum systems with
finite-dimensional state spaces'',
{\em Phys. Rev. A} {\bf 66}, 2, 022317 (2002);
quant-ph/0109064.

\item {\bf [Nielsen 02 a]}:
M. A. Nielsen,
``A simple formula for the average gate fidelity of a quantum dynamical operation'',
{\em Phys. Lett. A} {\bf 303}, 4, 249-252 (2002).

\item {\bf [Nielsen 02 b]}:
M. A. Nielsen,
``Rules for a complex quantum world'',
{\em Sci. Am.} {\bf ?}, 11, ?-? (2002);
Spanish version:
``Reglas para un mundo cu\'{a}ntico complejo'',
{\em Investigaci\'{o}n y Ciencia} 316, 44-53 (2003).
Reprinted in {\bf [Cabello 03 a]}, pp.~30-37.

\item {\bf [Nielsen 03]}:
M. A. Nielsen,
``Quantum computation by measurement and quantum memory'',
{\em Phys. Lett. A} {\bf 308}, 2-3, 96-100 (2003);
quant-ph/0108020.

\item {\bf [Nielsen-Dawson-Dodd-(+6) 03]}:
M. A. Nielsen, C. M. Dawson, J. L. Dodd,
A. Gilchrist, D. Mortimer, T. J. Osborne,
M. J. Bremner, A. W. Harrow, \& A. Hines,
``Quantum dynamics as a physical resource'',
{\em Phys. Rev. A} {\bf 67}, 5, 052301 (2003).

\item {\bf [Nielsen 04]}:
M. A. Nielsen,
``Optical quantum computation using cluster states'',
{\em Phys. Rev. Lett.} {\bf 93}, 4, 040503 (2004);
quant-ph/0402005.

\item {\bf [Nielsen-Dawson 04]}:
M. A. Nielsen, \& C. M. Dawson,
``Fault-tolerant quantum computation with cluster states'',
quant-ph/0405134.

\item {\bf [Nielsen-Petz 04]}:
M. A. Nielsen, \& D. Petz,
``A simple proof of the strong subadditivity inequality'',
quant-ph/0408130.

\item {\bf [Nikolic 01]}:
H. Nikolic,
`Comment on ``Classical interventions in quantum systems II.
Relativistic invariance''\,',
{\em Phys. Rev. A} {\bf 64}, 6, 066101 (2001);
quant-ph/0109120.
Comment on {\bf [Peres 00 c]}.
Reply: {\bf [Peres 01 b]}.

\item {\bf [Nikolic 04]}:
H. Nikolic,
``Bohmian particle trajectories in relativistic bosonic quantum field
theory'',
{\em Found. Phys. Lett.} (2004);
quant-ph/0208185.

\item {\bf [Nishimura-Ozawa 01]}:
H. Nishimura, \& M. Ozawa,
``Quantum subroutine problem and the robustness of quantum
complexity classes'',
quant-ph/0107089.

\item {\bf [Nishimura-Yamakami 04]}:
H. Nishimura, \& T. Yamakami,
``Polynomial time quantum computation with advice'',
{\em Information Processing Lett.} {\bf 90}, 195-204 (2004);
quant-ph/0305100.

\item {\bf [Nishioka-Ishizuka-Hasegawa 01]}:
T. Nishioka, H. Ishizuka, \& T. Hasegawa,
`\,``Circular type'' quantum key distribution',
quant-ph/0106083.

\item {\bf [Nishioka-Hasegawa-Ishizuka-(+2) 04]}:
T. Nishioka, T. Hasegawa, H. Ishizuka,
K. Imafuku, \& H. Imai,
``How much security does Y-00 protocol provide us?'',
{\em Phys. Lett. A} {\bf 327}, 1, 28-32 (2004).

\item {\bf [Niskanen-Nakahara-Salomaa 03]}:
A. O. Niskanen, M. Nakahara, \& M. M. Salomaa,
``Realization of arbitrary gates in holonomic quantum computation'',
{\em Phys. Rev. A} {\bf 67}, 1, 012319 (2003).

\item {\bf [Niskanen-Vartiainen-Salomaa 03]}:
A. O. Niskanen, J. J. Vartiainen, \& M. M. Salomaa,
``Optimal multiqubit operations for Josephson charge qubits'',
{\em Phys. Rev. Lett.} {\bf 90}, 19, 197901 (2003).

\item {\bf [Nistic\`{o}-Concetta Romania 94]}:
G. Nistic\`{o}, \& M. Concetta Romania,
``Knowledge about noncommuting quantum observables by means of
Einstein-Podolsky-Rosen correlations'',
{\em J. Math. Phys.} {\bf 35}, 9, 4534-4546 (1994).

\item {\bf [Nistic\`{o} 97]}:
G. Nistic\`{o},
``\,`Counterfactual' interpretation of the quantum measurement process'',
{\em Found. Phys. Lett.} {\bf 10}, 4, 371-381 (1997).

\item {\bf [Nistic\`{o} 99]}:
G. Nistic\`{o},
``Consistency conditions for probabilities of quantum histories'',
{\em Found. Phys.} {\bf 29}, 2, 221-240 (1999).

\item {\bf [Nistic\`{o} 00]}:
G. Nistic\`{o},
`Efficacy of non-locality theorems ``without inequalities''
for pairs of spin 1/2 particles',
quant-ph/0012012.
See {\bf [Nistic\`{o} 01]}.

\item {\bf [Nistic\`{o}-Beneduci 01]}:
G. Nistic\`{o}, \& R. Beneduci,
``Individual consistency of 2-events quantum histories'',
quant-ph/0101041.

\item {\bf [Nistic\`{o} 01]}:
G. Nistic\`{o},
``Impossibility of Greenberger-Horne-Shimony-Zeilinger's
theorems for pairs of spin-1/2 particles'',
{\em Phys. Lett. A} {\bf 281}, 5-6, 273-277 (2001).
See {\bf [Nistic\`{o} 00]}.

\item {\bf [Nistic\`{o}-Beneduci 02]}:
G. Nistic\`{o}, \& R. Beneduci,
``Self-decoherence criterion of consistency for quantum histories'',
{\em Phys. Lett. A} {\bf 299}, 5-6, 433-440 (2002).

\item {\bf [Nistic\`{o} 03]}:
G. Nistic\`{o},
``Conceptual analysis of quantum history theory'',
{\em Phys. Essays} {\bf 16}, 1, ?-? (2003).
quant-ph/0310135.

\item {\bf [Nistic\`{o} 04]}:
G. Nistic\`{o},
``Detection of incompatible properties without erasure'',
quant-ph/0409092.

\item {\bf [Niu-Griffiths 98 a]}:
C. Niu, \& R. B. Griffiths,
``Optimal copying of one quantum bit'',
{\em Phys. Rev. A} {\bf 58}, 6, 4377-4393 (1998).

\item {\bf [Niu-Griffiths 99]}:
C. Niu, \& R. B. Griffiths,
``Two-qubit copying machine for economical quantum eavesdropping'',
{\em Phys. Rev. A} {\bf 60}, 4, 2764-2776 (1999);
quant-ph/9810008.

\item {\bf [Niu-Wang 04]}:
W. Niu, \& A. M. Wang,
``Separability criterion and local information in separable states'',
{\em Phys. Lett. A} {\bf 327}, 2-3, 103-106 (2004).

\item {\bf [Niwa-Matsumoto-Imai 99]}:
J. Niwa, K. Matsumoto, \& H. Imai,
``General-purpose parallel simulator for quantum computing'',
{\em Phys. Rev. A} {\bf 66}, 6, 062317 (2002).

\item {\bf [del Noce 95]}:
C. del Noce,
``An algorithm for finding Bell-type inequalities'',
{\em Found. Phys. Lett.} {\bf 8}, 3, 213-229 (1995).

\item {\bf [Noel-Stroud 95]}:
M. W. Noel, \& C. R. Stroud, Jr.,
``Young's double-slit interferometrywithin an atom'',
{\em Phys. Rev. Lett.} {\bf 75}, 7, 1252-1255 (1995).

\item {\bf [Nogami-Toyama-Dijk 00]}:
Y. Nogami, F. M. Toyama, \& W. van Dijk,
``Bohmian description of a decaying quantum system'',
{\em Phys. Lett. A} {\bf 270}, 6, 279-287 (2000);
quant-ph/0005109.

\item {\bf [Nogueira-dos Aidos-Caldeira-Domingos 92]}:
F. Nogueira, F. D. dos Aidos, M. H. Caldeira, \& J. M. Domingos,
``Expectation values and non-commuting operators'',
{\em Eur. J. Phys.} {\bf 13}, 6, 284-285 (1992).

\item {\bf [Nogueira-Walborn-P\'{a}dua-Monken 02]}:
W. A. T. Nogueira, S. P. Walborn, S. P\'{a}dua, \& C. H. Monken,
``Spatial antibunching of photons with parametric down-conversion'',
{\em Phys. Rev. A} {\bf 66}, 5, 053810 (2002).

\item {\bf [Nogueira-Walborn-P\'{a}dua-Monken 04]}:
W. A. T. Nogueira, S. P. Walborn, S. P\'{a}dua, \& C. H. Monken,
``Generation of a two-photon singlet beam'',
{\em Phys. Rev. Lett.} {\bf 92}, 4, 043602 (2004).

\item {\bf [Nogues-Rauschenbeutel-Osnaghi-(+3) 99]}:
G. Nogues, A. Rauschenbeutel, S. Osnaghi,
M. Brune, J.-M. Raimond, \& S. Haroche,
``Seeing a single photon without destroying it'',
{\em Nature} {\bf 400}, 6741, 239-242 (1999).
See {\bf [Grangier 99]}, {\bf [Physics World 99]}.

\item {\bf [Nogues-Rauschenbeutel-Osnaghi-(+6) 00]}:
G. Nogues, A. Rauschenbeutel, S. Osnaghi, P. Bertet, M. Brune,
J.-M. Raimond, S. Haroche, L. G. Lutterbach, \& L. Davidovich,
``Measurement of a negative value for the Wigner
function of radiation'',
{\em Phys. Rev. A} {\bf 62}, 5, 054101 (2000).

\item {\bf [Noh-Hong 98]}:
T. G. Noh, \& C. K. Hong,
``Interaction-free measurement based on nonclassical fourth-order interference'',
{\em Quantum Semiclass. Opt.} {\bf 10}, 4, 637-641 (1998).

\item {\bf [Norris 00]}:
C. Norris,
{\em Quantum theory and the flight from realism},
Routledge, London, 2000.

\item {\bf [Novais-Castro Neto 04]}:
E. Novais, \& A. H. Castro Neto,
``Nuclear spin qubits in a pseudospin quantum chain'',
{\em Phys. Rev. A} {\bf 69}, 6, 062312 (2004).

\item {\bf [Nowakowski 99]}:
M. Nowakowski,
``Bell's locality condition confronting perfect
anti-correlations'',
{\em Phys. Lett. A} {\bf 260}, 3-4, 169-174 (1999).

\item {\bf [Nuri 98]}:
V. Z. Nuri,
``Local, deterministic hidden variable theories based on a loophole in
Bell's theorem'',
quant-ph/9808008.

\item {\bf [Nussinov 98]}:
S. Nussinov,
``Realistic experiments for measuring the wave function of a single
particle'',
{\em Found. Phys.} {\bf 28}, 6, 865-880 (1998).


\newpage

\subsection{}


\item {\bf [Obada-Hessian 04]}:
A.-S. F. Obada, \& H. A. Hessian,
``Entanglement generation and entropy growth due to intrinsic decoherence in the Jaynes–Cummings model'',
{\em J. Optical Soci. Am. B} {\bf 21}, 1535-1542 (2004).

\item {\bf [de Obaldia-Shimony-Wittel 88]}:
E. de Obaldia, A. Shimony, \& F. Wittel,
``Amplification of Belinfante's argument for the nonexistence
of dispersion-free states'',
{\em Found. Phys.} {\bf 18}, 10, 1013-1021 (1988).

\item {\bf [Obenland-Despain 98]}:
K. M. Obenland, \& A. M. Despain,
``A parallel quantum computer simulator'',
quant-ph/9804039.

\item {\bf [Oberbeck-Curson-Simmons-(+4) 02]}:
L. Oberbeck, N. J. Curson, M. Y. Simmons,
R. Brenner, A. R. Hamilton, S. R. Schofield, \& R. G. Clark,
``Encapsulation of phosphorus dopants in silicon for the fabrication of a
quantum computer'',
{\em Appl. Phys. Lett.} {\bf 81}, ?, 3197-? (2002).

\item {\bf [Oblak-Mikkelsen-Tittel-(+5) 03]}:
D. Oblak, J. K. Mikkelsen, W. Tittel, A. K. Vershovski,
J. L. Sorensen, P. G. Petrov, C. L. Garrido Alzar, \& E. S. Polzik,
``Quantum noise limited interferometric measurement of atomic noise:
Towards spin squeezing on the Cs clock transition'',
quant-ph/0312165.

\item {\bf [O'Brien-Pryde-White-(+2) 03]}:
J. L. O'Brien, G. J. Pryde, A. G. White,
T. C. Ralph, \& D. Branning,
``Demonstration of an all-optical quantum controlled-NOT gate'',
{\em Nature} {\bf 426}, 6964, 264-267 (2003);
quant-ph/0403062.

\item {\bf [O'Brien-Pryde-Gilchrist-(+4) 04]}:
J. L. O'Brien, G. J. Pryde, A. Gilchrist,
D. F. V. James, N. K. Langford, T. C. Ralph, \& A. G. White,
``Quantum process tomography of a controlled-NOT gate'',
quant-ph/0402166.

\item {\bf [O'Brien-Pryde-White-Ralph 04]}:
J. L. O'Brien, G. J. Pryde, A. G. White, \& T. C. Ralph,
``High-fidelity $Z$-measurement error correction of optical qubits'',
quant-ph/0408064.

\item {\bf [Ochs 72]}:
W. Ochs,
``On Gudder's hidden-variable theorems'',
{\em Nuovo Cimento B} {\bf 10}, 1, 172-184 (1972).

\item {\bf [O'Connell-Zuo 03]}:
R. F. O'Connell, \& J. Zuo,
``Effect of an external field on decoherence'',
{\em Phys. Rev. A} {\bf 67}, 6, 062107 (2003);
quant-ph/0311020.

\item {\bf [O'Connell 03 a]}:
R. F. O'Connell,
``The equation of motion of an electron'',
{\em Phys. Lett. A} {\bf 313}, ?, 491-? (2003);
quant-ph/0311021.

\item {\bf [O'Connell 03 b]}:
R. F. O'Connell,
``Wigner distribution function approach to dissipative problems in quantum
mechanics with emphasis on decoherence and measurement theory'',
Authors: R.F. O'Connell
{\em J. Opt. B: Quantum Semiclass. Opt.} {\bf 5}, ?, S349-? (2003);
quant-ph/0311084.

\item {\bf [O'Connor-Wootters 01]}:
K. M. O'Connor, \& W. K. Wootters,
``Entangled rings'',
{\em Phys. Rev. A} {\bf 63}, 5, 052302 (2001);
quant-ph/0009041.

\item {\bf [Oemrawsingh-Aiello-Eliel-(+2) 04]}:
S. S. R. Oemrawsingh, A. Aiello, E. R. Eliel,
G. Nienhuis, \& J. P. Woerdman,
``How to observe high-dimensional two-photon entanglement with only two detectors'',
{\em Phys. Rev. Lett.} {\bf 92}, 21, 217901 (2004);
quant-ph/0401148.

\item {\bf [Ogata 03]}:
Y. Ogata,
``Decoherence-free algebra'',
{\em Phys. Lett. A} {\bf 314}, 1-2, 19-22 (2003).

\item {\bf [\"{O}gren 83]}:
M. \"{O}gren,
``Some evaluations of Bell's inequality for particles of arbitrary spin'',
{\em Phys. Rev. D} {\bf 27}, 8, 1766-1773 (1983).

\item {\bf [Oh-Ahn-Hwang 00]}:
J. H. Oh, D. Ahn, \& S. W. Hwang,
``Optically driven qubits in artificial molecules'',
{\em Phys. Rev. A} {\bf 62}, 5, 052306 (2000).

\item {\bf [Oh-Lee-Lee 02]}:
S. Oh, S. Lee, \& H.-W. Lee,
``Fidelity of quantum teleportation through noisy channels'',
{\em Phys. Rev. A} {\bf 66}, 2, 022316 (2002).

\item {\bf [Oh-Kim 04]}:
S. Oh, \& J. Kim,
``Entanglement of electron spins of noninteracting electron gases'',
{\em Phys. Rev. A} {\bf 69}, 5, 054305 (2004).

\item {\bf [Ohshima 00]}:
T. Ohshima,
``All-optical electron spin quantum computer with ancilla
bits for operations in each coupled-dot cell'',
{\em Phys. Rev. A} {\bf 62}, 6, 062316 (2000).

\item {\bf [Ohya 98]}:
M. Ohya,
``A mathematical foundation of quantum information and quantum computer
---On quantum mutual entropy and entanglement---'',
quant-ph/9808051.

\item {\bf [Ohya-Volovich 03]}:
M. Ohya, \& I. V. Volovich,
``New quantum algorithm for studying NP-complete problems'',
{\em Rep. Math. Phys.} {\bf 52}, 1, 25-33 (2003);
quant-ph/0406216.

\item {\bf [Oi 03]}:
D. K. L. Oi,
``Interference of quantum channels'',
{\em Phys. Rev. Lett.} {\bf 91}, 6, 067902 (2003).

\item {\bf [Olavo 00]}:
L. S. F. Olavo,
``Foundations of quantum mechanics: Connection with stochastic processes'',
{\em Phys. Rev. A} {\bf 61}, 5, 052109 (2000).
Comment: {\bf [Alonso-Muga-Sala Mayato 01]}.

\item {\bf [Olaya Castro-Johnson-Quiroga 03 a]}:
A. Olaya-Castro, N. F. Johnson, \& L. Quiroga,
``Ultrafast deterministic generation of entanglement in a time-dependent
asymmetric two-qubit-cavity system'',
quant-ph/0311111.

\item {\bf [Olaya Castro-Johnson-Quiroga 03 b]}:
A. Olaya-Castro, N. F. Johnson, \& L. Quiroga,
``Dynamics of quantum correlations and linear entropy in a
multi-qubit-cavity system'',
quant-ph/0311181.

\item {\bf [Olaya Castro-Johnson 04]}:
A. Olaya-Castro, \& N. F. Johnson,
``Quantum information processing in nanostructures'',
in {\em Handbook of theoretical and computational nanotechnology};
quant-ph/0406133.

\item {\bf [Olaya Castro-Johnson-Quiroga 04]}:
A. Olaya-Castro, N. F. Johnson, \& L. Quiroga,
``A robust one-step catalytic machine for high fidelity anti-cloning and
$W$-state generation in a multi-qubit system'',
quant-ph/0409104.

\item {\bf [Olivares-Paris-Bonifacio 03]}:
S. Olivares, M. G. A. Paris, \& R. Bonifacio,
``Teleportation improvement by inconclusive photon subtraction'',
{\em Phys. Rev. A} {\bf 67}, 3, 032314 (2003);
quant-ph/0209140.

\item {\bf [Olivares-Paris 03]}:
S. Olivares, \& M. G. A. Paris,
``Binary optical communication in single-mode and entangled quantum noisy channels'',
quant-ph/0309096.

\item {\bf [Olivares-Paris-Rossi 03]}:
S. Olivares, M. G. A. Paris, \& A. R. Rossi,
``Optimized teleportation in Gaussian noisy channels'',
quant-ph/0309097.

\item {\bf [Olivares-Paris 04]}:
S. Olivares, \& M. G. A. Paris,
``Degaussification of twin-beam and nonlocality in the phase space'',
quant-ph/0406030.

\item {\bf [de Oliveira-Mizrahi-Dodonov 99]}:
M. C. de Oliveira, S. S. Mizrahi, \& V. V. Dodonov,
``Information transfer in the course of a quantum interaction'',
{\em J. Opt. B: Quantum Semiclass. Opt.} {\bf 1}, 5, 610-617 (1999);
quant-ph/9909045.

\item {\bf [de Oliveira-Munro 00]}:
M. C. de Oliveira, \& W. J. Munro,
``Quantum computation with mesoscopic superposition states'',
{\em Phys. Rev. A} {\bf 61}, 4, 042309 (2000);
quant-ph/0001018.

\item {\bf [de Oliveira-Milburn 02]}:
M. C. de Oliveira, \& G. J. Milburn,
``Discrete teleportation protocol of continuum spectra field states'',
{\em Phys. Rev. A} {\bf 65}, 3, 032304 (2002);
quant-ph/0204062.

\item {\bf [de Oliveira-da Silva-Mizrahi 02]}:
M. C. de Oliveira, L. F. da Silva, \& S. S. Mizrahi,
``Controlling quantum entanglement through photocounts'',
{\em Phys. Rev. A} {\bf 65}, 6, 062314 (2002);
quant-ph/0204061.

\item {\bf [de Oliveira 03]}:
M. C. de Oliveira,
``Teleportation of a Bose-Einstein condensate state by controlled elastic
collisions'',
{\em Phys. Rev. A} {\bf 67}, 2, 022307 (2003);
quant-ph/0212118.

\item {\bf [de Oliveira-Munro 03]}:
M. C. de Oliveira, \& W. J. Munro,
``Nonclassicality and information exchange in deterministic entanglement
formation'',
{\em Phys. Lett. A};
quant-ph/0205151.

\item {\bf [de Oliveira 04]}:
M. C. de Oliveira,
``On the P-representable subset of all bipartite Gaussian separable states'',
{\em Phys. Rev. A};
quant-ph/0401056.

\item {\bf [Oliver-Stroud 89]}:
B. J. Oliver, \& C. R. Stroud, Jr.,
``Predictions of violations of Bell's inequality in an
8-port homodyne detector'',
{\em Phys. Lett. A} {\bf 135}, 8-9, 407-410 (1989).

\item {\bf [Oliver-Yamaguchi-Yamamoto 02]}:
W. D. Oliver, F. Yamaguchi, \& Y. Yamamoto,
``Electron entanglement via a quantum dot'',
{\em Phys. Rev. Lett.} {\bf 88}, 3, 037901 (2002);
quant-ph/0107084.

\item {\bf [Olkhovsky-Recami 92]}:
V. S. Olkhovsky, \& E. Recami,
``Recent developments in the time analysis of tunneling
processes'',
{\em Phys. Rep.} {\bf 214}, 6, 339-356 (1992).

\item {\bf [Ollerenshaw-Lidar-Kay 03]}:
J. E. Ollerenshaw, D. A. Lidar, \& L. E. Kay,
``Magnetic resonance realization of decoherence-free quantum computation'',
{\em Phys. Rev. Lett.} {\bf 91}, 21, 217904 (2003);
quant-ph/0302175.

\item {\bf [Ollivier-Zurek 02]}:
H. Ollivier, \& W. H. Zurek,
``Quantum discord: A measure of the quantumness of correlations'',
{\em Phys. Rev. Lett.} {\bf 88}, 1, 017901 (2002);
quant-ph/0105072.

\item {\bf [Ollivier-Poulin-Zurek 03]}:
H. Ollivier, D. Poulin, \& W. H. Zurek,
``Emergence of objective properties from subjective quantum states:
Environment as a witness'',
quant-ph/0307229.
See {\bf [Ollivier-Poulin-Zurek 04]}.

\item {\bf [Ollivier-Tillich 04]}:
H. Ollivier, \& J.-P. Tillich,
``Quantum convolutional codes: Fundamentals'',
quant-ph/0401134.

\item {\bf [Ollivier-Poulin-Zurek 04]}:
H. Ollivier, D. Poulin, \& W. H. Zurek,
``Environment as a witness: Selective proliferation of information and
emergence of objectivity'',
quant-ph/0408125.
See {\bf [Ollivier-Poulin-Zurek 03]}.

\item {\bf [Omar-Paunkovi\'{c}-Bose-Vedral 02]}:
Y. Omar, N. Paunkovi\'{c}, S. Bose, \& V. Vedral,
``Spin-space entanglement transfer and quantum statistics'',
{\em Phys. Rev. A} {\bf 65}, 6, 062305 (2002);
quant-ph/0105120.

\item {\bf [Omar-Paunkovi\'{c}-Sheridan-Bose 04]}:
Y. Omar, N. Paunkovi\'{c}, L. Sheridan, \& S. Bose,
``Quantum walk on a line with two entangled particles'',
quant-ph/0411065.

\item {\bf [Omn\`{e}s 88 a]}:
R. Omn\`{e}s,
``Logical reformulation of quantum mechanics. I. Foundations'',
{\em J. Stat. Phys.} {\bf 53}, 3-4, 893-932 (1988).

\item {\bf [Omn\`{e}s 88 b]}:
R. Omn\`{e}s,
``Logical reformulation of quantum
mechanics. II. Interferences and the Einstein-Podolsky-Rosen experiment'',
{\em J. Stat. Phys.} {\bf 53}, 3-4, 933-955 (1988).

\item {\bf [Omn\`{e}s 88 c]}:
R. Omn\`{e}s,
``Logical reformulation of quantum
mechanics. III. Classical limit and irreversibility'',
{\em J. Stat. Phys.} {\bf 53}, 3-4, 957-975 (1988).

\item {\bf [Omn\`{e}s 89]}:
R. Omn\`{e}s,
``Logical reformulation of quantum
mechanics. IV. Projectors in semiclassical physics'',
{\em J. Stat. Phys.} {\bf 57}, 1-2, 357-382 (1989).

\item {\bf [Omn\`{e}s 90 a]}:
R. Omn\`{e}s,
``A consistent interpretation of quantum mechanics'',
in M. Cini, \& J. M. L\'{e}vy-Leblond (eds.),
{\em Quantum theory without reduction},
Adam Hilger, Bristol, 1990, pp.~27-48.

\item {\bf [Omn\`{e}s 90 b]}:
R. Omn\`{e}s,
``From Hilbert space to common sense:
A synthesis of recent progress in the interpretation of quantum mechanics'',
{\em Ann. Phys.} {\bf 201}, 2, 354-447 (1990).

\item {\bf [Omn\`{e}s 90 c]}:
R. Omn\`{e}s,
``Some progress in measurement theory:
The logical interpretation of quantum mechanics'',
in {\bf [Zurek 90]}, pp.~495-512.

\item {\bf [Omn\`{e}s 92]}:
R. Omn\`{e}s,
``Consistent interpretations of quantum mechanics'',
{\em Rev. Mod. Phys.} {\bf 64}, 2, 339-382 (1992).

\item {\bf [Omn\`{e}s 94 a]}:
R. Omn\`{e}s,
{\em The interpretation of quantum mechanics},
Princeton University Press, Princeton, New Jersey, 1994.
Review: {\bf [Griffiths 95]}.

\item {\bf [Omn\`{e}s 94 b]}:
R. Omn\`{e}s,
{\em Philosophie de la science contemporaine},
Editions Gallimard, Paris, 1994.
English version: {\em Quantum philosophy.
Understanding and interpreting contemporary science},
Princeton University Press, Princeton, New Jersey, 1994.
Review: {\bf [Scalettar 99]}, {\bf [Greenberger 01]}.

\item {\bf [Omn\`{e}s 95]}:
R. Omn\`{e}s,
``Une nouvelle interpr\'{e}tation de la m\'{e}canique quantique'',
{\em La Recherche} {\bf 26}, 280, 50-56 (1995).
Spanish version: ``Una nueva interpretaci\'{o}n de la
mec\'{a}nica cu\'{a}ntica'',
{\em Mundo Cient\'{\i}fico} {\bf 15}, 163, 1034-1040 (1995).

\item {\bf [Omn\`{e}s 97]}:
R. Omn\`{e}s,
``General theory of the decoherence effect in quantum mechanics'',
{\em Phys. Rev. A} {\bf 56}, 5, 3383-3394 (1997).

\item {\bf [Omn\`{e}s 98 a]}:
R. Omn\`{e}s,
``Comment on `Is the EPR paradox really a paradox?'\,'',
{\em Eur. J. Phys.} {\bf 20}, 1, L17 (1999).
Comment on {\bf [Tartaglia 98]}.

\item {\bf [Omn\`{e}s 98 b]}:
R. Omn\`{e}s,
``Theory of the decoherence effect'',
{\em Fortschr. Phys.} {\bf 46}, 6-8, 771-777 (1998).

\item {\bf [Omn\`{e}s 99 a]}:
R. Omn\`{e}s,
{\em Understanding quantum mechanics},
Princeton University Press, Princeton, New Jersey, 1999.
Review: {\bf [Scalettar 99]}.

\item {\bf [Omn\`{e}s 99 b]}:
R. Omn\`{e}s,
``Recent advances in the consistency of interpretation'',
in {\bf [Greenberger-Reiter-Zeilinger 99]}, pp.~103-111.

\item {\bf [Omn\`{e}s 00]}:
R. Omn\`{e}s,
``Quantum dialogue: The making of a revolution'',
{\em Eur. J. Phys.} {\bf 21}, 2, 200-201 (2000).
Review of {\bf [Beller 99]}.

\item {\bf [Omn\`{e}s 01]}:
R. Omn\`{e}s,
``Decoherence: An irreversible process'',
quant-ph/0106006.

\item {\bf [Omn\`{e}s 02]}:
R. Omn\`{e}s,
``Decoherence, irreversibility, and selection by decoherence of exclusive quantum states
with definite probabilities'',
{\em Phys. Rev. A} {\bf 65}, 5, 052119 (2002);
quant-ph/0304100.

\item {\bf [Onofrio-Viola 98]}:
R. Onofrio, \& L. Viola,
``Dynamics of decoherence in continuous atom-optical quantum
nondemolition measurements'',
{\em Phys. Rev. A} {\bf 58}, 1, 69-76 (1998);
quant-ph/9805060.

\item {\bf [Onofrio-Presilla 00]}:
R. Onofrio, \& Presilla,
`Comment on the paper ``Decoherence and the theory
of continuous quantum measurements'', by M. B. Mensky',
quant-ph/0009013.
Comment on {\bf [Mensky 98 a]}.

\item {\bf [Opatrn\'{y}-Kurizki 98]}:
T. Opatrn\'{y}, \& G. Kurizki,
``Optimization approach to entanglement distillation'',
{\em Phys. Rev. A} {\bf 60}, 1, 167-172 (1999);
quant-ph/9811090.

\item {\bf [Opatrn\'{y}-Kurizki-Welsch 00]}:
T. Opatrn\'{y}, G. Kurizki, \& D.-G. Welsch,
``Improvement on teleportation of continuous variables
by photon subtraction via conditional measurement'',
{\em Phys. Rev. A} {\bf 61}, 3, 032302 (2000);
quant-ph/9907048.

\item {\bf [Opatrn\'{y}-Kurizki 00]}:
T. Opatrn\'{y}, \& G. Kurizki,
``On the possibility of quantum computation based on photon exchange interactions'',
{\em Fortschr. Phys.} {\bf 48}, 9-11 (Special issue:
Experimental proposals for quantum computation), 1125-1131 (2000);
quant-ph/0003010.
See {\bf [Franson 00 b]}.

\item {\bf [Opatrn\'{y}-Kurizki 01]}:
T. Opatrn\'{y}, \& G. Kurizki,
``Matter-wave entanglement and teleportation by molecular dissociation and
collisions'',
{\em Phys. Rev. Lett.} {\bf 86}, 14, 3180-3183 (2001);
quant-ph/0009121.

\item {\bf [Opatrn\'{y}-Korolkova-Leuchs 02]}:
T. Opatrn\'{y}, N. Korolkova, \& G. Leuchs,
``Mode structure and photon number correlations in squeezed quantum pulses'',
{\em Phys. Rev. A} {\bf 66}, 5, 053813 (2002).

\item {\bf [Opatrn\'{y}-Deb-Kurizki 03]}:
T. Opatrn\'{y}, B. Deb, \& G. Kurizki,
``Proposal for translational entanglement of dipole-dipole interacting atoms
in optical lattices'',
{\em Phys. Rev. Lett.} {\bf 90}, 25, 250404 (2003);
quant-ph/0305107.

\item {\bf [Oppenheim-Reznik-Unruh 98]}:
J. Oppenheim, B. Reznik, \& W. G. Unruh,
``Time as an observable'',
in {\em 10th Max Born Symp.\ ``Quantum Future'' (Wroclaw, Poland, 1997)},
Springer-Verlag, Berlin, 1998;
quant-ph/9807058.

\item {\bf [Oppenheim-Reznik-Unruh 00]}:
J. Oppenheim, B. Reznik, \& W. G. Unruh,
``When does a measurement or event occur?'',
{\em Found. Phys. Lett.} {\bf 13}, 2, 107-118 (2000);
quant-ph/9805064.

\item {\bf [Oppenheim-Horodecki-Horodecki-Horodecki 01]}:
J. Oppenheim, M. Horodecki, P. Horodecki, \& R. Horodecki,
``Quantifying non-locality through heat engines and distributed negentropy'',
quant-ph/0112074.

\item {\bf [Oppenheim-Reznik-Unruh 02]}:
J. Oppenheim, B. Reznik, \& W. G. Unruh,
``Temporal ordering in quantum mechanics'',
{\em J. Phys. A} {\bf 35}, ?, 7641-7652 (2002);
quant-ph/0003130.

\item {\bf [Oppenheim-Horodecki-Horodecki-Horodecki 02]}:
J. Oppenheim, M. Horodecki, P. Horodecki, \& R. Horodecki,
``Thermodynamical approach to quantifying quantum correlations'',
{\em Phys. Rev. Lett.} {\bf 89}, 18, 180402 (2002);
quant-ph/0112074.

\item {\bf [Oppenheim-Horodecki-Horodecki 02]}:
J. Oppenheim, M. Horodecki, \& R. Horodecki,
``Are there phase transitions in information space?'',
{\em Phys. Rev. Lett.} {\bf 90}, 1, 010404 (2003);
quant-ph/0207169

\item {\bf [Oppenheim-Horodecki-Horodecki-(+2) 03]}:
J. Oppenheim, K. Horodecki, M. Horodecki,
P. Horodecki, \& R. Horodecki,
``Mutually exclusive aspects of information carried by physical systems:
Complementarity between local and nonlocal information'',
{\em Phys. Rev. A} {\bf 68}, 2, 022307 (2003).

\item {\bf [Oppenheim-Horodecki 03]}:
J. Oppenheim, \& M. Horodecki,
``How to reuse a one-time pad and other notes on authentication,
encryption and protection of quantum information'',
quant-ph/0306161.

\item {\bf [Oriols-Mart\'{\i}n-Su\~{n}e 96]}:
X. Oriols, F. Mart\'{\i}n, \& J. Su\~{n}e,
``Implications of the noncrossing property of Bohm
trajectories in one-dimensional tunneling configurations'',
{\em Phys. Rev. A} {\bf 54}, 4, 2594-2604 (1996).

\item {\bf [Orlov 99]}:
Y. F. Orlov,
``Origin of quantum indeterminism and irreversibility of measurements'',
{\em Phys. Rev. Lett.} {\bf 82}, 2, 243-246 (1999).

\item {\bf [Orlov 02 a]}:
Y. F. Orlov,
``Classical counterexamples to Bell's inequalities'',
{\em Phys. Rev. A} {\bf 65}, 4, 042106 (2002).

\item {\bf [Orlov 02 b]}:
Y. F. Orlov,
``Quantumlike bits and logic gates based on classical oscillators'',
{\em Phys. Rev. A} {\bf 66}, 5, 052324 (2002).

\item {\bf [Orrit 02]}:
M. Orrit,
``Molecular entanglements'',
{\em Science} {\bf 298}, ?, 369-370 (2002).
See {\bf [Hettich et al. 02]}.

\item {\bf [Ortiz-Gubernatis-Knill-Laflamme 01]}:
G. Ortiz, J. E. Gubernatis, E. Knill, \& R. Laflamme,
``Quantum algorithms for fermionic simulations'',
{\em Phys. Rev. A} {\bf 64}, 2, 022319 (2001).
Erratum: {\em Phys. Rev. A} {\bf 65}, 2, 029902 (2002);
cond-mat/0012334.

\item {\bf [Ortiz-Gubernatis-Knill-Laflamme 02]}:
G. Ortiz, J. E. Gubernatis, E. Knill, \& R. Laflamme,
```Simulating fermions on a quantum computer'',
{\em Comput. Phys. Comm.} {\bf 146}, 3, 302-316 (2002).

\item {\bf [Ortiz-Somma-Barnum-(+2) 04]}:
G. Ortiz, R. Somma, H. Barnum, E. Knill, \& L. Viola,
``Entanglement as an observer-dependent concept: An application to
quantum phase transitions'',
{\em Condensed Matter Theor.} {\bf 19}, ?, ?-? (2004);
quant-ph/0403043.

\item {\bf [Ortoli-Pharabod 84]}:
S. Ortoli, \& J.-P. Pharabod,
{\em Le cantique des quantiques},
La D\'{e}couverte, Paris, 1984.
Spanish version:
{\em El c\'{a}ntico de la cu\'{a}ntica}
Gedisa, Barcelona, 2001.

\item {\bf [Orus-Latorre-Mart\'{\i}n Delgado 02]}:
R. Orus, J. I. Latorre, \& M. A. Mart\'{\i}n-Delgado,
``Systematic analysis of majorization in quantum algorithms'',
quant-ph/0212094.

\item {\bf [Orus-Latorre-Mart\'{\i}n Delgado 03]}:
R. Orus, J. I. Latorre, \& M. A. Mart\'{\i}n-Delgado,
`` Natural majorization of the Quantum Fourier Transformation in
phase-estimation algorithms'',
{\em Quant. Inf. Proc.} {\bf 4}, ?, 283-302 (2003);
quant-ph/0206134.

\item {\bf [Orus-Latorre 03]}:
R. Orus, \& J. I. Latorre,
``Universality of entanglement and quantum computation complexity'',
{\em Phys. Rev. A};
quant-ph/0311017.

\item {\bf [Orus-Tarrach 04]}:
R. Orus, \& R. Tarrach,
``Weakly-entangled states are dense and robust'',
{\em Phys. Rev. A};
quant-ph/0404100.

\item {\bf [Osaki-Ban 03]}:
M. Osaki, \& M. Ban,
``Security evaluation of the quantum key distribution system with two-mode
squeezed states'',
{\em Phys. Rev. A} {\bf 68}, 2, 022325 (2003).

\item {\bf [Osawa-Nagaoka 00]}:
S. Osawa, \& H. Nagaoka,
``Numerical experiments on the capacity of quantum
channel with entangled input states'',
quant-ph/0007115.

\item {\bf [Osborne 00 a]}:
I. S. Osborne,
``Physics: Engineered quantum entanglement'',
{\em Science} {\bf 287}, 5462, 2377 (2000).
See {\bf [Sackett-Kielpinski-King-(+8) 00]}.

\item {\bf [Osborne 00 b]}:
I. S. Osborne,
``Applied physics: To catch one photon'',
{\em Science} {\bf 289}, 5476, 13 (2000).
See {\bf [Shields {\em et al.} 00]}.

\item {\bf [Osborne 00 c]}:
I. S. Osborne,
``Applied physics: Single photons from diamonds'',
{\em Science} {\bf 289}, 5478, 363 (2000).
See {\bf [Kurtsiefer-Mayer-Zarda-Weinfurter 00]}.

\item {\bf [Osborne 00 d]}:
I. S. Osborne,
``Physics: Heaven for quantum computation'',
{\em Science} {\bf 289}, 5484, 1433 (2000).
See {\bf [Bacon-Kempe-Lidar-Whaley 00]},
{\bf [Beige-Braun-Tregenna-Knight 00]}.

\item {\bf [Osborne 00 e]}:
I. S. Osborne,
``Physics: Wave-particle detection'',
{\em Science} {\bf 290}, 5492, 673 (2000).
See {\bf [Foster-Orozco-Castro Beltran-Carmichael 00]}.

\item {\bf [Osborne 02]}:
I. S. Osborne,
``Entanglement for rank-2 mixed states'',
quant-ph/0203087.

\item {\bf [Osborne-Nielsen 01]}:
T. J. Osborne, \& M. A. Nielsen,
``Entanglement, quantum phase transitions, and density matrix renormalization'',
quant-ph/0109024.

\item {\bf [Osborne-Nielsen 02]}:
T. J. Osborne, \& M. A. Nielsen,
``Entanglement in a simple quantum phase transition'',
{\em Phys. Rev. A} {\bf 66}, 3, 032110 (2002);
quant-ph/0202162.

\item {\bf [Osborne-Linden 03]}:
T. J. Osborne, \& N. Linden,
``The propagation of quantum information through a spin system'',
quant-ph/0312141.

\item {\bf [Osborne-Severini 04]}:
T. J. Osborne, \& S. Severini,
``Quantum algorithms and covering spaces'',
quant-ph/0403127.

\item {\bf [Osenda-Huang-Kais 03]}:
O. Osenda, Z. Huang, \& S. Kais,
``Tuning the entanglement for a one-dimensional magnetic system with
anisotropic coupling and impurities'',
{\em Phys. Rev. A} {\bf 67}, 6, 062321 (2003).

\item {\bf [Oshima 01]}:
K. Oshima,
``Driving Hamiltonian in a quantum search problem'',
quant-ph/0111021.

\item {\bf [Oshima-Azuma 03]}:
K. Oshima, \& K. Azuma,
``Proper magnetic fields for nonadiabatic geometric quantum gates in NMR'',
{\em Phys. Rev. A} {\bf 68}, 3, 034304 (2003).

\item {\bf [Osnaghi-Bertet-Auffeves-(+4) 01]}:
S. Osnaghi, P. Bertet, A. Auffeves,
P. Maioli, M. Brune, J. M. Raimond, \& S. Haroche,
``Coherent control of an atomic collision in a cavity'',
{\em Phys. Rev. Lett.} {\bf 87}, 3, 037902 (2001);
quant-ph/0105063.

\item {\bf [Osterloh-Amico-Falci-Fazio 02]}:
A. Osterloh, L. Amico, G. Falci, \& R. Fazio,
``Scaling of entanglement close to a quantum phase transition'',
{\em Nature} {\bf 416}, 6881, 608-610 (2002).

\item {\bf [Ottaviani-Vitali-Artoni-(+2) 03]}:
C. Ottaviani, D. Vitali, M. Artoni,
F. Cataliotti, \& P. Tombesi,
``Polarization qubit phase gate in driven atomic media'',
{\em Phys. Rev. Lett.} {\bf 90}, 19, 197902 (2003).

\item {\bf [Otte-Mahler 00]}:
A. Otte, \& G. Mahler,
``Adapted-operator representations: Selective versus collective
properties of quantum networks'',
{\em Phys. Rev. A} {\bf 62}, 1, 012303 (2000),
quant-ph/0101138.

\item {\bf [Ou-Hong-Mandel 87]}:
Z. Y. Ou, C. K. Hong, \& L. Mandel,
``Relation between input and output states for a beam splitter'',
{\em Opt. Comm.} {\bf 63}, 2, 118-122 (1987).

\item {\bf [Ou-Mandel 88 a]}:
Z. Y. Ou, \& L. Mandel,
``Violation of Bell's inequality
and classical probability in a two-photon correlation experiment'',
{\em Phys. Rev. Lett.} {\bf 61}, 1, 50-53 (1988).

\item {\bf [Ou-Mandel 88 b]}:
Z. Y. Ou, \& L. Mandel,
``Obervation of spatial quantum beating with separed photodetectors'',
{\em Phys. Rev. Lett.} {\bf 61}, 1, 54-57 (1988).

\item {\bf [Ou-Zou-Wang-Mandel 90]}:
Z. Y. Ou, X. Y. Zou, L. J. Wang, \& L. Mandel,
``Observation of nonlocal interference in separated photon channels'',
{\em Phys. Rev. Lett.} {\bf 65}, 3, 321-324 (1990).

\item {\bf [Ou-Wang-Zou-Mandel 90]}:
Z. Y. Ou, L. J. Wang, X. Y. Zou, \& L. Mandel,
``Evidence of phase memory in two-photon down conversion
through entanglement with the vacuum'',
{\em Phys. Rev. A} {\bf 41}, 1, 566-568 (1990).

\item {\bf [Ou-Pereira-Kimble-Peng 92]}:
Z. Y. Ou, S. F. Pereira, H. J. Kimble, \& K. C. Peng,
``Realization of the Einstein-Podolsky-Rosen paradox for continuous
variables'',
{\em Phys. Rev. Lett.} {\bf 68}, 25, 3663-3666 (1992).

\item {\bf [Ou 96]}:
Z. Y. Ou,
``Quantum multi-particle interference due to a single-photon
state'',
{\em Quantum Semiclass. Opt.} {\bf 8}, 2, 315-322 (1996).

\item {\bf [Ou 97]}:
Z. Y. Ou,
``Nonlocal correlation in the realization of a quantum eraser'',
{\em Phys. Lett. A} {\bf 226}, 6, 323-326 (1997).

\item {\bf [Ozawa 98 a]}:
M. Ozawa,
``Quantum nondemolition monitoring of universal quantum computers'',
{\em Phys. Rev. Lett.} {\bf 80}, 3, 631-634 (1998);
quant-ph/9704028.

\item {\bf [Ozawa 98 b]}:
M. Ozawa,
``Quantum state reduction: An operational approach'',
{\em Fortschr. Phys.} {\bf 46}, 6-8, 615-625 (1998);
quant-ph/9706027.

\item {\bf [Ozawa 98 c]}:
M. Ozawa,
``Controlling quantum state reduction'',
quant-ph/9805033.

\item {\bf [Ozawa 98 d]}:
M. Ozawa,
``Quantum Turing machines: Local transition, preparation, measurement, and
halting'',
{\em 4th Int.\ Conf.\ on Quantum
Communication, Measurement and Computing (Evanston, Illinois, 1998)};
quant-ph/9809038.

\item {\bf [Ozawa 98 e]}:
M. Ozawa,
``Measurability and computability'',
quant-ph/9809048.

\item {\bf [Ozawa-Nishimura 98]}:
M. Ozawa, \& H. Nishimura,
``Local transition functions of quantum Turing machines'',
quant-ph/9811069.

\item {\bf [Ozawa 00 a]}:
M. Ozawa,
``Entanglement measures and the Hilbert-Schmidt distance'',
{\em Phys. Lett. A} {\bf 268}, 3, 158-160 (2000);
quant-ph/0002036.

\item {\bf [Ozawa 00 b]}:
M. Ozawa,
``Measurements of nondegenerate discrete observables'',
{\em Phys. Rev. A} {\bf 62}, 6, 062101 (2000);
quant-ph/0003033.

\item {\bf [Ozawa 00 c]}:
M. Ozawa,
``Operational characterization of simultaneous measurements in quantum
mechanics'',
{\em Phys. Lett. A} {\bf 275}, 1-2, 5-11 (2000).

\item {\bf [Ozawa 01 a]}:
M. Ozawa,
``Operations, disturbance, and simultaneous measurability'',
{\em Phys. Rev. A} {\bf 63}, 3, 032109 (2001);
quant-ph/0005054.

\item {\bf [Ozawa 01 b]}:
M. Ozawa,
``Controlling quantum state reductions'',
{\em Phys. Lett. A} {\bf 282}, 6, 336-342 (2001).

\item {\bf [Ozawa 01 c]}:
M. Ozawa,
``Phase-creation algorithm to solve an extended Deutsch problem
by a quantum computer'',
{\em Phys. Rev. A} {\bf 63}, 5, 052312 (2001).

\item {\bf [Ozawa 01 d]}:
M. Ozawa,
``Position measuring interactions and the
Heisenberg uncertainty principle'',
quant-ph/0107001.

\item {\bf [Ozawa 01 e]}:
M. Ozawa,
``Quantum measurement, information, and completely positive maps'',
{\em 5th Int.\ Conf.\ on Quantum Communication, Measurement and
Computing (Capri, Italy, 2000)};
quant-ph/0107090.

\item {\bf [Ozawa 02 a]}:
M. Ozawa,
``Conservation laws, uncertainty relations,
and quantum limits of measurements'',
{\em Phys. Rev. Lett.} {\bf 88}, 5, 050402 (2002);
quant-ph/0112154.

\item {\bf [Ozawa 02 b]}:
M. Ozawa,
``Position measuring interactions and the Heisenberg
uncertainty principle'',
{\em Phys. Lett. A} {\bf 299}, 1, 1-7 (2002).

\item {\bf [Ozawa 02 c]}:
M. Ozawa,
``Conservative quantum computing'',
{\em Phys. Rev. Lett.} {\bf 89}, 5, 057902 (2002);
quant-ph/0112179.
Comment: {\bf [Lidar 03]}.
Reply: {\bf [Ozawa 03 b]}.

\item {\bf [Ozawa 03 a]}:
M. Ozawa,
``Universally valid reformulation of the Heisenberg uncertainty principle on
noise and disturbance in measurement'',
{\em Phys. Rev. A} {\bf 67}, 4, 042105 (2003).

\item {\bf [Ozawa 03 b]}:
M. Ozawa,
``Ozawa replies'',
{\em Phys. Rev. Lett.} {\bf 91}, 8, 089802 (2003).
Reply to {\bf [Lidar 03]}.
See {\bf [Ozawa 02 c]}.

\item {\bf [Ozawa 03 c]}:
M. Ozawa,
``Uncertainty relations for joint measurements of noncommuting observables'',
{\em A Satellite Workshop to EQIS'03:
Non-locality of Quantum Mechanics and Statistical Inference (Kyoto, 2003)},
quant-ph/0310070.

\item {\bf [Ozawa 03 d]}:
M. Ozawa,
``Uncertainty principle for quantum instruments and computing'',
{\em Int. J. Quantum Inf.} {\bf 1}, ?, 569-588 (2003);
quant-ph/0310071.

\item {\bf [Ozawa 03 e]}:
M. Ozawa,
``Perfect correlations between noncommuting observables'',
quant-ph/0310072.

\item {\bf [Ozawa 04]}:
M. Ozawa,
``Universal uncertainty principle and quantum state control under
conservation laws'',
quant-ph/0411074.

\item {\bf [\"{O}zdemir-Miranowicz-Koashi-Imoto 01]}:
S. K. \"{O}zdemir, A. Miranowicz, M. Koashi, \& N. Imoto,
``Quantum-scissors device for optical state truncation: A proposal for
practical realization'',
{\em Phys. Rev. A} {\bf 64}, 6, 063818 (2001).

\item {\bf [\"{O}zdemir-Miranowicz-Koashi-Imoto 02]}:
S. K. \"{O}zdemir, A. Miranowicz, M. Koashi, \& N. Imoto,
``Pulse-mode quantum projection synthesis: Effects of mode mismatch on optical
state truncation and preparation'',
{\em Phys. Rev. A} {\bf 66}, 5, 053809 (2002).

\item {\bf [\"{O}zdemir-Shimamura-Morikoshi-Imoto 02]}:
S. K. \"{O}zdemir, J. Shimamura, F. Morikoshi, \& N. Imoto,
``Samaritan's dilemma: Classical and quantum strategies in welfare game'',
quant-ph/0311074.

\item {\bf [\"{O}zdemir-Shimamura-Imoto 04]}:
S. K. \"{O}zdemir, J. Shimamura, \& N. Imoto,
``Quantum advantage does not survive in the presence of a corrupt source:
Optimal strategies in simultaneous move games'',
{\em Phys. Lett. A} {\bf 325}, 2, 104-111 (2004);
quant-ph/0402038.

\item {\bf [Ozhigov 97]}:
Y. Ozhigov,
``Quantum computers cannot speed up iterated applications of a
black box'',
quant-ph/9712051.

\item {\bf [Ozhigov 99]}:
Y. Ozhigov,
``Lower bounds of a quantum search for an extreme point'',
{\em Proc. R. Soc. Lond. A} {\bf 455}, 1986, 2165-2172 (1999).

\item {\bf [Ozhigov-Fedichkin 02]}:
Y. Ozhigov, \& L. Fedichkin,
``Quantum computer with fixed interaction is universal'',
quant-ph/0202030.

\item {\bf [Ozhigov 02 a]}:
Y. Ozhigov,
``Realization of the quantum Fourier transform and simulation of a wave
function on a quantum computer with fixed continuous interaction'',
{\em JETP Lett.} {\bf 76}, 641-? (2002).

\item {\bf [Ozhigov 02 b]}:
Y. Ozhigov,
``How behavior of systems with sparse spectrum can be predicted on a quantum
computer'',
{\em JETP Lett.} {\bf 76}, 675-? (2002).

\item {\bf [Ozhigov-Fedichkin 03]}:
Y. Ozhigov, \& L. Fedichkin,
``A quantum computer with fixed interaction is universal'',
{\em JETP Lett.} {\bf 77}, 328-? (2003).

\item {\bf [Ozorio de Almeida-Vallejos-Saraceno 04]}:
A. M. Ozorio de Almeida, R. O. Vallejos, \& M. Saraceno,
``Pure state correlations: Chords in phase space'',
quant-ph/0410129.


\newpage

\subsection{}


\item {\bf [Pablo Norman-Ruiz Altaba 00]}:
B. Pablo-Norman, \& M. Ruiz-Altaba,
``Noise in Grover's quantum search algorithm'',
{\em Phys. Rev. A} {\bf 61}, 1, 012301 (2000).

\item {\bf [Pachos-Zanardi-Rasetti 00]}:
J. Pachos, P. Zanardi, \& M. Rasetti,
``Non-Abelian Berry connections for quantum computation'',
{\em Phys. Rev. A} {\bf 61}, 1, 010305(R) (2000).

\item {\bf [Pachos-Chountasis 00]}:
J. Pachos, \& S. Chountasis,
``Optical holonomic quantum computer'',
{\em Phys. Rev. A} {\bf 62}, 5, 052318 (2000);
quant-ph/9912093.

\item {\bf [Pachos-Zinardi 00]}:
J. Pachos, \& P. Zinardi,
``Quantum holonomies for quantum computing'',
quant-ph/0007110.

\item {\bf [Pachos-Walther 02]}:
J. Pachos, \& H. Walther,
``Quantum computation with trapped ions in an optical cavity'',
{\em Phys. Rev. Lett.} {\bf 89}, 18, 187903 (2002);
quant-ph/0111088.

\item {\bf [Pachos 02 a]}:
J. Pachos,
``Quantum computation by geometrical means'',
in {\bf [Lomonaco-Brandt 02]} 245-250;
quant-ph/0003150.

\item {\bf [Pachos 02 b]}:
J. Pachos,
``Topological features in ion-trap holonomic computation'',
{\em Phys. Rev. A} {\bf 66}, 4, 042318 (2002).

\item {\bf [Pachos 02 c]}:
J. Pachos,
``Quantum computation by geometrical means'',
{\em Quantum computation and information (Washington, 2000)},
American Mathematical Society, Providence, Rhode Island, 2002, pp.~245-250.

\item {\bf [Pachos-Solano 02]}:
J. Pachos, \& E. Solano,
``Generation and degree of entanglement in a relativistic formulation'',
quant-ph/0203065.

\item {\bf [Pachos-Vedral 03]}:
J. K. Pachos, \& V. Vedral,
``Topological quantum gates with quantum dots'',
quant-ph/0302077.

\item {\bf [Pachos-Knight 03]}:
J. K. Pachos, \& P. L. Knight,
``Quantum computation with a one-dimensional optical lattice'',
{\em Phys. Rev. Lett.} {\bf 91}, 10, 107902 (2003);
quant-ph/0301084.

\item {\bf [Pachos-Beige 04]}:
J. K. Pachos, \& A. Beige,
``Decoherence-free dynamical and geometrical entangling phase gates'',
{\em Phys. Rev. A} {\bf 69}, 3, 033817 (2004);
quant-ph/0309180.

\item {\bf [Pachos-Plenio 04]}:
J. K. Pachos, \& M. B. Plenio,
``Three-spin interactions in optical lattices and criticality in cluster Hamiltonians'',
{\em Phys. Rev. Lett.} {\bf 93}, 5, 056402 (2004);
quant-ph/0401106.

\item {\bf [Pachos-Rico 04]}:
J. K. Pachos, \& E. Rico,
``Effective three-body interactions in triangular optical lattices'',
quant-ph/0404048.

\item {\bf [Paganelli-de Pasquale-Giampaolo 02]}:
S. Paganelli, F. de Pasquale, \& S. M. Giampaolo,
``Decoherence slowing down in a symmetry-broken environment'',
{\em Phys. Rev. A} {\bf 66}, 5, 052317 (2002).

\item {\bf [Page 82]}:
D. N. Page,
``The Einstein-Podolsky-Rosen physical reality is
completely described by quantum mechanics'',
{\em Phys. Lett. A} {\bf 91}, 2, 57-60 (1982).
Comments: {\bf [Todorov 83]}, {\bf [Kunstatter-Trainor 84 a]}.
See {\bf [Bitbol 83]}.

\item {\bf [Page 02]}:
D. N. Page,
``Quantum mechanics as a simple generalization of classical mechanics'',
quant-ph/0204015.

\item {\bf [Pagonis-Redhead-Clifton 91]}:
C. Pagonis, M. L. G. Redhead, \& R. K. Clifton,
``The breakdown of quantum non-locality in the classical limit'',
{\em Phys. Lett. A} {\bf 155}, 8-9, 441-444 (1991).

\item {\bf [Pagonis-Clifton 92]}:
C. Pagonis, \& R. K. Clifton,
``Hardy's nonlocality theorem for $n$ spin-$\frac{1}{2}$ particles'',
{\em Phys. Lett. A} {\bf 168}, 2, 100-102 (1992).
See {\bf [Wu-Xie 96]}.

\item {\bf [Pagonis 92]}:
C. Pagonis,
``Empty waves: No necessarily effective'',
{\em Phys. Lett. A} {\bf 169}, 3, 219-221 (1992).
Comment on {\bf [Hardy 92 c]}.
Reply: {\bf [Hardy 92 e]}.

\item {\bf [Pagonis 93]}:
C. Pagonis,
``The quantum theory of motion: An account of
the de Broglie-Bohm causal interpretation of quantum mechanics'',
{\em Nature} {\bf 364}, 6436, 398 (1993).
Review of {\bf [Holland 93]}.

\item {\bf [Pagonis-Clifton 95]}:
C. Pagonis, \& R. K. Clifton,
``Unremarkable contextualism: Dispositions in the Bohm theory'',
{\em Found. Phys.} {\bf 25}, 2, 281-296 (1995).

\item {\bf [Pagonis-Redhead-La Rivi\`{e}re 96]}:
C. Pagonis, M. L. G. Redhead, \& P. La Rivi\`{e}re,
``EPR, relativity, and the GHZ experiment'',
in R. K. Clifton (ed.),
{\em Perspectives on quantum reality: Non-relativistic, relativistic,
and field-theoretic (London, Western Ontario, Canada, 1994)},
Kluwer Academic, Dordrecht, Holland, 1996, pp.~?-?.
See {\bf [Clifton-Pagonis-Pitowsky 92]}.

\item {\bf [Pahlke-Mathis 03]}:
K. Pahlke, \& W. Mathis,
``Tolerant-gate implementation for the trapped-ion quantum-information
processor'',
{\em Phys. Rev. A} {\bf 67}, 6, 062305 (2003).

\item {\bf [Pais 79]}:
A. Pais,
``Einstein and the quantum theory'',
{\em Rev. Mod. Phys.} {\bf 51}, 4, 863-910 (1979).

\item {\bf [Pais 82]}:
A. Pais,
{\em ``Subtle is the Lord\ldots''. The science and the life
of Albert Einstein},
Oxford University Press, New York, 1982.
Spanish version: {\em ``El Se\~{n}or es sutil\ldots''.
La ciencia y la vida de Albert Einstein}, Ariel, Barcelona, 1984.

\item {\bf [Pais 86]}:
A. Pais,
{\em Inward bound. Of matter and forces in the physical world},
Oxford University Press, New York, 1986.

\item {\bf [Pais 91]}:
A. Pais,
{\em Niel's Bohr times, in physics, philosophy, and polity},
Oxford University Press, Oxford, 1991.

\item {\bf [Pais 00]}:
A. Pais,
{\em The genius of science.
A portrait gallery of twentieth-century physicists},
Oxford University Press, Oxford, 2000.
Review: {\bf [Bignami 00]}, {\bf [Schweber 00]},
{\bf [Phipps 00]}, {\bf [Bernstein 01]}, {\bf [Kox 02]}.

\item {\bf [Paladino-Faoro-Falci-Fazio 02]}:
E. Paladino, L. Faoro, G. Falci, \& R. Fazio,
``Decoherence and $1/f$ noise in Josephson qubits'',
{\em Phys. Rev. Lett.} {\bf 88}, 22, 228304 (2002).

\item {\bf [Palao-Kosloff 02]}:
J. P. Palao, \& R. Kosloff,
``Quantum computing by an optimal control algorithm for unitary transformations'',
{\em Phys. Rev. Lett.} {\bf 89}, 18, 188301 (2002);
quant-ph/0204101.

\item {\bf [Palao-Kosloff 03]}:
J. P. Palao, \& R. Kosloff,
``Optimal control theory for unitary transformations'',
{\em Phys. Rev. A} {\bf 68}, 6, 062308 (2003).

\item {\bf [Palma-Suominen-Ekert 96]}:
G. M. Palma, K. Suominen, \& A. K. Ekert,
``Quantum computers and dissipation'',
{\em Proc. R. Soc. Lond. A} {\bf 452}, 1946, 567-584 (1996).

\item {\bf [Pan-Zeilinger 98]}:
J.-W. Pan, \& A. Zeilinger,
``Greenberger-Horne-Zeilinger-state analyzer'',
{\em Phys. Rev. A} {\bf 57}, 3, 2208-2211 (1998).

\item {\bf [Pan-Bouwmeester-Weinfurter-Zeilinger 98]}:
J.-W. Pan, D. Bouwmeester, H. Weinfurter, \& A. Zeilinger,
``Experimental entanglement swapping:
Entangling photons that never interacted'',
{\em Phys. Rev. Lett.} {\bf 80}, 18, 3891-3894 (1998).
Comment: {\bf [Kok-Braunstein 99]}.

\item {\bf [Pan-Bouwmeester-Daniell-(+2) 00]}:
J.-W. Pan, D. Bouwmeester, M. Daniell, H. Weinfurter, \& A. Zeilinger,
``Experimental test of quantum nonlocality in
three-photon Greenberger-Horne-Zeilinger entanglement'',
{\em Nature} {\bf 403}, 6769, 515-519 (2000).
See {\bf [Bouwmeester-Pan-Daniell-(+2) 99]}.

\item {\bf [Pan-Simon-Bruckner-Zeilinger 01]}:
J.-W. Pan, C. Simon, \v{C}. Brukner, \& A. Zeilinger,
``Feasible entanglement purification for quantum communication'',
{\em Nature} {\bf 410}, 6832, 1067-1070 (2001);
quant-ph/0012026.

\item {\bf [Pan-Daniell-Gasparoni-(+2) 01]}:
J.-W. Pan, M. Daniell, S. Gasparoni,
G. Weihs, \& A. Zeilinger,
``Experimental demonstration of four-photon entanglement
and high-fidelity teleportation'',
{\em Phys. Rev. Lett.} {\bf 86}, 20, 4435-4438 (2001);
quant-ph/0104047.

\item {\bf [Pan-Zeilinger 02]}:
J.-W. Pan, \& A. Zeilinger,
``Multi-photon entanglement and quantum non-locality'',
in {\bf [Bertlmann-Zeilinger 02]}, pp.~225-240.

\item {\bf [Pan-Gasparoni-Aspelmeyer-(+2) 03]}:
J.-W. Pan, S. Gasparoni, M. Aspelmeyer, T. Jennewein, \& A. Zeilinger,
``Experimental realization of freely propagating teleported qubits'',
{\em Nature} {\bf 421}, 6924, 721-725 (2003).

\item {\bf [Pan-Gasparoni-Ursin-(+2) 03]}:
J.-W. Pan, S. Gasparoni, R. Ursin,
G. Weihs, \& A. Zeilinger,
``Experimental entanglement purification of arbitrary unknown states'',
{\em Nature} {\bf 423}, 6938, 417-422 (2003).

\item {\bf [Panov 02]}:
A. D. Panov,
``Inverse quantum Zeno effect in quantum oscillations'',
{\em Phys. Lett. A} {\bf 298}, 5-6, 295-300 (2002).

\item {\bf [Papadakos 02]}:
N. P. Papadakos,
``Quantum information theory and applications to quantum cryptography'',
quant-ph/0201057.

\item {\bf [Papaliolios 67]}:
C. Papaliolios,
``Experimental test of a hidden-variable quantum theory'',
{\em Phys. Rev. Lett.} {\bf 18}, 15, 622-625 (1967).

\item {\bf [Paris 98]}:
M. G. A. Paris,
``Entanglement and visibility at the output of a Mach-Zehnder interferometer'',
{\em Phys. Rev. A};
quant-ph/9811078.

\item {\bf [Paris 99]}:
M. G. A. Paris,
``Optical qubit by conditional interferometry'',
quant-ph/9909075.

\item {\bf [Paris 00]}:
M. G. A. Paris,
``Increasing the visibility of multiphoton entanglement'',
{\em Fortschr. Phys.} {\bf 48}, 5-7, 511-515 (2000).

\item {\bf [Paris-Plenio-Bose-(+2) 00]}:
M. G. A. Paris, M. B. Plenio, S. Bose,
D. Jonathan, \& G. M. D'Ariano,
``Optical Bell measurement by Fock filtering'',
{\em Phys. Lett. A} {\bf 273}, 3, 153-158 (2000);
quant-ph/9911036.

\item {\bf [Paris-D'Ariano-Sacchi 01]}:
M. G. A. Paris, G. M. D'Ariano, \& M. F. Sacchi,
``Maximum-likelihood method in quantum estimation'',
quant-ph/0101071.

\item {\bf [Paris 01]}:
M. G. A. Paris,
``A robust verification of the quantum nature of light'',
quant-ph/0102044.

\item {\bf [Paris 02]}:
M. G. A. Paris,
``Optimized quantum nondemolition measurement of a field quadrature'',
{\em Phys. Rev. A} {\bf 65}, 1, 012110 (2002);
quant-ph/0203097.

\item {\bf [Paris-Cola-Bonifacio 02]}:
M. G. A. Paris, M. Cola, \& R. Bonifacio,
``Remote state preparation and teleportation in phase space'',
{\em J. Opt. B: Quantum Semiclass. Opt.};
quant-ph/0209168.

\item {\bf [Paris-Cola-Bonifacio 03]}:
M. G. A. Paris, M. Cola, \& R. Bonifacio,
``Quantum-state engineering assisted by entanglement'',
{\em Phys. Rev. A} {\bf 67}, 4, 042104 (2003).

\item {\bf [Paris-D'Ariano-Lo Presti-Perinotti 03]}:
M. G. A. Paris, G. M. D'Ariano, P. Lo Presti, \& P. Perinotti,
``About the use of entanglement in the optical
implementation of quantum information processing'',
{\em Fortschr. Phys.} {\bf 51}, ?, 449-? (2003);
quant-ph/0210031.

\item {\bf [Paris-Illuminati-Serafini-De Siena 03]}:
M. A. G. Paris, F. Illuminati, A. Serafini, \& S. De Siena,
``Purity of Gaussian states: Measurement schemes and time evolution in
noisy channels'',
{\em Phys. Rev. A} {\bf 68}, 1, 012314 (2003);
quant-ph/0304059.

\item {\bf [Park 68 a]}:
J. L. Park,
``Quantum theoretical concepts of measurement: Part I'',
{\em Philos. Sci.} {\bf 35}, ?, 205-231 (1968).

\item {\bf [Park 68 b]}:
J. L. Park,
``Quantum theoretical concepts of measurement: Part II'',
{\em Philos. Sci.} {\bf 35}, ?, 389-411 (1968).

\item {\bf [Park-Margenau 68]}:
J. L. Park, \& H. Margenau,
``Simultaneous measurability in quantum theory'',
{\em Int. J. Theor. Phys.} {\bf 1}, 3, 211-283 (1968).
See {\bf [von Neumann 31]}, {\bf [Neumark 54]}.

\item {\bf [Park-Margenau 71]}:
J. L. Park, \& H. Margenau,
``The logic of noncommutability of quantum mechanical operators
and its empirical consequences'', in
W. Yourgrau, \& A. van der Merwe (eds.),
{\em Perspectives in quantum theory:
Essays in honor of Alfred Land\'e},
M.\ I.\ T.\ Press, Cambridge, Massachusetts, 1971.

\item {\bf [Park-Band 71]}:
J. L. Park, \& W. Band,
``A general theory of empirical
state determination in quantum physics: Part I'',
{\em Found. Phys.} {\bf 1}, 3, 211-225 (1971).
See {\bf [Band-Park 71]} (II).

\item {\bf [Park-Band 80]}:
J. L. Park, \& W. Band,
``Simultaneous measurement,
phase-space distributions and quantum state determination'',
{\em Annalen der Physik} {\bf 7}, ?, 189-199 (1980).

\item {\bf [Park-Band 92]}:
J. L. Park, \& W. Band,
``Preparation and measurements in quantum physics'',
{\em Found. Phys.} {\bf 22}, 5, 657-668 (1992).

\item {\bf [Parker-Rijmen 01]}:
M. G. Parker, \& V. Rijmen,
``The quantum entanglement of binary and bipolar sequences'',
{\em SETA01 (Bergen, 2001)};
quant-ph/0107106.

\item {\bf [Parker-Plenio 00]}:
S. Parker, \& M. B. Plenio,
``Efficient factorization with a single pure qubit
and $\log N$ mixed qubits'',
{\em Phys. Rev. Lett.} {\bf 85}, 14, 3049-3052 (2000);
quant-ph/0001066.

\item {\bf [Parker-Bose-Plenio 00]}:
S. Parker, S. Bose, \& M. B. Plenio,
``Entanglement quantification and purification in continuous-variable
systems'',
{\em Phys. Rev. A} {\bf 61}, 3, 032305 (2000);
quant-ph/9906098.

\item {\bf [Parker-Plenio 02]}:
S. Parker, \& M. B. Plenio,
``Entanglement simulations of Shor's algorithm'',
{\em Proc.\ ESF QIT Conf.\ Quantum Information: Theory, Experiment and Perspectives
(Gdansk, Poland, 2001)}, {\em J. Mod. Opt.} {\bf 49}, 8, 1325-1353 (2002);
quant-ph/0102136.

\item {\bf [Parkins-Marte-Zoller-Kimble 93]}:
A. S. Parkins, P. Marte, P. Zoller, \& H. J. Kimble,
``Synthesis of arbitrary quantum states via adiabatic transfer
of Zeeman coherence'',
{\em Phys. Rev. Lett.} {\bf 71}, 19, 3095-3098 (1993).

\item {\bf [Parkins-Kimble 99]}:
A. S. Parkins, \& H. J. Kimble,
``Proposal for teleportation of the wave function of a massive
particle'',
quant-ph/9909021.

\item {\bf [Parkins-Kimble 00]}:
A. S. Parkins, \& H. J. Kimble,
``Position-momentum Einstein-Podolsky-Rosen state of distantly
separated trapped atoms'',
{\em Phys. Rev. A} {\bf 61}, 5, 052104 (2000);
quant-ph/9907049.

\item {\bf [Parkins 00]}:
A. S. Parkins,
``Proposal for preparing delocalised mesoscopic
states of material oscillators'',
{\em J. Opt. B: Quantum Semiclass. Opt.};
quant-ph/0006113.

\item {\bf [Parkins-Larsabal 01]}:
A. S. Parkins, \& E. Larsabal,
``Preparation and light-mediated distribution of motional state
entanglement'',
{\em Phys. Rev. A} {\bf 63}, 1, 012304 (2001);
quant-ph/0008072.

\item {\bf [Parrondo 01]}:
J. M. R. Parrondo,
``Juegos cu\'{a}nticos'',
{\em Investigaci\'{o}n y Ciencia} 303, 82-83 (2001).
Reprinted in {\bf [Cabello 03 a]}, pp.~44-45.

\item {\bf [Particle Data Group 94]}:
Particle Data Group,
``Review of particle properties'',
{\em Phys. Rev. D} {\bf 50}, 3, Part I, 1173-1825 (1994).
Summary tables of particle properties published every two years.

\item {\bf [Partovi 04]}:
M. H. Partovi,
``Universal measure of entanglement'',
{\em Phys. Rev. Lett.} {\bf 92}, 7, 077904 (2004).

\item {\bf [Pascazio-Reignier 87]}:
S. Pascazio, \& J. Reignier,
``On the emission
lifetimes in atomic cascade tests of the Bell inequality'',
{\em Phys. Lett. A} {\bf 126}, 3, 163-167 (1987).

\item {\bf [Pascazio-Namiki-Badurek-Rauch 93]}:
S. Pascazio, M. Namiki, G. Badurek, \& H. Rauch
``Quantum Zeno effect with neutron spin'',
{\em Phys. Lett. A} {\bf 179}, 3, 155-160 (1993).

\item {\bf [Pascazio 03]}:
S. Pascazio,
``On noise-induced superselection rules'',
{\em J. Mod. Opt.};
quant-ph/0311195.

\item {\bf [Pashkin-Yamamoto-Astafiev-(+3) 03]}:
Y. A. Pashkin, T. Yamamoto, O. Astafiev,
Y. Nakamura, D. V. Averin, \& J. S. Tsai,
``Quantum oscillations in two coupled charge qubits'',
{\em Nature} {\bf 421}, 6925, 823-826 (2003).

\item {\bf [Passon 04]}:
O. Passon,
``How to teach quantum mechanics'',
quant-ph/0404128.

\item {\bf [Patel 98]}:
A. Patel,
``On how to produce entangled states violating Bell's
inequalities in quantum theory'',
quant-ph/9810071.

\item {\bf [Patel 99]}:
A. Patel,
``What is quantum computation?'',
quant-ph/9909082.

\item {\bf [Patel 01 a]}:
A. Patel,
``Quantum database search can do without sorting'',
{\em Phys. Rev. A} {\bf 64}, 3, 034303 (2001);
quant-ph/0012149.

\item {\bf [Patel 01 b]}:
A. Patel,
``Why genetic information processing could have a
quantum basis'',
{\em J. Biosciences} {\bf 26}, 145-151 (2001);
quant-ph/0105001.

\item {\bf [Patel-Raghunathan-Rungta 04]}:
A. Patel, K. S. Raghunathan, \& P. Rungta,
``Quantum random walks do not need coin toss'',
quant-ph/0405128.

\item {\bf [Paternostro-Kim-Ham 03]}:
M. Paternostro, M. S. Kim, \& B. S. Ham,
``Generation of entangled coherent states via cross-phase-modulation in a
double electromagnetically induced transparency regime'',
{\em Phys. Rev. A} {\bf 67}, 2, 023811 (2003).

\item {\bf [Paternostro-Falci-Kim-Palma 03]}:
M. Paternostro, G. Falci, M. S. Kim, \& G. M. Palma,
``Entanglement between two superconducting qubits
via interaction with nonclassical radiation'',
{\em Phys. Rev. B} {\bf 69}, 21, 214502 (2004);
quant-ph/0307163.

\item {\bf [Paternostro-Son-Kim-(+2) 04]}:
M. Paternostro, W. Son, M. S. Kim, G. Falci, \& G. M. Palma,
``Dynamical entanglement-transfer for quantum information networks'',
quant-ph/0403126.

\item {\bf [Paternostro-Son-Kim 04]}:
M. Paternostro, W. Son, \& M. S. Kim,
``Complete conditions for entanglement transfer'',
{\em Phys. Rev. Lett.};
quant-ph/0403180.

\item {\bf [Paternostro-Kim-Knight 04]}:
M. Paternostro, M. S. Kim, \& P. L. Knight,
``Vibrational coherent quantum computation'',
quant-ph/0405099.

\item {\bf [Paternostro-Palma-Kim-Falci 04]}:
M. Paternostro, G. M. Palma, M. S. Kim, \& G. Falci,
``Quantum state transfer in imperfect artificial spin networks'',
quant-ph/0407058.

\item {\bf [Paternostro-Kim 04]}:
M. Paternostro, \& M. S. Kim,
``Qubit state guidance without feedback'',
quant-ph/0410026.

\item {\bf [Paterson-Piper-Schack 04]}:
K. G. Paterson, F. Piper, \& R. Schack,
``Why quantum cryptography?'',
quant-ph/0406147.

\item {\bf [Pati 98 a]}:
A. K. Pati,
``Uncertainty, non-locality and Bell's inequality'',
quant-ph/9802026.

\item {\bf [Pati 98 b]}:
A. K. Pati,
``Testing Bell's inequality using Aharonov-Casher effect'',
{\em Phys. Rev. A} {\bf 58}, 1, R1-R3 (1998);
quant-ph/9804005.

\item {\bf [Pati 98 c]}:
A. K. Pati,
``Fast quantum search algorithm and bounds on it'',
quant-ph/9807067.

\item {\bf [Pati-Lawande 98]}:
A. K. Pati, \& S. V. Lawande,
``Geometry of the Hilbert space and the quantum Zeno effect'',
{\em Phys. Rev. A} {\bf 58}, 2, 831-835 (1998).

\item {\bf [Pati 99]}:
A. K. Pati,
``Quantum superposition of multiple clones and the novel cloning
machine'',
{\em Phys. Rev. Lett.} {\bf 83}, 14, 2849-2852 (1999).

\item {\bf [Pati 00 a]}:
A. K. Pati,
``Assisted cloning and orthogonal complementing of an unknown state'',
{\em Phys. Rev. A} {\bf 61}, 2, 022308 (2000).

\item {\bf [Pati 00 b]}:
A. K. Pati,
``Probabilistic exact cloning and probabilistic no-signalling'',
{\em Phys. Lett. A} {\bf 270}, 3-4, 103-107 (2000);
quant-ph/9908017.

\item {\bf [Pati-Braunstein 00 a]}:
A. K. Pati, \& S. L. Braunstein,
``Impossibility of deleting an unknown quantum state'',
{\em Nature} {\bf 404}, 6774, 164-165 (2000);
quant-ph/9911090.
See {\bf [Zurek 00 a]}, {\bf [Pati-Braunstein 00 b]}.
Comment: {\bf [Bhagwat-Khandekar-Menon-(+2) 00]}.

\item {\bf [Pati-Braunstein-Lloyd 00]}:
A. K. Pati, S. L. Braunstein, \& S. Lloyd,
``Quantum searching with continuous variables'',
quant-ph/0002082.

\item {\bf [Pati 00 c]}:
A. K. Pati,
``Existence of the Schmidt decomposition for tripartite systems'',
{\em Phys. Lett. A} {\bf 278}, 3, 118-122 (2000);
quant-ph/9911073.

\item {\bf [Pati 00 d]}:
A. K. Pati,
``Entropy decrease in quantum Zeno effect'',
quant-ph/0006089.

\item {\bf [Pati-Braunstein 00 b]}:
A. K. Pati, \& S. L. Braunstein,
``Quantum no-deleting principle and some of its implications'',
quant-ph/0007121.
See {\bf [Pati-Braunstein 00 a]}.

\item {\bf [Pati 01 a]}:
A. K. Pati,
``Minimum classical bit for remote preparation and measurement
of a qubit'',
{\em Phys. Rev. A} {\bf 63}, 1, 014302 (2001);
quant-ph/9907022.

\item {\bf [Pati 01 b]}:
A. K. Pati,
``Quantum cobweb: Remote shared-entangling of an unknown quantum
state'',
quant-ph/0101049.

\item {\bf [Pati-\.{Z}ukowski 01]}:
A. K. Pati, \& M. \.{Z}ukowski,
``Interference due to coherence swapping'',
quant-ph/0102041.

\item {\bf [Pati-Jain-Mitra-Ramanna 02]}:
A. K. Pati, S. R. Jain, A. Mitra, \& R. Ramanna,
``Quantum limit on computational time and speed'',
{\em Phys. Lett. A} {\bf 301}, 3-4, 125-129 (2002).

\item {\bf [Pati 02 a]}:
A. K. Pati,
``Remote state preparation and measurement of single photon'',
invited talk in {\em Sixth Int.\ Conf.\ on Photonics-2002 (Mumbai, India, 2002)};
quant-ph/0212164.

\item {\bf [Pati 02 b]}:
A. K. Pati,
``General impossible operations in quantum information'',
{\em Phys. Rev. A} {\bf 66}, 6, 062319 (2002);
quant-ph/0111153.

\item {\bf [Pati 03 a]}:
A. K. Pati,
``Relative phase change during quantum operation'',
talk in {\em Feynman Festival (University of
Maryland, College Park, Maryland, 2002)};
quant-ph/0301101.

\item {\bf [Pati-Braunstein 03 a]}:
A. K. Pati, \& S. L. Braunstein,
``Quantum mechanical universal constructor'',
quant-ph/0303124.

\item {\bf [Pati-Braunstein 03 b]}:
A. K. Pati, \& S. L. Braunstein,
``Quantum deleting and signalling'',
{\em Phys. Lett. A} {\bf 315}, 3-4, 208-212 (2003);
quant-ph/0305145.

\item {\bf [Pati 03 b]}:
A. K. Pati,
`Quantifying entanglement in ``Experimental entanglement of four
particles''\',
quant-ph/0311031.
See {\bf [Sackett-Kielpinski-King-(+8) 00]}.

\item {\bf [Pati 04]}:
A. K. Pati,
`Quantifying entanglement in ``experimental entanglement of four particles''\,',
{\em Phys. Lett. A} {\bf 322}, 5-6 301-304 (2004).

\item {\bf [Patil 98]}:
S. H. Patil,
``Quantum mechanical description of the Stern-Gerlach experiment'',
{\em Eur. J. Phys.} {\bf 19}, 1, 25-30 (1998).

\item {\bf [Paty 95]}:
M. Paty,
``The nature of Einstein's objections to the
Copenhagen interpretation of quantum mechanics'',
{\em Found. Phys.} {\bf 25}, 1, 183-204 (1995).

\item {\bf [Paty 99]}:
M. Paty,
``Are quantum systems physical objects with physical
properties?'',
{\em Eur. J. Phys.} {\bf 20}, 6, 373-388 (1999).

\item {\bf [Pa\v{s}kauskas-You 01]}:
R. Pa\v{s}kauskas, \& L. You,
``Quantum correlations in two-boson wavefunctions'',
{\em Phys. Rev. A} {\bf 64}, 4, 042310 (2001);
quant-ph/0106117.

\item {\bf [Paul 85]}:
H. Paul,
``Einstein-Podolsky-Rosen paradox and reality of
individual physical properties'',
{\em Am. J. Phys.} {\bf 53}, 4, 318-319 (1985).

\item {\bf [Paul-Pavi\v{c}i\'{c} 96]}:
H. Paul, \& M. Pavi\v{c}i\'{c},
``Resonance interaction-free measurement'',
{\em Int. J. Theor. Phys.} {\bf 35}, 10, 2085-2091 (1996).

\item {\bf [Paul-Pavi\v{c}i\'{c} 97]}:
H. Paul, \& M. Pavi\v{c}i\'{c},
``Nonclassical interaction-free
detection of objects in a monolithic total-internal-reflection
resonator'',
{\em J. Opt. Soc. Am. B} {\bf 14}, 6, 1275-1279 (1997);
quant-ph/9908023.

\item {\bf [Paul-Pavi\v{c}i\'{c} 98]}:
H. Paul, \& M. Pavi\v{c}i\'{c},
``Realistic interaction-free detection of objects in a
resonator'',
{\em Found. Phys.} {\bf 28}, 6, 959-970 (1998);
quant-ph/9906102.

\item {\bf [Paul 98 a]}:
H. Paul,
`Interference and ``which way'' information',
{\em Int. J. Theor. Phys.} {\bf 37}, 1, 511-518 (1998).

\item {\bf [Paul 98 b]}:
H. Paul,
``Interference destroyed by entanglement'',
{\em Phys. World} {\bf 11}, 11, 24-25 (1998).
See {\bf [D\"{u}rr-Nonn-Rempe 98]}.

\item {\bf [Paul 90]}:
W. Paul,
``Electromagnetic traps for charged and neutral particles'',
{\em Rev. Mod. Phys.} {\bf 62}, 3, 531-540 (1990).
Reprinted in {\bf [Macchiavello-Palma-Zeilinger 00]}, pp.~320-329.

\item {\bf [Pauli 26]}:
W. Pauli,
``\"{U}ber den Zusammenhang des Abschlusses der
Elektronengruppen im Atom mit der Komplexstruktur der Spektren'',
{\em Zeitschrift f\"{u}r Physik} {\bf 31}, 765-? (1926).

\item {\bf [Pauli 27 a]}:
W. Pauli,
``\"{U}ber Gasentartung und Paramagnetismus'',
{\em Zeitschrift f\"{u}r Physik} {\bf 43}, 81-102 (1927).

\item {\bf [Pauli 27 b]}:
W. Pauli,
``Zur Quantenmechanik des magnetischen Elektrons'',
{\em Zeitschrift f\"{u}r Physik} {\bf 43}, 601-623 (1927).
Reprinted in {\bf [Biedenharn-van Dam 65]}, pp.~14-36.

\item {\bf [Pauli 33]}:
W. Pauli,
``Die allgemeinen Prinzipien der Wellenmechanik'',
in Handbuch der Physik, vol. 24/1,
Springer-Verlag, Berlin, 1933, pp.~771-938.

\item {\bf [Pauli 58]}:
W. Pauli,
{\em Handbuch der Physik, vol. 5, Part 1:
Prinzipien der Quantentheorie},
Springer-Verlag, Berlin, 1958.
English version: {\em General principles of quantum theory},
Springer-Verlag, Berlin, 1980.

\item {\bf [Pauli 64]}:
W. Pauli,
{\em Collected scientific papers by Wolfgang Pauli},
R. Kronig, \& V. Weisskopf (eds.),
Wiley, New York, 1964.

\item {\bf [Pauli 79]}:
W. Pauli (edited by K. von Meyenn),
{\em Wissenschaftlicher Briefwechsel mit
Bohr, Einstein, Heisenberg u.a. /
Scientific correspondence with Bohr,
Einstein, Heisenberg a.o.
Band I/Volume I: 1919-1929},
Springer-Verlag, New York, 1979.
See {\bf [Pauli 85]} (II),
{\bf [Pauli 93]} (III), {\bf [Pauli 97]} (IV-I),
{\bf [Pauli 98]} (IV-II), {\bf [Pauli 01]} (IV-III).

\item {\bf [Pauli 80]}:
W. Pauli (edited by P. Achuthan, \& K. Venkatesan),
{\em General principles of quantum mechanics},
Springer-Verlag, Berlin, 1980.

\item {\bf [Pauli 85]}:
W. Pauli (edited by K. V. Meyenn, A. Hermann, \& V. F. Weisskopff),
{\em Wissenschaftlicher Briefwechsel mit
Bohr, Einstein, Heisenberg u.a. /
Scientific correspondence with Bohr,
Einstein, Heisenberg a.o.
Band II/Volume II: 1930-1939},
Springer-Verlag, New York, 1985.
See {\bf [Pauli 79]} (I),
{\bf [Pauli 93]} (III), {\bf [Pauli 97]} (IV-I),
{\bf [Pauli 98]} (IV-II), {\bf [Pauli 01]} (IV-III).

\item {\bf [Pauli 93]}:
W. Pauli (edited by K. von Meyenn),
{\em Wissenschaftlicher Briefwechsel mit Bohr,
Einstein, Heisenberg u.a. /
Scientific correspondence with
Bohr, Einstein, Heisenberg a.o.
Band III/Volume III: 1940-1949},
Springer-Verlag, Berlin, 1993.
See {\bf [Pauli 79]} (I), {\bf [Pauli 85]} (II),
{\bf [Pauli 97]} (IV-I),
{\bf [Pauli 98]} (IV-II), {\bf [Pauli 01]} (IV-III).

\item {\bf [Pauli 94]}:
W. Pauli (edited by C. P. Enz, \& K. von Mayenn),
{\em Writings on physics and philosophy},
Springer-Verlag, New York, 1994.

\item {\bf [Pauli 97]}:
W. Pauli (edited by K. von Meyenn),
{\em Wissenschaftlicher Briefwechsel mit Bohr,
Einstein, Heisenberg u.a. /
Scientific correspondence with
Bohr, Einstein, Heisenberg a.o.
Band IV/Volume IV. Teil I/Part I: 1950-1952},
Springer-Verlag, Berlin, 1997.
See {\bf [Pauli 79]} (I), {\bf [Pauli 85]} (II),
{\bf [Pauli 93]} (III),
{\bf [Pauli 98]} (IV-II), {\bf [Pauli 01]} (IV-III).

\item {\bf [Pauli 98]}:
W. Pauli (edited by K. von Meyenn),
{\em Wissenschaftlicher Briefwechsel mit Bohr,
Einstein, Heisenberg u.a. /
Scientific correspondence with
Bohr, Einstein, Heisenberg a.o.
Band IV/Volume IV. Teil II/Part II: 1953-1954},
Springer-Verlag, Berlin, 1998.
See {\bf [Pauli 79]} (I), {\bf [Pauli 85]} (II),
{\bf [Pauli 93]} (III), {\bf [Pauli 97]} (IV-I),
{\bf [Pauli 01]} (IV-III).

\item {\bf [Pauli 00]}:
W. Pauli (edited by C. P. Enz),
{\em Wave mechanics (Pauli Lectures on Physics Volume 5)},
Dover, New York, 2000.

\item {\bf [Pauli 01]}:
W. Pauli (edited by K. von Meyenn),
{\em Wissenschaftlicher Briefwechsel mit Bohr,
Einstein, Heisenberg u.a. /
Scientific correspondence with
Bohr, Einstein, Heisenberg a.o.
Band IV/Volume IV. Teil III/Part III: 1955-1956},
Springer-Verlag, Berlin, 2001.
See {\bf [Pauli 79]} (I), {\bf [Pauli 85]} (II),
{\bf [Pauli 93]} (III), {\bf [Pauli 97]} (IV-I),
{\bf [Pauli 98]} (IV-II).

\item {\bf [Paunkovi\'{c}-Omar-Bose-Vedral 02]}:
N. Paunkovi\'{c}, Y. Omar, S. Bose, \& V. Vedral,
``Entanglement concentration using quantum statistics'',
{\em Phys. Rev. Lett.} {\bf 88}, 18, 187903 (2002);
quant-ph/0112004.

\item {\bf [Pavon 99]}:
M. Pavon,
``Derivation of the wave function collapse in the context of
Nelson's stochastic mechanics'',
{\em J. Math. Phys.} {\bf 40}, 11, 5565-5577 (1999);
quant-ph/9912015.

\item {\bf [Pavon 02]}:
M. Pavon,
``Quantum Schr\"{o}dinger bridges'',
in A. Rantzer, \& C. I. Byrnes (eds.),
{\em Directions in mathematical systems theory and optimization},
Springer-Verlag, Berlin, 2002, pp.~227-238.

\item {\bf [Pavi\v{c}i\'{c} 81]}:
M. Pavi\v{c}i\'{c},
``Kriticki osvrt na potpunost kvantne mehanike
[A critical review of the completeness of quantum mechanics]'',
M.\ Sc. thesis, University of Zagreb, 1981.

\item {\bf [Pavi\v{c}i\'{c} 83]}:
M. Pavi\v{c}i\'{c},
``The Einstein locality without the Bell inequality'',
in P. Weingartner (ed.),
{\em Abstracts of the 7th International Congress of Logic, Methodology, and Philosophy of Science}, Vol. 4,
J. Huttegger OHG, Salzburg (1983), pp.~161-164.

\item {\bf [Pavi\v{c}i\'{c} 86 a]}:
M. Pavi\v{c}i\'{c},
``Algebarsko-logicka struktura kvantno-mehanickih interpretacija
[Algebraico-logical structure of the interpretations of quantum mechanics]'',
Ph.\ D. thesis, University of Belgrade, 1986.

\item {\bf [Pavi\v{c}i\'{c} 86 b]}:
M. Pavi\v{c}i\'{c},
``When do position and momentum distributions determine the quantum mechanical state?'',
{\em Phys. Lett. A} {\bf 118}, ?, 5-7 (1986).

\item {\bf [Pavi\v{c}i\'{c} 87 a]}:
M. Pavi\v{c}i\'{c},
``Complex Gaussians and the Pauli non-uniqueness'',
{\em Phys. Lett. A} {\bf 122}, ?, 280-282 (1987).

\item {\bf [Pavi\v{c}i\'{c} 87 b]}:
M. Pavi\v{c}i\'{c},
``Minimal quantum logic with merged implications'',
{\em Int. J. Theor. Phys.} {\bf 26}, ?, 845-852 (1987).

\item {\bf [Pavi\v{c}i\'{c} 89]}:
M. Pavi\v{c}i\'{c},
``Unified quantum logic'',
{\em Found. Phys.} {\bf 19}, ?, 999-1016 (1989).

\item {\bf [Pavi\v{c}i\'{c} 90 a]}:
M. Pavi\v{c}i\'{c},
``A relative frequency criterion for the repeatability of quantum measurements'',
{\em Nuovo Cimento B} {\bf 105}, 1103-1112 (1990)
Erratum: {\em Nuovo Cimento B} {\bf 106}, 105-106 (1991).

\item {\bf [Pavi\v{c}i\'{c} 90 b]}:
M. Pavi\v{c}i\'{c},
``A theory of deduction for quantum mechanics'',
{\em Nuova Critica} {\bf I-II} (Nuova Serie), 13-14, 109-129 (1990).

\item {\bf [Pavi\v{c}i\'{c} 90 c]}:
M. Pavi\v{c}i\'{c},
``Quantum Malus Law for composite systems as a hidden-variable theory'',
{\em Phys. Rev. D} {\bf 42}, ?, 3594-3595 (1990).

\item {\bf [Pavi\v{c}i\'{c} 90 d]}:
M. Pavi\v{c}i\'{c},
``There is a formal difference between the Copenhagen and the statistical interpretation of quantum mechanics'',
in J. Mizerski, A. Posiewnik, J. Pykacz, \& M. \.{Z}ukowski (eds.),
{\em Problems in quantum physics (Gdansk, 1989)},
World Scientific, Singapore, 1990, pp.~440-452.

\item {\bf [Pavi\v{c}i\'{c} 90 d]}:
M. Pavi\v{c}i\'{c},
``Epistemic vs. ontic frequency approach to quantum measurements'',
in {\em Symp.\ on the Foundations of Modern Physics (Turku, Finland, 1990)},
Report Series Turku-FTL-R183, 1990, pp.~144-147.

\item {\bf [Pavi\v{c}i\'{c} 92 a]}:
M. Pavi\v{c}i\'{c},
``A new axiomatization of unified quantum logic'',
{\em Int. J. Theor. Phys.} {\bf 31}, ? 1753-1766 (1992).

\item {\bf [Pavi\v{c}i\'{c} 92 b]}:
M. Pavi\v{c}i\'{c},
``Bibliography on quantum logics and related structures'',
{\em Int. J. Theor. Phys. 31}, ?, 373-461 (1992).

\item {\bf [Pavi\v{c}i\'{c} 92 c]}:
M. Pavi\v{c}i\'{c},
``In search for new observables'',
in A. van der Merwe, \& F. Selleri (eds.),
{\em Bell's theorem and the foundations of modern physics},
World Scientific, Singapore, 1992, pp.~371-374.

\item {\bf [Pavi\v{c}i\'{c} 93 a]}:
M. Pavi\v{c}i\'{c},
``On a formal difference between the individual and statistical interpretation of quantum mechanics'',
{\em Phys. Lett. A} 174, ?, 353-357 (1993).

\item {\bf [Pavi\v{c}i\'{c} 93 b]}:
M. Pavi\v{c}i\'{c},
``Probabilistic forcing in quantum logic'',
{\em Int. J. Theor. Phys.} {\bf 32}, ?, 1965-1979 (1993).

\item {\bf [Pavi\v{c}i\'{c} 94]}:
M. Pavi\v{c}i\'{c},
``Spin correlated interferometry for polarized and unpolarized photons on a beam splitter'',
{\em Phys. Rev. A} {\bf 50}, 4, 3486-3491 (1994).

\item {\bf [Pavi\v{c}i\'{c}-Summhammer 94]}:
M. Pavi\v{c}i\'{c}, \& J. Summhammer,
``Interferometry with two pairs of spin correlated photons'',
{\em Phys. Rev. Lett.} {\bf 73}, 24, 3191-3194 (1994).

\item {\bf [Pavi\v{c}i\'{c} 95 a]}:
M. Pavi\v{c}i\'{c},
``Preselection with certainty of photons in a singlet state from a set of independent photons'',
{\em Int. J. Theor. Phys.} {\bf 34}, ?, 1653-1665 (1995).

\item {\bf [Pavi\v{c}i\'{c} 95 b]}:
M. Pavi\v{c}i\'{c},
``Spin-correlated interferometry with beam-splitters preselection of
spin-correlated photons'',
{\em J. of the Optical Soc. Am. B} {\bf 12}, 5, 821-828 (1995);
quant-ph/9908024.

\item {\bf [Pavi\v{c}i\'{c} 95 c]}:
M. Pavi\v{c}i\'{c},
``Closure of the enhancement and detection loopholes in the Bell theorem
by fourth order interference with photons of different colours'',
{\em Phys. Lett. A} {\bf 209}, 5-6, 255-260 (1995).

\item {\bf [Pavi\v{c}i\'{c} 96 a]}:
M. Pavi\v{c}i\'{c},
`Resonance energy-exchange-free detection and
``welcher Weg'' experiment',
{\em Phys. Lett. A} {\bf 223}, 4, 241-245 (1996);
quant-ph/9907040.

\item {\bf [Pavi\v{c}i\'{c} 96 b]}:
M. Pavi\v{c}i\'{c},
``Preselected quantum optical correlations'',
in S. Jeffers, S. Roy, J.-P. Vigier, \& G. Hunter (eds.),
{\em The present status of the quantum theory of light},
Kluwer Academic, Dordrecht, Holland, 1996, pp.~311-322.

\item {\bf [Pavi\v{c}i\'{c} 96 c]}:
M. Pavi\v{c}i\'{c},
``Event-ready entanglement preparation'',
in F. de Martini, G. Denardo, \& Y. H. Shih (eds.),
{\em Quantum interferometry},
VCH Publishers, Weinheim, 1996, pp.~193-204;
quant-ph/9907038.

\item {\bf [Pavi\v{c}i\'{c} 96 d]}:
M. Pavi\v{c}i\'{c},
``Nonmaximal entaglement preparation'',
{\em EQEC'96--1996 European Quantum Electronics Conference},
Technical Digest, IEEE, Piscataway, New Jersey, 1996, p.~191.

\item {\bf [Pavi\v{c}i\'{c} 96 e]}:
M. Pavi\v{c}i\'{c},
``Preselected loophole-free Bell experiments'',
{\em Supplement to Optics \& Photonics News} {\bf 7}, 8, 181 (1996).

\item {\bf [Pavi\v{c}i\'{c} 97 a]}:
M. Pavi\v{c}i\'{c},
``Loophole-free four photon EPR experiment'',
{\em Phys. Lett. A} {\bf 224}, 4-5, 220-226 (1997).

\item {\bf [Pavi\v{c}i\'{c} 97 b]}:
M. Pavi\v{c}i\'{c},
``A method for reaching detection efficiencies
necessary for optical loophole-free Bell experiments'',
{\em Opt. Comm.} {\bf 142}, ?, 308-314 (1997).
quant-ph/9907039.

\item {\bf [Pavi\v{c}i\'{c} 98]}:
M. Pavi\v{c}i\'{c},
``Identity rule for classical and quantum theories'',
{\em Int. J. Theor. Phys.} {\bf 37}, 8, 2099-2104 (1998).

\item {\bf [Pavi\v{c}i\'{c}-Megill 99]}:
M. Pavi\v{c}i\'{c}, \& N. D. Megill,
``Non-orthomodular models for both quantum logic and standard classical logic:
Repercussions for quantum computers'',
{\em Helv. Phys. Acta} {\bf 72}, ?, 189-210 (1999);
quant-ph/9906101.

\item {\bf [Pavi\v{c}i\'{c} 00 a]}:
M. Pavi\v{c}i\'{c},
``Quantum logic for quantum computers'',
{\em Int. J. Theor. Phys.} {\bf 39}, ?, 813-825 (2000).

\item {\bf [Pavi\v{c}i\'{c} 00 b]}:
M. Pavi\v{c}i\'{c},
``Quantum simulators and quantum repeaters'',
{\em Fortschr. Phys.} {\bf 48}, 5-7, 497-503 (2000).

\item {\bf [Pavi\v{c}i\'{c} 00 c]}:
M. Pavi\v{c}i\'{c},
``A realistic interaction-free resonator'',
in P. Kumar, G. M. D'Ariano, \& O. Hirota (eds.),
{\em Quantum communication, measurement, and computing--Proc.\ of the
Fourth Int.\ Conf.\ on Quantum Communication, Measurement, and Computing (Evanston, Illinois, 1998)},
Academic/Plenum Publishers, New York, 2000, 527-531.

\item {\bf [Pavi\v{c}i\'{c} 00 d]}:
M. Pavi\v{c}i\'{c},
``Quantum logic for genuine quantum simulators'',
in E. Donkor, \& A. R. Pirich (eds.),
{\em Quantum computing. Proc.\ of SPIE} Vol. 4047, 2000, pp.~90-96.

\item {\bf [Pavi\v{c}i\'{c} 02]}:
M. Pavi\v{c}i\'{c},
``Quantum computers, discrete space, and entanglement'',
in N. Callaos, Y. He, \& J. A. P\'{e}rez-Peraza (eds.),
{\em SCI 2002, The 6th World Multiconference on Systemics, Cybernetics, and Informatics (Orlando, Florida, 2002)},
International Institute of Informatics and Systemics, Orlando, Florida, 2002, Vol. XVII, pp.~65-70.

\item {\bf [Pavi\v{c}i\'{c}-Merlet-McKay-Megill 04]}:
M. Pavi\v{c}i\'{c}, J.-P. Merlet, B. McKay, \& N. D. Megill,
``Kochen-Specker qubits'',
quant-ph/0409014.

\item {\bf [Payne-Deng 03]}:
M. G. Payne, \& L. Deng,
``Quantum entanglement of Fock states with perfectly efficient ultraslow
single-probe photon four-wave mixing'',
{\em Phys. Rev. Lett.} {\bf 91}, 12, 123602 (2003).

\item {\bf [Paz-Mahler 93]}:
J. P. Paz, \& G. Mahler,
``Proposed test for temporal Bell inequalities'',
{\em Phys. Rev. Lett.} {\bf 71}, 20, 3235-3239 (1993).

\item {\bf [Paz-Habib-Zurek 93]}:
J. P. Paz, S. Habib, \& W. H. Zurek,
``Reduction of the wave packet:
Preferred observable and decoherence time scale'',
{\em Phys. Rev. D} {\bf 47}, 2, 488-501 (1993).

\item {\bf [Paz-Zurek 93]}:
J. P. Paz, \& W. H. Zurek,
``Environment-induced decoherence, classicality,
and consistency of quantum histories'',
{\em Phys. Rev. D} {\bf 48}, 6, 2728-2738 (1993).

\item {\bf [Paz-Zurek 98]}:
J. P. Paz, \& W. H. Zurek,
``Continuous error correction'',
in D. P. DiVincenzo. E. Knill, R. Laflamme, \& W. H. Zurek (eds.),
{\em Quantum Coherence and Decoherence.
Proc.\ of the ITP Conf.\ (Santa Barbara, California, 1996)},
{\em Proc. R. Soc. Lond. A} {\bf 454}, 1969, 355-364 (1998);
quant-ph/9707049.

\item {\bf [Paz-Zurek 99]}:
J. P. Paz, \& W. H. Zurek,
``Quantum limit of decoherence:
Environment induced superselection of energy eigenstates'',
{\em Phys. Rev. Lett.} {\bf 82}, 26, 5181-5185 (1999);
quant-ph/9811026.

\item {\bf [Paz-Zurek 00]}:
J. P. Paz, \& W. H. Zurek,
``Environment-induced decoherence and the transition from quantum to
classical'',
quant-ph/0010011.

\item {\bf [Paz 01]}:
J. P. Paz,
``Quantum engineering: Protecting the quantum world'',
{\em Nature} {\bf 412}, 6850, 869-870 (2001).
See {\bf [Carvalho-Milman-de Matos Filho-Davidovich 01]}.

\item {\bf [Paz 02]}:
J. P. Paz,
``Discrete Wigner functions and the phase-space representation of quantum
teleportation'',
{\em Phys. Rev. A} {\bf 65}, 6, 062311 (2002).

\item {\bf [Paz-Roncaglia 03]}:
J. P. Paz, \& A. Roncaglia,
``Quantum gate arrays can be programmed to evaluate
the expectation value of any operator
{\em Phys. Rev. A} {\bf 68}, 5, 052316 (2003);
quant-ph/0306143.

\item {\bf [Paz-Roncaglia-Saraceno 04 a]}:
J. P. Paz, A. J. Roncaglia, \& M. Saraceno,
``Quantum algorithms for phase-space tomography'',
{\em Phys. Rev. A} {\bf 69}, 3, 032312 (2004);
quant-ph/0310126.

\item {\bf [Paz-Roncaglia-Saraceno 04 b]}:
J. P. Paz, A. J. Roncaglia, \& M. Saraceno,
``Qubits in phase space: Wigner function approach to quantum error
correction and the mean king problem'',
quant-ph/0410117.

\item {\bf [Peacock 92]}:
K. A. Peacock,
``Comment on `Tests of signal locality and
Einstein-Bell locality for multiparticle systems'\,'',
{\em Phys. Rev. Lett.} {\bf 69}, 18, 2733 (1992).
Comment on {\bf [Roy-Singh 91]}.

\item {\bf [Pearle 67]}:
P. M. Pearle,
``Alternative to the orthodox interpretation of quantum theory'',
{\em Am. J. Phys.} {\bf 35}, 8, 742-753 (1967).

\item {\bf [Pearle 70]}:
P. M. Pearle,
``Hidden-variable example based upon data rejection'',
{\em Phys. Rev. D} {\bf 2}, 8, 1418-1425 (1970).

\item {\bf [Pearle 76]}:
P. M. Pearle,
``Reduction of the state vector by a nonlinear Schr\"{o}dinger equation'',
{\em Phys. Rev. D} {\bf 13}, 4, 857-868 (1976).

\item {\bf [Pearle 79]}:
P. M. Pearle,
``Toward explaining why events occur'',
{\em Int. J. Theor. Phys.} {\bf 18}, 7, 489-518 (1979).

\item {\bf [Pearle 82]}:
P. M. Pearle,
``Might god toss coins?'',
{\em Found. Phys.} {\bf 12}, 3, 249-263 (1982).

\item {\bf [Pearle 84]}:
P. M. Pearle,
`Comment on ``Quantum measurement and stochastic processes''\,',
{\em Phys. Rev. Lett.} {\bf 53}, 18, 1775 (1984).
Comment on {\bf [Gisin 84 a]}.
Reply: {\bf [Gisin 84 b]}.

\item {\bf [Pearle 85]}:
P. M. Pearle,
``On the time it takes a state vector to reduce'',
{\em J. Stat. Phys.} {\bf 41}, 3-4, 719-727 (1985).

\item {\bf [Pearle 86 a]}:
P. M. Pearle,
``Stochastic dynamical reduction theories and
superluminal communication'',
{\em Phys. Rev. D} {\bf 33}, 8, 2240-2252 (1986).

\item {\bf [Pearle 86 b]}:
P. M. Pearle,
``?'',
in D. M. Greenberger (ed.),
{\em New techniques and ideas in quantum measurement theory.
Proc.\ of an international conference (New York, 1986),
Ann. N. Y. Acad. Sci.} {\bf 480}, 539-? (1986).

\item {\bf [Pearle 86 c]}:
P. M. Pearle,
``Models for reduction'',
in R. Penrose, \& C. Isham (eds.),
{\em Quantum concepts in space and time},
Clarendon Press, Oxford, 1986, pp.~84-108.

\item {\bf [Pearle 89]}:
P. M. Pearle,
``Combining stochastic dynamical state-vector
reduction with spontaneous localization'',
{\em Phys. Rev. A} {\bf 39}, 5, 2277-2289 (1989).

\item {\bf [Pearle 90]}:
P. M. Pearle,
``Toward a relativistic theory of statevector
reduction'',
in A. I. Miller (ed.),
{\em Sixty-two years of uncertainty: Historical,
philosophical and physical inquiries into the foundations of quantum mechanics.
Proc.\ from the Int. School of History of Science
(Erice, Italy, 1989)},
Plenum Press, New York, 1990, pp.~193-214.

\item {\bf [Pearle 91]}:
P. M. Pearle,
``?'',
in A. I. Fine, M. Forbes, \& L. Wessels (eds.),
{\em Proc.\ of the 1990 Biennial Meeting of the Philosophy of Science
Association}, East Lansing, Michigan, 1991, vol. 2, pp.~35-?.

\item {\bf [Pearle 92]}:
P. M. Pearle,
``Relativistic model for statevector reduction'',
in P. Cvitanovic (ed.),
{\em Quantum chaos---Quantum mesurement},
Kluwer Academic, Dordrecht, Holland, 1992, pp.~283-297.

\item {\bf [Pearle 93]}:
P. M. Pearle,
``Ways to describe dynamical state-vector reduction'',
{\em Phys. Rev. A} {\bf 48}, 2, 913-923 (1993).

\item {\bf [Pearle-Squires 94]}:
P. M. Pearle, \& E. J. Squires,
``Bound state excitation, nucleon decay experiments,
and models of wave function collapse'',
{\em Phys. Rev. Lett.} {\bf 73}, 1, 1-5 (1994).

\item {\bf [Pearle 97 a]}:
P. M. Pearle,
``True collapse and false collapse'',
in D. H. Feng, \& B. L. Hu (eds.),
{\em Proc.\ of the 4th Drexel Symp.\ on Quantum
Nonintegrability (Philadelphia, Pennsylvania, 1994)},
Int. Press, Cambridge, Massachusetts, 1997;
quant-ph/9805049.

\item {\bf [Pearle 97 b]}:
P. M. Pearle,
``Tales and tails and stuff and nonsense'',
in {\bf [Cohen-Horne-Stachel 97 a]}, pp.~143-156;
quant-ph/9805050.

\item {\bf [Pearle-Squires 98]}:
P. M. Pearle, \& E. J. Squires,
`Comment on ``DNA molecular cousin of Schr\"{o}dinger's cat:
A curious example of quantum measurement''\,',
{\em Phys. Rev. Lett.} {\bf 80}, 6, 1348 (1998).
Comment on {\bf [Home-Chattopadhyay 96]}.
Reply: {\bf [Home-Chattopadhyay 98]}.

\item {\bf [Pearle 99 a]}:
P. M. Pearle,
``Relativistic collapse model with tachyonic features'',
{\em Phys. Rev. A} {\bf 59}, 1, 80-101 (1999).

\item {\bf [Pearle 99 b]}:
P. M. Pearle,
``Collapse models'',
in F. Petruccione, \& H. P. Breuer (eds.),
{\em Open systems and measurement in relativistic quantum theory},
Springer-Verlag, Berlin, 1999;
quant-ph/9901077.

\item {\bf [Pearle-Ring-Collar-Avignone 99]}:
P. M. Pearle, J. Ring, J. I. Collar, \& F. T. Avignone III,
``CSL collapse model and spontaneous radiation: An update'',
{\em Found. Phys.} {\bf 29}, 3, 465-480 (1999);
quant-ph/0001041.

\item {\bf [Pearle 00]}:
P. M. Pearle,
``Wavefunction collapse and conservation laws'',
{\em Found. Phys.} {\bf 30}, 8, 1145-1160 (2000);
quant-ph/0004067.

\item {\bf [Pearle 02]}:
P. M. Pearle,
``Quantum reflections'',
{\em Am. J. Phys.} {\bf 70}, 3, 365-367 (2002).
Review of {\bf [Ellis-Amati 00]}.

\item {\bf [Pearle 03]}:
P. M. Pearle,
``Problems and aspects of energy-driven wavefunction collapse models'',
quant-ph/0310086.

\item {\bf [Peat 97]}:
F. D. Peat,
{\em Infinite potential: The life and times of David Bohm},
Addison-Wesley, Reading, Massachusetts, 1997.
Reviews: {\bf [Cushing 97]}, {\bf [Mermin 97 b]}.

\item {\bf [Peev-N\"{o}lle-Maurhardt-(+6) 04]}:
M. Peev, M. N\"{o}lle, O. Maurhardt,
T. Lor\"{u}nser, M. Suda, A. Poppe,
R. Ursin, A. Fedrizzi, \& A. Zeilinger,
``A novel protocol-authentication algorithm ruling out a man-in-the-middle
attack in quantum cryptography'',
in {\em Proc.\ of Foundations of Quantum Information" (Camerino, Italy, 2004)},
{\em Int. J. Quantum Inf.};
quant-ph/0407131.

\item {\bf [Pegg-Barnett 99]}:
D. T. Pegg, \& S. M. Barnett,
``Retrodiction in quantum optics'',
{\em J. Opt. B: Quantum Semiclass. Opt.} {\bf 1}, 4, 442-445
(1999).

\item {\bf [Pegg-Barnett-Jeffers 02 a]}:
D. T. Pegg, S. M. Barnett, \& J. Jeffers,
``Quantum retrodiction in open systems'',
{\em Phys. Rev. A} {\bf 66}, 2, 022106 (2002).

\item {\bf [Pegg-Barnett-Jeffers 02 b]}:
D. T. Pegg, S. M. Barnett, \& J. Jeffers,
``Quantum theory of preparation and measurement'',
{\em J. Mod. Opt.} {\bf 49}, 5-6, 913-924 (2002).

\item {\bf [Peierls 79]}:
R. Peierls,
{\em Surprises in theoretical physics},
Princeton University Press, Princeton, New Jersey, 1979.

\item {\bf [Peierls 85]}:
R. Peierls,
``Observations in quantum mechanics and the `collapse of
the wave function'\,'',
in P. J. Lahti, \& P. Mittelstaedt (eds.),
{\em Symp.\ on the Foundations of Modern
Physics: 50 Years of the Einstein-Podolsky-Rosen Experiment
(Joensuu, Finland, 1985)},
World Scientific, Singapore, 1985, pp.~187-196.

\item {\bf [Peierls 91]}:
R. Peierls,
``In defence of `measurement'\,'',
{\em Phys. World} {\bf 4}, 1, 19-20 (1991).
See {\bf [Bell 90]}.

\item {\bf [Pelliccia-Schettini-Sciarrino-(+2) 03]}:
D. Pelliccia, V. Schettini, F. Sciarrino,
C. Sias, \& F. De Martini,
``Contextual realization of the universal quantum cloning machine and of the
universal-NOT gate by quantum-injected optical parametric
amplification'',
{\em Phys. Rev. A} {\bf 68}, 4, 042306 (2003);
quant-ph/0302087.

\item {\bf [Pellizzari-Gardiner-Cirac-Zoller 95]}:
T. Pellizzari, S. A. Gardiner, J. I. Cirac, \& P. Zoller,
``Decoherence, continuous observation, and quantum
computing: A cavity QED model'',
{\em Phys. Rev. Lett.} {\bf 75}, 21, 3788-3791 (1995).

\item {\bf [Pellizzari 97]}:
T. Pellizzari,
``Quantum networking with optical fibres'',
{\em Phys. Rev. Lett.} {\bf 79}, 26, 5242-5245 (1997);
quant-ph/9707001.

\item {\bf [Pellizzari 98]}:
T. Pellizzari,
``Quantum computers, error-correction and networking:
Quantum optical approaches'',
in {\bf [Lo-Spiller-Popescu 98]}, pp.~270-310.

\item {\bf [Pelton-Santori-Solomon-(+2) 02]}:
M. Pelton, C. Santori, G. S. Solomon,
O. Benson, \& Y. Yamamoto,
``Triggered single photons and entangled photons from a quantum dot microcavity'',
{\em Eur. Phys. J. D} {\bf 18}, 2 (Special issue:
{\em Quantum interference and cryptographic keys:
Novel physics and advancing technologies (QUICK) (Corsica, 2001)}, 179-190 (2002).

\item {\bf [Peng-Parkins 02]}:
A. Peng, \& A. S. Parkins,
``Motion-light parametric amplifier and entanglement distributor'',
{\em Phys. Rev. A} {\bf 65}, 6, 062323 (2002);
quant-ph/0203136.

\item {\bf [Peng-Zhu-Fang-(+4) 00]}:
X. Peng, X. Zhu, X. Fang,
M. Feng, K. Gao, X. Yang, \& M. Liu,
``Preparation of pseudo-pure states by line-selective
pulses in nuclear magnetic resonance'',
quant-ph/0012038.

\item {\bf [Peng-Zhu-Fang-(+3) 01]}:
X. Peng, X. Zhu, X. Fang, M. Feng, M. Liu, \& K. Gao,
``Experimental implementation of Hogg's algorithm on a
three-quantum-bit NMR quantum computer'',
quant-ph/0108068.

\item {\bf [Peng-Zhu-Fang-(+3) 02]}:
X. Peng, X. Zhu, X. Fang,
M. Feng, M. Liu, \& K. Gao,
``Experimental implementation of Hogg's algorithm on a three-quantum-bit NMR
quantum computer'',
{\em Phys. Rev. A} {\bf 65}, 4, 042315 (2002).

\item {\bf [Peng-Zhu-Fang-(+4) 02]}:
X. Peng, X. Zhu, X. Fang,
M. Feng, X. Yang, M. Liu, \& K. Gao,
``Exhibition of the periodicity of quantum Fourier transformation
in nuclear magnetic resonance'',
quant-ph/0202010.

\item {\bf [Peng-Zhu-Fang-(+3) 03 a]}:
X. Peng, X. Zhu, X. Fang,
M. Feng, M. Liu, \& K. Gao,
``Experimental implementation of remote state
preparation by nuclear magnetic resonance'',
{\em Phys. Lett. A} {\bf 306}, 5-6, 271-276 (2003);
quant-ph/0202004.

\item {\bf [Peng-Zhu-Fang-(+3) 03 b]}:
X. Peng, X. Zhu, X. Fang,
M. Feng, M. Liu, \& K. Gao,
``An interferometric complementarity experiment
in a bulk nuclear magnetic resonance ensemble'',
{\em J. Phys. A} {\bf 36}, 10, 2555-2563 (2003);
quant-ph/0201146.

\item {\bf [Peng-Zhu-Fang-(+3) 04]}:
X. Peng, X. Zhu, X. Fang,
M. Feng, M. Liu, \& K. Gao,
`\,``Spectral Implementation'' for creating a labeled pseudo-pure state and the Bernstein–Vazirani algorithm
in a four-qubit nuclear magnetic resonance quantum processor',
{\em J. Chem. Phys.} {\bf 120}, 3579-3585 (2004);
quant-ph/0202008.

\item {\bf [Peng-Du-Suter 04]}:
X. Peng, J. Du, \& D. Suter,
``Quantum phase transition of ground-state entanglement in a Heisenberg
spin chain simulated in an NMR quantum computer'',
{\em Phys. Rev. A};
quant-ph/0411049.

\item {\bf [Pennini-Plastino-Plastino-Casas 02]}:
F. Pennini, A. Plastino, A. R. Plastino, \& M. Casas,
``How fundamental is the character of thermal uncertainty relations?'',
{\em Phys. Lett. A} {\bf 302}, 4, 156-162 (2002).

\item {\bf [Penrose-Rindler 84]}:
R. Penrose, \& W. Rindler,
{\em Spinors and
space-time. Volume 1. Two-spinor calculus and relativistic fields},
Cambridge University Press, Cambridge, 1984.

\item {\bf [Penrose-Rindler 86]}:
R. Penrose, \& W. Rindler,
{\em Spinors and
space-time. Volume 2. Spinor and twistor methods in space-time geometry},
Cambridge University Press, Cambridge, 1986.

\item {\bf [Penrose 87 a]}:
R. Penrose,
``Quantum physics and conscious thought'',
in {\bf [Hiley-Peat 87]}, pp.~105-120.

\item {\bf [Penrose 87 b]}:
R. Penrose,
``Newton, quantum theory and reality'',
in S. W. Hawking, \& W. Israel (eds.),
{\em Three hundred years of gravitation (Cambridge, 1987)},
Cambridge University Press, Cambridge, 1987, pp.~17-49.

\item {\bf [Penrose 89]}:
R. Penrose,
{\em The emperor's new mind: Concerning
computers, minds, and the laws of physics},
Oxford University Press, Oxford, 1989.
Spanish version: {\em La nueva mente del emperador},
Mondadori, Madrid, 1991.
Reviews: {\bf [Bennett 90]}, {\bf [Detlefsen 98]}.

\item {\bf [Penrose 93]}:
R. Penrose,
``Quantum non-locality and complex reality'',
in G. F. R. Ellis, A. Lanza, \& J. Miller (eds.),
{\em The renaissance of general relativity
and cosmology: A survey to celebrate the 65th birthday of Dennis Sciama},
Cambridge University Press, New York, 1993, pp.~314-325.

\item {\bf [Penrose 94 a]}:
R. Penrose,
``Non-locality and objectivity in quantum state reduction'',
in J. Anandan, \& J. L. Safko (eds.),
{\em Fundamental aspects of quantum theory},
World Scientific, Singapore, pp.~?-?.

\item {\bf [Penrose 94 b]}:
R. Penrose,
{\em Shadows of the mind: A search for the
missing science of consciousness},
Oxford University Press, Oxford, 1994.
Spanish version: {\em Las sombras de la mente.
Hacia una comprensi\'{o}n cient\'{\i}fica de la consciencia},
Mondadori, Barcelona, 1996.
Reviews: {\bf [Anderson 94]}, {\bf [Detlefsen 98]}.

\item {\bf [Penrose 95]}:
R. Penrose,
``Non-locality and objectivity in quantum state
reduction'', in J. S. Anandan, \& J. L. Safko (eds.),
{\em Quantum coherence and
reality. In celebration of the 60th birthday of Yakir Aharonov.
Int.\ Conf.\
on Fundamental Aspects of Quantum Theory (?, ?)},
World Scientific, Singapore, 1995, pp.~?-?.

\item {\bf [Penrose 96]}:
R. Penrose,
``On gravity's role in quantum state reduction'',
{\em General Relativity and Gravitation} {\bf 28}, 5, 581-609 (1996).

\item {\bf [Penrose 97 a]}:
R. Penrose,
{\em The large, the small and the human mind},
Cambridge University Press, Cambridge, 1997.
Reviews: {\bf [Anderson 97]}, {\bf [Detlefsen 98]}.

\item {\bf [Penrose 97 b]}:
R. Penrose,
``The mathematics of the electron's spin'',
{\em Eur. J. Phys.} {\bf 18}, 3, 164-168 (1997).

\item {\bf [Penrose 98]}:
R. Penrose,
``Quantum computation, entanglement and state reduction'',
in A. K. Ekert, R. Jozsa, \& R. Penrose (eds.),
{\em Quantum Computation: Theory and Experiment.
Proceedings of a Discussion Meeting held at the Royal
Society of London on 5 and 6 November 1997},
{\em Philos. Trans. R. Soc. Lond. A} {\bf 356}, 1743, 1927-1940
(1998).

\item {\bf [Penrose 00]}:
R. Penrose,
``On Bell non-locality without probabilities:
Some curious geometry'',
in {\bf [Ellis-Amati 00]}, pp.~1-27.
See {\bf [Zimba-Penrose 93]}.

\item {\bf [Penrose 00]}:
R. Penrose,
``John Bell, state reduction, and quanglement'',
in {\bf [Bertlmann-Zeilinger 02]}, pp.~319-332.

\item {\bf [de la Pe\~{n}a Auerbach 70]}:
L. de la Pe\~{n}a-Auerbach,
``Stochastic quantum mechanics for particles with spin'',
{\em Phys. Lett. A} {\bf 31}, ?, 403-404 (1970).

\item {\bf [de la Pe\~{n}a Auerbach-Cetto 71]}:
L. de la Pe\~{n}a-Auerbach, \& A. M. Cetto,
``Self interaction corrections
in a nonrelativistic theory of quantum mechanics'',
{\em Phys. Rev. D} {\bf 3}, 4, 795-800 (1971).

\item {\bf [de la Pe\~{n}a Auerbach-Cetto-Brody 72]}:
L. de la Pe\~{n}a-Auerbach, A. M. Cetto, \& T. A. Brody,
``On hidden-variable theories and Bell's inequality'',
{\em Lettere al Nuovo Cimento} {\bf 5}, 2, 177-181 (1972).

\item {\bf [de la Pe\~{n}a Auerbach-Cetto 75]}:
L. de la Pe\~{n}a-Auerbach, \& A. M. Cetto,
``Stochastic theory for classical and quantum mechanical systems'',
{\em Found. Phys.} {\bf 5}, 2, 355-370 (1975).

\item {\bf [de la Pe\~{n}a Auerbach-Cetto 82]}:
L. de la Pe\~{n}a-Auerbach, \& A. M. Cetto,
``Does quantum mechanics accept a stochastic support?'',
{\em Found. Phys.} {\bf 12}, ?, 1017-1037 (1982).
Reprinted in {\bf [Barut-van der Merwe-Vigier 84]}, pp.~93-113.

\item {\bf [de la Pe\~{n}a Auerbach-Santos 99]}:
L. de la Pe\~{n}a-Auerbach, \& E. Santos,
``Perturbation of the evolution of a quantum system induced by its
environment'',
{\em Phys. Lett. A} {\bf 259}, 2, 83-90 (1999).

\item {\bf [Percival 98]}:
I. C. Percival,
``Quantum transfer functions, weak nonlocality and relativity'',
{\em Phys. Lett. A} {\bf 244}, 6, 495-501 (1998);
quant-ph/980304.

\item {\bf [Percival 99]}:
I. C. Percival,
``Quantum measurement breaks Lorentz symmetry'',
quant-ph/9906005.

\item {\bf [Percival 00 a]}:
I. C. Percival,
``Cosmic quantum measurement'',
{\em Proc. R. Soc. Lond. A} {\bf 456}, 1993, 25-37 (2000);
quant-ph/9811089.

\item {\bf [Percival 00 b]}:
I. C. Percival,
``Speakable and unspeakable after John Bell'',
quant-ph/0012021.

\item {\bf [Percival 01]}:
I. C. Percival,
``Why do Bell experiments?'',
{\em Phys. Lett. A} {\bf 279}, 3-4, 105-109 (2001);
quant-ph/0008097.

\item {\bf [Pereira-Ou-Kimble 00]}:
S. F. Pereira, Z. Y. Ou, \& H. J. Kimble,
``Quantum communication with correlated nonclassical states'',
{\em Phys. Rev. A} {\bf 62}, 4, 042311 (2000);
quant-ph/0003094.

\item {\bf [Peres-Singer 60]}:
A. Peres, \& P. Singer,
``On possible experimental
tests for the paradox of Einstein, Podolsky and Rosen'',
{\em Nuovo Cimento} {\bf 15}, 6, 907-915 (1960).
See {\bf [Bohm-Aharonov 60]}.

\item {\bf [Peres-Rosen 64 a]}:
A. Peres, \& N. Rosen,
``Measurement of a quantum ensemble by a classical apparatus'',
{\em Ann. Phys.} {\bf 29}, ?, 366-? (1964).

\item {\bf [Peres-Rosen 64 b]}:
A. Peres, \& N. Rosen,
``Macroscopic bodies in quantum theory'',
{\em Phys. Rev.} {\bf 165}, 6B, B1486-B1488 (1964).

\item {\bf [Peres 74]}:
A. Peres,
``Quantum measurements are reversible'',
{\em Am. J. Phys.} {\bf 42}, 10, 886-891 (1974).
Comment: {\bf [van Heerden 75]}.
Reply: {\bf [Peres 75]}.

\item {\bf [Peres 75]}:
A. Peres,
``A single system has no state'',
{\em Am. J. Phys.} {\bf 43}, 11, 1015-1016 (1975).
Reply to {\bf [van Heerden 75]}.
See {\bf [Peres 74]}.

\item {\bf [Peres 78 a]}:
A. Peres,
``Unperformed experiments have no results'',
{\em Am. J. Phys.} {\bf 46}, 7, 745-747 (1978).
Reprinted in {\bf [Ballentine 88 b]}, pp.~100-?.

\item {\bf [Peres 78 b]}:
A. Peres,
``Pure states, mixtures, and compounds'',
in {\em Mathematical foundations of quantum theory},
Academic Press, New York, 1978, pp.~357-?.

\item {\bf [Peres 79]}:
A. Peres,
``Proposed test for complex versus quaternion
quantum theory'',
{\em Phys. Rev. Lett.} {\bf 42}, 11, 683-686 (1979).

\item {\bf [Peres 80 a]}:
A. Peres,
``Measurement of time by quantum clocks'',
{\em Am. J. Phys.} {\bf 48}, 7, 552-557 (1980).

\item {\bf [Peres 80 b]}:
A. Peres,
``Zeno paradox in quantum theory'',
{\em Am. J. Phys.} {\bf 48}, 11, 931-932 (1980).

\item {\bf [Peres 80 c]}:
A. Peres,
``The physicist's role in physical laws'',
{\em Found. Phys.} {\bf 10}, 7-8, 631-634 (1980).

\item {\bf [Peres 80 d]}:
A. Peres,
``Can we undo quantum measurements?'',
{\em Phys. Rev. D} {\bf 22}, 4, 879-883 (1980).
Reprinted in {\bf [Wheeler-Zurek 83]}, pp.~692-696.

\item {\bf [Peres 81]}:
A. Peres,
``Relativity, quantum theory, and statistical mechanics are
compatible'',
{\em Phys. Rev. D} {\bf 23}, 6, 1458-1459 (1981).

\item {\bf [Peres-Zurek 82]}:
A. Peres, \& W. H. Zurek,
``Is quantum theory universally valid?'',
{\em Am. J. Phys.} {\bf 50}, 9, 807-810 (1982).

\item {\bf [Peres 84 a]}:
A. Peres,
``What is a state vector?'',
{\em Am. J. Phys.}
{\bf 52}, 7, 644-650 (1984).

\item {\bf [Peres 84 b]}:
A. Peres,
``Ergodicity and mixing in quantum theory. I'',
{\em Phys. Rev. A} {\bf 30}, 1, 504-508 (1984).
See {\bf [Feingold-Moiseyev-Peres 84]} (II).

\item {\bf [Peres 84 c]}:
A. Peres,
``On quantum-mechanical automata'',
{\em Phys. Lett. A} {\bf 101}, 5-6, 249-250 (1984).

\item {\bf [Peres 84 d]}:
A. Peres,
``The classic paradoxes of quantum theory'',
{\em Found. Phys.} {\bf 14}, ?, 1131-? (1984).

\item {\bf [Peres 85 a]}:
A. Peres,
``Einstein, G\"{o}del, Bohr'',
{\em Found. Phys.} {\bf 15}, 2, 201-205 (1985).

\item {\bf [Peres-Wootters 85 b]}:
A. Peres, \& W. K. Wootters,
``Quantum measurements of finite duration'',
{\em Phys. Rev. D} {\bf 32}, 8, 1968-1974 (1985).

\item {\bf [Peres 85 b]}:
A. Peres,
``Reversible logic and quantum computers'',
{\em Phys. Rev. A} {\bf 32}, 6, 3266-3276 (1985).

\item {\bf [Peres 86 a]}:
A. Peres,
``When is a quantum measurement?'',
{\em Am. J. Phys.} {\bf 54}, 8, 688-692 (1986).
Comment: {\bf [Bussey 88]}.
Almost the same as {\bf [Peres 86 b]}.

\item {\bf [Peres 86 b]}:
A. Peres,
``When is a quantum measurement?'',
in D. M. Greenberger (ed.),
{\em New techniques and ideas in quantum measurement theory.
Proc.\ of an international conference (New York, 1986),
Ann. N. Y. Acad. Sci.} {\bf 480}, 438-448 (1986).
Almost the same as {\bf [Peres 86 a]}.

\item {\bf [Peres 86 c]}:
A. Peres,
``Existence of `free will' as a problem of physics'',
{\em Found. Phys.} {\bf 16}, 6, 573-584 (1986).
Reprinted in {\bf [Zurek-van der Merwe-Miller 88]}, pp.~592-602.

\item {\bf [Peres 86 d]}:
A. Peres,
``Semiclassical properties of Wigner functions'',
{\em Physica Scripta} {\bf 34}, 736-? (1986).

\item {\bf [Peres 88 a]}:
A. Peres,
``Schr\"{o}dinger's immortal cat'',
{\em Found. Phys.} {\bf 18}, ?, 57-76 (1988).

\item {\bf [Peres 88 b]}:
A. Peres,
``How to differentiate between non-orthogonal states'',
{\em Phys. Lett. A} {\bf 128}, 1-2, 19-24 (1988).
See {\bf [Ivanovic 87]}, {\bf [Dieks 88]}.

\item {\bf [Peres 88 c]}:
A. Peres,
``Quantum limitations on measurement of magnetic flux'',
{\em Phys. Rev. Lett.} {\bf 61}, 18, 2019-2021 (1988).

\item {\bf [Peres-Ron 88]}:
A. Peres, \& A. Ron,
``Cryptodeterminism and quantum theory'',
in A. van der Merwe, F. Selleri, \& G. Tarozzi (eds.),
{\em Microphysical reality and quantum formalism.
Proc.\ of an international conference (Urbino, Italy, 1985)},
Kluwer Academic, Dordrecht, Holland, 1988, vol. 2, pp.~115-123.

\item {\bf [Peres 89 a]}:
A. Peres,
``Quantum measurement with postselection'',
{\em Phys. Rev. Lett.} {\bf 62}, 19, 2326 (1989).
Comment on {\bf [Aharonov-Albert-Vaidman 88]}.
Reply: {\bf [Aharonov-Vaidman 89]}.
See {\bf [Leggett 89]},
{\bf [Duck-Stevenson-Sudarshan 89]}.

\item {\bf [Peres 89 b]}:
A. Peres,
``Do electrons exist?'',
{\em Phys. Essays} {\bf 2}, ?, 288-? (1989).

\item {\bf [Peres 89 c]}:
A. Peres,
``Quantum limited detectors for weak classical signals'',
{\em Phys. Rev. D} {\bf 39}, 10, 2943-2950 (1989).

\item {\bf [Peres 89 d]}:
A. Peres,
``Nonlinear variants of Schr\"{o}dinger's equation violate the second law
of thermodynamics'',
{\em Phys. Rev. Lett.} {\bf 63}, 10, 1114 (1989).
Comment on {\bf [Weinberg 89 b]}.
Reply: {\bf [Weinberg 89 c]}.

\item {\bf [Peres 89 c]}:
A. Peres,
``The logic of quantum nonseparability'',
in M. Kafatos (ed.),
{\em Bell's theorem, quantum theory, and conceptions of the universe.
Proc.\ of a workshop (George Mason University, 1988)},
Kluwer Academic, Dordrecht, Holland, 1989, pp.~51-60.
See {\bf [Peres 93 a]} (Sec. 7. 4).

\item {\bf [Peres 90 a]}:
A. Peres,
``Neumark's theorem and quantum inseparability'',
{\em Found. Phys.} {\bf 20}, 12, 1441-1453 (1990).
See {\bf [Neumark 54]}.

\item {\bf [Peres-Ron 90]}:
A. Peres, \& A. Ron,
`Incomplete ``collapse'' and partial quantum Zeno effect',
{\em Phys. Rev. A} {\bf 42}, 9, 5720-5722 (1990).

\item {\bf [Peres 90 b]}:
A. Peres,
``Incompatible results of quantum measurements'',
{\em Phys. Lett. A} {\bf 151}, 3-4, 107-108 (1990).
See {\bf [De Baere 96 a]}.

\item {\bf [Peres 90 c]}:
A. Peres,
``The grammar and syntax of quantum theory'',
in F. Cooperstock, L. P. Horwitz, \& J. Rosen (eds.),
{\em Developments in general relativity, astrophysics and quantum
theory},
{\em Ann. Phys. Soc. Israel} {\bf 9}, 255-267 (1990).

\item {\bf [Peres 90 d]}:
A. Peres,
``Consecutive quantum measurements'',
in M. Cini, \& J. M. L\'{e}vy-Leblond (eds.),
{\em Quantum theory without reduction},
Adam Hilger, Bristol, 1990, pp.~122-139.

\item {\bf [Peres 90 e]}:
A. Peres,
``Thermodynamic constraints on quantum axioms'',
in {\bf [Zurek 90]}, pp.~345-355.

\item {\bf [Peres-Wootters 91]}:
A. Peres, \& W. K. Wootters,
``Optimal detection of quantum information'',
{\em Phys. Rev. Lett.} {\bf 66}, 9, 1119-1122 (1991).
See {\bf [Ban-Yamazaki-Hirota 97]}.

\item {\bf [Peres 91 a]}:
A. Peres,
``Two simple proofs of the Kochen-Specker theorem'',
{\em J. Phys. A} {\bf 24}, 4, L175-L178 (1991).
See {\bf [Peres 93 a]} (Sec. 7. 3), {\bf [Kernaghan 94]},
{\bf [Cabello-Estebaranz-Garc\'{\i}a Alcaine 96 a]}.

\item {\bf [Peres 91 b]}:
A. Peres,
``Axiomatic quantum phenomenology'',
in P. J. Lahti, \& P. Mittelstaedt (eds.),
{\em Symp.\ on the Foundations of Modern Physics 1990.
Quantum Theory of Measurement and Related Philosophical Problems
(Joensuu, Finland, 1990)},
World Scientific, Singapore, 1991, pp.~317-331.
See {\bf [Peres 93 a]} (Chap.~2).

\item {\bf [Peres 92 a]}:
A. Peres,
``Recursive definition for elements of reality'',
{\em Found. Phys.} {\bf 22}, 3-4, 357-361 (1992).

\item {\bf [Peres 92 b]}:
A. Peres,
``Emergence of local realism in fuzzy observations
of correlated quantum systems'',
{\em Found. Phys.} {\bf 22}, 6, 819-828 (1992).

\item {\bf [Peres 92 c]}:
A. Peres,
``An experimental test for Gleason's theorem'',
{\em Phys. Lett. A} {\bf 163}, 4, 243-245 (1992).

\item {\bf [Peres 92 d]}:
A. Peres,
``Finite violation of a Bell inequality for arbitrarily large spin'',
{\em Phys. Rev. A} {\bf 46}, 7, 4413-4414 (1992).

\item {\bf [Peres 92 e]}:
A. Peres,
``Looking at the quantum world with classical eyes'',
in P. Cvitanovi\'{c}, I. Percival, \& A. Wirzba (eds.),
{\em Quantum chaos---Quantum measurements},
Kluwer Academic, Dordrecht, Holland, 1992, pp.~249-256.

\item {\bf [Peres 93 a]}:
A. Peres,
{\em Quantum theory: Concepts and methods},
Kluwer Academic, Dordrecht, Holland, 1993, 1995
(reprinted with corrections).
Reviews: {\bf [Sudbery 94]}, {\bf [Knight 94 b]},
{\bf [Caves 94]}, {\bf [Mayer 94]}, {\bf [Clifton 95 a]},
{\bf [Ballentine 95 b]}, {\bf [Mermin 97 a]}.

\item {\bf [Peres 93 b]}:
A. Peres,
``Quantum paradoxes and objectivity: An analysis of some paradoxes'',
in K. V. Laurikainen, \& C. Montonen (eds.),
{\em Proc.\ Symp.\ Foundations of Modern Physics 1992:
The Copenhagen Interpretation and Wolfgang Pauli (Helsinki, 1992)},
World Scientific, Singapore, 1993, pp.~57-78.

\item {\bf [Peres 93 c]}:
A. Peres,
``Storage and retrieval of quantum information'',
in {\em Workshop on Physics and Computation, PhysComp 92},
IEEE Computer Science Society Press,
Los Alamitos, California, 1993, pp.~155-158.

\item {\bf [Peres 94 a]}:
A. Peres,
``Time asymmetry in quantum mechanics: A retrodiction paradox'',
{\em Phys. Lett. A} {\bf 194}, 1-2, 21-25 (1994).
Comments: {\bf [Aharonov-Vaidman 95]}.
Reply: {\bf [Peres 95 d]}.

\item {\bf [Peres 94 b]}:
A. Peres,
``Classification of quantum paradoxes: Nonlocality vs.
contextuality'',
in L. Accardi (ed.),
{\em The interpretation of quantum theory: Where do we stand?},
Enciclopedia Italiana, 1994, pp.~117-135.

\item {\bf [Peres 95 a]}:
A. Peres,
``Relativistic quantum measurements'',
in D. M. Greenberger, \& A. Zeilinger (eds.),
{\em Fundamental problems in quantum theory:
A conference held in honor of professor John A. Wheeler,
Ann. N. Y. Acad. Sci.} {\bf 755}, 445-450 (1995).

\item {\bf [Peres 95 b]}:
A. Peres,
``Nonlocal effects in Fock space'',
{\em Phys. Rev. Lett.} {\bf 74}, 2, 4571 (1995).
Erratum: {\em Phys. Rev. Lett.} {\bf 76}, 12, 2205 (1996).
quant-ph/9501019.
See {\bf [Hardy 94]}.

\item {\bf [Peres 95 c]}:
A. Peres,
``Higher order Schmidt decompositions'',
{\em Phys. Lett. A} {\bf 202}, 1, 16-17 (1995);
quant-ph/9504006.
See {\bf [Elby-Bub 94]}.

\item {\bf [Peres 95 d]}:
A. Peres,
``Reply to the comment of Y. Aharonov and L. Vaidman on
`Time asymmetry in quantum mechanics: A retrodiction paradox'\,'',
{\em Phys. Lett. A} {\bf 203}, 2-3, 150-151 (1995);
quant-ph/9501005.
See {\bf [Peres 94 a]}, {\bf [Aharonov-Vaidman 95]}.

\item {\bf [Peres 96 a]}:
A. Peres,
``Nathan Rosen 1909-95'',
{\em Phys. World} {\bf 9}, 2, 49 (1996).
See {\bf [Peres 96 b]}, {\bf [Bergmann-Merzbacher-Peres 96]}.

\item {\bf [Peres 96 b]}:
A. Peres,
``Obituary: Nathan Rosen'',
{\em Found. Phys.} {\bf 26}, ?, ? (1996).

\item {\bf [Peres 96 c]}:
A. Peres,
``Generalized Kochen-Specker theorem'',
{\em Found. Phys.} {\bf 26}, 6, 807-812 (1996);
quant-ph/9510018.

\item {\bf [Peres 96 d]}:
A. Peres,
``Separability criterion for density matrices'',
{\em Phys. Rev. Lett.} {\bf 77}, 8, 1413-1415 (1996);
quant-ph/9604005.
See {\bf [Horodecki-Horodecki-Horodecki 96 c]},
{\bf [Wang 00 b]}.

\item {\bf [Peres 96 e]}:
A. Peres,
``Collective tests for quantum nonlocality'',
{\em Phys. Rev. A} {\bf 54}, 4, 2685-2689 (1996);
quant-ph/9603023.

\item {\bf [Peres 96 f]}:
A. Peres,
``Quantum cryptography with orthogonal states?'',
{\em Phys. Rev. Lett.} {\bf 77}, 15, 3264 (1996);
quant-ph/9509003.
Comment on {\bf [Goldenberg-Vaidman 95 a]}.
Reply: {\bf [Goldenberg-Vaidman 96]}.

\item {\bf [Peres 96 g]}:
A. Peres,
``Quaternionic quantum interferometry'',
in F. de Martini, G. Denardo, \& Y. H. Shih (eds.),
{\em Quantum interferometry},
VCH Publishers, New York, 1996, pp.~431-437;
quant-ph/9605024.

\item {\bf [Peres 96 h]}:
A. Peres,
``Error correction and symmetrization in quantum computers'',
in T. Toffoli, M. Biafore, \& J. Le\"{a}lo (eds.),
{\em PhysComp 96: Proc.\ 4th Workshop on Physics and Computation},
New England Complex Systems Institute,
Cambridge, Massachusetts, 1996, pp.~275-277.
quant-ph/9611046.

\item {\bf [Peres 96 i]}:
A. Peres,
``From E.P.R. to G.H.Z.: Conference highlights'',
in A. Mann, \& M. Revzen (eds.),
{\em The dilemma of Einstein, Podolsky and Rosen -- 60 years
later. An international symposium in honour of Nathan Rosen
(Haifa, Israel, 1995)},
{\em Ann. Phys. Soc. Israel} {\bf 12}, 305-314 (1996).

\item {\bf [Peres 96 j]}:
A. Peres,
``Quantum inseparability and free will'',
in U. Ketvel (ed.),
{\em Vastakohtien todellisuus},
University of Helsinki Press, Helsinki, 1996, pp.~117-121.

\item {\bf [Peres 97 a]}:
A. Peres,
``Quantum nonlocality and inseparability'',
in M. Ferrero, \& A. van der Merwe (eds.),
{\em New developments on fundamental problems
in quantum physics (Oviedo, Spain, 1996)},
Kluwer Academic, Dordrecht, Holland, 1997, pp.~301-310;
quant-ph/9609016.

\item {\bf [Peres 97 b]}:
A. Peres,
``Bell inequalities with postselection'',
in {\bf [Cohen-Horne-Stachel 97 b]}, pp.~191-196;
quant-ph/9512003.

\item {\bf [Peres 97 c]}:
A. Peres,
``Unitary dynamics for quantum codewords'',
in O. Hirota, A. S. Holevo (Kholevo), \& C. M. Caves (eds.),
{\em Proc.\ of symposium on quantum communication and measurement},
Plenum Press, New York, 1997, pp.~171-179;
quant-ph/9609015.

\item {\bf [Peres 97 d]}:
A. Peres,
``The ambivalent quantum observer'',
in D. H. Feng, \& B. L. Hu (eds.),
{\em Quantum classical correspondence},
Int. Press, Cambridge, Massachusetts, 1997, pp.~57-68.

\item {\bf [Peres 98 a]}:
A. Peres,
``Quantum entanglement: Criteria and collective tests'',
in E. B. Karlsson, \& E. Br\"{a}ndas (eds.),
{\em Proc.\ of the 104th Nobel Symp.\ ``Modern Studies of Basic Quantum Concepts and Phenomena'' (Gimo, Sweden, 1997)},
{\em Physica Scripta} {\bf T76}, 115-121 (1998);
quant-ph/9707026.

\item {\bf [Peres 98 b]}:
A. Peres,
``Quantum disentanglement and computation'',
{\em Superlatt. Microstruct.} {\bf 23}, 373-? (1998);
quant-ph/9707047.

\item {\bf [Peres 98 c]}:
A. Peres,
``Interpreting the quantum world'',
{\em Stud. Hist. Philos. Sci. Part B: Stud. Hist. Philos. Mod. Phys.}
{\bf 29}, 4, 611-620 (1998);
quant-ph/9711003.
Review of {\bf [Bub 97]}.
See {\bf [Bub 00]}.

\item {\bf [Peres 98 d]}:
A. Peres,
``Comparing the strengths of various Bell inequalities'',
quant-ph/9802022 (withdraw).

\item {\bf [Peres 98 e]}:
A. Peres,
``Book review. The quantum challenge: Modern
research on the foundations of quantum mechanics'',
{\em Am. J. Phys.} {\bf 66}, 5, 455 (1998).
Review of {\bf [Greenstein-Zajonc 98]}.

\item {\bf [Peres-Terno 98 a]}:
A. Peres, \& D. R. Terno,
``Optimal distinction between non-orthogonal quantum states'',
{\em J. Phys. A} {\bf 31}, 34, 7105-7112 (1998);
quant-ph/9804031.

\item {\bf [Peres-Terno 98 b]}:
A. Peres, \& D. R. Terno,
``Convex probability domain of generalized quantum measurements'',
{\em J. Phys. A} {\bf 31}, 38, L671-L675 (1998);
quant-ph/9806024.

\item {\bf [Peres 99 a]}:
A. Peres,
``All the Bell inequalities'',
{\em Found. Phys.} {\bf 29}, 4, 589-614 (1999);
quant-ph/9807017.

\item {\bf [Peres 99 b]}:
A. Peres,
``Error symmetrization in quantum computers'',
{\em Int. J. Theor. Phys.} {\bf 38}, 3, 799-806 (1999);
quant-ph/9605009.

\item {\bf [Peres 00 a]}:
A. Peres,
``Delayed choice for entanglement swapping'',
in V. Bu\v{z}zek, \& D. P. DiVincenzo (eds.),
{\em J. Mod. Opt.} {\bf 47}, 2-3 (Special issue:
Physics of quantum information), 139-143 (2000);
quant-ph/9904042.

\item {\bf [Peres 00 b]}:
A. Peres,
``Classical interventions in quantum systems.
I. The measuring process'',
{\em Phys. Rev. A} {\bf 61}, 2, 022116 (2000);
quant-ph/9906023.
See {\bf [Peres 00 c]} (II).

\item {\bf [Peres 00 c]}:
A. Peres,
``Classical interventions in quantum systems.
II. Relativistic invariance'',
{\em Phys. Rev. A} {\bf 61}, 2, 022117 (2000);
quant-ph/9906034.
See {\bf [Peres 00 b]} (I).
Comment: {\bf [Nikolic 01]}.
Reply: {\bf [Peres 01 b]}.

\item {\bf [Peres 00 d]}:
A. Peres,
``Bayesian analysis of Bell inequalities'',
{\em Fortschr. Phys.} {\bf 48}, 5-7, 531-535 (2000);
quant-ph/9905084.

\item {\bf [Peres 00 e]}:
A. Peres,
``Impossible things usually don't happen'',
{\em Phys. World} {\bf 13}, 5, 47-? (2000).

\item {\bf [Peres 00 f]}:
A. Peres,
``Opposite momenta lead to opposite directions'',
{\em Am. J. Phys.} {\bf 68}, 11, 991-992 (2000);
quant-ph/9910123.
See {\bf [Struyve-De Baere-De Neve-De Weirdt 04]}.

\item {\bf [Peres-Terno 01 a]}:
A. Peres, \& D. R. Terno,
``Hybrid classical-quantum dynamics'',
{\em Phys. Rev. A} {\bf 63}, 2, 022101 (2001);
quant-ph/0008068.

\item {\bf [Peres 01 a]}:
A. Peres,
``Karl Popper and the Copenhagen interpretation'',
{\em Stud. Hist. Philos. Sci. Part B:
Stud. Hist. Philos. Mod. Phys.} {\bf 32} (2001);
quant-ph/9910078.
See {\bf [Popper 56]}.

\item {\bf [Peres-Scudo 01]}:
A. Peres, \& P. F. Scudo,
``Entangled quantum states as direction indicators'',
{\em Phys. Rev. Lett.} {\bf 86}, 18, 4160-4162 (2001);
quant-ph/0010085.

\item {\bf [Peres-Scudo 01 b]}:
A. Peres, \& P. F. Scudo,
``Transmission of a Cartesian frame by a quantum system'',
{\em Phys. Rev. Lett.} {\bf 87}, 16, 167901 (2001);
quant-ph/0103149.

\item {\bf [Peres 01 b]}:
A. Peres,
`Reply to ``Comment on `Classical interventions in quantum systems.
II. Relativistic Invariance'\,''\,',
{\em Phys. Rev. A} {\bf 64}, 6, 066102 (2001).
Reply to {\bf [Nikolic 01]}.
See {\bf [Peres 00 c]}.

\item {\bf [Peres-Terno 02 a]}:
A. Peres, \& D. R. Terno,
``Lorentz transformations of open systems'',
{\em Proc.\ ESF QIT Conf.\ Quantum Information: Theory, Experiment and Perspectives
(Gdansk, Poland, 2001)}, {\em J. Mod. Opt.} {\bf 49}, 8, 1255-1261 (2002);
quant-ph/0106079.

\item {\bf [Peres-Scudo 02 a]}:
A. Peres, \& P. F. Scudo,
``Covariant quantum measurements may not be optimal'',
{\em Proc.\ ESF QIT Conf.\ Quantum Information: Theory, Experiment and Perspectives
(Gdansk, Poland, 2001)}, {\em J. Mod. Opt.} {\bf 49}, 8, ?-? (2002);
quant-ph/0107114.

\item {\bf [Peres-Scudo 02 b]}:
A. Peres, \& P. F. Scudo,
``Unspeakable quantum information'',
in A. Khrennikov (ed.),
{\em Quantum Theory: Reconsideration of Foundations (V\"{a}xj\"{o}, Sweden, 2001)},
V\"{a}xj\"{o} University Press, V\"{a}xj\"{o}, Sweden, 2002;
quant-ph/0201017.

\item {\bf [Peres-Scudo-Terno 02]}:
A. Peres, P. F. Scudo, \& D. R. Terno,
``Quantum entropy and special relativity'',
{\em Phys. Rev. Lett.} {\bf 88}, 23, 230402 (2002);
quant-ph/0203033.
Comment: {\bf [Czachor 03 b]}.

\item {\bf [Peres 02]}:
A. Peres,
``How the no-cloning theorem got its name'',
{\em Proc.\ of Quantum Interferometry IV};
quant-ph/0205076.

\item {\bf [Peres-Terno 03]}:
A. Peres, \& D. R. Terno,
``Relativistic Doppler effect in quantum communication'',
in M. Ferrero (ed.),
{\em Proc. of Quantum Information: Conceptual Foundations,
Developments and Perspectives (Oviedo, Spain, 2002)},
{\em J. Mod. Opt.} {\bf 50}, 6-7, 1165-1173 (2003);
quant-ph/0208128.

\item {\bf [Peres 03 a]}:
A. Peres,
``What's wrong with these observables?'',
{\em Found. Phys.} {\bf 33}, 10, 1543-1547 (2003);
quant-ph/0207020.

\item {\bf [Peres 03 b]}:
A. Peres,
``What is actually teleported?'',
{\em IBM J. Res. Dev.};
quant-ph/0304158.

\item {\bf [Peres 03 c]}:
A. Peres,
``Einstein, Podolsky, Rosen, and Shannon'',
quant-ph/0310010.

\item {\bf [Peres 03 d]}:
A. Peres,
``Finite precision measurement nullifies Euclid's postulates'',
quant-ph/0310035.
Comment on {\bf [Meyer 99 b]}.

\item {\bf [Peres-Terno 03]}:
A. Peres, \& D. R. Terno,
``Quantum information and special relativity'',
{\em Int. J. Quant. Inf.} {\bf 1}, ?, 225-? (2003);
quant-ph/0301065

\item {\bf [Peres-Terno 04]}:
A. Peres, \& D. R. Terno,
``Quantum information and relativity theory'',
{\em Rev. Mod. Phys.} {\bf 76}, 1, 93-123 (2004);
quant-ph/0212023.

\item {\bf [Peres 04 a]}:
A. Peres,
``I am the cat who walks by himself'',
physics/0404085.

\item {\bf [Peres 04 b]}:
A. Peres,
``Quantum information and general relativity'',
in
{\em Proc.\ of Quantum Optics for Quantum Information Processing (Rome, 2004)}
{\em Fortschr. Phys.};
quant-ph/0405127.

\item {\bf [P\'{e}rez Garc\'{\i}a 04]}:
D. P\'{e}rez Garc\'{\i}a,
``Deciding separability with a fixed error'',
{\em Phys. Lett. A} {\bf 330}, 3-4, 149-154 (2004);
quant-ph/0407247.

\item {\bf [P\'{e}rez-Curty-Santos-Garc\'{\i}a Fern\'{a}ndez 03]}:
E. P\'{e}rez, M. Curty, D. J. Santos, \& P. Garc\'{\i}a-Fern\'{a}ndez,
``Quantum authentication with unitary coding sets'',
in M. Ferrero (ed.),
{\em Proc. of Quantum Information: Conceptual Foundations,
Developments and Perspectives (Oviedo, Spain, 2002)},
{\em J. Mod. Opt.} {\bf 50}, 6-7, 1035-1047 (2003).

\item {\bf [P\'{e}rez Su\'{a}rez-Santos 04]}:
M. P\'{e}rez-Su\'{a}rez, \& D. J. Santos,
{\em Procesado de informaci\'{o}n con sistemas cu\'{a}nticos},
Universidad de Vigo, Vigo, Spain, 2004.

\item {\bf [P\'{e}rez Garc\'{\i}a-Gonzalo-P\'{e}rez D\'{\i}az 92]}:
V. M. P\'{e}rez-Garc\'{\i}a, I. Gonzalo, \& J. L. P\'{e}rez-D\'{\i}az,
``Theory of the stability of the quantum chiral state'',
{\em Phys. Lett. A} {\bf 167}, 4, 377-382 (1992).

\item {\bf [Pe\v{r}ina-Haderka-Soubusta 01]}:
J. Pe\v{r}ina, Jr., O. Haderka, \& J. Soubusta,
``Quantum cryptography using a photon source based on
postselection from entangled two-photon states'',
{\em Phys. Rev. A} {\bf 64}, 5, 052305 (2001);
quant-ph/0107086.

\item {\bf [Perrie-Duncan-Beyer-Kleinpoppen 85]}:
W. Perrie, A. J. Duncan, H. J. Beyer, \& H. Kleinpoppen,
``Polarization correlation of the two photons emitted by
metastable atomic deuterium: A test of Bell's inequality'',
{\em Phys. Rev. Lett.} {\bf 54}, 16, 1790-1793 (1985).
Erratum: {\em Phys. Rev. Lett.} {\bf 54}, 24, 2647 (1985).

\item {\bf [Peruzzi-Rimini 98]}:
G. Peruzzi, \& A. Rimini,
``Incompatible and contradictory retrodictions in the history
approach to quantum mechanics'',
{\em Found. Phys. Lett.} {\bf 11}, 2, 201-207 (1998).
See {\bf [Kent 00 b]}.

\item {\bf [Peruzzi-Rimini 00]}:
G. Peruzzi, \& A. Rimini,
``Compoundation invariance and Bohmian mechanics'',
{\em Found. Phys.} {\bf 30}, 9, 1445-1472 (2000).

\item {\bf [Peters-Altepeter-Branning-(+3) 04]}:
N. A. Peters, J. B. Altepeter, D. A. Branning,
E. R. Jeffrey, T.-C. Wei, \& P. G. Kwiat,
``Maximally entangled mixed states: Creation and concentration'',
{\em Phys. Rev. Lett.} {\bf 92}, 13, 133601 (2004);
quant-ph/0308003.

\item {\bf [Peters-Wei-Kwiat 04]}:
N. A. Peters, T.-C. Wei, \& P. G. Kwiat,
``Mixed state sensitivity of several quantum information benchmarks'',
quant-ph/0407172.

\item {\bf [Petersen 63]}:
A. Petersen,
``The philosophy of Niels Bohr'',
{\em Bulletin of the Atomic Scientists}, September 1963, pp.~8-14.

\item {\bf [Petersen-Madsen-M{\o}lmer 04]}:
V. Petersen, L. B. Madsen, \& K. M{\o}lmer,
``Magnetometry with entangled atomic samples'',
quant-ph/0409202.

\item {\bf [Petroni-Morato 00]}:
N. C. Petroni, \& L. M. Morato,
``Entangled states in stochastic mechanics'',
{\em J. Phys. A} {\bf 33}, 33, 5833-5848 (2000).

\item {\bf [Petrosky-Tasaki-Prigogine 90]}:
T. Petrosky, S. Tasaki, \& I. Prigogine,
``Quantum Zeno effect'',
{\em Phys. Lett. A} {\bf 151}, 3-4, 109-113 (1990).

\item {\bf [Petrosky-Ordonez-Prigogine 00]}:
T. Petrosky, G. Ordonez, \& I. Prigogine,
``Quantum transitions and nonlocality'',
{\em Phys. Rev. A} {\bf 62}, 4, 042106 (2000).

\item {\bf [Petrosyan 02]}:
D. Petrosyan, \& G. Kurizki,
``Scalable solid-state quantum processor using subradiant two-atom states'',
{\em Phys. Rev. Lett.} {\bf 89}, 20, 207902 (2002).

\item {\bf [Petrosyan-Kurizki-Shapiro 03]}:
D. Petrosyan, G. Kurizki, \& M. Shapiro,
``Entanglement transfer from dissociated molecules to photons'',
{\em Phys. Rev. A} {\bf 67}, 1, 012318 (2003);
quant-ph/0111093.

\item {\bf [Pfau-Sp\"{a}lter-Kurtsiefer-(+2) 94]}:
T. Pfau, S. Sp\"{a}lter, C. Kurtsiefer, C. Ekstrom \& J. Mlynek,
``Loss of spatial coherence by a single spontaneous emission'',
{\em Phys. Rev. Lett.} {\bf 73}, 9, 1223-1226 (1994).

\item {\bf [Pfau-Kurtsiefer-Mlynek 96]}:
T. Pfau, C. Kurtsiefer, \& J. Mlynek,
``Double-slit experiments with correlted atom-photon states'',
{\em Quantum Semiclass. Opt.} {\bf 8}, 3, 665-671 (1996).

\item {\bf [Philipp-Hess 02]}:
W. Philipp, \& K. Hess,
``A local mathematical model for EPR-experiments'',
quant-ph/0212085.

\item {\bf [Philippidis-Dewdney-Hiley 79]}:
C. Philippidis,
C. Dewdney, \& B. J. Hiley,
``Quantum interference and the quantum potential'',
{\em Nuovo Cimento B} {\bf 52}, 1, 15-28 (1979).

\item {\bf [Phillips 98]}:
W. D. Phillips,
``Laser cooling and trapping of neutral atoms'',
{\em Rev. Mod. Phys.} {\bf 70}, 3, 721-741 (1998).

\item {\bf [Phipps 00]}:
T. E. Phipps, Jr.,
``Book review. The genius of science: A portrait gallery
of 20th century physicists'',
{\em Found. Phys.} {\bf 30}, 8, 1321-1323 (2000).
Review of {\bf [Pais 00]}.

\item {\bf [Physics World 99]}:
Pysics World,
``First for single photons'',
{\em Phys. World} {\bf 12}, 8, 5 (1999).
See {\bf [Nogues-Rauschenbeutel-Osnaghi-(+3) 99]}.

\item {\bf [Phoenix 93]}:
S. J. D. Phoenix,
``Quantum cryptography without conjugate coding'',
{\em Phys. Rev. A} {\bf 48}, 1, 96-102 (1993).

\item {\bf [Phoenix-Barnett 93]}:
S. J. D. Phoenix, \& S. M. Barnett,
``Non-local interatomic correlations in the micromaser'',
{\em J. Mod. Opt.} {\bf 40}, 6, 979-983 (1993).

\item {\bf [Phoenix-Barnett-Townsend-Blow 95]}:
S. J. D. Phoenix, S. M. Barnett, P. D. Townsend, \& K. J. Blow,
``Multi-user quantum cryptography on optical networks'',
{\em J. Mod. Opt.} {\bf 42}, 6, 1155-1163 (1995).
See {\bf [Townsend-Smith 93]}, {\bf [Townsend-Blow 93]},
{\bf [Townsend-Phoenix-Blow-Barnett 94]},
{\bf [Barnett-Phoenix 94]}.

\item {\bf [Phoenix-Townsend 95]}:
S. J. D. Phoenix, \& P. D. Townsend,
``Quantum cryptography:
How to beat the code breakers using quantum mechanics'',
{\em Contemp. Phys.} {\bf 36}, 3, 165-195 (1995).

\item {\bf [Phoenix-Barnett 98]}:
S. J. D. Phoenix, \& S. M. Barnett,
``Method for key distribution using quantum cryptography'',
patent US5764765, 1998.

\item {\bf [Phoenix-Barnett-Chefles 00]}:
S. J. D. Phoenix, S. M. Barnett, \& A. Chefles,
``Three-state quantum cryptography'',
in V. Bu\v{z}zek, \& D. P. DiVincenzo (eds.),
{\em J. Mod. Opt.} {\bf 47}, 2-3 (Special issue:
Physics of quantum information), 507-516 (2000).

\item {\bf [Piccioni-Mehlhop-Wright 93]}:
O. Piccioni, W. Mehlhop, \& B. Wright,
``The EPR without Bell Inequalities'',
in A. van der Merwe, \& F. Selleri (eds.),
{\em Bell's theorem and the foundations of modern physics.
Proc.\ of an international
conference (Cesena, Italy, 1991)},
World Scientific, Singapore, 1993, pp.~375-387.

\item {\bf [Piechocinska 00]}:
B. Piechocinska,
``Information erasure'',
{\em Phys. Rev. A} {\bf 61}, 6, 062314 (2000).

\item {\bf [Piermarocchi-Chen-Sham-Steel 02]}:
C. Piermarocchi, P. Chen, L. J. Sham, \& D. G. Steel,
``Optical RKKY interaction between charged semiconductor quantum dots'',
{\em Phys. Rev. Lett.} {\bf 89}, 16, 167402 (2002).

\item {\bf [Pines 93]}:
D. Pines,
``David Bohm 1917-92'',
{\em Phys. World} {\bf 6}, 3, 67 (1993).

\item {\bf [Pinkse-Fischer-Maunz-Rempe 00]}:
P. W. H. Pinkse, T. Fischer, P. Maunz, \& G. Rempe,
``Trapping an atom with single photons'',
{\em Nature} {\bf 404}, 6776, 365-368 (2000).
See {\bf [Zoller 00]}.

\item {\bf [Piotrowski-Sladkowski 01]}:
E. W. Piotrowski, \& J. Sladkowski,
``Quantum bargaining games'',
quant-ph/0106140.

\item {\bf [Piotrowski-Sladkowski 02]}:
E. W. Piotrowski, \& J. Sladkowski,
``Trading by quantum rules--Quantum anthropic principle'',
quant-ph/0201045.

\item {\bf [Pipkin 78]}:
F. M. Pipkin,
``Atomic physics tests of the basic concepts in quantum mechanics'',
in D. R. Bates, \& B. Bederson (eds.),
{\em Advances in Atomic and Molecular Physics},
Academic Press, New York, 1978, vol. 14, pp.~281-340.

\item {\bf [Pirandola-Mancini-Vitali-Tombesi 03 a]}:
S. Pirandola, S. Mancini, D. Vitali, \& P. Tombesi,
``Continuous variable entanglement by radiation pressure'',
quant-ph/0302017.

\item {\bf [Pirandola-Mancini-Vitali-Tombesi 03 b]}:
S. Pirandola, S. Mancini, D. Vitali, \& P. Tombesi,
``Continuous-variable entanglement and quantum-state teleportation
between optical and macroscopic vibrational modes through radiation pressure'',
{\em Phys. Rev. A} {\bf 68}, 6, 062317 (2003);
quant-ph/0309078.

\item {\bf [Pirandola-Mancini-Vitali-Tombesi 03 c]}:
S. Pirandola, S. Mancini, D. Vitali, \& P. Tombesi,
``Light reflection upon a movable mirror as a paradigm for continuous
variable teleportation network'',
in R. Bonifacio, B. G. Englert, \& M. G. A. Paris (eds.),
{\em Mysteries, Puzzles and Paradoxes in Quantum
Mechanics (Garda Lake, Italy, 2003)}, {\em J. Mod. Opt.};
quant-ph/0311186.

\item {\bf [Pirandola-Mancini-Vitali-Tombesi 04]}:
S. Pirandola, S. Mancini, D. Vitali, \& P. Tombesi,
``Constructing finite dimensional codes with optical continuous variables'',
quant-ph/0402202.

\item {\bf [Pirandola 04]}:
S. Pirandola,
``A quantum teleportation game'',
quant-ph/0407248.

\item {\bf [Pirandola-Mancini-Vitali 04]}:
S. Pirandola, S. Mancini, \& D. Vitali,
``Conditioning two-party quantum teleportation within three-party quantum
channel'',
quant-ph/0410213.

\item {\bf [Piron 72]}:
C. Piron,
``Survey of general quantum physics'',
{\em Found. Phys.} {\bf 2}, 4, 287-314 (1972).
Reprinted in {\bf [Hooker 75]}, pp.~513-544.

\item {\bf [Piron 76]}:
C. Piron,
{\em Foundations of quantum physics},
Addison-Wesley, Reading, Massachusetts, 1976.

\item {\bf [Pironio 03]}:
S. Pironio,
``Violations of Bell inequalities as lower bounds on the communication cost of nonlocal correlations'',
{\em Phys. Rev. A} {\bf 68}, 6, 062102 (2003);
quant-ph/0304176.

\item {\bf [Pitowsky 82 a]}:
I. Pitowsky,
``Resolution of the Einstein-Podolsky-Rosen and Bell paradoxes'',
{\em Phys. Rev. Lett.} {\bf 48}, 19, 1299-1302 (1982).
Comments: {\bf [Mermin 82]}, {\bf [Macdonald 82]}.
Reply: {\bf [Pitowsky 82 b]}.

\item {\bf [Pitowsky 82 b]}:
I. Pitowsky,
``Pitowsky responds'',
{\em Phys. Rev. Lett.} {\bf 49}, 16, 1216 (1982).
Reply to {\bf [Mermin 82]}, {\bf [Macdonald 82]}.
See {\bf [Pitowsky 82 a]}.

\item {\bf [Pitowsky 82 c]}:
I. Pitowsky,
``Substitution and truth in quantum logic'',
{\em Philos. Sci.} {\bf 49}, 3, 380-401 (1982).

\item {\bf [Pitowsky 83 a]}:
I. Pitowsky,
``Deterministic model of spin and statistics'',
{\em Phys. Rev. D} {\bf 27}, 10, 2316-2326 (1983).

\item {\bf [Pitowsky 83 b]}:
I. Pitowsky,
``The logic of fundamental processes, non measurable sets and quantum mechanics'',
Ph.\ D. thesis, University of Western Ontario, 1983.

\item {\bf [Pitowsky 85]}:
I. Pitowsky,
``Quantum mechanics and value definiteness'',
{\em Philos. Sci.} {\bf 52}, 154-156 (1985).

\item {\bf [Pitowsky 86 a]}:
I. Pitowsky,
``A phase space model of quantum mechanics in which all operator commute'',
in L. M. Roth, \& A. Inomata (eds.),
{\em Fundamental questions in quantum mechanics},
Gordon and Breach, New York, 1986, pp.~241-252.

\item {\bf [Pitowsky 86 b]}:
I. Pitowsky,
``The range of quantum probability'',
{\em. J. Math. Phys.} {\bf 27}, ?, 1556-1565 (1986).

\item {\bf [Pitowsky 89 a]}:
I. Pitowsky,
{\em Quantum probability-quantum logic},
Springer-Verlag, Berlin, 1989.

\item {\bf [Pitowsky 89 b]}:
I. Pitowsky,
``From George Boole to John Bell---The
origins of Bell's inequality'',
in M. Kafatos (ed.),
{\em Bell's theorem, quantum theory,
and conceptions of the universe.
Proc.\ of a workshop (George Mason University, 1988)},
Kluwer Academic, Dordrecht, Holland, 1989, pp.~37-49.

\item {\bf [Pitowsky 91 a]}:
I. Pitowsky,
``Bohm's quantum potential and quantum gravity'',
{\em Found. Phys.} {\bf 21}, ?, 343-352 (1991).

\item {\bf [Pitowsky 91 b]}:
I. Pitowsky,
``The relativity of quantum predictions'',
{\em Phys. Lett. A} {\bf 156}, 3-4, 137-139 (1991).
Comment: {\bf [Herbut 92]}.
See {\bf [Pitowsky 92]}.

\item {\bf [Pitowsky 91 c]}:
I. Pitowsky,
``Correlation polytopes: Their geometry and complexity'',
{\em Math. Program. A} {\bf 50}, 395-414 (1991).

\item {\bf [Pitowsky 92]}:
I. Pitowsky,
``Reply to Herbut's comment on `The
relativity of quantum predictions'\,'',
{\em Phys. Lett. A} {\bf 166}, 3-4, 292 (1992).
Reply to {\bf [Herbut 92]}.
See {\bf [Pitowsky 91 b]}.

\item {\bf [Pitowsky 94]}:
I. Pitowsky,
`George Boole's ``Conditions of Possible Experience''\,',
{\em Brit. J. Philos. Sci.} {\bf 45}, ?, 95-125 (1994).

\item {\bf [Pitowsky 98 a]}:
I. Pitowsky,
``Infinite and finite Gleason's theorems and the logic of indeterminacy'',
{\em J. Math. Phys.} {\bf 39}, 1, 218-228 (1998).

\item {\bf [Pitowsky 98 b]}:
I. Pitowsky,
``Constructive quantum mechanics, Gleason's theorem and the indeterminacy relations'',
{\em Tatra Mountains Mathematical Publications} {\bf 15}, 1-10 (1998).

\item {\bf [Pitowsky 98 c]}:
I. Pitowsky,
``Feasible histories, maximum entropy and a MinMax choice criterion'',
{\em Phys. Lett. A} {\bf 247}, 1-2, 9-13 (1998).

\item {\bf [Pitowsky-Svozil 00]}:
I. Pitowsky, \& K. Svozil,
``Boole-Bell type inequalities for the
Greenberger-Horne-Zeilinger and 3-3 cases'',
quant-ph/0011060.

\item {\bf [Pitowsky-Svozil 01]}:
I. Pitowsky, \& K. Svozil,
``Optimal tests of quantum nonlocality'',
{\em Phys. Rev. A} {\bf 64}, 1, 014102 (2001).

\item {\bf [Pitowsky 02 a]}:
I. Pitowsky,
``Range theorems for quantum probability and entanglement'',
in A. Khrennikov (ed.),
{\em Quantum Theory: Reconsideration of Foundations (V\"{a}xj\"{o}, Sweden, 2001)},
V\"{a}xj\"{o} University Press, V\"{a}xj\"{o}, Sweden, 2002, pp.~299-308;
quant-ph/0112068.

\item {\bf [Pitowsky 02 b]}:
I. Pitowsky,
``Quantum speed-up of computations'',
{\em Philos. Sci.} {\bf 69}, S168-S177 (2002).

\item {\bf [Pitowsky 02 c]}:
I. Pitowsky,
``Most Bell operators do not significantly violate locality'',
quant-ph/0202053.

\item {\bf [Pitowsky 02 d]}:
I. Pitowsky,
``Betting on the outcomes of measurements:
A bayesian theory of quantum probability'' (2002),
quant-ph/0208121,
PITT-PHIL-SCI00000766.

\item {\bf [Pitowsky 04]}:
I. Pitowsky,
``Macroscopic objects in quantum mechanics: A combinatorial approach'',
{\em Phys. Rev. A};
quant-ph/0404051.

\item {\bf [Pittenger 99]}:
A. O. Pittenger,
{\em An introduction to quantum computing algorithms},
Birkhauser, Berlin, 1999.

\item {\bf [Pittenger-Rubin 00 a]}:
A. O. Pittenger, \& M. H. Rubin,
``Separability and Fourier representations of density matrices'',
{\em Phys. Rev. A} {\bf 62}, 3, 032313 (2000);
quant-ph/9912116.

\item {\bf [Pittenger-Rubin 00 b]}:
A. O. Pittenger, \& M. H. Rubin,
``Complete separability and Fourier representations of $n$-qubit states'',
{\em Phys. Rev. A} {\bf 62}, 4, 042306 (2000).

\item {\bf [Pittenger-Rubin 00 c]}:
A. O. Pittenger, \& M. H. Rubin,
``Separability and Fourier representations of density matrices'',
quant-ph/0001014.

\item {\bf [Pittenger-Rubin 00 d]}:
A. O. Pittenger, \& M. H. Rubin,
``Note on separability of the Werner states in arbitrary
dimensions'',
quant-ph/0001110.

\item {\bf [Pittenger-Rubin 02]}:
A. O. Pittenger, \& M. H. Rubin,
``Convexity and the separability problem of quantum
mechanical density matrices'',
{\em Linear Algebra Appl.} {\bf 346}, 47-71 (2002);
quant-ph/0103038.

\item {\bf [Pittenger-Rubin 03 a]}:
A. O. Pittenger, \& M. H. Rubin,
``Geometry of entanglement witnesses and local detection of entanglement'',
{\em Phys. Rev. A} {\bf 67}, 1, 012327 (2003).

\item {\bf [Pittenger-Rubin 03 b]}:
A. O. Pittenger, \& M. H. Rubin,
``Mutually unbiased bases, generalized spin matrices and separability'',
quant-ph/0308142.

\item {\bf [Pittman-Shih-Sergienko-Rubin 95]}:
T. B. Pittman, Y. H. Shih, A. V. Sergienko, \& M. H. Rubin,
``Experimental tests of Bell's inequalities based on
space-time and spin variables'',
{\em Phys. Rev. A} {\bf 51}, 5, 3495-3498 (1995).

\item {\bf [Pittman 95]}:
T. B. Pittman,
``On the use of double entanglement in four-photon experiments'',
{\em Phys. Lett. A} {\bf 204}, 3-4, 193-197 (1995).

\item {\bf [Pittman-Shih-Strekalov-Sergienko 95]}:
T. B. Pittman, Y. H. Shih, D. V. Strekalov, \& A. V. Sergienko,
``Optical imaging by means of two-photon quantum entanglement'',
{\em Phys. Rev. A} {\bf 52}, 5, R3429-R3432 (1995).

\item {\bf [Pittman-Strekalov-Migdall-(+3) 96]}:
T. B. Pittman, D. V. Strekalov, A. Migdall, M. H. Rubin,
A. V. Sergienko, \& Y. H. Shih,
``Can two-photon interference be considered the
interference of two photons?'',
{\em Phys. Rev. Lett.} {\bf 77}, 10, 1917-1920 (1996).

\item {\bf [Pittman-Shih-Strekalov-(+2) 96]}:
T. B. Pittman, Y. H. Shih, D. V. Strekalov,
A. V. Sergienko, \& M. H. Rubin,
``Multi-particle interferometry based on double entangled states'',
in D. Han, K. Peng, Y. S. Kim, \& V. I. Man'ko (eds.),
{\em 4th Int.\ Conf.\ on Squeezed States and Uncertainty Relations
(Taiyuan, Shanxi, China, 1995)},
NASA, Greenbelt, Maryland, 1996, pp.~139-144.

\item {\bf [Pittman-Jacobs-Franson 01 a]}:
T. B. Pittman, B. C. Jacobs, \& J. D. Franson,
``Probabilistic quantum logic operations using polarizing beam splitters'',
{\em Phys. Rev. A} {\bf 64}, 6, 062311 (2001);
quant-ph/0107091.

\item {\bf [Pittman-Jacobs-Franson 01 b]}:
T. B. Pittman, B. C. Jacobs, \& J. D. Franson,
``Demonstration of non-deterministic quantum
logic operations using linear optical elements'',
quant-ph/0109128.

\item {\bf [Pittman-Jacobs-Franson 02 a]}:
T. B. Pittman, B. C. Jacobs, \& J. D. Franson,
``Single photons on pseudodemand from stored parametric down-conversion '',
{\em Phys. Rev. A} {\bf 66}, 4, 042303 (2002).

\item {\bf [Pittman-Jacobs-Franson 02 b]}:
T. B. Pittman, B. C. Jacobs, \& J. D. Franson,
``Demonstration of feed-forward control for linear optics quantum computation'',
{\em Phys. Rev. A} {\bf 66}, 5, 052305 (2002);
quant-ph/0204142.

\item {\bf [Pittman-Franson 02]}:
T. B. Pittman, \& J. D. Franson,
``Cyclical quantum memory for photonic qubits'',
{\em Phys. Rev. A} {\bf 66}, 6, 062302 (2002).

\item {\bf [Pittman-Franson 03]}:
T. B. Pittman, \& J. D. Franson,
``Violation of Bell's inequality with photons from independent sources'',
{\em Phys. Rev. Lett.} {\bf 90}, 24, 240401 (2003);
quant-ph/0301169.

\item {\bf [Pittman-Fitch-Jacobs-Franson 03]}:
T. B. Pittman, M. J. Fitch, B. C Jacobs, \& J. D. Franson,
``Experimental controlled-NOT logic gate for single photons in
the coincidence basis'',
{\em Phys. Rev. A} {\bf 68}, 3, 032316 (2003);
quant-ph/0303095.

\item {\bf [Pittman-Donegan-Fitch-(+5) 03]}:
T. B. Pittman, M. M. Donegan, M. J. Fitch,
B. C. Jacobs, J. D. Franson, P. Kok, H. Lee, \& J. P. Dowling,
``Heralded two-photon entanglement from probabilistic quantum logic
operations on multiple parametric down-conversion sources'',
quant-ph/0303113.

\item {\bf [Pittman-Jacobs-Franson 04 a]}:
T. B. Pittman, B. C Jacobs, \& J. D. Franson,
``Probabilistic quantum encoder for single-photon qubits'',
{\em Phys. Rev. A} {\bf 69}, 4, 042306 (2004);
quant-ph/0312097.

\item {\bf [Pittman-Jacobs-Franson 04 b]}:
T. B. Pittman, B. C. Jacobs, \& J. D. Franson,
``Experimental demonstration of a quantum circuit using linear optics
gates'',
quant-ph/0404059.

\item {\bf [Pittman-Jacobs-Franson 04 c]}:
T. B. Pittman, B. C. Jacobs, \& J. D. Franson,
``Quantum computing using linear optics'',
quant-ph/0406192.

\item {\bf [Pittman-Jacobs-Franson 04 d]}:
T. B. Pittman, B. C. Jacobs, \& J. D. Franson,
``Heralding single photons from pulsed parametric down-conversion'',
quant-ph/0408093.

\item {\bf [Plaga 00]}:
R. Plaga,
`An extension of ``Popper's experiment'' can test
interpretations of quantum mechanics',
{\em Found. Phys. Lett.} {\bf 13}, 5, 461-476 (2000);
quant-ph/0010030.

\item {\bf [Planat-Rosu-Perrine-Saniga 04]}:
M. R. P. Planat, H. Rosu, S. Perrine, \& M. Saniga,
``Finite algebraic geometrical structures underlying mutually unbiased
quantum measurements'',
quant-ph/0409081.

\item {\bf [Planck 00]}:
M. Planck,
``Zur Theorie der Gesetzes der Energieverteilung im
Normalspektrum'',
{\em Verhandlungen der Deutschen Physikalischen Gesellschaft}
{\bf 2}, 237-245 (1900).

\item {\bf [Planck 48]}:
M. Planck,
{\em Wissenschaftliche Selbstbiographie},
?, ?, ?.
English version:
{\em Scientific autobiography and other papers},
Greenwood Press, London, 1948.
Spanish version:
{\em Autobiograf\'{\i}a cient\'{\i}fica y \'{u}ltimos escritos},
Nivola Ediciones, Madrid, 2000.

\item {\bf [Planck 69]}:
M. Planck,
{\em Die Quantenhypothese},
{\em Dokumente der Naturwissenschaft}, Ernst Battenberg Verlag, M\"{u}nchen, 1969.

\item {\bf [Plastina-Fazio-Palma 01]}:
F. Plastina, R. Fazio, \& G. M. Palma,
``Macroscopic entanglement in Josephson nanocircuits'',
{\em Phys. Rev. B} {\bf 64}, 11, 113306 (2001).

\item {\bf [Plastina-Falci 03]}:
F. Plastina, \& G. Falci,
``Communicating Josephson qubits'',
{\em Phys. Rev. B} {\bf 67}, 22, 224514 (2003).

\item {\bf [Plastino-Casas-Plastino 01]}:
A. R. Plastino, M. Casas, \& A. Plastino,
``Bohmian quantum theory of motion for particles with
position-dependent effective mass'',
{\em Phys. Lett. A} {\bf 281}, 5-6, 297-304 (2001).

\item {\bf [Plastino-Rigo-Casas-(+2) 02]}:
A. R. Plastino, A. Rigo, M. Casas, F. Garc\'{\i}as \& A. Plastino,
``Aplicaci\'{o}n del formalismo mec\'{a}nico cu\'{a}ntico supersim\'{e}trico a
sistemas con masa efectiva dependiente de la posici\'{o}n'',
in C. Mataix, \& A. Rivadulla (eds.),
{\em F\'{\i}sica cu\'{a}ntica y realidad.
Quantum physics and reality (Madrid, 2000)},
Editorial Complutense, Madrid, 2002, pp.~345-355.

\item {\bf [Platt 92]}:
D. E. Platt,
``A modern analysis of the Stern-Gerlach experiment'',
{\em Am. J. Phys.} {\bf 60}, 4, 306-308 (1992).

\item {\bf [Platzman-Dykman 99]}:
P. M. Platzman, \& M. I. Dykman,
``Quantum computing with electrons
floating on liquid helium'',
{\em Science} {\bf 284}, 5422, 1967-1969 (1999).

\item {\bf [Plenio-Knight 96]}:
M. B. Plenio, \& P. L. Knight,
``Realistic lower bounds for the factorization time of large numbers
on a quantum computer'',
{\em Phys. Rev. A} {\bf 53}, 5, 2986-2990 (1996).
See {\bf [Plenio-Knight 97]}.

\item {\bf [Plenio-Vedral-Knight 96]}:
M. B. Plenio, V. Vedral, \& P. L. Knight,
``Computers and communication in the quantum world'',
{\em Phys. World} {\bf 9}, 10, 19-20 (1996).

\item {\bf [Plenio-Vedral-Knight 97 a]}:
M. B. Plenio, V. Vedral, \& P. L. Knight,
``Quantum error correction in the presence of spontaneous emission'',
{\em Phys. Rev. A} {\bf 55}, 1, 67-71 (1997).

\item {\bf [Plenio-Vedral-Knight 97 b]}:
M. B. Plenio, V. Vedral, \& P. L. Knight,
``Conditional generation of error syndromes
in fault-tolerant error correction'',
{\em Phys. Rev. A} {\bf 55}, 6, 4593-4596 (1997);
quant-ph/9608028.

\item {\bf [Plenio-Knight 97 a]}:
M. B. Plenio, \& P. L. Knight,
``Decoherence limits to quantum factoring'',
in M. Ferrero, \& A. van der Merwe (eds.),
{\em New developments on fundamental problems in quantum physics
(Oviedo, Spain, 1996)},
Kluwer Academic, Dordrecht, Holland, 1997, pp.~311-316.
See {\bf [Plenio-Knight 96, 97 b]}.

\item {\bf [Plenio-Knight 97 b]}:
M. B. Plenio, \& P. L. Knight,
``Decoherence limits to quantum computation using trapped ions'',
{\em Proc. R. Soc. Lond. A} {\bf 453}, 1965, 2017-2041 (1997).
See {\bf [Plenio-Knight 96, 97 a]}.

\item {\bf [Plenio 98]}:
M. B. Plenio,
``Quantum optics'',
{\em Contemp. Phys.} {\bf 39}, 2, 137-139 (1998).
Review of {\bf [Scully-Zubairy 97]}.

\item {\bf [Plenio-Vedral 98]}:
M. B. Plenio, \& V. Vedral,
``Teleportation, entanglement and thermodynamics in the quantum world'',
{\em Contemp. Phys.} {\bf 39}, 6, 431-446 (1998).
See quant-ph/9804075.

\item {\bf [Plenio-Fern\'{a}ndez Huelga-Beige-Knight 98]}:
M. B. Plenio, S. G. Fern\'{a}ndez Huelga, A. Beige, \& P. L. Knight,
``Cavity loss induced generation of entangled atoms'',
quant-ph/9811003.

\item {\bf [Plenio 99]}:
M. B. Plenio,
``The Holevo bound and Landauer's principle'',
{\em Phys. Lett. A} {\bf 263}, 4-6, 281-284 (1999);
quant-ph/9910086.

\item {\bf [Plenio-Virmani-Papadopoulos 00]}:
M. B. Plenio, S. Virmani, \& P. Papadopoulos,
``Operator monotones, the reduction criterion
and the relative entropy'',
{\em J. Phys. A} {\bf 33}, 22, L193-L197 (2000);
quant-ph/0002075.

\item {\bf [Plenio 00 a]}:
M. B. Plenio,
``Quantum mechanics II'',
lecture notes, Imperial College, London, 2000;
http://www.lsr.ph.ic.ac.uk/TQO/People/Plenio/lectures.ps.

\item {\bf [Plenio 00 b]}:
M. B. Plenio,
``Equivalent classes of closed three-level systems'',
{\em Phys. Rev. A} {\bf 62}, 1, 015802 (2000);
quant-ph/0103109.

\item {\bf [Plenio-Vitelli 01]}:
M. B. Plenio, \& V. Vitelli,
``The physics of forgetting: Landauer's erasure principle
and information theory'',
{\em Contemp. Phys.} {\bf 42}, 1, 25-60 (2001);
quant-ph/0103108.

\item {\bf [Plenio-Vedral 01]}:
M. B. Plenio, \& V. Vedral,
``Bounds on relative entropy of entanglement for multi-party systems'',
in S. Popescu, N. Linden, \& R. Jozsa (eds.),
{\em J. Phys. A} {\bf 34}, 35
(Special issue: Quantum information and computation), 6997-7002 (2001);
quant-ph/0010080.

\item {\bf [Plenio-Fern\'{a}ndez Huelga 02]}:
M. B. Plenio, \& S. G. Fern\'{a}ndez Huelga,
``Entangled light from white noise'',
{\em Phys. Rev. Lett.} {\bf 88}, 19, 197901 (2002),
quant-ph/0110009.

\item {\bf [Plenio-Hartley-Eisert 04]}:
M. B. Plenio, J. Hartley, \& J. Eisert,
``Dynamics and manipulation of entanglement in coupled harmonic systems
with many degrees of freedom'',
quant-ph/0402004.

\item {\bf [Plenio-Eisert-Dreissig-Cramer 04]}:
M. B. Plenio, J. Eisert, J. Dreissig, \& M. Cramer,
``Geometric entropy in harmonic lattice systems'',
quant-ph/0405142.

\item {\bf [Plenio-Semiao 04]}:
M. B. Plenio, \& F. L. Semiao,
``High fidelity transfer of quantum information in translation invariant
quantum chains'',
quant-ph/0407034.

\item {\bf [Plesch-Bu\v{z}zek 03 a]}:
M. Plesch, \& V. Bu\v{z}zek,
``Entangled graphs: Bipartite entanglement in multiqubit systems'',
{\em Phys. Rev. A} {\bf 67}, 1, 012322 (2003);
quant-ph/0211020.

\item {\bf [Plesch-Bu\v{z}zek 03 b]}:
M. Plesch, \& V. Bu\v{z}zek,
``Entangled graphs. II. Classical correlations in multi-qubit entangled
systems'',
{\em Phys. Rev. A} {\bf 68}, 1, 012313 (2003);
quant-ph/0306001.

\item {\bf [Plesch-Novotny-Dzurakova-Bu\v{z}zek 03 b]}:
M. Plesch, J. Novotny, Z. Dzurakova, \& V. Bu\v{z}zek,
``Controlling bi-partite entanglement in multi-qubit systems'',
{\em J. Phys. A};
quant-ph/0311069.

\item {\bf [Plotnitsky 03]}:
A. Plotnitsky,
``Mysteries without mysticism and correlations without correlata:
On quantum knowledge and knowledge in general
{\em Found. Phys.} {\bf 33}, 11, 1649-1689 (2003).

\item {\bf [Plourde-Zhang-Whaley-(+7) 04]}:
B. L. T. Plourde, J. Zhang, K. B. Whaley,
F. K. Wilhelm, T. L. Robertson, T. Hime,
S. Linzen, P. A. Reichardt, C.-E. Wu, \& J. Clarke,
``Entangling flux qubits with a bipolar dynamic inductance'',
quant-ph/0406049.

\item {\bf [Podoshvedov 02]}:
S. A. Podoshvedov,
``Entangled state generation and modulation of vacuum squeezing by classical
Optical fields in birefringent fiber'',
{\em JETP Lett.} {\bf 75}, ?, 503-? (2002).

\item {\bf [Polachic-Rangacharyulu-van den Bergb-(+11) 04]}:
C. Polachic, C. Rangacharyulu, A. M. van den Bergb,
S. Hamiehb, M. N. Harakeh, M. Hunyadi,
M. A. de Huub, H. J. W\"{o}rtche, J. Heysec,
C. B\"{a}umer, D. Frekersd, S. Rakers,
J. A. Brookee, P. Buschf,
``Polarization correlations of $^1S_0$ proton pairs as tests of
hidden-variable theories'',
{\em Phys. Lett. A} {\bf 323}, 3-4, 176-181 (2004).

\item {\bf [Polchinski 91]}:
J. Polchinski,
``Weinberg's nonlinear quantum mechanics and
the Einstein-Podolsky-Rosen paradox'',
{\em Phys. Rev. Lett.} {\bf 66}, 4, 397-400 (1991).
See {\bf [Weinberg 89 a, b, c, d]}, {\bf [Gisin 90]}.
Comment: {\bf [Mielnik 00]}.

\item {\bf [Polkinghorne-Ralph 99]}:
R. E. S. Polkinghorne, \& T. C. Ralph,
``Continuous variable entanglement swapping'',
{\em Phys. Rev. Lett.} {\bf 83}, 11, 2095-2099 (1999);
quant-ph/9906066.

\item {\bf [Pollatsek 01]}:
H. Pollatsek,
``Quantum error correction: Classic group theory meets a quantum challenge'',
{\em Am. Math. Monthly} {\bf 108}, 10, 932-962 (2001).

\item {\bf [Pollatsek-Ruskai 03]}:
H. Pollatsek, \& M. B. Ruskai,
``Permutationally invariant codes for quantum error correction'',
{\em Lin. Alg. Appl.};
quant-ph/0304153.

\item {\bf [Polley 99]}:
L. Polley,
``Quantum-mechanical probability from the symmetries of two-state
systems'',
quant-ph/9906124.
See {\bf [Deutsch 99]},
{\bf [Finkelstein 99 c]},
{\bf [Summhammer 99]},
{\bf [Barnum-Caves-Finkelstein-(+2) 00]},
{\bf [Polley 01]}.

\item {\bf [Polley 01]}:
L. Polley,
``Position eigenstates and the statistical axiom of quantum
mechanics'',
in {\em Foundations of Probability and Physics (V\"{a}xj\"{o}, Sweden, 2000)};
quant-ph/0102113.

\item {\bf [Polyakov-Chou-Felinto-Kimble 04]}:
S. V. Polyakov, C. W. Chou, D. Felinto, \& H. J. Kimble,
``Temporal dynamics of photon pairs generated by an atomic ensemble'',
quant-ph/0406050.

\item {\bf [Polzik 00]}:
E. S. Polzik,
``Einstein-Podolsky-Rosen-correlated atomic ensembles'',
{\em Phys. Rev. A} {\bf 59}, 6, 4202-4205 (1999).

\item {\bf [Pomeransky 03]}:
A. A. Pomeransky,
``Strong superadditivity of the entanglement of formation follows from its
additivity'',
{\em Phys. Rev. A} {\bf 68}, 3, 032317 (2003).

\item {\bf [Pomeransky-Shepelyansky 04]}:
A. A. Pomeransky, \& D. L. Shepelyansky,
``Quantum computation of the Anderson transition in the presence of imperfections'',
{\em Phys. Rev. A} {\bf 69}, 1, 014302 (2004).

\item {\bf [Pompili 00]}:
A. Pompili,
``?'',
{\em Eur. J. Phys. C} {\bf 14}, ?, 469-? (2000).

\item {\bf [Ponomarenko-Wolf 01]}:
S. A. Ponomarenko, \& E. Wolf,
``Correlations in open quantum systems and associated
uncertainty relations'',
{\em Phys. Rev. A} {\bf 63}, 6, 062106 (2001).

\item {\bf [de Ponte-de Oliveira-Moussa 03]}:
M. A. de Ponte, M. C. de Oliveira, \& M. H. Y. Moussa,
``Decoherence in strongly coupled quantum oscillators'',
quant-ph/0309082.

\item {\bf [Pope-Drummond-Munro 00]}:
D. T. Pope, P. D. Drummond, \& W. J. Munro,
``Disagreement between correlations of quantum mechanics and
stochastic electrodynamics in the damped parametric oscillator'',
{\em Phys. Rev. A} {\bf 62}, 4, 042108 (2000);
quant-ph/0003131.

\item {\bf [Pope-Milburn 03]}:
D. T. Pope, \& G. J. Milburn,
``Multipartite entanglement and quantum state exchange'',
{\em Phys. Rev. A} {\bf 67}, 5, 052107 (2003);
quant-ph/0208098.

\item {\bf [Pope-Milburn 04]}:
D. T. Pope, \& G. J. Milburn,
``Bitwise Bell inequality violations for an entangled state
involving $2N$ ions'',
{em Phys. Rev. A};
quant-ph/0403091.

\item {\bf [Popescu-Ionicioiu 04]}:
A. E. Popescu, \& R. Ionicioiu,
``All-electrical quantum computation with mobile spin qubits'',
{\em Phys. Rev. B} {\bf 69}, 24, 245422 (2004).

\item {\bf [Popescu-Rohrlich 92 a]}:
S. Popescu, \& D. Rohrlich,
``Generic quantum nonlocality'',
{\em Phys. Lett. A} {\bf 166}, 5-6, 293-297 (1992).

\item {\bf [Popescu-Rohrlich 92 b]}:
S. Popescu, \& D. Rohrlich,
``Which states violate Bell's inequality maximally?'',
{\em Phys. Lett. A} {\bf 169}, 6, 411-414 (1992).

\item {\bf [Popescu-Rohrlich 94]}:
S. Popescu, \& D. Rohrlich,
``Quantum nonlocality as an axiom'',
{\em Found. Phys.} {\bf 24}, 3, 379-385 (1994).
See {\bf [Rohrlich-Popescu 96]}.

\item {\bf [Popescu-Vaidman 94]}:
S. Popescu, \& L. Vaidman,
``Causality constraints on nonlocal quantum measurements'',
{\em Phys. Rev. A} {\bf 49}, 6, 4331-4338 (1994);
hep-th/9306087.

\item {\bf [Popescu 94]}:
S. Popescu,
``Bell's inequalities versus teleportation: What is nonlocality?'',
{\em Phys. Rev. Lett.} {\bf 72}, 6, 797-799 (1994).

\item {\bf [Popescu 95 a]}:
S. Popescu,
``An optical method for teleportation'',
quant-ph/9501020.

\item {\bf [Popescu 95 b]}:
S. Popescu,
`Bell's inequalities and density matrices:
Revealing ``hidden'' nonlocality',
{\em Phys. Rev. Lett.} {\bf 74}, 14, 2619-2622 (1995);
quant-ph/9502005.

\item {\bf [Popescu 96]}:
S. Popescu,
``More powerful tests of nonlocality by sequences of measurements'',
in A. Mann, \& M. Revzen (eds.),
{\em The dilemma of Einstein, Podolsky and Rosen -- 60 years
later. An international symposium in honour of Nathan Rosen
(Haifa, Israel, 1995)},
{\em Ann. Phys. Soc. Israel} {\bf 12}, 144-151 (1996).

\item {\bf [Popescu-Rohrlich 97 a]}:
S. Popescu, \& D. Rohrlich,
``The relativistic EPR argument'',
in {\bf [Cohen-Horne-Stachel 97 b]};
quant-ph/9605004.

\item {\bf [Popescu-Rohrlich 97 b]}:
S. Popescu, \& D. Rohrlich,
``Thermodynamics and the measure of entanglement'',
{\em Phys. Rev. A} {\bf 56}, 5, R3319-R3321 (1997);
quant-ph/9610044.

\item {\bf [Popescu-Hardy-\.{Z}ukowski 97]}:
S. Popescu, L. Hardy, \& M. \.{Z}ukowski,
``Revisiting Bell's theorem for a class of down-conversion experiments'',
{\em Phys. Rev. A} {\bf 56}, 6, R4353-R4356 (1997);
quant-ph/9612030.

\item {\bf [Popescu-Rohrlich 97 b]}:
S. Popescu, \& D. Rohrlich,
``Causality and nonlocality as axioms for quantum mechanics'',
{\em Causality and Locality in Modern
Physics and Astronomy: Open Questions and Possible Solutions
(York University, Toronto, 1997)};
quant-ph/9709026.

\item {\bf [Popescu-Rohrlich 98]}:
S. Popescu, \& D. Rohrlich,
``The joy of entanglement'',
in {\bf [Lo-Spiller-Popescu 98]}, pp.~29-48.

\item {\bf [Popescu 99]}:
S. Popescu,
``What is quantum computation?'',
in J.-M. Frere, M. Henneaux, A. Sevrin, \& P. Spindel (eds.),
{\em Fundamental interactions: From symmetries to black holes,
Proc. of the conference in honour of Francois Englert, (Universite Libre de Bruxelles, 1999)}.

\item {\bf [Popescu-Gisin 00]}:
S. Popescu, \& N. Gisin,
``Quantum measurements and non-locality'',
in H.-P. Breuer, \& F. Petrucione (eds.),
{\em Relativistic quantum measurement and decoherence},
Springer-Verlag, New York, 2000.

\item {\bf [Popescu-Linden-Jozsa 01]}:
S. Popescu, N. Linden, \& R. Jozsa,
``Quantum information and computation'',
in S. Popescu, N. Linden, \& R. Jozsa (eds.),
{\em J. Phys. A} {\bf 34}, 35
(Special issue: Quantum information and computation), 6723-6724 (2001).

\item {\bf [Popescu-Groisman-Massar 04]}:
S. Popescu, B. Groisman, \& S. Massar,
``Lower bound on the number of Toffoli gates in a classical reversible
circuit through quantum information concepts'',
quant-ph/0407035.

\item {\bf [Popper 56]}:
K. R. Popper,
{\em Quantum theory and the schism in physics.
From the Postscript to the Logic of scientific discovery}, ?, 1956;
Routledge, London, 1992.
Spanish version: {\em Post scriptum a la L\'{o}gica de la
investigaci\'{o}n cient\'{\i}fica. Volumen III.
Teor\'{\i}a cu\'{a}ntica y el cisma en f\'{\i}sica},
Tecnos, Madrid, 1985
(contains a preface from 1982 entitled ``Sobre una interpretaci\'{o}n
realista y de sentido com\'{u}n de la teor\'{\i}a cu\'{a}ntica'').
See {\bf [Peres 99 e]}.

\item {\bf [Poppe-Fedrizzi-Loruenser-(+9) 04]}:
A. Poppe, A. Fedrizzi, T. Loruenser,
O. Maurhardt, R. Ursin, H. R. Boehm,
M. Peev, M. Suda, C. Kurtsiefer, H. Weinfurter,
T. Jennewein, \& A. Zeilinger
``Practical quantum key distribution with polarization-entangled photons'',
quant-ph/0404115.

\item {\bf [Popper 67]}:
K. R. Popper,
``Quantum mechanics without the observer'', in
M. Bunge (ed.),
{\em Quantum theory and reality},
Springer-Verlag, New York, 1967.

\item {\bf [Popper 84]}:
K. R. Popper,
``A critical note on the greatest days of quantum theory'',
in {\bf [Barut-van der Merwe-Vigier 84]}, pp.~49-54.

\item {\bf [Porras-Cirac 04]}:
D. Porras, \& J. I. Cirac,
``Effective quantum spin systems with trapped ions'',
{\em Phys. Rev. Lett.} {\bf 92}, 20, 207901 (2004);
quant-ph/0401102.

\item {\bf [Pospiech 00]}:
G. Pospiech,
``Information---the fundamental notion of quantum theory'',
quant-ph/0002009.

\item {\bf [Post 01]}:
E. J. Post,
``Book review. Quantum dialogue: The making of a revolution'',
{\em Found. Phys.} {\bf 31}, 11 (2001).
Review of {\bf [Beller 99]}.

\item {\bf [P\"{o}tting-Lee-Schmitt-(+3) 00]}:
S. P\"{o}tting, E. S. Lee, W. Schmitt, I. Rumyantsev,
B. Mohring, \& P. Meystre,
``Quantum coherence and interaction-free measurements'',
{\em Phys. Rev. A} {\bf 62}, 6, 060101(R) (2000).

\item {\bf [Potel-Mu\~{n}oz Ale\~{n}ar-Barranco-Vigezzi 02]}:
G. Potel, M. Mu\~{n}oz Ale\~{n}ar, F. Barranco, \& E. Vigezzi,
``Stability properties of $|\psi|^2$ in Bohmian dynamics'',
{\em Phys. Lett. A} {\bf 299}, 2-3, 125-130 (2002).
See {\bf [Barranco-Potel-Mu\~{n}oz Ale\~{n}ar-Bienvenido 02]}.

\item {\bf [Potel-Barranco-Cruz Barrios-G\'{o}mez Camacho 04]}:
G. Potel, F. Barranco, S. Cruz-Barrios, \& J. G\'{o}mez-Camacho,
``Quantum mechanical description of Stern-Gerlach experiments'',
quant-ph/0409206.

\item {\bf [Potting-Lee-Schmitt-(+3) 00]}:
S. Potting, E. S. Lee, W. Schmitt,
I. Rumyantsev, B. Mohring, \& P. Meystre,
``Quantum coherence and interaction-free measurements'',
quant-ph/0007014.

\item {\bf [Potvin 99]}:
G. Potvin,
``Dualist interpretation of quantum mechanics'',
quant-ph/9908019.

\item {\bf [Poulin 02]}:
D. Poulin,
``Classicality of quantum information processing'',
{\em Phys. Rev. A} {\bf 65}, 4, 042319 (2002);
quant-ph/0108102.

\item {\bf [Poulin-Blume Kohout 03]}:
D. Poulin, \& R. Blume-Kohout,
``Compatibility of quantum states'',
{\em Phys. Rev. A} {\bf 67}, 1, 010101 (2003);
quant-ph/0205033.

\item {\bf [Poulin-Laflamme-Milburn-Paz 03]}:
D. Poulin, R. Laflamme, G. J. Milburn, \& J. P. Paz,
``Testing integrability with a single bit of quantum information'',
{\em Phys. Rev. A} {\bf 68}, 2, 022302 (2003);
quant-ph/0303042.
Comment: {\bf [Wiseman 03 c]}.

\item {\bf [Poulin-Blume Kohout-Laflamme-Ollivier 03]}:
D. Poulin, R. Blume-Kohout, R. Laflamme, \& H. Ollivier,
``Exponential speed-up with a single bit of quantum information: Testing
the quantum butterfly effect'',
quant-ph/0310038.

\item {\bf [Poulin 04]}:
D. Poulin,
``Macroscopic observables'',
quant-ph/0403212.

\item {\bf [Power-Percival 00]}:
W. L. Power, \& I. C. Percival,
``Decoherence of quantum wavepackets due to interaction with conformal
spacetime fluctuations'',
{\em Proc. R. Soc. Lond. A} {\bf 456}, 1996, 955-968 (2000);
quant-ph/9811059.

\item {\bf [Poyatos-Walser-Cirac-(+2) 96]}:
J. F. Poyatos, R. Walser, J. I. Cirac, P. Zoller, \& R. Blatt,
``Motion tomography of a single trapped ion'',
{\em Phys. Rev. A} {\bf 53}, 4, R1966-R1969 (1996).

\item {\bf [Poyatos-Cirac-Zoller 96]}:
J. F. Poyatos, J. I. Cirac, \& P. Zoller,
``Quantum reservoir engineering with laser cooled trapped ions'',
{\em Phys. Rev. Lett.} {\bf 77}, 23, 4728-4731 (1996).

\item {\bf [Poyatos-Cirac-Zoller 97 a]}:
J. F. Poyatos, J. I. Cirac, \& P. Zoller,
``Complete characterization of a quantum process:
The two-bit quantum gate'',
{\em Phys. Rev. Lett.} {\bf 78}, 2, 390-393 (1997).

\item {\bf [Poyatos-Cirac-Zoller 97 b]}:
J. F. Poyatos, J. I. Cirac, \& P. Zoller,
`Quantum gates with ``hot'' trapped ions',
quant-ph/9712012.

\item {\bf [Poyatos-Cirac-Zoller 99]}:
J. F. Poyatos, J. I. Cirac, \& P. Zoller,
``From classical to quantum computers.
Quantum computations with trapped ions'',
{\em Physica Scripta}, {\bf ?}, ?-? (1999).

\item {\bf [Poyatos-Cirac-Zoller 00]}:
J. F. Poyatos, J. I. Cirac, \& P. Zoller,
``Schemes of quantum computations with trapped ions'',
{\em Fortschr. Phys.} {\bf 48}, 9-11 (Special issue: Experimental proposals for quantum computation), 785-799 (2000).

\item {\bf [Power-Tan-Wilkens 97]}:
W. L. Power, S. M. Tan, \& M. Wilkens,
``A scheme for the simultaneous measurement of atomic position and
momentum'',
{\em J. Mod. Opt.} {\bf 44}, 11-12 (Special issue: Quantum
state preparation and measurement), 2591-2605 (1997).

\item {\bf [Pradhan-Anantram-Wang 00]}:
P. Pradhan, M. P. Anantram, \& K. L. Wang,
``Quantum computation by optically coupled steady
atoms/quantum-dots inside a quantum electro-dynamic cavity'',
quant-ph/0002006.

\item {\bf [Prager 01]}:
T. Prager,
``A necessary and sufficient condition for
optimal decompositions'',
quant-ph/0106030.

\item {\bf [Pratt-Eberly 01]}:
J. S. Pratt, \& J. H. Eberly,
``Qubit cross talk and entanglement decay'',
{\em Phys. Rev. B} {\bf 64}, 19, 195314 (2001).

\item {\bf [Pravia-Chen-Yepez-Cory 02]}:
M. A. Pravia, Z. Chen, J. Yepez, \& D. G. Cory,
``Towards a NMR implementation of a quantum lattice gas algorithm'',
{\em Proc. of the Quantum Computation for Physical Modeling
Workshop 2000 (North Falmouth, Massachusetts)},
{\em Comput. Phys. Comm.} {\bf 146}, 3, 339-344 (2002).

\item {\bf [Pravia-Chen-Yepez-Cory 03]}:
M. A. Pravia, Z. Chen, J. Yepez, \& D. G. Cory,
``Experimental demonstration of quantum lattice gas computation'',
quant-ph/0303183.

\item {\bf [Pravia-Boulant-Emerson-(+4) 03]}:
M. A. Pravia, N. Boulant, J. Emerson,
A. Farid, E. M. Fortunato, T. F. Havel, \& D. G. Cory,
``Robust control of quantum information'',
quant-ph/0307062.

\item {\bf [Praxmeyer-Englert-W\'{o}dkiewicz 04]}:
L. Praxmeyer, B.-G. Englert, \& K. W\'{o}dkiewicz,
``Violation of Bell's inequality for continuous variables'',
{\em Quantum Information with Atoms, Ions and Photons (La Tuile, Italy, 2004)};
quant-ph/0406172.

\item {\bf [Presilla-Onofrio-Tambini 96]}:
C. Presilla, R. Onofrio, \& U. Tambini,
``Measurement quantum mechanics
and experiments on quantum Zeno effect'',
{\em Ann. Phys} {\bf 248}, ?, 95-121 (1996).

\item {\bf [Preskill 98 a]}:
J. Preskill,
``Reliable quantum computers'',
in D. P. DiVincenzo. E. Knill, R. Laflamme, \& W. H. Zurek (eds.),
{\em Quantum Coherence and Decoherence.
Proc.\ of the ITP Conf.\ (Santa Barbara, California, 1996)},
{\em Proc. R. Soc. Lond. A} {\bf 454}, 1969, 385-410 (1998);
quant-ph/9705031.
See {\bf [Preskill 98 b]}.

\item {\bf [Preskill 98 b]}:
J. Preskill,
``Quantum computing: Pro and con'',
in D. P. DiVincenzo. E. Knill, R. Laflamme, \& W. H. Zurek (eds.),
{\em Quantum Coherence and Decoherence.
Proc.\ of the ITP Conf.\ (Santa Barbara, California, 1996)},
{\em Proc. R. Soc. Lond. A} {\bf 454}, 1969, 469-486 (1998);
quant-ph/9705032.
See {\bf [Preskill 98 a]}.

\item {\bf [Preskill 98 c]}:
J. Preskill,
``Robust solutions to hard problems'',
{\em Nature} {\bf 391}, 6668, 631-632 (1998).

\item {\bf [Preskill 98 d]}:
J. Preskill,
``Fault-tolerant quantum computation'',
in {\bf [Lo-Spiller-Popescu 98]}, pp.~213-269;
quant-ph/9712048.

\item {\bf [Preskill 99 a]}:
J. Preskill,
``Unstable door to the future of computing'',
{\em Nature} {\bf 398}, 6723, 118-119 (1999).
Review of {\bf [Milburn 98]}.

\item {\bf [Preskill 99 b]}:
J. Preskill,
``Plug-in quantum software'',
{\em Nature} {\bf 402}, 6760, 357-358 (1999).
See {\bf [Gottesman-Chuang 99]}.

\item {\bf [Preskill 99 c]}:
J. Preskill,
``Battling decoherence: The fault-tolerant quantum computer'',
{\em Phys. Today} {\bf 52}, 6, 24, 30 (1999).

\item {\bf [Preskill 99 d]}:
J. Preskill,
``Lecture notes for the course
`Information for Physics 219/Computer Science 219,
Quantum Computation'\,'',
Caltech, Pasadena, California, 1999;
www.theory.caltech.edu/people/preskill/ph229.

\item {\bf [Preskill 00 a]}:
J. Preskill,
``Quantum information and physics: Some future directions'',
in V. Bu\v{z}zek, \& D. P. DiVincenzo (eds.),
{\em J. Mod. Opt.} {\bf 47}, 2-3 (Special issue:
Physics of quantum information), 127-137 (2000).

\item {\bf [Preskill 00 b]}:
J. Preskill,
``Quantum clock synchronization and quantum error correction'',
quant-ph/0010098.

\item {\bf [Prestage-Dick-Maleki 89]}:
J. D. Prestage, G. J. Dick, \& L. Maleki,
``New ion trap for frequency standard applications'',
{\em J. Appl. Phys.} {\bf 66}, 3, 1013-1017 (1989).
Reprinted in {\bf [Macchiavello-Palma-Zeilinger 00]}, pp.~334-338.

\item {\bf [Prezhdo-Brooksby 01]}:
O. V. Prezhdo, \& C. Brooksby,
``Quantum backreaction through the Bohmian particle'',
{\em Phys. Rev. Lett.} {\bf 86}, 15, 3215-3219 (2001).

\item {\bf [Price-Somaroo-Dunlop-(+2) 99]}:
M. D. Price, S. S. Somaroo, A. E. Dunlop, T. F. Havel, \& D. G. Cory,
``Generalized methods for the development of quantum
logic gates for an NMR quantum information processor'',
{\em Phys. Rev. A} {\bf 60}, 4, 2777-2780 (1999).

\item {\bf [Price-Havel-Cory 00]}:
M. D. Price, T. F. Havel, \& D. G. Cory,
``Multiqubit logic gates in NMR quantum computing'',
{\em New J. Phys.} {\bf 2}, 10.1-10.9 (2000).

\item {\bf [Price-Chissick 77]}:
W. C. Price, \& S. S. Chissick,
{\em The uncertainty principle and foundations of quantum mechanics:
A fifty years' survey},
John Wiley \& Sons, London, 1977.

\item {\bf [Primas 81]}:
H. Primas,
{\em Chemistry, quantum mechanics and reductionism},
Springer-Verlag, Berlin, 1981.

\item {\bf [Primas 87]}:
H. Primas,
``Contextual quantum objects and their ontic interpretation'',
in P. J. Lahti, \& P. Mittelstaedt (eds.),
{\em Symp. on the Foundations of Modern Physics, 1987.
The Copenhagen Interpretation 60 Years after the Como Lecture}, pp.~251-275.

\item {\bf [Primas 90 a]}:
H. Primas,
``Mathematical and philosophical questions in the theory of open and
macroscopic quantum systems'',
in A. I. Miller (ed.),
{\em Sixty-two years of uncertainty: Historical, philosophical and
physical inquiries into the foundations of quantum mechanics.
Proc.\ from the Int. School of History of Science
(Erice, Italy, 1989)},
Plenum Press, New York, 1990, pp.~?-?.

\item {\bf [Primas 90 b]}:
H. Primas,
``Induced nonlinear time evolution of open quantum objects'',
in A. I. Miller (ed.),
{\em Sixty-two years of uncertainty: Historical, philosophical and
physical inquiries into the foundations of quantum mechanics.
Proc.\ from the Int. School of History of Science
(Erice, Italy, 1989)},
Plenum Press, New York, 1990, pp.~259-280.

\item {\bf [Primas 90 c]}:
H. Primas,
``The measurement process in the individual interpretation of
quantum mechanics'',
in M. Cini, \& J. M. L\'{e}vy-Leblond (eds.),
{\em Quantum theory without reduction},
Adam Hilger, Bristol, 1990, pp.~49-68.

\item {\bf [Privman-Vagner-Kventsel 98]}:
V. Priman, I. D. Vagner, \& G. Kventsel,
``Quantum computation in quantum-Hall systems'',
{\em Phys. Lett. A} {\bf 239}, 3, 141-146 (1998).

\item {\bf [Privman-Mozyrsky-Vagner 02]}:
V. Privman, D. Mozyrsky, \& I. D. Vagner,
``Quantum computing with spin qubits in semiconductor structures'',
{\em Proc. of the Quantum Computation for Physical Modeling
Workshop 2000 (North Falmouth, Massachusetts)},
{\em Comput. Phys. Comm.} {\bf 146}, 3, 331-338 (2002);
cond-mat/0102308.

\item {\bf [Privman 02]}:
V. Privman,
``Short-time decoherence and deviation from pure quantum states'',
{\em Mod. Phys. Lett. B} {\bf 16}, ?, 459-465 (2002);
cond-mat/0203039.

\item {\bf [Proos-Zalka 03]}:
J. Proos \& C. Zalka,
``Shor's discrete logarithm quantum algorithm for elliptic curves'',
{\em Quant. Inf. Comp.} {\bf 3}, 4, 317-344 (2003);
quant-ph/0301141.

\item {\bf [Prosen-Znidaric 01]}:
T. Prosen, \& M. Znidaric,
``Can quantum chaos enhance the stability of quantum computation?'',
{\em J. Phys. A} {\bf 34}, 47, L681-L687 (2001).

\item {\bf [Protopopescu-Barhen 02]}:
V. Protopopescu, \& J. Barhen,
``Solving a class of continuous global
optimization problems using quantum algorithms'',
{\em Phys. Lett. A} {\bf 296}, 1, 9-14 (2002).

\item {\bf [Protopopescu-Perez-D'Helon-Schmulen 03]}:
V. Protopopescu, R. Perez, C. D'Helon, \& J. Schmulen,
``Robust control of decoherence in realistic one-qubit quantum gates'',
{\em J. Phys. A}, {\bf 36}, 8, 2175-? (2003);
quant-ph/0202141.

\item {\bf [Protsenko-Reymond-Schlosser-Grangier 02 a]}:
I. E. Protsenko, G. Reymond, N. Schlosser, \& P. Grangier,
``Operation of a quantum phase gate using neutral atoms in microscopic dipole
traps'',
{\em Phys. Rev. A} {\bf 65}, 5, 052301 (2002).

\item {\bf [Protsenko-Reymond-Schlosser-Grangier 02 b]}:
I. E. Protsenko, G. Reymond, N. Schlosser, \& P. Grangier,
``Conditional quantum logic using two atomic qubits'',
{\em Phys. Rev. A} {\bf 66}, 6, 062306 (2002).

\item {\bf [Pryde-White 03]}:
G. J. Pryde, \& A. G. White,
``Creation of maximally entangled photon-number states using optical fiber multiports'',
{\em Phys. Rev. A} {\bf 68}, 5, 052315 (2003);
quant-ph/0304135.

\item {\bf [Pryde-O'Brien-White-(+2) 04]}:
G. J. Pryde, J. L. O'Brien, A. G. White,
S. D. Bartlett, \& T. C. Ralph,
``Measuring a photonic qubit without destroying it'',
{\em Phys. Rev. Lett.} {\bf 92}, 19, 190402 (2004);
quant-ph/0312048.
Comment: {\bf [Kok-Munro 04]}.

\item {\bf [Pryde-O'Brien-White-Bartlett 04]}:
G. J. Pryde, J. L. O'Brien, A. G. White, \& S. D. Bartlett,
``Demonstrating quantum nonlocality without entanglement'',
quant-ph/0410203.

\item {\bf [Pryor-Flatt\'{e} 03]}:
C. E. Pryor, \& M. E. Flatt\'{e},
``Accuracy of circular polarization as a
measure of spin polarization in quantum dot qubits'',
{\em Phys. Rev. Lett.} {\bf 91}, 25, 257901 (2003).

\item {\bf [Pu-Meystre 00]}:
H. Pu, \& P. Meystre,
``Creating macroscopic atomic Einstein-Podolsky-Rosen
states from Bose-Einstein condensates'',
{\em Phys. Rev. Lett.} {\bf 85}, 19, 3987-3990 (2000).

\item {\bf [Pulmannov\'{a} 02]}:
S. Pulmannov\'{a},
``Hidden variables and Bell inequalities on quantum logics'',
{\em Found. Phys.} {\bf 32}, 2, 193-216 (2002).

\item {\bf [Purdue 02]}:
P. Purdue,
``Analysis of a quantum nondemolition speed-meter interferometer'',
{\em Phys. Rev. D} {\bf 66}, 2, 022001 (2002).

\item {\bf [Putnam 57]}:
H. Putnam,
``Three-valued logic'',
{\em Phil. Studies} {\bf 8}, ?, 73-80 (1957).

\item {\bf [Putnam 61]}:
H. Putnam,
``Comments on the paper of David Sharp'',
{\em Philos. Sci.} {\bf 28}, ?, 234-237 (1961).
Comment on {\bf [Sharp 61]}.

\item {\bf [Putnam 69]}:
H. Putnam,
``Is logic empirical?'', in
R. S. Cohen, \& M. W. Wartofsky (eds.),
{\em Boston studies in the philosophy of science. Vol. 5},
Reidel, Dordrecht, 1969, pp.~216-241.

\item {\bf [Putnam 74]}:
H. Putnam,
``How to think quantum-logically'',
{\em Synthese} {\bf 29}, ?, 55-61 (1974).
Reprinted in {\bf [Suppes 76]}, pp.~?-?.

\item {\bf [Putnam 81]}:
H. Putnam,
``Quantum mechanics and the observer'',
{\em Erkenntnis} {\bf 16}, ?, 193-219 (1981).

\item {\bf [Pykacz-Santos 90]}:
J. Pykacz, \& E. Santos,
``Constructive approach to
logics of physical systems: Application to EPR case'',
{\em Int. J. Theor. Phys.} {\bf 29}, 10, 1041-1058 (1990).

\item {\bf [Pykacz-Santos 91]}:
J. Pykacz, \& E. Santos,
``Hidden variables in
quantum logic approach reexamined'',
{\em J. Math. Phys.} {\bf 32}, 5, 1287-1292 (1991).

\item {\bf [Pykacz 93]}:
J. Pykacz,
``Hidden variables and Bell-type inequalities in quantum logic approach
to foundations of quantum mechanics'',
in A. van der Merwe, \& F. Selleri (eds.),
{\em Bell's theorem and the foundations of modern physics.
Proc.\ of an international
conference (Cesena, Italy, 1991)},
World Scientific, Singapore, 1993, pp.~397-401.

\item {\bf [Pykacz-Santos 95]}:
J. Pykacz, \& E. Santos,
``Bell-type inequalities on tensors products of quantum logics'',
{\em Found. Phys. Lett.} {\bf 8}, 3, 205-212 (1995).

\item {\bf [Pykacz 98]}:
J. Pykacz,
``Conjunctions, disjunctions, and Bell-type inequalities in
orthoalgebras'',
{\em Int. J. Theor. Phys.} {\bf 37}, 3, 1171-? (1998).


\newpage

\subsection{}


\item {\bf [Qamar-Zhu-Zubairy 03]}:
S. Qamar, S.-Y. Zhu, \& M. S. Zubairy,
``Teleportation of an atomic momentum state'',
{\em Phys. Rev. A} {\bf 67}, 4, 042318 (2003).

\item {\bf [Qiao-Ruda-Zhen 01]}:
B. Qiao, H. E. Ruda, \& X. H. Zhen,
``Quantum computing using an open system and projected subspace'',
{\em J. Phys. A};
quant-ph/0110002.

\item {\bf [Qiao-Ruda-Wang 02]}:
B. Qiao, H. E. Ruda, \& J. Wang,
``Multiqubit computing and error-avoiding codes in subspace using quantum dots'',
{\em J. Appl. Phys.} {\bf 91}, ?, 2524-? (2002).

\item {\bf [Qiao-Ruda-Zhan 02]}:
B. Qiao, H. E. Ruda, \& M. S. Zhan,
``Two-qubit quantum computing in a projected subspace'',
{\em Phys. Rev. A} {\bf 65}, 4, 042325 (2002);
quant-ph/0110027.

\item {\bf [Qiao-Ruda-Chang 02]}:
B. Qiao, H. E. Ruda, \& J. F. Chang,
``Using sequences of pulses to control coherence in an open quantum computing
system'',
{\em J. Appl. Phys.} {\bf 91}, ?, 9368-? (2002).

\item {\bf [QinetiQ 02]}:
{\em QinetiQ},
www.qinetiq.com.

\item {\bf [Qiu 02 a]}:
D. Qiu,
``Some analogies between quantum cloning and quantum deleting'',
{\em Phys. Rev. A} {\bf 65}, 5, 052303 (2002).

\item {\bf [Qiu 02 b]}:
D. Qiu,
``Combinations of probabilistic and approximate quantum cloning and deleting'',
{\em Phys. Rev. A} {\bf 65}, 5, 052329 (2002).

\item {\bf [Qiu 02 c]}:
D. Qiu,
``Non-optimal universal quantum deleting machine'',
{\em Phys. Lett. A} {\bf 301}, 3-4, 105-111 (2002).

\item {\bf [Qiu 02 d]}:
D. Qiu,
``Upper bound on the success probability for unambiguous discrimination'',
{\em Phys. Lett. A} {\bf 303}, 2-3, 140-146 (2002).

\item {\bf [Qiu 03 a]}:
D. Qiu,
``Some general probabilistic quantum cloning and deleting machines'',
{\em Phys. Lett. A} {\bf 308}, 5-6, 335-342 (2003).

\item {\bf [Qiu 03 b]}:
D. Qiu,
``Optimum unambiguous discrimination between subsets of quantum states'',
{\em Phys. Lett. A} {\bf 309}, 3-4, 189-197 (2003).

\item {\bf [Quan-Chaojing 02]}:
Z. Quan, \& T. Chaojing,
``Simple proof of the unconditional security of the Bennett 1992 quantum key
distribution protocol'',
{\em Phys. Rev. A} {\bf 65}, 6, 062301 (2002).

\item {\bf [Quiroga-Johnson 99]}:
L. Quiroga, \& N. F. Johnson,
``Entangled Bell and Greenberger-Horne-Zeilinger states of
excitons in coupled quantum dots'',
{\em Phys. Rev. Lett.} {\bf 83}, 11, 2270-2273 (1999);
cond-mat/9901201.

\item {\bf [Qureshi 03]}:
T. Qureshi,
``Popper's experiment, Copenhagen interpretation and nonlocality'',
quant-ph/0301123.

\item {\bf [Qureshi 04]}:
T. Qureshi,
``Understanding Popper's experiment'',
quant-ph/0405057.


\newpage

\subsection{}


\item {\bf [Rabenstein-Sverdlov-Averin 04]}:
K. Rabenstein, V. A. Sverdlov, \& D. V. Averin,
``Qubit decoherence by Gaussian low-frequency noise'',
{\em JETP Lett.} {\bf 79}, 646-649 (2004).

\item {\bf [Rae 86]}:
A. I. M. Rae,
{\em Quantum physics: Illusion or reality?},
Cambridge University Press, Cambridge, 1986.
Spanish version:
{\em F\'{\i}sica cu\'{a}ntica: ?`Ilusi\'{o}n o realidad?},
Alianza, Madrid, 1988.

\item {\bf [Rae 92]}:
A. I. M. Rae,
``Comment on `Does quantum mechanics violate
the Bell inequalities?'\,'',
{\em Phys. Rev. Lett.} {\bf 68}, 17, 2700.
Comment on {\bf [Santos 91 b]}.

\item {\bf [Rae 93]}:
A. I. M. Rae,
``The measurement problem and the quantum mechanics of
macroscopic objects'',
in A. van der Merwe, \& F. Selleri (eds.),
{\em Bell's theorem and the foundations of modern physics.
Proc.\ of an international
conference (Cesena, Italy, 1991)},
World Scientific, Singapore, 1993, pp.~402-406.

\item {\bf [Rae 97]}:
A. I. M. Rae,
``From cats to computers'',
{\em Nature} {\bf 389}, 6652, 686-687 (1997).
Review of {\bf [Milburn 97]}, {\bf [Hannabuss 97]},
{\bf [Bub 97]}.

\item {\bf [Rae 98]}:
A. I. M. Rae,
``Book review. The infamous boundary'',
{\em Stud. Hist. Philos. Sci. Part B: Stud. Hist. Philos. Mod. Phys.}
{\bf 29}, 2, 281-286 (1998).
Review of {\bf [Wick 95]}.

\item {\bf [Rae 99]}:
A. I. M. Rae,
``Quantum physics: Waves, particles and fullerenes'',
{\em Nature} {\bf 401}, 6754, 651-653 (1999).
See {\bf [Arndt-Nairz-Vos Andreae-(+3) 99]}.

\item {\bf [Rae 01]}:
A. I. M. Rae,
``In search of perfection'',
{\em Nature} {\bf 389}, 6839, 741-742 (2001).
Review of {\bf [Schwinger 01]}.

\item {\bf [De Raedt-Hams-Michielsen-(+2) 00]}:
H. De Raedt, A. Hams, K. Michielsen,
S. Miyashita, \& K. Saito,
``On the problem of programming quantum computers'',
quant-ph/0008015.

\item {\bf [De Raedt-Michielsen-De Raedt-Miyashita 01]}:
H. De Raedt, K. Michielsen, K. De Raedt, \& S. Miyashita,
``Number partitioning on a quantum computer'',
{\em Phys. Lett. A} {\bf 290}, 5-6, 227-233 (2001);
quant-ph/0010018.

\item {\bf [De Raedt-De Raedt-Michielsen 04]}:
K. De Raedt, H. De Raedt, \& K. Michielsen,
``Deterministic event-based simulation of quantum interference'',
quant-ph/0409213.

\item {\bf [Raginsky 01]}:
M. Raginsky,
``A fidelity measure for quantum channels'',
{\em Phys. Lett. A} {\bf 290}, 1-2, 11-18 (2001);
quant-ph/0107108.

\item {\bf [Raginsky 02]}:
M. Raginsky,
``Strictly contractive quantum channels and physically realizable quantum
computers'',
{\em Phys. Rev. A} {\bf 65}, 3, 032306 (2002);
quant-ph/0105141.

\item {\bf [Rahn-Doherty-Mabuchi 02]}:
B. Rahn, A. C. Doherty, \& H. Mabuchi,
``Exact performance of concatenated quantum codes'',
{\em Phys. Rev. A} {\bf 66}, 3, 032304 (2002);
quant-ph/0111003.

\item {\bf [Rai-Rai 00 a]}:
J. Rai, \& S. Rai,
``Entangled states of $N$ identical particles'',
quant-ph/0003055.

\item {\bf [Rai-Rai 00 b]}:
S. Rai, \& J. Rai,
``Group-theoretical structure of the entangled states of $N$
identical particles'',
quant-ph/0006107.

\item {\bf [Raimond-Brune-Haroche 01]}:
J. M. Raimond, M. Brune, \& S. Haroche,
``Colloquium: Manipulating quantum entanglement with
atoms and photons in a cavity'',
{\em Rev. Mod. Phys.} 73, 3, 565-582 (2001).

\item {\bf [Rains-Hardin-Shor-Sloane 97]}:
E. M. Rains, R. H. Hardin, P. W. Shor, \& N. J. A. Sloane,
``Nonadditive quantum code'',
{\em Phys. Rev. Lett.} {\bf 79}, 5, 953-954 (1997).

\item {\bf [Rains 97]}:
E. M. Rains,
``Entanglement purification via separable superoperators'',
quant-ph/9707002.

\item {\bf [Rains 99 a]}:
E. M. Rains,
``Rigorous treatment of distillable entanglement'',
{\em Phys. Rev. A} {\bf 60}, 1, 173-178 (1999);
quant-ph/9809078.

\item {\bf [Rains 99 b]}:
E. M. Rains,
``Bound on distillable entanglement'',
{\em Phys. Rev. A} {\bf 60}, 1, 179-184 (1999).
Erratum: {\em Phys. Rev. A} {\bf 63}, 1, 019902(E) (2001).
quant-ph/9809082.

\item {\bf [Rains 00]}:
E. M. Rains,
``A semidefinite program for distillable entanglement'',
quant-ph/0008047.

\item {\bf [Raiteri 98]}:
A. Raiteri,
``A realistic interpretation of the density matrix'',
quant-ph/9812011.

\item {\bf [Rajagopal 99]}:
A. K. Rajagopal,
``Quantum entanglement and the maximum-entropy
states from the Jaynes principle'',
{\em Phys. Rev. A} {\bf 60}, 6, 4338-4340 (1999);
quant-ph/9903083.

\item {\bf [Rajagopal-Jensen-Cummings 99]}:
A. K. Rajagopal, K. L Jensen, \& F. W. Cummings,
``Quantum entangled supercorrelated states in the Jaynes-Cummings model'',
{\em Phys. Lett. A} {\bf 259}, 3-4, 285-290 (1999).

\item {\bf [Rajagopal-Rendell 01 a]}:
A. K. Rajagopal, \& R. W. Rendell,
``Decoherence, correlation, and entanglement in a pair of coupled quantum
dissipative oscillators'',
{\em Phys. Rev. A} {\bf 63}, 2, 022116 (2001);
quant-ph/0009003.

\item {\bf [Rajagopal-Rendell 01 b]}:
A. K. Rajagopal, \& R. W. Rendell,
``Two qubits in the Dirac representation'',
{\em Phys. Rev. A} {\bf 64}, 2, 024303 (2001);
quant-ph/0006129.

\item {\bf [Rajagopal-Rendell 01 c]}:
A. K. Rajagopal, \& R. W. Rendell,
``Dissipative quantum theory:
Implications for quantum entanglement'',
quant-ph/0106050.

\item {\bf [Rajagopal-Rendell 02 a]}:
A. K. Rajagopal, \& R. W. Rendell,
``Robust and fragile entanglement of three qubits: Relation to permutation
symmetry'',
{\em Phys. Rev. A} {\bf 65}, 3, 032328 (2002);
quant-ph/0104122.

\item {\bf [Rajagopal-Rendell 02 b]}:
A. K. Rajagopal, \& R. W. Rendell,
``Separability and correlations in composite states based on entropy methods'',
{\em Phys. Rev. A} {\bf 66}, 2, 022104 (2002);
quant-ph/0203124.

\item {\bf [Rallan-Vedral 03]}:
L. Rallan, \& V. Vedral,
``Energy requirements for quantum data compression and $1-1$ coding'',
{\em Phys. Rev. A} {\bf 68}, 4, 042309 (2003);
quant-ph/0208039.

\item {\bf [Ralph-Clark-Everitt-(+3) 03]}:
J. F. Ralph, T. D. Clark, M. J. Everitt,
H. Prance, P. Stiffell, \& R. J. Prance,
``Characterising a solid state qubit via environmental noise'',
{\em Phys. Lett. A} {\bf 317}, 3-4, 199-205 (2003).

\item {\bf [Ralph-Clark-Spiller-Munro 04]}:
J. F. Ralph, T. D. Clark, T. P. Spiller, \& W. J. Munro,
``Entanglement generation in persistent current qubits'',
cond-mat/0401358.

\item {\bf [Ralph-Lam 98]}:
T. C. Ralph, \& P. K. Lam,
``Teleportation with bright squeezed light'',
{\em Phys. Rev. Lett.} {\bf 81}, 25, 5668-5671 (1998).

\item {\bf [Ralph 98]}:
T. C. Ralph,
``All optical quantum teleportation'',
quant-ph/9812021.

\item {\bf [Ralph-Lam-Polkinghorne 99]}:
T. C. Ralph, P. K. Lam, \& R. E. S. Polkinghorne,
``Characterizing teleportation in optics'',
{\em J. Opt. B: Quantum Semiclass. Opt.} {\bf 1}, 4, 483-489
(1999).

\item {\bf [Ralph 00 a]}:
T. C. Ralph,
``Continuous variable quantum cryptography'',
{\em Phys. Rev. A} {\bf 61}, 1, 010303(R) (2000);
quant-ph/9907073.

\item {\bf [Ralph 00 b]}:
T. C. Ralph,
``Mach-Zehnder interferometer and the teleporter'',
{\em Phys. Rev. A} {\bf 61}, 4, 044301 (2000);
quant-ph/9906072.

\item {\bf [Ralph-Munro-Polkinghorne 00]}:
T. C. Ralph, W. J. Munro, \& R. E. S. Polkinghorne,
``Proposal for the measurement of Bell-type correlations
from continuous variables'',
{\em Phys. Rev. Lett.} {\bf 85}, 10, 2035-2039 (2000);
quant-ph/0006057.
Comment: {\bf [Banaszek-Walmsley-W\'{o}dkiewicz 00]}.
Reply: {\bf [Ralph-Munro 01]}.

\item {\bf [Ralph 00 c]}:
T. C. Ralph,
``Teleportation criteria: Form and significance'',
in H. Carmichael, R. Glauber, \& M. O. Scully (eds.),
{\em Dan Walls Memorial},
Springer-Verlag, New York, 2000;
quant-ph/0004093.

\item {\bf [Ralph 00 d]}:
T. C. Ralph,
``Security of continuous-variable quantum cryptography'',
{\em Phys. Rev. A} {\bf 62}, 6, 062306 (2000);
quant-ph/0007024.

\item {\bf [Ralph 00 e]}:
T. C. Ralph,
``Franson-type interferometer for intensity fluctuations'',
{\em J. Opt. B: Quantum Semiclass. Opt.} {\bf 2}, 6, L31-L34 (2000).

\item {\bf [Ralph-Munro 01]}:
T. C. Ralph, \& W. J. Munro,
`Reply to Comment on ``Proposal for the measurement
of Bell-type correlations from continuous variables''\,',
quant-ph/0104092.
Reply to {\bf [Banaszek-Walmsley-W\'{o}dkiewicz 00]}.
See {\bf [Ralph-Munro-Polkinghorne 00]}.

\item {\bf [Ralph 01 a]}:
T. C. Ralph,
``Quantum key distribution with continuous variables in optics'',
in S. L. Braunstein, \& A. K. Pati (eds.),
{\em Quantum information theory with continuous variables},
quant-ph/0109096.

\item {\bf [Ralph 01 b]}:
T. C. Ralph,
``Coherent superposition states as quantum rulers'',
quant-ph/0109106.

\item {\bf [Ralph-White-Munro-Milburn 02]}:
T. C. Ralph, A. G. White, W. J. Munro, \& G. J. Milburn,
``Simple scheme for efficient linear optics quantum gates'',
{\em Phys. Rev. A} {\bf 65}, 1, 012314 (2002);
quant-ph/0108049.

\item {\bf [Ralph-Munro-Milburn 01]}:
T. C. Ralph, W. J. Munro, \& G. J. Milburn,
``Quantum computation with coherent states,
linear interactions and superposed resources'',
quant-ph/0110115.

\item {\bf [Ralph 02 a]}:
T. C. Ralph,
``Interferometric tests of teleportation'',
{\em Phys. Rev. A} {\bf 65}, 1, 012319 (2002);
quant-ph/0012109.

\item {\bf [Ralph 02 b]}:
T. C. Ralph,
``Coherent superposition states as quantum rulers'',
{\em Phys. Rev. A} {\bf 65}, 4, 042313 (2002).

\item {\bf [Ralph-Langford-Bell-White 02]}:
T. C. Ralph, N. K. Langford, T. B. Bell, \& A. G. White,
``Linear optical controlled-NOT gate in the coincidence basis'',
{\em Phys. Rev. A} {\bf 65}, 6, 062324 (2002);
quant-ph/0112088.

\item {\bf [Ralph-Huntington 02]}:
T. C. Ralph, \& E. H. Huntington,
``Unconditional continuous-variable dense coding'',
{\em Phys. Rev. A} {\bf 66}, 4, 042321 (2002).

\item {\bf [Ralph-Gilchrist-Milburn-(+2) 03]}:
T. C. Ralph, A. Gilchrist, G. J. Milburn, W. J. Munro, \& S. Glancy,
``Quantum computation with optical coherent states'',
quant-ph/0306004.

\item {\bf [Ralph 03]}:
T. C. Ralph,
``Proposal for a simple quantum error correction test gate in linear optics'',
{\em J. of Selected Topics in Quantum Electronics};
quant-ph/0306190.

\item {\bf [Ralph 04]}:
T. C. Ralph,
``Scaling of multiple postselected quantum gates in optics'',
{\em Phys. Rev. A} {\bf 70}, 1, 012312 (2004).

\item {\bf [Ramanathan-Cho-Cappellaro-(+2) 03]}:
C. Ramanathan, H. Cho, P. Cappellaro,
G. S. Boutis, \& D. G. Cory,
``Encoding multiple quantum coherences in non-commuting bases'',
{\em Chem. Phys. Lett.} {\bf 369}, 311-317 (2003);
quant-ph/0408167.

\item {\bf [Ramanathan-Boulant-Chen-(+3) 04]}:
C. Ramanathan, N. Boulant, Z. Chen,
D. G. Cory, I. Chuang, \& M. Steffen,
``NMR quantum information processing'',
quant-ph/0408166.

\item {\bf [Rangan-Bloch-Monroe-Bucksbaum 04]}:
C. Rangan, A. M. Bloch, C. Monroe, \& P. H. Bucksbaum,
``Control of trapped-ion quantum states with optical pulses'',
{\em Phys. Rev. Lett.} {\bf 92}, 11, 113004 (2004).

\item {\bf [Rao 03]}:
M. V. P. Rao,
``Solving a hidden subgroup problem using the adiabatic quantum-computing
paradigm'',
{\em Phys. Rev. A} {\bf 67}, 5, 052306 (2003).

\item {\bf [Raptis-Zapatrin 00]}:
I. Raptis, \& R. R. Zapatrin,
``Decomposition of pure states of a quantum register'',
quant-ph/0010104.
Comment: {\bf [Vlasov 00 b]}.

\item {\bf [Raptis 01]}:
I. Raptis,
``Sheafifying consistent histories'',
quant-ph/0107037.

\item {\bf [Rarity-Tapster 97]}:
J. G. Rarity, \& P. R. Tapster,
``Quantum interference: Experiments and applications'',
in P. L. Knight, B. Stoicheff, \& D. Walls (eds.),
{\em Highlight in Quantum Optics},
{\em Philos. Trans. R. Soc. Lond. A} {\bf 355}, 1733, 2267-2277 (1997).

\item {\bf [Rarity-Tapster 90 a]}:
J. G. Rarity, \& P. R. Tapster,
``Experimental violation of Bell's inequality based on phase and momentum'',
{\em Phys. Rev. Lett.} {\bf 64}, 21, 2495-2498 (1990).

\item {\bf [Rarity-Tapster 90 b]}:
J. G. Rarity, \& P. R. Tapster,
``Two-color
photons and nonlocality in fourth-order interference'',
{\em Phys. Rev. A} {\bf 41}, 9, 5139-5146 (1990).

\item {\bf [Rarity-Tapster-Jakeman-(+4) 90]}:
J. G. Rarity, P. R. Tapster, E. jakeman, T. Larchuk, R. A. Campos, M. C. Teich,
\& B. E. A. Saleh,
``Two-photon interference in a Mach-Zehnder interferometer'',
{\em Phys. Rev. Lett.} {\bf 65}, 11, 1348-1351 (1990).

\item {\bf [Rarity-Owens-Tapster 94]}:
J. G. Rarity, P. C. M. Owens, \& P. R. Tapster,
``Quantum random-number generation and key sharing'',
in S. M. Barnett, A. K. Ekert, \& S. J. D. Phoenix (eds.),
{\em J. Mod. Opt.} {\bf 41}, 12 (Special issue: Quantum
communication), 2435-2444 (1994).

\item {\bf [Rarity-Tapster 95]}:
J. G. Rarity, \& P. R. Tapster,
``Correlated photon pair optical communications system'',
patent US5418905, 1995.

\item {\bf [Rarity-Tapster 97]}:
J. G. Rarity, \& P. R. Tapster,
``Interference and
violation on Bell's inequalities using separate sources'',
in M. Ferrero, \& A. van der Merwe (eds.),
{\em New developments on fundamental problems in quantum physics
(Oviedo, Spain, 1996)},
Kluwer Academic, Dordrecht, Holland, 1997, pp.~335-345.

\item {\bf [Rarity-Tapster 99]}:
J. G. Rarity, \& P. R. Tapster,
``Three-particle entanglement from entangled photon pairs and a weak coherent state'',
{\em Phys. Rev. A} {\bf 59}, 1, R35-R38 (1999).

\item {\bf [Rarity-Tapster-Gorman-Knight 02]}:
J. G. Rarity, P. R. Tapster, P. M. Gorman, \& P. L. Knight,
``Ground to satellite secure key exchange using
quantum cryptography'',
{\em New J. Phys.} {\bf 4}, 82.1-82.21 (2002).

\item {\bf [Rasel-Oberthaler-Batelaan-(+2) 95]}:
E. M. Rasel, M. K. Oberthaler, H. Batelaan, J. Schmiedmayer, \& A. Zeilinger,
``Atom wave interferometry with diffraction gratings of light'',
{\em Phys. Rev. Lett.} {\bf 75}, 14, 2633-2637 (1995).

\item {\bf [Rastall 81]}:
P. Rastall,
``The meaning of the Bell inequalities'',
{\em Phys. Lett. A} {\bf 86}, 2, 85-86 (1981).

\item {\bf [Rastall 82]}:
P. Rastall,
``The Bell inequalities and nonlocality'',
{\em Phys. Lett. A} {\bf 87}, 2, 6, 279-280 (1982).

\item {\bf [Rastall 85]}:
P. Rastall,
``Locality, Bell's theorem, and quantum mechanics'',
{\em Found. Phys.} {\bf 15}, 9, 963-972 (1985).
Comment: {\bf [Stapp 85 b]}.
See {\bf [Bedford-Stapp 89]}.

\item {\bf [Rastegin 01 a]}:
A. E. Rastegin,
``Some bounds for quantum copying'',
quant-ph/0108014.

\item {\bf [Rastegin 01 b]}:
A. E. Rastegin,
``Some bounds for quantum copying with multiple copies'',
quant-ph/0111085.

\item {\bf [Rastegin 02]}:
A. E. Rastegin,
``Relative error of state-dependent cloning'',
{\em Phys. Rev. A} {\bf 66}, 4, 042304 (2002).

\item {\bf [Rastegin 03 a]}:
A. E. Rastegin,
``Upper bound on the global fidelity for mixed-state cloning'',
{\em Phys. Rev. A} {\bf 67}, 1, 012305 (2003).

\item {\bf [Rastegin 03 b]}:
A. E. Rastegin,
``Global-fidelity limits of state-dependent cloning of mixed states'',
{\em Phys. Rev. A} {\bf 68}, 3, 032303 (2003);
quant-ph/0301132.

\item {\bf [Rau 00]}:
A. R. P. Rau,
``Manipulating two-spin coherences and qubit pairs'',
{\em Phys. Rev. A} {\bf 61}, 3, 032301 (2000).

\item {\bf [Rau-Dunningham-Burnett 03]}:
A. V. Rau, J. A. Dunningham, \& K. Burnett,
``Measurement-induced relative-position localization through entanglement'',
{\em Science} {\bf 301}, 5636, 1081-1084 (2003).

\item {\bf [Rauch-Treimer-Bonse 74]}:
H. Rauch, W. Treimer, \& U. Bonse,
``Test of a single crystal neutron inerferometer'',
{\em Phys. Lett. A} {\bf 47}, 5, 369-371 (1974).

\item {\bf [Rauch-Summhammer 84]}:
H. Rauch, \& J. Summhammer,
``Static versus time-dependent absortion in neutron interferometry'',
{\em Phys. Lett. A} {\bf 104}, 1, 44-46 (1984).

\item {\bf [Rauch 86]}:
H. Rauch,
``Neutron interferometric tests of quantum mechanics'',
{\em Contemp. Phys.} {\bf 27}, 4, 345-360 (1986).

\item {\bf [Rauch 88]}:
H. Rauch,
``Neutron interferometric tests of quantum mechanics'',
{\em Helv. Phys. Acta} {\bf 61}, 5, 589-610 (1988).

\item {\bf [Rauch-Vigier 90]}:
H. Rauch, \& J.-P. Vigier,
`Proposed neutron interferometry test of Einstein's ``einweg''
assumption in the Bohr-Einstein controversy',
{\em Phys. Lett. A} {\bf 151}, 6-7, 269-275 (1990).

\item {\bf [Rauch-Summhammer-Zawisky-Jericha 90]}:
H. Rauch, J. Summhammer, M. Zawisky, \& E. Jericha,
``Low-contrast and low-counting-rate measurements in neutron interferometry",
{\em Phys. Rev. A} {\bf 42}, 7, 3726-3732 (1990).
Comments: {\bf [Unnerstall 90]}, {\bf [Lerner 91]}.
Reply: {\bf [Rauch-Vigier 91]}.

\item {\bf [Rauch-Vigier 91]}:
H. Rauch, \& J.-P. Vigier,
``Reply to comment on `Proposed neutron interferometry test of
Einstein's ``einweg'' assumption in the Bohr-Einstein controversy''\,'\,'',
{\em Phys. Lett. A} {\bf 157}, 4-5, 311-313 (1991).
Reply to: {\bf [Lerner 90]}.
See {\bf [Rauch-Vigier 90]}.

\item {\bf [Rauch 93 a]}:
H. Rauch,
``Phase space coupling in interference and EPR experiments'',
{\em Phys. Lett. A} {\bf 173}, 3, 240-242 (1993).

\item {\bf [Rauch 93 b]}:
H. Rauch,
``Neutron interferometry'',
{\em Science} {\bf 262}, 5138, 1384-1385 (1993).

\item {\bf [Rauch 00]}:
H. Rauch,
``Reality in neutron interference experiments'',
in {\bf [Ellis-Amati 00]}, pp.~28-68.

\item {\bf [Rauch-Werner 00]}:
H. Rauch, \& S. A. Werner,
{\em Neutron interferometry: Lessons in experimental quantum
mechanics},
Oxford University Press, New York, 2000.
Review: {\bf [Silverman 02]}.

\item {\bf [Rauch 02]}:
H. Rauch,
``Towards more quantum complete neutron experiments'',
in {\bf [Bertlmann-Zeilinger 02]}, pp.~351-374.

\item {\bf [Rauschenbeutel-Nogues-Osnaghi-(+4) 00]}:
A. Rauschenbeutel, G. Nogues, S. Osnaghi,
P. Bertet, M. Brune, J.-M. Raimond, \& S. Haroche,
``Step-by-step engineered multiparticle entanglement'',
{\em Science} {\bf 288}, 5473, 2024-2028 (2000).

\item {\bf [Rauschenbeutel-Bertet-Osnaghi-(+4) 01]}:
A. Rauschenbeutel, P. Bertet, S. Osnaghi,
G. Nogues, M. Brune, J. M. Raimond, \& S. Haroche,
``Controlled entanglement of two field modes in a
cavity quantum electrodynamics experiment'',
{\em Phys. Rev. A} {\bf 64}, 5, 050301(R) (2001);
quant-ph/0105062.

\item {\bf [Raussendorf-Briegel 01 a]}:
R. Raussendorf, \& H.-J. Briegel,
``A one-way quantum computer'',
{\em Phys. Rev. Lett.} {\bf 86}, 22, 5188-5191 (2001);
quant-ph/0010033.

\item {\bf [Raussendorf-Briegel 01 b]}:
R. Raussendorf, \& H.-J. Briegel,
``Computational model underlying the one-way quantum computer'',
quant-ph/0108067.

\item {\bf [Raussendorf-Browne-Briegel 02]}:
R. Raussendorf, D. E. Browne, \& H.-J. Briegel,
``The one-way quantum computer---A non-network model of quantum
computation'',
{\em Proc.\ ESF QIT Conf.\ Quantum Information: Theory, Experiment and Perspectives
(Gdansk, Poland, 2001)}, {\em J. Mod. Opt.} {\bf 49}, 8, 1299-1306 (2002);
quant-ph/0108118.

\item {\bf [Raussendorf-Browne-Briegel 03]}:
R. Raussendorf, D. E. Browne, \& H.-J. Briegel,
``Measurement-based quantum computation on cluster states'',
{\em Phys. Rev. A} {\bf 68}, 2, 022312 (2003);
quant-ph/0301052.

\item {\bf [Raussendorf-Bravyi-Harrington 04]}:
R. Raussendorf, S. Bravyi, \& J. Harrington,
``Long-range quantum entanglement in noisy cluster states'',
quant-ph/0407255.

\item {\bf [Raymer 97]}:
M. G. Raymer,
``Measuring the quantum mechanical wave function'',
{\em Contemp. Phys.} {\bf 38}, 5, 343-355 (1997).

\item {\bf [Raymer-Funk-Sanders-de Guise 03]}:
M. G. Raymer, A. C. Funk, B. C. Sanders, \& H. de Guise,
``Separability criterion for separate quantum systems'',
{\em Phys. Rev. A} {\bf 67}, 5, 052104 (2003).

\item {\bf [Raynal-L\"{u}tkenhaus-van Enk 03]}:
P. Raynal, N. L\"{u}tkenhaus, \& S. J. van Enk,
``Reduction theorems for optimal unambiguous state discrimination of density
matrices'',
{\em Phys. Rev. A} {\bf 68}, 2, 022308 (2003);
quant-ph/0304179.

\item {\bf [Raz 99]}:
R. Raz,
``Exponential separation of quantum and classical communication complexity'',
in {\em Proc.\ 31st Annual ACM Symp.\ on Theory of Computing (1999)},
ACM Press, New York, 1999, pp.~358-367.

\item {\bf [Razmi 98]}:
H. Razmi,
``An alternative commutation relation between position and
momentum operators of massless particles'',
quant-ph/9811071.

\item {\bf [Razmi-Golshani 98 a]}:
H. Razmi, \& M. Golshani,
`Bell's theorem and ``classical'' probability theory',
quant-ph/9811072.

\item {\bf [Razmi-Golshani 98 b]}:
H. Razmi,\& M. Golshani,
``Locality is an unnecessary assumption of Bell's theorem'',
quant-ph/9812029.

\item {\bf [Razmi 03]}:
H. Razmi,
``On the role of locality condition in Bell's theorem'',
{\em Int. J. Quant. Inf.};
quant-ph/0301008.

\item {\bf [R\c{e}bilas 04]}:
K. R\c{e}bilas,
``On the Unnikrishnan resolution of the EPR puzzle'',
{\em Found. Phys. Lett.} {\bf 17}, 3, 277-286 (2004).

\item {\bf [Recati-Calarco-Zanardi-(+2) 02]}:
A. Recati, T. Calarco, P. Zanardi, J. I. Cirac, \& P. Zoller,
``Holonomic quantum computation with neutral atoms'',
{\em Phys. Rev. A} {\bf 66}, 3, 0302309 (2002);
quant-ph/0204030.

\item {\bf [Recher-Loss 03]}:
P. Recher, \& D. Loss,
``Dynamical Coulomb blockade and spin-entangled electrons'',
{\em Phys. Rev. Lett.} {\bf 91}, 26, 267003 (2003).

\item {\bf [Reck-Zeilinger-Bernstein-Bertani 94 a]}:
M. Reck, A. Zeilinger, H. J. Bernstein, \& P. Bertani,
``Experimental realization of any discrete unitary operator'',
{\em Phys. Rev. Lett.} {\bf 73}, 1, 58-61 (1994).

\item {\bf [Reck-Zeilinger-Bernstein-Bertani 94 b]}:
M. Reck, A. Zeilinger, H. J. Bernstein, \& P. Bertani,
``How to build any discrete unitary operator in your laboratory'',
in {\em Proc.\ 5th EQEC, ?, ?}, 1994, pp.~43-44.

\item {\bf [R\'{e}dei 98]}:
M. R\'{e}dei,
{\em Quantum logic in algebraic approach},
Kluwer Academic, Dordrecht, Holland, 1998.
Review: {\bf [Gudder 98]}.

\item {\bf [Redhead 81]}:
M. L. G. Redhead,
``Experimental tests of the sum rule'',
{\em Philos. Sci.} {\bf 48}, 1, 50-64 (1981).

\item {\bf [Redhead 87]}:
M. L. G. Redhead,
{\em Incompletness, nonlocality, and realism},
Oxford University Press, New York, 1987.

\item {\bf [Redhead 92]}:
M. L. G. Redhead,
``Propensities, correlations, and metaphysics'',
{\em Found. Phys.} {\bf 22}, 3, 381-394 (1992).

\item {\bf [Redhead-La Rivi\`{e}re 97]}:
M. L. G. Redhead, \& P. La Rivi\`{e}re,
``The relativistic EPR argument'',
in {\bf [Cohen-Horne-Stachel 97 b]}.

\item {\bf [Redhead-Wagner 98]}:
M. L. G. Redhead, \& F. Wagner,
``Unified treatment of EPR and Bell arguments in algebraic
quantum field theory'',
{\em Found. Phys. Lett.} {\bf 11}, 2, 111-125 (1998);
quant-ph/9802010.

\item {\bf [Redhead 01]}:
M. L. G. Redhead,
``The tangled story of nonlocality'',
in {\bf [Rusell-Clayton-Wegter McNelly-Polkinghorne 01]}, pp.~141-158.

\item {\bf [Reeder-Clifton 95]}:
N. Reeder, \& R. K. Clifton,
``Uniqueness of
prime factorizations of linear operators in quantum mechanics'',
{\em Phys. Lett. A} {\bf 204}, 3-4, 198-204 (1995).

\item {\bf [\v{R}eh\'{a}\v{c}ek-Pe\v{r}ina-Facchi-(+2) 99]}:
J. \v{R}eh\'{a}\v{c}ek, J. Pe\v{r}ina, P. Facchi, S. Pascazio, \& L. Mista,
``Quantum Zeno effect in a probed downconversion process'',
quant-ph/9911018.

\item {\bf [\v{R}eh\'{a}\v{c}ek-Hradil-Je\v{z}ek 00]}:
J. \v{R}eh\'{a}\v{c}ek, Z. Hradil, \& M. Je\v{z}ek,
``Iterative algorithm for reconstruction of entangled states'',
quant-ph/0009093.

\item {\bf [\v{R}eh\'{a}\v{c}ek-Hradil-Fiur\'{a}\v{s}ek-Brukner 01]}:
J. \v{R}eh\'{a}\v{c}ek, Z. Hradil, J. Fiur\'{a}\v{s}ek, \& \v{C}. Brukner,
``Designing optimum completely positive maps for quantum teleportation'',
{\em Phys. Rev. A} {\bf 64}, 6, 060301 (2001);
quant-ph/0105119.

\item {\bf [\v{R}eh\'{a}\v{c}ek-Hradil 02]}:
J. \v{R}eh\'{a}\v{c}ek, \& Z. Hradil,
``Invariant information and quantum state estimation'',
{\em Phys. Rev. Lett.} {\bf 88}, 13, 130401 (2002).

\item {\bf [\v{R}eh\'{a}\v{c}ek-Hradil 03]}:
J. \v{R}eh\'{a}\v{c}ek, \& Z. Hradil,
``Quantification of entanglement by means of convergent iterations'',
{\em Phys. Rev. Lett.} {\bf 90}, 12, 127904 (2003).

\item {\bf [\v{R}eh\'{a}\v{c}ek-Englert-Kaszlikowski 04 a]}:
J. \v{R}eh\'{a}\v{c}ek, B.-G. Englert, \& D. Kaszlikowski,
``Minimal qubit tomography'',
quant-ph/0405084.

\item {\bf [\v{R}eh\'{a}\v{c}ek-Englert-Kaszlikowski 04 b]}:
J. \v{R}eh\'{a}\v{c}ek, B.-G. Englert, \& D. Kaszlikowski,
``Iterative procedure for computing accessible information in quantum
communication'',
quant-ph/0408134.

\item {\bf [Reichenbach 44]}:
H. Reichenbach,
{\em Philosophical foundations of quantum mechanics},
University of California Press, Los Angeles, 1944.

\item {\bf [Reid-Walls 86]}:
M. D. Reid, \& D. F. Walls,
``Violations of classical inequalities in quantum optics'',
{\em Phys. Rev. A} {\bf 34}, 2, 1260-1276 (1986).

\item {\bf [Reid-Munro 92]}:
M. D. Reid, \& W. J. Munro,
``Macroscopic boson states exhibiting the Greenberger-Horne-Zeilinger
contradiction with local realism'',
{\em Phys. Rev. Lett.} {\bf 69}, 7, 997-1001 (1992).

\item {\bf [Reid 99]}:
M. D. Reid,
``Quantum cryptography using continuous variable Einstein-Podolsky-Rosen
correlations and quadrature phase amplitude measurements'',
quant-ph/9909030.

\item {\bf [Reid 00 a]}:
M. D. Reid,
``Macroscopic elements of reality and the Einstein-Podolsky-Rosen
paradox'',
{\em Quantum. Semiclass. Opt.} {\bf 9}, 3, 489-499 (1997).

\item {\bf [Reid 00 b]}:
M. D. Reid,
``Incompatibility of macroscopic local realism with quantum mechanics
in measurements with macroscopic uncertainties'',
{\em Phys. Rev. Lett.} {\bf 84}, 13, 2765-2769 (2000);
quant-ph/0010023.

\item {\bf [Reid 00 c]}:
M. D. Reid,
``Violations of Bell inequalities for measurements with macroscopic
uncertainties: What it means to violate macroscopic local realism'',
{\em Phys. Rev. A} {\bf 62}, 2, 022110 (2000);
quant-ph/0009016.

\item {\bf [Reid 00 d]}:
M. D. Reid,
``Quantum cryptography with a predetermined key,
using continuous-variable Einstein-Podolsky-Rosen correlations'',
{\em Phys. Rev. A} {\bf 62}, 6, 062308 (2000);
quant-ph/9909030.

\item {\bf [Reid 01 a]}:
M. D. Reid,
``New tests of macroscopic local realism using
continuous variable measurements'',
quant-ph/0101050.

\item {\bf [Reid 01 b]}:
M. D. Reid,
``Proposal to test quantum mechanics against macroscopic
realism using continuous variable entanglement:
A definitive signature of a Schr\"{o}dinger cat'',
quant-ph/0101051.

\item {\bf [Reid 01 c]}:
M. D. Reid,
``Inseparability criteria for demonstration of the
Einstein-Podolsky-Rosen gedanken experiment'',
quant-ph/0103142.

\item {\bf [Reid 01 d]}:
M. D. Reid,
``The Einstein-Podolsky-Rosen paradox and entanglement 1:
Signatures of EPR correlations for continuous variables'',
quant-ph/0112038.

\item {\bf [Reid 01 e]}:
M. D. Reid,
``The Einstein-Podolsky-Rosen paradox and entanglement 2:
Application to proof of security for continuous variable quantum cryptography'',
quant-ph/0112039.

\item {\bf [Reid-Munro-De Martini 02]}:
M. D. Reid, W. J. Munro, \& F. De Martini,
``Violation of multiparticle Bell inequalities for low- and high-flux
parametric amplification using both vacuum and entangled input states'',
{\em Phys. Rev. A} {\bf 66}, 3, 033801 (2002);
quant-ph/0104139.

\item {\bf [Reina-Quiroga-Johnson 00 a]}:
J. H. Reina, L. Quiroga, \& N. F. Johnson,
``Quantum entanglement and information processing via
excitons in optically driven quantum dots'',
{\em Phys. Rev. A} {\bf 62}, 1, 012305 (2000);
quant-ph/9911123.

\item {\bf [Reina-Quiroga-Johnson 00 b]}:
J. H. Reina, L. Quiroga, \& N. F. Johnson,
``NMR-based nanostructure switch for quantum logic'',
quant-ph/0003014.

\item {\bf [Reina-Quiroga-Johnson 00 c]}:
J. H. Reina, L. Quiroga, \& N. F. Johnson,
``Quantum information processing in semiconductor
nanostructures'',
quant-ph/0009035.

\item {\bf [Reina-Johnson 01]}:
J. H. Reina, \& N. F. Johnson,
``Quantum teleportation in a solid-state system'',
{\em Phys. Rev. A} {\bf 63}, 1, 012303 (2001).

\item {\bf [Reina-Quiroga-Johnson 02]}:
J. H. Reina, L. Quiroga, \& N. F. Johnson,
``Decoherence of quantum registers'',
{\em Phys. Rev. A} {\bf 65}, 3, 032326 (2002);
quant-ph/0105029.

\item {\bf [Reina-Beausoleil-Spiller-Munro 04]}:
J. H. Reina, R. G. Beausoleil, T. P. Spiller, \& W. J. Munro,
``Radiative corrections and quantum gates in molecular systems'',
quant-ph/0408066.

\item {\bf [Reimpell-Werner 03]}:
M. Reimpell, \& R. F. Werner,
``Iterative optimization of quantum error correcting codes'',
quant-ph/0307138.

\item {\bf [Reinisch 99]}:
G. Reinisch,
``Stern-Gerlach experiment as the pioneer - and probably the simplest -
quantum entanglement test?'',
{\em Phys. Lett. A} {\bf 259}, 6, 427-430 (1999).

\item {\bf [Rekdal-Skagerstam-Knight 03]}:
P. K. Rekdal, B.-S. K. Skagerstam, \& P. L. Knight,
``On the preparation of pure states in resonant microcavities'',
quant-ph/0301148.

\item {\bf [Rembieli\'{n}ski 80]}:
J. Rembieli\'{n}ski,
``?'',
{\em Phys. Lett. A} {\bf 78}, ?, 33-? (1980).

\item {\bf [Rembieli\'{n}ski 97]}:
J. Rembieli\'{n}ski,
``?'',
{\em Int. J. Mod. Phys.} {\bf 12}, ?, 1677-? (1997);
hep-th/9607232.

\item {\bf [Rembieli\'{n}ski-Caban 97]}:
J. Rembieli\'{n}ski, \& P. Caban,
``The preferred frame and Poincar\'{e} symmetry'',
in H.-D. Doebner, W. Scherer, \& P. Nattermann (eds.),
{\em Physical applications and mathematical aspects of geometry,
groups and algebras},
World Scientific, Singapore, 1997, pp.~349-?;
hep-th/9612072.

\item {\bf [Rembieli\'{n}ski 00]}:
J. Rembieli\'{n}ski,
``Superluminal phenomena and the quantum preferred frame'',
quant-ph/00010026.

\item {\bf [Rembieli\'{n}ski-Smolinski 02]}:
J. Rembieli\'{n}ski, \& K. A. Smolinski,
``Einstein-Podolsky-Rosen correlations of spin measurements in two moving
inertial frames'',
{\em Phys. Rev. A} {\bf 66}, 5, 052114 (2002).

\item {\bf [Ren-Guo-Li-Guo 04]}:
X.-F. Ren, G.-P. Guo, J. Li, \& G.-C. Guo,
``Entanglement of the Hermite-Gaussian modes states of photons'',
quant-ph/0407225.

\item {\bf [Renes-Blume Kohout-Scott-Caves 03]}:
J. M. Renes, R. Blume-Kohout, A. J. Scott, \& C. M. Caves,
``Symmetric informationally complete quantum measurements'',
quant-ph/0310075.

\item {\bf [Renes 03]}:
J. M. Renes,
``Quantum key distribution using the trine ensemble'',
quant-ph/0311106.

\item {\bf [Renes 04 a]}:
J. M. Renes,
``Spherical code key distribution protocols for qubits'',
quant-ph/0402135.

\item {\bf [Renes 04 b]}:
J. M. Renes,
``Equiangular spherical codes in quantum cryptography'',
quant-ph/0409043.

\item {\bf [Reninger 60]}:
M. Renninger,
``?'',
{\em Zeitschrift f\"{u}r Physik} {\bf 158}, ?, 417-? (1960).

\item {\bf [Resch-Lundeen-Steinberg 01]}:
K. J. Resch, J. S. Lundeen, \& A. M. Steinberg,
``Experimental observation of nonclassical effects on
single-photon detection rates'',
{\em Phys. Rev. A} {\bf 63}, 2, 020102(R) (2001);
quant-ph/0006056.

\item {\bf [Resch-Lundeen-Steinberg 02 a]}:
K. J. Resch, J. S. Lundeen, \& A. M. Steinberg,
``Quantum state preparation and conditional coherence'',
{\em Phys. Rev. Lett.} {\bf 88}, 11, 113601 (2002).

\item {\bf [Resch-Lundeen-Steinberg 02 b]}:
K. J. Resch, J. S. Lundeen, \& A. M. Steinberg,
``Conditional-phase switch at the single-photon level'',
{\em Phys. Rev. Lett.} {\bf 89}, 3, 037904 (2002).

\item {\bf [Resch-Lundeen-Steinberg 02 c]}:
K. J. Resch, J. S. Lundeen, \& A. M. Steinberg,
``Practical creation and detection of polarization Bell states using parametric down-conversion'',
{\em Proc.\ Solvay Conference};
quant-ph/0204034.

\item {\bf [Resch-Steinberg 03]}:
K. J. Resch, \& A. M. Steinberg,
``Extracting joint weak values with local, single-particle measurements'',
quant-ph/0310113.

\item {\bf [Resch-Lundeen-Steinberg 04]}:
K. J. Resch, J. S. Lundeen, \& A. M. Steinberg,
``Experimental realization of the quantum box problem'',
{\em Phys. Lett. A} {\bf 324}, 2-3, 125-131 (2004);
quant-ph/0310091.

\item {\bf [Retamal-Guerra 99]}:
J. C. Retamal, \& E. S. Guerra,
``A protocol to preserve quantum coherence in cavity QED'',
quant-ph/9905035.

\item {\bf [Retamal-Zagury 01]}:
J. C. Retamal, \& N. Zagury,
``Stability of quantum states under dissipation'',
{\em Phys. Rev. A} {\bf 63}, 3, 032106 (2001).

\item {\bf [Retzker-Cirac-Reznik 04]}:
A. Retzker, J. I. Cirac, \& B. Reznik,
``Detection of vacuum entanglement in a linear ion trap'',
quant-ph/0408059.

\item {\bf [Revzen-Mann 96]}:
M. Revzen, \& A. Mann,
``Bell's inequality for a single particle'',
{\em Found. Phys.} {\bf 26}, 6, 847-850 (1996).
See {\bf [Revzen-Mann 97]}.

\item {\bf [Revzen-Mann 97]}:
M. Revzen, \& A. Mann,
``Bell's inequality for a particle'',
in M. Ferrero, \& A. van der Merwe (eds.),
{\em New developments on fundamental problems in quantum physics
(Oviedo, Spain, 1996)},
Kluwer Academic, Dordrecht, Holland, 1997, pp.~347-354.
See {\bf [Revzen-Mann 96]}.

\item {\bf [Revzen-Lokaj\'{\i}\v{c}ek-Mann 97]}:
M. Revzen, M. Lokaj\'{\i}\v{c}ek, \& A. Mann,
``Bell's inequality and operator's noncommutativity'',
{\em Quantum Semiclass. Opt.} {\bf 9}, 3, 501-506 (1997).

\item {\bf [Revzen-Mann 03]}:
M. Revzen, \& A. Mann,
``Relative phase determination and clocks synchronization via shared entanglement'',
{\em Phys. Lett. A} {\bf 312}, 1-2, 11-15 (2003).

\item {\bf [Rezakhani-Siadatnejad-Ghaderi 03]}:
A. T. Rezakhani, S. Siadatnejad, \& A. H. Ghaderi,
``Separability in asymmetric phase-covariant cloning'',
quant-ph/0312024.

\item {\bf [Reznik-Aharonov 95]}:
B. Reznik, \& Y. Aharonov,
``Time-symmetric formulation of quantum mechanics'',
{\em Phys. Rev. A} {\bf 52}, 4, 2538-2550 (1995).

\item {\bf [Reznik 96]}:
B. Reznik,
``Unitary evolution between pure and mixed states'',
{\em Phys. Rev. Lett.} {\bf 76}, 8, 1192-1195 (1996).

\item {\bf [Reznik-Aharonov-Groisman 02]}:
B. Reznik, Y. Aharonov, \& B. Groisman,
``Remote operations and interactions for systems of arbitrary-dimensional
Hilbert space: State-operator approach'',
{\em Phys. Rev. A} {\bf 65}, 3, 032312 (2002);
quant-ph/0107143.

\item {\bf [Reznik 96]}:
B. Reznik,
``Remote generalized measurements (POVMs) require non-maximal entanglement'',
quant-ph/0203055.

\item {\bf [Reznik 02]}:
B. Reznik,
``Entanglement from the vacuum'',
{\em Found. Phys.};
quant-ph/0212044.

\item {\bf [Reznik-Retzker-Silman 03]}:
B. Reznik, A. Retzker, \& J. Silman,
``Violation of Bell's inequalities in the vacuum'',
quant-ph/0310058.

\item {\bf [Ribeiro-Milman-Mosseri 04]}:
P. Ribeiro, P. Milman, \& R. Mosseri,
``Aperiodic quantum random walks'',
quant-ph/0406071.

\item {\bf [Ribordy-Gautier-Gisin-(+2) 98]}:
G. Ribordy, J-D. Gautier, N. Gisin, O. Guinnard, \& H. Zbinden,
``Automated `plug \& play' quantum key distribution'',
{\em Electron. Lett.} {\bf 34}, 22, 2116-2117 (1998);
quant-ph/9812052.

\item {\bf [Ribordy-Gautier-Gisin-(+2) 00]}:
G. Ribordy, J-D. Gautier, N. Gisin, O. Guinnard, \& H. Zbinden,
``Fast and user-friendly quantum key distribution'',
in V. Bu\v{z}zek, \& D. P. DiVincenzo (eds.),
{\em J. Mod. Opt.} {\bf 47}, 2-3 (Special issue:
Physics of quantum information), 517-531 (2000);
quant-ph/9905056.

\item {\bf [Ribordy-Gisin-Zbinden 00]}:
G. Ribordy, N. Gisin, \& H. Zbinden,
``Quantum key distribution'',
{\bf [Macchiavello-Palma-Zeilinger 00]}, pp.~235-239.

\item {\bf [Ribordy-Brendel-Gautier-(+2) 01]}:
G. Ribordy, J. Brendel, J-D. Gautier, N. Gisin, \& H. Zbinden,
``Long-distance entanglement-based quantum key distribution'',
{\em Phys. Rev. A} {\bf 63}, 1, 012309 (2001);
quant-ph/0008039.

\item {\bf [Ricci-Sciarrino-Sias-De Martini 03]}:
M. Ricci, F. Sciarrino, C. Sias, \& F. De Martini,
``Teleportation scheme implementing contextually the universal optimal
quantum cloning machine and the universal not gate. Complete experimental
realization'',
quant-ph/0310021.
See {\bf [Ricci-Sciarrino-Sias-De Martini 04]}.

\item {\bf [Ricci-Sciarrino-Sias-De Martini 04]}:
M. Ricci, F. Sciarrino, C. Sias, \& F. De Martini,
``Teleportation scheme implementing the universal optimal quantum cloning
machine and the universal NOT gate'',
{\em Phys. Rev. Lett.} {\bf 92}, 4, 047901 (2004);
quant-ph/0304070.
See {\bf [Ricci-Sciarrino-Sias-De Martini 03]}.

\item {\bf [Ricci-Martini-Cerf-(+3) 04]}:
M. Ricci, F. De Martini, N. J. Cerf,
R. Filip, J. Fiur\'{a}\v{s}ek, \& C. Macchiavello,
``Experimental purification of single qubits'',
quant-ph/0403118.

\item {\bf [Rice 97 a]}:
D. A. Rice,
``A geometric approach to nonlocality in the Bohm
model of quantum mechanics'',
{\em Am. J. Phys.} {\bf 65}, 2, 144-147 (1997).

\item {\bf [Rice 97 b]}:
D. A. Rice,
``Nonlocality and conservation laws in hidden variable theories'',
{\em Found. Phys.} {\bf 27}, 10, 1345-1353 (1997).

\item {\bf [Rice-Sanders 98]}:
D. A. Rice, \& B. C. Sanders,
``Complementarity and entangled coherent states'',
{\em Quantum Semiclass. Opt.} {\bf 10}, 3, L41-L47 (1998).

\item {\bf [Rice-Jaeger-Sanders 00]}:
D. A. Rice, G. Jaeger, \& B. C. Sanders,
``Two-coherent-state interferometry'',
{\em Phys. Rev. A} {\bf 62}, 1, 012101 (2000).

\item {\bf [Riebe-H\"{a}ffner-Roos-(+8) 04]}:
M. Riebe, H. H\"{a}ffner, C. F. Roos,
W. H\"{a}nsel, J. Benhelm, G. P. T. Lancaster,
T. W. K\"{o}rber, C. Becher, F. Schmidt-Kaler,
D. F. V. James, \& R. Blatt,
``Deterministic quantum teleportation with atoms'',
{\em Nature} {\bf 429}, ?, 734-737 (2004).
See {\bf [Barrett-Chiaverini-Schaetz-(+8) 04]}.

\item {\bf [de Riedmatten-Marcikic-Tittel-(+2) 02]}:
H. de Riedmatten, I. Marcikic, W. Tittel, H. Zbinden, \& N. Gisin,
``Quantum interference with photon pairs created in spatially separated sources'',
{\em Phys. Rev. A} {\bf 67}, 2, 022301 (2003);
quant-ph/0208174.

\item {\bf [de Riedmatten-Scarani-Marcikic-(+4) 03]}:
H. de Riedmatten, V. Scarani, I. Marcikic,
A. Ac\'{\i}n, W. Tittel, H. Zbinden, \& N. Gisin,
``Two independent photon pairs versus four-photon entangled states in
parametric down conversion'',
quant-ph/0310167.

\item {\bf [de Riedmatten-Marcikic-Tittel-(+3) 04]}:
H. de Riedmatten, I. Marcikic, W. Tittel,
H. Zbinden, D. Collins, \& N. Gisin,
``Long distance quantum teleportation in a quantum relay configuration'',
{\em Phys. Rev. Lett.} {\bf 92}, 4, 047904 (2004);
quant-ph/0309218.

\item {\bf [de Riedmatten-Marcikic-Scarani-(+3) 04]}:
H. de Riedmatten, I. Marcikic, V. Scarani,
W. Tittel, H. Zbinden, \& N. Gisin,
``Tailoring photonic entanglement in high-dimensional Hilbert spaces'',
{\em Phys. Rev. A} {\bf 69}, 5, 050304 (2004);
quant-ph/0309058.

\item {\bf [de Riedmatten-Marcikic-van Houwelingen-(+3) 04]}:
H. de Riedmatten, I. Marcikic, J. A. W. van Houwelingen,
W. Tittel, H. Zbinden, \& N. Gisin,
``Long distance entanglement swapping with photons from separated sources'',
quant-ph/0409093.

\item {\bf [Rieffel-Polak 98]}:
E. G. Rieffel, \& Polak,
``An introduction to quantum computing for non-physicists'',
submitted to {\em ACM Computing Surveys};
quant-ph/9809016.

\item {\bf [Rigden 00]}:
J. S. Rigden,
``\,`Geons, black holes, and quantum foam: A life in physics'
by John Archibald Wheeler with Kenneth Ford'',
{\em Am. J. Phys.} {\bf 68}, 6, 585 (2000).
Review of {\bf [Wheeler-Ford 98]}.

\item {\bf [Rigo-Plastino-Plastino-Casas 00]}:
A. Rigo, A. R. Plastino, A. Plastino, \& M. Casas,
``On the inference of entangled states:
A maximum-entropy-minimum-norm approach'',
{\em Phys. Lett. A} {\bf 270}, 1-2, 1-9 (2000).

\item {\bf [Rigolin 02]}:
G. Rigolin,
``Uncertainty relations for entangled states '',
{\em Found. Phys. Lett.} {\bf 15}, 3, 293-298 (2002).
quant-ph/0105057.

\item {\bf [Rigolin-Escobar 04]}:
G. Rigolin, \& C. O. Escobar,
``Lower bounds on the entanglement of formation for general Gaussian
states'',
{\em Phys. Rev. A} {\bf 69}, 1, 012307 (2004);
quant-ph/0307023.

\item {\bf [Rigolin 04]}:
G. Rigolin,
``Thermal entanglement in the two-qubit Heisenberg XYZ model'',
{\em Int. J. Quant. Inf.} {\bf 2}, ?, 393-? (2004);
quant-ph/0311185.

\item {\bf [De Rinaldis-D'Amico-Biolatti-(+3) 02]}:
S. De Rinaldis, I. D'Amico, E. Biolatti,
R. Rinaldi, R. Cingolani, \& F. Rossi,
``Intrinsic exciton-exciton coupling in GaN-based quantum dots: Application to
solid-state quantum computing'',
{\em Phys. Rev. B} {\bf 65}, 8, 081309 (2002).

\item {\bf [De Rinaldis 04]}:
S. De Rinaldis,
``Distinguishability of complete and unextendible product bases'',
{\em Phys. Rev. A} {\bf 70}, 2, 022309 (2004);
quant-ph/0304027.

\item {\bf [Ringo 97]}:
R. Ringo,
``A possible Einstein-Podolsky-Rosen probe of the
momentum-position uncertainty relation'',
quant-ph/9712006.

\item {\bf [Risco Delgado 97]}:
R. Risco Delgado,
``Localidad y no localidad bajo
recientes resultados te\'{o}ricos y experimentales'',
Ph.\ D. thesis, Universidad de Sevilla, 1997.

\item {\bf [Risco Delgado 02]}:
R. Risco Delgado,
``Bell's inequalities and indeterminism'',
quant-ph/0202099.

\item {\bf [Rivest-Shamir-Adleman 78]}:
R. L. Rivest, A. Shamir, L. M. Adleman,
``?'',
{\em Communications of the ACM} {\bf 21}, 120-? (1978).

\item {\bf [Roa-Delgado-Fuentes Guridi 03]}:
L. Roa, A. Delgado, \& I. Fuentes-Guridi,
``Optimal conclusive teleportation of quantum states'',
{\em Phys. Rev. A} {\bf 68}, 2, 022310 (2003).

\item {\bf [Roa-Retamal-Saavedra 02]}:
L. Roa, J. C. Retamal, \& C. Saavedra,
``Quantum-state discrimination'',
{\em Phys. Rev. A} {\bf 66}, 1, 012103 (2002).

\item {\bf [Robertson 29]}:
H. P. Robertson,
``The uncertainty principle'',
{\em Phys. Rev.} {\bf 34}, 1, 163-164 (1929).
Reprinted in {\bf [Wheeler-Zurek 83]}, pp.~127-128.

\item {\bf [Robinson 82]}:
A. L. Robinson,
``Quantum mechanics passes another test'',
{\em Science} {\bf 217}, 4558, 435-436 (1982).

\item {\bf [Robinson 83]}:
A. L. Robinson,
``Loophole closed in quantum mechanics test'',
{\em Science} {\bf 219}, 4580, 40-41 (1983).

\item {\bf [Robinson 86]}:
A. L. Robinson,
``Testing superposition in quantum mechanics'',
{\em Science} {\bf 231}, 4744, 1370-1372 (1986).

\item {\bf [Rodionov-Chirkin 04]}:
A. V. Rodionov, \& A. S. Chirkin,
``Entangled photon states in consecutive nonlinear optical interactions'',
{\em JETP Lett.} {\bf 79}, 253-256 (2004).
Erratum: {\em JETP Lett.} {\bf 79}, 582 (2004).

\item {\bf [Roehrig 98]}:
H. Roehrig,
``An upper bound for searching an ordered list'',
quant-ph/9812061.

\item {\bf [Rohrlich-Popescu 96]}:
D. Rohrlich, \& S. Popescu,
``Nonlocality as an axiom for quantum theory'',
in A. Mann, \& M. Revzen (eds.),
{\em The dilemma of Einstein, Podolsky and Rosen -- 60 years
later. An international symposium in honour of Nathan Rosen
(Haifa, Israel, 1995)},
{\em Ann. Phys. Soc. Israel} {\bf 12}, 152-156 (1996);
quant-ph/9508009.
See {\bf [Popescu-Rohrlich 94]}.

\item {\bf [Rohrlich 01]}:
D. Rohrlich,
``Thermodynamical analogues in quantum information theory'',
{\em Optics and Spectroscopy} {\bf 91}, ?, 394-? (2001).
quant-ph/0107026.

\item {\bf [Rohrlich 83]}:
F. Rohrlich,
``Facing quantum mechanical reality'',
{\em Science} {\bf 221}, 4617, 1251-1255 (1983).

\item {\bf [Rohrlich 85]}:
F. Rohrlich,
in ``Reality and the quantum theory'',
{\em Phys. Today} {\bf 38}, 11, 11-15 (1985).
Comment on {\bf [Mermin 85]}.

\item {\bf [Rojo 99]}:
A. G. Rojo,
``El jard\'{\i}n de los mundos que se ramifican: Borges y la Mec\'{a}nica Cu\'{a}ntica'',
in {\em Borges en 10 miradas},
Fundaci\'{o}n El Libro, Buenos Aires, 1999.

\item {\bf [Rojo 03]}:
A. G. Rojo,
``Optimally squeezed spin states'',
{\em Phys. Rev. A} {\bf 68}, 1, 013807 (2003).

\item {\bf [Roko-Bu\v{z}zek-Chouha-Hillery 03]}:
M. Roko, V. Bu\v{z}zek, P. R. Chouha, \& M. Hillery,
``Generalized measurements via a programmable quantum processor'',
{\em Phys. Rev. A} {\bf 68}, 6, 062302 (2003).

\item {\bf [Roland-Cerf 02]}:
J, Roland, \& N, J. Cerf,
``Quantum search by local adiabatic evolution'',
{\em Phys. Rev. A} {\bf 65}, 4, 042308 (2002);
quant-ph/0107015.

\item {\bf [Roland-Cerf 03 a]}:
J. Roland, \& N. J. Cerf,
``Quantum-circuit model of Hamiltonian search algorithms'',
{\em Phys. Rev. A} {\bf 68}, 6, 062311 (2003);
quant-ph/0302138.

\item {\bf [Roland-Cerf 03 b]}:
J. Roland, \& N. J. Cerf,
``Adiabatic quantum search algorithm for structured problems'',
{\em Phys. Rev. A} {\bf 68}, 6, 062312 (2003);
quant-ph/0304039.

\item {\bf [Rolston-Phillips 02]}:
S. L. Rolston, \& W. D. Phillips,
``Nonlinear and quantum atom optics'',
{\em Nature} {\bf 416}, 6877, 219-224 (2002).

\item {\bf [Romer 91]}:
R. H. Romer,
``Editorial: John S. Bell (1928-1990), the man
who proved Einstein was right'',
{\em Am. J. Phys.} {\bf 59}, 4, 299-300 (1991).

\item {\bf [Romero-Roa-Retamal-Saavedra 02]}:
J. L. Romero, L. Roa, J. C. Retamal, \& C. Saavedra,
``Entanglement purification in cavity QED using local operations'',
{\em Phys. Rev. A} {\bf 65}, 5, 052319 (2002).

\item {\bf [Roos-Riebe-H\"{a}ffner-(+6) 04]}:
C. F. Roos, M. Riebe, H. H\"{a}ffner,
W. H\"{a}nsel, J. Benhelm, G. P. T. Lancaster,
C. Becher, F. Schmidt-Kaler, \& R. Blatt,
``Control and measurement of three-qubit entangled states'',
{\em Science} {\bf 304}, ?, 1478-1481 (2004).

\item {\bf [Roos-Lancaster-Riebe-(+7) 04]}:
C. F. Roos, G. P. T. Lancaster, M. Riebe,
H. H\"{a}ffner, W. H\"{a}nsel, S. Gulde,
C. Becher, J. Eschner, F. Schmidt-Kaler, \& R. Blatt,
``Bell states of atoms with ultralong lifetimes and their tomographic state analysis'',
{\em Phys. Rev. Lett.} {\bf 92}, 22, 220402 (2004);
quant-ph/0307210.

\item {\bf [Roos-M{\o}lmer 04]}:
I. Roos, \& K. M{\o}lmer,
``Quantum computing with an inhomogeneously broadened ensemble of ions:
 Suppression of errors from detuning variations by specially adapted pulses and coherent population trapping'',
{\em Phys. Rev. A} {\bf 69}, 2, 022321 (2004);
quant-ph/0305060.

\item {\bf [Roscilde-Verrucchi-Fubini-(+2) 04]}:
T. Roscilde, P. Verrucchi, A. Fubini,
S. Haas, \& V. Tognetti,
``Studying quantum spin systems through entanglement estimators'',
{\em Phys. Rev. Lett.} {\bf 93}, 16, 167203 (2004);
cond-mat/0404403.

\item {\bf [R\"{o}seberg 94]}:
U. R\"{o}seberg,
``Hidden historicity: The challenge of Bohr's philosophical
thought'',
in {\bf [Faye-Folse 94]}, pp.~325-343.

\item {\bf [R\"{o}seberg 95]}:
U. R\"{o}seberg,
``Did they just misunderstood each other?'',
in K. Gavroglu, J. Stachel, \& M. W. Wartofsky (eds.),
{\em Physics, philosophy and the scientific community},
Kluwer Academic, Dordrecht, Holland, 1995, pp.~105-123.

\item {\bf [Rosen 84]}:
N. Rosen,
``A semiclassical interpretation of wave mechanics'',
{\em Found. Phys.} {\bf 14}, 7, 579-605 (1984).

\item {\bf [Rosen 85]}:
N. Rosen,
``Quantum mechanics and reality'',
in P. J. Lahti, \& P. Mittelstaedt (eds.),
{\em Symp.\ on the Foundations of Modern
Physics: 50 Years of the Einstein-Podolsky-Rosen Experiment
(Joensuu, Finland, 1985)},
World Scientific, Singapore, 1985, pp.~17-33.

\item {\bf [Rosen 94]}:
N. Rosen,
``Bell's theorem and quantum mechanics'',
{\em Am. J. Phys.} {\bf 62}, 2, 109-110 (1994).

\item {\bf [Rosen 96]}:
N. Rosen,
``Some reflections on the EPR work'',
in A. Mann, \& M. Revzen (eds.),
{\em The dilemma of Einstein, Podolsky and Rosen -- 60 years
later. An international symposium in honour of Nathan Rosen
(Haifa, Israel, 1995)},
{\em Ann. Phys. Soc. Israel} {\bf 12}, 23-26 (1996).

\item {\bf [Rosenblum-Kuttner 02]}:
B. Rosenblum, \& F. Kuttner,
``The observer in the quantum experiment'',
{\em Found. Phys.} {\bf 32}, 8, 1273-1293 (2002);
quant-ph/0011086.

\item {\bf [Rosinger 04 a]}:
E. E. Rosinger,
{\em Basics of quantum computation (Part 1)};
quant-ph/0407064.

\item {\bf [Rosinger 04 b]}:
E. E. Rosinger,
`What is wrong with von Neumann's theorem on ``no hidden variables''\,',
quant-ph/0408191.

\item {\bf [Ro\v{s}ko-Bu\v{z}zek-Chouha-Hillery 03]}:
M. Ro\v{s}ko, V. Bu\v{z}zek, P. R. Chouha, \& M. Hillery,
``Generalized measurements via programmable quantum processor'',
quant-ph/0311172.

\item {\bf [Ross 98]}:
D. A. Ross,
``A modification of Grover's algorithm as a fast database search'',
quant-ph/9807078.

\item {\bf [Rossi-Paris 04]}:
A. R. Rossi, \& M. G. A. Paris,
``About distillability of depolarized states'',
quant-ph/0401167.

\item {\bf [Rossi-Olivares-Paris 04 a]}:
A. R. Rossi, S. Olivares, \& M. G. A. Paris,
``Degradation of continuous variable entanglement in a phase-sensitive
environment'',
in R. Bonifacio, B. G. Englert, \& M. G. A. Paris (eds.),
{\em Mysteries, Puzzles and Paradoxes in Quantum
Mechanics (Garda Lake, Italy, 2003)},
{\em J. Mod. Opt.};
quant-ph/0401172.

\item {\bf [Rossi-Olivares-Paris 04 b]}:
A. R. Rossi, S. Olivares, \& M. G. A. Paris,
``Photon statistic without counting photons'',
quant-ph/0405139.

\item {\bf [Rossignoli-Canosa 02]}:
R. Rossignoli, \& N. Canosa,
``Generalized entropic criterion for separability'',
{\em Phys. Rev. A} {\bf 66}, 4, 042306 (2002).

\item {\bf [Rossignoli-Canosa 03]}:
R. Rossignoli, \& N. Canosa,
``Violation of majorization relations in entangled states and its detection by
means of generalized entropic forms'',
{\em Phys. Rev. A} {\bf 67}, 4, 042302 (2003).

\item {\bf [Rothman-Sudarshan 00]}:
A. Rothman, \& E. C. G. Sudarshan,
``Hidden variables or positive probabilities?'',
quant-ph/0004109.

\item {\bf [Rotter 00]}:
I. Rotter,
``Branch points in the complex plane and information
loss in quantum systems at high level density'',
quant-ph/0011008.

\item {\bf [Rotter 01]}:
I. Rotter,
``Correlations in quantum systems and branch points in the complex
plane'',
quant-ph/0107018.

\item {\bf [Roussel 88]}:
P. Roussel,
``Some properties of the functions satisfying
Bell's inequalities in relation to quantum mechanics'',
in A. van der Merwe, F. Selleri,
\& G. Tarozzi (eds.),
{\em Microphysical reality and quantum formalism.
Proc.\ of an international conference (Urbino, Italy, 1985)},
Kluwer Academic, Dordrecht, Holland, 1988, vol. 2, pp.~421-432.

\item {\bf [Rowe-Kielpinski-Meyer-(+4) 01]}:
M. A. Rowe, D. Kielpinski, V. Meyer, C. A. Sackett,
 W. M. Itano, C. Monroe, \& D. J. Wineland,
``Experimental violation of a Bell's inequality with
efficient detection'',
{\em Nature} {\bf 409}, 6822, 791-794 (2001).
See {\bf [Sackett-Kielpinski-King-(+8) 00]},
{\bf [Grangier 01 a]}, {\bf [Vaidman 01 b]}, {\bf [Santos 01]}.

\item {\bf [Roy-Sen-Sen 01]}:
A. Roy, A. Sen, \& U. Sen,
``Is it possible to clone using an arbitrary blank state?'',
{\em Phys. Lett. A} {\bf 286}, 1, 1-3 (2001);
quant-ph/0103093.

\item {\bf [Roy 98 a]}:
B. Roy,
``Nonclassical properties of the real and imaginary nonlinear
Schr\"{o}dinger cat states'',
{\em Phys. Lett. A} {\bf 249}, 1-2, 25-29 (1998).

\item {\bf [Roy-Singh 78]}:
S. M. Roy, \& V. Singh,
``Experimental tests of
quantum mechanics versus local hidden variable theories'',
{\em J. Phys. A} {\bf 11}, 8, L167-L171 (1978).

\item {\bf [Roy-Singh 79]}:
S. M. Roy, \& V. Singh,
``Completeness of tests of local hidden variable theories'',
{\em J. Phys. A} {\bf 12}, 7, 1003-1009 (1979).

\item {\bf [Roy-Singh 89]}:
S. M. Roy, \& V. Singh,
``Hidden variable theories
without non-local signalling and their experimental tests'',
{\em Phys. Lett. A} {\bf 139}, 9, 437-441 (1989).

\item {\bf [Roy-Singh 91]}:
S. M. Roy, \& V. Singh,
``Tests of signal locality and
Einstein-Bell locality for multiparticle systems'',
{\em Phys. Rev. Lett.} {\bf 67}, 20, 2761-2764 (1991).
Comment: {\bf [Peacock 92]}.
See {\bf [Mermin 90 c]}.

\item {\bf [Roy-Singh 93]}:
S. M. Roy, \& V. Singh,
``Quantum violation of
stochastic noncontextual hidden-variable theories'',
{\em Phys. Rev. A} {\bf 48}, 4, 3379-3381 (1993).

\item {\bf [Roy-Singh 99]}:
S. M. Roy, \& V. Singh,
``Maximally realistic causal quantum mechanics'',
{\em Phys. Lett. A} {\bf 255}, 4-6, 201-208 (1999);
quant-ph/9811041.

\item {\bf [Roy 98 b]}:
S. M. Roy,
``Maximally causal quantum mechanics'',
{\em Pramana};
quant-ph/9811047.

\item {\bf [Roy 99]}:
S. M. Roy,
``Contextual deterministic quantum mechanics'',
quant-ph/9908062.

\item {\bf [Roy 04]}:
S. M. Roy,
``$N$-partite separability inequalities exponentially stronger than local
reality inequalities'',
quant-ph/0401048.

\item {\bf [Royer 85]}:
A. Royer,
``Measurement of the Wigner function'',
{\em Phys. Rev. Lett.} {\bf 55}, 25, 2745-2748 (1985).

\item {\bf [Royer 89]}:
A. Royer,
``Measurements of quantum state and the Wigner function'',
{\em Found. Phys.} {\bf 19}, 1, 3-32 (1989).

\item {\bf [Royer 94]}:
A. Royer,
``Reversible quantum measurements on a spin
$\frac{1}{2}$ and measuring the state of a single system'',
{\em Phys. Rev. Lett.} {\bf 73}, 7, 913-917 (1994).

\item {\bf [Ruark 35]}:
A. E. Ruark,
``Is the quantum mechanical description of physical reality complete?'',
{\em Phys. Rev.} {\bf 48}, 5, 466-467 (1935).

\item {\bf [Rubin 01 a]}:
M. A. Rubin,
``Locality in the Everett interpretation
of Heisenberg-picture quantum mechanics'',
{\em Found. Phys. Lett.} {\bf 14}, 4, 301-322 (2001);
quant-ph/0103079.

\item {\bf [Rubin 01]}:
M. A. Rubin,
``Exponentially fast quantum search for a specified number of
targets'',
quant-ph/0104082.

\item {\bf [Rubin 02]}:
M. A. Rubin,
``Locality in the Everett interpretation
of quantum field theory'',
{\em Found. Phys.} {\bf 32}, 10, 1495-1523 (2001).

\item {\bf [Rubin 03]}:
M. A. Rubin,
``Relative frequency and probability in the Everett interpretation
of Heisenberg-picture quantum mechanics'',
{\em Found. Phys.} {\bf 33}, 3, 379-405 (2003).

\item {\bf [Rubin 04]}:
M. A. Rubin,
``There is no basis ambiguity in Everett quantum mechanics'',
{\em Found. Phys. Lett.} {\bf 17}, 4, 323-341 (2004);
quant-ph/0310186.

\item {\bf [Rubin-Shih 92]}:
M. H. Rubin, \& Y. H. Shih,
``Models of a two-photon
Einstein-Podolsky-Rosen interference experiment'',
{\em Phys. Rev. A} {\bf 45}, 11, 8138-8147 (1992).

\item {\bf [Rubin 00 a]}:
M. H. Rubin,
``Entanglement and state preparation'',
{\em Phys. Rev. A} {\bf 61}, 2, 022311 (2000);
quant-ph/9901071.

\item {\bf [Rubin 00 b]}:
M. H. Rubin,
``Measurement of entanglement states and state preparation'',
{\em Fortschr. Phys.} {\bf 48}, 5-7, 473-479 (2000).

\item {\bf [Rubin 02]}:
M. H. Rubin,
``Locality in the Everett interpretation of quantum field theory'',
quant-ph/0204024.

\item {\bf [Rudolph 96]}:
O. Rudolph,
``On the consistent effect histories approach to quantum mechanics'',
{\em J. Math. Phys.} {\bf 37}, 11, 5368-5379 (1996).

\item {\bf [Rudolph 99]}:
O. Rudolph,
``Consistent histories and POV measurements'',
{\em Helv. Phys. Acta} {\bf 72}, 1, 23-70 (1999).

\item {\bf [Rudolph 03]}:
O. Rudolph,
``Some properties of the computable cross-norm criterion for separability'',
{\em Phys. Rev. A} {\bf 67}, 3, 032312 (2003).

\item {\bf [Rudolph 00]}:
O. Rudolph,
``A separability criterion for density operators'',
{\em J. Phys. A} {\bf 33}, 21, 3951-3955 (2000);
quant-ph/0002026.

\item {\bf [Rudolph-Wright 00]}:
O. Rudolph, \& J. D. M. Wright,
``On unentangled Gleason theorems for quantum information theory'',
{\em Lett. Math. Phys.} {\bf 52}, 3, 239-245 (2000);
quant-ph/0004036.

\item {\bf [Rudolph 01 a]}:
O. Rudolph,
``A uniqueness theorem for entanglement measures'',
{\em J. Math. Phys.} {\bf 42}, 6, 2507-2512 (2001).
See {\bf [Donald-Horodecki-Rudolph 02]}.

\item {\bf [Rudolph 01 b]}:
O. Rudolph,
``A new class of entanglement measures'',
{\em J. Math. Phys.} {\bf 42}, 11, 5306-5314 (2001);
math-ph/0005011.

\item {\bf [Rudolph 02 a]}:
O. Rudolph,
``Some properties of the computable cross norm criterion for separability'',
{\em Phys. Rev. A};
quant-ph/0212047.

\item {\bf [Rudolph 02 b]}:
O. Rudolph,
``Further results on the cross norm criterion for separability'',
quant-ph/0202121.

\item {\bf [Rudolph 03]}:
O. Rudolph,
``On the cross norm criterion for separability'',
{\em J. Phys. A} {\bf 36}, 21, 5825 (2003).

\item {\bf [Rudolph 04 a]}:
O. Rudolph,
``On extremal quantum states of composite systems with fixed marginals'',
{\em J. Math. Phys.};
quant-ph/0406021.

\item {\bf [Rudolph 04 b]}:
O. Rudolph,
``Some aspects of separability revisited'',
{\em Phys. Lett. A} {\bf 321}, 4, 239-243 (2004).

\item {\bf [Rudolph-Sanders 01]}:
T. Rudolph, \& B. C. Sanders,
``Requirement of optical coherence for
continuous-variable quantum teleportation'',
{\em Phys. Rev. Lett.} {\bf 87}, 7, 077903 (2001);
quant-ph/0103147.
Comment: {\bf [Wiseman 01]}.
See {\bf [Wiseman 03 a]}.

\item {\bf [Rudolph-Pan 01]}:
T. Rudolph, \& J.-W. Pan,
``A simple gate for linear optics quantum computing'',
quant-ph/0108056.

\item {\bf [Rudolph 02]}:
T. Rudolph,
``The laws of physics and cryptographic security'',
quant-ph/0202143.

\item {\bf [Rudolph-Spekkens-Turner 03]}:
T. Rudolph, R. W. Spekkens, \& P. S. Turner,
``Unambiguous discrimination of mixed states'',
{\em Phys. Rev. A} {\bf 68}, 1, 010301 (2003);
quant-ph/0303071.

\item {\bf [Rudolph-Grover 03]}:
T. Rudolph, \& L. Grover,
``Quantum communication complexity of establishing a shared reference frame'',
{\em Phys. Rev. Lett.} {\bf 91}, 21, 217905 (2003);
quant-ph/0306017.

\item {\bf [Rudolph-Spekkens 03]}:
T. Rudolph, \& R. W. Spekkens,
``Quantum state targeting'',
quant-ph/0310060.

\item {\bf [Ruetsche 95]}:
L. Ruetsche,
``Measurement error and the Albert-Loewer problem'',
{\em Found. Phys. Lett.} {\bf 8}, 4, 327-344 (1995).
See {\bf [Albert-Loewer 91]}.

\item {\bf [Rungta-Munro-Nemoto-(+3) 00]}:
P. Rungta, W. J. Munro, K. Nemoto,
P. Deuar, G. J. Milburn, \& C. M. Caves,
``Qudit entanglement'',
quant-ph/0001075.

\item {\bf [Rungta-Bu\v{z}ek-Caves-(+2) 01]}:
P. Rungta, V. Bu\v{z}ek, C. M. Caves,
M. Hillery, \& G. J. Milburn,
``Universal state inversion and concurrence in arbitrary dimensions'',
{\em Phys. Rev. A} {\bf 64}, 4, 042315 (2001);
quant-ph/0102040.

\item {\bf [Rungta-Caves 03]}:
P. Rungta, \& C. M. Caves,
``Concurrence-based entanglement measures for isotropic states'',
{\em Phys. Rev. A} {\bf 67}, 1, 012307 (2003).

\item {\bf [Rupertsberger 98]}:
H. Rupertsberger,
``Bohm's interpretation of quantum mechanics and
the reconstruction of the probability distribution'',
quant-ph/9804074.

\item {\bf [Rupp-Hunzinger-Sorg 02]}:
S. Rupp, S. Hunzinger, \& M. Sorg,
``Exchange degeneracy of
relativistic two-particle quantum states'',
{\em Found. Phys.} {\bf 32}, 5, 705-750 (2002).

\item {\bf [Ruseckas-Kaulaky 01 a]}:
J. Ruseckas, \& B. Kaulakys,
``Real measurements and the quantum Zeno effect'',
{\em Phys. Rev. A} {\bf 63}, 6, 062103 (2001);
quant-ph/0105138.

\item {\bf [Ruseckas-Kaulakys 01 b]}:
J. Ruseckas, \& B. Kaulakys,
``Time problem in quantum mechanics and weak measurements'',
{\em Phys. Lett. A} {\bf 287}, 5-6, 297-303 (2001);
quant-ph/0202156.

\item {\bf [Ruseckas 01]}:
J. Ruseckas,
``Influence of the finite duration of the measurement on the quantum Zeno effect'',
{\em Phys. Lett. A} {\bf 291}, 4-5, 185-189 (2001);
quant-ph/0202157.

\item {\bf [Ruseckas 01]}:
J. Ruseckas,
``Possibility of the tunneling time determination'',
{\em Phys. Rev. A};
quant-ph/0101136.

\item {\bf [Ruseckas-Kaulakys 04]}:
J. Ruseckas, \& B. Kaulakys,
``General expression for the quantum Zeno and anti-Zeno effects'',
{\em Phys. Rev. A};
quant-ph/0403123.

\item {\bf [Ruseckas-Kaulakys 04]}:
J. Ruseckas, \& B. Kaulakys,
``Time problem in quantum mechanics and its analysis by the concept of
weak measurement'',
{\em Lithuanian J. Phys.} {\bf 44}, 161-182 (2004);
quant-ph/0409006.

\item {\bf [Rusell-Clayton-Wegter McNelly-Polkinghorne 01]}:
R. J. Rusell, P. Clayton, K. Wegter-McNelly, \& J. Polkinghorne (eds.),
{\em Quantum Mechanics. Perspectives on Divine Action. Volume 5
(Imperial College, London, 1999, and Vatican Observatory, Castel Gandolfo, Italy, 2000)},
Vatican Observatory Publications, Vatican City State-Center for Theology and Natural Sciences,
Berkeley, California, Specola Vaticana, Citt\`{a} del Vaticano, 2001.

\item {\bf [Ruskai 00]}:
M. B. Ruskai,
``Pauli exchange errors in quantum computation'',
{\em Phys. Rev. Lett.} {\bf 85}, 1, 194-197 (2000);
quant-ph/9906114.
See {\bf [Ruskai 02 c]}.

\item {\bf [Ruskai 02 a]}:
M. B. Ruskai,
``Inequalities for quantum entropy: A review with conditions
for equality'',
{\em J. Math. Phys.} {\bf 43}, 9, 4358-4375 (2002).

\item {\bf [Ruskai 02 b]}:
M. B. Ruskai,
``Comments on adiabatic quantum algorithms'',
quant-ph/0203127.

\item {\bf [Ruskai 02 c]}:
M. B. Ruskai,
``Pauli exchange and quantum error correction'',
in {\bf [Lomonaco-Brandt 02]} 251-263;
quant-ph/0006008.
Extended version of {\bf [Ruskai 00]}.

\item {\bf [Ruskai 02 d]}:
M. B. Ruskai,
``Entanglement breaking channels'',
quant-ph/0207100.
See {\bf [Horodecki-Shor-Ruskai 03]}, {\bf [Ruskai 03]}.

\item {\bf [Ruskai 02 e]}:
M. B. Ruskai,
``Pauli exchange and quantum error correction'',
{\em Quantum computation and information (Washington, 2000)},
American Mathematical Society, Providence, Rhode Island, 2002, pp.~251-263.

\item {\bf [Ruskai 03]}:
M. B. Ruskai,
``Qubit entanglement breaking channels'',
{\em Rev. Math. Phys.};
quant-ph/0302032.
See {\bf [Ruskai 02 d]}, {\bf [Horodecki-Shor-Ruskai 03]}.

\item {\bf [Ruskai 04]}:
M. B. Ruskai,
``Lieb's simple proof of concavity of ${\rm Tr} \left(A^p K^*
B^{(1-p) K}\right)$ and remarks on related inequalities'',
quant-ph/0404126.

\item {\bf [Ruskov-Korotkov 02]}:
R. Ruskov, \& A. N. Korotkov,
``Quantum feedback control of a solid-state qubit'',
{\bf Phys. Rev. B} {\bf 66}, 4, 041401 (2002).

\item {\bf [Ruskov-Korotkov 03]}:
R. Ruskov, \& A. N. Korotkov,
``Entanglement of solid-state qubits by measurement'',
{\em Phys. Rev. B} {\bf 67}, 24, 241305 (2003).

\item {\bf [Rutherford-Grobe 98]}:
G. H. Rutherford, \& R. Grobe,
`Comment on ``Stern-Gerlach effect for electron beams''\,',
{\em Phys. Rev. Lett.} {\bf 81}, 21, 4772 (1998).
Comment on {\bf [Batelaan-Gay-Schwendiman 97]}.
Reply: {\bf [Batelaan-Gay 98]}.

\item {\bf [Ryff 86 a]}:
L. C. B. Ryff,
``Bell's inequalities and their relevance to the
problem of nonlocality in quantum mechanics'',
{\em Phys. Lett. A} {\bf 118}, 1, 3-4 (1986).

\item {\bf [Ryff 86 b]}:
L. C. B. Ryff,
``A comparison between realistic and orthodox interpretations of a
possible experiments based on two recent tests of quantum mechanics'',
{\em Phys. Lett. A} {\bf 119}, 1, 1-2 (1986).

\item {\bf [Ryff 92]}:
``A proposal for testing indirect detection'',
{\em Phys. Lett. A} {\bf 170}, 4, 259-264 (1992).

\item {\bf [Ryff 95]}:
L. C. B. Ryff,
``Interference, distinguishability, and apparent
contradiction in an experiment of induced coherence'',
{\em Phys. Rev. A} {\bf 52}, 4, 2591-2596 (1995).

\item {\bf [Ryff 96]}:
L. C. B. Ryff,
``A proposal for testing local realism without using assumptions
related to hidden variable states'',
in D. Han, K. Peng, Y. S. Kim, \& V. I. Man'ko (eds.),
{\em 4th Int.\ Conf.\ on Squeezed States and Uncertainty Relations
(Taiyuan, Shanxi, China, 1995)},
NASA, Greenbelt, Maryland, 1996, pp.~151-158.

\item {\bf [Ryff 97 a]}:
L. C. B. Ryff,
``Bell and Greenberger, Horne, and Zeilinger theorems revisited'',
{\em Am. J. Phys.} {\bf 65}, 12, 1197-1199 (1997).

\item {\bf [Ryff 97 b]}:
L. C. B. Ryff,
``The strange behavior of entangled photons'',
{\em Found. Phys. Lett.} {\bf 10}, 3, 207-220 (1997).

\item {\bf [Ryff 98]}:
L. C. B. Ryff,
``Two-photon interference without intrinsic
indistinguishability'',
{\em Quantum Semiclass. Opt.} {\bf 10}, 2, 409-414 (1998).

\item {\bf [Ryff 99]}:
L. C. B. Ryff,
``Einstein, Podolsky, and Rosen correlations, quantum teleportation,
entanglement swapping, and special relativity'',
{\em Phys. Rev. A} {\bf 60}, 6, 5083-5086 (1999).

\item {\bf [Ryff-Monken 99]}:
L. C. B. Ryff, \& C. H. Monken,
``Nonlocal `interaction-free' reduction of the state vector'',
{\em J. Opt. B Quantum Semiclass. Opt.} {\bf 1}, 3, 345-348
(1999).

\item {\bf [Ryff-Souto Ribeiro 01]}:
L. C. B. Ryff, \& P. H. Souto Ribeiro,
``Mach-Zehnder interferometer for a two-photon wave packet'',
{\em Phys. Rev. A} {\bf 63}, 2, 023801 (2001).

\item {\bf [Ryff 01]}:
L. C. B. Ryff,
`Comment on ``Two-photon Franson-type experiment and local
realism''\,',
{\em Phys. Rev. Lett.} {\bf 86}, 8, 1908 (2001).
Comment on {\bf [Aerts-Kwiat-Larsson-\.{Z}ukowski 99]}.
Reply: {\bf [Aerts-Kwiat-Larsson-\.{Z}ukowski 01]}.

\item {\bf [Ryff 03]}:
L. C. B. Ryff,
`Comment on ``Experimental nonlocality proof of quantum teleportation and
entanglement swapping''\,',
quant-ph/0303082.
Comment on {\bf [Jennewein-Weihs-Pan-Zeilinger 02]}.
Reply: {\bf [Jennewein-Weihs-Pan-Zeilinger 03]}.


\newpage

\subsection{}


\item {\bf [Saavedra-Gheri-T\"{o}rm\"{a}-(+2) 00]}:
C. Saavedra, K. M. Gheri, P. T\"{o}rm\"{a}, J. I. Cirac, \& P. Zoller,
``Controlled source of entangled photonic qubits'',
{\em Phys. Rev. A} {\bf 61}, 6, 062311 (2000).

\item {\bf [Sacchi 01]}:
M. F. Sacchi,
``Characterizing a universal cloning machine
by maximum-likelihood estimation'',
{\em Phys. Rev. A} {\bf 64}, 2, 022106 (2001);
quant-ph/0010004.

\item {\bf [Sachs 88]}:
M. Sachs,
{\em Einstein versus Bohr},
Open Court, La Salle, Illinois, 1988.

\item {\bf [Sackett-Kielpinski-King-(+8) 00]}:
C. A. Sackett, D. Kielpinski, B. E. King, C. Langer, V. Meyer, C. J. Myatt,
M. Rowe, Q. A. Turchette, W. M. Itano, D. J. Wineland, \& C. Monroe,
``Experimental entanglement of four particles'',
{\em Nature} {\bf 404}, 6775, 256-259 (2000).
See {\bf [Blatt 00]}, {\bf [Osborne 00 a]},
{\bf [Rowe-Kielpinski-Meyer-(+4) 01]}, {\bf [Pati 03 b]}.

\item {\bf [Saffman-Walker 02]}:
M. Saffman, \& T. G. Walker,
``Creating single-atom and single-photon sources from entangled atomic
ensembles'',
{\em Phys. Rev. A} {\bf 66}, 6, 065403 (2002).

\item {\bf [Safonov 02]}:
V. L. Safonov,
``Continuous unitary transformations'',
quant-ph/0202095.

\item {\bf [Safronova-Williams-Clark 03]}:
M. S. Safronova, C. J. Williams, \& C. W. Clark,
``Optimizing the fast Rydberg quantum gate'',
{\em Phys. Rev. A} {\bf 67}, 4, 040303 (2003).

\item {\bf [Saito-Thorwart-Tanaka-(+4) 04]}:
S. Saito, M. Thorwart, H. Tanaka,
M. Ueda, H. Nakano, K. Semba, \& H. Takayanagi,
``Multiphoton transitions in a macroscopic quantum two-state system'',
{\em Phys. Rev. Lett.} {\bf 93}, 3, 037001 (2004).

\item {\bf [Sakaguchi-Ozawa-Amano-Fukumi 99]}:
U. Sakaguchi, H. Ozawa, C. Amano, \& T. Fukumi,
``Microscopic analogs of the Greenberger-Horne-Zeilinger
experiment on an NMR quantum computer'',
{\em Phys. Rev. A} {\bf 60}, 3, 1906-1911 (1999).

\item {\bf [Sakaguchi-Ozawa-Fukumi 00]}:
U. Sakaguchi, H. Ozawa, \& T. Fukumi,
``Method for effective pure states with any number of spins'',
{\em Phys. Rev. A} {\bf 61}, 4, 042313 (2000).

\item {\bf [Sakurai 67]}:
J. J. Sakurai,
{\em Advanced quantum mechanics},
Addison-Wesley, Reading, Massachusetts, 1967.

\item {\bf [Sakurai 85]}:
J. J. Sakurai, \& S. F. Tuan (ed.),
{\em Modern quantum mechanics},
Benjamin/Cummings, Menlo Park, California, 1985.
Revised edition: Addison-Wesley, Reading, Massachusetts, 1994.

\item {\bf [Salas-Sanz 01]}:
P. J. Salas, \& A. L. Sanz,
``Computaci\'{o}n cu\'{a}ntica: Una revoluci\'{o}n en el tratamiento de
la informaci\'{o}n'',
{\em Revista Espa\~{n}ola de F\'{\i}sica} {\bf 15}, 3, 20-28 (2001).

\item {\bf [Salas-Sanz 01 b]}:
P. J. Salas, \& A. L. Sanz,
``Numerical simulation of decoherence-control processes in quantum computers'',
{\em ICQI (Rochester, 2001)};
quant-ph/0110106.

\item {\bf [Salas-Sanz 02]}:
P. J. Salas, \& A. L. Sanz,
``Numerical simulation of information recovery in quantum computers'',
{\em Phys. Rev. A} {\bf 66}, 2, 022302 (2002).

\item {\bf [Salas-Sanz 04 a]}:
P. J. Salas, \& A. L. Sanz,
``Effect of ancilla's structure on quantum error correction using the
7-qubit Calderbank-Shor-Steane code'',
quant-ph/0405012.

\item {\bf [Salas-Sanz 04 b]}:
P. J. Salas, \& A. L. Sanz,
``Error threshold estimation by means of the [[7,1,3]] CSS quantum code'',
{\em Int. J. Quantum. Inf.};
quant-ph/0411042.

\item {\bf [Saleh-Popescu-Teich 96]}:
B. E. A. Saleh, S. Popescu, \& M. C. Teich,
``Generalized entangled-photon imaging'',
{\em Proc. of LEOS'96, 9th Annual Meeting IEEE Laser and Electro-Optics Society}
vol. 1, pp.~362-? (1996).

\item {\bf [Salgado-S\'{a}nchez G\'{o}mez 00]}:
D. Salgado, \& J. L. S\'{a}nchez G\'{o}mez,
``On the compatibility between the Markov property and the quantum
jump'',
quant-ph/0004023.

\item {\bf [Salgado-S\'{a}nchez G\'{o}mez 01 a]}:
D. Salgado, \& J. L. S\'{a}nchez G\'{o}mez,
``QSES's and the quantum jump'',
{\em Proc.\ 3rd Workshop on
Mysteries, Puzzles, and Paradoxes in Quantum Mechanics (Gargnano,
Italy, 2000)},
{\em Zeitschrift f\"{u}r Naturforschung A} {\bf 56}, 228-229 (2001);
quant-ph/0106116.

\item {\bf [Salgado-S\'{a}nchez G\'{o}mez 01 b]}:
D. Salgado, \& J. L. S\'{a}nchez G\'{o}mez,
``Lindblad evolution and stochasticity: The case
of a two-level system'',
quant-ph/0106151.

\item {\bf [Salgado 01]}:
D. Salgado,
``Decoherencia y entrelazamiento'',
{\em Revista Espa\~{n}ola de F\'{\i}sica} {\bf 15}, 4, 52 (2001).

\item {\bf [Salgado-S\'{a}nchez G\'{o}mez 02 a]}:
D. Salgado, \& J. L. S\'{a}nchez G\'{o}mez,
``Expressing stochastic unravellings using random evolution operators'',
{\em J. Opt. B: Quantum Semiclass. Opt.} {\bf 4}, ?, S458-? (2002).

\item {\bf [Salgado-S\'{a}nchez G\'{o}mez 02 b]}:
D. Salgado, \& J. L. S\'{a}nchez G\'{o}mez,
``Some comments on three suggested postulates for quantum theory'',
{\em Found. Phys. Lett.} {\bf 15}, 3, 209-227 (2002);
quant-ph/0106158.

\item {\bf [Salgado-S\'{a}nchez G\'{o}mez 02 c]}:
D. Salgado, \& J. L. S\'{a}nchez G\'{o}mez,
``Generalized model-independent approach to intrinsic decoherence'',
quant-ph/0204141.

\item {\bf [Salgado-S\'{a}nchez G\'{o}mez 02 d]}:
D. Salgado, \& J. L. S\'{a}nchez G\'{o}mez,
`Comment on
``Dynamics of open quantum systems initially entangled with environment:
Beyond the Kraus representation''\,',
quant-ph/0211164.
Comment on {\bf [\v{S}telmachovi\v{c}-Bu\v{z}ek 01]}.

\item {\bf [Salgado-S\'{a}nchez G\'{o}mez 02 d]}:
D. Salgado, \& J. L. S\'{a}nchez G\'{o}mez,
``Expressing decoherence with spectral and stochastic methods'',
quant-ph/0212042.

\item {\bf [Salgado-S\'{a}nchez G\'{o}mez 03]}:
D. Salgado, \& J. L. S\'{a}nchez-G\'{o}mez,
``Damped quantum interference using stochastic calculus'',
in M. Ferrero (ed.),
{\em Proc. of Quantum Information: Conceptual Foundations,
Developments and Perspectives (Oviedo, Spain, 2002)},
{\em J. Mod. Opt.} {\bf 50}, 6-7, 975-980 (2003).

\item {\bf [Salgado-S\'{a}nchez G\'{o}mez 04 a]}:
D. Salgado, \& J. L. S\'{a}nchez-G\'{o}mez,
``A formula for the Bloch vector of some Lindblad quantum systems'',
{\em Phys. Lett. A} {\bf 323}, 5-6, 365-373 (2004);
quant-ph/0308066.

\item {\bf [Salgado 04]}:
D. Salgado,
``Evoluci\'{o}n estoc\'{a}stica en espacios de Hilbert:
Aspectos fundamentales y aplicaciones en sistemas cu\'{a}nticos abiertos'',
Ph.\ D. thesis, Universidad Aut\'{o}noma de Madrid, 2004.

\item {\bf [Salgado-S\'{a}nchez G\'{o}mez 04 b]}:
D. Salgado, J. L. S\'{a}nchez-G\'{o}mez, \& M. Ferrero,
``A simple proof of the Jamiolkowski criterion for complete positivity of
linear maps'',
math-ph/0406010.

\item {\bf [Samal-Ghose 02]}:
M. K. Samal, \& P. Ghose,
``Grover's search algorithm and the quantum measurement problem'',
quant-ph/0202176.

\item {\bf [Samuelsson-Sukhorukov-Buttiker 03 a]}:
P. Samuelsson, E. Sukhorukov, \& M. Buttiker,
``Orbital entanglement and violation of Bell inequalities in the presence
of dephasing'',
{\em Turkish J. Phys.} (special issue on {\em Quantum Computation at the Atomic Scale});
cond-mat/0309540.

\item {\bf [Samuelsson-Sukhorukov-Buttiker 03 b]}:
P. Samuelsson, E. V. Sukhorukov, \& M. Buttiker,
``Orbital entanglement and violation of Bell inequalities in mesoscopic
conductors'',
{\em Phys. Rev. Lett.} {\bf 91}, 15, 157002 (2003).

\item {\bf [S\'{a}nchez G\'{o}mez-S\'{a}nchez Ron 83]}:
J. L. S\'{a}nchez G\'{o}mez, \& J. M. S\'{a}nchez Ron,
``Quantum mechanics and macroscopic separability: A critical review'',
{\em Anales de F\'{\i}sica A} {\bf 79}, 2, 85-94 (1983).

\item {\bf [S\'{a}nchez G\'{o}mez-S\'{a}nchez Ron 85]}:
J. L. S\'{a}nchez G\'{o}mez, \& J. M. S\'{a}nchez Ron,
``Quantum mechanics and macroscopic separability II: Further comments'',
{\em Anales de F\'{\i}sica A} {\bf 81}, 1, 40-43 (1985).

\item {\bf [S\'{a}nchez G\'{o}mez-Rosales 93]}:
J. L. S\'{a}nchez G\'{o}mez, \& J. L. Rosales,
``Quantum fluctuations and classical stochastic gravity'',
in A. van der Merwe, \& F. Selleri (eds.),
{\em Bell's theorem and the foundations of modern physics.
Proc.\ of an international
conference (Cesena, Italy, 1991)},
World Scientific, Singapore, 1993, pp.~407-411.

\item {\bf [S\'{a}nchez G\'{o}mez 96]}:
J. L. S\'{a}nchez G\'{o}mez,
``Decoherencia: Aspectos generales y aplicaci\'{o}n a la
cosmolog\'{\i}a cu\'{a}ntica'',
in M. Ferrero, A. Fern\'{a}ndez Ra\~{n}ada, J. L. S\'{a}nchez G\'{o}mez,
\& E. Santos (eds.),
{\em Fundamentos de la F\'{\i}sica Cu\'{a}ntica (San Lorenzo de El
Escorial, Spain, 1995)},
Editorial Complutense, Madrid, 1996, pp.~241-257.

\item {\bf [S\'{a}nchez G\'{o}mez-Unturbe 99]}:
J. L. S\'{a}nchez G\'{o}mez, \& J. Unturbe,
``Gravitationally induced decoherence in macroscopic systems'',
{\em Found. Phys. Lett.} {\bf 12}, 3, 233-250 (1999).

\item {\bf [S\'{a}nchez G\'{o}mez 00 a]}:
J. L. S\'{a}nchez G\'{o}mez,
``La interpretaci\'{o}n de la teor\'{\i}a cu\'{a}ntica:
Un debate permanente'',
{\em Revista Espa\~{n}ola de F\'{\i}sica} {\bf 14}, 1, 10-16 (2000).

\item {\bf [S\'{a}nchez G\'{o}mez 00 b]}:
J. L. S\'{a}nchez G\'{o}mez,
``La interpretaci\'{o}n actual de la teor\'{\i}a cu\'{a}ntica:
De los muchos universos a las historias consistentes'',
{\em Arbor} {\bf 167}, 659-660, 475-488 (2000).

\item {\bf [S\'{a}nchez Ron 01]}:
J. M. S\'{a}nchez Ron,
{\em Historia de la f\'{\i}sica cu\'{a}ntica.
I. El per\'{\i}odo fundacional (1860-1926)},
Cr\'{\i}tica, Barcelona, 2001.

\item {\bf [S\'{a}nchez Ruiz 93]}:
J. S\'{a}nchez Ruiz,
``Entropic uncertainty and certainty relations for complementary observables'',
{\em Phys. Lett. A} {\bf 173}, ?, 233-239 (1993).

\item {\bf [S\'{a}nchez Ruiz 94]}:
J. S\'{a}nchez Ruiz,
``States of minimal joint uncertainty for complementary observables
in three-dimensional Hilbert space'',
{\em J. Phys. A} {\bf 27}, ?, L843-L846 (1994).

\item {\bf [S\'{a}nchez Ruiz 95]}:
J. S\'{a}nchez Ruiz,
``Improved bounds in the entropic uncertainty and certainty relations
for complementary observables'',
{\em Phys. Lett. A} {\bf 2o1}, ?, 125-131 (1998).

\item {\bf [S\'{a}nchez Ruiz 98 a]}:
J. S\'{a}nchez Ruiz,
``Position-momentum entropic relation and complementarity in single-slit
and double-slit experiments'',
{\em Phys. Rev. A} {\bf 57}, 3, 1519-1525 (1998).

\item {\bf [S\'{a}nchez Ruiz 98 b]}:
J. S\'{a}nchez Ruiz,
``Optimal entropic uncertainty relation in two-dimensional Hilbert space'',
{\em Phys. Lett. A} {\bf 244}, 4, 189-195 (1998).

\item {\bf [S\'{a}nchez Ruiz 01]}:
J. S\'{a}nchez Ruiz,
``Informaci\'{o}n, incertidumbre y complementariedad'',
{\em Revista Espa\~{n}ola de F\'{\i}sica} {\bf 15}, 4, 28-33 (2001).

\item {\bf [S\'{a}nchez Ruiz 02]}:
J. S\'{a}nchez Ruiz,
``Informaci\'{o}n, incertidumbre y complementariedad'',
in C. Mataix, \& A. Rivadulla (eds.),
{\em F\'{\i}sica cu\'{a}ntica y realidad.
Quantum physics and reality (Madrid, 2000)},
Editorial Complutense, Madrid, 2002, pp.~267-283.

\item {\bf [S\'{a}nchez Soto 00]}:
L. L. S\'{a}nchez-Soto,
``El tao de la no interacci\'{o}n'',
{\em Revista Espa\~{n}ola de F\'{\i}sica} {\bf 14}, 4, 20-25 (2000).

\item {\bf [S\'{a}nchez Soto-Monz\'{o}n-Yonte-Cari\~{n}ena 01]}:
L. L. S\'{a}nchez-Soto, J. J. Monz\'{o}n, T. Yonte, \& J. F. Cari\~{n}ena,
``Simple trace criterion for classification of multilayers'',
{\em Opt. Lett.};
quant-ph/0108028.

\item {\bf [S\'{a}nchez Soto-Delgado-Klimov-Bj{\o}rk 02]}:
L. L. S\'{a}nchez-Soto, J. Delgado, A. B. Klimov, \& G. Bj{\o}rk,
``Description of entanglement in terms of quantum phase'',
{\em Phys. Rev. A} {\bf 66}, 4, 042112 (2002);
quant-ph/0202100.

\item {\bf [S\'{a}nchez Soto-Cari\~{n}ena-Barriuso-Monz\'{o}n 04]}:
L. L. S\'{a}nchez-Soto, J. F. Cari\~{n}ena, A. G. Barriuso, \& J. J. Monz\'{o}n,
``Vectorlike representation of one-dimensional scattering'',
{\em Festschrift in honor of Alberto Galindo},
World Scientific, Singapore, 2004;
quant-ph/0411081.

\item {\bf [Sanaka-Kawahara-Kuga 00]}:
K. Sanaka, K. Kawahara, \& T. Kuga,
``New high-efficiency source of photon pairs
for engineering quantum entanglement'',
quant-ph/0012028.

\item {\bf [Sanaka-Kawahara-Kuga 01]}:
K. Sanaka, K. Kawahara, \& T. Kuga,
``Franson-type experiment realizes two-qubit quantum logic gate'',
quant-ph/0108001.

\item {\bf [Sanaka-Kawahara-Kuga 02]}:
K. Sanaka, K. Kawahara, \& T. Kuga,
``Experimental probabilistic manipulation of down-converted photon pairs using
unbalanced interferometers'',
{\em Phys. Rev. A} {\bf 66}, 4, 040301 (2002).

\item {\bf [Sanaka-Jennewein-Pan-(+2) 04]}:
K. Sanaka, T. Jennewein, J.-W. Pan,
K. Resch, \& A. Zeilinger,
``Experimental nonlinear sign shift for scalable quantum computation'',
{\em Phys. Rev. Lett.} {\bf 92}, 1, 017902 (2004);
quant-ph/0308134.

\item {\bf [Sancho-Fern\'{a}ndez Huelga 00]}:
J. M. G. Sancho, \& S. G. Fern\'{a}ndez Huelga,
``Measuring the entanglement of bipartite pure states'',
{\em Phys. Rev. A} {\bf 61}, 4, 042303 (2000);
quant-ph/9910041.

\item {\bf [Sancho 97]}:
P. Sancho,
``Wave and particle pictures of correlated systems'',
{\em Found. Phys. Lett.} {\bf 10}, 3, 235-244 (1997).

\item {\bf [Sancho 01]}:
P. Sancho,
``The gedanken double-slit experiment with extended incoherent
sources'',
{\em Phys. Lett. A} {\bf 281}, 5-6, 289-296 (2001).

\item {\bf [Sancho 02]}:
P. Sancho,
``Popper's experiment revisited'',
{\em Found. Phys.} {\bf 32}, 5, 789-805 (2002).

\item {\bf [Sanders-Milburn 89]}:
B. C. Sanders, \& G. J. Milburn,
``Complementarity in a quantum nondemolition measurement'',
{\em Phys. Rev. A} {\bf 39}, 2, 694-702 (1989).

\item {\bf [Sanders 92]}:
B. C. Sanders,
``Entangled coherent states'',
{\em Phys. Rev. A} {\bf 45}, 9, 6811-6815 (1992).
Erratum: {\em Phys. Rev. A} {\bf 46}, 5, 2966 (1992).

\item {\bf [Sanders-Bartlett-Tregenna-Knight 03]}:
B. C. Sanders, S. D. Bartlett, B. Tregenna, \& P. L. Knight,
``Quantum quincunx in cavity quantum electrodynamics'',
{\em Phys. Rev. A} {\bf 67}, 4, 042305 (2003).

\item {\bf [Sanders-Bartlett-Rudolph-Knight 03]}:
B. C. Sanders, S. D. Bartlett, T. Rudolph, \& P. L. Knight,
``Photon number superselection and the entangled coherent state
representation'',
{\em Phys. Rev. A} {\bf 68}, 4, 042329 (2003);
quant-ph/0306076.

\item {\bf [Sanders-Kim-Holton 99 a]}:
G. D. Sanders, K. W. Kim, \& W. C. Holton,
``Quantum computing with complex instruction sets'',
{\em Phys. Rev. A} {\bf 59}, 2, 1098-1101 (1999).

\item {\bf [Sanders-Kim-Holton 99 b]}:
G. D. Sanders, K. W. Kim, \& W. C. Holton,
``Optically driven quantum-dot quantum computer'',
{\em Phys. Rev. A} {\bf 60}, 5, 4146-414 (1999);
quant-ph/9909070.

\item {\bf [Sagi 03]}:
Y. Sagi,
``Scheme for generating Greenberger-Horne-Zeilinger-type
states of $n$ photons'',
{\em Phys. Rev. A} {\bf 68}, 4, 042320 (2003);
quant-ph/0307097.

\item {\bf [Saguia-Sarandy 03]}:
A. Saguia, \& M. S. Sarandy,
``Entanglement in the one-dimensional Kondo necklace model'',
{\em Phys. Rev. A} {\bf 67}, 1, 012315 (2003).

\item {\bf [Sagioro-Olindo-Monken-P\'{a}dua 04]}:
M. A. Sagioro, C. Olindo, C. H. Monken, \& S. P\'{a}dua,
``Time control of two-photon interference'',
{\em Phys. Rev. A} {\bf 69}, 5, 053817 (2004).

\item {\bf [Sanpera-Macchiavella 97]}:
A. Sanpera, \& C. Macchiavello,
``Quantum privacy amplification: A secure method for cryptography'',
in M. Ferrero, \& A. van der Merwe (eds.),
{\em New developments on fundamental problems in quantum
physics (Oviedo, Spain, 1996)},
Kluwer Academic, Dordrecht, Holland, 1997, pp.~361-367.

\item {\bf [Sanpera-Tarrach-Vidal 97]}:
A. Sanpera, R. Tarrach, \& G. Vidal,
``Quantum separability, time reversal and canonical decompositions'',
quant-ph/9707041.

\item {\bf [Sanpera-Tarrach-Vidal 98 a]}:
A. Sanpera, R. Tarrach, \& G. Vidal,
``Local description of quantum inseparability'',
{\em Phys. Rev. A} {\bf 58}, 2, 826-830 (1998).
See {\bf [Lockhart-Steiner-Gerlach 00]}.

\item {\bf [Sanpera-Tarrach-Vidal 98 b]}:
A. Sanpera, R. Tarrach, \& G. Vidal,
``Quantum inseparability as local pseudomixture'',
quant-ph/9801024.

\item {\bf [Sanpera-Bru\ss-Lewenstein 01]}:
A. Sanpera, D. Bru\ss, \& M. Lewenstein,
``Schmidt number witnesses and bound entanglement'',
{\em Phys. Rev. A} {\bf 63}, 5, 050301(R) (2001);
quant-ph/0009109.

\item {\bf [San'Annat 00]}:
A. S. San'Annat,
``Indistinguishability and nonlocality in
Einstein-Podolsky-Rosen experiment'',
quant-ph/0003047.

\item {\bf [Santana-Khanna-Revzen 02]}:
A. E. Santana, F. C. Khanna, \& M. Revzen,
``Entropy of entangled states and $SU(1,1)$ and $SU(2)$ symmetries'',
{\em Phys. Rev. A} {\bf 65}, 3, 032119 (2002).

\item {\bf [Santori-Fattal-Vu\v{c}kovi\'{c}-(+2) 02]}:
C. Santori, D. Fattal, J. Vu\v{c}kovi\'{c}, G. S. Solomon, \& Y. Yamamoto,
``Indistinguishable photons from a single-photon device'',
{\em Nature} {\bf 419}, 6907, 594-597 (2002).

\item {\bf [Santos 84 a]}:
E. Santos,
``Microscopic and macroscopic Bell inequalities'',
{\em Phys. Lett. A} {\bf 101}, 8, 379-382 (1984).

\item {\bf [Santos 84 b]}:
E. Santos,
``Comment on `Proposed molecular test of local hidden-variables theories'\,'',
{\em Phys. Rev. A} {\bf 30}, 4, 2128-2129 (1984).
Comment on {\bf [Lo-Shimony 81]}.
See {\bf [Shimony 84 b]}.

\item {\bf [Santos 86]}:
E. Santos,
``The Bell inequalities as tests of classical logic'',
{\em Phys. Lett. A} {\bf 115}, 8, 363-365 (1986).

\item {\bf [Santos 88 a]}:
E. Santos,
``El teorema de Bell: Realismo local frente a mec\'{a}nica cu\'{a}ntica'',
{\em Revista Espa\~{n}ola de F\'{\i}sica} {\bf 2}, 2, 38-43 (1988).

\item {\bf [Santos 88 b]}:
E. Santos,
``The search for hidden variables in quantum mechanics'',
in F. Selleri (ed.),
{\em Quantum mechanics versus local realism: The
Einstein-Podolsky-Rosen paradox},
Plenum Press, New York, 1988, pp.~365-390.

\item {\bf [Santos 90]}:
E. Santos,
in `Can demolition of the ``elements of reality''
proceed on schedule?',
{\em Phys. Today} {\bf 43}, 12, 11-13 (1990).
Comment on {\bf [Mermin 90 a]}.

\item {\bf [Santos 91 a]}:
E. Santos,
``Interpretation of the quantum formalism and Bell's theorem'',
{\em Found. Phys.} {\bf 21}, 2, 221-241 (1991).

\item {\bf [Santos 91 b]}:
E. Santos,
``Does quantum mechanics violate the Bell inequalities?'',
{\em Phys. Rev. Lett.} {\bf 66}, 11, 1388-1390 (1991).
Erratum: {\em Phys. Rev. Lett.} {\bf 66}, 24, 3227.
Comments: {\bf [Rae 92]}, {\bf [Ben Aryeh-Postan 92 b]}.
Counterexample: {\bf [Jones-Adelberger 94]}.
See {\bf [Santos 92 b]}.

\item {\bf [Santos 92 a]}:
E. Santos,
``Comment on `Nonlocality of a single photon'\,'',
{\em Phys. Rev. Lett.} {\bf 68}, 6, 894 (1992).
Comment on {\bf [Tan-Walls-Collett 91]}.

\item {\bf [Santos 92 b]}:
E. Santos,
``Reply to comments on `Does quantum
mechanics violate the Bell inequalities?'\,'',
{\em Phys. Rev. Lett.} {\bf 68}, 17, 2702-2703 (1992).
Reply to {\bf [Rae 92]}, {\bf [Ben Aryeh-Postan 92 b]}.

\item {\bf [Santos 92 c]}:
E. Santos,
``Critical analysis of the empirical tests of local
hidden-variable theories'',
{\em Phys. Rev. A} {\bf 46}, 7, 3646-3656 (1992).

\item {\bf [Santos 93]}:
E. Santos,
``Does quantum mechanics predict violations of the Bell inequalities?'',
in A. van der Merwe, \& F. Selleri (eds.),
{\em Bell's theorem and the foundations of modern physics.
Proc.\ of an international
conference (Cesena, Italy, 1991)},
World Scientific, Singapore, 1993, pp.~412-421.

\item {\bf [Santos 95 a]}:
E. Santos,
``Foundations of quantum physics: Present and future'',
{\em Apeiron} {\bf 2}, 4, 108-111 (1995).

\item {\bf [Santos 95 b]}:
E. Santos,
``Constraints for the violation of the Bell
inequality in Einstein-Podolsky-Rosen-Bohm experiments'',
{\em Phys. Lett. A} {\bf 200}, 1, 1-6 (1995).

\item {\bf [Santos 95 c]}:
E. Santos,
``Comment on `Quantum mechanics and Bell's
inequalities'\,'', preprint, 1995.
Comment on {\bf [Jones-Adelberger 94]}.

\item {\bf [Santos 96 a]}:
E. Santos,
``Unreliability of performed tests of Bell's
inequality using parametric down-converted photons'',
{\em Phys. Lett. A} {\bf 212}, 1-2, 10-14 (1996).

\item {\bf [Santos 96 b]}:
E. Santos,
``On the conjecture that Bell's theorem is not true'',
in A. Mann, \& M. Revzen (eds.),
{\em The dilemma of Einstein, Podolsky and Rosen -- 60 years
later. An international symposium in honour of Nathan Rosen
(Haifa, Israel, 1995)},
{\em Ann. Phys. Soc. Israel} {\bf 12}, 127-131 (1996).

\item {\bf [Santos 96 c]}:
E. Santos,
``Experimentos con pares de fotones producidos
en cristales no lineales'',
in M. Ferrero, A. Fern\'{a}ndez Ra\~{n}ada, J. L. S\'{a}nchez
G\'{o}mez, \& E. Santos (eds.),
{\em Fundamentos de la F\'{\i}sica Cu\'{a}ntica
(San Lorenzo de El Escorial, Spain, 1995)},
Editorial Complutense, Madrid, 1996, pp.~63-81.

\item {\bf [Santos 97 a]}:
E. Santos,
``Effect of perturbing the internal structure in
atom interference: The relevance of photon dressing'',
{\em Phys. Lett. A} {\bf 228}, 4-5, 232-238 (1997).

\item {\bf [Santos 97 b]}:
E. Santos,
``The program of local hidden variables'',
in M. Ferrero, \& A. van der Merwe (eds.),
{\em New developments on fundamental problems
in quantum physics (Oviedo, Spain, 1996)},
Kluwer Academic, Dordrecht, Holland, 1997, pp.~369-379.

\item {\bf [Santos 98]}:
E. Santos,
``Is quantum mechanics really strange?'', to be
published in a book in honor to L. J. Boya.

\item {\bf [Santos-Casado-Fern\'{a}ndez Rueda-(+2) 99]}:
E. Santos. A. Casado, A. Fern\'{a}ndez Rueda,
J. Mart\'{\i}nez, \& R. Risco Delgado,
``Wigner function description of entangled photon pairs produced in
nonlinear crystals'',
in A. Ballesteros, F. J. Herranz, J. Negro,
L. M. Nieto, \& C. M. Pere\~{n}a (eds.),
{\em Proc.\ of the First Int. Workshop Symmetries in
Quantum Mechanics and Quantum Optics (Burgos, Spain, 1998)},
Universidad de Burgos, 1999, pp.~353-368.

\item {\bf [Santos-Ferrero 00]}:
E. Santos, \& M. Ferrero,
``Linear entropy and Bell inequalities'',
{\em Phys. Rev. A} {\bf 62}, 2, 024101 (2000).

\item {\bf [Santos 01]}:
E. Santos,
``Quantum mechanics vs. local realism, is that the question?'',
quant-ph/0103062.
See {\bf [Rowe-Kielpinski-Meyer-(+4) 01]}, {\bf [Vaidman 01 b]}.

\item {\bf [Santos 02]}:
E. Santos,
``What is entanglement?'',
quant-ph/0204020.

\item {\bf [Santos 04 a]}:
E. Santos,
``Entropy inequalities and Bell inequalities for two-qubit systems'',
{\em Phys. Rev. A} {\bf 69}, 2, 022305 (2004);
quant-ph/0301171.
Comment: {\bf [Jakobczyk 04]}.

\item {\bf [Santos 04 b]}:
E. Santos,
``Proposed optical tests of Bell's inequalities not resting upon the fair
sampling assumption'',
{\em Phys. Lett. A} {\em 327}, {\bf 1} 33-37 (2004);
quant-ph/0401003.

\item {\bf [Santos 04 c]}:
E. Santos,
``Bell's theorem and the experiments: Increasing empirical support to
local realism'',
quant-ph/0410193.

\item {\bf [Santos-Bru\ss\, 01]}:
L. Santos, \& D. Bru\ss,
``Reversible quantum teleportation in an optical lattice'',
in S. Popescu, N. Linden, \& R. Jozsa (eds.),
{\em J. Phys. A} {\bf 34}, 35
(Special issue: Quantum information and computation), 7003-7016 (2001);
quant-ph/9908041.

\item {\bf [Santos-Escobar 98]}:
L. F. Santos, \& C. O. Escobar,
``A beable interpretation of the GRW spontaneous collapse model'',
quant-ph/9810019.

\item {\bf [Santos-Escobar 99]}:
L. F. Santos, \& C. O. Escobar,
``Convergences in the measurement problem in quantum mechanics'',
quant-ph/9912005.

\item {\bf [Santos-Escobar 01]}:
L. F. Santos, \& C. O. Escobar,
``A proposed solution to the tail problem of dynamical reduction
models'',
{\em Phys. Lett. A} {\bf 278}, 6, 315-318 (2001).

\item {\bf [Santos 03 b]}:
L. F. Santos,
``Entanglement in quantum computers described by the $XXZ$ model with defects'',
{\em Phys. Rev. A} {\bf 67}, 6, 062306 (2003).

\item {\bf [Santos-Rigolin-Escobar 04]}:
L. F. Santos, G. Rigolin, \& C. O. Escobar,
``Entanglement versus chaos in disordered spin chains'',
{\em Phys. Rev. A} {\bf 69}, 4, 042304 (2004).

\item {\bf [Santos-Dykman-Shapiro-Izrailev 04]}:
L. F. Santos, M. I. Dykman, M. Shapiro, \& F. M. Izrailev,
``Strong many-particle localization and quantum computing with perpetually
coupled qubits'',
quant-ph/0405013.
See {\bf [Dykman-Izrailev-Santos-Shapiro 04]}.

\item {\bf [Santos-Lutterbach-Davidovich 01]}:
M. F. Santos, L. G. Lutterbach, \& L. Davidovich,
``Probing entanglement in phase space:
Signature of GHZ states in the Wigner function'',
{\em J. Opt. B: Quantum Semiclass. Opt.} {\bf 3}, 1, S55-S59 (2001).

\item {\bf [Santos-Milman-Khoury-Souto Ribeiro 01]}:
M. F. Santos, P. Milman, A. Z. Khoury, \& P. H. Souto Ribeiro,
``Measurement of the degree of polarization entanglement
through position interference'',
quant-ph/0102023.

\item {\bf [Sanz-S\'{a}nchez G\'{o}mez 87]}:
A. L. Sanz, \& J. L. S\'{a}nchez G\'{o}mez,
``On photon rescattering in the atomic-cascade experimental tests of Bell
inequalities'',
{\em Europhys. Lett.} {\bf 3}, 5, 519-522 (1987).

\item {\bf [Sanz-S\'{a}nchez G\'{o}mez 90]}:
A. L. Sanz, \& J. L. S\'{a}nchez G\'{o}mez,
``The classical, high-spin limit of the Bell-Mermin inequality'',
{\em Anales de F\'{\i}sica A} {\bf 86}, 1, 77-84 (1990).

\item {\bf [Sanz 90]}:
A. L. Sanz,
``Algunas cuestiones de fundamentaci\'{o}n de la
teor\'{\i}a cu\'{a}ntica: Separabilidad y aspectos estoc\'{a}sticos'',
Ph.\ D. thesis, Universidad Aut\'{o}noma de Madrid, 1990.

\item {\bf [Sanz-Borondo-Bastiaans 03]}:
A. S. Sanz, F. Borondo, \& M. J. Bastiaans,
``On the loss of coherence in double-slit diffraction experiments'',
quant-ph/0310095.

\item {\bf [Sanz-Borondo 03]}:
A. S. Sanz, \& F. Borondo,
``A Bohmian view on quantum decoherence'',
quant-ph/0310096.

\item {\bf [Saraga-Loss 03]}:
D. S. Saraga, \& D. Loss,
``Spin-entangled currents created by a triple quantum dot'',
{\em Phys. Rev. Lett.} {\bf 90}, 16, 166803 (2003).

\item {\bf [Saraga-Altshuler-Loss-Westervelt 04]}:
D. S. Saraga, B. L. Altshuler, D. Loss, \& R. M. Westervelt,
``Coulomb scattering in a 2D interacting electron gas and production of EPR pairs'',
{\em Phys. Rev. Lett.} {\bf 92}, 24, 246803 (2004).

\item {\bf [Sarandy-Lidar 04]}:
M. S. Sarandy, \& D. A. Lidar,
``The adiabatic approximation in open quantum systems'',
{\em Phys. Rev. A};
quant-ph/0404147.

\item {\bf [Sarovar-Ahn-Jacobs-Milburn 04]}:
M. Sarovar, C. Ahn, K. Jacobs, \& G. J. Milburn,
``A practical scheme for error control using feedback'',
{\em Phys. Rev. A} {\bf 69}, 5, 052324 (2004);
quant-ph/0402017.

\item {\bf [Sasada-Okamoto 03]}:
H. Sasada, \& M. Okamoto,
``Transverse-mode beam splitter of a light beam and its application to quantum
cryptography'',
{\em Phys. Rev. A} {\bf 68}, 1, 012323 (2003).
Erratum: {\em Phys. Rev. A} {\bf 69}, 3, 039901 (2004).

\item {\bf [Sasaki-Kato-Izutsu-Hirota 98]}:
M. Sasaki, K. Kato, M. Izutsu \& O. Hirota,
``Quantum channels showing superadditivity in channel capacity'',
{\em Phys. Rev. A} {\bf 58}, 1, 146-158 (1998).

\item {\bf [Sasaki-Barnett-Jozsa-(+2) 99]}:
M. Sasaki, S. M. Barnett, R. Jozsa, M. Osaki, \& O. Hirota,
``Accessible information and optimal strategies for real
symmetrical quantum sources'',
{\em Phys. Rev. A} {\bf 59}, 5, 3325-3335 (1999);
quant-ph/9812062.

\item {\bf [Sasaki-Carlini-Chefles 01]}:
M. Sasaki, A. Carlini, \& A. Chefles,
``Optimal phase estimation and square root measurement'',
in S. Popescu, N. Linden, \& R. Jozsa (eds.),
{\em J. Phys. A} {\bf 34}, 35
(Special issue: Quantum information and computation), 7017-7028 (2001);
quant-ph/0011057.

\item {\bf [Sasaki-Carlini-Jozsa 01]}:
M. Sasaki, A. Carlini, \& R. Jozsa,
``Quantum template matching'',
{\em Phys. Rev. A} {\bf 64}, 2, 022317 (2001);
quant-ph/0102020.

\item {\bf [Sasaki-Ban-Barnett 02]}:
M. Sasaki, M. Ban, \& S. M. Barnett,
``Optimal parameter estimation of a depolarizing channel'',
{\em Phys. Rev. A} {\bf 66}, 2, 022308 (2002);
quant-ph/0203113.

\item {\bf [Sasaki-Carlini 02]}:
M. Sasaki, \& A. Carlini,
``Quantum learning and universal quantum matching machine'',
{\em Phys. Rev. A} {\bf 66}, 2, 022303 (2002);
quant-ph/0202173.

\item {\bf [Sasura-Bu\v{z}zek 01 a]}:
M. Sasura, \& V. Bu\v{z}zek,
``Multiparticle entanglement with quantum logic networks:
Application to cold trapped ions'',
{\em Phys. Rev. A} {\bf 64}, 1, 012305 (2001);
quant-ph/0103067.

\item {\bf [Sasura-Bu\v{z}ek 01 b]}:
M. Sasura, \& V. Bu\v{z}ek,
``Cold trapped ions as quantum information processors'',
{\em J. Mod. Opt.};
quant-ph/0112041.

\item {\bf [Sasura-Steane 02]}:
M. Sasura, \& A. M. Steane,
``Realistic fast quantum gates with hot trapped ions'',
quant-ph/0212005.

\item {\bf [Sasura-Steane 03]}:
M. Sasura, \& A. M. Steane,
``Fast quantum logic by selective displacement of hot trapped ions'',
{\em Phys. Rev. A} {\bf 67}, 6, 062318 (2003).

\item {\bf [Sasura-Steane 04]}:
M. Sasura, \& A. M. Steane,
``Further quantum-gate methods using selective displacement of trapped
ions'',
quant-ph/0402054.

\item {\bf [Satinover 02 a]}:
J. Satinover,
``Decoherence-free subspaces in supersymmetric oscillator networks'',
quant-ph/0211172.

\item {\bf [Satinover 02 b]}:
J. Satinover,
``Modeling decoherence and decoherence-free subspaces in complex
environments'',
quant-ph/0212003.

\item {\bf [Saunders 93]}:
S. Saunders,
``Decoherence, relative states, and evolutionary adaptation'',
{\em Found. Phys.} {\bf 23}, 12, 1553-1595 (1993).

\item {\bf [Saunders 95]}:
S. Saunders,
``Time, quantum mechanics, and decoherence'',
{\em Synthese} {\bf 102}, ?, 235-66 (1995);
PITT-PHIL-SCI00000463.

\item {\bf [Saunders 98]}:
S. Saunders,
``Time, quantum mechanics, and probability'',
{\em Synthese} {\bf 114}, ?, 373-404 (1998);
quant-ph/0111047.

\item {\bf [Savinien-Taron-Tarrach 00]}:
J. Savinien, J. Taron, \& R. Tarrach,
``Triviality of GHZ operators of higher spin'',
{\em J. Phys. A} {\bf 33}, 48, L493-L495 (2000);
quant-ph/0007069.
See {\bf [Cabello 01 a]}.

\item {\bf [Sawicki 90]}:
M. Sawicki,
in `Can demolition of the ``elements of reality'' proceed on schedule?',
{\em Phys. Today} {\bf 43}, 12, 11-13 (1990).
Comment on {\bf [Mermin 90 a]}.

\item {\bf [Sazonova-Singh 01 a]}:
Z. S. Sazonova, \& R. Singh,
``Detection and correction of errors with
quantum tomography'',
quant-ph/0106146.

\item {\bf [Sazonova-Singh 01 b]}:
Z. S. Sazonova, \& R. Singh,
``Optical quantum computer based on RDS crystal'',
quant-ph/0108065.

\item {\bf [Sazonova-Singh 01 c]}:
Z. S. Sazonova, \& R. Singh,
``CNOT operator and its similar matrices in quantum computation'',
quant-ph/0110051.

\item {\bf [Scalettar 99]}:
R. Scalettar,
``Understanding quantum mechanics. Quantum philosophy:
Understanding and interpreting contemporary science'',
{\em Phys. Today} {\bf 52}, 12, 62-64 (1999).
Review of {\bf [Omn\`{e}s 94 b, 99 a]}.

\item {\bf [Scarcelli-Valencia-Gompers-Shih 03]}:
G. Scarcelli, A. Valencia, S. Gompers, \& Y. Shih,
``Remote spectral measurement using entangled photons'',
{\em App. Phys. Lett.} {\bf 83}, 26, 5560-? (2003);
quant-ph/0407164.

\item {\bf [Scarani-Suarez 98]}:
V. Scarani, \& A. Suarez,
``Introducing quantum mechanics: One-particle interferences'',
{\em Am. J. Phys.} {\bf 66}, 8, 718-721 (1998).

\item {\bf [Scarani 98]}:
V. Scarani,
``Quantum computing'',
{\em Am. J. Phys.} {\bf 66}, 11, 956-960 (1998);
quant-ph/9804044.

\item {\bf [Scarani-Tittel-Zbinden-Gisin 00]}:
V. Scarani, W. Tittel, H. Zbinden, \& N. Gisin,
``The speed of quantum information and the preferred frame:
Analysis of experimental data'',
{\em Phys. Lett. A} {\bf 276}, 1-4, 1-7 (2000);
quant-ph/0007008.

\item {\bf [Scarani-Gisin 01 a]}:
V. Scarani, \& N. Gisin,
``Quantum communication between $N$ partners and Bell's inequalities'',
{\em Phys. Rev. Lett.} {\bf 87}, 11, 117901 (2001);
quant-ph/0101110.
See {\bf [Scarani-Gisin 02]}.

\item {\bf [Scarani-Gisin 01 b]}:
V. Scarani, \& N. Gisin,
``Spectral decomposition of Bell's operators for qubits'',
quant-ph/0103068.

\item {\bf [Scarani-Gisin 01 c]}:
V. Scarani, \& N. Gisin,
``Loss of EPR correlations leads to signaling'',
quant-ph/0110074.

\item {\bf [Scarani-Gisin 02]}:
V. Scarani, \& N. Gisin,
``Quantum key distribution between $N$ partners: Optimal eavesdropping and
Bell's inequalities'',
{\em Phys. Rev. A} {\bf 65}, 1, 012311 (2002);
quant-ph/0104016.
See {\bf [Scarani-Gisin 01 a]}.

\item {\bf [Scarani-Ziman-Stelmachovic-(+2) 02]}:
V. Scarani, M. Ziman, P. Stelmachovic, N. Gisin, \& V. Bu\v{z}zek,
``Thermalizing quantum machines: Dissipation and entanglement'',
{\em Phys. Rev. Lett.} {\bf 88}, 9, 097905 (2002);
quant-ph/0110088.

\item {\bf [Scarani-Gisin 02]}:
V. Scarani, \& N. Gisin,
``Superluminal influences, hidden variables, and signaling'',
{\em Phys. Lett. A} {\bf 295}, 4, 167-174 (2002).

\item {\bf [Scarani-Ac\'{\i}n-Ribordy-Gisin 04]}:
V. Scarani, A. Ac\'{\i}n, G. Ribordy, \& N. Gisin,
``Quantum cryptography protocols robust against photon number splitting attacks for weak laser pulse implementations'',
{\em Phys. Rev. Lett.} {\bf 92}, 5, 057901 (2004);
quant-ph/0211131.

\item {\bf [Scarani-Gisin-Popescu 03]}:
V. Scarani, N. Gisin, \& S. Popescu,
``Energy-time entanglement of quasi-particles in solid-state devices'',
cond-mat/0307385.

\item {\bf [Scarani 03]}:
V. Scarani,
``Classical and quantum: Some mutual clarifications'',
{\em Proc.\ of the Workshop ``Multiscale Methods in Quantum
Mechanics'' (Rome, 2002)};
quant-ph/0309113.

\item {\bf [Scarani-Ac\'{\i}n-Schenck-Aspelmeyer 04]}:
V. Scarani, A. Ac\'{\i}n, E. Schenck, \& M. Aspelmeyer,
``The non-locality of cluster states'',
quant-ph/0405119.

\item {\bf [Scarani-de Riedmatten-Marcikic-(+2) 04]}:
V. Scarani, H. de Riedmatten, I. Marcikic,
H. Zbinden, \& N. Gisin,
``Four-photon correction in two-photon Bell experiments'',
quant-ph/0407189.

\item {\bf [Scarani-Gisin 04]}:
V. Scarani, \& N. Gisin,
``Superluminal hidden communication as the underlying mechanism for
quantum correlations: constraining models'',
quant-ph/0410025.

\item {\bf [Scarola-Park-Das Sarma 03]}:
V. W. Scarola, K. Park, \& S. Das Sarma,
``Pseudospin quantum computation in semiconductor nanostructures'',
{\em Phys. Rev. Lett.};
cond-mat/0304225.

\item {\bf [Scarola-Park-Das Sarma 03]}:
V. W. Scarola, K. Park, \& S. Das Sarma,
``Chirality in quantum computation with spin cluster qubits'',
{\em Phys. Rev. Lett.} {\bf 93}, 12, 120503 (2004);
cond-mat/0403444.

\item {\bf [Scarl 94]}:
D. Scarl,
``Quantum measurement'',
{\em Phys. Today} {\bf 47}, 1, 46-47 (1994).
Review of {\bf [Braginsky-Khalili 92]}.

\item {\bf [Schnabel-Bowen-Treps-(+3) 03]}:
R. Schnabel, W. P. Bowen, N. Treps,
T. C. Ralph, H.-A. Bachor, \& P. K. Lam,
``Stokes-operator-squeezed continuous-variable polarization states'',
{\em Phys. Rev. A} {\bf 67}, 1, 012316 (2003).

\item {\bf [Schnabel-Bowen-Treps-(+4) 03]}:
R. Schnabel, W. P. Bowen, N. Treps,
B. Buchler, T. C. Ralph, P. K. Lam, \& H.-A. Bachor,
``Optical experiments beyond the quantum limit: Squeezing, entanglement, and
teleportation'',
{\em Opt. Spectrosc.} {\bf 94}, 651 (2003).

\item {\bf [Schachner 03]}:
G. Schachner,
``The structure of Bell inequalities'',
quant-ph/0312117.

\item {\bf [Schack 98]}:
R. Schack,
``Using a quantum computer to investigate quantum chaos'',
{\em Phys. Rev. A} {\bf 57}, 3, 1634-1635 (1998).

\item {\bf [Schack-Caves 99 a]}:
R. Schack, \& C. M. Caves,
``Classical model for bulk-ensemble NMR quantum computation'',
{\em Phys. Rev. A} {\bf 60}, 6, 4354-4362 (1999).

\item {\bf [Schack-Caves 99 b]}:
R. Schack, \& C. M. Caves,
``Explicit product ensembles for separable quantum states'',
in V. Bu\v{z}zek, \& D. P. DiVincenzo (eds.),
{\em J. Mod. Opt.} {\bf 47}, 2-3 (Special issue:
Physics of quantum information), 387-399 (2000).

\item {\bf [Schack-Brun-Caves 00]}:
R. Schack, T. A. Brun, \& C. M. Caves,
``Quantum Bayes rule'',
{\em Phys. Rev. A} {\bf 64}, 1, 014305 (2001);
quant-ph/0008113.

\item {\bf [Schack 03]}:
R. Schack
``Quantum theory from four of Hardy's axioms'',
{\em Found. Phys.} {\bf 33}, 10, 1461-1468 (2003).
See {\bf [Hardy 01, 02 a]}.

\item {\bf [Schaetz-Barrett-Leibfried-(+6) 04]}:
T. Schaetz, M. D. Barrett, D. Leibfried,
J. Chiaverini, J. Britton, W. M. Itano,
J. D. Jost, C. Langer, \& D. J. Wineland,
``Quantum dense coding with atomic qubits'',
{\em Phys. Rev. Lett.} {\bf 93}, 4, 040505 (2004).

\item {\bf [Schafir 97]}:
R. L. Schafir,
``Comment on `Entangled entanglement'\,'',
{\em Phys. Rev. A} {\bf 56}, 5, 4335 (1997).
Comment on {\bf [Krenn-Zeilinger 96]}.
Reply: {\bf [Krenn-Zeilinger 97]}.
See {\bf [Krenn-Zeilinger 95]},
{\bf [Cereceda 97 c]}.

\item {\bf [Schafir 98 a]}:
R. L. Schafir,
``Nonlocality in the many-worlds and
consistent-histories interpretations'',
{\em Found. Phys.} {\bf 28}, 2, 157-166 (1998).
See {\bf [Griffiths 99 a]}.

\item {\bf [Schafir 98 b]}:
R. L. Schafir,
``Comment on `Consistent histories and quantum reasoning'\,'',
{\em Phys. Rev. A} {\bf 58}, 4, 3353-3355 (1998).
Comment on {\bf [Griffiths 96]}.
Repy: {\bf [Griffiths 98 b]}.

\item {\bf [Schafir 02 a]}:
R. L. Schafir,
``Comment on an alleged refutation of non-locality'',
quant-ph/0201065.
Comment on {\bf [Deutsch-Hayden 00]}.

\item {\bf [Schafir 02 b]}:
R. L. Schafir,
``Can a quantum measurement be cancelled in a very short period of time?'',
quant-ph/0201080.

\item {\bf [Schafir 02 c]}:
R. L. Schafir,
``Comment on a suggested Kochen-Specker test'',
quant-ph/0208127.
Comment on {\bf [Simon-\.{Z}ukowski-Weinfurter-Zeilinger 00]}.

\item {\bf [Schatten 93]}:
K. H. Schatten,
``Hidden-variable model for the Bohm Einstein-Podolsky-Rosen experiment'',
{\em Phys. Rev. A} {\bf 48}, 1, 103-104 (1993).

\item {\bf [Schauer 94 a]}:
D. L. Schauer,
``Comment on `Quantum mechanics, local
realistic theories, and Lorentz-invariant realistic theories'\,'',
{\em Phys. Rev. Lett.} {\bf 72}, 5, 782 (1994).
Comment on {\bf [Hardy 92 a]}.

\item {\bf [Schauer 94 b]}:
D. L. Schauer,
``Bell's inequalities between independent particles'',
{\em Found. Phys. Lett.} {\bf 7}, 4, 373-378 (1994).

\item {\bf [Scheel-Kn\"{o}ll-Opatrn\'{y}-Welsch 00 a]}:
S. Scheel, L. Kn\"{o}ll, T. Opatrn\'{y}, \& D.-G. Welsch,
``Entanglement transformation at absorbing and amplifying four-port
devices'',
{\em Phys. Rev. A} {\bf 62}, 4, 43803 (2000);
quant-ph/0004003.

\item {\bf [Scheel-Kn\"{o}ll-Opatrn\'{y}-Welsch 00 b]}:
S. Scheel, L. Kn\"{o}ll, T. Opatrn\'{y}, \& D.-G. Welsch,
``Entanglement transformation at dielectric four-port devices'',
quant-ph/0010053.

\item {\bf [Scheel-Welsch 01]}:
S. Scheel, \& D.-G. Welsch,
``Entanglement generation and degradation by passive optical devices'',
{\em Phys. Rev. A} {\bf 64}, 6, 063811 (2001);
quant-ph/0103167.

\item {\bf [Scheel-Eisert-Knight-Plenio 03]}:
S. Scheel, J. Eisert, P. L. Knight, \& M. B. Plenio,
``Hot entanglement in a simple dynamical model'',
in M. Ferrero (ed.),
{\em Proc. of Quantum Information: Conceptual Foundations,
Developments and Perspectives (Oviedo, Spain, 2002)},
{\em J. Mod. Opt.} {\bf 50}, 6-7, 881-889 (2003).

\item {\bf [Scheel-Nemoto-Munro-Knight 03]}:
S. Scheel, K. Nemoto, W. J. Munro, \& P. L. Knight,
``Measurement-induced nonlinearity in linear optics'',
{\em Phys. Rev. A} {\bf 68}, 3, 032310 (2003);
quant-ph/0305082.

\item {\bf [Scheel-L\"{u}tkenhaus 04]}:
S. Scheel, \& N. L\"{u}tkenhaus,
``Upper bounds on success probabilities in linear optics'',
{\em New J. Phys.} {\bf 6}, 51,-? (2004);
quant-ph/0403103.

\item {\bf [Scheel-Pachos-Hinds-Knight 04]}:
S. Scheel, J. Pachos, E. A. Hinds, \& P. L. Knight,
``Quantum gates and decoherence'',
quant-ph/0403152.

\item {\bf [Scheel 04]}:
S. Scheel,
``Scaling of success probabilities for linear optics gates'',
quant-ph/0410014.

\item {\bf [Scheibe 73]}:
E. Scheibe,
{\em The logical analysis of quantum mechanics},
Pergamon Press, Oxford, 1973.

\item {\bf [Scheibe 93]}:
E. Scheibe,
``Three remarks concerning Bell's inequality'',
in A. van der Merwe, \& F. Selleri (eds.),
{\em Bell's theorem and the foundations of modern physics.
Proc.\ of an international
conference (Cesena, Italy, 1991)},
World Scientific, Singapore, 1993, pp.~428-435.

\item {\bf [Schelpe-Kent-Munro-Spiller 03]}:
B. Schelpe, A. Kent, W. Munro, \& T. Spiller,
``Inferring superposition and entanglement in evolving systems from
measurements in a single basis'',
{\em Phys. Rev. A} {\bf 67}, 5, 052316 (2003).

\item {\bf [Schenkel-Persaud-Park-(+6) 02]}:
T. Schenkel, A. Persaud, S. J. Park,
J. Meijer, J. R. Kingsley, J. W. McDonald, J. P. Holder, J. Bokor, \& D. H. Schneider,
``Single ion implantation for solid state quantum computer development'',
{\em J. Vac. Sci. Technol. B} {\bf 20}, ?, 2819-? (2002).

\item {\bf [Schenkel-Persaud-Park-(+7) 03]}:
T. Schenkel, A. Persaud, S. J. Park,
J. Nilsson, J. Bokor, J. A. Liddle,
R. Keller, D. H. Schneider, D. W. Cheng, \& D. E. Humphries,
``Solid state quantum computer development in silicon with single ion implantation'',
{\em J. Appl. Phys.} {\bf 94}, 7017-7024 (2003).

\item {\bf [Schenzle 96]}:
A. Schenzle,
``Illusion or reality: The measurement process in quantum
optics'',
{\em Contemp. Phys.} {\bf 37}, 4, 303-320 (1996).

\item {\bf [Scherer-Soklakov-Schack 04]}:
A. Scherer, A. N. Soklakov, \& R. Schack,
``A simple necessary decoherence condition for a set of histories'',
{\em Phys. Lett. A} {\bf 326}, 5-6, 307-314 (2004);
quant-ph/0401132.
See {\bf [Scherer-Soklakov 04]}.

\item {\bf [Scherer-Soklakov 04]}:
A. Scherer, \& A. N. Soklakov,
``Initial states and decoherence of histories'',
quant-ph/0405080.

\item {\bf [Schiff 49]}:
L. I. Schiff,
{\em Quantum mechanics},
McGraw-Hill, New-York, 1949 (1st edition), 1968 (3rd edition).
English-Spanish version of the 2nd edition:
{\em Mec\'{a}nica cu\'{a}ntica},
McGraw-Hill (?), New-York (?), ?.

\item {\bf [Schirmer-Girardeau-Leahy 00]}:
S. G. Schirmer, M. D. Girardeau, \& J. V. Leahy,
``Efficient algorithm for optimal control of mixed-state quantum systems'',
{\em Phys. Rev. A} {\bf 61}, 1, 012101 (2000);
quant-ph/9906089.

\item {\bf [Schirmer-Greentree-Ramakrishna-Rabitz 01]}:
S. G. Schirmer, A. D. Greentree, V. Ramakrishna, \& H. Rabitz,
``Quantum control using sequences of simple
control pulses'',
quant-ph/0105155.

\item {\bf [Schirmer-Zhang-Leahy 03]}:
S. G. Schirmer, T. Zhang, \& J. V. Leahy,
``Orbits of quantum states and geometry of Bloch vectors for $N$-level
systems'',
quant-ph/0308004.

\item {\bf [Schirmer-Kolli-Oi 04]}:
S. G. Schirmer, A. Kolli, \& D. K. L. Oi,
``Experimental Hamiltonian identification for controlled two-level systems'',
{\em Phys. Rev. A} {\bf 69}, 5, 050306 (2004).

\item {\bf [Schlegel 71]}:
R. Schlegel,
``The Einstein-Podolsky-Rosen paradox'',
{\em Am. J. Phys.} {\bf 39}, 4, 458 (1971).
Comment on {\bf [Hooker 70]}.
See {\bf [Erlichson 72]}.

\item {\bf [Schleich 99]}:
W. P. Schleich,
``Quantum control: Sculpting a wavepacket'',
{\em Nature} {\bf 397}, 6716, 207-208 (1999).
See {\bf [Weinacht-Ahn-Bucksbaum 99]}.

\item {\bf [Schleich 00]}:
W. P. Schleich,
``Quantum physics: Engineering decoherence'',
{\em Nature} {\bf 403}, 6767, 256-257 (2000).
See {\bf [Myatt-King-Turchette-(+5)]}.

\item {\bf [Schliemann-Cirac-Ku\'s-(+2) 01]}:
J. Schliemann, J. I. Cirac, M. Ku\'s,
M. Lewenstein, \& D. Loss,
``Quantum correlations in two-fermion systems'',
{\em Phys. Rev. A} {\bf 64}, 2, 022303 (2001);
quant-ph/0012094.

\item {\bf [Schliemann-Khaetskii-Loss 02]}:
J. Schliemann, A. V. Khaetskii, \& D. Loss,
``Spin decay and quantum parallelism'',
{\em Phys. Rev. B} {\bf 66}, 24, 245303 (2002).

\item {\bf [Schliemann 03]}:
J. Schliemann,
``Entanglement in SU(2)-invariant quantum spin systems'',
{\em Phys. Rev. A} {\bf 68}, 1, 012309 (2003).

\item {\bf [Schlienz-Mahler 95]}:
J. Schlienz, \& G. Mahler,
``Description of entanglement'',
{\em Phys. Rev. A} {\bf 52}, 6, 4396-4404 (1995).

\item {\bf [Schlienz-Mahler 96]}:
J. Schlienz, \& G. Mahler,
``The maximal entangled three-particle state is unique'',
{\em Phys. Lett. A} {\bf 224}, 1-2, 39-44 (1996).

\item {\bf [Schlingemann-Werner 02]}:
D. Schlingemann, \& R. F. Werner,
``Quantum error-correcting codes associated with graphs'',
{\em Phys. Rev. A} {\bf 65}, 1, 012308 (2002);
quant-ph/0012111.

\item {\bf [Schlingemann 01]}:
D. Schlingemann,
``Stabilizer codes can be realized as graph codes'',
quant-ph/0111080.

\item {\bf [Schlingemann 03]}:
D. Schlingemann,
``Cluster states, algorithms and graphs'',
quant-ph/0305170.

\item {\bf [Schlosshauer-Fine 03]}:
M. Schlosshauer, \& A. Fine,
``On Zurek's derivation of the Born rule'',
{\em Found. Phys.};
quant-ph/0312058.
See {\bf [Zurek 03 b]}.

\item {\bf [Schlosshauer 05]}:
M. Schlosshauer,
``Decoherence, the measurement problem, and interpretations of quantum
mechanics'',
{\em Rev. Mod. Phys.} (2005);
quant-ph/0312059.

\item {\bf [Schmidt 07 a]}:
E. Schmidt,
``Zur Theorie der linearen und nichtlinearen Integralgleichungen. I. Teil:
Entwicklung willk\"{u}rlicher Funktionen nach Systemen vorgeschriebener'',
{\em Math. Ann.} {\bf 63}, 433-476 (1907).

\item {\bf [Schmidt 07 b]}:
E. Schmidt,
``Zur Theorie der linearen und nichtlinearen Integralgleichungen. Zweite
Abhandlung: Aufl\"{o}sung der allgemeinen linearen Integralgleichung'',
{\em Math. Ann.} {\bf 64}, 161-174 (1907).

\item {\bf [Schmidt Kaler-H\"{a}ffner-Riebe-(+7) 03]}:
F. Schmidt-Kaler, H. H\"{a}ffner, M. Riebe,
S. Gulde, G. P. T. Lancaster, T. Deuschle,
C. Becher, C. F. Roos, J. Eschner, \& R. Blatt,
``Realization of the Cirac-Zoller controlled-NOT quantum gate'',
{\em Nature} {\bf 422}, 6930, 408-411 (2003).
More details: {\bf [Schmidt Kaler-H\"{a}ffner-Riebe-(+7) 03]}.

\item {\bf [Schmidt Kaler-H\"{a}ffner-Gulde-(+5) 03]}:
F. Schmidt-Kaler, H. H\"{a}ffner, S. Gulde,
M. Riebe, G. Lancaster, J. Eschner,
C. Becher, \& R. Blatt,
``Quantized phase shifts and a dispersive universal quantum gate'',
quant-ph/0307211.

\item {\bf [Schmidt Kaler-H\"{a}ffner-Riebe-(+7) 03]}:
F. Schmidt-Kaler, H. H\"{a}ffner, M. Riebe,
G. P. T. Lancaster, T. Deuschle, C. Becher,
W. H\"{a}nsel, J. Eschner, C. F. Roos, \& R. Blatt,
``How to realize a universal quantum gate with trapped ions'',
{\em Appl. Phys. B} {\bf 77}, 789-796 (2003);
quant-ph/0312162.
See {\bf [Schmidt Kaler-H\"{a}ffner-Riebe-(+7) 03]}.

\item {\bf [Schneider-LaPuma 02]}:
M. B. Schneider, \& I. A. LaPuma,
``A simple experiment for discussion of quantum
interference and which-way measurement'',
{\em Am. J. Phys.} {\bf 70}, 3, 266-271 (2002).

\item {\bf [Schneider-Wiseman-Munro-Milburn 98]}:
S. Schneider, H. M. Wiseman, W. J. Munro, \& G. J. Milburn,
``Measurement and state preparation via ion trap quantum computing'',
{\em Fortschr. Phys.} {\bf 46}, 4-5, 391-399 (1998);
quant-ph/9709042.
See {\bf [D'Helon-Milburn 96]}.

\item {\bf [Schneider-James-Milburn 98]}:
S. Schneider, D. F. V. James, \& G. J. Milburnaper.
`Method of quantum computation with ``hot'' trapped ions',
quant-ph/9808012.

\item {\bf [Schneider-Milburn 98]}:
S. Schneider, \& G. J. Milburn,
``Decoherence and fidelity in ion traps with fluctuating
trap parameters'',
quant-ph/9812044.

\item {\bf [Schneider-James-Milburn 00]}:
S. Schneider, D. F. V. James, \& G. J. Milburn,
``Quantum controlled-NOT gate wit `hot' trapped ions'',
in V. Bu\v{z}zek, \& D. P. DiVincenzo (eds.),
{\em J. Mod. Opt.} {\bf 47}, 2-3 (Special issue:
Physics of quantum information), 499-505 (2000).

\item {\bf [Schneider-Milburn 02]}:
S. Schneider, \& G. J. Milburn,
``Entanglement in the steady state of a collective-angular-momentum (Dicke)
model'',
{\em Phys. Rev. A} {\bf 65}, 4, 042107 (2002);
quant-ph/0112042.

\item {\bf [Schneiderman-Stanley-Aravind 02]}:
J. F. Schneiderman, M. E. Stanley, \& P. K. Aravind,
``A pseudo-simulation of Shor's quantum factoring algorithm'',
quant-ph/0206101.

\item {\bf [Schoen-Beige 01]}:
C. Schoen, \& A. Beige,
``An analysis of a two-atom double-slit experiment
based on environment-induced measurements'',
{\em Phys. Rev. A};
quant-ph/0104076.

\item {\bf [Schonhammer 00]}:
K. Schonhammer,
``Stern-Gerlach measurements with arbitrary spin:
Quantum combs'',
{\em Am. J. Phys.} {\bf 68}, 1, 48-54 (2000).

\item {\bf [Schori-Julsgaard-Sorensen-Polzik 02]}:
C. Schori, B. Julsgaard, J. L. Sorensen, \& E. S. Polzik,
``Recording quantum properties of light in a long-lived atomic spin state:
Towards quantum memory'',
{\em Phys. Rev. Lett.} {\bf 89}, 5, 057903 (2002).

\item {\bf [Schori-Sorensen-Polzik 02]}:
C. Schori, J. L. Sorensen, \& E. S. Polzik,
``Narrow-band frequency tunable light source of continuous quadrature
entanglement'',
{\em Phys. Rev. A} {\bf 66}, 3, 033802 (2002).

\item {\bf [Schrader 00]}:
R. Schrader,
``On a quantum version of Shannon's conditional entropy'',
quant-ph/0003048.

\item {\bf [Schr\"{o}dinger 27]}:
E. Schr\"{o}dinger,
{\em Abhandlungen zur Wellenmechanik}, ?, ?, 1927.
English version: {\em Collected papers on wave
mechanics (together with his four lectures on wave mechanics)},
Chelsea, Nueva York, 1982 (3rd edition, enlarged).

\item {\bf [Schr\"{o}dinger 35 a]}:
E. Schr\"{o}dinger,
``Die gegenw\"{a}rtige Situation in der Quantenmechanik'',
{\em Naturwissenschaften} {\bf 23}, 48, 807-812; 49, 823-828;
50, 844-849 (1935).
English version: J. D. Trimmer,
``The present situation in quantum mechanics:
A translation of Schr\"{o}dinger's `cat paradox'\,'',
{\em Proc.\ Am. Philos. Soc.} {\bf 124}, 5, 323-338 (1980).
Reprinted in {\bf [Wheeler-Zurek 83]}, pp.~152-167.

\item {\bf [Schr\"{o}dinger 35 b]}:
E. Schr\"{o}dinger,
``Discussion of probability relations between separated systems'',
{\em Proc.\ Camb. Philos. Soc.} {\bf 31}, 555-563 (1935).

\item {\bf [Schr\"{o}dinger 36]}:
E. Schr\"{o}dinger,
``Probability relations between separated systems'',
{\em Proc.\ Camb. Philos. Soc.} {\bf 32}, 446-452 (1936).

\item {\bf [Schr\"{o}dinger 52]}:
E. Schr\"{o}dinger,
``Are there quantum jumps?'',
{\em Brit. J. Philos. Sci.} {\bf 3}, ?, 109-123; ? 233-247 (1952).

\item {\bf [Schr\"{o}dinger 53]}:
E. Schr\"{o}dinger,
``What is matter?'',
{\em Sci. Am.} September 1953, pp.~52-56.

\item {\bf [Schr\"{o}dinger 95]}:
E. Schr\"{o}dinger,
{\em The interpretation of quantum mechanics:
Dublin seminars (1949-1955) and other unpublished essays},
Ox Bow Press, Woodbridge, Connecticut, 1995.

\item {\bf [Schroeck 96]}:
F. E. Schroeck, Jr.,
{\em Quantum mechanics on phase space},
Kluwer Academic, Dordrecht, Holland, 1996.
Review: {\bf [Landsman 99]}.

\item {\bf [Schroer 03]}:
B. Schroer,
``Pascual Jordan, his contributions to quantum mechanics and his legacy in
contemporary local quantum physics'',
hep-th/0303241.

\item {\bf [Schuch-Siewert 03 a]}:
N. Schuch, \& J. Siewert,
``Natural two-qubit gate for quantum computation using the $XY$ interaction'',
{\em Phys. Rev. A} {\bf 67}, 3, 032301 (2003);
quant-ph/0209035.

\item {\bf [Schuch-Siewert 03 b]}:
N. Schuch, \& J. Siewert,
``Programmable networks for quantum algorithms'',
{\em Phys. Rev. Lett.} {\bf 91}, 2, 027902 (2003).

\item {\bf [Schuch-Verstraete-Cirac 04 a]}:
N. Schuch, F. Verstraete, \& J. I. Cirac,
``Nonlocal resources in the presence of superselection rules'',
{\em Phys. Rev. Lett.} {\bf 92}, 8, 087904 (2004);
quant-ph/0310124.

\item {\bf [Schuch-Verstraete-Cirac 04 b]}:
N. Schuch, F. Verstraete, \& J. I. Cirac,
``Quantum entanglement theory in the presence of superselection rules'',
quant-ph/0404079.

\item {\bf [Schulman 69]}:
L. S. Schulman,
``Measurement and interference'',
{\em Am. J. Phys.} {\bf 37}, 12, 1220-1221 (1969).

\item {\bf [Schulman-Mozyrsky 97]}:
L. S. Schulman, \& D. Mozyrsky,
``Measure of entanglement'',
quant-ph/9701030.

\item {\bf [Schulman 98 a]}:
L. S. Schulman,
``Bounds of decoherence and error'',
{\em Phys. Rev. A} {\bf 57}, 2, 840-844 (1998).

\item {\bf [Schulman 98 b]}:
L. S. Schulman,
``Continuous and pulsed observations in the quantum Zeno effect'',
{\em Phys. Rev. A} {\bf 57}, 3, 1509-1515 (1998).

\item {\bf [Schulman-Vazirani 98]}:
L. S. Schulman, \& U. Vazirani,
``Scalable NMR quantum computation'',
quant-ph/9804060.

\item {\bf [Schulman 99]}:
L. S. Schulman,
``Book review. Decoherence and quantum measurements'',
{\em Found. Phys.} {\bf 29}, 11, 1807-1810 (1999).
Review of {\bf [Namiki-Pascazio-Nakazato 97]}.

\item {\bf [Schulman 01]}:
L. S. Schulman,
``Jump time and passage time: the duration of a quantum transition'',
to appear in J. G. Muga, R. Sala Mayato, \& I. L. Egusquiza,
{\em Time in quantum mechanics},
Springer-Verlag;
quant-ph/0103151.

\item {\bf [Schulz-Steinh\"{u}bl-Weber-(+6) 03]}:
O. Schulz, R. Steinh\"{u}bl, M. Weber,
B.-G. Englert, C. Kurtsiefer, \& H. Weinfurter,
``Ascertaining the values of $\sigma_x$, $\sigma_y$, and $\sigma_z$ of a
polarization qubit'',
{\em Phys. Rev. Lett.} {\bf 90}, 17, 177901 (2003);
quant-ph/0209127.

\item {\bf [Schumacher 90]}:
B. W. Schumacher,
``Information from quantum measurements'',
in {\bf [Zurek 90]}, pp.~29-37.

\item {\bf [Schumacher 91]}:
B. W. Schumacher,
``Information and quantum nonseparability'',
{\em Phys. Rev. A} {\bf 44}, 11, 7047-7052 (1991).

\item {\bf [Schumacher 95]}:
B. W. Schumacher,
``Quantum coding'',
{\em Phys. Rev. A} {\bf 51}, 4, 2738-2747 (1995).

\item {\bf [Schumacher-Nielsen 96]}:
B. W. Schumacher, \& M. A. Nielsen,
``Quantum data processing and error correction'',
quant-ph/9604022.

\item {\bf [Schumacher 96]}:
B. W. Schumacher,
``Sending entanglement through noisy quantum channels'',
{\em Phys. Rev. A} {\bf 54}, 4, 2614-2628 (1996);
quant-ph/9604023.
Reprinted in {\bf [Macchiavello-Palma-Zeilinger 00]}, pp.~180-194.

\item {\bf [Schumacher-Westmoreland-Wootters 96]}:
B. W. Schumacher, M. Westmoreland, \& W. K. Wootters,
``Limitation on the amount of accessible information in a quantum channel'',
{\em Phys. Rev. Lett.} {\bf 76}, 18, 3452-3455 (1996).

\item {\bf [Schumacher-Westmoreland 97]}:
B. W. Schumacher, \& M. D. Westmoreland,
``Sending classical information via noisy quantum channels'',
{\em Phys. Rev. A} {\bf 56}, 1, 131-138 (1997).
Reprinted in {\bf [Macchiavello-Palma-Zeilinger 00]}, pp.~199-206.
See {\bf [Hausladen-Jozsa-Schumacher-(+2) 96]}.

\item {\bf [Schumacher-Westmoreland 98]}:
B. W. Schumacher, \& M. D. Westmoreland,
``Quantum privacy and quantum coherence'',
{\em Phys. Rev. Lett.} {\bf 80}, 25, 5695-5697 (1998);
quant-ph/9709058.

\item {\bf [Schumacher-Westmoreland 99]}:
B. W. Schumacher, \& M. D. Westmoreland,
``Optimal signal ensembles'',
{\em Phys. Rev. A} {\bf 63}, 2, 022308 (2001).
quant-ph/9912122.

\item {\bf [Schumacher-Westmoreland 01 a]}:
B. W. Schumacher, \& M. D. Westmoreland,
``Indeterminate-length quantum coding'',
{\em Phys. Rev. A} {\bf 64}, 4, 042304 (2001);
quant-ph/0011014.

\item {\bf [Schumacher-Westmoreland 01 b]}:
B. W. Schumacher, \& M. D. Westmoreland,
``Approximate quantum error correction'',
quant-ph/0112106.

\item {\bf [Schumacher-Westmoreland 02 a]}:
B. W. Schumacher, \& M. D. Westmoreland,
``Entanglement and perfect quantum error correction'',
{\em J. Math. Phys.} {\bf 43}, 9, 4279-4285 (2002);
quant-ph/0201061.

\item {\bf [Schumacher-Westmoreland 02 b]}:
B. W. Schumacher, \& M. D. Westmoreland,
``Relative entropy in quantum information theory'',
in {\bf [Lomonaco-Brandt 02]} 265-289;
quant-ph/0004045.

\item {\bf [Schumacher-Werner 04]}:
B. W. Schumacher, \& R. F. Werner,
``Reversible quantum cellular automata'',
quant-ph/0405174.

\item {\bf [Schumacher-Westmoreland 02 b]}:
B. W. Schumacher, \& M. D. Westmoreland,
``Locality and information transfer in quantum operations'',
quant-ph/0406223.

\item {\bf [Schumann 00]}:
R. H. Schumann,
``Quantum information theory'',
M.\ Sc. thesis, University of Stellenbosch, South Africa, 2000;
quant-ph/0010060.

\item {\bf [Schust-Mattes-Sorg 04]}:
P. Schust, M. Mattes, \& M. Sorg,
``Quantum entanglement in relativistic three-particle systems'',
{\em Found. Phys.} {\bf 34}, 1, 99-167 (2004).

\item {\bf [Schutzhold 03]}:
R. Schutzhold,
``Pattern recognition on a quantum computer'',
{\em Phys. Rev. A} {\bf 67}, 6, 062311 (2003).

\item {\bf [Schwarzschild 96]}:
B. Schwarzschild,
``Labs demonstrate logic gates for quantum computation'',
{\em Phys. Today} {\bf 49}, 3, 21-23 (1996).

\item {\bf [Schweber 00]}:
S. S. Schweber,
``The genius of physics: A portrait gallery'',
{\em Phys. Today} {\bf 53}, 9, 61-62 (2000).
Review of {\bf [Pais 00]}.

\item {\bf [Schweber 01]}:
S. S. Schweber,
``The historical development of quantum theory,
volume 6: The completion of quantum mechanics 1926-1941'',
{\em Phys. Today} {\bf 54}, 11, 56-? (2001).
Review of {\bf [Mehra-Rechenberg 00 a, b]}.

\item {\bf [Schwindt-Kwiat-Englert 99]}:
P. D. D. Schwindt, P. G. Kwiat, \& B.-G. Englert,
``Quantitative wave-particle duality and nonerasing quantum erasure'',
{\em Phys. Rev. A} {\bf 60}, 6, 4285-4290 (1999);
quant-ph/9908072.

\item {\bf [Schwinger-Scully-Englert 88]}:
J. Schwinger, M. O. Scully, \& B.-G. Englert,
``Is spin coherence like Humpty-Dumpty? II. General theory'',
{\em Zeitschrift f\"{u}r Physik D} {\bf 10}, 2-3, 135-144 (1988).
See {\bf [Englert-Schwinger-Scully 88]} (I),
{\bf [Scully-Englert-Schwinger 89]} (III).

\item {\bf [Schwinger 01]}:
J. Schwinger,
{\em Quantum mechanics: Symbolism of atomic measurements},
B.-G. Englert (ed.),
Spinger-Verlag, Berlin, 2001.
Review: {\bf [Rae 01]}, {\bf [Greenberger 03]}.

\item {\bf [Sciarrino-Lombardi-Milani-De Martini 02]}:
F. Sciarrino, E. Lombardi, G. Milani, \& F. De Martini,
``Delayed-choice entanglement swapping with vacuum--one-photon quantum states'',
{\em Phys. Rev. A} {\bf 66}, 2, 024309 (2002);
quant-ph/0201019.

\item {\bf [Sciarrino-Sias-Ricci-De Martini 02]}:
F. Sciarrino, C. Sias, M. Ricci, \& F. De Martini,
``Teleportation of the U-NOT gate and quantum cloning: A computational
network and experimental realization'',
quant-ph/0311160.

\item {\bf [Sciarrino-Sias-Ricci-De Martini 04]}:
F. Sciarrino, C. Sias, M. Ricci, \& F. De Martini,
``Quantum cloning and universal NOT gate by teleportation'',
{\em Phys. Lett. A} {\bf 323}, 1-2, 34-39 (2004).

\item {\bf [Sciarrino-De Martini-Bu\v{z}zek 04]}:
F. Sciarrino, F. De Martini, \& V. Bu\v{z}zek,
``Realization of the optimal universal quantum entangler'',
{\em Phys. Rev. A};
quant-ph/0410224.

\item {\bf [Sciarrino-Sias-Ricci-De Martini 04]}:
F. Sciarrino, C. Sias, M. Ricci, \& F. De Martini,
``Realization of universal optimal quantum machines by projective
operators and stochastic maps'',
quant-ph/0407176.

\item {\bf [Scott-Caves 03]}:
A. J. Scott, \& C. M. Caves,
``Entangling power of the quantum baker's map'',
{\em J. Phys. A} {\bf 36}, 36, 9553–9576 (2003);
quant-ph/0305046.

\item {\bf [Scott 04]}:
A. J. Scott,
``Multipartite entanglement, quantum-error-correcting codes,
 and entangling power of quantum evolutions'',
{\em Phys. Rev. A} {\bf 69}, 5, 052330 (2004).

\item {\bf [Scully-Dr\"{u}hl 82]}:
M. O. Scully, \& K. Dr\"{u}hl,
`Quantum eraser: A proposed photon correlation experiment concerning
observation and ``delayed choice'' in quantum mechanics',
{\em Phys. Rev. A} {\bf 25}, 4, 2208-2213 (1982).

\item {\bf [Scully-Cohen 86]}:
M. O. Scully, \& L. Cohen,
``EPRB and the Wigner distribution for spin-$\frac{1}{2}$ particles'',
in D. M. Greenberger (ed.),
{\em New techniques and ideas in quantum measurement theory.
Proc.\ of an international conference (New York, 1986),
Ann. N. Y. Acad. Sci.} {\bf 480}, 115-117 (1986).

\item {\bf [Scully-Walther 89]}:
M. O. Scully, \& H. Walther,
``Quantum optical test of observation and complementarity in quantum mechanics'',
{\em Phys. Rev. A} {\bf 39}, 10, 5229-5236 (1989).

\item {\bf [Scully-Englert-Schwinger 89]}:
M. O. Scully, B.-G. Englert, \& J. Schwinger,
``Spin coherence and Humpty-Dumpty. III. The effects of observation'',
{\em Phys. Rev. A} {\bf 40}, 4, 1775-1784 (1989).
See {\bf [Englert-Schwinger-Scully 88]} (I),
{\bf [Schwinger-Scully-Englert 88]} (II).

\item {\bf [Scully-Englert-Walther 91]}:
M. O. Scully, B.-G. Englert, \& H. Walther,
``Quantum optical tests of complementarity'',
{\em Nature} {\bf 351}, 6322, 111-116 (1991).
See {\bf [Englert-Scully-Walther 94]}, {\bf [Mohrhoff 96]},
{\bf [Scully-Zubairy 97]} (Chap.~20).

\item {\bf [Scully-Bednar-Rostovtsev-Zhu 97]}:
M. O. Scully, C. J. Bednar, Y. Rostovtsev, \& S.-Y. Zhu,
``Counter-counter-intuitive quantum coherence effects'',
in P. L. Knight, B. Stoicheff, \& D. Walls (eds.),
{\em Highlight in Quantum Optics},
{\em Philos. Trans. R. Soc. Lond. A} {\bf 355}, 1733, 2305-2311 (1997).

\item {\bf [Scully-Zubairy 97]}:
M. O. Scully, \& M. S. Zubairy,
{\em Quantum optics},
Cambridge University Press, Cambridge, 1997.
Reviews: {\bf [Milonni 98]}, {\bf [Plenio 98]}, {\bf [Walmsley 99]},
{\bf [Englert 99]}.

\item {\bf [Scully-Walther 98]}:
M. O. Scully, \& H. Walther,
``An operational analysis of quantum eraser and delayed choice'',
{\em Found. Phys.} {\bf 28}, 3, 399-413 (1998).

\item {\bf [Scully-Aharonov-Englert 99]}:
M. O. Scully, Y. Aharonov, \& B.-G. Englert,
``On the locality and reality of Einstein-Podolsky-Rosen correlations'',
in R. Bonifacio (ed.),
{\em Mysteries, Puzzles, and Paradoxes in Quantum Mechanics (Gargnano, Italy, 1998)},
American Institute of Physics, Woodbury, New York, 1999, pp.~47-68.

\item {\bf [Scully-Englert-Bednar 99]}:
M. O. Scully, B.-G. Englert, \& C. J. Bednar,
``Two-photon scheme for detecting the Bell basis using atomic coherence'',
{\em Phys. Rev. Lett.} {\bf 83}, 21, 4433-4436 (1999).

\item {\bf [Scully-Zubairy 01]}:
M. O. Scully, \& M. S. Zubairy,
``Quantum search protocol for an atomic array'',
{\em Phys. Rev. A} {\bf 64}, 2, 022304 (2001).
Comment:{\bf [Coffey 02]}.
Reply: {\bf [Scully-Zubairy 02 b]}.

\item {\bf [Scully-Zubairy 02 a]}:
M. O. Scully, \& M. S. Zubairy,
``Cavity QED implementation of the discrete quantum Fourier transform'',
{\em Phys. Rev. A} {\bf 65}, 5, 052324 (2002).

\item {\bf [Scully-Zubairy 02 b]}:
M. O. Scully, \& M. S. Zubairy,
`Reply to ``Comment on `Quantum search protocol for an atomic array'\,''\,',
{\em Phys. Rev. A} {\bf 66}, 5, 056301 (2002).
Reply to {\bf [Coffey 02]}.
See {\bf [Scully-Zubairy 01]}.

\item {\bf [Scully-Zubairy-Agarwal-Walther 03]}:
M. O. Scully, M. S. Zubairy, G. S. Agarwal, \& H. Walther,
``Extracting work from a single heat bath via vanishing quantum coherence'',
{\em Science} {\bf 299}, ?, 862-? (2003).

\item {\bf [Scutaru 95]}:
H. Scutaru,
``Lower bound for mutual information of a quantum channel'',
{\em Phys. Rev. Lett.} {\bf 75}, 5, 773-776 (1995).

\item {\bf [Scutaru 03]}:
H. Scutaru,
``On the separability of pure states'',
{\em Proc. Rom. Acad. Ser. A} {\bf 4}, 3, 183-188 (2003).
See {\bf [Wang 02 b]}.

\item {\bf [Seevinck-Uffink 02]}:
M. P. Seevinck, \& J. Uffink,
``Sufficient conditions for three-particle entanglement and their tests in
recent experiments'',
{\em Phys. Rev. A} {\bf 65}, 1, 012107 (2002);
quant-ph/0107072.

\item {\bf [Seevinck-Svetlichny 02]}:
M. P. Seevinck, \& G. Svetlichny,
``Bell-type inequalities for partial separability in $N$-particle
systems and quantum mechanical violations'',
{\em Phys. Rev. Lett.} {\bf 89}, 6, 060401 (2002);
quant-ph/0201046.
See {\bf [Collins-Gisin-Popescu-(+2) 02]}.

\item {\bf [Seevinck 02]}:
M. P. Seevinck,
``Entanglement, local hidden variables and Bell-inequalities. An
investigation in multi-partite quantum mechanics'',
Masters Thesis, University of Utrecht, 2002.

\item {\bf [Seevinck 04]}:
M. P. Seevinck
``Holism, physical theories and quantum mechanics'',
quant-ph/0402047.

\item {\bf [Segev-Milonni-Babb-Chiao 00]}:
B. Segev, P. W. Milonni, J. F. Babb, \& R. Y. Chiao,
``Quantum noise and superluminal propagation'',
{\em Phys. Rev. A} {\bf 62}, 2, 022114 (2000);
quant-ph/0004047.

\item {\bf [Segev-Milonni-Babb-Chiao 00]}:
B. Segev,
``Causality and propagation in the Wigner, Husimi, Glauber,
and Kirkwood phase-space representations'',
{\em Phys. Rev. A} {\bf 63}, 5, 052114 (2001).

\item {\bf [Segre 01]}:
G. Segre,
``Algorithmic information theoretic issues in quantum mechanics'',
Ph.\ D. thesis;
quant-ph/0110018.

\item {\bf [Segre 02]}:
G. Segre,
``The no cloning theorem versus the second law of thermodynamics'',
quant-ph/0202109.

\item {\bf [Segre 04]}:
G. Segre,
``A remark about the Mermin-Squires Music Hall's interludium'',
quant-ph/0403087.

\item {\bf [Sehat-S\"{o}derholm-Bj\"{o}rk-(+3) 04]}:
A. Sehat, J. S\"{o}derholm, G. Bj\"{o}rk,
P. Espinoza, A. B. Klimov, \& L. L. S\'{a}nchez-Soto,
``Quantum polarization properties of two-mode number states'',
quant-ph/0407184.

\item {\bf [Seife 97]}:
C. Seife,
``Physics: Flaw found in a quantum code'',
{\em Science} {\bf 276}, 5315, 1034 (1997).

\item {\bf [Seife 00 a]}:
C. Seife,
``Neuroscience: Cold numbers unmake the quantum mind'',
{\em Science} {\bf 287}, 5454, 791 (2000).

\item {\bf [Seife 00 b]}:
C. Seife,
``Quantum mechanics: `Spooky action' passes a relativistic test'',
{\em Science} {\bf 287}, 5460, 1909-1910 (2000).

\item {\bf [Seife 02]}:
C. Seife,
``The quandary of quantum information'',
{\em Science} {\bf 293}, ?, 2026-2027 (2001).

\item {\bf [Sekatski 03]}:
S. K. Sekatski,
``Preparation of entangled spin states for a free
electron and nucleus by resonance laser photoionization'',
{\em JETP Lett} {\bf 78}, 405-407 (2003).

\item {\bf [Selleri 78]}:
F. Selleri,
``On the consequences of Einstein locality'',
{\em Found. Phys.} {\bf 8}, 1-2, 103-116 (1978).

\item {\bf [Selleri-Tarozzi 81]}:
F. Selleri, \& G. Tarozzi,
``Quantum mechanics reality and separability'',
{\em Rivista del Nuovo Cimento} {\bf 4}, 2, 1-53 (1981).

\item {\bf [Selleri 82]}:
F. Selleri,
``Generalized EPR-paradox'',
{\em Found. Phys.} {\bf 12}, 7, 645-659 (1982).

\item {\bf [Selleri 85]}:
F. Selleri,
``Local realistic photon models and EPR-type experiments'',
{\em Phys. Lett. A} {\bf 108}, 4, 197-202 (1985).

\item {\bf [Selleri-Tarozzi 86]}:
F. Selleri, \& G. Tarozzi,
``Why quantum mechanics is incompatible with Einstein locality'',
{\em Phys. Lett. A} {\bf 119}, 3, 101-104 (1986).

\item {\bf [Selleri 88]}:
F. Selleri,
``History of the Einstein-Podolsky-Rosen paradox'',
in F. Selleri (ed.),
{\em Quantum mechanics versus local realism: The
Einstein-Podolsky-Rosen paradox},
Plenum Press, New York, 1988, pp.~1-61.

\item {\bf [Selleri 90]}:
F. Selleri,
{\em Quantum paradoxes and physical reality},
Kluwer Academic, Dordrecht, Holland, 1990.

\item {\bf [Selleri 97]}:
F. Selleri,
``Incompatibility between local realism and
quantum mechanics for pairs of neutral kaons'',
{\em Phys. Rev. A} {\bf 56}, 5, 3493-3506 (1997).

\item {\bf [Selleri 02]}:
F. Selleri,
``Weak and strong Bell type inequalities'',
in C. Mataix, \& A. Rivadulla (eds.),
{\em F\'{\i}sica cu\'{a}ntica y realidad.
Quantum physics and reality (Madrid, 2000)},
Editorial Complutense, Madrid, 2002, pp.~231-247.

\item {\bf [Selleri 04]}:
F. Selleri,
``Towards a nonlinear quantum physics'',
{\em Found. Phys.} {\bf 34}, 4, 705-710 (2004).
Review of {\bf [Croca 03]}.

\item {\bf [Semi\~{a}o-Vidiella Barranco-Roversi 01 a]}:
F. L. Semi\~{a}o, A. Vidiella-Barranco, \& J. A. Roversi,
``Entanglement between motional states of a single trapped ion and light'',
{\em Phys. Rev. A} {\bf 64}, 2, 024305 (2001);
quant-ph/0101046.

\item {\bf [Semi\~{a}o-Vidiella Barranco-Roversi 01 b]}:
F. L. Semi\~{a}o, A. Vidiella-Barranco, \& J. A. Roversi,
``A proposal of quantum logic gates using cold trapped ions in a cavity'',
{\em Phys. Lett. A} {\bf 299}, 5-6, 423-426 (2002).
Erratum: {\em Phys. Lett. A} {\bf 306}, 1, 62 (2002).
quant-ph/0112044.

\item {\bf [Semi\~{a}o-Vidiella Barranco 03]}:
F. L. Semiao, \& A. Vidiella-Barranco,
``Schr\"{o}dinger-cat states and entanglement in cavity QED with trapped ions'',
quant-ph/0312056.

\item {\bf [Sen 91]}:
A. Sen,
`Comment on ``Quantum mysteries revisited'', by N.
David Mermin [{\em Am. J. Phys.} {\bf 58}, 731-734 (1990)]',
{\em Am. J. Phys.}
{\bf 59}, 8, 761 (1991).
Comment on {\bf [Mermin 90 b]}.

\item {\bf [Sen De-Sen 02]}:
A. Sen De, \& U. Sen,
``On irreversible bit-encoding between two separable states'',
quant-ph/0203007.

\item {\bf [Sen De-Sen-\.{Z}ukowski 02]}:
A. Sen De, U. Sen, \& M. \.{Z}ukowski,
``Functional Bell inequalities can serve as a stronger entanglement witness
than conventional Bell inequalities'',
{\em Phys. Rev. A} {\bf 66}, 6, 062318 (2002);
quant-ph/0206165.

\item {\bf [Sen De-Sen 03]}:
A. Sen De, \& U. Sen,
``Can there be quantum correlations in a mixture of two separable states?'',
in M. Ferrero (ed.),
{\em Proc. of Quantum Information: Conceptual Foundations,
Developments and Perspectives (Oviedo, Spain, 2002)},
{\em J. Mod. Opt.} {\bf 50}, 6-7, 981-985 (2003).

\item {\bf [Sen De-Sen-\.{Z}ukowski 03 a]}:
A. Sen De, U. Sen, \& M. \.{Z}ukowski,
``Unified criterion for security of secret sharing in terms of violation
of Bell inequality'',
{\em Phys. Rev. A} {\bf 68}, 3, 032309 (2003);
quant-ph/0302156.

\item {\bf [Sen De-Sen-\.{Z}ukowski 03 b]}:
A. Sen De, U. Sen, \& M. \.{Z}ukowski,
``Output state in multiple entanglement swapping'',
{\em Phys. Rev. A} {\bf 68}, 6, 062301 (2003).

\item {\bf [Sen De-Sen-Wie\'{s}niak-(+2) 03]}:
A. Sen De, U. Sen, M. Wie\'{s}niak, D. Kaszlikowski, \& M. \.{Z}ukowski,
``Multiqubit $W$ states lead to stronger nonclassicality than Greenberger-Horne-Zeilinger states'',
{\em Phys. Rev. A} {\bf 68}, 6, 062306 (2003);
quant-ph/0211023.

\item {\bf [Sen De-Sen-\.{Z}ukowski 03 c]}:
A. Sen De, U. Sen, \& M. \.{Z}ukowski,
``Characteristics of the output in a quantum repeater of arbitrary length'',
quant-ph/0305044.

\item {\bf [Sen De-Sen-Brukner-(+2) 03]}:
A. Sen De, U. Sen, \v{C}. Brukner, V. Bu\v{z}zek, \& M. \.{Z}ukowski,
``Entanglement swapping of noisy states: A kind of superadditivity in
nonclassicality'',
quant-ph/0311194.

\item {\bf [Sen De-Sen-Lewenstein 04]}:
A. Sen De, U. Sen, \& M. Lewenstein,
``Nonergodicity of entanglement and its complementary behavior to
magnetization in infinite spin chain'',
quant-ph/0407114.

\item {\bf [Sen 96]}:
S. Sen,
``Average entropy of quantum subsystems'',
{\em Phys. Rev. Lett.} {\bf 77}, 1, 1-3 (1996).

\item {\bf [Senitzky 94]}:
I. R. Senitzky,
``Classical interpretation of `frustrated two-photon creation via interference'\,'',
{\em Phys. Rev. Lett.} {\bf 73}, 22, 3040 (1994).
Comment on {\bf [Herzog-Rarity-Weinfurter-Zeilinger 94 a]}.
Reply: {\bf [Herzog-Rarity-Weinfurter-Zeilinger 94 b]}.

\item {\bf [Serafini-De Siena-Illuminati-Paris 03]}:
A. Serafini, S. De Siena, F. Illuminati, \& M. G. A. Paris,
``Minimum decoherence cat-like states in Gaussian noisy channels'',
quant-ph/0310005.

\item {\bf [Serafini-De Siena-Illuminati 03]}:
A. Serafini, S. De Siena, \& F. Illuminati,
``Decoherence of number states in phase-sensitive reservoirs'',
quant-ph/0312055.

\item {\bf [Serafini-Illuminati-Paris-De Siena 04]}:
A. Serafini, F. Illuminati, M. G. A. Paris, \& S. De Siena,
``Entanglement and purity of two-mode Gaussian states in noisy channels'',
{\em Phys. Rev. A} {\bf 69}, 2, 022318 (2004);
quant-ph/0310087.

\item {\bf [Serafini-Eisert-Wolf 04]}:
A. Serafini, J. Eisert, \& M. M. Wolf,
``Multiplicativity of maximal output purities of Gaussian channels under
Gaussian inputs'',
quant-ph/0406065.

\item {\bf [Sergienko-Shih-Pittman-Rubin, 96]}:
A. V. Sergienko, Y. H. Shih, T. D. Pittman, \& M. H. Rubin,
``Two-photon entanglement and EPR experiments using type-II
spontaneous parametric down conversion'',
in D. Han, K. Peng, Y. S. Kim, \& V. I. Man'ko (eds.),
{\em 4th Int.\ Conf.\ on Squeezed States and Uncertainty Relations
(Taiyuan, Shanxi, China, 1995)},
NASA, Greenbelt, Maryland, 1996, pp.~159-164.

\item {\bf [Sergienko-Atat\"{u}re-Walton-(+3) 99]}:
A. V. Sergienko, M. Atat\"{u}re, Z. Walton,
G. Jaeger, B. E. A. Saleh, \& M. C. Teich,
``Quantum cryptography using femtosecond-pulsed parametric down-conversion'',
{\em Phys. Rev. A} {\bf 60}, 4, R2622-R2625 (1999).

\item {\bf [Serra-Villas B\^{o}as-de Almeida-Moussa 02]}:
R. M. Serra, C. J. Villas-B\^{o}as, N. G. de Almeida, \& M. H. Y. Moussa,
``High-fidelity teleportation of entanglements of running-wave field states'',
{\em J. Opt. B: Quantum Semiclass. Opt.} {\bf 4}, 5, 316-325 (2002);
quant-ph/0204057.

\item {\bf [Service 99]}:
R. F. Service,
``Quantum mechanics: Quantum computing makes solid progress'',
{\em Science} {\bf 284}, 5415, 722-723 (1999).

\item {\bf [Shabani-Lidar 04]}:
A. Shabani, \& D. A. Lidar,
``Completely positive post-Markovian master equation via a measurement
approach'',
quant-ph/0404077.

\item {\bf [Shafiee-Golshani 99 a]}:
A. Shafiee, \& M. Golshani,
``Single-particle Bell-type inequality'',
quant-ph/9907017.

\item {\bf [Shafiee-Golshani 99 b]}:
A. Shafiee, \& M. Golshani,
``The possibility of factorizable contextual hidden variable
theories'',
quant-ph/9907060.

\item {\bf [Shafiee-Golshani 02 a]}:
A. Shafiee, \& M. Golshani,
``On the significance of Bell's locality condition'',
{\em Ann. Found. L. de Broglie} {\bf 27}, 1, 101-111 (2002);
quant-ph/0002060.

\item {\bf [Shafiee-Golshani 02 b]}:
A. Shafiee, \& M. Golshani,
``Bell's theorem and chemical potential'',
{\em J. Phys. A} {\bf 35}, ?, 8627-? (2002);
quant-ph/0207054.

\item {\bf [Shafiee-Golshani 03]}:
A. Shafiee, \& M. Golshani,
``A Bell-type inequality for real experiments and the relevance of fair
sampling assumption'',
quant-ph/0305110.

\item {\bf [Shafiee 04]}:
A. Shafiee,
``A CH-type inequality for real experiments'',
{\em J. Mod. Opt.};
quant-ph/0401022.

\item {\bf [Shafiee-Golshani 04]}:
A. Shafiee, \& M. Golshani,
``Complete correlation, detection loophole and Bell's theorem'',
{\em Fortschr. Phys.};
quant-ph/0406042.

\item {\bf [Shafiee-Safinejad-Naqsh 04]}:
A. Shafiee, F. Safinejad, \& F. Naqsh,
``Information and the Brukner-Zeilinger interpretation of quantum
mechanics: A critical investigation'',
quant-ph/0407198.

\item {\bf [Shafiee 02]}:
F. Shafiee,
``Entangled quantum networks'',
quant-ph/0203010.

\item {\bf [Shahriar 00]}:
M. S. Shahriar,
``Phase locking of remote clocks using quantum entanglement'',
quant-ph/0010007.

\item {\bf [Shahriar-Hemmer-Lloyd-(+2) 02]}:
M. S. Shahriar, P. R. Hemmer, S. Lloyd,
P. S. Bhatia, \& A. E. Craig,
``Solid-state quantum computing using spectral holes'',
{\em Phys. Rev. A} {\bf 66}, 3, 032301 (2002);
quant-ph/0007074.

\item {\bf [Shalyt Margolin-Strazhev-Tregubovich 03]}:
A. E. Shalyt-Margolin, V. I. Strazhev, \& A. Y. Tregubovich,
``On geometric realization of quantum computations in an externally driven
four-level system'',
{\em Opt. Spectrosc.} {\bf 94}, 730 (2003).

\item {\bf [Shan 99]}:
G. Shan,
``Quantum superluminal communication does not result in the causal
loop'',
quant-ph/9906113.

\item {\bf [Shan 04]}:
G. Shan,
``Quantum collapse, consciousness and superluminal communication'',
{\em Found. Phys. Lett.} {\bf 17}, 2, 167-182 (2004).

\item {\bf [Shannon 48]}:
C. E. Shannon,
``A mathematical theory of communication'',
{\em Bell Syst. Tech. J.} {\bf 27}, 379-423 and 623-656 (1948).

\item {\bf [Shannon 49 a]}:
C. E. Shannon,
{\em A mathematical theory of communication},
University of Illinois Press, Urbana, 1949.

\item {\bf [Shannon 49 b]}:
C. E. Shannon,
``Communication theory of secrecy systems'',
{\em Bell Syst. Tech. J.} {\bf 28}, 656-715 (1949).

\item {\bf [Shapira-Mozes-Biham 03]}:
D. Shapira, S. Mozes, \& O. Biham,
``Effect of unitary noise on Grover's quantum search algorithm'',
{\em Phys. Rev. A} {\bf 67}, 4, 042301 (2003).

\item {\bf [Shapira-Biham-Bracken-Hackett 03]}:
D. Shapira, O. Biham, A. J. Bracken, \& M. Hackett,
``One-dimensional quantum walk with unitary noise'',
{\em Phys. Rev. A} {\bf 68}, 6, 062315 (2003).

\item {\bf [Shapiro-Spanner-Ivanov 03]}:
E. A. Shapiro, M. Spanner, \& M. Y. Ivanov,
``Quantum logic approach to wave packet control'',
{\em Phys. Rev. Lett.} {\bf 91}, 23, 237901 (2003).

\item {\bf [Shapiro-Wong 00]}:
J. H. Shapiro, \& N. C. Wong,
``An ultrabright narrowband source of polarization-entangled photon pairs'',
{\em J. Opt. B: Quantum Semiclass. Opt.} {\bf 2}, 1, L1-L4 (2000).

\item {\bf [Shapiro 02]}:
J. H. Shapiro,
``Architectures for long-distance quantum teleportation'',
{\em New J. Phys} {\bf 4}, 47.1-47.18 (2002).

\item {\bf [Shapiro 03]}:
J. H. Shapiro,
``Near-field turbulence effects on quantum-key distribution'',
{\em Phys. Rev. A} {\bf 67}, 2, 022309 (2003).

\item {\bf [Shapiro-Brumer 02]}:
M. Shapiro, \& P. Brumer,
``S-matrix approach to the construction of decoherence-free subspaces'',
{\em Phys. Rev. A} {\bf 66}, 5, 052308 (2002).

\item {\bf [Sharf-Havel-Cory 00 a]}:
Y. Sharf, T. F. Havel, \& D. G. Cory,
``Spatially encoded pseudopure states for
NMR quantum-information processing'',
{\em Phys. Rev. A} {\bf 62}, 5, 052314 (2000);
quant-ph/0005076.

\item {\bf [Sharf-Havel-Cory 00 b]}:
Y. Sharf, T. F. Havel, \& D. G. Cory,
``Quantum codes for controlling coherent evolution'',
submitted to {\em J. Chem. Phys.};
quant-ph/0004029.

\item {\bf [Sharf-Cory-Somaroo-(+3) 00]}:
Y. Sharf, D. G. Cory, S. S. Somaroo, T. F. Havel,
E. Knill, \& R. Laflamme,
``A study of quantum error correction by geometric algebra and
liquid-state NMR spectroscopy'',
{\em Molecular Phys.} {\bf 98}, 17, 1347-1363 (2000);
quant-ph/0004030.

\item {\bf [Sharma 03]}:
S. S. Sharma,
``Tripartite GHZ state generation with trapped ion in an optical cavity'',
{\em Phys. Lett. A} {\bf 311}, 2-3, 111-114 (2003).

\item {\bf [Sharma-Sharma 04]}:
S. S. Sharma, \& N. K. Sharma,
``Intrinsic decoherence effects on tripartite GHZ state generation using a
trapped ion coupled to an optical cavity'',
quant-ph/0411045.

\item {\bf [Sharp 61]}:
D. H. Sharp,
``The Einstein-Podolsky-Rosen paradox re-examined'',
{\em Philos. Sci.} {\bf 28}, ?, 225-233 (1961).
Comment: {\bf [Putnam 61]}.

\item {\bf [Sharp-Shanks 89]}:
W. D. Sharp, \& N. Shanks,
``The curious quantum statistics in the
interval between measurements'',
{\em Phys. Lett. A} {\bf 138}, 9, 451-453 (1989).

\item {\bf [Sharp-Shanks 93]}:
W. D. Sharp, \& N. Shanks,
``The rise and fall of time-symmetrized quantum mechanics'',
{\em Philos. Sci.} {\bf 60}, ?, 488-499 (1993).

\item {\bf [Shelly Sharma-Sharma 01]}:
S. Shelly Sharma, \& N. K. Sharma,
``Quantum mutual entropy for two-level ion in a
q-analog trap'',
quant-ph/0105060.

\item {\bf [Shende-Markov-Bullock 04]}:
V. V. Shende, I. L. Markov, \& S. S. Bullock,
``Minimal universal two-qubit controlled-NOT-based circuits'',
{\em Phys. Rev. A} {\bf 69}, 6, 062321 (2004);
quant-ph/0308033.

\item {\bf [Shende-Bullock-Markov 04]}:
V. V. Shende, S. S. Bullock, \& I. L. Markov,
``Recognizing small-circuit structure in two-qubit operators'',
{\em Phys. Rev. A} {\bf 70}, 1, 012310 (2004).

\item {\bf [Shenker-Hemmo 02 a]}:
O. R. Shenker, \& M. Hemmo,
``Quantum decoherence and the approach to equilibrium (Part 1)'' (2002),
PITT-PHIL-SCI00000881.
See {\bf [Shenker-Hemmo 02 b]} (II).

\item {\bf [Shenker-Hemmo 02 b]}:
O. R. Shenker, \& M. Hemmo,
``Quantum decoherence and the approach to equilibrium (Part 2)'' (2002),
PITT-PHIL-SCI00000882.
See {\bf [Shenker-Hemmo 02 b]} (II).

\item {\bf [Shenker-Hemmo 03]}:
O. R. Shenker, \& M. Hemmo,
``The Von Neumann entropy: A Reply to
Henderson'',
PITT-PHIL-SCI00001232.

\item {\bf [Shenvi-Kempe-Whaley 03]}:
N. Shenvi, J. Kempe, \& K. B. Whaley,
``Quantum random-walk search algorithm'',
{\em Phys. Rev. A} {\bf 67}, 5, 052307 (2003).

\item {\bf [Shenvi-Brown-Whaley 03]}:
N. Shenvi, K. R. Brown, \& K. B. Whaley,
``Effects of a random noisy oracle on search algorithm complexity'',
{\em Phys. Rev. A} {\bf 68}, 5, 052313 (2003);
quant-ph/0304138.

\item {\bf [Sherry-Sudarshan 78]}:
T. N. Sherry, \& E. C. G. Sudarshan,
``Interaction between classical and quantum systems: A new approach to quantum
measurement. I'',
{\em Phys. Rev. D} {\bf 18}, 12, 4580-4589 (1978).

\item {\bf [Sherson-Julsgaard-Polzik 04]}:
J. Sherson, B. Julsgaard, \& E. S. Polzik,
``Distant entanglement of macroscopic gas samples'',
quant-ph/0408146.

\item {\bf [Sherson-M{\o}lmer 04]}:
J. Sherson, \& K. M{\o}lmer,
``Entanglement of large atomic samples: A Gaussian state analysis'',
quant-ph/0410119.

\item {\bf [Sherwin-Imamoglu-Montroy 99]}:
M. S. Sherwin, A. Imamoglu, \& T. Montroy,
``Quantum computation with quantum dots and
terahertz cavity quantum electrodynamics'',
{\em Phys. Rev. A} {\bf 60}, 5, 3508-3514 (1999).

\item {\bf [Shi-Guo 00]}:
B.-S. Shi, \& G.-C. Guo,
``One-to-one and one-to-any quantum key distribution by improving
long-distance frequency-division interferometer'',
{\em J. Mod. Opt.} {\bf 46}, 6, 1011-1015 (1999).

\item {\bf [Shi-Jiang-Guo 00 a]}:
B.-S. Shi, Y.-K. Jiang, \& G.-C. Guo,
``Probabilistic teleportation of two-particle entangled state'',
{\em Phys. Lett. A} {\bf 268}, 3, 161-164 (2000);
quant-ph/0006001.

\item {\bf [Shi-Jiang-Guo 00 b]}:
B.-S. Shi, Y.-K. Jiang, \& G.-C. Guo,
``Manipulating the frequency entangled states by
acoustic-optical-modulator'',
{\em Phys. Rev. A};
quant-ph/0005065.

\item {\bf [Shi-Jiang-Guo 00 c]}:
B.-S. Shi, Y.-K. Jiang, \& G.-C. Guo,
``Optimal entanglement purification via entanglement swapping'',
{\em Phys. Rev. A} {\bf 62}, 5, 054301 (2000);
quant-ph/0005125.

\item {\bf [Shi-Jiang-Guo 00 b]}:
B.-S. Shi, Y.-K. Jiang, \& G.-C. Guo,
``Quantum key distribution using different-frequency photons'',
{\em Appl. Phys. B} {\bf 70}, ?, 415-417 (2000);
quant-ph/0006038.

\item {\bf [Shi-Li-Liu-(+2) 01]}:
B.-S. Shi, J. Li, J.-M. Liu, X.-F. Fan, \& G.-C. Guo,
``Quantum key distribution and quantum authentication
based on entangled state'',
{\em Phys. Lett. A} {\bf 281}, 2-3, 83-87 (2001);
quant-ph/0102058.

\item {\bf [Shi-Tomita 02 a]}:
B.-S. Shi, \& A. Tomita,
``Teleportation of an unknown state by $W$ state'',
{\em Phys. Lett. A} {\bf 296}, 4-5, 161-164 (2002).
Comment: {\bf [Joo-Park 02]}.
Reply: {\bf [Shi-Tomita 02 b]}.

\item {\bf [Shi-Tomita 02 b]}:
B.-S. Shi, \& A. Tomita,
``Reply to `Comment on: Teleportation of an unknown state
by $W$ state'\,'',
{\em Phys. Lett. A} {\bf 300}, 4-5, 538-539 (2002).
Reply to {\bf [Joo-Park 02]}.
See {\bf [Shi-Tomita 02 a]}.

\item {\bf [Shi-Tomita 02 c]}:
B.-S. Shi, \& A. Tomita,
``Remote state preparation of an entangled state'',
{\em J. Opt. B: Quantum Semiclass. Opt.} {\bf 4}, 6, 380-382 (2002).

\item {\bf [Shi-Tomita 03 a]}:
B.-S. Shi, \& A. Tomita,
``Generation of a pulsed polarization entangled-photon pair via a two-crystal
geometry'',
{\em Phys. Rev. A} {\bf 67}, 4, 043804 (2003).

\item {\bf [Shi-Tomita 03 b]}:
B.-S. Shi, \& A. Tomita,
``Generation of a pulsed polarization entangled photon pair using a Sagnac interferometer'',
{\em Phys. Rev. A} {\bf 69}, 1, 013803 (2004).

\item {\bf [Shi-Du 01 a]}:
M. Shi, \& J. Du,
``Remarks on the Lewenstein-Sanpera decomposition'',
{\em Phys. Lett. A} {\bf 285}, 5-6, 263-266 (2001);
quant-ph/0012145.
See {\bf [Lewenstein-Sanpera 98]}.

\item {\bf [Shi-Du 01 b]}:
M. Shi, \& J. Du,
``Boundary of the set of separable states'',
quant-ph/0103016.

\item {\bf [Shi 97]}:
Y. Shi,
``Non-partial reality'',
quant-ph/9705047.

\item {\bf [Shi 98]}:
Y. Shi,
``Early gedanken experiments of quantum mechanics revisited'',
quant-ph/9811050.

\item {\bf [Shi 99]}:
Y. Shi,
``Remarks on universal quantum computer'',
quant-ph/9908074.

\item {\bf [Shi 00]}:
Y. Shi,
``Quantum lower bound for sorting'',
quant-ph/0009091.
See {\bf [H\o{}yer-Neerbek-Shi 01]}.

\item {\bf [Shi 02 a]}:
Y. Shi,
``All entanglements in a multiparticle state'',
quant-ph/0201079.

\item {\bf [Shi 02 b]}:
Y. Shi,
``Quantum entanglement in condensed matter systems'',
quant-ph/0204058.

\item {\bf [Shi 03 a]}:
Y. Shi,
``Quantum entanglement of identical particles'',
{\em Phys. Rev. A} {\bf 67}, 2, 024301 (2003).

\item {\bf [Shi 03 b]}:
Y. Shi,
``Quantum disentanglement in long-range orders and spontaneous symmetry breaking'',
{\em Phys. Lett. A} {\bf 209}, 3-4, 254-261 (2003).

\item {\bf [Shi 04]}:
Y. Shi,
``Optical generation and quantitative characterizations of electron-hole entanglement'',
{\em Phys. Rev. A} {\bf 69}, 3, 032318 (2004).

\item {\bf [Shi-Wu 04]}:
Y. Shi, \& Y.-S. Wu,
``Perturbative formulation and nonadiabatic corrections in adiabatic quantum-computing schemes'',
{\em Phys. Rev. A} {\bf 69}, 2, 024301 (2004).

\item {\bf [Shields {\em et al.} 00]}:
?. Shields {\em et al.},
``?'',
{\em Appl. Phys. Lett.} {\bf 76}, ?, 3673-? (2000).
See {\bf [Osborne 00 b]}.

\item {\bf [Shields 02]}:
A. Shields,
``Quantum logic with light, glass, and mirrors'',
{\em Science} {\bf 297}, 1821-1822 (2002).

\item {\bf [Shifren-Akis-Ferry 00]}:
L. Shifren, R. Akis, \& D. K. Ferry,
``Correspondence between quantum and classical motion: comparing
Bohmian mechanics with a smoothed effective potential approach'',
{\em Phys. Lett. A} {\bf 274}, 1-2, 75-83 (2000).

\item {\bf [Shih-Alley 88]}:
Y. H. Shih, \& C. O. Alley,
``New type of Einstein-Podolsky-Rosen-Bohm experiment using pairs
of light quanta produced by optical parametric down conversion'',
{\em Phys. Rev. Lett.} {\bf 61}, 26, 2921-2924 (1988).

\item {\bf [Shih-Rubin 93]}:
Y. H. Shih, \& M. H. Rubin,
``Four photon interference experiment for the testing of the
Greenberger-Horne-Zeilinger theorem'',
{\em Phys. Lett. A} {\bf 182}, 1, 16-22 (1993).

\item {\bf [Shih-Strekalov-Pittman-Rubin 98]}:
Y. H. Shih, D. V. Strekalov, T. D. Pittman, \& M. H. Rubin,
``Why two-photon but not two photons?'',
{\em Fortschr. Phys.} {\bf 46}, 6-8, 627-641 (1998).

\item {\bf [Shih-Kim 00]}:
Y. H. Shih, \& Y.-H. Kim,
``Experimental realization of Popper's
experiment---Violation of the uncertainty principle?'',
{\em Fortschr. Phys.} {\bf 48}, 5-7, 463-471 (2000).

\item {\bf [Shimamura-\"{O}zdemir-Morikoshi-Imoto 03]}:
J. Shimamura, \c{S}. K. \"{O}zdemir, F. Morikoshi, \& N. Imoto,
``Quantum and classical correlations between players in game theory'',
quant-ph/0311104.

\item {\bf [Shimamura-\"{O}zdemir-Morikoshi-Imoto 04]}:
J. Shimamura, \c{S}. K. \"{O}zdemir, F. Morikoshi, \& N. Imoto,
``Entangled states that cannot reproduce original classical games in their
quantum version
quant-ph/0404151.

\item {\bf [Shimizu-Miyadera-Ukena 01]}:
A. Shimizu, T. Miyadera, \& A. Ukena,
``Decoherence of anomalously-fluctuating states of finite macroscopic systems'',
{\em Proc.\ 7th Int.\ Symp.\ Foundations of Quantum Mechanics (Tokyo, 2001)},
World Scientific, Singapore, 2001;
quant-ph/0109082.

\item {\bf [Shimizu-Imoto-Mukai 99]}:
K. Shimizu, N. Imoto, \& T. Mukai,
``Dense coding in photonic quantum communication
with enhanced information capacity'',
{\em Phys. Rev. A} {\bf 59}, 2, 1092-1097 (1999).
Erratum: {\em Phys. Rev. A} {\bf 60}, 2, 1731 (1999).

\item {\bf [Shimizu-Imoto 99]}:
K. Shimizu, \& N. Imoto,
``Communication channels secured from eavesdropping
via transmission of photonic Bell states'',
{\em Phys. Rev. A} {\bf 60}, 1, 157-166 (1999).

\item {\bf [Shimizu-Imoto 00]}:
K. Shimizu, \& N. Imoto,
``Single-photon-interference communication equivalent to
Bell-state-basis cryptographic quantum communication'',
{\em Phys. Rev. A} {\bf 62}, 5, 054303 (2000).

\item {\bf [Shimizu-Imoto 02 a]}:
K. Shimizu, \& N. Imoto,
``Fault-tolerant simple quantum-bit commitment unbreakable by individual
attacks'',
{\em Phys. Rev. A} {\bf 65}, 3, 032324 (2002).

\item {\bf [Shimizu-Imoto 02 b]}:
K. Shimizu, \& N. Imoto,
``Communication channels analogous to one out of two oblivious transfers based
on quantum uncertainty'',
{\em Phys. Rev. A} {\bf 66}, 5, 052316 (2002).
See: {\bf [Shimizu-Imoto 03]} (II).

\item {\bf [Shimizu-Miyadera-Ukena 02]}:
K. Shimizu, T. Miyadera, \& A. Ukena,
``Stability of quantum states of finite macroscopic systems'',
{\em Proc. of the Japan-Italy Joint Waseda Workshop on
``Fundamental Problems in Quantum Mechanics'' (Tokyo, 2001)},
World Scientific, Singapore, 2002;
quant-ph/0202096.

\item {\bf [Shimizu-Miyadera 02]}:
K. Shimizu, \& T. Miyadera,
``Stability of quantum states of finite macroscopic systems against
classical noises, perturbations from environments, and local measurements'',
quant-ph/0203106.

\item {\bf [Shimizu-Imoto 03]}:
K. Shimizu, \& N. Imoto,
``Communication channels analogous to one out of two oblivious transfers based
on quantum uncertainty. II. Closing EPR-type loopholes'',
{\em Phys. Rev. A} {\bf 67}, 3, 034301 (2003).
See: {\bf [Shimizu-Imoto 02 b]} (I).

\item {\bf [Shimizu-Edamatsu-Itoh 03]}:
R. Shimizu, K. Edamatsu, \& T. Itoh,
``Quantum diffraction and interference of spatially correlated photon pairs
generated by spontaneous parametric down-conversion'',
{\em Phys. Rev. A} {\bf 67}, 4, 041805 (2003).

\item {\bf [Shimoni-Shapira-Biham 04]}:
Y. Shimoni, D. Shapira, \& O. Biham,
``Characterization of pure quantum states of multiple qubits
using the Groverian entanglement measure'',
{\em Phys. Rev. A} {\bf 69}, 6, 062303 (2004).

\item {\bf [Shimono 02]}:
T. Shimono,
``Lower bound for entanglement cost of antisymmetric states'',
quant-ph/0203039.

\item {\bf [Shimony 63]}:
A. Shimony,
``Role of the observer in quantum theory'',
{\em Am. J. Phys.} {\bf 31}, 10, 755-773 (1963).
Reprinted in {\bf [Shimony 93]}, pp.~3-33.

\item {\bf [Shimony 71]}:
A. Shimony,
``Experimental test of local hidden-variable theories'',
in {\bf [d'Espagnat 71]}, pp.~182-194.
Reprinted in {\bf [Shimony 93]}, pp.~77-89.

\item {\bf [Shimony 74]}:
A. Shimony,
``Approximate measurement in quantum mechanics II'',
{\em Phys. Rev. D} {\bf 9}, 8, 2321-2323 (1974).

\item {\bf [Shimony-Horne-Clauser 76]}:
A. Shimony, M. A. Horne, \& J. F. Clauser,
``?'',
{\em Epistemological Lett.} no. 13 (Oct. 1976), 1-8.
Reprinted in {\bf [Bell-Shimony-Horne-Clauser 85]}.
See {\bf [Bell 76 a, b, 77]},
{\bf [Shimony-Horne-Clauser 78]}.

\item {\bf [Shimony 78]}:
A. Shimony,
``?'',
{\em Epistemological Lett.} no. 18 (? 1978), 1-3.

\item {\bf [Shimony-Horne-Clauser 78]}:
A. Shimony, M. A. Horne, \& J. F. Clauser,
``?'',
{\em Epistemological Lett.} no. 18 (? 1978), ?-?.
Reprinted in {\bf [Bell-Shimony-Horne-Clauser 85]}.
See {\bf [Bell 76 a, b, 77]},
{\bf [Shimony-Horne-Clauser 76]}.

\item {\bf [Shimony 79]}:
A. Shimony,
``Proposed neutron interferometer test of some
nonlinear variants of wave mechanics'',
{\em Phys. Rev. A} {\bf 20}, 2, 394-396 (1979).

\item {\bf [Shimony 81]}:
A. Shimony,
``Critique of the papers of Fine and Suppes'',
in P. D. Asquith, \& R. N. Giere (eds.),
{\em Proc.\ of the 1980 Biennial Meeting of
the Philosophy of Science Association},
East Lansing, Michigan, 1981, vol. 2, pp.~572-580.

\item {\bf [Shimony 84 a]}:
A. Shimony,
``Contextual hidden variables theories and Bell's inequalities'',
{\em Brit. J. Philos. Sci.} {\bf 35}, 1, 25-45 (1984).
Reprinted in {\bf [Shimony 93]}, pp.~104-129 (with a comment on GHZ).

\item {\bf [Shimony 84 b]}:
A. Shimony,
``Response to `Comment on ``Proposed
molecular test of local hidden-variables theories''\,'\,'',
{\em Phys. Rev. A} {\bf 30}, 4, 2130-2131 (1984).
Reply to {\bf [Santos 84 b]}.
See {\bf [Lo-Shimony 81]}.

\item {\bf [Shimony 84 c]}:
A. Shimony,
``Controllable and uncontrollable nonlocality'',
in S. Kamefuchi et al. (eds.),
{\em Foundations of quantum mechanics in the light of new technology
(Kokubunji, Tokyo, 1983)},
Physical Society of Japan, Tokyo, 1984, pp.~225-230.
Reprinted in {\bf [Shimony 93]}, pp.~130-139 (with a comment).

\item {\bf [Shimony 86]}:
A. Shimony,
``Events and processes in the quantum world'',
in R. Penrose, \& C. Isham (eds.),
{\em Quantum concepts in space and time},
Clarendon Press, Oxford, 1986, pp.~182-203.
Reprinted in {\bf [Shimony 93]}, pp.~140-162.

\item {\bf [Shimony 88 a]}:
A. Shimony,
``The reality of the quantum world'',
{\em Sci. Am.} {\bf 258}, 1, 36-43 (1988).
Reprinted in
{\bf [Rusell-Clayton-Wegter McNelly-Polkinghorne 01]}, pp.~3-16.
Spanish version: ``Realidad del mundo cu\'{a}ntico'',
{\em Investigaci\'{o}n y Ciencia} 138, 28-35 (1988).
Reprinted in {\bf [Cabello 97 c]}, pp.~28-35.

\item {\bf [Shimony 88 b]}:
A. Shimony,
``Physical and philosophical issues in the Bohr-Einstein debate'',
in H. Feshbach, T. Matsui, \& A. Oleson (eds.),
{\em Niels Bohr: Physics and the world},
Harwood, GmbH, 1988.
Reprinted in {\bf [Shimony 93]}, pp.~171-187
(with a comment on some works of Howard and Fine).

\item {\bf [Shimony 90]}:
A. Shimony,
``An exposition of Bell's theorem'',
in A. I. Miller (ed.),
{\em Sixty-two years of uncertainty: Historical, philosophical and physical
inquiries into the foundations of quantum mechanics.
Proc.\ Int. School
of History of Science (Erice, Italy, 1989)},
Plenum Press, New York, 1990, pp.~?-?.
Reprinted in {\bf [Shimony 93]}, pp.~90-103 (with a comment on Stapp).

\item {\bf [Shimony 91]}:
A. Shimony,
``John Stewart Bell: 1928-1990'',
{\em Found. Phys.} {\bf 21}, 5, 509-511 (1991).

\item {\bf [Shimony 93]}:
A. Shimony,
{\em Search for a naturalistic world view.
Vol II: Natural science and metaphysics},
Cambridge University Press, New York, 1993.
Review: {\bf [Ziman 94]}.

\item {\bf [Shimony 95]}:
A. Shimony,
``Degree of entanglement'',
in D. M. Greenberger, \& A. Zeilinger (eds.),
{\em Fundamental problems in quantum theory: A
conference held in honor of professor John A. Wheeler,
Ann. N. Y. Acad. Sci.} {\bf 755}, 675-679 (1995).

\item {\bf [Shimony 96 a]}:
A. Shimony,
``Philosophical reflections on EPR'',
in A. Mann, \& M. Revzen (eds.),
{\em The dilemma of Einstein, Podolsky and Rosen -- 60 years
later. An international symposium in honour of Nathan Rosen
(Haifa, Israel, 1995)},
{\em Ann. Phys. Soc. Israel} {\bf 12}, 27-41 (1996).

\item {\bf [Shimony 96 b]}:
A. Shimony,
``Measures of entanglement'',
in A. Mann, \& M. Revzen (eds.),
{\em The dilemma of Einstein, Podolsky and Rosen -- 60 years
later. An international symposium in honour of Nathan Rosen
(Haifa, Israel, 1995)},
{\em Ann. Phys. Soc. Israel} {\bf 12}, 163-176 (1996).

\item {\bf [Shimony 98]}:
A. Shimony,
``On ensembles that are both pre- and post-selected'',
{\em Fortschr. Phys.} {\bf 46}, 6-8, 725-728 (1998).

\item {\bf [Shimony 98]}:
A. Shimony,
``Philosophical and experimental perspectives on quantum physics
(6th Vienna Circle lecture)'',
in {\bf [Greenberger-Reiter-Zeilinger 99]}, pp.~1-18.

\item {\bf [Shimony 00]}:
A. Shimony,
``New aspects of Bell's theorem'',
in {\bf [Ellis-Amati 00]}, pp.~136-164.

\item {\bf [Shimony-Stein 01]}:
A. Shimony, \& H. Stein,
`Comment on ``Nonlocal character of quantum theory'' by
Henry P. Stapp [Am. J. Phys. {\bf 65} (4), 300-304 (1997)]',
{\em Am. J. Phys.} {\bf 69}, 8, 848-853 (2001).
Comment on {\bf [Stapp 97 a]}.
Reply: {\bf [Stapp 01 a]}.

\item {\bf [Shimony 02]}:
A. Shimony,
``John S. Bell: Some reminiscences and reflections'',
in {\bf [Bertlmann-Zeilinger 02]}, pp.~51-60.

\item {\bf [Shimony 04]}:
A. Shimony,
`An analysis of Stapp's ``A Bell-type theorem without hidden variables''\,',
quant-ph/0404121.

\item {\bf [Shirai 98]}:
H. Shirai,
``Reinterpretation of quantum mechanics based on the statistical
interpretation'',
{\em Found. Phys.} {\bf 28}, 11, 1633-1661 (1998).

\item {\bf [Shirokov 98]}:
M. I. Shirokov,
``Spin state determination using a Stern-Gerlach device'',
{\em Found. Phys.} {\bf 28}, 6, 985-998 (1998).

\item {\bf [Shnirman-Sch\"{o}n-Hermon 97]}:
A. Shnirman, G. Sch\"{o}n, \& Z. Hermon,
``Quantum manipulations od small Josephson junctions'',
{\em Phys. Rev. Lett.} {\bf 79}, 12, 2371-2374 (1997).
Reprinted in {\bf [Macchiavello-Palma-Zeilinger 00]}, pp.~391-394.

\item {\bf [Shojai-Shojai 01]}:
A. Shojai, \& F. Shojai,
``About some problems raised by the relativistic form of
De-Broglie--Bohm theory of pilot wave'',
{\em Physica Scripta} {\bf ?}, ?-?;
quant-ph/0109025.

\item {\bf [Shor 94]}:
P. W. Shor,
``Algorithms for quantum computation: Discrete logarithms and factoring'',
in S. Goldwasser (ed.),
{\em Proc.\ of the 35th Annual Symp.\ on the Foundations of Computer
Science (Santa Fe, New Mexico, 1994)},
IEEE Computer Science Society Press,
Los Alamitos, California, 1994, pp.~124-134.
Enlarged version: {\bf [Shor 97]}.

\item {\bf [Shor-Zurek-Chuang-Laflamme 95]}:
P. W. Shor, W. H. Zurek, I. L. Chuang, \& R. Laflamme,
``?'',
{\em Science} {\bf 270}, ?, 1633 (1995).

\item {\bf [Shor 95]}:
P. W. Shor,
``Scheme for reducing decoherence in quantum computer memory'',
{\em Phys. Rev. A} {\bf 52}, 4, R2493-R2496 (1995).
Reprinted in {\bf [Macchiavello-Palma-Zeilinger 00]}, pp.~134-137.

\item {\bf [Shor 96]}:
P. W. Shor,
``Fault-tolerant quantum computation'',
in ? (ed.),
{\em Proc.\ of the 37h Annual Symp.\ on the Foundations of Computer
Science (?, ?, 1994)},
IEEE Computer Science Society Press,
Los Alamitos, California, ?, pp.~56-65;
quant-ph/9605011.

\item {\bf [Shor 97]}:
P. W. Shor,
``Polynomial-time algorithms for prime factorization and discrete
logarithms on a quantum computer'',
{\em SIAM J. Comp.} {\bf 26}, 5, 1484-1509 (1997);
quant-ph/9508027.
Enlarged version of {\bf [Shor 94]}.

\item {\bf [Shor-Preskill 00]}:
P. W. Shor, \& J. Preskill,
``Simple proof of security of the BB84 quantum key distribution
protocol'',
{\em Phys. Rev. Lett.} {\bf 85}, 2, 441-444 (2000);
quant-ph/0003004.

\item {\bf [Shor-Smolin-Thapliyal 00]}:
P. W. Shor, J. A. Smolin, \& A. V. Thapliyal,
``Superactivation of bound entanglement'',
quant-ph/0005117.

\item {\bf [Shor 00]}:
P. W. Shor,
``On the number of elements needed in a POVM
attaining the accessible information'',
quant-ph/0009077.

\item {\bf [Shor-Smolin-Terhal 00]}:
P. W. Shor, J. A. Smolin, \& B. M. Terhal,
``Nonadditivity of bipartite distillable entanglement follows from
a conjecture on bound entangled Werner states'',
{\em Phys. Rev. Lett.} {\bf 86}, 12, 2681-2684 (2001);
quant-ph/0010054.

\item {\bf [Shor 02 a]}:
P. W. Shor,
``Additivity of the classical capacity of entanglement-breaking quantum
channels'',
{\em J. Math. Phys.} {\bf 43}, 9, 4334-4340 (2002);
quant-ph/0201149.

\item {\bf [Shor 02 b]}:
P. W. Shor,
``Introduction to quantum algorithms'',
in {\bf [Lomonaco 02 a]}, pp.~143-159;
quant-ph/0005003.

\item {\bf [Shor-Smolin-Thapliyal 03]}:
P. W. Shor, J. A. Smolin, \& A. V. Thapliyal,
``Superactivation of bound entanglement'',
{\em Phys. Rev. Lett.} {\bf 90}, 10, 107901 (2003).

\item {\bf [Shor 03 a]}:
P. W. Shor,
``Capacities of quantum channels and how to find them'',
talk for {\em 18th Int.\ Symp.\ on
Mathematical Programming (Copenhagen, Denmark, 2003)}
quant-ph/0304102.

\item {\bf [Shor 03 b]}:
P. W. Shor,
``Equivalence of additivity questions in quantum information theory'',
quant-ph/0305035.

\item {\bf [Shor 04]}:
P. W. Shor,
``The classical capacity achievable by a quantum channel assisted by
limited entanglement'',
{\em Festschrift celebrating Alexander S. Holevo's 60'th birthday};
 quant-ph/0402129.

\item {\bf [Short 00]}:
A. J. Short,
``Comment on `Experimental realization of Popper's experiment:
Violation of the uncertainty principle?'\,'',
quant-ph/0005063.
Comment on {\bf [Kim-Shih 99]}.

\item {\bf [Short 01]}:
A. J. Short,
``Popper's experiment and conditional uncertainty relations'',
{\em Found. Phys. Lett.} {\bf 14}, 3, 275-284 (2001).

\item {\bf [Shore-Knight 93]}:
B. W. Shore, \& P. L. Knight,
``The Jaynes-Cummings model'',
{\em J. Mod. Opt.} {\bf 40}, 7, 1195-1238 (1993).

\item {\bf [Shull-Atwood-Arthur-Horne 80]}:
C. G. Shull, D. K. Atwood,
J. Arthur, \& M. A. Horne,
``Search for a nonlinear variant of the Schr\"{o}dinger equation by
neutron interferometry'',
{\em Phys. Rev. Lett.} {\bf 44}, 12, 765-768 (1980).
See {\bf [G\"{a}hler-Klein-Zeilinger 81]}.

\item {\bf [Siegfried 00]}:
T. Siegfried,
{\em The bit and the pendulum. From quantum computing to
M theory-- The new physics of information},
Wiley, New York, 2000.

\item {\bf [Siewert-Fazio 01]}:
J. Siewert, \& R. Fazio,
``Quantum algorithms for Josephson networks'',
{\em Phys. Rev. Lett.} {\bf 87}, 25, 257905 (2001).

\item {\bf [Siewert-Fazio 02]}:
J. Siewert, \& R. Fazio,
``Implementation of the Deutsch-Jozsa algorithm with Josephson charge qubits'',
{\em J. Mod. Opt.};
quant-ph/0112135.

\item {\bf [Sica 99 a]}:
L. Sica,
``Bell's inequalities I: An explanation for their experimental
violation'',
{\em Opt. Comm.} {\bf 170}, 1, 55-60 (1999);
quant-ph/0101087.
See {\bf [Sica 99 b]} (II).

\item {\bf [Sica 99 b]}:
L. Sica,
``Bell's inequalities II: Logical loophole in their interpretation'',
{\em Opt. Comm.} {\bf 170}, 1, 61-66 (1999);
quant-ph/0101094.
See {\bf [Sica 99 a]} (I).

\item {\bf [Sica 02]}:
L. Sica,
``Correlations for a new Bell's inequality experiment'',
{\em Found. Phys. Lett.} {\bf 15}, 5, 473-486 (2002).

\item {\bf [Sifel-Hohenester 03]}:
C. Sifel, \& U. Hohenester,
``A turnstile electron-spin entangler in semiconductors'',
{\em Appl. Phys. Lett.} {\bf 83}, 153 (2003).

\item {\bf [Silberhorn-Lam-Weiss-(+3) 01]}:
C. Silberhorn, P. K. Lam, O. Weiss,
F. Koenig, N. Korolkova, \& G. Leuchs,
``Generation of continuous variable Einstein-Podolsky-Rosen entanglement
via the Kerr nonlinearity in an optical fibre'',
{\em Phys. Rev. Lett.};
quant-ph/0103002.

\item {\bf [Silberhorn-Korolkova-Leuchs 02]}:
C. Silberhorn, N. Korolkova, \& G. Leuchs,
``Quantum key distribution with bright entangled beams'',
{\em Phys. Rev. Lett.} {\bf 88}, 16, 167902 (2002);
quant-ph/0109009.

\item {\bf [Silberhorn-Ralph-L\"{u}tkenhaus-Leuchs 02]}:
C. Silberhorn, T. C. Ralph, N. L\"{u}tkenhaus, \& G. Leuchs,
``Continuous variable quantum cryptography: Beating the 3 dB loss limit'',
{\em Phys. Rev. Lett.} {\bf 89}, 16, 167901 (2002);
quant-ph/0204064.

\item {\bf [Silverman 02]}:
M. P. Silverman,
``Neutron interferometry: Lessons in experimental quantum mechanics'',
{\em Am. J. Phys.} {\bf 70}, 12, 1272-1274 (2002).
Review of {\bf [Rauch-Werner 00]}.

\item {\bf [Silvestrini-Stodolsky 01]}:
P. Silvestrini, \& L. Stodolsky,
``Demonstration of macroscopic coherence and decoherence
by adiabatic inversion, application to the SQUID'',
{\em Phys. Lett. A} {\bf 280}, 1-2, 17-22 (2001);
cond-mat/0004472.

\item {\bf [Silvestrov-Schomerus-Beenakker 01]}:
P. G. Silvestrov, H. Schomerus, \& C. W. J. Beenakker,
``Limits to error correction in quantum chaos'',
{\em Phys. Rev. Lett.} {\bf 86}, 22, 5192-5195 (2001);
quant-ph/0012119.

\item {\bf [Simmonds-Lang-Hite-(+3) 04]}:
R. W. Simmonds, K. M. Lang, D. A. Hite,
S. Nam, D. P. Pappas, \& J. M. Martinis,
``Decoherence in Josephson phase qubits from junction resonators'',
{\em Phys. Rev. Lett.} {\bf 93}, 7, 077003 (2004).

\item {\bf [Simon-Weihs-Zeilinger 99]}:
C. Simon, G. Weihs, \& A. Zeilinger,
``Quantum cloning and signaling'',
{\em Acta Phys. Slov.} {\bf 49}, ?, 755-760 (1999).

\item {\bf [Simon-Weihs-Zeilinger 00 a]}:
C. Simon, G. Weihs, \& A. Zeilinger,
``Optimal quantum cloning via stimulated emission'',
{\em Phys. Rev. Lett.} {\bf 84}, 13, 2993-2996 (2000),
quant-ph/9910048.

\item {\bf [Simon-Weihs-Zeilinger 00 b]}:
C. Simon, G. Weihs, \& A. Zeilinger,
``Optimal quantum cloning and universal NOT without quantum gates'',
in V. Bu\v{z}zek, \& D. P. DiVincenzo (eds.),
{\em J. Mod. Opt.} {\bf 47}, 2-3 (Special issue:
Physics of quantum information), 233-246 (2000).

\item {\bf [Simon-\.{Z}ukowski-Weinfurter-Zeilinger 00]}:
C. Simon, M. \.{Z}ukowski, H. Weinfurter, \& A. Zeilinger,
`Feasible ``Kochen-Specker'' experiment with single particles',
{\em Phys. Rev. Lett.} {\bf 85}, 9, 1783-1786 (2000);
quant-ph/0009074.
Comment: {\bf [Schafir 02 c]}.
See {\bf [Huang-Li-Zhang-(+2) 03]}.

\item {\bf [Simon 00 a]}:
C. Simon,
``The foundations of quantum information and feasible experiments'',
Ph.\ D. thesis, University of Vienna, 2000;
quant-ph/0103057.

\item {\bf [Simon 02]}:
C. Simon,
``Natural entanglement in Bose-Einstein condensates'',
{\em Phys. Rev. A} {\bf 66}, 5, 052323 (2002);
quant-ph/0110114.

\item {\bf [Simon-Kempe 02]}:
C. Simon, \& J. Kempe,
``Robustness of multiparty entanglement'',
{\em Phys. Rev. A} {\bf 65}, 5, 052327 (2002);
quant-ph/0109102.

\item {\bf [Simon-Pan 02]}:
C. Simon, \& J.-W. Pan,
``Polarization entanglement purification using spatial entanglement'',
{\em Phys. Rev. Lett.} {\bf 89}, 25, 257901 (2002).

\item {\bf [Simon-Irvine 03]}:
C. Simon, \& W. T. M. Irvine,
``Robust long-distance entanglement and a loophole-free Bell test with ions
and photons'',
{\em Phys. Rev. Lett.} {\bf 91}, 11, 110405 (2003);
quant-ph/0303023.

\item {\bf [Simon-Jaksch 04]}:
C. Simon, \& D. Jaksch,
``Could energy decoherence due to quantum gravity be observed?'',
quant-ph/0406007.

\item {\bf [Simon-Poizat 04]}:
C. Simon, \& J.-P. Poizat,
``Creating single time-bin entangled photon pairs'',
quant-ph/0409100.

\item {\bf [Simon 04]}:
C. Simon,
`Comment on ``Separability of quantum states and the violation of
Bell-type inequalities''\,',
{\em Phys. Rev. A};
quant-ph/0410032.
Comment on {\bf [Loubenets 04 a]}.

\item {\bf [Simon 94]}:
D. R. Simon,
``On the power of quantum computation'',
in {\em Proc.\ of the 35th Annual IEEE Symp.\ on the
Foundation of Computer Science (1995)},
IEEE Computer Science Society Press, Los Alamitos, California, 1994, p.~119.
Full version {\bf [Simon 97]}.

\item {\bf [Simon 97]}:
D. R. Simon,
``On the power of quantum computation'',
{\em SIAM J. Comput.} {\bf 26}, ?-? (1997).
Full version of {\bf [Simon 94]}.

\item {\bf [Simon 00 b]}:
R. Simon,
``Peres-Horodecki separability criterion for continuous variable
systems'',
{\em Phys. Rev. Lett.} {\bf 84}, 12, 2726-2729 (2000);
quant-ph/9909044.

\item {\bf [Simon-Brukner-Zeilinger 01]}:
C. Simon, \v{C}. Brukner, \& A. Zeilinger,
``Hidden-variable theorems for real experiments'',
{\em Phys. Rev. Lett.} {\bf 86}, 20, 4427-4430 (2001);
quant-ph/0006043.
See {\bf [Larsson 02 a]}.

\item {\bf [Simon-Bu\v{z}zek-Gisin 01]}:
C. Simon, V. Bu\v{z}zek, \& N. Gisin,
``The no-signaling condition and quantum dynamics'',
{\em Phys. Rev. Lett.} {\bf 87}, 17, 170405 (2001);
quant-ph/0102125.
Comment: {\bf [Bona 03]}.
Reply: {\bf [Simon-Bu\v{z}zek-Gisin 03]}.

\item {\bf [Simon-Pan 01]}:
C. Simon, \& J.-W. Pan,
``Polarization entanglement purification using spatial entanglement'',
quant-ph/0108063.

\item {\bf [Simon-Bu\v{z}zek-Gisin 03]}:
C. Simon, V. Bu\v{z}zek, \& N. Gisin,
``Simon, Bu\v{z}zek, and Gisin reply'',
{\em Phys. Rev. Lett.} {\bf 90}, 20, 208902 (2003).
Reply to {\bf [Bona 03]}.
See {\bf [Simon-Bu\v{z}zek-Gisin 01]}.

\item {\bf [Simon-Bouwmeester 03]}:
C. Simon, \& D. Bouwmeester,
``Theory of an entanglement laser'',
{\em Phys. Rev. Lett.} {\bf 91}, 5, 053601 (2003).
quant-ph/0302023.

\item {\bf [Simon-Platzman 00]}:
S. H. Simon, \& P. M. Platzman,
`Fundamental limit on ``interaction-free'' measurements',
{\em Phys. Rev. A} {\bf 61}, 5, 052103 (2000);
quant-ph/9905050.

\item {\bf [Simonius 78]}:
M. Simonius,
``Spontaneous symmetry breaking and blocking of metastable
states'',
{\em Phys. Rev. Lett.} {\bf 40}, 15, 980-983 (1978).

\item {\bf [Simonius 93]}:
M. Simonius,
``Measurement in quantum mechanics: From probabilities
to objective events'',
{\em Helv. Phys. Acta} {\bf 66}, 7, 727-742 (1993);
quant-ph/9811074.

\item {\bf [Simonius 97]}:
M. Simonius,
``Quantum interferometry, measurement and objectivity:
Some basic features revisited'',
in M. Ferrero, \& A. van der Merwe (eds.),
{\em New developments on fundamental problems in quantum
physics (Oviedo, Spain, 1996)},
Kluwer Academic, Dordrecht, Holland, 1997, pp.~395-401.

\item {\bf [Sinatra-Roch-Vigneron-(+3) 97]}:
A. Sinatra, J.-F. Roch, K. Vigneron, P. Grelu, J.-P. Poizat, \& P.
Grangier,
``Quantum non-demolition measurements using cold atoms in an
optical cavity'',
{\em J. Mod. Opt.} {\bf 44}, 10 (Special issue:
Fundamentals of quantum optics IV), 1967-1984 (1997).

\item {\bf [Singer-Stulpe 90]}:
M. Singer, \& W. Stulpe,
``Informational
incompletness of the observables $S_x$, $S_y$, $S_z$ for a spin-1 systems'',
{\em Found. Phys.} {\bf 20}, 4, 471-472 (1990).

\item {\bf [Singh 01]}:
C. Singh,
``Student understanding of quantum mechanics'',
{\em Am. J. Phys.} {\bf 69}, 8, 885-895 (2001).

\item {\bf [Singh-Srikanth 03]}:
S. K. Singh, \& R. Srikanth,
``Generalized quantum secret sharing'',
quant-ph/0307200.

\item {\bf [Singh 04]}:
S. K. Singh,
``Combinatorial approaches in quantum information theory'',
M.\ Sc. thesis, Department of Mathematics, IIT Kharagpur, India, 2004;
quant-ph/0405089.

\item {\bf [Singh-Srikanth 04]}:
S. K. Singh, \& R. Srikanth,
``Quantum seals'',
quant-ph/0410017.

\item {\bf [Sinolecka-\.{Z}yczkowski-Ku\'{s} 01]}:
M. M. Sinolecka, K. \.{Z}yczkowski, \& M. Ku\'{s},
``Manifolds of interconvertible pure states'',
quant-ph/0110082.

\item {\bf [Six 77]}:
J. Six,
``Test of the non-separability of the $K^{0}\overline{K}^{0}$ system'',
in J. L. Lopes, \& M. Paty (eds.),
{\em Quantum mechanics a half century later},
Reidel, Dordrecht, Holland, 1977, pp.~391-396.

\item {\bf [Skantzos-Saad-Kabashima 03]}:
N. S. Skantzos, D. Saad, \& Y. Kabashima,
``Analysis of common attacks in public-key cryptosystems based on low-density parity-check codes'',
{\bf Phys. Rev. E} {\bf 68}, 5, 056125 (2003).

\item {\bf [\v{S}karja-Bort\v{s}nik-L\"{o}ffler-Walther 99]}:
M. \v{S}karja, N. M. Bort\v{s}nik, M. L\"{o}ffler, \& H. Walther,
``Quantum interference and atom-atom entanglement in a two-mode, two-cavity micromaser'',
{\em Phys. Rev. A} {\bf 60}, 4, 3229-3232 (1999).

\item {\bf [Skinner-Davenport-Kane 03]}:
A. J. Skinner, M. E. Davenport, \& B. E. Kane,
``Hydrogenic spin quantum computing in silicon: A digital approach'',
{\em Phys. Rev. Lett.} {\bf 90}, 8, 087901 (2003).

\item {\bf [Skulimowski 02]}:
M. Skulimowski,
``Construction of time covariant POV measures'',
{\em Phys. Lett. A} {\bf 297}, 3-4, 129-136 (2002).

\item {\bf [Skyrms 82]}:
B. Skyrms,
``Counterfactual definiteness and local causation'',
{\em Philos. Sci.} {\bf 49}, 1, 43-50 (1982).

\item {\bf [Piotrowski-Sladkowski 03]}:
E. W. Piotrowski, \& J. Sladkowski,
``The next stage: Quantum game theory'',
quant-ph/0308027.

\item {\bf [Slater 98]}:
P. B. Slater,
`Sensitivity of estimates of the volume of the set of separable states to
the choice of ``uniform'' distributions',
quant-ph/9809042.
See {\bf [\.{Z}yczkowski-Horodecki-Sanpera-Lewenstein 98]}.

\item {\bf [Slater 99]}:
P. B. Slater,
``{\em A priori} probabilities of separable quantum states'',
{\em J. Phys. A} {\bf 32}, 28, 5261-5275 (1999);
quant-ph/9810026.
See {\bf [\.{Z}yczkowski-Horodecki-Sanpera-Lewenstein 98]}.

\item {\bf [Slater 00 a]}:
P. B. Slater,
``Information-theoretic analysis of two-level quantum systems'',
quant-ph/0002063.

\item {\bf [Slater 00 b]}:
P. B. Slater,
``Bures geometry of the three-level quantum systems'',
quant-ph/0008069.

\item {\bf [Slater 01]}:
P. B. Slater,
``Increased efficiency of quantum state estimation
using {\em non-separable} measurements'',
in S. Popescu, N. Linden, \& R. Jozsa (eds.),
{\em J. Phys. A} {\bf 34}, 35
(Special issue: Quantum information and computation), 7029-7046 (2001);
quant-ph/0006009.

\item {\bf [Slater 02 a]}:
P. B. Slater,
``Two coupled qubits are classically correlated with an exact
Bures probability of $2^(1/2)/24$'',
quant-ph/0203088.

\item {\bf [Slater 02 b]}:
P. B. Slater,
``A priori probability that a qubit-qutrit pair is unentangled'',
{\em Quant. Inf. Proc.};
quant-ph/0211150.

\item {\bf [Slater 03 a]}:
P. B. Slater,
``A priori probabilities of triseparable and biseparable Eggeling-Werner
states'',
quant-ph/0306053.

\item {\bf [Slater 03 b]}:
P. B. Slater,
``Silver mean conjectures for 15-d volumes and 14-d hyperareas of the
separable two-qubit systems'',
quant-ph/0308037.

\item {\bf [Slater 04 a]}:
P. B. Slater,
``Volumes and hyperareas of the spaces of separable and nonseparable
qubit-qutrit systems: Initial numerical analyses'',
quant-ph/0405114.

\item {\bf [Slater 04 b]}:
P. B. Slater,
``Separability analyses of two-qubit density matrices'',
quant-ph/0408157.

\item {\bf [Slater 04 c]}:
P. B. Slater,
``Qubit-qutrit separability probabilities'',
quant-ph/0410238.

\item {\bf [Slavnov 00]}:
D. A. Slavnov,
``Quantum mechanics with the permitted hidden parameters'',
quant-ph/0010069.

\item {\bf [Sleator-Weinfurter 95]}:
T. Sleator, \& H. Weinfurter,
``Realizable universal quantum logic gates'',
{\em Phys. Rev. Lett.} {\bf 74}, 20, 4087-4090 (1995).

\item {\bf [\'{S}liwa-Banaszek 03]}:
C. \'{S}liwa, \& K. Banaszek,
``Conditional preparation of maximal polarization entanglement'',
{\em Phys. Rev. A} {\bf 67}, 3, 030101(R) (2003),
quant-ph/0207117.

\item {\bf [\'{S}liwa 03]}:
C. \'{S}liwa,
``Symmetries of the Bell correlation inequalities'',
{\em Phys. Lett. A} {\bf 317}, 3-4, 165-168 (2003);
quant-ph/0305190.

\item {\bf [Slutsky-Rao-Sun-Fainman 98]}:
B. A. Slutsky, R. Rao, P. Sun, \& Y. Fainman,
``Security of quantum cryptography against individual attacks'',
{\em Phys. Rev. A} {\bf 57}, 4, 2383-2398 (1998).

\item {\bf [Smirnov 03]}:
A. Y. Smirnov,
``Decoherence and relaxation of a quantum bit in the presence of Rabi
oscillations'',
{\em Phys. Rev. B} {\bf 67}, 15, 155104 (2003).

\item {\bf [Smith 01]}:
A. Smith,
``Multi-party quantum computation'',
M.\ Sc. thesis, 2001;
quant-ph/0111030.

\item {\bf [Smith 00]}:
A. D. Smith,
``Quantum secret sharing for general access structures'',
quant-ph/0001087.

\item {\bf [Smith 03]}:
D. Smith,
``Algebraic partial Boolean algebras'',
{\em J. Phys. A} {\bf 36}, 13, 3899–3910 (2003).

\item {\bf [Smith-Reiner-Orozco-(+2) 02]}:
W. P. Smith, J. E. Reiner, L. A. Orozco,
S. Kuhr, \& H. M. Wiseman,
``Capture and release of a conditional state of a cavity QED system by quantum
feedback'',
{\em Phys. Rev. Lett.} {\bf 89}, 13, 133601 (2002).

\item {\bf [Smolin-DiVincenzo 97]}:
J. A. Smolin, \& D. P. DiVincenzo,
``Five two-bit quantum gates are sufficient
to implement the quantum Fredkin gate'',
{\em Phys. Rev. A} {\bf 53}, 4, 2855-2856 (1996).

\item {\bf [Smolin 01]}:
J. A. Smolin,
``Four-party unlockable bound entangled state'',
{\em Phys. Rev. A} {\bf 63}, 3, 032306 (2001);
quant-ph/0001001.

\item {\bf [Smolin 03]}:
J. A. Smolin,
``Can quantum cryptography imply quantum mechanics?'',
quant-ph/0310067.
See {\bf [Halvorson-Bub 03]}.

\item {\bf [Smolin 04]}:
J. A. Smolin,
``The continuous variable quantum teleportation controversy'',
quant-ph/0407009.

\item {\bf [Snigirev-Kohn-Snigireva-Lengeler 96]}:
A. Snigirev, V. Kohn, I. Snigireva, \& B. Lengeler,
``A compound refractive lens for focusing high-energy X-rays'',
{\em Nature} {\bf 384}, 6604, 49-51 (1996).

\item {\bf [Sobottka 00]}:
S. Sobottka,
``Quantum theory -- Interpretation, formulation, inspiration'',
{\em Phys. Today} {\bf 53}, 9, ? (2000).
Comment on {\bf [Fuchs-Peres 00 a]}.
Reply: {\bf [Fuchs-Peres 00 b]}.

\item {\bf [Socolovsky 03]}:
M. Socolovsky,
``Bell inequality, nonlocality and analyticity'',
{\em Phys. Lett. A} {\bf 316}, 1-2, 10-16 (2003).

\item {\bf [Sohma-Hirota 00]}:
M. Sohma, \& O. Hirota,
``Binary discretization for quantum continuous channels'',
{\em Phys. Rev. A} {\bf 62}, 5, 052312 (2000).
See {\bf [Hastings 96]}.

\item {\bf [Sohma-Hirota 01]}:
M. Sohma, \& O. Hirota,
``Information capacity formula of quantum optical channels'',
quant-ph/0105042.

\item {\bf [Sohma-Hirota 03]}:
M. Sohma, \& O. Hirota,
``Capacity of a channel assisted by two-mode squeezed states'',
{\em Phys. Rev. A} {\bf 68}, 2, 022303 (2003).

\item {\bf [Soklakov-Schack 00]}:
A. N. Soklakov, \& R. Schack,
``Preparation information and optimal decompositions for mixed quantum
states'',
quant-ph/0002027.

\item {\bf [Soklakov-Schack 02]}:
A. N. Soklakov, \& R. Schack,
``Decoherence and linear entropy increase in the quantum baker's map'',
{\em Phys. Rev. E} {\bf 66}, 3, 036212 (2002);
quant-ph/0107071.

\item {\bf [Soklakov-Schack 03]}:
A. N. Soklakov, \& R. Schack,
``Information dynamics in cavity QED'',
{\em Phys. Rev. A} {\bf 67}, 3, 033804 (2003);
quant-ph/0210024.

\item {\bf [Soklakov-Schack 04 a]}:
A. N. Soklakov, \& R. Schack,
``Efficient state preparation for a register of quantum bits'',
quant-ph/0408045.

\item {\bf [Soklakov-Schack 04 b]}:
A. N. Soklakov, \& R. Schack,
``State preparation based on Grover's algorithm in the presence of global
information about the state'',
quant-ph/0411010.

\item {\bf [Sokolov-Kolobov-Gatti-Lugiato 00]}:
I. V. Sokolov, M. I. Kolobov, A. Gatti, \& L. A. Lugiato,
``Quantum holographic teleportation'',
{\em Opt. Comm.} {\bf 193}, 1-6, 175-180 (2001);
quant-ph/0007026.

\item {\bf [Sokolovski-Liu 01]}:
D. Sokolovski, \& Y. Liu,
``Quantum histories and measurements'',
{\em Phys. Lett. A} {\bf 281}, 4, 207-212 (2001).

\item {\bf [Sola-Malinovsky-Santamar\'{\i}a 04]}:
I. R. Sola, V. S. Malinovsky, and J. Santamar\'{\i}a,
``Implementing quantum gates on oriented optical isomers'',
{\em Journal of Chem. Phys.} {\bf 120}, 10955-10960 (2004).

\item {\bf [Solano-de Matos Filho-Zagury 99]}:
E. Solano, R. L. de Matos Filho, \& N. Zagury,
``Deterministic Bell states and measurement of the motional
state of two trapped ions'',
{\em Phys. Rev. A} {\bf 59}, 4, R2539-R2543 (1999).

\item {\bf [Solano-Cesar-de Matos Filho-Zagury 99]}:
E. Solano, C. L. Cesar, R. L. de Matos Filho, \& N. Zagury,
``Reliable teleportation in trapped ions'',
{\em Eur. Phys. J. D} {\bf 13}, 1, 121-128 (2001).

\item {\bf [Solano-Santos-Milman 01]}:
E. Solano, M. F. Santos, \& P. Milman,
``Quantum phase gate with a selective interaction'',
{\em Phys. Rev. A} {\bf 64}, 2, 024304 (2001).

\item {\bf [Solano-de Matos Filho-Zagury 02]}:
E. Solano, R. L. de Matos Filho, \& N. Zagury,
``Entangled coherent states and squeezing in $N$ trapped ions'',
in R. Bonifacio, \& D. Vitali (eds.),
{\em Mysteries, Puzzles and Paradoxes in Quantum Mechanics IV:
Quantum Interference Phenomena (Gargnano, Italy, 2001)},
{\em J. Opt. B: Quantum Semiclass. Opt.} {\bf 4}, 4, S324-S327 (2002);
quant-ph/0202072.

\item {\bf [Solano-Agarwal-Walther 03 a]}:
E. Solano, G. S. Agarwal, \& H. Walther,
``Strong-driving-assisted multipartite entanglement in cavity QED'',
{\em Phys. Rev. Lett.} {\bf 90}, 2, 027903 (2003);
quant-ph/0202071.

\item {\bf [Solano-Agarwal-Walther 03 b]}:
E. Solano, G. S. Agarwal, \& H. Walther,
``Generalized Schr\"{o}dinger cat states in cavity QED'',
{\em Opt. Spectrosc.} {\bf 94}, 805 (2003).

\item {\bf [Solano 03]}:
E. Solano,
``Selective interactions in trapped ions: state reconstruction and quantum logic'',
quant-ph/0310007.

\item {\bf [Solinas-Zanardi-Zanghi-Rossi 03 a]}:
P. Solinas, P. Zanardi, N. Zanghi, \& F. Rossi,
``Semiconductor-based geometrical quantum gates'',
{\em Phys. Rev. B} {\bf 67}, 12, 121307 (2003).

\item {\bf [Solinas-Zanardi-Zanghi-Rossi 03 b]}:
P. Solinas, P. Zanardi, N. Zanghi, \& F. Rossi,
``Nonadiabatic geometrical quantum gates in semiconductor quantum dots'',
{\em Phys. Rev. A} {\bf 67}, 5, 052309 (2003).

\item {\bf [Solinas-Zanardi-Zanghi-Rossi 03 c]}:
P. Solinas, P. Zanardi, N. Zanghi, \& F. Rossi,
``Holonomic quantum gates: A semiconductor-based implementation'',
{\em Phys. Rev. A} {\bf 67}, 6, 062315 (2003).

\item {\bf [Solinas-Zanardi-Zanghi 03]}:
P. Solinas, P. Zanardi, \& N. Zanghi,
``Robustness of non-abelian holonomic quantum gates against parametric
noise'',
quant-ph/0312109.

\item {\bf [Solinas-Zanghi 04]}:
P. Solinas, \& N. Zanghi,
``Noise cancellation effect in quantum systems'',
quant-ph/0409020.

\item {\bf [Soljanin 01]}:
E. Soljanin,
``Compressing mixed-state sources by sending classical information'',
quant-ph/0110130.

\item {\bf [Solli-McCormick-Ropers-(+3) 03]}:
D. R. Solli, C. F. McCormick, C. Ropers,
J. J. Morehead, R. Y. Chiao, \& J. M. Hickmann,
``Demonstration of superluminal effects in an absorptionless,
non-reflective system'',
{\em Phys. Rev. Lett.};
quant-ph/0308135.

\item {\bf [Solli-McCormick-Chiao-(+2) 03]}:
D. R. Solli, C. F. McCormick, R. Y. Chiao,
S. Popescu, \& J. M. Hickmann,
``Fast light, slow light, and phase singularities: A connection to
generalized weak values'',
{\em Phys. Rev. Lett.};
quant-ph/0310048.

\item {\bf [Solvay 28]}:
?,
{\em Electrons et photons. Rapports et discussions du cinqui\`{e}me
conseil de physique tenu \`{a} Bruxelles du 24 au 29 Octobre 1927
sous les auspices de l'Institut International de Physique Solvay},
Gauthier-Villars, Paris, 1928.

\item {\bf [Somaroo-Tseng-Havel-(+2) 99]}:
S. S. Somaroo, C. H. Tseng, T. F. Havel, R. Laflamme, \& D. G. Cory,
``Quantum simulations on a quantum computer'',
{\em Phys. Rev. Lett.} {\bf 82}, 26, 5381-5384 (1999);
quant-ph/9905045.

\item {\bf [Somma-Ortiz-Gubernatis-(+2) 02]}:
R. Somma, G. Ortiz, J. E. Gubernatis,
E. Knill, \& R. Laflamme,
``Simulating physical phenomena by quantum networks'',
{\em Phys. Rev. A} {\bf 65}, 4, 042323 (2002);
quant-ph/0108146.

\item {\bf [Somma-Ortiz-Barnum-(+2) 04]}:
R. Somma, G. Ortiz, H. Barnum,
E. Knill, \& L. Viola,
``Nature and measure of entanglement in quantum phase transitions'',
quant-ph/0403035.

\item {\bf [Sommers-\.{Z}yczkowski 03]}:
H.-J. Sommers, \& K. \.{Z}yczkowski,
``Bures volume of the set of mixed quantum states'',
{\em J. Phys. A} {\bf 36}, 39, 10083-10100 (2003);
quant-ph/0304041.

\item {\bf [Sommers-\.{Z}yczkowski 04]}:
H.-J. Sommers, \& K. \.{Z}yczkowski,
``Statistical properties of random density matrices'',
quant-ph/0405031.

\item {\bf [Son-Lee-Kim-Park 01]}:
W. Son, J. Lee, M. S. Kim, \& Y.-J. Park,
``Conclusive teleportation of a $d$-dimensional unknown state'',
{\em Phys. Rev. A} {\bf 64}, 6, 064304 (2001);
quant-ph/0012092.

\item {\bf [Son-Kim-Lee-Ahn 02]}:
W. Son, M. S. Kim, J. Lee, \& D. Ahn,
``Entanglement transfer from continuous variables to qubits'',
{\em J. Mod. Opt.} {\bf 49}, 10, 1739-1746 (2002);
quant-ph/0110151.

\item {\bf [Son-Lee-Kim 03]}:
W. Son, J. Lee, \& M. S. Kim,
``$D$-outcome measurement for nonlocality test'',
quant-ph/0309193.

\item {\bf [Sonego 96]}:
S. Sonego,
``Hidden variables, wholeness, and the two-hole experiment'',
{\em Eur. J. Phys.} {\bf 17}, 3, 118-124 (1996).

\item {\bf [Song 04]}:
D. Song,
``Secure key distribution by swapping quantum entanglement'',
{\em Phys. Rev. A} {\bf 69}, 3, 034301 (2004);
quant-ph/0305168.
See {\bf [Cabello 00 c, 01 b, e]}.

\item {\bf [Song-Klappenecker 03]}:
G. Song, \& A. Klappenecker,
``The simplified Toffoli gate implementation by Margolus is optimal'',
quant-ph/0312225.

\item {\bf [Song-Jiang-Shi-Guo 99]}:
K.-H. Song, Y. Jiang, B. Shi, \& G.-C. Guo,
``Establishment of multi-particle entanglement between particles located at
different nodes of a communication network'',
{\em Phys. Lett. A} {\bf 264}, 4, 261-264 (1999).

\item {\bf [Song-Jiang-Shi-Guo 00]}:
K.-H. Song, Y. Jiang, B. Shi, \& G.-C. Guo,
``Preparation and applications of the disentanglement-free state'',
{\em J. Opt. B: Quantum Semiclass. Opt.} {\bf 2}, 3, 221-224 (2000).

\item {\bf [Song-Zhang 01]}:
K.-H. Song, \& W.-J. Zhang,
``Proposal for teleporting an entangled coherent state
via the dispersive atom-cavity-field interaction'',
{\em Phys. Lett. A} {\bf 290}, 5-6, 214-218 (2001).

\item {\bf [Song 00]}:
P. H. Song,
``Time evolution of Grover's algorithm with imperfections'',
quant-ph/0010075.

\item {\bf [Song-Shepelyansky 01]}:
P. H. Song, \& D. L. Shepelyansky,
``Quantum computing of quantum chaos and imperfection effects'',
{\em Phys. Rev. Lett.} {\bf 86}, 10, 2162-2165 (2001).

\item {\bf [Song-Zhu 02]}:
T.-Q. Song, \& Y.-J. Zhu,
``$n$-particle entangled states in the $n$-mode Fock space'',
{\em Mod. Phys. Lett. B} {\bf 16}, 17, 631-636 (2002).

\item {\bf [Song 03]}:
T.-Q. Song,
``Teleportation of quantum states with continuous variables'',
{\em Phys. Lett. A} {\bf 316}, 6, 363-368 (2003).

\item {\bf [Song-Yang-Cao 04 a]}:
W. Song, M. Yang, \& Z.-L. Cao,
``Strategies for state-dependent quantum deleting'',
{\em Phys. Lett. A} {\bf 327}, 2-3, 123-128 (2004).

\item {\bf [Song-Yang-Cao 04 b]}:
W. Song, M. Yang, \& Z.-L. Cao,
``Probabilistic implementation of universal Hadamard and unitary gates'',
{\em Phys. Lett. A} {\bf 330}, 3-4, 155-160 (2004);
quant-ph/0404141.

\item {\bf [Song-Yang-Cao 04 c]}:
W. Song, M. Yang, \& Z.-L. Cao,
``Scheme for the implementation of optimal phase-covariant quantum cloning for equatorial qubits'',
{\em Phys. Lett. A} {\bf 331}, 1-2, 34-38 (2004).

\item {\bf [Soo-Lin 04]}:
C. Soo, \& C. C. Y. Lin,
``Wigner rotations, Bell states, and Lorentz invariance of entanglement
and von Neumann entropy'',
{\em Int. J. Quant. Inf.} {\bf 2}, ?, 183-200 (2004);
quant-ph/0307107.

\item {\bf [S{\o}rensen 02]}:
A. S. S{\o}rensen,
``Bogoliubov theory of entanglement in a Bose-Einstein condensate'',
{\em Phys. Rev. A} {\bf 65}, 4, 043610 (2002).

\item {\bf [S{\o}rensen-M{\o}lmer 03 a]}:
A. S. S{\o}rensen, \& K. M{\o}lmer,
``Probabilistic generation of entanglement in optical cavities'',
{\em Phys. Rev. Lett.} {\bf 90}, 12, 127903 (2003).

\item {\bf [S{\o}rensen-M{\o}lmer 03 b]}:
A. S. S{\o}rensen, \& K. M{\o}lmer,
``Measurement induced entanglement and quantum computation with atoms in
optical cavities'',
{\em Phys. Rev. Lett.} {\bf 91}, 9, 097905 (2003).

\item {\bf [Sornborger-Stewart 99]}:
A. T. Sornborger, \& E. D. Stewart,
``Higher-order methods for simulations on quantum computers'',
{\em Phys. Rev. A} {\bf 60}, 3, 1956-1965 (1999).

\item {\bf [Soubusta-Cernoch-Fiur\'{a}\v{s}ek-Du\v{s}ek 04]}:
J. Soubusta, A. Cernoch, J. Fiur\'{a}\v{s}ek, \& M. Du\v{s}ek,
``Experimental realization of a programmable quantum-state discriminator
and a phase-covariant quantum multimeter'',
{\em Phys. Rev. A} {\bf 69}, 5, 052321 (2004).
quant-ph/0401166.

\item {\bf [Soucek 01]}:
J. Soucek,
``The new interpretation of quantum mechanics and hidden parameters'',
quant-ph/0107117.

\item {\bf [de Sousa-Das Sarma 03]}:
R. de Sousa, \& S. Das Sarma,
``Electron spin coherence in semiconductors: Considerations for a spin-based
solid-state quantum computer architecture'',
{\em Phys. Rev. B} {\bf 67}, 3, 033301 (2003).
Publisher's note: {\em Phys. Rev. B} {\bf 67}, 3, 039901 (2003).

\item {\bf [de Sousa-Delgado-Das Sarma 04]}:
R. de Sousa, J. D. Delgado, \& S. Das Sarma,
``Silicon quantum computation based on magnetic dipolar coupling'',
{\em Phys. Rev. A};
cond-mat/0311403.

\item {\bf [Souto Ribeiro-P\'{a}dua-Monken 00]}:
P. H. Souto Ribeiro, S. P\'{a}dua, \& C. H. Monken,
``Quantum erasure by transverse indistinguishability'',
quant-ph/0009120.

\item {\bf [Specker 60]}:
E. P. Specker,
``Die Logik nicht gleichzeitig entscheidbarer Aussagen'',
{\em Dialectica} {\bf 14}, 54-55, 239-246 (1960).
Reprinted in {\bf [Specker 90]}, pp.~175-182.
English version: ``The logic of propositions which are not
simultaneously decidable'',
in {\bf [Hooker 75]}, pp.~135-140.

\item {\bf [Specker 90]}:
E. P. Specker,
{\em Selecta},
Birkh\"{a}user Verlag, Basel, Switzerland, 1990.

\item {\bf [Spedalieri 01]}:
F. M. Spedalieri,
``Local deterministic transformations of three-qubit pure states'',
quant-ph/0110179.

\item {\bf [Spedalieri-Lee-Florescu-(+3) 04]}:
F. M. Spedalieri, H. Lee, M. Florescu,
K. T. Kapale, U. Yurtsever, \& J. P. Dowling,
``Exploiting the quantum Zeno effect to beat photon loss in linear optical
quantum information processors'',
quant-ph/0408026.

\item {\bf [Spedalieri 04]}:
F. M. Spedalieri,
``Quantum key distribution without reference frame alignment: Exploiting
photon orbital angular momentum'',
quant-ph/0409057.

\item {\bf [Spekkens-Sipe 01 a]}:
R. W. Spekkens, \& J. E. Sipe,
``Non-orthogonal core projectors for modal interpretations of quantum
mechanics'',
{\em Found. Phys.} {\bf 31}, 10, 1403-1430 (2001);
quant-ph/0003092.

\item {\bf [Spekkens-Sipe 01 b]}:
R. W. Spekkens, \& J. E. Sipe,
``A modal interpretation of quantum mechanics based on a principle
of entropy minimization'',
{\em Found. Phys.} {\bf 31}, 10, 1431-1464 (2001).

\item {\bf [Spekkens-Rudolph 01]}:
R. W. Spekkens, \& T. Rudolph,
``Optimization of coherent attacks in generalizations of the BB84
quantum bit commitment protocol'',
quant-ph/0107042.

\item {\bf [Spekkens-Rudolph 02]}:
R. W. Spekkens, \& T. Rudolph,
``Degrees of concealment and bindingness in quantum bit commitment protocols'',
{\em Phys. Rev. A} {\bf 65}, 1, 012310 (2002);
quant-ph/0106019.

\item {\bf [Spekkens-Rudolph 02]}:
R. W. Spekkens, \& T. Rudolph,
``Quantum protocol for cheat-sensitive weak coin flipping'',
{\em Phys. Rev. Lett.} {\bf 89}, 22, 227901 (2002);
quant-ph/0202118.

\item {\bf [Spekkens 04 a]}:
R. W. Spekkens,
``In defense of the epistemic view of quantum states: A toy theory'',
based on a talk given at the
{\em Rob Clifton Memorial Conference (College Park, Maryland, 2003)};
quant-ph/0401052.

\item {\bf [Spekkens 04 b]}:
R. W. Spekkens,
``Contextuality for preparations, transformations, and unsharp
measurements'',
quant-ph/0406166.

\item {\bf [Spiller-Clarck-Prance-Prance 90]}:
T. P. Spiller, T. D. Clarck, R. J. Prance, \& H. Prance,
``Barrier traversal time in the quantum potential picture'',
{\em Europhys. Lett.} {\bf 12}, 1, 1-4 (1990).

\item {\bf [Spiller 98]}:
T. P. Spiller,
``Basic elements of quantum information technology'',
in {\bf [Lo-Spiller-Popescu 98]}, pp.~1-28.

\item {\bf [Spiller 00]}:
T. P. Spiller,
``Superconducting circuits for quantum computing'',
{\em Fortschr. Phys.} {\bf 48}, 9-11 (Special issue: Experimental proposals for quantum computation), 1075-1094 (2000).

\item {\bf [Spreeuw 98]}:
R. J. C. Spreeuw,
``A classical analogy of entanglement'',
{\em Found. Phys.} {\bf 28}, 3, 361-374 (1998).

\item {\bf [Spreeuw 01]}:
R. J. C. Spreeuw,
``Classical wave-optics analogy of quantum information
processing'',
{\em Phys. Rev. A} {\bf 63}, 6, 062302 (2001);
quant-ph/0009066.

\item {\bf [Sprinzak-Buks-Heiblum-Shtrikman 99]}:
D. Sprinzak, E. Buks, M. Heiblum, \& H. Shtrikman,
``Controlled dephasing via phase detection of electrons:
Demonstration of Bohr's complementarity principle'',
cond-mat/9907162.

\item {\bf [S{\o}rensen-M{\o}lmer 98 a]}:
A. S. S{\o}rensen, \& K. M{\o}lmer,
``Error-free quantum communication through noisy channels'',
{\em Phys. Rev. A} {\bf 58}, 4, 2745-2749 (1998);
quant-ph/9810042.

\item {\bf [S{\o}rensen-M{\o}lmer 99]}:
A. S. S{\o}rensen, \& K. M{\o}lmer,
``Quantum computation with ions in thermal motion'',
{\em Phys. Rev. Lett.} {\bf 82}, 9, 1971-1974 (1999);
quant-ph/9810039.

\item {\bf [S{\o}rensen-M{\o}lmer 00 a]}:
A. S. S{\o}rensen, \& K. M{\o}lmer,
``Entanglement and quantum computation with ions in thermal motion'',
{\em Phys. Rev. A} {\bf 62}, 2, 022311 (2000).

\item {\bf [S{\o}rensen-M{\o}lmer 00 b]}:
A. S. S{\o}rensen, \& K. M{\o}lmer,
``Ion trap quantum computer with bichromatic light'',
{\em Fortschr. Phys.} {\bf 48}, 9-11 (Special issue:
Experimental proposals for quantum computation), 811-821 (2000).

\item {\bf [S{\o}rensen-Duan-Cirac-Zoller 01]}:
A. S. S{\o}rensen, L.-M. Duan, J. I. Cirac, \& P. Zoller,
``Many-particle entanglement with Bose-Einstein condensates'',
{\em Nature} {\bf 409}, 6816, 63-65 (2001);
quant-ph/0006111.
See {\bf [Bigelow 01]}.

\item {\bf [S{\o}rensen-M{\o}lmer 01]}:
A. S. S{\o}rensen, \& K. M{\o}lmer,
``Entanglement and extreme spin squeezing'',
{\em Phys. Rev. Lett.} {\bf 86}, 20, 4431-4434 (2001).

\item {\bf [S{\o}rensen-M{\o}lmer 02]}:
A. S. S{\o}rensen, \& K. M{\o}lmer,
``Entangling atoms in bad cavities'',
{\em Phys. Rev. A} {\bf 66}, 2, 022314 (2002);
quant-ph/0202073.

\item {\bf [S{\o}rensen-M{\o}lmer 03]}:
A. S. S{\o}rensen, \& K. M{\o}lmer,
``Measurement induced entanglement and quantum computation with atoms in
optical cavities'',
quant-ph/0304008.

\item {\bf [Squires 86]}:
E. J. Squires,
{\em The mystery of the quantum world},
Adam Hilger, Bristol, 1986; Institute of Physics, Bristol, 1994 (2nd edition).

\item {\bf [Squires 87 a]}:
E. J. Squires,
``Many views of one world---an interpretation of quantum theory'',
{\em Eur. J. Phys.} {\bf 8}, 3, 171-173 (1987).
See {\bf [Squires 87 b]}, {\bf [Whitaker 89]}.

\item {\bf [Squires 87 b]}:
E. J. Squires,
``?'',
{\em New Scientist} {\bf 1580}, ?, 79-?.
See {\bf [Squires 87 a]}, {\bf [Whitaker 89]}.

\item {\bf [Squires 88]}:
E. J. Squires,
``Non-self-adjoint observables'',
{\em Phys. Lett. A} {\bf 130}, 4-5, 192-193 (1988).

\item {\bf [Squires 90]}:
E. J. Squires,
``An attempt to understand the many-worlds interpretation of
quantum theory'',
in M. Cini, \& J. M. L\'{e}vy-Leblond (eds.),
{\em Quantum theory without reduction},
Adam Hilger, Bristol, 1990, pp.~151-160.

\item {\bf [Squires 91]}:
E. J. Squires,
``Wavefunction collapse and ultraviolet photons'',
{\em Phys. Lett. A} {\bf 158}, 9, 431-432 (1991).

\item {\bf [Squires 92 a]}:
E. J. Squires,
``Quantum challenge'',
{\em Phys. World} {\bf 4}, 1, 19-20 (1991).

\item {\bf [Squires 92 b]}:
E. J. Squires,
``Explicit collapse and superluminal signals'',
{\em Phys. Lett. A} {\bf 163}, 5-6, 356-358 (1992).

\item {\bf [Squires 93]}:
E. J. Squires,
``Consciousness in the quantum world'',
{\em Contemp. Phys.} {\bf 34}, 6, 329-331 (1993).
Review of {\bf [Stapp 93 b]}.

\item {\bf [Squires 94 a]}:
E. J. Squires,
``Has quantum nonlocality been
experimentally verified?'',
{\em Found. Phys. Lett.} {\bf 7}, 4, 353-363 (1994).

\item {\bf [Squires-Hardy-Brown 94]}:
E. J. Squires, L. Hardy, \& H. R. Brown,
``Non-locality from an analogue of the quantum Zeno effect'',
{\em Stud. Hist. Philos. Sci.} {\bf 25}, 3, 425-435 (1994).

\item {\bf [Srikanth 01 a]}:
R. Srikanth,
``On a generalized peaceful coexistence of special relativity and quantum mechanics'',
{\em Phys. Lett. A} {\bf 292}, 3, 161-165 (2001);

\item {\bf [Srikanth 01 b]}:
R. Srikanth,
``A limit of the complementarity principle'',
quant-ph/0102009.

\item {\bf [Srikanth 01 c]}:
R. Srikanth,
``The hidden cost of quantum teleportation and remote state
preparation'',
quant-ph/0104081.

\item {\bf [Srikanth 01 d]}:
R. Srikanth,
``Classical communication via the Einstein-Podolsky-Rosen channel alone:
A proposed experimental test'',
quant-ph/0109148.

\item {\bf [Srikanth 01 e]}:
R. Srikanth,
``Single particle nonlocality: A proposed experimental test'',
quant-ph/0110132.

\item {\bf [Srikanth 01 f]}:
R. Srikanth,
``Securing quantum bit commitment through reverse quantum communication'',
quant-ph/0112172.

\item {\bf [Srikanth 03 a]}:
R. Srikanth,
``A computational model for quantum measurement'',
{\em Quant. Inf. Proc.} {\bf 2}, 3, 153-199 (2003);
quant-ph/0302160.

\item {\bf [Srikanth 03 b]}:
R. Srikanth,
``Quantum bit commitment with a composite evidence'',
quant-ph/0306155.

\item {\bf [Srivastava-Vitiello-Widom 98]}:
Y. N. Srivastava, G. Vitiello, \& A. Widom,
``Quantum measurements, information and entropy production'',
quant-ph/9810095.

\item {\bf [Staanum-Drewsen 02]}:
P. Staanum, \& M. Drewsen,
``Trapped-ion quantum logic utilizing position-dependent ac Stark shifts'',
{\em Phys. Rev. A} {\bf 66}, 4, 040302 (2002).

\item {\bf [Staanum-Drewsen-M{\o}lmer 04]}:
P. Staanum, M. Drewsen, \& K. M{\o}lmer,
``Geometric quantum gate for trapped ions based on optical dipole forces
induced by Gaussian laser beams'',
quant-ph/0406186.

\item {\bf [Stace-Barnes 02]}:
T. M. Stace, \& C. H. W. Barnes,
``Effect of stochastic noise on quantum-state transfer'',
{\em Phys. Rev. A} {\bf 65}, 6, 062308 (2002);
quant-ph/0204083.

\item {\bf [Stace-Barrett 04]}:
T. M. Stace, \& S. D. Barrett,
``Continuous quantum measurement: Inelastic tunneling and lack of current oscillations'',
{\em Phys. Rev. Lett.} {\bf 92}, 13, 136802 (2004).

\item {\bf [Stadler-Hofer 02]}:
R. Stadler, \& W. A. Hofer,
``Comment on `Matter-wave interferometer for large molecules'\,'',
quant-ph/0204012.
Comment on {\bf [Brezger-Hackerm\"{u}ller-Uttenthaler-(+3) 02]}.

\item {\bf [Stairs 78]}:
A. Stairs,
``Quantum mechanics, logic and reality'',
Ph.\ D. thesis, University of Western Ontario, Canada, 1978.

\item {\bf [Stairs 79]}:
A. Stairs,
``On Arthur Fine's interpretation of quantum
mechanics'',
{\em Synthese} {\bf 42}, 1, 91-100 (1979).

\item {\bf [Stairs 82]}:
A. Stairs,
``Quantum logic and the L\"{u}ders rule'',
{\em Philos. Sci.} {\bf 49}, ?, 422-436 (1982).

\item {\bf [Stairs 83 a]}:
A. Stairs,
``On the logic of pairs of quantum systems'',
{\em Synthese} {\bf 56}, ?, 47-60 (1983).

\item {\bf [Stairs 83 b]}:
A. Stairs,
``Quantum logic, realism, and value definiteness'',
{\em Philos. Sci.} {\bf 50}, 4, 578-602 (1983).

\item {\bf [Stairs 84]}:
A. Stairs,
``Sailing into the Charybdis: Van Fraassen on Bell's theorem'',
{\em Synthese} {\bf 61}, ?, 351-359 (1984).
See {\bf [van Fraassen 82]}.

\item {\bf [Stapp 71]}:
H. P. Stapp,
``S-matrix interpretation of quantum theory'',
{\em Phys. Rev. D} {\bf 3}, 6, 1303-1320 (1971).

\item {\bf [Stapp 72]}:
H. P. Stapp,
``The Copenhagen interpretation'',
{\em Am. J. Phys.} {\bf 40}, 8, 1098-1116 (1972).
Reprinted in {\bf [Stapp 93 b]}, pp.~49-78.
Comment: {\bf [Ballentine 74]}. Reply: {\bf [Stapp 74]}.

\item {\bf [Stapp 74]}:
H. P. Stapp,
``Reply to Ballentine's comments'',
{\em Am. J. Phys.} {\bf 42}, 1, 83-85 (1974).
Reply to {\bf [Ballentine 74]}. See {\bf [Stapp 72]}.

\item {\bf [Stapp 75]}:
H. P. Stapp,
``Bell's theorem and world process'',
{\em Nuovo Cimento B} {\bf 29}, 2, 270-276 (1975).

\item {\bf [Stapp 77 a]}:
H. P. Stapp,
``Are superluminal connections necessary?'',
{\em Nuovo Cimento B} {\bf 40}, 1, 191-205 (1977).

\item {\bf [Stapp 77 b]}:
H. P. Stapp,
``Theory of reality'',
{\em Found. Phys.} {\bf 7}, ?, 313-323 (1977).

\item {\bf [Stapp 79]}:
H. P. Stapp,
``Whitehedian approach to quantum theory and
the generalized Bell's theorem'',
{\em Found. Phys.} {\bf 9}, 1-2, 1-25 (1979).

\item {\bf [Stapp 80]}:
H. P. Stapp,
``Locality and reality'',
{\em Found. Phys.} {\bf 10}, 9-10, 767-795 (1980).

\item {\bf [Stapp 82 a]}:
H. P. Stapp,
``Mind, matter, and quantum mechanics'',
{\em Found. Phys.} {\bf 12}, 4, 363-399 (1982).
Reprinted in {\bf [Stapp 93 b]}, pp.~79-116.

\item {\bf [Stapp 82 b]}:
H. P. Stapp,
``Bell's theorem as a nonlocality property of quantum theory'',
{\em Phys. Rev. Lett.} {\bf 49}, 20, 1470-1474 (1982).
See {\bf [Fine 82 a]}.

\item {\bf [Stapp 85 a]}:
H. P. Stapp,
``Bell's theorem and the foundations of quantum physics'',
{\em Am. J. Phys.} {\bf 53}, 4, 306-317 (1985).
Comments: {\bf [Guy-Deltete 88]}, {\bf [Fellows 88]}.
Reply: {\bf [Stapp 88 c, d]}.

\item {\bf [Stapp 85 b]}:
H. P. Stapp,
``Comments on `Locality, Bell's theorem, and quantum mechanics'\,'',
{\em Found. Phys.} {\bf 15}, 9, 973-976 (1985).
Comment on {\bf [Rastall 85]}.

\item {\bf [Stapp 85 c]}:
H. P. Stapp,
``On the unification of quantum theory and classical physics'',
in P. J. Lahti, \& P. Mittelstaedt (eds.),
{\em Symp.\ on the Foundations of Modern
Physics: 50 Years of the Einstein-Podolsky-Rosen Experiment
(Joensuu, Finland, 1985)},
World Scientific, Singapore, 1985, pp.~213-222.

\item {\bf [Stapp 85 d]}:
H. P. Stapp,
``EPR: What has it taught us?'',
in P. J. Lahti, \& P. Mittelstaedt (eds.),
{\em Symp.\ on the Foundations of Modern
Physics: 50 Years of the Einstein-Podolsky-Rosen Experiment
(Joensuu, Finland, 1985)},
World Scientific, Singapore, 1985, pp.~637-652.

\item {\bf [Stapp 87]}:
H. P. Stapp,
``Light as foundation of being'',
in {\bf [Hiley-Peat 87]}, pp.~255-266.

\item {\bf [Stapp 88 a]}:
H. P. Stapp,
``Quantum nonlocality'',
{\em Found. Phys.} {\bf 18}, ?, 427-448 (1988).

\item {\bf [Stapp 88 b]}:
H. P. Stapp,
``Spacetime and future quantum theory'',
{\em Found. Phys.} {\bf 18}, ?, 833-849 (1988).

\item {\bf [Stapp 88 c]}:
H. P. Stapp,
``Reply to `Note on ``Bell's theorem and the foundations of quantum
physics''\,'[{\em Am. J. Phys.} {\bf 56}, 565 (1988)]'',
{\em Am. J. Phys.} {\bf 56}, 6, 567 (1988).
Reply to {\bf [Guy-Deltete 88]}.
See {\bf [Stapp 85 a]}.

\item {\bf [Stapp 88 d]}:
H. P. Stapp,
``Reply to `Comment on ``Bell's theorem and
the foundations of quantum physics''\,'[{\em Am. J. Phys.}
{\bf 56}, 567 (1988)]'',
{\em Am. J. Phys.} {\bf 56}, 6, 568-569 (1988).
Reply to {\bf [Fellows 88]}.
See {\bf [Stapp 85 a]}.

\item {\bf [Stapp 88 e]}:
H. P. Stapp,
``Are faster-than-light influences necessary?'',
in F. Selleri (ed.),
{\em Quantum mechanics versus local realism:
The Einstein-Podolsky-Rosen paradox},
Plenum Press, New York, 1988, pp.~63-85.

\item {\bf [Stapp 88 f]}:
H. P. Stapp,
``Einstein locality, EPR locality, and the
significance for science of the nonlocal character of quantum theory'',
in A. van der Merwe, F. Selleri, \& G. Tarozzi (eds.),
{\em Microphysical reality and quantum formalism.
Proc.\ of an international conference (Urbino, Italy, 1985)},
Kluwer Academic, Dordrecht, Holland, 1988, vol. 2, pp.~367-378.

\item {\bf [Stapp 89 a]}:
H. P. Stapp,
``Quantum nonlocality and the description of nature'',
in J. T. Cushing, \& E. McMullin (eds.),
{\em Philosophical consequences of
quantum theory: Reflections on Bell's theorem},
University of Notre Dame Press, Notre
Dame, Indiana, 1989, pp.~?-?.

\item {\bf [Stapp 89 b]}:
H. P. Stapp,
``Comments on `Quantum theory does not require
action at a distance'\,'',
{\em Found. Phys. Lett.} {\bf 2}, 1, 9-13 (1989).
Comment on {\bf [Kraus 89]}.

\item {\bf [Stapp 90]}:
H. P. Stapp,
``Comments on `Nonlocal influences and possible worlds'\,'',
{\em Brit. J. Philos. Sci.} {\bf 41}, 1, 59-72 (1990).
Comment on {\bf [Clifton-Butterfield-Redhead 90]}.
See {\bf [Dickson 93]}.

\item {\bf [Stapp 91]}:
H. P. Stapp,
``EPR and Bell's theorem: A critical review'',
{\em Found. Phys.} {\bf 21}, 1, 1-23 (1991).

\item {\bf [Stapp 92]}:
H. P. Stapp,
``Noise-induced reduction of wave packets and
faster-than-light influences'',
{\em Phys. Rev. A} {\bf 46}, 11, 6860-6868 (1992).

\item {\bf [Stapp 93 a]}:
H. P. Stapp,
``Significance of an experiment of the
Greenberger-Horne-Zeilinger kind'',
{\em Phys. Rev. A} {\bf 47}, 2, 847-853 (1993).
See {\bf [Vaidman 98 d]}.

\item {\bf [Stapp 93 b]}:
H. P. Stapp,
{\em Mind, matter, and quantum mechanics},
Springer-Verlag, New York, 1993.
Review: {\bf [Squires 93]}, {\bf [Greenberger 94 b]}.

\item {\bf [Stapp 94 a]}:
H. P. Stapp,
``Strong versions of Bell's theorem'',
{\em Phys. Rev. A} {\bf 49}, 5, Part A, 3182-3187 (1994).

\item {\bf [Stapp 94 b]}:
H. P. Stapp,
``Reply to `Stapp's algebraic argument for nonlocality'\,'',
{\em Phys. Rev. A} {\bf 49}, 5, Part B, 4257-4260 (1994).
Comment on {\bf [Dickson-Clifton 94]}.

\item {\bf [Stapp 94 c]}:
H. P. Stapp,
``The undivided universe: An ontological
interpretation of quantum theory'',
{\em Am. J. Phys.} {\bf 62}, 10, 958-960 (1994).
Review of {\bf [Bohm-Hiley 93]}.

\item {\bf [Stapp 94 d]}:
H. P. Stapp,
``Comments on `Interpretations of quantum mechanics, joint measurements
of incompatible observables, and counterfactual definitness'\,'',
{\em Found. Phys.} {\bf 24}, 12, 1665-1669 (1994).
Comment on {\bf [De Muynck-de Baere-Martens 94]}.

\item {\bf [Stapp 97 a]}:
H. P. Stapp,
``Nonlocal character of quantum theory'',
{\em Am. J. Phys.} {\bf 65}, 4, 300-304 (1997).
See {\bf [Unruh 97]}.
Comment: {\bf [Mermin 97 c]}, {\bf [Finkelstein 98 b]}.
Reply: {\bf [Stapp 97 b, 01 a]}.
See {\bf [Mermin 98 b]}, {\bf [Stapp 97 e]},
{\bf [Mashkevich 98 a]}, {\bf [Shimony-Stein 01]}.

\item {\bf [Stapp 97 b]}:
H. P. Stapp,
``Mermin's suggestion and the nature of
Bohr's action-at-a-distance influence'',
quant-ph/9711060. First reply to {\bf [Mermin 98 b]}.
Reply: {\bf [Mermin 97 c]}.

\item {\bf [Stapp 97 c]}:
H. P. Stapp,
``On quantum theories of the mind'',
quant-ph/9711064.

\item {\bf [Stapp 97 d]}:
H. P. Stapp,
``Nonlocality and Bohr's reply to EPR'',
quant-ph/9712036. Second reply to {\bf [Mermin 98 b]}.
See {\bf [Stapp 97 a, b]}.

\item {\bf [Stapp 97 e]}:
H. P. Stapp,
``Reply to Unruh'',
quant-ph/9712043.
See {\bf [Unruh 97]}, {\bf [Stapp 98 a]}.

\item {\bf [Stapp 98 a]}:
H. P. Stapp,
``Comments on Unruh's paper'',
quant-ph/9801056.
See {\bf [Unruh 97]}, {\bf [Stapp 97 e, 98 b]}.

\item {\bf [Stapp 98 b]}:
H. P. Stapp,
``Is quantum mechanics non-local?'',
quant-ph/9805047.

\item {\bf [Stapp 98 c]}:
H. P. Stapp,
``Meaning of counterfactual statements in quantum physics'',
{\em Am. J. Phys.} {\bf 66}, 11, 924-926 (1998).
Reply to {\bf [Mermin 98 b]}.
See {\bf [Unruh 97]}, {\bf [Stapp 97 e, 98 a]}.

\item {\bf [Stapp 99 a]}:
H. P. Stapp,
``Quantum ontologies and mind-matter synthesis'',
in P. Blanchard, \& A. Jadczyk (eds.),
{\em Quantum Future},
Springer-Verlag, Berlin, 1999;
quant-ph/9905053.

\item {\bf [Stapp 99 b]}:
H. P. Stapp,
``Attention, intention, and will in quantum physics'',
{\em J. Conscious Studies}, July 1999;
quant-ph/9905054.

\item {\bf [Stapp 99 c]}:
H. P. Stapp,
``Nonlocality, counterfactuals, and consistent histories'',
quant-ph/9905055.

\item {\bf [Stapp 99 d]}:
H. P. Stapp,
``Comment on `Nonlocality
counterfactuals, and quantum mechanics'\,'',
{\em Phys. Rev. A} {\bf 60}, 3, 2595-2598 (1999).
Comment on {\bf [Unruh 99 a]}.
Reply: {\bf [Unruh 99 b]}.

\item {\bf [Stapp 00 b]}:
H. P. Stapp,
``From Einstein nonlocality to Von Neumann reality'',
quant-ph/0003064.

\item {\bf [Stapp 00 c]}:
H. P. Stapp,
``Decoherence, quantum Zeno effect, and the efficacy of mental
effort'',
quant-ph/0003065.

\item {\bf [Stapp 00 d]}:
H. P. Stapp,
``From quantum nonlocality to mind-brain interaction'',
submitted to {\em Proc. R. Soc. Lond. A};
quant-ph/0009062.

\item {\bf [Stapp 00 e]}:
H. P. Stapp,
``The importance of quantum decoherence in brain processes'',
submitted to {\em Phys. Rev. E};
quant-ph/0010029.

\item {\bf [Stapp 00 f]}:
H. P. Stapp,
``Bell's theorem without hidden variables'',
quant-ph/0010047.

\item {\bf [Stapp 01 a]}:
H. P. Stapp,
`Response to ``Comment on `Nonlocal character of quantum theory'\,''
by Abner Shimony and Howard Stein
[Am. J. Phys. {\bf 69} (8), 848-853 (2001)]',
{\em Am. J. Phys.} {\bf 69}, 8, 854-859 (2001);
quant-ph/0010086.
Reply to {\bf [Shimony-Stein 01]}.
See {\bf [Stapp 97 a]}.

\item {\bf [Stapp 01 b]}:
H. P. Stapp,
``Von Neumann's formulation of quantum theory and
the role of mind in nature'',
quant-ph/0101118.

\item {\bf [Stapp 01 c]}:
H. P. Stapp,
``Quantum theory and the role of mind in nature'',
{\em Found. Phys} {\bf 31}, 10, 1465-1499 (2001);
quant-ph/0103043.
Revised version of {\bf [Stapp 01 a]}.
Comment: {\bf [Mohrhoff 02 a]}.
Reply: {\bf [Stapp 02]}.

\item {\bf [Stapp 01 d]}:
H. P. Stapp,
``The basis problem in many worlds theories'',
quant-ph/0110148.

\item {\bf [Stapp 02]}:
H. P. Stapp,
``The 18-fold way'',
{\em Found. Phys.} {\bf 32}, 2, 255-266 (2002);
quant-ph/0108092.
Reply to {\bf [Mohrhoff 02 a]}.
See {\bf [Stapp 01 c]}.

\item {\bf [Stapp 03]}:
H. P. Stapp,
``Correspondence and analyticity'',
{\em Publications of RIMS},
quant-ph/0312013.

\item {\bf [Stapp 04 a]}:
H. P. Stapp,
``A Bell-type theorem without hidden variables'',
{\em Am. J. Phys.} {\bf 72}, 1, 30-33 (2004);
quant-ph/0205096.

\item {\bf [Stapp 04 b]}:
H. P. Stapp,
``Comments on Shimony's analysis'',
{\em Found. Phys.} (Festschrift in honor of Asher Peres);
quant-ph/0404169.

\item {\bf [Stay 00]}:
M. Stay,
``Artificial orbitals and a solution to Grover's problem'',
quant-ph/0010065.

\item {\bf [Steane 96 a]}:
A. M. Steane,
``Error correction codes in quantum theory'',
{\em Phys. Rev. Lett.} {\bf 77}, 5, 793-797 (1996).
Reprinted in {\bf [Macchiavello-Palma-Zeilinger 00]}, pp.~138-142.

\item {\bf [Steane 96 b]}:
A. M. Steane,
``Multiparticle interference and quantum error correction'',
{\em Proc. R. Soc. Lond. A} {\bf 452}, 1954, 2551-2577 (1996);
quant-ph/9601029.

\item {\bf [Steane 96 c]}:
A. M. Steane,
``Simple quantum error-correcting codes'',
{\em Phys. Rev. A} {\bf 54}, 6, 4741-4751 (1996);
quant-ph/9605021.

\item {\bf [Steane 97 a]}:
A. M. Steane,
``The ion trap quantum information processor'',
{\em Appl. Phys. B} {\bf 64}, ?, 623-? (1997);
quant-ph/9608011.

\item {\bf [Steane 97 b]}:
A. M. Steane,
``Active stabilization, quantum computation, and quantum state
synthesis'',
{\em Phys. Rev. Lett.} {\bf 78}, 11, 2252-2255 (1997);
quant-ph/9611027.

\item {\bf [Steane 98 a]}:
A. M. Steane,
``Space, time, parallelism and noise
requirements for reliable quantum computing'',
{\em Fortsch. Phys.} {\bf 46}, ?, 443-458 (1998).
quant-ph/9708021.

\item {\bf [Steane 98 b]}:
A. M. Steane,
``Quantum computing'',
{\em Rep. Prog. Phys.} {\bf 61}, 2, 117-173 (1998);
quant-ph/9708022.

\item {\bf [Steane 98 c]}:
A. M. Steane,
``Enlargement of Calderbank Shor Steane quantum codes'',
submitted to {\em IEEE Trans. Inf. Theory};
quant-ph/9802061.

\item {\bf [Steane 98 d]}:
A. M. Steane,
``Quantum error correction'',
in {\bf [Lo-Spiller-Popescu 98]}, pp.~184-212.

\item {\bf [Steane 98 e]}:
A. M. Steane,
``Introduction to quantum error correction'',
in A. K. Ekert, R. Jozsa, \& R. Penrose (eds.),
{\em Quantum Computation: Theory and Experiment.
Proceedings of a Discussion Meeting held at the Royal
Society of London on 5 and 6 November 1997},
{\em Philos. Trans. R. Soc. Lond. A} {\bf 356}, 1743, 1739-17587 (1998).

\item {\bf [Steane 98 e]}:
A. M. Steane,
``Space, time, parallelism and noise requirements
for reliable quantum computing'',
{\em Fortschr. Phys.} {\bf 46}, 4-5, 443-457 (1998);
quant-ph/9708021.

\item {\bf [Steane 99 a]}:
A. M. Steane,
``Efficient fault-tolerant quantum computing'',
{\em Nature} {\bf 399}, 6732, 124-126 (1999);
quant-ph/9809054.

\item {\bf [Steane 99 b]}:
A. M. Steane,
``Bit of a hype'',
{\em Phys. World} {\bf 12}, 9, 17-18 (1999).

\item {\bf [Steane 99 c]}:
A. M. Steane,
``Feynman's spirit lives on in computing'',
{\em Phys. World} {\bf 12}, 6, 48-49 (1999).
Review of {\bf [Hey 99]}.

\item {\bf [Steane 99 d]}:
A. M. Steane,
``Quantum Reed-Muller codes'',
{\em IEEE Trans. Inf. Theory} {\bf 45}, ?, 1701-1703 (1999);
quant-ph/9608026.

\item {\bf [Steane-van Dam 00]}:
A. M. Steane, \& W. van Dam,
``Physicists triumph at `Guess my number'\,'',
{\em Phys. Today} {\bf 53}, 2, 35-39 (2000).

\item {\bf [Steane-Roos-Stevens-(+4) 00]}:
A. M. Steane, C. F. Roos, D. Stevens, A. Mundt,
D. Leibfried, F. Schmidt-Kaler, \& R. Blatt,
``Speed of ion-trap quantum-information processors'',
{\em Phys. Rev. A} {\bf 62}, 4, 042305 (2000);
quant-ph/0003087.

\item {\bf [Steane 00]}:
A. M. Steane,
``A quantum computer only needs one universe'',
quant-ph/0003084.

\item {\bf [Steane-Lucas 00]}:
A. M. Steane, \& D. M. Lucas,
``Quantum computing with trapped ions, atoms and light'',
{\em Fortschr. Phys.} {\bf 48}, 9-11 (Special issue:
Experimental proposals for quantum computation), 839-858 (2000);
quant-ph/0004053.

\item {\bf [Steane 01]}:
A. M. Steane,
``Quantum computing and error correction'',
in A. Gonis, \& P. E. A. Turchi (eds.),
{\em Decoherence and its implications in quantum
computation and information transfer},
IOS Press, Amsterdam, 2001, pp.~284-298;
quant-ph/0304016.

\item {\bf [Steane 02 a]}:
A. M. Steane,
``Fast fault-tolerant filter for quantum codewords'',
quant-ph/0202036.

\item {\bf [Steane 02 b]}:
A. M. Steane,
``Quantum computer architecture for fast entropy extraction'',
quant-ph/0203047.

\item {\bf [Steane 02 c]}:
A. M. Steane,
``Overhead and noise threshold of fault-tolerant quantum error correction'',
quant-ph/0207119.

\item {\bf [Steane-Ibinson 03]}:
A. M. Steane, \& B. Ibinson,
``Fault-tolerant logical gate networks for CSS codes'',
quant-ph/0311014.

\item {\bf [Steck-Jacobs-Mabuchi-(+2) 04]}:
D. A. Steck, K. Jacobs, H. Mabuchi, T. Bhattacharya, \& S. Habib,
``Quantum feedback control of atomic motion in an optical cavity'',
{\em Phys. Rev. Lett.} {\bf 92}, 22, 223004 (2004).

\item {\bf [Steeb-Hardy 04]}:
W.-H. Steeb, \& Y. Hardy,
{\em Problems and solutions in quantum computing and quantum information},
World Scientific, Singapore, 2004.

\item {\bf [Stefanov-Gisin-Guinnard-(+2) 00]}:
A. Stefanov, N. Gisin, O. Guinnard, L. Guinnard, \& H. Zbinden,
``Optical quantum random number generator'',
{\em J. Mod. Opt.} {\bf 47}, 4, 595-598 (2000);
quant-ph/9907006.

\item {\bf [Stefanov-Zbinden-Gisin-Suarez 02]}:
A. Stefanov, H. Zbinden, N. Gisin, \& A. Suarez,
``Quantum correlations with spacelike separated beam splitters in motion:
Experimental test of multisimultaneity'',
{\em Phys. Rev. Lett.} {\bf 88}, 12, 120404 (2002);
quant-ph/0110117.
Comment: {\bf [Costa de Beauregard 01]}.

\item {\bf [Stefanov-Zbinden-Gisin-Suarez 03]}:
A. Stefanov, H. Zbinden, N. Gisin, \& A. Suarez,
``Quantum entanglement with acousto-optic modulators: Two-photon beats and
Bell experiments with moving beam splitters'',
{\em Phys. Rev. A} {\bf 67}, 4, 042115 (2003).

\item {\bf [Stefanovich 02]}:
E. V. Stefanovich,
``Is Minkowski space-time compatible
with quantum mechanics?'',
{\em Found. Phys.} {\bf 32}, 5, 673-703 (2002).

\item {\bf [Steffen-van Dam-Hogg-(+2) 03]}:
M. Steffen, W. van Dam, T. Hogg, G. Breyta. \& I. L. Chuang,
``Experimental implementation of an adiabatic quantum optimization
algorithm'',
{\em Phys. Rev. Lett.} {\bf 90}, 6, 067903 (2003);
quant-ph/0302057.

\item {\bf [Steffen-Martinis-Chuang 03]}:
M. Steffen, J. M. Martinis, \& I. L. Chuang,
``Accurate control of Josephson phase qubits'',
{\em Phys. Rev. B} {\bf 68}, 22, 224518 (2003).

\item {\bf [Steinbach-Gerry 98]}:
J. Steinbach, \& C. C. Gerry,
``Efficient scheme for the deterministic maximal
entanglement of $N$ trapped ions'',
{\em Phys. Rev. Lett.} {\bf 81}, 25, 5528-5531 (1998).

\item {\bf [Steinbach-Twamley 98]}:
J. Steinbach, \& J. Twamley,
``Motional quantum error correction'',
in V. Bu\v{z}zek, \& D. P. DiVincenzo (eds.),
{\em J. Mod. Opt.} {\bf 47}, 2-3 (Special issue:
Physics of quantum information), 453-485 (2000);
quant-ph/9811011.

\item {\bf [Steinberg-Kwiat-Chiao 92]}:
A. M. Steinberg, P. G. Kwiat, \& R. Y. Chiao,
``Dispersion cancellation and high-resolution time measurements
in a fourth-order optical interferometer'',
{\em Phys. Rev. A} {\bf 45}, 9, 6659-6665 (1992).

\item {\bf [Steinberg-Kwiat-Chiao 94]}:
A. M. Steinberg, P. G. Kwiat, \& R. Y. Chiao,
``Hidden and unhidden information in quantum tunneling'',
{\em Found. Phys. Lett.} {\bf 7}, 3, 223-239 (1994).

\item {\bf [Steinberg 97]}:
A. M. Steinberg,
``Can a falling tree make a noise in two
forests at the same time?'',
to be published in {\em Proc.\ Obsc. Unr. Conf.};
quant-ph/9710046.

\item {\bf [Steinberg 98]}:
A. M. Steinberg,
``Single-particle nonlocality and conditional measurements'',
{\em Found. Phys.} {\bf 28}, 3, 385-398 (1998).

\item {\bf [Steinberg 03]}:
A. M. Steinberg,
``Speakable and unspeakable, past and future'',
in J. D. Barrow, P. C. W. Davies, \& C. L. Harper, Jr. (eds.),
{\em Science and ultimate reality: Quantum theory, cosmology and complexity},
Cambridge University Press, Cambridge, 2003;
quant-ph/0302003.

\item {\bf [Steiner 00]}:
M. Steiner,
``Towards quantifying non-local information transfer:
Finite-bit non-locality'',
{\em Phys. Lett. A} {\bf 270}, 5, 239-244 (2000).

\item {\bf [Steiner-Rendell 00]}:
M. Steiner, \& R. W. Rendell,
``Entanglement of a double dot with a quantum point contact'',
{\em Phys. Rev. A} {\bf 63}, 5, 052304 (2001);
quant-ph/0009047.

\item {\bf [Steiner 03]}:
M. Steiner,
``Generalized robustness of entanglement'',
{\em Phys. Rev. A} {\bf 67}, 5, 054305 (2003);
quant-ph/0304009.

\item {\bf [Steinmann 96]}:
O. Steinmann,
``The EPR bingo'',
{\em Helv. Phys. Acta} {\bf 69}, 5-6, 702-705 (1996).

\item {\bf [Steinwandt-Janzing-Beth 01]}:
R. Steinwandt, D. Janzing, \& T. Beth,
``On using quantum protocols to detect traffic analysis'',
to appear in {\em Quant. Inform. and Comp.};
quant-ph/0106100.

\item {\bf [\v{S}telmachovi\v{c}-Bu\v{z}ek 01]}:
P. \v{S}telmachovi\v{c}, \& V. Bu\v{z}ek,
``Dynamics of open quantum systems initially entangled with environment:
Beyond the Kraus representation'',
{\em Phys. Rev. A} {\bf 64}, 6, 062106 (2001)];
quant-ph/0108136.
Comment: {\bf [Salgado-S\'{a}nchez G\'{o}mez 02 d]}.

\item {\bf [\v{S}telmachovi\v{c}-Bu\v{z}ek 03]}:
P. \v{S}telmachovi\v{c}, \& V. Bu\v{z}ek,
``Quantum information approach to the Ising model'',
{\em Phys. Rev. A};
quant-ph/0312154 (extended version with more appendices).

\item {\bf [Stenger 95]}:
V. J. Stenger,
{\em The unconscious quantum: Metaphysics
in modern physics and cosmology},
Prometheus Books, Amherst, New York, 1995.

\item {\bf [Stenholm 85]}:
S. Stenholm (ed.),
{\em Lasers in applied and
fundamental research}, Adam Hilger, Bristol, 1985.

\item {\bf [Stenholm 92]}:
S. Stenholm,
``Simultaneous measurements of conjugate variables'',
{\em Ann. Phys.} {\bf 218}, 2, 233-254 (1992).

\item {\bf [Stenholm-Wilkens 97]}:
S. Stenholm, \& M. Wilkens,
``Jumps in quantum theory'',
{\em Contemp. Phys.} {\bf 38}, 4, 257-268 (1997).

\item {\bf [Stenholm-Bardroff 98]}:
S. Stenholm, \& P. J. Bardroff,
``Teleportation of $N$-dimensional states'',
{\em Phys. Rev. A} {\bf 58}, 6, 4373-4376 (1998).

\item {\bf [Stenholm 00 a]}:
S. Stenholm,
``Observations and quantum information'',
in V. Bu\v{z}zek, \& D. P. DiVincenzo (eds.),
{\em J. Mod. Opt.} {\bf 47}, 2-3 (Special issue:
Physics of quantum information), 311-324 (2000).

\item {\bf [Stenholm 00 b]}:
S. Stenholm,
``From physics to philosophy'',
{\em Eur. J. Phys.} {\bf 21}, 2, 199 (2000).
Review of {\bf [Butterfield-Pagonis 00]}.

\item {\bf [Stenholm-Andersson 00]}:
S. Stenholm, \& E. Andersson,
``Shared access to quantum information'',
{\em Phys. Rev. A} {\bf 62}, 4, 044301 (2000).

\item {\bf [Stenholm 02]}:
S. Stenholm,
``Are there measurements?'',
in {\bf [Bertlmann-Zeilinger 02]}, pp.~185-198.

\item {\bf [Stepanenko-Bonesteel-DiVincenzo-(+2) 03]}:
D. Stepanenko, N. E. Bonesteel, D. P. DiVincenzo,
G. Burkard, \& D. Loss,
``Spin-orbit coupling and time-reversal symmetry in quantum gates'',
{\em Phys. Rev. B} {\bf 68}, 11, 115306 (2003).

\item {\bf [Stern-Aharonov-Imry 90]}:
A. Stern, Y. Aharonov, \& Y. Imry,
``Phase uncertainty and loss of interference: A general picture'',
{\em Phys. Rev. A} {\bf 41}, 7, 3436-3448 (1990).

\item {\bf [Steuernagel-Vaccaro 95]}:
O. Steuernagel, \& J. A. Vaccaro,
``Reconstructing the density operator via simple projectors'',
{\em Phys. Rev. Lett.} {\bf 75}, 18, 3201-3205 (1995).

\item {\bf [Steuernagel 99]}:
O. Steuernagel,
``Uncertainty is complementary to complementarity'',
quant-ph/9908011.

\item {\bf [Stevens-Brochard-Steane 98]}:
D. Stevens, J. Brochard, \& A. M. Steane,
``Simple experimental methods for trapped-ion quantum processors'',
{\em Phys. Rev. A} {\bf 58}, 4, 2750-2759 (1998);
quant-ph/9802059.

\item {\bf [Stevenson-Thompson-Shields-(+4) 02]}:
R. M. Stevenson, R. M. Thompson, A. J. Shields,
I. Farrer, B. E. Kardynal, D. A. Ritchie, \& M. Pepper,
``Quantum dots as a photon source for passive quantum key encoding'',
{\em Phys. Rev. B} {\bf 66}, 8, 081302 (2002).

\item {\bf [Stobi\'{n}ska-W\'{o}dkiewicz 04]}:
M. Stobi\'{n}ska, \& K. W\'{o}dkiewicz,
``Witnessing entanglement with second-order interference'',
quant-ph/0408047.

\item {\bf [Stocks-Redhead 96]}:
A. C. Stocks, \& M. L. G. Redhead,
``A value rule for non-maximal observables'',
{\em Found. Phys. Lett.} {\bf 9}, 2, 109-119 (1996).

\item {\bf [Stockton-Armen-Mabuchi 02]}:
J. K. Stockton, M. Armen, \& H. Mabuchi,
``Programmable logic devices in experimental quantum optics'',
{\em J. Opt. Soc. Am. B} {\bf 19}, ?, 3019-? (2002).

\item {\bf [Stockton-Geremia-Doherty-Mabuchi 03]}:
J. K. Stockton, J. M. Geremia, A. C. Doherty, \& H. Mabuchi,
``Characterizing the entanglement of symmetric many-particle spin-$\frac{1}{2}$
systems'',
{\em Phys. Rev. A} {\bf 67}, 2, 022112 (2003);
quant-ph/0210117.

\item {\bf [Stoeltzner 02]}:
M. Stoeltzner,
``Bell, Bohm, and von Neumann: Some philosophical inequalities
concerning no-go theorems and the axiomatic method'',
in T. Placek, \& J. N. Butterfield (eds.),
{\em Modality, probability, and Bell's theorem},
Kluwer Academic, Dordrecht, Holland, 2002;
PITT-PHIL-SCI00000494.

\item {\bf [Stollsteimer-Mahler 01]}:
M. Stollsteimer, \& G. Mahler,
``Suppression of arbitrary internal coupling in a quantum register'',
{\em Phys. Rev. A} {\bf 64}, 5, 052301 (2001);
quant-ph/0107059.

\item {\bf [Stomphorst 01]}:
R. G. Stomphorst,
``Tunneling times and excited state interactions between
chromophores'',
Ph.\ D. thesis, Wageningen Universiteit, Holland, 2001.

\item {\bf [Stomphorst 02]}:
R. G. Stomphorst,
``Transmission and reflection in a double potential well:
Doing it the Bohmian way'',
{\em Phys. Lett. A} {\bf 292}, 4-5, 213-221 (2002);
quant-ph/0111045.

\item {\bf [Storcz-Wilhelm 03]}:
M. J. Storcz, \& F. K. Wilhelm,
``Decoherence and gate performance of coupled solid-state qubits'',
{\em Phys. Rev. A} {\bf 67}, 4, 042319 (2003).
Publisher's note:
{\em Phys. Rev. A} {\bf 67}, 5, 059901 (2003).

\item {\bf [Storey-Tan-Collett-Walls 94 a]}:
E. P. Storey, S. M. Tan. M. J. Collett, \& D. F. Walls,
``Path detection and the uncertainty principle'',
{\em Nature} {\bf 367}, 6464, 626-628 (1994).
See {\bf [Englert-Fearn-Scully-Walther 94]},
{\bf [Englert-Scully-Walther 95]},
{\bf [Storey-Tan-Collett-Walls 94 b, 95]},
{\bf [Wiseman-Harrison 95]}.

\item {\bf [Storey-Tan-Collett-Walls 94 b]}:
E. P. Storey, S. M. Tan. M. J. Collett, \& D. F. Walls,
``?'',
in F. De Martini, G. Denardo, \& A. Zeilinger,
{\em Quantum interferometry},
World Scientific, Singapore, 1994, pp.~120-129.
See {\bf [Englert-Fearn-Scully-Walther 94]},
{\bf [Storey-Tan-Collett-Walls 94 a, 95]},
{\bf [Englert-Scully-Walther 95]},
{\bf [Wiseman-Harrison 95]}.

\item {\bf [Storey-Tan-Collett-Walls 95]}:
E. P. Storey, S. M. Tan. M. J. Collett, \& D. F. Walls,
``Complementarity and uncertainty'',
{\em Nature} {\bf 375}, 6530, 367-368 (1995).
See {\bf [Englert-Fearn-Scully-Walther 94]},
{\bf [Englert-Scully-Walther 95]},
{\bf [Storey-Tan-Collett-Walls 94 a, b]},
{\bf [Wiseman-Harrison 95]}.

\item {\bf [Strassner-Witte 00]}:
J. Strassner, \& C. Witte,
``Entanglement optimization for pairs of qubits'',
quant-ph/0012126.

\item {\bf [Strauch-Johnson-Dragt-(+3) 03]}:
F. W. Strauch, P. R. Johnson, A. J. Dragt, C. J. Lobb,
J. R. Anderson, \& F. C. Wellstood,
``Quantum logic gates for coupled superconducting phase qubits'',
{\em Phys. Rev. Lett.} {\bf 91}, 16, 167005 (2003);
quant-ph/0303002.

\item {\bf [Streater-Wightman 64]}:
R. F. Streater, \& A. S. Wightman,
{\em PCT, spin and statistics, and all that}, Benjamin, New York, 1964.

\item {\bf [Strekalov-Pittman-Sergienko-(+2) 96]}:
D. V. Strekalov, T. P. Pittman, A. V. Sergienko, Y. H. Shih, \& P. G. Kwiat,
``Postselection-free energy-time entanglement'',
{\em Phys. Rev. A} {\bf 54}, 1, R1-R4 (1996).

\item {\bf [Strekalov-Pittman-Shih 98]}:
D. V. Strekalov, T. B. Pittman, \& Y. H. Shih,
``What we can learn about single photons in a
two-photon interference experiment'',
{\em Phys. Rev. A} {\bf 57}, 1, 567-570 (1998).

\item {\bf [Strekalov-Kim-Shih 99]}:
D. V. Strekalov, Y. Kim, \& Y. H. Shih,
``Experimental study of a subsystem in an entangled two-photon state'',
{\em Phys. Rev. A} {\bf 60}, 4, 2685-2688 (1999);
quant-ph/9811060.

\item {\bf [Stroke 95]}:
H. H. Stroke (ed.),
{\em The Physical Review: The first
hundred years. A selection of seminal papers and comentaries},
American Institute of Physics Press, Woodbury, New York, 1995.

\item {\bf [Strunz-Di\'{o}si-Gisin-Yu 99]}:
W. T. Strunz, L. Di\'{o}si, N. Gisin, \& T. Yu,
``Quantum trajectories for Brownian motion'',
{\em Phys. Rev. Lett.} {\bf 83}, 24, 4909-4913 (1999).

\item {\bf [Strunz-Haake-Braun 02]}:
W. T. Strunz, F. Haake, \& D. Braun,
``Unversality of decoherence in the macroworld'',
quant-ph/0204129.

\item {\bf [Struyve-De Baere 02]}:
W. Struyve, \& W. De Baere,
``Comments on some recently proposed experiments that should
distinguish Bohmian mechanics from quantum mechanics'',
in A. Khrennikov (ed.),
{\em Quantum Theory: Reconsideration of Foundations (V\"{a}xj\"{o}, Sweden, 2001)},
V\"{a}xj\"{o} University Press, V\"{a}xj\"{o}, Sweden, 2002;
quant-ph/0108038.

\item {\bf [Struyve-De Baere-De Neve-De Weirdt 03]}:
W. Struyve, W. De Baere, J. De Neve, \& S. De Weirdt,
``Comment on `Bohmian prediction about a two double-slit
experiment and its disagreement with standard quantum mechanics'\,'',
{\em J. Phys. A} {\bf 36}, 5, 1525-1530 (2003).
Comment on {\bf [Golshani-Akhavan 01 a]}.

\item {\bf [Struyve-De Baere-De Neve-De Weirdt 04]}:
W. Struyve, W. De Baere, J. De Neve, \& S. De Weirdt,
`On Peres' statement ``opposite momenta lead to opposite directions'',
decaying systems and optical imaging',
{\em Found. Phys.};
quant-ph/0401095.
See {\bf [Peres 00 f]}.

\item {\bf [Stuart 91]}:
C. I. J. M. Stuart,
``Inconsistency of the Copenhagen interpretation'',
{\em Found. Phys.} {\bf 21}, 5, 591-622 (1991).

\item {\bf [Stucki-Ribordy-Stefanov-(+3) 01]}:
D. Stucki, G. Ribordy, A. Stefanov, H. Zbinden, J. G.
Rarity, \& T. Walls,
``Photon counting for quantum key distribution
with Peltier cooled InGaAs/InP APD's'',
{\em J. Mod. Opt.} {\bf 48}, ?, 1967-1981 (2001);
quant-ph/0106007.

\item {\bf [Stucki-Gisin-Guinnard-(+2) 02]}:
D. Stucki, N. Gisin, O. Guinnard, G. Ribordy, \&, H. Zbinden,
``Quantum key distribution over 67 km with a plug \& play system'',
{\em New J. Phys} {\bf 4}, 41.1-41.8 (2002);
quant-ph/0203118.

\item {\bf [Stulpe-Swat 01]}:
W. Stulpe, \& M. Swat,
``Quantum states as probability measures'',
{\em Found. Phys. Lett.} {\bf 14}, 3, 285-293 (2001).

\item {\bf [Sturzu 01]}:
I. Sturzu,
``State transformations after quantum fuzzy measurements'',
{\em Phys. Rev. A} {\bf 64}, 5, 054101 (2001);
quant-ph/0204110.

\item {\bf [Sturzu 02]}:
I. Sturzu,
``Topics on the stochastical treatment of an open quantum system'',
quant-ph/0204014.

\item {\bf [Styer 96]}:
D. F. Styer,
``Common misconceptions regarding quantum mechanics'',
{\em Am. J. Phys.} {\bf 64}, 1, 31-34 (1996).
Erratum: {\em Am. J. Phys.} {\bf 64}, 9, 1202 (1996).

\item {\bf [Styer 00 a]}:
D. F. Styer,
{\em The strange world of quantum mechanics},
Cambridge University Press, Cambridge, 2000.
Review: {\bf [Holland 00]}, {\em [Jagannathan 02]}.

\item {\bf [Styer 00 b]}:
D. Styer,
``Quantum theory -- Interpretation, formulation, inspiration'',
{\em Phys. Today} {\bf 53}, 9, ? (2000).
Comment on {\bf [Fuchs-Peres 00 a]}.
Reply: {\bf [Fuchs-Peres 00 b]}.

\item {\bf [Styer-Balkin-Becker-(+10) 02]}:
D. F. Styer, M. S. Balkin, K. M. Becker,
M. R. Burns, C. E. Dudley, S. T. Forth,
J. S. Gaumer, M. A. Kramer, D. C. Oertel,
L. H. Park, M. T. Rinkoski, C. T. Smith, \& T. D. Wotherspoon,
``Nine formulations of quantum mechanics'',
{\em Am. J. Phys.} {\bf 70}, 3, 288-297 (2002).
Comment: {\bf [Tong 02]}.

\item {\bf [Suarez-Scarani 97]}:
A. Suarez, \& V. Scarani,
``Does entanglement depend on the timing of the impacts at the beam-splitters'',
{\em Phys. Lett. A} {\bf 232}, 1-2, 9-14 (1997);
quant-ph/9704038.

\item {\bf [Suarez 97]}:
A. Suarez,
``Relativistic nonlocality in an experiment with 2 non-before impacts'',
{\em Phys. Lett. A} {\bf 236}, 5-6, 383-390 (1997);
quant-ph/9712049.

\item {\bf [Suarez 98]}:
A. Suarez,
``Quantum mechanics versus multisimultaneity
in interferometer-series experiments'',
{\em Phys. Lett. A} {\bf 250}, 1-3, 39-47 (1998);
quant-ph/9812009.

\item {\bf [Suarez 00 a]}:
A. Suarez,
``Quantum mechanics versus multisimultaneity in experiments
with acousto-optic choice-devices'',
{\em Phys. Lett. A} {\bf 269}, 5-6, 293-302 (2000).
See {\bf [Suarez 00 b]}.

\item {\bf [Suarez 00 b]}:
A. Suarez,
``Preferred frame versus multisimultaneity:
Meaning and relevance of a forthcoming experiment'',
quant-ph/0006053.
See {\bf [Suarez 00 a]}.

\item {\bf [Suarez 01 a]}:
A. Suarez,
``Imbedding nonlocality in a relativistic chronology'',
quant-ph/0108045.

\item {\bf [Suarez 01 b]}:
A. Suarez,
``Is there a real time ordering behind the nonlocal correlations?'',
quant-ph/0110124.

\item {\bf [Suarez 02]}:
A. Suarez,
``Bell's theorem and Bohr's principle that the measurement must be classical'',
quant-ph/0202134.

\item {\bf [Suarez 03]}:
A. Suarez,
``Entanglement and time'',
quant-ph/0311004.

\item {\bf [Subrahmanyam 04 a]}:
V. Subrahmanyam,
``Quantum entanglement in Heisenberg antiferromagnets'',
{\em Phys. Rev. A} {\bf 69}, 2, 022311 (2004).

\item {\bf [Subrahmanyam 04 b]}:
V. Subrahmanyam,
``Entanglement dynamics and quantum-state transport in spin chains'',
{\em Phys. Rev. A} {\bf 69}, 3, 034304 (2004).

\item {\bf [Sudarshan 01]}:
E. C. G. Sudarshan,
``Quantum theory and probabilities'',
quant-ph/0109159.

\item {\bf [Sudbery 93]}:
A. Sudbery,
``Instant teleportation'',
{\em Nature} {\bf 362}, 6421, 586-587 (1993).

\item {\bf [Sudbery 94]}:
A. Sudbery,
``?'',
{\em Phys. World} {\bf 7}, 4, ?-?, (1994).
Review of {\bf [Peres 93 a]}.

\item {\bf [Sudbery 95]}:
A. Sudbery,
``Constructing reality'',
{\em Nature} {\bf 375}, 6533, 644 (1995).
Review of {\bf [Gribbin 95]}.

\item {\bf [Sudbery 97]}:
A. Sudbery,
``The fastest way from A to B'',
{\em Nature} {\bf 390}, 6660, 551-552 (1997).
See {\bf [Bouwmeester-Pan-Mattle-(+3) 97]}.

\item {\bf [Sudbery 00 a]}:
A. Sudbery,
``Diese verdammte quantenspringerei'',
quant-ph/0011082.

\item {\bf [Sudbery 00 b]}:
A. Sudbery,
``Why am I me? and why is my world so classical?'',
quant-ph/0011084.

\item {\bf [Sudbery 01]}:
A. Sudbery,
``On local invariants of pure three-qubit states'',
{\em J. Phys. A} {\bf 34}, 3, 643-652 (2001).

\item {\bf [Sudbery-Szulc 04]}:
A. Sudbery, \& J. Szulc,
``Compatibility of subsystem states'',
quant-ph/0407227.

\item {\bf [Sulcs-Oppy-Gilbert 00]}:
S. Sulcs, G. Oppy, \& B. C. Gilbert,
``The rejection of local realism is premature'',
{\em Found. Phys. Lett.} {\bf 13}, 6, 521-541 (2000).

\item {\bf [Sullivan-Citrin 04]}:
D. M. Sullivan, \& D. S. Citrin,
``Time-domain simulation of a universal quantum gate'',
{\em J. Appl. Phys.} {\bf 96}, 1540-1546 (2004).

\item {\bf [Summers-Werner 87 a]}:
S. Summers, \& R. Werner,
``Bell's inequalities and quantum field theory. I. General setting'',
{\em J. Math. Phys.} {\bf 28}, 10, 2440-2447 (1987).
See {\bf [Summers-Werner 87 b]} (II).

\item {\bf [Summers-Werner 87 b]}:
S. Summers, \& R. Werner,
``Bell's inequalities and quantum field theory. II.
Bell's inequalities are maximally violated in the vacuum'',
{\em J. Math. Phys.} {\bf 28}, 10, 2448-2456 (1987).
See {\bf [Summers-Werner 87 a]} (I).

\item {\bf [Summhammer-Badurek-Rauch-Kischko 82]}:
J. Summhammer, G. Badurek, H. Rauch, \& U. Kischko,
``Explicit experimental verification of quantum spin-state superposition'',
{\em Phys. Lett. A} {\bf 90}, 3, 110-112 (1982).
See {\bf [Badurek-Rauch-Summhammer 83]},
{\bf [Badurek-Rauch-Tuppinger 86]}.

\item {\bf [Summhammer-Rauch-Tuppinger 87]}:
J. Summhammer, H. Rauch, \& D. Tuppinger,
``Stochastic and deterministic absorption in neutron-interference experiments'',
{\em Phys. Rev. A} {\bf 36}, 9, 4447-4455 (1987).

\item {\bf [Summhammer 97]}:
J. Summhammer,
``Factoring and Fourier transformation with a Mach-Zehnder interferometer'',
{\em Phys. Rev. A} {\bf 56}, 5, 4324-4326 (1997);
quant-ph/9708047.
See {\bf [Clauser-Dowling 96]}.

\item {\bf [Summhammer 99]}:
J. Summhammer,
``Maximum predictive power and the superposition principle'',
quant-ph/9910039.

\item {\bf [Summhammer 01]}:
J. Summhammer,
``Structure of probabilistic information and quantum laws'',
in {\em Foundations of Probability and Physics (V\"{a}xj\"{o}, Sweden, 2000)};
quant-ph/0102099.

\item {\bf [Sun-Mazurenko-Fainman 95]}:
P. C. Sun, Y. Mazurenko, \& Y. Fainman,
``Long-distance frequency-division interferometer
for communication and quantum cryptography'',
{\em Opt. Lett.} {\bf 20}, 1062-1063 (1995).

\item {\bf [Sun-Zhan-Liu 98]}:
C. P. Sun, H. Zhan, \& X. F. Liu,
``Decoherence and relevant universality in quantum algorithms via a
dynamic theory for quantum measurement'',
{\em Phys. Rev. A} {\bf 58}, 3, 1810-1821 (1998).

\item {\bf [Sun-Zhou-Yu-Liu 01]}:
C. P. Sun, D. L. Zhou, S. X. Yu, \& X. F. Liu,
``Quantum decoherence from adiabatic entanglement
with external one or a few degrees of freedom'',
{\em Eur. Phys. J. D} {\bf 13}, 1, 145-155 (2001).

\item {\bf [Sun-Yi-You 03]}:
C. P. Sun, S. Yi, \& L. You,
``Decoherence of collective atomic spin states due to inhomogeneous coupling'',
{\em Phys. Rev. A} {\bf 67}, 6, 063815 (2003).

\item {\bf [Sun-Li-Liu 03]}:
C. P. Sun, Y. Li, \& X. F. Liu,
``Quasi-spin-wave quantum memories with a dynamical symmetry'',
{\em Phys. Rev. Lett.} {\bf 91}, 14, 147903 (2003).

\item {\bf [Sun-Zhang-Feng-Ying 02]}:
X. Sun, S. Zhang, Y. Feng, \& M. Ying,
``Mathematical nature of and a family of lower bounds for the success
probability of unambiguous discrimination'',
{\em Phys. Rev. A} {\bf 65}, 4, 044306 (2002).

\item {\bf [Sun-Hillery-Bergou 01]}:
Y. Sun, M. Hillery, \& J. A. Bergou,
``Optimum unambiguous discrimination between linearly independent
nonorthogonal quantum states and its optical realization'',
{\em Phys. Rev. A} {\bf 64}, 2, 022311 (2001);
quant-ph/0012131.

\item {\bf [Sun-Bergou-Hillery 02]}:
Y. Sun, J. A. Bergou, \& M. Hillery,
``Optimum unambiguous discrimination between subsets
of nonorthogonal quantum states'',
{\em Phys. Rev. A} {\bf 66}, 3, 032315 (2002);
quant-ph/0112051.

\item {\bf [Sun-Chen-Chen 03]}:
Y. Sun, Y. Chen, \& H. Chen,
``Thermal entanglement in the two-qubit Heisenberg $XY$ model under a nonuniform
external magnetic field'',
{\em Phys. Rev. A} {\bf 68}, 4, 044301 (2003).

\item {\bf [Suppes 76]}:
P. Suppes (ed.),
{\em Logic and probability in quantum mechanics},
Reidel, Dordrecht, Holland, 1976.

\item {\bf [Suppes 80]}:
P. Suppes (ed.),
{\em Studies in the foundations of quantum mechanics},
East Lansing, Michigan, 1980.

\item {\bf [Suppes-Zanotti 91]}:
P. Suppes, \& M. Zanotti,
``New Bell-type
inequalities for $N > 4$ necessary for existence of a hidden variable'',
{\em Found. Phys. Lett.} {\bf 4}, 1, 101-107 (1991).

\item {\bf [Suter-Lim 02]}:
D. Suter, \& K. Lim,
``Scalable architecture for spin-based quantum computers with a single type of
gate'',
{\em Phys. Rev. A} {\bf 65}, 5, 052309 (2002).

\item {\bf [Sutherland 00]}:
R. I. Sutherland,
``A suggestive way of deriving the quantum probability rule'',
{\em Found. Phys. Lett.} {\bf 13}, 4, 379-386 (2000).

\item {\bf [Svetlichny 87]}:
G. Svetlichny,
``Distinguishing three-body from two-body nonseparability by a Bell-type inequality'',
{\em Phys. Rev. D} {\bf 35}, 10, 3066-3069 (1987).

\item {\bf [Svetlichny-Redhead-Brown-Butterfield 88]}:
G. Svetlichny, M. L. G. Redhead, H. Brown, \& J. N. Butterfield,
``Do the Bell inequalities require the existence
of joint probability distributions?'',
{\em Philos. Sci.} {\bf 55}, 2, 387-401 (1988).

\item {\bf [Svetlichny 90]}:
G. Svetlichny,
``On the inverse of the EPR problem: Quantum is classical'',
{\em Found. Phys.} {\bf 20}, 6, 635-650 (1990).

\item {\bf [Svetlichny 98]}:
G. Svetlichny,
``Quantum formalism with state-collapse and superluminal communication'',
{\em Found. Phys.} {\bf 28}, 2, 131-155 (1998).

\item {\bf [Svetlichny 00]}:
G. Svetlichny,
``The space-time origin of quantum mechanics: Covering law'',
{\em Found. Phys.} {\bf 30}, 11, 1819-1847 (2001).

\item {\bf [Svetlichny 03 a]}:
G. Svetlichny,
``Causality implies formal state collapse'',
{\em Found. Phys.} {\bf 33}, 4, 641-655 (2003);
quant-ph/0207180.

\item {\bf [Svetlichny 03 b]}:
G. Svetlichny,
``Non-linear quantum mechanics and high energy cosmic rays'',
hep-th/0305100.

\item {\bf [Svetlichny 04 a]}:
G. Svetlichny,
``On linearity of separating multi-particle differential Schr\"{o}dinger
operators for identical particles'',
{\em J. Math. Phys.};
quant-ph/0308001.

\item {\bf [Svetlichny 04 b]}:
G. Svetlichny,
``Informal resource letter - Nonlinear quantum mechanics on arXiv up to
August 2004'',
quant-ph/0410036.

\item {\bf [Svetlichny 04 c]}:
G. Svetlichny,
``Amplification of acausal effects in nonlinear quantum mechanics under
localization'',
quant-ph/0410186.

\item {\bf [Svetlichny 04 d]}:
G. Svetlichny,
``Nonlinear quantum mechanics at the Planck scale'',
quant-ph/0410230.

\item {\bf [Svore-Terhal-DiVincenzo 04]}:
K. M. Svore, B. M. Terhal, \& D. P. DiVincenzo,
``Local fault-tolerant quantum computation'',
quant-ph/0410047.

\item {\bf [Svozil 94]}:
K. Svozil,
{\em Randomness and undecidability in physics},
World Scientific, Singapore, 1994.

\item {\bf [Svozil 95]}:
K. Svozil,
``Consistent use of paradoxes in deriving
constraints on the dynamics of physical systems and of no-go theorems'',
preprint, 1995.

\item {\bf [Svozil-Tkadlec 96]}:
K. Svozil, \& J. Tkadlec,
``Greechie diagrams, nonexistence of measures in quantum logics,
and Kochen-Specker-type constructions'',
{\em J. Math. Phys.} {\bf 37}, 11, 5380-5401 (1996).

\item {\bf [Svozil 96]}:
K. Svozil,
``Experimental realization of any discrete operator'',
quant-ph/9612010.

\item {\bf [Svozil 98 a]}:
K. Svozil,
``Analogues of quantum complementarity in the theory of automata'',
{\em Stud. Hist. Philos. Sci. Part B: Stud. Hist. Philos. Mod. Phys.}
{\bf 29}, 1, 61-80 (1998).

\item {\bf [Svozil 98 b]}:
K. Svozil,
{\em Quantum logic},
Springer-Verlag, Singapore, 1998.

\item {\bf [Svozil 99]}:
K. Svozil,
``Testing quantum contextuality'',
quant-ph/9907015.

\item {\bf [Svozil 00 a]}:
K. Svozil,
``Quantum interfaces'',
quant-ph/0001064.

\item {\bf [Svozil 00 b]}:
K. Svozil,
``The information interpretation of quantum mechanics'',
quant-ph/0006033.

\item {\bf [Svozil 00 c]}:
K. Svozil,
``On possible extensions of quantum mechanics'',
quant-ph/0012066.

\item {\bf [Svozil 01 a]}:
K. Svozil,
``What is wrong with SLASH?'',
quant-ph/0103166.

\item {\bf [Svozil 01 b]}:
K. Svozil,
``Irreducibility of $n$-ary quantum information'',
quant-ph/0111113.

\item {\bf [Svozil 02 a]}:
K. Svozil,
``Conventions in relativity theory and quantum mechanics'',
{\em Found. Phys.} {\bf 32}, 4, 479-502 (2002);
quant-ph/0110054.

\item {\bf [Svozil 02 b]}:
K. Svozil,
``Quantum information in base $n$ defined by state partitions'',
{\em Phys. Rev. A} {\bf 66}, 4, 044306 (2002).
Erratum: {\em Phys. Rev. A} {\bf 68}, 5, 059901 (2003);
quant-ph/0205031.

\item {\bf [Svozil 04 a]}:
K. Svozil,
``Eutactic quantum codes'',
{\em Phys. Rev. A} {\bf 69}, 3, 034303 (2004);
quant-ph/0304056.

\item {\bf [Svozil 04 b]}:
K. Svozil,
``Context translation and quantum information via state partitions'',
{\em J. Mod. Opt.} {\bf 15}, ? ({\em Garda Workshop 2003}), 811-819 (2004);
quant-ph/0308110.

\item {\bf [Svozil 04 c]}:
K. Svozil,
``Farewell to quantum contextuality?'',
quant-ph/0401112.

\item {\bf [Svozil 04 d]}:
K. Svozil,
``Single particle interferometric analogues of multipartite entanglement'',
quant-ph/0401113.

\item {\bf [Svozil 04 e]}:
K. Svozil,
``Against contextuality, for context translation'',
quant-ph/0406014.

\item {\bf [Swift-Wright 80]}:
A. R. Swift, \& R. Wright,
``Generalized Stern-Gerlach experiments and the observability of
arbitrary spin operators'',
{\em J. Math. Phys.} {\bf 21}, 1, 77-82 (1980).

\item {\bf [Sylju\aa sen 03]}:
O. F. Sylju\aa sen,
``Entanglement and spontaneous symmetry breaking in quantum spin models'',
{\em Phys. Rev. A} {\bf 68}, 6, 060301 (2003).

\item {\bf [Synak-Horodecki 04]}:
B. Synak, \& M. Horodecki,
``Classical information deficit and monotonicity on local operations'',
quant-ph/0403167.

\item {\bf [Synak-Horodecki 04]}:
B. Synak, K. Horodecki, \& M. Horodecki,
``Bounds on localisable information via semidefinite programming'',
quant-ph/0405149.

\item {\bf [Sypher-Brereton-Wiseman-(+2) 02]}:
D. R. Sypher, I. M. Brereton, H. M. Wiseman,
B. L. Hollis, \& B. C. Travaglione,
``Read-only-memory-based quantum computation: Experimental explorations using
nuclear magnetic resonance and future prospects'',
{\em Phys. Rev. A} {\bf 66}, 1, 012306 (2002);
quant-ph/0112127.

\item {\bf [Szab\'{o} 98 a]}:
L. E. Szabo,
``Quantum structures do not exist in reality'',
{\em Int. J. Theor. Phys.} {\bf 37}, 1, 449-456 (1998).

\item {\bf [Szab\'{o} 98 b]}:
L. E. Szab\'{o},
``Complete resolution of the EPR-Bell paradox'',
quant-ph/9806074.

\item {\bf [Szab\'{o} 00 a]}:
L. E. Szab\'{o},
``On Fine's resolution of the EPR-Bell problem'',
{\em Found. Phys.} {\bf 30}, 11, 1891-1909 (2000);
quant-ph/0002030.

\item {\bf [Szab\'{o} 00 b]}:
L. E. Szab\'{o},
``A continuum local hidden variable model for the EPR
experiment'',
quant-ph/0012042.

\item {\bf [Szab\'{o} 01]}:
L. E. Szab\'{o},
``Critical reflections on quantum probability theory'',
in M. R\'{e}dei, \& M. St\"{o}ltzner (eds.),
{\em John von Neumann and the foundations of quantum physics},
Kluwer Academic, Dordrecht, Holland, 2001, pp.~?-?.

\item {\bf [Szab\'{o}-Fine 02]}:
L. E. Szab\'{o}, \& A. Fine,
``A local hidden variable theory for the GHZ experiment'',
{\em Phys. Lett. A} {\bf 295}, 5-6, 229-240 (2002).
Erratum: {\em Phys. Lett. A} {\bf 302}, 5-6, 345 (2002);
quant-ph/0007102.


\newpage

\subsection{}


\item {\bf [Tahan-Friesen-Joynt 02]}:
C. Tahan, M. Friesen, \& R. Joynt,
``Decoherence of electron spin qubits in Si-based quantum computers'',
{\em Phys. Rev. B} {\bf 66}, 3, 035314 (2002).

\item {\bf [Takabayashi 54]}:
T. Takabayashi,
``The formulation of quantum mechanics in terms of ensemble in phase space'',
{\em Prog. Theor. Phys. (Japan)} {\bf 11}, 4-5, 341-373 (1954).

\item {\bf [Takahashi-Honda-Tanaka-(+3) 99]}:
Y. Takahashi, K. Honda, N. Tanaka, K. Toyoda, K. Ishikawa, \& T. Yabuzaki,
``Quantum nondemolition measurement of spin via
the paramagnetic Faraday rotation'',
{\em Phys. Rev. A} {\bf 60}, 6, 4974-4979 (1999).

\item {\bf [Takeoka-Ban-Sasaki 02]}:
M. Takeoka, M. Ban, \& M. Sasaki,
``Quantum channel of continuous variable teleportation
and nonclassicality of quantum states'',
{\em J. Opt. B: Quantum Semiclass. Opt.} {\bf 4}, 2, 114-122 (2002);
quant-ph/0110031.

\item {\bf [Takeoka-Ban-Sasaki 03 a]}:
M. Takeoka, M. Ban, \& M. Sasaki,
``Practical scheme for optimal measurement in quantum interferometric devices'',
{\em Phys. Lett. A} {\bf 313}, 1-2, 16-20 (2003).

\item {\bf [Takeoka-Sasaki-Ban 03]}:
M. Takeoka, M. Sasaki, \& M. Ban,
``Continuous variable teleportation as a quantum channel'',
{\em Opt. Spectrosc.} {\bf 94}, 675 (2003).

\item {\bf [Takeoka-Ban-Sasaki 03 b]}:
M. Takeoka, M. Ban, \& M. Sasaki,
``Unambiguous quantum-state filtering'',
{\em Phys. Rev. A} {\bf 68}, 1, 012307 (2003).

\item {\bf [Takeoka-Fujiwara-Mizuno-Sasaki 04]}:
M. Takeoka, M. Fujiwara, J. Mizuno, \& M. Sasaki,
``Implementation of generalized quantum measurements: Superadditive quantum coding,
accessible information extraction, and classical capacity limit'',
{\em Phys. Rev. A} {\bf 69}, 5, 052329 (2004);
quant-ph/0306034.

\item {\bf [Takeoka-Sasaki-van Loock-L\"{u}tkenhaus 04]}:
M. Takeoka, M. Sasaki, P. van Loock, \& N. L\"{u}tkenhaus,
``Implementation of projective measurements with linear optics and
continuous photon counting'',
quant-ph/0410133.

\item {\bf [Takesue-Inoue 04]}:
H. Takesue, \& K. Inoue,
``Generation of polarization entangled photon pairs and violation of
Bell's inequality using spontaneous four-wave mixing in fiber loop'',
quant-ph/0408032.

\item {\bf [Takeuchi 00 a]}:
S. Takeuchi,
``Analysis of errors in linear-optics quantum computation'',
{\em Phys. Rev. A} {\bf 61}, 5, 052302 (2000).

\item {\bf [Takeuchi 00 b]}:
S. Takeuchi,
``Experimental demonstration of a three-qubit quantum computation
algorithm using a single photon and linear optics'',
{\em Phys. Rev. A} {\bf 62}, 3, 032301 (2000).

\item {\bf [Tamaki-Koashi-Imoto 03 a]}:
K. Tamaki, M. Koashi, \& N. Imoto,
``Security of the Bennett 1992 quantum-key distribution protocol against
individual attack over a realistic channel',
{\em Phys. Rev. A} {\bf 67}, 3, 032310 (2003);
quant-ph/0212161.

\item {\bf [Tamaki-Koashi-Imoto 03 b]}:
K. Tamaki, M. Koashi, \& N. Imoto,
``Unconditionally secure key distribution based on two nonorthogonal states'',
{\em Phys. Rev. Lett.} {\bf 90}, 16, 167904 (2003);
quant-ph/0212162.

\item {\bf [Tamaki-L\"{u}tkenhaus 04]}:
K. Tamaki, \& N. L\"{u}tkenhaus,
``Unconditional security of the Bennett 1992 quantum key-distribution protocol
over a lossy and noisy channel'',
{\em Phys. Rev. A} {\bf 69}, 3, 032316 (2004);
quant-ph/0308048.

\item {\bf [Tan-Jeffers-Barnett-Pegg 03]}:
E.-K. Tan, J. Jeffers, S. M. Barnett, \& D. T. Pegg,
``Retrodictive states and two-photon quantum imaging'',
{\em Eur. Phys. J. D} {\bf 22}, ?, 495-? (2003);
quant-ph/0303099.

\item {\bf [Tan-Walls-Collett 91]}:
S. M. Tan, D. F. Walls, \& M. J. Collett,
``Nonlocality of a single photon'',
{\em Phys. Rev. Lett.} {\bf 66}, 3, 252-255 (1991).
Comment: {\bf [Santos 92 a]}.

\item {\bf [Tan 99]}:
S. M. Tan,
``Confirming entanglement in continuous variable quantum teleportation'',
{\em Phys. Rev. A} {\bf 60}, 4, 2752-2758 (1999).

\item {\bf [Tanaka 02]}:
A. Tanaka,
``Semiclassical theory of weak values'',
{\em Phys. Lett. A} {\bf 297}, 5-6, 307-312 (2002);
quant-ph/0203149.

\item {\bf [Tanamoto 00 a]}:
T. Tanamoto,
``Quantum gates by coupled asymmetric quantum dots and
controlled-NOT-gate operation'',
{\em Phys. Rev. A} {\bf 61}, 2, 022305 (2000).

\item {\bf [Tanamoto 00 b]}:
T. Tanamoto,
``Quantum gates by coupled quantum dots and measurement procedure in field-effect-transistor structure'',
{\em Fortschr. Phys.} {\bf 48}, 9-11 (Special issue: Experimental proposals for quantum computation), 1005-1021 (2000).

\item {\bf [Tanamoto 01]}:
T. Tanamoto,
``One- and two-dimensional $N$-qubit systems in capacitively coupled quantum
dots'',
{\em Phys. Rev. A} {\bf 64}, 6, 062306 (2001);
quant-ph/0009030.

\item {\bf [Tanimura-Hayashi-Nakahara 04]}:
S. Tanimura, D. Hayashi, \& M. Nakahara,
``Exact solutions of holonomic quantum computation'',
{\em Phys. Lett. A} {\bf 325}, 3-4, 199-205 (2004).

\item {\bf [Tannous-Langlois 99]}:
C. Tannous, \& J. Langlois,
``Comment on: Breakdown of Bohr's correspondence principle by:
Bo Gao Phys. Rev. Lett. {\bf 83}, 4225 (1999)'',
quant-ph/0102042.
Comment on {\bf [Gao 99]}.

\item {\bf [Tanona 02]}:
S. Tanona,
``Idealization and formalism in Bohr's approach to quantum theory'' (2002),
PITT-PHIL-SCI00000880.

\item {\bf [Tanzilli-Tittel-De Riedmatten-(+5) 02]}:
S. Tanzilli, W. Tittel, H. De Riedmatten,
H. Zbinden, P. Baldi, M. De Micheli, D. B. Ostrowsky, \& N. Gisin,
``PPLN waveguide for quantum communication'',
{\em Eur. Phys. J. D} {\bf 18}, 2 (Special issue:
{\em Quantum interference and cryptographic keys:
Novel physics and advancing technologies (QUICK) (Corsica, 2001)}, 155-160 (2002);
quant-ph/0107125.

\item {\bf [Tapster-Rarity-Owens 94]}:
P. R. Tapster, J. G. Rarity, \& P. C. M. Owens,
``Violation of Bell's inequality over 4 km of optical fiber'',
{\em Phys. Rev. Lett.} {\bf 73}, 14, 1923-1926 (1994).

\item {\bf [Tarasov 01]}:
V. E. Tarasov,
``Open $n$-qubit system as a quantum computer with four-valued logic'',
quant-ph/0112023.

\item {\bf [Tarasov 02 a]}:
V. E. Tarasov,
``Quantum computer with mixed states and four-valued logic'',
{\em J. Phys. A} {\bf 35}, ?, 5207-5235 (2002);
quant-ph/0312131.

\item {\bf [Tarasov 02 b]}:
V. E. Tarasov,
``Quantum computation by quantum operations on mixed states'',
quant-ph/0201033.

\item {\bf [Tarozzi 93]}:
G. Tarozzi,
``Bell's theorem and the conflict between
the two basic principles of quantum mechanics'',
in A. van der Merwe, \& F. Selleri (eds.),
{\em Bell's theorem and the foundations of modern physics.
Proc.\ of an international
conference (Cesena, Italy, 1991)},
World Scientific, Singapore, 1993, pp.~448-457.

\item {\bf [Tarrach 97]}:
R. Tarrach,
``What when Gleason's theorem fails?'',
in M. Ferrero, \& A. van der Merwe (eds.),
{\em New developments on fundamental problems
in quantum physics (Oviedo, Spain, 1996)},
Kluwer Academic, Dordrecht, Holland, 1997, pp.~415-419.

\item {\bf [Tarrach-Vidal 99]}:
R. Tarrach, \& G. Vidal,
``Universality of optimal measurements'',
{\em Phys. Rev. A} {\bf 60}, 5, R3339-R3342 (1999);
quant-ph/9907098.

\item {\bf [Tarrach 00]}:
R. Tarrach,
``Una teor\'{\i}a determinante'',
{\em ABC-El Cultural}, 20 Dec. 2000, pp.~64-65.

\item {\bf [Tartaglia 98]}:
A. Tartaglia,
``Is the EPR paradox really a paradox?'',
{\em Eur. J. Phys.} {\bf 19}, 3, 307-311 (1998);
quant-ph/9805074.
Comment: {\bf [Omn\`{e}s 98]}.
See: {\bf [Chibeni 01]}.

\item {\bf [Taubes 96 a]}:
G. Taubes,
``Information science: All together for quantum computing'',
{\em Science} {\bf 273}, 5279, 1164 (1996).

\item {\bf [Taubes 96 b]}:
G. Taubes,
``Physics: Atomic mouse probes the lifetime of a quantum cat'',
{\em Science} {\bf 274}, 5293, 1615 (1996).

\item {\bf [Taubes 96 c]}:
G. Taubes,
``Quantum mechanics: To send data, physicists resort to quantum voodoo'',
{\em Science} {\bf 274}, 5287, 504-505 (1996).

\item {\bf [Taubes 97]}:
G. Taubes,
``Computing: Putting a quantum computer to work in a cup of coffee'',
{\em Science} {\bf 275}, 5298, 307-309 (1997).
See {\bf [Gershenfeld-Chuang 97]}.

\item {\bf [Tatara-Garc\'{\i}a 03]}:
G. Tatara, \& N. Garc\'{\i}a,
``Quantum toys for quantum computing: Persistent currents controlled by the
spin Josephson effect'',
{\em Phys. Rev. Lett.} {\bf 91}, 7, 076806 (2003).

\item {\bf [Taylor-Marcus-Lukin 03]}:
J. M. Taylor, C. M. Marcus, \& M. D. Lukin,
``Long-lived memory for mesoscopic quantum bits'',
{\em Phys. Rev. Lett.} {\bf 90}, 20, 206803 (2003).

\item {\bf [Taylor-Imamoglu-Lukin 03]}:
J. M. Taylor, A. Imamoglu, \& M. D. Lukin,
``Controlling a mesoscopic spin environment by quantum bit manipulation'',
{\em Phys. Rev. Lett.} {\bf 91}, 24, 246802 (2003).

\item {\bf [Taylor 01]}:
J. R. Taylor,
``Book review. Introduction to quantum mechanics'',
{\em Found. Phys.} {\bf 31}, 3, 561-563 (2001).
Review of {\bf [Griffiths 95]}.

\item {\bf [Tegmark 98]}:
M. Tegmark,
``The interpretation of quantum mechanics: Many worlds or many words?'',
{\em Fortschr. Phys.} {\bf 46}, 6-8, 855-862 (1998).

\item {\bf [Tegmark 00]}:
M. Tegmark,
``The importance of quantum decoherence in brain processes'',
{\em Phys. Rev. E} {\bf 61}, 4, 4194-4206 (2000);
quant-ph/9907009.

\item {\bf [Tegmark-Wheeler 01]}:
M. Tegmark, \& J. A. Wheeler,
``100 years of the quantum'',
{\em Sci. Am.} {\bf 284}, 2, 68-76 (2001);
quant-ph/0101077.
Spanish version: ``Cien a\~{n}os de misterios cu\'{a}nticos'',
{\em Investigaci\'{o}n y Ciencia} 295, 48-56 (2001).

\item {\bf [Tejada-Garg-Gider-(+3) 96]}:
J. Tejada, A. Garg, S. Gider,
D. D. Awschalom, D. P. DiVincenzo, \& D. Loss,
``Does macroscopic quantum coherence occur in ferritin?'',
{\em Science} {\bf 272}, 5260, 424-426 (1996).

\item {\bf [Tejero 02]}:
C. Tejero,
``In pursuit of intelectual beauty'',
{\em Spain Gourmetour. Food, Wine \& Travel Magazine}
55 (January-April 2002), pp.~96-100.
French version:
``A la recherche de la beaut\'{e} intellectuelle'',
{\em Spain Gourmetour. Revue de la gastronomie, des vins et du tourisme}
45 (Janvier-Avril 2002), pp.~96-100.
German version:
``Auf der Suche nach intellektueller sch\"{o}nheit'',
{\em Spain Gourmetour. Magazin f\"{u}r Essen, Wein und Reisen}
32 (Januar-April 2002), pp.~96-100.

\item {\bf [Teklemariam-Fortunato-Pravia-(+2) 01]}:
G. Teklemariam, E. M. Fortunato, M. A. Pravia,
T. F. Havel, \& D. G. Cory,
``NMR analog of the quantum disentanglement eraser'',
{\em Phys. Rev. Lett.} {\bf 86}, 26, 5845-5849 (2001);
quant-ph/0104131.

\item {\bf [Teklemariam-Fortunato-Pravia-(+5) 02]}:
G. Teklemariam, E. M. Fortunato, M. A. Pravia,
Y. Sharf, T. F. Havel, D. G. Cory, A. Bhattaharyya, \& J. Hou,
``Quantum erasers and probing classifications of entanglement via nuclear
magnetic resonance'',
{\em Phys. Rev. A} {\bf 66}, 1, 012309 (2002).

\item {\bf [Teklemariam-Fortunato-L\'{o}pez-(+4) 03]}:
G. Teklemariam, E. M. Fortunato, C. C. L\'{o}pez,
J. Emerson, J. P. Paz, T. F. Havel, \& D. G. Cory,
``Method for modeling decoherence on a quantum-information processor'',
{\em Phys. Rev. A} {\bf 67}, 6, 062316 (2003);
quant-ph/0303115.

\item {\bf [Telcordia Technologies 02]}:
{\em Telcordia Technologies},
www.telcordia.com.

\item {\bf [Telegdi 03]}:
V. L. Telegdi,
``No time to be brief: A scientific biography of Wolfgang Pauli'',
{\em Phys. Today} {\bf 56}, 11, ?-? (2003).

\item {\bf [Teodorescu Frumosu-Jaeger 03}:
M. Teodorescu-Frumosu, \& G. Jaeger,
``Quantum Lorentz-group invariants of $n$-qubit systems'',
{\em Phys. Rev. A} {\bf 67}, 5, 052305 (2003).

\item {\bf [Teplitsky-Gea Banacloche 03]}:
M. Teplitsky, \& J. Gea-Banacloche,
``Addendum: Extracting an entangled state of $n-t$ qubits from an $n$-qubit
entangled state after errors at $t$ sites'',
{\em Phys. Rev. A} {\bf 67}, 1, 014307 (2003).
Addendum to {\bf [Yang-Gea Banacloche 01 b]}.

\item {\bf [Terashima-Ueda 02]}:
H. Terashima, \& M. Ueda,
``Einstein-Podolsky-Rosen correlation seen from moving observers'',
quant-ph/0204138.

\item {\bf [Terashima-Ueda 04]}:
H. Terashima, \& M. Ueda,
``Einstein-Podolsky-Rosen correlation in a gravitational field'',
{\em Phys. Rev. A} {\bf 69}, 3, 032113 (2004);
quant-ph/0307114.

\item {\bf [Terhal-Smolin 97]}:
B. M. Terhal, \& J. A. Smolin,
``Superfast quantum
algorithms for coin weighing and binary search problems'',
quant-ph/9705041.

\item {\bf [Terhal-Smolin 98]}:
B. M. Terhal, \& J. A. Smolin,
``Single quantum querying of a database'',
{\em Phys. Rev. A} {\bf 58}, 3, 1822-1826 (1998).

\item {\bf [Terhal-Chuang-DiVincenzo-(+2) 99]}:
B. M. Terhal, I. L. Chuang, D. P. DiVincenzo, M. Grassl, \& J. A. Smolin,
``Simulating quantum operations with mixed environments'',
{\em Phys. Rev. A} {\bf 60}, 2, 881-885 (1999).

\item {\bf [Terhal 99]}:
B. M. Terhal,
``Quantum algorithms and quantum entanglement'',
Ph.\ D. thesis, University of Amsterdam, 1999.

\item {\bf [Terhal-DiVincenzo 00]}:
B. M. Terhal, \& D. P. DiVincenzo,
``Problem of equilibration and the computation of correlation
functions on a quantum computer'',
{\em Phys. Rev. A} {\bf 61}, 2, 022301 (2000);
quant-ph/9810063.

\item {\bf [Terhal-Horodecki 00]}:
B. M. Terhal, \& P. Horodecki,
``Schmidt number for density matrices'',
{\em Phys. Rev. A} {\bf 61}, 4, 040301(R) (2000);
quant-ph/9911117.

\item {\bf [Terhal 00]}:
B. M. Terhal,
``Bell inequalities and the separability criterion'',
{\em Phys. Lett. A} {\bf 271}, 5-6, 319-326 (2000);
quant-ph/9911057.

\item {\bf [Terhal-Vollbrecht 00]}:
B. M. Terhal, \& K. G. H. Vollbrecht,
``Entanglement of formation for isotropic states'',
{\em Phys. Rev. Lett.} {\bf 85}, 12, 2625-2628 (2000);
quant-ph/0005062.

\item {\bf [Terhal-DiVincenzo-Leung 01]}:
B. M. Terhal, D. P. DiVincenzo, \& D. W. Leung,
``Hiding bits in Bell states'',
{\em Phys. Rev. Lett.} {\bf 86}, 25, 5807-5810 (2001);
quant-ph/0011042.
See {\bf [DiVincenzo-Leung-Terhal 02]}.

\item {\bf [Terhal 01]}:
B. M. Terhal,
``A family of indecomposable positive linear maps based on
entangled quantum states'',
{\em Linear Algebra Appl.} {\bf 323}, ?, 61-73 (2001).

\item {\bf [Terhal 02]}:
B. M. Terhal,
``Detecting quantum entanglement'',
{\em Theoret. Comput. Sci.} {\bf 287}, 1, 313-335 (2002);
quant-ph/0101032.

\item {\bf [Terhal-DiVincenzo 02 a]}:
B. M. Terhal, \& D. P. DiVincenzo,
``Classical simulation of noninteracting-fermion quantum circuits'',
{\em Phys. Rev. A} {\bf 65}, 3, 032325 (2002);
quant-ph/0108010.

\item {\bf [Terhal-Horodecki-Leung-DiVincenzo 02]}:
B. M. Terhal, M. Horodecki, D. W. Leung, D. P. DiVincenzo,
``The entanglement of purification'',
{\em J. Math. Phys.} {\bf 43}, 9, 4286-4298 (2002);
quant-ph/0202044.

\item {\bf [Terhal-DiVincenzo 02 b]}:
B. M. Terhal, \& D. P. DiVincenzo,
``Adaptive quantum computation, constant depth quantum circuits and
Arthur-Merlin games'',
{\em Quant. Inf. Comp.};
quant-ph/0205133.

\item {\bf [Terhal-Wolf-Doherty 03]}:
B. M. Terhal, M. M. Wolf, \& A. C. Doherty,
``Quantum entanglement: A modern perspective'',
{\em Phys. Today} {\bf 56}, 4, ?-? (2003).

\item {\bf [Terhal-Doherty-Schwab 03]}:
B. M. Terhal, A. C. Doherty, \& D. Schwab,
``Symmetric extensions of quantum states and local hidden variable theories'',
{\em Phys. Rev. Lett.} {\bf 90}, 15, 157903 (2003).

\item {\bf [Terhal 03]}:
B. M. Terhal,
``Is entanglement monogamous?'',
{\em IBM J. Res. Dev.};
quant-ph/0307120.

\item {\bf [Terhal-Burkard 04]}:
B. M. Terhal, \& G. Burkard,
``Fault-tolerant quantum computation for local non-Markovian noise'',
quant-ph/0402104.

\item {\bf [Terno 99]}:
D. R. Terno,
``Nonlinear operations in quantum-information theory'',
{\em Phys. Rev. A} {\bf 59}, 5, 3320-3324 (1999);
quant-ph/9811036.

\item {\bf [Terno 01]}:
D. R. Terno,
`Comment on ``Counterfactual entanglement and
non-local correlations in separable states''\,',
{\em Phys. Rev. A} {\bf 63}, 1, 016101 (2001);
quant-ph/9912002.
Comment on {\bf [Cohen 99 a]}.
Reply: {\bf [Cohen 01]}.

\item {\bf [Terno 02]}:
D. R. Terno,
``Extension of quantum information theory to curved spacetimes'',
in A. Khrennikov (ed.),
{\em Quantum Theory: Reconsideration of Foundations (V\"{a}xj\"{o}, Sweden, 2001)},
V\"{a}xj\"{o} University Press, V\"{a}xj\"{o}, Sweden, 2002;
quant-ph/0111144.

\item {\bf [Terno 04 a]}:
D. R. Terno,
``Entropy, holography and the second law'',
{\em Phys. Rev. Lett.} {\bf 93}, 5, 051303 (2004).

\item {\bf [Terno 04 b]}:
D. R. Terno,
``Inconsistency of quantum--classical dynamics, and what it implies'',
quant-ph/0402092.

\item {\bf [Terra Cunha 98]}:
M. O. Terra Cunha,
``What is surrealistic about Bohm trajectories?'',
quant-ph/9809006.
See {\bf [Englert-Scully-S\"{u}ssmann-Walther 93 a]}.

\item {\bf [Terra Cunha-Man'ko-Scully 01]}:
M. O. Terra Cunha, V.I. Man'ko, \& M.O. Scully,
``Quasiprobability and probability distributions for spin-1/2
states'',
{\em Found. Phys. Lett.} {\bf 14}, 2, 103-117 (2001);
quant-ph/0102038.

\item {\bf [Terra Cunha-Nemes 02]}:
M. O. Terra Cunha, \& M. C. Nemes,
``Analysing a complementarity experiment
on the quantum-classical boundary'',
{\em Phys. Lett. A} {\bf 305}, 6, 313-321 (2002).

\item {\bf [Terra Cunha-Nemes 04]}:
M. O. Terra Cunha, \& M. C. Nemes,
``Towards an understanding of decoherence in ion traps'',
{\em Phys. Lett. A} {\bf 329}, 6, 409-413 (2004).

\item {\bf [Terra Cunha-Geraij Mokarzel-Peixoto de Faria-Nemes 04]}:
M. O. Terra Cunha, S. Geraij Mokarzel, J. G. Peixoto de Faria, \& M. C. Nemes,
``Timescales for decoherence and dissipation: reasons for the success of
master equations'',
quant-ph/0409061.

\item {\bf [Terraneo-Georgeot-Shepelyansky 02]}:
M. Terraneo, B. Georgeot, \& D. L. Shepelyansky,
``Strange attractor simulated on a quantum computer'',
quant-ph/0203062.

\item {\bf [Terraneo-Shepelyansky 04]}:
M. Terraneo, \& D. L. Shepelyansky,
``Dynamical localization and repeated measurements in a quantum computation process'',
{\em Phys. Rev. Lett.} {\bf 92}, 3, 037902 (2004).

\item {\bf [Tesch-de Vivie Riedle 02]}:
C. M. Tesch, \& R. de Vivie-Riedle,
``Quantum computation with vibrationally excited molecules'',
{\em Phys. Rev. Lett.} {\bf 89}, 15, 157901 (2002).

\item {\bf [Tesche 00]}:
C. Tesche,
``Quantum mechanics:
Enhanced: Schr\"{o}dinger's cat is out of the hat'',
{\em Science} {\bf 290}, 5492, 720-721 (2000).
See {\bf [van der Wal-ter Haar-Wilhelm-(+5) 00]}.

\item {\bf [Tessier-Deutsch-Delgado-Fuentes Guridi 03]}:
T. E. Tessier, I. H. Deutsch, A. Delgado, \& I. Fuentes-Guridi,
``Entanglement sharing in the two-atom Tavis-Cummings model'',
{\em Phys. Rev. A} {\bf 68}, 6, 062316 (2003).

\item {\bf [Tessieri-Wilkie 03]}:
L. Tessieri, \& J. Wilkie,
``Decoherence in a spin–spin-bath model with environmental self-interaction'',
{\em J. Phys. A} {\bf 36}, 49, 12305-12327 (2003).

\item {\bf [Tessieri-Deutsch-Caves 04]}:
T. E. Tessier, I. H. Deutsch, \& C. M. Caves,
``Efficient classical-communication-assisted local simulation of n-qubit
GHZ correlations'',
quant-ph/0407133.

\item {\bf [Tewari-Hariharan 97]}:
S. P. Tewari, \& P. Hariharan,
``Generation of entangled four-photon states by parametric
down-conversion'',
{\em J. Mod. Opt.} {\bf 44}, 3, 543-553 (1997).

\item {\bf [Teufel-Berndl-D\"{u}rr-(+2) 96]}:
S. Teufel, K. Berndl, D. D\"{u}rr, S. Goldstein, \& N. Zangh\`{\i},
``Locality and causality in hidden-variables models of quantum theory'',
{\em Phys. Rev. A} {\bf 56}, 2, 1217-1227 (1997);
quant-ph/9609005.

\item {\bf [Thaheld 00]}:
F. H. Thaheld,
``Comment on `Quantum superpositions and definite perceptions:
Envisaging new experimental tests'
[{\em Phys. Lett. A} {\bf 262} (1999) 1-14]'',
{\em Phys. Lett. A} {\bf 273}, 4, 232-234 (2000).
Comment on {\bf [Ghirardi 99 b]}.

\item {\bf [Thankappan-Menon 97]}:
V. K. Thankappan, \& R. K. Menon,
``EPR elements of physical reality in quantum mechanics'',
{\em Int. J. Mod. Phys. A} {\bf 12}, 29, 5289-5303 (1997).

\item {\bf [Thapliyal 99]}:
A. V. Thapliyal,
``Multipartite pure-state entanglement'',
{\em Phys. Rev. A} {\bf 59}, 5, 3336-3342 (1999);
quant-ph/9811091.

\item {\bf [Thapliyal-Smolin 03]}:
A. V. Thapliyal, \& J. A. Smolin,
``Multipartite entanglement gambling: The power of asymptotic state
transformations assisted by a sublinear amount of quantum communication'',
{\em Phys. Rev. A} {\bf 68}, 6, 062324 (2003);
quant-ph/0212098.

\item {\bf [Thew-Munro 01 a]}:
R. T. Thew, \& W. J. Munro,
``Entanglement manipulation and concentration'',
{\em Phys. Rev. A} {\bf 63}, 3, 030302(R) (2001);
quant-ph/0012049.

\item {\bf [Thew-Munro 01 b]}:
R. T. Thew, \& W. J. Munro,
``Mixed state entanglement:
Manipulating polarization-entangled photons'',
{\em Phys. Rev. A} {\bf 64}, 2, 022320 (2001);
quant-ph/0103005.

\item {\bf [Thew-Nemoto-White-Munro 02]}:
R. T. Thew, K. Nemoto, A. G. White, \& W. J. Munro,
``Qudit quantum-state tomography'',
{\em Phys. Rev. A} {\bf 66}, 1, 012303 (2002);
quant-ph/0201052.

\item {\bf [Thew-Tanzilli-Tittel 02]}:
R. T. Thew, S. Tanzilli, W. Tittel, H. Zbinden, \& N. Gisin,
``Experimental investigation of the robustness of partially entangled photons over 11 km'',
{\em Phys. Rev. A} {\bf 66}, 6, 062304 (2002);
quant-ph/0203067.

\item {\bf [Thew-Ac\'{\i}n-Zbinden-Gisin 03]}:
R. T. Thew, A. Ac\'{\i}n, H. Zbinden, \& N. Gisin,
``Entangled qutrits for quantum communication'',
{\em Quant. Inf. Comm.};
quant-ph/0307122.

\item {\bf [Thew-Ac\'{\i}n-Zbinden-Gisin 04]}:
R. T. Thew, A. Ac\'{\i}n, H. Zbinden, \& N. Gisin,
``A Bell-type test of energy-time entangled qutrits'',
{\em Phys. Rev. Lett.} {\bf 93}, 1, 010503 (2004);
quant-ph/0402048.

\item {\bf [Thew-Tanzilli-Ac\'{\i}n-(+2) 04]}:
R. T. Thew, S. Tanzilli, A. Ac\'{\i}n, H. Zbinden, \& N. Gisin,
``Energy-time entangled qutrits: Bell tests and quantum communication'',
{\em Proc.\ QCMC 2004 (Glasgow)};
quant-ph/0409111.

\item {\bf [Thompson 97]}:
C. H. Thompson,
`Timing, ``accidentals'' and other artifacts in EPR experiments',
quant-ph/9711044.

\item {\bf [Thompson 99]}:
C. H. Thompson,
``Rotational invariance, phase relationships
and the quantum entanglement illusion'',
quant-ph/9912082.

\item {\bf [Thompson-Rempe-Kimble 92]}:
R. J. Thompson, G. Rempe, \& H. J. Kimble,
``Observation of normal-mode splitting for an atom in an optical cavity'',
{\em Phys. Rev. Lett.} {\bf 68}, 8, 1132-1135 (1992).

\item {\bf [Thorne-Drever-Caves-(+2) 78]}:
K. S. Thorne,
R. W. P. Drever, C. M. Caves, M. Zimmermann, \& V. D. Sandberg,
``Quantum nondemolition measurements of harmonic oscillators'',
{\em Phys. Rev. Lett.} {\bf 40}, 11, 667-671 (1978).

\item {\bf [Thorwart-Hanggi 02]}:
M. Thorwart, \& P. Hanggi,
``Decoherence and dissipation during a quantum [small-caps XOR] gate operation'',
{\em Phys. Rev. A} {\bf 65}, 1, 012309 (2002).

\item {\bf [Tian-Lloyd 00]}:
L. Tian, \& S. Lloyd,
``Resonant cancellation of off-resonant effects
in a multilevel qubit'',
{\em Phys. Rev. A} {\bf 62}, 5, 050301(R) (2000);
quant-ph/0003034.

\item {\bf [Tian-Lloyd-Orlando 02]}:
L. Tian, S. Lloyd, \& T. P. Orlando,
``Decoherence and relaxation of a superconducting quantum bit during
measurement'',
{\em Phys. Rev. B} {\bf 65}, 14, 144516 (2002).

\item {\bf [Tian-Lloyd-Orlando 03]}:
L. Tian, S. Lloyd, \& T. P. Orlando,
``Projective measurement scheme for solid-state qubits'',
{\em Phys. Rev. B} {\bf 67}, 22, 220505 (2003);
quant-ph/0310083.

\item {\bf [Tian-Zoller 03]}:
L. Tian, \& P. Zoller,
``Quantum computing with atomic Josephson junction arrays'',
quant-ph/0306085.

\item {\bf [Tian-Rabl-Blatt-Zoller 04]}:
L. Tian, P. Rabl, R. Blatt, \& P. Zoller,
``Interfacing quantum-optical and solid-state qubits'',
{\em Phys. Rev. Lett.} {\bf 92}, 24, 247902 (2004);
quant-ph/0310057.

\item {\bf [Tian-Zoller 04]}:
L. Tian, \& P. Zoller,
``Coupled ion-nanomechanical systems'',
quant-ph/0407020.

\item {\bf [Tikochinsky 88 a]}:
Y. Tikochinsky,
``?'',
{\em J. Math. Phys.} {\bf 29}, ?, 398-? (1988).
See {\bf [Tikochinsky-Gull 00]}.

\item {\bf [Tikochinsky 88 b]}:
Y. Tikochinsky,
``On the generalized multiplication and addition of complex numbers'',
{\em Int. J. Theor. Phys.} {\bf 27}, 2, 398-399 (1988).
See {\bf [Tikochinsky-Gull 00]}.

\item {\bf [Tikochinsky-Gull 00]}:
Y. Tikochinsky, \& S. F. Gull,
``Consistency, amplitudes and probabilities in quantum theory'',
{\em J. Phys. A} {\bf 33}, 31, 5615-5618 (2000).
See {\bf [Tikochinsky 88 a, b]}, {\bf [Caticha 98 a]}.

\item {\bf [Tilma-Byrd-Sudarshan 02]}:
T. Tilma, M. Byrd, \& E. C. G. Sudarshan,
``A parametrization of
bipartite systems based on $\rm SU(4)$ Euler angles'',
{\em J. Phys. A} {\bf 35}, 48, 10445-10465 (2002).

\item {\bf [Timpson 01]}:
C. G. Timpson,
``On the supposed conceptual inadequacy of the Shannon information'' (2001),
quant-ph/0112178,
PITT-PHIL-SCI00000529.
See {\bf [Brukner-Zeilinger 01 a]}.

\item {\bf [Timpson-Brown 02]}:
C. G. Timpson, \& H. R. Brown,
``Entanglement and relativity'',
in R. Lupacchini, \& V. Fano (eds.),
{\em Understanding physical knowledge},
Department of Philosophy, University of Bologna, 2002;
quant-ph/0212140.

\item {\bf [Timpson-Brown 04]}:
C. G. Timpson, \& H. R. Brown,
``Proper and improper separability'',
quant-ph/0402094.

\item {\bf [Ting 99]}:
J. Ting,
``Noise effects on one-Pauli channels'',
{\em Phys. Lett. A} {\bf 259}, 5, 349-354 (1999).
quant-ph/9906120.

\item {\bf [Ting-Feng Li-Zhi Xi 04]}:
G. Ting, Y. Feng-Li, \& W. Zhi-Xi,
``Probabilistically cloning and quantum computation'',
{\em Chinese Phys. Lett.} {\bf 21}, 995 (2004).
quant-ph/0406104.

\item {\bf [Tipler 86]}:
F. J. Tipler,
``The many-worlds interpretation of quantum mechanics in quantum cosmology'',
in R. Penrose, \& C. Isham (eds.),
{\em Quantum concepts in space and time},
Clarendon Press, Oxford, 1986, pp.~?-?.

\item {\bf [Tipler 00]}:
F. J. Tipler,
``Does quantum nonlocality exist?
Bell's theorem and the many-worlds interpretation'',
quant-ph/0003146.

\item {\bf [Tisza 99]}:
L. Tisza,
``Unreasonable and `unreasonable' in quantum mechanics?'',
{\em Found. Phys.} {\bf 29}, 3, 491-496 (1999).

\item {\bf [Tittel-Brendel-Gisin-(+3) 97]}:
W. Tittel, J. Brendel, N. Gisin, T. Herzog, \& H. Zbinden,
``Non-local two-photon correlations
using interferometers physically separated by 35 meters'',
{\em Europhys. Lett.} {\bf 40} 6, 595-600 (1997);
quant-ph/9703023.

\item {\bf [Tittel-Ribordy-Gisin 98]}:
W. Tittel, G. Ribordy, \& N. Gisin,
``Quantum cryptography'',
{\em Phys. World} {\bf 11}, 3, 41-45 (1998).
Reprinted in {\bf [Macchiavello-Palma-Zeilinger 00]}, pp.~240-244.

\item {\bf [Tittel-Brendel-Gisin-(+3) 98]}:
W. Tittel, J. Brendel, B. Gisin, T. Herzog,
H. Zbinden, \& N. Gisin,
``Experimental demonstration
of quantum-correlations over more than 10 kilometers'',
{\em Phys. Rev. A} {\bf 57}, 5, 3229-3232 (1998);
quant-ph/9707042.

\item {\bf [Tittel-Brendel-Zbinden-Gisin 98]}:
W. Tittel, J. Brendel, H. Zbinden, \& N. Gisin,
``Violation of Bell inequalities by photons more than 10 km apart'',
{\em Phys. Rev. Lett.} {\bf 81}, 17, 3563-3566 (1998);
quant-ph/9806043.

\item {\bf [Tittel-Brendel-Gisin-Zbinden 99]}:
W. Tittel, J. Brendel, N. Gisin, \& H. Zbinden,
``Long-distance Bell-type tests using energy-time entangled photons'',
{\em Phys. Rev. A} {\bf 59}, 6, 4150-4163 (1999);
quant-ph/9809025.

\item {\bf [Tittel-Zbinden-Gisin 99]}:
W. Tittel, H. Zbinden, \& N. Gisin,
``Quantum secret sharing using pseudo-GHZ states'',
quant-ph/9912035.

\item {\bf [Tittel-Brendel-Zbinden-Gisin 00]}:
W. Tittel, J. Brendel, H. Zbinden, \& N. Gisin,
``Quantum cryptography using entangled photons in energy-time Bell
states'',
{\em Phys. Rev. Lett.} {\bf 84}, 20, 4737-4740 (2000);
quant-ph/9911109.

\item {\bf [Tittel-Weihs 01]}:
W. Tittel, \& G. Weihs,
``Photonic entanglement for fundamental tests and quantum communication'',
submitted to {\em Quant. Inf. Comp.};
quant-ph/0107156.

\item {\bf [Tiwari 02 a]}:
S. C. Tiwari,
`Comment on ``Experimental test of nonlocal quantum correlation
in relativistic configurations''\,',
{\em Phys. Rev. A} {\bf 65}, 1, 016101 (2002).
Comment on {\bf [Zbinden-Brendel-Gisin-Tittel 01]}.

\item {\bf [Tiwari 02 b]}:
S. C. Tiwari,
``Relativity, entanglement and the physical reality of the photon'',
{\em J. Opt. B: Quantum Semiclass. Opt.} {\bf 4}, 2, S39-S46 (2002).

\item {\bf [Tkadlec 98]}:
J. Tkadlec,
``Greechie diagrams of small quantum logics with small state spaces'',
{\em Int. J. Theor. Phys.} {\bf 37}, 1, 203-210 (1998).

\item {\bf [Todorov 83]}:
N. S. Todorov,
``On the Einstein-Podolsky-Rosen argument:
Its recent critique and its meaning'',
{\em Phys. Lett. A} {\bf 97}, 3, 91-94 (1983).
Comment on {\bf [Page 82]}.
See {\bf [Kunstatter-Trainor 84 a]}.

\item {\bf [Tokunaga 01]}:
Y. Tokunaga,
``Quantum coin flipping with arbitrary small bias is impossible'',
quant-ph/0108026.

\item {\bf [de Toledo Piza, \& M. C. Nemes 01]}:
A. F. R. de Toledo Piza, \& M. C. Nemes,
``Manipulating decay rates by entanglement and the Zeno effect'',
{\em Phys. Lett. A} {\bf 290}, 1-2, 6-10 (2001).

\item {\bf [Tolkunov-Privman 04]}:
D. Tolkunov, \& V. Privman,
``Short-time decoherence for general system-environment interactions'',
{\em Phys. Rev. A} {\bf 69}, 6, 062309 (2004).

\item {\bf [Tombesi-Giovannetti-Vitali 00]}:
P. Tombesi, V. Giovannetti, \& D. Vitali,
``Quantum state protection using all-optical feedback'',
in H. Carmichael, R. Glauber, \& M. O. Scully (eds.),
{\em Dan Walls Memorial},
Springer-Verlag, New York, 2000;
quant-ph/0004027.

\item {\bf [Tombesi-Hirota 01]}:
P. Tombesi, \& O. Hirota (eds.),
{\em Quantum communication, computing, and measurement 3},
Kluwer Academic, Dordrecht, Holland, 2001.

\item {\bf [Tomita-Hirota 00]}:
A. Tomita, \& O. Hirota,
``Security of classical noise-based cryptography'',
{\em J. Opt. B: Quantum Semiclass. Opt.} {\bf 2}, 6, 705-710 (2000);
quant-ph/0002044.

\item {\bf [Tomita 00]}:
A. Tomita,
``Complete Bell state measurement with controlled
photon absorption and quantum interference'',
quant-ph/0006093.

\item {\bf [Tomita 01]}:
A. Tomita,
``Complete Bell state measurement with a solid state device'',
{\em Phys. Lett. A} {\bf 282}, 6, 331-335 (2001).

\item {\bf [Tommasini 01 a]}:
D. Tommasini,
``Quantum electrodynamics is free from the Einstein-Podolsky-Rosen paradox'',
quant-ph/0110059.

\item {\bf [Tommasini 01 b]}:
D. Tommasini,
``Quantum theory finally reconciled with special relativity'',
quant-ph/0111083.

\item {\bf [Tommasini 03]}:
D. Tommasini,
``Photon uncertainty solves the Einstein-Podolsky-Rosen paradox'',
{\em Opt. Spectrosc.} {\bf 94}, 741 (2003);
quant-ph/0202175,
PITT-PHIL-SCI00000598.

\item {\bf [Toner-Bacon 03]}:
B. F. Toner, \& D. Bacon,
``The communication cost of simulating Bell correlations'',
{\em Phys. Rev. Lett.} {\bf 91}, 18, 187904 (2003);
quant-ph/0304076.

\item {\bf [Tong 02]}:
C. Tong,
``Nine formulations of quantum mechanics'',
{\em Am. J. Phys.} {\bf 70}, 7, 664 (2002).
Comment on {\bf [Styer-Balkin-Becker-(+10) 02]}.

\item {\bf [Tong-Sj\"{o}qvist-Kwek-(+2) 02]}:
D. M. Tong, E. Sj\"{o}qvist, L. C. Kwek,
C. H. Oh, \& M. Ericsson,
``Relation between geometric phases of entangled bipartite systems and their
subsystems'',
{\em Phys. Rev. A} {\bf 68}, 2, 022106 (2003).

\item {\bf [Tong-Kuang 00]}:
Z.-Y. Tong, \& L.-M. Kuang,
``Broadcasting of entanglement in three-particle GHZ
state via quantum copying'',
{\em Chin. Phys. Lett.};
quant-ph/0005070.

\item {\bf [Tong-Kuang 02]}:
Z.-Y. Tong, \& L.-M. Kuang,
``Entanglement preserving in quantum copying of three-qubit entangled
state'',
{\em Commun. Theor. Phys.} {\bf 38}, 541-546 (2002);
quant-ph/0402074.

\item {\bf [Tong-Chen-Kwek-(+2) 03]}:
D. M. Tong, J.-L. Chen, L. C. Kwek,
C. H. Lai, \& C. H. Oh,
``General formalism of Hamiltonians for realizing a prescribed evolution of a qubit'',
{\em Phys. Rev. A} {\bf 68}, 6, 062307 (2003);
quant-ph/0406185.

\item {\bf [Torgerson-Branning-Mandel 95]}:
J. R. Torgerson, D. Branning, \& L. Mandel,
``A method for demonstrating violation of local realism with a two-photon
downconverter without use of Bell iniequalities'',
{\em Appl. Phys. B} {\bf 60}, 2-3, 267-269 (1995).
See {\bf [Torgerson-Branning-Monken-Mandel 95]}.

\item {\bf [Torgerson-Branning-Monken-Mandel 95]}:
J. R. Torgerson, D. Branning, C. H. Monken, \& L. Mandel,
``Experimental demonstration of the violation
of local realism without Bell inequalities'',
{\em Phys. Lett. A} {\bf 204}, 5-6, 323-328 (1995).
See {\bf [Torgerson-Branning-Mandel 95]}, {\bf [Garuccio 95 b]},
{\bf [Boschi-De Martini-Di Giuseppe 97]},
{\bf [Di Giuseppe-De Martini-Boschi 97]}.
Comment: {\bf [Cabello-Santos 96]}.
Reply: {\bf [Torgerson-Branning-Monken-Mandel 96]}.

\item {\bf [Torgerson-Branning-Monken-Mandel 96]}:
J. R. Torgerson, D. Branning, C. H. Monken, \& L. Mandel,
``Reply to the comment by Cabello and Santos on `Experimental demonstration
of the violation of local realism without Bell inequalities'\,'',
{\em Phys. Lett. A} {\bf 214}, 5-6, 319-320 (1996).
Reply to {\bf [Cabello-Santos 96]}.
See {\bf [Torgerson-Branning-Monken-Mandel 95]},
{\bf [Garuccio 95 b]}.

\item {\bf [Torma-Jex-Schleich 02]}:
P. Torma, I. Jex, \& W. P. Schleich,
``Localization and diffusion in Ising-type quantum networks'',
{\em Phys. Rev. A} {\bf 65}, 5, 052110 (2002).

\item {\bf [Torres-Deyanova-Torner-Molina Terriza 03]}:
J. P. Torres, Y. Deyanova, L. Torner, \& G. Molina-Terriza,
``Preparation of engineered two-photon entangled states for multidimensional
quantum information'',
{\em Phys. Rev. A} {\bf 67}, 5, 052313 (2003).

\item {\bf [Torres-Alexandrescu-Torner 03]}:
J. P. Torres, A. Alexandrescu, \& L. Torner,
``Quantum spiral bandwidth of entangled two-photon states'',
{\em Phys. Rev. A} {\bf 68}, 5, 050301 (2003).

\item {\bf [T\"{o}rnquist 81]}:
N. A. T\"{o}rnquist,
``Suggestion for Einstein-Podolsky-Rosen using reactions like
$e^+e^-\to \Lambda \kern 1pt \bar \Lambda \to \pi ^-p\,\pi ^+ \bar p$'',
{\em Found. Phys.} {\bf 11}, 1-2, 171-177 (1981).

\item {\bf [T\"{o}rnquist 86]}:
N. A. T\"{o}rnquist,
``Bell's inequalities as triangle inequalities for cross-sections'',
{\em Europhys. Lett.} {\bf 1}, 8, 377-380 (1986).

\item {\bf [de la Torre-Catuogno-Ferrando 89]}:
A. C. de la Torre, P. Catuogno, \& S. Ferrando,
``Uncertainty and nonseparability'',
{\em Found. Phys. Lett.} {\bf 2}, 3, 235-244 (1989).

\item {\bf [de la Torre 94 a]}:
A. C. de la Torre,
``The concept of transmission in quantum mechanics'',
{\em Found. Phys. Lett.} {\bf 7}, 2, 143-159 (1994).

\item {\bf [de la Torre 94 b]}:
A. C. de la Torre,
``Contextuality in quantum systems'',
{\em Am. J. Phys.} {\bf 62}, 9, 808-812 (1994).
See {\bf [de la Torre-Dotson 96]}.

\item {\bf [de la Torre-Dotson 96]}:
A. C. de la Torre, \& A. C. Dotson,
``An entangled opinion on the interpretation of quantum mechanics'',
{\em Am. J. Phys.} {\bf 64}, 2, 174 (1996).
See {\bf [de la Torre 94 a]}.

\item {\bf [de la Torre-Iguain 98]}:
A. C. de la Torre, \& J. L. Iguain,
``Manifest and concealed correlations in quantum mechanics'',
{\em Eur. J. Phys.} {\bf 19}, 6, 563-569 (1998).

\item {\bf [de la Torre 99]}:
A. C. de la Torre,
``A one-dimensional lattice model for a quantum mechanical free particle'',
quant-ph/9905031.

\item {\bf [de la Torre-Daleo 99]}:
A. C. de la Torre, \& A. Daleo,
``Quantum mechanics as a classical field theory'',
quant-ph/9905032.

\item {\bf [de la Torre-Daleo 00]}:
A. C. de la Torre, \& A. Daleo,
``A one-dimensional lattice model for a quantum mechanical free
particle'',
{\em Eur. Phys. J. D} {\bf 8}, 2, 165-168 (2000).

\item {\bf [de la Torre-Daleo-Garc\'{\i}a Mata 00]}:
A. C. de la Torre, A. Daleo, \& I. Garc\'{\i}a-Mata,
``The photon-box Bohr-Einstein debate demythologized'',
{\em Eur. J. Phys.} {\bf 21}, 3, 253-260 (2000).
Comment: {\bf [Hnizdo 02]}.
Reply: {\bf [de la Torre-Daleo-Garc\'{\i}a Mata 02]}.

\item {\bf [de la Torre-Daleo-Garc\'{\i}a Mata 02]}:
A. C. de la Torre, A. Daleo, \& I. Garc\'{\i}a-Mata,
``The photon-box Bohr-Einstein debate
demythologized. Reply to comment'',
{\em Eur. J. Phys.} {\bf 23}, 4, L15-L16 (2002).
Reply to {\bf [Hnizdo 02]}.

\item {\bf [de la Torre 02]}:
A. C. de la Torre,
``Relativity of representations in quantum mechanics'',
{\em Am. J. Phys.} {\bf 70}, 3, 298-300 (2002);
quant-ph/0109129.

\item {\bf [de la Torre-Goyeneche 03]}:
A. C. de la Torre, \& D. Goyeneche,
``Quantum mechanics in finite-dimensional Hilbert space'',
{\em Am. J. Phys.} {\bf 71}, 1, 49-54 (2003).
See {\bf [de la Torre-Goyeneche 04]}.

\item {\bf [de la Torre-Goyeneche 04]}:
A. C. de la Torre, \& D. Goyeneche,
``Finite dimensional Hilbert space'',
{\em Am. J. Phys.} {\bf 72}, 7, 855 (2004).
See {\bf [de la Torre-Goyeneche 03]}.

\item {\bf [de la Torre 04 a]}:
A. C. de la Torre,
``Understanding light quanta: First quantization of the free
electromagnetic field'',
quant-ph/0410171.

\item {\bf [de la Torre 04 b]}:
A. C. de la Torre,
``Understanding light quanta: The photon'',
quant-ph/0410179.

\item {\bf [T\"{o}rm\"{a} 98]}:
P. T\"{o}rm\"{a},
``Transitions in quantum networks'',
{\em Phys. Rev. Lett.} {\bf 81}, 11, 2185-2189 (1998).

\item {\bf [Toscano-de Matos Filho-Davidovich 04]}:
F. Toscano, R. L. de Matos Filho, \& L. Davidovich,
``Decoherence and the quantum-classical limit in the presence of chaos'',
quant-ph/0410146.

\item {\bf [Toschek-Wunderlich 01]}:
P. E. Toschek, \& C. Wunderlich,
``What does an observed quantum system reveal to its observer?'',
{\em Eur. Phys. J. D} {\bf 14}, 3, 387-396 (2001);
quant-ph/0009021.

\item {\bf [T\'{o}th-Lent 01]}:
G. T\'{o}th, \& C. S. Lent,
``Quantum computing with quantum-dot cellular automata'',
{\em Phys. Rev. A} {\bf 63}, 5, 052315 (2001).

\item {\bf [T\'{o}th-Simon-Cirac 03]}:
G. T\'{o}th, C. Simon, \& J. I. Cirac,
``Entanglement detection based on interference and particle counting'',
{\em Phys. Rev. A} {\bf 68}, 6, 062310 (2003);
quant-ph/0306086.

\item {\bf [T\'{o}th 04]}:
G. T\'{o}th,
``Entanglement detection in optical lattices of bosonic atoms with
collective measurements'',
{\em Phys. Rev. A} {\bf 69}, 5, 052327 (2004);
quant-ph/0310039.

\item {\bf [T\'{o}th-G\"{u}hne 04]}:
G. T\'{o}th, \& O. G\"{u}hne,
``Detecting genuine multipartite entanglement with two local measurements'',
quant-ph/0405165.

\item {\bf [T\'{o}th-Cirac 04]}:
G. T\'{o}th, \& J. I. Cirac,
``Entanglement witnesses in spin models'',
quant-ph/0406061.

\item {\bf [Toulmin 70]}:
S. Toulmin (ed.),
{\em Physical reality}, Harper and Row,
Evanstone \& London, 1970.

\item {\bf [Townsend-Rarity-Tapster 93 a]}:
P. D. Townsend, J. G. Rarity, \& P. R. Tapster,
``Single photon interference in a 10 km long optical fiber interferometer'',
{\em Electron. Lett.} {\bf 29}, ?, 634-639 (1993).

\item {\bf [Townsend-Rarity-Tapster 93 b]}:
P. D. Townsend, J. G. Rarity, \& P. R. Tapster,
``Enhanced single photon fringe visibility in a 10 km-long
prototype quantum cryptography channel'',
{\em Electron. Lett.} {\bf 29}, 14, 1291-1293 (1993).

\item {\bf [Townsend-Smith 93]}:
P. D. Townsend, \& D. W. Smith,
``?'',
patent application EP93307120.1, 1993.
See {\bf [Townsend-Blow 93]},
{\bf [Townsend-Phoenix-Blow-Barnett 94]},
{\bf [Barnett-Phoenix 94]},
{\bf [Phoenix-Barnett-Townsend-Blow 95]}.

\item {\bf [Townsend-Blow 93]}:
P. D. Townsend, \& K. J. Blow,
``?'',
patent application EP93307121.9, 1993.
See {\bf [Townsend-Smith 93]},
{\bf [Townsend-Phoenix-Blow-Barnett 94]},
{\bf [Barnett-Phoenix 94]},
{\bf [Phoenix-Barnett-Townsend-Blow 95]}.

\item {\bf [Townsend 94]}:
P. D. Townsend,
``Secure key distribution system based on quantum cryptography'',
{\em Electron. Lett.} {\bf 30}, ?, 809-811 (1994).

\item {\bf [Townsend-Phoenix-Blow-Barnett 94]}:
P. D. Townsend, S. J. D. Phoenix, K. J. Blow, \& S. M. Barnett,
``Design of QC systems for passive optical networks'',
{\em Electron. Lett.} {\bf 30}, ?, 1875-1876 (1994).
See {\bf [Townsend-Smith 93]}, {\bf [Townsend-Blow 93]},
{\bf [Barnett-Phoenix 94]},
{\bf [Phoenix-Barnett-Townsend-Blow 95]}.

\item {\bf [Townsend-Thompson 94]}:
P. D. Townsend, \& I. Thompson,
``A quantum key distribution channel based on optical fibre'',
in S. M. Barnett, A. K. Ekert, \& S. J. D. Phoenix (eds.)
{\em J. Mod. Opt.} {\bf 41}, 12
(Special issue: Quantum communication), 2425-2433 (1994).

\item {\bf [Townsend 97 a]}:
P. D. Townsend,
``Quantum cryptography on multiuser optical fibre networks'',
{\em Nature} {\bf 385}, 6611, 47-49 (1997).

\item {\bf [Townsend 97 b]}:
P. D. Townsend,
``Quantum cryptography'',
patent EP776558A1, 1997.
See {\bf [Townsend 99]}.

\item {\bf [Townsend 97 c]}:
P. D. Townsend,
``Method and apparatus for polarization-insensitive
quantum cryptography'',
patent WO9744936A1, 1997.
See {\bf [Townsend 00]}.

\item {\bf [Townsend 97 d]}:
P. D. Townsend,
``System and method for key distribution using quantum
cryptography'',
patent US5675648, 1997.

\item {\bf [Townsend 97 e]}:
P. D. Townsend,
``Simultaneous quantum cryptographic key distribution and
conventional data transmission over installed fibre using WDM'',
{\em Electron. Lett.} {\bf 33}, ?, 188-190 (1997).

\item {\bf [Townsend 98 a]}:
P. D. Townsend,
``Experimental investigation of the performance limits for
first telecommunications-window quantum cryptography systems'',
{\em IEEE Photonics Technol. Lett.} {\bf 10}, 1048-1050 (1998).

\item {\bf [Townsend 98 b]}:
P. D. Townsend,
``Quantum cryptography on optical fiber networks'',
{\em Opt. Fiber Technol.: Mater., Devices Syst.} {\bf 4}, 345-370 (1998).

\item {\bf [Townsend-Smith 98 a]}:
P. D. Townsend, \& D. W. Smith,
``Key distribution in a multiple access network using quantum
cryptography'',
patent US5768378, 1998.
See {\bf [Townsend-Smith 98 b]}.

\item {\bf [Townsend-Smith 98 b]}:
P. D. Townsend, \& D. W. Smith,
``Key distribution in a multiple access network using quantum
cryptography'',
patent EP717895B1, 1998.
See {\bf [Townsend-Smith 98 a]}.

\item {\bf [Townsend-Blow 98]}:
P. D. Townsend, \& K. J. Blow,
``System and method for key distribution using quantum
cryptography'',
patent US5850441, 1998.
See {\bf [Townsend-Blow 99]}.

\item {\bf [Townsend-Blow 99]}:
P. D. Townsend, \& K. J. Blow,
``System and method for key distribution using quantum
cryptography'',
patent EP717896B1, 1999.
See {\bf [Townsend-Blow 98]}.

\item {\bf [Townsend 99]}:
P. D. Townsend,
``Quantum cryptography'',
patent US5953421, 1999.
See {\bf [Townsend 97 b]}.

\item {\bf [Townsend 00]}:
P. D. Townsend,
``Method and apparatus for polarization-insensitive
quantum cryptography'',
patent EP972373A1, 2000.
See {\bf [Townsend 97 c]}.

\item {\bf [Toyota 03]}:
N. Toyota,
``Quantization of the stag hunt game and the Nash equilibrilum'',
quant-ph/0307029.

\item {\bf [Tran-Pfister 02]}:
N.-K. Tran, \& O. Pfister,
``Quantum teleportation with close-to-maximal entanglement from a beam
splitter'',
{\em Phys. Rev. A} {\bf 65}, 5, 052313 (2002);
quant-ph/0107073.

\item {\bf [Traub-Wozniakowski 01]}:
J. F. Traub, \& H. Wozniakowski,
``Path integration on a quantum computer'',
quant-ph/0109113.

\item {\bf [Travaglione-Milburn 01]}:
B. C. Travaglione, \& G. J. Milburn,
``Generation of eigenstates using the phase-estimation algorithm'',
{\em Phys. Rev. A} {\bf 63}, 3, 032301 (2001).

\item {\bf [Travaglione-Nielsen-Wiseman-Ambainis 01]}:
B. C. Travaglione, M. A. Nielsen, H. M. Wiseman, \& A. Ambainis,
``ROM-based computation: Quantum versus classical'',
quant-ph/0109016.

\item {\bf [Travaglione-Milburn 02 a]}:
B. C. Travaglione, \& G. J. Milburn,
``Implementing the quantum random walk'',
{\em Phys. Rev. A} {\bf 65}, 3, 032310 (2002).

\item {\bf [Travaglione-Milburn 02 b]}:
B. C. Travaglione, \& G. J. Milburn,
``Preparing encoded states in an oscillator'',
{\em Phys. Rev. A} {\bf 66}, 5, 052322 (2002).

\item {\bf [Treacy 03]}:
P. B. Treacy,
``Role of locality in Einstein-Podolsky-Rosen correlations and teleportation'',
{\em Phys. Rev. A} {\bf 67}, 1, 014101 (2003).

\item {\bf [Tregenna-Beige-Knight 02]}:
B. Tregenna, A. Beige, \& P. L. Knight,
``Quantum computing in a macroscopic dark period'',
{\em Phys. Rev. A} {\bf 65}, 3, 032305 (2002);
quant-ph/0109006.

\item {\bf [Treiman 99]}:
S. Treiman,
{\em The odd quantum},
Princeton University Press, Princeton, New Jersey, 1999.
Reviews: {\bf [March 00]}, {\bf [von Baeyer 00]},
{\bf [Holstein 01]}, {\bf [Borcherds 03]}.

\item {\bf [Treussart-Alleaume-Le Floc'h 02]}:
F. Treussart, R. Alleaume, V. Le Floc'h,
L. T. Xiao, J.-M. Courty, \& J.-F. Roch,
``Direct measurement of the photon statistics
of a triggered single photon source'',
{\em Phys. Rev. Lett.} {\bf 89}, 9, 093601 (2002).

\item {\bf [Treutlein-Hommelhoff-Steinmetz-(+2) 04]}:
P. Treutlein, P. Hommelhoff, T. Steinmetz,
T. W. Hänsch, \& J. Reichel,
``Coherence in microchip traps'',
{\em Phys. Rev. Lett.} {\bf 92}, 20, 203005 (2004).

\item {\bf [Trifonov-Bjork-Soderholm 01]}:
A. Trifonov, G. Bjork, \& J. Soderholm,
``Simultaneous minimum-uncertainty measurement of
discrete-valued complementary observables'',
{\em Phys. Rev. Lett.} {\bf 86}, 20, 4423-4426 (2001);
quant-ph/0012124.

\item {\bf [Trifonov-Bj\"{o}rk-S\"{o}derholm-Tsegaye 02]}:
A. Trifonov, G. Bj\"{o}rk, J. S\"{o}derholm, \& T. Tsegaye,
``Comprehensive experimental test of quantum erasure'',
{\em Eur. Phys. J. D} {\bf 18}, 2 (Special issue:
{\em Quantum interference and cryptographic keys:
Novel physics and advancing technologies (QUICK) (Corsica, 2001)}, 251-258 (2002);
quant-ph/0009097.

\item {\bf [Trifonov 03]}:
D. A. Trifonov,
``On the position uncertainty measure on the circle'',
{\em J. Phys. A} {\bf 36}, 47, 11873-11879 (2003).

\item {\bf [Troiani-Hohenester-Molinari 00]}:
F. Troiani, U. Hohenester, \& E. Molinari,
``Exploiting exciton-exciton interactions in semiconductor
quantum dots for quantum-information processing'',
{\em Phys. Rev. B} {\bf 62}, 4, R2263-R2266 (2000);
quant-ph/0005064.

\item {\bf [Troiani-Hohenester-Molinari 02]}:
F. Troiani, U. Hohenester, \& E. Molinari,
``Electron-hole localization in coupled quantum dots'',
{\em Phys. Rev. B} {\bf 65}, 16, 161301 (2002).

\item {\bf [Troiani-Hohenester-Molinari 03]}:
F. Troiani, E. Molinari, \& U. Hohenester,
``High-finesse optical quantum gates for electron spins in artificial
molecules'',
{\em Phys. Rev. Lett.} {\bf 90}, 20, 206802 (2003).

\item {\bf [Trojek-Schmid-Bourennane-(+2) 04]}:
P. Trojek, C. Schmid, M. Bourennane,
H. Weinfurter, \& C. Kurtsiefer,
``Compact source of polarization-entangled photon pairs'',
{\em Opt. Express} {\bf 12}, 276-281 (2004).

\item {\bf [Trugenberger 00]}:
C. A. Trugenberger,
``Probabilistic quantum memories'',
{\em Phys. Rev. Lett.} {\bf 87}, 6, 067901 (2001).
Comment: {\bf [Brun-Klauck-Nayak-Zalka 03]}.

\item {\bf [Trugenberger 02]}:
C. A. Trugenberger,
``Quantum optimization for combinatorial searches'',
{\em New J. Phys.} {\bf 4}, 26.1-26.7 (2002);
quant-ph/0107081.
Comment: {\bf [Zalka-Brun 02]}.

\item {\bf [Trump-Bru\ss-Lewenstein 01]}:
C. Trump, D. Bru\ss, \& M. Lewenstein,
``Realistic teleportation with linear optical elements'',
{\em Phys. Lett. A} {\bf 279}, 1-2, 7-11 (2001);
quant-ph/0010113.

\item {\bf [Tsallis-Lloyd-Baranger 00]}:
C. Tsallis, S. Lloyd, \& M. Baranger,
``Generalization of the Peres criterion for local realism
through nonextensive entropy'',
quant-ph/0007112.

\item {\bf [Tsallis-Prato-Anteneodo 02]}:
C. Tsallis, D. Prato, \& C. Anteneodo,
``Separable-entangled frontier in a bipartite harmonic system'',
quant-ph/0202077.

\item {\bf [Tsegaye-Bj\"{o}rk-Atat\"{u}re-(+3) 00]}:
T. Tsegaye, G. Bj\"{o}rk, M. Atat\"{u}re, A. V. Sergienko,
B. E. A. Saleh, \& M. C. Teich,
``Complementarity and quantum erasure with entangled-photon states'',
{\em Phys. Rev. A} {\bf 62}, 3, 032106 (2000).

\item {\bf [Tsegaye-S\"{o}derholm-Atat\"{u}re-(+5) 00]}:
T. Tsegaye, J. S\"{o}derholm, M. Atat\"{u}re,
A. Trifonov, G. Bj\"{o}rk, A. V. Sergienko,
B. E. A. Saleh, \& M. C. Teich,
``Experimental demonstration of three mutually orthogonal
polarization states of entangled photons'',
{\em Phys. Rev. Lett.} {\bf 85}, 24, 5013-5017 (2000).

\item {\bf [Tseng-Somaroo-Sharf-(+4) 00 a]}:
C. H. Tseng, S. Somaroo, Y. Sharf, E. Knill,
R. Laflamme, T. F. Havel, \& D. G. Cory,
``Quantum simulation of a three-body-interaction Hamiltonian
on an NMR quantum computer'',
{\em Phys. Rev. A} {\bf 61}, 1, 012302 (2000);
quant-ph/9908012.

\item {\bf [Tseng-Somaroo-Sharf-(+4) 00 b]}:
C. H. Tseng, S. Somaroo, Y. Sharf, E. Knill,
R. Laflamme, T. F. Havel, \& D. G. Cory,
``Quantum simulation with natural decoherence'',
{\em Phys. Rev. A} {\bf 62}, 3, 032309 (2000).

\item {\bf [Tsomokos-Chong-Vourdas 04]}:
D. I. Tsomokos, C. C. Chong, \& A. Vourdas,
``Entanglement of distant electron interference experiments'',
{\em Phys. Rev. A} {\bf 69}, 1, 013810 (2004);
quant-ph/0404099.

\item {\bf [Tsujino-Takeuchi-Sasaki 02]}:
K. Tsujino, S. Takeuchi, \& K. Sasaki,
``Detailed analysis of the fidelity of quantum teleportation using photons:
Considering real experimental parameters'',
{\em Phys. Rev. A} {\bf 66}, 4, 042314 (2002).

\item {\bf [Tsurumaru 04]}:
T. Tsurumaru,
``An implementable protocol of quantum bit string commitment'',
quant-ph/0407174.

\item {\bf [Tu-Long 01]}:
C.-C. Tu, \& G. L. Long,
``Chen and Diao's quantum search algorithm is not exponentially fast'',
submitted to {\em Chinese Phys. Lett.}
quant-ph/0110098.
See {\bf [Chen-Diao 00 b]}.

\item {\bf [Tucci 98]}:
R. R. Tucci,
``A rudimentary quantum compiler'',
quant-ph/9805015.

\item {\bf [Tucci 99]}:
R. R. Tucci,
``Quantum entanglement and conditional information transmission'',
quant-ph/9909041.

\item {\bf [Tucci 00 a]}:
R. R. Tucci,
``Quantum computer as an inference engine'',
quant-ph/0004028.

\item {\bf [Tucci 00 b]}:
R. R. Tucci,
``Separability of density matrices and conditional information
transmission'',
quant-ph/0005119.

\item {\bf [Tucci 00 c]}:
R. R. Tucci,
``Entanglement of formation and conditional information
transmission'',
quant-ph/0010041.

\item {\bf [Tucci 01 a]}:
R. R. Tucci,
``Relaxation method for calculating quantum entanglement'',
quant-ph/0101123.

\item {\bf [Tucci 01 b]}:
R. R. Tucci,
``Entanglement of Bell mixtures of two qubits'',
quant-ph/0103040.

\item {\bf [Tucci 02]}:
R. R. Tucci,
``Entanglement of distillation and conditional mutual information'',
quant-ph/0202144.


\item {\bf [Tumulka 04 a]}:
R. Tumulka,
``Understanding Bohmian mechanics: A dialogue'',
{\em Am. J. Phys.} {\bf 72}, 9, 1220-1226 (2004);
quant-ph/0408113.

\item {\bf [Tumulka 04 b]}:
R. Tumulka,
``A relativistic version of the Ghirardi-Rimini-Weber model'',
quant-ph/0406094.

\item {\bf [Turchette-Hood-Lange-(+2) 95]}:
Q. A. Turchette, C. J. Wood, W. Lange
H. Mabuchi, \& H. J. Kimble,
``Measurement of conditional phase shifts for quantum logic'',
{\em Phys. Rev. Lett.} {\bf 75}, 25, 4710-4713 (1995);
Reprinted in {\bf [Macchiavello-Palma-Zeilinger 00]}, pp.~290-293.

\item {\bf [Turchette-Wood-King-(+5) 98]}:
Q. A. Turchette, C. S. Wood, B. E. King, C. J. Myatt,
D. Leibfried, W. M. Itano, C. Monroe, \& D. J. Wineland,
``Deterministic entanglement of two trapped ions'',
{\em Phys. Rev. Lett.} {\bf 81}, 17, 3631-3634 (1998);
quant-ph/9806012.
Reprinted in {\bf [Macchiavello-Palma-Zeilinger 00]}, pp.~359-362.

\item {\bf [Turner 68]}:
J. E. Turner,
``Violation of the quantum ordering of
propositions in hidden-variable theories'',
{\em J. Math. Phys.} {\bf 9}, 9, 1411-1415 (1968).

\item {\bf [Tuszynski-Brown-Hawrylak 98]}:
J. A. Tuszynski, J. A. Brown, \& P. Hawrylak,
``Dielectric polarization, electrical conduction, information
processing and quantum computation in microtubules. Are
they plausible?'',
in A. K. Ekert, R. Jozsa, \& R. Penrose (eds.),
{\em Quantum Computation: Theory and Experiment.
Proceedings of a Discussion Meeting held at the Royal
Society of London on 5 and 6 November 1997},
{\em Philos. Trans. R. Soc. Lond. A} {\bf 356}, 1743, 1897-1926
(1998).

\item {\bf [Twamley 00]}:
J. Twamley,
``A hidden shift quantum algorithm'',
{\em J. Phys. A} {\bf 33}, 48, 8973-8980 (2000).

\item {\bf [Twamley 03]}:
J. Twamley,
``Quantum-cellular-automata quantum computing with endohedral fullerenes'',
{\em Phys. Rev. A} {\bf 67}, 5, 052318 (2003).

\item {\bf [Tworzydlo-Beenakker 02]}:
J. Tworzydlo, \& C. W. J. Beenakker,
``Quantum optical communication rates through an amplifying random medium'',
{\em Phys. Rev. Lett.} {\bf 89}, 4, 043902 (2002);
quant-ph/0203094.

\item {\bf [Tyapkin-Vinduska 91]}:
A. A. Tyapkin, \& M. Vinduska,
``The geometrical aspects of the Bell inequalities'',
{\em Found. Phys.} {\bf 21}, 2, 185-195 (1991).

\item {\bf [Tyc-Sanders 02]}:
T. Tyc, \& B. C. Sanders,
``How to share a continuous-variable quantum secret by optical interferometry'',
{\em Phys. Rev. A} {\bf 65}, 4, 042310 (2002);
quant-ph/0107074.

\item {\bf [Tyc-Rowe-Sanders 03]}:
T. Tyc, D. J. Rowe, \& B. C. Sanders,
``Efficient sharing of a continuous-variable quantum secret'',
{\em J. Phys. A} {\bf 36}, 27, 7625–7637 (2003).

\item {\bf [Tyc-Sanders 04]}:
T. Tyc, \& B. C. Sanders,
``Operational formulation of homodyne detection'',
{\em J. Phys. A} {\bf 37}, ?, 7341-? (2004);
quant-ph/0404090.

\item {\bf [Tyson 03]}:
J. E. Tyson,
``Operator-Schmidt decompositions and the Fourier transform,
with applications to the operator-Schmidt numbers of unitaries'',
{\em J. Phys. A} {\bf 36}, 39, 10101-10114 (2003).


\newpage

\subsection{}


\item {\bf [Uchiyama-Aihara 02]}:
C. Uchiyama, \& M. Aihara,
``Multipulse control of decoherence'',
{\em Phys. Rev. A} {\bf 66}, 3, 032313 (2002).

\item {\bf [Uchiyama 97]}:
F. Uchiyama,
``Generalized Bell inequality in two neutral kaon systems'',
{\em Phys. Lett. A} {\bf 231}, 4-5, 295-298 (1997).

\item {\bf [Uchiyama 95]}:
S. Uchiyama,
``Local reality: Can it exist in the EPR-Bohm {\em gedanken} experiment?'',
{\em Found. Phys.} {\bf 25}, 11, 1561-1575 (1995).

\item {\bf [Ueda-Kitagawa 92]}:
M. Ueda, \& M. Kitagawa,
``Reversibility in quantum measurement processes'',
{\em Phys. Rev. Lett.} {\bf 68}, 23, 3424-3427 (1992).

\item {\bf [Uffink-Hilgevoord 84]}:
J. Uffink, \& J. Hilgevoord,
``New bounds for the uncertainty principle'',
{\em Phys. Lett. A} {\bf 105}, ?, 474-? (1984).

\item {\bf [Uffink 85]}:
J. Uffink,
``Verification of the uncertainty principle in neutron interferometry'',
{\em Phys. Lett. A} {\bf 108}, ?, 59-? (1985).

\item {\bf [Uffink-Hilgevoord 85 a]}:
J. Uffink, \& J. Hilgevoord,
``Uncertainty principle and uncertainty relations'',
{\em Found. Phys.} {\bf 15}, ?, 925-? (1985).

\item {\bf [Uffink-Hilgevoord 85 b]}:
J. Uffink, \& J. Hilgevoord,
``More certainty about the uncertainty principle'',
{\em Eur. J. Phys.} {\bf ?}, ?, 165-? (1985).

\item {\bf [Uffink-Hilgevoord 88 a]}:
J. Uffink, \& J. Hilgevoord,
``The mathematical expression of the uncertainty principle'',
in A. van der Merwe {\em et al.} (eds.),
{\em Microphysical reality and quantum description},
Kluwer Academic, Dordrecht, 1988, pp.~?-?.

\item {\bf [Uffink-Hilgevoord 88 b]}:
J. Uffink, \& J. Hilgevoord,
``Interference and distinguishability in quantum mechanics'',
{\em Physica B} {\bf 151}, 309-313 (1988);
quant-ph/9904003.

\item {\bf [Uffink-Hilgevoord 89 a]}:
J. Uffink, \& J. Hilgevoord,
``Spacetime symmetries and the uncertainty principle'',
{\em Nuclear Phys. B} {\bf 6}, 246-249 (1989).

\item {\bf [Uffink-Hilgevoord 89 b]}:
J. Uffink, \& J. Hilgevoord,
``Comments on `Violation of Heisenberg's uncertainty
relations\ldots' by Z. Mari\'{c}, K. Popper and J. P. Vigier'',
{\em Found. Phys. Lett.} {\bf 2}, ?, 403-? (1989).

\item {\bf [Uffink-Hilgevoord 90]}:
J. Uffink, \& J. Hilgevoord,
``A new view on the uncertainty principle'',
in A. I. Miller (ed.),
{\em Sixty-two years of uncertainty:
Historical, philosophical and physical inquiries into the foundations of quantum
mechanics.
Proc.\ Int. School of History of Science (Erice, Italy, 1989)},
Plenum Press, New York, 1990, pp.~121-139.

\item {\bf [Uffink 90]}:
J. Uffink,
``Measures of uncertainty and the uncertainty principle'',
Ph.\ D. thesis, University of Utrecht, 1990.

\item {\bf [Uffink 91]}:
J. Uffink,
``Uncertainty in prediction and in inference'',
{\em Found. Phys.} {\bf 21}, ?, 323-341 (1991).

\item {\bf [Uffink 93]}:
J. Uffink,
``The rate of evolution of a quantum state'',
{\em Am. J. Phys.} {\bf 61}, ?, 935-? (1993).
See {\bf [Vaidman 92]}.

\item {\bf [Uffink 94 a]}:
J. Uffink,
``Two new kinds of uncertainty relations'',
{\em Proc.\ of the Third
Int.\ Workshop on Squeezed States and Uncertainty Relations},
NASA Conf.\ Publications 3270, ?, 1994, pp.~155-161.

\item {\bf [Uffink 94 b]}:
J. Uffink,
``The joint measurement problem'',
{\em Int. J.Theor. Phys.} {\bf 33}, 199-212 (1994).

\item {\bf [Uffink 99]}:
J. Uffink,
``How to protect the interpretation of the
wave function against protective measurements'',
{\em Phys. Rev. A} {\bf 60}, 5, 3474-3481 (1999).

\item {\bf [Uffink 02]}:
J. Uffink,
``Quadratic Bell inequalities as tests for multipartite entanglement'',
{\em Phys. Rev. Lett.} {\bf 88}, 23, 230406 (2002);
quant-ph/0201070.

\item {\bf [Uhlmann 99]}:
A. Uhlmann,
``Quantum channels of the Einstein-Podolski-Rosen kind'',
in {\em 12th Max Born Symp.\ ``Fine de siecle'' (Wroclaw, Poland, 1998)};
quant-ph/9901027.

\item {\bf [Uhlmann 00 a]}:
A. Uhlmann,
``Operators and maps affiliated to EPR channels'',
in
{\em Trends in quantum mechanics (Goslar, 1998)},
World Scientific, River Edge, New Jersey, 2000, pp.~138-145.

\item {\bf [Uhlmann 00 b]}:
A. Uhlmann,
``Fidelity and concurrence of conjugated states'',
{\em Phys. Rev. A} {\bf 62}, 3, 032307 (2000);
quant-ph/9909060.

\item {\bf [Uhlmann 01]}:
A. Uhlmann,
``On 1-qubit channels'',
in S. Popescu, N. Linden, \& R. Jozsa (eds.),
{\em J. Phys. A} {\bf 34}, 35
(Special issue: Quantum information and computation), 7047-7056 (2001);
quant-ph/0011106.

\item {\bf [Uhlmann 03]}:
A. Uhlmann,
``Antilinearity in bipartite quantum systems and imperfect quantum
teleportation'',
in W. Freudenberg (ed.),
{\em Quantum probability and white noise analysis},
World Scientific, Singapore, 2003;
quant-ph/0407244.

\item {\bf [Ukena-Shimizu 04]}:
A. Ukena, \& A. Shimizu,
``Appearance and stability of anomalously fluctuating states in Shor's factoring algorithm'',
{\em Phys. Rev. A} {\bf 69}, 2, 022301 (2004).

\item {\bf [Ulfbeck-Bohr 01]}:
O. Ulfbeck, \& A. Bohr,
``Genuine fortuitousness. Where did that click come from?'',
{\em Found. Phys.} {\bf 31}, 5, 757-774 (2001).
See {\bf [Bohr-Mottelson-Ulfbeck 04]}.

\item {\bf [Um-Hong-Yeon 98]}:
C. Um, S. Hong, \& K. Yeon,
``Analysis of Mach-Zehnder interferometer with two degenerated
optical parametric amplifiers'',
{\em Phys. Lett. A} {\bf 244}, 4, 201-210 (1998).

\item {\bf [Unanyan-Vitanov-Bergmann 01]}:
R. G. Unanyan, N. V. Vitanov, \& K. Bergmann,
``Preparation of entangled states by adiabatic passage'',
{\em Phys. Rev. Lett.} {\bf 87}, 13, 137902 (2001).

\item {\bf [Unanyan-Fleischhauer 02]}:
R. G. Unanyan, \& M. Fleischhauer,
``Efficient and robust entanglement generation in a many-particle
system with resonant dipole-dipole interactions'',
{\em Phys. Rev. A} {\bf 66}, 3, 032109 (2002);
quant-ph/0111095.

\item {\bf [Unanyan-Fleischhauer-Vitanov-Bergmann 02]}:
R. G. Unanyan, M. Fleischhauer, N. V. Vitanov, \& K. Bergmann,
``Entanglement generation by adiabatic navigation in the space of symmetric
multiparticle states'',
{\em Phys. Rev. A} {\bf 66}, 4, 042101 (2002);
quant-ph/0205118.

\item {\bf [Unanyan-Fleischhauer 03]}:
R. G. Unanyan, \& M. Fleischhauer,
``Decoherence-free generation of many-particle entanglement by adiabatic
ground-state transitions'',
{\em Phys. Rev. Lett.} {\bf 90}, 13, 133601 (2003);
quant-ph/0208144.

\item {\bf [Ungar 02]}:
A. A. Ungar,
``The density matrix for mixed state qubits and
hyperbolic geometry'',
{\em Quant. Inf. Comput.} {\bf 2}, 6, 513-514 (2002).

\item {\bf [Unnerstall 90]}:
T. Unnerstall,
``A comment on the Rauch-Vigier experiments on neutron
interferometry'',
{\em Phys. Lett. A} {\bf 151}, 6-7, 263-268 (1990).
Comment on {\bf [Rauch-Vigier 90]}.

\item {\bf [Unnikrishnan 00 a]}:
C. S. Unnikrishnan,
``Popper's experiment, uncertainty principle,
signal locality and momentum conservation'',
{\em Found. Phys. Lett.} {\bf 13}, 2, 197-200 (2000).

\item {\bf [Unnikrishnan 02]}:
C. S. Unnikrishnan,
``Is the quantum mechanical
description of physical reality complete? Proposed resolution of
the EPR puzzle'',
{\em Found. Phys. Lett.} {\bf 15}, 1, 1-25 (2002);
quant-ph/0005103.
See {\bf [d'Espagnat 03]}.

\item {\bf [Unruh 78]}:
W. G. Unruh,
``Analysis of quantum-nondemolition measurement'',
{\em Phys. Rev. D} {\bf 18}, 6, 1764-1772 (1978).

\item {\bf [Unruh 79]}:
W. G. Unruh,
``Quantum nondemolition and gravity-wave detection'',
{\em Phys. Rev. D} {\bf 19}, 10, 2888-2896 (1979).

\item {\bf [Unruh-Zurek 89]}:
W. G. Unruh, \& W. H. Zurek,
``Reduction of a wave packet in quantum Brownian motion'',
{\em Phys. Rev. D} {\bf 40}, 4, 1071-1094 (1989).

\item {\bf [Unruh 95]}:
W. G. Unruh,
``?'',
{\em Phys. Rev. A} {\bf 51}, ?, 992 (1995).

\item {\bf [Unruh 97]}:
W. G. Unruh,
``Is quantum mechanics non-local?'',
quant-ph/9710032.
See {\bf [Stapp 97 a, e, 98 a]}.

\item {\bf [Unruh 99 a]}:
W. G. Unruh,
``Nonlocality, counterfactuals, and quantum mechanics'',
{\em Phys. Rev. A} {\bf 59}, 1, 126-130 (1999).
Comment: {\bf [Stapp 99 d]}.
Reply: {\bf [Unruh 99 b]}.

\item {\bf [Unruh 99 b]}:
W. G. Unruh,
``Reply to `Comment on ``Nonlocality
counterfactuals, and quantum mechanics''\,'\,'',
{\em Phys. Rev. A} {\bf 60}, 3, 2599-2600 (1999).
Reply to {\bf [Stapp 99 d]}.
See {\bf [Unruh 99 a]}.

\item {\bf [U'Ren-Banaszek-Walmsley 03]}:
A. B. U'Ren, K. Banaszek, \& I. A. Walmsley,
``Photon engineering for quantum information processing'',
quant-ph/0305192.

\item {\bf [Usami-Nambu-Ishizaka-(+4) 01]}:
K. Usami, Y. Nambu, S. Ishizaka,
T. Hiroshima, Y. Tsuda, K. Matsumoto, \& K. Nakamura,
``Restoration of entanglement by spectral filters'',
quant-ph/0107121.

\item {\bf [Usami-Nambu-Tsuda-(+2) 03]}:
K. Usami, Y. Nambu, Y. Tsuda,
K. Matsumoto, \& K. Nakamura,
``Accuracy of quantum-state estimation utilizing Akaike's information
criterion'',
{\em Phys. Rev. A} {\bf 68}, 2, 022314 (2003);
quant-ph/0306083.

\item {\bf [Usera 00]}:
J. I. Usera,
``An approach to measurement by quantum-stochastic-parameter
averaged Bohmian mechanics'',
quant-ph/0001054.

\item {\bf [Usmanov-Ioffe 04]}:
R. A. Usmanov, \& L. B. Ioffe,
``Theoretical investigation of a protected quantum bit in
a small Josephson junction array with tetrahedral symmetry'',
{\em Phys. Rev. B} {\bf 69}, 21, 214513 (2004).

\item {\bf [Usuda-Takumi-Hata-Hirota 99]}:
T. S. Usuda, I. Takumi, M. Hata, \& O. Hirota,
``Minimum error detection of classical linear code sending through a quantum
channel'',
{\em Phys. Lett. A} {\bf 256}, 2-3, 104-108 (1999).

\item {\bf [Usuda-Usami-Takumi-Hata 02]}:
T. S. Usuda, S. Usami, I. Takumi, \& M. Hata,
``Superadditivity in capacity of quantum channel for $q$-ary
linearly dependent real symmetric-state signals'',
{\em Phys. Lett. A} {\bf 305}, 3-4, 125-134 (2002).

\item {\bf [Utsunomiya-Master-Yamamoto 04]}:
S. Utsunomiya, C. P. Master, \& Y. Yamamoto,
``Algorithm-based analysis of collective decoherence in quantum
computation'',
quant-ph/0408162.


\newpage

\subsection{}


\item {\bf [Vaccaro-Steuernagel-Fern\'{a}ndez Huelga 01]}:
J. A. Vaccaro, O. Steuernagel, \& S. G. Fern\'{a}ndez Huelga,
``A class of symmetric controlled quantum operations'',
in S. Popescu, N. Linden, \& R. Jozsa (eds.),
{\em J. Phys. A} {\bf 34}, 35
(Special issue: Quantum information and computation), 7057-7066 (2001);
quant-ph/0102015.

\item {\bf [Vaccaro-Anselmi-Wiseman 03]}:
J. A. Vaccaro, F. Anselmi, \& H. M. Wiseman,
``Entanglement of identical particles and reference phase uncertainty'',
{\em Int. J. Quant. Inf.} {\bf 1}, 4, 427-441 (2003);
quant-ph/0311028.

\item {\bf [Vaidman 87]}:
L. Vaidman,
``?'',
Ph.\ D. thesis, Tel-Aviv University, 1987.

\item {\bf [Vaidman-Aharonov-Albert 87]}:
L. Vaidman, Y. Aharonov, \& D. Z. Albert,
``How to ascertain the values of $\sigma_x$, $\sigma_y$, $\sigma_z$
of a spin-$\frac{1}{2}$ particle'',
{\em Phys. Rev. Lett.} {\bf 58}, 14, 1385-1387 (1987).
See {\bf [Mermin 95 b]}.

\item {\bf [Vaidman 88]}:
L. Vaidman,
``Meaning and measurability of nonlocal quantum states'',
in A. van der Merwe, F. Selleri, \& G. Tarozzi (eds.),
{\em Microphysical reality and quantum formalism.
Proc.\ of an international conference (Urbino, Italy, 1985)},
Kluwer Academic, Dordrecht, Holland, 1988, vol. 1, pp.~81-88.

\item {\bf [Vaidman 91]}:
L. Vaidman,
``A quantum time machine'',
{\em Found. Phys.} {\bf 21}, ?, 947-? (1991).

\item {\bf [Vaidman 92]}:
L. Vaidman,
``Minimum time for the evolution to an orthogonal quantum state'',
{\em Am. J. Phys.} {\bf 60}, ?, 182–183 (1992).
See {\bf [Uffink 93]}.

\item {\bf [Vaidman 93]}:
L. Vaidman,
`Lorentz-invariant ``elements of reality'' and
the joint measurability of commuting observables',
{\em Phys. Rev. Lett.} {\bf 70}, 22, 3369-3372 (1993);
hep-th/9305162.
See {\bf [Cohen-Hiley 95 a, 96]}.

\item {\bf [Vaidman 94 a]}:
L. Vaidman,
``Teleportation of quantum states'',
{\em Phys. Rev. A} {\bf 49}, 2, 1473-1476 (1994).

\item {\bf [Vaidman 94 b]}:
L. Vaidman,
``How to detect an excited atom without disturbing it'',
in D. Han, Y. S. Kim, N. H. Rubin, Y. Shin, \& W. W. Zachary (eds.),
{\em Squeezed states and uncertainty relations},
NASA Conf.\ Publ.\ 3270, ?, 1994, p.~299-?.
See {\bf [Elitzur-Vaidman 93 a, b]}, {\bf [Vaidman 94 c]}.

\item {\bf [Vaidman 94 c]}:
L. Vaidman,
``On the realization of interaction-free measurements'',
{\em Quantum Opt.} {\bf 6}, 3, 119-126 (1994).
See {\bf [Elitzur-Vaidman 93 a, b]}, {\bf [Vaidman 94 b]}.

\item {\bf [Vaidman 94 d]}:
L. Vaidman,
``On the paradoxical aspects of new quantum experiments'',
in D. Hull, M. Forbes, \& R. Burian (eds.),
{\em Proc.\ of the 1994
Biennial Meeting of the Philosophy of Science Association},
East Lansing, Michigan, 1994, vol. 1, pp.~211-?;
PITT-PHIL-SCI00000564.

\item {\bf [Vaidman 95 a]}:
L. Vaidman,
``Measurement of nonlocal variables without breaking causality'',
in J. S. Anandan, \& J. L. Safko (eds.),
{\em Quantum coherence and reality. In celebration of the 60th birthday
of Yakir Aharonov. Int.\
Conf.\ on Fundamental Aspects of Quantum Theory (?, ?)},
World Scientific, Singapore, 1995, pp.~?-?;
hep-th/9306086.

\item {\bf [Vaidman 95 b]}:
L. Vaidman,
``Nonlocality of a single photon revisited again'',
{\em Phys. Rev. Lett.} {\bf 75}, 10, 2063 (1995);
quant-ph/9501003.
Comment on {\bf [Hardy 94]}.
Reply: {\bf [Hardy 95 b]}.

\item {\bf [Vaidman 95 c]}:
L. Vaidman,
``Nonlocal measurements and teleporation of quantum states'',
in M. Ferrero, \& A. van der Merwe (eds.),
{\em Fundamental problems in quantum physics.
Proc.\ of an international symposium (Oviedo, Spain, 1993)},
Kluwer Academic, Dordrecht, Holland, 1995, pp.~347-356.

\item {\bf [Vaidman 96 a]}:
L. Vaidman,
``Weak-measurement elements of reality'',
{\em Found. Phys.} {\bf 26}, 7, 895-906 (1996);
quant-ph/9601005.
See {\bf [Kastner 98 b]}.

\item {\bf [Vaidman-Goldenberg-Wiesner 96]}:
L. Vaidman, L. Goldenberg, \& S. Wiesner,
``Error prevention scheme with four particles'',
{\em Phys. Rev. A} {\bf 54}, 3, R1745-R1748 (1996);
quant-ph/9603031.

\item {\bf [Vaidman 96 b]}:
L. Vaidman,
``Emergence of weak values'',
in F. de Martini, G. Denardo, \& Y. H. Shih (eds.),
{\em Quantum interferometry},
VCH Publishers, New York, 1996, pp.~?-?;
quant-ph/9607023.

\item {\bf [Vaidman 96 c]}:
L. Vaidman,
``On schizophrenic experiences of the neutron or why we should
believe in the many-worlds interpretation of quantum theory'',
quant-ph/9609006.

\item {\bf [Vaidman 96 d]}:
L. Vaidman,
``Defending time-symmetrized quantum theory'',
quant-ph/9609007.

\item {\bf [Vaidman 96 e]}:
L. Vaidman,
``Interaction-free measurements'',
quant-ph/9610033.

\item {\bf [Vaidman 97]}:
L. Vaidman,
``The analysis of Hardy's experiment revisited'',
quant-ph/9703018.
See {\bf [Cohen-Hiley 95 a]}.

\item {\bf [Vaidman 98 a]}:
L. Vaidman,
``Validity of the Aharonov-Bergmann-Lebowitz rule'',
{\em Phys. Rev. A} {\bf 57}, 3, 2251-2253 (1998);
quant-ph/9703001.
See {\bf [Cohen 98]}.

\item {\bf [Vaidman 98 b]}:
L. Vaidman,
``Time-symmetrized counterfactuals in quantum theory'',
quant-ph/9802042, quant-ph/9807075.

\item {\bf [Vaidman 98 c]}:
L. Vaidman,
``Teleportation: Dream or reality?'',
invited talk in the conference
{\em Mysteries, Puzzles, and Paradoxes in Quantum Mechanics},
quant-ph/9810089.

\item {\bf [Vaidman 98 d]}:
L. Vaidman,
``?'',
{\em Int. Stud. Philos. Sci.} {\bf 12}, ?, ?-? (1998).

\item {\bf [Vaidman 98 e]}:
L. Vaidman,
``Time-symmetrized quantum theory'',
{\em Fortschr. Phys.} {\bf 46}, 6-8, 729-739 (1998);
quant-ph/9710036.

\item {\bf [Vaidman-Belkind 98]}:
L. Vaidman, \& O. Belkind,
``Strict bounds of Franson inequality'',
{\em Phys. Rev. A} {\bf 57}, 3, 1583-1585 (1998);
quant-ph/9707060.

\item {\bf [Vaidman-Yoran 99]}:
L. Vaidman, \& N. Yoran,
``Methods for reliable teleportation'',
{\em Phys. Rev. A} {\bf 59}, 1, 116-125 (1999);
quant-ph/9808040.
See {\bf [L\"{u}tkenhaus-Calsamiglia-Suominen 99]}.

\item {\bf [Vaidman 99 a]}:
L. Vaidman,
``Defending time-symmetrized quantum counterfactuals'',
{\em Stud. Hist. Philos. Sci. Part B: Stud. Hist. Philos. Mod. Phys.}
{\bf 30}, 3, 373-397 (1999);
quant-ph/9811092.
See {\bf [Kastner 98 a, b, 99 b, c]}, {\bf [Vaidman 99 c]}.

\item {\bf [Vaidman 99 b]}:
L. Vaidman,
``Variations on the theme of the Greenberger-Horne-Zeilinger proof'',
{\em Found. Phys.} {\bf 29}, 4, 615-630 (1999);
quant-ph/9808022.

\item {\bf [Vaidman 99 c]}:
L. Vaidman,
``Time-symmetrized counterfactuals in quantum theory'',
{\em Found. Phys.} {\bf 29}, 5, 755-766 (1999);
quant-ph/9807075.
See {\bf [Kastner 98 a, b, 99 b, c]}, {\bf [Vaidman 99 a, d]}.

\item {\bf [Vaidman 99 d]}:
L. Vaidman,
``The meaning of elements of reality and quantum counterfactuals:
Reply to Kastner'',
{\em Found. Phys.} {\bf 29}, 6, 865-876 (1999);
quant-ph/9903095.
See {\bf [Kastner 98 a, b, 99 b, c]}, {\bf [Vaidman 99 a, c]}.

\item {\bf [Vaidman 00 a]}:
L. Vaidman,
``Discussion: Byrne and Hall on Everett and Chalmers'',
quant-ph/0001057.

\item {\bf [Vaidman 00 b]}:
L. Vaidman,
``Are interaction-free measurements interaction free?'',
contribution to the {\em ICQO 2000, Raubichi, Belarus};
quant-ph/0006077.

\item {\bf [Vaidman 01 a]}:
L. Vaidman,
``The paradoxes of the interaction-free measurement'',
contribution to
{\em Mysteries and Paradoxes in Quantum Mechanics (Gargnano, Italy,
2000)},
{\em Zeitschrift f\"{u}r Naturforschung};
quant-ph/0102049.

\item {\bf [Vaidman 01 b]}:
L. Vaidman,
``Tests of Bell inequalities'',
{\em Phys. Lett. A} {\bf 286}, 4, 241-244 (2001);
quant-ph/0102139; quant-ph/0107057.
See {\bf [Rowe-Kielpinski-Meyer-(+4) 01]}, {\bf [Santos 01]}.

\item {\bf [Vaidman 02 a]}:
L. Vaidman,
``Probability and the many-worlds interpretation of quantum theory'',
in A. Khrennikov (ed.),
{\em Quantum Theory: Reconsideration of Foundations (V\"{a}xj\"{o}, Sweden, 2001)},
V\"{a}xj\"{o} University Press, V\"{a}xj\"{o}, Sweden, 2002;
quant-ph/0111072.

\item {\bf [Vaidman 02 b]}:
L. Vaidman,
``An impossible necklace'',
in {\bf [Bertlmann-Zeilinger 02]}, pp.~221-223.

\item {\bf [Vaidman-Mitrani 04]}:
L. Vaidman, \& Z. Mitrani,
``Qubit versus bit for measuring an integral of a classical field'',
{\em Phys. Rev. Lett.} {\bf 92}, 21, 217902 (2004);
quant-ph/0212165.

\item {\bf [Vaidman 03 a]}:
L. Vaidman,
``Instantaneous measurement of nonlocal variables'',
{\em Phys. Rev. Lett.} {\bf 90}, 1, 010402 (2003);
quant-ph/0111124.

\item {\bf [Vaidman 03 b]}:
L. Vaidman,
``The meaning of the interaction-free measurements'',
{\em Found. Phys.} {\bf 33}, 3, 491-510 (2003);
quant-ph/0103081.

\item {\bf [Vaidman 03 c]}:
L. Vaidman,
``Discussion: Time-symmetric quantum counterfactuals'',
PITT-PHIL-SCI00001108.

\item {\bf [Vaidman 03 d]}:
L. Vaidman,
``The reality in Bohmian quantum mechanics or can you kill with an empty
wave bullet?'',
contribution to the Jim Cushing Memorial, {\em Found. Phys.};
quant-ph/0312227.

\item {\bf [Vaidman-Kalev 04]}:
L. Vaidman, \& A. Kalev,
``Measurement of an integral of a classical field with a single quantum
particle'',
quant-ph/0406024.

\item {\bf [Vala-Whaley 02]}:
J. Vala, \& K. B. Whaley,
``Encoded universality for generalized anisotropic exchange Hamiltonians'',
{\em Phys. Rev. A} {\bf 66}, 2, 022304 (2002);
quant-ph/0204016.

\item {\bf [Vala-Amitay-Zhan-(+2) 02]}:
J. Vala, Z. Amitay, B. Zhang, S. R. Leone, \& R. Kosloff,
``Experimental implementation of the Deutsch-Jozsa algorithm for three-qubit
functions using pure coherent molecular superpositions'',
{\em Phys. Rev. A} {\bf 66}, 6, 062316 (2002);
quant-ph/0107058.

\item {\bf [Valdats 00]}:
M. Valdats,
``The class of languages recognizable by 1-way quantum finite
automata is not closed under union'',
quant-ph/0001005.
See {\bf [Ambainis-Kikusts-Valdats 00]}.

\item {\bf [Valdenebro 02]}:
A. G. Valdenebro,
``Assumptions underlying Bell's inequalities'',
{\em Eur. J. Phys.} {\bf 23}, 5, 569-577 (2002);
quant-ph/0208161.

\item {\bf [Valencia-Chekhova-Trifonov-Shih 02]}:
A. Valencia, M. V. Chekhova, A. Trifonov, \& Y. Shih,
``Entangled two-photon wave packet in a dispersive medium'',
{\em Phys. Rev. Lett.} {\bf 88}, 18, 183601 (2002);
quant-ph/0407201.

\item {\bf [Valencia-Scarcelli-Shih 04]}:
A. Valencia, G. Scarcelli, \& Y. Shih,
``Distant clock synchronization using entangled photon pairs'',
{\em App. Phys. Lett.};
quant-ph/0407204.

\item {\bf [Valentini 01]}:
A. Valentini,
``Signal-locality and subquantum information in
deterministic hidden-variables theories'',
quant-ph/0112151.

\item {\bf [Valentini 02 a]}:
A. Valentini,
``Signal-locality in hidden-variables theories'',
{\em Phys. Lett. A} {\bf 297}, 5-6, 273-278 (2002);
quant-ph/0106098.

\item {\bf [Valentini 02 b]}:
A. Valentini,
``Subquantum information and computation'',
in R. Ghosh (ed.),
{\em Proc.\ of the 2nd Winter Institute on Foundations of Quantum Theory and Quantum Optics:
Quantum Information Processing (Bangalore, India, 2002)}, pp.~?-?;
quant-ph/0203049.

\item {\bf [Valentini 04]}:
A. Valentini,
``Universal signature of non-quantum systems'',
{\em Phys. Lett. A} {\bf 332}, 3-4, 187-193 (2004);
quant-ph/0309107.

\item {\bf [Valentini-Westman 04]}:
A. Valentini, \& H. Westman,
``Dynamical origin of quantum probabilities'',
quant-ph/0403034.

\item {\bf [Valiant 02]}:
L. G. Valiant,
``Quantum circuits that can be simulated classically in polynomial time'',
{\em Siam J. Comput.} {\bf 31}, 1229 (2002).

\item {\bf [Valiev-Kokin 99]}:
K. A. Valiev, \& A. A. Kokin,
``Solid-state NMR quantum computer with individual
access to qubits and some its ensemble developments'',
quant-ph/9909008.

\item {\bf [Vatan-Williams 04 a]}:
F. Vatan, \& C. P. Williams,
``Optimal quantum circuits for general two-qubit gates'',
{\em Phys. Rev. A} {\bf 69}, 3, 032315 (2004);
quant-ph/0308006.

\item {\bf [Vatan-Williams 04 b]}:
F. Vatan, \& C. P. Williams,
``Realization of a general three-qubit quantum gate'',
quant-ph/0401178.

\item {\bf [Van Assche-Iblisdir-Cerf 04]}:
G. Van Assche, S. Iblisdir, \& N. J. Cerf,
``Secure coherent-state quantum key distribution protocols with efficient
reconciliation'',
quant-ph/0410031.

\item {\bf [van Dam 98 a]}:
W. van Dam,
``Quantum oracle interrogation: Getting all information for almost half the price'', Wim van Dam,
in {\em Proc.\ of the 39th Annual IEEE Symposium on Foundations of Computer Science (FOCS'98)}, 362-367;
quant-ph/9805006.

\item {\bf [van Dam 98 b]}:
W. van Dam,
``Two classical queries versus one quantum query'',
quant-ph/9806090.

\item {\bf [van Dam 00]}:
W. van Dam,
``Quantum algorithms for weighing matrices and quadratic
residues'',
quant-ph/0008059.

\item {\bf [van Dam-Hallgren 00]}:
W. van Dam, \& S. Hallgren,
``Efficient quantum algorithms for shifted quadratic character
problems'',
quant-ph/0011067.

\item {\bf [van Dam-Hayden 02 a]}:
W. van Dam, \& P. Hayden,
``Embezzling entangled quantum states'',
quant-ph/0201041.

\item {\bf [van Dam-Hayden 02 b]}:
W. van Dam, \& P. Hayden,
``Renyi-entropic bounds on quantum communication'',
quant-ph/0204093.

\item {\bf [van Dam-Seroussi 02]}:
W. van Dam, \& G. Seroussi,
``Efficient quantum algorithms for estimating Gauss sums'',
quant-ph/0207131.

\item {\bf [van Dam 02]}:
W. van Dam,
``On quantum computation theory'',
Ph.\ D. thesis, University of Amsterdam, 2002.

\item {\bf [van Dam-Hallgren-Ip 02]}:
W. van Dam, S. Hallgren, \& L. Ip,
``Quantum algorithms for some hidden shift problems'',
{\em Proc.\ ACM-SIAM Symp.\ on Discrete Algorithms}, 489-498 (2003);
quant-ph/0211140.

\item {\bf [van Dam-Hayden 03]}:
W. van Dam, \& P. Hayden,
``Universal entanglement transformations without communication'',
{\em Phys. Rev. A} {\bf 67}, 6, 060302 (2003).

\item {\bf [van Dam-Grunwald-Gill 03]}:
W. van Dam, P. Grunwald, \& R. D. Gill,
``The statistical strength of nonlocality proofs'',
quant-ph/0307125.

\item {\bf [van Dam 03 a]}:
W. van Dam,
``Think nonlocally''
{\em Am. Scientist} {\bf 91}, 3, ? (2003).
Review of {\bf [Aczel 02]}.

\item {\bf [van Dam 03 b]}:
W. van Dam,
`Comment on ``Quantum identification schemes with entanglements''\,',
{\em Phys. Rev. A} {\bf 68}, 2, 026301 (2003);
quant-ph/0307126.
Comment on {\bf [Mihara 02]}.

\item {\bf [van Dam 04]}:
W. van Dam,
``Quantum computing and zeroes of zeta functions'',
quant-ph/0405081.

\item {\bf [van der Merwe 84]}:
A. van der Merwe,
`Editorial postscript to ``The evolution of the ideas of Louis de Broglie on the interpretation of
wave mechanics''\,',
in {\bf [Barut-van der Merwe-Vigier 84]}, pp.~34-41.
See {\bf [Lochak 84]}.

\item {\bf [Van den Nest-Dehaene-De Moor 04 a]}:
M. Van den Nest, J. Dehaene, \& B. De Moor,
``Graphical description of the action of local Clifford transformations on graph states'',
{\em Phys. Rev. A} {\bf 69}, 2, 022316 (2004);
quant-ph/0308151.

\item {\bf [Van den Nest-Dehaene-De Moor 04 b]}:
M. Van den Nest, J. Dehaene, \& B. De Moor,
``Local invariants of stabilizer codes'',
{\em Phys. Rev. A};
quant-ph/0404106.

\item {\bf [Van den Nest-Dehaene-De Moor 04 c]}:
M. Van den Nest, J. Dehaene, \& B. De Moor,
``An efficient algorithm to recognize local Clifford equivalence of graph
states'',
{\em Phys. Rev. A};
quant-ph/0405023.

\item {\bf [Van den Nest-De Moor 04]}:
M. Van den Nest, \& B. De Moor,
``The invariants of the local Clifford group'',
quant-ph/0410035.

\item {\bf [van der Waerden 67]}:
B. L. van der Waerden (ed.),
{\em Sources of quantum mechanics},
North-Holland, Amsterdam, 1967;
Dover, New York, 1968.

\item {\bf [van der Wal-ter Haar-Wilhelm-(+5) 00]}:
C. H. van der Wal, A. C. J. ter Haar, F. K. Wilhelm, R. N. Schouten,
C. J. P. M. Harmans, T. P. Orlando, S. Lloyd, \& J. E. Mooij,
``Quantum superposition of macroscopic persistent-current states'',
{\em Science} {\bf 290}, 5492, 773-777 (2000).
See {\bf [Tesche 00]}.

\item {\bf [van der Wal-Kouwenhoven 99]}:
C. van der Wal, \& L. Kouwenhoven,
``Solid start for solid-state quantum bits'',
{\em Phys. World} {\bf 12}, 7, 21-22 (1999).

\item {\bf [van der Wal-Eisaman-Andre-(+4) 03]}:
C. H. van der Wal, M. D. Eisaman, A. Andre,
R. L. Walsworth, D. F. Phillips,
A. S. Zibrov, \& M. D. Lukin,
``Atomic memory for correlated photon states'',
{\em Science} {\bf 301}, ?, 196-? (2003).

\item {\bf [Van Dyck 03]}:
M. Van Dyck,
``The roles of one thought experiment in interpreting
quantum mechanics. Werner Heisenberg meets Thomas Kuhn'',
PITTPHILSCI 1158 (2003).

\item {\bf [van Enk-Cirac-Zoller 97 a]}:
S. J. van Enk, J. I. Cirac, \& P. Zoller,
``Ideal quantum communication over noisy channels:
A quantum optical implementation'',
{\em Phys. Rev. Lett.} {\bf 78}, 22, 4293-4296 (1997).

\item {\bf [van Enk-Cirac-Zoller 97 b]}:
S. J. van Enk, J. I. Cirac, \& P. Zoller,
``Purifying two-bit quantum gates and joint measurements in cavity QED'',
{\em Phys. Rev. Lett.} {\bf 79}, 25, 5178-5181 (1997);
quant-ph/9708032.

\item {\bf [van Enk-Cirac-Zoller-(+2) 97]}:
S. J. van Enk, J. I. Cirac, P. Zoller, H. J. Kimble, \& H. Mabuchi,
``Quantum state transfer in a quantum network: A quantum optical implementation'',
{\em J. Mod. Opt.} {\bf 44}, 10 (Special issue:
Fundamentals of quantum optics IV), 1727-1736 (1997).
See {\bf [van Enk-Cirac-Zoller-(+2) 98]}.

\item {\bf [van Enk-Cirac-Zoller-(+2) 98]}:
S. J. van Enk, J. I. Cirac, P. Zoller, H. J. Kimble, \& H. Mabuchi,
``Transmission of quantum information in a quantum network:
A quantum optical implementation'',
{\em Fortschr. Phys.} {\bf 46}, 6-8, 689-695 (1998).
See {\bf [van Enk-Cirac-Zoller-(+2) 97]}.

\item {\bf [van Enk 98]}:
S. J. van Enk,
``No-cloning and superluminal signaling'',
quant-ph/9803030.
See {\bf [Westmoreland-Schumacher 98]}.

\item {\bf [van Enk-Cirac-Zoller 98]}:
S. J. van Enk, J. I. Cirac, \& P. Zoller,
``Photonic channels for quantum communication'',
{\em Science} {\bf 279}, 5348, 205-208 (1998).
Reprinted in {\bf [Macchiavello-Palma-Zeilinger 00]}, pp.~229-232.

\item {\bf [van Enk-Kimble-Cirac-Zoller 99]}:
S. J. van Enk, H. J. Kimble, J. I. Cirac, \& P. Zoller,
``Quantum communication with dark photons'',
{\em Phys. Rev. A} {\bf 59}, 4, 2659-2664 (1999);
quant-ph/9805003.

\item {\bf [van Enk 99]}:
S. J. van Enk,
``Discrete formulation of teleportation of continuous variables'',
{\em Phys. Rev. A} {\bf 60}, 6, 5095-5097 (1999);
quant-ph/9905081.

\item {\bf [van Enk 00]}:
S. J. van Enk,
``Quantum and classical game strategies'',
{\em Phys. Rev. Lett.} {\bf 84}, 4, 789 (2000).
Comment on {\bf [Meyer 99 a]}.

\item {\bf [van Enk-Hirota 01]}:
S. J. van Enk, \& O. Hirota,
``Entangled coherent states: Teleportation and decoherence'',
{\em Phys. Rev. A} {\bf 64}, 2, 022313 (2001);
quant-ph/0012086.

\item {\bf [van Enk 01]}:
S. J. van Enk,
``The physical meaning of phase and its
importance for quantum teleportation'',
submitted to {\em J. Mod. Opt.};
quant-ph/0102004.

\item {\bf [van Enk-Fuchs 02 a]}:
S. J. van Enk, \& C. A. Fuchs,
``Quantum state of an ideal propagating laser field'',
{\em Phys. Rev. Lett.} {\bf 88}, 2, 027902 (2002);
quant-ph/0104036.
See {\bf [Wiseman 03 a]}.

\item {\bf [van Enk-Fuchs 02 b]}:
S. J. van Enk, \& C. A. Fuchs,
``The quantum state of a propagating laser field'',
{\em Quant. Inf. Comp.} {\bf 2}, ?, 151-? (2002);
quant-ph/0111157.

\item {\bf [van Enk-Pike 02]}:
S. J. van Enk, \& R. Pike,
``Classical rules in quantum games'',
{\em Phys. Rev. A} {\bf 66}, 2, 024306 (2002);
quant-ph/0203133.

\item {\bf [van Enk 02 a]}:
S. J. van Enk,
``Phase measurements with weak reference pulses'',
{\em Phys. Rev. A} {\bf 66}, 4, 042308 (2002);
quant-ph/0207142.

\item {\bf [van Enk 02 b]}:
S. J. van Enk,
``Unambiguous state discrimination of coherent states with linear optics:
Application to quantum cryptography'',
{\em Phys. Rev. A} {\bf 66}, 4, 042313 (2002);
quant-ph/0207138.

\item {\bf [van Enk 03 a]}:
S. J. van Enk,
``Entanglement of electromagnetic fields'',
{\em Phys. Rev. A} {\bf 67}, 2, 022303 (2003).

\item {\bf [van Enk 03 b]}:
S. J. van Enk,
``Odd numbers of photons and teleportation'',
{\em Phys. Rev. A} {\bf 67}, 2, 022318 (2003).

\item {\bf [van Enk 03 c]}:
S. J. van Enk,
``Entanglement capabilities in infinite dimensions: Multidimensional entangled
coherent states'',
{\em Phys. Rev. Lett.} {\bf 91}, 1, 017902 (2003);
quant-ph/0302086.

\item {\bf [van Enk-Kimble 03]}:
S. J. van Enk, \& H. J. Kimble,
`Reply I to ``Comment on `Some implications of the quantum nature of laser
fields for quantum computations'\,''\,',
{\em Phys. Rev. A} {\bf 68}, 4, 046302 (2003).
Reply to: {\bf [Itano 03]}.
See: {\bf [Gea Banacloche 02 a, 03]}.

\item {\bf [van Enk-Rudolph 03]}:
S. J. van Enk, \& T. Rudolph,
``On continuous-variable entanglement with and without phase references'',
quant-ph/0303096.

\item {\bf [van Enk 04 a]}:
S. J. van Enk,
``Entangled states of light'',
quant-ph/0403119.

\item {\bf [van Enk 04 b]}:
S. J. van Enk,
``Reference frames and refbits'',
quant-ph/0410083.

\item {\bf [van Fraassen 72]}:
B. C. van Fraassen,
``A formal approach to the philosophy of science'',
in R. G. Colodny (ed.),
{\em Paradigms and paradoxes. The
philosophical challenge of the quantum domain},
University of Pittsburgh Press, Pittsburgh, 1972, pp.~303-366.

\item {\bf [van Fraassen 73]}:
B. C. van Fraassen,
``Semantic analysis of quantum logic'',
in {\bf [Hooker 73]}, pp.~80-113.

\item {\bf [van Fraassen 74 a]}:
B. C. van Fraassen,
``The Einstein-Podolsky-Rosen paradox'',
{\em Synthese} {\bf 29}, ?, 291-309 (1974).

\item {\bf [van Fraassen 74 b]}:
B. C. van Fraassen,
``The labyrinth of quantum logic'',
in R. S. Cohen, \& M. W. Wartofsky (eds.),
{\em Boston studies in the philosophy of sciences. Vol. 5},
Reidel, Dordrecht, Holland, 1974, pp.~72-102.
Reprinted in {\bf [Hooker 75]}, pp.~577-607.

\item {\bf [van Fraassen 79]}:
B. C. van Fraassen,
``Hidden variables and the modal
interpretation of quantum theory'',
{\em Synthese} {\bf 42}, 1, 155-165 (1979).

\item {\bf [van Fraassen 81]}:
B. C. van Fraassen,
``A modal interpretation of quantum mechanics'',
in {\bf [Beltrametti-van Fraassen 81]}, pp.~?-?.

\item {\bf [van Fraassen 82]}:
B. C. van Fraassen,
``The Charybdis of realism: Epistemological implications of Bell's inequality'',
{\em Synthese} {\bf 52}, ?, 25-38 (1982).
See {\bf [Stairs 84]}.

\item {\bf [van Fraassen 85]}:
B. C. van Fraassen,
``EPR: When is a correlation not a mystery?'',
in P. J. Lahti, \& P. Mittelstaedt (eds.),
{\em Symp.\ on the Foundations of Modern
Physics: 50 Years of the Einstein-Podolsky-Rosen Experiment
(Joensuu, Finland, 1985)},
World Scientific, Singapore, 1985, pp.~113-128.

\item {\bf [van Fraassen 91 a]}:
B. C. van Fraassen,
{\em Quantum mechanics: An empiricist view},
Clarendon Press, Oxford, 1991.

\item {\bf [van Fraassen 91 b]}:
B. C. van Fraassen,
``The modal interpretation of quantum mechanics'',
in P. J. Lahti, \& P. Mittelstaedt (eds.),
{\em Symp.\ on the Foundations of Modern Physics 1990.
Quantum Theory of Measurement and Related Philosophical Problems
(Joensuu, Finland, 1990)},
World Scientific, Singapore, 1991, pp.~?-?.

\item {\bf [van Fraassen 98]}:
B. C. van Fraassen,
``Book review. Interpreting the quantum world'',
{\em Found. Phys.} {\bf 28}, 4, 683-689 (1998).
Review of {\bf [Bub 97]}.

\item {\bf [van Heerden 75]}:
P. J. van Heerden,
``Single systems do have eigenstates'',
{\em Am. J. Phys.} {\bf 43}, 11, 114-115 (1975).
Comment on {\bf [Peres 74]}.
Reply: {\bf [Peres 75]}.

\item {\bf [van Kampen 88]}:
N. G. van Kampen,
``Ten theorems about quantum mechanical measurements'',
{\em Physica A} {\bf 153}, ?, 97-113 (1988).

\item {\bf [van Kampen 90]}:
N. G. van Kampen,
``Quantum criticism'',
{\em Phys. World} {\bf 3}, 10, 20 (1990).

\item {\bf [van Kampen 91]}:
N. G. van Kampen,
``Mystery of quantum measurement'',
{\em Phys. World} {\bf 4}, 12, 16-17 (1991).

\item {\bf [van Loock-Braunstein 00 a]}:
P. van Loock, \& S. L. Braunstein,
``Unconditional teleportation of continuous-variable entanglement'',
{\em Phys. Rev. A} {\bf 61}, 1, 010302(R) (2000);
quant-ph/9906075.

\item {\bf [van Loock-Braunstein 00 b]}:
P. van Loock, \& S. L. Braunstein,
``Multipartite entanglement for continuous variables:
A quantum teleportation network'',
{\em Phys. Rev. Lett.} {\bf 84}, 15, 3482-3485 (2000);
quant-ph/9906021.

\item {\bf [van Loock-Braunstein-Kimble 00]}:
P. van Loock, S. L. Braunstein, \& H. J. Kimble,
``Broadband teleportation'',
{\em Phys. Rev. A} {\bf 62}, 2, 022309 (2000).

\item {\bf [van Loock-Braunstein 00 c]}:
P. van Loock, \& S. L. Braunstein,
``Greenberger-Horne-Zeilinger nonlocality in phase space'',
{\em Phys. Rev. A} {\bf 63}, 2, 022106 (2001);
quant-ph/0006029.

\item {\bf [van Loock-Braunstein 01]}:
P. van Loock, \& S. L. Braunstein,
``Telecloning of continuous quantum variables'',
{\em Phys. Rev. Lett.} {\bf 87}, 24, 247901 (2001);
quant-ph/0012063.

\item {\bf [van Loock-Furusawa 03]}:
P. van Loock, \& A. Furusawa,
``Detecting genuine multipartite continuous-variable entanglement'',
{\em Phys. Rev. A} {\bf 67}, 5, 052315 (2003);
quant-ph/0212052.

\item {\bf [van Loock-L\"{u}tkenhaus 04]}:
P. van Loock, \& N. L\"{u}tkenhaus,
``Simple criteria for the implementation of projective measurements with linear optics'',
{\em Phys. Rev. A} {\bf 69}, 1, 012302 (2004);
quant-ph/0304057.

\item {\bf [van Tonder 04]}:
A. van Tonder,
``A lambda calculus for quantum computation'',
{\em SIAM J. Comp.} {\bf 33}, ?, 1109-1135 (2004).

\item {\bf [van Velsen-Tworzydlo-Beenakker 03]}:
J. L. van Velsen, J. Tworzydlo, \& C. W. J. Beenakker,
``Scattering theory of plasmon-assisted entanglement transfer and distillation'',
{\em Phys. Rev. A} {\bf 68}, 4, 043807 (2003).

\item {\bf [van Velsen-Beenakker 04]}:
J. L. van Velsen, \& C. W. J. Beenakker,
``Transition from pure-state to mixed-state entanglement by random scattering'',
quant-ph/0403093.

\item {\bf [Vandersypen-Yannoni-Sherwood-Chuang 99]}:
L. M. K. Vandersypen, C. S. Yannoni, M. H. Sherwood, \& I. L. Chuang,
``Realization of effective pure states for bulk quantum computation'',
{\em Phys. Rev. Lett.} {\bf 83}, 15, 3085-3088 (1999);
quant-ph/9905041.

\item {\bf [Vandersypen-Steffen-Sherwood-(+3) 00]}:
L. M. K. Vandersypen, M. Steffen, M. H. Sherwood,
C. S. Yannoni, G. Breyta, \& I. L. Chuang,
``Implementation of a three-quantum-bit search algorithm'',
{\em Appl. Phys. Lett.} {\bf 76}, 5, 646-648 (2000);
quant-ph/9910075.

\item {\bf [Vandersypen-Steffen-Breyta-(+3) 00]}:
L. M. K. Vandersypen, M. Steffen, G. Breyta,
C. S. Yannoni, R. Cleve, \& I. L. Chuang,
``Experimental realization of an order-finding algorithm with
an NMR quantum computer'',
{\em Phys. Rev. Lett.} {\bf 85}, 25, 5452-5455 (2000);
quant-ph/0007017.

\item {\bf [Vandersypen-Yannoni-Chuang 00]}:
L. M. K. Vandersypen, C. S. Yannoni, \& I. L. Chuang,
``Liquid state NMR quantum computing computation'',
in ?. Grant, \& ?. Harris (eds.),
{\em Encyclopedia of NMR};
quant-ph/0012108.

\item {\bf [Vandersypen-Steffen-Breyta-(+3) 01]}:
L. M. K. Vandersypen, M. Steffen, G. Breyta,
C. S. Yannoni, M. H. Sherwood, \& I. L. Chuang,
``Experimental realization of Shor's quantum factoring algorithm
using nuclear magnetic resonance'',
{\em Nature} {\bf 414}, 6866, 883-887 (2001);
quant-ph/0112176.

\item {\bf [Vandersypen-Chuang 04]}:
L. M. K. Vandersypen, \& and I. L. Chuang,
``NMR techniques for quantum control and computation'',
{\em Rev. Mod. Phys.};
quant-ph/0404064.

\item {\bf [Varadarajan 62]}:
V. S. Varadarajan,
``Probability in physics and a
theorem on simultaneous observability'',
{\em Comm. on Pure and Appl. Math.} {\bf 15}, 2, 189-217 (1962).
Erratum: {\em Comm. on Pure and Appl. Math.} {\bf 18}, 4, 757 (1965).
Also in {\bf [Hooker 75]}, pp.~171-204 (incomplete),
the rest is in {\bf [Hooker 79]}, pp.~xvii-xix.

\item {\bf [Varadarajan 68]}:
V. S. Varadarajan,
{\em Geometry of quantum theory},
Van Nostrand, Princeton, New Jersey, 1968 (1st edition, 2 vols.),
Springer-Verlag, New York, 1985 (2nd edition, 1 vol.).

\item {\bf [Vartiainen-Niskanen-Nakahara-Salomaa 04]}:
J. J. Vartiainen, A. O. Niskanen, M. Nakahara, \& M. M. Salomaa,
``Implementing Shor's algorithm on Josephson charge qubits'',
{\em Phys. Rev. A} {\bf 70}, 1, 012319 (2004);
quant-ph/0308171.

\item {\bf [Vazirani 98]}:
U. Vazirani,
``On the power of quantum computation'',
in A. K. Ekert, R. Jozsa, \& R. Penrose (eds.),
{\em Quantum Computation: Theory and Experiment.
Proceedings of a Discussion Meeting held at the Royal
Society of London on 5 and 6 November 1997},
{\em Philos. Trans. R. Soc. Lond. A} {\bf 356}, 1743, 1759-1768 (1998).

\item {\bf [Vaziri-Weihs-Zeilinger 01]}:
A. Vaziri, G. Weihs, \& A. Zeilinger,
``Superpositions of the orbital angular momentum for
applications in quantum experiments'',
{\em J. Opt. B: Quantum Semiclass. Opt.} {\bf 4}, 2, S47-S51 (2002);
quant-ph/0111033.

\item {\bf [Vazirani-Weihs-Zeilinger 02]}:
A. Vaziri, G. Weihs, \& A. Zeilinger,
``Experimental two-photon, three-dimensional entanglement for quantum
communication'',
{\em Phys. Rev. Lett.} {\bf 89}, 24, 240401 (2002).

\item {\bf [Vazirani 02]}:
U. Vazirani,
``A survey of quantum complexity theory'',
in {\bf [Lomonaco 02 a]}, pp.~193-217.

\item {\bf [Vazirani-Pan-Jennewein-(+2) 03]}:
A. Vaziri, J.-W. Pan, T. Jennewein, G. Weihs, \& A. Zeilinger,
``Concentration of higher dimensional entanglement:
Qutrits of photon orbital angular momentum'',
{\em Phys. Rev. Lett.} {\bf 91}, 22, 227902 (2003);
quant-ph/0303003.

\item {\bf [Vecchi 00]}:
I. Vecchi,
``Interference of macroscopic superpositions'',
quant-ph/0007117.

\item {\bf [Vecchi 01]}:
I. Vecchi,
``Is entanglement observer-dependent?'',
quant-ph/0106003.

\item {\bf [Vedral-Barenco-Ekert 96]}:
V. Vedral, A. Barenco, \& A. K. Ekert,
``Quantum networks for elementary arithmetic operations'',
{\em Phys. Rev. A} {\bf 54}, 1, 147-153 (1996);
quant-ph/9511018.

\item {\bf [Vedral-Rippin-Plenio 97]}:
V. Vedral, M. A. Rippin, \& M. B. Plenio,
``Quantum correlations, local interactions and error correction'',
{\em J. Mod. Opt.} {\bf 44}, 11-12 (Special issue: Quantum
state preparation and measurement), 2185-2205 (1997).

\item {\bf [Vedral-Plenio-Rippin-Knight 97]}:
V. Vedral, M. B. Plenio, M. A. Rippin, \& P. L. Knight,
``Quantifying entanglement'',
{\em Phys. Rev. Lett.} {\bf 78}, 12, 2275-2279 (1997);
quant-ph/9702027.

\item {\bf [Vedral-Plenio-Jacobs-Knight 97]}:
V. Vedral, M. B. Plenio, K. Jacobs, \& P. L. Knight,
``Statistical inference, distinguishability of quantum states, and
quantum entanglement'',
{\em Phys. Rev. A} {\bf 56}, 6, 4452-4455 (1997);
quant-ph/9703025.

\item {\bf [Vedral-Plenio 98 a]}:
V. Vedral, \& M. B. Plenio,
``Entanglement measures and purification procedures'',
{\em Phys. Rev. A} {\bf 57}, 3, 1619-1633 (1998);
quant-ph/9707035.

\item {\bf [Vedral-Plenio 98 b]}:
V. Vedral, \& M. B. Plenio,
``Basics of quantum computation'',
quant-ph/9802065.

\item {\bf [Vedral 99]}:
V. Vedral,
``On bound entanglement assisted distillation'',
{\em Phys. Lett. A} {\bf 262}, 2-3, 121-124 (1999);
quant-ph/9908047.

\item {\bf [Vedral 00]}:
V. Vedral,
``Landauer's erasure, error correction and entanglement'',
{\em Proc. R. Soc. Lond. A} {\bf 456}, 1996, 969-984 (2000).

\item {\bf [Vedral 01]}:
V. Vedral,
``Foundations and interpretation of quantum mechanics'',
{\em Prog. Quant. Electron.} {\bf 25}, 4, 191 (2001).
Review of {\bf [Auletta 00]}.

\item {\bf [Vedral 02 a]}:
V. Vedral,
``The role of relative entropy in quantum information theory'',
{\em Rev. Mod. Phys.} {\bf 74}, 1, 197-234 (2002);
quant-ph/0102094.

\item {\bf [Vedral-Kashefi 02]}:
V. Vedral, \& E. Kashefi,
``Uniqueness of the entanglement measure for bipartite pure states and
thermodynamics'',
{\em Phys. Rev. Lett.} {\bf 89}, 3, 037903 (2002);
quant-ph/0112137.

\item {\bf [Vedral 02 b]}:
V. Vedral,
``Untangling quantum mechanics. [Review of] Entanglement:
The greatest mystery in physics, by Amir Aczel'',
{\em Nature} {\bf 420}, 6913, 271 (2002).
Review of {\bf [Aczel 02]}.

\item {\bf [Vedral 03 a]}:
V. Vedral,
``Classical correlations and entanglement in quantum measurements'',
{\em Phys. Rev. Lett.} {\bf 90}, 5, 050401 (2003).

\item {\bf [Vedral 03 b]}:
V. Vedral,
``Entanglement hits the big time'',
{\em Nature} {\bf 425}, ?, 28-29 (2002).
See {\bf [Ghosh-Rosenbaum-Aeppli-Coppersmith 03]}.

\item {\bf [Vedral 03 c]}:
V. Vedral,
``Mean field approximations and multipartite thermal correlations'',
quant-ph/0312104.

\item {\bf [Vedral 04 a]}:
V. Vedral,
``High temperature macroscopic entanglement'',
quant-ph/0405102.

\item {\bf [Vedral 04 b]}:
V. Vedral,
``The Meissner effect and massive particles as witnesses of macroscopic
entanglement'',
quant-ph/0410021.

\item {\bf [Velleman 98]}:
D. J. Velleman,
``Probability and quantum mechanics'',
{\em Am. J. Phys.} {\bf 66}, 11, 967-969 (1998).


\item {\bf [Ventura-Mart\'{\i}nez 98 a]}:
D. Ventura \& A. Mart\'{\i}nez,
``A quantum computational learning algorithm'',
quant-ph/9807052.

\item {\bf [Ventura-Mart\'{\i}nez 98 b]}:
D. Ventura \& A. Mart\'{\i}nez,
``Quantum associative memory'',
quant-ph/9807053.

\item {\bf [Ventura-Mart\'{\i}nez 98 c]}:
D. Ventura \& A. Mart\'{\i}nez,
``Initializing the amplitude distribution of a quantum state'',
quant-ph/9807054.

\item {\bf [Venugopalan 97]}:
A. Venugopalan,
``Decoherence and Schr\"{o}dinger-cat states in
a Stern-Gerlach-type experiment'',
{\em Phys. Rev. A} {\bf 56}, 5, 4307-4310 (1997).

\item {\bf [Venugopalan 00]}:
A. Venugopalan,
``Pointer states via decoherence in a quantum measurement'',
{\em Phys. Rev. A} {\bf 61}, 1, 012102 (2000);
quant-ph/9909005.

\item {\bf [Vera 00]}:
J. Vera,
``Entrevista con Juan Ignacio Cirac'',
{\em Muy Interesante}, 235, 206-208 (2000).

\item {\bf [Verch-Werner 04]}:
R. Verch, \& R. F. Werner,
``Distillability and positivity of partial transposes in general quantum
field systems'',
quant-ph/0403089.

\item {\bf [Verhulst-Liivak-Sherwood-(+2) 01]}:
A. S. Verhulst, O. Liivak, M. H. Sherwood,
H.-M. Vieth, \& I. L. Chuang,
``Non-thermal nuclear magnetic resonance quantum computing using
hyperpolarized xenon'',
{\em Appl. Phys. Lett.} {\bf 79}, ?, 2480-? (2001).

\item {\bf [Vermaas 94]}:
P. E. Vermaas,
`Comment on ``Getting contextual and
nonlocal elements-of-reality the easy way'',
by Rob Clifton [{\em Am. J. Phys.} {\bf 61}, 443-447 (1993)]',
{\em Am. J. Phys.} {\bf 62}, 7, 658-660 (1994).
Comment on {\bf [Clifton 93]}.

\item {\bf [Vermaas-Dieks 95]}:
P. E. Vermaas, \& D. Dieks,
``The modal interpretation of quantum mechanics and its generalization
to density operators'',
{\em Found. Phys.} {\bf 25}, 1, 145-158 (1995).

\item {\bf [Vermaas 96]}:
P. E. Vermaas,
``Unique transition probabilities in the modal interpretation'',
{\em Stud. Hist. Philos. Sci. Part B: Stud. Hist. Philos. Mod. Phys.}
{\bf 27}, 2, 133-159 (1996).

\item {\bf [Vermaas 97]}:
P. E. Vermaas,
``A no-go theorem for joint property
ascriptions in modal interpretations of quantum mechanics'',
{\em Phys. Rev. Lett.} {\bf 78}, 11, 2033-2037 (1997).

\item {\bf [Vermaas 99 a]}:
P. E. Vermaas,
``Two no-go theorems for modal interpretations of quantum mechanics'',
{\em Stud. Hist. Philos. Sci. Part B: Stud. Hist. Philos. Mod. Phys.}
{\bf 30}, 3, 403-431 (1999).

\item {\bf [Vermaas 99 b]}:
P. E. Vermaas,
{\em A philosopher's understanding of quantum mechanics.
Possibilities and impossibilities of a modal interpretation},
Cambridge University Press, Cambridge, 1999.

\item {\bf [Vernac-Pinard-Giacobino 01]}:
L. Vernac, M. Pinard, \& E. Giacobino,
``Quantum state transfer from light beams to atomic ensembles'',
{\em Eur. Phys. J. D} {\bf 17}, 1, 125-136 (2001).

\item {\bf [Vernam 26 a]}:
G. S. Vernam,
``Cipher printing telegraph systems'',
{\em Transactions of the AIEE} {\bf 45}, 295-301 (1926).
See {\bf [Leung 00]}.

\item {\bf [Vernam 26 b]}:
G. S. Vernam,
``Cipher printing telegraph systems for secret wire and radio
telegraphic communications'',
{\em J. Am. Inst. Electr. Eng.} {\bf 55}, 109-115 (1926).
See {\bf [Leung 00]}.

\item {\bf [Vershik-Cirel'son 92]}:
A. M. Vershik, \& B. S. Cirel'son [Tsirelson],
`Formulation of Bell type problems, and ``noncommutative" convex geometry'',\',
{\em Advances in Soviet Math.} {\bf 9}, 95-114 (1992).

\item {\bf [Verstraete-Dehaene-De Moor 00]}:
F. Verstraete, J. Dehaene, \& B. De Moor,
``Local filtering operations on two qubits'',
{\em Phys. Rev. A} {\bf 62}, 1, 010101(R) (2001);
quant-ph/0011111.

\item {\bf [Verstraete-Audenaert-De Moor 01]}:
F. Verstraete, K. M. R. Audenaert, \& B. De Moor,
``Maximally entangled mixed states of two qubits'',
{\em Phys. Rev. A} {\bf 64}, 1, 012316 (2001);
quant-ph/0011110.

\item {\bf [Verstraete-Audenaert-Dehaene-De Moor 01]}:
F. Verstraete, K. M. R. Audenaert, J. Dehaene, \& B. De Moor,
``A comparison of the entanglement measures negativity and concurrence'',
{\em J. Phys. A} {\bf 34}, 47, 10327-10332 (2001);
quant-ph/0108021.

\item {\bf [Verstraete-Dehaene-De Moor 01]}:
F. Verstraete, J. Dehaene, \& B. De Moor,
``Normal forms, entanglement monotones and
optimal filtration of multipartite quantum systems'',
quant-ph/0105090.

\item {\bf [Verstraete-Dehaene-De Moor 02 a]}:
F. Verstraete, J. Dehaene, \& B. De Moor,
``On the geometry of entangled states'',
{\em Proc.\ ESF QIT Conf.\ Quantum Information: Theory, Experiment and Perspectives
(Gdansk, Poland, 2001)}, {\em J. Mod. Opt.} {\bf 49}, 8, 1277-1287 (2002);
quant-ph/0107155.

\item {\bf [Verstraete-Dehaene-De Moor 02 b]}:
F. Verstraete, J. Dehaene, \& B. De Moor,
``Lorentz singular-value decomposition and its applications to pure states of
three qubits'',
{\em Phys. Rev. A} {\bf 65}, 3, 032308 (2002);
quant-ph/0108043.

\item {\bf [Verstraete-Dehaene-De Moor-Verschelde 02]}:
F. Verstraete, J. Dehaene, B. De Moor, \& H. Verschelde,
``Four qubits can be entangled in nine different ways'',
{\em Phys. Rev. A} {\bf 65}, 5, 052112 (2002);
quant-ph/0109033.

\item {\bf [Verstraete-Verschelde 02 a]}:
F. Verstraete, \& H. Verschelde,
``Fidelity of mixed states of two qubits'',
{\em Phys. Rev. A} {\bf 66}, 2, 022307 (2002);
quant-ph/0203073.
See {\bf [Verstraete-Verschelde 03 a]}.

\item {\bf [Verstraete-Wolf 02]}:
F. Verstraete, \& M. M. Wolf,
``Entanglement versus Bell violations and their
behavior under local filtering operations'',
{\em Phys. Rev. Lett.} {\bf 89}, 17, 170401 (2002);
quant-ph/0112012.

\item {\bf [Verstraete-Verschelde 02 b]}:
F. Verstraete, \& H. Verschelde,
``On one-qubit channels'',
quant-ph/0202124.

\item {\bf [Verstraete-Verschelde 03 a]}:
F. Verstraete, \& H. Verschelde,
``Optimal teleportation with a mixed state of two qubits'',
{\em Phys. Rev. Lett.} {\bf 90}, 9, 097901 (2003);
quant-ph/0303007.
See {\bf [Verstraete-Verschelde 02 a]}.

\item {\bf [Verstraete-Verschelde 03 b]}:
F. Verstraete, \& H. Verschelde,
``Optimal teleportation with a mixed state of two qubits'',
{\em Phys. Rev. Lett.} {\bf 90}, 9, 097901 (2003).

\item {\bf [Verstraete-Cirac 03 a]}:
F. Verstraete, \& J. I. Cirac,
``Quantum nonlocality in the presence of superselection rules and data hiding
protocols'',
{\em Phys. Rev. Lett.} {\bf 91}, 1, 010404 (2003);
quant-ph/0302039.

\item {\bf [Verstraete-Dehaene-De Moor 03]}:
F. Verstraete, J. Dehaene, \& B. De Moor,
``Normal forms and entanglement measures for multipartite quantum states'',
{\em Phys. Rev. A} {\bf 68}, 1, 012103 (2003).

\item {\bf [Verstraete-Cirac 03 b]}:
F. Verstraete, \& J. I. Cirac,
``Valence bond solids for quantum computation'',
quant-ph/0311130.

\item {\bf [Verstraete-Popp-Cirac 04]}:
F. Verstraete, M. Popp, \& J. I. Cirac,
``Entanglement versus correlations in spin systems'',
{\em Phys. Rev. Lett.} {\bf 92}, 2, 027901 (2004);
quant-ph/0307009.

\item {\bf [Verstraete-Mart\'{\i}n Delgado-Cirac 04]}:
F. Verstraete, M. A. Mart\'{\i}n-Delgado, \& J. I. Cirac,
``Diverging entanglement length in gapped quantum spin systems'',
{\em Phys. Rev. Lett.} {\bf 92}, 8, 087201 (2004);
quant-ph/0311087.

\item {\bf [Verstraete-Porras-Cirac 04]}:
F. Verstraete, D. Porras, \& J. I. Cirac,
``DMRG and periodic boundary conditions: A quantum information perspective'',
cond-mat/0404706.

\item {\bf [Verstraete-Cirac-Latorre-(+2) 04]}:
F. Verstraete, J. I. Cirac, J. I. Latorre,
E. Rico, \& M. M. Wolf,
``Renormalization group transformations on quantum states'',
quant-ph/0410227.

\item {\bf [Vervoort 00]}:
L. Vervoort,
``Bell's theorem and nonlinear systems'',
{\em Europhys. Lett.} {\bf 50}, 2, 142-147 (2000);
quant-ph/0003057.

\item {\bf [Viamontes-Markov-Hayes 04]}:
G. F. Viamontes, I. L. Markov, \& J. P. Hayes,
``Is quantum search practical?'',
quant-ph/0405001.

\item {\bf [Vianna-Rabelo-Monken 03]}:
R. O. Vianna, W. R. M. Rabelo, \& C. H. Monken,
``The semi-quantum computer'',
{\em Int. J. Quant. Inf.} {\bf 1}, 2, 279-288 (2003);
quant-ph/0304085.

\item {\bf [Vidal 98]}:
G. Vidal,
``On the characterization of entanglement'',
quant-ph/9807077.

\item {\bf [Vidal-Tarrach 98]}:
G. Vidal, \& R. Tarrach,
``Robustness of entanglement'',
{\em Phys. Rev. A} {\bf 59}, 1, 141-155 (1999);
quant-ph/9806094.

\item {\bf [Vidal-Latorre-Pascual-Tarrach 99]}:
G. Vidal, J. I. Latorre, P. Pascual, \& R. Tarrach,
``Optimal minimal measurements of mixed states'',
{\em Phys. Rev. A} {\bf 60}, 1, 126-135 (1999);
quant-ph/9812068.

\item {\bf [Vidal 99 a]}:
G. Vidal,
``Entanglement of pure states for a single copy'',
{\em Phys. Rev. Lett.} {\bf 83}, 5, 1046-1049 (1999);
quant-ph/9902033.

\item {\bf [Vidal 99 b]}:
G. Vidal,
``Quantum states and entanglement'',
Ph.\ D. thesis, Universitat de Barcelona, 1999.

\item {\bf [Vidal 00 a]}:
G. Vidal,
``Entanglement monotones'',
in V. Bu\v{z}zek, \& D. P. DiVincenzo (eds.),
{\em J. Mod. Opt.} {\bf 47}, 2-3 (Special issue:
Physics of quantum information), 355-376 (2000);
quant-ph/9807077.

\item {\bf [Vidal-D\"{u}r-Cirac 00]}:
G. Vidal, W. D\"{u}r, \& J. I. Cirac,
``Reversible combination of inequivalent kinds of multipartite entanglement'',
{\em Phys. Rev. Lett.} {\bf 85}, 3, 658-661 (2000);
quant-ph/0004009.

\item {\bf [Vidal-Jonathan-Nielsen 00]}:
G. Vidal, D. Jonathan, \& M. A. Nielsen,
``Approximate transformations and robust manipulation of
bipartite pure-state entanglement'',
{\em Phys. Rev. A} {\bf 62}, 1, 012304 (2000);
quant-ph/9910099.

\item {\bf [Vidal 00 b]}:
G. Vidal,
``Optimal local preparation of an arbitrary mixed state
of two qubits: Closed expression for the single-copy case'',
{\em Phys. Rev. A} {\bf 62}, 6, 062315 (2000);
quant-ph/0003002.

\item {\bf [Vidal-Cirac 00]}:
G. Vidal, \& J. I. Cirac,
``Storage of quantum dynamics on quantum states:
A quasi-perfect programmable quantum gate'',
quant-ph/0012067.
See {\bf [Vidal-Masanes-Cirac 02]}.

\item {\bf [Vidal-Cirac 01 a]}:
G. Vidal, \& J. I. Cirac,
``Irreversibility in asymptotic manipulations of entanglement'',
{\em Phys. Rev. Lett.} {\bf 86}, 25, 5803-5806 (2001);
quant-ph/0102036.

\item {\bf [Vidal-Cirac 01 b]}:
G. Vidal, \& J. I. Cirac,
``When only two thirds of the entanglement can be distilled'',
quant-ph/0107051.

\item {\bf [Vidal-Cirac 01 c]}:
G. Vidal, \& J. I. Cirac,
``Optimal simulation of nonlocal Hamiltonians
using local operations and classical communication'',
quant-ph/0108076.

\item {\bf [Vidal-D\"{u}r-Cirac 01]}:
G. Vidal, W. D\"{u}r, \& J. I. Cirac,
``Entanglement cost of antisymmetric states'',
quant-ph/0112131.

\item {\bf [Vidal-Cirac 02 a]}:
G. Vidal, \& J. I. Cirac,
``Irreversibility in asymptotic manipulations of a distillable entangled state'',
{\em Phys. Rev. A} {\bf 65}, 1, 012323 (2002).

\item {\bf [Vidal-Masanes-Cirac 02]}:
G. Vidal, L. Masanes, \& J. I. Cirac,
``Storing quantum dynamics in quantum states: A stochastic programmable gate'',
{\em Phys. Rev. Lett.} {\bf 88}, 4, 047905 (2002);
extension of {\bf [Vidal-Cirac 00]};
quant-ph/0102037.

\item {\bf [Vidal-Werner 02]}:
G. Vidal, \& R. F. Werner,
``Computable measure of entanglement'',
{\em Phys. Rev. A} {\bf 65}, 3, 032314 (2002);
quant-ph/0102117.

\item {\bf [Vidal-Cirac 02 b]}:
G. Vidal, \& J. I. Cirac,
``Catalysis in nonlocal quantum operations'',
{\em Phys. Rev. Lett.} {\bf 88}, 16, 167903 (2002);
quant-ph/0108077.

\item {\bf [Vidal-Hammerer-Cirac 02]}:
G. Vidal, K. Hammerer, \& J. I. Cirac,
``Interaction cost of nonlocal gates'',
{\em Phys. Rev. Lett.} {\bf 88}, 23, 237902 (2002);
quant-ph/0112168.
See {\bf [Haselgrove-Nielsen-Osborne 03 b]}.

\item {\bf [Vidal-D\"{u}r-Cirac 02]}:
G. Vidal, W. D\"{u}r, \& J. I. Cirac,
``Entanglement cost of bipartite mixed states'',
{\em Phys. Rev. Lett.} {\bf 89}, 2, 027901 (2002).

\item {\bf [Vidal-Cirac 02]}:
G. Vidal, \& J. I. Cirac,
``Nonlocal Hamiltonian simulation assisted by local operations and classical
communication'',
{\em Phys. Rev. A} {\bf 66}, 2, 022315 (2002).

\item {\bf [Vidal 02]}:
G. Vidal,
``On the continuity of asymptotic measures of entanglement'',
quant-ph/0203107.

\item {\bf [Vidal-Latorre-Rico-Kitaev 02]}:
G. Vidal, J. I. Latorre, E. Rico, \& A. Kitaev,
``Entanglement in quantum critical phenomena'',
{\em Phys. Rev. Lett.} {\bf 90}, 22, 227902 (2003).

\item {\bf [Vidal 03]}:
G. Vidal,
``Efficient classical simulation of slightly entangled quantum computations'',
{\em Phys. Rev. Lett.} {\bf 91}, 14, 147902 (2003);
quant-ph/0301063.

\item {\bf [Vidal-Dawson 04]}:
G. Vidal, \& C. M. Dawson,
``Universal quantum circuit for two-qubit
transformations with three controlled-NOT gates'',
{\em Phys. Rev. A} {\bf 69}, 1, 010301 (2004);
quant-ph/0307177.

\item {\bf [Vidal 04]}:
G. Vidal,
``Efficient simulation of one-dimensional quantum many-body systems'',
{\em Phys. Rev. Lett.} {\bf 93}, 4, 040502 (2004);
quant-ph/0310089.

\item {\bf [Vidal-Mosseri-Dukelsky 03]}:
J. Vidal, R. Mosseri, \& J. Dukelsky,
``Entanglement in a first order quantum phase transition'',
cond-mat/0312130.

\item {\bf [Vidal-Palacios-Mosseri 04]}:
J. Vidal, G. Palacios, \& R. Mosseri,
``Entanglement in a second-order quantum phase transition'',
{\em Phys. Rev. A} {\bf 69}, 2, 022107 (2004);
cond-mat/0305573

\item {\bf [Vidiella Barranco 99]}:
A. Vidiella-Barranco,
``Entanglement and nonextensive statistics'',
{\em Phys. Lett. A} {\bf 260}, 5, 335-339 (1999);
quant-ph/9909057.

\item {\bf [Vidiella Barranco-Moya Cessa 01]}:
A. Vidiella-Barranco, \& H. Moya-Cessa,
``Nonextensive approach to decoherence in quantum mechanics'',
{\em Phys. Lett. A} {\bf 279}, 1-2, 56-60 (2001).

\item {\bf [Vigier-Dewdney-Holland-Kyprianidis 87]}:
J.-P. Vigier, C. Dewdney, P. R. Holland, \& A. Kyprianidis,
``Causal particle trajectories and the interpretation of quantum mechanics'',
in {\bf [Hiley-Peat 87]}, pp.~169-204.

\item {\bf [Vigier 94]}:
J.-P. Vigier,
``Possible test of the reality of superluminal phase waves and
particle phase space motions in the Einstein-de Broglie-Bohm causal
stochastic interpretation of quantum mechanics'',
{\em Found. Phys.} {\bf 24}, 1, 61-83 (1994).

\item {\bf [Villas B\^{o}as-de Almeida-Moussa 99]}:
C. J. Villas-B\^{o}as, N. G. de Almeida, \& M. H. Y. Moussa,
``Teleportation of a zero- and one-photon
running-wave state by projection synthesis'',
{\em Phys. Rev. A} {\bf 60}, 4, 2759-2763 (1999).

\item {\bf [Villoresi-Tamburini-Aspelmeyer-(+6) 04]}:
P. Villoresi, F. Tamburini, M. Aspelmeyer,
T. Jennewein, R. Ursin, C. Pernechele,
G. Bianco, A. Zeilinger, \& C. Barbieri,
``Space-to-ground quantum-communication using an optical ground station: a
feasibility study'',
quant-ph/0408067.

\item {\bf [Vinduska 92]}:
M. Vinduska,
``Information-theoretic Bell inequalities and
the relative measure of probability'',
{\em Found. Phys.} {\bf 22}, 3, 343-355 (1992).

\item {\bf [Vink 93]}:
J. Vink,
``Quantum mechanics in terms of discrete beables'',
{\em Phys. Rev. A} {\bf 48}, 3, 1808-1818 (1993).

\item {\bf [Viola-Lloyd 98 a]}:
L. Viola, \& S. Lloyd,
``Dynamical suppression of
decoherence in two-state quantum systems'',
{\em Phys. Rev. A} {\bf 58}, 4, 2733-2744 (1998);
quant-ph/9803057.

\item {\bf [Viola-Lloyd 98 b]}:
L. Viola, \& S. Lloyd,
``Decoherence control in quantum information processing: Simple models'',
{\em 4th Int.\ Conf.\ on Quantum Communication, Measurement and
Computing (Evanston, Illinois, 1998)};
quant-ph/9809058.

\item {\bf [Viola-Knill-Lloyd 99]}:
L. Viola, E. Knill, \& S. Lloyd,
``Dynamical decoupling of open quantum systems'',
{\em Phys. Rev. Lett.} {\bf 82}, 12, 2417-2421 (1999);
quant-ph/9809071.

\item {\bf [Viola-Lloyd-Knill 99]}:
L. Viola, S. Lloyd, \& E. Knill,
``Universal control of decoupled quantum systems'',
{\em Phys. Rev. Lett.} {\bf 83}, 23, 4888-4891 (1999);
quant-ph/9906094.

\item {\bf [Viola-Knill-Lloyd 00]}:
L. Viola, E. Knill, \& S. Lloyd,
``Dynamical generation of noiseless quantum subsystems'',
{\em Phys. Rev. Lett.} {\bf 85}, 16, 3520-3523 (2000).

\item {\bf [Viola-Knill-Laflamme 01]}:
L. Viola, E. Knill, \& R. Laflamme,
``Constructing qubits in physical systems'',
in S. Popescu, N. Linden, \& R. Jozsa (eds.),
{\em J. Phys. A} {\bf 34}, 35
(Special issue: Quantum information and computation), 7067-7080 (2001);
quant-ph/0101090.

\item {\bf [Viola-Fortunato-Pravia-(+3) 01]}:
L. Viola, E. M. Fortunato, M. A. Pravia,
E. Knill, R. Laflamme, \& D. G. Cory,
``Experimental realization of
noiseless subsystems for quantum information processing'',
{\em Science} {\bf 293}, ?, 2059-2063 (2001);
quant-ph/0210057.

\item {\bf [Viola 02]}:
L. Viola,
``Quantum control via encoded dynamical decoupling'',
{\em Phys. Rev. A} {\bf 66}, 1, 012307 (2002);
quant-ph/0111167.

\item {\bf [Viola-Knill 03 a]}:
L. Viola, \& E. Knill,
``Robust dynamical decoupling of quantum systems with bounded controls'',
{\em Phys. Rev. Lett.} {\bf 90}, 3, 037901 (2003).

\item {\bf [Viola-Onofrio 03]}:
L. Viola, \& R. Onofrio,
``Contractive Schr\"{o}dinger cat states for a free mass'',
{\em New J. Phys.} {\bf 5}, 5.1-5.21, (2003);
quant-ph/0303031.

\item {\bf [Viola-Knill 03 b]}:
L. Viola, \& E. Knill,
``Verification procedures for quantum noiseless subsystems'',
{\em Phys. Rev. A} {\bf 68}, 3, 032311 (2003);
quant-ph/0303165.

\item {\bf [Viola-Barnum-Knill-(+2) 04]}:
L. Viola, H. Barnum, E. Knill, G. Ortiz, \& R. Somma,
``Entanglement beyond subsystems'',
{\em Proc.\ of the Coding Theory and Quantum Computing Workshop},
American mathematical Society, 2004;
quant-ph/0403044.

\item {\bf [Viola 04]}:
L. Viola,
``Advances in decoherence control'',
{\em Proc.\ of the XXXIV Winter Colloquium on the Physics of Quantum Electronics (Snowbird, 2004)},
{\em J. Mod. Opt.},
quant-ph/0404038.

\item {\bf [Vion-Aassime-Cottet-(+5) 02]}:
D. Vion, A. Aassime, A. Cottet,
P. Joyez, H. Pothier, C. Urbina, D. Esteve, \& M. H. Devoret,
``Manipulating the quantum state of an electrical circuit'',
{\em Science} {\bf 296}, ?, 886-889 (2002).

\item {\bf [Virmani-Plenio 00]}:
S. Virmani, \& M. B. Plenio,
``Ordering states with entanglement measures'',
{\em Phys. Lett. A} {\bf 268}, 1-2, 31-34 (2000);
quant-ph/9911119.

\item {\bf [Virmani-Sacchi-Plenio-Markham 01]}:
S. Virmani, M. F. Sacchi, M. B. Plenio, \& D. Markham,
``Optimal local discrimination of two multipartite pure states'',
{\em Phys. Lett. A} {\bf 288}, 2, 62-68 (2001);
quant-ph/0102073.

\item {\bf [Virmani-Plenio 03]}:
S. Virmani, \& M. B. Plenio,
``Construction of extremal local positive-operator-valued measures under
symmetry'',
{\em Phys. Rev. A} {\bf 67}, 6, 062308 (2003);
quant-ph/0212020.

\item {\bf [Virmani-Fern\'{a}ndez Huelga-Plenio 04]}:
S. Virmani, S. G. Fern\'{a}ndez Huelga, \& M. B. Plenio,
``Classical simulatability, entanglement breaking, and quantum computation
thresholds'',
quant-ph/0408076.

\item {\bf [Visser-Eliel-Nienhuis 02]}:
J. Visser, E. R. Eliel, \& G. Nienhuis,
``Polarization entanglement in a crystal with threefold symmetry'',
{\em Phys. Rev. A} {\bf 66}, 3, 033814 (2002).

\item {\bf [Vitali-Tombesi-Milburn 97]}:
D. Vitali, P. Tombesi, \& G. J. Milburn,
``Protecting Schr\"{o}dinger cat states using feedback'',
{\em J. Mod. Opt.} {\bf 44}, 11-12 (Special issue: Quantum
state preparation and measurement), 2033-2041 (1997).

\item {\bf [Vitali-Tombesi 99]}:
D. Vitali, \& P. Tombesi,
``Using parity kicks for decoherence control'',
{\em Phys. Rev. A} {\bf 59}, 6, 4178-4186 (1999).

\item {\bf [Vitali-Fortunato-Tombesi-De Martini 00]}:
D. Vitali, M. Fortunato, P. Tombesi, \& F. De Martini,
``Generating entangled Schr\"{o}dinger cat states within a parametric
oscillator'',
{\em Fortschr. Phys.} {\bf 48}, 5-7, 437-446 (2000).

\item {\bf [Vitali-Fortunato-Tombesi 00]}:
D. Vitali, M. Fortunato, \& P. Tombesi,
``Complete quantum teleportation with a Kerr nonlinearity'',
{\em Phys. Rev. Lett.} {\bf 85}, 2, 445-448 (2000);
quant-ph/0003082.

\item {\bf [Vitali-Tombesi 02]}:
D. Vitali, \& P. Tombesi,
``Heating and decoherence suppression using decoupling techniques'',
{\em Phys. Rev. A} {\bf 65}, 1, 012305 (2002);
quant-ph/0108007.

\item {\bf [Vitali 02]}:
D. Vitali,
``Decoupling methods for heating and decoherence control'',
in R. Bonifacio, \& D. Vitali (eds.),
{\em Mysteries, Puzzles and Paradoxes in Quantum Mechanics IV:
Quantum Interference Phenomena (Gargnano, Italy, 2001)},
{\em J. Opt. B: Quantum Semiclass. Opt.} {\bf 4}, 4, S337-S344 (2002).

\item {\bf [Vitanyi 99]}:
P. Vitanyi,
``Three approaches to the quantitative definition of information
in an individual pure quantum state'',
in {\em Proc.\ 15th IEEE Conf.\ Computational Complexity, 2000};
quant-ph/9907035.

\item {\bf [Vlasov 99 a]}:
A. Y. Vlasov,
``Quantum gates and Clifford algebras'',
quant-ph/9907079.

\item {\bf [Vlasov 99 b]}:
A. Y. Vlasov,
``Error correction with euclidean qubits'',
quant-ph/9911074.

\item {\bf [Vlasov 00 a]}:
A. Y. Vlasov,
``Algebra, logic and qubits: Quantum abacus'',
quant-ph/0001100.

\item {\bf [Vlasov 00 b]}:
A. Y. Vlasov,
``Comment on `Decomposition of pure states of a quantum
register'\,'',
quant-ph/0011017.
Comment on {\bf [Raptis-Zapatrin 00]}.

\item {\bf [Vlasov 01 a]}:
A. Y. Vlasov,
``Clifford algebras and universal sets of quantum gates'',
{\em Phys. Rev. A} {\bf 63}, 5, 054302 (2001);
quant-ph/0010071.

\item {\bf [Vlasov 01 b]}:
A. Y. Vlasov,
``Probabilities, tensors and qubits'',
quant-ph/0104126.

\item {\bf [Vlasov 01 c]}:
A. Y. Vlasov,
``Clifford algebras, noncommutative tori and universal quantum computers'',
quant-ph/0109010.

\item {\bf [Vlasov 02]}:
A. Y. Vlasov,
``Noncommutative tori and universal sets of nonbinary quantum gates'',
{\em J. Math. Phys.} {\bf 43}, 6, 2959-2964 (2002).

\item {\bf [Vlasov 03 a]}:
A. Y. Vlasov,
``Quantum processors and controllers'',
quant-ph/0301147.

\item {\bf [Vlasov 03 b]}:
A. Y. Vlasov,
``From Fermat's last theorem to the quantum computer'',
quant-ph/0307019.

\item {\bf [Vlasov 03 c]}:
A. Y. Vlasov,
``Von Neumann quantum processors'',
{\em 304th WEH Seminar ``Elementary Quantum Processors'' (2003)},
quant-ph/0311196.

\item {\bf [Vlasov 04]}:
A. Y. Vlasov,
``Stairway quantum computer'',
quant-ph/0409216.

\item {\bf [Vogel-Akulin-Schleich 93]}:
K. Vogel, V. M. Akulin, \& W. P. Schleich,
``Quantum state engineering of the radiation field'',
{\em Phys. Rev. Lett.} {\bf 71}, 12, 1816-1819 (1993).

\item {\bf [Vogel 00 a]}:
W. Vogel,
``Nonclassical states: An observable criterion''.
{\em Phys. Rev. Lett.} {\bf 84}, 9, 1849-1852 (2000).
Comment: {\bf [Di\'{o}si 00]}.
Reply: {\bf [Vogel 00 b]}.

\item {\bf [Vogel 00 b]}:
W. Vogel,
``Vogel replies''.
{\em Phys. Rev. Lett.} {\bf 85}, 13, 2842 (2000).
Reply to {\bf [Di\'{o}si 00]}.
See {\bf [Vogel 00 a]}.

\item {\bf [Vogels-Xu-Ketterle 02]}:
J. M. Vogels, K. Xu, \& W. Ketterle,
``Generation of macroscopic pair-correlated atomic
beams by four wave mixing in Bose-Einstein condensates'',
{\em Phys. Rev. Lett.} {\bf 89}, 2, 020401 (2002).

\item {\bf [Vogt 89]}:
A. Vogt,
``Bell's theorem and Mermin's gedanken experiment'',
in M. Kafatos (ed.),
{\em Bell's theorem, quantum theory, and
conceptions of the universe.
Proc.\ of a workshop (George Mason University, 1988)},
Kluwer Academic, Dordrecht, Holland, 1989, pp.~61-63.

\item {\bf [Vollbrecht-Werner 99]}:
K. G. H. Vollbrecht, \& R. F. Werner,
``Why two qubits are special'',
quant-ph/9910064.

\item {\bf [Vollbrecht-Werner 01]}:
K. G. H. Vollbrecht, \& R. F. Werner,
``Entanglement measures under symmetry'',
{\em Phys. Rev. A} {\bf 64}, 6, 062307 (2001);
quant-ph/0010095.

\item {\bf [Vollbrecht-Wolf 02 a]}:
K. G. H. Vollbrecht, \& M. M. Wolf,
``Activating distillation with an infinitesimal amount of bound entanglement'',
{\em Phys. Rev. Lett.} {\bf 88}, 24, 247901 (2002).

\item {\bf [Vollbrecht-Wolf 02 b]}:
K. G. H. Vollbrecht, \& M. M. Wolf,
``Conditional entropies and their relation to entanglement criteria'',
{\em J. Math. Phys.} {\bf 43}, 9, 4299-4306 (2002);
quant-ph/0202058.

\item {\bf [Vollbrecht-Wolf 02 c]}:
K. G. H. Vollbrecht, \& M. M. Wolf,
``Activating NPPT distillation with an infinitesimal amount of bound entanglement'',
quant-ph/0201103.

\item {\bf [Vollbrecht-Wolf 03]}:
K. G. H. Vollbrecht, \& M. M. Wolf,
``Efficient distillation beyond qubits'',
{\em Phys. Rev. A} {\bf 67}, 1, 012303 (2003).

\item {\bf [Vollbrecht-Werner-Wolf 04]}:
K. G. H. Vollbrecht, R. F. Werner, \& M. M. Wolf,
``Irreversibility of entanglement distillation for a class of symmetric states'',
{\em Phys. Rev. A} {\bf 69}, 6, 062304 (2004);
quant-ph/0301072.

\item {\bf [Vollbrecht-Verstraete 04]}:
K. G. H. Vollbrecht, \& F. Verstraete,
``Interpolation of recurrence and hashing entanglement distillation
protocols'',
quant-ph/0404111.

\item {\bf [Vollbrecht-Solano-Cirac 04]}:
K. G. H. Vollbrecht, E. Solano, \& J. I. Cirac,
``Ensemble quantum computation with atoms in periodic potentials'',
quant-ph/0405014.

\item {\bf [Volovich 99]}:
I. V. Volovich,
``Atomic quantum computer'',
quant-ph/9911062.

\item {\bf [Volovich-Volovich 00]}:
I. V. Volovich, \& Y. Volovich,
``Bell's theorem and random variables'',
quant-ph/0009058.

\item {\bf [Volovich 00 a]}:
I. V. Volovich,
``Bell's theorem and locality in space'',
quant-ph/0012010.

\item {\bf [Volovich 00 b]}:
I. V. Volovich,
``An attack to quantum cryptography from space'',
quant-ph/0012054.

\item {\bf [Volovich 01 a]}:
I. V. Volovich,
``Quantum information in space and time'',
quant-ph/0108073.

\item {\bf [Volovich-Volovich 01]}:
I. V. Volovich, \& Y. I. Volovich,
``On classical and quantum cryptography'',
quant-ph/0108133.

\item {\bf [Volovich 01 b]}:
I. V. Volovich,
``Quantum computing and Shor's factoring algorithm'',
quant-ph/0109004.

\item {\bf [Volz-Kurtsiefer-Weinfurter 01]}:
J. Volz, C. Kurtsiefer, \& H. Weinfurter,
``Compact all-solid-state source of polarization-entangled photon pairs'',
{\em Appl. Phys. Lett.} {\bf 79}, 6, 869-871 (2001).

\item {\bf [Vorojtsov-Mucciolo-Baranger 04]}:
S. Vorojtsov, E. R. Mucciolo, \& H. U. Baranger,
``Spin qubits in multielectron quantum dots'',
{\em Phys. Rev. B} {\bf 69}, 11, 115329 (2004).

\item {\bf [Voronov 01]}:
V. V. Voronov,
``The use of NMR for solving the problems of quantum computers'',
quant-ph/0102021.

\item {\bf [Voss 96]}:
D. Voss,
``Approaching the quantum gate'',
{\em Science} {\bf 271}, 5246, 162 (1996).

\item {\bf [von Baeyer 00]}:
H. C. von Baeyer,
``Quantum mechanics at the end of the 20th century'',
{\em Science} {\bf 287}, 5460, 1935 (2000).
Review of {\bf [Treiman 99]}.

\item {\bf [von Baeyer 01]}:
H. C. von Baeyer,
``In the beginning was the bit'',
{\em New Scientist} (17 February) 2001.

\item {\bf [von Borzeszkowski-Mensky 00]}:
H.-H. von Borzeszkowski, \& M. B. Mensky,
``EPR effect in gravitational field: Nature of non-locality'',
{\em Phys. Lett. A} {\bf 269}, 4, 197-203 (2000);
quant-ph/0007085.

\item {\bf [von Borzeszkowski 00]}:
H.-H. von Borzeszkowski,
``Book review. Quantum measurements and decoherence: Models and phenomenology'',
{\em Found. Phys.} {\bf 30}, 11, 1991-1993 (2000).
Review of {\bf [Mensky 00]}.

\item {\bf [von Borzeszkowski-Mensky 01]}:
H.-H. von Borzeszkowski, \& M. B. Mensky,
``Gravitational effects on entangled states and
interferometer with entangled atoms'',
{\em Phys. Lett. A} {\bf 286}, 2-3, 102-106 (2001).

\item {\bf [Von Korff-Kempe 04]}:
J. Von Korff, \& J. Kempe,
``Quantum color-coding is better'',
quant-ph/0405086.

\item {\bf [von Meyenn-Schucking 01]}:
K. von Meyenn, \& E. Schucking,
``Wolfgang Pauli'',
{\em Phys. Today} {\bf 54}, 2, ?-? (2001).

\item {\bf [von Neumann 31]}:
J. von Neumann,
``\"{U}ber Funktionen von Funktionaloperatoren'',
{\em Ann. of Math.} {\bf 32}, 191-226 (1931).
Reprinted in
{\bf [von Neumann 61]}, vol. 2, pp.~177-212.
See {\bf [von Neumann 32]},
Chap.~II, Sec.~10, pp.~124-127 in the Spanish version; {\bf [Neumark 54]},
{\bf [Park-Margenau 68]}.

\item {\bf [von Neumann 32]}:
J. von Neumann,
{\em Mathematische Grundlagen der Quantenmechanik},
Springer-Verlag, Berlin, 1932.
Spanish version:
{\em Fundamentos matem\'{a}ticos de la mec\'{a}nica cu\'{a}ntica},
C. S. I. C., Madrid, 1949, 1991.
English version: {\em Mathematical foundations of quantum mechanics},
Princeton University Press, Princeton, New Jersey, 1955.

\item {\bf [von Neumann 61]}:
J. von Neumann,
{\em Collected works},
A. H. Taub (ed.),
Pergamon Press, New York, 1961.

\item {\bf [von Weizs\"{a}cker 71]}:
C. F. von Weizs\"{a}cker,
``The Copenhagen interpretation'',
in E. Bastin (ed.),
{\em Quantum theory and beyond},
Cambridge University Press, Cambridge, 1971, pp.~25-32.

\item {\bf [Vourdas 02]}:
A. Vourdas,
``Coding with finite quantum systems'',
{\em Phys. Rev. A} {\bf 65}, 4, 042321 (2002).

\item {\bf [Vourdas 03]}:
A. Vourdas,
``Factorization in finite quantum systems'',
{\em J. Phys. A} {\bf 36}, 20, 5645-5653 (2003).

\item {\bf [Vourdas-Tsomokos-Chong 04]}:
A. Vourdas, D. I. Tsomokos, \& C. C. Chong,
``Photon-induced entanglement of distant mesoscopic SQUID rings'',
in {\em Macroscopic Quantum Coherence and Computing
(Napoli, Italy, 2004)},
Kluwer, Dordrecht, Holland, 2004;
quant-ph/0406059.

\item {\bf [Vrijen-Yablonovitch-Wang-(+5) 00]}:
R. Vrijen, E. Yablonovitch, K. Wang, H. W. Jiang,
A. Balandin, V. Roychowdhury, T. Mor, \& D. P. DiVincenzo,
``Electron-spin-resonance transistors for quantum computing
in silicon-germanium heterostructures'',
{\em Phys. Rev. A} {\bf 62}, 1, 012306 (2000);
quant-ph/9905096.

\item {\bf [Vuji\v{c}i\'{c}-Herbut 84]}:
M. Vuji\v{c}i\'{c}, \& F. Herbut,
``A quantum-mechanical theory of distant correlations'',
{\em J. Math. Phys.} {\bf 25}, 7, 2253-2259 (1984).

\item {\bf [Vukics-Janszky-Kobayashi 02]}:
A. Vukics, J. Janszky, \& T. Kobayashi,
``Nonideal teleportation in coherent-state basis'',
{\em Phys. Rev. A} {\bf 66}, 2, 023809 (2002).

\item {\bf [Vu\v{s}kovi\'{c}-Arsenovi\'{c}-Bo\v{z}i\'{c} 02]}:
L. Vu\v{s}kovi\'{c}, D. Arsenovi\'{c}, \& M. Bo\v{z}i\'{c},
``Non-classical behavior of atoms in an interferometer'',
{\em Found. Phys.} {\bf 32}, 9, 1329-1346 (2002).
quant-ph/0105129.


\newpage

\subsection{}


\item {\bf [Wagh 99]}:
A. G. Wagh,
``Neutron interferometry was the first to correlate
complementarity and `which-way' information'',
{\em Phys. Lett. A} {\bf 259}, 2, 81-82 (1999).

\item {\bf [Waks-Zeevi-Yamamoto 02]}:
E. Waks, A. Zeevi, \& Y. Yamamoto,
``Security of quantum key distribution with entangled photons against
individual attacks'',
{\em Phys. Rev. A} {\bf 65}, 5, 052310 (2002);
quant-ph/0012078.

\item {\bf [Waks-Santori-Yamamoto 02]}:
E. Waks, C. Santori, \& Y. Yamamoto,
``Security aspects of quantum key distribution with sub-Poisson light'',
{\em Phys. Rev. A} {\bf 66}, 4, 042315 (2002).

\item {\bf [Waks-Inoue-Santori-(+4) 02]}:
E. Waks, K. Inoue, C. Santori,
D. Fattal, J. Vuckovic, G. S. Solomon, \& Y. Yamamoto,
``Secure communication: Quantum cryptography with a photon turnstile'',
{\em Nature} {\bf 420}, 6917, 762 (2002).

\item {\bf [Waks-Diamanti-Sanders-(+2) 03]}:
E. Waks, E. Diamanti, B. C. Sanders,
S. D. Bartlett, \& Y. Yamamoto,
``Direct observation of non-classical photon statistics in parametric
downconversion'',
quant-ph/0307162.

\item {\bf [Waks-Inoue-Diamanti-Yamamoto 03]}:
E. Waks, K. Inoue, E. Diamanti, \& Y. Yamamoto,
``High efficiency photon number detection for quantum information
processing'',
quant-ph/0308054.

\item {\bf [Waks-Diamanti-Yamamoto 03]}:
E. Waks, E. Diamanti, \& Y. Yamamoto,
``Generation of photon number states'',
quant-ph/0308055.

\item {\bf [Walborn-Terra Cunha-P\'{a}dua-Monken 02]}:
S. P. Walborn, M. O. Terra Cunha, S. P\'{a}dua, \& C. H. Monken,
``Double-slit quantum eraser'',
{\em Phys. Rev. A} {\bf 65}, 3, 033818 (2002);
quant-ph/0106078.
See {\bf [Walborn-Terra Cunha-P\'{a}dua-Monken 03]}.

\item {\bf [Walborn-Terra Cunha-P\'{a}dua-Monken 03]}:
S. P. Walborn, M. O. Terra Cunha, S. P\'{a}dua, \& C. H. Monken,
``Quantum erasure'',
{\em Am. Scientist} {\bf 91}, 336-343 (2003).
Spanish version: ``Borrado cu\'{a}ntico'',
{\em Investigaci\'{o}n y Ciencia} 329, 59-67 (2004).
See {\bf [Walborn-Terra Cunha-P\'{a}dua-Monken 02]}.

\item {\bf [Walborn-de Oliveira-P\'{a}dua-Monken 03]}:
S. P. Walborn, A. N. de Oliveira, S. P\'{a}dua, \& C. H. Monken,
``Multimode Hong-Ou-Mandel interference'',
{\em Phys. Rev. Lett.} {\bf 90}, 14, 143601 (2003).

\item {\bf [Walborn-P\'{a}dua-Monken 03]}:
S. P. Walborn, S. P\'{a}dua, \& C. H. Monken,
``Hyperentanglement-assisted Bell-state analysis'',
{\em Phys. Rev. A} {\bf 68}, 4, 042313 (2003);
quant-ph/0307212.

\item {\bf [Walborn-de Oliveira-Thebaldi-Monken 04]}:
S. P. Walborn, A. N. de Oliveira, R. S. Thebaldi, \& C. H. Monken,
``Entanglement and conservation of orbital angular momentum in spontaneous parametric down-conversion'',
{\em Phys. Rev. A} {\bf 69}, 2, 023811 (2004).

\item {\bf [Walborn-P\'{a}dua-Monken 04]}:
S. P. Walborn, S. P\'{a}dua, \& C. H. Monken,
``Conservation and entanglement of Hermite-Gaussian modes in parametric
down-conversion'',
quant-ph/0407216.

\item {\bf [Walgate-Short-Hardy-Vedral 00]}:
J. Walgate, A. J. Short, L. Hardy, \& V. Vedral,
``Local distinguishability of multipartite orthogonal quantum states'',
{\em Phys. Rev. Lett.} {\bf 85}, 23, 4972-4975 (2000);
quant-ph/0007098.
See {\bf [Chen-Li 03 a]}.

\item {\bf [Walgate-Hardy 02]}:
J. Walgate, \& L. Hardy,
``Nonlocality, asymmetry, and distinguishing bipartite states'',
{\em Phys. Rev. Lett.} {\bf 89}, 14, 147901 (2002);
quant-ph/0202034.

\item {\bf [Wallace 01 a]}:
D. Wallace,
``Simple computer model for the quantum Zeno effect'',
{\em Phys. Rev. A} {\bf 63}, 2, 022109 (2001).

\item {\bf [Wallace 01 b]}:
D. Wallace,
``Worlds in the Everett interpretation'',
to appear in {\em Stud. Hist. Philos. Sci. Part B:
Stud. Hist. Philos. Mod. Phys.};
quant-ph/0103092.

\item {\bf [Wallace 01 c]}:
D. Wallace,
``Everett and structure'' (2001),
quant-ph/0107144,
PITT-PHIL-SCI00000681.

\item {\bf [Wallace 03]}:
D. Wallace,
``Quantum probability and decision theory, revisited'',
{\em Stud. Hist. Philos. Sci. Part B: Stud. Hist. Philos. Mod. Phys.};
quant-ph/0303050,
PITT-PHIL-SCI00000885, PITT-PHIL-SCI00001030.

\item {\bf [Wallace 96]}:
P. R. Wallace,
{\em Paradox lost. Images of the quantum},
Springer-Verlag, New York, 1996.

\item {\bf [Wallach 02]}:
N. R. Wallach,
``An unentangled Gleason's theorem'',
in {\bf [Lomonaco-Brandt 02]} 291-298;
quant-ph/0002058.

\item {\bf [Wallentowitz-de Matos Filho-Vogel 97]}:
S. Wallentowitz, R. L. de Matos Filho, \& W. Vogel,
``Determination of entangled quantum states of a trapped atom'',
{\em Phys. Rev. A} {\bf 56}, 2, 1205-1211 (1997).

\item {\bf [Wallentowitz-Walmsley-Eberly 00]}:
S. Wallentowitz, I. A. Walmsley, \& J. H. Eberly,
``How big is a quantum computer?'',
quant-ph/0009069.

\item {\bf [Walls-Milburn 95]}:
D. F. Walls, \& G. J. Milburn,
{\em Quantum optics},
Springer-Verlag, New York, 1995.

\item {\bf [Walmsley 99]}:
I. A. Walmsley,
``Quantum optics by Marlan O. Scully and M. Suhail Zubairy'',
{\em Am. J. Phys.} {\bf 67}, 7, 648 (1999).
Review of {\bf [Scully-Zubairy 97]}.

\item {\bf [Walther 92]}:
H. Walther,
``Experiments on cavity quantum electrodynamics'',
{\em Phys. Rep.} {\bf 219}, 3-6, 263-281 (1992).

\item {\bf [Walther 97]}:
H. Walther,
``Entanglement and superposition states in the micromaser'',
in P. L. Knight, B. Stoicheff, \& D. Walls (eds.),
{\em Highlight in Quantum Optics},
{\em Philos. Trans. R. Soc. Lond. A} {\bf 355}, 1733, 2343-2351 (1997).

\item {\bf [Walther 98]}:
H. Walther,
``Single atom experiments in cavities and traps'',
in D. P. DiVincenzo. E. Knill, R. Laflamme, \& W. H. Zurek (eds.),
{\em Quantum Coherence and Decoherence.
Proc.\ of the ITP Conf.\ (Santa Barbara, California, 1996)},
{\em Proc. R. Soc. Lond. A} {\bf 454}, 1969, 431-445 (1998).

\item {\bf [Walther-Pan-Aspelmeyer-(+3) 03]}:
P. Walther, J.-W. Pan, M. Aspelmeyer,
R. Ursin, S. Gasparoni, \& A. Zeilinger
``De Broglie wavelength of a nonlocal four-photon'',
quant-ph/0312197.

\item {\bf [Walther-Zeilinger 03]}:
P. Walther, \& A. Zeilinger,
``Experimental realization of a photonic Bell-state analyzer'',
quant-ph/0410244.

\item {\bf [Walther-Fry 97]}:
T. Walther, \& E. S. Fry,
``An experimental realization of Bohm's spin-$\frac{1}{2}$ particle EPR
{\em gedanken} experiment'',
in M. Ferrero, \& A. van der Merwe (eds.),
{\em New developments on fundamental problems
in quantum physics (Oviedo, Spain, 1996)},
Kluwer Academic, Dordrecht, Holland, 1997, pp.~431-439.

\item {\bf [Walton-Sergienko-Atat\"{u}re 01]}:
Z. D. Walton, A. V. Sergienko, M. Atat\"{u}re,
B. E. A. Saleh, \& M. C. Teich,
``Performance of photon-pair quantum key distribution systems'',
quant-ph/0103145.

\item {\bf [Walton-Abouraddy-Sergienko-(+2) 03 a]}:
Z. D. Walton, A. F. Abouraddy, A. V. Sergienko,
B. E. A. Saleh, \& M. C. Teich,
``One-way entangled-photon autocompensating quantum cryptography'',
{\em Phys. Rev. A} {\bf 67}, 6, 062309 (2003);
quant-ph/0207167.

\item {\bf [Walton-Abouraddy-Sergienko-(+2) 03 b]}:
Z. D. Walton, A. F. Abouraddy, A. V. Sergienko,
B. E. A. Saleh, \& M. C. Teich,
``Decoherence-free subspaces in quantum key distribution'',
{\em Phys. Rev. Lett.} {\bf 91}, 8, 087901 (2003);
quant-ph/0304075.

\item {\bf [Walton-Booth-Sergienko-(+2) 03]}:
Z. D. Walton, M. C. Booth, A. V. Sergienko,
B. E. A. Saleh, \& M. C. Teich,
``Controllable frequency entanglement via auto-phase-matched spontaneous
parametric down-conversion'',
{\em Phys. Rev. A} {\bf 67}, 5, 053810 (2003).

\item {\bf [Walton-Sergienko-Levitin-(+2) 03]}:
Z. D. Walton, A. V. Sergienko, L. B. Levitin,
B. E. A. Saleh, \& M. C. Teich,
``Symmetric autocompensating quantum key distribution'',
quant-ph/0309050.

\item {\bf [Walton-Sergienko-Saleh-Teich 04]}:
Z. D. Walton, A. V. Sergienko, B. E. A. Saleh, \& M. C. Teich,
``Generating polarization-entangled photon pairs with arbitrary joint
spectrum'',
quant-ph/0405021.

\item {\bf [Wan 80]}:
K.-K. Wan,
``Superselection rules, quantum measurement and the Schr\"{o}dinger's cat'',
{\em Canadian J. Phys.} {\bf 58}, ?, 976-982 (1980).

\item {\bf [Wan-Bradshaw-Trueman-Harrison 98]}:
K.-K. Wan, J. Bradshaw, C. Trueman, \& F. E. Harrison,
``Classical systems, standard quantum systems, and
mixed quantum systems in Hilbert space'',
{\em Found. Phys.} {\bf 28}, 12, 1739-1784 (1998).

\item {\bf [Wang 00 a]}:
A. M. Wang,
``Improving relative entropy of entanglement'',
quant-ph/0001023.
See {\bf [Wang 02]}.

\item {\bf [Wang 00 b]}:
A. M. Wang,
``Eigenvalues, Peres' separable condition and entanglement'',
quant-ph/0002073.
See {\bf [Peres 96 c]}.

\item {\bf [Wang 00 c]}:
A. M. Wang,
``New construction of an universal quantum network and its
applications'',
quant-ph/0006122.

\item {\bf [Wang 00 d]}:
A. M. Wang,
``Generalization of the entanglement of formation for multi-party
systems'',
quant-ph/0011040.

\item {\bf [Wang 00 e]}:
A. M. Wang,
``Bounds on the generalized entanglement of formation for multi-party
systems'',
quant-ph/0011091.

\item {\bf [Wang 00 f]}:
A. M. Wang,
`` Improved relative entropy of entanglement for multi-party
systems'',
quant-ph/0012029.

\item {\bf [Wang 02 a]}:
A. M. Wang,
``Modification of relative entropy of entanglement'',
quant-ph/0203093.
See {\bf [Wang 00 a]}.

\item {\bf [Wang 02 b]}:
A. M. Wang,
``Separability criterion for pure states in multipartite and high dimensional systems'',
quant-ph/0207136.
See {\bf [Scutaru 03]}.

\item {\bf [Wang 04 a]}:
A. M. Wang,
``Generalized GHZ-class and W-class concurrence and entanglement vectors
of the multipartite systems consisting of qubits'',
quant-ph/0406114.

\item {\bf [Wang-Ruda-Bi 01]}:
J. Wang, H. E. Ruda, \& Q. Bi,
``Decoherence of quantum registers in the weak-coupling limit'',
{\em Phys. Lett. A} {\bf 294}, 1, 6-12 (2002).

\item {\bf [Wang-Kais 04]}:
J. Wang, \& S. Kais,
``Scaling of entanglement at a quantum phase transition for a two-dimensional array of quantum dots'',
{\em Phys. Rev. A} {\bf 70}, 2, 022301 (2004).

\item {\bf [Wang-Wiseman-Milburn 04]}:
J. Wang, H. M. Wiseman, \& G. J. Milburn,
``Dynamical creation of entanglement by homodyne-mediated feedback'',
quant-ph/0409154.

\item {\bf [Wang-Li-Zheng 03]}:
L. Wang, S.-S. Li, \& H.-Z. Zheng,
``Conditions for the local manipulation of tripartite Gaussian states'',
{\em Phys. Rev. A} {\bf 67}, 6, 062317 (2003).

\item {\bf [Wang-Li-Yang-(+3) 03]}:
L. Wang, S.-S. Li, F.-H. Yang,
Z.-C. Niu, S.-L. Feng, \& H.-Z. Zheng,
``Conditions for the local manipulation of all bipartite Gaussian states'',
{\em Phys. Rev. A} {\bf 68}, 2, 020302 (2003).

\item {\bf [Wang-Zou-Mandel 91 a]}:
L. J. Wang, X. Y. Zou, \& L. Mandel,
``Experimental test of the de Broglie guided-wave theory for photons'',
{\em Phys. Rev. Lett.} {\bf 66}, 9, 1111-1114 (1991).
Comment: {\bf [Holland-Vigier 91]}.
Reply: {\bf [Wang-Zou-Mandel 91 b]}.

\item {\bf [Wang-Zou-Mandel 91 b]}:
L. J. Wang, X. Y. Zou, \& L. Mandel,
``Wang. Zou, and Mandel reply'',
{\em Phys. Rev. Lett.} {\bf 67}, 3, 402 (1991).
Reply to {\bf [Holland-Vigier 91]}.
See {\bf [Wang-Zou-Mandel 91 a]}.

\item {\bf [Wang-Hong-Friberg 01]}:
L. J. Wang, C. K. Hong, \& S. R. Friberg,
``Generation of correlated photons via four-wave mixing in optical fibres'',
{\em J. Opt. B: Quantum Semiclass. Opt.} {\bf 3}, 5, 346-352 (2001).

\item {\bf [Wang-Kwek-Oh 00]}:
X.-B. Wang, L. C. Kwek, \& C. H. Oh,
``Quantum roulette: An extended quantum strategy'',
{\em Phys. Lett. A} {\bf 278}, 1-2, 44-46 (2000).

\item {\bf [Wang 03]}:
X.-B. Wang,
``Security proof for quantum key distribution in decoherence-free subspace'',
quant-ph/0308092.

\item {\bf [Wang-Shi-Tomita-Matsumoto 04]}
X.-B. Wang, B. S. Shi, A. Tomita, \& K. Matsumoto,
``Quantum entanglement swapping with spontaneous parametric down-conversion'',
{\em Phys. Rev. A} {\bf 69}, 1, 014303 (2004).

\item {\bf [Wang 04 b]}:
X.-B. Wang,
``Quantum key distribution with two-qubit quantum codes'',
{\em Phys. Rev. Lett.} {\bf 92}, 7, 077902 (2004).

\item {\bf [Wang 04 c]}:
X.-B. Wang,
``Quantum error-rejection code with spontaneous parametric down-conversion'',
{\em Phys. Rev. A} {\bf 69}, 2, 022320 (2004).

\item {\bf [Wang-S{\o}rensen-M{\o}lmer 00]}:
X.-G. Wang, A. S. S{\o}rensen, \& K. M{\o}lmer,
``Multibit gates for quantum computing'',
{\em Phys. Rev. Lett.} {\bf 86}, 17, 3907-3910 (2001);
quant-ph/0012055.

\item {\bf [Wang 01 a]}:
X.-G. Wang,
``Effects of anisotropy on thermal entanglement'',
{\em Phys. Lett. A} {\bf 281}, 2-3, 101-104 (2001).

\item {\bf [Wang 01 b]}:
X.-G. Wang,
``Quantum teleportation of entangled coherent states'',
{\em Phys. Rev. A} {\bf 64}, 2, 022302 (2001);
quant-ph/0102048.

\item {\bf [Wang-Fu-Solomon 01]}:
X.-G. Wang, H. Fu, \& A. I. Solomon,
``Thermal entanglement in three-qubit Heisenberg models'',
quant-ph/0105075.

\item {\bf [Wang 01 c]}:
X.-G. Wang,
``Continuous-variable and hybrid quantum gates'',
{\em J. Phys. A} {\bf 34}, 44, 9577-9584 (2001).

\item {\bf [Wang 01 d]}:
X.-G. Wang,
``Entanglement in the quantum Heisenberg $XY$ model'',
quant-ph/0101013.

\item {\bf [Wang 01 e]}:
X.-G. Wang,
``Maximal entanglement of nonorthorgonal states'',
quant-ph/0102011.

\item {\bf [Wang 01 f]}:
X.-G. Wang,
``Violation of Bell inequality for thermal states
of interaction qubits via a multi-qubit Heisenberg model'',
quant-ph/0111165.

\item {\bf [Wang-Sanders 02]}:
X. Wang, \& B. C. Sanders,
``Multipartite entangled coherent states'',
{\em Phys. Rev. A} {\bf 65}, 1, 012303 (2002);
quant-ph/0104011.

\item {\bf [Wang 02 a]}:
X. Wang,
``Bipartite entangled non-orthogonal states'',
{\em J. Phys. A} {\bf 35}, 1, 165-174 (2002).

\item {\bf [Wang-M{\o}lmer 02]}:
X.-G. Wang, \& K. M{\o}lmer,
``Pairwise entanglement in symmetric multi-qubit systems'',
{\em Eur. Phys. J. D} {\bf 18}, 3, 385-391 (2002);
quant-ph/0106145.

\item {\bf [Wang-Zanardi 02 a]}:
X.-G. Wang, \& P. Zanardi,
``Quantum entanglement and Bell inequalities in Heisenberg spin chains'',
{\em Phys. Lett. A} {\bf 301}, 1-2, 1-6 (2002);
quant-ph/0202108.

\item {\bf [Wang-Zanardi 02 b]}:
X.-G. Wang, \& P. Zanardi,
``Quantum entanglement of unitary operators on bipartite systems'',
{\em Phys. Rev. A} {\bf 66}, 4, 044303 (2002).

\item {\bf [Wang 02 b]}:
X.-G. Wang,
``Thermal and ground-state entanglement in Heisenberg $XX$ qubit rings'',
{\em Phys. Rev. A} {\bf 66}, 3, 034302 (2002);
quant-ph/0203141.

\item {\bf [Wang 02 c]}:
X.-G. Wang,
``Violation of the Bell inequality for thermal states
of interaction qubits investigated via a multi-qubit Heisenberg model'',
{\em New J. Phys.} {\bf 4}, 11.1-11.9 (2002).

\item {\bf [Wang 02 d]}:
X.-G. Wang,
``Threshold temperature for pairwise and many-particle thermal entanglement in
the isotropic Heisenberg model'',
{\em Phys. Rev. A} {\bf 66}, 4, 044305 (2002).

\item {\bf [Wang-Feng-Sanders 03]}:
X.-G. Wang, M. Feng, \& B. C. Sanders,
``Multipartite entangled states in coupled quantum dots and cavity QED'',
{\em Phys. Rev. A} {\bf 67}, 2, 022302 (2003).

\item {\bf [Wang-Sanders-Berry 03]}:
X.-G. Wang, B. C. Sanders, \& D. W. Berry,
``Entangling power and operator entanglement in qudit systems'',
{\em Phys. Rev. A} {\bf 67}, 4, 042323 (2003).

\item {\bf [Wang-Sanders 03 a]}:
X.-G. Wang, \& B. C. Sanders,
``Entanglement capability of a self-inverse Hamiltonian evolution'',
{\em Phys. Rev. A} {\bf 68}, 1, 014301 (2003).

\item {\bf [Wang-Sanders 03 b]}:
X.-G. Wang, \& B. C. Sanders,
``Spin squeezing and pairwise entanglement for symmetric multiqubit states'',
{\em Phys. Rev. A} {\bf 68}, 1, 012101 (2003).

\item {\bf [Wang-Sanders 03 c]}:
X.-G. Wang, \& B. C. Sanders,
``Relations between bosonic quadrature squeezing and atomic spin squeezing'',
{\em Phys. Rev. A} {\bf 68}, 3, 033821 (2003).

\item {\bf [Wang-Ghose-Sanders-Hu 04]}:
X.-G. Wang, S. Ghose, B. C. Sanders, \& B. Hu,
``Entanglement as a signature of quantum chaos'',
{\em Phys. Rev. E} {\bf 70}, 1, 016217 (2004);
quant-ph/0312047.

\item {\bf [Wang 04 d]}
X.-G. Wang,
``Analytical results for entanglement in the five-qubit anisotropic Heisenberg model'',
{\em Phys. Lett. A} {\bf 329}, 6, 439-444 (2004).

\item {\bf [Wang 04 e]}
X.-G. Wang,
``Entanglement and spin squeezing in the three-qubit transverse Ising model'',
{\em Phys. Lett. A} {\bf 331}, 3-4, 164-169 (2004).

\item {\bf [Wang-Sanders 04]}
X.-G. Wang, \& B. C. Sanders,
``Canonical entanglement for two indistinguishable particle'',
quant-ph/0409200.

\item {\bf [Wang-Kuang 03]}:
Y.-H. Wang, \& L.-M. Kuang,
``Nonmaximally-entangled-state quantum photolithography'',
{\em J. Opt. B: Quantum Semiclass. Opt.} {\bf 5}, ?, 405-408 (2003);
quant-ph/0402059.

\item {\bf [Wang-Zhang-Li-Guo 00]}:
Z. Wang, C. Zhang, C.-F. Li, \& G.-C. Guo,
``Universal quantum entanglement concentration gate'',
quant-ph/0001060.

\item {\bf [Wang-Feng-Gong-Xu 01]}:
Z.-Y. Wang, X.-L. Feng, S.-Q. Gong, \& Z.-Z. Xu,
``Atomic-state teleportation by using a quantum switch'',
{\em Phys. Rev. A} {\bf 63}, 6, 062308 (2001).

\item {\bf [Warren-Gershenfeld-Chuang 97]}:
W. S. Warren, N. Gershenfeld, \& I. L. Chuang,
``The usefulness of NMR quantum computing'',
{\em Science} {\bf 277}, ?, 1688-1690 (1997).

\item {\bf [Warszawski-Wiseman-Mabuchi 02]}:
P. Warszawski, H. M. Wiseman, \& H. Mabuchi,
``Quantum trajectories for realistic detection'',
{\em Phys. Rev. A};
quant-ph/0112129.

\item {\bf [Warszawski-Gambetta-Wiseman 04]}:
P. Warszawski, J. Gambetta, \& H. M. Wiseman,
``Dynamical parameter estimation using realistic photodetection'',
{\em Phys. Rev. A} {\bf 69}, 4, 042104 (2004).

\item {\bf [Watanabe 04]}:
T. Watanabe,
``Locality and orthomodular structure of compound systems'',
{\em J. Math. Phys.} {\bf 45}, ?, 1795-1803 (2004).

\item {\bf [Watrous 00]}:
J. Watrous,
``Quantum algorithms for solvable groups'',
quant-ph/0011023.

\item {\bf [Watrous 01]}:
J. Watrous,
``Quantum simulations of classical random
walks and undirected graph connectivity'',
{\em J. Comp. Syst. Sci.} {\bf 62}, 2, 376–391 (2001).

\item {\bf [Watrous 04]}:
J. Watrous,
``Many copies may be required for entanglement distillation'',
{\em Phys. Rev. Lett.} {\bf 93}, 1, 010502 (2004);
quant-ph/0312123.

\item {\bf [Watson 95]}:
A. Watson,
``Quantum physics:
`Eraser' rubs out information to reveal light's dual nature'',
{\em Science} {\bf 270}, 5238, 913-914 (1995).

\item {\bf [Watson 96]}:
A. Watson,
``Quantum mechanics:
Physicists trap photons and count them one by one'',
{\em Science} {\bf 272}, 5258, 34 (1996).

\item {\bf [Watson 97 a]}:
A. Watson,
``Quantum spookiness wins, Einstein loses in photon test'',
{\em Science} {\bf 277}, 5325, 481 (1997).

\item {\bf [Watson 97 b]}:
A. Watson,
``Quantum mechanics: Teleportation beams up a photon's state'',
{\em Science} {\bf 278}, 5345, 1881-1882 (1997).

\item {\bf [Watson 98 a]}:
A. Watson,
``Quantum computing: Fighting corruption in the quantum world'',
{\em Science} {\bf 281}, 5384, 1781 (1998).

\item {\bf [Watson 98 b]}:
A. Watson,
``Communications: Quantum encryption takes first step to orbit'',
{\em Science} {\bf 282}, 5389, 605-606 (1998).

\item {\bf [Watson 99 a]}:
A. Watson,
``Quantum mechanics: Entangled trio to put nonlocality to the test'',
{\em Science} {\bf 283}, 5407, 1429 (1999).
See {\bf [Bouwmeester-Pan-Daniell-(+2) 99]},
{\bf [Pan-Bouwmeester-Daniell-(+2) 00]}.

\item {\bf [Watson 99 b]}:
A. Watson,
``Quantum mechanics: Physicists tame a single photon'',
{\em Science} {\bf 285}, 5426, 307-309 (1999).

\item {\bf [Wawer-Keller-Liebman-Mahler 98]}:
R. Wawer, M. Keller, A. Liebman, \& G. Mahler,
``Quantum Zenon effects in composite systems'',
{\em Eur. Phys. J. D} {\bf 1}, 1, 15-28 (1998).

\item {\bf [Weedbrook-Lance-Bowen-(+3) 04]}:
C. Weedbrook, A. M. Lance, W. P. Bowen,
T. Symul, T. C. Ralph, \& P. K. Lam,
``Quantum cryptography without switching'',
{\em Phys. Rev. Lett.} {\bf 93}, 17, 170504 (2004);
quant-ph/0405105.

\item {\bf [Wehner-de Wolf 04]}:
S. Wehner, \& R. de Wolf,
``Improved lower bounds for locally decodable codes and private
information retrieval'',
quant-ph/0403140.

\item {\bf [Wei-Luo-Sun-(+2) 04]}:
D. Wei, J. Luo, X. Sun,
X. Zeng \& M. Liu,
``Three-qubit quantum error-correction scheme based on quantum cloning'',
{\em Phys. Lett. A} {\bf 329}, 4-5, 294-297 (2004).

\item {\bf [Wei-Lei 00]}:
L. F. Wei, \& X. L. Lei,
``Quantum controlled-NOT gates with a single trapped ion'',
{\em J. Opt. B: Quantum Semiclass. Opt.} {\bf 2}, 5, 581-583 (2000).

\item {\bf [Wei-Liu-Lei 02]}:
L. F. Wei, S. Y. Liu, \& X. L. Lei,
``Quantum computation with two-level trapped cold ions beyond Lamb-Dicke limit'',
{\em Phys. Rev. A} {\bf 65}, 6, 062316 (2002).

\item {\bf [Wei-Liu-Nori 04]}:
L. F. Wei, Y.-X. Liu, \& F. Nori,
``Testing Bell's inequality in a capacitively coupled Josephson circuit'',
quant-ph/0408089.

\item {\bf [Wei-Yang-Luo-(+4) 01]}:
D. Wei, X. Yang, J. Luo,
X. Sun, X. Zeng, M. Liu, \& S. Ding,
``Experimental realization of 7-qubit universal perfect
controlled-NOT and controlled square-root NOT gates'',
quant-ph/0109002.

\item {\bf [Wei-Xue-Morgera 98]}:
H. Wei, X. Xue, \& S. D. Morgera,
``Single molecule magnetic resonance and quantum computation'',
quant-ph/9807057.

\item {\bf [Wei-Goldbart 02]}:
T.-C. Wei, \& P. M. Goldbart,
`` Geometric measure of entanglement for multipartite quantum states'',
quant-ph/0212030.

\item {\bf [Wei-Nemoto-Goldbart-(+3) 03]}:
T.-C. Wei, K. Nemoto, P. M. Goldbart,
P. G. Kwiat, W. J. Munro, \& F. Verstraete,
``Maximal entanglement versus entropy for mixed quantum states'',
{\em Phys. Rev. A} {\bf 67}, 2, 022110 (2003).

\item {\bf [Wei-Goldbart 03 a]}:
T.-C. Wei, \& P. M. Goldbart,
``Application of the geometric measure of entanglement to three-qubit
mixed states'',
quant-ph/0303158.

\item {\bf [Wei-Goldbart 03 b]}:
T.-C. Wei, \& P. M. Goldbart,
``Geometric measure of entanglement and applications to bipartite and
multipartite quantum states'',
{\em Phys. Rev. A} {\bf 68}, 4, 042307 (2003);
quant-ph/0307219.

\item {\bf [Wei-Altepeter-Goldbart-Munro 04]}:
T.-C. Wei, J. B. Altepeter, P. M. Goldbart, \& W. J. Munro,
``Measures of entanglement in multipartite bound entangled states'',
{\em Phys. Rev. A} {\bf 70}, ?, ? (2004);
quant-ph/0308031.

\item {\bf [Wei-Ericsson-Goldbart-Munro 04]}:
T.-C. Wei, M. Ericsson, P. M. Goldbart, \& W. J. Munro,
``Connections between relative entropy of entanglement and geometric
measure of entanglement'',
quant-ph/0405002.

\item {\bf [Weigert 92]}:
S. Weigert,
``Pauli problem for a spin of arbitrary length: A
simple method to determine its wave function'',
{\em Phys. Rev. A} {\bf 45}, 11, 7688-7696 (1992).

\item {\bf [Weigert 98]}:
S. Weigert,
``Reconstruction of spin states and its conceptual implications'',
quant-ph/9809065.

\item {\bf [Weigert 01]}:
S. Weigert,
``Quantum parametric resonance'',
quant-ph/0106138.

\item {\bf [Weigert 03 a]}:
S. Weigert,
``The quantum way to diagonalize hermitean matrices'',
{\em Fortschr. Phys.} {\bf 51}, ?, 248-? (2003);
quant-ph/0305140.

\item {\bf [Weigert 03 b]}:
S. Weigert,
``Quantum search for zeros of polynomials'',
quant-ph/0305177.

\item {\bf [Weigert-Busch 03]}:
S. Weigert \& P. Busch,
``L\"{u}ders theorem for coherent-state POVMs'',
quant-ph/0308035.

\item {\bf [Weihs-Reck-Weinfurter-Zeilinger 96 a]}:
G. Weihs, M. Reck, H.
Weinfurter, \& A. Zeilinger,
``Two-photon interference in optical fiber multiports'',
{\em Phys. Rev. A} {\bf 54}, 1, 893-897 (1996).

\item {\bf [Weihs-Reck-Weinfurter-Zeilinger 96 a]}:
G. Weihs, M. Reck, H.
Weinfurter, \& A. Zeilinger,
``All-fiber three-path Mach-Zehnder interferometer'',
{\em Opt. Lett.} {\bf 21}, ?, 302-304 (1996).

\item {\bf [Weihs-Weinfurter-Zeilinger 97]}:
G. Weihs, H. Weinfurter, \& A. Zeilinger,
``Towards a long distance Bell-experiment with independent observers'',
in {\bf [Cohen-Horne-Stachel 97 a]}, pp.~271-280.

\item {\bf [Weihs-Jennewein-Simon-(+3) 98]}:
G. Weihs, T. Jennewein, C. Simon, H. Weinfurter, \& A. Zeilinger,
``Violation of Bell's inequality under strict Einstein locality conditions'',
{\em Phys. Rev. Lett.} {\bf 81}, 23, 5039-5043 (1998);
quant-ph/9810080.
See {\bf [Aspect 99]}, {\bf [Weihs-Jennewein-Simon-(+3) 99]}.

\item {\bf [Weihs-Jennewein-Simon-(+3) 99]}:
G. Weihs, T. Jennewein, C. Simon, H. Weinfurter, \& A. Zeilinger,
``A Bell experiment under strict Einstein locality conditions'',
in {\bf [Greenberger-Reiter-Zeilinger 99]}, pp.~?-?.
See {\bf [Weihs-Jennewein-Simon-(+3) 98]}.

\item {\bf [Weihs 02]}:
G. Weihs,
``Bell's theorem for space-like separation'',
in {\bf [Bertlmann-Zeilinger 02]}, pp.~155-162.

\item {\bf [Weinacht-Ahn-Bucksbaum 99]}:
T. C. Weinacht, J. Ahn, \& P. H. Bucksbaum,
``Controlling the shape of a quantum wavefunction'',
{\em Nature} {\bf 397}, 6716, 233-235 (1999).
See {\bf [Schleich 99]}.

\item {\bf [Weinberg 89 a]}:
S. Weinberg,
``Testing quantum mechanics'',
{\em Ann. Phys.} {\bf 194}, 2, 336-386 (1989).

\item {\bf [Weinberg 89 b]}:
S. Weinberg,
``Precision tests of quantum mechanics'',
{\em Phys. Rev. Lett.} {\bf 62}, 5, 485-488 (1989).
See {\bf [Gisin 90]}, {\bf [Polchinski 91]},
{\bf [Mielnik 00]}.

\item {\bf [Weinberg 89 c]}:
S. Weinberg,
``Weinberg replies'',
{\em Phys. Rev. Lett.} {\bf 63}, 10, 1115 (1989).
Reply to {\bf [Peres 89 d]}.
See {\bf [Weinberg 89 b]}.

\item {\bf [Weinberg 89 d]}:
S. Weinberg,
``?'',
{\em Nucl. Phys. B (Proc.\ Suppl.)} {\bf 6}, ?, 67-? (1989).

\item {\bf [Weinberg 92]}:
S. Weinberg,
{\em Dreams of a final theory:
The scientist's search for the ultimate laws of nature},
?, ?, 1992; Vintage, New York, 1993.
Spanish version: {\em El sue\~{n}o de una teor\'{\i}a final},
Cr\'{\i}tica, Barcelona, 1994.

\item {\bf [Weinfurter 94]}:
H. Weinfurter,
``Experimental Bell-state analysis'',
{\em Europhys. Lett.} {\bf 25}, 8, 559-564 (1994).

\item {\bf [Weinfurter-Herzog-Kwiat-(+3) 95]}:
H. Weinfurter, T. J. Herzog, P. G. Kwiat, J. G. Rarity,
A. Zeilinger, \& M. \.{Z}ukowski,
``Frustrated downconversion: Virtual or real photons?'',
{\em Ann. N. Y. Acad. Sci.} {\bf 755}, 61-72 (1995).

\item {\bf [Weinfurter-\.{Z}ukowski 01]}:
H. Weinfurter, \& M. \.{Z}ukowski,
``Four-photon entanglement from down-conversion'',
{\em Phys. Rev. A} {\bf 64}, 1, 010102(R) (2001);
quant-ph/0103049.

\item {\bf [Weinstein-Lloyd-Cory 99]}:
Y. S. Weinstein, S. Lloyd, \& D. G. Cory,
``Implementation of the quantum Fourier transform'',
quant-ph/9906059.

\item {\bf [Weinstein-Pravia-Fortunato-(+2) 01]}:
Y. S. Weinstein, M. A. Pravia, E. M. Fortunato,
S. Lloyd, \& D. G. Cory,
``Implementation of the quantum Fourier transform'',
{\em Phys. Rev. Lett.} {\bf 86}, 9, 1889-1891 (2001).

\item {\bf [Weinstein-Lloyd-Emerson-Cory 99]}:
Y. S. Weinstein, S. Lloyd, J. Emerson, \& D. G. Cory,
``Experimental implementation of the quantum Baker's map'',
{\em Phys. Rev. Lett.} {\bf 89}, 15, 157902 (2002).

\item {\bf [Weinstein-Hellberg 04]}:
Y. S. Weinstein, \& C. S. Hellberg,
``Quantum-cellular-automata pseudorandom maps'',
{\em Phys. Rev. A} {\bf 69}, 6, 062301 (2004).

\item {\bf [Weinstein-Havel-Emerson-(+4) 04]}:
Y. S. Weinstein, T. F. Havel, J. Emerson,
N. Boulant, M. Saraceno, S. Lloyd, \& D. G. Cory,
``Quantum process tomography of the quantum Fourier transform'',
{\em J. Chem. Phys.};
quant-ph/0406239.

\item {\bf [Weissman 99]}:
M. B. Weissman,
``Emergent measure-dependent probabilities from modified
quantum dynamics without state-vector reduction'',
{\em Found. Phys. Lett.} {\bf 12}, ?, 407-426 (1999);
quant-ph/9906127.

\item {\bf [Wellard-Hollenberg-Pauli 02]}:
C. Wellard, L. C. L. Hollenberg, \& H. C. Pauli,
``Nonadiabatic controlled-NOT gate for the Kane solid-state
quantum computer'',
{\em Phys. Rev. A} {\bf 65}, 3, 032303 (2002).

\item {\bf [Wellard-Orus 03]}:
C. Wellard, \& R. Orus,
``Quantum phase transitions in anti-ferromagnetic planar cubic lattices'',
quant-ph/0401144.

\item {\bf [Wellens-Ku\'{s} 01]}:
T. Wellens, \& M. Ku\'{s},
``Separable approximation for mixed states of composite quantum systems'',
{\em Phys. Rev. A} {\bf 64}, 5, 052302 (2001);
quant-ph/0104098.

\item {\bf [Welsch-Scheel-Chizhov 01]}:
D.-G. Welsch, S. Scheel, \& A. V. Chizhov,
``Entanglement degradation in
continuous-variable quantum teleportation'',
contribution to {\em 7th Int.\
Conf.\ on Squeezed States and Uncertainty Relations ICCSUR
VII (Boston, 2001)};
quant-ph/0105111.

\item {\bf [Wen 02]}:
X. G. Wen,
``Quantum order: a quantum entanglement of many particles'',
{\em Phys. Lett. A} {\bf 300}, 2-3, 175-181 (2002).

\item {\bf [Wenger-Hafezi-Grosshans-(+2) 03]}:
J. Wenger, M. Hafezi, F. Grosshans,
R. Tualle-Brouri, \& P. Grangier,
``Maximal violation of Bell inequalities using continuous-variable
measurements'',
{\em Phys. Rev. A} {\bf 67}, 1, 012105 (2003).

\item {\bf [Wenger-Tualle Brouri-Grangier 04]}:
J. Wenger, R. Tualle-Brouri, \& P. Grangier,
``Non-Gaussian statistics from individual pulses of squeezed light'',
{\em Phys. Rev. Lett.};
quant-ph/0402192.

\item {\bf [Wenger-Tualle Brouri-Grangier 04 b]}:
J. Wenger, R. Tualle-Brouri, \& P. Grangier,
``Pulsed homodyne measurements of femtosecond squeezed pulses generated by
single-pass parametric deamplification'',
{\em Opt. Lett.};
quant-ph/0402193.

\item {\bf [Wenger-Fiur\'{a}\v{s}ek-Tualle Brouri 04]}:
J. Wenger, J. Fiur\'{a}\v{s}ek, R. Tualle-Brouri,
N. J. Cerf, \& P. Grangier,
``Pulsed squeezed vacuum characterization without homodyning'',
quant-ph/0403234.

\item {\bf [Wenger-Ourjoumtsev-Tualle Brouri-Grangier 04]}:
J. Wenger, A. Ourjoumtsev, R. Tualle-Brouri, \& P. Grangier,
``Time-resolved homodyne characterization of individual
quadrature-entangled pulses'',
quant-ph/0409211.

\item {\bf [Werbos-Dolmatova 00]}:
P. J. Werbos, \& L. Dolmatova,
``The backwards-time interpretation of
quantum mechanics - Revisited with experiment'',
quant-ph/0008036.

\item {\bf [Werner-Milburn 93]}:
M. J. Werner, \& G. J. Milburn,
``Eavesdropping using quantum nondemolition measurements'',
{\em Phys. Rev. A} {\bf 47}, 1, 639-641 (1993).

\item {\bf [Werner 86]}:
R. Werner,
``A generalization of stochastic mechanics and its relation to
quantum mechanics'',
{\em Phys. Rev. D} {\bf 34}, 2, 463-469 (1986).

\item {\bf [Werner 89]}:
R. F. Werner,
``Quantum states with Einstein-Podolsky-Rosen correlations
admitting a hidden-variable model'',
{\em Phys. Rev. A} {\bf 40}, 8, 4277-4281 (1989).

\item {\bf [Werner 98]}:
R. F. Werner,
``Optimal cloning of pure states'',
{\em Phys. Rev. A} {\bf 58}, 3, 1827-1832 (1998);
quant-ph/9804001.

\item {\bf [Werner 99]}:
R. F. Werner,
``EPR states for von Neumann algebras'',
quant-ph/9910077.

\item {\bf [Werner-Wolf 00]}:
R. F. Werner, \& M. M. Wolf,
``Bell's inequalities for states with positive partial
transpose'',
{\em Phys. Rev. A} {\bf 61}, 6, 062102 (2000);
quant-ph/9910063.

\item {\bf [Werner-Vollbrecht 00]}:
R. F. Werner, \& K. G. H. Vollbrecht,
``A counterexample to a conjectured entanglement inequality'',
quant-ph/0006046.

\item {\bf [Werner-Wolf 01 a]}:
R. F. Werner, \& M. M. Wolf,
``Bound entangled Gaussian states'',
{\em Phys. Rev. Lett.} {\bf 86}, 16, 3658-3661 (2001);
quant-ph/0009118.

\item {\bf [Werner 01 a]}:
R. F. Werner,
``All teleportation and dense coding schemes'',
in S. Popescu, N. Linden, \& R. Jozsa (eds.),
{\em J. Phys. A} {\bf 34}, 35
(Special issue: Quantum information and computation), 7081-7094 (2001);
quant-ph/0003070.

\item {\bf [Werner 01 b]}:
R. F. Werner,
``Quantum information theory -- An invitation'',
to appear in {\em Quantum information -- An introduction to basic
theoretical concepts and experiments},
Springer-Verlag, New York;
quant-ph/0101061.

\item {\bf [Werner-Wolf 01 b]}:
R. F. Werner, \& M. M. Wolf,
``All-multipartite Bell-correlation inequalities
for two dichotomic observables per site'',
{\em Phys. Rev. A} {\bf 64}, 3, 032112 (2001);
quant-ph/0102024.

\item {\bf [Werner-Wolf 01 c]}:
R. F. Werner, \& M. M. Wolf,
``Bell inequalities and entanglement'',
{\em Quant. Inf. Comp.} {\bf 1}, 1, 1-? (2001);
quant-ph/0107093.

\item {\bf [Werner-Holevo 02]}:
R. F. Werner, \& A. S. Holevo (Kholevo),
``Counterexample to an additivity conjecture for output purity of quantum
channels'',
{\em J. Math. Phys.} {\bf 43}, 9, 4353-4357 (2002);
quant-ph/0203003.

\item {\bf [Werner 04]}:
R. F. Werner,
``The uncertainty relation for joint measurement of position and momentum'',
for A. S. Holevo Festschrift,
quant-ph/0405184.

\item {\bf [Wesenberg-M{\o}lmer 02]}:
J. Wesenberg, \& K. M{\o}lmer,
``Mixed collective states of many spins'',
{\em Phys. Rev. A} {\bf 65}, 6, 062304 (2002);
quant-ph/0201043.

\item {\bf [Wesenberg-M{\o}lmer 03]}:
J. Wesenberg, \& K. M{\o}lmer,
``Robust quantum gates and a bus architecture for quantum computing with
rare-earth-ion-doped crystals'',
{\em Phys. Rev. A} {\bf 68}, 1, 012320 (2003).

\item {\bf [Westmoreland-Schumacher 98]}:
M. D. Westmoreland, \& B. W. Schumacher,
``Quantum entanglement and the nonexistence of superluminal signals'',
quant-ph/9801014.
Comments: {\bf [Mashkevich 98 b]}, {\bf [van Enk 98]}.

\item {\bf [Wharton 98]}:
W. R. Wharton,
``Backward causation and the EPR paradox'',
quant-ph/9810060.

\item {\bf [Wheeler 57]}:
J. A. Wheeler,
`Assessment of Everett's ``relative state''
formulation of quantum theory',
{\em Rev. Mod. Phys.} {\bf 29}, 3, 463-465 (1957).
Reprinted in {\bf [DeWitt-Graham 73]}.
See {\bf [Everett 57 b]}.

\item {\bf [Wheeler 78]}:
J. A. Wheeler,
``?'',
in A. R. Marlow (ed.),
{\em Mathematical Foundations of Quantum Theory
(Loyola University, New Orleans, 1977)},
Academic Press, New York, 1978, pp.~9-48.

\item {\bf [Wheeler 81]}:
J. A. Wheeler,
``Delayed-choice experiments and the Bohr-Einstein dialog'',
{\em American Philosophical Society and the Royal Society: Papers
read at a meeting, June 5, 1980},
Am. Philosophical Society, Philadelphia, Pennsylvania, 1981, pp.~9-40.
Extracts reprinted in {\bf [Wheeler-Zurek 83]}, pp.~182-213.

\item {\bf [Wheeler 83]}:
J. A. Wheeler,
`On recognizing ``Law without law''\,',
{\em Am. J. Phys.} {\bf 51}, ?, 398-? (1983).

\item {\bf [Wheeler-Zurek 83]}:
J. A. Wheeler, \& W. H. Zurek (eds.),
{\em Quantum theory and measurement},
Princeton University Press, Princeton, New Jersey, 1983.

\item {\bf [Wheeler 90]}:
J. A. Wheeler,
``Information, physics, quantum: The search for links'',
in {\bf [Zurek 90]}, pp.~1-28.
Reproduced in T. Hey (ed.),
{\em Feynman and computation},
Perseus, Reading, Massachusetts, 1999, pp.~?-?.

\item {\bf [Wheeler 95]}:
J. A. Wheeler,
``Toward `It from bit'\,'',
in J. Anandan, \& J. L. Safko (eds.),
{\em Quantum coherence and reality. In celebration of the 60th
birthday of Yakir Aharonov.
Int.\ Conf.\ on Fundamental Aspects of Quantum Theory (?, ?)},
World Scientific, Singapore, 1995, pp.~?-?.

\item {\bf [Wheeler-Ford 98]}:
J. A. Wheeler, \& K. W. Ford,
{\em Geons, black holes, and quantum foam: A life in physics},
W. W. Norton, New York, 1998.
Reviews: {\bf [Wilczek 98]}, {\bf [Goldhaber 99]}, {\bf [Rigden 00]}.

\item {\bf [Whitaker 96]}:
A. Whitaker,
{\em Einstein, Bohr and the quantum dilemma},
Cambridge University Press, Cambridge, 1996.
Review: {\bf [Knight 97]}.

\item {\bf [Whitaker 98 a]}:
A. Whitaker,
``John Bell and the most profound discovery of science'',
{\em Phys. World} {\bf 11}, 12, 29-34 (1998).

\item {\bf [Whitaker 02]}:
A. Whitaker,
``John Bell in Belfast: Early years and education'',
in {\bf [Bertlmann-Zeilinger 02]}, pp.~7-20.

\item {\bf [Whitaker-Singh 81]}:
M. A. B. Whitaker, \& I. Singh,
``Use of reduced density matrix for the EPR paradox'',
{\em Phys. Lett. A} {\bf 87}, 1-2, 9-10 (1981).
Comment on {\bf [Cantrell-Scully 78]}.
See {\bf [Whitaker-Singh 82]}.

\item {\bf [Whitaker-Singh 82]}:
M. A. B. Whitaker, \& I. Singh,
``Interpretations
of quantum mechanics and some claimed resolutions of the EPR paradox'',
{\em J. Phys. A} {\bf 15}, 8, 2377-2382 (1982).
See {\bf [Cantrell-Scully 78]}, {\bf [Whitaker-Singh 81]}.

\item {\bf [Whitaker 85]}:
M. A. B. Whitaker,
``The relative states and many-worlds
interpretations of quantum mechanics and the EPR problem'',
{\em J. Phys. A} {\bf 18}, 2, 253-264 (1985).

\item {\bf [Whitaker 89]}:
M. A. B. Whitaker,
``On Squires' many-views interpretation of quantum theory'',
{\em Eur. J. Phys.} {\bf 10}, 1, 73-74 (1989).
See {\bf [Squires 87 a, b]}.

\item {\bf [Whitaker-Dennison 88]}:
M. A. B. Whitaker, \& J. G. Dennison,
``Quantum theory, EPR experiments and locality'',
{\em Phys. Lett. A} {\bf 133}, 9, 466-470 (1988).

\item {\bf [Whitaker 93]}:
M. A. B. Whitaker,
``The quantum watched-pot paradox and
its influence on fundamental aspects of quantum theory'',
in A. van der Merwe, \& F. Selleri (eds.),
{\em Bell's theorem and the foundations of modern physics.
Proc.\ of an international
conference (Cesena, Italy, 1991)},
World Scientific, Singapore, 1993, pp.~470-472.

\item {\bf [Whitaker 98 b]}:
M. A. B. Whitaker,
``Interaction-free measurement and the quantum Zeno effect'',
{\em Phys. Lett. A} {\bf 244}, 6, 502-506 (1998).

\item {\bf [Whitaker 04]}:
M. A. B. Whitaker,
``The EPR paper and Bohr's response: A re-assessment'',
{\em Found. Phys.} {\bf 34}, 9, 1305-1340 (2004).

\item {\bf [White-Mitchell-Nairz-Kwiat 98]}:
A. G. White, J. R. Mitchell, O. Nairz, \& P. G. Kwiat,
`\,``Interaction-free'' imaging',
{\em Phys. Rev. A} {\bf 58}, 1, 605-613 (1998);
quant-ph/9803060.

\item {\bf [White-Kwiat-James 99]}:
A. G. White, P. G. Kwiat, \& D. F. V. James,
`\,``Interaction-free'' measurements of quantum objects?',
in R. Bonifacio (ed.),
{\em Mysteries, Puzzles, and Paradoxes in Quantum Mechanics (Gargnano, Italy, 1998)},
American Institute of Physics, Woodbury, New York, 1999, pp.~268-271.

\item {\bf [White-James-Eberhard-Kwiat 99]}:
A. G. White, D. F. V. James, P. H. Eberhard, \& P. G. Kwiat,
``Nonmaximally entangled states:
Production, characterization, and utilization'',
{\em Phys. Rev. Lett.} {\bf 83}, 16, 3103-3107 (1999);
quant-ph/9908081.

\item {\bf [White-James-Munro-Kwiat 02]}:
A. G. White, D. F. V. James, W. J. Munro, \& P. G. Kwiat,
``Exploring Hilbert space: Accurate characterization of quantum information'',
{\em Phys. Rev. A} {\bf 65}, 1, 012301 (2002);
quant-ph/0108088.

\item {\bf [White-Gilchrist, G. J. Pryde-(+3) 03]}:
A. G. White, A. Gilchrist, G. J. Pryde,
J. L. O'Brien, M. J. Bremner, \& N. K. Langford,
``Measuring controlled-NOT and two-qubit gate operation'',
quant-ph/0308115.

\item {\bf [Wichmann 63]}:
E. H. Wichmann,
``Density matrices arising from incomplete measurements'',
{\em J. Math. Phys.} {\bf 4}, 7, 884-896 (1963).

\item {\bf [Wick 95]}:
D. Wick,
{\em The infamous boundary. Seven decades of
heresy in quantum physics},
Birkh\"{a}uer, Boston, Massachusetts, 1995;
Copernicus Books, Springer-Verlag, New York, 1996.
Reviews: {\bf [Jack 96]}, {\bf [Rae 98]}.

\item {\bf [Wick-Wightman-Wigner 52]}:
G. C. Wick, A. S. Wightman, \& E. P. Wigner,
``The intrinsic parity of elementary particles'',
{\em Phys. Rev.} {\bf 88}, 1, 101-105 (1952).

\item {\bf [Wickes-Alley-Jakubowicz 81]}:
W. C. Wickes, C. O. Alley, \& O. Jakubowicz,
`A ``delayed-choice'' quantum mechanics experiment',
University of Maryland preprint.
Reprinted in {\bf [Wheeler-Zurek 83]}, pp.~457-461 (1981).

\item {\bf [Wickramasekara-Bohm 03]}:
S. Wickramasekara, \& A. Bohm,
``On Einstein Causality and Time Asymmetry in Quantum Physics'',
{\em J. Phys. A} {\bf 35}, ?, L715-? (2003);
quant-ph/0302056.

\item {\bf [Wielinga-Sanders 93]}:
B. Wielinga, \& B. C. Sanders,
``Entangled coherent states with variable weighting'',
{\em J. Mod. Opt.} {\bf 40}, 10, 1923-1937 (1993).

\item {\bf [Wiesner 83]}:
S. Wiesner,
``Conjugate coding'',
{\em SIGACT News} {\bf 15}, 1, 23-27 (1983).

\item {\bf [Wiesner 96]}:
S. Wiesner,
``Simulations of many-body quantum systems by a quantum
computer'',
quant-ph/9603028.
See {\bf [Abrams-Lloyd 97]}.

\item {\bf [Wightman 95]}:
A. S. Wightman,
``Superselection rules; old and new'',
{\em Nuovo Cimento B} {\bf 110}, 5-6, 751-769 (1995).
Also in C. Di Castro, F. Guerra, \& G. Jona-Lasinio (eds.),
{\em Proc.\ of the Int.\ Conf.\ on
Mesoscopic Physics and Fundamental Problems in Quantum Mechanics,
(Rome, 1994)}.

\item {\bf [Wigner 32]}:
E. P. Wigner,
``On the quantum correction for thermodynamic equilibrium'',
{\em Phys. Rev.} {\bf 40}, 5, 749-759 (1932).

\item {\bf [Wigner 52]}:
E. P. Wigner,
``Die Messung quantenmechanischer Operatoren'',
{\em Zeitschrift f\"{u}r Physik} {\bf 133}, ?, 101-108 (1952).

\item {\bf [Wigner 61 a]}:
E. P. Wigner,
``Remarks on the mind-body question'',
in I. J. Good (ed.),
{\em The scientist speculates. An antologhy of partly-baked ideas},
Heinemann, London, 1961 (Basic Books, New York, 1962), pp.~284-302;
Basic Books, New York, 1962, pp.~284-302.
Reprinted in {\bf [Wigner 67]}, pp.~171-184.
Reprinted in {\bf [Wheeler-Zurek 83]}, pp.~168-181.

\item {\bf [Wigner 61 b]}:
E. P. Wigner,
``The probability of the existence of a self-reproducing unit'',
in
{\em The logic of personal knowledge. Essays in honor of Michael Polanyi},
Routledge and Kegan Paul, London, 1961.
Reprinted in {\bf [Wigner 67]}.

\item {\bf [Wigner 63]}:
E. P. Wigner,
``The problem of measurement'',
{\em Am. J. Phys.} {\bf 31}, 1, 6-15 (1963).
Reprinted in {\bf [Wigner 67]}, pp.~153-170, and
in {\bf [Wheeler-Zurek 83]}, pp.~324-341.

\item {\bf [Wigner 67]}:
E. P. Wigner,
{\em Symmetries and reflections}, Indiana
University Press, Bloomington, Indiana, 1967.

\item {\bf [Wigner 70]}:
E. P. Wigner,
``On hidden variables and quantum mechanical probabilities'',
{\em Am. J. Phys.} {\bf 38}, 8, 1005-1009 (1970).
See {\bf [Clauser 71 a, b]}, {\bf [Wigner 71 a]}.

\item {\bf [Wigner 71 a]}:
E. P. Wigner,
``Rejoinder'',
{\em Am. J. Phys.} {\bf 39}, 9, 1097-1098 (1971).
See {\bf [Wigner 70]}, {\bf [Clauser 71 a, b]}.

\item {\bf [Wigner 71 b]}:
E. P. Wigner,
``The subject of our discussions'',
in {\bf [d'Espagnat 71]}, pp.~1-19.

\item {\bf [Wigner 73]}:
E. P. Wigner,
``Epistemological perspective on quantum theory'',
in C. A. Hooker (ed.),
{\em Contemporary
research in the foundations and philosophy of quantum theory
Proc. of a Conf. held at
the University of Western Ontario, London, Canada},
Reidel, Dordrecht, Holland, 1973, pp.~369-385.

\item {\bf [Wigner 76]}:
E. P. Wigner,
``Interpretation of quantum mechanics'',
lectures originally given in the Physics Department of
Princeton University during 1976, as revised for publication, 1981,
{\bf [Wheeler-Zurek 83]}, pp.~260-314.

\item {\bf [Wigner 79]}:
E. P. Wigner,
``Quantum-mechanical distribution functions revisited'',
in W. Yourgrau, \& A. van der Merwe (eds.),
{\em Perspectives in quantum theory},
Dover, New York, 1979, pp.~25-36.

\item {\bf [Wigner 83]}:
E. P. Wigner,
``Review of the quantum measurement problem'',
in P. Meystre, \& M. O. Scully (eds.),
{\em Quantum optics, experimental gravity, and measurement
theory},
Plenum Press, New York, 1983, pp.~43-63.

\item {\bf [Wigner 97 a]}:
E. P. Wigner,
{\em The collected works of Eugene Paul Wigner.
Part A: The scientific papers.
Vol. 3. Part I: Particles and fields.
Part II: Foundations of quantum mechanics},
A. S. Wightman (ed. and annotations of part I), \&
A. Shimony (annotations of part II),
Springer-Verlag, Berlin, 1997.
Review: {\bf [Faris 99]}.

\item {\bf [Wigner 97 b]}:
E. P. Wigner,
{\em The collected works of Eugene Paul Wigner.
Part B: Historical, philosophical, and socio-political papers.
Vol. 6: Philosophical reflections and syntheses},
J. Mehra (ed.), \& G. G. Emch (annotations),
Springer-Verlag, Berlin, 1997.
Review: {\bf [Esfeld 99]}.

\item {\bf [Wilce 04]}:
A. Wilce,
``Topological test spaces'',
{\em Int. J. Theor. Phys.};
quant-ph/0405178.

\item {\bf [Wilczek 98]}:
F. Wilczek,
``Physics: The long life of a thoughtful teacher'',
{\em Science} {\bf 282}, 5396, 1998 (1998).
Review of {\bf [Wheeler-Ford 98]}.

\item {\bf [Wilczek 00]}:
F. Wilczek,
``What is quantum theory?'',
{\em Phys. Today} {\bf 53}, 6, 11-12 (2000).

\item {\bf [Wildfeuer-Schiller 03]}:
C. Wildfeuer, \& D. H. Schiller,
``Generation of entangled $N$-photon states in a two-mode Jaynes-Cummings model'',
{\em Phys. Rev. A} {\bf 67}, 5, 053801 (2003);
quant-ph/0210138.

\item {\bf [Wilhelm 03]}:
F. K. Wilhelm,
``Asymptotic von Neumann measurement strategy for solid-state qubits'',
{\em Phys. Rev. B} {\bf 68}, 6, 060503 (2003).

\item {\bf [Williams-Clearwater 98]}:
C. P. Williams, \& S. H. Clearwater,
{\em Explorations in quantum computing},
Springer-Verlag, New York, 1998.
Review: {\bf [Bennett 99]}.

\item {\bf [Williams 99 a]}:
C. P. Williams (ed.),
{\em Quantum computing and quantum communications},
Springer-Verlag, New York, 1999.
Review: {\bf [Gudder 99 c]}.

\item {\bf [Williams 99 b]}:
C. P. Williams,
{\em Ultimate zero and one: Computing at the quantum frontier},
Copernicus Books, Springer-Verlag, New York, 1999.
Review: {\bf [Gudder 00 b]}.

\item {\bf [Williams 98]}:
R. S. Williams,
``Computing in the 21st century: Nanocircuitry, defect
tolerance and quantum logic'',
in A. K. Ekert, R. Jozsa, \& R. Penrose (eds.),
{\em Quantum Computation: Theory and Experiment.
Proceedings of a Discussion Meeting held at the Royal
Society of London on 5 and 6 November 1997},
{\em Philos. Trans. R. Soc. Lond. A} {\bf 356}, 1743, 1783-1792 (1998).

\item {\bf [Williamson-Vedral 02]}:
M. Williamson, \& V. Vedral,
``Eavesdropping on practical quantum cryptography'',
{\em J. Mod. Opt.};
quant-ph/0211155.

\item {\bf [Wilson-Jeong-Kim 01]}:
D. Wilson, H. Jeong, \& M. S. Kim,
``Quantum nonlocality for a mixed entangled coherent state'',
quant-ph/0109121.

\item {\bf [Wilson-Bushev-Eschner-(+4) 03]}:
M. A. Wilson, P. Bushev, J. Eschner,
F. Schmidt-Kaler, C. Becher, R. Blatt, \& U. Dorner,
``Vacuum-field level shifts in a single trapped ion mediated by a single
distant mirror'',
{\em Phys. Rev. Lett.};
quant-ph/0306159.

\item {\bf [Wineland-Bollinger-Itano-(+2) 92]}:
D. J. Wineland, J. J. Bollinger, W. M. Itano,
F. L. Moore, \& D. J. Heinzen,
``Spin squeezing and reduced quantum noise in spectroscopy'',
{\em Phys. Rev. A} {\bf 46}, 11, R6797-R6800 (1992).

\item {\bf [Wineland-Monroe-Itano-(+5) 98]}:
D. J. Wineland, C. Monroe, W. M. Itano, B. E. King,
D. Leibfried, D. M. Meekhof, C. Myatt, \& C. Wood,
``Experimental primer on the trapped ion quantum computer'',
{\em Fortschr. Phys.} {\bf 46}, 4-5, 363-390 (1998).

\item {\bf [Wineland-Monroe-Meekhof-(+7) 98]}:
D. J. Wineland, C. Monroe, D. M. Meekhof, B. E. King,
D. Leibfried, W. M. Itano, J. C. Bergquist, D. Berkeland,
J. J. Bollinger, \& J. Miller,
``Quantum state manipulation of trapped atomic ions'',
in D. P. DiVincenzo. E. Knill, R. Laflamme, \& W. H. Zurek (eds.),
{\em Quantum Coherence and Decoherence.
Proc.\ of the ITP Conf.\ (Santa Barbara, California, 1996)},
{\em Proc. R. Soc. Lond. A} {\bf 454}, 1969, 411-429 (1998).

\item {\bf [Wineland-Barrett-Britton-(+9) 02]}:
D. J. Wineland, M. Barrett, J. Britton,
J. Chiaverini, B. DeMarco, W. M. Itano,
B. Jelenkovi\'{c}, C. Langer,
D. Leibfried, V. Meyer, T. Rosenband, \& T. Sch\"{a}tz,
``Quantum information processing with trapped ions''`,
{\em Proc.\ of the Discussion Meeting on Practical Realisations of Quantum
Information Processing, held at the Royal Society (London, 2002)},
quant-ph/0212079.

\item {\bf [Winter 99]};
A. Winter,
``Coding theorems of quantum information theory'',
Ph.\ D. thesis, Bielefeld University, 1999;
quant-ph/9907077.

\item {\bf [Winter 01 a]}:
A. Winter,
``On the fidelity of two pure states'',
in S. Popescu, N. Linden, \& R. Jozsa (eds.),
{\em J. Phys. A} {\bf 34}, 35
(Special issue: Quantum information and computation), 7095-7102 (2001);
quant-ph/0011053.

\item {\bf [Winter-Massar 00]};
A. Winter, \& S. Massar,
``Compression of quantum-measurement operations'',
{\em Phys. Rev. A} {\bf 64}, 1, 012311 (2001);
quant-ph/0012128.

\item {\bf [Winter 01 b]}:
A. Winter,
``Extrinsic and intrinsic data in quantum measurements:
Asymptotic convex decomposition of positive operator valued measures'',
quant-ph/0109050.

\item {\bf [Winter 02]}:
A. Winter,
``Scalable programmable quantum gates and a new aspect of the additivity
problem for the classical capacity of quantum channels'',
{\em J. Math. Phys.} {\bf 43}, 9, 4341-4352 (2002);
quant-ph/0108066.

\item {\bf [Winter 04 a]}:
A. Winter,
``Quantum and classical message identification via quantum channels'',
to Holevo's 60th birthday,
quant-ph/0401060.

\item {\bf [Winter 04 b]}:
A. Winter,
``Identification via quantum channels in the presence of prior correlation
and feedback'',
quant-ph/0403203.

\item {\bf [Wiseman-Harrison 95]}:
H. M. Wiseman, \& F. E. Harrison,
``Uncertainty over complementarity?'',
{\em Nature} {\bf 377}, 6550, 584 (1995).
See {\bf [Englert-Fearn-Scully-Walther 94]},
{\bf [Storey-Tan-Collett-Walls 94 a, b, 95]},
{\bf [Englert-Scully-Walther 95]}.

\item {\bf [Wiseman 96]}:
H. M. Wiseman,
``Quantum trajectories and quantum measurement theory'',
{\em Quantum Semiclass. Opt.} {\bf 8}, 205-222 (1996);
quant-ph/0302080.

\item {\bf [Wiseman-Harrison-Collet-(+3) 97]}:
H. M. Wiseman, F. E. Harrison, M. J. Collet,
S. M. Tan, D. F. Walls, \& R. B. Killip,
``Nonlocal momentum transfer in {\em welcher Weg} measurements'',
{\em Phys. Rev. A} {\bf 56}, 1, 55-75 (1997).

\item {\bf [Wiseman 98 a]}:
H. M. Wiseman,
``Bohmian analysis of momentum transfer in welcher Weg measurements'',
{\em Phys. Rev. A} {\bf 58}, 3, 1740-1756 (1998).

\item {\bf [Wiseman 98 b]}:
H. M. Wiseman,
``Extending Heisenberg's measurement-disturbance relation
to the twin-slit case'',
{\em Found. Phys.} {\bf 28}, 11, 1619-1632 (1998).

\item {\bf [Wiseman-M{\o}lmer 00]}:
H. M. Wiseman, \& K. M{\o}lmer,
``Induced coherence with and without induced emission'',
{\em Phys. Lett. A} {\bf 270}, 5, 245-248 (2000).

\item {\bf [Wiseman-Hollis 00]}:
H. M. Wiseman, \& B. L. Hollis,
``Space-bounded computation: Quantum is better than classical'',
quant-ph/0009054.

\item {\bf [Wiseman-Vaccaro 01]}:
H. M. Wiseman, \& J. A. Vaccaro,
``Inequivalence of pure state ensembles for open quantum systems:
The preferred ensembles are those that are physically realizable'',
{\em Phys. Rev. Lett} {\bf 87}, 24, 240402 (2001);
quant-ph/0112115.

\item {\bf [Wiseman 01]}:
H. M. Wiseman,
`Comment on ``Requirement of optical coherence for continuous--
variable quantum teleportation'' by Terry Rudolph and Barry C.
Sanders',
quant-ph/0104004.
Comment on {\bf [Rudolph-Sanders 01]}.

\item {\bf [Wiseman 02]}:
H. M. Wiseman,
``Weak values, quantum trajectories, and the Stony-Brook cavity QED experiment'',
{\em Phys. Rev. A};
quant-ph/0112116.

\item {\bf [Wiseman 03 a]}:
H. M. Wiseman,
``Optical coherence and teleportation: Why a laser is a clock, not a
quantum channel'',
{\em Proc.\ SPIE Conf.\ Fluctuations and Noise (Santa Fe, New Mexico, 2003)};
quant-ph/0303116.
See {\bf [Rudolph-Sanders 01]}, {\bf [van Enk-Fuchs 02 a]}.

\item {\bf [Wiseman 03 b]}:
H. M. Wiseman,
`Directly observing momentum transfer in twin-slit ``which-way'' experiments',
{\em Phys. Lett. A} {\bf 311}, 4-5, 285-291 (2003).

\item {\bf [Wiseman-Vaccaro 03]}:
H. M. Wiseman, \& J. A. Vaccaro,
``Entanglement of indistinguishable particles shared between two parties'',
{\em Phys. Rev. Lett.} {\bf 91}, 9, 097902 (2003);
quant-ph/0210002.

\item {\bf [Wiseman 03 c]}:
H. M. Wiseman,
`Comment on ``Testing integrability with a single bit of quantum
information''\,',
quant-ph/0305153.
Comment on {\bf [Poulin-Laflamme-Milburn-Paz 03]}.

\item {\bf [Wiseman-Bartlett-Vaccaro 03]}:
H. M. Wiseman, S. D. Bartlett, \& J. A. Vaccaro,
``Ferreting out the fluffy bunnies: Entanglement constrained by
generalized superselection rules'',
{\em Proc.\ 16th Int.\ Conf.\ on Laser Spectroscopy (2003)};
quant-ph/0309046.

\item {\bf [Wiseman 04]}:
H. M. Wiseman,
``Defending continuous variable teleportation: Why a laser is a clock, not
a quantum channel'',
{\em J. Opt. B: Quantum Semiclass. Opt..};
quant-ph/0403137.

\item {\bf [Wiseman-Milburn 04]}:
H. M. Wiseman, \& G. J. Milburn,
``All-optical versus electro-optical quantum-limited feedback'',
quant-ph/0409050.

\item {\bf [Witmer 67]}:
W. Witmer,
``Interpretation of quantum mechanics and the future of physics'',
{\em Am. J. Phys.} {\bf 35}, 1, 40-? (1967).

\item {\bf [Witte-Trucks 99]}:
C. Witte, \& M. Trucks,
``A new entanglement measure induced by the Hilbert-Schmidt norm'',
{\em Phys. Lett. A} {\bf 257}, 1-2, 14-20 (1999);
quant-ph/9811027.

\item {\bf [Wocjan-R\"{o}tteler-Janzing-Beth 01]}:
P. Wocjan, M. R\"{o}tteler, D. Janzing, \& T. Beth,
``Universal simulation of Hamiltonians using a finite set of control operations'',
quant-ph/0109063.

\item {\bf [Wocjan-Rotteler-Janzing-Beth 02]}:
P. Wocjan, M. Rotteler, D. Janzing, \& T. Beth,
``Simulating Hamiltonians in quantum networks: Efficient schemes and
complexity bounds'',
{\em Phys. Rev. A} {\bf 65}, 4, 042309 (2002);
quant-ph/0109088.

\item {\bf [Wocjan-Beth 04]}:
P. Wocjan, \& T. Beth,
``New construction of mutually unbiased bases in square dimensions'',
quant-ph/0407081.

\item {\bf [W\'{o}dkiewicz-Scully 90]}:
K. W\'{o}dkiewicz, \& M. O. Scully,
``Weinberg's nonlinear wave mechanics'',
{\em Phys. Rev. A} {\bf 42}, 9, 5111-5116 (1990).

\item {\bf [W\'{o}dkiewicz 92]}:
K. W\'{o}dkiewicz,
``On the Bell's inequalities for arbitrary spin and many particles'',
in T. D. Black, M. Mart\'{\i}n Nieto, H. S.
Pilloff, M. O. Scully, \& R. M. Sinclair (eds.),
{\em Foundations of quantum
mechanics. Workshop (Santa Fe, New Mexico, 1991)},
World Scientific, Singapore, 1992, pp.~276-282.

\item {\bf [W\'{o}dkiewicz-Wang-Eberly 93 a]}:
K. W\'{o}dkiewicz,
L. Wang, \& J. H. Eberly,
``Perfect correlations of three-particle entangled states in cavity QED'',
{\em Phys. Rev. A} {\bf 47}, 4, Part B, 3280-3284 (1993).
See {\bf [W\'{o}dkiewicz-Wang-Eberly 93 b]}.

\item {\bf [W\'{o}dkiewicz-Wang-Eberly 93 b]}:
K. W\'{o}dkiewicz,
L. Wang, \& J. H. Eberly,
``?'',
in F. Ehlotzky (ed.),
{\em Fundamentals of quantum optics III},
Springer-Verlag, Berlin, 1993, pp.~306-?.
See {\bf [W\'{o}dkiewicz-Wang-Eberly 93 a]}.

\item {\bf [W\'{o}dkiewicz 94]}:
K. W\'{o}dkiewicz,
``?'',
{\em Acta Phys. Pol. A} {\bf 86}, ?, 223-? (1994).

\item {\bf [W\'{o}dkiewicz 95 a]}:
K. W\'{o}dkiewicz,
``Randomness, nonlocality, and information in entangled correlations'',
{\em Phys. Rev. A} {\bf 52}, 5, 3503-3510 (1995).

\item {\bf [W\'{o}dkiewicz 95 b]}:
K. W\'{o}dkiewicz,
``Non-local and local ghost fields in quantum correlations'',
{\em Contemp. Phys.} {\bf 36}, 3, 139-147 (1995).

\item {\bf [W\'{o}dkiewicz 02]}:
K. W\'{o}dkiewicz,
``Nonlocality of the Schr\"{o}odinger cat'',
{\em New J. Phys.} {\bf 2}, 21.1-21.8 (2002).

\item {\bf [Woerdeman 03]}:
H. J. Woerdeman,
``Checking $2 \times M$ quantum separability via semidefinite programming'',
{\em Phys. Rev. A} {\bf 67}, 1, 010303 (2003);
quant-ph/0301058.

\item {\bf [Wojcik 03]}:
A. Wojcik,
`Eavesdropping on the ``ping-pong'' quantum communication protocol',
{\em Phys. Rev. Lett.} {\bf 90}, 15, 157901 (2003).
Comment: {\bf [Zhang 04]}.

\item {\bf [Wojcik-Grudka 03]}:
A. Wojcik, \& A. Grudka,
`Comment I on ``Dense coding in entangled states''\,',
{\em Phys. Rev. A} {\bf 68}, 1, 016301 (2003);
Comment on: {\bf [Lee-Ahn-Hwang 02]}.

\item {\bf [Wojcik-Grudka-Chhajlany 03]}:
A. Wojcik, A. Grudka, \& R. W. Chhajlany,
``Generation of inequivalent generalized Bell bases'',
quant-ph/0305034.

\item {\bf [Wojcik-Dorfman 03]}:
D. K. Wojcik, \& J. R. Dorfman,
``Diffusive-ballistic crossover in 1D quantum walks'',
{\em Phys. Rev. Lett.} {\bf 90}, 23, 230602 (2003).

\item {\bf [Wolf-Eisert-Plenio 03]}:
M. M. Wolf, J. Eisert, \& M. B. Plenio,
``Entangling power of passive optical elements'',
{\em Phys. Rev. Lett.} {\bf 90}, 4, 047904 (2003).

\item {\bf [Wolf-Giedke-Kr\"{u}ger-Werner-Cirac 03]}:
M. M. Wolf, G. Giedke, O. Kr\"{u}ger,
R. F. Werner, \& J. I. Cirac,
``Gaussian entanglement of formation'',
{\em Phys. Rev. A} {\bf 69}, 5, 052320 (2004);
quant-ph/0306177.

\item {\bf [Wolf-Cirac 03]}:
M. M. Wolf, \& J. I. Cirac,
``Entanglement frustration for Gaussian states on symmetric graphs'',
quant-ph/0307060.

\item {\bf [Wolf-Verstraete-Cirac 03]}:
M. M. Wolf, F. Verstraete, \& J. I. Cirac,
``Entanglement and frustration in ordered systems'',
{\em Proc.\ QIT-EQIS'03};
quant-ph/0311051.

\item {\bf [Wolf-Verstraete-Cirac 04]}:
M. M. Wolf, F. Verstraete, \& J. I. Cirac,
``Entanglement frustration for Gaussian states on symmetric graphs'',
{\em Phys. Rev. Lett.} {\bf 92}, 8, 087903 (2004).

\item {\bf [de Wolf 02]}:
R. de Wolf,
``Quantum communication and complexity'',
{\em Theoret. Comput. Sci.} {\bf 287}, 1, 337-353 (2002).

\item {\bf [de Wolf 03]}:
R. de Wolf,
``Nondeterministic quantum query and communication complexities'',
{\em Siam J. Comput.} {\bf 32}, 681 (2003).

\item {\bf [Wong-Christensen 00]}:
A. Wong, \& N. Christensen,
``A potential multipartide entanglement measure'',
quant-ph/0010052.

\item {\bf [Woodward-Br\"{u}schweiler 00]}:
F. M. Woodward, \& R. Br\"{u}schweiler,
``Solution of the Deutsch-Josza problem by NMR ensemble
computing without sensitivity scaling'',
quant-ph/0006024.

\item {\bf [Wootters-Zurek 79]}:
W. K. Wootters, \& W. H. Zurek,
``Complementarity in the double-slit experiment:
Quantum nonseparability and a
quantitative statement of Bohr's principle'',
{\em Phys. Rev. D} {\bf 19}, 2, 473-484 (1979).

\item {\bf [Wootters 80]}:
W. K. Wootters,
``The acquisition of information from quantum measurements'',
Ph.\ D. thesis, University of Texas at Austin, 1980.

\item {\bf [Wootters-Zurek 82]}:
W. K. Wootters, \& W. H. Zurek,
``A single quantum cannot be cloned'',
{\em Nature} {\bf 299}, 5886, 802-803 (1982).
Comment: {\bf [Bussey 83]}.
Reply: {\bf [Wootters-Zurek 83]}.
See {\bf [Mandel 83]}.

\item {\bf [Wootters-Zurek 83]}:
W. K. Wootters, \& W. H. Zurek,
``Wootters and Zurek reply'',
{\em Nature} {\bf 304}, 5922, 188-189 (1983).
Reply to {\bf [Bussey 83]}.

\item {\bf [Wootters 88]}:
W. K. Wootters,
``Quantum mechanics without probability amplitudes'',
in {\bf [Zurek-van der Merwe-Miller 88]}, pp.~507-521.

\item {\bf [Wootters-Fields 89]}:
W. K. Wootters, \& B. D. Fields,
``Optimal state-determination by mutually unbiased measurements'',
{\em Ann. Phys.} {\bf 191}, 2, 363-381 (1989).

\item {\bf [Wootters 90 a]}:
W. K. Wootters,
``Random quantum states'',
{\em Found. Phys.} {\bf 20}, ?, 1365-1378 (1990).

\item {\bf [Wootters 90 b]}:
W. K. Wootters,
``Local accessibility of quantum states'',
in {\bf [Zurek 90]}, pp.~39-46.

\item {\bf [Wootters 94]}:
W. K. Wootters,
``Is time asymmetry logically prior to quantum mechanics?'',
in J. Halliwell, J. P\'{e}rez Mercader, \& W. Zurek (eds.),
{\em Physical origins of time asymmetry},
Cambridge University Press, Cambridge, 1994.

\item {\bf [Wootters 98 a]}:
W. K. Wootters,
``Entanglement of formation of an arbitrary state of two qubits'',
{\em Phys. Rev. Lett.} {\bf 80}, 10, 2245-2248 (1998);
quant-ph/9709029.

\item {\bf [Wootters 98 b]}:
W. K. Wootters,
``Quantum entanglement as a quantifiable resource'',
in A. K. Ekert, R. Jozsa, \& R. Penrose (eds.),
{\em Quantum Computation: Theory and Experiment.
Proceedings of a Discussion Meeting held at the Royal
Society of London on 5 and 6 November 1997},
{\em Philos. Trans. R. Soc. Lond. A} {\bf 356}, 1743, 1717-1731 (1998).

\item {\bf [Wootters 01]}:
W. K. Wootters,
``Entanglement of formation and concurrence'',
{\em Quant. Inf. Comp.} {\bf 1}, 1, 27-? (2001).

\item {\bf [Wootters 02 a]}:
W. K. Wootters,
``Parallel transport in an entangled ring'',
{\em J. Math. Phys.} {\bf 43}, 9, 4307-4325 (2002);
quant-ph/0202048.

\item {\bf [Wootters 02 b]}:
W. K. Wootters,
``Entangled chains'',
in {\bf [Lomonaco-Brandt 02]} 299-310.

\item {\bf [Wootters 02 c]}:
W. K. Wootters,
``Entangled chains'',
{\em Contemp. Math.} {\bf 305}, 299-? (2002);
quant-ph/0001114.

\item {\bf [Wootters 03 a]}:
W. K. Wootters,
``Why things fall'',
{\em Found. Phys.} {\bf 33}, 10, 1549-1557 (2003).

\item {\bf [Wootters 03 b]}:
W. K. Wootters,
``Picturing qubits in phase space'',
{\em IBM J. Res. Dev.};
quant-ph/0306135.

\item {\bf [Wu-Shaknov 50]}:
C. S. Wu, \& I. Shaknov,
``The angular correlaton of scattered annihilation radiation'',
{\em Phys. Rev.} {\bf 77}, 1, 136 (1950).

\item {\bf [Wu-Sprung 99]}:
H. Wu, \& D. W. L. Sprung,
``Quantum chaos in terms of Bohm trajectories'',
{\em Phys. Lett. A} {\bf 261}, 3-4, 150-157 (1999).

\item {\bf [Wu 00]}:
J. W. Wu,
``Violation of Bell's inequalities and two-mode
quantum-optical state measurement'',
{\em Phys. Rev. A} {\bf 61}, 2, 022111 (2000).

\item {\bf [Wu-Byrd-Lidar 01]}:
L.-A. Wu, M. S. Byrd, \& D. A. Lidar,
``Testing pairing models on a quantum computer'',
quant-ph/0108110.

\item {\bf [Wu-Lidar 01]}:
L.-A. Wu, \& D. A. Lidar,
``Qubits as parafermions'',
quant-ph/0109078.

\item {\bf [Wu-Lidar 02 a]}:
L.-A. Wu, \& D. A. Lidar,
``Power of anisotropic exchange interactions: Universality and efficient codes
for quantum computing'',
{\em Phys. Rev. A} {\bf 65}, 4, 042318 (2002);
quant-ph/0103039.

\item {\bf [Wu-Lidar 02 b]}:
L.-A. Wu, \& D. A. Lidar,
``Creating decoherence-free subspaces using strong and fast pulses'',
{\em Phys. Rev. Lett.} {\bf 88}, 20, 207902 (2002);
quant-ph/0112144.

\item {\bf [Wu-Byrd-Lidar 02 a]}:
L.-A. Wu, M. S. Byrd, \& D. A. Lidar,
``Polynomial-time simulation of pairing models on a quantum computer'',
{\em Phys. Rev. Lett.} {\bf 89}, 5, 057904 (2002).
Erratum: {\em Phys. Rev. Lett.} {\bf 89}, 13, 139901 (2002).
Publisher's note: {\em Phys. Rev. Lett.} {\bf 89}, 19, 199902 (2002).
Comment: {\bf [Dukelsky-Roman-Sierra 03]}.

\item {\bf [Wu-Byrd-Lidar 02 b]}:
L.-A. Wu, M. S. Byrd, \& D. A. Lidar,
``Efficient universal leakage elimination for physical and encoded qubits'',
{\em Phys. Rev. Lett.} {\bf 89}, 12, 127901 (2002);
quant-ph/0202168.

\item {\bf [Wu-Lidar 02 c]}:
L.-A. Wu, \& D. A. Lidar,
``Qubits as parafermions'',
{\em J. Math. Phys.} {\bf 43}, 9, 4506-4525 (2002).

\item {\bf [Wu-Lidar 02 d]}:
L.-A. Wu, \& D. A. Lidar,
``Universal quantum logic from Zeeman and anisotropic exchange interactions'',
{\em Phys. Rev. A} {\bf 66}, 6, 062314 (2002).

\item {\bf [Wu-Lidar 03]}:
L.-A. Wu, \& D. A. Lidar,
``Universal quantum computation using exchange interactions and measurements
of single- and two-spin observables'',
{\em Phys. Rev. A} {\bf 67}, 5, 050303 (2003).

\item {\bf [Wu-Byrd-Lidar 03]}:
L.-A. Wu, M. S. Byrd, \& D. A. Lidar,
``Wu et al. reply'',
{\em Phys. Rev. Lett.} {\bf 90}, 24, 249804 (2003);
quant-ph/0305159.
Reply to {\bf [Dukelsky-Roman-Sierra 03]}.
See {\bf [Wu-Byrd-Lidar 02 a]}.

\item {\bf [Wu-Lidar 03]}:
L.-A. Wu, \& D. A. Lidar,
``Dressed qubits'',
{\em Phys. Rev. Lett.} {\bf 91}, 9, 097904 (2003);
quant-ph/0303129.

\item {\bf [Wu-Lo-Lidar 03]}:
L.-A. Wu, H.-K. Lo, \& D. A. Lidar,
``Simple solution to loss and decoherence in optical fibers'',
quant-ph/0307178.

\item {\bf [Wu-Lidar-Friesen 03]}:
L.-A. Wu, D. A. Lidar, \& M. Friesen,
``One-spin quantum logic gates from the exchange interaction and a global
magnetic field'',
{\em Phys. Rev. Lett.} {\bf 93}, 3, 030501 (2004);
quant-ph/0310094.

\item {\bf [Wu-Lidar-Friesen 04]}:
L.-A. Wu, D. A. Lidar, \& S. Schneider,
``Long-range entanglement generation via frequent measurements'',
quant-ph/0402209.

\item {\bf [Wu-Sarandy-Lidar 04]}:
L.-A. Wu, M. S. Sarandy, \& D. A. Lidar,
``Quantum phase transitions and bipartite entanglement'',
quant-ph/0407056.

\item {\bf [Wu-Chen-Zhang 00]}:
S. Wu, X. Chen, \& Y. Zhang,
``A necessary and sufficient criterion for multipartite separable
states'',
{\em Phys. Lett. A} {\bf 275}, 4, 244-249 (2000);
quant-ph/0006058.

\item {\bf [Wu-Zhang 00]}:
S. Wu, \& Y. Zhang,
``Calculating the relative entropy of entanglement'',
quant-ph/0004018.

\item {\bf [Wu-Zhang 01]}:
S. Wu \&, Y. Zhang,
``Multipartite pure-state entanglement and the
generalized Greenberger-Horne-Zeilinger states'',
{\em Phys. Rev. A} {\bf 63}, 1, 012308 (2001);
quant-ph/0004020.

\item {\bf [Wu-Anandan 02]}:
S. Wu, \& J. Anandan,
``Some aspects of separability'',
{\em Phys. Lett. A} {\bf 297}, 1-2, 4-8 (2002);
quant-ph/0112177.

\item {\bf [Wu 04]}:
S. Wu,
``The convex sum of product states for a separable state'',
{\em Phys. Lett.} {\bf 321}, 5-6, 301-307 (2004).

\item {\bf [Wu-Wu 04]}:
W. Wu, \& L.-A. Wu,
``Two new types of quantum states generated via higher powers of Bogoliubov's transformation'',
{\em J. Math. Phys.} {\bf 45}, ?, 1752-1761 (2004).

\item {\bf [Wu-Xie 96]}:
X. Wu, \& R. Xie,
``Hardy's nonlocality theorem for three spin-half particles'',
{\em Phys. Lett. A} {\bf 211}, 3, 129-133 (1996).
See {\bf [Pagonis-Clifton 92]}.

\item {\bf [Wu-Xie-Huang-Hsia 96]}:
X. Wu, R. Xie, X. Huang, \& Y. Hsia,
``Local realism violations in two-particle interferometry'',
{\em Phys. Rev. A} {\bf 53}, 4, R1927-R1930 (1996).
Comment: {\bf [Cereceda 97 b]}.
See: {\bf [Cereceda 99 c]}, {\bf [Cabello 00 b]}.

\item {\bf [Wu-Huang 00]}:
X. Wu, \& W. Huang,
``New Bell inequality for the optical test of local hidden-variables
model'',
{\em Phys. Lett. A} {\bf 276}, 1-4, 12-15 (2000).

\item {\bf [Wu-Zong-Pang 00]}:
X.-H. Wu, H.-S. Zong, \& H.-R. Pang,
``Hardy's theorem for Greenberger-Horne-Zeilinger states'',
{\em Phys. Lett. A} {\bf 276}, 5-6, 221-224 (2000).

\item {\bf [Wu-Zong-Pang-Wang 01 a]}:
X.-H. Wu, H.-S. Zong, H.-R. Pang, \& F. Wang,
``A new Bell inequality for two spin-1 particle system'',
{\em Phys. Lett. A} {\bf 281}, 4, 203-206 (2001).

\item {\bf [Wu-Zong-Pang-Wang 01 b]}:
X.-H. Wu, H.-S. Zong, H.-R. Pang, \& F. Wang,
``Bell inequality for Werner states'',
{\em Phys. Rev. A} {\bf 64}, 2, 022103 (2001);
quant-ph/0103087.

\item {\bf [Wu-Zong 03]}:
X.-H. Wu, \& H.-S. Zong,
``A new Bell inequality for three spin-half particle system'',
{\em Phys. Lett. A} {\bf 307}, 5-6, 262-264 (2003).

\item {\bf [Wu-Zong 03]}:
X.-H. Wu, \& H.-S. Zong,
``Violation of local realism by a system with $N$ spin-1/2 particles'',
{\em Phys. Rev. A} {\bf 68}, 3, 032102 (2003).

\item {\bf [Wu-Zong 04]}:
X.-H. Wu, \& H.-S. Zong,
``A new constraint for all Bell inequalities'',
{\em Phys. Lett. A} {\bf 332}, 5-6, 350-354 (2004).

\item {\bf [Wu 01]}:
Y. Wu,
``Elaboration of the Ahn-Weinacht-Bucksbaum scheme for information storage
or retrieval through a quantum phase with a single operation'',
{\em Phys. Rev. A} {\bf 63}, 5, 052303 (2001).

\item {\bf [Wu 99]}:
Y. L. Wu,
`Comment on ``Quantum telegraph: Is it possible?'' by
B. B. Kadomtsev' (Phys. Lett. A {\bf 210} (1996) 371)',
{\em Phys. Lett. A} {\bf 255}, 3, 119-121 (1999).
Comment on {\bf [Kadomtsev 96]}.
Reply: {\bf [Kadomtsev 99]}.

\item {\bf [Wu-Payne-Hagley-Deng 04]}:
Y. Wu, M. G. Payne, E. W. Hagley, \& L. Deng,
``Preparation of multiparty entangled states using pairwise perfectly
efficient single-probe photon four-wave mixing'',
{\em Phys. Rev. A} {\bf 69}, 6, 063803 (2004).

\item {\bf [Wunderlich-Balzer-Toschek 01]}:
C. Wunderlich, C. Balzer, \& P. E. Toschek,
``Evolution of an atom impeded by measurement: The quantum Zeno effect'',
in {\em 3rd Workshop Puzzles, and Paradoxes in Quantum Mechanics)},
{\em Zeitschrift f\"{u}r Naturforschung A} {\bf 56}, 160-164 (2001);
quant-ph/0108040.


\newpage

\subsection{}


\item {\bf [Xi-Hao-Chen-Yue 02]}:
X.-Q. Xi, S.-R. Hao, W.-X. Chen, \& R.-H. Yue,
``Impurity entanglement in three-qubit Heisenberg $XX$ chain'',
{\em Phys. Lett. A} {\bf 297}, 5-6, 291-299 (2002).

\item {\bf [Xi-Chen-Hao-Yue 02]}:
X.-Q. Xi, W.-X. Chen, S.-R. Hao, \& R.-H. Yue,
``Pairwise thermal entanglement in the $n$-qubit ($n \le 5$) Heisenberg
$XX$ chain'',
{\em Phys. Lett. A} {\bf 300}, 6, 567 (2002).

\item {\bf [Xia-Guo 03]}:
Y.-J. Xia, \& G.-C. Guo,
``Squeezing and entanglement in continuous variable systems'',
quant-ph/0307194.

\item {\bf [Xiang Bin-Keiji-Akihisa 01]}:
W. Xiang-Bin, M. Keiji, \& T. Akihisa,
``Detecting the inseparability and distillability of
continuous variable states in Fock space'',
{\em Phys. Rev. Lett.} {\bf 87}, 13, 137903 (2001).

\item {\bf [Xiang Bin 01 a]}:
W. Xiang-Bin,
``A fully efficient secure quantum cryptography protocol'',
quant-ph/0110053.

\item {\bf [Xiang Bin 01 b]}:
W. Xiang-Bin,
``On the role of coherent attacks in a type of
strategic problem related to quantum key distribution'',
quant-ph/0110073.
See {\bf [Cirac-Gisin 97]}.

\item {\bf [Xiang Bin 02 a]}:
W. Xiang-Bin,
``Theorem for the beam-splitter entangler'',
{\em Phys. Rev. A} {\bf 66}, 2, 024303 (2002);
quant-ph/0204039.

\item {\bf [Xiang Bin-Yong-Keiji 02]}:
W. Xiang-Bin, H. W. Yong, \& M. Keiji,
``Arbitrary classical privacy amplification
can be used for the quantum key distribution'',
quant-ph/0201027.

\item {\bf [Xiang Bin 02 b]}:
W. Xiang-Bin,
``Properties of a beam-splitter entangler with Gaussian input states'',
{\em Phys. Rev. A} {\bf 66}, 6, 064304 (2002);
quant-ph/0204082.

\item {\bf [Xiang Bin 03]}:
W. Xiang-Bin,
``Possibility of producing the event-ready two-photon polarization entangled
state with normal photon detectors'',
{\em Phys. Rev. A} {\bf 68}, 4, 042304 (2003);
quant-ph/0208166.

\item {\bf [Xiang Bin-Heng 03]}:
W. Xiang-Bin, \& F. Heng,
``Entanglement concentration by ordinary linear optical devices without postselection'',
{\em Phys. Rev. A} {\bf 68}, 6, 060302 (2003).

\item {\bf [Xiao-Long-Yan-Sun 02]}:
L. Xiao, G. L. Long, H.-Y. Yan, \& Y. Sun,
``Experimental realization of the Br\"{u}schweiler's algorithm in a
homonuclear system'',
{\em J. Chem. Phys.} {\bf 117}, 3310-? (2002);
quant-ph/0112161.

\item {\bf [Xiao-Long 02]}:
L. Xiao, \& G. L. Long,
``Fetching marked items from an unsorted database in NMR ensemble computing'',
{\em Phys. Rev. A} {\bf 66}, 5, 052320 (2002);
quant-ph/0112162.

\item {\bf [Xiao-Long-Deng-Pan 04]}:
L. Xiao, G. L. Long, F.-G. Deng, \& J.-W. Pan,
``Efficient multi-party quantum secret sharing schemes'',
{\em Phys. Rev. A} {\bf 69}, 5, 052307 (2004);
quant-ph/0405179.

\item {\bf [Xiao-Lin-Gao-(+3) 04]}:
Y.-F. Xiao, X.-M. Lin, J. Gao,
Y. Yang, Z.-F. Han, \& G.-C. Guo,
``Realizing quantum controlled phase flip through cavity-QED'',
quant-ph/0408033.

\item {\bf [Xu-Zhou-Zhou-Nie 02]}:
T. Xu, Y. Zhou, Y. Zhou, \& Q. Nie,
``Entanglement of one-dimensional spin chains'',
{\em Phys. Lett. A} {\bf 298}, 4, 219-224 (2002).

\item {\bf [Xue-Du-Zhou-(+2) 03]}:
F. Xue, J. Du, X. Zhou, R. Han, \& J. Wu,
``Experimentally obtaining the likeness of two unknown quantum states on
an NMR quantum information processor'',
{\em Chin. Phys. Lett.} {\bf 20}, 10, 1669-? (2003);
quant-ph/0204049.

\item {\bf [Xue-Chen-Shi-(+3) 03]}:
F. Xue, Z.-B. Chen, M. Shi, X. Zhou, J. Du, \& R. Han,
``Architecture of a deterministic quantum central processing unit'',
{\em Phys. Lett. A} {\bf 312}, 5-6, 301-306 (2003);
quant-ph/0207032.

\item {\bf [Xue-Li-Zhang-Guo 01]}:
P. Xue, C.-F. Li, Y.-S. Zhang, \& G.-C. Guo,
``Three-party quantum communication complexity via entangled tripartite pure states'',
{\em J. Opt. B: Quantum Semiclass. Opt.} {\bf 4}, 4, 219-222 (2001).

\item {\bf [Xue-Huang-Zhang-(+2) 01]}:
P. Xue, Y.-F. Huang, Y.-S. Zhang, C.-F. Li, \& G.-C. Guo,
``Reducing the communication complexity with quantum entanglement'',
{\em Phys. Rev. A} {\bf 64}, 3, 032304 (2001);
quant-ph/0012052.

\item {\bf [Xue-Li-Guo 01]}:
P. Xue, C.-F. Li, \& G.-C. Guo,
``Efficient quantum-key-distribution scheme
with nonmaximally entangled states'',
{\em Phys. Rev. A} {\bf 64}, 3, 032305 (2001);
quant-ph/0101043.
See {\bf [Xue-Li-Guo 02]}.

\item {\bf [Xue-Li-Guo 02]}:
P. Xue, C.-F. Li, \& G.-C. Guo,
``Conditional efficient multiuser quantum cryptography network'',
{\em Phys. Rev. A} {\bf 65}, 2, 022317 (2002).

\item {\bf [Xue-Li-Guo 02]}:
P. Xue, C.-F. Li, \& G.-C. Guo,
`Addendum to ``Efficient quantum-key-distribution scheme with nonmaximally
entangled states''\,',
{\em Phys. Rev. A} {\bf 65}, 3, 034302 (2002).
See {\bf [Xue-Li-Guo 01]}.

\item {\bf [Xue-Guo 03 a]}:
P. Xue, \& G.-C. Guo,
``Scheme for preparation of multipartite entanglement of atomic ensembles'',
{\em Phys. Rev. A} {\bf 67}, 3, 034302 (2003);
quant-ph/0205176.

\item {\bf [Xue-Guo 03 b]}:
P. Xue, \& G.-C. Guo,
``Efficient scheme for multipartite entanglement and quantum information
processing using atomic ensembles`'',
{\em Phys. Lett. A} {\bf 319}, 3-4, 225-232 (2003).

\item {\bf [Xue-Guo 04]}:
P. Xue, \& G.-C. Guo,
``Nondeterministic scheme for preparation of nonmaximal entanglement between two atomic ensembles'',
{\em J. Opt. Soc. Amer. B} {\bf 21}, ?, 1358-1363 (2004).

\item {\bf [Xue-Han-Yu-(+2) 04]}:
P. Xue, C. Han, B. Yu,
X.-M. Lin, \& G.-C. Guo,
``Entanglement preparation and quantum communication with atoms in optical cavities'',
{\em Phys. Rev. A} {\bf 69}, 5, 052318 (2004).


\newpage

\subsection{}


\item {\bf [Yaffe 82]}:
L. G. Yaffe,
``Large $N$ limits as classical mechanics'',
{\em Rev. Mod. Phys.} {\bf 54}, 2, 407-435 (1982).

\item {\bf [Yam 97]}:
P. Yam,
``Bringing Schr\"{o}dinger's cat to life'',
{\em Sci. Am.} {\bf 276}, 6, 124-129 (1997).
Spanish version: ``La frontera entre lo
cu\'{a}ntico y lo cl\'{a}sico'',
{\em Investigaci\'{o}n y Ciencia} 251, 18-24 (1997).
Reprinted in {\bf [Cabello 97 c]}, pp.~84-90.

\item {\bf [Yamaguchi-Master-Yamamoto 00]}:
F. Yamaguchi, C. P. Master, \& Y. Yamamoto,
``Concurrent quantum computation'',
quant-ph/0005128.

\item {\bf [Yamaguchi-Milman-Brune-(+2) 02]}:
F. Yamaguchi, P. Milman, M. Brune,
J. M. Raimond, \& S. Haroche,
``Quantum search with two-atom collisions in cavity QED'',
{\em Phys. Rev. A} {\bf 66}, 1, 010302 (2002);
quant-ph/0203146.

\item {\bf [Yamakami 02]}:
T. Yamakami,
``Quantum optimization problems'',
quant-ph/0204010.

\item {\bf [Yamamoto-Koashi-Imoto 01]}:
T. Yamamoto, M. Koashi, \& N. Imoto,
``Concentration and purification scheme for two partially
entangled photon pairs'',
{\em Phys. Rev. A} {\bf 64}, 1, 012303 (2001);
quant-ph/0101042.

\item {\bf [Yamamoto-Tamaki-Koashi-Imoto 01]}:
T. Yamamoto, K. Tamaki, M. Koashi, \& N. Imoto,
``Polarization-entangled $W$ state using parametric down-conversion'',
{\em Phys. Rev. A} {\bf 66}, 6, 064301 (2002).

\item {\bf [Yamamoto-Koashi-\"{O}zdemir-Imoto 03]}:
T. Yamamoto, M. Koashi, S. K. \"{O}zdemir, \& N. Imoto,
``Experimental extraction of an entangled photon pair from two identically
decohered pairs'',
{\em Nature} {\bf 421}, 6921, 343-346 (2003).

\item {\bf [Yamamoto-Shimamura-\"{O}zdemir-(+2) 04]}:
T. Yamamoto, J. Shimamura, S. K. \"{O}zdemir,
M. Koashi, \& N. Imoto,
``Distribution of a qubit over collective-noise channel assisted with one
additional qubit'',
quant-ph/0407046.

\item {\bf [Yamamoto-Haus 86]}:
Y. Yamamoto, \& H. A. Haus,
``Preparation,
measurement and information capacity of optical quantum states'',
{\em Rev. Mod. Phys.} {\bf 58}, 4, 1001-1029 (1986).

\item {\bf [Yamasaki-Kobayashi-Imai 03]}:
T. Yamasaki, H. Kobayashi, \& H. Imai,
``Analysis of absorbing times of quantum walks'',
{\em Phys. Rev. A} {\bf 68}, 1, 012302 (2003).

\item {\bf [Yan-Wang 03]}:
F. Yan \& D. Wang,
``Probabilistic and controlled teleportation of unknown quantum states'',
{\em Phys. Lett. A} {\bf 316}, 5, 297-303 (2003).

\item {\bf [Yan-Tan-Yang 02]}:
F.-L. Yan, H.-G. Tan, \& L.-G. Yang,
``Probabilistic teleportation of two-particle state of general formation'',
{\em Commun. Theor. Phys.} {\bf 32}, 6, 649-654 (2002).

\item {\bf [Yang-Guo 99]}:
C.-P. Yang, \& G.-C. Guo,
``Disentanglement-free state of two pairs of two-level atoms'',
{\em Phys. Rev. A} {\bf 59}, 6, 4217-4222 (1999).

\item {\bf [Yang-Gea Banacloche 01 a]}:
C.-P. Yang, \& J. Gea-Banacloche,
``Three-qubit quantum error-correction scheme for collective decoherence'',
{\em Phys. Rev. A} {\bf 63}, 2, 022311 (2001).

\item {\bf [Yang-Gea Banacloche 01 b]}:
C.-P. Yang, \& J. Gea-Banacloche,
``Extracting an entangled state of $n-t$ qubits
from an $n$-qubit entangled state after errors at $t$ sites'',
{\em Phys. Rev. A} {\bf 64}, 3, 032309 (2001).
Addendum: {\bf [Teplitsky-Gea Banacloche 03]}.

\item {\bf [Yang-Gea Banacloche 01 c]}:
C.-P. Yang, \& J. Gea-Banacloche,
``Teleportation of rotations and receiver-encoded secret sharing'',
{\em J. Opt. B: Quantum Semiclass. Opt.} {\bf 3}, 6, 407-412 (2001);
quant-ph/0107100.

\item {\bf [Yang-Chu-Han 02]}:
C.-P. Yang, S.-I. Chu, \& S. Han,
``Error-prevention scheme
for protecting three-qubit quantum information'',
{\em Phys. Lett. A} {\bf 299}, 1, 31-37 (2002);
quant-ph/0112141.

\item {\bf [Yang-Chu-Han 02]}:
C.-P. Yang, S.-I. Chu, \& S. Han,
``Error-prevention scheme with two pairs of qubits'',
{\em Phys. Rev. A} {\bf 66}, 3, 034301 (2002).

\item {\bf [Yang-Chu-Han 03]}:
C.-P. Yang, S.-I. Chu, \& S. Han,
``Possible realization of entanglement, logical gates, and quantum-information
transfer with superconducting-quantum-interference-device qubits in cavity
QED'',
{\em Phys. Rev. A} {\bf 67}, 4, 042311 (2003).

\item {\bf [Yang-Chu-Han 04]}:
C.-P. Yang, S.-I. Chu, \& S. Han,
``Quantum information transfer and entanglement with SQUID qubits in cavity QED:
A dark-state scheme with tolerance for nonuniform device parameter'',
{\em Phys. Rev. Lett.} {\bf 92}, 11, 117902 (2004).

\item {\bf [Yang-Chen 04]}:
D. Yang, \& Y.-X. Chen,
``Mixture of multiple copies of maximally entangled states is quasipure'',
{\em Phys. Rev. A} {\bf 69}, 2, 024302 (2004).

\item {\bf [Yang 01]}:
L. Yang,
``Quantum theory cannot forbid superluminal signaling'',
quant-ph/0103154.

\item {\bf [Yang-Song-Cao 04]}:
M. Yang, W. Song, \& Z.-L. Cao,
``Entanglement swapping without joint measurement'',
quant-ph/0408172.

\item {\bf [Yang-Wei-Luo-Miao 02]}:
X. Yang, D. Wei, J. Luo, \& X. Miao,
``Modification and realization of Br\"{u}schweiler's search'',
{\em Phys. Rev. A} {\bf 66}, 4, 042305 (2002).

\item {\bf [Yannoni-Sherwood-Vandersypen-(+3) 99]}:
C. S. Yannoni, M. H. Sherwood, L. M. K. Vandersypen,
D. C. Miller, M. G. Kubinec, \& I. L. Chuang,
``Nuclear magnetic resonance quantum computing using liquid crystal
solvents'',
{\em Appl. Phys. Lett.} {\bf 75}, 22, 3563-3565 (1999);
quant-ph/9907063.

\item {\bf [Yao 79]}:
A. C.-C. Yao,
``?'',
in {\em Proc.\ of the 11th Annual ACM
Symposium on Theory of Computing (STOC 1978)}
ACM Press, New York, 1979), pp.~209-213.

\item {\bf [Yao-Liu-Sham 04]}:
W. Yao, R. Liu, \& L. J. Sham,
``Nanodot-cavity electrodynamics and photon entanglement'',
{\em Phys. Rev. Lett.} {\bf 92}, 21, 217402 (2004).

\item {\bf [Yasuda 99]}:
K. Yasuda,
``Direct determination of the quantum-mechanical density matrix:
Parquet theory'',
{\em Phys. Rev. A} {\bf 59}, 6, 4133-4149 (1999).

\item {\bf [Ye-Yao-Guo 02]}:
L. Ye, C.-M. Yao, \& G.-C. Guo,
``The entanglement purification for entangled multi-particle
states'',
{\em J. Opt. B: Quantum Semiclass. Opt.} {\bf 4}, 3, 215-217 (2002).

\item {\bf [Ye-Guo 02]}:
L. Ye, \& G.-C. Guo,
``Scheme for the generation of three-atom Greenberger-Horne-Zeilinger states
and teleportation of entangled atomic states'',
{\em J. Opt. Soc. Am. B} {\bf 20}, ?, 97-? (2002).

\item {\bf [Ye-Guo 04]}:
L. Ye, \& G.-C. Guo,
``Scheme for entanglement concentration of atomic entangled states in cavity QED'',
{\em Phys. Lett. A} {\bf 327}, 4, 284-289 (2004);
Liu Ye and Guang-Can Guo

\item {\bf [Ye-Zhang-Guo 04 a]}:
M.-Y. Ye, Y.-S. Zhang, \& G.-C. Guo,
``Faithful remote state preparation using finite classical bits and a
non-maximally entangled state'',
{\em Phys. Rev. A} {\bf 69}, 2, 022310 (2004);
quant-ph/0307027.
See {\bf [Berry 04]}.

\item {\bf [Ye-Zhang-Guo 04 b]}:
M.-Y. Ye, Y.-S. Zhang, \& G.-C. Guo,
``Super controlled gates and controlled gates in two-qubit gate
simulations'',
quant-ph/0407108.

\item {\bf [Ye-Zhang 04]}:
P. Ye, \& Y.-Z. Zheng,
``Entanglement capabilities of non-local Hamiltonians with maximally entangled ancillary particles'',
{\em Phys. Lett. A} {\bf 328}, 4-5, 284-288 (2004).

\item {\bf [Yelin-Wang 03]}:
S. F. Yelin, \& B. C. Wang,
``Time-frequency bases for BB84 protocol'',
quant-ph/0309105.

\item {\bf [Yeo 02]}:
Y. Yeo,
``Teleportation via thermally entangled states of a two-qubit Heisenberg $XX$
chain'',
{\em Phys. Rev. A} {\bf 66}, 6, 062312 (2002).

\item {\bf [Yeo 03 a]}:
Y. Yeo,
``Teleportation via thermally entangled state of a two-qubit Heisenberg XX chain'',
{\em Phys. Lett. A} {\bf 309}, 3-4, 215-217 (2003);
quant-ph/0302101.

\item {\bf [Yeo 03 b]}:
Y. Yeo,
``Quantum channels with correlated noise and entanglement teleportation'',
{\em Phys. Rev. A} {\bf 67}, 5, 054304 (2003).

\item {\bf [Yeo-Skeen 03]}:
Y. Yeo, \& A. Skeen,
``Time-correlated quantum amplitude-damping channel'',
{\em Phys. Rev. A} {\bf 67}, 6, 064301 (2003).

\item {\bf [Yeo 03 c]}:
Y. Yeo,
``Studying the thermally entangled state of a three-qubit Heisenberg $XX$ ring
via quantum teleportation'',
{\em Phys. Rev. A} {\bf 68}, 2, 022316 (2003).

\item {\bf [Yeo 03 d]}:
Y. Yeo,
``Quantum teleportation using three-particle entanglement'',
quant-ph/0302030.

\item {\bf [Yeo 03 e]}:
Y. Yeo,
``Entanglement teleportation using three-qubit entanglement'',
quant-ph/0302203.

\item {\bf [Yi-Sun 99]}:
X. X. Yi, \& C. P. Sun,
``Factoring the unitary evolution operator and quantifying entanglement'',
{\em Phys. Lett. A} {\bf 262}, 4-5, 287-295 (1999);
quant-ph/0003024.

\item {\bf [Yi-Guo 00]}:
X. X. Yi, \& G. C. Guo,
``Distillability of two entanglement particles
transmitted through a noisy channel'',
{\em Phys. Rev. A} {\bf 62}, 6, 062312 (2000).

\item {\bf [Yi-Jin-Zhou 01]}:
X. X. Yi, G. R. Jin, \& D. L. Zhou,
``Creating Bell states and decoherence effects in quantum dots system'',
{\em Phys. Rev. A} {\bf 63}, 6, 062307 (2001);
quant-ph/0011058.

\item {\bf [Yi-Cui-Wang 03]}:
X. X. Yi, H. T. Cui, \& X. G. Wang,
``Dynamics of the entanglement rate in the presence of decoherence'',
{\em Phys. Lett. A} {\bf 306}, 5-6, 285-290 (2003).

\item {\bf [Yi-Su-You 03]}:
X. X. Yi, X. H. Su, \& L. You,
``Conditional quantum phase gate between two 3-state atoms'',
{\em Phys. Rev. Lett.} {\bf 90}, 9, 097902 (2003).

\item {\bf [Yi-Yu-Zhou-Song 03]}:
X. X. Yi, C. S. Yu, L. Zhou, \& H. S. Song,
``Noise-assisted preparation of entangled atoms'',
{\em Phys. Rev. A} {\bf 68}, 5, 052304 (2003).

\item {\bf [Yimsiriwattana-Lomonaco 04 a]}:
A. Yimsiriwattana, \& S. J. Lomonaco, Jr.,
``Generalized GHZ state and distributed quantum computing'',
quant-ph/0402148.

\item {\bf [Yimsiriwattana-Lomonaco 04 b]}:
A. Yimsiriwattana, \& S. J. Lomonaco, Jr.,
``Distributed quantum computing: A distributed Shor algorithm'',
quant-ph/0403146.

\item {\bf [Ying 02 a]}:
M. Ying,
``Wootters-Zurek quantum-copying machine:
The higher-dimensional case'',
{\em Phys. Lett. A} {\bf 299}, 2-3, 107-115 (2002).

\item {\bf [Ying 02 b]}:
M. Ying,
``Universal quantum-copying machines:
A sufficient and necessary condition'',
{\em Phys. Lett. A} {\bf 302}, 1, 1-7 (2002).

\item {\bf [Yoran-Reznik 03]}:
N. Yoran, \& B. Reznik,
``Deterministic linear optics quantum computation with single photon qubits'',
{\em Phys. Rev. Lett.} {\bf 91}, 3, 037903 (2003).

\item {\bf [Yoshizawa-Kaji-Tsuchida03]}:
A. Yoshizawa, R. Kaji, \& H. Tsuchida,
``After-pulse-discarding in single-photon detection to reduce bit errors in
quantum key distribution'',
{\em Opt. Express} {\bf 11}, 1303-? (2003).

\item {\bf [You-Tsai-Nori 02]}:
J. Q. You, J. S. Tsai, \& F. Nori,
``Scalable quantum computing with Josephson charge qubits'',
{\em Phys. Rev. Lett.} {\bf 89}, 19, 197902 (2002);
cond-mat/0306209.

\item {\bf [You-Tsai-Nori 03]}:
J. Q. You, J. S. Tsai, \& F. Nori,
``Controllable manipulation and entanglement of macroscopic quantum states in
coupled charge qubits'',
{\em Phys. Rev. B} {\bf 68}, 2, 024510 (2003).

\item {\bf [You-Nori 03]}:
J. Q. You, \& F. Nori,
``Quantum information processing with superconducting qubits in a microwave
field'',
{\em Phys. Rev. B} {\bf 68}, 6, 064509 (2003).

\item {\bf [You-Chapman 00]}:
L. You, \& M. S. Chapman,
``Quantum entanglement using trapped atomic spins'',
{\em Phys. Rev. A} {\bf 62}, 5, 052302 (2000);
quant-ph/0002029.

\item {\bf [You 03]}:
L. You,
``Creating maximally entangled atomic states in a Bose-Einstein condensate'',
{\em Phys. Rev. Lett.} {\bf 90}, 3, 030402 (2003).

\item {\bf [You-Yi-Su 03]}:
L. You, X. X. Yi, \& X. H. Su,
``Quantum logic between atoms inside a high-$Q$ optical cavity'',
{\em Phys. Rev. A} {\bf 67}, 3, 032308 (2003).

\item {\bf [Youssef 95]}:
S. Youssef,
``Is complex probability theory consistent with Bell's theorem?'',
{\em Phys. Lett. A} {\bf 204}, 3-4, 181-187 (1995).

\item {\bf [Yu-Zhou-Zhang-(+2) 04]}:
B. Yu, Z.-W. Zhou, Y. Zhang,
G.-Y. Xiang, \& G.-C. Guo,
``Robust high-fidelity teleportation of an atomic state through the detection of cavity decay'',
{\em Phys. Rev. A} {\bf 70}, 1, 014302 (2004);
quant-ph/0311046.

\item {\bf [Yu-Song 04]}:
C.-S. Yu, \& H.-S. Song,
``Free entanglement measure of multiparticle quantum states'',
{\em Phys. Lett. A} {\bf 330}, 5, 377-383 (2004).

\item {\bf [Yu-Sun 00]}:
S. Yu, \& C. Sun,
``Canonical quantum teleportation'',
{\em Phys. Rev. A} {\bf 61}, 2, 022310 (2000);
quant-ph/0001052.

\item {\bf [Yu-Chen-Pan-Zhang 03]}:
S. Yu, Z.-B. Chen, J.-W. Pan, \& Y.-D. Zhang,
``Classifying $N$-qubit entanglement via Bell's inequalities'',
{\em Phys. Rev. Lett.} {\bf 90}, 8, 080401 (2003).

\item {\bf [Yu-Pan-Chen-Zhang 03]}:
S. Yu, J.-W. Pan, Z.-B. Chen, \& Y.-D. Zhang,
``Comprehensive test of entanglement for two-level systems
via the indeterminacy relationship'',
{\em Phys. Rev. Lett.} {\bf 91}, 21, 217903 (2003).

\item {\bf [Yu-Percival 02]}:
T. Yu, \& I. C. Percival,
``Generalized quantum measurement'',
{\em Phys. Lett. A} {\bf 294}, 2, 59-65 (2002).

\item {\bf [Yu-Eberly 02]}:
T. Yu, \& J. H. Eberly,
``Phonon decoherence of quantum entanglement: Robust and fragile states'',
{\em Phys. Rev. B} {\bf 66}, 19, 193306 (2002).

\item {\bf [Yu-Brown-Chuang 04]}:
T. M. Yu, K. R. Brown, \& I. L. Chuang,
``Bounds on the entanglability of thermal states in liquid-state nuclear
magnetic resonance'',
quant-ph/0409170.

\item {\bf [Yu-Han-Chu-(+2) 02]}:
Y. Yu, S. Han, X. Chu, S.-I. Chu, \& Z. Wang,
``Coherent temporal oscillations of macroscopic quantum states in a Josephson
junction'',
{\em Science} {\bf 296}, ?, 889-? (2002).

\item {\bf [Yu-Eberly 03]}:
T. Yu, \& J. H. Eberly,
``Qubit disentanglement and decoherence via dephasing'',
{\em Phys. Rev. B} {\bf 68}, 16, 165322 (2003).

\item {\bf [Yu-Nakada-Lee-(+5) 04]}:
Y. Yu, D. Nakada, J. C. Lee,
B. Singh, D. S. Crankshaw, T. P. Orlando,
K. K. Berggren, \& W. D. Oliver,
``Energy relaxation time between macroscopic quantum levels in a superconducting persistent-current qubit'',
{\em Phys. Rev. Lett.} {\bf 92}, 11, 117904 (2004).

\item {\bf [Yu-Feng-Zhan 02]}:
Y.-F. Yu, J. Feng, \& M.-S. Zhan,
``Multi-output programmable quantum processor'',
{\em Phys. Rev. A} {\bf 66}, 5, 052310 (2002).

\item {\bf [Yu-Feng-Zhan 03 a]}:
Y.-F. Yu, J. Feng, \& M.-S. Zhan,
``Preparing remotely two instances of quantum state'',
{\em Phys. Lett. A} {\bf 310}, 5-6, 329-332 (2003).

\item {\bf [Yu-Feng-Zhan 03 b]}:
Y.-F. Yu, J. Feng, \& M.-S. Zhan,
``Remote information concentration by a Greenberger-Horne-Zeilinger state and
by a bound entangled state'',
{\em Phys. Rev. A} {\bf 68}, 2, 024303 (2003).

\item {\bf [Yuasa-Nakazato-Takazawa 03]}:
K. Yuasa, H. Nakazato, \& T. Takazawa,
``Purification of quantum state through Zeno-like measurements'',
{\em Proc. Waseda International Symposium on Fundamental Physics: New Perspectives in Quantum
Physics (Tokyo, 2002)},
{\em J. Phys. Soc. Japan} {\bf 72} Suppl. C, 34-37 (2003);
quant-ph/0402183.

\item {\bf [Yuasa-Nakazato-Takazawa 04]}:
K. Yuasa, H. Nakazato, \& M. Unoki,
``Entanglement purification through Zeno-like measurements'',
{\em J. Mod. Opt.};
quant-ph/0402184.

\item {\bf [Yung-Leung-Bose 03]}:
M.-H. Yung, D. W. Leung, \& S. Bose,
``An exact effective two-qubit gate in a chain of three spins'',
quant-ph/0312105.

\item {\bf [Yuen 86 a]}:
H. P. Yuen,
``Quantum amplification and the detection of empty de Broglie waves'',
{\em Phys. Lett. A} {\bf 113}, 8, 401-404 (1986).

\item {\bf [Yuen 86 b]}:
H. P. Yuen,
``Amplification of quantum states and noiseless photon amplifiers'',
{\em Phys. Lett. A} {\bf 113}, 8, 405-407 (1986).

\item {\bf [Yuen-Ozawa 93]}:
H. P. Yuen, \& M. Ozawa,
``Ultimate information carrying limit of quantum systems'',
{\em Phys. Rev. Lett.} {\bf 70}, 4, 363-366 (1993).

\item {\bf [Yuen 96]}:
H. P. Yuen,
``Quantum amplifiers, quantum duplicators and quantum cryptography'',
{\em Quantum Semiclass. Opt.} {\bf 8}, 4, 939-949 (1996).

\item {\bf [Yuen-Kim 98]}:
H. P. Yuen, \& A. M. Kim,
``Classical noise-based cryptography similar to two-state quantum cryptography'',
{\em Phys. Lett. A} {\bf 241}, 3, 135-138 (1998).

\item {\bf [Yuen 00 a]}:
H. P. Yuen,
``No-clone smartcard via quantum memory'',
{\em Phys. Lett. A} {\bf 265}, 3, 173-177 (2000).

\item {\bf [Yuen 00 b]}:
H. P. Yuen,
``Unconditionally secure quantum bit commitment is possible'',
quant-ph/0006109.

\item {\bf [Yuen 00 c]}:
H. P. Yuen,
``Anonymous-key quantum cryptography and unconditionally
secure quantum bit commitment'',
quant-ph/0009113.

\item {\bf [Yuen 01 a]}:
H. P. Yuen,
``Unconditional security in quantum bit commitment'',
quant-ph/0106001.
See {\bf [Yuen 01 c]}.

\item {\bf [Yuen 01 b]}:
H. P. Yuen,
``Communication and measurement with squeezed states'',
in P. D. Drummond, \& Z. Ficek (eds.),
{\em Quantum squeezing},
Springer-Verlag, New York, 2001;
quant-ph/0109054.

\item {\bf [Yuen 01 c]}:
H. P. Yuen,
``How unconditionally secure quantum bit commitment is possible'',
quant-ph/0109055.
See {\bf [Yuen 01 a]}.

\item {\bf [Yukalov 03 a]}:
V. I. Yukalov,
``Entanglement measure for composite systems '',
{\em Phys. Rev. Lett.} {\bf 90}, 16, 167905 (2003);
quant-ph/0309015.

\item {\bf [Yukalov 03 b]}:
V. I. Yukalov,
``Quantifying entanglement production of quantum operations'',
{\em Phys. Rev. A} {\bf 68}, 2, 022109 (2003);
quant-ph/0309054.

\item {\bf [Yura 03]}:
F. Yura,
``Entanglement cost of three-level antisymmetric states'',
{\em J. Phys. A} {\bf 36}, 15, L237-L242 (2003).

\item {\bf [Yurke-Stoler 92 a]}:
B. Yurke, \& D. Stoler,
``Einstein-Podolsky-Rosen
effects from independent particle sources'',
{\em Phys. Rev. Lett.} {\bf 68}, 9, 1251-1254 (1992).

\item {\bf [Yurke-Stoler 92 b]}:
B. Yurke, \& D. Stoler,
``Bell's-inequality experiments using independent-particle sources'',
{\em Phys. Rev. A} {\bf 46}, 5, 2229-2234 (1992).

\item {\bf [Yurke-Stoler 93]}:
B. Yurke, \& D. Stoler,
``Using the Pauli exclusion principle to exhibit local-realism violations
in overlapping interferometers'',
{\em Phys. Rev. A} {\bf 47}, 3, 1704-1707 (1993).

\item {\bf [Yurke-Stoler 95]}:
B. Yurke, \& D. Stoler,
``Bell's-inequality experiment
employing four harmonic oscillators'',
{\em Phys. Rev. A} {\bf 51}, 5, 3437-3444 (1995).

\item {\bf [Yurke-Stoler 97]}:
B. Yurke, \& D. Stoler,
``Observing local realism
violations with a combination of sensitive and insensitive detectors'',
{\em Phys. Rev. Lett.} {\bf 79}, 25, 4941-4945 (1997).

\item {\bf [Yurke-Hillery-Stoler 99]}:
B. Yurke, M. Hillery, \& D. Stoler,
``Position-momentum local-realism violation of the Hardy type'',
{\em Phys. Rev. A} {\bf 60}, 5, 3444-3447 (1999);
quant-ph/9909042.

\item {\bf [Yurtsever-Dowling 02]}:
U. Yurtsever, \& J. P. Dowling,
``Lorentz-invariant look at quantum clock-synchronization protocols based on
distributed entanglement'',
{\em Phys. Rev. A} {\bf 65}, 5, 052317 (2002);
quant-ph/0010097.

\item {\bf [Yurtsever 01]}:
U. Yurtsever,
``Interferometry with entangled atoms'',
quant-ph/0102006.


\newpage

\subsection{}


\item {\bf [Zachar-Alter 91]}:
O. Zachar, \& O. Alter,
``Interpretation of the curious
results of the new quantum formalism of pre- and post-selected systems'',
{\em Found. Phys.} {\bf 21}, 7, 803-820 (1991).
See {\bf [Hu 90]}.

\item {\bf [Zadoyan-Kohen-Lidar-Apkarian 01]}:
R. Zadoyan, D. Kohen, D. A. Lidar, \& V. A. Apkarian,
``The manipulation of massive ro-vibronic superpositions using time-frequency-resolved
coherent anti-Stokes Raman scattering (TFRCARS): From quantum control to quantum computing'',
{\em Chem. Phys.} {\bf 266}, ?, 323-351 (2001);
physics/0102091.

\item {\bf [Zagury-de Toledo Piza 94]}:
N. Zagury, \& A. F. R. de Toledo Piza,
``Large correlation effects of small perturbations by preselection
and postselection of states'',
{\em Phys. Rev. A} {\bf 50}, 4, 2908-2914 (1994).

\item {\bf [Zajonc-Wang-Zou-Mandel 91]}:
A. G. Zajonc, L. J. Wang, X. Y. Zou, \& L. Mandel,
``Quantum eraser'',
{\em Nature} {\bf 353}, 6344, 507-508 (1991).

\item {\bf [Zak-Williams 98]}:
M. Zak, \& C. P. Williams,
``Quantum neural nets'',
{\em Int. J. Theor. Phys.} {\bf 37}, 2, 651-684 (1998).

\item {\bf [Zalka 98 a]}:
C. Zalka,
``Efficient simulation of quantum systems by quantum computers'',
{\em Fortschr. Phys.} {\bf 46}, 6-8, 877-879 (1998).

\item {\bf [Zalka 98 b]}:
C. Zalka,
``Simulating quantum systems on a quantum computer'',
in D. P. DiVincenzo, E. Knill, R. Laflamme, \& W. H. Zurek (eds.),
{\em Quantum Coherence and Decoherence.
Proc.\ of the ITP Conf.\ (Santa Barbara, California, 1996)},
{\em Proc. R. Soc. Lond. A} {\bf 454}, 1969, 313-322 (1998).

\item {\bf [Zalka 98 c]}:
C. Zalka,
``Fast versions of Shor's quantum factoring algorithm'',
quant-ph/9806040.

\item {\bf [Zalka 98 d]}:
C. Zalka,
``An introduction to quantum computers'',
quant-ph/9811006.

\item {\bf [Zalka 99]}:
C. Zalka,
``Grover's quantum searching algorithm is optimal'',
{\em Phys. Rev. A} {\bf 60}, 4, 2746-2751 (1999);
quant-ph/9711070.

\item {\bf [Zalka 00]}:
C. Zalka,
``Using Grover's quantum algorithm for searching
actual databases'',
{\em Phys. Rev. A} {\bf 62}, 5, 052305 (2000);
quant-ph/9901068.

\item {\bf [Zalka 01]}:
C. Zalka,
`Comment on ``Stable quantum computation of unstable classical chaos''\,',
quant-ph/0110019.
Comment on {\bf [Georgeot-Shepelyansky 01 b]}.
See {\bf [Di\'{o}si 01 b]}.

\item {\bf [Zalka-Rieffel 02]}:
C. Zalka, \& E. Rieffel,
``Quantum operations that cannot be implemented using a small mixed environment'',
{\em J. Math. Phys.} {\bf 43}, 9, 4376-4381 (2002);
quant-ph/0111084.

\item {\bf [Zalka-Brun 02]}:
C. Zalka, \& T. A. Brun,
`Comment on ``Quantum optimization for combinatorial
searches''\,',
{\em New J. Phys.} {\bf 4} C1.1-C1.2 (2002);
Comment on {\bf [Trugenberger 02]}.

\item {\bf [Zalka 02]}:
C. Zalka,
``Introduction to quantum computers and quantum algorithms'',
in R. Arvieu \& S. Weigert (eds.),
{\em IX Seminaire Rhodanien de Physique (Dolomieu, France, 2001)},
?, ?, 2002;
quant-ph/0305053.

\item {\bf [Zanardi 97]}:
P. Zanardi,
``Dissipative dynamics in a quantum register'',
{\em Phys. Rev. A} {\bf 56}, 6, 4445-4451 (1997);
quant-ph/9708042.

\item {\bf [Zanardi-Rasetti 97 a]}:
P. Zanardi, \& M. Rasetti,
``Noiseless quantum codes'',
{\em Phys. Rev. Lett.} {\bf 79}, 17, 3306-3309 (1997);
quant-ph/9705044.

\item {\bf [Zanardi-Rasetti 97 b]}:
P. Zanardi, \& M. Rasetti,
``Error avoiding quantum codes'',
{\em Mod. Phys. Lett. B} {\bf 25}, ?, 1085-? (1997);
quant-ph/9710041.

\item {\bf [Zanardi-Rasetti 97 c]}:
P. Zanardi, \& M. Rasetti,
``Comment on: `Preserving coherence in quantum computation
by pairing quantum bits'\,'',
quant-ph/9710002.
Comment on {\bf [Duan-Guo 97 a]}.

\item {\bf [Zanardi 98 a]}:
P. Zanardi,
``Dissipation and decoherence in a quantum register'',
{\em Phys. Rev. A} {\bf 57}, 5, 3276-3284 (1998);
quant-ph/9705045.

\item {\bf [Zanardi 98 b]}:
P. Zanardi,
``Quantum cloning in $d$ dimensions'',
{\em Phys. Rev. A} {\bf 58}, 5, 3484-3490 (1998);
quant-ph/9804011.

\item {\bf [Zanardi-Rossi 98]}:
P. Zanardi, \& F. Rossi,
``Quantum information in semiconductors:
Noiseless encoding in a quantum-dot array'',
{\em Phys. Rev. Lett.} {\bf 81}, 23, 4752-4755 (1998);
quant-ph/9804016.

\item {\bf [Zanardi-Rossi 99]}:
P. Zanardi, \& F. Rossi,
``Subdecoherent information encoding in a quantum-dot array'',
{\em Phys. Rev. B} {\bf 59}, 12, 8170-8181 (1999);
quant-ph/9808036.

\item {\bf [Zanardi 99 a]}:
P. Zanardi,
``Symmetrizing evolutions'',
{\em Phys. Lett. A} {\bf 258}, 2-3, 77-82 (1999);
quant-ph/9809064.

\item {\bf [Zanardi 99 b]}:
P. Zanardi,
``Computation on an error-avoiding quantum code and symmetrization'',
{\em Phys. Rev. A} {\bf 60}, 2, R729-R732 (1999);
quant-ph/9901047.

\item {\bf [Zanardi-Rasetti 99]}:
P. Zanardi, \& M. Rasetti,
``Holonomic quantum computation'',
{\em Phys. Lett. A} {\bf 264}, 2-3, 94-99 (1999);
quant-ph/9904011.

\item {\bf [Zanardi-Zalka-Faoro 00]}:
P. Zanardi, C. Zalka, \& L. Faoro,
``Entangling power of quantum evolutions'',
{\em Phys. Rev. A} {\bf 62}, 3, 030301(R) (2000);
quant-ph/0005031.

\item {\bf [Zanardi 00]}:
P. Zanardi,
``Entanglement of quantum evolutions'',
quant-ph/0010074.

\item {\bf [Zanardi 01 a]}:
P. Zanardi,
``Stabilizing quantum information'',
{\em Phys. Rev. A} {\bf 63}, 1, 012301 (2001);
quant-ph/9910016.

\item {\bf [Zanardi 01 b]}:
P. Zanardi,
``Virtual quantum subsystems'',
{\em Phys. Rev. Lett.} {\bf 87}, 7, 077901 (2001).

\item {\bf [Zanardi-Wang 02]}:
P. Zanardi, \& X. Wang,
``Fermionic entanglement in itinerant systems'',
quant-ph/0201028.

\item {\bf [Zanardi 02 a]}:
P. Zanardi,
``Quantum entanglement in fermionic lattices'',
{\em Phys. Rev. A} {\bf 65}, 4, 042101 (2002).

\item {\bf [Zanardi 02 b]}:
P. Zanardi,
``Stabilization of quantum information:
A unified dynamical-algebraic approach'',
in A. V. Averin, B. Ruggiero, \& P. Silvestrini (eds.),
{\em Macroscopic Quantum Coherence and Quantum Computing (Napoli, Italy, 2000},
Kluwer Academic/Plenum Publisher, New York, pp.~?-?;
quant-ph/0203008.

\item {\bf [Zanardi-Lloyd 03]}:
P. Zanardi, \& S. Lloyd,
``Topological protection and quantum noiseless subsystems'',
{\em Phys. Rev. Lett.} {\bf 90}, 6, 067902 (2003);
quant-ph/0208132.

\item {\bf [Zanardi 03]}:
P. Zanardi,
``Bipartite mode entanglement of bosonic condensates on tunneling graphs'',
{\em Phys. Rev. A} {\bf 67}, 5, 054301 (2003).

\item {\bf [Zanardi-Lloyd 04]}:
P. Zanardi, \& S. Lloyd,
``Universal control of quantum subspaces and subsystems'',
{\em Phys. Rev. A} {\bf 69}, 2, 022313 (2004);
quant-ph/0305013.

\item {\bf [Zanardi-Lidar-Lloyd 04]}:
P. Zanardi, D. A. Lidar, \& S. Lloyd,
``Quantum tensor product structures are observable induced'',
{\em Phys. Rev. Lett.} {\bf 92}, 6, 060402 (2004);
quant-ph/0308043.

\item {\bf [Zanardi-Lidar 04]}:
P. Zanardi, \& D. Lidar,
``Purity and state fidelity of quantum channels'',
{\em Phys. Rev. A} {\bf 70}, 1, 012315 (2004);
quant-ph/0403074.

\item {\bf [Zanghi-Tumulka 03]}:
N. Zanghi, \& R. Tumulka,
``John Bell across space and time'',
{\em Am. Scientist} {\bf 91}, 5, 461-462 (2003);
quant-ph/0309020.
Review of {\bf [Bertlmann-Zeilinger 02]}.

\item {\bf [Zhai-Li-Gao 04]}:
Z. Zhai, Y. Li, \& J. Gao,
``Entanglement characteristics of subharmonic modes reflected from a cavity for type-II second-harmonic generation'',
{\em Phys. Rev. A} {\bf 69}, 4, 044301 (2004).

\item {\bf [Zhangx-Kasai-Hayasaka 04]}:
Y. Zhangx, K. Kasai, \& K. Hayasaka,
``Single-beam noise characteristics of quantum correlated twin beams'',
{\em J. Opt. Soc. Am. B};
quant-ph/0402119.

\item {\bf [Zhang-Ye-Guo 04]}:
Y.-S. Zhang, M.-Y. Ye, \& G.-C. Guo,
``Conditions for optimal construction of two-qubit non-local gates'',
quant-ph/0411058.

\item {\bf [Zapatrin 03]}:
R. R. Zapatrin,
``Combinatorial topology of multipartite entangled states'',
in M. Ferrero (ed.),
{\em Proc. of Quantum Information: Conceptual Foundations,
Developments and Perspectives (Oviedo, Spain, 2002)},
{\em J. Mod. Opt.} {\bf 50}, 6-7, 891-899 (2003).

\item {\bf [Zarda-Chiangga-Jennewein-Weinfurter 99]}:
P. Zarda, S. Chiangga, T. Jennewein, \& H. Weinfurter,
``Quantum mechanics and secret communication'',
in {\bf [Greenberger-Reiter-Zeilinger 99]}, pp.~271-273.

\item {\bf [Zazunov-Shumeiko-Bratus'-(+2) 03]}:
A. Zazunov, V. S. Shumeiko, E. N. Bratus',
J. Lantz, \& G. Wendin,
``Andreev level qubit'',
{\em Phys. Rev. Lett.} {\bf 90}, 8, 087003 (2003).

\item {\bf [Zbinden-Gautier-Gisin-(+3) 97]}:
H. Zbinden, J.D. Gautier, N. Gisin,
B. Huttner, A. Muller, \& W. Tittel,
``Interferometry with Faraday mirrors for quantum cryptography'',
{\em Elect. Lett.} {\bf 33}, 586-588 (1997);
quant-ph/9703024.

\item {\bf [Zbinden 98]}:
H. Zbinden,
``Experimental quantum cryptography'',
in {\bf [Lo-Spiller-Popescu 98]}, pp.~120-142.

\item {\bf [Zbinden-Gisin-Huttner-(+2) 98]}:
H. Zbinden, N. Gisin, B. Huttner,
A. Muller, \& W. Tittel,
``Practical aspects of quantum cryptographic key distribution'',
{\em J. Cryptology} {\bf 11}, 1-14 (1998).
Reprinted in {\bf [Macchiavello-Palma-Zeilinger 00]}, pp.~245-258.

\item {\bf [Zbinden-Brendel-Gisin-Tittel 01]}:
H. Zbinden, J. Brendel, N. Gisin, \& W. Tittel,
``Experimental test of non-local quantum correlation in
relativistic configurations'',
{\em Phys. Rev. A} {\bf 63}, 2, 022111 (2001);
quant-ph/0007009.
Comment: {\bf [Tiwari 02 a]}.

\item {\bf [Zbinden-Brendel-Tittel-Gisin 01]}:
H. Zbinden, J. Brendel, W. Tittel, \& N. Gisin,
``Experimental test of relativistic quantum state
collapse with moving reference frames'',
in S. Popescu, N. Linden, \& R. Jozsa (eds.),
{\em J. Phys. A} {\bf 34}, 35
(Special issue: Quantum information and computation), 7103-7110 (2001);
quant-ph/0002031.

\item {\bf [Zeh 70]}:
H. D. Zeh,
``On the interpretation of measurement in quantum theory'',
{\em Found. Phys.} {\bf 1}, 1, 69-76 (1970).

\item {\bf [Zeh 71]}:
H. D. Zeh,
``On the irreversibility of time and observation in quantum theory'',
in {\bf [d'Espagnat 71]}, pp.~?-?.

\item {\bf [Zeh 73]}:
H. D. Zeh,
``Toward a quantum theory of observation'',
{\em Found. Phys.} {\bf 3}, ?, 109-? (1973);
quant-ph/0306151.

\item {\bf [Zeh 79]}:
H. D. Zeh,
``Quantum theory and time asymmetry'',
{\em Found. Phys.} {\bf 9}, ?, 803-? (1979);
quant-ph/0307013.

\item {\bf [Zeh 90]}:
H. D. Zeh,
``Quantum measurements and entropy'',
in {\bf [Zurek 90]}, pp.~405-422.

\item {\bf [Zeh 93]}:
H. D. Zeh,
``There are no quantum jumps, nor are there particles!'',
{\em Phys. Lett. A} {\bf 172}, 4, 189-192 (1993).

\item {\bf [Zeh 97]}:
H. D. Zeh,
``What is achieved by decoherence?'',
in M. Ferrero, \& A. van der Merwe (eds.),
{\em New developments on fundamental problems in
quantum physics (Oviedo, Spain, 1996)},
Kluwer Academic, Dordrecht, Holland, 1997, pp.~441-451.

\item {\bf [Zeh 99 a]}:
H. D. Zeh,
``Why Bohm's quantum theory?'',
{\em Found. phys. Lett.} {\bf 12}, 2, 197-200 (1999);
quant-ph/9812059.
Comment on {\bf [Goldstein 98 a]}.

\item {\bf [Zeh 99 b]}:
H. D. Zeh,
``The meaning of decoherence'',
quant-ph/9905004.

\item {\bf [Zeh 00]}:
H. D. Zeh,
``The problem of conscious observation in quantum
mechanical description'',
{\em Found. Phys. Lett.} {\bf 13}, 3, 221-234 (2000);
quant-ph/9908084.

\item {\bf [Zeh 02]}:
H. D. Zeh,
``The wave function: It or bit?'',
quant-ph/0204088.

\item {\bf [Zeier-Grassl-Beth 04]}:
R. Zeier, M. Grassl, \& T. Beth,
``Gate simulation and lower bounds on the simulation time'',
quant-ph/0403082.

\item {\bf [Zeilinger 81]}:
A. Zeilinger,
``General properties of lossless beam splitters in interferometry'',
{\em Am. J. Phys.} {\bf 49}, 9, 882-883 (1981).

\item {\bf [Zeilinger 86]}:
A. Zeilinger,
``Testing Bell's inequalities with periodic switching'',
{\em Phys. Lett. A} {\bf 118}, 1, 1-2 (1986).

\item {\bf [Zeilinger 90]}:
A. Zeilinger,
``Experiment and quantum measurement theory'',
in M. Cini, \& J. M. L\'{e}vy-Leblond (eds.),
{\em Quantum theory without reduction},
Adam Hilger, Bristol, 1990, pp.~9-26.

\item {\bf [Zeilinger-Horne-Greenberger 92]}:
A. Zeilinger, M. A. Horne, \& D. M. Greenberger,
``Higher-order quantum entanglement'',
in D. Han, Y. S. Kim, \& W. W. Zachary (eds.),
{\em Proc.\ of Squeezed States and Quantum Uncertainty},
NASA Conf. Publ., 1992, pp.~3135-?.

\item {\bf [Zeilinger-Bernstein-Greenberger-(+2) 93]}:
A. Zeilinger, H. J. Bernstein,
D. M. Greenberger, M. A. Horne, \& M. \.{Z}ukowski,
``Possible realization of the GHZ state using parametric down-conversion'',
in H. Ezawa, \& Y. Muruyama (eds.),
{\em Quantum control and measurement},
Elsevier, Amsterdam, 1993, pp.~?-?.

\item {\bf [Zeilinger 94]}:
A. Zeilinger,
``Probing higher dimensions of Hilbert space in experiment'',
{\em Acta Phys. Pol.} {\bf 85}, 4, 717-723 (1994).

\item {\bf [Zeilinger-Bernstein-Horne 94]}:
A. Zeilinger, H. J. Bernstein, \& M. A. Horne,
``Information transfer with two-state two-particle quantum systems'',
in S. M. Barnett, A. K. Ekert, \& S. J. D. Phoenix (eds.),
{\em J. Mod. Opt.} {\bf 41}, 12 (Special issue: Quantum
communication), 2375-2384 (1994).

\item {\bf [Zeilinger-\.{Z}ukowski-Horne-(+2) 94]}:
A. Zeilinger, M. \.{Z}ukowski, M. A. Horne,
H. J. Bernstein, \& D. M. Greenberger,
``Einstein-Podolsky-Rosen correlations in higher dimensions'',
in F. De Martini, \& A. Zeilinger (eds.),
{\em Quantum interferometry},
World scientific, Singapore, 1994, pp.~?-?.

\item {\bf [Zeilinger-Horne-Weinfurter 95]}:
A. Zeilinger, M. A. Horne, \& H. Weinfurter,
``Fundamental problems of quantum theory'',
{\em Ann. N. Y. Acad. Sci.} {\bf 755}, 91-? (1995).

\item {\bf [Zeilinger-\.{Z}ukowski-Horne-(+2) 95]}:
A. Zeilinger, M. \.{Z}ukowski, M. A. Horne,
H. J. Bernstein, \& D. M. Greenberger,
``Einstein-Podolsky-Rosen correlations in higher dimensions'',
in J. Anandan, \& J. L. Safko (eds.),
{\em Quantum coherence and reality.
In celebration of the 60th birthday of Yakir Aharonov.
Int.\ Conf.\ on Fundamental Aspects of Quantum Theory (?, ?)},
World Scientific, Singapore, 1995, pp.~?-?.

\item {\bf [Zeilinger 96 a]}:
A. Zeilinger,
``Einstein-Podolsky-Rosen interferometry'',
in A. Mann, \& M. Revzen (eds.),
{\em The dilemma of Einstein, Podolsky and Rosen -- 60 years
later. An international symposium in honour of Nathan Rosen
(Haifa, Israel, 1995)},
{\em Ann. Phys. Soc. Israel} {\bf 12}, 57-72 (1996).

\item {\bf [Zeilinger 96 b]}:
A. Zeilinger,
``On the interprtation and philosophical foundation of quantum mechanics'',
in U. Ketvel et al. (eds), {\em Vastakohtien todellisuus},
Helsinky University Press, 1996, pp.~?-?.

\item {\bf [Zeilinger-Horne-Weinfurter-\.{Z}ukowski 97]}:
A. Zeilinger, M. A.
Horne, H. Weinfurter, \& M. \.{Z}ukowski,
``Three-particle Greenberger-Horne-Zeilinger
states from two entangled pairs'',
{\em Phys. Rev. Lett.} {\bf 78}, 16, 3031-3034 (1997).

\item {\bf [Zeilinger 97 a]}:
A. Zeilinger,
``Get set for the quantum revolution'',
{\em Phys. World} {\bf 9}, 9, 54-55 (1997).
Review of {\bf [Milburn 97]}.

\item {\bf [Zeilinger 97 b]}:
A. Zeilinger,
``Quantum teleportation and the non-locality of information'',
in P. L. Knight, B. Stoicheff, \& D. Walls (eds.),
{\em Highlight in Quantum Optics},
{\em Philos. Trans. R. Soc. Lond. A} {\bf 355}, 1733, 2401-2404 (1997).

\item {\bf [Zeilinger 98 a]}:
A. Zeilinger,
``Teleportation: Who was first?'',
{\em Phys. World} {\bf 11}, 3, 24 (1998).
See {\bf [De Martini 98 a]}.

\item {\bf [Zeilinger 98 b]}:
A. Zeilinger,
``Fundamentals of quantum information'',
{\em Phys. World} {\bf 11}, 3, 35-40 (1998).

\item {\bf [Zeilinger 98 c]}:
A. Zeilinger,
``Quantum entanglement: A fundamental concept finding its applications'',
{\em Physica Scripta} {\bf T76}, 203-209 (1998).
Reprinted in {\bf [Macchiavello-Palma-Zeilinger 00]}, pp.~12-18.

\item {\bf [Zeilinger 99 a]}:
A. Zeilinger,
``A foundational principle for quantum mechanics'',
{\em Found. Phys.} {\bf 29}, 4, 631-644 (1999).

\item {\bf [Zeilinger 99 b]}:
A. Zeilinger,
``Albert Einstein: Philosopher-scientist'',
{\em Nature} {\bf 398}, 6724, 210-211 (1999).
Review of {\bf [Bohr 49]}, {\bf [Einstein 49]}.

\item {\bf [Zeilinger 99 c]}:
A. Zeilinger,
``Experiment and the foundations of quantum physics'',
{\em Rev. Mod. Phys.} {\bf 71}, 2, S288-S297 (1999).
Reprinted in {\bf [Macchiavello-Palma-Zeilinger 00]}, pp.~19-28.

\item {\bf [Zeilinger 00 a]}:
A. Zeilinger,
``Quantum computing:
Quantum entangled bits step closer to IT'',
{\em Science} {\bf 289}, 5478, 405-406 (2000).

\item {\bf [Zeilinger 00 b]}:
A. Zeilinger,
``The quantum centennial'',
{\em Nature} {\bf 408}, 6813, 639-641 (2000).

\item {\bf [Zeilinger 00 c]}:
A. Zeilinger,
``Quantum teleportation'',
{\em Sci. Am.} {\bf 282}, 5, ?-? (2000);
Spanish version:
``Teletransporte cu\'{a}ntico'',
{\em Investigaci\'{o}n y Ciencia} 285, 58-67 (2000).
Reprinted in {\bf [Cabello 03 a]}, pp.~46-54.

\item {\bf [Zeilinger 00 d]}:
A. Zeilinger,
``Introductory concepts'',
{\bf [Macchiavello-Palma-Zeilinger 00]}, p.~3.

\item {\bf [Zeilinger 02]}:
A. Zeilinger,
``Bell's theorem, information and quantum physics'',
in {\bf [Bertlmann-Zeilinger 02]}, pp.~241-256.

\item {\bf [Zeilinger 03]}:
A. Zeilinger,
{\em Einsteins Schleier.
Die neue Welt der Quantenphysik},
C. H. Beck, M\"{u}nchen, 2003.

\item {\bf [Zeng-Liu-Li-Long 01]}:
B. Zeng, X. S. Liu, Y. S. Li, \& G. L. Long,
``General schemes for multi-particle $d$-dimensional
cat-like state teleportation'',
quant-ph/0104102.

\item {\bf [Zeng-Zhang 02]}:
B. Zeng, \& P. Zhang,
``Remote-state preparation in higher dimension and the parallelizable manifold
$S^{n-1}$'',
{\em Phys. Rev. A} {\bf 65}, 2, 022316 (2002);
quant-ph/0105088.

\item {\bf [Zeng-Zhai-Xu 02]}:
B. Zeng, H. Zhai, \& Z. Xu,
``Entanglement properties of some fractional quantum Hall liquids'',
{\em Phys. Rev. A} {\bf 66}, 4, 042324 (2002).

\item {\bf [Zeng-Zhou-Xu-Sun 03]}:
B. Zeng, D.L. Zhou, Z. Xu, \& C.P. Sun,
``Quantum teleportation using cluster states'',
quant-ph/0304165.

\item {\bf [Zeng-Zhou-Zhang-(+2) 03]}:
B. Zeng, D. L. Zhou, P. Zhang,
Z. Xu, \& L. You,
``Criterion for testing multiparticle negative-partial-transpose entanglement'',
{\em Phys. Rev. A} {\bf 68}, 4, 042316 (2003).

\item {\bf [Zeng 98 a]}:
G. Zeng,
``Improvement of quantum key distribution protocols'',
quant-ph/9810021.

\item {\bf [Zeng-Wang 98]}:
G. Zeng, \& X. Wang,
``Attacks of BB84 protocol in quantum cryptography'',
quant-ph/9812022.
See {\bf [Zeng 98 b]}.

\item {\bf [Zeng 98 b]}:
G. Zeng,
``A simple eavesdropping strategy of BB84 protocol'',
quant-ph/9812064.
See {\bf [Zeng-Wang 98]}.

\item {\bf [Zeng-Zhang 00]}:
G. Zeng, \& W. Zhang,
``Identity verification in quantum key distribution'',
{\em Phys. Rev. A} {\bf 61}, 2, 022303 (2000).

\item {\bf [Zeng 00]}:
G. Zeng,
``Quantum key distribution based on Greenberger-Horne-Zeilinger state'',
quant-ph/0001044.

\item {\bf [Zeng-Wang-Wang 00]}:
G. Zeng, Z. Wang, \& X. Wang,
``Quantum key distribution relied on trusted information center'',
quant-ph/0001045.

\item {\bf [Zeng-Guo 00]}:
G. Zeng, \& G.-C. Guo,
``Quantum authentication protocol'',
quant-ph/0001046.

\item {\bf [Zeng-Keitel 02 a]}:
G. Zeng, \& C. H. Keitel,
``Arbitrated quantum-signature scheme'',
{\em Phys. Rev. A} {\bf 65}, 4, 042312 (2002);
quant-ph/0109007.

\item {\bf [Zeng-Keitel 02 b]}:
G. Zeng, \& C. H. Keitel,
``Inhibiting decoherence via ancilla processes'',
{\em Phys. Rev. A} {\bf 66}, 2, 022306 (2002);
quant-ph/0203012.

\item {\bf [Zeng-Saavedra-Keitel 02]}:
G. Zeng, C. Saavedra, \& C. H. Keitel,
``Asymmetrical quantum cryptographic algorithm'',
quant-ph/0202021.

\item {\bf [Zeng-Kuang 00 a]}:
H.-S. Zeng, \& L.-M. Kuang,
``Preparation of GHZ states via Grover's quantum searching
algorithm'',
{\em Chin. Phys. Lett.} {\bf 17}, ?, 410-? (2000);
quant-ph/0005002.

\item {\bf [Zeng-Kuang 00 b]}:
H.-S. Zeng, \& L.-M. Kuang,
``Efficient state preparation via ion trap quantum computing
and quantum searching algorithm'',
{\em Comm. Theor. Phys.} {\bf 33}, ?, 11-? (2000);
quant-ph/0005036.

\item {\bf [Zeng-Kuang-Gao 04]}:
H.-S. Zeng, L.-M. Kuang, \& K.-L. Gao
``Coherent manipulation of motional states of trapped ions'',
{\em J. Opt. B: Quantum Semiclass. Opt.} {\bf 6}, 269-275 (2004);
quant-ph/0405136.

\item {\bf [Zeng-Zhu-Pei 02]}:
J.-Y. Zeng, H.-B. Zhu, \& S.-Y. Pei,
``Multiparticle reduced density matrix and a useful kind of entangled state
for quantum teleportation'',
{\em Phys. Rev. A} {\bf 65}, 5, 052307 (2002).

\item {\bf [Zeng-Bi-Guo-Ruda 03]}:
X.-H. Zeng, Q. Bi, G.-C. Guo, \& H. E. Ruda,
``Decoherence free in subspace using Na@C$_{60}$ as quantum qubit'',
{\em Phys. Lett. A} {\bf 313}, 1-2, 21-28 (2003).

\item {\bf [Zhang 04]}:
C. Zhang,
``Preparation of polarization-entangled mixed states of two photons'',
{\em Phys. Rev. A} {\bf 69}, 1, 014304 (2004);
quant-ph/0107145.

\item {\bf [Zhang-Li-Guo 99 a]}:
C.-W. Zhang, C.-F. Li, \& G.-C. Guo,
``General strategies for discrimination of quantum states'',
{\em Phys. Lett. A} {\bf 261}, 1-2, 25-29 (1999);
quant-ph/9908001.

\item {\bf [Zhang-Li-Guo 99 b]}:
C.-W. Zhang, C.-F. Li, \& G.-C. Guo,
``Realizing probabilistic identification and cloning of quantum
states. II Multiparticles system'',
quant-ph/9908002.
See {\bf [Zhang-Wang-Li-Guo 00]} (I).

\item {\bf [Zhang-Li-Guo 99 c]}:
C.-W. Zhang, C.-F. Li, \& G.-C. Guo,
``No-concentrating theorem of pure entangled states'',
quant-ph/9910018.

\item {\bf [Zhang-Wang-Li-Guo 00]}:
C.-W. Zhang, Z.-Y. Wang, C.-F. Li, \& G.-C. Guo,
``Realizing probabilistic identification and cloning of quantum
states via universal quantum logic gates'',
{\em Phys. Rev. A} {\bf 61}, 6, 062310 (2000);
quant-ph/9907097.
See {\bf [Zhang-Li-Guo 99 b]} (II).

\item {\bf [Zhang-Li-Wang-Guo 00]}:
C.-W. Zhang, C.-F. Li, Z.-Y. Wang, \& G.-C. Guo,
``Probabilistic quantum cloning via
Greenberger-Horne-Zeilinger states'',
{\em Phys. Rev. A} {\bf 62}, 4, 042302 (2000);
quant-ph/0001081.

\item {\bf [Zhang-Li-Guo 00 a]}:
C.-W. Zhang, C.-F. Li, \& G.-C. Guo,
``Conditions for manipulation of a set of entangled pure states'',
quant-ph/0002068.

\item {\bf [Zhang-Li-Guo 00 b]}:
C.-W. Zhang, C.-F. Li, \& G.-C. Guo,
``Quantum clone and states estimation for $n$-state system'',
{\em Phys. Lett. A} {\bf 271}, 1-2, 31-34 (2000);
quant-ph/9908003.

\item {\bf [Zhang-Li-Guo 00 c]}:
C.-W. Zhang, C.-F. Li, \& G.-C. Guo,
``Quantum authentication using entangled state'',
quant-ph/0008044.


\item {\bf [Zhang-Li-Guo 01 a]}:
C.-W. Zhang, C.-F. Li, \& G.-C. Guo,
``Comment on `Quantum key distribution without alternative
measurements' [Phys. Rev. A {\bf 61}, 052312 (2000)]'',
{\em Phys. Rev. A} {\bf 63}, 3, 036301 (20001);
quant-ph/0009042.
Comment on {\bf [Cabello 00 c]}.
Reply: {\bf [Cabello 01 b, e]}.

\item {\bf [Zhang 04]}:
C.-W. Zhang,
``Entanglement concentration of individual photon pairs via linear optical
logic'',
{\em Quant. Inf. Comp.} {\bf 4}, ?, 196-? (2004);
quant-ph/0104054.

\item {\bf [Zhang-Peng 00]}:
J. Zhang, \& K. Peng,
``Quantum teleportation and dense coding by means of bright
amplitude-squeezed light and direct measurement of a Bell state'',
{\em Phys. Rev. A} {\bf 62}, 6, 064302 (2000).

\item {\bf [Zhang-Xie-Peng 01 a]}:
J. Zhang, C. Xie, \& K. Peng,
``Quantum teleportation for continuous variables by means of
a phase sensitive nondegenerate optical parametric amplifier'',
{\em Phys. Lett. A} {\bf 287}, 1-2, 7-11 (2001).

\item {\bf [Zhang-Xie-Peng 01 b]}:
J. Zhang, C. Xie, \& K. Peng,
``Quantum switch for continuous variable teleportation'',
{\em J. Opt. B: Quantum Semiclass. Opt.} {\bf 3}, 5, 293-297 (2001);
quant-ph/0105073.

\item {\bf [Zhang-Lu-Shan-Deng 02]}:
J. Zhang, Z. Lu, L. Shan, \& Z. Deng,
``Realization of generalized quantum searching using nuclear magnetic
resonance'',
{\em Phys. Rev. A} {\bf 65}, 3, 034301 (2002);
quant-ph/0110076.

\item {\bf [Zhang-Xie-Peng 02 a]}:
J. Zhang, C. Xie, \& K. Peng,
``Entanglement swapping using nondegenerate optical parametric amplifier'',
{\em Phys. Lett. A} {\bf 299}, 5-6, 427-432 (2002).

\item {\bf [Zhang-Xie-Peng 02 b]}:
J. Zhang, C. Xie, \& K. Peng,
``Controlled dense coding for continuous variables using three-particle
entangled states'',
{\em Phys. Rev. A} {\bf 66}, 3, 032318 (2002).

\item {\bf [Zhang-Lu-Shan-Deng 02 a]}:
J. Zhang, Z. Lu, L. Shan, \& Z. Deng,
``Experimental study of quantum decoherence
using nuclear magnetic resonance'',
quant-ph/0202146.

\item {\bf [Zhang-Lu-Shan-Deng 02 b]}:
J. Zhang, Z. Lu, L. Shan, \& Z. Deng,
``Simulating decoherence behavior of a system in
entangled state using nuclear magnetic resonance'',
quant-ph/0204113.

\item {\bf [Zhang-Lu-Shan-Deng 02 c]}:
J. Zhang, Z. Lu, L. Shan, \& Z. Deng,
``Synthesizing NMR analogs of Einstein-Podolsky-Rosen states using the
generalized Grover's algorithm'',
{\em Phys. Rev. A} {\bf 66}, 4, 044308 (2002);
quant-ph/0205180.

\item {\bf [Zhang-Xie-Peng 02]}:
J. Zhang, C. Xie, \& K. Peng,
``Quantum entanglement and squeezing of the quadrature difference of bright
light fields'',
{\em Phys. Rev. A} {\bf 66}, 4, 042319 (2002).

\item {\bf [Zhang-Lu 03]}:
J. Zhang, \& Z. Lu,
``Similarity between Grover's quantum search algorithm and classical two-body
collisions'',
{\em Am. J. Phys.} {\bf 71}, 1, 83-86 (2003);
quant-ph/0110077.

\item {\bf [Zhang 03]}:
J. Zhang,
``Einstein-Podolsky-Rosen sideband entanglement in broadband squeezed light'',
{\em Phys. Rev. A} {\bf 67}, 5, 054302 (2003).

\item {\bf [Zhang-Vala-Sastry-Whaley 03 a]}:
J. Zhang, J. Vala, S. Sastry, \& K. B. Whaley,
``Geometric theory of nonlocal two-qubit operations'',
{\em Phys. Rev. A} {\bf 67}, 4, 042313 (2003).

\item {\bf [Zhang-Vala-Sastry-Whaley 03 b]}:
J. Zhang, J. Vala, S. Sastry, \& K. B. Whaley,
``Exact two-qubit universal quantum circuit'',
{\em Phys. Rev. Lett.} {\bf 91}, 2, 027903 (2003);
quant-ph/0212109.

\item {\bf [Zhang-Peng-Braunstein 03]}:
J. Zhang, K. Peng, \& S. L. Braunstein,
``Quantum-state transfer from light to macroscopic oscillators'',
{\em Phys. Rev. A} {\bf 68}, 1, 013808 (2003).

\item {\bf [Zhang-Vala-Sastry-Whaley 04]}:
J. Zhang, J. Vala, S. Sastry, \& K. B. Whaley,
``Minimum construction of two-qubit quantum operations'',
{\em Phys. Rev. Lett.} {\bf 93}, 2, 020502 (2004);
quant-ph/0312193.

\item {\bf [Zhang-You 03]}:
M. Zhang, \& L. You,
``Quantum Zeno subspace and entangled Bose-Einstein condensates'',
{\em Phys. Rev. Lett.} {\bf 91}, 23, 230404 (2003);
cond-mat/0310638.

\item {\bf [Zhang-Xue-Zhao-Xie 02]}:
P. Zhang, Q.-K. Xue, X.-G. Zhao, \& X. C. Xie,
``Coulomb-enhanced dynamic localization and Bell-state generation in coupled
quantum dots'',
{\em Phys. Rev. A} {\bf 66}, 2, 022117 (2002).

\item {\bf [Zhang-Liu-Sun 02]}:
P. Zhang, X. F. Liu, \& C. P. Sun,
``An quantum approach of measurement based on the Zurek's triple model'',
quant-ph/0204121.

\item {\bf [Zhang-Chan-Xue-Zhao 03]}:
P. Zhang, C. K. Chan, Q.-K. Xue, \& X.-G. Zhao,
``Quantum entanglement of excitons in coupled quantum dots'',
{\em Phys. Rev. A} {\bf 67}, 1, 012312 (2003).

\item {\bf [Zhang-Xue-Zhao-Xie 04]}:
P. Zhang, Q.-K. Xue, X.-G. Zhao, \& X. C. Xie,
``Generation of spatially separated spin entanglement in a triple-quantum-dot system'',
{\em Phys. Rev. A} {\bf 69}, 4, 042307 (2004).

\item {\bf [Zhang-Ying 02]}:
S. Zhang, \& M. Ying,
``Set discrimination of quantum states'',
{\em Phys. Rev. A} {\bf 65}, 6, 062322 (2002).

\item {\bf [Zhang-Goh-Chou-(+3) 03]}:
T. C. Zhang, K. W. Goh, C. W. Chou, P. Lodahl, \& H. J. Kimble,
``Quantum teleportation of light beams'',
{\em Phys. Rev. A} {\bf 67}, 3, 033802 (2003).

\item {\bf [Zhang-Li-Li-(+2) 00]}:
Y. Zhang, C.-F. Li, W. Li, Y. Huang, \& G.-C. Guo,
``Optical realization of quantum gambling machine'',
quant-ph/0001008.

\item {\bf [Zhang-Kasai-Hayasaka 04]}:
Y. Zhang, K. Kasai, \& K. Hayasaka,
``Quantum channel using photon number correlated twin beams'',
{\em Opt. Express} {\bf 11}, 3592-3597 (2003);
quant-ph/0401033.

\item {\bf [Zhang-Zhou-Yu-Guo 04]}:
Y. Zhang, Z.-W. Zhou, B. Yu, \& G.-C. Guo,
``Concatenating bang-bang control with decoherence-free subspaces for
quantum computation'',
{\em Phys. Rev. A} {\bf 69}, 4, 042315 (2004);
quant-ph/0311067.

\item {\bf [Zhang-Li-Guo 01 b]}:
Y.-S. Zhang, C.-F. Li, \& G.-C. Guo,
``Quantum key distribution via quantum encryption'',
{\em Phys. Rev. A} {\bf 64}, 2, 024302 (2001);
quant-ph/0011034.

\item {\bf [Zhang-Huang-Li-Guo 03]}:
Y.-S. Zhang, Y.-F. Huang, C.-F. Li, \& G.-C. Guo,
``Experimental preparation of the Werner state via spontaneous parametric
down-conversion'',
{\em Phys. Rev. A} {\bf 66}, 6, 062315 (2002).

\item {\bf [Zhang 04]}:
Z. Zhang,
`Comment on ``Eavesdropping on the ping-pong communication protocol''\,',
quant-ph/0402022.
Comment on {\bf [Wojcik 03]}.

\item {\bf [Zhao-Chen 02]}:
Y. Zhao, \& G. H. Chen,
``Density matrix negativity for two oscillators in an Agarwal bath'',
{\em Phys. Rev. E} {\bf 65}, 5, 056120 (2002).

\item {\bf [Zhao-Zanardi-Chen 04]}:
Y. Zhao, P. Zanardi, \& G. Chen,
``Quantum entanglement and the self-trapping transition in polaronic
systems'',
quant-ph/0407080.

\item {\bf [Zhao-Pan-Zhan 01]}:
Z. Zhao, J.-W. Pan, \& M. S. Zhan,
``Practical scheme for entanglement concentration'',
{\em Phys. Rev. A} {\bf 64}, 1, 014301 (2001);
quant-ph/0104039.

\item {\bf [Zhao-Yang-Chen-(+2) 02]}:
Z. Zhao, T. Yang, Z.-B. Chen,
J. Du, \& J.-W. Pan,
``Deterministic and highly efficient quantum cryptography with entangled
photon pairs'',
quant-ph/0211098.

\item {\bf [Zhao-Yang-Chen-(+2) 03]}:
Z. Zhao, T. Yang, Y.-A. Chen,
A.-N. Zhang, \& J.-W. Pan,
``Experimental realization of entanglement concentration and a quantum
repeater'',
{\em Phys. Rev. Lett.} {\bf 90}, 20, 207901 (2003);
quant-ph/0301118.

\item {\bf [Zhao-Yang-Chen-(+3) 03]}:
Z. Zhao, T. Yang, Y.-A. Chen, A.-N. Zhang, \.{Z}ukowski, \& J.-W. Pan,
``Experimental violation of local realism by four-photon Greenberger-Horne-Zeilinger entanglement'',
{\em Phys. Rev. Lett.} {\bf 91}, 18, 180401 (2003);
quant-ph/0302137.

\item {\bf [Zhao-Chen-Zhang-(+3) 04]}:
Z. Zhao, Y.-A. Chen, A.-N. Zhang,
T. Yang, H. Briegel, \& J.-W. Pan,
``Experimental demonstration of five-photon entanglement and
open-destination teleportation'',
{\em Nature} {\bf 430}, ?, 54-58 (2004);
quant-ph/0402096.

\item {\bf [Zhao-Zhang-Chen-(+4) 04]}:
Z. Zhao, A.-N. Zhang, Y.-A. Chen,
H. Zhang, J.-F. Du, T. Yang, \& J.-W. Pan,
``Experimental demonstration of a non-destructive controlled-NOT quantum
gate for two independent photon-qubits'',
quant-ph/0404129.

\item {\bf [Zheng-Guo 97 a]}:
S. Zheng, \& G.-C. Guo,
``Teleportation of an unknown atomic state through the Raman atom-cavity-field'',
{\em Phys. Lett. A} {\bf 232}, 3-4, 171-174 (1997).

\item {\bf [Zheng-Guo 97 b]}:
S. Zheng, \& G.-C. Guo,
``Teleportation of superpositions of macroscopic states of a cavity field'',
{\em Phys. Lett. A} {\bf 236}, 3, 180-182 (1997).

\item {\bf [Zheng 01]}:
S.-B. Zheng,
``One-step synthesis of multiatom Greenberger-Horne-Zeilinger states'',
{\em Phys. Rev. Lett.} {\bf 87}, 23, 230404 (2001).
Comment: {\bf [Agarwal-Puri-Singh 01]}.

\item {\bf [Zheng 02 a]}:
S.-B. Zheng,
``Fast scheme for the generation of entangled states of multiple hot trapped
ions'',
{\em Phys. Rev. A} {\bf 65}, 5, 051804 (2002).

\item {\bf [Zheng 02 b]}:
S.-B. Zheng,
``Quantum nonlocality for a three-particle nonmaximally entangled state
without inequalities'',
{\em Phys. Rev. A} {\bf 66}, 1, 014103 (2002).
Comment: {\bf [Cereceda 03 c]}.

\item {\bf [Zheng 02 c]}:
S.-B. Zheng,
``Establishment of multiparticle maximal entanglement
from nonmaximally entangled pairs in a communication network'',
{\em J. Opt. B: Quantum Semiclass. Opt.} {\bf 4}, 5, 235-239 (2002).

\item {\bf [Zheng 02 d]}:
S.-B. Zheng,
``Quantum-information processing and multiatom-entanglement engineering with a
thermal cavity'',
{\em Phys. Rev. A} {\bf 66}, 6, 060303 (2002).

\item {\bf [Zheng 03]}:
S.-B. Zheng,
``Quantum logic gates for hot ions without a speed limitation'',
{\em Phys. Rev. Lett.} {\bf 90}, 21, 217901 (2003).

\item {\bf [Zhong 04 a]}:
S.-B. Zheng,
``Scheme for approximate conditional teleportation of an
unknown atomic state without the Bell-state measurement'',
{\em Phys. Rev. A} {\bf 69}, 6, 064302 (2004).

\item {\bf [Zhong 04 b]}:
Z.-Z. Zhong,
``Exact entanglement bases and general bound entanglement'',
quant-ph/0409080.

\item {\bf [Zhou-Tao-Shen-Liang 02]}:
B. Zhou, R. Tao, S.-Q. Shen, \& J.-Q. Liang,
``Quantum computing of molecular magnet Mn$_{12}$'',
{\em Phys. Rev. A} {\bf 66}, 1, 010301 (2002).

\item {\bf [Zhou-Zeng 03]}:
C. Zhou, \& H. Zeng,
``Time-division single-photon Sagnac interferometer for quantum key
distribution'',
{\em Appl. Phys. Lett.} {\bf 82}, 832-? (2003).

\item {\bf [Zhou-Wu-Chen-Zeng 03]}:
C. Zhou, G. Wu, X. Chen, \& H. Zeng,
`\,``Plug and play'' quantum key distribution system with differential phase
shift',
{\em Appl. Phys. Lett.} {\bf 83}, 1692-? (2003).

\item {\bf [Zhou-Zhang-Sun 01]}:
D. L. Zhou, P. Zhang, \& C. P. Sun,
``High order quantum decoherence via
multi-particle amplitude for boson system'',
quant-ph/0105122.
See {\bf [Zhou-Zhang-Sun 02]}.

\item {\bf [Zhou-Zhang-Sun 02]}:
D. L. Zhou, P. Zhang, \& C. P. Sun,
``Understanding destruction of nth-order quantum
coherence in terms of multi-path interference'',
quant-ph/0203082.
See {\bf [Zhou-Zhang-Sun 01]}.

\item {\bf [Zhou-Zeng-Xu-Sun 03]}:
D. L. Zhou, B. Zeng, Z. Xu, \& C. P. Sun,
``Quantum computation based on $d$-level cluster state'',
{\em Phys. Rev. A} {\bf 68}, 6, 062303 (2003);
quant-ph/0304054.

\item {\bf [Zhou-Zhang 00]}:
J. Zhou, \& Y. Zhang,
``Teleportation scheme of $S$-level quantum pure states by two-level
EPRs'',
quant-ph/0004095.

\item {\bf [Zhou-Hou-Wu-Zhang 00]}:
J. Zhou, G. Hou, S. Wu, \& Y. Zhang,
``Controlled quantum teleportaion'',
quant-ph/0006030.

\item {\bf [Zhou-Hou-Zhang 01]}:
J. Zhou, G. Hou, \& Y. Zhang,
``Teleportation scheme of $S$-level quantum pure states
by two-level Einstein-Podolsky-Rosen states'',
{\em Phys. Rev. A} {\bf 64}, 1, 012301 (2001).

\item {\bf [Zhou-Song-Luo 01]}:
L. Zhou, H. S. Song, \& Y. X. Luo,
``Role of degenerate atomic levels in the
entanglement and the decoherence'',
quant-ph/0105092.

\item {\bf [Zhou-Kuang 03]}:
L. Zhou, \& L.-M. Kuang,
``Proposal for optically realizing a quantum game'',
{\em Phys. Lett. A} {\bf 315}, 6, 426-430 (2003);
quant-ph/0402161.

\item {\bf [Zhou-Song-Guo-Li 03]}:
L. Zhou, H. S. Song, Y. Q. Guo, \& C. Li,
``Enhanced thermal entanglement in an anisotropic Heisenberg $XYZ$ chain'',
{\em Phys. Rev. A} {\bf 68}, 2, 024301 (2003).

\item {\bf [Zhou-Leung-Chuang 00]}:
X. Zhou, D. W. Leung, \& I. L. Chuang,
``Quantum logic gate constructions with one-bit
`teleportation'\,'',
quant-ph/0002039.

\item {\bf [Zhou-Leung-Chuang 00]}:
X. Zhou, D. W. Leung, \& I. L. Chuang,
``Quantum algorithms which accept hot qubit inputs'',
quant-ph/9906112.

\item {\bf [Zhou-Leung-Chuang 00]}:
X. Zhou, D. W. Leung, \& I. L. Chuang,
``Methodology for quantum logic gate construction'',
{\em Phys. Rev. A} {\bf 62}, 5, 052316 (2000).

\item {\bf [Zhou-Zhou-Feldman-Guo 01 a]}:
X. Zhou, Z.-W. Zhou, M. J. Feldman, \& G.-C. Guo,
``Nondistortion quantum interrogation using
Einstein-Podolsky-Rosen entangled photons'',
{\em Phys. Rev. A} {\bf 64}, 4, 044103 (2001);
quant-ph/0103097.

\item {\bf [Zhou-Zhou-Guo-Feldman 01]}:
X. Zhou, Z.-W. Zhou, G.-C. Guo, \& M. J. Feldman,
``High efficiency nondistortion quantum interrogation
of atoms in quantum superpositions'',
quant-ph/0105023.

\item {\bf [Zhou-Wulf-Zhou-(+3) 04]}:
X. Zhou, M. Wulf, Z. Zhou, G.-C. Guo, \& M. J. Feldman,
``Dispersive manipulation of paired superconducting qubits'',
{\em Phys. Rev. A} {\bf 69}, 3, 030301 (2004).

\item {\bf [Zhou-Yu-Zhan 02]}:
X.-Q. Zhou, Y.-F. Yu, \& M.-S. Zhan,
``Generating
multi-photon entangled state in a scalable way'',
{\em Phys. Lett. A} {\bf 297}, 5-6, 285-290 (2002).

\item {\bf [Zhou-Zhou-Guo-Feldman 02]}:
X. Zhou, Z.-W. Zhou, G.-C. Guo, \& M. J. Feldman,
``Quantum computation with untunable couplings'',
{\em Phys. Rev. Lett.} {\bf 89}, 19, 197903 (2002).

\item {\bf [Zhou-Chu-Han 02]}:
Z. Zhou, S.-I. Chu, \& S. Han,
``Quantum computing with superconducting devices: A three-level SQUID qubit'',
{\em Phys. Rev. B} {\bf 66}, 5, 054527 (2002).

\item {\bf [Zhou-Guo 00 a]}:
Z.-W. Zhou, \& G.-C. Guo,
``Disentanglement and inseparability correlation in a two-qubit
system'',
{\em Phys. Rev. A} {\bf 61}, 3, 032108 (2000);
quant-ph/9911063.

\item {\bf [Zhou-Guo 00 b]}:
Z.-W. Zhou, \& G.-C. Guo,
``Efficient scheme for two-atom entanglement and quantum information
processing in cavity QED'',
{\em Phys. Rev. Lett.} {\bf 85}, 11, 2392-2395 (2000).

\item {\bf [Zhou-Guo 00 c]}:
Z.-W. Zhou, \& G.-C. Guo,
``Basic limitations for entanglement catalysis'',
quant-ph/0005005.

\item {\bf [Zhou-Zhou-Feldman-Guo 01 b]}:
Z.-W. Zhou, X. Zhou, M. J. Feldman, \& G.-C. Guo,
``Conditions for nondistortion interrogation of quantum system'',
quant-ph/0103102.

\item {\bf [Zhou-Zhou-Lin-(+2) 02]}:
Z.-W. Zhou, X. Zhou, X. D. Lin, M. J. Feldman, \& G.-C. Guo,
``A general framework of nondistortion quantum interrogation'',
quant-ph/0203114.

\item {\bf [Zhou-Yu-Zhou-(+2) 04]}:
Z.-W. Zhou, B. Yu, X. Zhou, M. J. Feldman, \& G.-C. Guo,
``Scalable fault-tolerant quantum computation in decoherence-free subspaces'',
{\em Phys. Rev. Lett.} {\bf 93}, 1, 010501 (2004);
quant-ph/0407241.

\item {\bf [Zhu-Zanardi 04]}:
S.-L. Zhu, \& P. Zanardi,
``Geometric quantum gates robust against stochastic control errors'',
quant-ph/0407177.

\item {\bf [Zhukov-Maslennikov-Chekhova 02]}:
A. A. Zhukov, G. A. Maslennikov, \& M. V. Chekhova,
``Operationalistic orthogonality condition for single-mode biphotons (qutrits)'',
{\em JETP Lett.} {\bf 76}, 10, 596-599 (2002).

\item {\bf [Zhu-Wang 02 a]}:
S.-L. Zhu, \& Z. D. Wang,
``Implementation of universal quantum gates
based on nonadiabatic geometric phases'',
{\em Phys. Rev. Lett.} {\bf 89}, 9, 097902 (2002).
Erratum: {\em Phys. Rev. Lett.} {\bf 89}, 28, 289901 (2002).

\item {\bf [Zhu-Wang 02 b]}:
S.-L. Zhu, \& Z. D. Wang,
``Geometric phase shift in quantum computation using superconducting
nanocircuits: Nonadiabatic effects'',
{\em Phys. Rev. A} {\bf 66}, 4, 042322 (2002).

\item {\bf [Zhu-Wang 03]}:
S.-L. Zhu, \& Z. D. Wang,
``Universal quantum gates based on a pair of orthogonal cyclic states:
Application to NMR systems'',
{\em Phys. Rev. A} {\bf 67}, 2, 022319 (2003);
quant-ph/0210027.

\item {\bf [Zhu-Wang-Yang 03]}:
S.-L. Zhu, Z. D. Wang, \& K. Yang,
``Quantum-information processing using Josephson junctions coupled through
cavities'',
{\em Phys. Rev. A} {\bf 68}, 3, 034303 (2003);
quant-ph/0304173.

\item {\bf [Zhu-Wang-Zanardi 04]}:
S.-L. Zhu, Z. D. Wang, \& P. Zanardi,
``Geometric quantum computation and multi-qubit entanglement with
superconducting charge qubits inside a cavity'',
quant-ph/0403004.

\item {\bf [Zhu-Fang-Feng-(+3) 00]}:
X. Zhu, X. Fang, M. Feng, F. Du, K. Gao, \& X. Mao,
``Experimental realization of a highly structured search
algorithm'',
{\em Physica D} {\bf 156}, 1-2, 179-185 (2001);
quant-ph/0007040.

\item {\bf [Zhu-Fang-Peng-(+3) 00]}:
X. Zhu, X. Fang, X. Peng, M. Feng, K. Gao, \& F. Du,
``Experimental test of complementarity by
nuclear magnetic resonance techniques'',
quant-ph/0011094.

\item {\bf [Zibrov-Matsko-Kocharovskaya-(+3) 02]}:
A. S. Zibrov, A. B. Matsko, O. Kocharovskaya,
Y. V. Rostovtsev, G. R. Welch, \& M. O. Scully,
``Transporting and time reversing light via atomic coherence'',
{\em Phys. Rev. Lett.} {\bf 88}, 10, 103601 (2002).

\item {\bf [Zierler-Schlessinger 65]}:
N. Zierler, \& M. Schlessinger,
``Boolean embeddings of orthomodular sets and quantum logic'',
{\em Duke, Math. J.} {\bf 32}, 251-262 (1965).

\item {\bf [Ziman 94]}:
J. Ziman,
``Attacking the theory of knowledge from several directions'',
{\em Phys. Today} {\bf 47}, 3, 57 (1994).
Review of {\bf [Shimony 93]}.

\item {\bf [Ziman-Bu\v{z}ek 00]}:
M. Ziman, \& V. Bu\v{z}ek,
``Equally distant, partially entangled alphabet states
for quantum channels'',
{\em Phys. Rev. A} {\bf 62}, 5, 052301 (2000);
quant-ph/0009075.

\item {\bf [Ziman-Stelmachovic-Bu\v{z}zek 01]}:
M. Ziman, P. Stelmachovic, \& V. Bu\v{z}zek,
``On the local unitary equivalence of states of multi-partite systems'',
quant-ph/0107016.

\item {\bf [Ziman-Stelmachovic-Bu\v{z}zek-(+3) 02]}:
M. Ziman, P. Stelmachovic, V. Bu\v{z}zek,
M. Hillery, V. Scarani, \& N. Gisin,
``Diluting quantum information: An analysis of information transfer in
system-reservoir interactions'',
{\em Phys. Rev. A} {\bf 65}, 4, 042105 (2002);
quant-ph/0110164.

\item {\bf [Ziman-Bu\v{z}ek 03]}:
M. Ziman, \& V. Bu\v{z}zek,
``Correlation-assisted quantum communication'',
{\em Phys. Rev. A} {\bf 67}, 4, 042321 (2003).

\item {\bf [Ziman-Plesch-Bu\v{z}ek 04]}:
M. Ziman, M. Plesch, \& V. Bu\v{z}zek,
``Reconstruction of superoperators from incomplete measurements'',
quant-ph/0406088.

\item {\bf [Ziman-Stelmachovic-Bu\v{z}ek 04]}:
M. Ziman, P. Stelmachovic, \& V. Bu\v{z}zek,
``Description of quantum dynamics of open systems based on collision-like
models'',
{\em Open Syst. Inf. Dynamics};
quant-ph/0410161.

\item {\bf [Zimba 93]}:
J. R. Zimba,
``Finitary proofs of contextuality and non-locality using
Majorana representation of spin-$\frac{3}{2}$ states'',
M.\ Sc. thesis, Oxford University, 1993.

\item {\bf [Zimba-Penrose 93]}:
J. R. Zimba, \& R. Penrose,
``On Bell non-locality without probabilities: More curious geometry'',
{\em Stud. Hist. Philos. Sci.} {\bf 24}, 5, 697-720 (1993).
See {\bf [Massad 98]}, {\bf [Massad-Aravind 99]}, {\bf [Penrose 00]}.

\item {\bf [Zimba 98]}:
J. R. Zimba,
``Simple realism and canonically conjugate observables in
non-relativistic quantum mechanics'',
{\em Found. Phys. Lett.} {\bf 11}, 6, 503-533 (1998).

\item {\bf [Zimba-Clifton 98]}:
J. R. Zimba, \& R. K. Clifton,
``Valuations on functionally closed sets of quantum mechanical observables
and von Neumann's `no-hidden-variables' theorem'',
in {\bf [Dieks-Vermaas 98]}, pp.~69-101;
quant-ph/9802038.

\item {\bf [Zimba 00]}:
J. R. Zimba,
``Quantum-mechanical uncertainty and the stability of
incompatibility'',
{\em Found. Phys.} {\bf 30}, 2, 179-203 (2000).

\item {\bf [Zippilli-Vitali-Tombesi-Raimond 03]}:
S. Zippilli, D. Vitali, P. Tombesi, \& J.-M. Raimond,
``Scheme for decoherence control in microwave cavities'',
{\em Phys. Rev. A} {\bf 67}, 5, 052101 (2003).

\item {\bf [Zisis 00]}:
M. Zisis,
``Approximate hidden variables'',
{\em Found. Phys.} {\bf 30}, 7, 971-1000 (2000).

\item {\bf [Zoller 00]}:
P. Zoller,
``Quantum optics: Tricks with a single photon'',
{\em Nature} {\bf 404}, 6776, 340-341 (2000).
See {\bf [Pinkse-Fischer-Maunz-Rempe 00]}.

\item {\bf [Zoller-Cirac-Duan-Garc\'{\i}a Ripoll 04]}:
P. Zoller, J. I. Cirac, L.-M. Duan, \& J. J. Garc\'{\i}a-Ripoll,
``Implementing quantum information processing with atoms, ions and photons'',
in D. Est\`{e}ve, J. M. Raimond, \& J. Dalibard (eds.),
{\em Quantum entanglement and information processing.
Proc.\ of the Les Houches Summer School, Session 79},
Elsevier, Amsterdam, 2004;
quant-ph/0405025.

\item {\bf [Zorin 04]}:
A. B. Zorin,
``Josephson charge-phase qubit with radio frequency readout: Coupling and decoherence'',
{\em J. Exp. Theor. Phys.} {\bf 98}, ?, 1250-1261 (2004).

\item {\bf [Zou-Shao 99]}:
J. Zou, \& B. Shao,
``Quantum coherence in a mesoscopic Josephson junction'',
{\em Phys. Lett. A} {\bf 256}, 5-6, 375-378 (1999).

\item {\bf [Zou-Pahlke-Mathis 01 a]}:
X. B. Zou, K. Pahlke, \& W. Mathis,
``Generation of entangled states of two traveling modes
for fixed number of photon'',
quant-ph/0110149.

\item {\bf [Zou-Pahlke-Mathis 01 b]}:
X. B. Zou, K. Pahlke, \& W. Mathis,
``The feasible generation of entangled photon
states by using linear optical elements'',
quant-ph/0110168.

\item {\bf [Zou-Pahlke-Mathis 01 c]}:
X. B. Zou, K. Pahlke, \& W. Mathis,
``Teleportation implementation of non-deterministic
quantum logic operations by using linear optical elements'',
quant-ph/0110177.

\item {\bf [Zou-Pahlke-Mathis 01 d]}:
X. B. Zou, K. Pahlke, \& W. Mathis,
``The generation of an entangled four-photon state from two pairs of entangled
two-photon states by using linear optical elements'',
quant-ph/0111014.

\item {\bf [Zou-Pahlke-Mathis 01 e]}:
X. B. Zou, K. Pahlke, \& W. Mathis,
``The entanglement concentration and purification
for entangled polarization photon state'',
quant-ph/0111049.

\item {\bf [Zou-Pahlke-Mathis 01 f]}:
X. B. Zou, K. Pahlke, \& W. Mathis,
``Continuous variable quantum cloning via the cavity QED with trapped ions'',
quant-ph/0112036.

\item {\bf [Zou-Pahlke-Mathis 01 g]}:
X. B. Zou, K. Pahlke, \& W. Mathis,
``Quantum computation with trapped ions in strongly detuned optical cavity'',
quant-ph/0112057.

\item {\bf [Zou-Pahlke-Mathis 02 a]}:
X. B. Zou, K. Pahlke, \& W. Mathis,
``Generation of multi-photon entangled states by
using linear optical elements and a projective measurement'',
{\em Phys. Lett. A} {\bf 306}, 1, 10-13 (2002).

\item {\bf [Zou-Pahlke-Mathis 02 b]}:
X. B. Zou, K. Pahlke, \& W. Mathis,
``The generation of an entangled four-photon $W$-state by using linear optical elements'',
quant-ph/0202090.

\item {\bf [Zou-Pahlke-Mathis 02 c]}:
X. B. Zou, K. Pahlke, \& W. Mathis,
``Linear optical implementation of a single-mode quantum filter and generation
of multiphoton polarization entangled states'',
{\em Phys. Rev. A} {\bf 66}, 6, 064302 (2002).

\item {\bf [Zou-Pahlke-Mathis 03 a]}:
X.-B. Zou, K. Pahlke, \& W. Mathis,
``The non-deterministic quantum logic operation and the teleportation of the
superposition of the vacuum state and the single-photon state via parametric amplifiers'',
{\em Phys. Lett. A} {\bf 311}, 4-5, 271-276 (2003);
quant-ph/0303006.

\item {\bf [Zou-Pahlke-Mathis 03 b]}:
X.-B. Zou, K. Pahlke, \& W. Mathis,
``Generation of entangled states of two three-level atoms in cavity QED'',
quant-ph/0303015.

\item {\bf [Zou-Pahlke-Mathis 04]}:
X.-B. Zou, K. Pahlke, \& W. Mathis,
``Conditional generation of quantum entanglement of many distant atoms via multifold coincidence detection'',
{\em Phys. Rev. A} {\bf 69}, 1, 013811 (2004).

\item {\bf [Zou-Wang-Mandel 91]}:
X. Y. Zou, L. J. Wang, \& L. Mandel,
``Induced coherence and indistinguishability in optical interference'',
{\em Phys. Rev. Lett.} {\bf 67}, 3, 318-321 (1991).

\item {\bf [Zou-Pahlke-Mathis 02 a]}:
X. Zou, K. Pahlke, \& W. Mathis,
``Generation of two-mode nonclassical states and a quantum-phase-gate
operation in trapped-ion cavity QED'',
{\em Phys. Rev. A} {\bf 65}, 6, 064303 (2002);
quant-ph/0201032.

\item {\bf [Zou-Pahlke-Mathis 02 b]}:
X. Zou, K. Pahlke, \& W. Mathis,
``Teleportation implementation of nondeterministic quantum logic operations by
using linear optical elements'',
{\em Phys. Rev. A} {\bf 65}, 6, 064305 (2002).

\item {\bf [Zou-Pahlke-Mathis 02 c]}:
X. Zou, K. Pahlke, \& W. Mathis,
``Generation of entangled photon states by using linear optical elements'',
{\em Phys. Rev. A} {\bf 66}, 1, 014102 (2002);
quant-ph/0204070.

\item {\bf [Zou-Pahlke-Mathis 02 d]}:
X. Zou, K. Pahlke, \& W. Mathis,
``Generation of an entangled four-photon $W$ state'',
{\em Phys. Rev. A} {\bf 66}, 4, 044302 (2002).

\item {\bf [Zou-Pahlke-Mathis 02 e]}:
X. Zou, K. Pahlke, \& W. Mathis,
``Quantum phase-gate implementation for trapped ions in thermal motion'',
{\em Phys. Rev. A} {\bf 66}, 4, 044307 (2002).

\item {\bf [Zou-Pahlke-Mathis 03 a]}:
X. Zou, K. Pahlke, \& W. Mathis,
``Scheme for the implementation of a universal quantum cloning machine via
cavity-assisted atomic collisions in cavity QED'',
{\em Phys. Rev. A} {\bf 67}, 2, 024304 (2003).

\item {\bf [Zou-Pahlke-Mathis 03 b]}:
X. Zou, K. Pahlke, \& W. Mathis,
``Generation of an entangled state of two three-level atoms in cavity QED'',
{\em Phys. Rev. A} {\bf 67}, 4, 044301 (2003).

\item {\bf [Zou-Pahlke-Mathis 03 c]}:
X. Zou, K. Pahlke, \& W. Mathis,
``Conditional generation of the Greenberger-Horne-Zeilinger state of four
distant atoms via cavity decay'',
{\em Phys. Rev. A} {\bf 68}, 2, 024302 (2003).

\item {\bf [Zou-Pahlke-Mathis 03 d]}:
X. Zou, K. Pahlke, \& W. Mathis,
``Generation of arbitrary superpositions of the Dicke states of excitons in
optically driven quantum dots'',
{\em Phys. Rev. A} {\bf 68}, 3, 034306 (2003).

\item {\bf [Zou-Lee-Kim-(+2) 02]}:
X.-B. Zou, H.-W. Lee, J. Kim, J.-W. Lee, \& E. K. Lee,
``Nonlocality of Hardy type in experiments using independent particle sources'',
{\em Phys. Lett. A} {\bf 305}, 6, 354-358 (2002).

\item {\bf [Zrenner-Beham-Stufler-(+3) 02]}:
A. Zrenner, E. Beham, S. Stufler,
F. Findeis, M. Bichler, \& G. Abstreiter,
``Coherent properties of a two-level system based on a quantum-dot photodiode'',
{\em Nature} {\bf 418}, 6898, 612-614 (2002).

\item {\bf [\.{Z}ukowski 90]}:
M. \.{Z}ukowski,
``Two-particle spatial quantum beats: Feasible test of Bell's inequalities'',
{\em Phys. Lett. A} {\bf 150}, 3-4, 136-142 (1990).

\item {\bf [\.{Z}ukowski-Zeilinger 91]}:
M. \.{Z}ukowski, \& A. Zeilinger,
``Test of the
Bell inequality based on phase and linear momentum as well as spin'',
{\em Phys. Lett. A} {\bf 155}, 2-3, 69-72 (1991).

\item {\bf [\.{Z}ukowski 91 a]}:
M. \.{Z}ukowski,
`Definite values for observables versus quantum predictions:
A ``GHZ-like'' test',
{\em Phys. Lett. A} {\bf 157}, 4-5, 198-202 (1991).

\item {\bf [\.{Z}ukowski 91 b]}:
M. \.{Z}ukowski,
``GHZ correlations in quadrature phase measurements'',
{\em Phys. Lett. A} {\bf 157}, 4-5, 203-208 (1991).

\item {\bf [\.{Z}ukowski 93]}:
M. \.{Z}ukowski,
``Bell theorem involving all settings of measuring apparatus'',
{\em Phys. Lett. A} {\bf 177}, 4-5, 290-296 (1993).

\item {\bf [\.{Z}ukowski-Zeilinger-Horne-Ekert 93]}:
M. \.{Z}ukowski, A. Zeilinger, M. A. Horne, \& A. K. Ekert,
`\,``Event-ready-detectors'' Bell experiment via entanglement swapping',
{\em Phys. Rev. Lett.} {\bf 71}, 26, 4287-4290 (1993).

\item {\bf [\.{Z}ukowski-Zeilinger-Weinfurter 95]}:
M. \.{Z}ukowski, A. Zeilinger, \& H. Weinfurter,
``Entangling photons radiated by independent pulsed sources'',
in D. M. Greenberger, \& A. Zeilinger (eds.),
{\em Fundamental problems in quantum theory:
A conference held in honor of professor John A. Wheeler,
Ann. N. Y. Acad. Sci.} {\bf 755}, 91-102 (1995).

\item {\bf [\.{Z}ukowski-Zeilinger-Horne-Ekert 95]}:
M. \.{Z}ukowski, A. Zeilinger, M. A. Horne, \& A. K. Ekert,
`Extensions of Bell theorem: Experiment involving
independent sources in ``event-ready'' configuration',
in M. Ferrero, \& A. van der Merwe (eds.),
{\em Fundamental problems in quantum physics.
Proc.\ of an international symposium (Oviedo, Spain, 1993)},
Kluwer Academic, Dordrecht, Holland, 1995, pp.~363-373.

\item {\bf [\.{Z}ukowski-Zeilinger-Horne 97]}:
M. \.{Z}ukowski, A. Zeilinger, \& M. A. Horne,
``Realizable higher-dimensional two-particle entanglements via
multiport beam splitters'',
{\em Phys. Rev. A} {\bf 55}, 4, 2564-2579 (1997).

\item {\bf [\.{Z}ukowski-Kaszlikowski 97]}:
M. \.{Z}ukowski, \& D. Kaszlikowski,
``Critical visibility for $N$-particle Greenberger-Horne-Zeilinger
correlations to violate local realism'',
{\em Phys. Rev. A} {\bf 56}, 3, R1682-R1685 (1997);
quant-ph/9910045.

\item {\bf [\.{Z}ukowski-Horodecki-Horodecki-Horodecki 98]}:
M. \.{Z}ukowski, R. Horodecki, M. Horodecki, \& P. Horodecki,
``Generalized quantum measurements and local realism'',
{\em Phys. Rev. A} {\bf 58}, 3, 1694-1698 (1998);
quant-ph/9608035.

\item {\bf [\.{Z}ukowski-Zeilinger-Horne-Weinfurter 98]}:
M. \.{Z}ukowski, A. Zeilinger, M. A. Horne, \& H. Weinfurter,
``Quest for GHZ states'',
{\em Acta Phys. Pol. A} {\bf 93}, 1, 187-194 (1998).

\item {\bf [\.{Z}ukowski-Kaszlikowski-Baturo-Larsson 99]}:
M. \.{Z}ukowski, D. Kaszlikowski, A. Baturo, \& J.-\AA. Larsson,
``Strengthening the Bell theorem:
Conditions to falsify local realism in an experiment'',
quant-ph/9910058.

\item {\bf [\.{Z}ukowski-Kaszlikowski 99 a]}:
M. \.{Z}ukowski, \& D. Kaszlikowski,
``Greenberger-Horne-Zeilinger paradoxes with symmetric
multiport beam splitters'',
{\em Phys. Rev. A} {\bf 59}, 5, 3200-3203 (1999);
quant-ph/9911039.

\item {\bf [\.{Z}ukowski-Kaszlikowski 99 b]}:
M. \.{Z}ukowski, \& D. Kaszlikowski,
``Greenberger-Horne-Zeilinger paradox for three tritters'',
in {\bf [Greenberger-Reiter-Zeilinger 99]}, pp.~275-278.

\item {\bf [\.{Z}ukowski-Kaszlikowski-Santos 99]}:
M. \.{Z}ukowski, D. Kaszlikowski, \& E. Santos,
``Irrelevance of photon events distinguishability
in a class of Bell experiments'',
{\em Phys. Rev. A} {\bf 60}, 4, R2614-R2617 (1999);
quant-ph/9911027.

\item {\bf [\.{Z}ukowski-Zeilinger-Horne-Weinfurter 99]}:
M. \.{Z}ukowski, A. Zeilinger, M. A. Horne, \& H. Weinfurter,
``Independent photons and entanglement: A short overview'',
{\em Int. J. Theor. Phys.} {\bf 38}, ?, 501-517 (1999).

\item {\bf [\.{Z}ukowski-Kaszlikowski 99 c]}:
M. \.{Z}ukowski, \& D. Kaszlikowski,
``Faster than light Bell telephone of Michalski transmits only noise'',
quant-ph/9911028.

\item {\bf [\.{Z}ukowski 00 a]}:
M. \.{Z}ukowski,
``Violations of local realism in multiphoton interference experiments'',
{\em Phys. Rev. A} {\bf 61}, 2, 022109 (2000);
quant-ph/9811013.
See {\bf [Bouwmeester-Pan-Daniell-(+2) 98]}.

\item {\bf [\.{Z}ukowski-Kaszlikowski 00 a]}:
M. \.{Z}ukowski, \& D. Kaszlikowski,
``Entanglement swapping with PDC sources'',
{\em Acta Phys. Slovaca} {\bf 49}, 4, 621-626 (2000);
quant-ph/9911045.

\item {\bf [\.{Z}ukowski-Kaszlikowski 00 b]}:
M. \.{Z}ukowski, \& D. Kaszlikowski,
``Bell theorem for some quantum interferometric experiments'',
{\em Fortschr. Phys.} {\bf 48}, 5-7, 489-496 (2000).

\item {\bf [\.{Z}ukowski 00 b]}:
M. \.{Z}ukowski,
``Bell theorem for the nonclassical part of
the quantum teleportation process'',
{\em Phys. Rev. A} {\bf 62}, 3, 032101 (2000);
quant-ph/9912029.

\item {\bf [\.{Z}ukowski-Brukner 02]}:
M. \.{Z}ukowski, \& \v{C}. Brukner,
``Bell's theorem for general $N$-qubit states'',
{\em Phys. Rev. Lett.} {\bf 88}, 21, 210401 (2002);
quant-ph/0102039.

\item {\bf [\.{Z}ukowski-Brukner-Laskowski-Wie\'{s}niak 02]}:
M. \.{Z}ukowski, \v{C}. Brukner, W. Laskowski, \& M. Wie\'{s}niak,
``Do all pure entangled states violate Bell's inequalities for correlation
functions?'',
{\em Phys. Rev. Lett.} {\bf 88}, 21, 210402 (2002);
quant-ph/0110095.

\item {\bf [\.{Z}ukowski 03]}:
M. \.{Z}ukowski,
``Some news about Bell inequalities'',
in M. Ferrero (ed.),
{\em Proc. of Quantum Information: Conceptual Foundations,
Developments and Perspectives (Oviedo, Spain, 2002)},
{\em J. Mod. Opt.} {\bf 50}, 6-7, 1151-1163 (2003).

\item {\bf [Zubairy 98]}:
M. S. Zubairy,
``Quantum teleportation of a field state'',
{\em Phys. Rev. A} {\bf 58}, 6, 4368-4372 (1998).

\item {\bf [Zubairy-Kim-Scully 03]}:
M. S. Zubairy, M. Kim, \& M. O. Scully,
``Cavity-QED-based quantum phase gate'',
{\em Phys. Rev. A} {\bf 68}, 3, 033820 (2003).

\item {\bf [Zubairy-Agarwal-Scully 04]}:
M. S. Zubairy, G. S. Agarwal, \& M. O. Scully,
``Quantum disentanglement eraser: A cavity QED implementation'',
{\em Phys. Rev. A} {\bf 70}, 1, 012316 (2004).

\item {\bf [Zurek 81 a]}:
W. H. Zurek,
``Pointer basis of quantum apparatus: Into what
mixture does the wave packet collapse?'',
{\em Phys. Rev. D} {\bf 24}, 6, 1516-1525 (1981).

\item {\bf [Zurek 81 b]}:
W. H. Zurek,
``Information transfer in quantum measurements:
Irreversibility and amplification'',
lecture at {\em NATO ASI (Bad Windsheim, 1981)};
quant-ph/0111137.

\item {\bf [Zurek 82]}:
W. H. Zurek,
``Environment-induced superselection rules'',
{\em Phys. Rev. D} {\bf 26}, 8, 1862-1880 (1982).

\item {\bf [Zurek 83]}:
W. H. Zurek,
``Information transfer in quantum
measurements: Irreversibility and amplification'',
in P. Meystre, \& M. O. Scully (eds.),
{\em Quantum optics, experimental gravitation and measurement theory},
Plenum Press, New York, 1983, pp.~87-116.

\item {\bf [Zurek-van der Merwe-Miller 88]}:
W. H. Zurek, A. van der Merwe, \& W. A. Miller (eds.),
{\em Between quantum and cosmos.
Studies and essays in honor of John Archibald Wheeler},
Princeton University Press, Princeton, New Jersey, 1988.

\item {\bf [Zurek 90]}:
W. H. Zurek (ed.),
{\em Complexity, entropy, and the physics of information (Santa
Fe, New Mexico, 1989)},
Santa Fe Institute of Studies in the Science of Complexity,
Addison-Wesley, Redwood City, California, 1990.

\item {\bf [Zurek 91 a]}:
W. H. Zurek,
``Algorithmic information content, Church-Turing thesis,
physical entropy, and Maxwell's demon'',
in W. H. Zurek (ed.),
{\em Complexity, entropy, and the physics of information
(Santa Fe, New Mexico, 1989)},
Santa Fe Institute of Studies in the Science of Complexity,
Addison-Wesley, Redwood City, California, 1991, pp.~73-89.

\item {\bf [Zurek 91 b]}:
W. H. Zurek,
``Decoherence and the transition from
quantum to classical'',
{\em Phys. Today} {\bf 44}, 10, 36-44 (1991).
Comments in {\bf [Zurek 93 a]}.

\item {\bf [Zurek 93 a]}:
W. H. Zurek,
``Negotiating the tricky border between quantum and classical'',
{\em Phys. Today} {\bf 46}, 4, 13-90 (1991).
Comments and reply to {\bf [Zurek 91 b]}.

\item {\bf [Zurek-Paz 93 a]}:
W. H. Zurek, \& J. P. Paz,
``Coherent states via decoherence'',
{\em Phys. Rev. Lett.} {\bf 70}, 9, 1187-1190 (1993).

\item {\bf [Zurek 93 b]}:
W. H. Zurek,
``Preferred states, predictabilty, classicality,
and the environment-induced decoherence'',
{\em Prog. Theor. Phys.} {\bf 89}, 2, 281-312 (1993).

\item {\bf [Zurek-Paz 93 b]}:
W. H. Zurek, \& J. P. Paz,
``Environment-induced decoherence, classicality,
and consistency of quantum histories'',
{\em Phys. Rev. D} {\bf 48}, 6, 2728-2738 (1993).

\item {\bf [Zurek-Paz 93 c]}:
W. H. Zurek, \& J. P. Paz,
``Decoherence, chaos, the quantum and the classical'',
{\em Nuovo Cimento B} {\bf 110}, 5-6, 611-624 (1995).
Also in C. Di Castro, F. Guerra, \& G. Jona-Lasinio (eds.),
{\em Proc.\ of the Int.\ Conf.\ on Mesoscopic Physics
and Fundamental Problems in Quantum Mechanics (Rome, 1994)}.

\item {\bf [Zurek-Habib-Paz 93]}:
W. H. Zurek, S. Habib, \& J. P. Paz,
``Coherent states via decoherence'',
{\em Phys. Rev. Lett.} {\bf 70}, 9, 1187-1190 (1993).

\item {\bf [Zurek-Paz 94]}:
W. H. Zurek, \& J. P. Paz,
``Decoherence, chaos, and the second law'',
{\em Phys. Rev. Lett.} {\bf 72}, 16, 2508-2511 (1994).
Comment: {\bf [Casati-Chirikov 95]}.
Reply: {\bf [Zurek-Paz 95 b]}.

\item {\bf [Zurek-Paz 95 a]}:
W. H. Zurek, \& J. P. Paz,
``?'',
{\em Physica D} {\bf 83}, ?, 300-? (1995).

\item {\bf [Zurek-Paz 95 b]}:
W. H. Zurek, \& J. P. Paz,
``Zurek and Paz reply'',
{\em Phys. Rev. Lett.} {\bf 75}, 2, 351 (1995).
Reply to {\bf [Casati-Chirikov 95]}.
See {\bf [Zurek-Paz 94]}.

\item {\bf [Zurek-Laflamme 96]}:
W. H. Zurek, \& R. Laflamme,
``Quantum logical operations on encoded qubits'',
{\em Phys. Rev. Lett.} {\bf 77}, 22, 4683-4686 (1996).

\item {\bf [Zurek 97]}:
W. H. Zurek,
``Probing quantum origins of the classical'',
{\em Phys. World} {\bf 10}, 1, 24-25.
See {\bf [Davidovich-Brune-Raimond-Haroche 96]},
{\bf [Brune-Hagley-Dreyer-(+5) 96]}.

\item {\bf [Zurek 98 a]}:
W. H. Zurek,
``Decoherence, einselection, and the existential interpretation
(the rough guide)'',
in A. K. Ekert, R. Jozsa, \& R. Penrose (eds.),
{\em Quantum Computation: Theory and Experiment.
Proceedings of a Discussion Meeting held at the Royal
Society of London on 5 and 6 November 1997},
{\em Philos. Trans. R. Soc. Lond. A} {\bf 356}, 1743, 1793-1821 (1998);
quant-ph/9805065.

\item {\bf [Zurek 98 b]}:
W. H. Zurek,
``Algorithmic randomness, physical entropy, measurements,
and the demon of choice'',
quant-ph/9807007.

\item {\bf [Zurek-Paz 99]}:
W. H. Zurek, \& J. P. Paz,
``Why we don't need quantum planetary dynamics:
Decoherence and the correspondence principle for
chaotic systems'',
in {\bf [Greenberger-Reiter-Zeilinger 99]}, pp.~167-177.

\item {\bf [Zurek 00 a]}:
W. H. Zurek,
``Quantum cloning: Schr\"{o}dinger's sheep'',
{\em Nature} {\bf 404}, 6774, 130-131 (2000).
See {\bf [Pati-Braunstein 00 a]}.

\item {\bf [Zurek 00 b]}:
W. H. Zurek,
``Einselection and decoherence from an information theory
perspective'',
{\em Annalen der Physik} {\bf 9}, 5, 853-862 (2000);
quant-ph/0011039.

\item {\bf [Zurek 01]}:
W. H. Zurek,
``Sub-Planck structure in phase space and its relevance
for quantum decoherence'',
{\em Nature} {\bf 412}, 6848, 712-717 (2001).
See {\bf [Albrecht 01]}.

\item {\bf [Zurek 02]}:
W. H. Zurek,
``Decoherence and the transition from quantum to classical---Revisited'',
{\em Los Alamos Science} 27, 2-25 (2002);
quant-ph/0306072.

\item {\bf [Zurek 03 a]}:
W. H. Zurek,
``Quantum discord and Maxwell's demons'',
{\em Phys. Rev. A} {\bf 67}, 1, 012320 (2003);
quant-ph/0202123 and quant-ph/0301127.

\item {\bf [Zurek 03 b]}:
W. H. Zurek,
``Environment-assisted invariance, entanglement, and probabilities in quantum
physics'',
{\em Phys. Rev. Lett.} {\bf 90}, 12, 120404 (2003).
See {\bf [Schlosshauer-Fine 03]}, {\bf [Mohrhoff 04 c]}.

\item {\bf [Zurek 03 c]}:
W. H. Zurek,
``Decoherence, einselection, and the quantum
origins of the classical'',
{\em Rev. Mod. Phys.} {\bf 75}, 3, 715-775 (2003);
quant-ph/0105127.

\item {\bf [Zurek 03 d]}:
W. H. Zurek,
``Reduction of the wavepacket: How long does it take?'',
quant-ph/0302044.

\item {\bf [Zurek 03 e]}:
W. H. Zurek,
``Quantum Darwinism and envariance'',
quant-ph/0308163.

\item {\bf [Zurek-Cucchietti-Paz 03]}:
W. H. Zurek, F. M. Cucchietti, \& J. P. Paz,
``Gaussian decoherence from random spin environments'',
quant-ph/0312207.

\item {\bf [Zurek 04]}:
W. H. Zurek,
``Probabilities from envariance'',
quant-ph/0405161.

\item {\bf [Zweifel 74]}:
P. F. Zweifel,
``Measurement in quantum mechanics and the EPR paradox'',
{\em Int. J. Theor. Phys.} {\bf 10}, 1, 67-72 (1974).

\item {\bf [\.{Z}yczkowski-Horodecki-Sanpera-Lewenstein 98]}:
K. \.{Z}yczkowski, P. Horodecki, A. Sanpera, \& M. Lewenstein,
``On the volume of the set of separable states'',
{\em Phys. Rev. A} {\bf 58}, 2, 883-892 (1998);
quant-ph/9804024.
See {\bf [Slater 98, 99]}, {\bf [\.{Z}yczkowski 99]} (II).

\item {\bf [\.{Z}yczkowski 99]}:
K. \.{Z}yczkowski,
``Volume of the set of separable states. II'',
{\em Phys. Rev. A} {\bf 60}, 5, 3496-3507 (1999).
See {\bf [\.{Z}yczkowski-Horodecki-Sanpera-Lewenstein 98]} (I).

\item {\bf [\.{Z}yczkowski-Sommers 01]}:
K. \.{Z}yczkowski, \& H.-J. Sommers,
``Induced measures in the space of mixed quantum states'',
in S. Popescu, N. Linden, \& R. Jozsa (eds.),
{\em J. Phys. A} {\bf 34}, 35
(Special issue: Quantum information and computation), 7111-7125 (2001);
quant-ph/0012101.

\item {\bf [\.{Z}yczkowski-Bengtsson 01]}:
K. \.{Z}yczkowski, \& I. Bengtsson,
``Relativity of pure states entanglement'',
quant-ph/0103027.

\item {\bf [\.{Z}yczkowski-Horodecki-Horodecki-Horodecki 02]}:
K. \.{Z}yczkowski, P. Horodecki, M. Horodecki, \& R. Horodecki,
``Dynamics of quantum entanglement'',
{\em Phys. Rev. A} {\bf 65}, 1, 012101 (2002);
quant-ph/0008115.

\item {\bf [\.{Z}yczkowski-Sommers 03 a]}:
K. \.{Z}yczkowski, \& H.-J. Sommers,
``Hilbert-Schmidt volume of the set of mixed quantum states'',
{\em J. Phys. A} {\bf 36}, 39, 10115–10130 (2003);
quant-ph/0302197.

\item {\bf [\.{Z}yczkowski 03]}:
K. \.{Z}yczkowski,
``Renyi extrapolation of Shannon entropy'',
quant-ph/0305062.

\item {\bf [\.{Z}yczkowski-Sommers 03 b]}:
K. \.{Z}yczkowski, \& H.-J. Sommers,
``Average fidelity between random quantum states'',
quant-ph/0311117.

\item {\bf [\.{Z}yczkowski-Bengtsson 04]}:
K. \.{Z}yczkowski, \& I. Bengtsson,
``On duality between quantum maps and quantum states'',
quant-ph/0401119.


\end{enumerate}




\end{document}